\pdfoutput=1
\documentclass[oneside,english,sort&compress,oneside]{yalephd}
\usepackage[T1]{fontenc}
\usepackage[latin9]{inputenc}
\usepackage{babel}
\usepackage{varioref}
\usepackage{units}
\usepackage{amsmath}
\usepackage{amssymb}
\usepackage{wasysym}
\usepackage{graphicx}
\usepackage{esint}
\usepackage[numbers]{natbib}
\PassOptionsToPackage{normalem}{ulem}
\usepackage{ulem}
\usepackage[unicode=true,pdfusetitle,
 bookmarks=true,bookmarksnumbered=false,bookmarksopen=false,
 breaklinks=false,pdfborder={0 0 1},backref=false,colorlinks=false]
 {hyperref}

\makeatletter

\providecommand{\LyX}{\texorpdfstring%
  {L\kern-.1667em\lower.25em\hbox{Y}\kern-.125emX\@}
  {LyX}}
\providecommand{\tabularnewline}{\\}

\newenvironment{lyxlist}[1]
{\begin{list}{}
{\settowidth{\labelwidth}{#1}
 \setlength{\leftmargin}{\labelwidth}
 \addtolength{\leftmargin}{\labelsep}
 }}
{\end{list}}

\usepackage[figure]{hypcap}

\usepackage{placeins}

\makeatother

\begin{document}

\title{Persistent Currents in Normal Metal Rings}

\author{William Ennis Shanks}

\advisor{Jack Harris}

\date{May 2011}
\begin{abstract}
One striking phenomenon of mesoscopic physics is the ability of a
resistive ring to sustain a constant electrical current while in thermal
equilibrium and in the absence of an external excitation. The observability
of persistent currents normal metal rings was first predicted in 1983
\citep{buttiker1983josephson}. Subseqently, these persistent currents
have been studied experimentally several times but with conflicting
results due in part to the difficulty of the measurements.

In this work, I present measurements of persistent currents in normal
metal rings performed with cantilever torsional magnetometry. With
this technique, the typical persistent current (the component that
varies randomly from ring to ring) was measured with high sensitivity.
I report measured magnitudes ($\apprle1\,\text{pA}$) over two orders
of magnitude smaller than observed in previous studies. These measurements
extend the range of temperature and magnetic field over which the
typical current has been observed. The wide magnetic field range allowed
us to study the effect of magnetic field penetrating the ring. It
also enabled the recording of many independent measurements of the
current magnitude in a single sample. These independent measurements
are necessary to characterize the persistent current magnitude because
it is a random quantity. From these measurements of the persistent
current, I also characterize the parametric dependence of the typical
current on sample orientation and number of rings. 

In addition to presenting the experimental results, I thoroughly review
the theory of the typical persistent current in the diffusive regime.
I begin with the simplest model and build up to the case appropriate
for the samples studied in our experiments. I also present in detail
the experimental apparatus used to measure the persistent currents.
\end{abstract}
\maketitle
\tableofcontents{}

\listoffigures

\listoftables

\frontmattersection{Acknowledgments}

I wish to begin this text by acknowledging my advisor Jack Harris
to whom I am very grateful. I feel fortunate to have been involved
in the beginnings of the persistent currents experiment in the Harris
Lab. When I decided to attend Yale for my graduate work, I knew that
I wanted to do something involving electronic systems behaving quantum
mechanically, but I did not even know Jack, in his first year as a
professor at the time, was at Yale. Fortuitously, I happened across
the Harris Lab website in time to work for Jack over the summer and
found both the lab and the persistent current project to be perfect
fits. Over the past six years, Jack has displayed all the traits of
a great advisor: patience, striking physical intuition, strong managerial
skills, and general wisdom.

Continuing the discussion of my good fortune as a graduate student,
I must also acknowledge Ania Bleszynski Jayich with whom I built the
persistent current experiment. While I had worked in a couple experimental
labs over the previous few years, it was my time with Ania during
which I learned to be an experimentalist. Her strong work ethic, openness
to new ideas, and positive attitude in the face of challenges have
had great influences on me (I hope!).

Jack's good judgment of character was exemplified by the lab's early
roster. It is hard to imagine a lab environment with better camaraderie
than the early Harris Lab composed of myself, Ben Zwickl, Andrew Jayich,
Jeff Thompson, and Ania. Amongst us all there was a shared vision
of how a lab should operate, which allowed work to progress smoothly.
I will remember Ben for his positive attitude towards all things including
Notre Dame during the Charlie Weis era and Mathematica's ability to
perform any task, Andrew for his infectious enthusiasm for {}``taking
data'' (and doing pretty much everything else really), and Jeff for
his massive intellect and healthy skepticism of authority.

I worked directly with several other Harris Lab members on the persistent
current experiment. Sofia Magkiriadou wrote the basic LabVIEW shell
on which the vi's that recorded all of the persistent current data
were based. She also had a great sense of humor and was a pleasure
to work with. Bruno Peaudecerf helped build several electronic components
and brought a meticulous attention to detail to calculations of the
cantilever frequency shift due to the persistent current. Dustin Ngo,
my successor as graduate student on the persistent current experiment,
contributed most directly to the work discussed here through his characterization
of the fiber-and-lens version of the cantilever detection set up.
Dustin has a wide-ranging intellectual curiosity, and I expect him
to shed further light on the persistent current phenomena in the coming
years. 

Lastly, I worked closely over my last five months in the lab with
Manuel Castellanos Beltran. Manuel's greatest contribution to this
text is that he read it in its entirety and has probably corrected
more errors in it that I have myself. Most of my work with Manuel
was devoted to the fabrication and measurement of single ring persistent
current samples, the discussion of which goes beyond the scope of
this text. I will remember Manuel for the alacrity with which he attacked
all problems and for his humility, perhaps best exemplified by his
tendency to say he understands nothing while actually understanding
almost everything.

Though he did not work on persistent currents, Jack Sankey's ability
to identify the puzzling aspects of any problem led him to contribute
to our understanding of the persistent current measurement. Jack also
impressed upon me a healthy respect for Python and the open source
revolution. This text would not have been written in \LyX{} without
him. I would also like to acknowledge the indirect contributions of
other Harris Lab members Brian Yang, Nathan Flowers-Jacobs, Jason
Merrill, and Matthew Harrison.

Many people outside of the Harris Lab contributed to the work discussed
here. First, there are my direct collaborators and co-authors, Eran
Ginossar, Leonid Glazman, Felix von Oppen, and Rob Ilic. The first
three were responsible for understanding the effects of a strong magnetic
field on the persistent current. Rob's vast knowledge of fabrication
of silicon micromechanical devices was invaluable when developing
the persistent current sample recipe. Dan Prober provided insight,
equipment, and old theses relevant to the transport measurements used
to characterize the electron phase coherence length. In addition to
sharing his keen insight into mesoscopic physics, Michel Devoret allowed
us to use his clean aluminum evaporator for depositing persistent
current samples. Luigi Frunzio also provided help with the evaporator
as well as with some early stages of the persistent current sample
fabrication. Several members of the labs of Michel Devoret and Rob
Schoelkopf provided assistance with sample preparation and characterization,
in particular Markus Brink, Nico Bergeal, and Julie Wyatt. Finally,
while mentioning people who worked in the Becton Engineering and Applied
Science Center, I'd also like to acknowledge Eric Akkermans, from
whose course and textbook I base much of my understanding of mesoscopics.
I have also useful exchanges with John Mamin, Martino Poggio, and
Hendrik Bluhm.

Within Sloane Physics Laboratory, I have been surrounded by many bright,
helpful people over the past six years. Dave DeMille, Sean Barrett,
Dan McKinsey, and members of their labs have provided equipment and
occasional advice. I am grateful to Sid Cahn for reminding me not
to hurt myself in the machine shop and for several loans of liquid
helium. Many people in the administrative offices upstairs, including
John Fox, SJ Compton, Lilian Snipes, and Julie Murphy, made it easy
to focus on research rather than paperwork. Outside of Yale, the hospitality
of Evan and Colleen at the Inn at City Lights in Ithaca made the long
trips to the Cornell NanoScale Facility for sample fabrication much
more bearable.

This text was written primarily on the wide screen television of Tera
Gahlsdorf. I am thankful for her daily company and support throughout
the writing process.

I thank my parents, Bill and Jean, and my sister Amanda for their
unflagging support over many years. My family has always encouraged
my pursuit of physics despite not understanding it, and I likely would
have taken a different path long ago without their faith in me. Over
the long period between when I finished a complete draft of this work
and when I finally completed the ultimate version for print, my father
would ask regularly to be sent the finished copy when it was ready
despite knowing that he would likely not understand much of it. To
me, this work will always be a symbol of the love of my family, without
which it could not have been created.

Finally, I thank you, the reader, for being brave enough to open this
document. I have struggled to make it as perfect as I could, working
on it well after the submission date and stopping only when I discovered
that my edits had reached a steady state in which they introduced
as many errors as they corrected.

This work was performed in part at the Cornell NanoScale Facility,
a member of the National Nanotechnology Infrastructure Network, which
is supported by the National Science Foundation (Grant ECS-0335765).

\frontmattersection{Glossary}

Many symbols are employed in this work. The most used ones are given
below along with the quantities the represent. Some symbols are used
multiple times to represent different quantities when that symbol
is the most fitting choice for each quantity and the chance of ambiguity
is low. These different quantities are given in the order in which
they appear and are separated by semicolons. References to the equation
or section where the quantity is first introduced are given in parentheses.
\begin{lyxlist}{00.00.0000}
\item [{$\left\langle \ldots\right\rangle $}] Average taken over disorder
(Sections \ref{sec:CHPCTh_DiffusiveRegime}, \ref{sub:CHPCTh_FluxThroughMetal})
\item [{$\overline{\ldots\vphantom{k}}$}] Average over a quantity other
than disorder (Section \ref{sub:CHPCTh_FiniteCrossSection})
\item [{$\alpha$}] Derivative of the normalized cantilever mode shape
$U$ with respect to the normalized displacement $\eta$ (Eq. \ref{eq:CHTorsMagn_Alpha})
\item [{$\beta$}] Magnetic field frequency (Section \ref{sub:ChData_SigProcDescription})
\item [{$\beta_{1}$}] Magnetic field frequency of the first harmonic of
the persistent current (Section \ref{sub:ChData_SigProcDescription})
\item [{$\gamma$}] Geometric factor relating the toroidal magnetic field
to the uniform one applied in experiment (Section \ref{sub:ChData_StatUncertainty})
\item [{$\Delta_{1}$}] Mean energy level spacing of a single transverse
channel of a ring (Section \ref{sub:CHPCTh_1DPerfFinT})
\item [{$\Delta_{1,M}$}] Mean energy level spacing of a single transverse
channel averaged over all channels in a ring (Section \ref{sub:CHPCTh_FiniteCrossSection})
\item [{$\Delta_{M}$}] Mean energy level spacing of a ring with a finite
cross-section (Eq. \ref{eq:CHPCTh_MultiChannelMeanLevelSpacing})
\item [{$\Delta f$}] Cantilever frequency shift (Section \ref{sec:CHTorsMagn_deltaFZeroDrive})
\item [{$\varepsilon$}] Energy of electrons within a ring
\item [{$\varepsilon_{\perp}$}] Transverse eigenvalues of the persistent
current diffusion constant (Eq. \ref{eq:CHPCTh_EperpToroidal})
\item [{$\varepsilon_{F}$}] Fermi energy (Section \ref{sub:CHPCTh_1DPerfFinT})
\item [{$\eta$}] Ratio of the statistical uncertainty in a quantity found
from a number of measurements $M_{\text{eff}}$ to the expected value
of that quantity
\item [{$\theta$}] Angle between magnetic field and cantilever beam axis
(Section \ref{sec:CHTorsMagn_deltaFZeroDrive})
\item [{$\kappa_{i}$}] The $i^{th}$ cumulant of a statistical distribution
(Eq. \ref{eq:AppCumul_CumulantDef})
\item [{$\lambda$}] Optical wavelength (Section \ref{sub:CHExpSetup_CantileverFiberInterferometer})
\item [{$\lambda_{0}$}] First order scale factor for the magnitude of
the persistent current due to electron-electron interactions (Eq.
\ref{eq:CHPCTh_Fee}); mean optical wavelength (Fig. \ref{fig:CHExpSetup_FiberCantileverFringe})
\item [{$\lambda_{\text{eff}}$}] Renormalized scale factor for the magnitude
of the persistent current due to electron-electron interactions (Eq.
\ref{eq:CHPCTh_Iee})
\item [{$\mu$}] Magnetic moment (Section \ref{sec:CHTorsMagn_deltaFZeroDrive})
\item [{$\nu$}] Density of states (Section \ref{sub:CHPCTh_1DPerfFinT})
\item [{$\nu_{0}$}] Density of states in the absence of disorder (Section
\ref{sub:PCTh_DisorderIntro})
\item [{$\tau$}] Torque (Eq. \ref{eq:TorqueRingGeneralFieldwithTheta})
\item [{$\phi$}] Magnetic flux through a ring (Section \ref{sub:CHPCTh_1DRingSingleLevelSolutions})
\item [{$\phi_{0}$}] Magnetic flux quantum $h/e$ 
\item [{$\Omega$}] Grand canonical free energy (Eq. \ref{eq:CHPCTh_GrandPotential})
\item [{$\omega$}] Angular frequency (usually $2\pi f$) (Section \ref{sec:CHTorsMagn_CantileverSHO})
\item [{$A$}] Area enclosed by a ring (typically the mean area in the
case of a ring with finite linewidth) (Section \ref{sec:CHTorsMagn_deltaFZeroDrive})
\item [{$B$}] Magnetic field
\item [{$B_{c}$}] Superconducting critical field (Eq. \ref{eq:CHSensitivity_BCSBc})
\item [{$B_{c,p}$}] Persistent current correlation field for the $p^{th}$
harmonic (Eq. \ref{eq:CHPCTh_BcpToroidalField})
\item [{$B_{M}$}] Toroidal magnetic field (Section \ref{sub:CHPCTh_FluxThroughMetal})
\item [{$b_{L}$}] Lorentzian function (Section \ref{sub:PCTh_DisorderIntro})
\item [{$C_{1}^{(0)}$}] Zero temperature persistent current autocorrelation
in energy (Eq. \ref{eq:CHPCTh_CurrentCurrentCorC10Def})
\item [{$c_{p}^{0}$}] Normalized $p^{th}$ harmonic of the zero temperature
persistent current autocorrelation in energy (Eq. \ref{eq:CHPCTh_cp0})
\item [{$c_{p}^{T}$}] Normalized square of the $p^{th}$ harmonic of the
temperature dependent persistent current magnitude (Eq. \ref{eq:CHPCTh_C10ZeemanT})
\item [{$D$}] Diffusion constant (Eq. \ref{eq:AppGrFu_DiffusionEquation})
\item [{$dI_{p}$}] Estimated $p^{th}$ harmonic of the derivative of the
persistent current with respect to field including a correction for
finite cantilever amplitude of motion (Eq. \ref{eq:ChData_dIpA})
\item [{$E$}] Energy of the ring-cantilever system
\item [{$E_{c}$}] Correlation scale of the energy levels of a ring in
the diffusive regime (Eq. \ref{eq:ChPCTh_ThoulessEnergy})
\item [{$E_{n}$}] Longitudinal eigenvalues of the persistent current diffusion
equation (Eq. \ref{eq:CHPCTh_DiffusonCooperonEigenvalues})
\item [{$E_{SO}$}] Spin-orbit energy scale (Section \ref{sub:CHPCTh_SpinOrbit})
\item [{$E_{Z}$}] Zeeman energy (Section \ref{sub:CHPCTh_Zeeman})
\item [{$e$}] Absolute value of the electron charge
\item [{$F_{p}$}] Helper function defined in the calculation of the persistent
current autocorrelation function (Eq. \ref{eq:CHPCTh_Fp})
\item [{$f$}] Fermi-Dirac distribution function; cantilever frequency
(Eq. \ref{eq:CHPCTh_PCthermalSum})
\item [{$G$}] Gain (Eq. \ref{eq:ChSensitivity_vPiezoComplexAmp}); Green
function (Appendix \ref{cha:AppGrFu_})
\item [{$g_{1}$}] Normalized temperature dependence of an ideal, one-dimensional
ring (Eq. \ref{eq:CHPCTh_g11DPerfectRing})
\item [{$g_{D}$}] Normalized temperature dependence of the typical current
of a diffusive ring (Eq. \ref{eq:CHPCTh_gDintegral})
\item [{$g_{M}$}] Normalized temperature dependence of an ideal ring of
finite cross-section (Eq. \ref{eq:CHPCTh_gM3DPerfectRing})
\item [{$H_{1}^{(0)}$}] Helper function defined in the calculation of
the persistent current autocorrelation function (Eq. \ref{eq:CHPCTh_H1Def})
\item [{$h$}] Planck's constant
\item [{$I$}] Persistent current (Eq. \ref{eq:CHPCTh_PCthermalSum})
\item [{$I'$}] Magnetic field derivative of the persistent current scaled
by $1/2\pi\beta_{1}$ so that the first harmonic amplitude should
have the same magnitude as the persistent current (Eq. \ref{eq:ChData_dIPrimeScaling})
\item [{$I_{0}$}] Characteristic magnitude $ev_{F}/L$ of the persistent
current in an ideal, one-dimensional ring (Eq. \ref{eq:CHPCTh_I0})
\item [{$I^{\text{can}}$}] Contribution to the average persistent current
present in the canonical ensemble but no the grand canonical ensemble
(Section \ref{sub:CHPCTh_AvgSingleParticle})
\item [{$I^{ee}$}] Contribution to the average persistent current due
to electron-electron interactions (Section \ref{sub:CHPCTh_AvgInteraction})
\item [{$i$}] Single energy level persistent current (Section \ref{sub:CHPCTh_1DSingleTotalCurrent})
\item [{$K_{p}$}] Normalized and scaled autocorrelation in magnetic field
of the $p^{th}$ harmonic of the persistent current (Eq. \ref{eq:ChData_KpCorr})
\item [{$k$}] Cantilever spring constant (Eq. \ref{eq:ChTorsMagn_springK})
\item [{$k_{B}$}] Boltzmann's constant
\item [{$k_{F,M}$}] Effective Fermi wave vector of one channel as a function
of channel indices in a three dimensional ring (Eq. \ref{eq:CHPCTh_kFM})
\item [{$L$}] Ring circumference (Fig. \ref{fig:CHPCTh_IdealRing}); wire
length (Appendix \ref{cha:AppTransport_})
\item [{$L_{\phi}$}] Phase coherence length of the electron (Section \ref{sec:AppTransport_phaseCoherence})
\item [{$l$}] Cantilever length (Fig. \ref{fig:LabeledUnflexedCantilever})
\item [{$l_{e}$}] Elastic mean free path of the electron (Eq. \ref{eq:AppGrFu_GDisorderAverage})
\item [{$L_{so}$}] Spin-orbit scattering length (Eq. \ref{eq:CHPCTh_ESO})
\item [{$M$}] Number of transverse channels in a three-dimensional ring
(Eq. \ref{eq:CHPCTh_MTransverse})
\item [{$M_{\text{eff}}$}] Effective number of transverse channels in
the diffusive regime (Eq. \ref{eq:CHPCTh_Meff}); effective number
of independent measurements contained in a trace of persistent current
versus magnetic field (Eq. \ref{eq:ChData_EffectiveNumberIndependent})
\item [{$m$}] Mass
\item [{$N$}] Number of rings in an array (Eq. \ref{eq:CHSensitivity_OptDimDeltaFreqPC})
\item [{$P$}] Optical power (Eq. \ref{eq:CHExpSetup_PInterferometerFull})
\item [{$p$}] Harmonic index of frequency shift or persistent current
(Eq. \ref{eq:CHPCTh_PCHarmonics})
\item [{$p_{\text{zero}}$}] Lowest value of the harmonic index $p$ for
which the finite drive correction results in no cantilever frequency
shift (Eq. \ref{eq:ChData_pzero})
\item [{$Q$}] Mechanical quality factor of a cantilever (Eq. \ref{eq:CHTorsMagn_Fdamping})
\item [{$R$}] Optical reflection coefficient (Eq. \ref{eq:CHExpSetup_PInterferometerFull});
electrical resistance
\item [{$S$}] Power spectral density (Section \ref{sub:CHSensitivity_ForceNoise})
\item [{$\mathcal{S}_{pc}$}] Ratio of persistent current signal to measurement
noise (Eq. \ref{eq:ChSensitivity_Sensitivity})
\item [{$T$}] Temperature; optical transmission coefficient (Eq. \ref{eq:CHExpSetup_PInterferometerFull})
\item [{$T_{b}$}] Temperature of the refrigerator in thermometry measurements
(Section \ref{sub:CHSensitivity_ThermometrySection})
\item [{$T_{c}$}] Superconducting transition temperature (Eq. \ref{eq:CHSensitivity_BCSBc})
\item [{$T_{e}$}] Electron temperature in thermometry measurements (Eq.
\ref{eq:CHSensitivity_BCSBc})
\item [{$T_{n}$}] Brownian motion noise temperature of the cantilever
(Eq. \ref{sub:CHSensitivity_MeasureBrownianMotion})
\item [{$T_{p}$}] Characteristic temperature of the $p^{th}$ harmonic
of the persistent current (Eqs. \ref{eq:CHPCTh_PerfectRingTp} and
\ref{eq:CHPCTh_TpDiffusive})
\item [{$t$}] Time; thickness (cantilever, ring, or wire where appropriate)
depending context
\item [{$t_{r}$}] Ring thickness (denoted by $t$ when there is no ambiguity)
\item [{typ}] Superscript used to denote the square root of the square
of a quantity averaged over disorder (e.g. $I^{\text{typ}}=\sqrt{\langle I^{2}\rangle}$)
(Eq. \ref{eq:CHPCTh_IpTyp})
\item [{$U_{m}$}] Normalized cantilever flexural mode shape for mode $m$
(Eq. \ref{eq:CHTorsMagn_UmCantileverMode})
\item [{$V$}] Disorder potential in a metal ring; voltage
\item [{$w$}] Width (cantilever, ring, or wire where appropriate)
\item [{$w_{r}$}] Ring linewidth (denoted by $w$ when there is no ambiguity)
\item [{$w_{w}$}] Wire linewidth (denoted by $w$ when there is no ambiguity)
\item [{$x$}] Displacement of the cantilever tip from its equilibrium
position (Eq. \ref{eq:CantileverRingdown})
\item [{$x_{0}$}] Distance from optical fiber to cantilever equilibrium
position (Fig. \ref{fig:CHExpSetup_CantileverFiberInterferometer})
\item [{$x_{1}$}] Displacement of the cantilever at the point addressed
by the fiber (Section \ref{sub:CHExpSetup_CantileverFiberInterferometer})
\item [{$x_{f,\max}$}] Amplitude of displacement of the cantilever at
the point addressed by the fiber (Eq. \ref{sub:CHExpSetup_CantileverFiberInterferometer})
\item [{$x_{\max}$}] Amplitude of displacement at the tip of an oscillating
cantilever (Eq. \ref{eq:CHTorsMagn_K1pertHamActAng})
\item [{$z$}] Expression composed various energy scales and introduced
for convenience in Chapter \ref{cha:CHMeso_} (Eq. \ref{eq:CHPCTh_z});
distance to a point on the cantilever from the base (Fig. \ref{fig:Flexed-cantilever-schematic})
\item [{$z_{f}$}] Distance from the base to the point on the cantilever
where the laser is incident (Section \ref{sub:CHExpSetup_CantileverFiberInterferometer})\end{lyxlist}

\frontmatterend

\chapter{\label{cha:Introduction_}Introduction}

The main goals of my graduate research were to develop a high quality
cantilever torsional magnetometer and most importantly to apply this
magnetometer to the measurement of persistent currents in normal metal
rings. In this text, I describe the success that I and my lab mates
had in achieving these goals. I document in detail each step of this
process including the theoretical description of persistent currents,
the nanofabrication of persistent current samples, the physical implementation
of the magnetometer, and the analysis of our torsional magnetometry
measurements. The results which I report here both add to an existing
body of research on persistent currents in normal metal rings and
open up the possibility of new measurements using torsional magnetometry.

In this text, I use {}``persistent current'' to mean an electrical
current flowing around a ring which is constant in time and which
is not driven by a power source external to the ring. As such, it
is a thermal equilibrium property of the ring system. While persistent
currents are usually associated with superconductors and atomic and
molecular quantum systems, quantum mechanics allows, surprisingly,
for such a current to exist in a resistive material as well. For the
persistent current in a ring of resistive material to be measurable,
the ring must be cold ($\apprle1\,\text{K}$) and small ($\apprle1\,\text{\ensuremath{\mu}m}$).

The specific focus of my experimental work is what I refer to in this
text as the {}``typical'' persistent current $I^{\text{typ}}$ in
normal metal rings. In practice, the atomic lattice of a metal contains
many microscopic defects upon which electrons scatter. The typical
persistent current depends sensitively on the microscopic details
of the metal's disorder and varies seemingly randomly in magnitude
and sign across an ensemble of rings fabricated with the same dimensions.
As a random quantity, the typical persistent current must be characterized
by its typical root-mean-square magnitude $I^{\text{typ}}=\sqrt{\langle I^{2}\rangle}$
averaged across an ensemble of rings with the same dimensions but
different microscopic arrangements of defects (here $\langle\ldots\rangle$
denotes averaging over all possible realizations of disorder with
the same density of scatterers). The typical persistent current may
be contrasted with a related quantity, the average persistent current
$I^{\text{avg}}=\langle I\rangle$, which is the same for each ring
in an ensemble.%
\footnote{This distinction between the typical and average components of the
persistent current is somewhat artificial because for any single ring
or array of rings there is only one persistent current signal (i.e.
the typical and average values of the persistent current are not measured
separately). However, the distinction is useful for discussing the
measurements of this text and their implications for the study of
persistent currents. As discussed in Chapter \ref{cha:CHMeso_}, our
measurements can not distinguish between the typical and average values
of the persistent current because, at the strong magnetic fields employed
in our measurements, the signal from an array of rings with moderate
variation in ring radius would appear random. In Chapter \ref{cha:CHMeso_},
different physical mechanisms (electron-electron interactions, simple
quantum interference in a diffusive system, flux-dependence of the
density of states in an isolated system, etc.) are identified as being
responsible for the leading contributions to the average and typical
values of the persistent current in a metal ring. Each of these mechanisms
is of theoretical and experimental interest. At strong magnetic fields,
all contributions to the average current are expected to be strongly
suppressed. I use the typical/average distinction in order to signal
to other researchers in the field of persistent currents that our
measurements (according to my interpretation of them) mainly provide
insight into one persistent current mechanism, namely that of non-interacting
electrons in the diffusive regime, without addressing the others.
I use {}``typical persistent current'' as a shorthand for {}``the
typical magnitude of the fluctuations of the persistent current of
non-interacting electrons in the diffusive regime'' because I know
of no other contribution to the fluctuations of the persistent current
that is not negligible compared to this contribution.%
}

At the beginning of this project, two measurements of the typical
persistent current in normal metal rings had been performed, one in
1991 and the other in 2001 \citep{chandrasekhar1991magnetic,jariwala2001diamagnetic}.
Despite the fact that in both cases gold rings of similar size were
studied using SQUID magnetometers, the reported magnitudes of the
persistent current for the two experiments were inconsistent with
each other. The first experiment observed persistent currents two
orders of magnitude larger than what was expected by theory and an
order of magnitude larger than what was observed in the second experiment.
The second measurement had reasonable agreement with theory.%
\footnote{Because the sample in the second experiment was observed to have a
diffusion constant $D$ larger than that of the first experiment,
the magnitude of the current expected using the analysis of Chapter
\ref{cha:CHMeso_} was an order of magnitude larger than in the first
experiment.%
} Adding to the intrigue surrounding persistent currents, measurements
of the average persistent current performed between 1990 and 2002
observed currents that differed from theory in both magnitude and
sign \citep{levy1990magnetization,reulet1995dynamic,jariwala2001diamagnetic,deblock2002acelectric,deblock2002diamagnetic}.
For completeness, I note that the typical persistent current was also
studied in semiconductor rings and found to agree roughly with theory
\citep{mailly1993experimental,rabaud2001persistent}. Additionally,
the typical current was measured again in gold rings at about the
same time as the measurements discussed here and was found to agree
well with theory \citep{bluhm2009persistent}.%
\footnote{In this brief introduction, I leave out the experiment of Ref. \citealp{kleemans2007oscillatory}
which studied persistent currents in a different parameter regime
(the ballistic regime with few electrons per ring). It is notable
that this measurement also used a torsional magnetometry technique,
though not one employing a high quality factor micromechanical device.%
}

The measurements which I report greatly expand the range of parameters
over which the typical persistent current has been measured. We studied
aluminum in which normal state persistent currents had not previously
been investigated. Across all samples (Tables \ref{tab:ChData_CLs}
and \ref{tab:ChData_Rings}), the range of observed current magnitudes
span three orders of magnitude from $\sim0.7\,\text{pA}$ to $\sim700\,\text{pA}$
whereas previous measurements had observed currents spanning less
than two orders of magnitude and had never observed currents smaller
than $\sim200\,\text{pA}$. The dimensions of the samples discussed
here varied sufficiently to possess characteristic temperatures $T_{p}$
(see Table \ref{tab:ChData_RingResults}) differing by a factor of
$\sim8$. No previous experiment had studied the temperature dependence
of co-deposited samples with significantly different characteristic
temperatures. Our measurements increase the maximum temperature at
which the typical current has been observed in normal metal rings
from $500\,\text{mK}$ to $2.5\,\text{K}$. We observe hundreds of
oscillations of the persistent current, which is roughly sinusoidal
in applied magnetic field, whereas previous measurements had measured
only a few oscillations, and we raise the maximum field at which the
typical persistent current has been observed from $\sim10\,\text{mT}$
to $8.4\,\text{T}$.

These many additions to the existing body of work on persistent currents
are made possible in large part by the torsional magnetometry measurement
technique. All of the previous measurements of the typical persistent
current had been performed using SQUID magnetometers. The torsional
magnetometry measurement increases in sensitivity with applied magnetic
field and is thus complementary to the SQUID measurement which requires
a weak applied magnetic field. The large magnetic field range afforded
by torsional magnetometry enabled the study of the magnetic field
correlation of the persistent current oscillation. This correlation
had never been studied previously.

The torsional magnetometry measurement technique permits the study
of persistent currents in a quiet electromagnetic environment. All
previous measurements of both the typical and average persistent current
had employed either a SQUID magnetometer or a superconducting resonator.
In the SQUID magnetometry measurements, the high frequency Josephson
current present in the measurement SQUID coupled inductively to the
ring sample.%
\footnote{In principle, the high frequency components of the magnetic flux produced
by the SQUID could have been shielded from the sample \citep{proberprivate}.%
} The superconducting resonators used to measure persistent currents
were also coupled inductively to the rings and possessed resonant
frequencies greater than $100\,\text{MHz}$. In attempting to explain
the experimental results obtained using these techniques, several
theoretical studies of non-equilibrium effects due to the presence
of high frequency electromagnetic radiation found that these effects
could mimic the equilibrium persistent current \citep{trivedi1988mesoscopic,efetov1991dynamic,janssen1993linearresponse,marchesoni1993persistent,kravtsov1993directcurrent,kravtsov1993disorderinduced,aronov1993nonlinear,reulet1994acconductivity,genkin1994resonance,galperin1996nonequilibrium,berman1997quantum,kopietz1998nonlinear,keller2000angularmomentum,yudson2001limitsof,chalaev2002aharonovbohm,yudson2003electron,arrachea2004dcresponse,matos-abiague2005photoinduced,dajka2006theinfluence,cohen2006rateof,kulik2007mesoscopic,moskalenko2007nonequilibrium}.
Thus the torsional magnetometry technique demonstrated here, which
does not necessitate the presence of high frequency electromagnetic
radiation, is advantageous because it allows the equilibrium persistent
current to be studied without the need to account for the possibility
of non-equilibrium effects.

Beyond non-equilibrium effects, persistent currents in the normal
state have been studied theoretically in many regimes (see chapter
\ref{cha:CHPrevWork}). Experimental progress has lagged behind. Besides
the work described in this text, only three experiments have been
reported in the last eight years \citep{kleemans2007oscillatory,bluhm2009persistent,koshnick2007fluctuation}.
The demonstration of cantilever torsional magnetometry as a powerful
alternative to SQUID and resonator based measurements is perhaps as
important a contribution to the field of persistent currents as our
experimental results themselves. Some possible follow up work to the
measurements described here is proposed in chapter \ref{cha:ChOutlook_}.

My main goal with this thesis is to provide a thorough account of
my experiment as well as a detailed theoretical description of persistent
currents. A secondary goal of mine is to present these topics beginning
at the level of a first year physics graduate student. Often in my
graduate tenure, I have been puzzled by an aspect of a published journal
article and dug up the thesis of the lead author only to be disappointed
by the lack of a clarification in its text. I hope that anyone searching
for additional explanation on any point in one of my published works
is aided by this text. If a section seems tedious, it is likely that
I struggled with its subject and want to dispel confusion from anyone
else dealing with it. I have tried to push the topics somewhat tangential
to persistent currents and their measurement into the appendices.

In chapter \ref{cha:CHMeso_}, I present a thorough theoretical description
of the basic persistent current phenomena. I begin with the simplest
possible model and build up to the regime relevant to the measurement
which I discuss in chapter \ref{cha:Data}. Chapter \ref{cha:CHMeso_}
draws heavily on the presentation given in Ref. \citep{akkermans2007mesoscopic}
and follows closely the results of Refs. \citealp{cheung1988isolated,cheung1988persistent,ambegaokar1990coherence,altshuler1991persistent,riedel1993mesoscopic,ginossar2010mesoscopic}
among others. I believe some small aspects of chapter \ref{cha:CHMeso_},
namely the discussion of the perfect three dimensional ring, the detailed
consideration of the effects of Zeeman splitting and spin-orbit scattering,
and the extension of the toroidal field model to the average persistent
current, to be novel. The theoretical machinery needed for chapter
\ref{cha:CHMeso_} is reviewed in Appendix \ref{cha:AppGrFu_}. In
chapter \ref{cha:CHPrevWork}, I review the persistent current literature
thoroughly, providing more detail to many of the points touched upon
in this introduction. 

In chapter \ref{cha:Cantilever-torsional-magnetometry}, I discuss
the physics of cantilever torsional magnetometry. This torsional magnetometry
technique was developed by myself, Ania Jayich, Jack Sankey, and Jack
Harris to take advantage of the unique magnetic properties of the
persistent current effect. In torsional magnetometry, one studies
a magnetic moment $\boldsymbol{\mu}$ through the torque $\boldsymbol{\tau}=\boldsymbol{\mu}\times\boldsymbol{B}$
that it experiences in a magnetic field $\boldsymbol{B}$. For our
cantilever torsional magnetometry technique, one infers the magnetic
properties of a sample by monitoring the spring constant of the cantilever
on which it is mounted. Because the persistent current depends strongly
on the orientation of the applied field, the frequency shift of the
cantilever varies quadratically in $\boldsymbol{B}$ (rather than
the typical linear dependence). This quadratic dependence on magnetic
field is what leads to the high sensitivity to persistent currents
which I discuss in chapter \ref{cha:CHSensitivity}. The torsional
magnetometry of persistent currents discussed in chapter \ref{cha:Cantilever-torsional-magnetometry}
makes use of some classical physics which is reviewed in Appendix
\ref{app:AppCanonPert}.

In chapter \ref{cha:CHExpSetup_} I present the experimental apparatus
and the procedures used to apply it to the persistent current measurement.
I was the first member of the Harris lab to work on the persistent
current project and began working in nearly an empty lab. Most of
equipment used in this experiment was acquired or designed by myself
and Ania Jayich under the direction of Jack Harris. The sample fabrication
process is also discussed in this chapter (with a more detailed recipe
given in Appendix \ref{app:AppSampFab}). The difficult process of
trial and error involved in developing a sample fabrication recipe
was done mostly by Ania Jayich along with Rob Ilic. Ania and I designed
the samples reported on in chapter \ref{cha:Data} together. We each
did roughly half of the cleanroom work for these samples. We took
all of the measurements together. Most of the LabVIEW routines used
in the measurements were programmed by me. 

In chapter \ref{cha:Data}, I outline the signal analysis procedure
developed by myself along with Ania Jayich and Jack Harris. I also
present analysis of all of our measurements of the persistent current.
All of the analysis presented in this chapter was performed by myself.%
\footnote{Ania also performed the analysis independently so we could check each
other for errors.%
} Transport measurements and analysis of a co-deposited wire are presented
in Appendix \ref{cha:AppTransport_}. These measurements allowed us
to check some properties of the samples independent of the persistent
current measurement. In chapter \ref{cha:ChOutlook_}, I discuss future
measurements which could build upon the results discussed in chapter
\ref{cha:Data}.

\chapter{\label{cha:CHMeso_}Review of theory relevant to mesoscopic persistent
currents in the normal state}

In this chapter we will review the basic theoretical results describing
persistent currents in small normal metal rings. We will begin with
the simplest possible model for the ring system and then refine it
to be applicable to three-dimensional rings of a normal metal in large
magnetic fields. We will acknowledge the original publications of
the results we summarize where appropriate but will reserve most discussion
of the persistent current literature for Chapter \ref{cha:CHPrevWork}.
In all cases, we assume that the ring dimensions are less than the
electron phase coherence length $L_{\phi}$ so that decoherence effects
may be ignored.

\section{\label{sec:CHPCTh_1DRing}The ideal one-dimensional ring}

The normal state persistent current studied in this dissertation is
an equilibrium property of the system. It is a quantum effect related
to the Aharonov-Bohm phase (discussed in Section \ref{sub:CHPCTh_1DRingSingleLevelSolutions})
picked up by a charged particle moving in a magnetic field. Although
it is possible for collective effects to contribute to an equilibrium
current, we will primarily be concerned with the current due to non-interacting
electrons (sometimes referred to as the {}``single-particle'' current).
This current arises from the fact that for any disorder configuration
of the ring each energy eigenstate possesses a non-zero expectation
value for its angular momentum. Since the electron is a charged particle,
this angular momentum has a current associated with it. These eigenstates
with finite angular momentum are analogous to the eigenstates of an
electron orbiting a nucleus which also possess finite angular momenta
for electrons outside of the \emph{s}-orbital. In principle, the calculation
of the current in a ring of many electrons boils down to calculating
the current associated with each energy eigenstate and then summing
up each of these currents with an appropriate weighting factor reflecting
the thermal population of that state.

\subsection{\label{sub:CHPCTh_1DRingSingleLevelSolutions}Solution of Schrödinger's
equation}

\begin{figure}

\begin{centering}
\includegraphics[width=0.35\paperwidth]{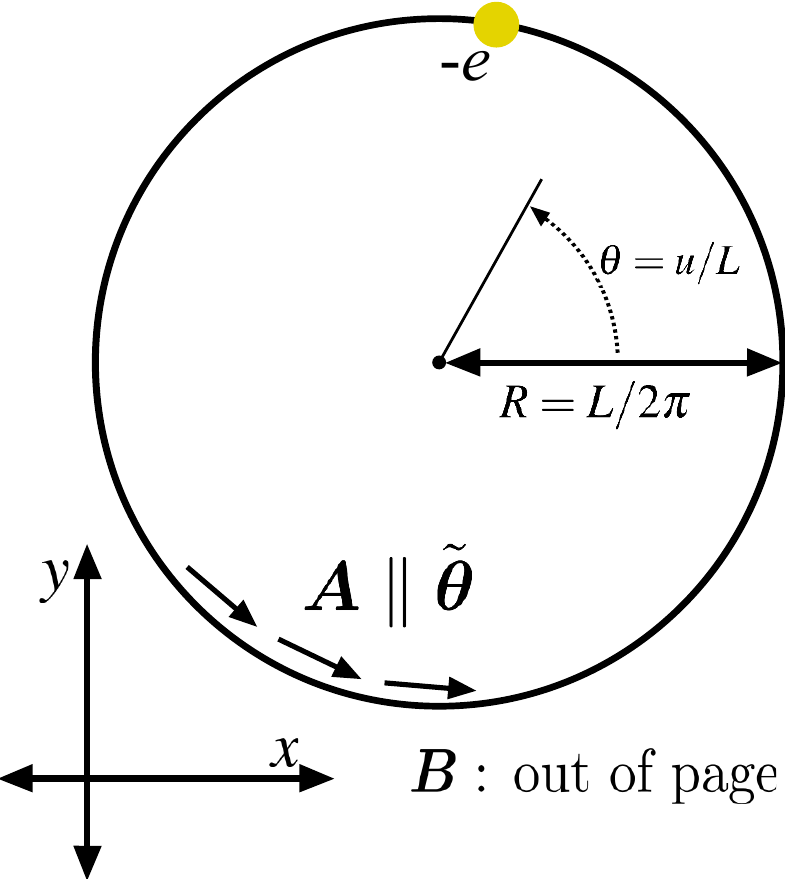}
\par\end{centering}

\caption[A perfect one-dimensional ring]{\label{fig:CHPCTh_IdealRing}A perfect one-dimensional ring. The
figure shows the orientation relative to an applied magnetic field
of a perfect, one-dimensional ring of radius $L/2\pi$ to which a
particle with charge $-e$ is confined. The applied magnetic field
$\boldsymbol{B}$ is parallel to the $\tilde{\boldsymbol{z}}$ direction
and points out of the page. The vector potential $\boldsymbol{A}$
chosen in the text is everywhere tangent to the ring and parallel
to the cylindrical unit vector $\tilde{\boldsymbol{\theta}}$.}

\end{figure}

We begin with the simplest possible model for the ring: a clean, one-dimensional
ring into which we put a single electron of mass $m$ and charge $-e$.
The task of describing this system is the textbook particle-in-a-box
problem from introductory quantum mechanics, here with periodic boundary
conditions. We parametrize the ring by the coordinate $u$ and call
its circumference $L$, as shown in Fig. \ref{fig:CHPCTh_IdealRing}.
The time-independent Schrödinger equation for the Hamiltonian $\hat{H}=\hat{P}^{2}/2m$
is 
\begin{equation}
-\frac{\hbar^{2}}{2m}\frac{d^{2}\psi_{n}}{du^{2}}=\varepsilon_{n}\psi_{n}\label{eq:CHPCTh_1DHamiltonian}
\end{equation}
with the wavefunction $\psi$ subject to the boundary condition
\begin{equation}
\psi\left(u+L\right)=\psi\left(u\right).\label{eq:CHPCTh_RingBoundCond}
\end{equation}
The eigenfunctions and eigenenergies indexed by integer $n$ are 
\begin{equation}
\psi_{n}\left(u\right)=\frac{1}{\sqrt{L}}\exp\left(2\pi in\frac{u}{L}\right)\label{eq:CHPCTh_1DIdealEigenfunctions}
\end{equation}
and
\[
\varepsilon_{n}=\frac{h^{2}}{2mL^{2}}n^{2}.
\]
We consider the effects of applying a constant, uniform magnetic field
$\boldsymbol{B}$ to the system. We choose coordinates so that the
ring lies in the $xy$-plane and is centered on the origin. We assume
that the magnetic field $\boldsymbol{B}=B\tilde{\boldsymbol{z}}$
is parallel to the z-axis.

In the presence of a magnetic field, the canonical momentum $\boldsymbol{P}_{\text{can}}$
for a particle of charge $-e$ and mass $m$ becomes $\boldsymbol{P}_{\text{can}}=\boldsymbol{P}_{\text{mech}}-e\boldsymbol{A}$
where $\boldsymbol{P}_{\text{mech}}=m\boldsymbol{v}$ is the particle's
mechanical momentum at velocity $\boldsymbol{v}$ and $\boldsymbol{A}$
is the vector potential of the magnetic field satisfying $\boldsymbol{B}=\nabla\times\boldsymbol{A}$.
Only the mechanical momentum contributes to the particle's energy.
Thus the Hamiltonian becomes $\hat{H}=(\hat{\boldsymbol{P}}+e\boldsymbol{A}(\hat{\boldsymbol{X}}))^{2}/2m$,
where $\hat{\boldsymbol{X}}$ is the position operator.%
\footnote{We will sometimes suppress the dependence of $\boldsymbol{A}$ on
the operator$\hat{\boldsymbol{X}}$ by expressing the vector potential
as an operator $\hat{\boldsymbol{A}}$.%
}

For $\boldsymbol{B}=B\tilde{\boldsymbol{z}}$, we can write $\boldsymbol{A}=(-B/2)(y\tilde{\boldsymbol{x}}-x\tilde{\boldsymbol{y}})$.
Defining cylindrical coordinates $(r,\theta)$ and corresponding unit
vectors $(\tilde{\boldsymbol{r}},\tilde{\boldsymbol{\theta}})$ satisfying
\[
x=r\cos\theta
\]
\[
y=r\sin\theta
\]
\[
r=\sqrt{x^{2}+y^{2}}
\]
\[
\theta=\tan^{-1}\left(y/x\right)
\]
\[
\tilde{\boldsymbol{r}}=\cos\theta\,\tilde{\boldsymbol{x}}+\sin\theta\,\tilde{\boldsymbol{y}}
\]
\[
\tilde{\boldsymbol{\theta}}=-\sin\theta\,\tilde{\boldsymbol{x}}+\cos\theta\,\tilde{\boldsymbol{y}},
\]
we can write $\boldsymbol{A}=Br\tilde{\boldsymbol{\theta}}/2$. Confined
to $r=L/2\pi$ along the ring, we can also write $\boldsymbol{A}=\frac{\phi}{L}\tilde{\boldsymbol{\theta}}$
where $\phi=\pi(L/2\pi)^{2}B$ is the flux enclosed by the ring. The
coordinate $u=L\theta/2\pi$ follows the ring in a circle about the
origin, and thus the derivative $\frac{d}{du}$, which is everywhere
tangent to the ring, is always parallel to $\tilde{\boldsymbol{\theta}}$.
Thus, in the presence of $\boldsymbol{B}$, the one-dimensional time-independent
Schrödinger equation parametrized by $u$ becomes
\[
\frac{1}{2m}\left(-i\hbar\frac{d\psi_{n}}{du}+e\frac{\phi}{L}\right)^{2}=\varepsilon_{n}\psi_{n}.
\]
The eigenfunctions $\psi_{n}$ are again given by Eq. \ref{eq:CHPCTh_1DIdealEigenfunctions}
with the eigenenergies now
\begin{align}
\varepsilon_{n} & =\frac{1}{2m}\left(\frac{2\pi\hbar}{L}n+e\frac{\phi}{L}\right)^{2}\nonumber \\
 & =\frac{h^{2}}{2mL^{2}}\left(n+\frac{\phi}{\phi_{0}}\right)^{2}\label{eq:CHPCTh_EnergyLevel1DPerfectRing}
\end{align}
where we have introduced the flux quantum $\phi_{0}=h/e$.

\begin{figure}
\begin{centering}
\includegraphics[width=0.6\paperwidth]{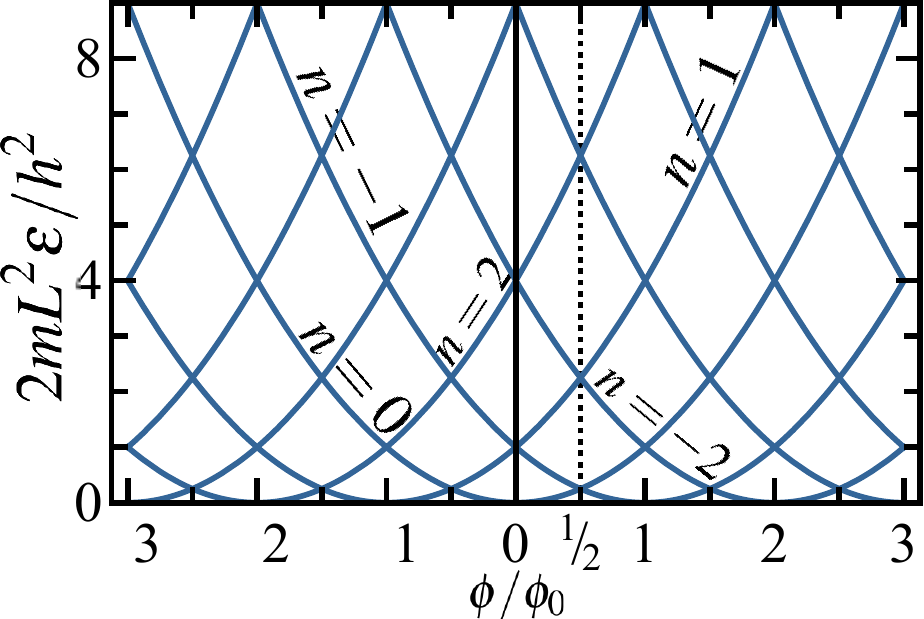}
\par\end{centering}

\caption[Eigenenergies of a perfect one-dimensional ring]{\label{fig:CHPCTh_EnergyLevelsSimple}Eigenenergies of a perfect
one-dimensional ring. The figure shows the first few eigenenergies
$\varepsilon_{n}$ of Eq. \ref{eq:CHPCTh_EnergyLevel1DPerfectRing}
plotted versus the normalized flux $\phi/\phi_{0}$. The flux periodicity
of the energy spectrum discussed in the text can clearly be seen.
The dashed line marks the point $\phi=\phi_{0}/2$. In the range $0<\phi<\phi_{0}/2$,
the energy levels are, in increasing order, $n=0,\,-1,\,+1,\,-2,\,+2,$
etc.}
\end{figure}

\begin{figure}

\begin{centering}
\includegraphics[width=0.6\paperwidth]{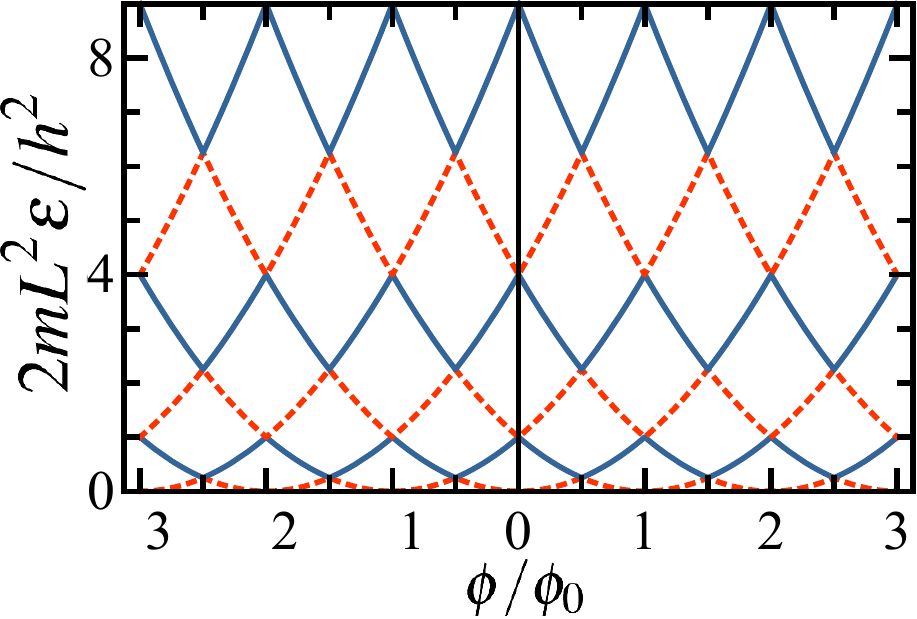}
\par\end{centering}

\caption[Energy levels of a perfect one-dimensional ring]{\label{fig:CHPCTh_EnergyLevelsSimpleReindexed}Energy levels of a
perfect one-dimensional ring. The figure shows the energy spectrum
of Fig. \ref{fig:CHPCTh_EnergyLevelsSimple} with each single energy
level drawn either with a solid or a dashed curve. Each energy level
has a kink at the points $\phi=m\phi_{0}/2$ for integer $m$ where
two of the eigenenergy parabolae intersect. These kinks are smoothed
out with the introduction of disorder to the ring.}
\end{figure}

The energies $\varepsilon_{n}$ are plotted versus flux $\phi$ in
Fig. \ref{fig:CHPCTh_EnergyLevelsSimple}. Although the eigenenergy
$\varepsilon_{n}$ of the $n^{th}$ eigenstate is a parabola, the
even spacing of the different $\varepsilon_{n}$ in $\phi$ produces
a spectrum that is periodic overall. We will discuss this flux periodicity
shortly. First, we note that in later sections it will often be preferable
to discuss the energy spectrum as a set of {}``levels'' rather than
a set of eigenenergies corresponding to particular eigenstates. By
energy levels, we mean $\varepsilon$ curves which are periodic in
flux. In Fig. \ref{fig:CHPCTh_EnergyLevelsSimpleReindexed}, the energy
spectrum of Fig. \ref{fig:CHPCTh_EnergyLevelsSimple} is replotted
with the energy levels, rather than the eigenenergies, indicated by
the alternating use of solid and dashed lines for each level. Since
the energy spectrum is symmetric under both $\phi\rightarrow-\phi$
and $\phi\rightarrow\phi+\phi_{0}$, the spectrum is totally specified
by its set of values in the range $0<\phi<\phi_{0}/2$. We could thus
label the energy levels by the index $n$ of the corresponding eigenenergy
over this range, in which case the energy levels are in increasing
order: $n=0,\,-1,\,+1,\,-2,\,+2,$ etc.

Rather than solving the Schrödinger equation with $\boldsymbol{A}$
included explicitly, it is possible to use a gauge transformation
to remove it from the Hamiltonian, restoring the original $\boldsymbol{B}=0$
Hamiltonian. In quantum mechanics, a gauge transformation involves
replacing $\boldsymbol{A}$ and $\psi$ by 
\[
\boldsymbol{A}'=\boldsymbol{A}+\nabla\Lambda
\]
and
\[
\psi'=\psi\exp\left(-i\frac{e\Lambda}{\hbar}\right)
\]
where $\Lambda$ is some smooth function of position. Since $\nabla\times\nabla\Lambda=0$
for all $\Lambda$, this modification to $\boldsymbol{A}$ does not
change $\boldsymbol{B}$. The product rule then gives 
\begin{align*}
\left(\hat{\boldsymbol{P}}+e\hat{\boldsymbol{A}}'\right)\psi' & =\left(-i\hbar\nabla\right)\psi'+\left(e\hat{\boldsymbol{A}}+e\nabla\Lambda\right)\psi'\\
 & =\left(-i\hbar\exp\left(-i\frac{e\Lambda}{\hbar}\right)\nabla\psi-e\left(\nabla\Lambda\right)\psi'\right)+\left(e\hat{\boldsymbol{A}}+e\nabla\Lambda\right)\psi'\\
 & =\exp\left(-i\frac{e\Lambda}{\hbar}\right)\left(-i\hbar\nabla+e\hat{\boldsymbol{A}}\right)\psi\\
 & =\exp\left(-i\frac{e\Lambda}{\hbar}\right)\left(\hat{\boldsymbol{P}}+e\hat{\boldsymbol{A}}\right)\psi.
\end{align*}

Often in quantum mechanics an overall phase factor such as the factor
$\exp(-ie\Lambda/\hbar)$ introduced in the gauge transformation has
no impact on the calculation of any physical observable and so the
vector potential can be freely shifted by $\nabla\Lambda$ without
consequence. However, Ehrenberg and Siday \citep{ehrenberg1949therefractive},
and later Aharonov and Bohm \citep{aharonov1959significance}, showed
that, when two trajectories enclosing a magnetic flux $\phi$ interfere,
the phase factor due the vector potential plays an important role,
shifting the overall phase of the interference by $2\pi\phi/\phi_{0}$.
A similar effect holds for eigenfunctions of the ideal ring which
we have been considering.

In the case of the ring, we can choose 
\begin{align*}
\Lambda & =-\frac{1}{2}RBu\\
 & =-\frac{1}{2}R^{2}B\theta\\
 & =-\phi\frac{\theta}{2\pi}.
\end{align*}
In cylindrical coordinates, $\nabla\Lambda=\tilde{\boldsymbol{r}}\frac{\partial\Lambda}{\partial r}+\tilde{\boldsymbol{\theta}}\frac{1}{r}\frac{\partial\Lambda}{\partial\theta}+\tilde{\boldsymbol{z}}\frac{\partial\Lambda}{\partial z}$,
so along the ring $\nabla\Lambda=-\frac{\phi}{L}\tilde{\boldsymbol{\theta}}$
and $\boldsymbol{A}'=0$. With this gauge transformation, we thus
return to the Hamiltonian given in Eq. \ref{eq:CHPCTh_1DHamiltonian}.
The new wavefunctions are 
\[
\psi'\left(u\right)=\exp\left(2\pi i\frac{\phi}{\phi_{0}}\frac{u}{L}\right)\psi\left(u\right).
\]
Using the original boundary condition of Eq. \ref{eq:CHPCTh_RingBoundCond},
the new boundary condition on $\psi'$ is
\begin{equation}
\psi'\left(u+L\right)=\exp\left(2\pi i\frac{\phi}{\phi_{0}}\right)\psi'\left(u\right).\label{eq:CHPCTh_BoundaryConditionGaugeTrans}
\end{equation}
This boundary condition, which determines which combinations of the
eigenfunctions given in Eq. \ref{eq:CHPCTh_1DIdealEigenfunctions}
are permitted, now depends periodically on $\phi$ with period $\phi_{0}$.
Since this is a general property of the solutions of the Hamiltonian,
it follows that \emph{all} properties of the system are periodic in
$\phi$ with period $\phi_{0}$. This property carries over in the
generalization to a disordered, three-dimensional ring and is one
of the key signatures of the persistent current.

The phrase {}``carries over'' is to be taken loosely here. When
the ring has a finite linewidth, the flux threading the ring is not
a well-defined quantity. In this case, we can define $\phi$ to be
the flux through the mean radius of the ring. The persistent current
is no longer strictly periodic in $\phi$, but its Fourier transform
with respect to $\phi$
\[
I\left(p\right)=\int_{-\infty}^{\infty}d\phi\, I\left(\phi\right)e^{-2\pi ip\phi/\phi_{0}}
\]
is peaked near $p=1$, with its peak width determined by the ring's
finite cross-section. 

Often the persistent current is discussed using an idealized {}``Aharonov-Bohm
flux'' $\phi$ threading the ring but not penetrating the linewidth
of the ring itself, similar to the arrangement presented by Refs.
\citep{ehrenberg1949therefractive} and \citep{aharonov1959significance}.
It is possible to attain all of the results given here for the one-dimensional,
perfect ring using this idealized flux. If the flux $\phi$ is taken
to be produced by a field $B$ threading a disk of radius $a$, then
the vector potential can be written as 
\begin{equation}
\boldsymbol{A}=\begin{cases}
Br\tilde{\boldsymbol{\theta}}/2, & r<a\\
\phi\tilde{\boldsymbol{\theta}}/2\pi r, & r>a
\end{cases}\label{eq:CHPCTh_VectorAABflux}
\end{equation}
which will give the same values for $\boldsymbol{A}$ along the ring
as we found above.

More generally, we can consider any arrangement of field $\boldsymbol{B}(\boldsymbol{r})$
for which $B=0$ everywhere inside the ring. Choosing the Lorenz gauge
for which $\nabla\cdot\boldsymbol{A}=0$ (a condition which the $\boldsymbol{A}$
defined in Eq. \ref{eq:CHPCTh_VectorAABflux} satisfies), the line
integral 
\[
\Lambda\left(\boldsymbol{r}\right)=-\int_{\boldsymbol{r}_{0}}^{\boldsymbol{r}}d\boldsymbol{r}'\cdot\boldsymbol{A}\left(\boldsymbol{r}'\right)
\]
is independent of the path of integration within the ring, as long
as the path does not encircle the ring. The choice of $\boldsymbol{r}_{0}$
is arbitrary and shifts $\Lambda$ by a constant with no physical
significance. If the path does encircle the ring, the line integral
can be decomposed into 
\[
\Lambda\left(\boldsymbol{r}\right)=-\int_{\boldsymbol{r}_{0}}^{\boldsymbol{r}}d\boldsymbol{r}'\cdot\boldsymbol{A}\left(\boldsymbol{r}'\right)-n\oint d\boldsymbol{r}'\cdot\boldsymbol{A}\left(\boldsymbol{r}'\right)
\]
where the first integral does not encircle the ring, the second integral
represents a closed loop encircling the ring, and $n$ is the number
of times that the original path encircled the ring. Since $\nabla\times\boldsymbol{A}=\boldsymbol{B}=0$
over the path of integration, Stokes' theorem can be applied to state
\begin{align*}
\oint d\boldsymbol{r}'\cdot\boldsymbol{A}\left(\boldsymbol{r}'\right) & =\oiint d\boldsymbol{S}\cdot\left(\nabla\times\boldsymbol{A}\right)\\
 & =\oiint d\boldsymbol{S}\cdot\boldsymbol{B}\\
 & =\phi
\end{align*}
where the surface integral is taken over the enclosed path and thus
is equal to the total flux $\phi$ threading the ring. If we take
this $\Lambda$ to be the function in the gauge transformation introduce
above, the transformed potential is $\boldsymbol{A}'=\boldsymbol{A}+\nabla\Lambda(\boldsymbol{r})=0$
and the magnetic field is eliminated from the Hamiltonian. Under the
gauge transformation, the boundary condition of Eq. \ref{eq:CHPCTh_RingBoundCond}
becomes 
\begin{align*}
\psi\left(u+L\right) & =\exp\left(2\pi i\left(\Lambda\left(\boldsymbol{r}+L\right)-\Lambda\left(\boldsymbol{r}\right)\right)/\phi_{0}\right)\psi\left(u\right)\\
 & =\exp\left(2\pi i\left(\oint d\boldsymbol{r}'\cdot\boldsymbol{A}\left(\boldsymbol{r}'\right)\right)/\phi_{0}\right)\psi\left(u\right)\\
 & =\exp\left(2\pi i\phi/\phi_{0}\right)\psi\left(u\right)
\end{align*}
where $\Lambda(\boldsymbol{r}+L)$ represents a line integral following
a path from $\boldsymbol{r}_{0}$ to point $\boldsymbol{r}$ plus
the integral representing one closed curve encircling the ring. Thus,
this more general treatment results in the same flux periodic boundary
condition as was given above in Eq. \ref{eq:CHPCTh_BoundaryConditionGaugeTrans}.
With this last, general form of the gauge transformation, we have
derived this flux periodicity without specifying the dimensionality
of the ring nor the potential energy term $V(\boldsymbol{r})$ of
the Hamiltonian. Thus, for this idealized Aharonov-Bohm flux threading
the ring but not penetrating the metal, the property of flux periodicity
carries over exactly from the perfect ring to the three-dimensional,
disordered ring. Due to the micrometer size scale necessary for the
persistent current to be observable, such an idealized flux would
be nearly impossible to impose in practice and is useful only as a
conceptual device (and as a first approximation for a ring with a
reasonable aspect ratio).

\FloatBarrier

\subsection{\label{sub:CHPCTh_1DSingleTotalCurrent}The single-level and total
current}

We now find the current associated with each eigenstate. A particle
moving at velocity $v$ around a ring of circumference $L$ makes
one round trip in time $\Delta t=L/v$. If the particle has charge
$-e$, then the average current, the charge passing any given point
of the ring per unit time averaged over one period, is $i=-e/\Delta t=-ev/L$.
As we stated above, the velocity of such a charged particle in a magnetic
field can be written as $v=P_{\text{mech}}/m=(P_{\text{can}}+eA)/m$.
Then for the system we have been considering, the velocity $v_{n}$
associated with the $n^{\text{th}}$ eigenstate as given in Eq. \ref{eq:CHPCTh_1DIdealEigenfunctions}
satisfies
\begin{align*}
v_{n}\psi_{n}\left(u\right) & =\frac{\left(P_{\text{can}}+eA\right)}{m}\psi_{n}\left(u\right)\\
 & =\frac{1}{m}\left(-i\hbar\frac{d}{du}+e\frac{\phi}{L}\right)\left(\frac{1}{\sqrt{L}}\exp\left(2\pi in\frac{u}{L}\right)\right)\\
 & =\frac{h}{mL}\left(n+\frac{\phi}{\phi_{0}}\right)\left(\frac{1}{\sqrt{L}}\exp\left(2\pi in\frac{u}{L}\right)\right),
\end{align*}
giving a current
\begin{align*}
i_{n} & =-\frac{e}{L}v_{n}\\
 & =-\frac{eh}{mL^{2}}\left(n+\frac{\phi}{\phi_{0}}\right).
\end{align*}
We note that in terms of the energy $\varepsilon_{n}$ given in Eq.
\ref{eq:CHPCTh_EnergyLevel1DPerfectRing} the current $i_{n}$ may
be written as
\[
i_{n}=-\frac{\partial\varepsilon_{n}}{\partial\phi}.
\]

Having described the single-particle states of the ideal ring, we
now consider the case of a ring filled with many electrons, as would
be the case in a real metal ring. We assume that the electrons are
non-interacting. This assumption is often appropriate due to charge
screening which suppresses long-range interactions within the metal
\citep{mattuck1992aguide}, but it is one which we will revisit later.
To find the total current $I$ at temperature $T$, we sum over the
contributions of each individual level with the weighting factor $f(\varepsilon,\mu,T)$
given by the Fermi-Dirac distribution function%
\footnote{This analysis is appropriate for the grand canonical ensemble in which
the ring can exchange electrons with a reservoir. Technically, such
a model is not accurate for the isolated ring system which we study.
However, it can be shown that for a gas of Fermions in the grand canonical
ensemble the typical fluctuations $\sqrt{\left\langle \Delta N^{2}\right\rangle }$
in particle number $N$ scale as $\sqrt{T}$ and so are negligible
at sufficiently low temperature. Many of the early and most cited
publications on persistent currents, such as Refs. \citep{cheung1988isolated,cheung1988persistent,cheung1989persistent,riedel1989persistent,riedel1993mesoscopic,ginossar2010mesoscopic},
use this model to calculate the current. We note that some authors
have taken issue with the use of the grand canonical ensemble \citep{yip1996persistent},
while others have argued that the calculation in the grand canonical
ensemble can be related to the one in the canonical ensemble by making
the chemical potential flux dependent so that as the energy levels
change with $\phi$ the number of occupied levels remains constant
\citep{altshuler1991persistent,schmid1991persistent,saminadayar2004equilibrium}.
We will touch on the difference between the canonical and grand canonical
ensemble again when discussing the average current in the diffusive
regime.%
}
\begin{align}
I & =\sum_{n}i_{n}f\left(\varepsilon_{n},\mu,T\right)\nonumber \\
 & =\sum_{n}\left(-\frac{\partial\varepsilon_{n}}{\partial\phi}\right)\left(1+\exp\left(-\frac{\left(\mu-\varepsilon_{n}\right)}{k_{B}T}\right)\right)^{-1}\nonumber \\
 & =-\frac{\partial\Omega}{\partial\phi}\label{eq:CHPCTh_PCthermalSum}
\end{align}
where
\[
\Omega=-k_{B}T\sum_{n}\left(\ln\left(1+\exp\left(\frac{\left(\mu-\varepsilon_{n}\right)}{k_{B}T}\right)\right)\right)
\]
is the grand potential of thermodynamics and $n$ also indexes both
spin states. We assume that the temperature is well below the Fermi
temperature so that the chemical potential $\mu\approx\varepsilon_{F}$
where $\varepsilon_{F}$ is the Fermi energy. In the limit of low
temperature $T\rightarrow0$, the Fermi-Dirac distribution function
$f(\varepsilon_{n},\varepsilon_{F},T)$ becomes $\theta(\varepsilon_{F}-\varepsilon_{n})$
where $\theta\left(x\right)$ is the Heaviside function equal to 1
for $x>0$ and 0 otherwise. Thus the sum in the first two lines of
Eq. \ref{eq:CHPCTh_PCthermalSum} becomes a sum over the single-level
currents of all energy levels below the Fermi level. 

Within the flux range $0<\phi<\phi_{0}/2$, the energy levels in ascending
order are 0, -1, +1, -2, +2, etc. as shown in Fig. \ref{fig:CHPCTh_EnergyLevelsSimple}.
For $-\phi_{0}/2<\phi<0$, the ordering of the levels $+n$ and $-n$
is reversed for each $n$. If we consider filling up all of the levels
to $n=\pm N$ as is appropriate for this flux range and $T=0$, we
will get a contribution $(-eh/mL^{2})\phi/\phi_{0}$ to the current
due to the $n=0$ level and a total contribution $2(-eh/mL^{2})\phi/\phi_{0}$
due to each pair $\pm n$. Denoting the total current due to all levels
$|n|\leq N$ by $I_{N}$ and accounting for the twofold degeneracy
of each level due to spin,%
\footnote{Note that, including spin degeneracy, $I_{N}$ is the current in a
ring filled with $4N+2$ electrons, and, in the notation used below,
$I_{N+n}$ is the current for a ring with $4N+2+n$ electrons.%
} we find
\[
I_{N}=-\frac{eh}{mL^{2}}\left(4N+2\right)\frac{\phi}{\phi_{0}}.
\]
Since we summed each pair $\pm n$, this result does not depend on
the ordering and is valid for $-\phi_{0}/2<\phi<\phi_{0}/2$. 

Over this range, the contributions of the next three electrons%
\footnote{Here we are switching back temporarily to the canonical ensemble.
In the grand canonical ensemble, each level is either doubly filled
or empty, so the expressions below for singly occupied levels ($I_{N+1}$
and $I_{N+3}$) are never applicable.%
} to the total current are $\frac{eh}{mL^{2}}(\sigma N-\phi/\phi_{0})$,
$\frac{eh}{mL^{2}}(\sigma N-\phi/\phi_{0})$, and $\frac{eh}{mL^{2}}(-\sigma N-\phi/\phi_{0})$
respectively with $\sigma=\text{sgn}(\phi)$ the sign of $\phi$.
The velocity of the $N^{\text{th}}$ level at $\phi=0$ is $hN/mL$.
Taking $N\gg1$ and denoting the magnitude of the velocity of the
highest filled as $v_{F}\approx hN/mL$, we can write 
\begin{align}
I_{N+0} & \approx-4\frac{ev_{F}}{L}\frac{\tilde{\phi}}{\phi_{0}}\label{eq:CHPCTh_IN0}\\
I_{N+1} & \approx\sigma\frac{ev_{F}}{L}-4\frac{ev_{F}}{L}\frac{\tilde{\phi}}{\phi_{0}}\label{eq:CHPCTh_IN1}\\
I_{N+2} & \approx2\sigma\frac{ev_{F}}{L}-4\frac{ev_{F}}{L}\frac{\tilde{\phi}}{\phi_{0}}\label{eq:CHPCTh_IN2}\\
I_{N+3} & \approx\sigma\frac{ev_{F}}{L}-4\frac{ev_{F}}{L}\frac{\tilde{\phi}}{\phi_{0}}\label{eq:CHPCTh_IN3}
\end{align}
where we have used the reduced flux $\tilde{\phi}=\phi-M\phi_{0}$
where $M$ is the integer required to satisfy $-\phi_{0}/2<\tilde{\phi}<\phi_{0}/2$.%
\footnote{More precisely, $M=\left\lfloor \left(\phi+\phi_{0}/2\right)/\phi_{0}\right\rfloor $
where $\left\lfloor z\right\rfloor $ denotes the nearest integer
less than $z$.%
} 

\begin{figure}
\centering{}\includegraphics[width=0.7\paperwidth]{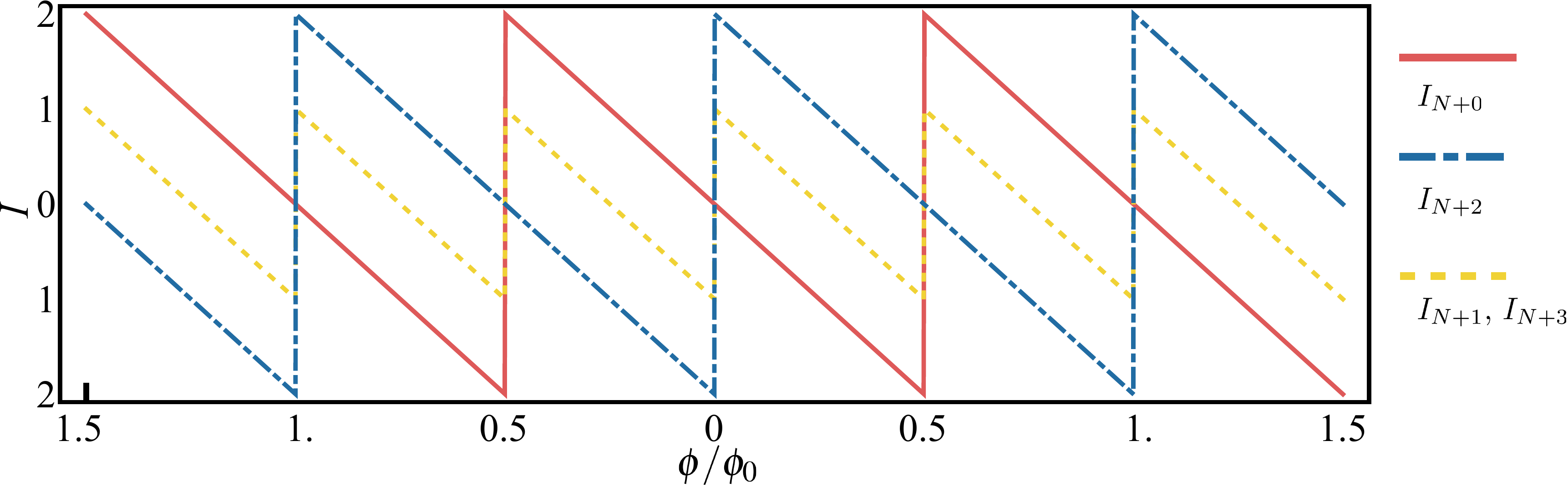}\caption[Current versus flux for a perfect one-dimensional ring]{\label{fig:PCTh_CurrentSimple}Current versus flux for a perfect
one-dimensional ring. The currents for different numbers of electrons
as given in Eqs. \ref{eq:CHPCTh_IN0} through \ref{eq:CHPCTh_IN3}
are plotted versus $\phi/\phi_{0}$. The current axis is scaled in
units of $I_{0}$. For each filling of the ring, the current has a
diamagnetic slope and is of the same order of magnitude. However,
the discontinuous jumps in the current occur at different values of
$\phi/\phi_{0}$ for different fillings. The jump at $\phi=0$ for
$I_{N+2}$ results in its odd Fourier components taking positive values
despite the fact that the slope is always negative (except at the
points of discontinuity). As is discussed below, in the presence of
finite temperature or disorder, higher harmonics are suppressed, meaning
that $I_{N+2}$ is dominated by its first harmonic and thus has a
paramagnetic slope at $\phi=0$.}
\end{figure}

The current for various numbers of electrons in the ring as given
in Eqs. \ref{eq:CHPCTh_IN0} through \ref{eq:CHPCTh_IN3} is plotted
against flux threading the ring in Fig. \ref{fig:PCTh_CurrentSimple}.
In all cases, the typical current magnitude is of order 
\begin{equation}
I_{0}=\frac{ev_{F}}{L}.\label{eq:CHPCTh_I0}
\end{equation}
Since this value has the same magnitude as that of the electron in
the highest occupied level, each additional electron has a strong
effect on the flux dependence of the current. These conclusions can
also be drawn from Fig. \ref{fig:CHPCTh_EnergyLevelsSimple}. Since
the single-level currents are equal to the slopes of the energy levels
$-\partial\varepsilon/\partial\phi$ and the slopes of successive
energy levels are anti-correlated, the current of each electron added
to the ring tends to cancel out the contribution of the previous electron
resulting in a total current with a magnitude of the order of the
current contribution from the highest energy level and with a flux
dependence that is sensitive to the number of electrons in the ring.

Taking the analysis one more step, we can calculate the harmonics
of the current. We note that the current is in all cases periodic
in $\phi$ and antisymmetric about $\phi=0$ and can thus be expanded
as 
\begin{equation}
I=\sum_{p}I_{p}\sin2\pi p\frac{\phi}{\phi_{0}}.\label{eq:CHPCTh_PCHarmonics}
\end{equation}
Using $I_{p}=\frac{4}{\phi_{0}}\int_{0}^{\phi_{0}/2}d\phi\, I(\phi)\sin(2\pi p\phi/\phi_{0})$,
$\int_{0}^{1/2}dx\, x\sin(2\pi px)=-(-1)^{p}/4\pi p$, and $\int_{0}^{1/2}dx\,\sin(2\pi px)=(1-(-1)^{p})/2\pi p$,
we find the harmonics of the current for the different fillings of
the energy levels to be
\begin{align}
I_{p,N+0} & =\frac{4}{\pi p}\left(-1\right)^{p}I_{0}\label{eq:CHPCTh_IpN0coeff}\\
I_{p,N+1} & =\frac{2}{\pi p}\left(1+\left(-1\right)^{p}\right)I_{0}\nonumber \\
I_{p,N+2} & =\frac{4}{\pi p}I_{0}\label{eq:CHPCTh_IpN2coeff}\\
I_{p,N+3} & =I_{p,N+1}.\nonumber 
\end{align}
Notably, for $p$ odd $I_{p,N+0}=-I_{p,N+2}$ and $I_{p,N+1}=I_{p,N+3}=0$,
whereas for $p$ even the harmonic is the same for each number of
electrons. For an ensemble of rings with a spread in the number of
electrons per ring greater than four, the ensemble average of the
harmonics $I_{p}$ with odd $p$ will be zero, while for even harmonics
it will be $4I_{0}/\pi p$.

Much of what we have seen for the perfect one-dimensional ring has
a counterpart in the case of the disordered, three-dimensional ring.
In that case, the perfectly anti-correlated single energy levels are
replaced by bands of energy levels correlated on an energy scale $E_{c}$.
Anti-correlation of successive bands (as well as lack of correlation
beyond a certain energy scale) causes the current contributions of
most bands to cancel out. This cancellation results in a typical total
current of the order of the current of the top-most filled band of
levels instead of the top-most filled energy level as we found for
the perfect, one-dimensional ring. Likewise, the sensitivity to electron
number of the odd harmonics of the current for the perfect ring can
be likened to the sensitivity of the disordered ring to its microscopic
disorder configuration.%
\footnote{The odd harmonics in a perfect three-dimensional ring are also sensitive
to the exact values of the cross-sectional dimensions. See Section
\ref{sub:CHPCTh_FiniteCrossSection}.%
} Averaging over the various possible disorder configurations also
results in the odd harmonics of the current vanishing. Survival of
the finite disorder average for the even harmonics is discussed in
\ref{sub:CHPCTh_AverageCurrent}.

\FloatBarrier

\section{\label{sub:CHPCTh_IdealRingFiniteTemperature}Effects of finite temperature,
finite cross-section, and the introduction of disorder on the persistent
current of the ideal one-dimensional ring}

\subsection{\label{sub:CHPCTh_1DPerfFinT}Finite temperature}

First we consider finite disorder which can be addressed using Eq.
\ref{eq:CHPCTh_PCthermalSum}. We begin by writing the grand canonical
potential as 
\begin{equation}
\Omega=-k_{B}T\int_{-\infty}^{\infty}d\varepsilon\,\nu\left(\varepsilon,\phi\right)\left(\ln\left(1+\exp\left(-\frac{\left(\varepsilon-\varepsilon_{F}\right)}{k_{B}T}\right)\right)\right)\label{eq:CHPCTh_PCOmega}
\end{equation}
where $\nu(\varepsilon,\phi)=\sum_{n}2\delta(\varepsilon_{n}(\phi)-\varepsilon)$
is the flux-dependent density of states. The factor of 2 has been
added to $\nu$ to account for spin degeneracy. We perform two integrations
by parts on the integral in Eq. \ref{eq:CHPCTh_PCOmega}. Because
$\nu(\varepsilon,\phi)$ and its integrals are zero for $\varepsilon<0$
and the $\ln$ term and its derivatives are zero at $\varepsilon=\infty$,
the boundary terms can be discarded. Using the Poisson summation formula
(see Eq. \ref{eq:AppMath_PCDOS}), the first term in the integral
in Eq. \ref{eq:CHPCTh_PCOmega} becomes
\begin{align}
\int_{0}^{\varepsilon}d\varepsilon'\int_{0}^{\varepsilon'}d\varepsilon''\,\nu\left(\varepsilon'',\phi\right) & =2\int_{0}^{\varepsilon}d\varepsilon'\int_{0}^{\varepsilon'}d\varepsilon''\,\sum_{p}\sqrt{\frac{2mL^{2}}{h^{2}\varepsilon''}}\cos\left(2\pi p\sqrt{\frac{2mL^{2}\varepsilon''}{h^{2}}}\right)e^{2\pi ip\phi/\phi_{0}}\nonumber \\
 & =4\int_{0}^{\varepsilon}d\varepsilon'\int_{0}^{\varepsilon'}d\varepsilon''\,\sum_{p>0}\sqrt{\frac{2mL^{2}}{h^{2}\varepsilon''}}\cos\left(2\pi p\sqrt{\frac{2mL^{2}\varepsilon''}{h^{2}}}\right)\cos\left(2\pi p\frac{\phi}{\phi_{0}}\right)\nonumber \\
 & \phantom{=}+\int_{0}^{\varepsilon}d\varepsilon'\int_{0}^{\varepsilon'}d\varepsilon''\,\sqrt{\frac{2mL^{2}}{h^{2}\varepsilon''}}\nonumber \\
 & =4\sum_{p>0}\cos\left(2\pi p\frac{\phi}{\phi_{0}}\right)\int_{0}^{\varepsilon}d\varepsilon'\,\frac{1}{\pi p}\sin\left(2\pi p\sqrt{\frac{2mL^{2}\varepsilon'}{h^{2}}}\right)\nonumber \\
 & \phantom{=}+\frac{8}{3}\sqrt{\frac{2mL^{2}}{h^{2}}}\varepsilon^{3/2}.\label{eq:CHPCTh_NuIntCleanEval}
\end{align}
Changing variables to $k=\sqrt{2m\varepsilon}/\hbar$, 
\begin{align*}
\int_{0}^{\varepsilon}d\varepsilon'\int_{0}^{\varepsilon'}d\varepsilon''\,\nu\left(\varepsilon'',\phi\right)-\frac{8}{3}\sqrt{\frac{2mL^{2}}{h^{2}}}\varepsilon^{3/2} & =4\sum_{p>0}\cos\left(2\pi p\frac{\phi}{\phi_{0}}\right)\int_{0}^{\varepsilon}d\varepsilon'\,\frac{1}{\pi p}\sin\left(2\pi p\sqrt{\frac{2mL^{2}\varepsilon'}{h^{2}}}\right)\\
 & =4\sum_{p>0}\cos\left(2\pi p\frac{\phi}{\phi_{0}}\right)\int_{0}^{k}dk'\,\frac{\hbar^{2}k'}{m}\frac{1}{\pi p}\sin\left(pk'L\right)\\
 & =4\sum_{p>0}\cos\left(2\pi p\frac{\phi}{\phi_{0}}\right)\frac{\hbar^{2}}{m\pi p}\int_{0}^{k}dk'\, k'\sin\left(pk'L\right).
\end{align*}
Integrating by parts and setting the constant of integration to zero
gives 
\begin{align*}
\int^{\varepsilon}d\varepsilon'\int^{\varepsilon'}d\varepsilon''\,\nu\left(\varepsilon'',\phi\right)-\frac{8}{3}\sqrt{\frac{2mL^{2}}{h^{2}}}\varepsilon^{3/2} & =4\sum_{p>0}\cos\left(2\pi p\frac{\phi}{\phi_{0}}\right)\frac{\hbar^{2}}{m\pi p}\left(\frac{\sin\left(pkL\right)}{p^{2}L^{2}}-k\frac{\cos\left(pkL\right)}{pL}\right)\\
 & =\sum_{p>0}\frac{4\hbar^{2}k}{m\pi p^{2}L}\cos\left(2\pi p\frac{\phi}{\phi_{0}}\right)\left(\frac{\sin\left(pkL\right)}{pkL}-\cos\left(pkL\right)\right).
\end{align*}
Meanwhile, the second term in the integral in Eq. \ref{eq:CHPCTh_PCOmega}
becomes

\begin{align*}
\partial_{\varepsilon}^{2}\left(\ln\left(1+\exp\left(-\frac{\left(\varepsilon-\varepsilon_{F}\right)}{k_{B}T}\right)\right)\right) & =\partial_{\varepsilon}\left(\frac{-1/k_{B}T}{1+\exp\left(\left(\varepsilon-\varepsilon_{F}\right)/k_{B}T\right)}\right)\\
 & =\left(\frac{1}{k_{B}T}\right)^{2}\left(\frac{\exp\left(\left(\varepsilon-\varepsilon_{F}\right)/k_{B}T\right)}{\left(1+\exp\left(\left(\varepsilon-\varepsilon_{F}\right)/k_{B}T\right)\right)^{2}}\right)\\
 & =\left(\frac{\text{sech}\left(\left(\varepsilon-\varepsilon_{F}\right)/2k_{B}T\right)}{2k_{B}T}\right)^{2}.
\end{align*}

\begin{figure}
\centering{}\includegraphics[width=0.6\paperwidth]{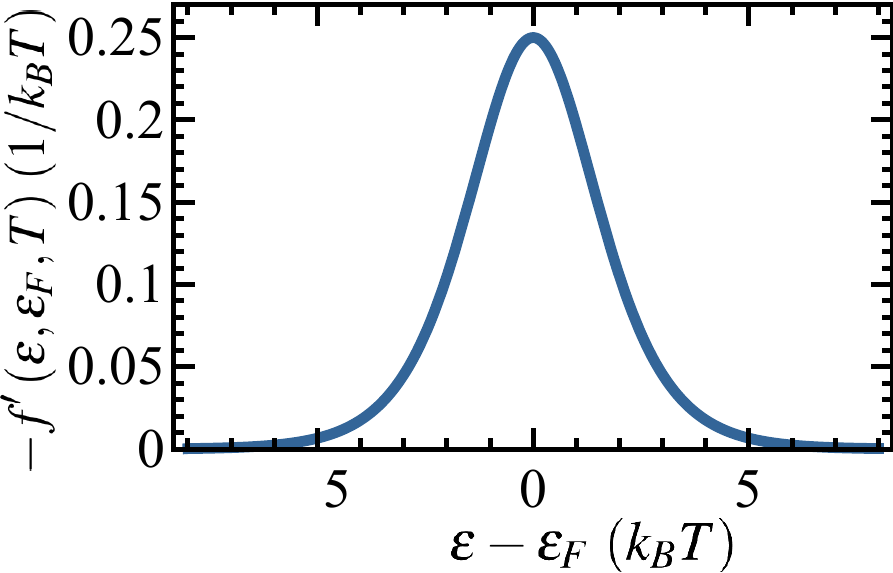}\caption[The thermal averaging function $y\left(\epsilon\right)$]{\label{fig:CHPCTh_ThermalAvgFunctionY}The thermal averaging function
$-f'(\varepsilon,\varepsilon_{F},T)$. The function $-f'(\varepsilon,\varepsilon_{F},T)$
represents the weighting of different energy components in the calculation
of the thermally averaged persistent current at finite temperature
$T$. The vertical axis is plotted in units of $1/k_{B}T$ while the
horizontal axis is centered on $\varepsilon_{F}$ and in units of
$k_{B}T$. The functional form of $-f'(\varepsilon,\varepsilon_{F},T)$
is given in Eq. \ref{eq:CHPCTh_ThermalAvgFunctionY}.}
\end{figure}

Using these results the persistent current is
\begin{align*}
I & =-\frac{\partial\Omega}{\partial\phi}\\
 & =-\sum_{p>0}\sin\left(2\pi p\frac{\phi}{\phi_{0}}\right)\int_{-\infty}^{\infty}d\varepsilon\,\frac{8\hbar^{2}k}{mpL\phi_{0}}\left(\frac{\sin\left(pkL\right)}{pkL}-\cos\left(pkL\right)\right)\frac{\text{sech}^{2}\left(\left(\varepsilon-\varepsilon_{F}\right)/2k_{B}T\right)}{4k_{B}T}.
\end{align*}
The last factor
\begin{align}
-f'\left(\varepsilon,\varepsilon_{F},T\right) & =\frac{\text{sech}^{2}\left(\left(\varepsilon-\varepsilon_{F}\right)/2k_{B}T\right)}{4k_{B}T}\label{eq:CHPCTh_ThermalAvgFunctionY}
\end{align}
is the negative derivative of the Fermi-Dirac distribution and is
peaked around the Fermi energy $\varepsilon_{F}$ with characteristic
width $k_{B}T$. The function $-f'(\varepsilon,\varepsilon_{F},T)$
is plotted in Fig. \ref{fig:CHPCTh_ThermalAvgFunctionY}. Typically,
the temperatures considered satisfy $k_{B}T\ll\varepsilon_{F}$, so
the factors of $k$ in the integral over $\varepsilon$ will be of
order $k_{F}=\sqrt{2m\varepsilon_{F}}/\hbar$. Since $k_{F}L\gg1$,
we can approximate the current as 
\begin{equation}
I\approx\sum_{p>0}\sin\left(2\pi p\frac{\phi}{\phi_{0}}\right)\int_{-\infty}^{\infty}d\varepsilon\,\frac{8\hbar^{2}k}{mpL\phi_{0}}\cos\left(pkL\right)\left(-f'\left(\varepsilon,\varepsilon_{F},T\right)\right).\label{eq:CHPCTh_1DCurrentThermalInt}
\end{equation}
The integral $-\int_{-\infty}^{\infty}d\varepsilon\, f'(\varepsilon,\varepsilon_{F},T)=1$
for all $T$, while the integrand is $1/k_{B}T$ at $\varepsilon=\varepsilon_{F}$.
Thus, as temperature decreases, this factor maintains a constant integral
while becoming more and more sharply peaked at $\varepsilon_{F}$
and can be represented by $\delta(\varepsilon-\varepsilon_{F})$ in
the limit $T\rightarrow0$. In this limit of $T\rightarrow0$, the
current becomes 
\begin{align}
I & =\sum_{p>0}\frac{8\hbar^{2}k_{F}}{mpL\phi_{0}}\cos\left(pk_{F}L\right)\sin\left(2\pi p\frac{\phi}{\phi_{0}}\right)\nonumber \\
 & =\sum_{p>0}\left(\frac{ev_{F}}{L}\right)\frac{4}{\pi p}\cos\left(pk_{F}L\right)\sin\left(2\pi p\frac{\phi}{\phi_{0}}\right).\label{eq:CHPCTh_1DCurrentZeroT}
\end{align}
If we take $k_{F}=2\pi(N+(n+1)/2)/L$ for $n=0,\,1$, we find 
\[
I=\sum_{p>0}I_{0}\frac{4}{\pi p}\left(-1\right)^{p\left(n+1\right)}\sin\left(2\pi p\frac{\phi}{\phi_{0}}\right),
\]
which has harmonic coefficients $I_{p}=I_{0}\frac{4}{\pi p}(-1)^{p(n+1)}$.
As expected, these coefficients match the $I_{p,N+2n}$ given in Eqs.
\ref{eq:CHPCTh_IpN0coeff} and \ref{eq:CHPCTh_IpN2coeff} for the
case of a fixed number of electrons at zero temperature.

We now evaluate the thermal averaging integral of Eq. \ref{eq:CHPCTh_1DCurrentThermalInt}.
To evaluate this integral, we rewrite the relation $k=\sqrt{2m\varepsilon}/\hbar$
as 
\begin{align}
k & =\frac{\sqrt{2m\varepsilon_{F}}}{\hbar}\sqrt{1+\left(\frac{\varepsilon}{\varepsilon_{F}}-1\right)}\nonumber \\
 & \approx k_{F}\left(1+\frac{1}{2}\left(\frac{\varepsilon}{\varepsilon_{F}}-1\right)-\frac{1}{8}\left(\frac{\varepsilon}{\varepsilon_{F}}-1\right)^{2}+\mathcal{O}\left(\left(\frac{\varepsilon}{\varepsilon_{F}}-1\right)^{3}\right)\right).\label{eq:CHPCTh_epsKrelation}
\end{align}
Assuming $k_{B}T\ll\varepsilon_{F}$, as noted above, the $f'(\varepsilon,\varepsilon_{F},T)$
factor in the integral will be sharply peaked around $\varepsilon=\varepsilon_{F}$
with width of order $k_{B}T$. Thus, the integrand is appreciable
only when $|\frac{\varepsilon}{\varepsilon_{F}}-1|\ll1$. We can thus
replace the prefactor $k$ in the integral by $k_{F}$. Using Eq.
\ref{eq:CHPCTh_EnergyLevel1DPerfectRing}, we can write $k_{F}\approx2\pi N/L$
where $N\gg1$ is the highest occupied level. Then using Eq. \ref{eq:CHPCTh_epsKrelation},
we have that the change of the argument of the $\cos(pkL)$ factor
in Eq. \ref{eq:CHPCTh_1DCurrentThermalInt} as $\varepsilon$ varies
from $\varepsilon_{F}$ to $\varepsilon_{F}+k_{B}T$ (i.e., over the
range over which the thermal averaging function in Fig. \ref{fig:CHPCTh_ThermalAvgFunctionY}
is appreciable) is 
\[
pkL-pk_{F}L\sim\pi N\left(\frac{k_{B}T}{\varepsilon_{F}}\right)-\frac{\pi N}{4}\left(\frac{k_{B}T}{\varepsilon_{F}}\right)^{2}.
\]
As this factor is in the argument of a cosine, variations small compared
to $\pi$ may be neglected. For $v_{F}=10^{6}\,\text{m/s}$, $T=1\,\text{K}$,
$L=1\,\text{\ensuremath{\mu}m}$, and $m=9.1\times10^{-31}$, we have
$N\sim10^{3}$ and $(k_{B}T/\varepsilon_{F})\sim10^{-5}$. Thus, we
can to good approximation drop all powers of $\varepsilon/\varepsilon_{F}$
beyond the first. With this linearization of $k$, Eq. \ref{eq:CHPCTh_1DCurrentThermalInt}
gives a current of 
\begin{align*}
I & =\sum_{p>0}\sin\left(2\pi p\frac{\phi}{\phi_{0}}\right)\int_{-\infty}^{\infty}d\varepsilon\,\frac{8\hbar^{2}k_{F}}{mpL\phi_{0}}\cos\left(pk_{F}L+\frac{pk_{F}L}{2\varepsilon_{F}}\left(\varepsilon-\varepsilon_{F}\right)\right)\left(-f'\left(\varepsilon,\varepsilon_{F},T\right)\right)\\
 & =\sum_{p>0}\sin\left(2\pi p\frac{\phi}{\phi_{0}}\right)\int_{-\infty}^{\infty}d\varepsilon\,\frac{4ev_{F}}{\pi pL}\cos\left(pk_{F}L+\frac{pk_{F}L}{2\varepsilon_{F}}\varepsilon\right)\frac{\text{sech}^{2}\left(\varepsilon/2k_{B}T\right)}{4k_{B}T}\\
 & =\sum_{p>0}\frac{4}{\pi p}I_{0}\sin\left(2\pi p\frac{\phi}{\phi_{0}}\right)\int_{-\infty}^{\infty}d\varepsilon\,\cos\left(pk_{F}L\right)\cos\left(\frac{pk_{F}L}{2\varepsilon_{F}}\varepsilon\right)\frac{\text{sech}^{2}\left(\varepsilon/2k_{B}T\right)}{4k_{B}T}\\
 & =\sum_{p>0}\frac{4}{\pi p}I_{0}\cos\left(pk_{F}L\right)\sin\left(2\pi p\frac{\phi}{\phi_{0}}\right)\int_{-\infty}^{\infty}d\varepsilon\,\exp\left(-i\frac{pk_{F}L}{2\varepsilon_{F}}\varepsilon\right)\frac{\text{sech}^{2}\left(\varepsilon/2k_{B}T\right)}{4k_{B}T}\\
 & =\sum_{p>0}\frac{4}{\pi p}I_{0}\cos\left(pk_{F}L\right)\sin\left(2\pi p\frac{\phi}{\phi_{0}}\right)\int_{-\infty}^{\infty}dx\,\frac{1}{2}\exp\left(-i\frac{4mpLk_{B}T}{\hbar^{2}k_{F}}x\right)\text{sech}^{2}\left(x\right)
\end{align*}
In several steps above, we have made use of the symmetry of $\text{sech}(x)$.
The last integral is just the Fourier transform of $\text{sech}^{2}(x)$
which is given in Eq. \ref{eq:AppMath_sechSquaredFourierTransform}.
Using this relation, the current can be written as 
\begin{equation}
I=\sum_{p>0}\left(\frac{4}{\pi p}I_{0}\cos\left(pk_{F}L\right)\right)\left(\frac{T/T_{p}}{\sinh\left(T/T_{p}\right)}\right)\sin\left(2\pi p\frac{\phi}{\phi_{0}}\right)\label{eq:CHPCTh_IperfectRing}
\end{equation}
where $T_{p}=\frac{1}{\pi pk_{B}}\frac{\hbar^{2}}{2m}\frac{k_{F}}{L}$.
We denote the normalized temperature dependence of the harmonics of
the single-channel ring as
\begin{equation}
g_{1}\left(T/T_{p}\right)=\frac{T/T_{p}}{\sinh\left(T/T_{p}\right)}.\label{eq:CHPCTh_g11DPerfectRing}
\end{equation}

If we denote the highest index $n$ in Eq. \ref{eq:CHPCTh_EnergyLevel1DPerfectRing}
for the energy levels $\varepsilon_{n}$ by $N$, then there are $2N$
levels between the lowest at $\varepsilon\approx0$ and $\varepsilon_{F}\approx h^{2}N^{2}/2mL^{2}$.
The mean level spacing $\Delta_{1}$ can be written as 
\begin{align*}
\Delta_{1} & =\varepsilon_{F}/2N\\
 & =\frac{\pi^{2}\hbar^{2}N}{mL^{2}}\\
 & =\frac{\pi\hbar^{2}k_{F}}{2mL}.
\end{align*}
The characteristic temperature of the $p^{th}$ harmonic of the current
can be written in terms of $\Delta_{1}$ as 
\begin{equation}
T_{p}=\frac{1}{k_{B}}\frac{\hbar^{2}k_{F}}{2\pi mpL}=\frac{1}{k_{B}}\frac{\Delta_{1}}{\pi^{2}p}.\label{eq:CHPCTh_PerfectRingTp}
\end{equation}
For $T>T_{p}$, the $p^{th}$ harmonic of the current decays exponentially
in $T$.

\begin{figure}
\centering{}\includegraphics[width=0.7\paperwidth]{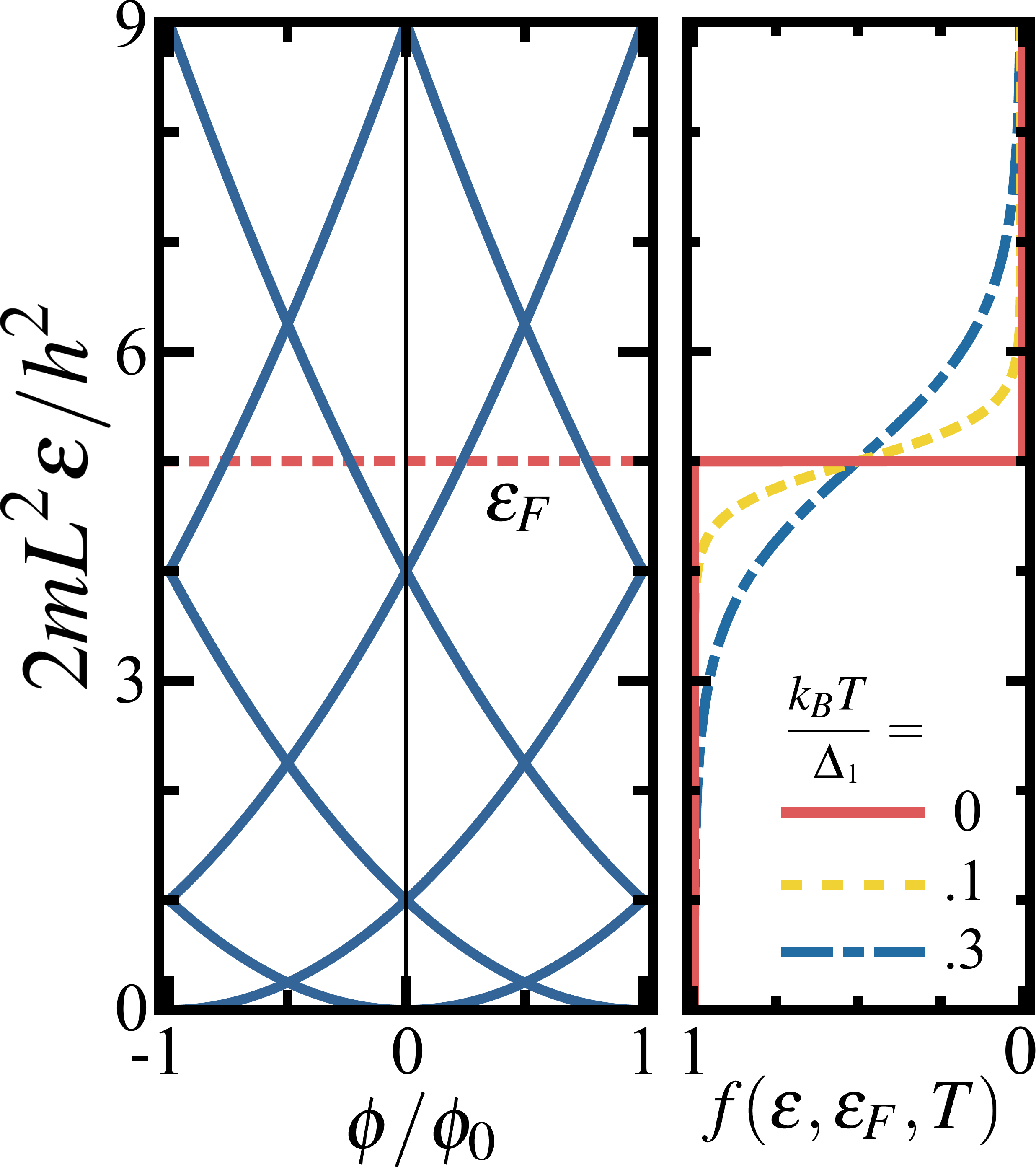}\caption[Spread of energy level occupancy at finite temperature in an ideal
ring]{\label{fig:CHPCTh_EnergyLevelsThermalDistribution}Spread of energy
level occupancy at finite temperature in an ideal ring. The left plot
shows the same energy levels as Fig. \ref{fig:CHPCTh_EnergyLevelsSimple}
while the right plot shows the Fermi-Dirac distribution function $f(\varepsilon,\varepsilon_{F},T)$
on the horizontal axis for the same vertical axis as the left plot.
This vertical energy axis is scaled by $2mL^{2}/h^{2}$ so that at
$\phi=0$ the $n^{th}$ energy level should be at $n^{2}$. In the
plot on the right, the Fermi-Dirac distribution is plotted for temperatures
representing $0$, $0.1\Delta_{1}/k_{B}$, and $0.3\Delta_{1}/k_{B}$
where $\Delta_{1}$ is the mean level spacing. For the plot, $2mL^{2}\varepsilon_{F}/h^{2}=5$
and $2mL^{2}\Delta_{1}/h^{2}\approx2.2$. These small values were
chosen for the figure for the sake of clarity, but the expressions
derived in the text are only appropriate when these values are $\gg1$.}
\end{figure}

As we noted when introducing Eq. \ref{eq:CHPCTh_PCthermalSum}, the
calculation of the current at finite temperature is performed by taking
a weighted sum of the single level currents $i_{n}$ with each energy
level $\varepsilon_{n}$ weighted by the Fermi-Dirac distribution
function $f(\varepsilon_{n},\varepsilon_{F},T)$. In Fig. \ref{fig:CHPCTh_EnergyLevelsThermalDistribution},
we plot the Fermi-Dirac distribution alongside the energy levels previously
shown in Fig. \ref{fig:CHPCTh_EnergyLevelsSimple} for three different
temperatures. As the temperature becomes an appreciable fraction of
the mean level spacing $\Delta_{1}$, adjacent energy levels begin
to have comparable thermal occupancy. We found in the previous section
that the current contributions of adjacent levels tend to cancel producing
at net current at $T=0$ of the same order as the current in the highest
occupied level. Thus, it is quite sensible to expect the current to
decay quickly once the spread in energy $k_{B}T$ of the occupied
levels becomes comparable to the spacing between levels and levels
with nearly opposite currents are given roughly equal weights. 

\begin{figure}
\begin{centering}
\includegraphics[width=0.6\paperwidth]{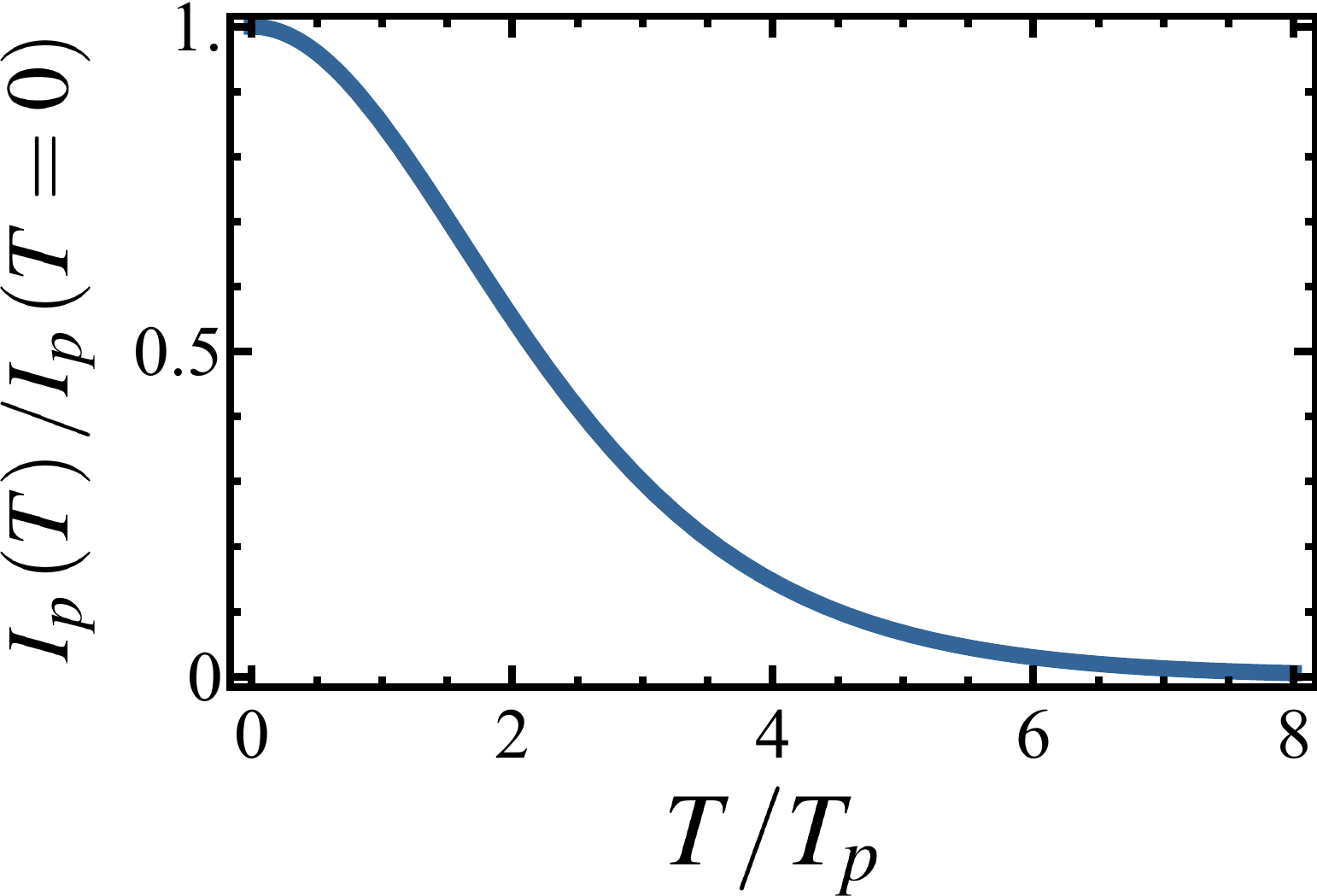}
\par\end{centering}

\caption[Temperature dependence of the harmonics of the persistent current
in a one-dimensional ring]{\label{fig:CHPCTh_IdealRingIpvsT}Temperature dependence of the harmonics
of the persistent current in a one-dimensional ring. The figure plots
the magnitude of the $p^{th}$ harmonic $I_{p}$ (see Eq. \ref{eq:CHPCTh_IpPerfectRing})
scaled by its value at $T=0$ as a function temperature $T$ scaled
by the characteristic temperature $T_{p}=\Delta_{1}/p\pi^{2}k_{B}$
of the $p^{th}$ harmonic (see Eq. \ref{eq:CHPCTh_PerfectRingTp}).
At low temperatures, the current is not affected by temperature, but,
once $k_{B}T$ reaches a value of $\sim\Delta_{1}/p\pi^{2}$, $I_{p}$
begins to decay quickly with $T/T_{p}$.}
\end{figure}

\begin{figure}

\begin{centering}
\includegraphics[width=0.7\paperwidth]{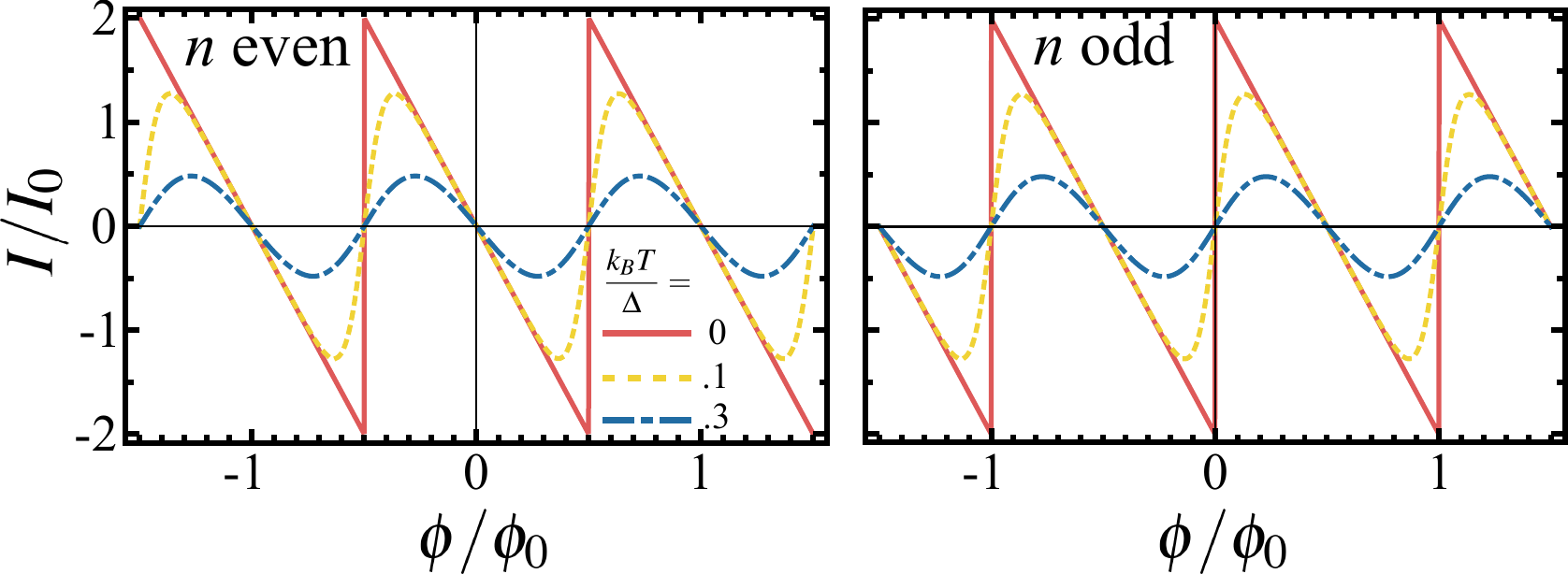}
\par\end{centering}

\caption[Persistent current in a one-dimensional ring versus flux for a series
of temperatures]{\label{fig:CHPCTh_IdealRingCurrentSeveralT}Persistent current in
a one-dimensional ring versus flux for a series of temperatures. The
figure shows $I$ of Eq. \ref{eq:CHPCTh_IperfectRing}, scaled by
$I_{0}=ev_{F}/L$, with $k_{F}L=2\pi(N+(n+1)/2)$ for $N$ and $n$
integers. The top panel corresponds to $n$ even and matches $I_{N+0}$
derived in Eq. \ref{eq:CHPCTh_IN0}, while the bottom panel corresponds
to $n$ odd and thus $I_{N+2}$ of Eq. \ref{eq:CHPCTh_IN2}. As temperature
is increased, the current transitions from a sawtooth to a sinusoid.
The three temperatures plotted ($T=0,$ $0.1\Delta_{1}$, and $0.3\Delta_{1}$
as indicated in the legend) are the same as those of Fig. \ref{fig:CHPCTh_EnergyLevelsThermalDistribution}.}
\end{figure}

Fig. \ref{fig:CHPCTh_IdealRingIpvsT} displays this decay by plotting
the magnitude of the $p^{th}$ harmonic 
\begin{equation}
I_{p}\left(T\right)=\left(\frac{4}{\pi p}I_{0}\cos\left(pk_{F}L\right)\right)\left(\frac{T/T_{p}}{\sinh\left(T/T_{p}\right)}\right)\label{eq:CHPCTh_IpPerfectRing}
\end{equation}
as a function of temperature. Fig. \ref{fig:CHPCTh_IdealRingCurrentSeveralT}
shows the decay in a different way by plotting the current $I$ given
in Eq. \ref{eq:CHPCTh_IperfectRing} as a function of normalized flux
$\phi/\phi_{0}$ for the same values of temperature as used in Fig.
\ref{fig:CHPCTh_EnergyLevelsThermalDistribution}. At finite temperature,
the discontinuities in $I(\phi)$ are rounded out. For temperatures
$T\apprge0.3\Delta_{1}/k_{B}$, the higher harmonics of the current
are suppressed resulting in a sinusoidal form for $I$. 

Similar results will hold for the case of the three-dimensional ring
in the diffusive regime. In that case, the current begins to decay
once the spread in energy of the occupied levels becomes comparable
to the correlation energy $E_{c}$ discussed in the previous section.
Likewise, the $p^{th}$ harmonic of the current decays on a characteristic
scale $T_{p}$ which scales as $p^{-1}$, meaning that for most achievable
temperatures the current is sinusoidal to a good approximation.

\FloatBarrier

\subsection{\label{sub:CHPCTh_FiniteCrossSection}Finite ring cross-section}

We now consider a three-dimensional ring with a finite cross-section.
Giving the ring a finite cross-section introduces additional energy
bands to the spectrum plotted in Fig. \ref{fig:CHPCTh_EnergyLevelsSimple}.
For this discussion, we ignore the effects of magnetic flux penetrating
the finite cross-section of the ring by considering the applied flux
$\phi$ to be an idealized Aharonov-Bohm flux threading the ring like
that introduced at the end of Section \ref{sub:CHPCTh_1DRingSingleLevelSolutions}.
We will describe one way of accounting for the magnetic flux penetrating
the metal in \ref{sub:CHPCTh_FluxThroughMetal}.

We denote the ring thickness by $t$ and its linewidth by $w$, with
the mean radius still $L/2\pi$. For simplicity, we use the vector
potential given in Eq. \ref{eq:CHPCTh_VectorAABflux}. For the function
\[
\Lambda\left(\boldsymbol{r}\right)=\int_{\boldsymbol{r}_{0}}^{\boldsymbol{r}}d\boldsymbol{r}'\cdot\boldsymbol{A}\left(\boldsymbol{r}'\right)
\]
we choose $\boldsymbol{r}_{0}=(L/2\pi-w/2)\tilde{\boldsymbol{x}}$.
For the point $\boldsymbol{r}=x\tilde{\boldsymbol{x}}+y\tilde{\boldsymbol{y}}+z\tilde{\boldsymbol{z}}$
in the argument of $\Lambda$, we can first integrate along the path
given by $\boldsymbol{r}_{0}+tz\tilde{\boldsymbol{z}}$ as $t$ goes
from 0 to 1 and then along the path from $\boldsymbol{r}_{0}+z\hat{\boldsymbol{z}}+t(\sqrt{x^{2}+y^{2}}-(L/2\pi-w/2))\tilde{\boldsymbol{x}}$
as $t$ goes from 0 to 1. Since $\boldsymbol{A}$ is parallel to $\tilde{\boldsymbol{\theta}}$
and $d\boldsymbol{r}'$ is parallel to $\tilde{\boldsymbol{z}}$ along
the first path and $\tilde{\boldsymbol{r}}$ along the second, both
of these integrals are zero. Finally, we integrate along the path
$\sqrt{x^{2}+y^{2}}(\cos(\theta')\hat{\boldsymbol{x}}+\sin(\theta')\tilde{\boldsymbol{y}})+z\tilde{\boldsymbol{z}}$
for $\theta'$ from 0 to $\theta=\tan^{-1}(y/x)$. This final path
integration gives
\begin{align*}
\Lambda\left(\boldsymbol{r}\right) & =\int_{0}^{\theta}\left(\sqrt{x^{2}+y^{2}}\tilde{\boldsymbol{\theta}}\, d\theta'\right)\cdot\left(\frac{\phi}{2\pi\sqrt{x^{2}+y^{2}}}\tilde{\boldsymbol{\theta}}\right)\\
 & =\int_{0}^{\theta}d\theta'\left(\frac{\phi}{2\pi}\right)\\
 & =\frac{\theta}{2\pi}\phi.
\end{align*}
The relation between the gauge transformed wave function $\psi^{\prime}$
and the untransformed $\psi$ is $\psi\prime=\exp(i\theta\phi/\phi_{0})\psi$.

Solving the time-independent Schrödinger equation 
\begin{equation}
-\frac{\hbar^{2}}{2m}\left(\nabla+i\frac{e}{\hbar}\boldsymbol{A}\right)^{2}\psi=\varepsilon\psi\label{eq:CHPCTh_3DSchrodingerVectorPotential}
\end{equation}
boils down to solving 
\[
\nabla^{2}\psi^{\prime}=-\frac{2m\varepsilon}{\hbar^{2}}\psi^{\prime}
\]
with $\psi'$ subject to the constraint that it goes smoothly to zero
at the edges of the ring and that $\psi'$ is a periodic function
of $\theta$ up to a factor of $\exp(i\theta\phi/\phi_{0})$. This
equation can be solved by standard separation of variables. Taking
$\psi'=P\left(r\right)\exp(i\theta\phi/\phi_{0})Q(\theta)Z(z)$, the
equation in cylindrical coordinates becomes
\[
\frac{1}{P\left(r\right)}\frac{1}{r}\frac{\partial}{\partial r}\left(r\frac{\partial P}{\partial r}\right)+\frac{e^{-i\theta\phi/\phi_{0}}}{Q\left(\theta\right)}\frac{1}{r^{2}}\frac{\partial^{2}}{\partial\theta^{2}}\left(e^{i\theta\phi/\phi_{0}}Q\left(\theta\right)\right)+\frac{1}{Z\left(z\right)}\frac{\partial^{2}Z}{\partial z^{2}}=-\frac{2m\varepsilon}{\hbar^{2}}.
\]

Since all of the $z$ dependence is contained in one term, that term
must be independent of $z$ and $\ddot{Z}=-k_{z}^{2}Z$ for some constant
$k_{z}$. The function $Z$ must then be some linear combination of
the functions $\exp(ik_{z}z)$ and $\exp(-ik_{z}z)$. Taking the $z$
coordinate of the ring to range from 0 to $t$, the boundary conditions
$Z(0)=Z(t)=0$ require $Z=\sin(k_{z}z)$ with $k_{z}=\pi n_{z}/t$
for a positive integer $n_{z}$ ($-n_{z}$ gives the same function
as $n_{z}$).

Writing $\varepsilon(n_{z})=\hbar^{2}k_{z}^{2}/2m$, the Schrödinger
equation can be rewritten as
\[
\frac{1}{P\left(r\right)}r\frac{\partial}{\partial r}\left(r\frac{\partial P}{\partial r}\right)+\frac{e^{-i\theta\phi/\phi_{0}}}{Q\left(\theta\right)}\frac{\partial^{2}}{\partial\theta^{2}}\left(e^{i\theta\phi/\phi_{0}}Q\left(\theta\right)\right)=-r^{2}\frac{2m\left(\varepsilon-\varepsilon\left(n_{z}\right)\right)}{\hbar^{2}}
\]
where now the $\theta$ dependence has been isolated to one term,
so that 
\[
\frac{\partial^{2}}{\partial\theta^{2}}\left(e^{i\theta\phi/\phi_{0}}Q\left(\theta\right)\right)=-a\left(\phi\right)e^{i\theta\phi/\phi_{0}}Q\left(\theta\right)
\]
must hold for some $a$. The boundary conditions on $Q(\theta)$ require
that it be periodic in $\theta$ with period $2\pi$. Thus $Q(\theta)$
must be some linear combination of terms of the form $\exp(in\theta)$.%
\footnote{We could denote $n$ by $n_{\theta}$ to follow the convention used
with the other indices. We use $n$ in order to emphasize the connection
with the index $n$ from the previous sections.%
} Each of these terms already satisfies the boundary on its own with
$a_{n}(\phi)=(n+\phi/\phi)^{2}$ for $Q=\exp(in\theta)$.

Schrödinger's equation can now be rewritten as
\begin{equation}
r^{2}\frac{\partial^{2}P}{\partial r^{2}}+r\frac{\partial P}{\partial r}+\left(\frac{2m\left(\varepsilon-\varepsilon\left(n_{z}\right)\right)}{\hbar^{2}}r^{2}-\left(n+\phi/\phi\right)^{2}\right)P=0\label{eq:CHPCTh_BesselDiffEqR}
\end{equation}
which bears solutions for $P$ that are linear combinations of $J_{n+\phi/\phi_{0}}(k_{r}(n_{r},n,\phi)r)$
and $Y_{n+\phi/\phi_{0}}(k_{r}(n_{r},n,\phi)r)$ (Bessel functions
of the first and second kind respectively) for some constant 
\[
k_{r}\left(n_{r},n,\phi\right)=\sqrt{\frac{2m\varepsilon_{r}}{\hbar^{2}}}
\]
where $\varepsilon_{r}=\varepsilon-\varepsilon(n_{z})$. The radial
boundary conditions can be stated as 
\[
J_{n+\phi/\phi_{0}}\left(k_{r}\left(n_{r},n,\phi\right)\left(L/2\pi-w/2\right)\right)+C_{n_{r}}Y_{n+\phi/\phi_{0}}\left(k_{r}\left(n_{r},n,\phi\right)\left(L/2\pi-w/2\right)\right)=0
\]
and 
\[
J_{n+\phi/\phi_{0}}\left(k_{r}\left(n_{r},n,\phi\right)\left(L/2\pi+w/2\right)\right)+C_{n_{r}}Y_{n+\phi/\phi_{0}}\left(k_{r}\left(n_{r},n,\phi\right)\left(L/2\pi+w/2\right)\right)=0
\]
for some constant $C_{n_{r}}$. These boundary conditions admit a
series of possible $k_{r}$'s which we index by $n_{r}$ (justifying
the notation introduced above). The $k_{r}$ satisfy the equation
\begin{equation}
J_{n+\phi/\phi_{0}}\left(x\left(1-\frac{\delta}{2}\right)\right)Y_{n+\phi/\phi_{0}}\left(x\left(1+\frac{\delta}{2}\right)\right)-J_{n+\phi/\phi_{0}}\left(x\left(1+\frac{\delta}{2}\right)\right)Y_{n+\phi/\phi_{0}}\left(x\left(1-\frac{\delta}{2}\right)\right)=0\label{eq:CHPCTh_PerfectRing3DRadialBC}
\end{equation}
where $x=k_{r}L/2\pi$ and $\delta=2\pi w/L$. 

In the limit of $2\pi w/L\ll1$, the values of $x$ satisfying Eq.
\ref{eq:CHPCTh_PerfectRing3DRadialBC} must be large since the arguments
of the different Bessel functions must be significantly different
for the two terms to cancel. For example, for $2\pi w/L=.3$ and $n=\phi/\phi_{0}=0$,
the lowest value of $x$ satisfying Eq. \ref{eq:CHPCTh_PerfectRing3DRadialBC}
is $\sim10$. In the limit of $x\gg\alpha^{2}$, the Bessel functions
take the asymptotic forms%
\footnote{The actual form for the coefficients (ignoring the $\sqrt{\pi/2x}$
term) is $\sum_{n}\frac{1}{n!}\left(\frac{-1}{8x}\right)^{n}\prod_{p,\text{odd}}^{n}\left(4\alpha^{2}-p^{2}\right)$
where the sum over $n$ runs over the odd integers for the first term
(the cosine term for $J_{\alpha}$) and the even integers for the
second term (the sine term for $J_{\alpha}$) and the index $p$ runs
over the first $n$ odd integers (the product is defined to be 1 for
$n=0$). In the text, we give the asymptotic forms keeping up to $n=1$.
In order to drop the $n=2$ term, it must hold that $\left(4\alpha^{2}-1\right)\left(4\alpha^{2}-9\right)$/$\left(2\left(8x\right)^{2}\right)\ll1$.
For $\alpha\gtrsim4$, the condition can be written as $\alpha^{4}\ll8x^{2}$.
For more discussion of the asymptotic form of the Bessel functions,
see e.g. Ref. \citealp{arfken2001mathematical}.%
}
\[
J_{\alpha}\left(x\right)\sim\sqrt{\frac{\pi}{2x}}\left(\cos\left(x-\frac{\pi\alpha}{2}-\frac{\pi}{4}\right)-\frac{4\alpha^{2}-1}{8x}\sin\left(x-\frac{\pi\alpha}{2}-\frac{\pi}{4}\right)\right)
\]
and
\[
Y_{\alpha}\left(x\right)\sim\sqrt{\frac{\pi}{2x}}\left(\sin\left(x-\frac{\pi\alpha}{2}-\frac{\pi}{4}\right)+\frac{4\alpha^{2}-1}{8x}\cos\left(x-\frac{\pi\alpha}{2}-\frac{\pi}{4}\right)\right).
\]
Taking 
\begin{align*}
\alpha & =n+\phi/\phi_{0},\\
\beta & =\pi\alpha/2+\pi/4,\\
x_{\pm} & =x\left(1\pm\frac{\delta}{2}\right),
\end{align*}
and $\delta\ll1$ so that $x\gg\alpha^{2}$, Eq. \ref{eq:CHPCTh_PerfectRing3DRadialBC}
can be written as 
\begin{align*}
0 & =\left(\cos\left(x_{-}-\beta\right)-\frac{4\alpha^{2}-1}{8x_{-}}\sin\left(x_{-}-\beta\right)\right)\left(\sin\left(x_{+}-\beta\right)+\frac{4\alpha^{2}-1}{8x_{+}}\cos\left(x_{+}-\beta\right)\right)\\
 & \phantom{=}-\left(\cos\left(x_{+}-\beta\right)-\frac{4\alpha^{2}-1}{8x_{+}}\sin\left(x_{+}-\beta\right)\right)\left(\sin\left(x_{-}-\beta\right)+\frac{4\alpha^{2}-1}{8x_{-}}\cos\left(x_{-}-\beta\right)\right)\\
 & \approx\cos\left(x_{-}-\beta\right)\sin\left(x_{+}-\beta\right)-\cos\left(x_{+}-\beta\right)\sin\left(x_{-}-\beta\right)\\
 & \phantom{\approxeq}+\frac{4\alpha^{2}-1}{8}\left(\frac{1}{x_{+}}-\frac{1}{x_{-}}\right)\left(\cos\left(x_{+}-\beta\right)\cos\left(x_{-}-\beta\right)+\sin\left(x_{-}-\beta\right)\sin\left(x_{+}-\beta\right)\right)
\end{align*}
Using the Eqs. \ref{eq:AppMath_TrigCosAB} and \ref{eq:AppMath_TrigSinAB},
the condition on $x$ becomes
\begin{align*}
0 & =\sin\left(x_{+}-x_{-}\right)+\frac{4\alpha^{2}-1}{8}\left(\frac{x_{-}-x_{+}}{x_{+}x_{-}}\right)\cos\left(x_{+}-x_{-}\right)\\
 & =\sin\left(x\delta\right)-\frac{4\alpha^{2}-1}{8}\left(\frac{x\delta}{x^{2}\left(1-\delta^{2}/4\right)}\right)\cos\left(x\delta\right)\\
 & \approx\sin\left(x\delta\right)-\frac{4\alpha^{2}-1}{8}\left(\frac{\delta}{x}\right)\cos\left(x\delta\right).
\end{align*}
From the conditions on $x$ and $\delta$, we know that the coefficient
of the second term is very small so that the sum of the two terms
goes to zero near where the first does at $x=n_{r}\pi/\delta$ for
$n_{r}$ a non-zero integer. Performing a Taylor expansion on $x$
about $n_{r}\pi/\delta$, we have to first order in $(x-\pi n_{r}/\delta)$
\begin{multline*}
\sin\left(x\delta\right)-\frac{4\alpha^{2}-1}{8}\left(\frac{\delta}{x}\right)\cos\left(x\delta\right)\approx\\
\left(-1\right)^{n_{r}}\delta\left(x-\frac{n_{r}\pi}{\delta}\right)-\frac{4\alpha^{2}-1}{8}\left(-1\right)^{n_{r}}\delta\left(\frac{\delta}{n_{r}\pi}-\left(\frac{\delta}{n_{r}\pi}\right)^{2}\left(x-\frac{n_{r}\pi}{\delta}\right)\right)
\end{multline*}
This expression is equal to zero for 
\begin{align*}
x & =\frac{n_{r}\pi}{\delta}+\frac{4\alpha^{2}-1}{8}\frac{\delta}{n_{r}\pi}\left(1+\frac{4\alpha^{2}-1}{8}\left(\frac{\delta}{n_{r}\pi}\right)^{2}\right)^{-1}\\
 & \approx\frac{n_{r}\pi}{\delta}+\frac{4\alpha^{2}-1}{8}\frac{\delta}{n_{r}\pi}
\end{align*}
where we have dropped terms of the order of $\delta^{3}$. Restoring
the expressions for $x$, $\alpha$ and $\delta$, we find the values
of $k_{r}$ are 
\[
k_{r}\left(n_{r},n,\phi\right)=\frac{\pi n_{r}}{w}-\frac{\pi}{2n_{r}}\frac{w}{L^{2}}+\frac{2\pi}{n_{r}}\frac{w}{L^{2}}\left(n+\frac{\phi}{\phi_{0}}\right)^{2}.
\]
Note that Eq. \ref{eq:CHPCTh_BesselDiffEqR} is unchanged for $r\rightarrow-r$
and thus for $k_{r}\rightarrow-k_{r}$ and $n_{r}\rightarrow-n_{r}$.
We can restrict the index $n_{r}$ to be a positive integer without
discarding any unique solutions.

Having obtained the form of the eigenfunctions of the Schrödinger
equation, we can now write down the eigenvalues. For
\begin{align}
\psi'\left(n_{r},n,\phi,n_{z}\right) & =\left(J_{n+\phi/\phi_{0}}\left(k_{r}\left(n_{r},n,\phi\right)r\right)+C\left(n_{r},n,\phi\right)Y_{n+\phi/\phi_{0}}\left(k_{r}\left(n_{r},n,\phi\right)r\right)\right)\nonumber \\
 & \phantom{=}\times\exp\left(i\left(n+\frac{\phi}{\phi_{0}}\right)\theta\right)\sin\left(\frac{\pi n_{z}z}{t}\right)\label{eq:CHPCTh_MultichannelEigenfunction}
\end{align}
with
\[
C\left(n_{r},n,\phi\right)=-\frac{J_{n+\phi/\phi_{0}}\left(k_{r}\left(n_{r},n,\phi\right)\left(\frac{L}{2\pi}-\frac{w}{2}\right)\right)}{Y_{n+\phi/\phi_{0}}\left(k_{r}\left(n_{r},n,\phi\right)\left(\frac{L}{2\pi}-\frac{w}{2}\right)\right)},
\]
we have 
\begin{align*}
-\frac{\hbar^{2}}{2m}\nabla^{2}\psi^{\prime} & =-\frac{\hbar^{2}}{2m}\left(\frac{1}{r}\frac{\partial}{\partial r}\left(r\frac{\partial\psi'}{\partial r}\right)+\frac{1}{r^{2}}\frac{\partial^{2}}{\partial\theta^{2}}\left(\psi'\right)+\frac{\partial^{2}\psi'}{\partial z^{2}}\right)\\
 & =\frac{\hbar^{2}}{2m}\left(k_{r}^{2}\left(n_{r},n,\phi\right)+k_{z}^{2}\right).
\end{align*}
In the limit, $\frac{2\pi w}{L}\ll1$ discussed in the preceding paragraph,
we can write the eigenenergy indexed by $(n_{r},n,n_{z})$ and parametrized
by $\phi$ as
\begin{align}
\varepsilon\left(n_{r},n,\phi,n_{z}\right) & =\frac{\hbar^{2}}{2m}\left(\left(\frac{\pi n_{r}}{w}-\frac{\pi}{2n_{r}}\frac{w}{L^{2}}+\frac{2\pi}{n_{r}}\frac{w}{L^{2}}\left(n+\frac{\phi}{\phi_{0}}\right)^{2}\right)^{2}+\left(\frac{\pi n_{z}}{t}\right)^{2}\right)\nonumber \\
 & \approx\frac{\hbar^{2}}{2m}\left(\frac{4\pi^{2}}{L^{2}}\left(n+\frac{\phi}{\phi_{0}}\right)^{2}+\left(\frac{\pi n_{r}}{w}\right)^{2}-\frac{\pi^{2}}{L^{2}}+\left(\frac{\pi n_{z}}{t}\right)^{2}\right)\nonumber \\
 & =\frac{h^{2}}{2mL^{2}}\left(n+\frac{\phi}{\phi_{0}}\right)^{2}+\frac{h^{2}n_{r}^{2}}{8mw^{2}}+\frac{h^{2}n_{z}^{2}}{8mt^{2}}-\frac{h^{2}}{8mL^{2}}.\label{eq:CHPCTh_3DPerfectRingEigenenergies}
\end{align}
The first term in Eq. \ref{eq:CHPCTh_3DPerfectRingEigenenergies}
matches Eq. \ref{eq:CHPCTh_EnergyLevel1DPerfectRing} for the energies
of the one-dimensional ring plotted in Fig. \ref{fig:CHPCTh_EnergyLevelsSimple}.
The other terms offset the energy levels of the one-dimensional ring
by an amount that depends on the indices $n_{r}$ and $n_{z}$ associated
with the transverse dimensions. We refer to the different sets of
eigenstates indexed by $(n_{r},n_{z})$ as {}``transverse modes,''
{}``subbands,'' or {}``transverse channels.'' We also see from
Eq. \ref{eq:CHPCTh_3DPerfectRingEigenenergies} that the {}``one-dimensional
limit'' is reached when $w$ and $t$ are sufficiently small so that
\[
\frac{h^{2}n_{r}^{2}}{8mw^{2}},\frac{h^{2}n_{z}^{2}}{8mt^{2}}>\varepsilon_{F}
\]
for $n_{r},n_{z}\geq2$. In this case, only the lowest transverse
mode $n_{r}=n_{z}=1$ is occupied.

\begin{figure}

\begin{centering}
\includegraphics[width=0.7\paperwidth]{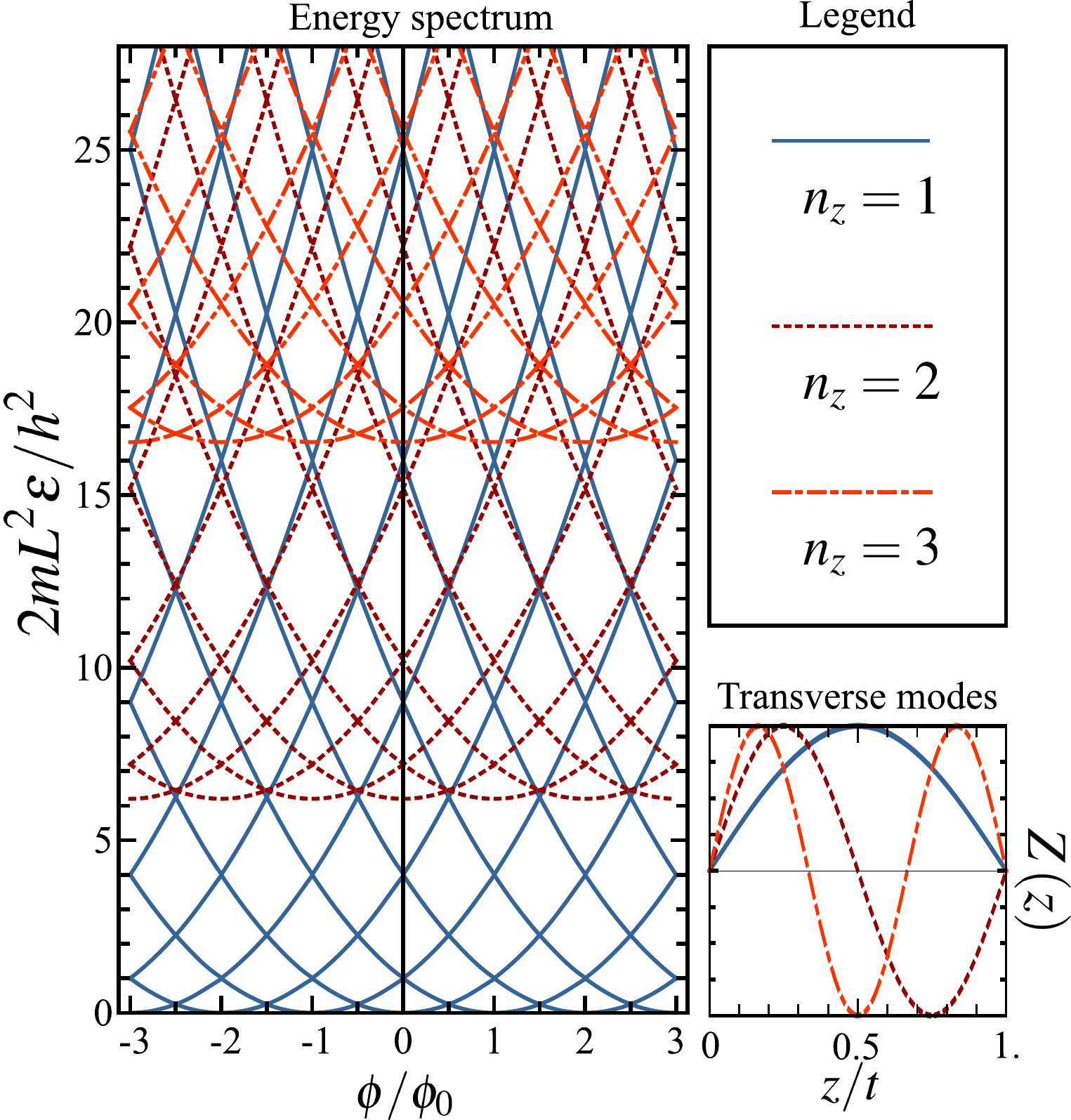}
\par\end{centering}

\caption[Energy levels of a perfect three-dimensional ring]{\label{fig:CHPCTh_3DPerfectRingEnergyLevels}Energy levels of a perfect
three-dimensional ring. The panel on the left plots the energy levels
of the three lowest transverse channels versus flux $\phi$ for $t/L=0.38$
and $w/L\leq0.16$ (so that only $n_{r}=1$ is shown). The three channels
plotted correspond to $n_{z}=1,2,3$ and $n_{r}=1$. The energy axis
has been shifted so that the lowest energy level within the ring is
at $\varepsilon=0$. All levels within one channel are plotted in
the same style as indicated by the legend on the right. Below the
legend, the transverse mode shape $Z\left(z\right)=\sin(n_{z}\pi z/t)$
along the $z$ direction is shown. When $w$ and $t$ are of comparable
dimensions and $\varepsilon_{F}$ is high enough to allow many modes
to be occupied, the energy spectrum becomes very dense.}
\end{figure}

When the one-dimensional limit does not hold, several transverse modes
will be occupied and the energy spectrum of Fig. \ref{fig:CHPCTh_EnergyLevelsSimple}
will be overlaid with additional sets of energy bands offset in $\varepsilon$
by amounts determined by $n_{r}$, $n_{z}$, $w$, and $t$ as described
in Eq. \ref{eq:CHPCTh_3DPerfectRingEigenenergies}. In Fig. \ref{fig:CHPCTh_3DPerfectRingEnergyLevels},
we plot the energy levels of Eq. \ref{eq:CHPCTh_3DPerfectRingEigenenergies}
versus flux $\phi$ for $t^{2}/L^{2}=1/12$ and $w^{2}/L^{2}\leq1/144$.
As can be seen in the figure, the energy spectrum begins to become
complicated as more channels become occupied.

Writing $x=\pi n_{r}/w$, $y=\pi n_{z}/w$, and $\varepsilon_{F}=\hbar^{2}k_{F}^{2}/2m$,
we can rewrite the condition
\[
\frac{h^{2}n_{r}^{2}}{8mw^{2}}+\frac{h^{2}n_{z}^{2}}{8mt^{2}}\leq\varepsilon_{F}
\]
for a channel $(n_{r},n_{z})$ to be occupied as 
\[
x^{2}+\left(\frac{w}{t}\right)^{2}y^{2}\leq k_{F}^{2}.
\]
The total number of occupied levels, $M$, can be found by summing
up all of the occupied levels
\begin{align*}
M & =\sum_{n_{r},n_{z}}1
\end{align*}
such that 
\[
\left(\frac{n_{r}}{w}\right)^{2}+\left(\frac{n_{z}}{t}\right)^{2}\leq\frac{8m\varepsilon_{F}}{h^{2}}.
\]
When $\varepsilon_{F}\gg\frac{h^{2}}{8mw^{2}},\frac{h^{2}}{8mt^{2}}$
(for which $M\gg1$), the sum over $n_{r}$ and $n_{z}$ can be approximated
by an integral
\[
M\approx\frac{1}{4}\left(\frac{w}{\pi}\right)^{2}\iint_{x^{2}+\left(\frac{w}{t}\right)^{2}y^{2}\leq k_{F}^{2}}dx\, dy
\]
where the factor of $1/4$ accounts for the fact that $n_{r},n_{z}>0$.
This integral is just the area of an ellipse with major and minor
diameters of $k_{F}$ and $\frac{t}{w}k_{F}$. Thus, 
\begin{align}
M & =\frac{1}{4}\left(\frac{w}{\pi}\right)^{2}\left(\pi\frac{t}{w}k_{F}^{2}\right)\nonumber \\
 & =\frac{1}{4\pi}wtk_{F}^{2}.\label{eq:CHPCTh_MTransverse}
\end{align}

We now consider the total persistent current in a ring of finite cross-section.
As we have seen, each channel $(n_{r},n_{z})$ possesses the same
energy spectrum as the one-dimensional ring considered in previous
sections with only the energy offset
\[
\varepsilon_{M}\left(n_{r},n_{z}\right)=\frac{h^{2}n_{r}^{2}}{8mw^{2}}+\frac{h^{2}n_{z}^{2}}{8mt^{2}}-\frac{h^{2}}{8mL^{2}}
\]
varying from channel to channel. Thus the total current of any single
channel is given by Eq. \ref{eq:CHPCTh_IperfectRing} with $k_{F}$
replaced by 
\begin{equation}
k_{F,M}\left(n_{r},n_{z}\right)=\sqrt{\frac{2m}{\hbar^{2}}\left(\varepsilon_{F}-\varepsilon_{M}\left(n_{r},n_{z}\right)\right)}.\label{eq:CHPCTh_kFM}
\end{equation}
The expression for the current becomes
\begin{align}
I_{M} & =\sum_{n_{r},n_{z}}\sum_{p>0}I_{p,M}\left(n_{r},n_{z}\right)\sin\left(2\pi p\frac{\phi}{\phi_{0}}\right)\nonumber \\
 & =\sum_{n_{r},n_{z}}\sum_{p>0}\left(\frac{4}{\pi p}\left(\frac{e\hbar}{mL}k_{F,M}\right)\cos\left(pk_{F,M}L\right)\right)\left(\frac{T/T_{p}\left(k_{F,M}\right)}{\sinh\left(T/T_{p}\left(k_{F,M}\right)\right)}\right)\sin\left(2\pi p\frac{\phi}{\phi_{0}}\right)\label{eq:CHPCTh_3DPerfectCurrent}
\end{align}
with 
\[
T_{p}\left(k_{F,M}\right)=\frac{1}{\pi pk_{B}}\frac{\hbar^{2}}{2m}\frac{k_{F,M}}{L}
\]

For the case of many occupied channels ($M\gg1$) and large aspect
ratio ($L\gg w,t$), the quantity $k_{F,M}L$ can depend strongly
on the dimensions of the ring and the channel indices $(n_{r},n_{z})$.
To see this, we note that for channels $(n_{r},n_{z})$ with $\varepsilon_{M}\ll\varepsilon_{F}$
\begin{align*}
k_{F,M}\left(n_{r}+1,n_{z}\right) & =\sqrt{\frac{2m}{\hbar^{2}}\left(\varepsilon_{F}-\varepsilon_{M}\left(n_{r},n_{z}\right)-\frac{h^{2}\left(2n_{r}+1\right)}{8mw^{2}}\right)}\\
 & \approx\sqrt{\frac{2m}{\hbar^{2}}\varepsilon_{F}}\left(1-\frac{1}{2}\frac{\pi^{2}\hbar^{2}\left(2n_{r}+1\right)}{2mw^{2}\varepsilon_{F}}\right)\\
 & \approx k_{F,M}\left(n_{r},n_{z}\right)-\frac{\pi^{2}\left(2n_{r}+1\right)}{2w^{2}k_{F}}.
\end{align*}
Thus, the variation in $k_{F,M}L$ satisfies
\begin{align*}
\Delta\left(k_{F,M}L\right) & =k_{F,M}\left(n_{r}+1,n_{z}\right)L-k_{F,M}\left(n_{r},n_{z}\right)L\\
 & =-\frac{\pi^{2}}{2}\left(2n_{r}+1\right)\left(\frac{L}{w}\right)\left(\frac{1}{k_{F}w}\right).
\end{align*}
For a ring with roughly symmetric cross-section, the condition $M\gg1$
implies $k_{F}w,k_{F}t\gg2\sqrt{\pi}$, so the expression above is
the product of a large quantity $L/w$ and a small quantity $(k_{F}w)^{-1}$.
(A similar expression is of course possible for the other channel
index by replacing $n_{r}$ with $n_{z}$ and $w$ with $t$). To
estimate $\Delta(k_{F,M}L)$, we use the typical values for a metal
of $v_{F}=10^{6}\,\text{m/s}$ and $m=10^{-30}\,\text{kg}$ and for
standard lithographically realizable dimensions $w=60\times10^{-9}\,\text{m}$
and $L=2\times10^{-6}\,\text{m}$. Then we have $k_{F}=mv_{F}/h=9\times10^{9}\,\text{m}^{-1}$,
$M\sim2\times10^{5}$, the number of electrons per channel $N_{M}\sim5\times10^{4}$
and
\[
\Delta\left(k_{F,M}L\right)=-0.3\left(2n_{r,z}+1\right).
\]
Thus the argument of the $\cos(pk_{F,M}L)$ factor in Eq. \ref{eq:CHPCTh_3DPerfectCurrent}
for the current of the multichannel ring can vary by $\gg\pi$ from
channel to channel with the exact magnitude of the variation depending
on $w/L$, $t/L$, $k_{F}$, and the channel indices $n_{r}$ and
$n_{z}$. This large variation leads to a lack of correlation in the
sign of the current in channels with similar values of $k_{F,M}$.
Similarly, when varying $w$, $t$, or $k_{F}$ over a small range
about its typical value, we expect $\cos(pk_{F,M}L)$ to vary between
-1 and 1. We would thus expect the average current of each channel
to be zero upon averaging over such a small range of $w$, $t$, or
$k_{F}$. Since the sign of the current in nearby channels is uncorrelated,
the total current should also average to zero.

Since the current $I_{M}$ varies strongly with the ring dimensions
and the Fermi level, we consider its typical value $\overline{I_{M}^{2}}$,
where the average $\overline{\ldots}$ is over some range of parameters
sufficient to produce a range of uncorrelated values for $\cos(pk_{F,M}L)$
between the different channels. In this case, each $I_{p,M}(n_{r},n_{z})$
in Eq. \ref{eq:CHPCTh_3DPerfectCurrent} is uncorrelated and we have
\begin{align*}
\overline{I_{M}^{2}} & =\overline{\left(\sum_{n_{r},n_{z}}\sum_{p>0}I_{p,M}\left(n_{r},n_{z}\right)\sin\left(2\pi p\frac{\phi}{\phi_{0}}\right)\right)^{2}}\\
 & =\sum_{n_{r},n_{z}}\sum_{p>0}\overline{I_{p,M}^{2}\left(n_{r},n_{z}\right)}\sin^{2}\left(2\pi p\frac{\phi}{\phi_{0}}\right)\\
 & =\sum_{p>0}\left(\frac{4}{\pi p}\frac{e\hbar}{mL}\right)^{2}\left(\sum_{n_{r},n_{z}}k_{F,M}^{2}\overline{\cos^{2}\left(pk_{F,M}L\right)}\left(\frac{T/T_{p}\left(k_{F,M}\right)}{\sinh\left(T/T_{p}\left(k_{F,M}\right)\right)}\right)^{2}\right)\sin^{2}\left(2\pi p\frac{\phi}{\phi_{0}}\right)\\
 & =\sum_{p>0}\frac{1}{2}\left(\frac{4}{\pi p}\frac{e\hbar}{mL}\right)^{2}\left(\sum_{n_{r},n_{z}}k_{F,M}^{2}\left(\frac{T/T_{p}\left(k_{F,M}\right)}{\sinh\left(T/T_{p}\left(k_{F,M}\right)\right)}\right)^{2}\right)\sin^{2}\left(2\pi p\frac{\phi}{\phi_{0}}\right)\\
 & =\sum_{p>0}\overline{I_{p,M}^{2}}\sin^{2}\left(2\pi p\frac{\phi}{\phi_{0}}\right)
\end{align*}
where we have averaged the $\cos^{2}$ term to $1/2$ and otherwise
left the expressions for $k_{F,M}$ unchanged, assuming that the average
over the small range of $k_{F,M}$ of these slowly varying expressions
is roughly equal to their value evaluated at the mean $k_{F,M}$.%
\footnote{Replacing $\cos^{2}\left(pk_{F,M}L\right)$ by $1/2$ is correct for
the grand canonical ensemble where $k_{F,M}$ can vary freely. For
the canonical ensemble, the $\cos\left(pk_{F,M}L\right)$ must be
$\pm1$ so $\cos^{2}\left(pk_{F,M}L\right)$ should be replaced by
$1$ in the average and in this case $\overline{I_{M}^{2}}$ is a
factor of 2 larger. Our temperature dependence is appropriate only
for the grand canonical ensemble.%
} The quantity $T_{p}(k_{F,M})$ is 
\begin{align*}
T_{p}\left(k_{F,M}\right) & =\frac{1}{\pi pk_{B}}\frac{\hbar^{2}}{2m}\frac{k_{F,M}\left(n_{r},n_{z}\right)}{L}\\
 & =\frac{k_{F,M}}{k_{F}}T_{p},
\end{align*}
where $T_{p}$ is the characteristic temperature of the single channel
ring with the same Fermi energy $\varepsilon_{F}$ as the three-dimensional
ring under consideration. In the limit $M\gg1$, the sum over channels
can be replaced by an integral. We first evaluate this integral at
$T=0$ to find
\begin{align*}
\sum_{n_{r},n_{z}}k_{F,M}^{2} & =\frac{1}{4}\int_{k_{r}^{2}+k_{z}^{2}\leq k_{F}^{2}}dk_{r}\, dk_{z}\,\left(\frac{wt}{\pi^{2}}\right)\left(k_{F}^{2}-k_{r}^{2}-k_{z}^{2}\right)\\
 & =\frac{wt}{4\pi^{2}}\int_{0}^{k_{F}}dk\int_{0}^{2\pi}d\theta\, k\left(k_{F}^{2}-k^{2}\right)\\
 & =\frac{wt}{2\pi}k_{F}^{4}\int_{0}^{1}dx\, x\left(1-x^{2}\right)\\
 & =\frac{wt}{2\pi}\frac{k_{F}^{4}}{4}.
\end{align*}
We can thus write the typical magnitude of the $p^{th}$ harmonic
of the current for the multichannel ring at $T=0$ as 
\begin{align}
I_{p,M}^{\text{typ}} & =\sqrt{\overline{I_{p,M}^{2}}}\label{eq:ChPCTh_TypCurrentMultichannel}\\
 & =\left(\frac{1}{4}\left(\frac{4}{\pi p}\right)^{2}\left(\frac{e\hbar}{mL}k_{F}\right)^{2}\left(\frac{wtk_{F}^{2}}{4\pi}\right)\right)^{1/2}\nonumber \\
 & =\frac{\sqrt{M}}{2}\left(\frac{4}{\pi p}I_{0}\right),\nonumber 
\end{align}
which is $\sqrt{M}/2$ times the magnitude of the single channel value
given in Eqs. \ref{eq:CHPCTh_IpN0coeff} and \ref{eq:CHPCTh_IpN2coeff}.
The typical current of the perfect ring grows as the number of transverse
channels is increased. 

\begin{figure}

\begin{centering}
\includegraphics[width=0.5\paperwidth]{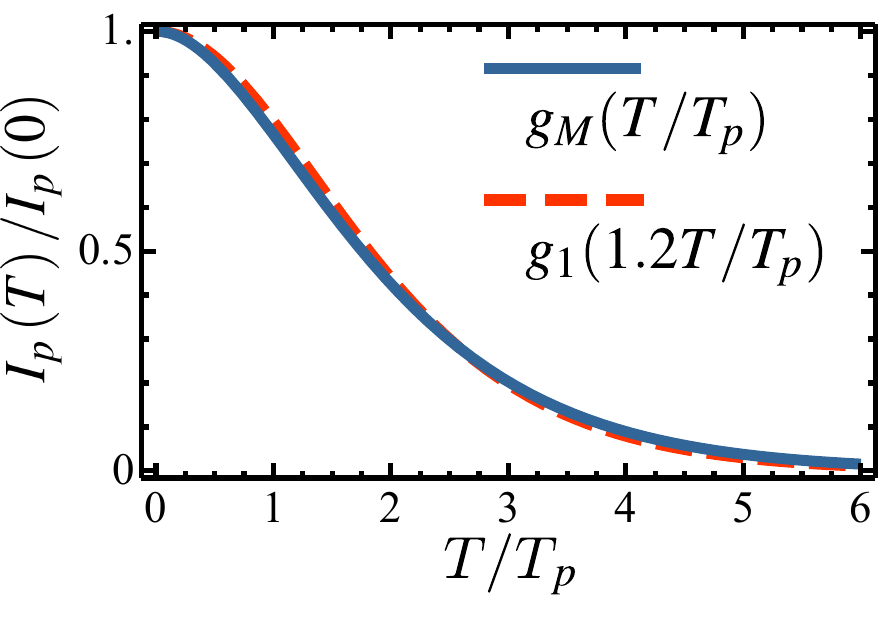}
\par\end{centering}

\caption[Temperature dependence of a perfect three-dimensional ring]{\label{fig:CHPCTh_3DperfectRingTempDependence}Temperature dependence
of a perfect three-dimensional ring. The solid curve gives the normalized
temperature dependence $g_{M}$ of the typical $p^{th}$ harmonic
of the current for the three-dimensional ring as given by Eq. \ref{eq:CHPCTh_gM3DPerfectRing}
in the text. The dashed curve gives the normalized magnitude $g_{1}$
of the $p^{th}$ harmonic of the current for the single channel ring
defined in Eq. \ref{eq:CHPCTh_g11DPerfectRing}. The temperature axis
is scaled by the characteristic temperature $T_{p}$, which was defined
in Eq. \ref{eq:CHPCTh_PerfectRingTp}. The argument of the single
channel ring function $g_{1}$ was scaled by an additional factor
of 1.2 so that the two curves would overlap.}
\end{figure}

In the case of finite temperature, the sum over channels becomes
\begin{align*}
\sum_{n_{r},n_{z}}k_{F,M}^{2}\left(\frac{T/T_{p}\left(k_{F,M}\right)}{\sinh\left(T/T_{p}\left(k_{F,M}\right)\right)}\right)^{2} & =\frac{wt}{4\pi^{2}}\int_{0}^{k_{F}}dk\int_{0}^{2\pi}d\theta\, kk_{F,M}^{2}\left(\frac{T}{k_{F,M}T_{p}/k_{F}}\right)^{2}\frac{1}{\sinh^{2}\left(\frac{T}{k_{F,M}T_{p}/k_{F}}\right)}\\
 & =\frac{wt}{4\pi^{2}}\int_{0}^{k_{F}}dk\int_{0}^{2\pi}d\theta\, k\frac{k_{F}^{2}\left(T/T_{p}\right)^{2}}{\sinh^{2}\left(\frac{T}{T_{p}}\left(1-\frac{k^{2}}{k_{F}^{2}}\right)^{-1/2}\right)}\\
 & =\left(\frac{wt}{2\pi}\frac{k_{F}^{4}}{4}\right)\left(4\left(\frac{T}{T_{p}}\right)^{2}\int_{0}^{1}dx\,\frac{x}{\sinh^{2}\left(\frac{T}{T_{p}}\left(1-x^{2}\right)^{-1/2}\right)}\right).
\end{align*}
For the single channel ring, we wrote $T_{p}$ in terms of the mean
level spacing $\Delta_{1}$ in Eq. \ref{eq:CHPCTh_PerfectRingTp}.
For the multichannel ring, we can calculate $\Delta_{1,M}$, the mean
level spacing within a single channel by averaging $\Delta_{1}$ over
all occupied channels. This averaging gives
\begin{align*}
\Delta_{1,M} & =\frac{\pi\hbar^{2}}{2mL}\left(\frac{1}{M}\sum_{n_{r},n_{z}}k_{F,M}\left(n_{r},n_{z}\right)\right)\\
 & =\frac{\pi\hbar^{2}}{2mL}\left(\frac{1}{M}\left(\frac{wt}{\pi^{2}}\right)\left(\frac{2\pi}{4}\right)\int_{0}^{k_{F}}dk\, k\sqrt{k_{F}^{2}-k^{2}}\right)\\
 & =\frac{\pi\hbar^{2}}{2mL}\left(\frac{1}{M}\left(\frac{wt}{4\pi}\right)2k_{F}^{3}\int_{0}^{1}dx\, x\sqrt{1-x^{2}}\right)\\
 & =\frac{\pi\hbar^{2}}{2mL}\left(2k_{F}\int_{0}^{\pi/2}d\theta\,\cos^{2}\theta\sin\theta\right)\\
 & =\frac{2}{3}\Delta_{1}.
\end{align*}
We could also define a characteristic temperature $T_{p,M}$ in analogy
to Eq. \ref{eq:CHPCTh_PerfectRingTp} by
\begin{align}
T_{p,M} & =\frac{1}{k_{B}}\frac{\Delta_{1,M}}{\pi^{2}p}\nonumber \\
 & =\left(\frac{2}{3}\right)\frac{1}{k_{B}}\frac{\Delta_{1}}{\pi^{2}p}\nonumber \\
 & =\left(\frac{2}{3}\right)T_{p}.\label{eq:CHPCTh_Tpm3DPerfectRing}
\end{align}
However, since this temperature scale differs from that of the single
channel ring by a simple numerical factor, we continue to use $T_{p}$
for simplicity. Defining 
\begin{equation}
g_{M}^{2}\left(\frac{T}{T_{p}}\right)=4\left(\frac{T}{T_{p}}\right)^{2}\int_{0}^{1}dx\,\frac{x}{\sinh^{2}\left(\frac{T}{T_{p}}\left(1-x^{2}\right)^{-1/2}\right)},\label{eq:CHPCTh_gM3DPerfectRing}
\end{equation}
the typical magnitude of the $p^{th}$ harmonic of the current for
the multichannel ring is
\begin{align}
I_{p,M}^{\text{typ}} & =\frac{\sqrt{M}}{2}\left(\frac{4}{\pi p}I_{0}\right)g_{M}\left(\frac{T}{T_{p}}\right).\label{eq:CHPCTh_3DTypCurrentFiniteT}
\end{align}
The function $g_{M}(y)$ does not have a closed analytic form.%
\footnote{At least, my efforts to find one were unsuccessful.%
} In Fig. \ref{fig:CHPCTh_3DperfectRingTempDependence}, we plot $g_{M}(T/T_{p})$
as well as $g_{1}(1.2T/T_{p})$. The function $g_{1}(T/T_{p})$ was
defined in Eq. \ref{eq:CHPCTh_g11DPerfectRing} to give the normalized
temperature dependence of the single channel ring. The factor of 1.2
was chosen so that the two curves agree over the range $T/T_{p}\apprle1$
shown. This agreement indicates that the temperature dependence of
the multichannel ring is determined by the single-channel level spacing
$\Delta_{1,M}$ rather than the average level spacing $\Delta_{M}$
across all levels 
\begin{align}
\Delta_{M} & =\frac{\Delta_{1,M}}{M}\label{eq:CHPCTh_MultiChannelMeanLevelSpacing}\\
 & =\frac{4\pi}{3}\frac{1}{wtk_{F}^{2}}\frac{\hbar v_{F}}{L},\nonumber 
\end{align}
which depends on the transverse dimensions of the ring. 

The correspondence between the temperature dependence of the single-channel
and multichannel rings is due to the fact that the levels within a
given channel are all anti-correlated as described in Section \ref{sub:CHPCTh_1DPerfFinT}
while we have taken different channels to be uncorrelated. From the
assumption of no correlation between different channels, it follows
that the current from levels in different channels does not add up
to a net average current and so does not cancel out with the thermal
broadening of the energy range over which levels are partially occupied. 

The decay with temperature of the multichannel ring is stronger than
the single channel ring by a factor of $\sim1.2$. To explain this
stronger decay, we note that, in the limit of many electrons, the
mean level spacing $\Delta_{1,M}(n_{r},n_{z})$ of the lowest indexed
channel $(n_{r}=1,n_{z}=1)$ is approximately the same as the mean
level spacing $\Delta_{1}$ of a single-channel ring with the same
Fermi energy. For higher indexed channels, the mean level spacing
$ $$\Delta_{1,M}(n_{r},n_{z})$ is smaller and so is the characteristic
temperature of decay for that channel. Since the current in the higher
channels decays more quickly than the single channel current, the
current summed over all channels does as well.

The results of this section for the multichannel ring provide a striking
contrast to those found for a ring in the diffusive limit. Here we
found that the typical magnitude of the current scales with the square-root
of the number of transverse channels. In the diffusive regime, the
current magnitude is independent of the number of transverse channels.
As we will see in the next section, disorder distorts the shape of
the individual energy levels. This distortion leads to correlation
between levels from different channels but near each other in energy
and destroys the anti-correlation of the levels within a single channel
because the energy scale $E_{c}$ of the correlation is less than
$\Delta_{1}$.%
\footnote{The relationship between $E_{c}$ and $\Delta_{1}$ is not obvious
at this point in our discussion. In Eq. \ref{eq:ChPCTh_ThoulessEnergy},
we will define $E_{c}=\hbar D/L^{2}$. The correlation energy $E_{c}$
can be rewritten in terms of $\Delta_{1}$ as $E_{c}=(2/\pi)(l_{e}/L)\Delta_{1}$
where the elastic mean free path $l_{e}$ is a measure of the strength
of the disorder (we note though that this definition of $E_{c}$ was
chosen for its notational simplicity. In Fig. \ref{fig:CHPCTh_H1CurrCurrCor},
it can be seen that the energy levels are actually correlated on a
scale of $\sim10E_{c}$). In the diffusive regime, $l_{e}\ll L$,
and thus $E_{c}\ll\Delta_{1}$.%
} Thus, the different channels no longer make large, uncorrelated contributions
to the current. In the presence of disorder the total current magnitude
is actually a fraction of the disorder-free single channel current.
As discussed in Section \ref{sub:CHPCTh_1DPerfFinT}, the temperature
dependence in the diffusive regime is set by the correlation scale
$E_{c}$. Since $E_{c}<\Delta_{1}$, the decay of the current with
temperature is also stronger in the diffusive regime. Similar results
to the ones discussed in this section were obtained in 1970 by Kulik
for a two-dimensional ring $(w\ll t,L)$ \citep{kulik1970fluxquantization}.
For a more detailed account of the finite cross-section ring, see
Ref. \citep{cheung1988isolated}.

\FloatBarrier

\subsection{\label{sub:PCTh_DisorderIntro}Introduction of disorder}

Before discussing the diffusive regime in the next section, we make
a few general remarks about the introduction of disorder to the perfect
ring which has been considered in this section.

First we examine the effect of disorder on two levels in the energy
spectrum (Figs. \ref{fig:CHPCTh_EnergyLevelsSimple}, \ref{fig:CHPCTh_EnergyLevelsThermalDistribution},
and \ref{fig:CHPCTh_3DPerfectRingEnergyLevels}). We label states
by $|\psi_{1}\rangle$ and $|\psi_{2}\rangle$ and their flux-dependent
energies by $\varepsilon_{1}(\phi)$ and $\varepsilon_{2}(\phi)$.
In the $[|\psi_{1}\rangle,|\psi_{2}\rangle]$ basis, we can write
the unperturbed Hamiltonian as 
\[
\hat{H}_{0}\left(\phi\right)=\left[\begin{array}{cc}
\varepsilon_{1}\left(\phi\right) & 0\\
0 & \varepsilon_{2}\left(\phi\right)
\end{array}\right].
\]
We consider the effect of introducing a spatially-dependent disorder
potential 
\[
\hat{V}=\left[\begin{array}{cc}
V_{11} & V_{12}\\
V_{12}^{*} & V_{22}
\end{array}\right]
\]
to the Hamiltonian $\hat{H}=\hat{H}_{0}+\hat{V}$, where 
\begin{align}
V_{ij} & =\left\langle \psi_{i}|\hat{V}|\psi_{j}\right\rangle \nonumber \\
 & =\int_{R-w/2}^{R+w/2}dr\int_{0}^{t}dz\int_{0}^{2\pi}d\theta\, r\psi_{i}^{\prime*}\left(n_{ri},n_{i},\phi,n_{zi}\right)V\left(r,\theta,z\right)\psi_{j}^{\prime}\left(n_{rj},n_{j},\phi,n_{zj}\right)\label{eq:CHPCTh_VijPertCrossterms}
\end{align}
where $\psi_{j}^{\prime}(n_{rj},n_{j},\phi,n_{zj})$ is the eigenfunction
(with $(r,\theta,z)$ dependence suppressed) given in Eq. \ref{eq:CHPCTh_MultichannelEigenfunction}.

The diagonal values $V_{11}$ and $V_{22}$ merely shift $\varepsilon_{1}$
and $\varepsilon_{2}$ and can be absorbed into these terms by taking
$\varepsilon_{i}+V_{ii}\rightarrow\varepsilon_{i}$. Dropping the
explicit use of $V_{ii}$, the Hamiltonian in the presence of disorder
is 
\[
\hat{H}\left(\phi\right)=\left[\begin{array}{cc}
\varepsilon_{1}\left(\phi\right) & V_{12}\\
V_{12}^{*} & \varepsilon_{2}\left(\phi\right)
\end{array}\right].
\]
In the presence of $\hat{V}$, the Hamiltonian is no longer diagonalized,
indicating that $\{|\psi_{1}\rangle,|\psi_{2}\rangle\}$ are no longer
the energy eigenstates. Instead $\hat{V}$ produces mixed eigenstates
which are linear combinations of the form $A|\psi_{1}\rangle+B|\psi_{2}\rangle$.
From Eq. \ref{eq:CHPCTh_VijPertCrossterms} and the orthogonality
of the different factors of the eigenstates as expressed in Eq. \ref{eq:CHPCTh_MultichannelEigenfunction},
some general conclusions can be reached about the effect of $\hat{V}$
on the energy levels. In order for $\hat{V}$ to mix two levels from
different orbitals $n_{1}\neq n_{2}$, it must have a non-zero Fourier
component $\widetilde{V}(n_{1}-n_{2},r,z)\neq0$ when expanded in
a Fourier series $V(r,\theta,z)=\sum_{n}\widetilde{V}(n,r,z)e^{in\theta}$.
Additionally, to mix levels from different channels $V(r,\theta,z)$
can not be uniform in $r$ and $z$, $V(r,\theta,z)\neq V(\theta)$,
since the transverse eigenfunctions are themselves orthogonal in $(r,z)$.

The new eigenenergies are found by solving 
\begin{align*}
0 & =\det\left[\hat{H}-\lambda\hat{I}\right]\\
 & =\left(\varepsilon_{1}-\lambda\right)\left(\varepsilon_{2}-\lambda\right)-\left|V_{12}\right|^{2}\\
 & =\lambda^{2}-\left(\varepsilon_{1}+\varepsilon_{2}\right)+\varepsilon_{1}\varepsilon_{2}-\left|V_{12}\right|^{2}.
\end{align*}
Setting $V^{2}=\left|V_{12}\right|^{2}$, the quadratic formula gives
the new eigenenergies $\lambda_{\pm}$ as 
\begin{align}
\lambda_{\pm} & =\frac{1}{2}\left(\varepsilon_{1}+\varepsilon_{2}\right)\pm\frac{1}{2}\sqrt{\left(\varepsilon_{1}+\varepsilon_{2}\right)^{2}-4\varepsilon_{1}\varepsilon_{2}+4V^{2}}\nonumber \\
 & =\frac{1}{2}\left(\varepsilon_{1}+\varepsilon_{2}\right)\pm\frac{1}{2}\sqrt{\left(\varepsilon_{1}-\varepsilon_{2}\right)^{2}+4V^{2}}.\label{eq:CHPCTh_lambdaAvoidedCrossing}
\end{align}
When $|\varepsilon_{1}-\varepsilon_{2}|\gg2V$, the new eigenenergies
match the original ones, with $\lambda_{+}\approx\max(\varepsilon_{1},\varepsilon_{2})$
and $\lambda_{-}\approx\min(\varepsilon_{1},\varepsilon_{2})$. When
$\varepsilon_{1}\approx\varepsilon_{2}=\varepsilon_{D}$, the new
eigenenergies are split from the degenerate value $\varepsilon_{D}$
by $2V$, $\lambda_{\pm}\approx\varepsilon_{D}\pm V$. When the $\varepsilon_{i}$
are functions of some parameter $\phi$ so that $|\varepsilon_{1}-\varepsilon_{2}|$
changes as well, the splitting of $(\lambda_{+},\lambda_{-})$ at
a point in $\phi$ where $|\varepsilon_{1}-\varepsilon_{2}|\rightarrow0$
is referred to as an {}``avoided crossing.'' Taking $\varepsilon_{2}>\varepsilon_{1}$,
we can write
\begin{align}
\lambda_{+} & =\varepsilon_{2}+\frac{1}{2}\left(\sqrt{\left(\varepsilon_{2}-\varepsilon_{1}\right)^{2}+4V^{2}}-\left(\varepsilon_{2}-\varepsilon_{1}\right)\right)\label{eq:ChPCTh_LambaPlusAC}\\
\lambda_{-} & =\varepsilon_{1}-\frac{1}{2}\left(\sqrt{\left(\varepsilon_{2}-\varepsilon_{1}\right)^{2}+4V^{2}}-\left(\varepsilon_{2}-\varepsilon_{1}\right)\right)\label{eq:CHPCTh_LambdaMinusAC}
\end{align}
from which it is easily seen that the deviations from $(\varepsilon_{1},\varepsilon_{2})$
are always $\leq V$. Since $V$ only changes the levels close to
each other, this two-level analysis can be extended to many levels
as long as $V$ is small enough that only two adjacent levels are
ever within $V$ of each other at a time (otherwise a treatment handling
the three or more relevant levels would be necessary).

\begin{figure}

\begin{centering}
\includegraphics[width=0.54\paperwidth]{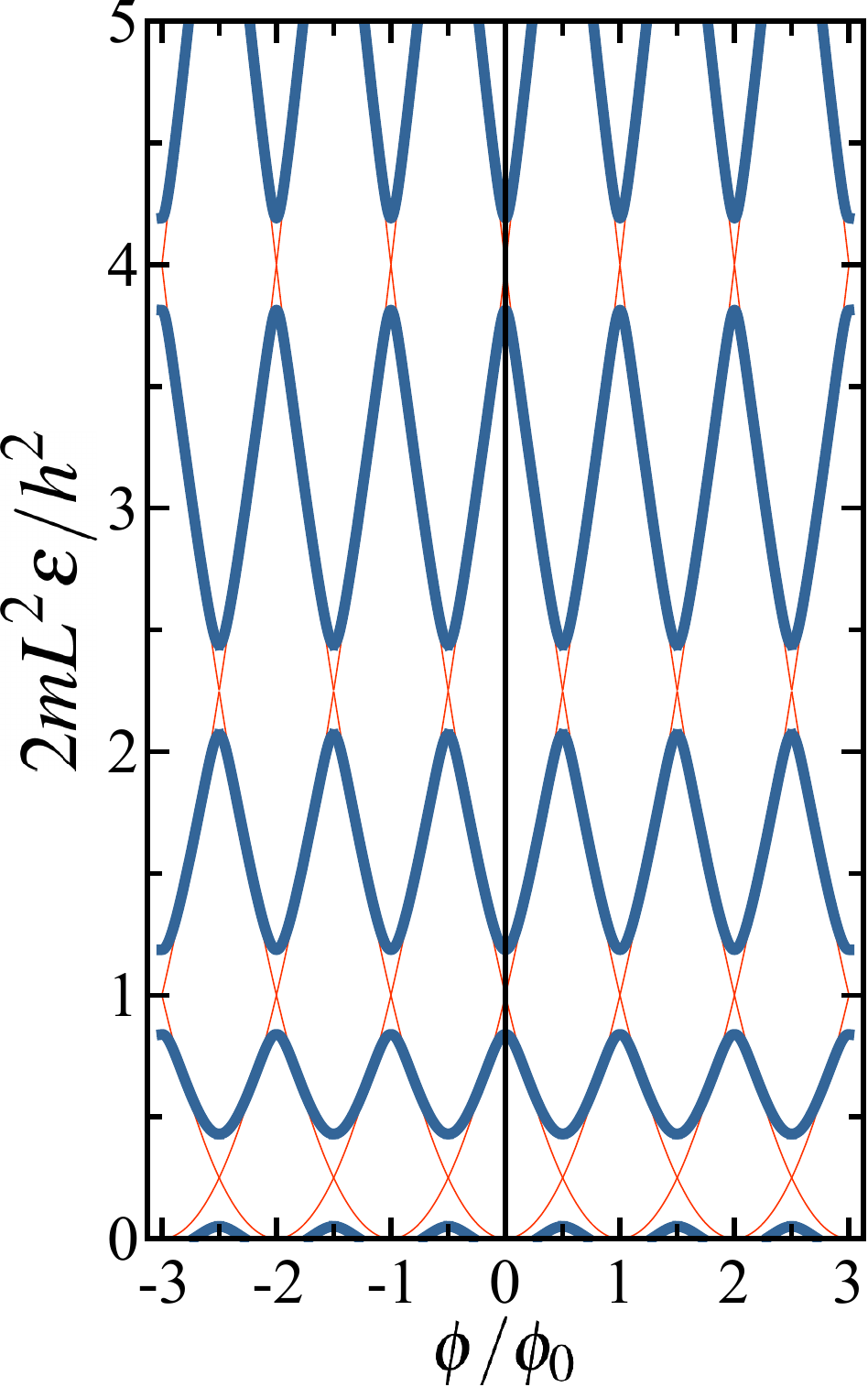}
\par\end{centering}

\caption[Energy spectrum of a disordered one-dimensional ring]{\label{fig:CHPCTh_1DDisorderedEnergyLevels}Energy spectrum of a
disordered one-dimensional ring. The thick lines show the energy levels
calculated using Eq. \ref{eq:CHPCTh_lambdaAvoidedCrossing} with $V_{ij}=0.2h^{2}/2mL^{2}$
for each pair of levels $i\neq j$ and $V_{ii}=0$ for each level
$i$. The thinner lines reproduce the unperturbed ($V=0$) spectrum
of the one-dimensional ring shown in Figs. \ref{fig:CHPCTh_EnergyLevelsSimple}
and \ref{fig:CHPCTh_EnergyLevelsSimpleReindexed}.}
\end{figure}

As a special case, we note that for $\varepsilon_{1,2}=(\pm n+\phi)^{2}$
(which is the case of the one-dimensional ring with $h^{2}/2mL^{2}$
and $\phi_{0}$ set to 1) the eigenenergies are 
\begin{align*}
\lambda_{\pm} & =\frac{1}{2}\left(\left(n+\phi\right)^{2}+\left(n-\phi\right)^{2}\right)\pm\frac{1}{2}\sqrt{\left(\left(n+\phi\right)^{2}-\left(n-\phi\right)^{2}\right)^{2}+4V^{2}}\\
 & =\left(\phi^{2}+n^{2}\right)\pm\sqrt{2n^{2}\phi^{2}+V^{2}}.
\end{align*}
Near the crossing at $\phi=0$, the eigenenergies become to lowest
order in $\phi$
\begin{align*}
\lambda_{\pm} & \approx\left(\phi^{2}+n^{2}\right)\pm\left(V+\frac{n^{2}\phi^{2}}{V}\right)\\
 & =n^{2}\pm V+\left(1\pm\frac{n^{2}}{V}\right)\phi^{2}.
\end{align*}
All energy level crossings%
\footnote{Pairs of levels degenerate at $\phi=m\phi_{0}/2$ with $m\neq0$ can
be put in this form by shifting $\phi$ by $-m\phi_{0}/2$.%
} for the one-dimensional ring are of this form and thus all crossings
in the spectrum are replaced by these quadratic avoided crossings,
as indicated in Fig. \ref{fig:CHPCTh_1DDisorderedEnergyLevels}. In
that figure, it can be seen that the kinks in the energy levels are
smoothed out while the $\phi_{0}$ periodicity of the spectrum is
maintained, as expected from the discussion in Section \ref{sub:CHPCTh_1DRingSingleLevelSolutions}.
It can also be seen that this smoothing reduces the range of energy
$\varepsilon$ and energy slope $\partial\varepsilon/\partial\phi$
experienced by each level. Since the current in a given level is proportional
to this slope, the disorder reduces the single level currents and
the total current as a result.

\begin{figure}
\begin{centering}
\includegraphics[width=0.52\paperwidth]{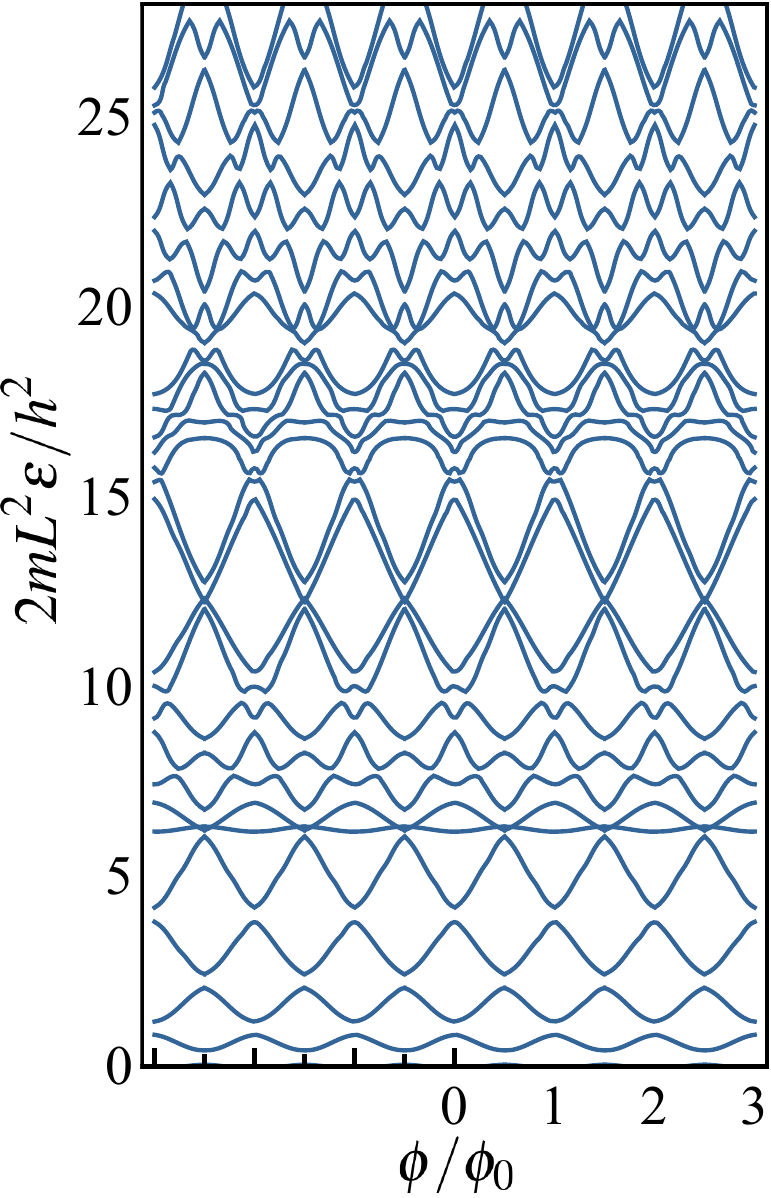}
\par\end{centering}

\caption[Energy spectrum of a disordered multichannel ring]{\label{fig:CHPCTh_3DDisorderedEnergyLevels}Energy spectrum of a
disordered multichannel ring. The figure shows the spectrum for a
ring with the same dimensions ($t/L=0.38$ and $w/L\leq0.16$) as
those used to make Fig. \ref{fig:CHPCTh_3DPerfectRingEnergyLevels}
but with an off-diagonal perturbation applied to break the level degeneracies.
Towards the top of the plot, the energy level density is higher and
the two-level limit is not strictly valid. Because of this fact, some
levels are pushed quite close to each other and appear to intersect.
These intersections can be viewed as very close avoided crossings
for which the finite thickness of the curves depicting adjacent energy
levels overlap.}
\end{figure}

In Fig. \ref{fig:CHPCTh_3DDisorderedEnergyLevels}, the energy spectrum
of the multichannel ring shown in Fig. \ref{fig:CHPCTh_3DPerfectRingEnergyLevels}
is replotted for the case of a disordered ring. The energy levels
were calculated by applying Eq. \ref{eq:CHPCTh_lambdaAvoidedCrossing}
to each pair of intersecting unperturbed levels.%
\footnote{More precisely, for two levels $\varepsilon_{1}$ and $\varepsilon_{2}$
which cross at $\phi_{c}$ and for which $\varepsilon_{1}<\varepsilon_{2}$
when $\phi<\phi_{c}$, the quantity
\begin{equation}
\frac{1}{2}\left(\sqrt{\left(\varepsilon_{2}-\varepsilon_{1}\right)^{2}+4V^{2}}-\left(\varepsilon_{2}-\varepsilon_{1}\right)\right)\label{eq:ChPCTh_PerturbationCorrect}
\end{equation}
was subtracted from level $\varepsilon_{1}$ and added to $\varepsilon_{2}$
in order to create two new perturbed levels (see Eqs. \ref{eq:ChPCTh_LambaPlusAC}
and \ref{eq:CHPCTh_LambdaMinusAC}). For levels undergoing multiple
crossings, this procedure was repeated for each crossing with a term
such as the one in Eq. \ref{eq:ChPCTh_PerturbationCorrect} above
being added or subtracted at each crossing. In calculating the new
levels, the terms given by Eq. \ref{eq:ChPCTh_PerturbationCorrect}
were calculated using the unperturbed levels. Thus the figure does
not include the second order effects that occur when two distant unperturbed
levels are shifted by other levels close enough to each other that
they should be subject to a repulsion away from each other as well.%
} The same disorder potential $V$ ($V_{ij}=0.2h^{2}/2mL^{2}$ for
each pair of levels $i\neq j$ and $V_{ii}=0$ for each level $i$)
as used in Fig. \ref{fig:CHPCTh_1DDisorderedEnergyLevels} was employed
for this figure so the low-lying levels ($\varepsilon<5h^{2}/2mL^{2}$)
of Fig. \ref{fig:CHPCTh_3DDisorderedEnergyLevels} reproduce Fig.
\ref{fig:CHPCTh_1DDisorderedEnergyLevels}. At higher energies the
density of levels increases. Each individual level represents the
result several avoided crossings and takes on quite a complicated,
but still periodic, flux dependence. As was the case for the single
channel ring of Fig. \ref{fig:CHPCTh_1DDisorderedEnergyLevels}, the
disorder smooths out the energy levels and so reproduces the associated
single level currents. 

As the ratio $V/\Delta_{M}$ of the disorder strength $V$ to the
level spacing $\Delta_{M}$ is increased to $\sim1$, the two-level
picture no longer remains valid. The disorder induced level repulsion
becomes strong enough to push one of the two repulsed levels into
a third nearby level. Since the avoided crossing derivation is quite
general, it is also valid for the level pushed into the third level
and they repel each other as well. When this multi-level repulsion
occurs for many levels, neighboring levels become correlated on an
energy scale set by the disorder. Also, the long-range anti-correlation
of the levels of each unperturbed channel noted previously is destroyed.
The effect of this energy level correlation on the total persistent
current is addressed in the next section.

It is possible to obtain a quantitative expression for the effect
of weak disorder within the framework introduced in this section for
the ideal ring. To show this, we use the result from Eq. \ref{eq:AppGrFu_DOSLorentzianConvolution}
for a Gaussian disorder potential characterized by an elastic scattering
time $\tau_{e}$. In the weak disorder limit $k(\varepsilon)l_{e}\gg1$
(where $l_{e}=v_{F}\tau_{e}$ is the elastic mean free path), the
disorder averaged density of states $\nu(\varepsilon)$ is related
to the density of states%
\footnote{In this section, we will use {}``0'' in subscripts to denote quantities
calculated previously for a ring without disorder.%
} $\nu_{0}(\varepsilon)$ in the absence of disorder by 
\[
\nu\left(\varepsilon\right)=\int_{0}^{\infty}d\varepsilon'\,\nu_{0}\left(\varepsilon^{\prime}\right)b_{L}\left(\varepsilon-\varepsilon^{\prime},\frac{\hbar}{2\tau_{e}}\right)
\]
where $b_{L}(\varepsilon,\delta)$ is the Lorentzian function
\[
b_{L}\left(\varepsilon,\delta\right)=\frac{1}{\pi}\frac{\delta}{\varepsilon^{2}+\delta^{2}}.
\]
In Section \ref{sub:CHPCTh_1DPerfFinT}, we argued that the grand
canonical potential $\Omega$ expressed in Eq. \ref{eq:CHPCTh_PCOmega}
could be rewritten using two integrations by parts (for which the
boundary terms could be dropped) as
\begin{equation}
\Omega=\int_{0}^{\infty}d\varepsilon\,\left(\int_{0}^{\varepsilon}d\varepsilon'\int_{0}^{\varepsilon'}d\varepsilon''\,\nu\left(\varepsilon'',\phi\right)\right)f'\left(\varepsilon,\varepsilon_{F},T\right).\label{eq:CHPCTh_GrandPotential}
\end{equation}
Because we have defined the energy levels to begin at $\varepsilon=0$,
we can freely extend the lower limit of integration of $\nu$ and
$\nu_{0}$ from 0 to $-\infty$ when it is convenient. We can rewrite
the integral of the disorder averaged density of states as
\begin{align*}
\int_{0}^{\varepsilon}d\varepsilon_{1}\,\nu\left(\varepsilon_{1}\right) & =\int_{-\infty}^{\varepsilon}d\varepsilon_{1}\int_{-\infty}^{\infty}d\varepsilon_{2}\,\nu_{0}\left(\varepsilon_{2}\right)b_{L}\left(\varepsilon_{1}-\varepsilon_{2},\frac{\hbar}{2\tau_{e}}\right)\\
 & =-\int_{-\infty}^{\infty}d\varepsilon_{2}\,\left(\int_{-\infty}^{\varepsilon_{2}}d\varepsilon_{3}\,\nu_{0}\left(\varepsilon_{3}\right)\right)\int_{-\infty}^{\varepsilon}d\varepsilon_{1}\,\left(-1\right)b_{L}^{\prime}\left(\varepsilon_{1}-\varepsilon_{2},\frac{\hbar}{2\tau_{e}}\right)\\
 & \phantom{=}+\left(\int_{-\infty}^{\varepsilon_{2}}d\varepsilon_{3}\nu_{0}\left(\varepsilon_{3}\right)\right)\left(\int_{-\infty}^{\varepsilon}d\varepsilon_{1}\, b_{L}\left(\varepsilon_{1}-\varepsilon_{2},\frac{\hbar}{2\tau_{e}}\right)\right)\Bigg|_{\varepsilon_{2}=-\infty}^{\infty}
\end{align*}
where we have used integration by parts with respect to $\varepsilon_{2}$.
Since $b_{L}(\varepsilon)\rightarrow0$ as $\varepsilon\rightarrow\pm\infty$,
the boundary term can be dropped. We then have
\begin{align}
\int^{\varepsilon}d\varepsilon_{1}\,\nu\left(\varepsilon_{1}\right) & =\int_{-\infty}^{\infty}d\varepsilon_{2}\,\left(\int_{-\infty}^{\varepsilon_{2}}d\varepsilon_{3}\,\nu_{0}\left(\varepsilon_{3}\right)\right)b_{L}\left(\varepsilon-\varepsilon_{2},\frac{\hbar}{2\tau_{e}}\right).\label{eq:CHPCTh_NuIntDisorderAverage}
\end{align}
This integration by parts did not make use of any properties of $\nu_{0}$
and can be repeated replacing $\nu_{0}(\varepsilon)$ with $\int_{-\infty}^{\varepsilon}d\varepsilon_{1}\,\nu_{0}(\varepsilon_{1})$
so that
\begin{equation}
\left(\int_{-\infty}^{\varepsilon}d\varepsilon_{1}\int_{-\infty}^{\varepsilon_{1}}d\varepsilon_{2}\,\nu\left(\varepsilon_{2},\phi\right)\right)=\int_{-\infty}^{\infty}d\varepsilon_{1}\,\left(\int_{-\infty}^{\varepsilon_{1}}d\varepsilon_{2}\int_{-\infty}^{\varepsilon_{2}}d\varepsilon_{3}\,\nu_{0}\left(\varepsilon_{3},\phi\right)\right)b_{L}\left(\varepsilon-\varepsilon_{1},\frac{\hbar}{2\tau_{e}}\right).\label{eq:CHPCTh_DOSCleanDisordered}
\end{equation}

At $T=0$, $f'(\varepsilon,\varepsilon_{F},T)=\delta(\varepsilon-\varepsilon_{F})$
and the persistent current is
\begin{align}
I & =-\frac{\partial\Omega}{\partial\phi}\nonumber \\
 & =-\frac{\partial}{\partial\phi}\left(\int_{-\infty}^{\varepsilon_{F}}d\varepsilon'\int_{-\infty}^{\varepsilon'}d\varepsilon''\,\nu\left(\varepsilon'',\phi\right)\right).\label{eq:CHPCTh_CurrentZeroTDOSform}
\end{align}
By Eq. \ref{eq:CHPCTh_DOSCleanDisordered}, the disorder average is
accomplished by convolving the clean ring result at energy $\varepsilon$
with the Lorentzian $b_{L}(\varepsilon_{F}-\varepsilon,\hbar/2\tau_{e})$.
Using the result of Eq. \ref{eq:CHPCTh_1DCurrentZeroT} for $T=0$,
we can write the disorder averaged current as 
\begin{align*}
I & =\int_{-\infty}^{\infty}d\varepsilon\, I_{1D,0}\left(\varepsilon\right)b_{L}\left(\varepsilon_{F}-\varepsilon,\frac{\hbar}{2\tau_{e}}\right)\\
 & =\sum_{p>0}\sin\left(2\pi p\frac{\phi}{\phi_{0}}\right)\frac{8\hbar^{2}}{mpL\phi_{0}}\int_{-\infty}^{\infty}d\varepsilon\, k\left(\varepsilon\right)\cos\left(pk\left(\varepsilon\right)L\right)b_{L}\left(\varepsilon_{F}-\varepsilon,\frac{\hbar}{2\tau_{e}}\right).
\end{align*}
The function $b_{L}(\varepsilon_{F}-\varepsilon,\hbar/2\tau_{e})$
is sharply peaked near the Fermi energy $\varepsilon_{F}$ so we can
once again use the expansion given in Eq. \ref{eq:CHPCTh_epsKrelation}
for $k(\varepsilon)$ about $\varepsilon_{F}$,
\[
k\left(\varepsilon\right)\approx k_{F}\left(1+\frac{1}{2}\left(\frac{\varepsilon}{\varepsilon_{F}}-1\right)-\frac{1}{8}\left(\frac{\varepsilon}{\varepsilon_{F}}-1\right)^{2}+\mathcal{O}\left(\left(\frac{\varepsilon}{\varepsilon_{F}}-1\right)^{3}\right)\right).
\]
The function $b_{L}(\varepsilon_{F}-\varepsilon,\hbar/2\tau_{e})$
is appreciable for $|\varepsilon-\varepsilon_{F}|\apprle\hbar/2\tau_{e}$.
Note that
\begin{align*}
\frac{\hbar}{2\tau_{e}} & =\frac{\hbar v_{F}}{2l_{e}}\\
 & =\frac{\hbar}{2l_{e}}\left(\frac{\hbar k_{F}}{m}\right)\\
 & =\frac{\varepsilon_{F}}{k_{F}l_{e}},
\end{align*}
so that over this same range with $\alpha\leq1$ 
\[
k\left(\varepsilon_{F}+\alpha\frac{\hbar}{2\tau_{e}}\right)\approx k_{F}\left(1+\frac{1}{2k_{F}l_{e}}\alpha-\frac{1}{8\left(k_{F}l_{e}\right)^{2}}\alpha^{2}\right).
\]
Since we are assuming weak disorder $k_{F}l_{e}\gg1$, it is sufficient
to replace the first factor of $k$ in the expression for $I$ by
$k_{F}$. We make the additional assumption that $L\apprge l_{e}$
so that $k_{F}l_{e}\gg L/l_{e}$. Only up to the first order term
in $\varepsilon-\varepsilon_{F}$ must be kept in the argument of
the cosine factor of the expression for $I$. Thus, we have 
\begin{align*}
 & \int_{-\infty}^{\infty}d\varepsilon\, k\left(\varepsilon\right)\cos\left(pk\left(\varepsilon\right)L\right)b_{L}\left(\varepsilon_{F}-\varepsilon,\frac{\hbar}{2\tau_{e}}\right)\\
 & \phantom{\int_{0}^{\infty}d\varepsilon}\approx k_{F}\int_{-\infty}^{\infty}d\varepsilon\,\cos\left(pk_{F}L+p\frac{k_{F}L}{2\varepsilon_{F}}\left(\varepsilon-\varepsilon_{F}\right)\right)b_{L}\left(\varepsilon_{F}-\varepsilon,\frac{\hbar}{2\tau_{e}}\right)\\
 & \phantom{\int_{0}^{\infty}d\varepsilon}\approx k_{F}\int_{-\infty}^{\infty}d\varepsilon\,\left(\cos\left(pk_{F}L\right)\cos\left(\frac{pk_{F}L}{2\varepsilon_{F}}\varepsilon\right)-\sin\left(pk_{F}L\right)\sin\left(\frac{pk_{F}L}{2\varepsilon_{F}}\varepsilon\right)\right)b_{L}\left(\varepsilon,\frac{\hbar}{2\tau_{e}}\right)\\
 & \phantom{\int_{0}^{\infty}d\varepsilon}=k_{F}\cos\left(pk_{F}L\right)\int_{-\infty}^{\infty}d\varepsilon\,\exp\left(i\frac{pk_{F}L}{2\varepsilon_{F}}\varepsilon\right)b_{L}\left(\varepsilon,\frac{\hbar}{2\tau_{e}}\right)
\end{align*}
where we made use of the narrowness and symmetry of $b_{L}(\varepsilon)$.
The last line is the Fourier transform of the Lorentzian which has
the well-known form
\begin{align}
\int_{-\infty}^{\infty}d\varepsilon\, k\left(\varepsilon\right)\cos\left(pk\left(\varepsilon\right)L\right)b_{L}\left(\varepsilon_{F}-\varepsilon,\frac{\hbar}{2\tau_{e}}\right) & =k_{F}\cos\left(pk_{F}L\right)\exp\left(-\frac{\left|p\right|\hbar k_{F}L}{4\varepsilon_{F}\tau_{e}}\right)\nonumber \\
 & =k_{F}\cos\left(pk_{F}L\right)\exp\left(-\frac{\left|p\right|L}{2l_{e}}\right).\label{eq:CHPCTh_FourierTransformLorentzian}
\end{align}
We can write the single-channel, disorder averaged persistent current
as
\begin{align}
I & =\sum_{p>0}\frac{8\hbar^{2}k_{F}}{mpL\phi_{0}}\cos\left(pk_{F}L\right)\exp\left(-p\frac{L}{2l_{e}}\right)\sin\left(2\pi p\frac{\phi}{\phi_{0}}\right)\nonumber \\
 & =\sum_{p>0}I_{p,0}\exp\left(-p\frac{L}{2l_{e}}\right)\sin\left(2\pi p\frac{\phi}{\phi_{0}}\right)\label{eq:CHPCTh_1DDisorderedCurrent}
\end{align}
where the coefficients $I_{p,0}$ are the harmonic amplitudes of the
current in the absence of disorder. Note that the suppression factors
have no dependence on $k_{F}$ so that the result in Eq. \ref{eq:CHPCTh_1DDisorderedCurrent}
remains valid at finite temperature just by using the temperature
dependent form for the $I_{p,0}(T)$ given in Eq. \ref{eq:CHPCTh_1DDisorderedCurrent}. 

Similarly, the typical harmonic magnitudes $I_{p,M,0}^{\text{typ}}$
given in Eq. \ref{eq:CHPCTh_3DTypCurrentFiniteT} for the multichannel
ring can be multiplied by the factor $\exp(-pL/2l_{e})$ to give the
disorder averaged result (since each single channel in the ring is
suppressed by this same factor). Recall that, in the absence of disorder
and for rings with dimensions similar to those studied in this text,
the $p^{th}$ harmonic $I_{p,M}$ of the multi-channel ring was found
to fluctuate strongly in sign and magnitude under small changes of
the Fermi wave vector $k_{F}$, thickness $t$, or width $w$. Because
of this strong dependence on dimensions, the current from nominally
identical rings could differ greatly in magnitude and sign due to
lithographic imperfections. To describe the current in the multi-channel
ring quantitatively, we introduced in Eq. \ref{eq:ChPCTh_TypCurrentMultichannel}
the typical magnitude $I_{p,M,0}^{\text{typ}}$ of the $p^{th}$ harmonic
of the current. This typical magnitude was found by averaging $I_{p,M,0}^{2}$
over small variations in $k_{F}$, $w$, or $t$ and then taking the
square root. To account for disorder, one replaces $I_{p,M,0}$ with
$I_{p,M,0}\exp(-pL/2l_{e})$. The calculation of the typical current
magnitude under variations of $k_{F}$, $w$, or $t$ proceeds as
before with this additional factor of $\exp(-pL/2l_{e})$. To be precise,
this quantity $I_{p,M,0}^{\text{typ}}\exp(-pL/2l_{e})$ is the result
of finding the disorder averaged current of a multichannel ring and
then finding its standard deviation over a small range of $k_{F}$,
$t$, or $w$. 

By taking the disorder average of the current first, we obtain the
exponential dependence on $L/l_{e}$. If it were possible to fabricate
many rings with precisely the same dimensions but different realizations
of the microscopic disorder, we would expect the average current in
all of these rings to be $I_{p,M,0}\exp(-pL/2l_{e})$ where the sign
and magnitude of the $I_{p,M,0}$ can be found exactly from Eq. \ref{eq:CHPCTh_3DPerfectCurrent}.
In practice, this calculation requires unrealistic precision in the
specification of the ring dimensions (namely $k_{F}$, $w$, and $t$).
For this reason, we calculate the typical current by averaging over
dimensions. It turns out that this quantity $I_{p,M}^{\text{typ}}$
is not especially useful for us. In \ref{sec:CHPCTh_DiffusiveRegime},
we consider the fluctuations of the persistent current with disorder
by averaging the square of the current over disorder (rather than
averaging the current over disorder and then taking the square average
over $k_{F}$, $w$, $t$). The typical current magnitude found in
this way does not depend sensitively on the ring dimensions, nor does
it decay exponentially in $L/l_{e}$. Consequently, it is much larger
than $I_{p,M}^{\text{typ}}$ for the rings studied experimentally
in this text for which $L/l_{e}\gg1$. The subtlety of this distinction
in the method of disorder averaging perhaps explains why the possibility
of measuring persistent currents experimentally was overlooked for
as long as it was.

The reason for the strong suppression of the current in $L/l_{e}$
is best described in terms of Green's functions (see Appendix \ref{cha:AppGrFu_}).
From Eq. \ref{eq:AppGrFu_DOSSpatialForm}, it is seen that the density
of states, which is intimately related to the persistent current (see
Eq. \ref{eq:CHPCTh_CurrentZeroTDOSform}), can be written as a spatial
average of Green's function amplitudes for closed paths within the
ring. The disorder average effectively throws away the contributions
of all closed paths in which the electron wavefunction is scattered
from one wavevector $\boldsymbol{k}$ to another $\boldsymbol{k}'$,
which happens when the electron travels an average distance $l_{e}$
(albeit with $|\boldsymbol{k}|=|\boldsymbol{k}'|$ unchanged because
the scattering is elastic). It was pointed out by Landauer, Büttiker,
and Imry that, because the electron maintains phase coherence over
these discarded trajectories, many of them produce sizable contributions
to the persistent current \citep{buttiker1983josephson}. We discuss
the contribution of these diffusive trajectories to the persistent
current in the next section.

\FloatBarrier

\section{\label{sec:CHPCTh_DiffusiveRegime}Persistent currents in the diffusive
regime}

With all of the framework for describing persistent currents established
in the previous sections, we consider the diffusive regime relevant
for metal rings. As discussed in Section \ref{sub:PCTh_DisorderIntro},
in the presence of disorder, the energy levels of the clean ring are
distorted and nearby levels repelled from each other. A ring with
a cross-section on the order of tens of nanometers can have on the
order of $M=10^{5}$ transverse channels resulting in relatively dense
channels. 

\begin{figure}
\begin{centering}
\includegraphics[width=0.7\paperwidth]{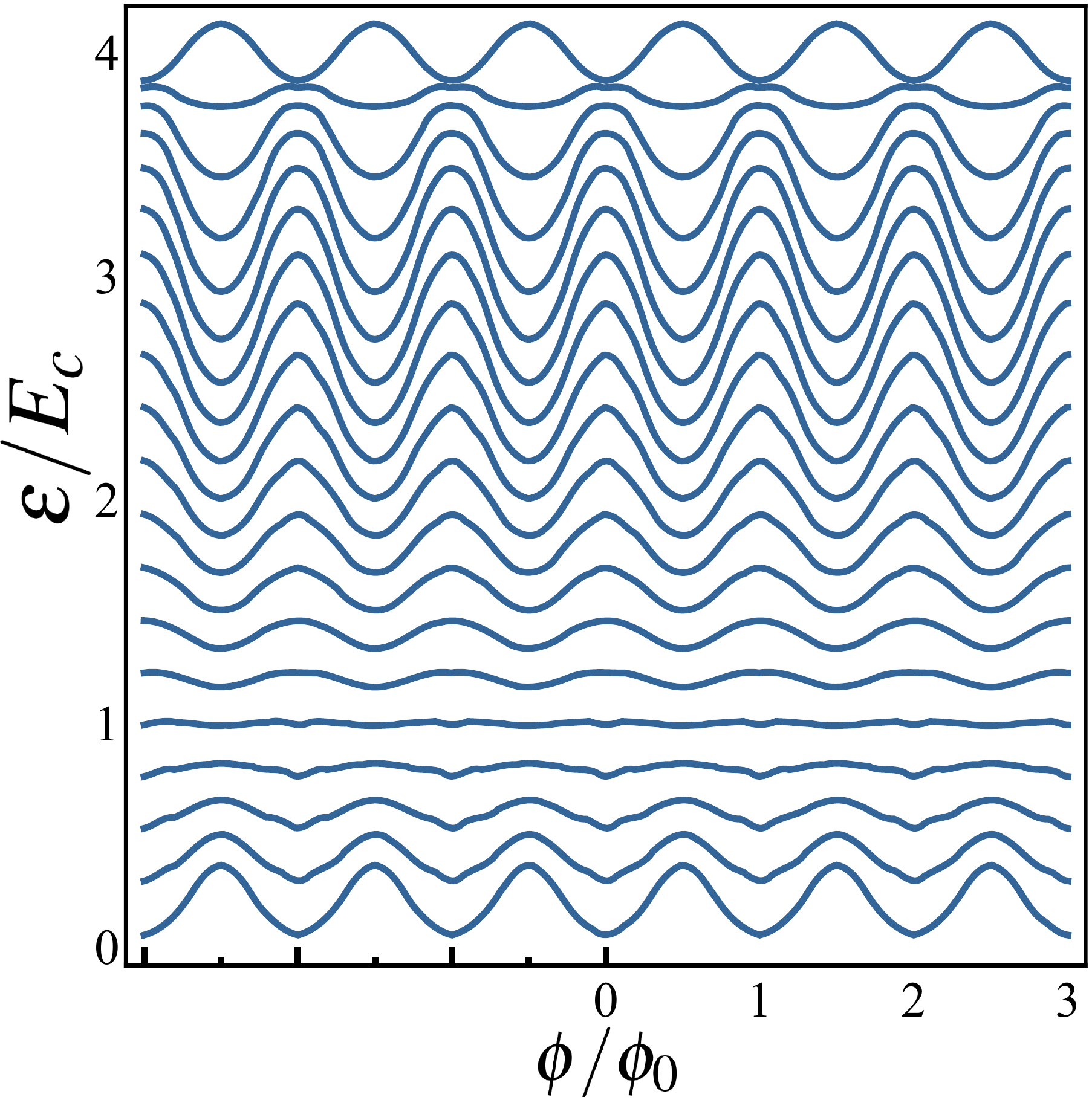}
\par\end{centering}

\caption[Drawing of energy level spectrum in the diffusive regime]{\label{fig:CHPCTh_DiffusiveEnergyLevels}Drawing of energy level
spectrum in the diffusive regime. The figure represents a cartoon
of the energy level spectrum for one disorder realization of a ring
in the diffusive regime. The spectrum was designed to have a correlation
scale of $M_{\text{eff}}=4$ energy levels. The energy axis is scaled
by the correlation energy while the flux axis is scaled by the flux
quantum.}
\end{figure}

As all of these dense channels repel each other strongly in the diffusive
regime, the flux-dependence of the energy levels is smoothed out leading
to a reduction in the current $i_{n}=-\partial\varepsilon_{n}/\partial\phi$
of each single level $\varepsilon_{n}$. However, this strong repulsion
of dense levels also results in a correlation in the features of neighboring
levels, as illustrated in Fig. \ref{fig:CHPCTh_DiffusiveEnergyLevels}.
Although the spectrum as a whole still consists of energy levels with
currents $\langle i_{n}\rangle=0$ which have no net disorder average,
this correlation means that a large number $M_{\text{eff}}\gg1$ of
levels add coherently to the total current $I=\sum_{n}i_{n}$. While
the current of most levels will cancel out, the net current will have
a typical magnitude $I^{\text{typ}}=\sqrt{\langle I^{2}\rangle}$
of $M_{\text{eff}}\sqrt{\langle i^{2}\rangle}$ rather than simply
$\sqrt{\langle i^{2}\rangle}$. It turns out the number of correlated
levels $M_{\text{eff}}=\frac{l_{e}}{L}M$.

\begin{figure}
\begin{centering}
\includegraphics[width=0.7\paperwidth]{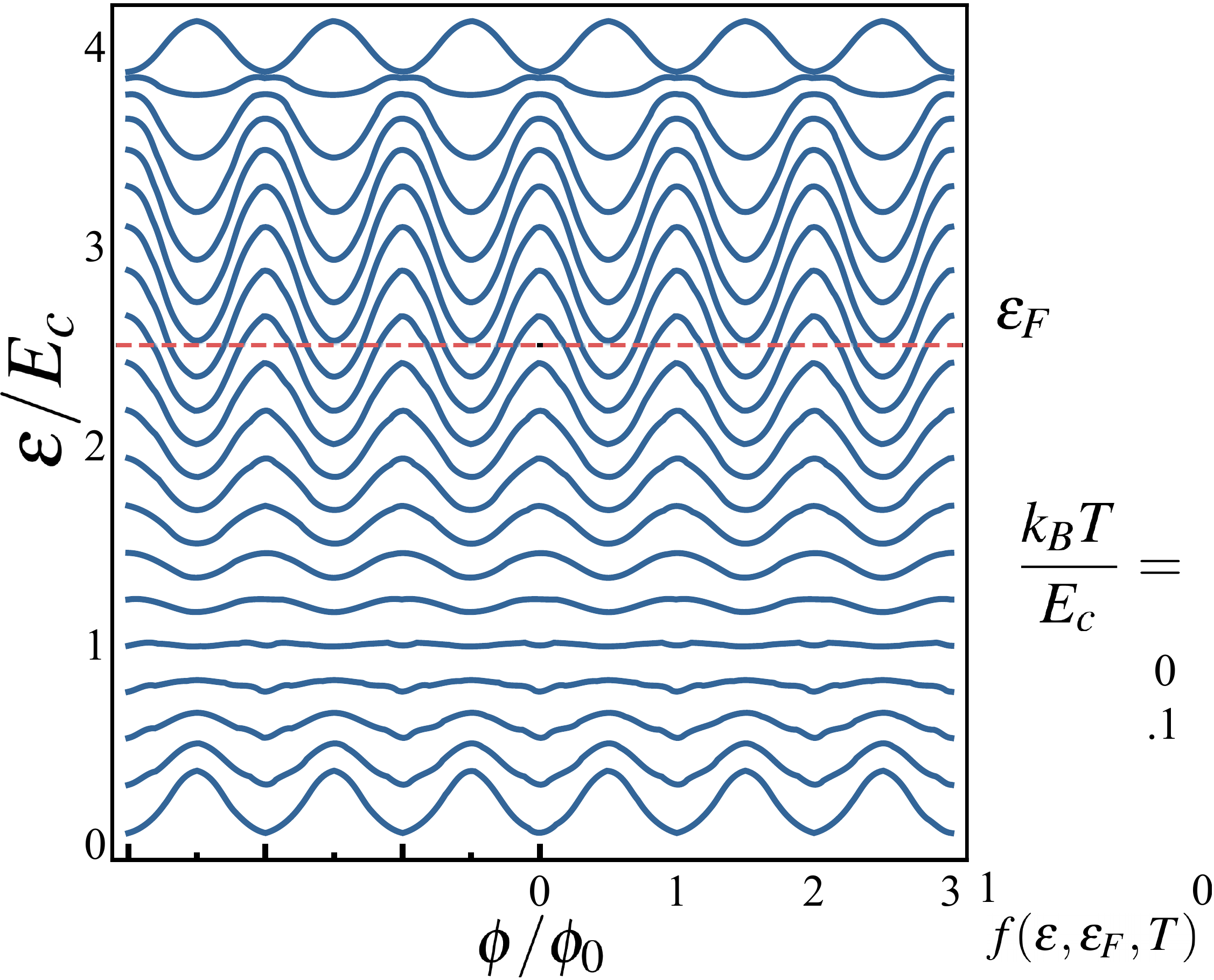}
\par\end{centering}

\caption[Drawing of distribution of occupancy of the energy level spectrum
in the diffusive regime]{\label{fig:CHPCTh_DiffusiveEnergyLevelsT}Drawing of distribution
of occupancy of the energy level spectrum in the diffusive regime.
The same spectrum as shown in Fig. \ref{fig:CHPCTh_DiffusiveEnergyLevels}
is replotted alongside a plot of the Fermi-Dirac distribution $f(\varepsilon,\varepsilon_{F},T)$
for three different temperatures ($0\times E_{c}/k_{B}$, $0.1\times E_{c}/k_{B}$,
$0.3\times E_{c}/k_{B}$). By the highest temperature shown the level
occupancy begins spread out over levels with different shapes from
those right at the Fermi level $\varepsilon_{F}$.}
\end{figure}

\begin{figure}
\begin{centering}
\includegraphics[width=0.7\paperwidth]{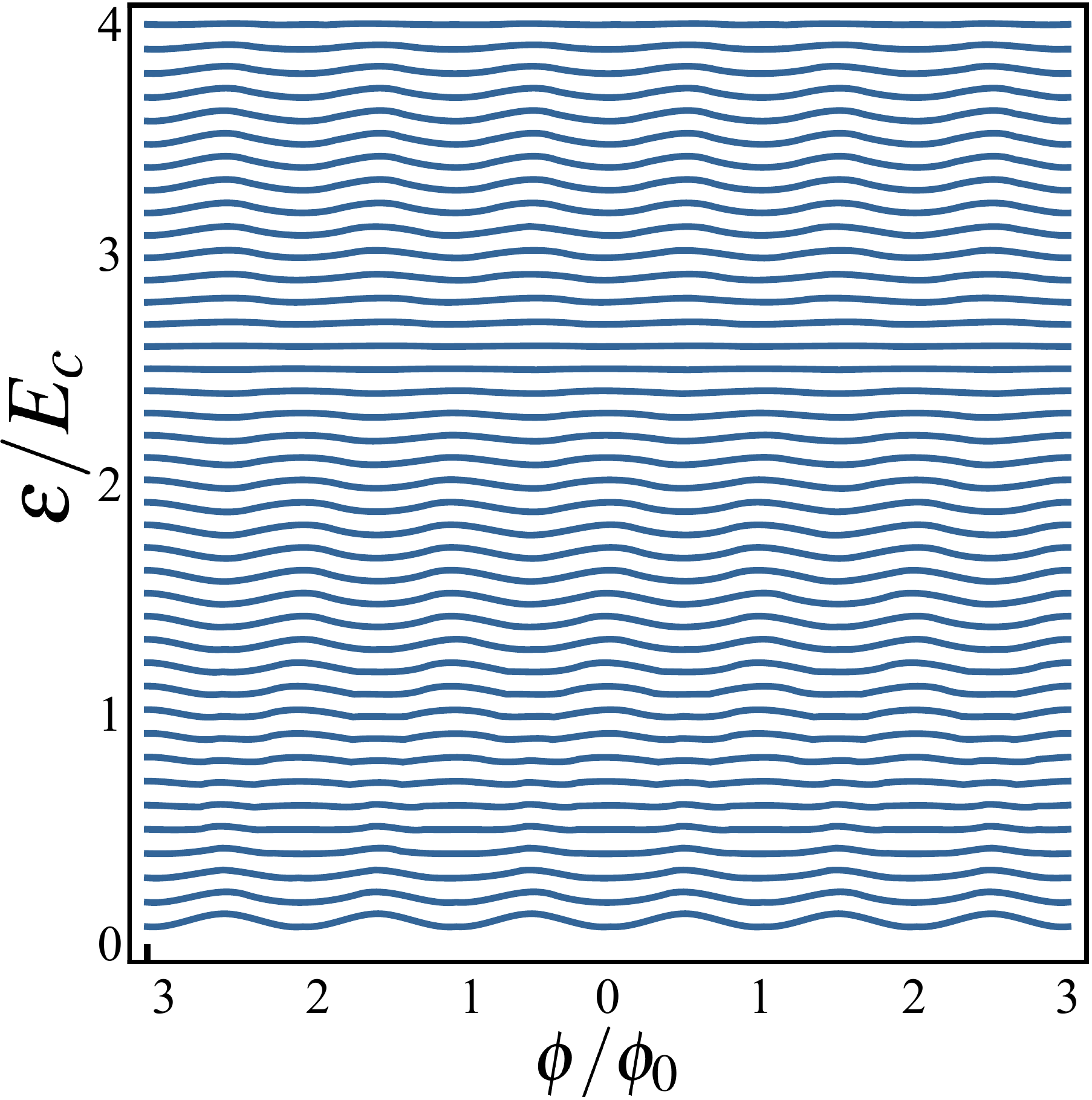}
\par\end{centering}

\caption[Alternate drawing of energy level spectrum in the diffusive regime]{\label{fig:CHPCTh_DiffusiveDenseEnergyLevels}Alternate drawing of
energy level spectrum in the diffusive regime. A spectrum is shown
for which eight neighboring energy levels are correlated. In contrast
to Fig. \ref{fig:CHPCTh_DiffusiveEnergyLevels}, this spectrum (for
the same correlation energy scale $E_{c}$) is denser and has less
steeply sloped energy curves. As discussed in the text, this spectrum
and that of Fig. \ref{fig:CHPCTh_DiffusiveEnergyLevels} have the
same typical current because the larger number of coherent levels
$M_{\text{eff}}=8$ is compensated for by the smaller single level
current $i=\partial\varepsilon/\partial\phi$. The typical current
of each spectrum could be calculated, for example, by finding $I_{N}=\sum_{n=1}^{N}i_{n}$
for a range of highest occupied levels $N$ and then averaging $I_{N}^{2}$
over $N$.}
\end{figure}

The correlation of energy levels means that levels within $\sim M_{\text{eff}}\Delta$
about the Fermi energy $\varepsilon_{F}$ all give the same contribution
to the current and thus the thermal occupancy, shown in Fig. \ref{fig:CHPCTh_DiffusiveEnergyLevelsT},
must be broadened to this energy scale $k_{B}T\sim M_{\text{eff}}\Delta$
rather than the smaller scale of the single level spacing $\Delta$
before the current begins to decay. Because the correlation in the
energy spectrum results from level repulsion, changing the density
of levels (e.g. by varying the ring cross-section and so the number
of transverse channels $M$) does not affect the typical current magnitude.
A greater density of levels means a greater number of correlated levels
contribute to $I^{\text{typ}}$. However, a greater density of levels
also means that each level $n$ is further flattened and so possesses
less current $i_{n}$. These two effects cancel each other out. An
indication of this phenomena is given in Fig. \ref{fig:CHPCTh_DiffusiveDenseEnergyLevels}
which shows an energy spectrum twice as dense as that shown in Fig.
\ref{fig:CHPCTh_DiffusiveEnergyLevels} but with a typical level slope
of half the magnitude of Fig. \ref{fig:CHPCTh_DiffusiveEnergyLevels}.
We will now discuss all of these effects more quantitatively.

\FloatBarrier

\subsection{\label{sub:CHPCTh_TypicalCurrent}Typical persistent current magnitude
in the diffusive regime}

To calculate the persistent current in the diffusive regime,%
\footnote{In addition to Refs. \citealp{riedel1993mesoscopic} and \citealp{ginossar2010mesoscopic},
I acknowledge private communication with Eran Ginossar and Felix von
Oppen in the composition of this derivation.%
} we return to the general formula 
\begin{align*}
I & =-\frac{\partial\Omega}{\partial\phi},
\end{align*}
which we rewrite using Eq. \ref{eq:CHPCTh_GrandPotential} and integration
by parts on $\varepsilon'$ as
\begin{align*}
I & =-\frac{\partial}{\partial\phi}\int_{0}^{\infty}d\varepsilon\,\left(\int_{0}^{\varepsilon}d\varepsilon'\int_{0}^{\varepsilon'}d\varepsilon''\,\nu\left(\varepsilon'',\phi\right)\right)f'\left(\varepsilon,\varepsilon_{F},T\right)\\
 & =-\frac{\partial}{\partial\phi}\int_{0}^{\infty}d\varepsilon\,\left(\int_{0}^{\varepsilon}d\varepsilon'\,\left(\varepsilon-\varepsilon'\right)\nu\left(\varepsilon',\phi\right)\right)f'\left(\varepsilon,\varepsilon_{F},T\right).
\end{align*}
In Section \ref{sub:PCTh_DisorderIntro}, we found that the current
averaged over disorder decays as $\exp(-L/2l_{e})$. This same conclusion
can also be reached by noting that Eq. \ref{eq:AppGrFu_DOSSpatialForm}
relates $\nu(\varepsilon)$ to a Green function and that according
to Eq. \ref{eq:AppGrFu_GDisorderAverage} this Green function decays
on a characteristic length scale $2l_{e}$ when averaged over disorder.
Since the average value of the current decays strongly, we focus here
on the typical magnitude $\sqrt{\langle I^{2}\rangle}$ of the fluctuations
of the current and revisit the average current in Section \ref{sub:CHPCTh_AverageCurrent}.

We begin by considering a ring threaded by an ideal Aharonov-Bohm
flux $\phi$ (see Section \ref{sub:CHPCTh_1DRingSingleLevelSolutions}).
We calculate the current-current correlation function
\begin{align}
\left\langle I\left(\phi\right)I\left(\phi'\right)\right\rangle  & =\frac{\partial^{2}}{\partial\phi\partial\phi'}\left\langle \Omega\left(\phi\right)\Omega\left(\phi'\right)\right\rangle \nonumber \\
 & =\int_{0}^{\infty}d\varepsilon_{1}\int_{0}^{\infty}d\varepsilon_{1}^{\prime}\,\left(f'\left(\varepsilon_{1}\right)f'\left(\varepsilon_{1}^{\prime}\right)\right)C_{1}^{\left(0\right)}\left(\varepsilon_{1},\phi;\varepsilon_{1}^{\prime},\phi'\right)\label{eq:ChPCTh_CurrentCurrentCorrelationDef}
\end{align}
where $f'(\varepsilon_{1})$ is short for $f'(\varepsilon_{1},\varepsilon_{F},T)$
and
\begin{equation}
C_{1}^{\left(0\right)}\left(\varepsilon_{1},\phi;\varepsilon_{1}^{\prime},\phi'\right)=\frac{\partial^{2}}{\partial\phi\partial\phi'}\int_{0}^{\varepsilon_{1}}d\varepsilon_{2}\int_{0}^{\varepsilon_{1}^{\prime}}d\varepsilon_{2}^{\prime}\,\left(\varepsilon_{1}-\varepsilon_{2}\right)\left(\varepsilon_{1}^{\prime}-\varepsilon_{2}^{\prime}\right)\left\langle \nu\left(\varepsilon_{2},\phi\right)\nu\left(\varepsilon_{2}^{\prime},\phi'\right)\right\rangle .\label{eq:CHPCTh_CurrentCurrentCorC10Def}
\end{equation}
As we have noted before the derivative of the Fermi-Dirac distribution
obeys $f'(\varepsilon,\varepsilon_{F},T)\rightarrow\delta(\varepsilon-\varepsilon_{F})$
as $T\rightarrow0$. Thus the current-current correlation function
at zero temperature is 
\[
\left\langle I\left(\phi\right)I\left(\phi'\right)\right\rangle =C_{1}^{\left(0\right)}\left(\varepsilon_{F},\phi;\varepsilon_{F},\phi'\right).
\]
The function $C_{1}^{(0)}(\varepsilon,\phi,\varepsilon',\phi')$ can
be thought of as the current-current correlation function at zero
temperature for a ring at two different Fermi energies $\varepsilon$
and $\varepsilon'$ and Aharonov-Bohm fluxes $\phi$ and $\phi'$:
\[
\left\langle I\left(\varepsilon,\phi\right)I\left(\varepsilon',\phi'\right)\right\rangle =C_{1}^{\left(0\right)}\left(\varepsilon,\phi;\varepsilon',\phi'\right).
\]
Because this quantity is rather abstract, we will use the more general-looking
syntax $C_{1}^{(0)}$ rather than the correlation function syntax
$\langle I(\phi)I(\phi')\rangle$. We first evaluate the current-current
correlation function at zero temperature and then consider the more
general case.

\subsubsection{\label{sub:CHPCTh_ZeroTDiffusiveCurrent}Zero temperature}

Because of the flux derivatives in Eq. \ref{eq:CHPCTh_CurrentCurrentCorC10Def},
we need only consider the flux-dependent portion of the disorder averaged
density of states correlation function. This flux dependent part of
the correlation function is specified by Eq. \ref{eq:AppGrFu_DOSFieldDependent},
\[
\left\langle \nu\left(\varepsilon,B\right)\nu\left(\varepsilon-\hbar\omega,B'\right)\right\rangle _{d,c}=2\left(\frac{1}{2\pi\hbar}\right)^{2}\sum_{\mp}\sum_{n}\text{Re}\left(\left(\frac{1}{i\omega+DE_{n}\left(B_{\mp}\right)}\right)^{2}\right),
\]
 and Eq. \ref{eq:AppGrFu_DiffusonCooperonEigenvalues},
\begin{equation}
\left(\nabla'+i\frac{e}{\hbar}\boldsymbol{A}_{\mp}\right)^{2}P_{d,c}\left(\boldsymbol{r},\boldsymbol{r}',\omega\right)=E_{n}^{d,c}\left(B_{\mp}\right)P_{d,c}\left(\boldsymbol{r},\boldsymbol{r}',\omega\right).\label{eq:CHPCTh_DiffusonCooperonEigenvalues}
\end{equation}
Since $\langle\nu(\varepsilon_{2},\phi)\nu(\varepsilon_{2}^{\prime},\phi')\rangle$
depends only on $|\varepsilon_{2}-\varepsilon_{2}^{\prime}|$, we
can rewrite this correlation function as
\begin{align}
 & C_{1}^{\left(0\right)}\left(\varepsilon_{1},\phi;\varepsilon_{1}^{\prime},\phi'\right)=\frac{\partial^{2}}{\partial\phi\partial\phi'}\int_{-\infty}^{\varepsilon_{1}}d\varepsilon_{2}\int_{-\infty}^{\varepsilon_{1}^{\prime}}d\varepsilon_{2}^{\prime}\,\left(\varepsilon_{1}-\varepsilon_{2}\right)\left(\varepsilon_{1}^{\prime}-\varepsilon_{2}^{\prime}\right)\left\langle \nu\left(\varepsilon_{2},\phi\right)\nu\left(\varepsilon_{2}^{\prime},\phi'\right)\right\rangle \nonumber \\
 & \phantom{C_{1}^{\left(0\right)}}=\frac{\partial^{2}}{\partial\phi\partial\phi'}\int_{-\infty}^{0}d\varepsilon_{2}\int_{-\infty}^{0}d\varepsilon_{2}^{\prime}\,\varepsilon_{2}\varepsilon_{2}^{\prime}\left\langle \nu\left(\varepsilon_{2}+\varepsilon_{1},\phi\right)\nu\left(\varepsilon_{2}^{\prime}+\varepsilon_{1}^{\prime},\phi'\right)\right\rangle \nonumber \\
 & \phantom{C_{1}^{\left(0\right)}}=\frac{\hbar}{8}\frac{\partial^{2}}{\partial\phi\partial\phi'}\int_{-\infty}^{0}d\varepsilon\int_{\varepsilon/\hbar}^{-\varepsilon/\hbar}d\omega\,\left(\varepsilon^{2}-\hbar^{2}\omega^{2}\right)\left\langle \nu\left(\frac{\varepsilon+\hbar\omega}{2}+\varepsilon_{1},\phi\right)\nu\left(\frac{\varepsilon-\hbar\omega}{2}+\varepsilon_{1}^{\prime},\phi'\right)\right\rangle \label{eq:CHPCTh_CurrentCurrentCorrelation}
\end{align}
where we have used the change of variables $\varepsilon=\varepsilon_{2}+\varepsilon_{2}^{\prime}$
and $\hbar\omega=\varepsilon_{2}-\varepsilon_{2}^{\prime}$ and again
extended the integrals to $-\infty$.

Before evaluating the integrals, we must evaluate the disorder averaged
density of states correlation function. For a ring subject to an ideal
Aharonov-Bohm flux $\phi$, Eq. \ref{eq:AppGrFu_DiffusonCooperonEigenvalues}
for the eigenvalues $E_{n}(\phi_{\pm})$ of the diffuson and cooperon
is identical to the Schrödinger equation, Eq. \ref{eq:CHPCTh_3DSchrodingerVectorPotential},
considered in Section \ref{sub:CHPCTh_FiniteCrossSection} with $\hbar^{2}/2m\rightarrow1$
and the boundary conditions changed to Eq. \ref{eq:AppGrFu_DiffusionBoundaryConditions},
\begin{equation}
\tilde{\boldsymbol{n}}\cdot\left(\nabla'+i\frac{e}{\hbar}\boldsymbol{A}_{\mp}\right)P_{d,c}\left(\boldsymbol{r},\boldsymbol{r}',\omega\right)=0.\label{eq:CHPCTh_DiffusonCooperonBoundaryCondition}
\end{equation}
The eigenvalues are thus given by Eq. \ref{eq:CHPCTh_3DPerfectRingEigenenergies}
except that the new boundary conditions change the indices allowed
for the transverse degrees of freedom. Using the same gauge considered
in that section, we have that $\tilde{\boldsymbol{n}}\cdot\boldsymbol{A}=0$
on all surfaces, and thus the boundary condition becomes
\[
\tilde{\boldsymbol{n}}\cdot\nabla'P_{d,c}\left(\boldsymbol{r},\boldsymbol{r}',\omega\right)=0
\]
 which admits (amongst others) solutions of the form $P_{d,c}(r,\theta,z)=P_{d,c}(\theta)$,
which are independent of $r$ and $z$. These functions correspond
to $n_{r}=n_{z}=0$, which were not allowed for Eq. \ref{eq:CHPCTh_3DPerfectRingEigenenergies}.%
\footnote{The new boundary condition also changes the form of the eigenfunctions.
For example, instead of $Z(z)=\sin(\pi n_{z}z/t)$ we would have $Z(z)=\cos(\pi n_{z}z/t)$.
However, the eigenvalue is the same in both cases. Since we are concerned
only with the eigenvalues and not the eigenfunctions here, we do not
write out the new form for the eigenfunctions. %
} The eigenvalues for the diffuson and cooperon are then%
\footnote{A few more steps are necessary to show that the extra constant offset
in Eq. \ref{eq:CHPCTh_3DPerfectRingEigenenergies} drops out when
the radial eigenvalues are recalculated for the new boundary conditions.%
}
\[
E_{n}\left(\phi_{\pm}\right)=\frac{\left(2\pi\right)^{2}}{L^{2}}\left(n+\frac{\phi}{\phi_{0}}\right)^{2}+\frac{\pi^{2}n_{r}^{2}}{2w^{2}}+\frac{\pi^{2}n_{z}^{2}}{2t^{2}}.
\]
We will write these eigenvalues as 
\begin{equation}
E_{n}\left(\phi_{\pm}\right)=\frac{\left(2\pi\right)^{2}}{L^{2}}\left(n+\frac{\phi_{\pm}}{\phi_{0}}\right)^{2}+\frac{1}{L^{2}}\varepsilon_{\perp}^{\pm}\label{eq:CHPCTh_EperpDef}
\end{equation}
where $\varepsilon_{\perp}^{\pm}$ stands in for all transverse eigenvalues
scaled by $1/L^{2}$. We can then write the density of states correlation
function as
\begin{align}
\left\langle \nu\left(\frac{\varepsilon+\hbar\omega}{2}+\varepsilon_{1},\phi\right)\nu\left(\frac{\varepsilon-\hbar\omega}{2}+\varepsilon_{1}^{\prime},\phi'\right)\right\rangle \hphantom{\text{Re}\left(\left(\frac{i\varepsilon_{1}-i\varepsilon_{1}^{\prime}}{\hbar}+\frac{\left(2\pi\right)^{2}D}{L^{2}}\left(n+\frac{\phi_{\pm}}{\phi_{0}}\right)^{2}+\frac{D\varepsilon_{\perp}}{L^{2}}\right)^{-2}\right)}\nonumber \\
=\frac{1}{2\pi^{2}\hbar^{2}}\sum_{\pm}\sum_{\varepsilon_{\perp}^{\pm}}\sum_{n}\text{Re}\left(\left(i\omega+i\left(\varepsilon_{1}-\varepsilon_{1}^{\prime}\right)+\frac{\left(2\pi\right)^{2}D}{L^{2}}\left(n+\frac{\phi_{\pm}}{\phi_{0}}\right)^{2}+\frac{D\varepsilon_{\perp}}{L^{2}}\right)^{-2}\right)\nonumber \\
=\frac{1}{2\pi^{2}}\sum_{\pm}\sum_{\varepsilon_{\perp}^{\pm}}\sum_{n}\text{Re}\left(\left(i\hbar\omega+i\left(\varepsilon_{1}-\varepsilon_{1}^{\prime}\right)+E_{c}\varepsilon_{\perp}+\left(2\pi\right)^{2}E_{c}\left(n+\frac{\phi_{\pm}}{\phi_{0}}\right)^{2}\right)^{-2}\right)\label{eq:CHPCTh_DOSCorrelation}
\end{align}
where we have introduced the energy scale 
\begin{equation}
E_{c}=\frac{\hbar D}{L^{2}}\label{eq:ChPCTh_ThoulessEnergy}
\end{equation}
known in the literature as the Thouless or correlation energy.%
\footnote{This energy scale is defined with various factors of 2 and $\pi$
by different authors.%
} Using the Poisson summation formula given in Eq. \ref{eq:AppMath_PoissonSummationFormula},
we can replace the sum over $n$ by a sum over $p$. Writing 
\begin{eqnarray*}
\alpha & = & i\hbar\omega+i\left(\varepsilon_{1}-\varepsilon_{1}^{\prime}\right)+E_{c}\varepsilon_{\perp},
\end{eqnarray*}
we have
\begin{align}
\sum_{n}\frac{1}{\left(\alpha+\left(2\pi\right)^{2}E_{c}\left(n+\frac{\phi_{\pm}}{\phi_{0}}\right)^{2}\right)^{2}} & =-\frac{\partial}{\partial\alpha}\sum_{n}\frac{1}{\alpha+4\pi^{2}E_{c}\left(n+\frac{\phi_{\pm}}{\phi_{0}}\right)^{2}}\nonumber \\
 & =-\frac{1}{2\sqrt{E_{c}}}\frac{\partial}{\partial\alpha}\sum_{p=-\infty}^{\infty}e^{2\pi ip\phi_{\pm}/\phi_{0}}\frac{1}{\sqrt{\alpha}}\int_{-\infty}^{\infty}d\phi\,\frac{1}{\pi}\frac{\frac{\phi_{0}\sqrt{\alpha}}{2\pi\sqrt{E_{c}}}e^{-2\pi ip\phi/\phi_{0}}}{\frac{\phi_{0}^{2}\alpha}{4\pi^{2}E_{c}}+\phi^{2}}\label{eq:CHPCTh_PoissonSumDiffusonCooperon}
\end{align}
where the last integral is now the Fourier transform of a Lorentzian.
Performing the Fourier transform gives
\begin{align}
\sum_{n}\frac{1}{\left(\alpha+\left(2\pi\right)^{2}E_{c}\left(n+\frac{\phi_{\pm}}{\phi_{0}}\right)^{2}\right)^{2}} & =-\frac{1}{2\sqrt{E_{c}}}\frac{\partial}{\partial\alpha}\sum_{p=-\infty}^{\infty}e^{2\pi ip\phi_{\pm}/\phi_{0}}\frac{1}{\sqrt{\alpha}}\exp\left(-\left|p\right|\sqrt{\frac{\alpha}{E_{c}}}\right)\nonumber \\
 & =\frac{1}{\sqrt{E_{c}}}\frac{i}{\hbar}\frac{\partial}{\partial\omega}\sum_{p=1}^{\infty}\cos\left(2\pi p\frac{\phi_{\pm}}{\phi_{0}}\right)\frac{1}{\sqrt{\alpha}}\exp\left(-p\sqrt{\frac{\alpha}{E_{c}}}\right)\label{eq:CHPCTh_DiffusonCooperonSum}
\end{align}
where in the last line we have prematurely dropped the $p=0$ term.
This term will drop out under the derivatives with respect to $\phi$
and $\phi'$ which will be performed when evaluating $C_{1}^{\left(0\right)}(\varepsilon_{1},\phi;\varepsilon_{1}^{\prime},\phi')$.

Putting the results of Eqs. \ref{eq:CHPCTh_DOSCorrelation} and \ref{eq:CHPCTh_DiffusonCooperonSum}
into Eq. \ref{eq:CHPCTh_CurrentCurrentCorrelation} for the current
current correlation function gives
\begin{align*}
 & C_{1}^{\left(0\right)}\left(\varepsilon_{1},\phi;\varepsilon_{1}^{\prime},\phi'\right)\\
 & \phantom{C_{1}^{\left(0\right)}}=\frac{\hbar}{8}\frac{\partial^{2}}{\partial\phi\partial\phi'}\int_{-\infty}^{0}d\varepsilon\int_{\varepsilon/\hbar}^{-\varepsilon/\hbar}d\omega\,\left(\varepsilon^{2}-\hbar^{2}\omega^{2}\right)\ldots\\
 & \phantom{C_{1}^{\left(0\right)}}\phantom{=}\ldots\times\frac{1}{2\pi^{2}}\sum_{\pm}\sum_{\varepsilon_{\perp}^{\pm}}\text{Re}\left(\frac{1}{\sqrt{E_{c}}}\frac{i}{\hbar}\frac{\partial}{\partial\omega}\sum_{p=1}^{\infty}\cos\left(2\pi p\frac{\phi_{\pm}}{\phi_{0}}\right)\frac{1}{\sqrt{\alpha}}\exp\left(-p\sqrt{\frac{\alpha}{E_{c}}}\right)\right)\\
 & \phantom{C_{1}^{\left(0\right)}}=\frac{1}{4}\sum_{p=1}^{\infty}\sum_{\pm}\sum_{\varepsilon_{\perp}^{\pm}}\left(\pm\frac{p^{2}}{\phi_{0}^{2}}\cos\left(2\pi p\frac{\phi_{\pm}}{\phi_{0}}\right)\right)\int_{-\infty}^{0}d\varepsilon\int_{\varepsilon/\hbar}^{-\varepsilon/\hbar}d\omega\,\left(\varepsilon^{2}-\hbar^{2}\omega^{2}\right)\ldots\\
 & \phantom{C_{1}^{\left(0\right)}}\phantom{=}\ldots\times\text{Im}\left(\frac{1}{\sqrt{E_{c}}}\frac{\partial}{\partial\omega}\frac{1}{\sqrt{\alpha}}\exp\left(-p\sqrt{\frac{\alpha}{E_{c}}}\right)\right)\\
 & \phantom{C_{1}^{\left(0\right)}}=\frac{\hbar^{2}}{2}\text{Im}\sum_{p=1}^{\infty}\sum_{\pm}\sum_{\varepsilon_{\perp}^{\pm}}\left(\pm\frac{p^{2}}{\phi_{0}^{2}}\cos\left(2\pi p\frac{\phi_{\pm}}{\phi_{0}}\right)\right)\int_{-\infty}^{0}d\varepsilon\int_{\varepsilon/\hbar}^{-\varepsilon/\hbar}d\omega\,\frac{\omega}{\sqrt{\alpha E_{c}}}\exp\left(-p\sqrt{\frac{\alpha}{E_{c}}}\right)\\
 & \phantom{C_{1}^{\left(0\right)}}=-\frac{\hbar^{2}}{2\sqrt{E_{c}}}\text{Im}\sum_{p=1}^{\infty}\sum_{\pm}\sum_{\varepsilon_{\perp}^{\pm}}\left(\pm\frac{p^{2}}{\phi_{0}^{2}}\cos\left(2\pi p\frac{\phi_{\pm}}{\phi_{0}}\right)\right)\int_{-\infty}^{0}d\varepsilon\int_{-\varepsilon/\hbar}^{\varepsilon/\hbar}d\omega\,\frac{\omega}{\sqrt{\alpha}}\exp\left(-p\sqrt{\frac{\alpha}{E_{c}}}\right)
\end{align*}
where in the next to last line we have used integration by parts on
$\omega$. To perform the final two integrals, we first use the change
of variables $x=\sqrt{\alpha}$ and the abbreviations $\sigma=i(\varepsilon_{1}-\varepsilon_{1}^{\prime})+E_{c}\varepsilon_{\perp}$
and $\kappa=p/\sqrt{E_{c}}$ to find
\begin{align*}
\int_{-\varepsilon/\hbar}^{\varepsilon/\hbar}d\omega\,\frac{\omega}{\sqrt{\alpha}}\exp\left(-p\sqrt{\frac{\alpha}{E_{c}}}\right) & =-\frac{2}{\hbar^{2}}\int_{\sqrt{-i\varepsilon+\sigma}}^{\sqrt{i\varepsilon+\sigma}}dx\,\left(x^{2}-\sigma\right)\exp\left(-\kappa x\right)\\
 & =-\frac{2}{\hbar^{2}}\int_{\sqrt{-i\varepsilon+\sigma}}^{\sqrt{i\varepsilon+\sigma}}dx\,\left(\frac{\partial^{2}}{\partial\kappa^{2}}-\sigma\right)\exp\left(-\kappa x\right)\\
 & =\frac{2}{\hbar^{2}}\left(\frac{\partial^{2}}{\partial\kappa^{2}}-\sigma\right)\sum_{\pm}\pm\frac{1}{\kappa}\exp\left(-\kappa\sqrt{\sigma\pm i\varepsilon}\right).
\end{align*}
The final integral over $\varepsilon$ involves only the last factor
of this expression. Using another change of variables $y=\sqrt{\sigma\pm i\varepsilon}$,
we have
\begin{align*}
\int_{-\infty}^{0}d\varepsilon\,\exp\left(-\kappa\sqrt{\sigma\pm i\varepsilon}\right) & =\int_{\sqrt{\sigma\mp i\left(\varepsilon_{1}+\varepsilon_{1}^{\prime}\right)}}^{\sqrt{\sigma}}dy\,\left(\mp2iy\right)\exp\left(-\kappa y\right)\\
 & =\pm2i\frac{\partial}{\partial\kappa}\int_{\sqrt{\sigma\mp i\infty}}^{\sqrt{\sigma}}dy\,\exp\left(-\kappa y\right)\\
 & \approx\mp2i\frac{\partial}{\partial\kappa}\left(\frac{1}{\kappa}\exp\left(-\kappa\sqrt{\sigma}\right)\right)
\end{align*}
where in the last line we dropped the term $\exp(-\kappa\sqrt{\sigma\mp i\infty})\rightarrow0$.
Because the derivatives of the Fermi-Dirac distribution function in
Eq. \ref{eq:ChPCTh_CurrentCurrentCorrelationDef} are peaked around
$\varepsilon_{F}\gg E_{c}$, this term will be proportional to approximately
$\exp(-\sqrt{2\varepsilon_{F}/\pi E_{c}})\approx0$. Together the
two integrals are
\begin{align*}
\int_{-\infty}^{0}d\varepsilon\int_{-\varepsilon/\hbar}^{\varepsilon/\hbar}d\omega\,\frac{\omega}{\sqrt{\alpha}}\exp\left(-\frac{p}{\pi}\sqrt{\frac{\alpha}{E_{c}}}\right) & =\frac{2}{\hbar^{2}}\left(\frac{\partial^{2}}{\partial\kappa^{2}}-\sigma\right)\sum_{\pm}\pm\frac{1}{\kappa}\left(\mp2i\frac{\partial}{\partial\kappa}\left(\frac{1}{\kappa}\exp\left(-\kappa\sqrt{\sigma}\right)\right)\right)\\
 & =-\frac{8i}{\hbar^{2}}\left(\frac{\partial^{2}}{\partial\kappa^{2}}-\sigma\right)\left(\frac{1}{\kappa}\frac{\partial}{\partial\kappa}\left(\frac{1}{\kappa}\exp\left(-\kappa\sqrt{\sigma}\right)\right)\right)\\
 & =-\frac{8i}{\hbar^{2}}\left(\frac{\partial^{2}}{\partial\kappa^{2}}-\sigma\right)\left(\left(-\frac{1}{\kappa^{3}}-\frac{\sqrt{\sigma}}{\kappa^{2}}\right)\exp\left(-\kappa\sqrt{\sigma}\right)\right)\\
 & =-\frac{8i}{\hbar^{2}}\left(\left(\frac{\sigma}{\kappa^{3}}+\frac{\sigma\sqrt{\sigma}}{\kappa^{2}}\right)\exp\left(-\kappa\sqrt{\sigma}\right)\right)\\
 & \phantom{=}-\frac{8i}{\hbar^{2}}\frac{\partial}{\partial\kappa}\left(\left(\frac{3}{\kappa^{4}}+\frac{\sqrt{\sigma}}{\kappa^{3}}+\frac{2\sqrt{\sigma}}{\kappa^{3}}+\frac{\sigma}{\kappa^{2}}\right)\exp\left(-\kappa\sqrt{\sigma}\right)\right)\\
 & =-\frac{8i}{\hbar^{2}}\exp\left(-\kappa\sqrt{\sigma}\right)\bigg(\frac{\sigma}{\kappa^{3}}+\frac{\sigma\sqrt{\sigma}}{\kappa^{2}}-\frac{12}{\kappa^{5}}-\frac{9\sqrt{\sigma}}{\kappa^{4}}-\frac{2\sigma}{\kappa^{3}}\\
 & \phantom{=-\frac{8i}{\hbar^{2}}\exp\left(-\kappa\sqrt{\sigma}\right)\bigg(}\ldots-\frac{3\sqrt{\sigma}}{\kappa^{4}}-\frac{3\sigma}{\kappa^{3}}-\frac{\sigma\sqrt{\sigma}}{\kappa^{2}}\bigg)\\
 & =-\frac{8i}{\hbar^{2}}\left(-\frac{4\sigma}{\kappa^{3}}-\frac{12\sqrt{\sigma}}{\kappa^{4}}-\frac{12}{\kappa^{5}}\right)\exp\left(-\kappa\sqrt{\sigma}\right).
\end{align*}
With this result for the integrals, we can write down a final expression
for the zero temperature current autocorrelation function:
\begin{align}
 & C_{1}^{\left(0\right)}\left(\varepsilon_{1},\phi;\varepsilon_{1}^{\prime},\phi'\right)\nonumber \\
 & \phantom{C_{1}^{\left(0\right)}}=\frac{\hbar^{2}}{2\sqrt{E_{c}}}\text{Im}\sum_{p=1}^{\infty}\sum_{\pm}\left(\pm\frac{p^{2}}{\phi_{0}^{2}}\cos\left(2\pi p\frac{\phi_{\pm}}{\phi_{0}}\right)\right)\frac{8i}{\hbar^{2}}\left(-\frac{4\sigma}{\kappa^{3}}-\frac{12\sqrt{\sigma}}{\kappa^{4}}-\frac{12}{\kappa^{5}}\right)\exp\left(-\kappa\sqrt{\sigma}\right)\nonumber \\
 & \phantom{C_{1}^{\left(0\right)}}=-16\frac{E_{c}^{2}}{\phi_{0}^{2}}\text{Re}\sum_{p=1}^{\infty}\sum_{\pm}\left(\pm\cos\left(2\pi p\frac{\phi_{\pm}}{\phi_{0}}\right)\right)\left(\frac{3}{p^{3}}+\frac{3\sqrt{z}}{p^{2}}+\frac{z}{p}\right)\exp\left(-p\sqrt{z}\right)\label{eq:CHPCTh_CurrCurrCorT0Simple}
\end{align}
where we have introduced the notation 
\begin{equation}
z=\varepsilon_{\perp}+\frac{i\left(\varepsilon_{1}-\varepsilon_{1}^{\prime}\right)}{E_{c}}.\label{eq:CHPCTh_z}
\end{equation}
For convenience, we define the function 
\begin{equation}
F_{p}\left(z\right)=\text{Re}\left[\left(\frac{3}{p^{3}}+\frac{3\sqrt{z}}{p^{2}}+\frac{z}{p}\right)\exp\left(-p\sqrt{z}\right)\right]\label{eq:CHPCTh_Fp}
\end{equation}
which allows us to write the zero temperature correlation function
as
\begin{align}
C_{1}^{\left(0\right)}\left(\varepsilon_{1},\phi;\varepsilon_{1}^{\prime},\phi'\right) & =-16\frac{E_{c}^{2}}{\phi_{0}^{2}}\sum_{\varepsilon_{\perp}^{\pm}}\sum_{p=1}^{\infty}\pm F_{p}\left(z\right)\cos\left(2\pi p\frac{\phi\pm\phi'}{\phi_{0}}\right)\label{eq:CHPCTh_CurrCurrCosCos}
\end{align}

We note that, with the definition of $\varepsilon_{\perp}$ in Eq.
\ref{eq:CHPCTh_EperpDef}, the exponential in $F_{p}(z)$ takes the
form $\exp(-p\sqrt{\frac{L^{2}\pi^{2}n_{r}^{2}}{2w^{2}}+\frac{L^{2}\pi^{2}n_{z}^{2}}{2t^{2}}+i\delta})$
where $\delta$ is a purely real number. For a high aspect ratio ring
$L\gg w,t$, this exponential is negligible for $n_{r},n_{z}>0$.
Thus we can discard these higher order transverse terms and use $\varepsilon_{\perp}$
to signify the first term in the sum with $n_{r}=n_{z}=0$. With the
system we have considered so far, this first transverse energy term
$\varepsilon_{\perp}$ is equal to zero. Despite this fact, we do
not drop the transverse energy $\varepsilon_{\perp}$ term from our
derivations. In Section \ref{sub:ChPCTh_DiffusiveRefinements}, we
will consider an effect, magnetic flux penetrating the metal of the
ring, which leads to a non-zero value for $\varepsilon_{\perp}$.
With this effect, $\varepsilon_{\perp}$ takes different values for
the diffuson and cooperon, so we also retain the $\pm$ notation to
distinguish these two terms where necessary. When only $\varepsilon_{\perp}=0$
is significant, we can write
\begin{align*}
\left\langle I\left(\varepsilon_{1},\phi\right)I\left(\varepsilon_{1}^{\prime},\phi'\right)\right\rangle  & =C_{1}^{\left(0\right)}\left(\varepsilon_{1},\phi;\varepsilon_{1}^{\prime},\phi'\right)\\
 & =32\frac{E_{c}^{2}}{\phi_{0}^{2}}\sum_{\varepsilon_{\perp}^{\pm}}\sum_{p=1}^{\infty}F_{p}\left(z\right)\left(\sin\left(2\pi p\frac{\phi}{\phi_{0}}\right)\sin\left(2\pi p\frac{\phi'}{\phi_{0}}\right)\right).
\end{align*}
As noted at the beginning of this section, setting $\varepsilon_{1}=\varepsilon_{1}^{\prime}=\varepsilon_{F}$
gives the zero temperature current-current correlation function
\begin{align*}
\left\langle I\left(\phi\right)I\left(\phi'\right)\right\rangle _{T=0} & =C_{1}^{\left(0\right)}\left(\varepsilon_{F},\phi;\varepsilon_{F},\phi'\right)\\
 & =96\frac{E_{c}^{2}}{\phi_{0}^{2}}\sum_{p=1}^{\infty}\frac{1}{p^{3}}\sin\left(2\pi p\frac{\phi}{\phi_{0}}\right)\sin\left(2\pi p\frac{\phi'}{\phi_{0}}\right)
\end{align*}
or in terms of the ring parameters
\[
\left\langle I\left(\phi\right)I\left(\phi'\right)\right\rangle _{T=0}=\left(1.11\frac{eD}{L^{2}}\right)^{2}\sum_{p=1}^{\infty}\frac{2}{p^{3}}\sin\left(2\pi p\frac{\phi}{\phi_{0}}\right)\sin\left(2\pi p\frac{\phi'}{\phi_{0}}\right)
\]
where the numerical factor is $2\sqrt{3}/\pi=1.11$. We also define
the typical magnitude $I_{p}^{\text{typ}}$ of the $p^{th}$ harmonic
(per spin) by
\begin{align}
I_{p}^{\text{typ}} & =\frac{4\sqrt{3}}{p^{1.5}}\frac{E_{c}}{\phi_{0}}\nonumber \\
 & =\frac{1.11}{p^{1.5}}\frac{eD}{L^{2}}\label{eq:CHPCTh_IpTyp}
\end{align}
so that
\begin{equation}
\left\langle I\left(\phi\right)I\left(\phi'\right)\right\rangle _{T=0}=\sum_{p=1}^{\infty}2\left(I_{p}^{\text{typ}}\right)^{2}\sin\left(2\pi p\frac{\phi}{\phi_{0}}\right)\sin\left(2\pi p\frac{\phi'}{\phi_{0}}\right).\label{eq:CHPCTh_IpSinSinDef}
\end{equation}

So far we have neglected the spin of the electron. Assuming the two
spin states are degenerate, we simply multiply $\nu(\varepsilon)$
by a factor of 2 which results in a factor of 4 for $C_{1}^{(0)}$.
We will neglect this factor of 2 for spin degeneracy until considering
spin effects more closely in Sections \ref{sub:CHPCTh_Zeeman} and
\ref{sub:CHPCTh_Zeeman}.

For convenience in plotting normalized curves, we define the normalized
correlation function $H_{1}^{(0)}(x)$ as
\begin{equation}
H_{1}^{\left(0\right)}\left(x\right)=\text{Re}\left[\left(1+\sqrt{x}+\frac{x}{3}\right)\exp\left(-\sqrt{x}\right)\right]\label{eq:CHPCTh_H1Def}
\end{equation}
so that the current-current correlation function $C_{1}^{(0)}(\varepsilon_{1},\phi;\varepsilon_{1}^{\prime},\phi')$
can be expanded as
\begin{align}
\left\langle I\left(\varepsilon_{1},\phi\right)I\left(\varepsilon_{1}^{\prime},\phi'\right)\right\rangle  & =C_{1}^{\left(0\right)}\left(\varepsilon_{1},\phi;\varepsilon_{1}^{\prime},\phi'\right)\nonumber \\
 & =3\times32\frac{E_{c}^{2}}{\phi_{0}^{2}}\sum_{p}\frac{H_{1}^{\left(0\right)}\left(p^{2}z\right)}{p^{3}}\sin\left(2\pi p\frac{\phi}{\phi_{0}}\right)\sin\left(2\pi p\frac{\phi'}{\phi_{0}}\right)\\
 & =\sum_{p=1}^{\infty}2\left(I_{p}^{\text{typ}}\right)^{2}H_{1}^{\left(0\right)}\left(p^{2}z\right)\sin\left(2\pi p\frac{\phi}{\phi_{0}}\right)\sin\left(2\pi p\frac{\phi'}{\phi_{0}}\right).\label{eq:CHPCTh_C1H1}
\end{align}
The function $H_{1}^{(0)}(p\sqrt{z})$ gives the normalized autocorrelation
function of the harmonics of the current at the energy difference
$\varepsilon_{1}-\varepsilon_{1}^{\prime}$. That is,
\[
H_{1}^{\left(0\right)}\left(p\sqrt{\varepsilon_{\perp}+i\frac{\delta\varepsilon}{E_{c}}}\right)=\frac{\left\langle I_{p}\left(\varepsilon_{1}\right)I_{p}\left(\varepsilon_{1}+\delta\varepsilon\right)\right\rangle }{\left\langle \left(I_{p}\left(\varepsilon_{1}\right)\right)^{2}\right\rangle }.
\]
\begin{figure}
\begin{centering}
\includegraphics[width=0.6\paperwidth]{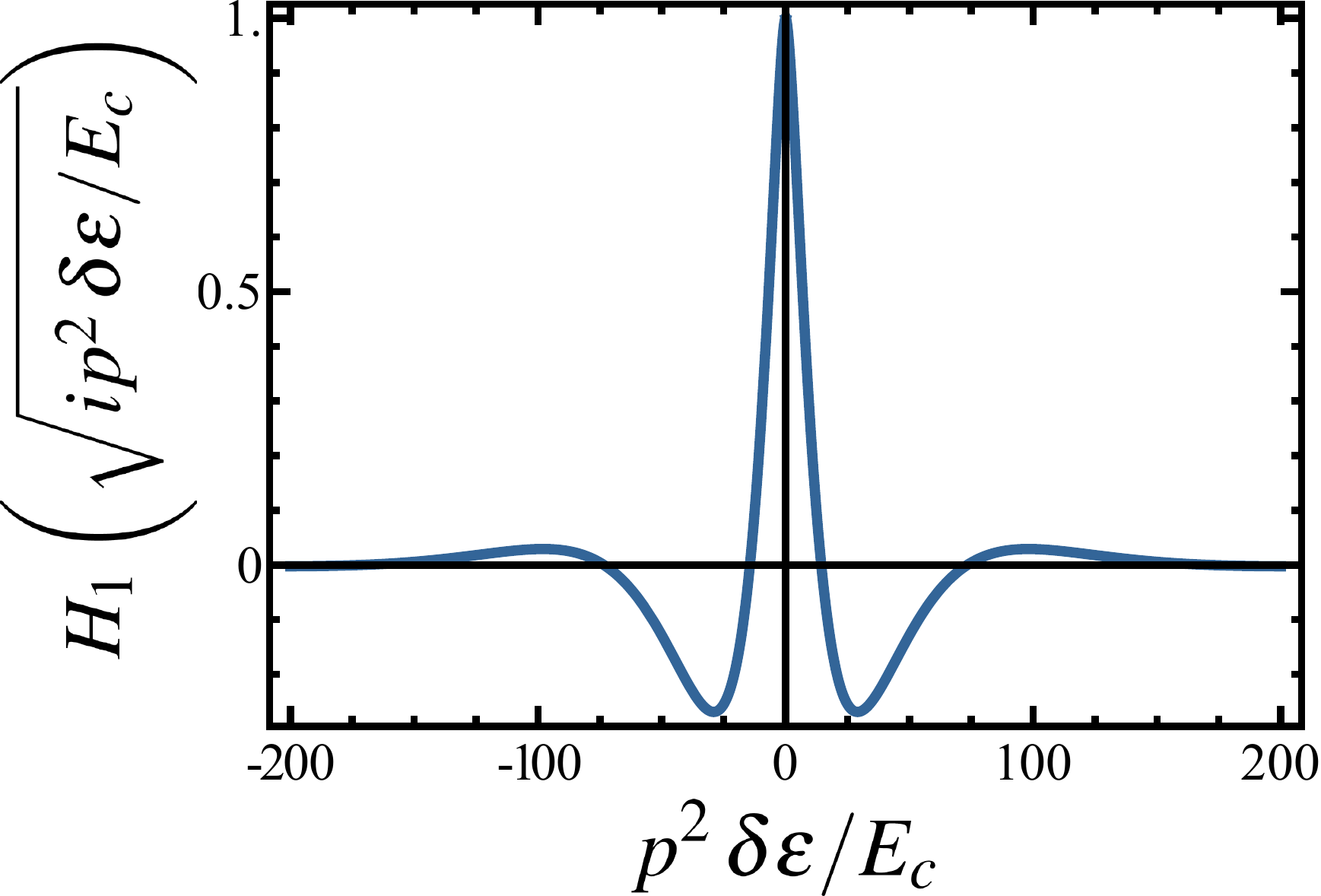}
\par\end{centering}

\caption[Current-current harmonic correlation function $H_{1}^{(0)}$ versus
energy difference]{\label{fig:CHPCTh_H1CurrCurrCor}Current-current harmonic correlation
function $H_{1}^{(0)}$ versus energy difference. The $p^{th}$ harmonic
of the total current is seen to be correlated over an energy range
$\delta\varepsilon\sim14E_{c}/p^{2}$. The correlation function shows
a sizable amount of anti-correlation before decaying significantly. }
\end{figure}
In Fig. \ref{fig:CHPCTh_H1CurrCurrCor}, we plot $H_{1}^{(0)}(\sqrt{ip^{2}\,\delta\varepsilon/E_{c}})$
versus $p^{2}\,\delta\varepsilon/E_{c}$. The figure shows that the
amplitude of the $p^{th}$ harmonic of the total current is correlated
over an energy range $\sim\pm14E_{c}/p^{2}$. Interestingly, it can
also be seen that the harmonics of the current possess some anti-correlation,
reminiscent of the perfect anti-correlation of the single level currents
of the ideal one-dimensional ring. For more discussion of $H_{1}^{(0)}$
see Ref. \citealp{riedel1993mesoscopic},%
\footnote{While mentioning this paper, I would also like to point out its discussion
of the sensitivity of the persistent current to the location of impurities.
It is argued that in the diffusive regime the autocorrelation function
of the current as a function of the location of a single scattering
center (with the rest of the disorder potential held fixed) decays
exponentially on the length scale $k_{F}^{-1}$. In other words, the
current is completely randomized when one scattering center within
the ring is moved by a distance of one Fermi wavelength.%
} where it is argued that an increase $\delta\varepsilon$ in the Fermi
energy equal to the single level spacing $\Delta_{M}$ is equivalent
to an increase in the number of electrons in the ring by 1. Among
the results of this argument is the conclusion that, for a large number
$M_{\text{eff}}$ of correlated levels and low harmonic index $p$,
the typical single level current harmonic magnitude $i_{p}^{\text{typ}}=\sqrt{\langle i_{p}^{2}\rangle}=p^{-1/2}\sqrt{2}\Delta_{M}/\phi_{0}$
scales with the single level spacing $\Delta_{M}$. This result is
reasonable given the conclusions drawn above that the single level
current is proportional to the slope of that level with respect to
flux and that adjacent levels are repelled from each other. That the
total current is larger than this figure reinforces the interpretation
given at the beginning of this section that a certain number of levels
are correlated. By comparing the single level current and the total
current
\[
\frac{I_{p}^{\text{typ}}}{i_{p}^{\text{typ}}}=\frac{8\sqrt{3}}{\sqrt{2}p}\frac{E_{c}}{\Delta_{M}},
\]
we can estimate that the number $M_{\text{eff}}$ of correlated levels
correlated for the $p^{th}$ harmonic is 
\begin{equation}
M_{\text{eff}}\sim10E_{c}/p\Delta_{M}.\label{eq:CHPCTh_Meff}
\end{equation}

With $D=v_{F}l_{e}/3$, we can see that the typical current magnitude
is approximately the single channel perfect ring current $I_{0}$
(Eq. \ref{eq:CHPCTh_I0}) reduced by the factor $l_{e}/L$. From the
sinusoidal factors, it can be seen that, although each harmonic has
a random sign, its phase with respect to flux is well defined. The
$p^{th}$ harmonic of the current always is zero for $\phi=N\phi_{0}/2p$
with $N$ an integer.

\FloatBarrier

\subsubsection{\label{sub:CHPCTh_TypicalIFiniteT}Finite temperature}

\begin{figure}
\begin{centering}
\includegraphics[width=0.5\paperwidth]{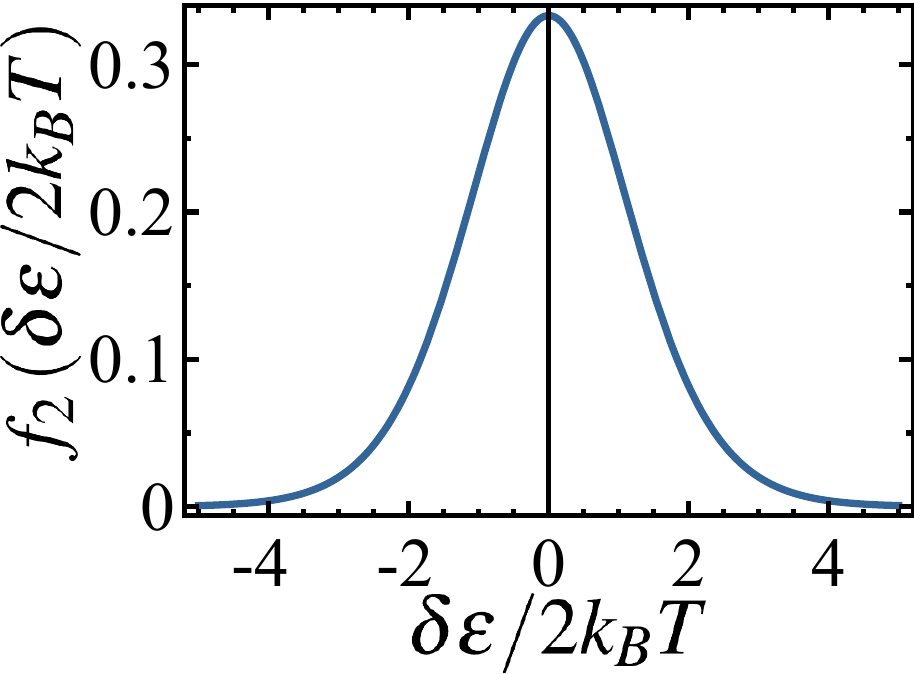}
\par\end{centering}

\caption[Temperature weighting function $f_{2}$ plotted against energy difference
$\delta\varepsilon$]{\label{fig:CHPCTh_f2TemperatureFunc}Temperature weighting function
$f_{2}$ plotted against energy difference $\delta\varepsilon$. The
figure shows the function $f_{2}$ of Eq. \ref{eq:CHPCTh_f2TemperatureFunc}
which provides the weighting given to the current-current correlation
function $H_{1}^{(0)}(ip^{2}\,\delta\varepsilon/E_{c})$ (Fig. \ref{fig:CHPCTh_H1CurrCurrCor})
in the integral of Eq. \ref{eq:CHPCTh_gDintegral} for the suppression
of $I_{p}^{\text{typ}}$ due to finite temperature (with $\varepsilon_{\perp}=0$).
As temperature increases, $f_{2}(\delta\varepsilon/2k_{B}T)$ becomes
broader in $\delta\varepsilon$ and less weight is given to the central
peak of $H_{1}^{(0)}(ip^{2}\,\delta\varepsilon/E_{c})$.}
\end{figure}

\begin{figure}
\begin{centering}
\includegraphics[width=0.5\paperwidth]{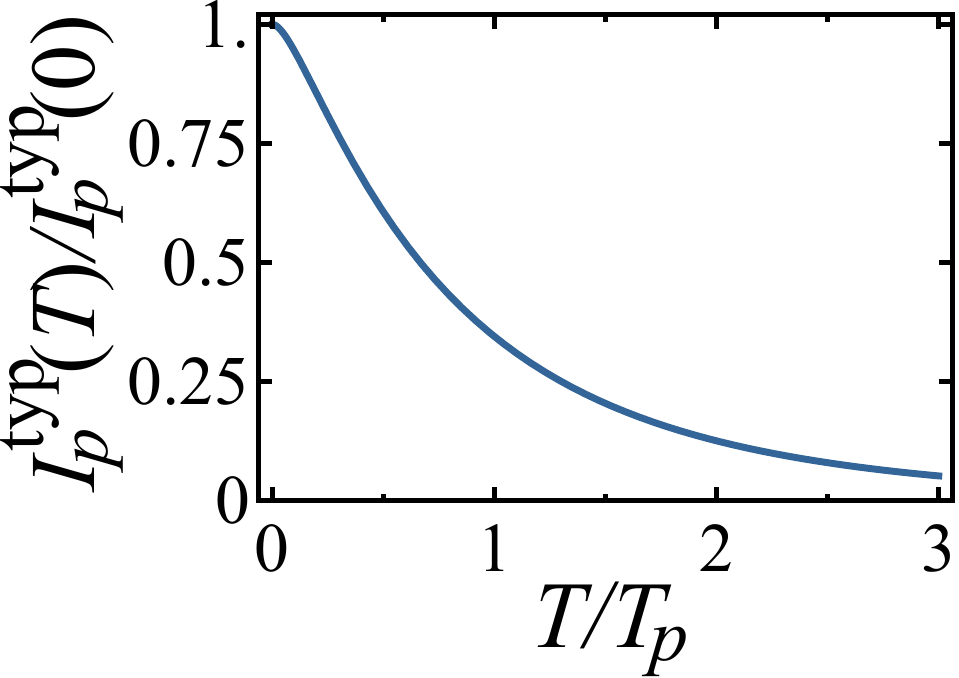}
\par\end{centering}

\caption[Temperature dependence of the persistent current in the diffusive
regime]{\label{fig:CHPCTh_IpTIpT0DiffusiveSimple}Temperature dependence
of the persistent current in the diffusive regime. The typical magnitude
$I_{p}^{\text{typ}}(T)$ of the $p^{th}$ harmonic of the current,
given in Eq. \ref{eq:CHPCTh_IpTypTBothSpins}, is shown as a function
of temperature. The vertical axis is normalized by the magnitude $I_{p}^{\text{typ}}\left(0\right)$
of the current at zero temperature, while the horizontal axis is normalized
by the characteristic temperature $T_{p}$ defined in Eq. \ref{eq:CHPCTh_TpDiffusive}.}
\end{figure}

We now evaluate the integrals of Eq. \ref{eq:ChPCTh_CurrentCurrentCorrelationDef}
in order to find the temperature dependence of the current-current
correlation function. From Eq. \ref{eq:CHPCTh_CurrCurrCorT0Simple},
we can see that $C_{1}^{(0)}(\varepsilon_{1},\phi;\varepsilon_{1}^{\prime},\phi')$
depends only on the difference $\varepsilon_{1}-\varepsilon_{1}^{\prime}$.
This dependence motivates using the change of variables of Eq. \ref{eq:CHPCTh_CurrentCurrentCorrelation}
on the integrals of Eq. \ref{eq:ChPCTh_CurrentCurrentCorrelationDef}.
Since $f'(\varepsilon,\varepsilon_{F},T)$ is well localized around
$\varepsilon_{F}\gg E_{c}$ and $C_{1}^{(0)}(\varepsilon_{1},\phi;\varepsilon_{1}^{\prime},\phi')$
decays for $|\varepsilon_{1}-\varepsilon_{1}^{\prime}|\gg E_{c}$,
we extend the lower bound of the $\varepsilon_{1}$ and $\varepsilon_{1}^{\prime}$
integrals to $-\infty$. Using Eq. \ref{eq:CHPCTh_ThermalAvgFunctionY}
for the form of $f'(\varepsilon,\varepsilon_{F},T)$, we can then
write
\begin{align*}
\left\langle I\left(\phi\right)I\left(\phi'\right)\right\rangle  & =\int_{0}^{\infty}d\varepsilon_{1}\int_{0}^{\infty}d\varepsilon_{1}^{\prime}\,\left(f'\left(\varepsilon_{1}\right)f'\left(\varepsilon_{1}^{\prime}\right)\right)C_{1}^{\left(0\right)}\left(\varepsilon_{1},\phi;\varepsilon_{1}^{\prime},\phi'\right)\\
 & \approx\int_{-\infty}^{\infty}d\varepsilon_{1}d\varepsilon_{1}^{\prime}\,\left(\frac{1}{4k_{B}T}\right)^{2}\text{sech}^{2}\left(\frac{\left(\varepsilon_{1}-\varepsilon_{F}\right)}{2k_{B}T}\right)\text{sech}^{2}\left(\frac{\left(\varepsilon_{1}^{\prime}-\varepsilon_{F}\right)}{2k_{B}T}\right)C_{1}^{\left(0\right)}\left(\varepsilon_{1},\phi;\varepsilon_{1}^{\prime},\phi'\right)\\
 & \approx\left(\frac{1}{4k_{B}T}\right)^{2}\int_{-\infty}^{\infty}d\varepsilon_{1}d\varepsilon_{1}^{\prime}\,\text{sech}^{2}\left(\frac{\varepsilon_{1}}{2k_{B}T}\right)\text{sech}^{2}\left(\frac{\varepsilon_{1}^{\prime}}{2k_{B}T}\right)C_{1}^{\left(0\right)}\left(\varepsilon_{1},\phi;\varepsilon_{1}^{\prime},\phi'\right)\\
 & =\frac{1}{4}\int_{-\infty}^{\infty}d\varepsilon\, C_{1}^{\left(0\right)}\left(\varepsilon_{1},\phi;\varepsilon_{1}^{\prime},\phi'\right)\int_{-\infty}^{\infty}d\sigma\,\text{sech}^{2}\left(\sigma+\frac{\varepsilon}{2}\right)\text{sech}^{2}\left(\sigma-\frac{\varepsilon}{2}\right)
\end{align*}
where we have used $\varepsilon=(\varepsilon_{1}-\varepsilon_{1}^{\prime})/2k_{B}T$
and $\sigma=(\varepsilon_{1}+\varepsilon_{1}^{\prime})/4k_{B}T$.
Using the expression for the integral over $\sigma$ given in Eq.
\ref{eq:AppMath_SechSechSigmaIntegral}, we find
\[
\left\langle I\left(\phi\right)I\left(\phi'\right)\right\rangle =\int_{-\infty}^{\infty}d\varepsilon\, f_{2}\left(\varepsilon\right)C_{1}^{\left(0\right)}\left(\varepsilon_{1},\phi;\varepsilon_{1}^{\prime},\phi'\right).
\]
with the thermal weighting function (see Fig. \ref{fig:CHPCTh_f2TemperatureFunc})
\begin{equation}
f_{2}\left(\varepsilon\right)=\frac{\varepsilon\cosh\varepsilon-\sinh\varepsilon}{\sinh^{3}\varepsilon}.\label{eq:CHPCTh_f2TemperatureFunc}
\end{equation}
Using Eq. \ref{eq:CHPCTh_CurrCurrCorT0Simple}, we can write the current-current
correlation function in terms of its harmonics as
\begin{equation}
\left\langle I\left(\phi\right)I\left(\phi'\right)\right\rangle =\sum_{p=1}^{\infty}\left(I_{p}^{\text{typ}}\right)^{2}\sum_{\pm}\left(\mp\cos\left(2\pi p\frac{\phi_{\pm}}{\phi_{0}}\right)\right)g_{D}\left(p^{2}\varepsilon_{\perp},20.8\frac{T}{T_{p}}\right)\label{eq:CHPCTh_CurrCurrCorTempDependence}
\end{equation}
where the normalized temperature dependence (see Fig. \ref{fig:CHPCTh_IpTIpT0DiffusiveSimple})
is 
\begin{equation}
g_{D}\left(x,y\right)=\int_{-\infty}^{\infty}d\varepsilon\, f_{2}\left(\varepsilon\right)H_{1}^{\left(0\right)}\left(x+iy\varepsilon\right)\label{eq:CHPCTh_gDintegral}
\end{equation}
and the characteristic temperature of decay is
\begin{align}
T_{p} & =\frac{10.4}{k_{B}}\frac{E_{c}}{p^{2}}\nonumber \\
 & =\frac{10.4}{k_{B}}\frac{\hbar D}{p^{2}L^{2}}.\label{eq:CHPCTh_TpDiffusive}
\end{align}

We can also define a temperature dependent form for the typical magnitude
$I_{p}^{\text{typ}}$ of the persistent current harmonics (keeping
only $\varepsilon_{\perp}=0$)
\begin{align}
I_{p}^{\text{typ}}\left(T\right) & =I_{p}^{\text{typ}}\left(T=0\right)\sqrt{g_{D}\left(0,20.8\frac{T}{T_{p}}\right)}\nonumber \\
 & =\frac{1.11}{p^{1.5}}\frac{eD}{L^{2}}\sqrt{g_{D}\left(0,20.8\frac{T}{T_{p}}\right)}.\label{eq:CHPCTh_IpTypTBothSpins}
\end{align}
As discussed in Section \ref{sec:AppMath_gDsumExp}, the function
$g_{D}(x,y)$ can also be expressed as the sum
\[
g_{D}\left(x,y\right)=\frac{\pi^{2}y^{2}}{12}\sum_{N=1}^{\infty}N\text{Re}\left(\exp\left(-\sqrt{x+\pi Ny}\right)\right).
\]
When $x=0$, this sum takes the simpler form 
\[
g_{D}\left(0,y\right)=\frac{\pi^{2}y^{2}}{12}\sum_{N=1}^{\infty}N\exp\left(-\sqrt{\pi Ny}\right).
\]
Over the experimentally relevant range of $y<50$, the temperature
dependence is roughly exponential (see Section \ref{sec:AppMath_gDsumExp})
with
\begin{equation}
g_{D}\left(0,y\right)\approx\exp\left(-0.096y\right),\label{eq:CHPCTh_gDExponential}
\end{equation}
meaning that the typical magnitude $I_{p}^{\text{typ}}$ decays exponentially
on the temperature scale $\sim T_{p}$. 

\FloatBarrier

\subsection{\label{sub:ChPCTh_DiffusiveRefinements}Refinements to the calculation
of the typical current}

In the previous section, we investigated the persistent current in
a ring threaded by an idealized Aharonov-Bohm flux. In practice, our
measurements were performed at high magnetic field where the flux
through the metal can not be ignored. The flux through the metal affects
the persistent current in two ways. First, it modifies the vector
potential $\boldsymbol{A}_{\pm}$ of Eq. \ref{eq:CHPCTh_DiffusonCooperonEigenvalues}
and so changes the eigenvalues $E_{n}^{d,c}(B_{\pm})$. The major
result of this modification is the suppression of the cooperon contribution
to $\langle I(\phi)I(\phi')\rangle$. Additionally, the Zeeman effect
lifts the spin degeneracy of the electrons. Spin degeneracy is also
lifted by spin orbit scattering, which is non-negligible in the samples
we measured (see Appendix \ref{cha:AppTransport_}). All of these
effects were considered recently in Ref. \citealp{ginossar2010mesoscopic}.

\subsubsection{\label{sub:CHPCTh_FluxThroughMetal}Flux through the metal of the
ring}

In Eq. \ref{eq:CHPCTh_IpSinSinDef}, we wrote the current-current
correlation function as a sum over harmonics. This form was a rearrangement
of a previous form (see e.g. Eq. \ref{eq:CHPCTh_CurrCurrCosCos})
involving the sum and difference of fluxes. These two forms are
\begin{align}
\left\langle I\left(\phi\right)I\left(\phi'\right)\right\rangle  & =\sum_{p=1}^{\infty}\left\langle I_{p}^{2}\right\rangle \left(\cos\left(2\pi\frac{\phi-\phi'}{\phi_{0}}\right)-\cos\left(2\pi\frac{\phi+\phi'}{\phi_{0}}\right)\right)\label{eq:CHPCTh_CurrCurrPhiForms}\\
 & =\sum_{p=1}^{\infty}2\left\langle I_{p}^{2}\right\rangle \sin\left(2\pi\frac{\phi}{\phi_{0}}\right)\sin\left(2\pi\frac{\phi'}{\phi_{0}}\right).\nonumber 
\end{align}
Looking back over the derivation of Section \ref{sub:CHPCTh_ZeroTDiffusiveCurrent},
the term involving $\phi_{+}=\phi+\phi'$ can be linked to the cooperon,
while the term involving $\phi_{-}=\phi-\phi'$ is due to the diffuson.
For $\phi=\phi'$, we have 
\[
\left\langle I^{2}\left(\phi\right)\right\rangle =\sum_{p}\left\langle I_{p}^{2}\right\rangle \left(1-\cos\left(2\pi\frac{2\phi}{\phi_{0}}\right)\right),
\]
from which it can be seen that the diffuson term provides a constant
contribution to typical magnitude of each harmonic while the cooperon
contribution oscillates with period $\phi_{0}/2$. The role of the
cooperon is to provide a definite phase reference for the oscillations
of $I(\phi)$ with respect to $\phi$. Because of the cooperon, the
average variance of the current $\langle I^{2}(N\phi_{0}/2)\rangle$
is zero at flux values $\phi=N\phi_{0}/2$ for all integer $N$. Consider
the quantity 
\[
\overline{\left\langle I^{2}\right\rangle }=\frac{1}{\phi_{0}}\int_{0}^{\phi_{0}}d\phi\,\left\langle I^{2}\left(\phi\right)\right\rangle ,
\]
where the bar $\overline{\ldots\vphantom{I}}$ denotes that the variance
of the current is averaged over one period $\phi_{0}$. The contribution
of the $p^{th}$ harmonic to $\overline{\langle I^{2}\rangle}$ can
be broken up into the contribution from the diffuson
\begin{align*}
\overline{\left\langle I_{p}^{2}\right\rangle }_{d} & =\frac{1}{\phi_{0}}\int_{0}^{\phi_{0}}d\phi\,\left\langle I_{p}^{2}\right\rangle \\
 & =\left\langle I_{p}^{2}\right\rangle 
\end{align*}
and the cooperon
\begin{align*}
\overline{\left\langle I_{p}^{2}\right\rangle }_{c} & =\frac{1}{\phi_{0}}\int_{0}^{\phi_{0}}d\phi\,\left\langle I_{p}^{2}\right\rangle \cos\left(2\pi\frac{2\phi}{\phi_{0}}\right)\\
 & =0.
\end{align*}
With this breakdown of the contributions to $\overline{\langle I^{2}\rangle}$,
it can be seen that all of the magnitude of the variance of the current
$\overline{\langle I^{2}\rangle}$ is due to the diffuson, while the
role of the cooperon is solely to modify the flux dependence of the
variance of the current.

We note that the suppression of the cooperon is consistent with the
randomization of the phases of the persistent current harmonics. We
can write the persistent current in the form
\begin{align*}
I\left(\phi\right) & =\sum_{p}\sqrt{2}I_{p}\sin\left(2\pi p\frac{\phi}{\phi_{0}}+\alpha_{p}\right)\\
 & =\sum_{p}\sqrt{2}I_{p}\left(\cos\left(\alpha_{p}\right)\sin\left(2\pi p\frac{\phi}{\phi_{0}}\right)+\sin\left(\alpha_{p}\right)\cos\left(2\pi p\frac{\phi}{\phi_{0}}\right)\right),
\end{align*}
where with the cooperon unsuppressed all phases $\alpha_{p}$ must
be zero and the factor of $\sqrt{2}$ is included so that $I_{p}$
is the same quantity as in Eq. \ref{eq:CHPCTh_CurrCurrPhiForms}.
The amplitude $I_{p}$ and phase $\alpha_{p}$ are assumed to vary
independently with disorder configuration and to be independent from
harmonic to harmonic:
\begin{align*}
\left\langle I_{p}I_{p'}\right\rangle  & =\left\langle I_{p}^{2}\right\rangle \delta_{pp'}\\
\left\langle \sin\left(\alpha_{p}\right)\sin\left(\alpha_{p'}\right)\right\rangle  & =\frac{\delta_{pp'}}{2}\\
\left\langle \cos\left(\alpha_{p}\right)\cos\left(\alpha_{p'}\right)\right\rangle  & =\frac{\delta_{pp'}}{2}\\
\left\langle \sin\left(\alpha_{p}\right)\cos\left(\alpha_{p'}\right)\right\rangle  & =0\\
\left\langle I_{p}f\left(\alpha_{p'}\right)\right\rangle  & =\left\langle I_{p}\right\rangle \left\langle f\left(\alpha_{p'}\right)\right\rangle 
\end{align*}
where $f$ is any function. With these assumptions, we can write
\begin{align*}
\left\langle I\left(\phi\right)I\left(\phi'\right)\right\rangle  & =\sum_{p}2\left\langle I_{p}^{2}\right\rangle \left(\frac{1}{2}\sin\left(2\pi p\frac{\phi}{\phi_{0}}\right)\sin\left(2\pi p\frac{\phi'}{\phi_{0}}\right)+\frac{1}{2}\cos\left(2\pi p\frac{\phi}{\phi_{0}}\right)\cos\left(2\pi p\frac{\phi'}{\phi_{0}}\right)\right)\\
 & =\sum_{p}\left\langle I_{p}^{2}\right\rangle \cos\left(2\pi p\frac{\phi-\phi'}{\phi_{0}}\right).
\end{align*}
This final expression matches Eq. \ref{eq:CHPCTh_CurrCurrPhiForms}
with the $\phi_{+}=\phi+\phi'$ term associated with the cooperon
dropped. We now consider the effects of magnetic flux penetrating
the metal of the ring, which leads to such a suppression of the cooperon.

In the experimental arrangement discussed in this text, the metal
ring is subjected to a uniform field applied at an angle relative
to the plane of the ring. Ideally, we would like to decompose the
field $\boldsymbol{B}=\boldsymbol{B}_{\phi}+\boldsymbol{B}_{M}$ into
the fields $\boldsymbol{B}_{\phi}$ threading flux through the ring
and $\boldsymbol{B}_{M}$ penetrating the metal, so that we can use
the results of the previous section with $\boldsymbol{B}_{\phi}$
leading to the dependence on $\phi$ in the results of that section
and $\boldsymbol{B}_{M}$ providing perturbative corrections to that
result. Such an approach is complicated, however, by the boundary
conditions given in Eq. \ref{eq:CHPCTh_DiffusonCooperonBoundaryCondition}
for the diffuson and cooperon. In principle, $\boldsymbol{B}_{M}$,
and thus the corresponding vector potential $\boldsymbol{A}_{M}$,
varies in a non-uniform manner throughout the ring. Because of the
boundary condition's dependence on the vector potential $\boldsymbol{A}$,
the diffuson and cooperon eigenfunctions depend on the perturbing
field $\boldsymbol{B}_{M}$, complicating the simple perturbation
theory approach.

In Ref. \citealp{ginossar2010mesoscopic}, the complications associated
with an arbitrary $\boldsymbol{B}_{M}$ are circumvented by choosing
a particular form of $\boldsymbol{B}_{M}$, the toroidal field, for
which $\boldsymbol{A}_{M}\cdot\tilde{\boldsymbol{n}}=0$ and thus
for which the boundary condition of Eq. \ref{eq:CHPCTh_DiffusonCooperonBoundaryCondition}
is independent of $\boldsymbol{B}_{M}$. The toroidal field is specified
by $\boldsymbol{B}_{M}=B_{M}\tilde{\boldsymbol{\theta}}$ within the
ring. While such a form for $\boldsymbol{B}_{M}$ is geometrically
quite distinct from the experimental arrangement, it does allow an
analytical solution for the current-current correlation function for
the case of a magnetic field inside the ring. As is argued in Ref.
\citealp{ginossar2010mesoscopic}, this analytical solution differs
from the solution for the field $\boldsymbol{B}_{M,\text{exp}}$ appropriate
for the experimental set-up only by a geometrical scaling $\alpha_{M}\sim1$
of the field strength as $\boldsymbol{B}_{M,\text{exp}}=\gamma\boldsymbol{B}_{M}$.

In order to solve for the current-current correlation function analytically
in the presence of the toroidal field, a couple other approximations
must be made. First, the ring is taken to have a circular cross-section
rather than a rectangular one. We keep the cross-sectional area of
the ring constant by taking the radius of the circular cross-section
to be $r_{T}=\sqrt{wt/\pi}$. Second, the high aspect ratio limit
is taken where the curvature of the ring can be ignored. In this limit,
the ring can be viewed as a cylinder centered along the $z$-axis.
The quantity $L\theta/2\pi$ then becomes $z$, with the condition
$\theta+2\pi=\theta$ becoming $z+L=z$ and $\boldsymbol{A}_{\phi}=\phi\tilde{\boldsymbol{\theta}}/2\pi r\rightarrow\phi\tilde{\boldsymbol{z}}/L$.
Eq. \ref{eq:CHPCTh_DiffusonCooperonEigenvalues} then becomes, for
$\boldsymbol{A}_{M}=0$, 
\begin{align*}
\left(\nabla'+i\frac{e}{\hbar}\boldsymbol{A}_{\mp}\right)^{2}P_{d,c}\left(\boldsymbol{r},\boldsymbol{r}',\omega\right) & =E_{n}^{d,c}\left(B_{\mp}\right)P_{d,c}\left(\boldsymbol{r},\boldsymbol{r}',\omega\right)\\
\left(\frac{\partial^{2}}{\partial x^{\prime2}}+\frac{\partial^{2}}{\partial y^{\prime2}}+\left(\frac{\partial}{\partial z'}+\frac{2\pi i}{L}\frac{\phi_{\mp}}{\phi_{0}}\right)^{2}\right)P_{d,c}\left(\boldsymbol{r},\boldsymbol{r}',\omega\right) & =E_{n}^{d,c}\left(B_{\mp}\right)P_{d,c}\left(\boldsymbol{r},\boldsymbol{r}',\omega\right),
\end{align*}
which once again separates to allow eigenfunctions of the form $P_{d,c}(\boldsymbol{r},\boldsymbol{r}',\omega)=X(x,y)\exp(2\pi inz/L)$
and has eigenvalues of the form of Eq. \ref{eq:CHPCTh_EperpDef}.
With this coordinate system, the toroidal field is $\boldsymbol{B}_{M}=B_{M}\hat{\boldsymbol{z}}$,
for which the vector potential can be written as 
\begin{align*}
\boldsymbol{A}_{M} & =-\frac{B_{M}}{2}y\tilde{\boldsymbol{x}}+\frac{B_{M}}{2}x\tilde{\boldsymbol{y}}.
\end{align*}
This form for $\boldsymbol{A}_{M}$ is always parallel to the surface
of the ring (it is always orthogonal to the vector $x\tilde{\boldsymbol{x}}+y\tilde{\boldsymbol{y}}$
normal to the ring surface). 

For the lowest transverse mode at $B_{M}=0$, the eigenvalue is $\varepsilon_{\perp}=0$
and the eigenfunction $X_{0}(x,y)=(\pi r_{T}^{2})^{-1/2}$ is a constant.
Treating the toroidal field as a perturbation with 
\begin{equation}
wtB_{M}\ll\phi_{0},\label{eq:CHPCTh_ToroidalCondition}
\end{equation}
we find the leading correction to the lowest transverse eigenvalue
to the cooperon (+) and diffuson (-) to be
\begin{align}
\varepsilon_{\perp}^{\pm} & =L^{2}\left\langle X_{0}\left|\left(\nabla+2\pi i\frac{\boldsymbol{A}_{M}^{\pm}}{\phi_{0}}\right)^{2}\right|X_{0}\right\rangle \nonumber \\
 & =L^{2}\oiint dxdy\, X_{0}^{2}\left(x,y\right)\left(\frac{\pi^{2}\left(B_{M}^{\pm}\right)^{2}y^{2}}{\phi_{0}^{2}}+\frac{\pi^{2}\left(B_{M}^{\pm}\right)^{2}x^{2}}{\phi_{0}^{2}}\right)\nonumber \\
 & =\frac{L^{2}}{wt}\int_{0}^{\sqrt{wt/\pi}}dr\int_{0}^{2\pi}d\theta\, r^{3}\frac{\pi^{2}\left(B_{M}^{\pm}\right)^{2}}{\phi_{0}^{2}}\nonumber \\
 & =\frac{\pi}{2}\frac{\left(B_{M}^{\pm}\right)^{2}wtL^{2}}{\phi_{0}^{2}}\label{eq:CHPCTh_EperpToroidal}
\end{align}
where the factor of $L^{2}$ comes from the scaling used when $\varepsilon_{\perp}$
was introduced in Eq. \ref{eq:CHPCTh_EperpDef}. 

\begin{figure}
\begin{centering}
\includegraphics[width=0.7\paperwidth]{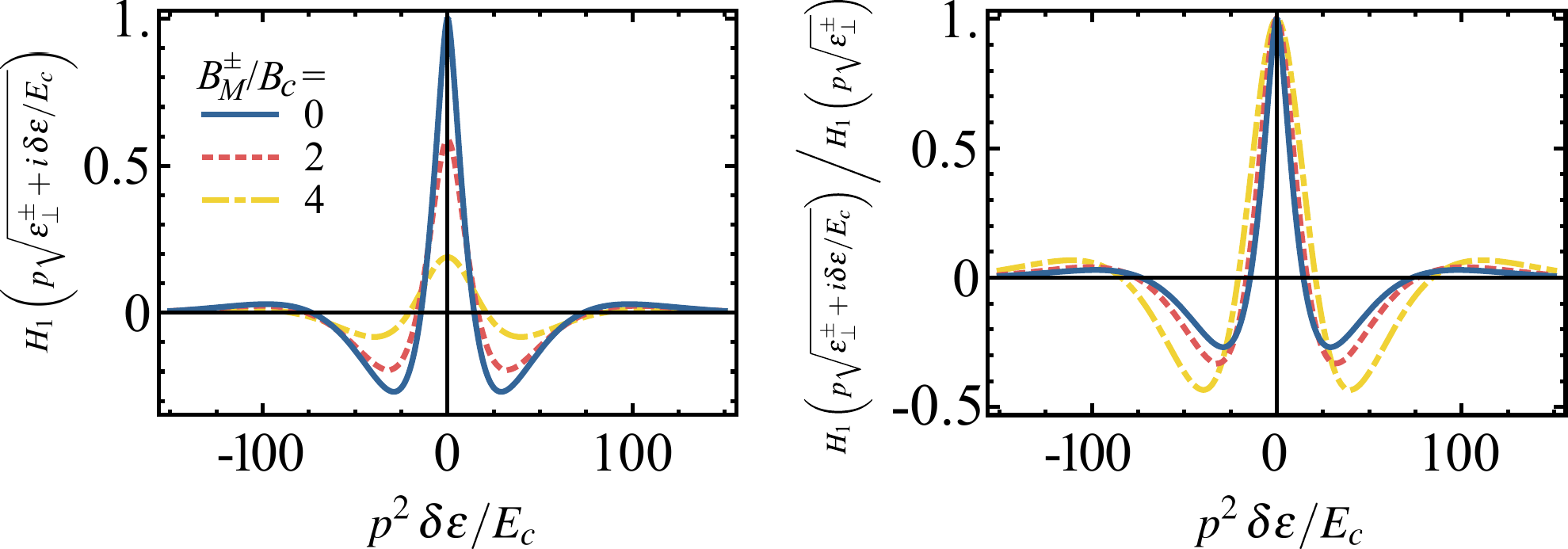}
\par\end{centering}

\caption[Current-current harmonic correlation function $H_{1}^{(0)}$ versus
energy difference for different values of the toroidal field $B_{M}^{\pm}$]{\label{fig:CHPCTh_H1CurrCurrCorBpm}Current-current harmonic correlation
function $H_{1}^{(0)}$ versus energy difference for different values
of the toroidal field $B_{M}^{\pm}$. The left plot shows $H_{1}^{(0)}(p\sqrt{\varepsilon_{\perp}^{\pm}+i\delta\varepsilon/E_{c}})$
versus the normalized energy difference $p^{2}\delta\varepsilon/E_{c}$
for different values of the toroidal field $B_{M}^{\pm}/B_{c,p}$.
The main effect of finite $B_{M}^{\pm}$ is an overall reduction of
the current-current correlation function, with otherwise minor effects
on $H_{1}^{(0)}$'s dependence on $\delta\varepsilon$, as seen in
the right plot where the same curves are scaled so that they are all
unity at $\delta\varepsilon=0$).}
\end{figure}

In the previous section, we retained the $\varepsilon_{\perp}$ in
most expressions despite concluding that for a high aspect ratio ring
we could take $\varepsilon_{\perp}=0$. We did this so that the relations
derived there would be valid for the case of the toroidal field with
$\varepsilon_{\perp}$ as given in Eq. \ref{eq:CHPCTh_EperpToroidal}.
Adapting Eq. \ref{eq:CHPCTh_CurrCurrCosCos} (and ignoring the factor
of 4 for spin degeneracy), we find that the current-current correlation
function in the presence of a toroidal field is
\[
C_{1}^{\left(0\right)}\left(\varepsilon_{1},\phi,B_{M};\varepsilon_{1}^{\prime},\phi',B_{M}^{\prime}\right)=16\frac{E_{c}^{2}}{\phi_{0}^{2}}\sum_{p=1}^{\infty}\sum_{\pm}F_{p}\left(z_{\pm}\right)\left(\mp\cos\left(2\pi p\frac{\phi\pm\phi'}{\phi_{0}}\right)\right)
\]
where 
\[
z_{\pm}=\varepsilon_{\perp}^{\pm}+\frac{i\left(\varepsilon_{1}-\varepsilon_{1}^{\prime}\right)}{E_{c}}.
\]
In Fig. \ref{fig:CHPCTh_H1CurrCurrCorBpm}, we plot the correlation
function $H_{1}^{(0)}$ defined in Eq. \ref{eq:CHPCTh_H1Def} and
shown previously in Fig. \ref{fig:CHPCTh_H1CurrCurrCor} now for different,
finite values of $B_{M}$. This function gives the normalized correlation
in energy of the $p^{th}$ harmonic of the current at zero temperature,
\[
H_{1}^{\left(0\right)}\left(p\sqrt{\varepsilon_{\perp}^{\pm}+i\frac{\delta\varepsilon}{E_{c}}}\right)=\frac{\left\langle I_{p}\left(\varepsilon_{1},B_{M}\right)I_{p}\left(\varepsilon_{1}+\delta\varepsilon,B_{M}^{\prime}\right)\right\rangle }{\left\langle \left(I_{p}\left(\varepsilon_{1},B_{M}^{\pm}=0\right)\right)^{2}\right\rangle }.
\]

\begin{figure}
\begin{centering}
\includegraphics[width=0.7\paperwidth]{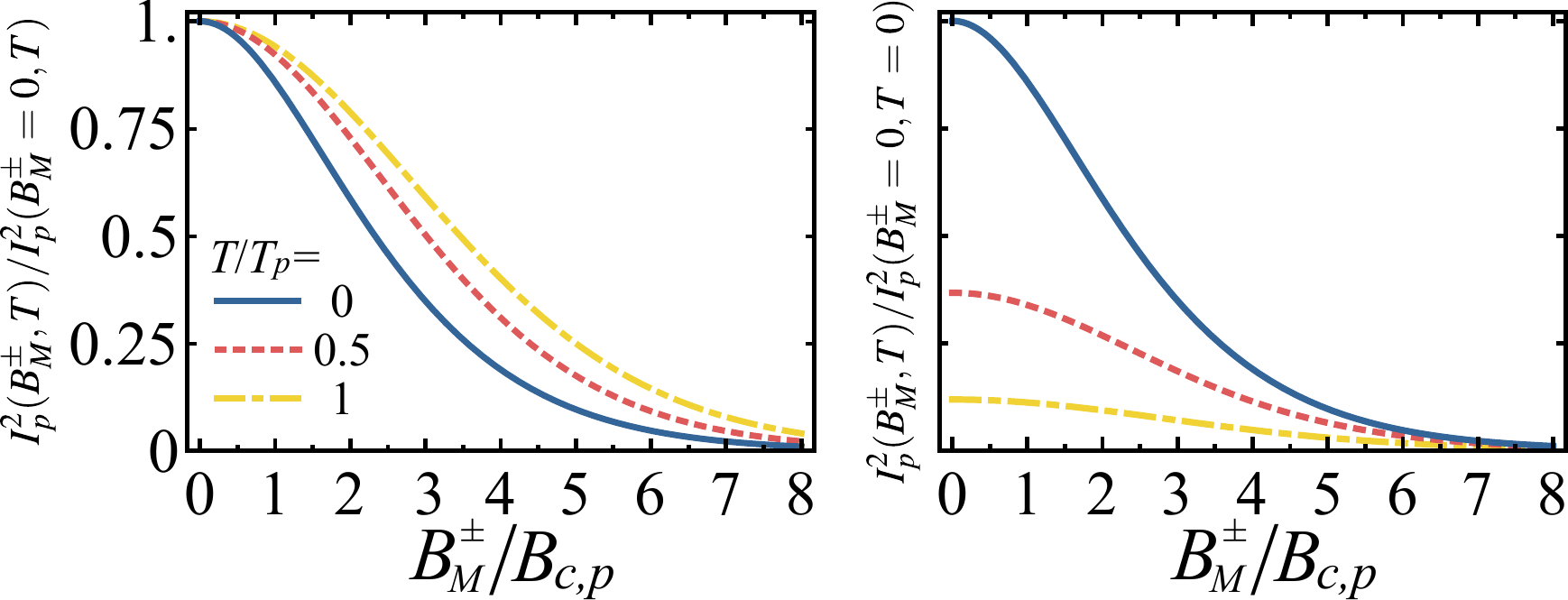}
\par\end{centering}

\caption[Current-current correlation function versus toroidal field]{\label{fig:CHPCTh_Bcorr}Current-current correlation function versus
toroidal field. The correlation function $I_{p}^{2}(B_{M}^{\pm},T)$
of the $p^{th}$ harmonic of the current is plotted against the scaled
toroidal field $B_{M}^{\pm}/B_{c,p}$ for three different scaled temperatures
$T=0,\,0.5T_{p},$ and $T_{p}$. The correlation function is scaled
by its value $I_{p}^{2}(B_{M}^{\pm}=0,T)$ at zero field and finite
temperature. In the inset, the same curves are shown scaled by the
zero-temperature, zero-field value $I_{p}^{2}(B_{M}^{\pm}=0,T=0)$
of the correlation function. The correlation function is seen to decay
on a field scale $B_{M}^{\pm}\sim3B_{c,p}$, indicating that the cooperon
is suppressed at large field. The diffuson is not suppressed with
absolute field but does have a finite correlation in field difference
$B_{M}^{-}=B_{M}-B_{M}^{\prime}$. Temperature has only a small effect
on the field dependence of $I_{p}^{2}(B_{M}^{\pm},T)$ in the temperature
range between 0 and $T_{p}$.}
\end{figure}

At finite temperature, the current-current correlation function, following
Eq. \ref{eq:CHPCTh_CurrCurrCorTempDependence}, is
\[
\left\langle I\left(\phi,B_{M}\right)I\left(\phi',B_{M}^{\prime}\right)\right\rangle =\sum_{p}\left(I_{p}^{\text{typ}}\right)^{2}\sum_{\pm}\left(\mp\cos\left(2\pi p\frac{\phi_{\pm}}{\phi_{0}}\right)\right)g_{D}\left(p^{2}\varepsilon_{\perp}^{\pm},20.8\frac{T}{T_{p}}\right).
\]
We can also write this function as 
\[
\left\langle I\left(\phi,B_{M}\right)I\left(\phi',B_{M}^{\prime}\right)\right\rangle =\sum_{p}\sum_{\pm}I_{p}^{2}\left(B_{M}^{\pm},T\right)\left(\mp\cos\left(2\pi p\frac{\phi_{\pm}}{\phi_{0}}\right)\right)
\]
with 
\[
I_{p}^{2}\left(B_{M}^{\pm},T\right)=\left(I_{p}^{\text{typ}}\right)^{2}g_{D}\left(p^{2}\varepsilon_{\perp}^{\pm},20.8\frac{T}{T_{p}}\right)
\]
the temperature dependent correlation function of the $p^{th}$ harmonic
with respect to the toroidal field $B_{M}$. In Fig. \ref{fig:CHPCTh_Bcorr},
the persistent current harmonic correlation function $I_{p}^{2}(B_{M}^{\pm},T)$
is plotted for several temperatures $T$ against $B_{M}^{\pm}$ scaled
by the field scale
\begin{equation}
B_{c,p}=\frac{1}{p}\sqrt{\frac{2}{\pi}}\frac{\phi_{0}}{L\sqrt{wt}}.\label{eq:CHPCTh_BcpToroidalField}
\end{equation}
The field scale $B_{c,p}$ is of the order of the field required to
thread flux $\phi_{0}/p$ through the shadow cast by the ring onto
the plane perpendicular to the applied magnetic field.%
\footnote{For typical ring dimensions, this projected area changes by no more
than a factor of three as the ring is rotated so this flux is of the
appropriate magnitude for all ring orientations.%
}

The persistent current correlation $I_{p}^{2}(B_{M}^{\pm},T)$ decays
over the toroidal field scale $B_{M}^{\pm}\sim3B_{c,p}$. For the
cooperon, this result means that the cooperon contribution to the
persistent current correlation function decays for $B_{M}\sim1.5B_{c}$.
For the diffuson, this decay means that the amplitude of the harmonics
of the current become uncorrelated on the scale $B_{M}-B_{M}^{\prime}\sim3B_{c,p}$.
For the experimentally relevant case of a large uniform magnetic field
applied at an angle $\theta$ with respect to the plane of the ring,
the quantities the threaded flux $\phi$ through the ring and the
toroidal field $B_{M}$ penetrating the metal are proportional and
related to each other through the aspect ratio of the ring and the
geometrical factor $\gamma$ mentioned above. As the applied magnetic
field $B$ is swept, the current undergoes oscillations of frequency
$\beta_{0}=\phi_{0}/(\pi R^{2}\sin\theta)$ as the component of $B$
perpendicular to the ring plane changes by enough to thread $\phi_{0}$
through it. At the same time, the field $B_{M}$ through the metal
of the ring is changing. Once $B_{M}$ changes by around 
\[
B_{c}\equiv B_{c,p=1},
\]
the oscillations of the current in $B$ are no longer correlated.
Effectively, the amplitude and phase of the oscillations of the persistent
are randomized on this field scale. Since the autocorrelation function
and the power spectral density are related by a Fourier transform,
we can rephrase this result, by stating that the flux through the
finite linewidth of the ring introduces a broadening of the persistent
current's magnetic field frequency $\beta$ of $\sim1/3B_{c}$.

While it is extremely useful for the correlation field $3B_{c}$ to
be larger than the period $1/\beta$ of the persistent current oscillation
so that the current signal has a well-defined magnetic field frequency
which can be distinguished from background noise, we note in passing
that the finite correlation of the persistent current also brings
benefits from an experimental point of view. By appeal to the ergodic
hypothesis, it can be argued that the finite correlation of the persistent
current oscillation leads to measurements separated by a field $>3B_{c}$
being statistically independent \citep{lee1985universal,lee1987universal}.
By measuring over a field range covering many $B_{c}$, a statistical
distribution of the persistent current can be built up from measurements
of only a single sample.%
\footnote{Additionally, the fact that different regions of magnetic field correspond
to independent samples means that every physical sample should be
capable of producing a signal of roughly the same magnitude. For SQUID
and superconducting resonator techniques, the measurement is confined
to a field region less than $3B_{c}$. For some samples, the current
will be small as it is being drawn from a probability distribution
centered around 0. For a measurement over a wide field range, it is
possible that the current will be small (compared to $I^{\text{typ}})$
in some field regions. However, it is also statistically unlikely
that the current will be small over the entirety of a region much
greater than $3B_{c}$. In fact, because the phase of the persistent
current becomes randomized, the persistent current becomes a random
phasor. If the underlying distribution of the current is the normal
distribution with standard deviation $I^{\text{typ}}$, then the amplitude
of each quadrature of the persistent current phasor should be drawn
from a normal distribution as well. The amplitude of the persistent
current phasor then follows the Rayleigh distribution, which is peaked
at $I^{\text{typ}}$. That is, the most probable amplitude of the
current is $I^{\text{typ}}$. The fact that each sample should produce
a measurable signal is useful when trying to {}``debug'' a measurement.%
}

\FloatBarrier

\subsubsection{\label{sub:CHPCTh_Zeeman}Zeeman splitting}

An applied magnetic field also affects the persistent current through
its interaction with the electron spins. A spin $\boldsymbol{\sigma}$
in a magnetic field $\boldsymbol{B}$ receives an energy shift $E_{Z}$
proportional to $\boldsymbol{\sigma}\cdot\boldsymbol{B}$. The degeneracy
of up and down spins is broken by this splitting. As the Zeeman energy
$E_{Z}$ is turned up, the entire sets of spin up and spin down energy
levels are shifted relative to each other. From Fig. \ref{fig:CHPCTh_H1CurrCurrCor},
it was seen that the energy levels in the diffusive ring are correlated
only over an energy range $\sim14E_{c}$. In the absence of Zeeman
splitting, the amplitude $I_{p\uparrow}$ of the $p^{th}$ harmonic
of the current associated with the spin up states is identical to
the amplitude $I_{p\downarrow}$ of the spin down states so that the
typical total amplitude of this harmonic is
\begin{align*}
\sqrt{\left\langle I_{p,\text{tot}}^{2}\right\rangle } & =\sqrt{\left\langle \left(I_{p\uparrow}+I_{p\downarrow}\right)^{2}\right\rangle }\\
 & =\sqrt{\left\langle \left(2I_{p\uparrow}\right)^{2}\right\rangle }\\
 & =2\sqrt{\left\langle I_{p\uparrow}^{2}\right\rangle },
\end{align*}
twice the typical current of the single spin set of levels. In the
presence of strong Zeeman splitting $E_{Z}\gg14E_{c}$, the current
associated with the spin up and spin down states is no longer correlated,
$\langle I_{p\uparrow}I_{p\downarrow}\rangle=0$. The typical harmonic
amplitude is then reduced by a factor of $\sqrt{2}$:
\begin{align*}
\sqrt{\left\langle I_{p,\text{tot}}^{2}\right\rangle } & =\sqrt{\left\langle \left(I_{p\uparrow}+I_{p\downarrow}\right)^{2}\right\rangle }\\
 & =\sqrt{\left\langle I_{p\uparrow}^{2}\right\rangle +\left\langle I_{p\downarrow}^{2}\right\rangle +2\left\langle I_{p\uparrow}I_{p\downarrow}\right\rangle }\\
 & =\sqrt{2}\sqrt{\left\langle I_{p\uparrow}^{2}\right\rangle },
\end{align*}
where we make use of the fact that the typical current of each set
of spin states is unaffected by Zeeman splitting.

In order to find the current in between these two extremes, we re-perform
the preceding calculation using the modified form of the density of
states correlation function $\langle\nu(\varepsilon_{1},B)\nu(\varepsilon_{1}^{\prime},B')\rangle$
given in Eq. \ref{eq:AppGrFu_NuNuEZ} which accounts for the Zeeman
splitting.%
\footnote{For simplicity, it is assumed that magnetic field used to calculate
the Zeeman energy is the same for both $B$ and $B'$. Such an approximation
is exact for any quantity calculated for which $B=B'$. Additionally,
it is accurate for finite $B_{M}^{-}=B_{M}-B_{M}^{\prime}\ll B_{M}$,
which is the regime of interest when calculating the field correlation
of the diffuson at large toroidal field. %
} The result for the current-current correlation function (Eq. \ref{eq:CHPCTh_CurrCurrCosCos})
is
\begin{multline}
C_{1}^{\left(0\right)}\left(\varepsilon_{1},\phi,B_{M};\varepsilon_{1}^{\prime},\phi',B_{M}^{\prime};E_{Z}\right)=\\
16\frac{E_{c}^{2}}{\phi_{0}^{2}}\sum_{p=1}^{\infty}\sum_{\pm}\left(\mp\cos\left(2\pi p\frac{\phi_{\pm}}{\phi_{0}}\right)\right)\left(2F_{p}\left(z\right)+F_{p}\left(z-2i\frac{E_{Z}}{E_{c}}\right)+F_{p}\left(z+2i\frac{E_{Z}}{E_{c}}\right)\right)\label{eq:CHPCTh_C10Zeeman}
\end{multline}
with $F_{p}$ given by Eq. \ref{eq:CHPCTh_Fp} and $z=\varepsilon_{\perp}^{\pm}+i(\varepsilon_{1}-\varepsilon_{1}^{\prime})/E_{c}$,
or, following Eqs. \ref{eq:CHPCTh_IpTyp} and \ref{eq:CHPCTh_C1H1},
\begin{align}
 & C_{1}^{\left(0\right)}\left(\varepsilon_{1},\phi,B_{M};\varepsilon_{1}^{\prime},\phi',B_{M}^{\prime};E_{Z}\right)\nonumber \\
 & \phantom{C_{1}^{\left(0\right)}}=\sum_{p=1}^{\infty}\left(I_{p}^{\text{typ}}\right)^{2}\sum_{\pm}\left(\mp\cos\left(2\pi p\frac{\phi_{\pm}}{\phi_{0}}\right)\right)c_{p}^{0}\left(\varepsilon_{1}-\varepsilon_{1}^{\prime},B_{M}^{\pm},E_{Z},E_{SO}=0\right)\nonumber \\
 & \phantom{C_{1}^{\left(0\right)}}=\sum_{p=1}^{\infty}\left(I_{p}^{\text{typ}}\right)^{2}\sum_{\pm}\left(\mp\cos\left(2\pi p\frac{\phi_{\pm}}{\phi_{0}}\right)\right)\times\ldots\nonumber \\
 & \phantom{C_{1}^{\left(0\right)}=\sum_{p=1}^{\infty}\left(I_{p}^{\text{typ}}\right)^{2}\sum_{\pm}}\times\left(2H_{1}^{(0)}\left(p^{2}z\right)+H_{1}^{(0)}\left(p^{2}z-2ip^{2}\frac{E_{Z}}{E_{c}}\right)+H_{1}^{(0)}\left(p^{2}z+2ip^{2}\frac{E_{Z}}{E_{c}}\right)\right),\label{eq:CHPCTh_cp0}
\end{align}
where we have introduced the notation $c_{p}^{0}(\varepsilon,B_{M}^{\pm},E_{Z},E_{SO})$
as an abbreviation for the normalized $p^{th}$ harmonic and $\pm$
component of $C_{1}^{(0)}$. We include a dependence on $E_{SO}$
in anticipation of the results of the next section. At finite temperature,
Eq. \ref{eq:CHPCTh_CurrCurrCorTempDependence} is modified to
\begin{align}
\left\langle I\left(\phi\right)I\left(\phi'\right)\right\rangle  & =\sum_{p=1}^{\infty}\left(I_{p}^{\text{typ}}\right)^{2}\sum_{\pm}\left(\mp\cos\left(2\pi p\frac{\phi_{\pm}}{\phi_{0}}\right)\right)c_{p}^{T}\left(T,B_{M}^{\pm},E_{Z},E_{SO}=0\right)\nonumber \\
 & =\sum_{p=1}^{\infty}\left(I_{p}^{\text{typ}}\right)^{2}\sum_{\pm}\left(\mp\cos\left(2\pi p\frac{\phi_{\pm}}{\phi_{0}}\right)\right)\times\ldots\nonumber \\
 & \phantom{=\sum_{p}\left(I_{p}^{\text{typ}}\right)^{2}\sum_{\pm}}\times\Bigg(2g_{D}\left(p^{2}\varepsilon_{\perp}^{\pm},20.8\frac{T}{T_{p}}\right)+\ldots\nonumber \\
 & \phantom{=\sum_{p}\left(I_{p}^{\text{typ}}\right)^{2}\sum_{\pm}\times\Bigg(}+g_{D}\left(p^{2}\left(\varepsilon_{\perp}^{\pm}-2i\frac{E_{Z}}{E_{c}}\right),20.8\frac{T}{T_{p}}\right)+\ldots\nonumber \\
 & \phantom{=\sum_{p}\left(I_{p}^{\text{typ}}\right)^{2}\sum_{\pm}\times\Bigg(}+g_{D}\left(p^{2}\left(\varepsilon_{\perp}^{\pm}+2i\frac{E_{Z}}{E_{c}}\right),20.8\frac{T}{T_{p}}\right)\Bigg)\label{eq:CHPCTh_C10ZeemanT}
\end{align}
where we have again introduced a normalized notation $c_{p}^{T}$
for the $p^{th}$ harmonic and $\pm$ component of the sum (note that
$c_{p}^{0}$ is a zero temperature correlation in energy, whereas
$c_{p}^{T}$ is a finite-temperature correlation in field $B_{M}^{\pm}$).

From Eqs. \ref{eq:CHPCTh_C10Zeeman} and \ref{eq:CHPCTh_C10ZeemanT},
it can be seen that in the presence of Zeeman splitting the zero-temperature
current-current correlation function $C_{1}^{(0)}(\varepsilon_{1},\phi;\varepsilon_{1}^{\prime},\phi')$
is modified from being from four times its value for spin-less particles
to being twice its value for spin-less particles plus two terms shifted
in energy difference $\delta\varepsilon=\varepsilon_{1}-\varepsilon_{1}^{\prime}$
by $\pm2E_{Z}$. These latter two terms represent the shifting in
energy of the up and down spin states due to the Zeeman effect. The
levels that are correlated are shifted to finite energy difference. 

\begin{figure}
\begin{centering}
\includegraphics[width=0.7\paperwidth]{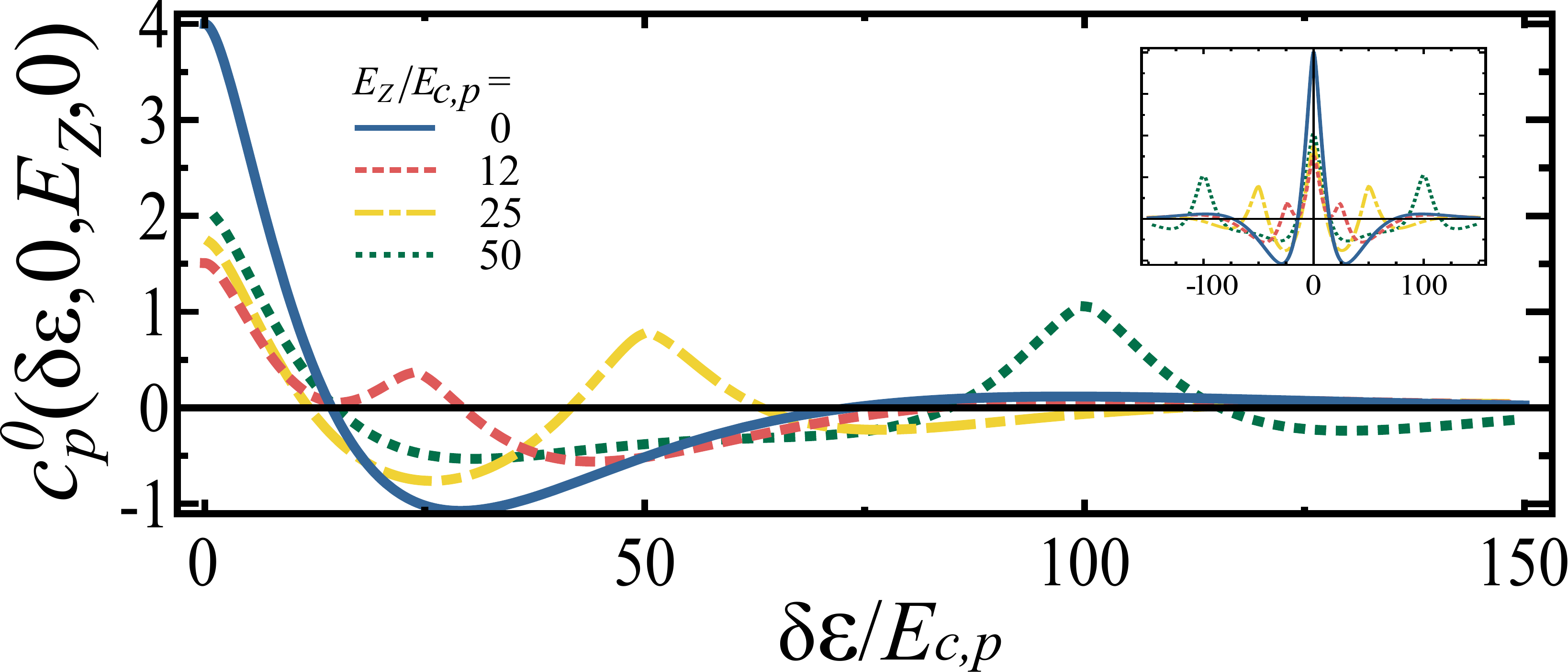}
\par\end{centering}

\caption[Zero-temperature current-current correlation function versus energy
difference at finite Zeeman splitting]{\label{fig:CHPCTh_C10EZ}Zero-temperature current-current correlation
function versus energy difference at finite Zeeman splitting. The
curves show the normalized current-current correlation function $c_{p}^{0}(\delta\varepsilon,B_{M}^{\pm},E_{Z},E_{SO})$
versus the normalized energy difference $\delta\varepsilon/E_{c}$
for values of the Zeeman energy $E_{Z}/E_{c}=$ 0, 12, 25, 50 (and
$B_{M}^{\pm}=E_{SO}=0$). For $E_{Z}=0$, the correlation function
is four times that of the spin-less case. At large $E_{Z}$, it is
reduced to half its value near $\delta\varepsilon=0$ and has additional
copies of the spin-less correlation function centered at $\pm2E_{Z}$.
At intermediate $E_{Z}$, these shifted correlation functions overlap
leading to a complicated form for the total correlation function.
Interestingly, the two intermediate curves have lower values for $c_{p}^{0}$
at $\delta\varepsilon=0$ than either the weak or strong $E_{Z}$
regimes. The inset shows the same curves plotted over a different
range of $\delta\varepsilon$. It is seen that $c_{p}^{0}$ remains
symmetric in $\delta\varepsilon$ at finite $E_{Z}$.}
\end{figure}

These shifts are shown in Fig. \ref{fig:CHPCTh_C10EZ} where the normalized
current-current correlation function $c_{p}^{0}(\delta\varepsilon,B_{M}^{\pm},E_{Z},E_{SO})$
is plotted versus $\delta\varepsilon/E_{c,p}$ for several values
of $E_{z}/E_{c,p}$ (and $B_{M}^{\pm}=E_{SO}=0$) with $E_{c,p}=E_{c}/p^{2}$.
Interestingly, due to the anti-correlation in $C_{1}^{(0)}(\varepsilon_{1},\phi;\varepsilon_{1}^{\prime},\phi')$
near $\varepsilon_{1}-\varepsilon_{1}^{\prime}\sim\pm25E_{c}$, the
value of $c_{p}^{0}$ for $\delta\varepsilon=0$ and $E_{Z}\sim\pm12E_{c,p}$
drops below 2, its value at $\delta\varepsilon=0$ in the limit of
large $E_{Z}$ where the levels of the two spin states are uncorrelated,
because the Zeeman splitting causes these anti-correlated levels to
overlap. We note that, for typical values of the diffusion constant
$D=0.02\,\text{m}^{2}/\text{s}$ and the ring circumference $L=2\,\mu\text{m}$,
the Zeeman energy $E_{Z}=2\mu_{B}B$ at $1\,\text{T}$ corresponds
to $E_{Z}\approx35E_{c}$.

\begin{figure}
\begin{centering}
\includegraphics[width=0.7\paperwidth]{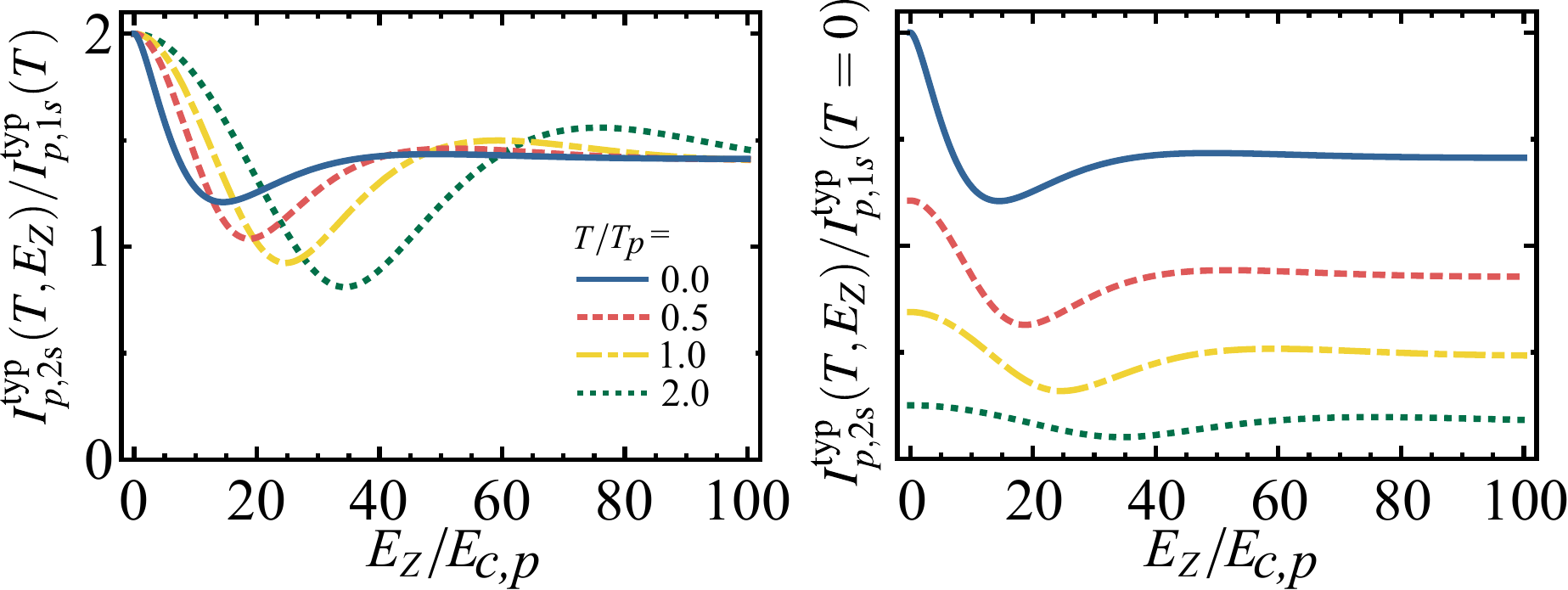}
\par\end{centering}

\caption[Typical magnitude of the persistent current harmonics versus Zeeman
energy at finite temperature]{\label{fig:CHPCTh_IITEZ}Typical magnitude of the persistent current
harmonics versus Zeeman energy at finite temperature. In the left
plot, the curves represent the typical magnitude of the current at
finite $T$ and $E_{Z}$ normalized by the single spin magnitude of
the current at finite $T$, so that all curves agree at $E_{Z}=0$.
The curves represent the normalized current magnitude for $T/T_{p}=0$,
0.5, 1, and 2. As $E_{Z}$ is increased, the typical magnitude of
the current decays from $2I_{p,1s}^{\text{typ}}(T)$ to $\sqrt{2}I_{p,1s}^{\text{typ}}(T)$
in an oscillatory fashion which indicates the anti-correlation of
the current in different regions of energy $\delta\varepsilon$. This
transition is seen to occur more slowly at higher temperatures. The
right plot shows, on the same scale and for the same series of temperatures,
the quantity $\sqrt{c_{p}^{T}(T,0,E_{Z},0)}$ which represents $I_{p,\text{2s}}^{\text{typ}}(T,E_{Z})/I_{p,1s}^{\text{typ}}(T=0)$.}
\end{figure}

The oscillation of the current-current correlation function's magnitude
near $\delta\varepsilon=0$ with $E_{Z}$ corresponds to an oscillation
in the typical magnitude of the current as a function of $E_{Z}$.
From Eq. \ref{eq:CHPCTh_C10ZeemanT} it can be seen that the quantity
$c_{p}^{T}(T,B_{M}^{\pm},E_{Z},E_{SO})$ gives the square magnitude
of the $p^{th}$ harmonic of the current at finite $T$, $B_{M}^{\pm}$,
$E_{Z}$, and $E_{SO}$, normalized by the single-spin value at $T=B_{M}^{\pm}=E_{Z}=E_{SO}=0$.
That is,
\[
c_{p}^{T}\left(T,B_{M}^{\pm},E_{Z},E_{SO}\right)=\frac{\left\langle I^{2}\left(T,B_{M}^{\pm},E_{Z},E_{SO}\right)\right\rangle _{\text{both spins}}}{\left\langle I^{2}\left(T=0,B_{M}^{\pm}\right)\right\rangle _{\text{single spin}}}.
\]
 In Fig. \ref{fig:CHPCTh_IITEZ}, the normalized typical magnitude
\[
\frac{I_{p,2s}^{\text{typ}}\left(T,E_{Z}\right)}{I_{p,1s}^{\text{typ}}\left(T\right)}=\sqrt{\frac{c_{p}^{T}\left(T,0,E_{Z},0\right)}{g_{D}\left(0,20.8\frac{T}{T_{p}}\right)}}
\]
of the $p^{th}$ harmonic of the current is plotted versus $E_{Z}$
for several values of temperature. The label $2s$ denotes a quantity
that accounts for the two spin states, while $1s$ ignores spin. The
label $1s$ could be applied to results of all preceding calculations
in the diffusive regime as up to now we have ignored spin. Whenever
a Zeeman or spin-orbit dependence is written, the two spin label $2s$
is implied since Zeeman splitting and spin orbit scattering are spin-dependent
effects. The magnitude of the current is seen to decay in an oscillatory
fashion from twice the single-spin value $I_{p,1s}^{\text{typ}}$
for the spin-degenerate case of $E_{Z}=0$ to $\sqrt{2}I_{p,1s}^{\text{typ}}$
in the uncorrelated spin regime of $E_{Z}\gg E_{c}$. The transition
between these two regimes occurs more slowly at higher temperature. 

\begin{figure}
\begin{centering}
\includegraphics[width=0.7\paperwidth]{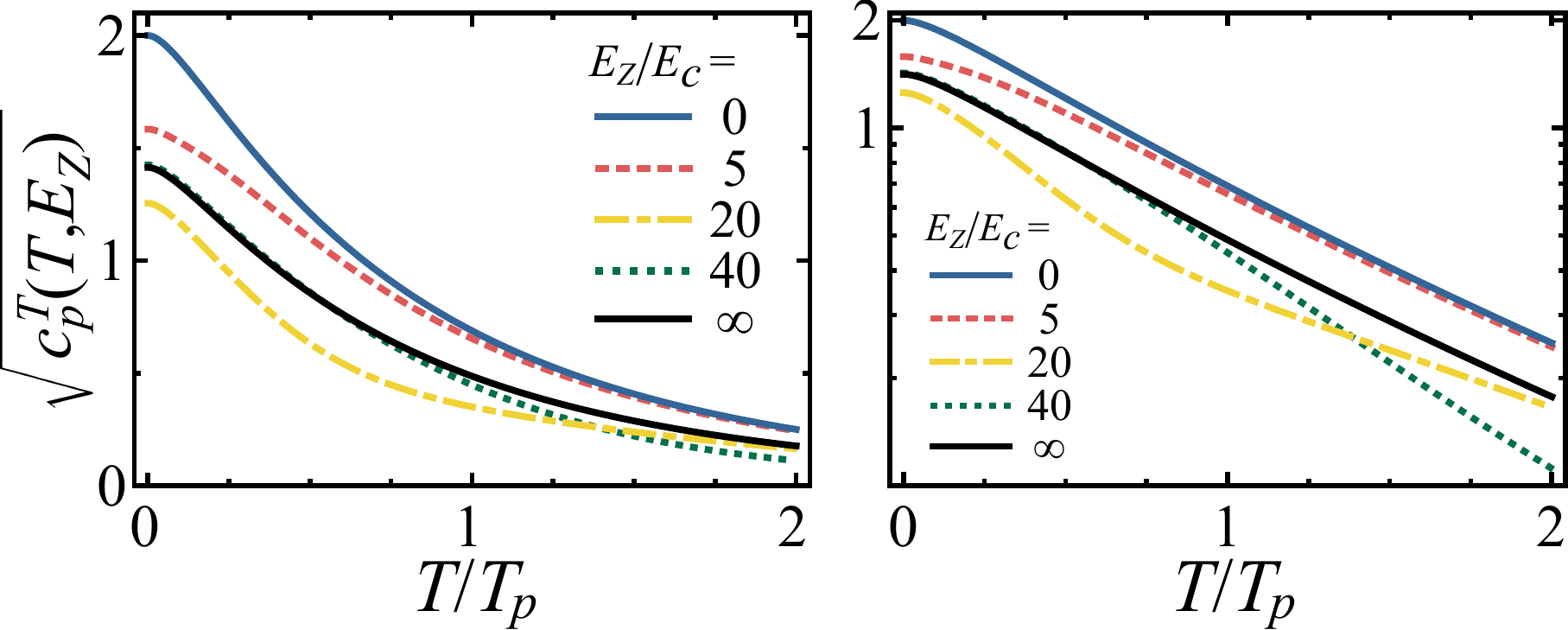}
\par\end{centering}

\caption[Typical magnitude of the $p^{th}$ harmonic of the current versus
temperature for finite Zeeman splitting]{\label{fig:CHPCTh_IIvsTEZseries}Typical magnitude of the $p^{th}$
harmonic of the current versus temperature for finite Zeeman splitting.
The two plots show the normalized typical magnitude of the $p^{th}$
harmonic of the current $\sqrt{c_{p}^{T}(T,0,E_{Z},0)}$ versus the
normalized temperature $T/T_{p}$ for the Zeeman splittings $E_{Z}/E_{c}=0$,
5, 20, 40, and $\infty$ on both linear (left plot) and log (right
plot) scales. The curve for $E_{Z}=0$ is exactly a factor of $\sqrt{2}$
larger than that for $E_{Z}=\infty$ for all values of $T$. For the
measurements discussed in this work, $E_{Z}$ was always greater than
$84E_{c}$.}
\end{figure}

\begin{figure}
\begin{centering}
\includegraphics[width=0.7\paperwidth]{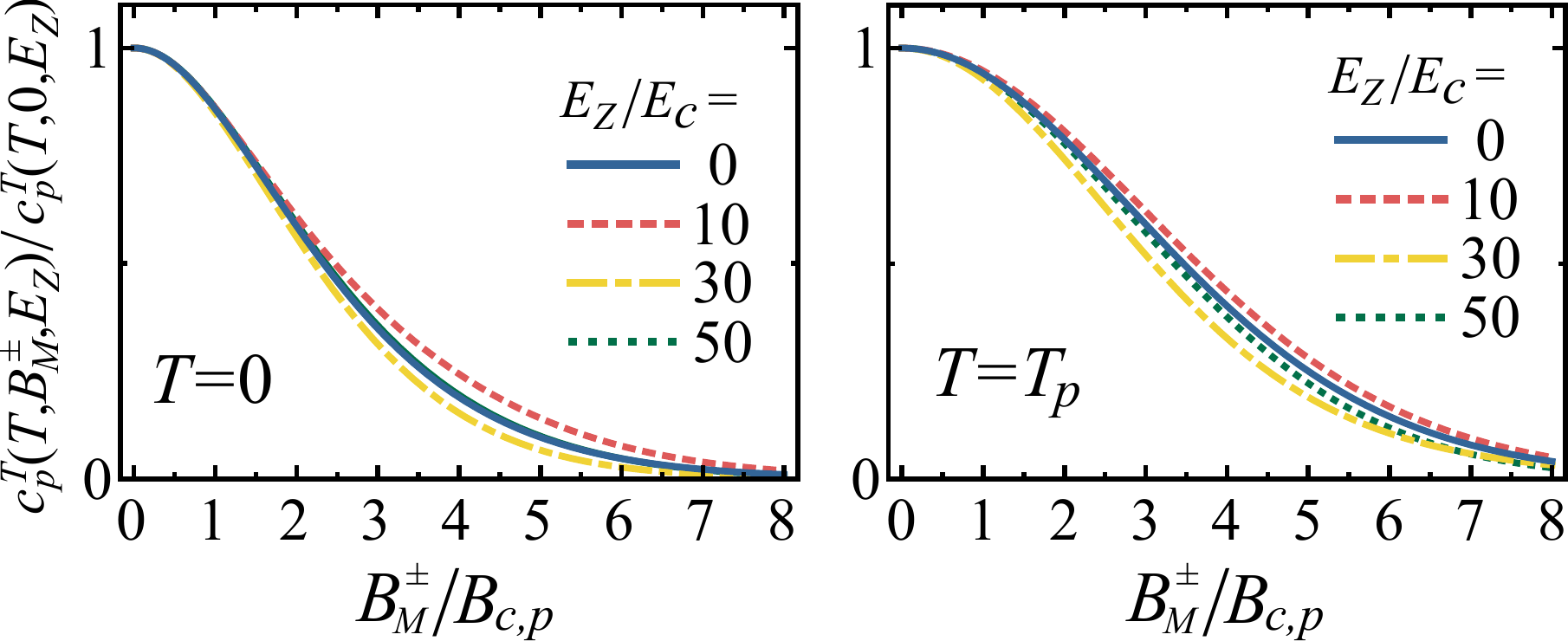}
\par\end{centering}

\caption[Current-current correlation function versus toroidal field at finite
Zeeman splitting]{\label{fig:CHPCTh_BcorrEZT}Current-current correlation function
versus toroidal field at finite Zeeman splitting. The curves show
the current-current correlation function $c_{p}^{T}(T,B_{M}^{\pm},E_{Z},E_{SO})$
at fixed values of $E_{Z}$ and $T$ (and $E_{SO}=0$) versus the
normalized toroidal field $B_{M}^{\pm}/B_{c,p}$. The curves display
the autocorrelation of the $p^{th}$ harmonic of the current as a
function of the toroidal field. Each curve has been normalized by
its value at $B_{M}^{\pm}$, which removes the overall suppression
of the current due to finite $E_{Z}$ and finite $T$. Qualitatively,
the curves are all very similar, indicating that the main role of
$E_{Z}$ and $T$ is to reduce the typical magnitude of the current-current
correlation without affecting its dependence on $B_{M}^{\pm}$.}
\end{figure}

The explanation for this interplay between the temperature and Zeeman
energy on the typical current magnitude can be deduced by consulting
Figs. \ref{fig:CHPCTh_f2TemperatureFunc} (showing the thermal weighting
function $f_{2}(\delta\varepsilon/2k_{B}T)$) and \ref{fig:CHPCTh_C10EZ}
(showing $c_{p}^{0}(\delta\varepsilon,0,E_{Z},0)$). The typical current
is found by integrating the product of these two functions and taking
the square root. As $E_{Z}$ is increased, the correlation close to
$\delta\varepsilon=0$ is decreased at the expense of greater correlation
at larger $\delta\varepsilon$ as the Zeeman split terms are shifted.
As temperature is increased, the range of $\delta\varepsilon$ given
an appreciable weight by $f_{2}$ increases and the Zeeman shifted
features in the correlation at larger values of $\delta\varepsilon$
contribute more to the integral. Physically speaking, as $E_{Z}$
is increased, the correlated orbitals of opposite spin are shifted
apart and the correlation becomes less significant. At finite temperature,
the occupancy of the levels is spread out, allowing the electrons
to occupy the displaced correlated levels. The large Zeeman splitting
regime requires both $E_{Z}\gg E_{c}$ and $E_{Z}\gg k_{B}T$. The
temperature dependence of the transition for weak to strong Zeeman
splitting also results in added features in the plot of the typical
current versus temperature, (Fig. \ref{fig:CHPCTh_IIvsTEZseries}).
Other than an overall suppression, the current-current correlation
as a function of toroidal field $B_{M}^{\pm}$ is only slightly affected
by $E_{Z}$ (Fig. \ref{fig:CHPCTh_BcorrEZT}).

In order to get a more quantitative estimate of the cross-over from
weak Zeeman splitting to strong Zeeman splitting at finite temperature,
we fit the normalized square magnitude of the current $(I_{p,2s}^{\text{typ}}\left(T,E_{Z}\right)/I_{p,1s}^{\text{typ}}\left(T\right))^{2}$
to 
\[
h\left(E_{Z}\right)=2+2\cos\left(2\pi\frac{E_{Z}}{E_{Z,\text{period}}\left(T\right)}\right)\exp\left(-\frac{E_{Z}}{E_{Z,\text{decay}}\left(T\right)}\right)
\]
at a series of temperatures. This fit function was chosen as a rough
approximation to the shape of the current magnitude's dependence on
$E_{Z}$ and provides a functional but inexact match. The extracted
values for $E_{Z,\text{decay}}(T)$ are roughly hyperbolic in $T$
with $E_{Z,\text{decay}}\left(T\right)\approx11E_{c}$ for $T\apprle0.3T_{p}$
and $E_{Z,\text{decay}}\left(T\right)\approx30E_{c}(T/T_{p})$ for
$.3T_{p}\apprle T\apprle3T_{p}$. With these results, we can state
that the weak Zeeman splitting regime is $E_{Z}\ll E_{Z,\text{CO}}$
while the strong Zeeman splitting regime is $E_{Z}\gg E_{Z,\text{CO}}$,
where the cross-over energy is 
\[
E_{Z,\text{CO}}=\max(11E_{c},30E_{c}(T/T_{p})).
\]
For sample CL17 with the smallest rings, the ratio $E_{Z}/E_{c}\approx84$
in the low range of the applied magnetic field which was $\sim3\,\text{T}$.
This range was measured at low temperature for which $E_{Z,\text{CO}}=11E_{c}$.
At the highest temperatures $T\sim5T_{p}$, the cross-over to the
strong Zeeman splitting limit is $E_{Z,\text{CO}}\sim150E_{c}$. Measurements
at this temperature occurred at fields greater than $5\,\text{T}$
for which $E_{Z}$ was close to this cross-over value. The majority
of the data was taken at larger magnetic fields and lower values of
$T/T_{p}$ (and, for the larger ring samples, smaller values of $E_{c}$)
for which $E_{Z}$ does exceed $E_{Z,\text{CO}}$ and the strong Zeeman
splitting limit applies. The main impact of the Zeeman energy introduced
by the large magnetic field used in the measurement is thus to reduce
the typical current magnitude by an overall factor of $1/\sqrt{2}$.

\subsubsection{\label{sub:CHPCTh_SpinOrbit}Spin-orbit scattering}

In addition to coupling to an external field through the Zeeman effect,
the spin degrees of freedom of electrons in a ring can also couple
directly to the disordered medium of the ring. This interaction can
be either in the form of a direct coupling of the electron spin to
magnetic impurities or an indirect coupling to the disorder potential
through the spin-orbit interaction. Both interactions can be important.
However, for measurements discussed in this text, high purity aluminum
was used to fabricate the rings in order to minimize the effects of
magnetic impurities. Additionally, it has previously been observed
in copper samples that for magnetic fields $B\apprge5k_{B}T/\mu_{B}$
magnetic impurities become polarized and spin-flip interactions are
suppressed \citep{pierre2002dephasing}. Most of our measurements,
those with $T\apprle1\,\text{K}$, were performed in this regime.
Moreover, no magnetic impurity has been observed to retain a localized
moment when dissolved into aluminum \citep{ashcroft1976solidstate}.
For these reasons, we neglect interactions with magnetic impurities.

The coupling of the spin degrees of freedom of an electron to the
disorder potential of its metallic host is relativistic in origin.
The disorder potential is essentially a spatially inhomogeneous electric
field through which the electron moves. In the rest frame of the moving
electron, the purely electric field is transformed into a mixture
of electric and magnetic fields. Thus the spin of the electron experiences
an effective magnetic field that varies with the velocity of the electron
and the strength of the disorder potential. A more detailed description
of the spin-orbit interaction is provided in Section \ref{sub:AppGrFu_Spin}.

The spin-orbit interaction mixes the spin up and spin down states.
The result of this mixing is that the spectrum of doubly degenerate
states is replaced by a non-degenerate spectrum which is twice as
dense. As was argued in the introduction to this section, in the diffusive
regime the magnitude of the persistent current should be independent
of the density of states because a denser spectrum leads to a greater
number of levels contributing to the current at the expense of a smaller
single-level current. Therefore, if we expect that with spin-orbit
scattering the typical persistent current magnitude is $2I^{\text{typ}}$,
twice the typical current of a spin-less system, we expect that in
the presence of a strong spin-orbit coupling the typical persistent
current magnitude will be reduced to $I^{\text{typ}}$, the same magnitude
as the spin-less case.

We now calculate the current in the presence of finite spin-orbit
interaction, confirming the two limiting cases just described. We
restrict our focus to the diffuson contribution to the persistent
current. Our measurements were performed at large magnetic fields
where the cooperon contribution to the typical current should be strongly
suppressed according to the analysis of Section \ref{sub:CHPCTh_FluxThroughMetal}.
The cooperon contribution is straightforward to calculate but takes
a slightly different form than the diffuson. Because the cooperon
involves reversed paths, it couples to different pairs of spins than
the diffuson. Namely, it mixes the $\uparrow\downarrow$ and $\downarrow\uparrow$
pairs split by the Zeeman effect (see Section \ref{sub:AppGrFu_Spin})
and consequently results in a more complicated expression for the
typical current in the presence of both finite Zeeman and spin-orbit
interactions.

As was the case for the Zeeman interaction in the preceding section,
to calculate the typical persistent current in the presence of finite
spin-orbit interaction we must recalculate the density of states correlation
function $\langle\nu(\varepsilon)\nu(\varepsilon')\rangle$. This
calculation is discussed in Section \ref{sub:AppGrFu_Spin} and the
result is given in Eq. \ref{eq:AppGrFu_NuNuZSO}. Repeating the calculation
of the current-current correlation function of the preceding sections
with this form for the density of states correlation function, we
find for the zero-temperature current-current correlation function

\begin{align*}
 & C_{1}^{\left(0\right)}\left(\varepsilon_{1},\phi,B_{M};\varepsilon_{1}^{\prime},\phi',B_{M}^{\prime};E_{Z},E_{SO}\right)=\\
 & \phantom{C_{1}^{\left(0\right)}}16\frac{E_{c}^{2}}{\phi_{0}^{2}}\sum_{p=1}^{\infty}\cos\left(2\pi p\frac{\phi-\phi'}{\phi_{0}}\right)\Bigg(F_{p}\left(z\right)+F_{p}\left(z+\frac{4}{3}\frac{E_{SO}}{E_{c}}\right)+F_{p}\left(z-2i\frac{E_{Z}}{E_{c}}+\frac{4}{3}\frac{E_{SO}}{E_{c}}\right)+\ldots\\
 & \phantom{C_{1}^{\left(0\right)}16\frac{E_{c}^{2}}{\phi_{0}^{2}}\sum_{p=1}^{\infty}\cos\left(2\pi p\frac{\phi-\phi'}{\phi_{0}}\right)\Bigg(}+F_{p}\left(z+2i\frac{E_{Z}}{E_{c}}+\frac{4}{3}\frac{E_{SO}}{E_{c}}\right)\Bigg)
\end{align*}
with $F_{p}$ given by Eq. \ref{eq:CHPCTh_Fp}, $z=\varepsilon_{\perp}^{-}+i(\varepsilon_{1}-\varepsilon_{1}^{\prime})/E_{c}$,
$\varepsilon_{\perp}^{-}$ given by Eq. \ref{eq:CHPCTh_EperpToroidal},
and the energy scale
\begin{equation}
E_{SO}=\frac{\hbar}{\tau_{SO}}=\frac{\hbar D}{L_{SO}^{2}}.\label{eq:CHPCTh_ESO}
\end{equation}
The spin-orbit scattering time $\tau_{SO}$ was introduced in Section
\ref{sub:AppGrFu_Spin}. Here we introduce the spin-orbit scattering
length $L_{SO}=\sqrt{\tau_{SO}}$ as well. Following Eqs. \ref{eq:CHPCTh_IpTyp},
\ref{eq:CHPCTh_C1H1}, and \ref{eq:CHPCTh_cp0}, we can also write
this correlation function in terms of the normalized zero-temperature
correlation function of the $p^{th}$ harmonic, $c_{p}^{0}$:
\begin{align}
 & C_{1}^{\left(0\right)}\left(\varepsilon_{1},\phi,B_{M};\varepsilon_{1}^{\prime},\phi',B_{M}^{\prime};E_{Z},E_{SO}\right)\nonumber \\
 & \phantom{C_{1}^{\left(0\right)}}=\sum_{p=1}^{\infty}\left(I_{p}^{\text{typ}}\right)^{2}\cos\left(2\pi p\frac{\phi-\phi'}{\phi_{0}}\right)c_{p}^{0}\left(\varepsilon_{1}-\varepsilon_{1}^{\prime},B_{M}^{\pm},E_{Z},E_{SO}\right)\nonumber \\
 & \phantom{C_{1}^{\left(0\right)}}=\sum_{p=1}^{\infty}\left(I_{p}^{\text{typ}}\right)^{2}\cos\left(2\pi p\frac{\phi-\phi'}{\phi_{0}}\right)\Bigg(H_{1}^{(0)}\left(p^{2}z\right)+H_{1}^{(0)}\left(p^{2}z+\frac{4}{3}\frac{E_{SO}}{E_{c,p}}\right)+\ldots\nonumber \\
 & \phantom{C_{1}^{\left(0\right)}=\sum_{p=1}^{\infty}\left(I_{p}^{\text{typ}}\right)^{2}\sum_{\pm}\left(\mp\cos\left(2\pi p\frac{\phi-\phi'}{\phi_{0}}\right)\right)\Bigg(}+H_{1}^{(0)}\left(p^{2}z+\frac{4}{3}\frac{E_{SO}}{E_{c,p}}-2i\frac{E_{Z}}{E_{c,p}}\right)+\ldots\nonumber \\
 & \phantom{C_{1}^{\left(0\right)}=\sum_{p=1}^{\infty}\left(I_{p}^{\text{typ}}\right)^{2}\sum_{\pm}\left(\mp\cos\left(2\pi p\frac{\phi-\phi'}{\phi_{0}}\right)\right)\Bigg(}+H_{1}^{(0)}\left(p^{2}z+\frac{4}{3}\frac{E_{SO}}{E_{c,p}}+2i\frac{E_{Z}}{E_{c,p}}\right)\Bigg),\label{eq:CHPCTh_C10BMEZESO}
\end{align}
where we use the notation
\[
E_{c,p}=\frac{E_{c}}{p^{2}}.
\]
At finite temperature, Eqs. \ref{eq:CHPCTh_CurrCurrCorTempDependence}
and \ref{eq:CHPCTh_C10ZeemanT} are modified to
\begin{align}
\left\langle I\left(\phi\right)I\left(\phi'\right)\right\rangle  & =\sum_{p=1}^{\infty}\left(I_{p}^{\text{typ}}\right)^{2}\cos\left(2\pi p\frac{\phi-\phi'}{\phi_{0}}\right)c_{p}^{T}\left(T,B_{M}^{-},E_{Z},E_{SO}\right)\nonumber \\
 & =\sum_{p=1}^{\infty}\left(I_{p}^{\text{typ}}\right)^{2}\cos\left(2\pi p\frac{\phi-\phi'}{\phi_{0}}\right)\times\ldots\nonumber \\
 & \phantom{=\sum_{p}\left(I_{p}^{\text{typ}}\right)^{2}\sum_{\pm}}\times\Bigg(g_{D}\left(p^{2}\varepsilon_{\perp}^{-},20.8\frac{T}{T_{p}}\right)+\ldots\nonumber \\
 & \phantom{=\sum_{p}\left(I_{p}^{\text{typ}}\right)^{2}\sum_{\pm}\times\Bigg(}+g_{D}\left(p^{2}\left(\varepsilon_{\perp}^{-}+\frac{4}{3}\frac{E_{SO}}{E_{c}}\right),20.8\frac{T}{T_{p}}\right)+\ldots\nonumber \\
 & \phantom{=\sum_{p}\left(I_{p}^{\text{typ}}\right)^{2}\sum_{\pm}\times\Bigg(}+g_{D}\left(p^{2}\left(\varepsilon_{\perp}^{-}+\frac{4}{3}\frac{E_{SO}}{E_{c}}-2i\frac{E_{Z}}{E_{c}}\right),20.8\frac{T}{T_{p}}\right)+\ldots\nonumber \\
 & \phantom{=\sum_{p}\left(I_{p}^{\text{typ}}\right)^{2}\sum_{\pm}\times\Bigg(}+g_{D}\left(p^{2}\left(\varepsilon_{\perp}^{-}+\frac{4}{3}\frac{E_{SO}}{E_{c}}+2i\frac{E_{Z}}{E_{c}}\right),20.8\frac{T}{T_{p}}\right)\Bigg),\label{eq:CHPCTh_IIFiniteTZSO}
\end{align}
where $c_{p}^{T}(T,B_{M}^{-},E_{Z},E_{SO})$ gives the normalized
magnitude of the autocorrelation of the $p^{th}$ harmonic of the
current at finite $T$, $B_{M}^{-}$, $E_{Z}$, and $E_{SO}$. The
typical magnitude of the $p^{th}$ harmonic of the current can be
written as 
\begin{equation}
I_{p}^{\text{typ}}\left(T,E_{Z},E_{SO}\right)=I_{p}^{\text{typ}}\sqrt{c_{p}^{T}\left(T,0,E_{Z},E_{SO}\right)}.\label{eq:ChPCTh_IpTypTEZESO}
\end{equation}
As this is the final form of the persistent current in the diffusive
limit presented in this text, we restate the following relevant expressions
for ease of reference:
\[
I_{p}^{\text{typ}}=\frac{1.11}{p^{1.5}}\frac{eD}{L^{2}}=\frac{4\sqrt{3}}{p^{1.5}}\frac{E_{c}}{\phi_{0}},
\]
\[
T_{p}=\frac{10.4}{k_{B}}\frac{\hbar D}{p^{2}L^{2}}=\frac{10.4}{p^{2}}\frac{E_{c}}{k_{B}},
\]
\[
E_{c}=\frac{\hbar D}{L^{2}},
\]
\[
\varepsilon_{\perp}^{-}=\frac{\pi}{2}\frac{\left(B_{M}^{-}\right)^{2}wtL^{2}}{\phi_{0}^{2}},
\]
\[
B_{c,p}=\frac{1}{p}\sqrt{\frac{2}{\pi}}\frac{\phi_{0}}{L\sqrt{wt}},
\]
and
\begin{align*}
g_{D}\left(x,y\right) & =\int_{-\infty}^{\infty}d\varepsilon\, f_{2}\left(\varepsilon\right)H_{1}^{\left(0\right)}\left(x+iy\varepsilon\right)\\
 & =\int_{-\infty}^{\infty}d\varepsilon\,\left(\frac{\varepsilon\cosh\varepsilon-\sinh\varepsilon}{\sinh^{3}\varepsilon}\right)\text{Re}\left[\left(1+\sqrt{x+iy\varepsilon}+\frac{x+iy\varepsilon}{3}\right)\exp\left(-\sqrt{x+iy\varepsilon}\right)\right]
\end{align*}
with $g_{D}\left(0,y\right)\approx\exp\left(-0.096y\right).$

\begin{figure}
\begin{centering}
\includegraphics[width=0.65\paperwidth]{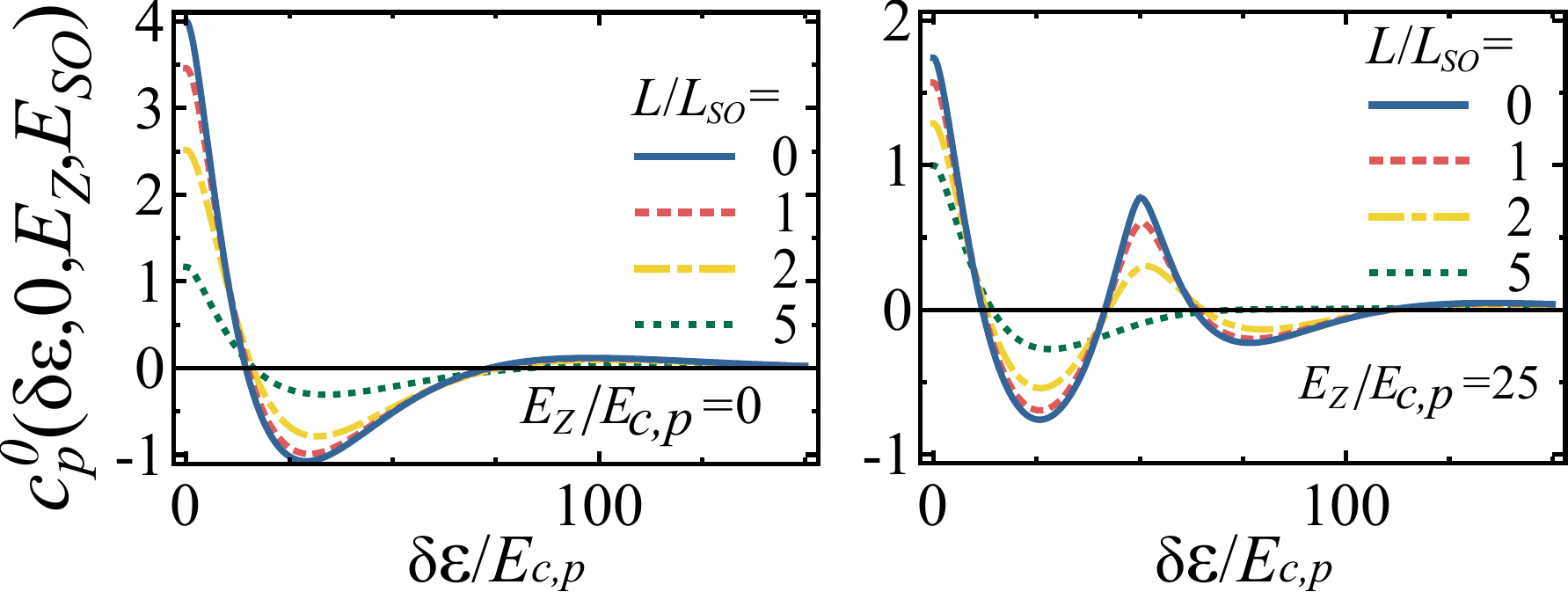}
\par\end{centering}

\caption[Zero-temperature current-current correlation function versus energy
difference at finite spin-orbit scattering and Zeeman splitting]{\label{fig:CHPCTh_cp0LSOEZ}Zero-temperature current-current correlation
function versus energy difference at finite spin-orbit scattering
and Zeeman splitting. The top graph shows $c_{p}^{T}(T,0,E_{Z},E_{SO})$
with $E_{Z}=0$ and $L/L_{SO}=0$, 1, 2, and 5. The bottom graph shows
the correlation function for the same series of values for $L/L_{SO}$
but with $E_{Z}=25E_{c}$. The figure can be meaningfully contrasted
with Figs. \ref{fig:CHPCTh_H1CurrCurrCorBpm} and \ref{fig:CHPCTh_C10EZ},
which show the same correlation function in different parameter regimes.
As $L/L_{SO}$ is increased with $E_{Z}=0$, the current-current correlation
function is reduced to one fourth of its value in the absence of spin-orbit
scattering but otherwise largely unchanged (left graph). For finite
$E_{Z}$, features at larger energy difference are reduced as the
modes coupling to the Zeeman interaction are suppressed (right graph).}
\end{figure}

\begin{figure}
\begin{centering}
\includegraphics[width=0.7\paperwidth]{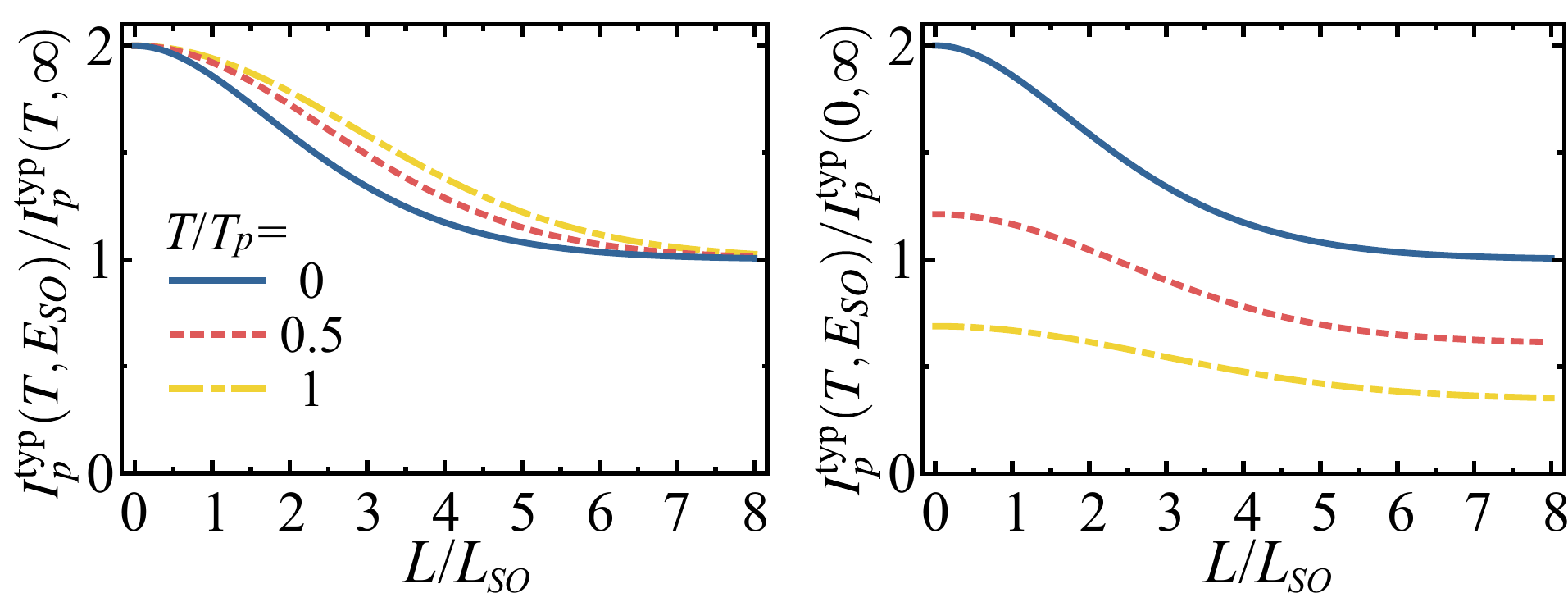}
\par\end{centering}

\caption[Typical persistent current magnitude versus spin-orbit scattering
strength]{\label{fig:CHPCTh_IvsLLSO}Typical persistent current magnitude versus
spin-orbit scattering strength. The function $\sqrt{c_{p}^{T}(T,0,0,E_{SO})/c_{p}^{T}(T,0,0,\infty)}$,
representing the magnitude $I_{p}^{\text{typ}}(T,E_{SO})/I_{p}^{\text{typ}}(T,\infty)$
of the $p^{th}$ harmonic of the persistent current at finite $T$
and $E_{SO}$ normalized by its magnitude in the strong spin-orbit
scattering ({}``spin-less'') limit, is plotted on the left against
$L/L_{SO}$. The normalized temperatures plotted are $T/T_{p}=0$,
0.5, and 1. It is seen that the current magnitude decays to the spin-less
value for $L/L_{SO}\sim4$, with the cross-over point increasing weakly
with temperature. The right plot shows the typical current magnitude
$I_{p}^{\text{typ}}(T,E_{SO})/I_{p}^{\text{typ}}(0,\infty)$ (or $\sqrt{c_{p}^{T}(T,0,0,E_{SO})/c_{p}^{T}(0,0,0,\infty)}$)
normalized by the zero temperature magnitude of the current so that
the decay with temperature can be seen.}
\end{figure}

\begin{figure}
\begin{centering}
\includegraphics[width=0.65\paperwidth]{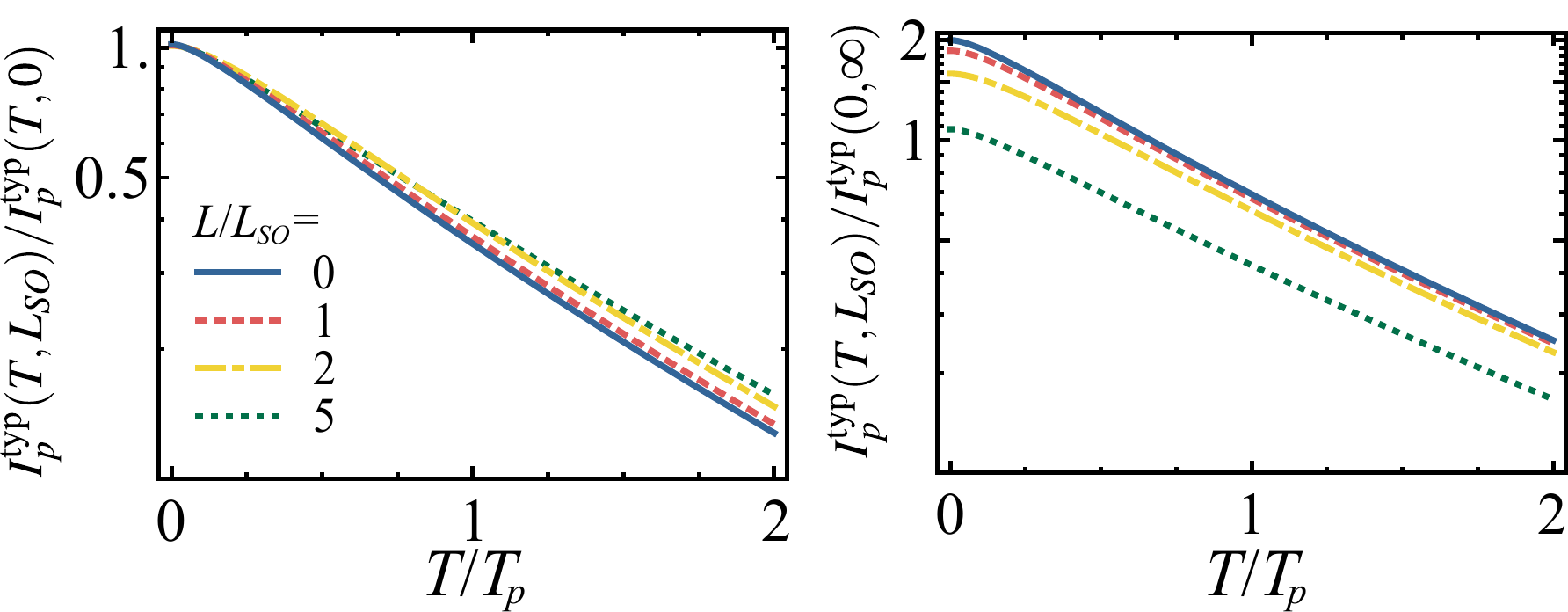}
\par\end{centering}

\caption[Typical persistent current magnitude versus temperature at finite
spin-orbit scattering strength]{\label{fig:CHPCTh_IvsTLSO}Typical persistent current magnitude versus
temperature at finite spin-orbit scattering strength. The left plot
shows $\sqrt{c_{p}^{T}(T,0,0,E_{SO})/c_{p}^{T}(T,0,0,0)}$, which
represents the typical magnitude $I_{p}^{\text{typ}}(T,L_{SO})/I_{p}^{\text{typ}}(T,0)$
of the current at finite $L_{SO}$ normalized by its magnitude in
the absence of spin-orbit scattering, versus normalized temperature
$T/T_{p}$. The curves correspond to $L/L_{SO}=0$, 1, 2, and 5. The
normalized magnitudes change very little with $L/L_{SO}\apprle5$.
In the right plot which shows $\sqrt{c_{p}^{T}(T,0,0,E_{SO})}$ (effectively
$I_{p}^{\text{typ}}(T,L_{SO})/I_{p}^{\text{typ}}(0,\infty)$), the
suppression of the current magnitude by a factor of 2 with increasing
$L/L_{SO}$ can be observed.}
\end{figure}

\begin{figure}
\begin{centering}
\includegraphics[width=0.65\paperwidth]{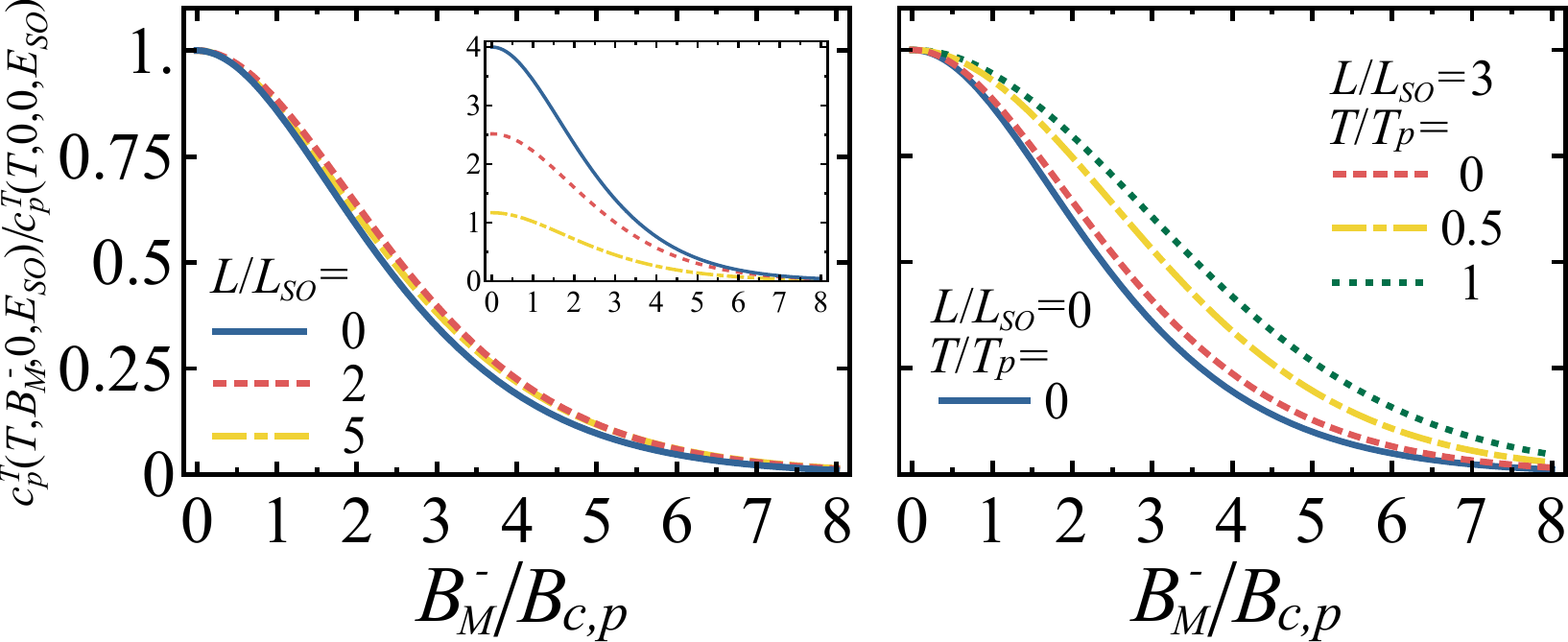}
\par\end{centering}

\caption[Current-current correlation function versus toroidal field in the
presence of spin orbit scattering]{\label{fig:CHPCTh_BcorrLSO}Current-current correlation function
versus toroidal field in the presence of spin orbit scattering. Both
the left and right plots show $c_{p}^{T}(T,B_{M}^{-},0,E_{SO})/c_{p}^{T}(T,0,0,E_{SO})$
versus the normalized toroidal field difference $B_{M}^{-}/B_{c,p}$.
The curves represent $\langle I_{p}(B_{M}+B_{M}^{-})I_{p}(B_{M})\rangle$
normalized by its $B_{M}^{-}=0$ value for different values of $L/L_{SO}$
and $T$. The left plot shows the $T=0$ case for $L/L_{SO}=0,$ 2,
and 5. The inset shows $c_{p}^{T}(0,B_{M}^{-},0,E_{SO})$ which corresponds
to $\langle I_{p}(B_{M}+B_{M}^{-})I_{p}(B_{M})\rangle$ at $T=0$
and finite $E_{SO}$ normalized by its value at $E_{SO}=\infty$.
Spin-orbit scattering has very little effect on the correlation in
$B_{M}^{-}$ other than to suppress the overall magnitude by a factor
of 4. In the right figure, the $T=0$, $L/L_{SO}=0$ case is replotted
along with three curves representing $L/L_{SO}=3$ and $T/T_{p}=0,$
0.5, and 1. This plot can be compared to Fig. \ref{fig:CHPCTh_Bcorr}
where a similar 33\% increase in the scale of correlation field is
observed with increasing $T/T_{p}$ from 0 to 1 in the absence of
spin-orbit scattering. Other than the overall suppression of the current
magnitude, spin-orbit scattering has negligible effect on the correlation
of the current with toroidal field.}
\end{figure}

Referring to each of the four terms summed together at fixed $p$
as {}``modes,'' we see that the spin-orbit interaction affects three
of the modes, including the two affected by Zeeman splitting, in exactly
the same fashion as the toroidal field does the spin-less mode. These
three modes are thus suppressed with $L/L_{SO}$ in a similar fashion
to how the current-current correlation was seen to be suppressed at
finite toroidal fields. This suppression, including the weakening
of features associated with Zeeman splitting, can be seen in Fig.
\ref{fig:CHPCTh_cp0LSOEZ} showing the normalized zero-temperature
current-current correlation function $c_{p}^{0}(\delta\varepsilon,0,E_{Z},E_{SO})$
versus $\delta\varepsilon$ and in Fig. \ref{fig:CHPCTh_IvsLLSO}
showing the typical magnitude of the $p^{th}$ current harmonic normalized
by its value in the absence of spin-orbit scattering $\sqrt{c_{p}^{T}(T,0,0,E_{SO})/c_{p}^{T}(T,0,0,0)}$
versus $L/L_{SO}$. With $E_{Z}=0$, the current-current correlation
function is mostly unchanged as a function of $\delta\varepsilon$
at finite $E_{SO}$ other than an overall suppression, leading to
a weak temperature dependence when $c_{p}^{T}(T,0,0,E_{SO})$ is normalized
by its value $c_{p}^{T}(T,0,0,0)$ in the absence of spin-orbit scattering.
This weak temperature dependence is also visible in Fig. \ref{fig:CHPCTh_IvsTLSO}
which shows the normalized magnitude $\sqrt{c_{p}^{T}(T,0,0,E_{SO})}$
of the $p^{th}$ harmonic of the current versus temperature. Similarly,
the main effect of spin-orbit scattering on the correlation $c_{p}^{T}(T,B_{M}^{-},0,E_{SO})$
of the current at finite $B_{M}^{-}$ is largely just to scale down
the magnitude of the correlation without affecting the toroidal field
dependence (Fig. \ref{fig:CHPCTh_BcorrLSO}).

\begin{figure}
\begin{centering}
\includegraphics[width=0.65\paperwidth]{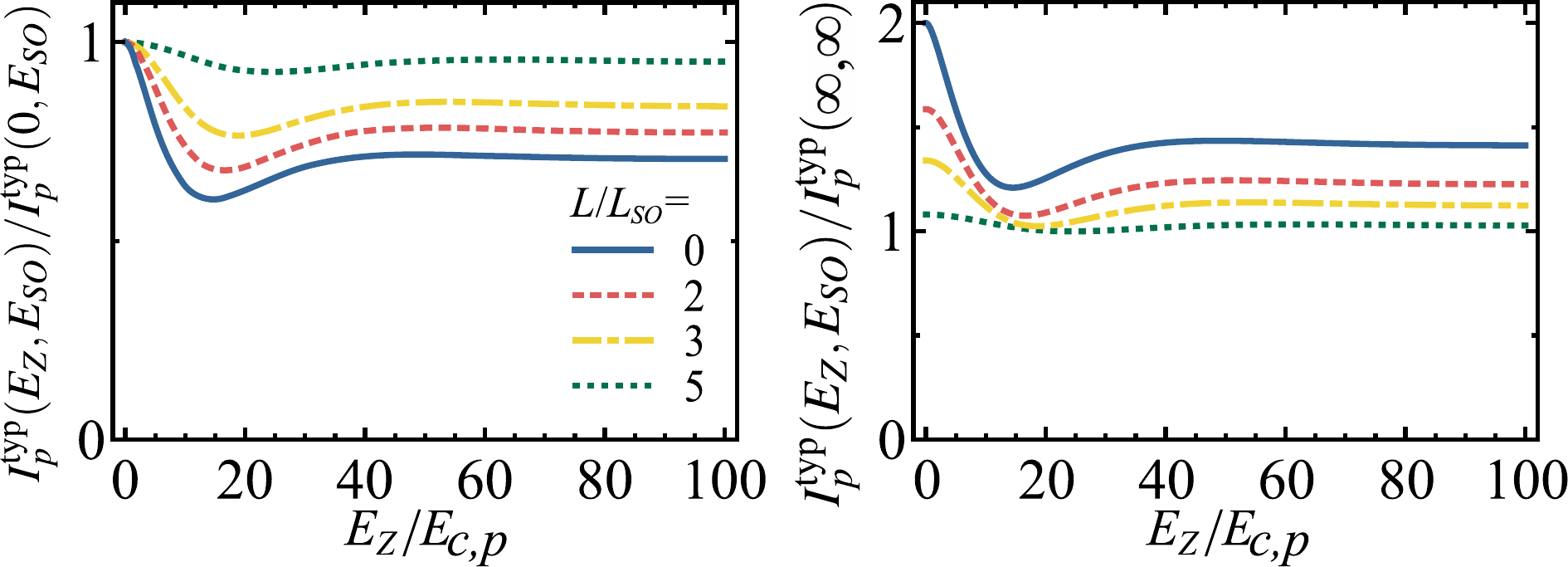}
\par\end{centering}

\caption[Typical persistent current magnitude versus Zeeman splitting with
finite spin-orbit scattering]{\label{fig:CHPCTh_IvsEZLLSO}Typical persistent current magnitude
versus Zeeman splitting with finite spin-orbit scattering. The left
plot shows the typical current magnitude $I_{p}^{\text{typ}}(E_{Z},E_{SO})/I_{p}^{\text{typ}}(0,E_{SO})$
(or $\sqrt{c_{p}^{0}(0,0,E_{Z},E_{SO})/c_{p}^{0}(0,0,0,E_{SO})}$)
at zero temperature versus the normalized Zeeman splitting $E_{Z}/E_{c,p}$.
The current magnitude has been normalized by its value at $E_{Z}=0$
so that all curves begin 1. As the spin-orbit scattering strength
is increased, the oscillatory feature in the current magnitude is
flattened out. The right plot shows the current magnitude $I_{p}^{\text{typ}}(E_{Z},E_{SO})/I_{p}^{\text{typ}}(\infty,\infty)$
normalized by the spin-less typical current. In the limit of strong
Zeeman splitting and no spin-orbit scattering, the persistent current
magnitude is reduced from $2I_{p}^{\text{typ}}$to $\sqrt{2}I_{p}^{\text{typ}}$,
where $I_{p}^{\text{typ}}$ is the spin-less typical current magnitude.
In the limit of strong spin-orbit scattering the current magnitude
is reduced to $I_{p}^{\text{typ}}$, independent of the strength $E_{Z}$
of the Zeeman splitting.}
\end{figure}

\begin{figure}
\begin{centering}
\includegraphics[width=0.7\paperwidth]{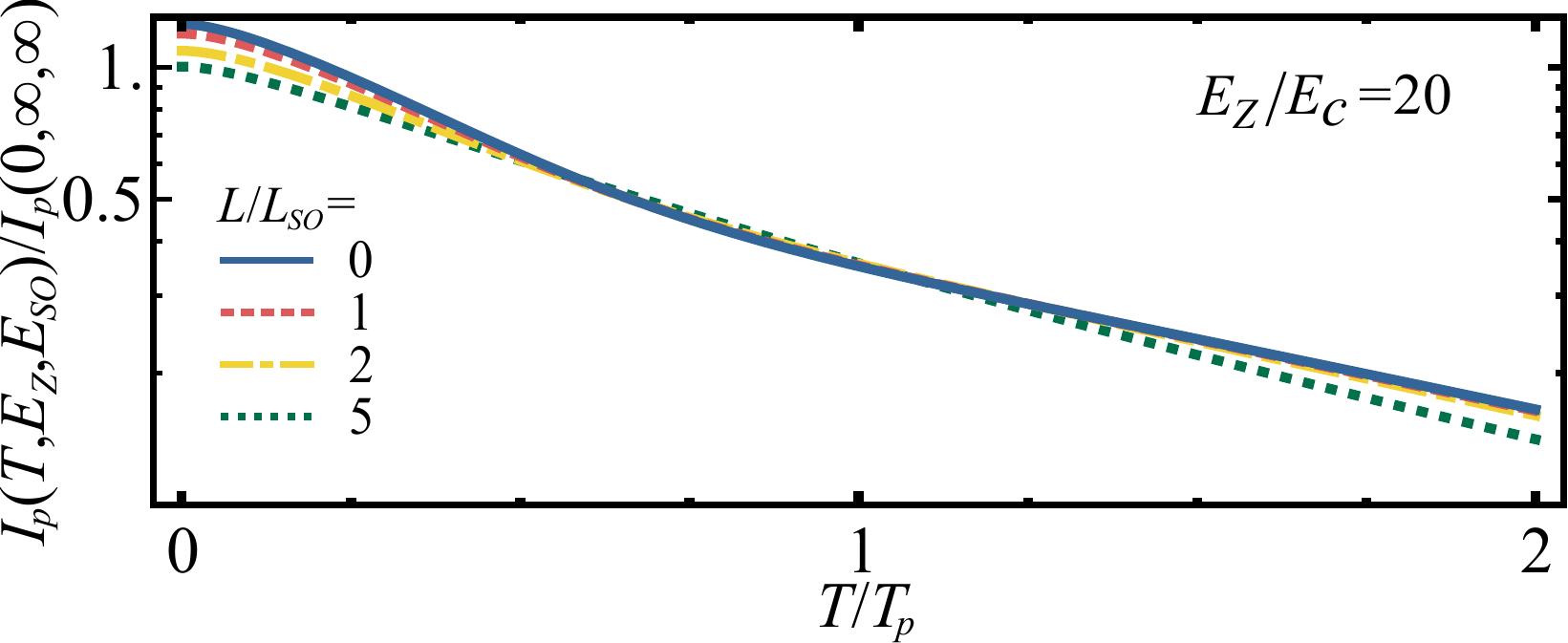}
\par\end{centering}

\caption[Typical persistent current magnitude versus temperature at finite
Zeeman splitting and spin-orbit scattering]{\label{fig:CHPCTh_IvsTEZLLSO}Typical persistent current magnitude
versus temperature at finite Zeeman splitting and spin-orbit scattering.
The normalized typical magnitude of the $p^{th}$ harmonic $I_{p}(T,E_{Z},E_{SO})/I_{p}(0,\infty,\infty)$
(or $\sqrt{c_{p}^{T}(T,0,E_{Z},E_{SO})/c_{p}^{T}(T,0,\infty,\infty)}$)
at finite Zeeman splitting and spin-orbit scattering is plotted versus
the normalized temperature $T/T_{p}$. The current magnitude, which
has been normalized by its value in the large Zeeman splitting, strong
spin-orbit scattering limit (the spin-less case), is shown for $E_{Z}/E_{c}=20$
and $L/L_{SO}=0$, 1, 2, and 5. The value of $E_{Z}$ was chosen because,
on a log scale, it produced the largest deviations from a straight
line in Fig. \ref{fig:CHPCTh_IIvsTEZseries} (which showed the current
magnitude's temperature dependence in the absence of spin-orbit scattering
for several values of $E_{Z}/E_{c}$). With increasing spin orbit
strength, it is seen that the persistent current magnitude decreases
slightly and loses much of its non-linearity on the log scale.}
\end{figure}

From all of these observations, we can summarize the effect of the
spin-orbit interaction on the persistent current as reducing the overall
magnitude of the current by correlating the spin and spatial degrees
of freedom and thus removing the spin degeneracy. In the strong spin-orbit
scattering limit, the current magnitude is reduced by a factor of
two and thus is the same as in the spin-less case. As indicated in
Fig. \ref{fig:CHPCTh_cp0LSOEZ}, the other major consequence of spin-orbit
scattering is to diminish the importance of the Zeeman splitting.
The reduction of the features associated with Zeeman splitting is
shown in Figs. \ref{fig:CHPCTh_IvsEZLLSO} and \ref{fig:CHPCTh_IvsTEZLLSO}.
In \ref{fig:CHPCTh_IvsEZLLSO}, the oscillatory feature of $I_{p}(E_{Z})$
versus $E_{Z}$ is seen to be flattened out by increasing $L/L_{SO}$.
Similarly, in Fig. \ref{fig:CHPCTh_IvsTEZLLSO}, the corresponding
oscillatory feature in $I_{p}(T)$ at finite $E_{Z}$ is suppressed
with increasing $L/L_{SO}$.

\FloatBarrier

\subsection{\label{sub:CHPCTh_AverageCurrent}Contributions to the average current
in the diffusive regime}

We now briefly describe two contributions to the average persistent
current and explain why these contributions are negligible for the
measurements discussed in the text. Addressing the average current
contributions is important because the measurements discussed in Chapter
\ref{cha:Data} were performed on arrays of $\sim10^{3}$ rings. The
total typical current in an array of $N$ rings due to a random current
contribution with typical value $I^{\text{typ}}=\sqrt{\langle I^{2}\rangle}$
is $\sqrt{N}I^{\text{typ}}$ per ring, while the total current in
the array due to an average current contribution $I^{\text{avg}}=\langle I\rangle$
per ring is $NI^{\text{avg}}$. Thus the average current contribution
to the total current signal could be a factor of $\sqrt{N}$ times
smaller than the contribution from the random component and still
contribute equally to the total current in an array.

The first contribution $I^{\text{can}}$ arises when one constrains
the number of particles in the ring to be a fixed number, which was
not done for most of the preceding calculations. It can be understood
as a single-level effect. The second contribution $I^{ee}$ results
from incorporating electron-electron interactions into the calculation
and thus is a collective effect. As discussed in Section \ref{cha:AppGrFu_},
the diffuson and cooperon depend on magnetic fields $B$ and $B'$
through $B^{-}=B-B'$ and $B^{+}=B+B'$ respectively. When considering
an average quantity, there is only one magnetic field $B=B'$. The
average persistent current is defined in terms of a magnetic field
derivative and thus has no diffuson contribution (since $B^{-}=0$
for average quantities). The gist of the argument for why these two
contributions may be neglected in our analysis is that they are derived
from the cooperon contribution to the current, which, as seen in Section
\ref{sub:CHPCTh_FluxThroughMetal}, is suppressed in the presence
of a strong magnetic field penetrating the diffusive medium of the
ring, as is the case in our measurements.

\subsubsection{\label{sub:CHPCTh_AvgSingleParticle}Single particle contribution}

We first discuss the contribution to the average current referred
to alternately as the {}``single-particle,'' {}``mesoscopic,''
{}``canonical,'' or {}``non-interacting electron'' contribution.
This contribution was first calculated independently by three different
groups \citep{altshuler1991persistent,schmid1991persistent,oppen1991average}.
Our discussion is based off of Ref. \citep{akkermans2007mesoscopic}
(which, in turn, was based on the three previously cited references).
We will refer to this current as $I^{\text{can}}$.

The starting point for the calculation is the observation that the
samples measured in experiments, consisting of metal rings on insulating
substrates, are more appropriately described by the canonical ensemble
in which the number $N$ of electrons is fixed rather than the grand
canonical ensemble in which the Fermi level $\varepsilon_{F}$ is
fixed and the particle number is allowed to fluctuate. One method
of correcting for this inaccuracy is to make the Fermi level $\varepsilon_{F}(\phi)$
depend on flux $\phi$ by enforcing the condition
\[
N=\int_{0}^{\varepsilon_{F}}d\varepsilon\,\nu\left(\varepsilon,\phi\right)
\]
for fixed $N$ and all $\phi$. Eq. \ref{eq:CHPCTh_PCthermalSum}
defining the persistent current then becomes
\[
I=-\frac{\partial\Omega\left(\varepsilon_{F}\left(\phi\right),\phi\right)}{\partial\phi}
\]
which now has an extra dependence upon $\phi$ through $\varepsilon_{F}(\phi)$.
Because of the arguments given in Section \ref{sub:CHPCTh_1DRingSingleLevelSolutions},
the quantity $\varepsilon(\phi)$ is periodic in flux with period
$\phi_{0}$.

Assuming that the variations of $\varepsilon_{F}(\phi)$ with $\phi$
are small compared to its average
\[
\varepsilon_{F}=\frac{1}{\phi_{0}}\int_{0}^{\phi_{0}}d\phi\,\varepsilon_{F}\left(\phi\right),
\]
we can expand the expression for $I$ to first order about $\varepsilon_{F}$
as
\[
I=-\frac{\partial}{\partial\phi}\left(\Omega\left(\varepsilon_{F},\phi\right)+\frac{\partial\Omega\left(\varepsilon_{F},\phi\right)}{\partial\varepsilon_{F}}\left(\varepsilon_{F}\left(\phi\right)-\varepsilon_{F}\right)\right).
\]
The first term calculated at constant $\varepsilon_{F}$ is simply
the quantity calculated in previous sections in the grand canonical
ensemble at constant $\varepsilon_{F}$. It was found to decay exponentially
in $L/l_{e}$ and thus was negligible in the diffusive regime. The
second term can be rewritten using a couple of thermodynamic relations.
First, we recall that at low temperature the Fermi level is equal
to the chemical potential $\mu\approx\varepsilon_{F}$ to lowest order
in temperature. The derivative of the grand potential with respect
to $\mu$ is
\[
\frac{\partial\Omega}{\partial\mu}=-N.
\]
Additionally, a change in the chemical potential at fixed particle
number $\delta\mu|_{N}$ can be related to a change in average particle
number at fixed chemical potential $\delta N|_{\mu}$ by 
\[
\delta\mu|_{N}=\left(\frac{\partial\mu}{\partial N}\right)_{N}\delta N|_{\mu}=\Delta_{M}\,\delta N|_{\mu}
\]
where $\Delta_{M}$ is the energy level spacing at the Fermi level,
roughly equal to the quantity given in Eq. \ref{eq:CHPCTh_MultiChannelMeanLevelSpacing}.
Using these two relations (and dropping the term already found to
be negligible in the diffusive regime), we have
\begin{align*}
I & =\Delta_{M}\frac{\partial N\left(\phi\right)}{\partial\phi}\left(N\left(\phi\right)-N\right)\\
 & =\frac{\Delta_{M}}{2}\frac{\partial}{\partial\phi}\left(N\left(\phi\right)-N\right)^{2},
\end{align*}
where the expressions are for fixed Fermi level with $N$ corresponding
to $\varepsilon_{F}$ and $N(\phi)$ to $\varepsilon_{F}(\phi)$.
We can write these quantities in terms of the density of states as
\[
N=\int_{0}^{\varepsilon_{F}}d\varepsilon\,\nu\left(\varepsilon\right)
\]
and
\[
N\left(\phi\right)=\int_{0}^{\varepsilon_{F}\left(\phi\right)}d\varepsilon\,\nu\left(\varepsilon\right)=\int_{0}^{\varepsilon_{F}}d\varepsilon\,\nu\left(\varepsilon,\phi\right).
\]

The disorder averaged current is
\[
I^{\text{can}}=\left\langle I\right\rangle =\frac{\Delta_{M}}{2}\left(\frac{\partial}{\partial\phi}\left\langle N^{2}\left(\phi\right)\right\rangle -2N\frac{\partial}{\partial\phi}\left\langle N\left(\phi\right)\right\rangle \right).
\]
In the absence of disorder, the number of levels below energy $\varepsilon$
\[
N_{0}\left(\varepsilon,\phi\right)=\int_{0}^{\varepsilon}d\varepsilon'\,\nu_{0}\left(\varepsilon',\phi\right)
\]
was calculated in Eq. \ref{eq:CHPCTh_NuIntCleanEval} as
\[
N_{0}\left(\varepsilon,\phi\right)=4\sum_{p>0}\cos\left(2\pi p\frac{\phi}{\phi_{0}}\right)\frac{1}{\pi p}\sin\left(2\pi p\sqrt{\frac{2mL^{2}\varepsilon}{h^{2}}}\right),
\]
where we have dropped the flux independent term. In Eq. \ref{eq:CHPCTh_NuIntDisorderAverage},
it was found that the disorder averaged number of levels $N(\varepsilon,\phi)=\langle N_{0}(\varepsilon,\phi)\rangle$
was the convolution of $N_{0}$ with a Lorentzian: 
\[
N\left(\varepsilon,\phi\right)=\int^{\varepsilon}d\varepsilon_{1}\,\nu\left(\varepsilon_{1},\phi\right)=\int_{0}^{\infty}d\varepsilon_{2}\, N_{0}\left(\varepsilon_{2},\phi\right)b_{L}\left(\varepsilon-\varepsilon_{2},\frac{\hbar}{2\tau_{e}}\right)
\]
with flux-dependent part of the n
\[
b_{L}\left(\varepsilon,\delta\right)=\frac{1}{\pi}\frac{\delta}{\varepsilon^{2}+\delta^{2}}.
\]
Taking $\varepsilon=\varepsilon_{F}$ and expanding $\varepsilon_{2}$
about $\varepsilon_{F}$, we have
\begin{align*}
N\left(\phi\right) & =\sum_{p}\frac{4}{\pi p}\cos\left(2\pi p\frac{\phi}{\phi_{0}}\right)\int_{0}^{\infty}d\varepsilon\,\sin\left(2\pi p\sqrt{\frac{2mL^{2}\varepsilon_{F}}{h^{2}}}\left(1+\frac{\varepsilon-\varepsilon_{F}}{2\varepsilon_{F}}\right)\right)b_{L}\left(\varepsilon_{F}-\varepsilon,\frac{\hbar}{2\tau_{e}}\right)\\
 & \approx\sum_{p}\frac{4}{\pi p}\cos\left(2\pi p\frac{\phi}{\phi_{0}}\right)\int_{-\infty}^{\infty}d\varepsilon\,\sin\left(pk\left(\varepsilon_{F}\right)L\left(1+\frac{\varepsilon}{2\varepsilon_{F}}\right)\right)b_{L}\left(\varepsilon,\frac{\hbar}{2\tau_{e}}\right)\\
 & =\sum_{p}\frac{4}{\pi p}\cos\left(2\pi p\frac{\phi}{\phi_{0}}\right)\Bigg(\sin\left(pk\left(\varepsilon_{F}\right)L\right)\int_{-\infty}^{\infty}d\varepsilon\,\cos\left(pk\left(\varepsilon_{F}\right)L\frac{\varepsilon}{2\varepsilon_{F}}\right)b_{L}\left(\varepsilon,\frac{\hbar}{2\tau_{e}}\right)+\ldots\\
 & \phantom{=\sum_{p}\frac{4}{\pi p}\cos\left(2\pi p\frac{\phi}{\phi_{0}}\right)\Bigg(}+\cos\left(pk\left(\varepsilon_{F}\right)L\right)\int_{-\infty}^{\infty}d\varepsilon\,\sin\left(pk\left(\varepsilon_{F}\right)L\frac{\varepsilon}{2\varepsilon_{F}}\right)b_{L}\left(\varepsilon,\frac{\hbar}{2\tau_{e}}\right)\Bigg)
\end{align*}
where we have that $\varepsilon_{F}\gg\hbar/2\tau_{e}$. The last
term is zero by symmetry. The other term is the Fourier transform
of the Lorentzian which was previously encountered in Eq. \ref{eq:CHPCTh_FourierTransformLorentzian}:
\begin{align*}
N\left(\phi\right) & =\sum_{p}\frac{4}{\pi p}\cos\left(2\pi p\frac{\phi}{\phi_{0}}\right)\sin\left(pk\left(\varepsilon_{F}\right)L\right)\int_{-\infty}^{\infty}d\varepsilon\,\cos\left(\pi pk\left(\varepsilon_{F}\right)L\frac{\varepsilon}{2\varepsilon_{F}}\right)b_{L}\left(\varepsilon,\frac{\hbar}{2\tau_{e}}\right)\\
 & =\sum_{p}\frac{4}{\pi p}\cos\left(2\pi p\frac{\phi}{\phi_{0}}\right)\sin\left(pk\left(\varepsilon_{F}\right)L\right)\exp\left(-\frac{\hbar}{2\tau_{e}}\frac{\pi pk\left(\varepsilon_{F}\right)L}{2\varepsilon_{F}}\right)\\
 & =\sum_{p}\frac{4}{\pi p}\cos\left(2\pi p\frac{\phi}{\phi_{0}}\right)\sin\left(pk\left(\varepsilon_{F}\right)L\right)\exp\left(-\frac{pL}{2l_{e}}\right).
\end{align*}
In the diffusive regime, this leads to a negligible contribution to
the current.

Dropping the $\langle N(\phi)\rangle$ term, we have 
\[
I^{\text{can}}=\frac{\Delta_{M}}{2}\frac{\partial}{\partial\phi}\left\langle N^{2}\left(\phi\right)\right\rangle .
\]
The second moment of $N(\phi)$ can be written as
\[
\left\langle N^{2}\left(\phi\right)\right\rangle =\int_{0}^{\varepsilon_{F}}d\varepsilon\int_{0}^{\varepsilon_{F}}d\varepsilon'\,\left\langle \nu\left(\varepsilon,\phi\right)\nu\left(\varepsilon',\phi\right)\right\rangle ,
\]
which has the cooperon as given in Eq. \ref{eq:AppGrFu_DOSFieldDependent}
as its leading flux dependent contribution (the diffuson is flux independent
since $B'=B$ in Eq. \ref{eq:AppGrFu_DOSFieldDependent} in this case).
Adapting Eqs. \ref{eq:CHPCTh_DOSCorrelation} and \ref{eq:CHPCTh_DiffusonCooperonSum},
we can write
\begin{align*}
\left\langle \nu\left(\varepsilon,\phi\right)\nu\left(\varepsilon',\phi\right)\right\rangle _{c} & =\frac{1}{2\pi^{2}}\sum_{n}\text{Re}\left(\left(i\hbar\omega+E_{c}\varepsilon_{\perp}+\left(2\pi\right)^{2}E_{c}\left(n+\frac{2\phi}{\phi_{0}}\right)^{2}\right)^{-2}\right)\\
 & =\frac{1}{2\pi^{2}}\frac{1}{\sqrt{E_{c}}}\frac{1}{\hbar}\text{Im}\frac{\partial}{\partial\omega}\sum_{p=1}^{\infty}\cos\left(4\pi p\frac{\phi}{\phi_{0}}\right)\frac{1}{\sqrt{i\hbar\omega+E_{c}\varepsilon_{\perp}}}\exp\left(-p\sqrt{\frac{i\hbar\omega+E_{c}\varepsilon_{\perp}}{E_{c}}}\right)
\end{align*}
with $\hbar\omega=\varepsilon-\varepsilon'$. Using the change of
variables $\varepsilon_{1}=\varepsilon+\varepsilon'$ and $\hbar\omega=\varepsilon-\varepsilon'$,
\begin{align*}
\left\langle N^{2}\left(\phi\right)\right\rangle _{c} & =\int_{0}^{\varepsilon_{F}}d\varepsilon\int_{0}^{\varepsilon_{F}}d\varepsilon'\,\left\langle \nu\left(\varepsilon,\phi\right)\nu\left(\varepsilon',\phi\right)\right\rangle _{c}\\
 & =\frac{1}{4\pi^{2}}\frac{1}{\sqrt{E_{c}}}\sum_{p=1}^{\infty}\cos\left(4\pi p\frac{\phi}{\phi_{0}}\right)\text{Im}\int_{0}^{2\varepsilon_{F}}d\varepsilon_{1}\int_{-\varepsilon_{1}/\hbar}^{\varepsilon_{1}/\hbar}d\omega\,\frac{\partial}{\partial\omega}\frac{\exp\left(-p\sqrt{i\hbar\omega/E_{c}+\varepsilon_{\perp}}\right)}{\sqrt{i\hbar\omega+E_{c}\varepsilon_{\perp}}}\\
 & =\frac{1}{4\pi^{2}}\frac{1}{\sqrt{E_{c}}}\sum_{p=1}^{\infty}\cos\left(4\pi p\frac{\phi}{\phi_{0}}\right)\text{Im}\int_{0}^{2\varepsilon_{F}}d\varepsilon_{1}\,\frac{\exp\left(-p\sqrt{i\hbar\omega/E_{c}+\varepsilon_{\perp}}\right)}{\sqrt{i\hbar\omega+E_{c}\varepsilon_{\perp}}}\Bigg|_{-\varepsilon_{1}\hbar}^{\varepsilon_{1}/\hbar}
\end{align*}
\begin{align*}
\phantom{\left\langle N^{2}\left(\phi\right)\right\rangle _{c}} & =\frac{1}{2\pi^{2}}\frac{1}{\sqrt{E_{c}}}\sum_{p=1}^{\infty}\cos\left(4\pi p\frac{\phi}{\phi_{0}}\right)\text{Im}\int_{0}^{2\varepsilon_{F}}d\varepsilon_{1}\,\frac{\exp\left(-p\sqrt{i\varepsilon_{1}/E_{c}+\varepsilon_{\perp}}\right)}{\sqrt{i\varepsilon_{1}+E_{c}\varepsilon_{\perp}}}
\end{align*}
where in the last line we used the fact that $\text{Im}(a-a^{*})=\text{Im}(2i\text{Im}(a))=2\text{Im}(a)$.
Performing the final integral, we have
\begin{align*}
\left\langle N^{2}\left(\phi\right)\right\rangle _{c} & =\frac{1}{2\pi^{2}}\frac{1}{\sqrt{E_{c}}}\sum_{p=1}^{\infty}\cos\left(4\pi p\frac{\phi}{\phi_{0}}\right)\text{Im}\left(\frac{2i\sqrt{E_{c}}}{p}\exp\left(-p\sqrt{\frac{i\varepsilon_{1}+E_{c}\varepsilon_{\perp}}{E_{c}}}\right)\Bigg|_{0}^{2\varepsilon_{F}}\right)\\
 & \approx-\frac{1}{\pi^{2}}\sum_{p=1}^{\infty}\frac{1}{p}\cos\left(4\pi p\frac{\phi}{\phi_{0}}\right)\text{Re}\left(\exp\left(-p\sqrt{\varepsilon_{\perp}}\right)\right)
\end{align*}
where we used the fact that $\varepsilon_{F}\gg E_{c}$ to drop the
oscillatory term. The average current is then
\begin{align}
I^{\text{can}} & =\frac{\Delta_{M}}{2}\frac{\partial}{\partial\phi}\left\langle N^{2}\left(\phi\right)\right\rangle \nonumber \\
 & =\frac{2}{\pi}\frac{\Delta_{M}}{\phi_{0}}\sum_{p=1}^{\infty}\sin\left(4\pi p\frac{\phi}{\phi_{0}}\right)\text{Re}\left[\exp\left(-p\sqrt{\varepsilon_{\perp}}\right)\right].\label{eq:CHPCTh_AverageCurrentSingleLevel}
\end{align}
From this expression, we see that the average current is of the order
of the single level current $\Delta_{M}/\phi_{0}$. Additionally,
the current is periodic with period $\phi_{0}/2$ and is paramagnetic
($\partial\langle I\rangle/\partial\phi>0$) at $\phi=0$. This result
could perhaps have been anticipated from Eqs. \ref{eq:CHPCTh_IN0}
through \ref{eq:CHPCTh_IN3} which predicted a paramagnetic current
for the even harmonics of the current in the ideal ring with a fixed
number of electrons. We have retained the transverse eigenvalues of
the cooperon with the term $\varepsilon_{\perp}$. Within the toroidal
field model, 
\begin{equation}
p^{2}\varepsilon_{\perp}=\left(\frac{2B_{M}}{B_{c,p}}\right)^{2}\label{eq:CHPCTh_ToroidalFieldEperp}
\end{equation}
where $B_{M}$ is the toroidal field strength and $B_{c,p}$ was given
in terms of the ring parameters in Eq. \ref{eq:CHPCTh_BcpToroidalField}.
Thus, this average current decays exponentially on the characteristic
field scale $B_{c,p}/2$. For this reason, we may neglect this term
as it is strongly suppressed at large field.

Since this contribution is negligible, we do not calculate its temperature
dependence or the effects of Zeeman splitting or spin-orbit scattering.
In passing, we note that the temperature dependence goes, very roughly,
as $T^{2}\exp(-\sqrt{k_{B}T/E_{c}})$ at low temperature and over
the range of experimental interest decays more strongly with temperature
than the interaction contribution (characteristic temperature $\sim3E_{c}/k_{B}$)
and the typical current contribution (characteristic temperature $\sim10E_{c}/k_{B}$
as seen in Eq. \ref{eq:CHPCTh_TpDiffusive}) \citep{ambegaokar1991comment,schmid1991persistent}.
To account for spin, the expression in Eq. \ref{eq:CHPCTh_AverageCurrentSingleLevel}
should be multiplied by 4 for the case of spin degeneracy. For Zeeman
splitting, the factor of $\text{Re}[\exp(-p\sqrt{\varepsilon_{\perp}})]$
should be replaced by 
\[
\text{Re}\left[2\exp\left(-p\sqrt{\varepsilon_{\perp}}\right)+\exp\left(-p\sqrt{\varepsilon_{\perp}+2i\frac{E_{Z}}{E_{c}}}\right)+\exp\left(-p\sqrt{\varepsilon_{\perp}-2i\frac{E_{Z}}{E_{c}}}\right)\right],
\]
and for spin orbit scattering the factor should be replaced by 
\[
\exp\left(-p\sqrt{\varepsilon_{\perp}}\right)+3\exp\left(-p\sqrt{\varepsilon_{\perp}+\frac{4}{3}\frac{E_{SO}}{E_{c}}}\right)
\]
following the procedure of Refs. \ref{sub:AppGrFu_Spin}, \ref{sub:CHPCTh_Zeeman}
and \ref{sub:CHPCTh_SpinOrbit}. As mentioned in Section \ref{sub:AppGrFu_Spin},
the expression accounting for both Zeeman splitting and spin-orbit,
using the eigenvalues given in the comment \vref{fn:AppGrFu_CooperonZeemanSO}
requires replacing $\text{Re}[\exp(-p\sqrt{\varepsilon_{\perp}})]$
by
\begin{align*}
 & \text{Re}\Bigg[2\exp\left(-p\sqrt{\varepsilon_{\perp}+\frac{4}{3}\frac{E_{SO}}{E_{c}}}\right)+\exp\left(-p\sqrt{\varepsilon_{\perp}+\frac{2}{3}\frac{E_{SO}}{E_{c}}+\sqrt{\left(\frac{2}{3}\frac{E_{SO}}{E_{c}}\right)^{2}-4\left(\frac{E_{Z}}{E_{c}}\right)^{2}}}\right)\\
 & +\exp\left(-p\sqrt{\varepsilon_{\perp}+\frac{2}{3}\frac{E_{SO}}{E_{c}}-\sqrt{\left(\frac{2}{3}\frac{E_{SO}}{E_{c}}\right)^{2}-4\left(\frac{E_{Z}}{E_{c}}\right)^{2}}}\right)\Bigg],
\end{align*}
from which it can be seen that this contribution is suppressed for
large Zeeman splitting and strong spin-orbit scattering even in the
absence of a toroidal field.

In principle, we should also consider the contribution of this canonical
current to the typical current. Such a contribution would involve
two fluxes $\phi$ and $\phi'$ and could thus contain a diffuson
which would survive at high field. However, we neglect this term as
we expect it to remain of order $\Delta_{M}/\phi_{0}$ which is smaller
than the contribution calculated in the previous sections by a factor
of $M_{\text{eff}}\sim10^{3}$. The small size of this correction
justifies the use of the grand canonical ensemble in the previous
sections.

\subsubsection{\label{sub:CHPCTh_AvgInteraction}Electron-electron contribution}

We now briefly discuss the contribution to the average current due
to electron-electron interactions, a contribution which was first
investigated in the normal state in Refs. \citealp{ambegaokar1990coherence,eckern1991coherence}.
Our discussion is once again based on Ref. \citealp{akkermans2007mesoscopic}.
We refer to this contribution to the average current as $I^{ee}$.

The interaction between electrons is handled by introducing a two-body
potential $U(\boldsymbol{r}-\boldsymbol{r}')$ describing the energy
associated with two electrons located at $\boldsymbol{r}$ and $\boldsymbol{r}'$.
It is typically taken to be short-ranged due to the tendency of the
electron gas in a metal to screen any net charge. For simplicity,
we work at zero temperature where the free energy is simply the sum
of the energies of the $N$ occupied states. Denoting by $\psi_{i}(\boldsymbol{r})$
the $i^{th}$ energy eigenfunction and by 
\[
n\left(\boldsymbol{r}\right)=\sum_{m=1}^{N}\left|\psi_{m}\left(\boldsymbol{r}\right)\right|^{2}
\]
the spatial density of electrons, the classical electro-static energy
due to the interaction for the $i^{th}$ eigenstate is
\begin{align*}
\delta\varepsilon_{i}^{ee,H} & =\int d\boldsymbol{r}\int d\boldsymbol{r}'\, U\left(\boldsymbol{r}-\boldsymbol{r}'\right)\left(n\left(\boldsymbol{r}\right)-\bar{n}\right)\left|\psi_{i}\left(\boldsymbol{r}'\right)\right|^{2}\\
 & =\sum_{m=1}^{N}\int d\boldsymbol{r}\int d\boldsymbol{r}'\, U\left(\boldsymbol{r}-\boldsymbol{r}'\right)\left(\left|\psi_{m}\left(\boldsymbol{r}\right)\right|^{2}-\bar{n}\right)\left|\psi_{i}\left(\boldsymbol{r}'\right)\right|^{2}
\end{align*}
where $\overline{n}$ is the electron density averaged over all of
the occupied states and over disorder. The average density $\overline{n}$
is assumed to give a uniformly neutral charge distribution when combined
with the positive ions making up the metal. This term does not take
the anti-symmetry of the wave function into account. Following the
standard second-quantization procedure of many-body physics, accounting
for the anti-symmetry of the wave function is accomplished by adding
the exchange term
\[
\delta\varepsilon_{i}^{ee,F}=-\sum_{m\neq i}^{N}\int d\boldsymbol{r}\int d\boldsymbol{r}'\, U\left(\boldsymbol{r}-\boldsymbol{r}'\right)\psi_{i}^{*}\left(\boldsymbol{r}\right)\psi_{m}\left(\boldsymbol{r}'\right)\psi_{m}^{*}\left(\boldsymbol{r}\right)\psi_{i}\left(\boldsymbol{r}'\right)
\]
to $\delta\varepsilon_{i}^{ee,H}$ where the sum over $m$ and $i$
is taken only over states of opposite spin.

To find the contribution of these interaction terms to the average
persistent current at zero temperature, we calculate the disorder
averaged contribution $F^{ee}$ of the interaction to the free energy
(the same as the energy at zero temperature) and take the derivative
with respect to flux:
\[
I^{ee}=-\frac{\partial\left\langle F^{ee}\right\rangle }{\partial\phi}.
\]
Referring to the $N^{th}$ level as the Fermi level with $\varepsilon_{N}=\varepsilon_{F}$
and making use of the non-local density of states $\nu(\boldsymbol{r},\boldsymbol{r}',\varepsilon)$
defined in Eq. \ref{eq:AppGrFu_NonLocalDOS}, we can write
\begin{align}
F^{ee} & =\sum_{i=1}^{N}\left(\delta\varepsilon_{i}^{ee,H}+\delta\varepsilon_{i}^{ee,F}\right)\nonumber \\
 & =\frac{1}{2}\int_{0}^{\varepsilon_{F}}d\varepsilon\int_{0}^{\varepsilon_{F}}d\varepsilon'\int d\boldsymbol{r}\int d\boldsymbol{r}'\, U\left(\boldsymbol{r}-\boldsymbol{r}'\right)\nu(\boldsymbol{r},\boldsymbol{r},\varepsilon)\nu(\boldsymbol{r}',\boldsymbol{r}',\varepsilon')\nonumber \\
 & \phantom{=}-\int_{0}^{\varepsilon_{F}}d\varepsilon\int_{0}^{\varepsilon_{F}}d\varepsilon'\int d\boldsymbol{r}\int d\boldsymbol{r}'\, U\left(\boldsymbol{r}-\boldsymbol{r}'\right)\nu(\boldsymbol{r},\boldsymbol{r}',\varepsilon)\nu(\boldsymbol{r}',\boldsymbol{r},\varepsilon')\label{eq:CHPCTh_FeeIntegrals}
\end{align}
where we have dropped the term like $\overline{n}\nu(\boldsymbol{r},\boldsymbol{r},\varepsilon)$
because we have seen previously (e.g. the preceding section) that
this term produces a contribution to the current which decays as $\exp(-pL/2l_{e})$.
For simplicity, we assume that the range of the interaction is very
short so that $U(\boldsymbol{r}-\boldsymbol{r}')\approx U\delta(\boldsymbol{r}-\boldsymbol{r}').$
In this case, we have
\begin{equation}
F^{ee}=-\frac{U}{2}\int_{0}^{\varepsilon_{F}}d\varepsilon\int_{0}^{\varepsilon_{F}}d\varepsilon'\int d\boldsymbol{r}\,\nu(\boldsymbol{r},\boldsymbol{r},\varepsilon)\nu(\boldsymbol{r},\boldsymbol{r},\varepsilon').\label{eq:CHPCTh_FeeIntegralForm}
\end{equation}

The cooperon and diffuson contributions to the disorder averaged product
$\langle\nu(\boldsymbol{r},\boldsymbol{r},\varepsilon)\nu(\boldsymbol{r}',\boldsymbol{r}',\varepsilon')\rangle$
were given as the integrand of Eq. \ref{eq:AppGrFu_DOSCorrelationIntegral}
(long-range contribution) and the expression of Eq. \ref{eq:AppGrFu_DOSCorrShortRange}
(short-range contribution). Since only the cooperon depends on $\phi$
(because we are considering $\langle\nu(\phi)\nu(\phi)\rangle$),
we only need to consider the cooperon term. For the long-range contribution,
setting $\boldsymbol{r}'=\boldsymbol{r}$ and then performing the
spatial integral leads to a contribution to Eq. \ref{eq:CHPCTh_FeeIntegralForm}
proportional to 
\[
\sum_{n,n'}\int d\boldsymbol{r}\,\frac{\phi_{n}^{*}\left(\boldsymbol{r}\right)\phi_{n}\left(\boldsymbol{r}\right)\phi_{n'}^{*}\left(\boldsymbol{r}\right)\phi_{n'}\left(\boldsymbol{r}\right)}{\left(i\omega+DE_{n}^{c}\right)\left(i\omega+DE_{n'}^{c}\right)}
\]
where we have used the eigenfunction expansion for $P_{c}(\boldsymbol{r},\boldsymbol{r}',\omega)$
introduced in Eq. \ref{eq:AppGrFu_DiffCoopEigenExpansion} and $\omega=(\varepsilon-\varepsilon')/\hbar$.
Since the terms in this sum have no reason to add coherently (there
is no reason for pairs of eigenfunctions to have appreciable magnitudes
over the same spatial region), we take this integral to be negligible.
Again using the eigenfunction expansion of Eq. \ref{eq:AppGrFu_DiffCoopEigenExpansion},
we can write the short-range contribution to Eq. \ref{eq:CHPCTh_FeeIntegralForm}
as
\begin{align*}
\int d\boldsymbol{r}\,\left\langle \nu(\boldsymbol{r},\boldsymbol{r},\varepsilon)\nu(\boldsymbol{r},\boldsymbol{r},\varepsilon')\right\rangle _{c} & =\text{Re}\left(\frac{\nu_{0}L^{d}}{\pi}\right)\int d\boldsymbol{r}\,\sum_{n}\frac{\left|\phi_{n}\left(\boldsymbol{r}\right)\right|^{2}}{i\omega+DE_{n}^{c}}\\
 & =\text{Re}\left(\frac{\nu_{0}L^{d}}{\pi}\right)\sum_{n}\frac{1}{i\omega+DE_{n}^{c}},
\end{align*}
leading to
\[
\left\langle F^{ee}\right\rangle =-\frac{U}{2}\int_{0}^{\varepsilon_{F}}d\varepsilon\int_{0}^{\varepsilon_{F}}d\varepsilon'\,\text{Re}\left(\frac{\nu_{0}L^{d}}{\pi}\right)\sum_{n}\frac{1}{i\omega+DE_{n}^{c}}.
\]
Using the Poisson summation given in Eqs. \ref{eq:CHPCTh_PoissonSumDiffusonCooperon}
and \ref{eq:CHPCTh_DiffusonCooperonSum} and setting $\lambda_{0}=U\nu_{0}L^{d}/2$,
we write
\begin{equation}
\left\langle F^{ee}\right\rangle =-\frac{\lambda_{0}}{\pi}\int_{0}^{\varepsilon_{F}}d\varepsilon\int_{0}^{\varepsilon_{F}}d\varepsilon'\,\text{Re}\frac{1}{\sqrt{E_{c}}}\sum_{p=1}^{\infty}\cos\left(4\pi p\frac{\phi}{\phi_{0}}\right)\frac{1}{\sqrt{i\hbar\omega+E_{c}\varepsilon_{\perp}}}\exp\left(-p\sqrt{\frac{i\hbar\omega+E_{c}\varepsilon_{\perp}}{E_{c}}}\right).\label{eq:CHPCTh_Fee}
\end{equation}
We use the standard change of variables $\varepsilon_{1}=\varepsilon+\varepsilon'$
and $\hbar\omega=\varepsilon-\varepsilon'$ to write
\begin{align*}
\left\langle F^{ee}\right\rangle  & =-\frac{\lambda_{0}}{\pi}\sum_{p=1}^{\infty}\cos\left(4\pi p\frac{\phi}{\phi_{0}}\right)\text{Re}\int_{0}^{2\varepsilon_{F}}d\varepsilon_{1}\int_{-\varepsilon_{1}/\hbar}^{\varepsilon_{1}/\hbar}d\omega\,\frac{\hbar}{2\sqrt{E_{c}}}\frac{1}{\sqrt{i\hbar\omega+E_{c}\varepsilon_{\perp}}}\exp\left(-p\sqrt{\frac{i\hbar\omega+E_{c}\varepsilon_{\perp}}{E_{c}}}\right)\\
 & =-\frac{\lambda_{0}}{\pi}\sum_{p=1}^{\infty}\cos\left(4\pi p\frac{\phi}{\phi_{0}}\right)\text{Re}\int_{0}^{2\varepsilon_{F}}d\varepsilon_{1}\,\frac{\hbar}{\sqrt{E_{c}}}\left(\frac{\sqrt{E_{c}}i}{p\hbar}\right)\exp\left(-p\sqrt{\frac{i\hbar\omega+E_{c}\varepsilon_{\perp}}{E_{c}}}\right)\Bigg|_{-\varepsilon_{1}/\hbar}^{\varepsilon_{1}/\hbar}\\
 & =-\frac{\lambda_{0}}{\pi}\sum_{p=1}^{\infty}\cos\left(4\pi p\frac{\phi}{\phi_{0}}\right)\text{Im}\int_{0}^{2\varepsilon_{F}}d\varepsilon_{1}\,\left(\frac{2}{p}\right)\exp\left(-p\sqrt{i\frac{\varepsilon_{1}}{E_{c}}+\varepsilon_{\perp}}\right).
\end{align*}
Using a couple changes of variables ($x=\varepsilon_{1}/E_{c}$, $y=ix+\varepsilon_{\perp}$,
and $z=p\sqrt{y}$), we can evaluate the integral to find
\begin{align*}
\left\langle F^{ee}\right\rangle  & =-\frac{\lambda_{0}}{\pi}E_{c}\sum_{p=1}^{\infty}\cos\left(4\pi p\frac{\phi}{\phi_{0}}\right)\text{Im}\int_{0}^{2\varepsilon_{F}/E_{c}}dx\,\left(\frac{2}{p}\right)\exp\left(-p\sqrt{ix+\varepsilon_{\perp}}\right)\\
 & =-\frac{\lambda_{0}}{\pi}E_{c}\sum_{p=1}^{\infty}\cos\left(4\pi p\frac{\phi}{\phi_{0}}\right)\text{Im}\int_{\varepsilon_{\perp}}^{2i\varepsilon_{F}/E_{c}+\varepsilon_{\perp}}dy\,\left(-i\right)\left(\frac{2}{p}\right)\exp\left(-p\sqrt{y}\right)\\
 & =\frac{\lambda_{0}}{\pi}E_{c}\sum_{p=1}^{\infty}\cos\left(4\pi p\frac{\phi}{\phi_{0}}\right)\text{Re}\int_{p\sqrt{\varepsilon_{\perp}}}^{p\sqrt{2i\varepsilon_{F}/E_{c}+\varepsilon_{\perp}}}dz\,\left(\frac{4}{p^{3}}\right)ze^{-z}.
\end{align*}
The final integral can be evaluated using integration by parts:
\begin{align*}
\int_{p\sqrt{\varepsilon_{\perp}}}^{p\sqrt{2i\varepsilon_{F}/E_{c}+\varepsilon_{\perp}}}dz\, ze^{-z} & =-\int_{p\sqrt{\varepsilon_{\perp}}}^{p\sqrt{2i\varepsilon_{F}/E_{c}+\varepsilon_{\perp}}}dz\,\left(-e^{-z}\right)+z\left(-e^{-z}\right)\Bigg|_{p\sqrt{\varepsilon_{\perp}}}^{p\sqrt{2i\varepsilon_{F}/E_{c}+\varepsilon_{\perp}}}\\
 & =-\left(1+z\right)e^{-z}\Bigg|_{p\sqrt{\varepsilon_{\perp}}}^{p\sqrt{2i\varepsilon_{F}/E_{c}+\varepsilon_{\perp}}}\\
 & \approx\left(1+p\sqrt{\varepsilon_{\perp}}\right)\exp\left(-p\sqrt{\varepsilon_{\perp}}\right).
\end{align*}
where the upper boundary term is dropped since it decays exponentially
with $\varepsilon_{F}/E_{c}\gg1$. Ultimately, the persistent current
due to the interaction is
\begin{align}
I^{ee} & =-\frac{\partial\left\langle F^{ee}\right\rangle }{\partial\phi}\nonumber \\
 & =16\lambda_{0}\frac{E_{c}}{\phi_{0}}\sum_{p}\sin\left(4\pi p\frac{\phi}{\phi_{0}}\right)\frac{1}{p^{2}}\left(1+p\sqrt{\varepsilon_{\perp}}\right)\exp\left(-p\sqrt{\varepsilon_{\perp}}\right).\label{eq:CHPCTh_IeeSimple}
\end{align}
Within the toroidal field model, we can write
\begin{equation}
I^{ee}=16\lambda_{0}\frac{E_{c}}{\phi_{0}}\sum_{p}\sin\left(4\pi p\frac{\phi}{\phi_{0}}\right)\frac{1}{p^{2}}\left(1+\frac{2B_{M}}{B_{c,p}}\right)\exp\left(-\frac{2B_{M}}{B_{c,p}}\right)\label{eq:CHPCTh_IeeToroidalFieldSimple}
\end{equation}
with $B_{c,p}$ given in terms of the ring parameters by Eq. \ref{eq:CHPCTh_BcpToroidalField}.

Before interpreting the expression in Eq. \ref{eq:CHPCTh_IeeToroidalFieldSimple}
for $I^{ee}$, we must state a few important caveats regarding the
way in which we handled the interaction \uline{$U(\boldsymbol{r}-\boldsymbol{r}')$}.
First, the assumption $U(\boldsymbol{r}-\boldsymbol{r}')\approx U\delta(\boldsymbol{r}-\boldsymbol{r}')$
of a local form for the interaction while fairly accurate is not necessary.
It is possible to evaluate the spatial integrals of Eq. \ref{eq:CHPCTh_FeeIntegrals}
using an analytical function for $U(\boldsymbol{r}-\boldsymbol{r}')$
(typically an exponentially screened $r^{-1}$ potential) and for
$\langle\nu(\boldsymbol{r},\boldsymbol{r},\varepsilon)\nu(\boldsymbol{r}',\boldsymbol{r}',\varepsilon')\rangle$
and $\langle\nu(\boldsymbol{r},\boldsymbol{r}',\varepsilon)\nu(\boldsymbol{r}',\boldsymbol{r},\varepsilon')\rangle$.
Doing so simply rescales the value of $\lambda_{0}$ in our derivation.
Additionally, our derivation consisted of finding the average contribution
of the interaction to the total energy using the eigenstates of the
system in the absence of the interaction, essentially a perturbative
calculation of the first order contribution of the interaction to
the total energy of the system. Taking higher orders leads to higher
order corrections in $U$. Carrying out the calculation to $n^{th}$
order leads to a term proportional to $U^{n}$ so that, very roughly,
the full calculation of the interaction contribution to the current
leads to a sum of terms similar to the geometric series $U+U^{2}+U^{3}+\ldots=U/(1-U)$
and essentially rescales $\lambda_{0}$ to $\lambda_{\text{eff}}$,
giving
\begin{equation}
I^{ee}=16\lambda_{\text{eff}}\frac{E_{c}}{\phi_{0}}\sum_{p}\sin\left(4\pi p\frac{\phi}{\phi_{0}}\right)\frac{1}{p^{2}}\left(1+\frac{2B_{M}}{B_{c,p}}\right)\exp\left(-\frac{2B_{M}}{B_{c,p}}\right).\label{eq:CHPCTh_Iee}
\end{equation}
We gloss over these details because, as discussed in Chapter \ref{cha:CHPrevWork},
the exact nature of the interaction and the correct value for $\lambda_{\text{eff}}$
is one of the outstanding questions in persistent current research.%
\footnote{Note that the comparison to a geometric series is provided just to
give a sense of the kind of effect that performing the full calculation
has on the prefactor in the expression for $I^{ee}$. The actual calculation
is not as simple as summing a geometric series.%
}

Theoretical and experimental values for $\lambda_{\text{eff}}$ have
fallen in the range of $\sim0.02-0.5$ \citep{ambegaokar1990coherence,ambegaokar1991comment,eckern1991coherence,bouchiat1991persistent,levy1991persistent,reulet1995dynamic,jariwala2001diamagnetic,deblock2002acelectric,deblock2002diamagnetic},
putting this current on roughly the same order of magnitude as the
typical current $I^{\text{typ}}\sim6E_{c}/\phi_{0}$ found in Section
\ref{sub:CHPCTh_TypicalCurrent}. Additionally, we note that this
current oscillates with a fundamental period of $\phi_{0}/2$ and
the sign of its slope $\partial I^{ee}/\partial\phi$ at zero flux
$\phi=0$ is determined by the sign of the interaction $\lambda_{\text{eff}}\propto U$.
For a repulsive interaction such as the Coulomb interaction, $U>0$
and the current is paramagnetic at low field, while for an attractive
interaction, such as the phonon-mediated interaction leading to superconductivity,
the current is diamagnetic at low field.

As noted at the outset of this section, for an array of $N$ rings
such as those measured in the experiments described in this text,
an average current $I^{ee}$ per ring of comparable size to the typical
current fluctuations $I^{\text{typ}}$ per ring should dominate the
total current in the array with $\sum I^{ee}/\sum I^{\text{typ}}\sim\sqrt{N}$
where the sums are over all the rings in the array. However, as can
be seen in Eq. \ref{eq:CHPCTh_Iee}, the contribution $I^{ee}$ to
the current becomes strongly suppressed on the field scale $\gamma B_{c,p}/2$,
which is typically $\sim10\,\text{mT}$.%
\footnote{This figure was calculated using Eq. \ref{eq:CHPCTh_BcpToroidalField},
sample dimensions similar to those given in Table \ref{tab:ChData_Rings},
and a geometrical factor of $\gamma$. It should be possible to fabricate
smaller rings for which $B_{c,p}/2\apprge50\,\text{mT}$.%
} The measurements discussed in this text were performed at fields
$>3\,\text{T}$ where this interaction contribution to the average
current should be negligible. The expression given in Eq. \ref{eq:CHPCTh_Iee}
is valid for low fields (Eq. \ref{eq:CHPCTh_ToroidalCondition}).
In Ref. \citealp{ginossar2010mesoscopic}, the high-field limit $wtB\gg\phi_{0}/2$
was considered. It was found that the prefactor of the $p^{th}$ harmonic
in Eq. \ref{eq:CHPCTh_Iee} becomes
\[
0.14\frac{1}{p^{2.5}}\sqrt{\frac{wt}{L}}\left(\frac{B_{M}}{\phi_{0}}\right)^{1/4}\left(1+2.75p\sqrt{\frac{L^{2}B_{M}}{\phi_{0}}}\right)\exp\left(-2.75p\sqrt{\frac{L^{2}B_{M}}{\phi_{0}}}\right),
\]
which gives currents of $\sim10^{-61}\lambda_{\text{eff}}E_{c}/\phi_{0}$
for typical ring parameters and $B_{M}=3\,\text{T}$ (assuming the
geometric factor $\gamma=1$). Despite the strong suppression with
magnetic field, the magnitude of the current given in Eq. \ref{eq:CHPCTh_Iee}
is actually large enough that it could be measured using the torsional
magnetometry technique discussed in Chapters \ref{cha:Cantilever-torsional-magnetometry}
and \ref{cha:CHSensitivity}. Such a measurement would need to be
performed at a field scale of $\sim\gamma B_{c,p}$ where Eq. \ref{eq:CHPCTh_Iee}
is valid and would require a large cantilever with $\sim10^{6}$ rings. 

To measure the average current in a large ensemble of rings, it is
important that the phases of the persistent current oscillations of
each individual ring remain synchronized. The frequency $\beta$ of
persistent current oscillation in magnetic field $B$ applied at angle
$\theta$ relative to the plane of the rings is given by 
\[
\beta=\frac{\pi R^{2}\sin\theta}{\phi_{0}/2}
\]
for the average current oscillation with period $\phi_{0}/2$. We
call the mean area of each ring in the ensemble $A_{0}$ and the mean
magnetic field frequency $\beta_{0}=2A_{0}\sin\theta/\phi_{0}$. In
any fabricated ensemble of rings, there will be ring-to-ring variations
in the dimensions and thus the area. We characterize these variations
by a normal distribution of ring areas with mean $A_{0}$ and standard
deviation $\alpha A_{0}$. This distribution in ring area results
in distribution of magnetic field frequencies with mean $\beta_{0}$
and standard deviation $\alpha\beta_{0}$. For an ensemble of $N$
rings with $N\gg1$, the total current signal $I^{\text{tot}}$ is
well approximated by $N$ times the ensemble-averaged current:
\[
I^{\text{tot}}\left(B\right)=N\sum_{p=1}^{\infty}I_{p}^{ee}\left(B\right)\int_{-\infty}^{\infty}d\beta\,\sin\left(2\pi p\beta B\right)\left(\frac{1}{\sqrt{2\pi\alpha^{2}\beta_{0}^{2}}}\exp\left(-\frac{\left(\beta-\beta_{0}\right)^{2}}{2\alpha^{2}\beta_{0}^{2}}\right)\right).
\]
The integral giving the ensemble-averaged oscillation is just the
imaginary part of the well-known Fourier transform of the Gaussian
function. Evaluating the Fourier transform gives
\begin{equation}
I^{\text{tot}}=N\sum_{p=1}^{\infty}I_{p}^{ee}\left(B\right)\sin\left(2\pi p\beta_{0}B\right)\exp\left(-2\pi^{2}p^{2}\alpha^{2}\beta_{0}^{2}B^{2}\right).\label{eq:CHPCTh_EnsembleSuppression}
\end{equation}
The ensemble average introduces a new field scale of suppression
\[
B_{\text{var},p}=\frac{1}{\sqrt{2}\pi p\alpha\beta_{0}}.
\]
The functions $\exp(-b^{2})$ and $(1+2b)\exp(-2b)$ are, very roughly,
equal for $b\apprle1$. Thus, which effect is more important to calculating
the average current in an ensemble of rings, the suppression of the
cooperon governed by Eq. \ref{eq:CHPCTh_Iee} or the dephasing of
the oscillation due to variations in persistent current frequency
governed by Eq. \ref{eq:CHPCTh_EnsembleSuppression}, depends on the
relative magnitudes of the field scales $\gamma B_{c,p}$ and $B_{\text{var},p}$,
with the shorter field scale being the more significant. Writing out
the ratio of the two field scales in terms of the ring parameters,
one finds
\[
\frac{B_{\text{var},p}}{\gamma B_{c,p}}=\frac{\sqrt{\pi}}{\alpha\gamma\sin\theta}\frac{\sqrt{wt}}{L}.
\]
Typically, $w\sim t\sim L/20$, $\gamma\sim$1, and $\sin\theta\sim1/\sqrt{2}$
so that $B_{\text{var},p}/B_{c,p}\approx1/8\alpha$. Typically, $\alpha<1/8$
for current electron-beam tolerances,%
\footnote{Note that $\alpha$ is the fractional variation of the mean ring area.
Calculating the area of each ring averages over the small variations
in radius within a single ring.%
} so that the suppression of the cooperon dominates over these ring-to-ring
variations. We note, however, that this dephasing of the average current
due to geometric variations would suppress its total magnitude on
a field scale of $\sim1\,\text{T}$ for $\alpha=0.01$ and ring dimensions
of the order of those studied in this text. Thus, even in the absence
of cooperon suppression any average contribution to the current should
be strongly dephased for the measurements discussed in Chapter \ref{cha:Data},
which were performed at fields greater than $3\,\text{T}$.

Since this interaction contribution to the persistent current is strongly
suppressed for the measurements discussed in this text, we do not
consider the effects of temperature or spin in any detail. Over the
temperature range of experimental interest, it can be shown that $I^{ee}$
decays roughly exponentially with a characteristic temperature of
$\sim3E_{c}/k_{B}$, which is slightly smaller than that of the typical
current contribution ($\sim10E_{c}/k_{B}$ as seen in Eq. \ref{eq:CHPCTh_TpDiffusive})
\citep{ambegaokar1991comment,schmid1991persistent}. Spin effects
are less straightforward than in the previous derivations due to the
exchange term in the Hartree-Fock model for the interaction. It can
be argued that spin-orbit scattering does not affect $I^{ee}$ because
the spin-dependence of the interaction (through the exchange term)
results in only the spin-orbit independent term of expressions like
Eq. \ref{eq:AppGrFu_NuNuZSO} contributing \citep{akkermans2007mesoscopic}.
Because of this spin dependence, the dependence of $I^{ee}$ on Zeeman
splitting takes on a slightly complicated form.

Finally, we address the effect of electron-electron interactions on
the typical fluctuations of the persistent current. While some have
argued the typical current in the presence of interactions could be
as large as $I^{ee,\text{typ}}\sim ev_{F}/L$ \citep{eckern1992persistent,eckern1993stochastic},
it has generally been accepted that interactions do not change the
form of the typical current derived in Section \ref{sub:CHPCTh_TypicalCurrent}
\citep{smith1992systematic,eckern1995normalpersistent,bussemaker1997effectivefieldtheory,houzet2010distribution}.

\chapter{\label{cha:CHPrevWork}Review of previous work on persistent currents
in normal metal rings}

In this chapter, we discuss previous theoretical and experimental
work on persistent currents in normal metal rings. By my accounting,
there have been over 450 papers published which investigate persistent
currents theoretically, while there have been nine published measurements
of persistent currents.%
\footnote{I have cataloged many of these publications at http://www.citeulike.org/user/willshanks/tags/persistent-current.%
} This large discrepancy is due in part to the difficulty of measuring
persistent currents and to the surprising nature of some of the early
experimental findings. Additionally, theorists have found that the
topology of the persistent current system grants access to several
different kinds of physical phenomena, many of which remain unstudied
experimentally.

We will first review the major theoretical results in chronological
order and then discuss the experiments. In order to put the theoretical
work into the proper context, we first give a brief summary of the
experiments. The earliest measurement of the typical persistent current
in a single normal metal ring reported current magnitudes over one
order of magnitude larger than that expected from Eq. \ref{eq:CHPCTh_CurrCurrCorTempDependence}
\citep{chandrasekhar1991magnetic}. This large magnitude was not reproduced
in later experiments \citep{jariwala2001diamagnetic,bluhm2009persistent,bleszynski-jayich2009persistent}.
The earliest measurement of the average current in an ensemble of
normal metal (copper) rings reported a current magnitude somewhat
smaller than expected from Eq. \ref{eq:CHPCTh_IeeToroidalFieldSimple}
using the $\lambda_{\text{eff}}$ corresponding to a repulsive screened
Coulomb interaction. Additionally, the low-field sign of the current
was found to be diamagnetic, corresponding to an attractive electron-electron
interaction. This result was unexpected as none of the materials studied
was thought to possess electrons with an attractive interaction. These
findings were confirmed by experiments on different materials \citep{reulet1995dynamic,deblock2002acelectric,deblock2002diamagnetic}.
A large portion of the theoretical literature has been devoted to
the explanation of the large currents found in the first measurement
of individual rings as well as the low-field sign of the average current.

\section{\label{sec:ChPrevWork_theory}Overview of the theoretical literature}

Various different papers have been cited as the earliest precursor
to the phenomenon of persistent currents in a ring of a diffusive
system. A reasonable choice is Bohr's early work in 1913 on the quantization
of electron orbits in atoms and molecules \citep{bohr1913ion}. All
of the calculations of the persistent current in the preceding chapter
depend critically on the quantization of electron orbits, as the Bohr-van
Leeuwen theorem states that classically the magnetization of any system
of charged particles in thermal equilibrium is zero \citep{ashcroft1976solidstate,dingle1952somemagnetic}. 

The next step towards the prediction of persistent currents in solid
state systems was taken by Ehrenfest, who in 1925 conjectured that
the anomalous diamagnetism of bismuth could be explained by electron
orbits enclosing several atoms in the crystal lattice \citep{ehrenfest1925opmerkingen,ehrenfest1929bemerkungen}.
This conjecture was developed by Landau in 1930 into his theory of
diamagnetism \citep{landau1930diamagnetismus,peierls1933zurtheorie}.
Landau's theory was further developed in the 1930's by, among others,
Peierls \citep{peierls1933zurtheorie2} and Blackman \citep{blackman1938onthe}
to explain the de Haas-van Alphen effect (which is described briefly
from an experimental perspective in \ref{sec:CHTorsMagn_Intro}).
Landau diamagnetism and the de Haas-van Alphen effect are both magnetization
effects arising from the orbital motion of conduction electrons in
a solid. In a sense, they can be thought of as the analogues to the
persistent current (ring topology) for a singly-connected topology.

The earliest studies of the magnetization due the orbital motion of
electrons in a ring structure were performed by Pauling in 1936 \citep{pauling1936thediamagnetic}
and London in 1937 \citep{london1937quantum}. Both of these authors
were concerned with the anisotropic magnetization of aromatic compounds
such as benzene. The large diamagnetic susceptibility perpendicular
to the hexagonal lattice of such compounds was attributed to the {}``ring
currents'' of electrons free to move in an orbit encircling one hexagonal
unit of the lattice. Shortly afterward in 1938, Hund published the
first quantum mechanical investigation of the magnetization of a metal
system with a ring topology \citep{hund1938rechnungen}. In 1952,
Dingle elaborated upon Hund's work \citep{dingle1952somemagnetic}.
Both works considered two and three dimensional cylinders and treated
the effect of the magnetic field perturbatively to first order.

The flux quantum periodicity of electron interference effects was
first noted by Ehrenberg and Siday in 1949 \citep{ehrenberg1949therefractive}
and again ten years later by Aharonov and Bohm \citep{aharonov1959significance}.
The flux quantum periodicity of all equilibrium properties of a system
with a ring topology was shown by Byers and Yang in 1961 \citep{byers1961theoretical}.
Following this work, several authors (including Refs. \citealp{bloch1965offdiagonal,bloch1968fluxquantization,schick1968fluxquantization,gunther1969fluxquantization})
made generalizations to the normal state while studying persistent
currents and flux quantization in superconducting rings. A particularly
noteworthy work along these lines was the 1970 paper of Kulik in which
he obtained the one-dimensional persistent current expression given
in Eq. \ref{eq:CHPCTh_IperfectRing} and argued for the exponential
suppression with disorder given in Eq. \ref{eq:CHPCTh_1DDisorderedCurrent}.

The most important work in the history of persistent current research
was the 1983 paper of Landauer, Büttiker, and Imry \citep{buttiker1983josephson}.
In the Landauer formalism for microscopic electron conduction developed
by one of the authors, transport through a narrow conductor with dimensions
smaller than the electron phase coherence length $L_{\phi}$ is decomposed
into the transport through many transverse channels each characterized
by a transmission coefficient $t$. These transverse channels, similar
to those discussed in \ref{sub:CHPCTh_FiniteCrossSection}, act as
one dimensional conductors in parallel with each other. Electrons
entering the conductor from a reservoir at chemical potential $\mu_{1}$
on the left side of the conductor are transmitted to the reservoir
at chemical potential $\mu_{2}<\mu_{1}$ on the right side with probability
$|t|^{2}$ and are otherwise scattered elastically back into the first
reservoir. In either case, the kinetic energy of the electron is conserved.
In this picture, the Joule heating usually associated with the transport
of electrons through a disordered conductor in the normal state occurs
not in the conductor itself but in the reservoirs as the electrons
transported from left to right equilibrate from a distribution associated
with $\mu_{1}$ to one associated with $\mu_{2}$ through inelastic
processes in the right reservoir (as the electrons travel over distances
much larger than $L_{\phi}$ within the reservoir itself). Likewise,
the energy loss associated with transport through a normal conductor
occurs in the left reservoir as the higher energy states associated
with the electrons transported from left to right must be repopulated.
This picture of transport in which energy loss occurs in the leads
led the authors to consider what would happen if the reservoirs were
eliminated by closing the conductor upon itself in a ring geometry.
The authors conjectured that it was possible that a realistic disordered
ring with dimensions larger than the elastic scattering length $l_{e}$
but smaller than the inelastic length $L_{\phi}$ could sustain a
significant persistent current. This conjecture was further developed
by the authors in subsequent works \citep{buttiker1984quantum,buttiker1986fluxsensitive,buttiker1986quantum}.
In 1983, the importance of the distinction between elastic and inelastic
scattering was just becoming appreciated with the experiments of Sharvin
and Sharvin in 1981%
\footnote{A similar experiment had previously been performed in 1974 \citep{shablo1974quantization},
but it had been interpreted in terms of superconducting fluctuations
rather than single electron coherence.%
} \citep{sharvin1981magneticflux} and Washburn, Webb, \emph{et al}.
in 1984 and 1985 \citep{umbach1984magnetoresistance,webb1985observation}.
A review of the advances leading to the understanding of the difference
between elastic and inelastic processes in disordered conductors,
an understanding which was critical to the serious consideration of
persistent currents in the normal state, was written by Washburn and
Webb \citep{washburn1986aharonovbohm}.

Following the work of Landauer, Büttiker, and Imry, Riedel and colleagues
performed several analytical calculations using Green function methods
and numerical simulations using the tight-binding model to study the
effect of disorder and finite temperature on the typical magnitude
of the persistent current in the ballistic and diffusive regimes \citep{cheung1988isolated,cheung1988persistent,riedel1989persistent,cheung1989persistent,cheung1989energyspectrum,riedel1991persistent}.
Results similar to those of \ref{sub:CHPCTh_TypicalCurrent} for the
typical current magnitude and temperature dependence in the diffusive
regime were first presented in Ref. \citealp{cheung1989persistent}
in 1989. This calculation was refined by Riedel and von Oppen in 1993
by including additional Green function diagrams \citep{riedel1993mesoscopic}.
In 2010, the applicability of this calculation was expanded to the
regime of strong magnetic fields and strong spin-orbit scattering
by Ginossar \emph{et al}. for analysis of the measurements discussed
in this text \citep{ginossar2010mesoscopic}.

Montambaux, Bouchiat \emph{et al}. performed additional early numerical
studies in 1990 which predicted that the ensemble average of the persistent
current for non-interacting electrons in the diffusive regime had
an appreciable contribution with a half flux quantum $\phi_{0}/2$
periodicity \citep{montambaux1990persistent,bouchiat1991persistent}.
This result was corroborated analytically by several different groups
which each derived the result given in \ref{sub:CHPCTh_AvgSingleParticle}
\citep{schmid1991persistent,oppen1991average,altshuler1991persistent,imry1991persistent}.
However, this result for non-interacting electrons has largely been
insignificant as it is predicted to be much smaller than the contribution
to the average persistent current due to electron-electron interactions
(see \ref{sub:CHPCTh_AvgInteraction}) first calculated by Ambegaokar
and Eckern in 1990 \citep{ambegaokar1990coherence,eckern1991coherence}.
The previously cited work of Ginossar \emph{et al}. extended the calculation
of both of these contributions to large magnetic fields, where both
contributions are strongly suppressed \citep{ginossar2010mesoscopic}.

The magnitude of the persistent current is a random quantity which
in the diffusive regime depends on the microscopic details of the
disorder configuration. The average and typical values of the persistent
current characterize the first two moments of the statistical distribution
from which the current is drawn. In 1992, Eckern and Schmid argued
that in the presence of electron-electron interactions the typical
persistent current would be $I^{\text{typ}}\sim ev_{F}/L$ (with $L$
the circumference of the ring), a factor of $L/l_{e}$ larger than
the value in the non-interacting case \citep{eckern1992persistent}.
They also calculated the higher-order cumulants of the current and
found that cumulants of order $2n+1$ and $2n+2$ with $n>1$ were
suppressed by a factor of $N^{-n}$ where $N$ is the number of electrons
in the ring. In the same year, Smith and Ambegaokar challenged Eckern
and Schmid's calculation of the typical current in the presence of
interactions (arguing that the leading contribution to the typical
current was that of the non-interacting electron case) but confirmed
the form of the suppression of the higher order cumulants \citep{smith1992systematic}.
Smith and Ambegaokar also questioned the conclusion that the persistent
current followed a Gaussian distribution, citing the possibility of
a long tail to the distribution. Subsequently, Eckern and Schmid published
a new calculation in which they addressed the criticism of Smith and
Ambegaokar and confirmed their earlier result for the typical current
in the presence of electron-electron interactions \citep{eckern1993stochastic}.
Later in 1995, Eckern and Schwab described the determination of the
typical current in the presence of electron-electron interactions
as an open question due to the ambiguities involved in the previously
published calculations \citep{eckern1995normalpersistent}. In 1994,
Cattaneo \emph{et al}. found that the persistent current follows a
Gaussian distribution in the idealized one dimensional limit \citep{cattaneo1994statistics}.
In 1997, Bussemaker and Kirkpatrick used the nonlinear sigma model
to confirm the results of Smith and Ambegaokar for the cumulants of
the persistent current in the presence of interactions \citep{bussemaker1997effectivefieldtheory}.
In 2010, Houzet used the nonlinear sigma model to derive the full
distribution function for persistent currents in the diffusive regime,
finding a Gaussian distribution both with and without electron-electron
interactions \citep{houzet2010distribution}. This result is valid
for values of the current $|I|\ll\sqrt{g}I^{\text{typ}}$ where $g=2\pi E_{c}/\Delta_{M}$
is the dimensionless conductance. Also in 2010, Danon and Brouwer
calculated the leading correction in $g$ to the third order correlation
function $\langle I(\phi)I(\phi')I(\phi'')\rangle$, finding a correction
of order $(I^{\text{typ}})^{3}/g$ \citep{danon2010nongaussian}.
For reference, we note that $g\sim2.5\times10^{4}$ for sample CL17
of Table \ref{tab:ChData_Rings}, resulting in $\langle I(\phi)I(\phi')I(\phi'')\rangle^{1/3}/I^{\text{typ}}\sim0.03$.
In related work, several authors have investigated the statistics
of the individual energy levels \citep{akkermans1992conductance,simons1993universalities,szafer1993universal,braun1994universal,montambaux1997spectral},
and Feldmann \emph{et al}. calculated the distribution of the flux
dependent density of states in the limit $l_{e}\ll L$ \citep{feldmann2000distribution}.

The preceding discussion touches upon all of the results directly
relevant to analyzing the measurements discussed in this text. Before
concluding this section, however, we give a brief overview of some
of the other aspects of persistent currents which have been investigated
theoretically. Excluding the measurements discussed in Chapter \ref{cha:Data},
only the magnitude, low-field susceptibility, temperature dependence,
and flux periodicity of the current have been measured. The theoretical
predictions described below can be viewed as additional motivation
for our measurements. Although our measurements do not test the predictions
directly, they demonstrate a new method of persistent current detection
and raise the possibility of observing many of these to date untested
hypotheses.

As mentioned above, much theoretical effort has been put into attempting
to explain the large current magnitude reported in early measurements.
One strategy has been to calculate the non-equilibrium currents associated
with rings the presence of an externally applied, time-varying electromagnetic
field \citep{trivedi1988mesoscopic,efetov1991dynamic,janssen1993linearresponse,marchesoni1993persistent,kravtsov1993directcurrent,kravtsov1993disorderinduced,aronov1993nonlinear,reulet1994acconductivity,genkin1994resonance,galperin1996nonequilibrium,berman1997quantum,kopietz1998nonlinear,keller2000angularmomentum,yudson2001limitsof,chalaev2002aharonovbohm,yudson2003electron,arrachea2004dcresponse,matos-abiague2005photoinduced,dajka2006theinfluence,cohen2006rateof,kulik2007mesoscopic,moskalenko2007nonequilibrium}.
A ring studied experimentally could potentially be subject to such
electromagnetic radiation through the back-action of the measurement
apparatus. Following the work of Ambegaokar and Eckern mentioned above
\citep{ambegaokar1990coherence,eckern1991coherence,eckern1992persistent,smith1992systematic,eckern1993stochastic},
many authors have investigated the possibility of additional contributions
to the persistent current due to electron-electron interactions \citep{muller-groeling1993interacting,kopietz1993universal,kopietz1993universal2,vignale1994comment,kopietz1994kopietz,altland1994comment,kopietz1994kopietz2,kato1994suppression,muller-groeling1994persistent,wendler1995persistent,berkovits1995significant,jeon1996persistent,giamarchi1995persistent,ramin1995electronelectron,li1996exactly,schmitteckert1996phasecoherence,stafford1997interactioninduced,schechter2003magnetic,eckern2004comment,schechter2004schechter,chau2004linearmagnetic}.
Others have considered the role of magnetic impurities, mostly finding
that they suppress the current \citep{yoshioka1993fluctuations,eckern1993persistent,schwab1996impurity,schwab1996impurity2,hausler1996persistent}.
However, Schwab and Eckern have identified a large contribution to
the average current due to magnetic impurities \citep{schwab1997persistent}.
More recently, Bary-Soroker and colleagues have predicted that a dilute
amount of magnetic impurities could totally quench superconductivity
in a metal while only weakly suppressing the effect of the attractive
BCS interaction on the persistent current (as calculated by Ambegaokar
and Eckern \citep{bary-soroker2008effectof,bary-soroker2009pairbreaking}).
Various other mechanisms, including effects related to the nature
of the disorder potential, have also been considered as possible explanations
of the large observed magnitude of the persistent current \citep{maynard1992apossible,buttiker1994characteristic,kirczenow1995whyare,unnikrishnan1996persistent,kusmartsev1997nelectron,stelzer2000mesoscopic,apel2000enhancement,tomita2002persistent,chowdhury2008largediamagnetic,chen2008persistent,feilhauer2010conductance}.

Several authors have considered the behavior of persistent currents
in systems beyond the simplest case of a ring in a uniform magnetic
field. Some studies have been done on the persistent current in rings
connected to leads \citep{buttiker1986fluxsensitive,akkermans1991relation,buttiker1994characteristic,bandopadhyay2004quantum,chowdhury2008largediamagnetic}
and in networks of many connected rings \citep{pascaud1997magnetization,pascaud1999persistent}.
Others have considered normal-superconducting hybrid rings \citep{imry1998mesoscopic,cayssol2003isolated}.
Additionally, it has been predicted that the interaction of a non-uniform
magnetic field with the spin of the electrons in a ring can lead to
a Berry phase and associated persistent current distinct from the
one considered in this text \citep{loss1992persistent,gao1993aharonovanandan,kawabata1999berryphase}.

The work discussed so far has had potential applications to persistent
currents in the diffusive regime appropriate for metal rings. Persistent
currents have also been studied extensively in the ballistic regime
and the localized regime (mainly in the framework of the Hubbard model
in, e.g., Refs. \citep{yu1992persistent,fujimoto1993persistent,giamarchi1995persistent,r?mer1995enhanced,schmitteckert1998fromthe,gambetti-cesare2002disorderinduced}).
The ballistic regime is of particular interest for this text because
it is realizable in semi-conductor heterostructures for which a current
magnitude similar to that observed in normal metal rings is possible.
It should be possible to integrate semiconductor rings onto cantilevers
and to employ the measurement technique discussed in Chapter \ref{cha:Cantilever-torsional-magnetometry}
to study their persistent currents. 

Many of the works cited above are also applicable to the ballistic
(or localized) case. Several interesting phenomena unique to the ballistic
case have also been studied. Many authors have considered rings with
integrated quantum dots, either within the ring or side-coupled to
it. Such systems have been identified as test-systems for studying
Kondo physics \citep{zvyagin1996persistent,zvyagin1996finitesize,ferrari1999kondoresonance,eckle2000kondoimpurity,kang2000mesoscopic,zvyagin2001comment,kang2001kangand,affleck2001detecting,hu2001mesoscopic,eckle2001kondoresonance,affleck2002comment,eckle2002eckleet,cho2001spinfluctuation,anda2002kondoeffect,luo2002magnetic,s?rensen2005kondoscreening},
adiabatic quantum pumping \citep{moskalets2003hiddenquantum,moskalets2003quantum,zhou2005geometric},
a noise-induced quantum phase transition \citep{tong2006signatures},
and quantum zero-point fluctuations \citep{cedraschi2000zeropoint,cedraschi2001zeropoint,semenov2010persistent}.
Ballistic rings have also been proposed as potential qubits for quantum
computation \citep{barone2002quantum,kulik2003persistent,kulik2003quantum,kulik2003quantum2,kulik2004spontaneous,zipper2006fluxqubit,szopa2006coherent,kurpas2007decoherence,giorgi2007quantum,kurpas2008coherent,szelag2008persistent,szopa2010fluxqubits}
and more broadly as model systems for the study of non-classical light,
quantum gambling, and the quantum Smoluchowski regime \citep{berman1997quantum,dajka2004persistent,dajka2005theinfluence,tsomokos2005electron,dajka2006theinfluence,paku&lstrok;a2007quantum,kurpas2007decoherence,kurpas2008coherent,kurpas2009entanglement}.
Quite recently, Bary-Soroker \emph{et al}. have studied the transition
of the persistent current from the ballistic to the diffusive regime
\citep{bary-soroker2010persistent}.

While the work discussed above is most directly applicable to the
measurements discussed in this text, persistent currents have also
been studied in other regimes. The study of persistent currents in
superconductors is too broad of an area of research to discuss here,
but we note that persistent currents have been measured experimentally
in rings of similar size to those measured in this text, both in rings
made entirely of superconducting material \citep{zhang1997susceptibility}
and in rings consisting of superconducting material interrupted by
Josephson junctions \citep{wal2000quantum}. In superconductors in
their normal state, persistent currents due to thermal and quantum
fluctuations have been studied theoretically \citep{ambegaokar1990nonlinear,oppen1992fluxperiodic,schwiete2009persistent},
with the persistent currents due to thermal fluctuations having also
been measured experimentally \citep{koshnick2007fluctuation}. Persistent
currents have been studied theoretically \citep{beenakker1991influence,pietilainen1993interactingelectron,viefers2004quantum}
and observed experimentally with torsional magnetometry \citep{kleemans2007oscillatory}
in few-electron quantum rings. Other regimes for which persistent
currents have been studied theoretically but as yet unprobed experimentally
include the Luttinger liquid \citep{loss1992parityeffects,loss1993absence,schmeltzer1993persistent},
the quantum Hall state \citep{avishai1993persistent,eliyahu1994mesoscopic,georgiev2004magneticmoment},
carbon nanotubes \citep{odintsov1999persistent,latil2003persistent,szopa2004coherence,marganska2005orbital},
graphene \citep{cotaescu2007signatures,moskalenko2007nonequilibrium,bellucci2010induced},
and topological insulators \citep{michetti2010bound}.

For reference, we conclude this summary of previous work by noting
that persistent currents in the normal state have been the subject
of several reviews and introductory articles \citep{levi1992experiments,eckern1994sinddauerstrome,imry1994recentdevelopments,eckern1995normalpersistent,montambaux1997spectral,montambaux1997spectral2,noat1999signature,kulik2003persistent,saminadayar2004equilibrium,viefers2004quantum,akkermans2007mesoscopic,bouchiat2008newclues,birge2009sensing,wilson2009sensitive,buttiker2010fromanderson}.

\section{Previous measurements of persistent currents}

We now review the measurements of persistent currents reported prior
to those discussed in this text. The experiments may be divided into
two types, measurements of the typical current and measurements of
the average current. We discuss these two types separately rather
than reviewing all of them in chronological order. Some of the measurements,
the first two in particular \citep{levy1990magnetization,chandrasekhar1991magnetic},
were published prior to the most plausible theoretical prediction
describing them. Rather than summarizing the interpretations of these
experiments given by the authors, we will discuss their results in
the context of the theoretical framework reviewed in Chapter \ref{cha:CHMeso_}.
We know of no proposed theory that accurately describes all of the
reported measurements. Besides the measurements discussed below, persistent
currents have been observed in superconducting rings above the superconducting
transition temperature \citep{koshnick2007fluctuation} and in self-assembled
InAs quantum rings \citep{kleemans2007oscillatory}.

\subsection{Measurements of the typical current}

The measurement of the typical persistent current with $h/e$ flux
periodicity was reported by Chandrasekhar \emph{et al}. in 1991 \citep{chandrasekhar1991magnetic}.
Measurements were performed on single gold rings with circumferences
$L$ of 7.5, 8.0, and $12.6\,\text{\ensuremath{\mu}m}$ at a base
temperature of $4.5\,\text{mK}$. The current in the two smaller rings
was observed to decrease with temperature, consistent with an exponential
decay with a characteristic temperature of $T_{0}=22\,\text{mK}$.
Using Eq. \ref{eq:CHPCTh_TpDiffusive} for the characteristic temperature
$T_{p=1}$ of a diffusive ring, we find values of $0.0156$ and $0.0178\,\text{m}^{2}/\text{s}$
for the diffusion constant $D=L^{2}k_{B}T_{0}/10.4\hbar$ in these
samples. Using $v_{F}=1.4\times10^{6}\,\text{m/s}$ for gold\citep{ashcroft1976solidstate},
these values of $D$ correspond to elastic mean free paths $l_{e}=3D/v_{F}$
of 34 and $38\,\text{nm}$ respectively, slightly smaller than the
value of $70\,\text{nm}$ measured by Chandrasekhar \emph{et al}.
in codeposited wires. From Eqs. \ref{eq:CHPCTh_IpTypTBothSpins} and
\ref{eq:CHPCTh_gDExponential}, the expected current magnitude $I=\sqrt{2}(1.11eD/L^{2})\,\exp(-T/T_{0})$
for both samples at $4.5\,\text{mK}$ is $57\,\text{pA}$. Using the
mean value of $D$ from the two small samples and the associated $T_{0}=10.4\hbar D/k_{B}L^{2}=8.3\,\text{mK}$,
the expected current magnitude for the for the $12.6\,\text{\ensuremath{\mu}m}$
sample at $4.5\,\text{mK}$ is $11\,\text{pA}$. All of these values
stand in striking contrast to the reported values of 6 and $30\,\text{nA}$
for the two smaller rings and $3\,\text{nA}$ for the larger one.
Due to the variability of the observed background in this experiment
and the substantial filtering of the data presented, it is difficult
to assign a signal to noise ratio to the measured persistent current
magnitude. However, I would estimate it to be no larger than 4.

Following the experiment of Chandrasekhar \emph{et al}., Mailly \emph{et
al}. measured persistent currents in a single ring etched into the
two dimensional electron gas of a GaAs/GaAlAs heterostructure \citep{mailly1993experimental,mailly1994persistent}.
The ring was etched with leads so that the ring's conductance and
magnetization could be measured simultaneously. The mean circumference
of the ring was $L\sim8.5\,\text{\ensuremath{\mu}m}$, while the elastic
mean free path measured via transport was $l_{e}\sim11\,\text{\ensuremath{\mu}m}$,
placing this sample in the ballistic regime $L<l_{e}$ described in
\ref{sec:CHPCTh_1DRing} and \ref{sub:CHPCTh_IdealRingFiniteTemperature}.
For this two dimensional ring, the number $M=2w/\lambda_{F}$ of transverse
channels was $\sim8$ for the reported width $w=160\,\text{nm}$ and
Fermi wavelength $\lambda_{F}=42\,\text{nm}$. Repeating the analysis
of \ref{sub:CHPCTh_FiniteCrossSection} for two dimensions, one can
show that the typical amplitude of the $p^{th}$ harmonic of the current
(the two dimensional analogue of Eq. \ref{eq:CHPCTh_3DTypCurrentFiniteT})
is
\begin{equation}
I_{p,2D,\text{ballistic}}^{\text{typ}}=2\times\sqrt{\frac{2M}{3}}\frac{4}{\pi p}I_{0}g_{M,2D}\left(\frac{T}{T_{p}}\right)\exp\left(-\frac{pL}{2l_{e}}\right)\label{eq:PrevWork_I2Dballistic}
\end{equation}
with
\[
g_{M,2D}\left(x\right)=\sqrt{\frac{3}{2}x^{2}\int_{0}^{1}dy\,\text{csch}^{2}\left(\frac{x}{\sqrt{1-y^{2}}}\right)}
\]
and $T_{p}$ the single-channel characteristic temperature of Eq.
\ref{eq:CHPCTh_PerfectRingTp}. Additionally, in the expression for
$I_{p,2D,\text{ballistic}}^{\text{typ}}$ we have included the correction
for finite elastic scattering length $l_{e}$ first introduced in
Eq. \ref{eq:CHPCTh_FourierTransformLorentzian} and a factor of 2
for spin. Using the reported value $v_{F}=2.6\times10^{5}\,\text{m/s}$
(and corresponding $T_{p=1}=37\,\text{mK}$), one finds an expected
value of $I_{p,2D,\text{ballistic}}^{\text{typ}}=18\,\text{nA}$ for
the fundamental $h/e$ periodic component of the current,%
\footnote{In the analysis of Refs. \citealp{mailly1993experimental,mailly1994persistent},
the expected current is listed as half this value with no mention
of including spin degeneracy.%
} which is a bit larger than the measured value $4\pm2\,\text{nA}$.
During the measurement, the disorder configuration of the ring changed
due to slow relaxation processes in the semiconductor. This effect
allowed measurements of the current to be made for a few different
realizations of disorder but could also have introduced some averaging
over disorder within a single measurement. The change of the disorder
configuration was monitored by measuring the $h/e$ transport signal
at the same time as the measurement of the persistent current.%
\footnote{Several theoretical works have investigated the effects of a transport
current on the behavior of the persistent current, including Refs.
\citealp{jayannavar1994persistent,jayannavar1995persistent,pareek1995effectof}.
It goes beyond the scope of this text to analyze the effects of the
transport current in the experiment of Mailly \emph{et al}.%
} The authors attributed some of the discrepancy in the current magnitude
to the small number of independent samples of the current magnitude.
The temperature dependence of the current was not studied. Due to
the large low frequency feature in the measured persistent current
spectrum of Mailly \emph{et al}., it is difficult to define a signal
to noise ratio from the results presented in Refs. \citealp{mailly1993experimental,mailly1994persistent},
but it appears to be no greater than 2.

In 2001, Jariwala \emph{et al}. published a study of the magnetization
of $N=30$ gold rings similar in dimensions ($L=8\,\text{\ensuremath{\mu}m}$)
to those studied individually by the same group in 1991 \citep{jariwala2001diamagnetic}.
Oscillatory currents with flux periodicities of both $h/e$ and $h/2e$
were observed. Focusing first on the $h/e$ signal, the observed characteristic
temperature $T_{h/e}$ was $166\,\text{mK}$, which corresponds to
a diffusion constant $D=L^{2}k_{B}T_{h/e}/10.4\hbar$ of $0.134\,\text{m}^{2}/\text{s}$
following Eq. \ref{eq:CHPCTh_TpDiffusive}. This diffusion constant
corresponds to a current $I_{h/e}=\sqrt{2}(1.11eD/L^{2})\,\exp(-T/T_{0})$
of $510\,\text{pA}$ at $5.5\,\text{mK}$, in reasonable agreement
with the measured value of $350\,\text{pA}$ per ring. This value
of the current per ring was obtained by scaling the total magnetization
by $1/\sqrt{N}$ because the current is expected to vary randomly
in sign from ring to ring. On a separate wire with the same cross-section
(thickness $t=60\,\text{nm}$) as the rings, the sheet resistance
$R_{\square}$ was measured to be $0.15\,\Omega/\square$, which by
Eq. \ref{eq:AppTransport_DiffusionConstant} corresponds to a diffusion
constant of $D=0.131\,\text{m}^{2}/\text{s}$,%
\footnote{This number differs from the value quoted in Ref. \citep{jariwala2001diamagnetic}.
The measured resistivity $\rho=tR_{\square}$ was $9\times10^{-9}\,\Omega\,\text{m}$.
The density of states per unit volume was $\eta=(2\Delta V)^{-1}=3.3\times10^{46}\,\text{J}^{-1}\text{m}^{-3}$
for the reported ring volume $V=wtL=5.76\times10^{-20}\,\text{m}^{-3}$
and mean level spacing $\Delta=k_{B}\times19\,\text{\ensuremath{\mu}K}$.
The value of $D$ given above was found using these numbers and Eq.
\ref{eq:AppTransport_DiffusionConstant}.%
} in agreement with the value inferred from the $h/e$ persistent current
signal. Due to the substantial filtering of the data presented, it
is difficult to assign a signal to noise ratio to the measured $h/e$
persistent current magnitude.

Analysis of the $h/2e$ signal from the array of gold rings is more
problematic. From the value of $D$ found for the $h/e$ signal, the
expected characteristic temperature of the second harmonic of the
typical current $T_{p=2}=T_{h/e}/4$ is $42\,\text{mK}$ while the
characteristic temperature $T_{ee}=3\hbar D/k_{B}L^{2}$ of the interaction
contribution to the average current discussed in \ref{sub:CHPCTh_AvgInteraction}
is $48\,\text{mK}$. In contrast, the decay of the $h/2e$ signal
observed by Jariwala \emph{et al}. displayed a characteristic temperature
of $T_{h/2e}=89\,\text{mK}$. The expected magnitude $I_{p=2}=(1.11eD/2^{1.5}L^{2})\,\exp(-T/T_{p=2})$
of the second harmonic of the typical current at $5.5\,\text{mK}$
was $115\,\text{pA}$, giving a typical contribution to the total
$h/2e$ current signal for the entire array of $\sqrt{30}\times115\,\text{pA}=0.63\,\text{nA}$.
Assuming a repulsive Coulomb interaction, the expected magnitude $I^{ee}=8\lambda_{\text{eff}}eD/\pi L^{2}\exp(-T/T_{ee})$
of the interaction contribution to the average $h/2e$ signal at $5.5\,\text{mK}$
was $41\,\text{pA}$, giving an average contribution to the total
$h/2e$ current signal for the entire array of $30\times41\,\text{pA}=1.22\,\text{nA}$.%
\footnote{\label{fn:CHPrevWork_lambdaEff}Here following Refs. \citealp{eckern1991coherence}
and \citealp{akkermans2007mesoscopic}, we use the formula $\lambda_{\text{eff}}=\ln(\varepsilon_{F}e^{\lambda_{0}^{-1}}/E_{c})$
with $\lambda_{0}=x^{2}\ln(1+4/x^{2})/8$ and $x=0.81\sqrt{(r_{s}/a_{0})(m^{*}/m)}$.
The quantity $r_{s}/a_{0}$ is the radius of a sphere with volume
equal to the volume of the ring divided by the number of conduction
electrons and normalized by the Bohr radius (see Table 1.1 of Ref.
\citealp{ashcroft1976solidstate}), $\varepsilon_{F}$ is the Fermi
energy (see Table 2.1 of Ref. \citealp{ashcroft1976solidstate}),
and $m^{*}/m$ is the ratio of the effective mass to the bare electron
mass (see Table 2.3 of Ref. \citealp{ashcroft1976solidstate}). This
expression is valid assuming a screened Coulomb repulsion between
the electrons. With the values from the tables in Ref. \citealp{ashcroft1976solidstate}
and the $E_{c}$ inferred from the $h/e$ data of the array of gold
rings, one finds $\lambda_{\text{eff}}=0.053$.%
} The expected values of $I_{p=2}$ and $I^{ee}$ are consistent with
the measured $1.98\,\text{nA}$ for the total $h/2e$ current signal
of the array at $5.5\,\text{mK}$. However, the sign of the total
$h/2e$ current was found to be diamagnetic at low field, while a
paramagnetic sign is expected for the Coulomb interaction. Although
the typical contribution to the $h/2e$ current has a random sign,
a diamagnetic fluctuation of $I_{p=2}$ of five times its typical
value is required to mask the expected paramagnetic value of $I^{ee}$
and match the data.%
\footnote{If one infers a value of $D$ from the observed $T_{h/2e}$ and either
the expression for $T_{p=2}$ or $T_{ee}$, one finds larger expected
values for the typical current $I_{p=2}\sim266\,\text{pA}$ ($D$
from $T_{p=2}$) and interaction current $I_{ee}\sim80\,\text{pA}$
($D$ from $T_{ee}$), giving $\sqrt{30}\times I_{p=2}=1.5\,\text{nA}$
and $30\times I_{ee}=2.4\,\text{nA}$. With these numbers it is slightly
more plausible that the observed sign could be due to insufficient
averaging of the typical fluctuations of the $h/2e$ current signal.
However, there is no theoretical justification for assuming a different
value for $D$ for the $h/2e$ current contributions compared to that
observed for the $h/e$ contribution.%
} The $h/2e$ signal was observed to decay consistently with the form
given in Eq. \ref{eq:CHPCTh_Iee} with $B_{c,p=1}=0.51\,\text{mT}$,
roughly a factor of 10 smaller than the value of $B_{c,p=1}$ expected
from Eq. \ref{eq:CHPCTh_BcpToroidalField}. This discrepancy could
possibly be due to the geometrical correction factor between the toroidal
field model and the orientation of the magnetic field (perpendicular
to the plane of the rings) in the experiment. As with the $h/e$ signal,
due to the substantial filtering of the data presented, it is difficult
to assign a signal to noise ratio to the measured $h/2e$ persistent
current magnitude.

Also in 2001, Rabaud \emph{et al}. reported measurements on an array
of 16 connected GaAs/GaAlAs rings similar to the single GaAs/GaAlAs
ring studied in 1993 by some of the same authors \citep{rabaud2001persistent}.
The ring array sample was fabricated in such a way that all of the
rings could be isolated from each other by applying a gate voltage.
For these rings, the perimeter $L=12\,\text{\ensuremath{\mu}m}$ was
slightly larger than the measured elastic mean free path $l_{e}=8\,\text{\ensuremath{\mu}m}$,
putting this sample closer to the diffusive regime than the previously
measured single semiconductor ring. Using the expression in Eq. \ref{eq:PrevWork_I2Dballistic}
for the current expected for the ballistic regime gives a much larger
typical current ($\sim25\,\text{nA}$) than the measured $0.35\,\text{nA}$
per ring at $20\,\text{mK}$. The expression for the diffusive ring
$I=2\times\sqrt{2}(1.11eD/L^{2})\,\exp(-T/T_{0})$ gives an expected
current of $4.3\,\text{nA}$ for $D=v_{F}l_{e}/2=1.26\,\text{m}^{2}/\text{s}$
and $T_{0}=10.4\hbar D/L^{2}=695\,\text{mK}$. The authors cited the
finite phase coherence length $L_{\phi}=20\,\text{\ensuremath{\mu}m}$
and the geometry of the square-shaped rings as possible explanations
of the smaller magnitude of the current. It is noteworthy that, because
of instabilities in the semiconductor leading to changes in the disorder
configuration, hundreds of independent measurements of the current
were able to be taken. As with the single semiconductor ring, the
instability of the disorder configuration could also have slightly
reduced the magnitude of the measured current. Although it is not
directly relevant to the measurements of this text, we note that the
measured magnitude of the current in the connected rings, $0.40\,\text{nA}$
per ring, was slightly larger than for the disconnected rings while
expected to be a factor of 0.58 smaller. The signal to noise ratio
was $\sim5$.

The four measurements described above constituted the entire body
of published experimental research on the typical persistent current
when the work detailed in this text was undertaken. Shortly before
the measurements of Chapter \ref{cha:Data} were published in 2009,
Bluhm \emph{et al}. published a third study of the typical current
in gold rings \citep{bluhm2008magnetic,bluhm2009persistent}. For
rings with $L=4.2\,\text{\ensuremath{\mu}m}$, the persistent current
was observed to decay exponentially on a scale of $380\,\text{mK}$,
in good agreement with the characteristic temperature $T_{p=1}=10.4\hbar D/k_{B}L^{2}=402\,\text{mK}$
expected for the value of the diffusion constant $D=0.09\,\text{m}^{2}/\text{s}$
found in transport measurements on codeposited wires. The typical
current magnitude $I_{h/e}=0.9\,\text{nA}$ observed in measurements
of 15 such rings at $150\,\text{mK}$ was also in good agreement with
the magnitude expected $I_{p=1}=\sqrt{2}1.11eD/L^{2}\exp(-T/T_{p=1})=0.88\,\text{nA}$
for this diffusion constant. Similar current magnitudes were observed
in measurements on four rings with $L=3.6\,\text{\ensuremath{\mu}m}$.
Due to the need to remove the ensemble-averaged background signal,
no inference of the average current could be made. It is noteworthy
that the base temperature of this measurement was an order of magnitude
larger than that of the earlier measurements due to heating of the
samples from the $10\,\text{GHz}$ Josephson current in the SQUID
detector pickup loop. The signal to noise ratio of this measurement
at base temperature was $\sim6$.

Summarizing the measurements of the typical persistent current prior
to the work of this text, we find that the two measurements in semiconductor
rings reported current magnitudes close to, but slightly smaller than,
that expected by theory while two of the three measurements on gold
rings reported $h/e$ currents in agreement with theory. All of these
measurements employed various forms of SQUID magnetometers. No generally
accepted explanation for the contradiction between the earliest measurement
on gold rings and the two subsequent measurements has been proposed.
Additionally, we note that at least in the experiment of Bluhm \emph{et
al}. the high frequency electromagnetic signal of the SQUID detector
was observed to have an effect on the sample. The absolute signal
to noise ratio in all of these experiments was quite small, ranging
from $\sim4-6$.

\subsection{Measurements of the average current}

The first measurement of persistent currents in normal metal rings
was reported by Levy \emph{et al}. in 1990 on an array of $10^{7}$
copper rings of circumference $L=2.2\,\text{\ensuremath{\mu}m}$ \citep{levy1990magnetization,levy1991persistent}.
The magnetization signal, which oscillated with an $h/2e$ flux periodicity,
was observed to decay exponentially with temperature on a characteristic
scale $T_{h/2e}=80\,\text{mK}$. Using the form for the characteristic
temperature $T_{ee}=3\hbar D/L^{2}$ of the interaction contribution
to the average current discussed in \ref{sub:CHPCTh_AvgInteraction},
one finds that the observed $T_{h/2e}$ corresponds to a diffusion
constant of $D=0.017\,\text{m}^{2}/\text{s}$, in good agreement with
the value found from transport measurements on codeposited wires.%
\footnote{Note that the value of $l_{e}$ was revised in Ref. \citealp{levy1991persistent}
from the value first reported in Ref. \citealp{levy1990magnetization}.
The revised value of $30\,\text{nm}$ corresponds to a diffusion constant
of $D=v_{F}l_{e}/3=0.016\,\text{m}^{2}/\text{s}$, using $v_{F}=1.57\times10^{6}\,\text{m/s}$
from Ref. \citealp{ashcroft1976solidstate}.%
} This value of $D$ leads to an expected average current due to the
repulsive Coulomb interaction $I^{ee}=(8\lambda_{\text{eff}}eD/\pi L^{2})\exp(-k_{B}T/T_{ee})$
of $71\,\text{pA}$ at $7\,\text{mK}$,%
\footnote{This estimate assumes $\lambda_{\text{eff}}=0.054$ following the
procedure described in note \vref{fn:CHPrevWork_lambdaEff}.%
} a bit smaller than the observed $400\,\text{pA}$. However, the low-field
sign of the current was found to be diamagnetic, while, as noted above
and in \ref{sub:CHPCTh_AvgInteraction}, a paramagnetic current is
expected for a repulsive electron-electron interaction. The signal
to noise ratio of this measurement was $\sim7$.

All of the measurements discussed so far were performed using SQUID
magnetometers operated at frequencies of order $10\,\text{Hz}$ (though
high frequency electromagnetic signals were still present due to the
Josephson oscillations of the SQUIDs). The remaining two studies were
performed by inductively coupling arrays of rings to a high Q superconducting
resonator. The resonant frequencies of the resonators were all over
$200\,\text{MHz}$. As noted in \ref{sec:ChPrevWork_theory} and by
the authors themselves, it is possible that the presence of high frequency
electromagnetic radiation could produce non-equilibrium effects in
the rings which behave similarly to the equilibrium persistent current,
complicating the analysis.

Two similar measurements of the persistent currents in arrays of $10^{5}$
GaAs/GaAlAs rings were reported first by Reulet \emph{et al}. in 1995
\citep{reulet1995dynamic} and later by Deblock \emph{et al}. in 2002
\citep{deblock2002acelectric}. The same group later noted (see note
16 in Ref. \citep{deblock2002diamagnetic}) that electric and magnetic
responses of the rings were not well separated in the earlier measurement
of Reulet \emph{et al}., so we focus on the results of Deblock \emph{et
al}. Qualitatively the results of the two measurements were similar,
with the second set of measurements reporting an $h/2e$ periodic
current $\sim5$ times smaller. For the measurements of Deblock, the
rings were in the intermediate regime between ballistic and diffusive
motion of the electrons with $L=5.2\,\text{\ensuremath{\mu}m}$ and
$l_{e}=3\,\text{\ensuremath{\mu}m}$. For the ideal ring, we found
in Eqs. \ref{eq:CHPCTh_IpN0coeff} and \ref{eq:CHPCTh_IpN2coeff}
a current magnitude $I_{p=2}$ which was independent of electron number
and could be written as $I_{p=2}=(4/\pi)\Delta_{1}/\phi_{0}$ where
$\Delta_{1}$ is the mean level spacing of the single channel ring
(including a factor of 2 for spin). The mean level spacing averaged
over all the transverse channels (in two dimensions) is $\Delta_{1,M}=\pi\Delta_{1}/4$.
Taking all transverse channels into account, one would then expect
an average current in the ballistic ring at zero temperature of $I_{p=2}=M(\Delta_{1}/\phi_{0})\exp(-L/l_{e})$.
However, using the values of Ref. \citep{deblock2002acelectric} ($v_{F}=2.2\times10^{5}\,\text{m/s}$
and $w=200\,\text{nm}$) gives an expected current of $\sim10\,\text{nA}$,
much greater than the observed $0.25\,\text{nA}$ per ring at $20\,\text{mK}$.
The expected average current $I^{\text{can}}=8\Delta_{M}/\pi\phi_{0}$
in the diffusive, non-interacting electron regime described by Eq.
\ref{eq:CHPCTh_AverageCurrentSingleLevel} predicts a current of $0.7\,\text{nA}$
at zero temperature (following Ref. \citealp{ambegaokar1991comment},
this contribution is not expected to be strongly attenuated for $k_{B}T/E_{c}\sim0.25$
as is the case for this measurement; this contribution is more significant
for the semiconductor rings than the metal ones because of the larger
mean level spacing). The expected average current contribution $I^{ee}=(8\lambda_{\text{eff}}eD/\pi L^{2})\exp(-k_{B}T/T_{ee})$
due to the repulsive Coulomb interaction is $0.36\,\text{nA}$.%
\footnote{The scale factor $\lambda_{\text{eff}}$ is once again calculated
following the procedure of note \vref{fn:CHPrevWork_lambdaEff}. In
the expression for $\lambda_{0}$ given there, we use $x=\lambda_{F}/2\pi\lambda_{s}$,
where $\lambda_{s}$ is the Thomas Fermi screening length specified
to be $16\,\text{nm}$ in Ref. \citep{deblock2002acelectric}. The
calculated value for $\lambda_{\text{eff}}$ is then 0.077.%
} So both the diffusive non-interacting and the interacting electron
contributions to the average current are on the same order as the
observed current. However, once again the low field sign of the current
was observed to be diamagnetic, while both of these contributions
correspond to paramagnetic behavior.

The final measurement of the average persistent current was reported
by Deblock \emph{et al}. in 2002 for an array of $10^{5}$ silver
rings with circumference $L=4\,\text{\ensuremath{\mu}m}$. The $h/2e$
current signal was observed to decay on a temperature scale of $39\,\text{mK}$,
corresponding to a diffusion constant of $D=k_{B}T_{ee}L^{2}/3\hbar=0.027\,\text{m}^{2}/\text{s}$
in rough agreement with the value ($0.018\,\text{m}^{2}/\text{s}$)
measured in codeposited wires. At $T=40\,\text{mK}$, the expected
current $I^{ee}=8\lambda_{\text{eff}}eD/\pi L^{2}\exp(-T/T_{ee})=13\,\text{pA}$
for the inferred diffusion constant is smaller than the observed current
$I_{h/2e}=330\,\text{pA}$.%
\footnote{This estimate assumes $\lambda_{\text{eff}}=0.053$ following the
procedure described in note \vref{fn:CHPrevWork_lambdaEff}.%
} Additionally, the low field sign of the current was observed to be
diamagnetic, in contradiction to the expectation of a repulsive Coulomb
interaction between the electrons.

Including the $h/2e$ signal measured by Jariwala \emph{et al}. in
the array of gold rings, the low field sign of the average persistent
current has been observed to be diamagnetic in Cu, Au, GaAs, and Ag
rings, in contradiction to the sign predicted both for non-interacting
electrons and for electrons with a repulsive interaction. As described
in \ref{sec:ChPrevWork_theory}, non-equilibrium effects and an attractive
electron-electron interaction have been proposed as possible explanations
of these observations. An attractive electron-electron interaction
would be surprising as none of these materials has been observed to
superconduct. Additionally, the measured current in each of the samples
was somewhat larger in magnitude than that expected for the current
due to the Coulomb interaction. These results highlight the need for
further study. While the measurements described in this text do not
detect the average persistent current, they demonstrate a measurement
technique that could be applied to the study of the average persistent
current. Finally, we note that the difficulty in interpreting the
experimental results for the average persistent currents results an
advantage on the measurements described in Chapter \ref{cha:Data}
over previous work. Because we operate in a regime of magnetic field
where all contributions to the average persistent current are strongly
suppressed, we can analyze our measurements, and our observed $h/2e$
signal in particular, assuming that the current is entirely due to
the non-interacting diffusive contribution and ignoring the poorly
understood average contributions.

\chapter{\label{cha:Cantilever-torsional-magnetometry}Cantilever torsional
magnetometry}

\section{\label{sec:CHTorsMagn_Intro}Introduction to cantilever torsional
magnetometry}

Cantilever torsional magnetometry has a long history and has been
applied to the study of many different physical problems. The origins
of torsional magnetometry can be traced back to the development of
the compass which took place in China no later than the eleventh century
. These early compasses took the form of a floating magnet in a bowl
of water or a magnetic needle suspended by a long, thin thread, a
form not unlike modern torsional magnetometers. In the thirteenth
century, Petrus Peregrinus wrote the earliest text to describe the
use of a freely pivoting magnet with a graduated circle to study the
properties of magnetic materials \citep{Peregrinus1904letter}. While
the compass is capable of measuring magnetic field direction, it was
not until the first of half of the nineteenth century that Gauss and
Weber devised a magnetometer capable of measuring magnetic field strength
on an absolute scale \citep{gubbins2007encyclopedia}. Again, the
instrument was a torsional magnetometer composed of a permanent magnet
suspended by a long, thin thread. Gauss and Weber's method required
measuring the frequency of oscillation of the suspended magnet when
it was angularly displaced from its equilibrium position. This technique,
used previously by von Humboldt to measure relative magnetic field
strength, is quite similar to that described in this chapter and used
in the measurements discussed in this text \citep{gauss1830intensitas}.

Following the work of Gauss and Weber, one of the most notable scientific
achievements involving torsional magnetometry was the precise measurement
of the de Haas-van Alphen effect, the most important element in the
experimental confirmation of the concept of the Fermi surface \citep{hoch1983akey}.
While the original measurements were performed with a Faraday balance
\citep{haas1930oscillations,shoenberg1936themagnetic,shoenberg1936themagnetic2},
much more precise data over a wider field range was obtained by Shoenberg
using a torsional magnetometer developed by Krishnan \citep{shoenberg1939themagnetic}.
In this case, torsional magnetometry takes advantage of a sample's
anisotropic magnetization by using a uniform magnetic field. The magnetic
field gradient of the Faraday balance becomes problematic when the
resulting variation in magnetic field across the sample is of the
same order of magnitude as the magnetic field scale of the experimental
features of interest \citep{shoenberg1982therutherford}. It is of
historical interest to note that the first measurements of the de
Haas-van Alphen effect in a material other than bismuth were performed
at Yale in Rooms 16 and 17 of Sloane Physics Laboratory \citep{wheeler2006theearly},%
\footnote{To be completely accurate and discreet, I note that I have only been
able to confirm that the helium liquefier of C. T. Lane was located
in Room 17 of Sloane Physics Laboratory, but it seems highly likely
that these cryogenic experiments would be performed in the same location
as the source of the cryogen.%
} the same location as the measurements discussed in this text, using
a Faraday balance \citep{marcus1947thedehaasvan,marcus1949magnetic}
and using a torsional magnetometer \citep{sydoriak1949magnetic,berlincourt1952thede}.
These measurements were important for demonstrating that the de Haas-van
Alphen effect was not a peculiarity of bismuth and led to numerous
experiments by other groups outside of Yale \citep{hoch1983akey}.
Although a variety of alternate magnetometers such as the superconducting
quantum interference device (SQUID), Hall effect probe, and cesium
vapor magnetometer have been developed, torsional magnetometry using
a sample suspended by a wire has continued to be used up to the present
day \citep{griffiths1951anoscillation,gradmann1976highsensitivity,schaapman2002amultipurpose},
including notably for the measurement of persistent currents in self-assembled
semiconductor quantum rings \citep{kleemans2007oscillatory}.

Cantilever magnetometry was first developed in the late 1980s, arising
out of advances over the previous two decades in micromachining. These
advances allowed for the fabrication of single crystal oscillators
with high quality factors \citep{petersen1982silicon,kleiman1985singlecrystal}
and led to the subsequent development of the atomic force microscope
\citep{binnig1986atomicforce}. The earliest applications of cantilever
magnetometry were magnetic force microscopy \citep{martin1987magnetic,sa?enz1987observation}
and studies of flux lattice melting in high temperature superconductors
by cantilever torsional magnetometry \citep{gammel1988evidence}.
Much of the pioneering work on cantilever magnetometry was performed
by Dan Rugar and collaborators at IBM for magnetic resonance force
microscopy \citep{rugar1992mechanical}, who have reported detection
of the magnetic moment of single electrons \citep{rugar2004singlespin}
and small ensembles of atomic nuclei \citep{degen2009nanoscale}.
Dovetailing nicely with our discussion of the de Haas-van Alphen effect
and its historical connection to the measurements discussed in this
text, another major highlight for cantilever torsional magnetometry
was the work done by my advisor Jack Harris to study magnetization
effects, including the de Haas-van Alphen effect, in two dimensional
electron gases \citep{harris1999integrated,harris2001magnetization}.

\section{\label{sec:CHTorsMagn_CantileverSHO}Cantilever as a simple harmonic
oscillator}

\begin{figure}
\begin{centering}
\includegraphics[width=0.7\paperwidth]{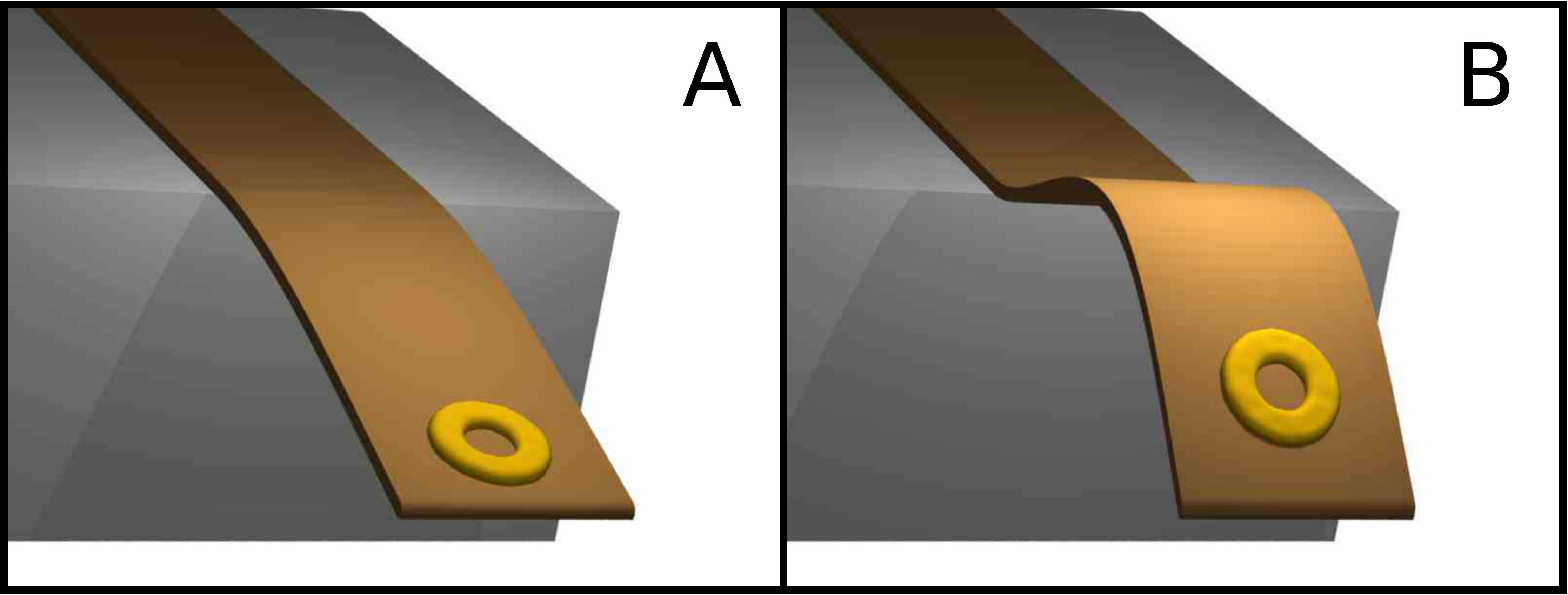}\caption[Cantilever mode shapes]{\label{fig:CantileverModeShapes}Cantilever mode shapes. Panel A
shows a cantilever with a normal metal ring deformed from its equilibrium
position according to its fundamental flexural mode of vibration,
while panel B shows the same cantilever deformed according to the
second flexural mode of vibration.}

\par\end{centering}

\end{figure}

A cantilever's motion can be decomposed into sets of indexed normal
modes, including flexural and torsional modes, each associated with
a particular deformation of the beam. The shapes of the first two
flexural modes of a cantilever are shown in Fig. \ref{fig:CantileverModeShapes}.
Focusing on the flexural modes, the beam's deformation consists in
a displacement at each point of the beam in the direction perpendicular
to the plane defined by the unflexed beam. The beam's deformation
varies as a function of position but follows a fixed functional form
which is scaled by an overall factor. The motion of each mode can
be treated as an independent, one-dimensional simple harmonic oscillator,
with this overall scaling (which can be written as the maximum displacement
at the cantilever tip) serving as the single degree of freedom \citep{cleland2003foundations}.
In this section, we will introduce many terms and symbols which will
be used throughout the text to discuss the cantilever motion explicitly.

\begin{figure}
\centering{}\includegraphics[width=0.5\paperwidth]{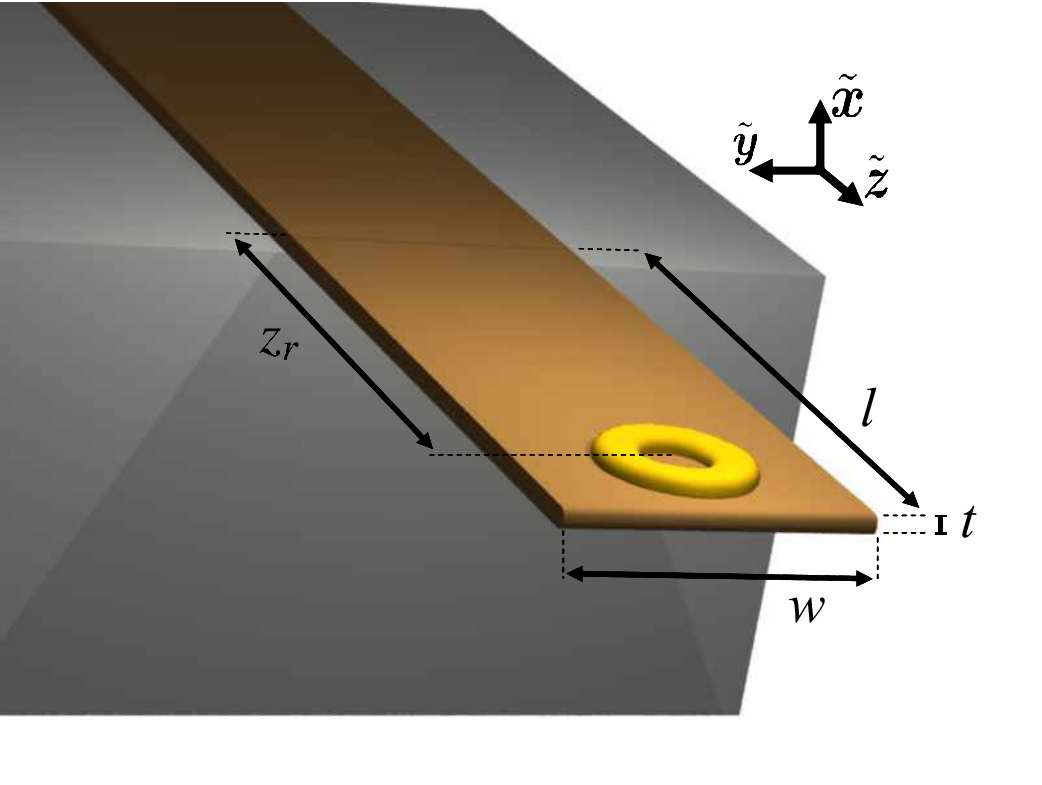}\caption[Labeled diagram of an unflexed cantilever]{\label{fig:LabeledUnflexedCantilever}Labeled diagram of an unflexed
cantilever. The cantilever has a length $l$, a width $w$, and a
thickness $t$. The cantilever's beam axis is parallel to the $z$-axis
with the wide surface of the cantilever parallel to the $yz$-plane.
The center of the normal metal ring sitting on the cantilever is located
a distance $z_{r}$ from the base of the cantilever. The ring is also
parallel to the $yz$-plane. A circulating current in the ring would
produce a magnetic moment pointed in the $\tilde{\boldsymbol{x}}$
direction.}
\end{figure}

To describe the motion of the cantilever quantitatively, we consider
a cantilever of length $l$ with its long dimension parallel to the
$z$-axis and with its wide face parallel to the $yz$-plane (see
Fig. \ref{fig:LabeledUnflexedCantilever}). When the cantilever undergoes
motion in its $m^{th}$ flexural mode, a point located a distance
$z$ from the cantilever base moves a distance $x_{z}=xU_{m}(z/l)$
in the $\tilde{\boldsymbol{x}}$ direction where $x$ is the amount
of displacement at the cantilever tip and $U_{m}$, the normalized
mode shape for the $m^{th}$ flexural mode, scales the $\tilde{\boldsymbol{x}}$
displacement appropriately for position $z$. All of the flexural
mode shapes are independent of $y$. The normalized mode shape is
given by 

\begin{equation}
U_{m}(\eta)=\frac{a_{m}(cos(\beta_{m}\eta)-cosh(\beta_{m}\eta))+sin(\beta_{m}\eta)-sinh(\beta_{m}\eta)}{a_{m}(cos(\beta_{m})-cosh(\beta_{m}))+sin(\beta_{m})-sinh(\beta_{m})}\label{eq:CHTorsMagn_UmCantileverMode}
\end{equation}
where $\beta_{m}$ and $a_{m}$ are constants associated with mode
$m$ \citep{cleland2003foundations}. The constant $\beta_{m}$ is
given by the solution of 
\begin{equation}
cos\beta_{m}cosh\beta_{m}+1=0,\label{eq:CHTorsMagn_BetaMDef}
\end{equation}
and
\[
a_{m}=-\frac{sin\beta_{m}+sinh\beta_{m}}{cos\beta_{m}+cosh\beta_{m}}.
\]

When the cantilever is elastically deformed, it experiences a restoring
force $F_{\text{restoring}}$ which can be modeled as a force acting
on the cantilever tip and obeying by Hooke's law $F_{\text{restoring}}=-kx$
for spring constant $k$. Additionally, when the cantilever moves,
it experiences a viscous damping force proportional to its velocity
\begin{equation}
F_{\text{damping}}=-m_{\text{eff}}\omega_{0}\dot{x}/Q,\label{eq:CHTorsMagn_Fdamping}
\end{equation}
where $\omega_{0}$ is the angular resonant frequency of the cantilever
and $Q$ is the cantilever mechanical quality factor. The equation
of motion for $x$ can be written as 

\begin{equation}
\ddot{x}+\frac{\omega_{0}}{Q}\dot{x}+\omega_{0}^{2}x=\frac{F(t)}{m_{\textrm{eff}}},\label{eq:CHTorsMagnCantEquationMotionTime}
\end{equation}
where $F(t)$ is the effective external force acting on the cantilever
tip at time $t$ and $m_{\textrm{eff}}$ is the cantilever effective
mass.%
\footnote{$t$ is such a natural symbol for both time and thickness that it
will be used for both in this document. Which quantity is denoted
by $t$ should be clear by context%
} In Eq. \ref{eq:CHTorsMagnCantEquationMotionTime}, the spring constant
has been eliminated using the expression for the angular resonant
frequency 
\begin{equation}
\omega_{0}=\sqrt{k/m_{\textrm{eff}}}.\label{eq:CHTorsMagn_FreqFromKandMeff}
\end{equation}
The cantilever spring constant $k$ can be defined as the product
\begin{equation}
k=\frac{Q}{m_{\text{eff}}}\frac{F\left(\omega_{0}\right)}{x\left(\omega_{0}\right)}\label{eq:ChTorsMagn_springK}
\end{equation}
where $F(\omega_{0})$ is the amplitude of an effective force applied
at the resonant frequency at the cantilever tip%
\footnote{Unlike a true mass-and-spring simple harmonic oscillator, it is not
possible to define the spring constants of the flexural modes of the
cantilever as the ratio of an applied static force to the corresponding
static displacement. The static deflection of the beam is described
by a different set of equations. Although we do not always write the
mode index $m$, the cantilever spring constant and resonant frequency
vary with $m$.%
} and $x(\omega_{0})$ is the amplitude of the cantilever response
at that frequency. The spring constant of the $m^{th}$ flexural mode
can be written in terms of the cantilever's mechanical parameters
as 
\begin{equation}
k_{m}=\frac{\beta_{m}^{4}}{48}E_{Y}\frac{wt^{3}}{l^{3}}\label{eq:CHTorsMagn_springKfromDimensions}
\end{equation}
where $E_{Y}$ is Young's modulus for the cantilever material and
$w$, $t$, and $l$ are the cantilever width, thickness, and length
respectively (see Fig. \ref{fig:LabeledUnflexedCantilever}). The
effective mass of the cantilever $m_{\textrm{eff}}$ satisfies 
\begin{equation}
m_{\textrm{eff}}=\frac{\rho}{4}wtl\label{eq:CHTorsMagn_MeffFromDimensions}
\end{equation}
where $\rho$ is the cantilever density. Using the relation $x(\omega)=\int_{-\infty}^{\infty}dt\, e^{-i\omega t}x(t)$
for the Fourier transform,%
\footnote{We will favor the ordinary frequency definition of the Fourier transform
and its inverse, namely $x(f)=\int_{-\infty}^{\infty}dt\, e^{-2\pi ift}x(t)$
and $x(t)=\int_{-\infty}^{\infty}df\, e^{2\pi ift}x(f)$, though as
above we will sometimes use the angular frequency form when convenient.
These two forms for the Fourier transform are totally equivalent under
the substitution $\omega=2\pi f$, though their inverse transforms
are not.%
} the equation of motion \eqref{eq:CHTorsMagnCantEquationMotionTime}
can be rewritten as $x(\omega)=G(\omega)F(\omega)$ where $G(\omega)$
is the cantilever transfer function given by
\begin{equation}
G(\omega)=\frac{1/m_{\textrm{eff}}}{\omega_{0}^{2}-\omega^{2}-i\omega_{0}\omega/Q}.\label{eq:GCantileverTransferFunction}
\end{equation}

From these relations we see that the cantilever motion is fully characterized
by the cantilever's dimensions and material properties and the quality
factor $Q$. The mechanical quality factor $Q$ describes the rate
of energy transfer from the cantilever's mechanical motion to its
environment. While the geometrical dimensions can be specified during
fabrication and the material properties are fixed and can be easily
looked up, a complete understanding of the mechanisms determining
the mechanical quality factor in micro-electromechanical systems is
lacking \citep{yasumura2000quality}. The mechanical quality factor
can be extracted from measurements of the cantilever's motion. For
example, the mechanical quality factor satisfies the relation $Q=\omega_{0}/\Delta\omega$,
where $\Delta\omega$ is the full width at half maximum of $|G(\omega)|^{2}$.
The transfer function $G(\omega)$ can be measured by monitoring the
cantilever amplitude as the frequency of a fixed amplitude excitation
is varied. Alternatively, in the absence of an applied force and the
limit of $Q\gg1$, the cantilever position as a function of time is
\begin{equation}
x(t)=x_{i}e^{-\omega_{0}t/2Q}\cos(\omega_{0}t),\label{eq:CantileverRingdown}
\end{equation}
where $x_{i}$ is the position of the cantilever at $t=0$.%
\footnote{Here, we have assumed that the cantilever has no kinetic energy at
$t=0$. This assumption can be removed at the cost of including an
additional phase factor $\phi$ in the argument of the cosine in Eq.
\eqref{eq:CantileverRingdown} satisfying $\tan(\phi)=-v_{0}/\omega_{0}x_{0}$
where $v_{0}$ is the value of $\dot{x}$ at $t=0$. For completeness,
we note here that the $Q\gg1$ limit mentioned above was used to drop
a factor of $\sqrt{1-1/4Q^{2}}$ in the argument of the cosine in
Eq. \eqref{eq:CantileverRingdown}.%
} In this case, the cantilever oscillates with an amplitude that decays
exponentially. By monitoring this decay and determining its characteristic
time scale $\tau=2Q/\omega_{0}$, the mechanical quality factor $Q$
can be extracted.

\FloatBarrier

\section{\label{sec:CHTorsMagn_deltaFZeroDrive}Cantilever frequency shift
due to persistent currents in the small amplitude limit}

For the measurements of cantilever motion discussed in this text,
the primary effect of placing a normal metal ring onto the end of
the cantilever is to produce a shift in the cantilever's resonant
frequency proportional to the persistent current in the ring.%
\footnote{One could also imagine using a time varying flux through the ring
to excite the cantilever resonantly and extracting the magnitude of
the persistent current from the amplitude of the cantilever's motion.
Similarly, one could vary the flux through the ring at twice the cantilever's
resonant frequency and use the influence of the ring on the cantilever
to amplify the cantilever motion parametrically.%
} In this section, we derive the change in the resonant frequency of
the cantilever due to the interaction of an integrated normal metal
ring with an applied magnetic field. We will ignore corrections due
to the finite extent of the cantilever's motion, which will be discussed
in \ref{sec:CHTorsMagn_FiniteDriveSection}.

To begin the derivation of the frequency shift, it is more convenient
to discuss the cantilever spring constant than the resonant frequency.
The cantilever spring constant represents a force gradient experienced
by the cantilever tip, $k=-\partial F/\partial x$, or equivalently
the curvature of the cantilever's potential energy $E_{\textrm{elastic}},$
$k=\partial^{2}E_{\textrm{elastic}}/\partial x^{2}$, at equilibrium.
The ring has its own potential energy $E_{\text{ring}}$ which depends
on its position and orientation in an applied magnetic field. When
the ring is integrated onto the cantilever as in Fig. \ref{fig:LabeledUnflexedCantilever},
the position of the ring becomes coupled to the position of the cantilever.
We can then write the total potential energy curvature (or spring
constant) as $k_{\text{tot}}=k+\Delta k$ with $\Delta k=\partial^{2}E_{\text{ring}}/\partial x^{2}$.
Using the relation (Eq. \ref{eq:CHTorsMagn_FreqFromKandMeff}) between
frequency and spring constant from the preceding section, we can write
the frequency of the cantilever-ring system as 
\begin{eqnarray*}
\omega_{0}+\Delta\omega & = & \sqrt{(k+\Delta k)/m_{\textrm{eff}}}\\
 & \approx & \omega_{0}\left(1+\frac{\Delta k}{2k}\right)
\end{eqnarray*}
so that $\Delta f/f_{0}\approx\Delta k/2k$ where $f_{0}$ and $\Delta f$
are the resonant frequency of the bare cantilever and the shift in
the resonant frequency due to the ring. The resonant frequency shift
of the cantilever due to the ring can be written as
\begin{equation}
\Delta f=\frac{f_{0}}{2k}\frac{\partial^{2}E_{\text{ring}}}{\partial x^{2}}.\label{eq:CHTorsMagn_FreqShiftdEdxGeneral}
\end{equation}

\begin{figure}[h]
\centering{}\includegraphics[bb=0bp 0bp 208bp 163bp,width=0.45\paperwidth]{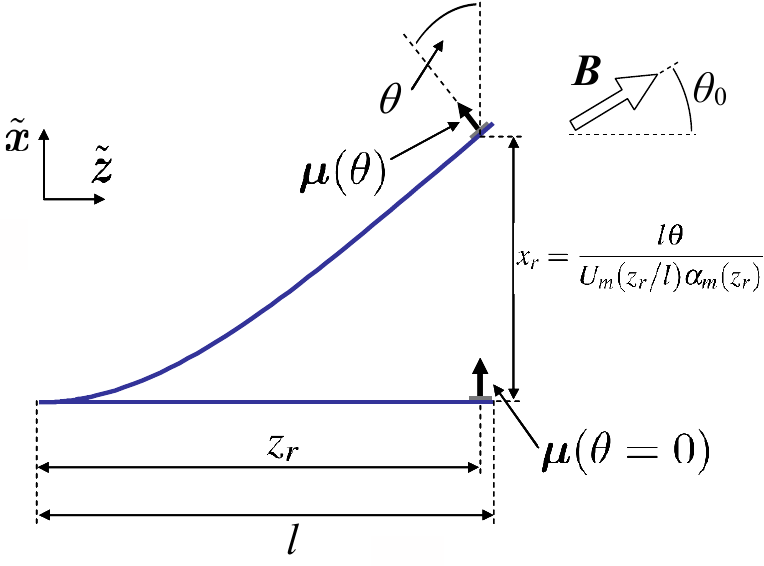}\caption[Flexed cantilever schematic]{\label{fig:Flexed-cantilever-schematic}Flexed cantilever schematic.
Profiles of a cantilever beam (purple lines) unflexed and flexed in
its fundamental mode shape are shown. The coordinate axes and cantilever
length $l$ are the same as those used in Figs. \ref{fig:CantileverModeShapes}
and \ref{fig:LabeledUnflexedCantilever}. A ring carrying a persistent
current which produces a magnetic moment $\boldsymbol{\mu}(\theta)$
is located a distance $z_{r}$ from the cantilever base and is represented
by a thick, solid arrow pointing in the direction of $\boldsymbol{\mu}(\theta)$.
When the cantilever flexes, the ring is displaced a distance $x_{r}$
along the $\boldsymbol{\hat{x}}$ direction. The ring is also tilted
an angle $\theta=\alpha_{m}(z_{r})x/l$. The factor $\alpha_{m}(z_{r})$
corrects for the bending of the beam. This factor would be unity and
independent of $z$ if the beam were totally rigid and merely pivoted
about its base. When the ring is tilted, the direction of $\boldsymbol{\mu}$
is rotated by the same angle $\theta.$ In the presence of a magnetic
field $\boldsymbol{B}$, the magnitude of $\boldsymbol{\mu}$ also
varies with $\theta$ due the change in magnetic flux through the
ring as described in the Chapter \ref{cha:CHMeso_}. The component
of $\boldsymbol{B}$ in the $xz$-plane, represented by a hollow arrow,
is oriented at an angle $\theta_{0}$ relative to the $\boldsymbol{\hat{z}}$
axis.}
\end{figure}

To describe the dependence of the ring's energy $E_{\text{ring}}$
on the motion of the cantilever, it is convenient to introduce a coordinate
representing the position of the ring. We use the angular deflection
of the ring $\theta$. We will see shortly that the angular deflection
of the ring couples to the magnitude of the magnetic field, while
the linear position of the ring couples to the magnetic field gradient.
In the measurements described in this text, a uniform magnetic field
was applied, so the angular deflection $\theta$ is the more natural
variable to use. In Fig. \ref{fig:Flexed-cantilever-schematic}, we
show a schematic, two-dimensional diagram of the cantilever depicted
in Figs. \ref{fig:CantileverModeShapes} and \ref{fig:LabeledUnflexedCantilever}.
We place the center of the ring a distance $z_{r}$ from the base
of the cantilever. The linear displacement of the ring from its equilibrium
position $x_{r}$ is given by $x_{r}=xU_{m}(z_{r})$. We can also
relate the linear displacement of the ring to its angular deflection
by 
\begin{align*}
\theta & =\alpha_{m}\left(z_{r}\right)\frac{x}{l}\\
 & =\frac{\alpha_{m}\left(z_{r}\right)}{U_{m}\left(z_{r}\right)}\frac{x_{r}}{l}
\end{align*}
where 
\begin{equation}
\alpha_{m}(z_{r})=\partial_{\eta}U_{m}\bigg|_{{\displaystyle {\displaystyle \eta=\frac{z_{r}}{l}}}}\label{eq:CHTorsMagn_Alpha}
\end{equation}
is the normalized, dimensionless derivative of the cantilever mode
shape. We will usually abbreviate $U_{m}\left(z_{r}\right)$ by $U$
and $\alpha_{m}(z_{r})$ by $\alpha$ when it is clear that we are
discussing a particular mode and particular $z_{r}$ coordinate on
the cantilever. With this definition, Eq. \ref{eq:CHTorsMagn_FreqShiftdEdxGeneral}
for the resonant frequency shift becomes
\begin{equation}
\Delta f=\frac{f_{0}}{2k}\left(\frac{\alpha_{m}\left(z_{r}\right)}{l}\right)^{2}\frac{\partial^{2}E_{\text{ring}}}{\partial\theta^{2}}.\label{eq:CHTorsMagn_FreqShiftdEdThetaGeneral}
\end{equation}

We will consider the motion of the cantilever-ring system in the presence
an arbitrary magnetic field $\boldsymbol{B}=B_{x}\tilde{\boldsymbol{x}}+B_{y}\tilde{\boldsymbol{y}}+B_{z}\tilde{\boldsymbol{z}}$
which could be spatially inhomogeneous. The ring has a mean radius
$r$ and so mean area $A=\pi r^{2}$ (we ignore finite linewidth effects
in the present discussion). We also define the vectorial form of the
ring area $\boldsymbol{A}=A(\cos\theta\tilde{\boldsymbol{x}}-\sin\theta\tilde{\boldsymbol{z}})$
which is orthogonal to the plane containing the ring. The magnetic
moment of the ring can be written as $\boldsymbol{\mu}=I\boldsymbol{A}$
where $I$ is the current in the ring \citep{purcell1985electricity}.
We assume that the persistent current is a function of the flux threading
the ring with period $\phi_{0}$ so that it can be expanded as 
\begin{equation}
I=\sum_{p}I_{p}\sin(2\pi p\phi_{\text{tot}}/\phi_{0}+\psi_{p})\label{eq:CHTorsMagnPCHarmonicExpansion}
\end{equation}
with the $I_{p}$ and $\psi_{p}$ constant but otherwise arbitrary
and $\phi_{\text{tot}}=\boldsymbol{A\cdot\boldsymbol{B}}$ the total
flux threading the mean radius of the ring.%
\footnote{For an ideal Aharonov-Bohm flux, the current should possess nonzero
Fourier coefficients only for integral values of $p$. However, the
finite magnetic field correlation of the persistent current discussed
in \ref{sub:CHPCTh_FluxThroughMetal} results in a broadening of the
magnetic flux frequency peaks of the persistent current oscillation.
The persistent current Fourier coefficients for $p$ close to each
integral value then take on nonzero values as well.%
}

The energy of a magnetic moment in an applied magnetic field can be
written as
\begin{equation}
E=-\intop^{\boldsymbol{B}}d\boldsymbol{B'}\cdot\boldsymbol{\mu}(\boldsymbol{B'})\label{eq:ChTorsMagn_EnergyMagMoment}
\end{equation}
where the lower bound of the integral produces an irrelevant shift
in the energy. Performing this integral for the case of the ring as
described above, we obtain
\begin{eqnarray}
E_{\textrm{ring}} & = & \sum_{p}E_{p,\textrm{ring}}\nonumber \\
 & = & \sum_{p}\frac{I_{p}\phi_{0}}{2\pi p}\cos\left(2\pi p\frac{A(B_{x}\cos\theta-B_{z}\sin\theta)}{\phi_{0}}+\psi_{p}\right).\label{eq:CHTorsMagn_EringExpansion}
\end{eqnarray}
Taking the angular derivative of $E_{\text{ring}}$, we find the torque
on the ring due to the magnetic field
\begin{align}
\tau_{ring}= & -\frac{\partial E_{\textrm{ring}}}{\partial\theta}\nonumber \\
= & \sum_{p}\begin{array}[t]{l}
{\displaystyle I_{p}A\sin\left(2\pi p\frac{A(B_{x}\cos\theta-B_{z}\sin\theta)}{\phi_{0}}+\psi_{p}\right)}\\
{\displaystyle \times\left(\frac{\partial B_{x}}{\partial\theta}\cos\theta-B_{x}\sin\theta-\frac{\partial B_{z}}{\partial\theta}\sin\theta-B_{z}\cos\theta\right).}
\end{array}\label{eq:TorqueRingGeneralFieldwithTheta}
\end{align}
Setting $\theta=0,$ Eq. \eqref{eq:TorqueRingGeneralFieldwithTheta}
reduces to 
\[
\tau_{\textrm{ring}}=\sum_{p}I_{p}A\sin\left(2\pi p\frac{\phi_{\text{tot}}}{\phi_{0}}+\psi_{p}\right)\left(U\frac{l}{\alpha}\frac{\partial B_{x}}{\partial x_{r}}-B_{z}\cos\theta\right),
\]
and follows the expected form for the torque and force on a fixed
magnetic moment
\[
\boldsymbol{\tau}=\boldsymbol{\mu\times B}+\frac{U_{m}\left(z\right)}{\alpha_{m}(z)}\boldsymbol{l\times}\left(\left(\boldsymbol{\mu\cdot\nabla}\right)\boldsymbol{B}\right)
\]
where the first term is the usual expression for the torque and the
second term is the usual expression for the force but converted into
a torque by the operation $[(U_{m}(z)/\alpha_{m}(z))\boldsymbol{l\times}]$
where $\boldsymbol{l}$ has a magnitude $l$ and points along the
cantilever beam axis (in the $\tilde{\boldsymbol{y}}$ direction in
Fig. \ref{fig:LabeledUnflexedCantilever}) \citep{sidles1995magnetic}. 

Performing a second derivative on Eq. \eqref{eq:TorqueRingGeneralFieldwithTheta}
and setting $\theta=0,$we obtain the cantilever resonant frequency
shift in the limit of zero drive
\begin{eqnarray}
\Delta f & = & \frac{f_{0}}{2k}\left(\frac{\alpha}{l}\right)^{2}\frac{\partial^{2}E_{\textrm{ring}}}{\partial\theta^{2}}\nonumber \\
 & = & -\frac{f_{0}}{2k}\left(\frac{\alpha}{l}\right)^{2}\sum_{p}\begin{array}[t]{c}
{\displaystyle I_{p}A\Bigg[\left(\frac{2\pi p}{\phi_{0}}A\right)\left(\frac{\partial B_{x}}{\partial\theta}-B_{z}\right)^{2}\cos\left(\frac{2\pi p}{\phi_{0}}AB_{x}+\psi_{p}\right)}\\
{\displaystyle +\left(-B_{x}+\frac{\partial^{2}B_{x}}{\partial\theta^{2}}-2\frac{\partial B_{z}}{\partial\theta}\right)\sin\left(\frac{2\pi p}{\phi_{0}}AB_{x}+\psi_{p}\right)\Bigg].}
\end{array}+\ldots\label{eq:FullZeroDriveFreqShiftIGradients}
\end{eqnarray}
We can rewrite this expression as 
\begin{equation}
\Delta f=-\frac{f_{0}}{2k}\left[\left(A\frac{\partial\mu}{\partial\phi}\right)\left(U\frac{\partial B_{x}}{\partial x}-\frac{\alpha}{l}B_{z}\right)^{2}+\mu\left(-\left(\frac{\alpha}{l}\right)^{2}B_{x}-2\frac{\alpha}{l}U\frac{\partial B_{z}}{\partial x}+U^{2}\frac{\partial^{2}B_{x}}{\partial x^{2}}\right)\right].\label{eq:FullZeroDriveFreqShiftMuGradients}
\end{equation}
The various terms in Eq. \eqref{eq:FullZeroDriveFreqShiftMuGradients}
have different physical origins. As discussed above, the frequency
shift is proportional to the sum of the torque and force gradients,
$\partial_{\theta}\tau_{y}$ and $\partial_{x}F_{x}$ respectively,
experienced by the ring. 

The terms proportional to $\mu$ in Eq. \eqref{eq:FullZeroDriveFreqShiftMuGradients}
represent the torque and force gradients felt by a rigid magnetic
moment of fixed magnitude whose orientation and position are coupled
to those of the cantilever. The $B_{x}$ term is usually the dominant
term in cantilever torsional magnetometry and represents the restoring
torque which tends to align $\boldsymbol{\mu}$ parallel to a uniform
$\boldsymbol{B}$. Similarly, the $\partial_{x}B_{z}$ term represents
the fact that, although a uniform $\boldsymbol{B}$ along $\tilde{\boldsymbol{z}}$
produces a uniform torque with no torque gradient for our chosen cantilever
and corresponding magnetic moment orientation, giving the $B_{z}$
a gradient along $\tilde{\boldsymbol{x}}$ allows the torque due to
the $\tilde{\boldsymbol{z}}$ component of $\boldsymbol{B}$ to produce
a torque gradient along $\tilde{\boldsymbol{x}}$. This term is usually
dominant in magnetic force microscopy. Finally, the $\partial_{x}^{2}B_{x}$
term corresponds to the force gradient arising from the force experienced
by a magnetic moment in a magnetic field gradient when that field
gradient itself is non-uniform. 

Unlike the terms in Eq. \eqref{eq:FullZeroDriveFreqShiftMuGradients}
proportional to $\mu$, the terms proportional to $\partial_{\phi}\mu$
represent the force gradient felt by the ring that arise from the
fact that the ring's magnetic moment itself varies with cantilever
position.%
\footnote{It will be shown presently that this term is the dominant contribution
to the persistent current signal in our detection scheme. Without
this term, measuring persistent currents would have been much more
difficult. I would like to add a few remarks regarding it to the historical
record. We overlooked this contribution for quite a long time in our
analysis of the persistent current signal. For the first couple of
years of work on the persistent current project, we experimented unsuccessfully
with various measurement schemes including conventional frequency
modulation magnetometry with $\boldsymbol{B}$ parallel to $\boldsymbol{A}$,
static deflection magnetometry with $\boldsymbol{B}$ orthogonal to
$\boldsymbol{A}$, and a variant of resonant force magnetometry in
which a time-varying flux through the ring is applied on resonance
with the cantilever as described in \citep{bleszynski-jayich2008highsensitivity}.
Ania Jayich and I independently discovered this flux-through-the-ring
dependent contribution to the cantilever frequency shift, without
much physical intuition on our parts, while following an exhortation
from Jack Harris to revisit the frequency shift derivation with a
more careful approach (which evolved into the analysis covered in
this section). Jack Sankey also played a role in leading us to perform
the derivation more carefully. The impetus for revisiting the frequency
shift derivation was the unexpectedly strong signal that we measured
with the rings in the superconducting state (see Appendix \ref{cha:AppSC_}).
In an earlier version of the experiment, the cantilevers were oriented
at $\theta_{0}=0^{\circ}$ while we attempted to thread flux through
the rings with a second, smaller superconducting coil which we had
wound ourselves. Expecting only the term proportional to $\mu$ (which
is 0 at $\theta_{0}=0^{\circ}$), we were initially quite surprised
by the large frequency shift signal which we observed at low magnetic
field. Once we understood its origin and implications, we removed
the second coil, introduced a known, non-zero amount of tilt $\theta_{0}$,
and performed the measurements described in this text.%
} The magnetic moment of the ring is modulated by the magnetic flux
threading it, and this flux changes when the cantilever moves. When
the cantilever moves, the flux through the ring can change in two
ways. First, it can change because the ring moves in the $\tilde{\boldsymbol{x}}$
direction and the magnitude of the magnetic field itself changes due
to a gradient, $\partial_{x}B_{x}$. Second, the flux can change because
the ring tilts slightly so that $\boldsymbol{A}$ picks up a $\tilde{\boldsymbol{z}}$
component and the component of $\boldsymbol{B}$ along $\tilde{\boldsymbol{z}}$,
$B_{z}$ now contributes to the flux. It is interesting to note that
for typical parameters used for measuring the persistent current,
$U=1$, $\alpha_{1}=1.377$, $l=150\,\mathrm{\mu}$m, and $B=9$ T,
the product $\alpha B/l=8.3\times10^{4}$ T/m is a bit less than two
orders of magnitude smaller than the largest magnetic field gradients,
$\partial_{x}B_{x}=4\times10^{6}$ T/m reported in the literature
\citep{degen2009nanoscale}.

Although the numbers seem promising for measuring persistent currents
in a magnetic field gradient,%
\footnote{Based on the numerical values for $\alpha B/l$ and $\partial_{x}B_{x}$
presented in the previous paragraph and the equivalence of these two
terms in the expression for the frequency shift given in Eq. \ref{eq:FullZeroDriveFreqShiftMuGradients},
any results derived for the magnitude of the persistent current signal
and its signal to noise ratio in the presence of a uniform field can
be translated over to the case of a field gradient. While the uniform
field measurement is more straightforward and was thus chosen for
the measurements discussed in this text, there are some effects, namely
those discussed in \ref{cha:CHMeso_} which involve the cooperon,
which are strongly suppressed at high field. Measuring these effects
with a uniform field would be highly difficult due to low fields required.
A magnetic field gradient measurement could be a better choice in
these cases since it can achieve the same ratio of frequency shift
to current as the strong uniform field measurement but can be conducted
at low field.%
} we will now drop the magnetic field gradient terms in order to focus
more closely on the experiments described in this text which were
all performed in a uniform magnetic field. We note that the $\tilde{\boldsymbol{y}}$
component of $\boldsymbol{B}$ does not appear in Eqs. \eqref{eq:FullZeroDriveFreqShiftIGradients}
and \eqref{eq:FullZeroDriveFreqShiftMuGradients}. Without loss of
generality, we will set $B_{y}=0$ (which also matches the experimental
arrangement discussed in this text). We can then write $\boldsymbol{B}=B\sin\theta_{0}\tilde{\boldsymbol{x}}+B\cos\theta_{0}\tilde{\boldsymbol{z}}$
with $\theta_{0}$ the angle between $\boldsymbol{B}$ and the unflexed
cantilever beam (see Fig. \eqref{fig:Flexed-cantilever-schematic}).
With these simplifications, Eq. \eqref{eq:FullZeroDriveFreqShiftIGradients}
reduces to
\begin{eqnarray}
\Delta f & = & -\frac{f_{0}}{2k}\left(\frac{\alpha}{l}\right)^{2}\sum_{p}\begin{array}[t]{c}
{\displaystyle I_{p}A\Bigg[\left(\frac{2\pi p}{\phi_{0}}A\right)B^{2}\cos^{2}\theta_{0}\cos\left(\frac{2\pi p}{\phi_{0}}AB\sin\theta_{0}+\psi_{p}\right)}\\
{\displaystyle -B\sin\theta_{0}\sin\left(\frac{2\pi p}{\phi_{0}}AB\sin\theta_{0}+\psi_{p}\right)\Bigg].}
\end{array}\label{eq:CHTorsMagn_ZeroDriveFreqShiftBothTerms}
\end{eqnarray}
The first term will dominate the second when the condition $B\gg\left[\left(\tan\theta_{0}/\cos\theta_{0}\right)\left(\phi_{0}/2\pi A\right)\right]$
is met. The first factor $\left(\tan\theta_{0}/\cos\theta_{0}\right)$
is no larger than $10$ as long as $\theta_{0}<70^{\circ}$. For a
typical ring size for measuring the normal state persistent current,
the second factor $\left(\phi_{0}/2\pi A\right)$ is on the order
of a millitesla. All of the measurements of the normal state persistent
current reported in this text were performed under the conditions
 $\theta_{0}\leq45^{\circ}$ and $B>1$ T. Retaining only the dominant
term, we arrive at the simplified expression for the frequency shift
due to the persistent current valid for all measurements reported
in this text (in the limit of zero drive)
\begin{eqnarray}
\Delta f & = & -\frac{f_{0}}{2k}\left(\frac{\alpha}{l}AB\cos\theta_{0}\right)^{2}\sum_{p}I_{p}\left(\frac{2\pi p}{\phi_{0}}\right)\cos\left(\frac{2\pi p}{\phi_{0}}AB\sin\theta_{0}+\psi_{p}\right)\nonumber \\
 & = & -\frac{f_{0}}{2k}\left(\frac{\alpha}{l}AB\cos\theta_{0}\right)^{2}\frac{\partial I}{\partial\phi}.\label{eq:ZeroDriveFreqShiftSimple}
\end{eqnarray}

\FloatBarrier

\section{\label{sec:CHTorsMagn_FiniteDriveSection}The effect of finite cantilever
oscillation amplitude on the frequency shift due to the persistent
current}

The cantilever frequency measurement is made by monitoring the cantilever
position as a function of time while the cantilever is driven at some
fixed amplitude and by then extracting the dominant frequency of the
resulting time trace. The main sources of noise in the measurement
of the cantilever position, the cantilever's Brownian motion and the
technical noise of the measurement system, are independent of the
cantilever's amplitude. The sensitivity to persistent current thus
increases with cantilever amplitude since the measurement signal increases
while the magnitude of the noise remains constant.%
\footnote{We ignore here the effects of cantilever non-linearity which lead
to an amplitude dependence of the cantilever frequency and can complicate
an analysis of the uncertainties involved in extracting the cantilever
frequency from a time trace of the cantilever position.%
} 

There is a limit to the effectiveness of increasing the cantilever
amplitude in the measurement of the frequency shift due to the persistent
current. This limit results from the fact that the flux through the
ring changes when the cantilever is displaced. For sufficiently large
displacements the flux through the ring varies by more than $\phi_{0}$
in one period of cantilever oscillation and the current in the ring,
which follows the form given by Eq. \ref{eq:CHTorsMagnPCHarmonicExpansion},
tends to be washed out when averaged over the cantilever motion. We
will now discuss this effect more quantitatively.

\begin{figure}
\begin{centering}
\includegraphics[width=0.7\paperwidth]{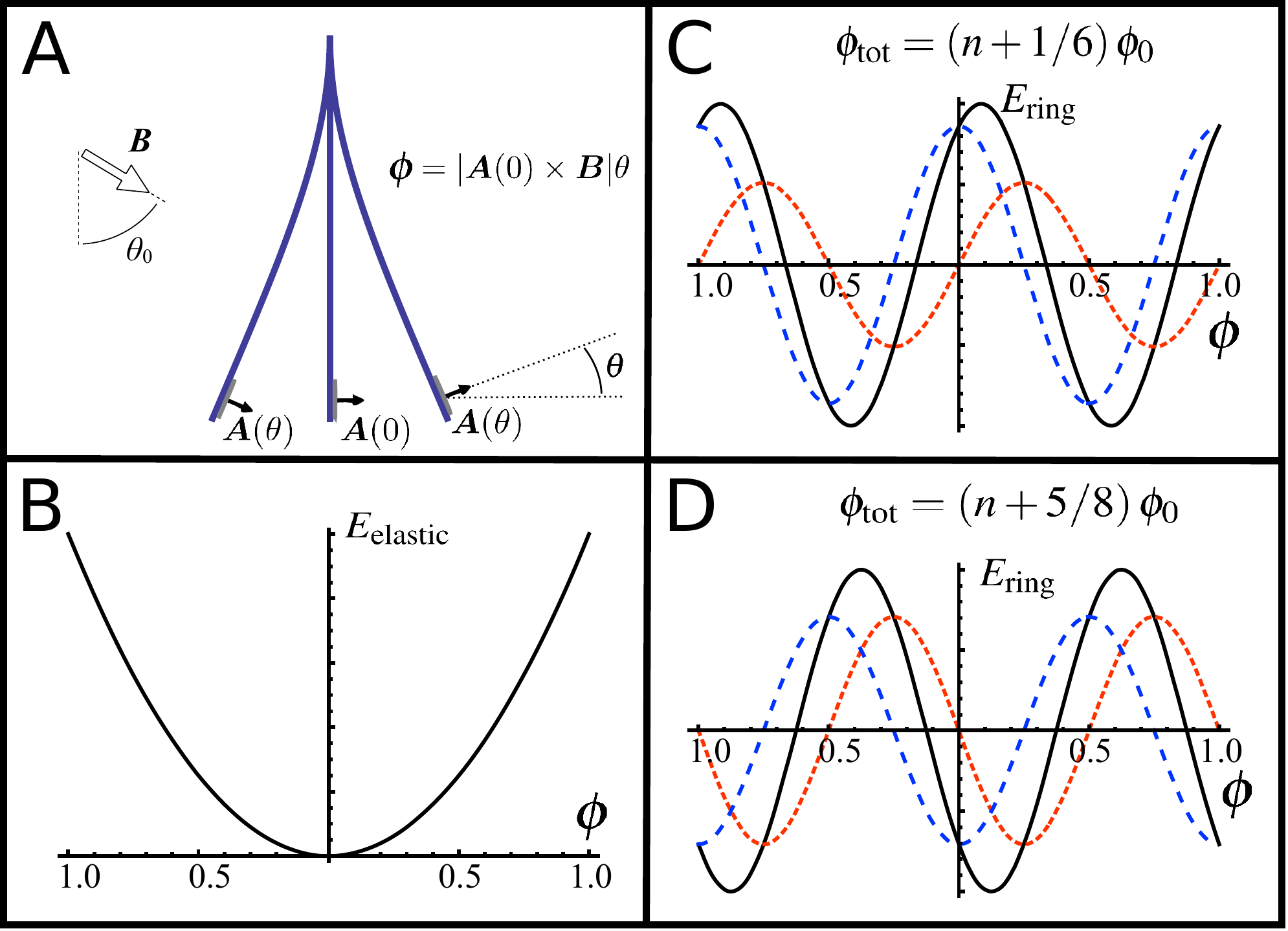}\caption[Potential energy landscapes of cantilever and ring]{\label{fig:CHTorsMagn_Potential-energy-landscapes}Potential energy
landscapes of cantilever and ring. Panel \textsf{A} shows a profile
of the cantilever beam in both flexed and unflexed positions. A side
view of the ring is shown near the end of the cantilever along with
an arrow indicating the direction of $\boldsymbol{A}$, the ring area
vector which always points normal to the plane containing the ring.
When the cantilever is flexed, the direction of $\boldsymbol{A}$
changes by $\theta$. Also indicated in this panel is the direction
of the vector $\boldsymbol{B}$ and the angle $\theta_{0}$ that it
makes relative to the unflexed cantilever beam. When the cantilever
flexes and $\boldsymbol{A}$ changes direction by $\theta$, the flux
through the ring changes by $\phi=|\boldsymbol{A}(0)\times\boldsymbol{B}|\theta$
from its value when the cantilever is unflexed. For fixed $\boldsymbol{B}$,
the flux $\phi$ is proportional the displacement of the cantilever.
In panel \textsf{B,} the potential energy of the bare cantilever $E_{\textrm{elastic}}$
is plotted versus the displacement of the cantilever, expressed as
$\phi$. This energy landscape displays the familiar parabolic form
for a simple harmonic oscillator. In panels \textsf{C }and\textsf{
D,} the potential energy of the ring $E_{\textrm{ring}}$ (solid black
curves) is shown for two different values of the total flux $\phi_{\textrm{tot}}$
through the ring when the cantilever is unflexed. For simplicity,
only the $p=1$ term of Eq. \ref{eq:CHTorsMagn_EringExpansion} is
shown, and $\psi_{p}$ is take to be 0. The values of $\phi_{\textrm{tot}}$
in panels \textsf{C }and\textsf{ D} are $(n+1/6)\phi_{0}$ and $(n+5/8)\phi_{0}$
respectively where $n$ is an integer. The scale of the horizontal
axes in panels \textsf{B, C }and\textsf{ D} is in units of the flux
quantum $\phi_{0}$. The scale of the vertical axes is arbitrary.
For the measurements reported in this text, the vertical scale of
panel \textsf{B }is typically seven or more orders of magnitude larger
than that of panels \textsf{C }and\textsf{ D}. Panels \textsf{C }and\textsf{
D} also show the components of $E_{\textrm{ring}}$ proportional to
$\sin2\pi\phi/\phi_{0}$ (red dotted line) and $\cos2\pi\phi/\phi_{0}$
(blue dashed line) described in Eq. \ref{eq:CHTorsMagnEpRingSinCos}.
The total value of $E_{\textrm{ring}}$ (solid black curves) is equal
to the sum of these two components.}

\par\end{centering}

\end{figure}

When the cantilever flexes and the ring undergoes a small angular
deflection $\theta$ from its equilibrium position, the amount of
flux threading the ring changes by $\phi=|\boldsymbol{A}(\theta=0)\times\boldsymbol{B}|\theta$.%
\footnote{We follow the convention that $x$, $\theta$, and $\phi$ represent
deviations in the different parameters due to the displacement of
the cantilever from its equilibrium position while $x_{0}$, $\theta_{0}$,
and $\phi_{\text{tot}}$ denote the values of those parameters at
the equilibrium position (the symbol $\phi_{0}$ is traditionally
reserved for the flux quantum). The symbols $x_{\max}$, $\theta_{\max}$,
and $\phi_{\max}$ represent the amplitudes of $x$, $\theta$, and
$\phi$ when the cantilever moves periodically.%
} Due to the proportionality between $\phi$ and $\theta$, we can
parametrize the motion of the cantilever and ring either in terms
of the angle $\theta$ or in terms of this change in flux $\phi$.
As the ring-cantilever system evolves in time, the cantilever position
moves through the potential energy landscape (see Fig. \ref{fig:CHTorsMagn_Potential-energy-landscapes})
given by the sum of the parabola corresponding to cantilever's energy
$E_{\text{elastic}}$ of elastic deformation plus the energy $E_{\text{ring}}$
of the ring in the presence of a magnetic field given in Eq. \ref{eq:CHTorsMagn_EringExpansion}.
For simplicity, we will consider only the $p^{\textrm{th}}$ term
from the sum in Eq. \ref{eq:CHTorsMagn_EringExpansion} defining the
energy of the ring. All of our analysis will be linear so that the
effects of the full ring potential can be found by summing over $p$. 

In Fig. \ref{fig:CHTorsMagn_Potential-energy-landscapes}, the potential
energy landscape of both the cantilever and the ring are plotted versus
$\phi$. The potential energy of the ring can be written as the sum
of terms proportional to $\cos(2\pi p\phi/\phi_{0})$ and $\sin(2\pi p\phi/\phi_{0})$
as
\begin{align}
E_{p,\textrm{ring}} & =\frac{I_{p}\phi_{0}}{2\pi p}\cos\left(2\pi p\frac{\phi_{\textrm{tot}}-\phi}{\phi_{0}}+\psi_{p}\right)\nonumber \\
 & =\frac{I_{p}\phi_{0}}{2\pi p}\Bigg(\cos\left(2\pi p\frac{\phi_{\textrm{tot}}}{\phi_{0}}+\psi_{p}\right)\cos\left(2\pi p\frac{\phi}{\phi_{0}}\right)+\ldots\nonumber \\
 & \phantom{=\frac{I_{p}\phi_{0}}{2\pi p}\Bigg(}+\sin\left(2\pi p\frac{\phi_{\textrm{tot}}}{\phi_{0}}+\psi_{p}\right)\sin\left(2\pi p\frac{\phi}{\phi_{0}}\right)\Bigg).\label{eq:CHTorsMagnEpRingSinCos}
\end{align}
where $\phi_{\text{tot}}$ is the total flux threading the mean radius
of the ring. These two terms are plotted individually in Fig. \ref{fig:CHTorsMagn_Potential-energy-landscapes}.
For cantilever motion traversing a sufficiently small range of $\phi$,
the cosine term (dashed blue line) merely adjusts the curvature of
the parabolic shape of $E_{\textrm{elastic}}$ while the sine term
shifts the minimum of $E_{\textrm{elastic}}$ without changing the
curvature. Since the frequency of a system with a parabolic potential
is given by the potential's curvature, these observations are sufficient
to determine the shift of the cantilever's resonant frequency due
to the ring as described in \ref{sec:CHTorsMagn_deltaFZeroDrive}.
However, once $\phi\sim\phi_{0}/2$, the parabolic and linear approximations
are no longer good enough and a more detailed analysis of the cantilever's
motion is needed.

To perform this more detailed analysis, we use perturbation theory
in the Hamilton-Jacobi formalism with action-angle variables. The
basic principles behind this technique are outlined in Appendix \ref{app:AppCanonPert}.
For a more detailed account, see Ref. \citep{goldstein2001classical}.
Our perturbative approach gives the shift in the frequency of the
cantilever's periodic free evolution due to interaction of the persistent
current ring with the magnetic field. The calculation ignores cantilever
damping (the term proportional to $\dot{x}$ in Eq. \ref{eq:CHTorsMagnCantEquationMotionTime})
and external, time dependent forces acting on the cantilever ($F(t)$
in Eq. \ref{eq:CHTorsMagnCantEquationMotionTime}). In the case of
the unperturbed simple harmonic oscillator driven on resonance, the
force of friction is exactly canceled by the resonant external driving
force so that the energy of the oscillator is constant in time. In
practice, we will measure the resonant frequency of the cantilever
by measuring its position while exciting it with such an external
driving force. We assume that even in the presence of the perturbation
the only appreciable effect of damping is to offset the energy added
to the oscillator by the external driving force. This assumption will
be justified by agreement between the expressions derived below and
our measurements. However, a quantitative analysis of the effect of
the perturbation on the cantilever's motion accounting for damping
and an external driving force would be worthwhile.

The full Hamiltonian of the cantilever-ring system in the presence
of a magnetic field is given by
\[
H=H_{0}+H_{1}
\]
with 
\[
H_{0}=E_{\textrm{ring}}=\frac{1}{2m_{\textrm{eff}}}\frac{\alpha^{2}p_{\theta}^{2}}{l^{2}}+\frac{1}{2}m_{\textrm{eff}}\omega_{0}^{2}\frac{l^{2}\theta^{2}}{\alpha^{2}}
\]
the Hamiltonian of the bare cantilever expressed in terms of $\theta$
and its canonically conjugate angular momentum $p_{\theta}$ and 
\[
H_{1}=E_{p,\textrm{ring}}=\frac{I_{p}\phi_{0}}{2\pi p}\cos\left(2\pi p\frac{AB\sin(\theta_{0}-\theta)}{\phi_{0}}+\psi_{p}\right)
\]
the $p^{\text{th}}$ harmonic of the perturbing Hamiltonian due to
the ring as written in Eq. \eqref{eq:CHTorsMagnEpRingSinCos} with
$\phi_{\text{tot}}+\phi=AB\sin(\theta_{0}-\theta)$. At this point,
we have not yet taken the small angle approximation for $\theta$
which results in $\phi_{a}\propto\theta$ as discussed qualitatively
above. We will make this approximation explicitly below. 

In order to find the shift in the resonant frequency of the system
due to $H_{1}$, we rewrite $H_{1}(\theta,p_{\theta})$ in terms of
the action-angle variables $(\eta,J)$ as $K_{1}(\eta,J)$ where $J$
is the action variable of the unperturbed system. In Appendix \ref{app:AppCanonPert},
we wrote the transformed perturbation Hamiltonian as $K_{1}(\eta,j)$
where $j$ is the action variable of the perturbed system. However,
$K_{1}$ is already first order in the perturbation parameter $\varepsilon$
and the two action variables $J$ and $j$ differ only by a term proportional
to $\varepsilon$. Thus using $J$ rather than $j$ leads to a correction
to the frequency shift that is second order in $\varepsilon$ and
can be discarded. Using Eq. \ref{eq:AppCanonPertqInTermsJEta}, we
find
\begin{equation}
K_{1}(\eta,J)=\frac{I_{p}\phi_{0}}{2\pi p}\cos\left(2\pi p\frac{AB}{\phi_{0}}\sin\left(\theta_{0}-\frac{\alpha}{l}\sqrt{\frac{J}{2\pi^{2}mf_{0}}}\sin2\pi\eta\right)+\psi_{p}\right)\label{eq:CHTorsMagn_K1pertHamActAng}
\end{equation}
where we take $J=kx_{\text{\ensuremath{\max}}}^{2}/2f_{0}$ with $x_{\text{\ensuremath{\max}}}$
the amplitude of motion of the tip of the cantilever. The first order
correction to the resonant frequency of the cantilever due to the
persistent current is given by Eq. \ref{eq:AppCanonPertFreqShiftFormula}
as 
\begin{eqnarray}
\Delta f & = & \frac{\partial}{\partial J}\int_{0}^{1}d\eta K_{1}\left(\eta,J\right).\label{eq:CHTorsMagn_FreqShiftetaJGeneric}
\end{eqnarray}
To simplify the analysis, we make the abbreviations 
\[
\begin{array}{cc}
M=2\pi p\frac{AB}{\phi_{0}}\sin\theta_{0},\, & F(J)=\frac{\alpha}{l}\sqrt{\frac{J}{2\pi^{2}mf_{0}}},\\
N=2\pi p\frac{AB}{\phi_{0}}\cos\theta_{0},\, & \varepsilon_{p}=\frac{I_{p}\phi_{0}}{2\pi p}
\end{array}.
\]
Using the identities in Eqs. \ref{eq:AppMath_TrigCosAB} and \ref{eq:AppMath_TrigSinAB},
we expand Eq. \ref{eq:CHTorsMagn_K1pertHamActAng} to
\[
K_{1}(\eta,J)=\varepsilon_{p}\Big[\begin{array}[t]{l}
\cos\psi_{p}\cos\left(M\cos\left(F(J)\sin2\pi\eta\right)\right)\cos\left(N\sin\left(F(J)\sin2\pi\eta\right)\right)\\
+\sin\psi_{p}\cos\left(M\cos\left(F(J)\sin2\pi\eta\right)\right)\sin\left(N\sin\left(F(J)\sin2\pi\eta\right)\right)\\
+\cos\psi_{p}\sin\left(M\cos\left(F(J)\sin2\pi\eta\right)\right)\sin\left(N\sin\left(F(J)\sin2\pi\eta\right)\right)\\
-\sin\psi_{p}\sin\left(M\cos\left(F(J)\sin2\pi\eta\right)\right)\cos\left(N\sin\left(F(J)\sin2\pi\eta\right)\right)\Big].
\end{array}
\]

The quantity $F(J)$ represents the amplitude of angular deflection
of the cantilever. Generally, this is a small quantity no larger than
a few degrees. We use the small angle approximations $\cos\theta\approx1-\theta^{2}/2$
and $\sin\theta\approx\theta$ to remove one level from the nested
series of sines and cosines operating on $\eta$ in the expression
for $K_{1}$. We keep to second order in $\theta$ because in Section
\ref{sec:CHTorsMagn_deltaFZeroDrive} we evaluated $\delta f$ in
the small amplitude limit by taking two derivatives of $E_{p,\text{ring}}$
with respect to $\theta$ before taking $\theta\rightarrow0$. 

With these approximations and the trigonometric identity of Eq. \ref{eq:AppMath_TrigSinSquared},
the Hamiltonian takes on the unwieldy form
\begin{eqnarray}
K_{1}(\eta,J) & = & \varepsilon_{p}\Big[\begin{array}[t]{l}
\cos\psi_{p}\cos\left(M-M\frac{F^{2}}{4}+M\frac{F^{2}}{4}\cos\left(4\pi\eta\right)\right)\cos\left(NF\sin2\pi\eta\right)\\
+\sin\psi_{p}\cos\left(M-M\frac{F^{2}}{4}+M\frac{F^{2}}{4}\cos\left(4\pi\eta\right)\right)\sin\left(NF\sin2\pi\eta\right)\\
+\cos\psi_{p}\sin\left(M-M\frac{F^{2}}{4}+M\frac{F^{2}}{4}\cos\left(4\pi\eta\right)\right)\sin\left(NF\sin2\pi\eta\right)\\
-\sin\psi_{p}\sin\left(M-M\frac{F^{2}}{4}+M\frac{F^{2}}{4}\cos\left(4\pi\eta\right)\right)\cos\left(NF\sin2\pi\eta\right)\Big]
\end{array}\nonumber \\
 & = & \varepsilon_{p}\Big[\begin{array}[t]{l}
\cos\psi_{p}\cos\left(M-M\frac{F^{2}}{4}\right)\cos\left(M\frac{F^{2}}{4}\cos\left(4\pi\eta\right)\right)\cos\left(NF\sin2\pi\eta\right)\\
-\cos\psi_{p}\sin\left(M-M\frac{F^{2}}{4}\right)\sin\left(M\frac{F^{2}}{4}\cos\left(4\pi\eta\right)\right)\cos\left(NF\sin2\pi\eta\right)\\
+\sin\psi_{p}\cos\left(M-M\frac{F^{2}}{4}\right)\cos\left(M\frac{F^{2}}{4}\cos\left(4\pi\eta\right)\right)\sin\left(NF\sin2\pi\eta\right)\\
-\sin\psi_{p}\sin\left(M-M\frac{F^{2}}{4}\right)\sin\left(M\frac{F^{2}}{4}\cos\left(4\pi\eta\right)\right)\sin\left(NF\sin2\pi\eta\right)\\
+\cos\psi_{p}\sin\left(M-M\frac{F^{2}}{4}\right)\cos\left(M\frac{F^{2}}{4}\cos\left(4\pi\eta\right)\right)\sin\left(NF\sin2\pi\eta\right)\\
+\cos\psi_{p}\cos\left(M-M\frac{F^{2}}{4}\right)\sin\left(M\frac{F^{2}}{4}\cos\left(4\pi\eta\right)\right)\sin\left(NF\sin2\pi\eta\right)\\
-\sin\psi_{p}\sin\left(M-M\frac{F^{2}}{4}\right)\cos\left(M\frac{F^{2}}{4}\cos\left(4\pi\eta\right)\right)\cos\left(NF\sin2\pi\eta\right)\\
-\sin\psi_{p}\cos\left(M-M\frac{F^{2}}{4}\right)\sin\left(M\frac{F^{2}}{4}\cos\left(4\pi\eta\right)\right)\cos\left(NF\sin2\pi\eta\right)\Big].
\end{array}\label{eq:CHTorsMagn_K1UnwieldyExpansion}
\end{eqnarray}
In order to simplify Eq. \ref{eq:CHTorsMagn_K1UnwieldyExpansion},
we use the Jacobi-Anger identity given in Eqs. \eqref{eq:AppMath_JacAngCS},
\eqref{eq:AppMath_JacAngSS}, \eqref{eq:AppMath_JacAngCC}, and \eqref{eq:AppMath_JacAngSC}
to expand out the nested trigonometric functions of $\eta$. Then
the integral over $\eta$ in \eqref{eq:CHTorsMagn_K1pertHamActAng}
can be evaluated. We have four different integrals to evaluate, which
give
\begin{align*}
\int_{0}^{1}d\eta\, & \cos\left(M\frac{F^{2}}{4}\cos\left(4\pi\eta\right)\right)\cos\left(NF\sin2\pi\eta\right)\\
 & =\int_{0}^{1}d\eta\,\begin{array}[t]{l}
{\displaystyle \left(\sum_{n=-\infty}^{\infty}(-1)^{n}J_{2n}\left(M\frac{F^{2}}{4}\right)\cos\left(8\pi n\eta\right)\right)\left(\sum_{m=-\infty}^{\infty}J_{2m}(NF)\cos\left(4\pi m\eta\right)\right)}\end{array}\\
 & =\sum_{n=-\infty}^{\infty}(-1)^{n}J_{2n}\left(M\frac{F^{2}}{4}\right)J_{4n}(NF),
\end{align*}
\begin{align*}
\int_{0}^{1}d\eta\, & \sin\left(M\frac{F^{2}}{4}\cos\left(4\pi\eta\right)\right)\cos\left(NF\sin2\pi\eta\right)\\
 & =\int_{0}^{1}d\eta\,\begin{array}[t]{l}
{\displaystyle \left(\sum_{n=-\infty}^{\infty}(-1)^{n+1}J_{2n-1}\left(M\frac{F^{2}}{4}\right)\cos\left(4\pi\left(2n-1\right)\eta\right)\right)}\end{array}\\
 & =\sum_{n=-\infty}^{\infty}(-1)^{n+1}J_{2n-1}\left(M\frac{F^{2}}{4}\right)J_{4n-2}(NF),
\end{align*}

\begin{align*}
\int_{0}^{1}d\eta\, & \cos\left(M\frac{F^{2}}{4}\cos\left(4\pi\eta\right)\right)\sin\left(NF\sin2\pi\eta\right)\\
 & =\int_{0}^{1}d\eta\,\begin{array}[t]{l}
{\displaystyle \left(\sum_{n=-\infty}^{\infty}(-1)^{n}J_{2n}\left(M\frac{F^{2}}{4}\right)\cos\left(8\pi n\eta\right)\right)}\\
{\displaystyle \times\left(\sum_{n=-\infty}^{\infty}J_{n}(NF)\sin\left(2\pi n\eta\right)\right)}
\end{array}\\
 & =0,
\end{align*}
and
\begin{align*}
\int_{0}^{1}d\eta\, & \sin\left(M\frac{F^{2}}{4}\cos\left(4\pi\eta\right)\right)\sin\left(NF\sin2\pi\eta\right)\\
 & =\int_{0}^{1}d\eta\,\begin{array}[t]{l}
{\displaystyle \left(\sum_{n=-\infty}^{\infty}(-1)^{n+1}J_{2n-1}\left(M\frac{F^{2}}{4}\right)\cos\left(4\pi\left(2n-1\right)\eta\right)\right)}\\
{\displaystyle \times\left(\sum_{n=-\infty}^{\infty}J_{n}(NF)\sin\left(2\pi n\eta\right)\right)}
\end{array}\\
 & =0,
\end{align*}
where we have made use of the relations given in Eqs. \eqref{eq:AppMath_TrigIntegrals1},
\eqref{eq:AppMath_TrigIntegrals}, and \eqref{eq:AppMath_TrigIntegrals2}.
Using these results and the trigonometric identities of Eqs. \eqref{eq:AppMath_TrigCosAB}
and \eqref{eq:AppMath_TrigSinAB}, we find for the integral of the
entire Hamiltonian in Eq. \eqref{eq:CHTorsMagn_K1UnwieldyExpansion}
\begin{align*}
\int_{0}^{1}d\eta\, K_{1}\left(\eta,J\right)= & \varepsilon_{p}\Bigg(\cos\left(M-M\frac{F^{2}}{4}+\psi_{p}\right)\left(\sum_{n=-\infty}^{\infty}(-1)^{n}J_{2n}\left(M\frac{F^{2}}{4}\right)J_{4n}(NF)\right)\\
 & \phantom{\varepsilon_{p}}+\sin\left(M-M\frac{F^{2}}{4}+\psi_{p}\right)\left(\sum_{n=-\infty}^{\infty}(-1)^{n}J_{2n-1}\left(M\frac{F^{2}}{4}\right)J_{4n-2}(NF)\right)\Bigg).
\end{align*}
The frequency shift of Eq. \ref{eq:CHTorsMagn_FreqShiftetaJGeneric}
becomes
\[
\frac{\Delta f}{\varepsilon_{p}}=\begin{array}[t]{l}
{\displaystyle \frac{MF^{2}}{4J}\sin Q\left(\sum_{n=-\infty}^{\infty}(-1)^{n}J_{2n}\left(\frac{MF^{2}}{4}\right)J_{4n}(NF)\right)}\\
{\displaystyle +\frac{NF}{4J}\cos Q\left(\sum_{n=-\infty}^{\infty}(-1)^{n}J_{2n}\left(\frac{MF^{2}}{4}\right)\left(J_{4n-1}(NF)-J_{4n+1}(NF)\right)\right)}\\
{\displaystyle +\frac{MF^{2}}{8J}\cos Q\left(\sum_{n=-\infty}^{\infty}(-1)^{n}\left(J_{2n-1}\left(\frac{MF^{2}}{4}\right)-J_{2n+1}\left(M\frac{F^{2}}{4}\right)\right)J_{4n}(NF)\right)}\\
+\ldots
\end{array}
\]
\[
\phantom{\frac{\Delta f}{\varepsilon_{p}}=}\begin{array}[t]{l}
{\displaystyle -\frac{MF^{2}}{4J}\cos Q\left(\sum_{n=-\infty}^{\infty}(-1)^{n}J_{2n-1}\left(\frac{MF^{2}}{4}\right)J_{4n-2}(NF)\right)}\\
{\displaystyle +\frac{NF}{4J}\sin Q\left(\sum_{n=-\infty}^{\infty}(-1)^{n}J_{2n-1}\left(\frac{MF^{2}}{4}\right)\left(J_{4n-3}(NF)-J_{4n-1}(NF)\right)\right)}\\
{\displaystyle +\frac{MF^{2}}{8J}\sin Q\left(\sum_{n=-\infty}^{\infty}(-1)^{n}\left(J_{2n-2}\left(\frac{MF^{2}}{4}\right)-J_{2n}\left(M\frac{F^{2}}{4}\right)\right)J_{4n-2}(NF)\right).}
\end{array}
\]
where we have used the identities $\partial_{J}F=F/2J$ and $J_{n}^{\prime}=(J_{n-1}-J_{n+1})/2$
and the abbreviation $Q=M-MF^{2}/4+\psi_{p}$.

To simplify the expression for the shift in the resonant frequency
of the cantilever, we need to make further approximations. We note
that $F=\alpha x_{\max}/l$ and $M=2\pi p\phi_{\text{tot}}/\phi_{0}$.
In the case of the experiment, $F\lesssim1.377\times(200\,\text{nm})/(150\,\text{\ensuremath{\mu}m})\approx2\times10^{-3}$
and $M\lesssim3\times10^{4}$ for $p=1$, $B=9\,\text{T}$ and $r=780\,\text{nm}$.
Thus, the relations $MF^{2}/4\ll1$ and $F^{2}/4\ll1$ both hold.
In the limit $x\ll1$, the Bessel functions obey $J_{n}(x)\propto x^{n}$.
Retaining only the lowest order terms (with $J_{0}(MF^{2}/4)\sim1$)
and taking $F^{2}/4\sim0$, we have
\[
\Delta f\approx\varepsilon_{p}\left(-\frac{NF}{2J}\cos\left(M+\psi_{p}\right)J_{1}(NF)+{\displaystyle \frac{MF^{2}}{4J}\sin\left(M+\psi_{p}\right)J_{0}(NF))}\right).
\]
Specifically, we have kept the $n=0$ term of the first two lines
of the previous expression for $\Delta f$ and allowed the $n=0$
and $n=1$ terms of the last line to cancel. All other terms in all
of the sums were dropped because they contain a factor $J_{n}(MF^{2}/4)$
with $\left|n\right|>0$. Restoring the full expressions for $\varepsilon_{p}$,
$M$, $N$, and $F$ and the sum over $p$ and writing $J=kx_{\max}^{2}/2f_{0}$,
we find
\begin{align*}
\Delta f=\frac{f_{0}}{2k}\sum_{p}I_{p}\Big(- & \frac{2\pi p}{\phi_{0}}\left(AB\cos\theta_{0}\frac{\alpha}{l}\right)^{2}\cos\left(2\pi p\frac{AB}{\phi_{0}}\sin\theta_{0}+\psi_{p}\right)\mathrm{jinc}\left(2\pi p\frac{AB}{\phi_{0}}\cos\theta_{0}\frac{\alpha}{l}x_{\text{\ensuremath{\max}}}\right)\\
 & +AB\sin\theta_{0}\left(\frac{\alpha}{l}\right)^{2}\sin\left(2\pi\frac{AB}{\phi_{0}}\sin\theta_{0}+\psi_{p}\right)J_{0}\left(2\pi p\frac{AB}{\phi_{0}}\cos\theta_{0}\frac{\alpha}{l}x_{\text{\ensuremath{\max}}}\right)\\
\phantom{\Delta f}=\frac{f_{0}}{2k}\sum_{p}I_{p}\Big(- & \frac{2\pi p}{\phi_{0}}\left(AB\cos\theta_{0}\frac{\alpha}{l}\right)^{2}\cos\left(2\pi p\frac{\phi_{\text{tot}}}{\phi_{0}}+\psi_{p}\right)\mathrm{jinc}\left(2\pi p\frac{\phi_{\max}}{\phi_{0}}\right)\\
 & +AB\sin\theta_{0}\left(\frac{\alpha}{l}\right)^{2}\sin\left(2\pi\frac{\phi_{\text{tot}}}{\phi_{0}}+\psi_{p}\right)J_{0}\left(2\pi p\frac{\phi_{\max}}{\phi_{0}}\right).
\end{align*}
where we define $\mathrm{jinc}(x)=2J_{1}(x)/x$ and use $\phi_{\max}=AB\cos\theta_{0}\alpha x_{\text{\ensuremath{\max}}}/l$
to represent amplitude of the change $\phi$ in flux. In the limit
$x_{\max}\rightarrow0$, this expression matches Eq. \eqref{eq:CHTorsMagn_ZeroDriveFreqShiftBothTerms}
as expected.%
\footnote{One might wonder why we did not drop the second order terms earlier
in the derivation. Doing so greatly simplifies the derivation. However,
it also misses the second term in the expression above and thus does
not match Eq. \eqref{eq:CHTorsMagn_ZeroDriveFreqShiftBothTerms} in
the limit of zero cantilever amplitude. Since either of the two terms
in Eq. \eqref{eq:CHTorsMagn_ZeroDriveFreqShiftBothTerms} can be dominant
depending on the parameters chosen, I find it more reassuring to follow
the derivation in a way that produces both of them.%
} As before, for the conditions of the experiment the first term dominates
and we can take 
\begin{align}
\Delta f & =-\frac{f_{0}}{2k}\sum_{p}\frac{2\pi p}{\phi_{0}}I_{p}\left(AB\cos\theta_{0}\frac{\alpha}{l}\right)^{2}\cos\left(2\pi p\frac{AB}{\phi_{0}}\sin\theta_{0}+\psi_{p}\right)\mathrm{jinc}\left(2\pi p\frac{AB}{\phi_{0}}\cos\theta_{0}\frac{\alpha}{l}x_{\text{\ensuremath{\max}}}\right)\nonumber \\
 & =-\frac{f_{0}}{2k}\sum_{p}\frac{2\pi p}{\phi_{0}}I_{p}\left(AB\cos\theta_{0}\frac{\alpha}{l}\right)^{2}\cos\left(2\pi p\frac{\phi_{\text{tot}}}{\phi_{0}}+\psi_{p}\right)\mathrm{jinc}\left(2\pi p\frac{\phi_{\max}}{\phi_{0}}\right)\label{eq:CHTorsMagn_FiniteAmpFreqShift}
\end{align}
as the shift in the resonant frequency of the cantilever due to the
persistent current for the case of finite cantilever amplitude.

\begin{figure}

\centering{}\includegraphics[width=0.5\paperwidth]{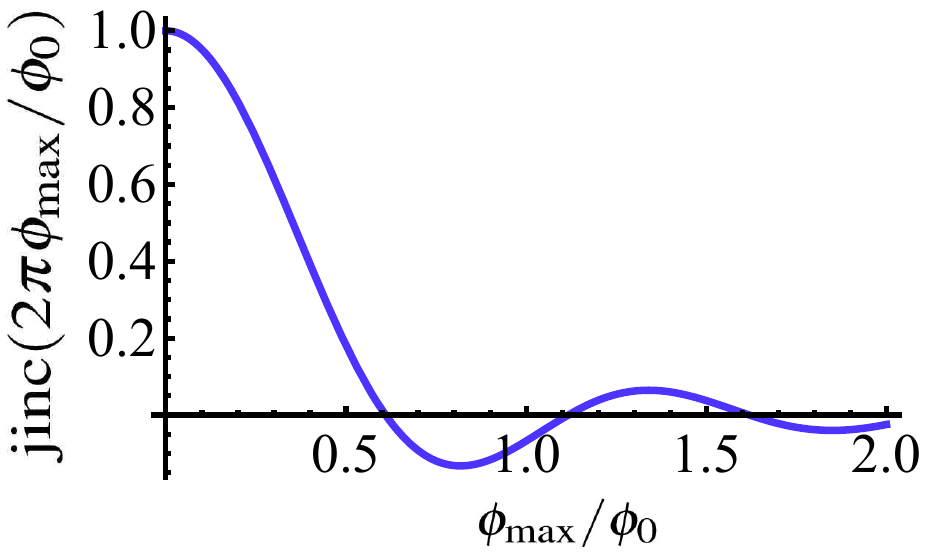}\caption[Persistent current signal suppression factor due to finite cantilever
amplitude]{\label{fig:CHTorsMagn_JincPlot}Persistent current signal suppression
factor due to finite cantilever amplitude. As the amplitude $\phi_{\max}=AB\cos\theta_{0}\alpha x_{\text{\ensuremath{\max}}}/l$
of the change in flux threading the ring during cantilever oscillation
increases, the resonant frequency shift due to the persistent current
is suppressed. The characteristic scale of this suppression is $\phi_{\max}=\phi_{0}/2p$
(in the figure $p=1$). The suppression has an oscillatory component,
but the subsequent peaks are successively smaller and never surpass
14\% of the value of the peak at $\phi_{a,\max}=0$.}
\end{figure}

In Fig. \eqref{fig:CHTorsMagn_JincPlot}, the suppression factor $\mathrm{jinc}(2\pi p\phi_{\max}/\phi_{0})$
is plotted versus $\phi_{\max}/\phi_{0}$ for $p=1$. The main consequence
of the form of the suppression factor is that during measurements
of the persistent current the amplitude $x_{\max}$ of the cantilever
tip must be kept small enough that $\phi_{\max}=AB\cos\theta_{0}\alpha x_{\max}/l\lesssim\phi_{0}/3p$
for the highest $p$ of interest. The suppression factor also has
an impact on the analysis of the measured frequency signal. At first,
it might appear that Eq. \eqref{eq:CHTorsMagn_FiniteAmpFreqShift}
states that the frequency shift of the cantilever is proportional
to the convolution of $\partial I/\partial\phi_{\text{tot}}$ with
the inverse Fourier transform of the $\mathrm{jinc}$ suppression
factor. However, note that this factor depends on both the transform
variable $p$ and the original variable $\phi_{\text{tot}}$.%
\footnote{It is $\phi_{a,\max}$ that appears in the $\text{jinc}$ factor,
but experimentally $\phi_{a,\max}$ and $\phi_{\text{tot}}$ are not
independent since they are both proportional to the applied magnetic
field $B$.%
} This dependency on $p$ and $\phi_{\text{tot}}$ complicates the
analysis. Dealing with this factor will be covered in more detail
when data analysis is discussed in \ref{sec:ChData_SigProc}.

Finally, we address one concern that might be raised regarding this
analysis. In deriving Eq. \ref{eq:CHTorsMagn_FiniteAmpFreqShift},
we have treated $I_{p}$ and $\psi_{p}$ as constants while in \ref{sub:CHPCTh_FluxThroughMetal}
it was shown that the persistent current oscillation has a finite
range of correlation in magnetic field due to the effect of magnetic
field penetrating the metal of the ring. One might wonder whether
$I_{p}$ and $\psi_{p}$ change as the ring tilts in the static magnetic
field and the flux $\phi_{\text{tot}}$ changes. We believe that such
changes in $I_{p}$ and $\psi_{p}$ should be small because, although
the tilting of the ring with the cantilever's motion results in changes
of $\phi_{\text{tot}}$ on the order of $\phi_{0}$, the change in
the flux $\phi_{M}$ actually penetrating the metal of the ring should
be small as long as the aspect ratio of the ring cross-section is
not extreme.

\chapter{\label{cha:CHExpSetup_}Experimental set-up and measurement}

\section{Cantilever sample fabrication}

\subsection{\label{sub:CHExpSetup_ThermometryCantilevers}Preliminary thermometry
cantilevers}

\begin{figure}
\centering{}\includegraphics[clip,width=0.5\paperwidth]{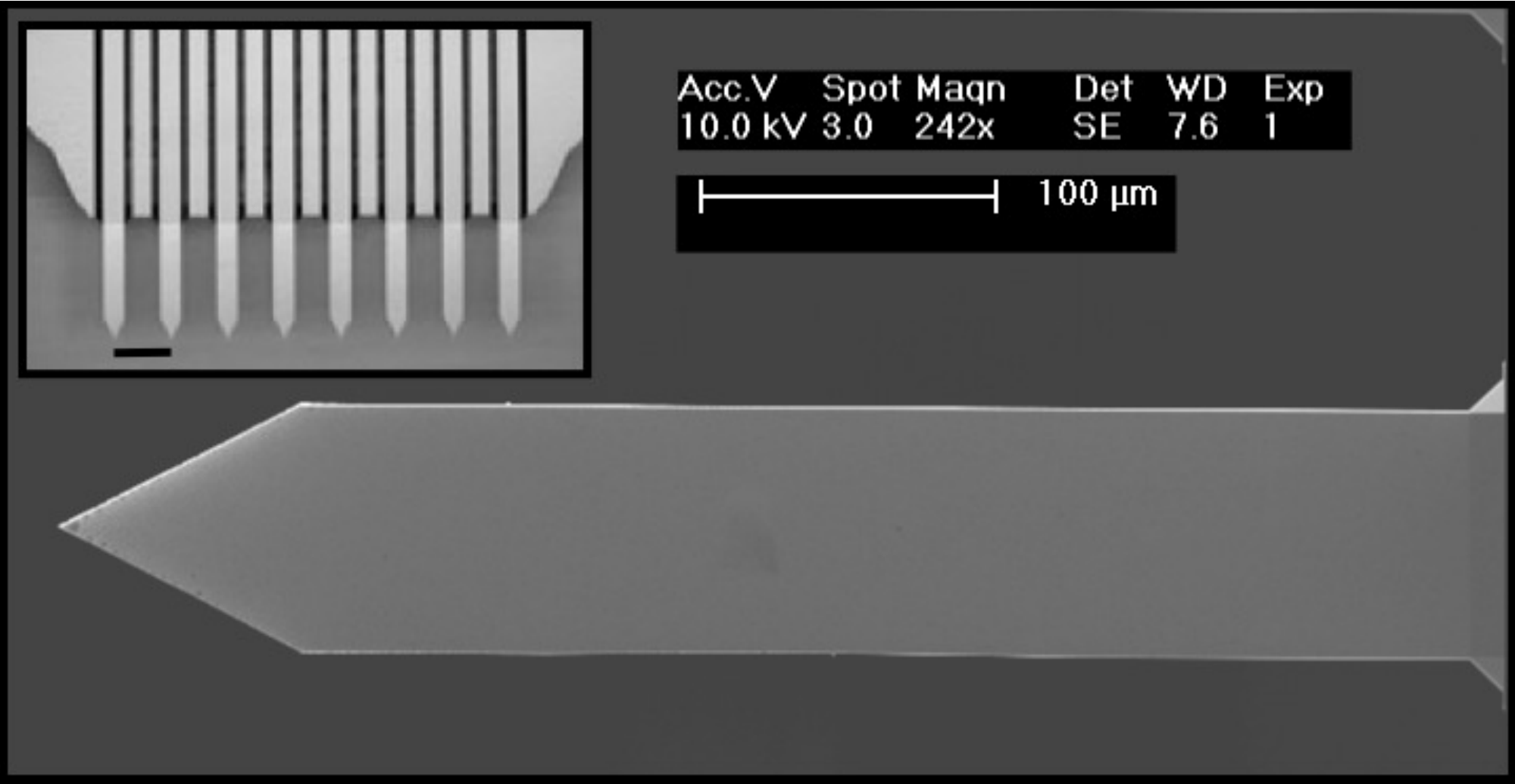}\caption[Bare cantilever used for Brownian motion measurements.]{\label{fig:CHExpSetup_BareArrows}Bare cantilever used for Brownian
motion measurements. The main figure shows the bare cantilever used
in Brownian motion thermometry measurements described in \ref{sub:CHSensitivity_ThermometrySection}.
The inset shows a typical Arrow$^{\text{TM}}$ TL8 chip with no samples
attached to the cantilever tips. The scale bar in the inset is 250
$\mu$m long.}
\end{figure}

For the preliminary experiment discussed in \ref{sub:CHSensitivity_ThermometrySection},
commercially available atomic force microscope (AFM) cantilevers were
used. In particular, we used the Arrow$^{\text{TM}}$ TL8 chip from
NanoWorld (NanoWorld, Neuchâtel, Switzerland). These are tipless cantilevers
with nominal dimensions of 500 $\mu$m by 100 $\mu$m by 1 $\mu$m
and typical spring constant and resonant frequency of 0.03 N/m and
6 kHz, respectively. A single chip contains eight cantilevers with
a center to center pitch of 250 $\mu$m between cantilevers. A picture
of a typical chip is shown in the inset of Fig. \ref{fig:CHExpSetup_BareArrows}.

\begin{figure}

\begin{centering}
\includegraphics[width=0.5\paperwidth]{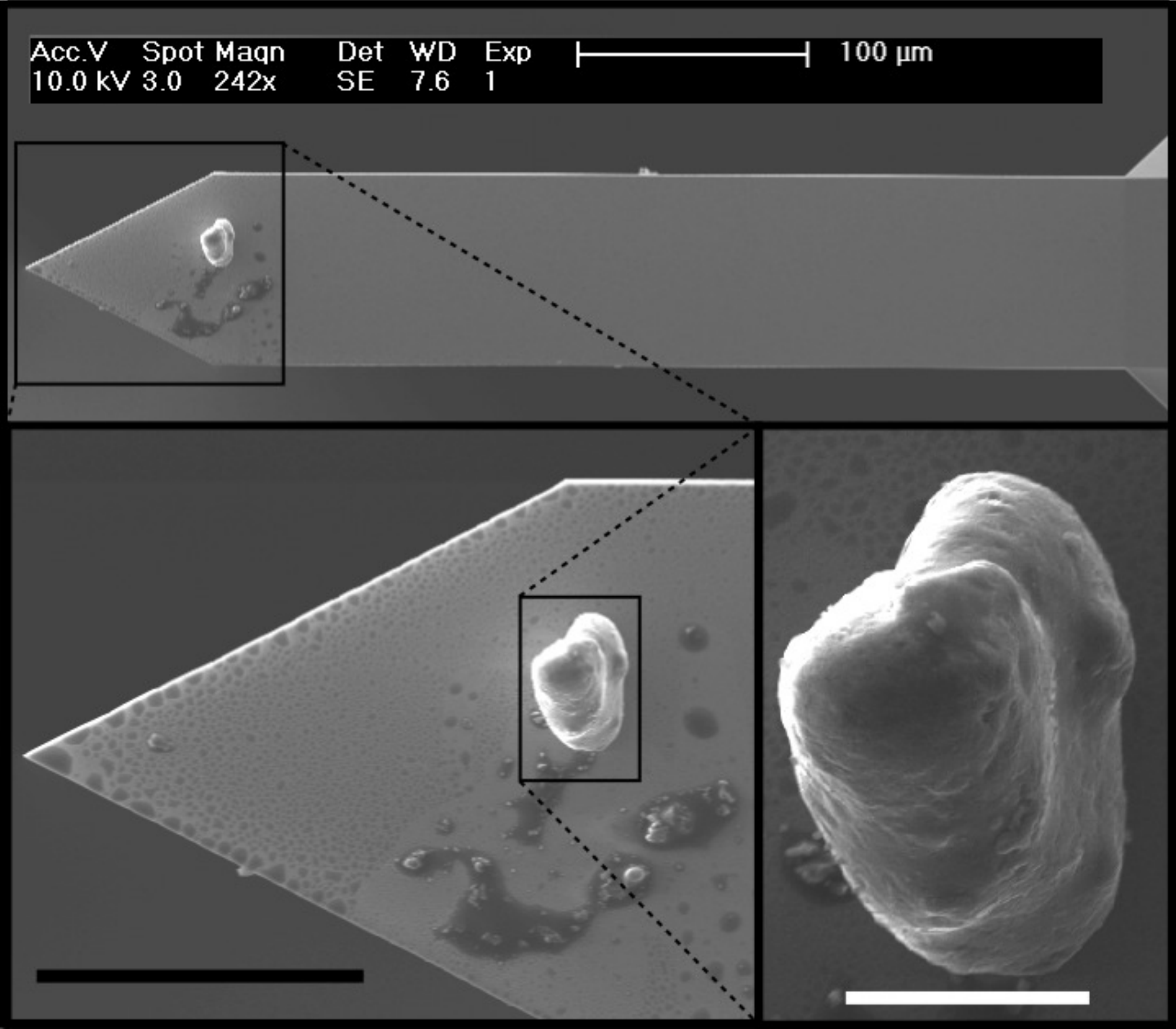}\caption[Cantilever with aluminum grain glued on tip]{\label{fig:CHExpSetup_AlGrainArrow}Cantilever with aluminum grain
glued on tip. The top panel shows the cantilever used for the aluminum
superconducting transition thermometry measurements described in \ref{sub:CHSensitivity_ThermometrySection}.
The lower two panels show successive magnified images of the aluminum
grain. The discolorations near the grain are puddles of epoxy and
smaller pieces of aluminum. The scale bar in the lower left panel
is 50 $\mu$m long, and the one in the lower right is 10 $\mu$m long.}

\par\end{centering}

\end{figure}

Various samples were glued to the ends of the arrow cantilevers using
thermally conductive Stycast$^{\textregistered}$ 2850 FT epoxy cured
with Catalyst 23LV (Henkel Emerson \& Cuming, Billerica, MA, USA).
In principle, up to eight different samples could be measured in one
cool down of the cryostat using an Arrow$^{\text{TM}}$ TL8 chip.
In practice, we only measured two cantilevers in any detail: the bare
cantilever shown in Fig. \ref{fig:CHExpSetup_BareArrows} used for
the Brownian motion thermometry measurements and the aluminum grain-mounted
cantilever shown in Fig. \ref{fig:CHExpSetup_AlGrainArrow} used for
the superconducting transition thermometry measurements. The aluminum
grain was taken from a 99.99\% pure aluminum shot sample (ESPI, Ashland,
OR, USA).

\FloatBarrier

\subsection{\label{sub:CHExpSetup_PersistentCurrentFabrication}Persistent current
cantilever-with-ring sample fabrication}

The persistent current samples discussed in this text were created
at the Cornell NanoScale Facility (CNF) located on the campus of Cornell
University in Ithaca, NY, USA. The majority of the preliminary work
determining dosages, exposure times, etc. was performed by Ania Jayich
with guidance from Rob Ilic of the CNF. Ania and I fabricated most
of the samples reported on in this text together.

\begin{figure}
\centering{}\includegraphics[width=0.5\paperwidth]{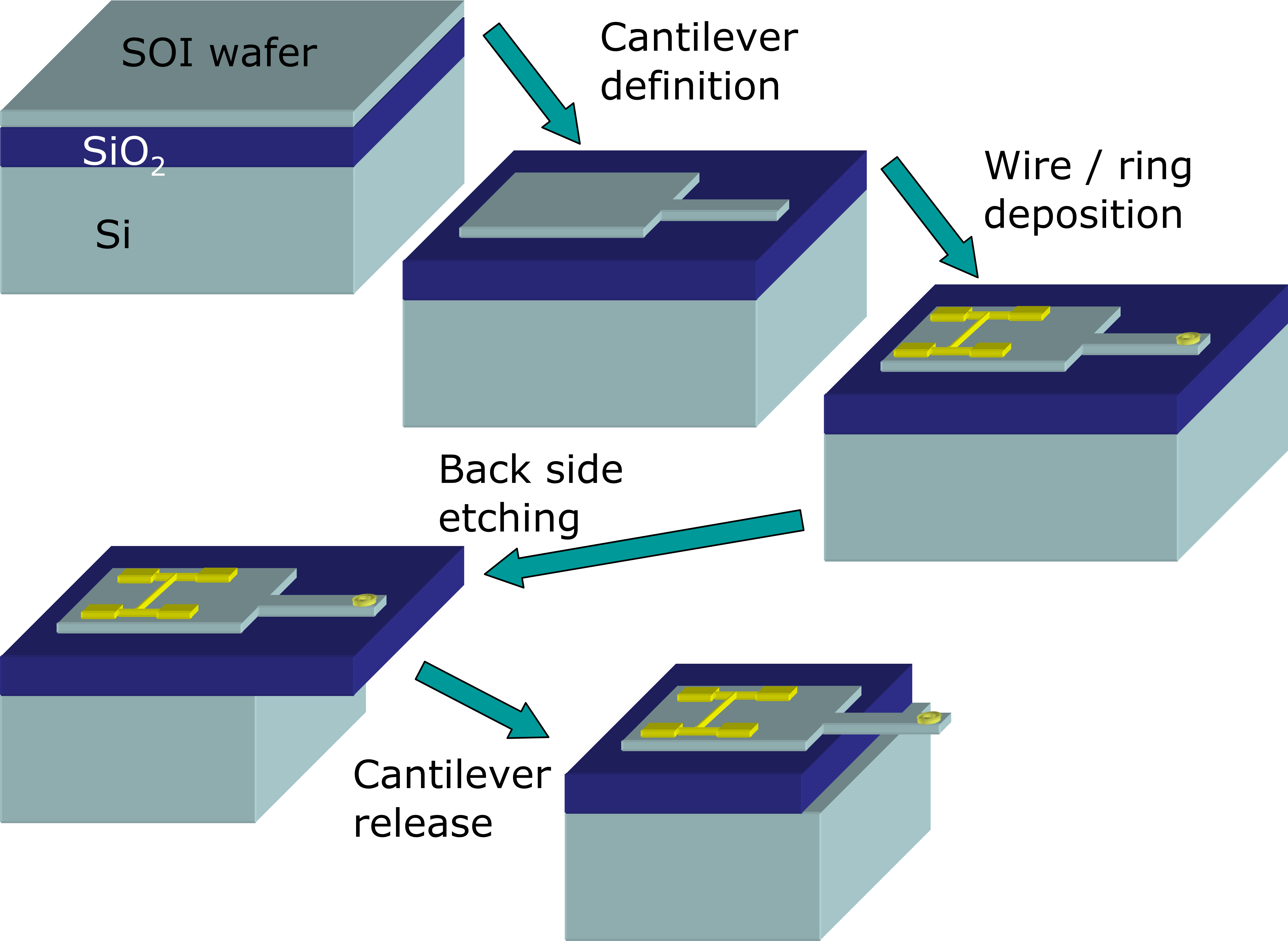}\caption[Persistent current sample fabrication schematic]{\label{fig:CHExpSetup_FabSchematic}Persistent current sample fabrication
schematic. \textbf{Cantilever definition:} photolithography and reactive
ion etch. \textbf{Wire / ring deposition:} electron beam lithography
and electron beam evaporation of aluminum. \textbf{Backside etching:}
photolithography and deep reactive ion etching. \textbf{Cantilever
release:} wet etch with buffered oxide etchant and critical point
dry.}
\end{figure}

An outline of the major steps in the persistent current sample fabrication
procedure is depicted in Fig. \ref{fig:CHExpSetup_FabSchematic}.
We begin with a clean silicon-on-insulator (SOI) wafer. The microelectronics
industry has developed etching procedures highly selective to silicon
and silicon dioxide. The insulator/oxide layer of the SOI wafer serves
as an etch stop for both frontside and backside processing during
the cantilever fabrication. The cantilevers are defined on the frontside
by using standard photolithography and a reactive ion etch (RIE) to
remove the surrounding frontside silicon. Then the rings and wires
are defined with electron beam lithography and created with standard
electron beam evaporation and lift-off. After this, the backside silicon
is removed by deep reactive ion etch (DRIE) of windowed regions defined
by another step of photolithography. Finally the cantilevers are released
with a wet etch of the silicon dioxide membrane and dried in a critical
point dryer. Images of fully fabricated persistent current samples
are shown in Fig. \ref{fig:CHExpSetup_PCSamples}. A more detailed
account of the fabrication procedure is given in Appendix \ref{app:AppSampFab}.

\begin{figure}

\centering{}\includegraphics[width=0.6\paperwidth]{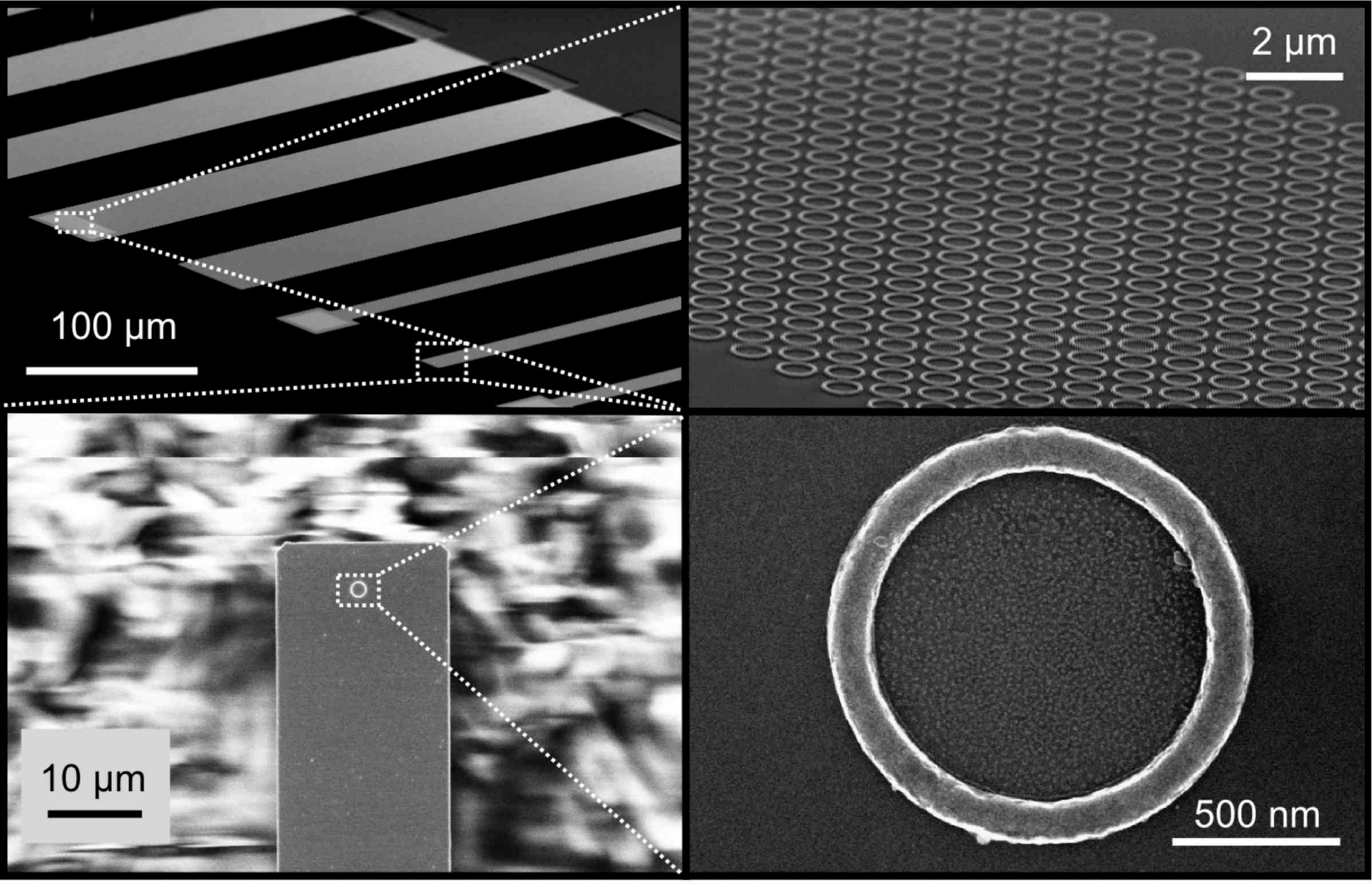}\caption[Scanning electron micrograph images of cantilevers with integrated
aluminum rings]{\label{fig:CHExpSetup_PCSamples}Scanning electron micrograph images
of cantilevers with integrated aluminum rings. The upper left panel
shows an angled view of several cantilevers on one cantilever chip.
The samples on the ends of the cantilevers alternate between arrays
and single rings. The other panels show magnified images of an array
of rings and a cantilever with a single ring as indicated by the dotted
lines in the figure. Some of the distortion in the array image is
due to cantilever vibration which is difficult to eliminate when imaging
released cantilevers. The samples shown in the lower two panels were
part of different chips but have similar dimensions to the regions
indicated in adjacent panels.}
\end{figure}

\FloatBarrier

\section{Experimental apparatus}

\subsection{\label{sub:ChSens_DewarFridge}Dewar and refrigerator}

\begin{figure}
\begin{centering}
\includegraphics[width=0.7\paperwidth]{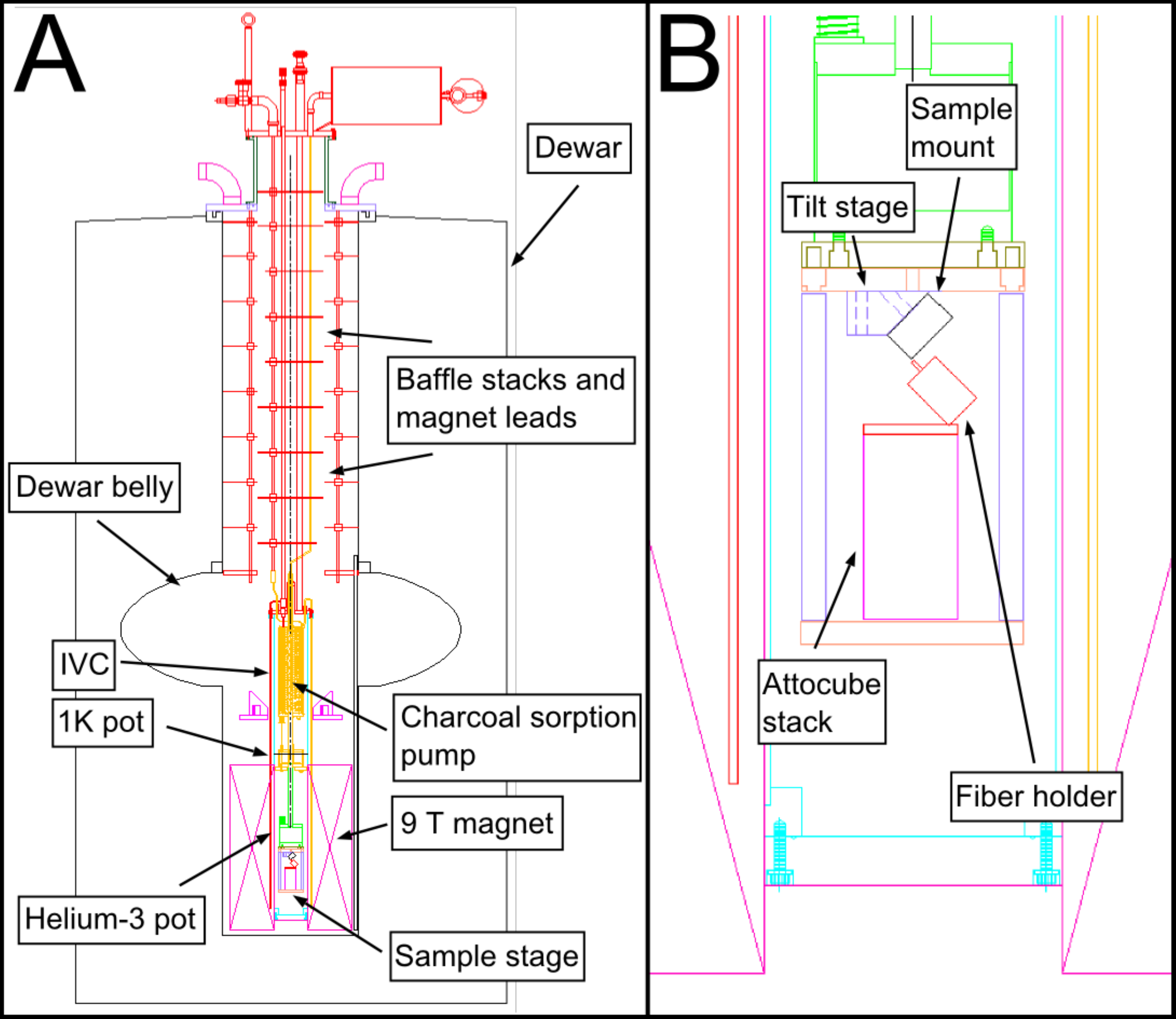}\caption[Schematic of Dewar and helium-3 refrigerator]{\label{fig:CHExpSetup_FridgeDewarSchematic}Schematic of Dewar and
helium-3 refrigerator. Panel A shows the entire Dewar and refrigerator
layout with the main elements labeled. The IVC is the inner vacuum
chamber. Panel B shows a close-up of the sample stage hanging from
the bottom of the refrigerator. The configuration for cantilevers
mounted at $45^{\circ}$ is shown.}

\par\end{centering}

\end{figure}

All experiments discussed in this text were performed in a helium-3
refrigerator (He-3-SSV, Janis Research Company, Inc., Wilmington,
MA, USA) inserted into a $70\,$L helium Dewar (Precision Cryogenic
Systems, Inc., Indianapolis, IN, USA) equipped with a $9\,$T magnet
(American Magnetics Inc., Oak Ridge, TN, USA).%
\footnote{The magnet was controlled with the Model 420 Power Supply Programmer
(American Magnetics Inc., Oak Ridge, TN, USA) and the 4Q050100PS Four
Quadrant Power Supply (American Magnetics Inc., Oak Ridge, TN, USA).%
} The magnet had a 3'' bore which fit snugly around the outside of
the refrigerator's inner vacuum chamber. The Dewar was mounted on
a felt-covered aluminum frame (1'' by 2'' cross-section aluminum
beams, 80/20 Inc., Columbia City, IN, USA) with four feet. Each foot
was supported by a stack of 2.5'' by 2.5'' by .25'' pieces of ultra-soft
polyurethane (McMaster-Carr, Elmhurst, IL, USA) and ribbed elastomer
(Vib-X-Pads, Vibrasciences, Branford, CT, USA) separated by similarly
sized pieces of aluminum. These stacks were the only form of vibration
isolation added to the cryostat.%
\footnote{It is possible that the structure of the cryostat naturally provides
some vibration isolation itself. The magnet also provided some vibration
reduction through eddy current damping.%
} A schematic of the refrigerator and Dewar is shown in Fig. \ref{fig:CHExpSetup_FridgeDewarSchematic}.

The refrigerator was outfitted with thermometers on its charcoal sorption
pump, $1\,$K pot, and helium-3 pot and with resistive heaters on
the charcoal sorption pump and helium-3 pot. An additional thermometer
(RX-202A, Lake Shore Cryotronics, Inc., Westerville, OH, USA) was
glued into each sample mounting piece with the sample with Stycast$^{\textregistered}$
2850 FT in order to achieve good thermal contact. This thermometer
was wired with the refrigerator's 0.005'' manganin wire. The thermometers
were measured and the heaters controlled by a piece of electronics
from Lake Shore (Model 340 Temperature Controller, Lake Shore Cryotronics,
Inc., Westerville, OH, USA). The usual mode of operation%
\footnote{For fine temperature control close to the refrigerator base temperature
and for temperatures above $4.2\,$K, the helium-3 pot heater was
used.%
} of the refrigerator after condensing the helium-3 was to leave the
$1\,$K pot's needle valve and exhaust valve open (so that it was
well anchored to $4\,$K) and to heat the charcoal sorption pump slightly
($\sim20\,$K) to control the sample stage temperature. A constant
sample temperature was maintained by a PID feedback loop controlling
the charcoal sorption pump heater while monitoring the sample stage
thermometer. The feedback loop was implemented by the Lake Shore controller.
The lowest temperatures were reached by allowing the $1\,$K pot to
fill completely, then closing the needle valve, and pumping on the
$1\,$K pot's exhaust line (using a SC 15 D scroll pump from Oerlikon
Leybold Vacuum GmbH, Cologne, Germany). A base temperature of $287\,$mK
was achieved by this method with the cantilever detection apparatus
mounted to the refrigerator. 

\FloatBarrier

\subsection{\label{sub:CHExpSetup_CantileverDetectionSetup}Cantilever detection
set-up}

The basic cantilever detection set-up is shown in Fig. \ref{fig:CHExpSetup_CantileverMeasurementSchematic}.
The cantilever motion is detected optically using a fiber-based interferometer
in a manner similar to that described in Refs. \citep{rugar1991mechanical},
\citep{albrecht1991frequency}, and \citep{albrecht1992lowtemperature}.
The cantilever signal is converted to a voltage signal using a photodiode-amplifier
package and measured with a lock-in amplifier. The cantilever is mounted
on a piezoelectric actuator which is driven in a phase-locked loop
by the local oscillator of the lock-in which uses the cantilever signal
as its clock. Various feedback circuits were used to control the laser
power and wavelength and the cantilever amplitude of motion. We will
now discuss each component of the set-up in more detail.

\subsubsection{\label{sub:CHExpSetup_LaserSetup}Laser source}

The laser source used in all persistent current measurements was a
$1550\,$nm fiber-coupled diode laser from JDS Uniphase (CQF935/66
26 $50\,$mW $1550\,$nm CW DFB Laser with PM fiber for WDM applications,
JDS Uniphase, Milpitas, CA, USA).%
\footnote{The thermometry experiments discussed in \ref{sub:CHSensitivity_ThermometrySection}
and preliminary attempts at measuring persistent currents were made
using a laser source from Thorlabs (S3FC $1550\,$nm DFB Benchtop
Laser Source, Thorlabs, Newton, NJ, USA). This laser source was less
tunable than the JDS Uniphase laser. We did not implement feedback
schemes for this laser's wavelength or power.%
} The laser was powered by a low noise current source from ILX (LDX-3620
Ultra low noise current source, ILX Lightwave, Bozeman, MT, USA).
The driving current of the laser was approximately $100\,$mA for
all measurements. During measurement, the LDX-3620 current source
powering the laser was operated in its constant power mode. In constant
power mode, the LDX-3620 current source stabilized its output drive
to the laser by feeding back on the output of the reference photodiode
(described below). The voltage output of the reference photodiode
package was connected through a $1\,$k$\Omega$ resistor to the LDX-3620
photodiode reference port, which required a current input. The laser
wavelength was tuned via its temperature using a thermoelectric cooler
mount (LM14S2 Universal 14-Pin Butterfly Laser Diode Mount, Thorlabs,
Newton, NJ, USA) and controller (TED200C Thermoelectric Temperature
Controller, Thorlabs, Newton, NJ, USA). The thermoelectric cooler
mount was equipped with a bias-T adapter which allowed for RF modulation
of the laser driving current. 

We employed a constant RF modulation of the laser current in order
to reduce optical feedback noise and optical interference noise \citep{arimoto1986optimum,ojima1986diodelaser,fukuma2005development}.
The $\sim1\,$MHz RF modulation signal was generated by a voltage-controlled
oscillator (ZX95-850W+, Mini-Circuits, Brooklyn, NY, USA) which passed
through a voltage-variable attenuator (ZX73-2500+, Mini-Circuits,
Brooklyn, NY, USA) and an amplifier (ZFL-1000VH2, Mini-Circuits, Brooklyn,
NY, USA). After the laser was first turned on, the RF components were
turned on%
\footnote{It is important to turn the laser on before the RF components to avoid
reverse biasing the laser diode.%
} and the voltage-controlled oscillator and voltage-variable attenuator
were tuned while watching the interferometer signal (to be described
below) on an oscilloscope (DPO 2014, Tektronix, Beaverton, OR, USA)
until the signal became quiet. The transition from noisy to quiet
interferometer signal was very sharp and typically reflected the reduction
of interference between unwanted reflectors in the fiber interferometer
path through the shortening of the laser coherence length. The shortened
coherence length used during measurements of the cantilever motion
was approximately $1\,\text{cm}$.

\subsubsection{Fiber optic components}

\begin{figure}
\centering{}\includegraphics[width=0.7\paperwidth]{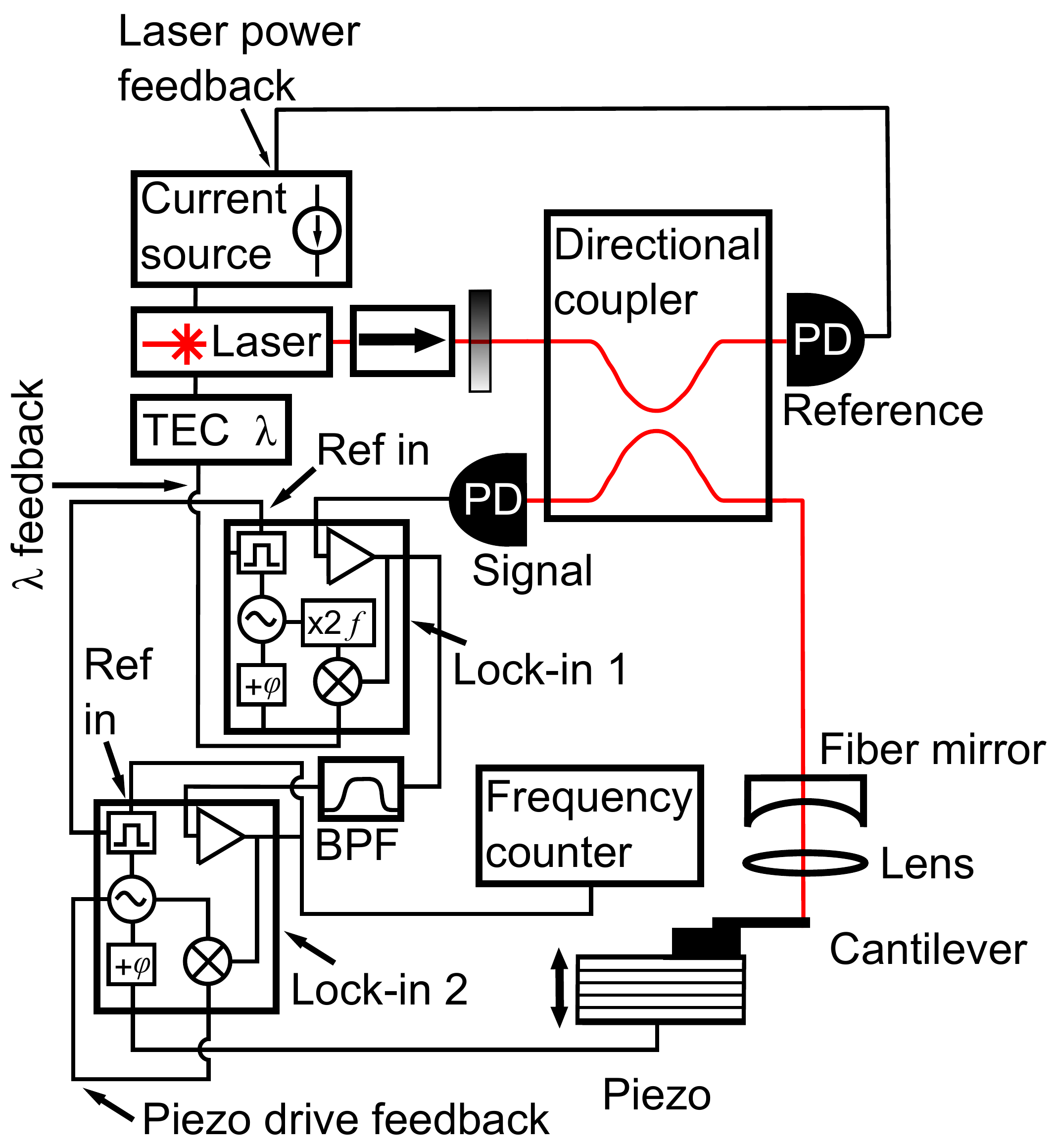}\caption[Cantilever measurement schematic]{\label{fig:CHExpSetup_CantileverMeasurementSchematic}Cantilever
measurement schematic. The figure displays the arrangement of optical
and electronic components during a typical measurement of the cantilever's
motion, such as during a persistent current measurement. The laser
first passes through an optical isolator and variable attenuator before
entering the 99:1 directional coupler. Two of the directional coupler
ports are connected to photodiodes (PD) while the other passes through
a feedthrough into the fridge and down to the cantilever. The reference
photodiode reading is used to stabilize the input laser power. The
signal photodiode reading is fed through two lock-in amplifiers and
a band pass filter (BPF). The second lock-in drives the piezo actuator
holding the cantilever in a phase-locked loop with the signal photodiode
as its reference. The outputs of the lock-in mixers (circles with
crosses) are to be understood schematically as containing both magnitude
and phase information which is processed externally by a computer
to implement the feedback loops on the cantilever amplitude through
the piezo drive and laser wavelength through the thermoelectric temperature
controller (TEC). More details of the cantilever measurement are given
in the text.}
\end{figure}

The optical beam path used to monitor the cantilever position is shown
in Fig. \ref{fig:CHExpSetup_CantileverMeasurementSchematic}. The
fiber-coupled output of the laser was connected in series first to
an optical isolator (4015SAFC, Thorlabs, Newton, NJ, USA) and then
to a variable attenuator (VOA50-FC, Thorlabs, Newton, NJ, USA). The
attenuator could be adjusted by hand to change the laser power between
measurements. The driving current of the laser was not varied to change
the laser power. The attenuator output was connected to the input
port of a 99:1 directional coupler (10202A-99-FC, Thorlabs, Newton,
NJ, USA). The through port of the directional coupler was connected
to a fiber-coupled photodiode package (2011-FC 200-kHz Front-End Photoreceiver,
New Focus, Santa Clara, CA, USA) which we refer to as the reference
photodiode. The coupled port of the directional coupler was connected
to a long fiber which was fed into the inner vacuum chamber (IVC)
of the cryostat and mounted with its end addressing the cantilever
to be detected. The details of this fiber are discussed more below.
The isolated port of the directional coupler (which is also the through
port from the perspective of the optical signal returning from the
cantilever) was connected to another photodiode package identical
to the reference photodiode's and is known as the signal photodiode
in this text.

The optical path from the directional coupler down to the cantilever
was composed of one single fiber (custom-ordered $8\,$m 9/125 bare
fiber pigtails with $5\,$m of furcation tubing, Fiber Instrument
Sales, Inc., Oriskany, NY, USA).%
\footnote{The optical fibers were made by Corning (SMF-28, Corning Inc., Corning,
NY, USA).%
} Early on, we attempted to use various commercial fiber vacuum feedthroughs.
However, all feedthroughs we used had sizable losses which seemed
to vary when the fiber was disconnected and reconnected. We chose
to do away with the fiber feedthrough element to simplify determination
of the incident laser power on the cantilever. The vacuum feedthrough
was achieved by epoxying the fiber into a 2'' long copper tube with
outer diameter chosen to match the inner diameter of a Swagelok$^{\textregistered}$
Ultra-Torr fitting (Swagelok, Solon, OH, USA). From outside to inside,
the optical fiber layers were a $3\,$mm yellow PVC jacket, a woven
layer of yellow aramid yarn, a $1.7\,$mm Teflon tube, a $250\,\mu$m
acrylate buffer coating, and the $9/125\,\mu$m fiber core/cladding.
In order to make the feedthrough the outer PVC, aramid yarn, and teflon
tube layers were terminated inside the copper tube.%
\footnote{Alternatively, the Teflon tube was terminated just after the the copper
tube, but a section of the teflon which would be inside of the copper
tube was cut away so that epoxy could seep into the tube. This arrangement
provided extra strain relief for the buffer coated fiber emerging
from the epoxy.%
} The PVC jacket of the fiber was then glued into place using 5-Minute$^{\textregistered}$
Epoxy (ITW Devcon, Danvers, MA, USA) to form a plug on the atmosphere
side of the copper tube. The tube was then turned atmosphere side
down and filled with Stycast$^{\textregistered}$1266 A/B epoxy (Henkel
Emerson \& Cuming, Billerica, MA, USA). 1266 was chosen for its low
viscosity, which allowed it to fill the small openings between different
fiber jacket layers.

\subsubsection{Cantilever detection fiber termination and mounting}

The preparation of the termination of the fiber directed at the cantilever
changed over the course of the experiment. In the earliest measurements
(such as the thermometry measurements discussed in \ref{sub:CHSensitivity_ThermometrySection}),
the fiber was epoxied into a thin copper tube and then cleaved. The
copper tube was epoxied to a piece of brass which was screwed down
to the Attocube positioners described below. The fiber tip was then
positioned $\sim100\,\mu$m away from the cantilever for measurement. 

\begin{figure}
\centering{}\includegraphics[width=0.5\paperwidth]{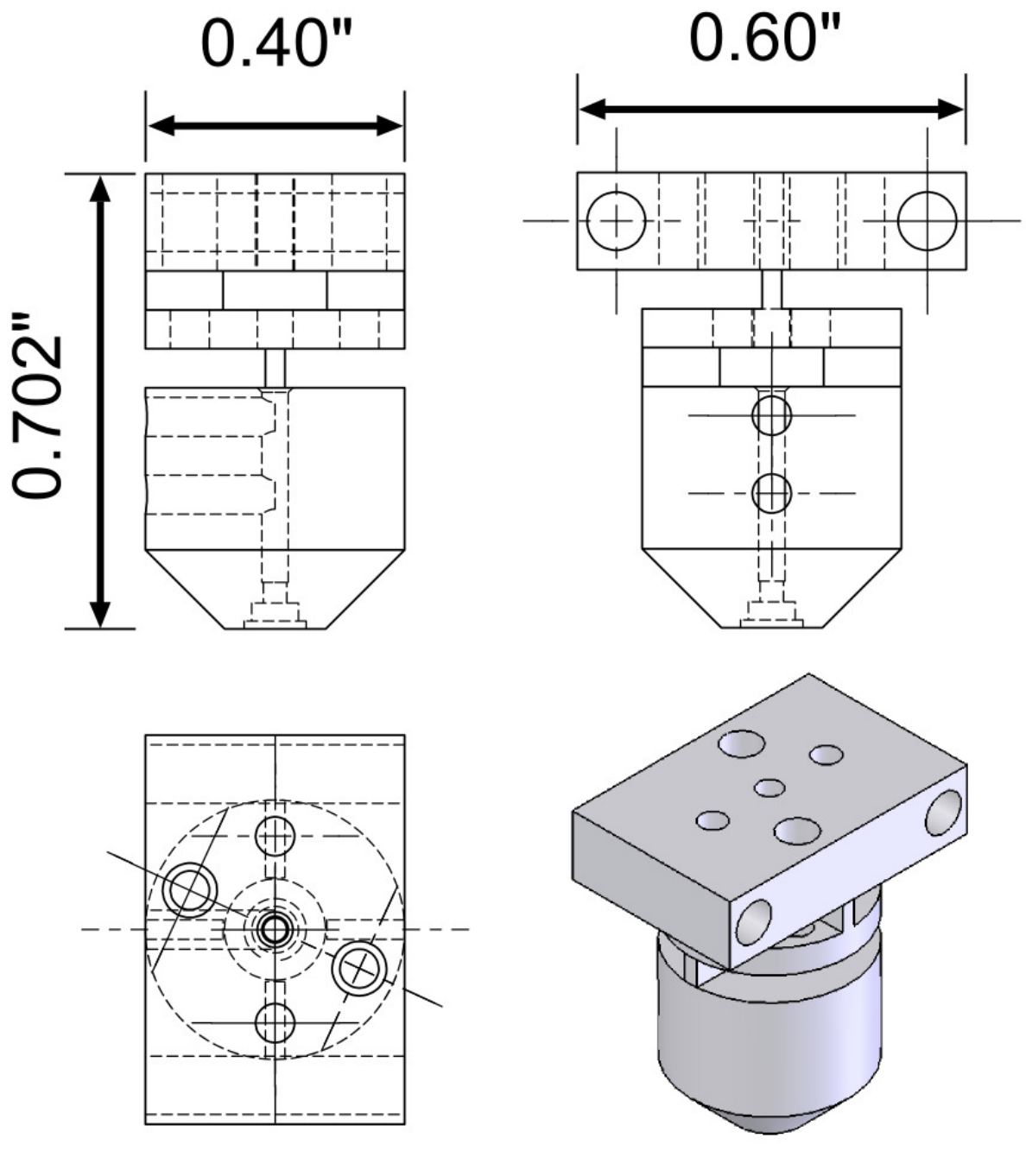}\caption[Optical fiber holder drawing]{\label{fig:CHExpSetup_FiberHolderDrawing}Optical fiber holder drawing.
The figure shows technical drawings of the fiber holder from three
different angles and a three dimensional rendition viewed from an
angle askew to the principle axes of symmetry. The drawings represent
the lens-mounted version of the fiber holder discussed in the text.
The pocket for the lens is located on the bottom of the part in the
top two drawings. The fiber holder for the Bragg reflector fiber looked
similar but did not have this pocket.}
\end{figure}

For the majority of the measurements discussed in this text, the fiber
was modified to include an integrated Bragg grating reflector (fabricated
by Avensys, Montreal, QC, Canada) with reflectivity $\sim$20\% located
$\sim8\,$mm from the cleaved end of the fiber. Again, the fiber tip
was positioned $\sim100\,\mu$m from the cantilever. However, the
effective cavity length with the Bragg reflector was $\sim8\,$mm
rather than $\sim100\,\mu$m.%
\footnote{One can think of the section of fiber between the Bragg reflector
and the fiber tip as playing the role of the {}``Lens'' in Fig.
\ref{fig:CHExpSetup_CantileverMeasurementSchematic} because it keeps
the light from diverging over most of the length between the Bragg
reflector and the cantilever. In the early cleaved fiber set-up, there
was no analogue to the lens.%
} This longer cavity length is useful for the wavelength tuning described
below. With the Bragg reflector fiber, the fiber holder design was
also changed. The fiber end was fed into a $14\,$mm long, $129\,\mu$m
inner diameter /$1\,$mm outer diameter borosilicate ferrule (Vitrocom,
Mountain Lakes, NJ, USA). One end of the ferrule had a tapered opening
to allow the fiber to be fed in. The buffer coating layer was carefully
stripped from the end of the fiber so that when the fiber was inserted
into the ferrule the tip of the fiber extended the desired distance
($\sim1\,$mm) from the ferrule while the end of the buffer coating
layer fit inside of the tapered ferrule lead-in.%
\footnote{It is very important that the buffer coating mate with the ferrule
lead-in. The bare fiber cladding (with the buffer coating removed)
is extremely brittle and will easily break at the epoxy joint if it
is not sufficiently strain-relieved.%
} The fiber was then glued into the ferrule using Stycast$^{\textregistered}$
2850 FT at the buffer coating end (and not on the cleaved tip side).
This ferrule was then slid into a brass holder piece similar to the
one shown in Fig. \ref{fig:CHExpSetup_FiberHolderDrawing} with the
cleaved fiber tip extending a bit past the end of the fiber holder.
In the center of the brass piece was a tube with a diameter 0.001''
wider than the ferrule (this tight tolerance%
\footnote{Great care must taken in cleaning and deburring the fiber holder due
to this tight tolerance, especially with the lens set-up described
below. It is nearly impossible to clean the inside of the fiber holder
without damaging the lens once the lens has been mounted.%
} was more important for the lens set-up described below). Two set
screws oriented perpendicularly to the axis of the ferrule were used
to hold the ferrule in place.

In the latest design of fiber holder with which I worked, an anti-reflection
coated aspheric lens (352140-C f=1.45 mm, NA=0.55, Unmounted Geltech
Aspheric Lens, Thorlabs, Newton, NJ, USA) was incorporated into the
fiber holder. The drawing shown in Fig. \ref{fig:CHExpSetup_FiberHolderDrawing}
depicts the fiber holder with a lens mount at one end. Other than
the cut-out for the lens mount, the only functional difference between
the lens-mounted fiber holder and the fiber holder for the Bragg reflector
fiber was the addition of a slight constriction of the cylinder inside
the fiber holder which prevented the ferrule from bumping into the
lens. This constriction was approximately one focal length away from
the lens. The lens was glued into the fiber holder with Stycast$^{\textregistered}$
2850 FT.%
\footnote{To prevent accidental smudging of the lens with epoxy, we applied
the epoxy with a bit of Kapton which we taped to thin wire which in
turn was taped to a three-dimensional translation stage with micrometer
screws (Ultra-align 561D, Newport Corporation, Irvine, CA, USA). The
Kapton tip was used to prevent scratching of the lens coating and
surface (which occurred with the wire tip). The wire tip was used
to make the object touching the lens extremely pliable and unlikely
to knock the lens out of its seat. The translation stage was used
to steady the epoxy applicator. Different translation stages were
used in different lens mounting attempts, and there were no stringent
specification requirements on stage performance. The epoxy was applied
while viewing the lens with a microscope.%
} A bare cleaved fiber (no Bragg reflector) was used in the lens set-up.%
\footnote{A higher reflectivity surface would boost the cantilever signal. We
experimented with gold coated fibers (Evaporated Coatings, Inc., Willow
Grove, PA, USA) but were not able to mount a coated fiber successfully
because the fibers had been stripped too far prior to being coated
and were very fragile as noted above. It is possible to obtain coated
fibers with less stripped coating. Our fibers were stripped so far
because we wanted to attempt to ground the coating to eliminate possible
electrostatic effects between the fiber tip and the cantilever. These
effects should not be a concern with the lens set-up.%
} With the lens set-up, the fiber holder was positioned with the lens
approximately two focal lengths ($\sim3\,$mm) away from the cantilever
for measurement. The lens set-up thus has a similar cavity length
to the Bragg reflector set-up but did not require that an object be
brought into close proximity with the cantilever. Interactions (e.g.
electrostatic) between the cantilever and a nearby surface can reduce
the mechanical quality factor and also modify the cantilever resonant
frequency.

\FloatBarrier

\subsubsection{Sample holder}

The sample holder stage used in the cantilever thermometry measurements
and preliminary persistent current measurements consisted of a top
plate and cylindrical bottom piece both machined out of brass. The
sample holder stage used in the persistent current measurements consisted
of a copper top plate, three brass%
\footnote{For some measurements, parts of the sample holder were machined from
copper nickel rather than brass in order to increase material resistivity
and reduce eddy current damping when ramping the magnetic field. No
significant change was observed using the copper nickel components
instead their brass counterparts.%
} rods, and a brass bottom plate screwed together in a fashion similar
to that shown in Fig. \ref{fig:CHExpSetup_FridgeDewarSchematic} (see
note in figure caption).

\begin{figure}

\centering{}\includegraphics[width=0.5\paperwidth]{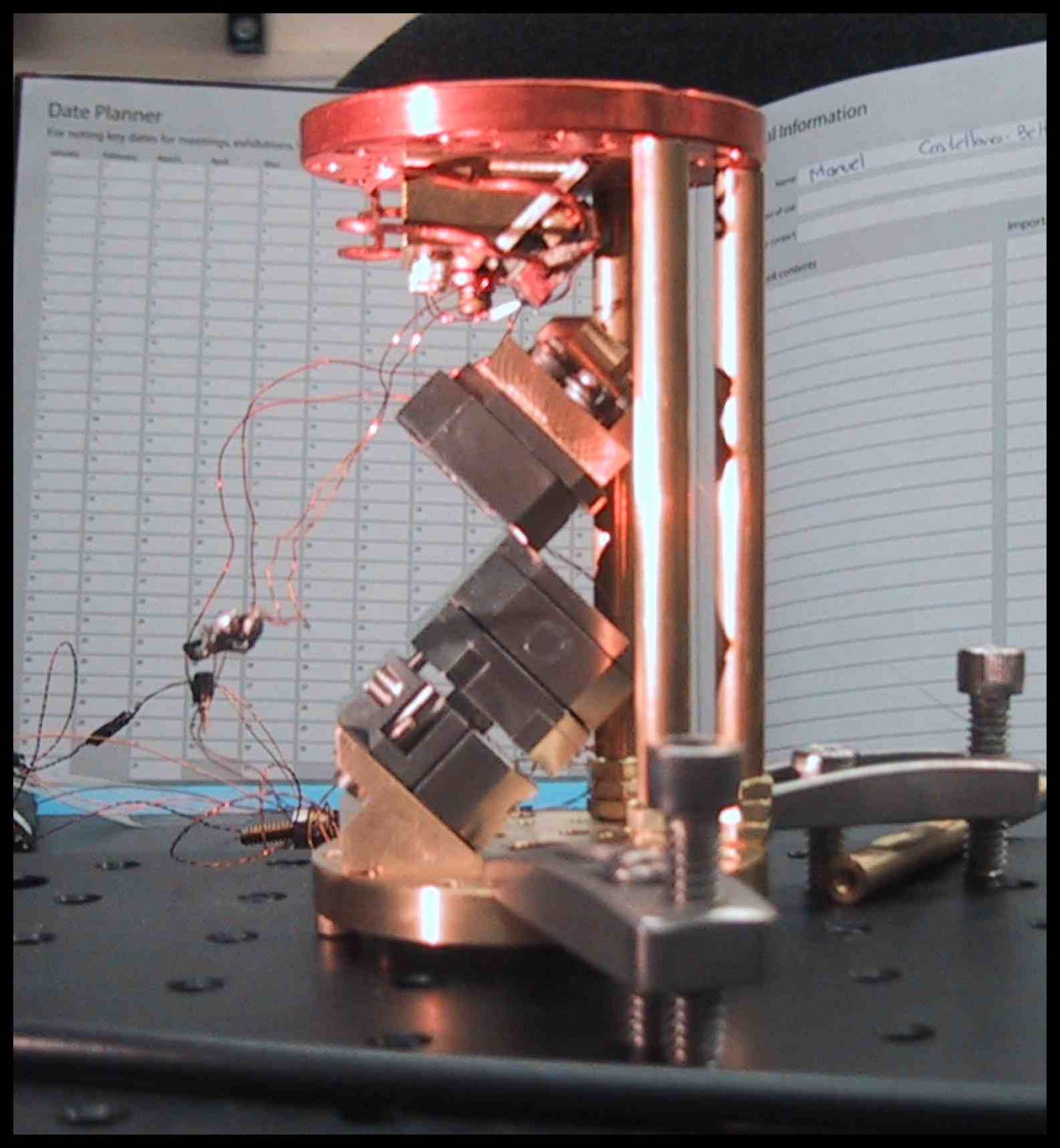}\caption[Persistent current sample holder stage photograph]{\label{fig:CHExpSetup_SampleStagePhoto}Persistent current sample
holder stage photograph. The image shows the lens-mounted fiber holder
seated on top of a set of three Attocube linear positioners. The Attocubes
are mounted on a $45^{\circ}$ tilt stage. The unusual stacking of
the Attocubes is necessary for them to fit within the inner vacuum
chamber while tilted at this angle. The $45^{\circ}$ sample tilt
stage is mounted to the copper top plate above the fiber holder. Insulating
spacers, a piezoelectric actuator, and a copper block for mounting
the cantilever (with thick heat sink leads) are mounted on the sample
tilt stage. During measurement three posts were arranged symmetrically
about the sample stage but were moved for the picture. The figure
gives a representative depiction of the sample stage used during measurement
with only a connector bracket used for interfacing the fridge wiring
with the sample stage thermometer, piezo, and Attocubes missing.}
\end{figure}

The fiber holder in each of its incarnations was screwed down to a
stack of Attocube linear positioners (2 ANPx101/LT/HV and 1 ANPz101/LT/HV,
Attocube systems AG, Munich, Germany).%
\footnote{Early measurements were performed using an earlier generation of Attocube
positioners (2 ANPx100/LT and 1 ANPz100/LT). At one point during the
persistent current measurements, one of the ANPx101 models stopped
working and one of the ANPx100 versions was used while it was repaired.%
} The Attocubes had $\sim6\,$mm of travel in each direction, allowing
access to an entire sample chip, and a minimum step size of $\sim40\,$nm.
For most of the persistent current measurements, the Attocubes were
stacked with their central axis parallel to gravity, as indicated
in panel B of Fig. \ref{fig:CHExpSetup_FridgeDewarSchematic}. When
the lens was incorporated into the fiber holder, an angled Attocube
mount was also added to the sample stage so that the central axis
of the Attocubes was parallel to the cantilevers as shown in Fig.
\ref{fig:CHExpSetup_SampleStagePhoto}. The Attocube wiring consisted
of 0.005'' copper wires from the top of the fridge down to the $4\,$K
plate at the top of the inner vacuum chamber, then heat-sunk superconducting
twisted pairs down to a connector bracket on the sample stage, and
finally short 32 AWG copper jumpers connected to the Attocubes. The
Attocubes were operated with a controller created by Attocube (ANC150/3,
Attocube systems AG, Munich, Germany).

Sample chips were held to a block of copper by means of a clip cut
out of a beryllium-copper gasket (9-78D-A Finger stock gaskets, Leader
Tech, Tampa, FL, USA), and epoxied to the copper block with Stycast$^{\textregistered}$
2850 FT. The sample stage thermometer was glued into a hole drilled
into the copper block. Thick (16 AWG) copper wires were silver soldered
to the copper block and to a copper bobbin which was screwed down
to top plate of the sample stage holder in order to ensure good thermal
contact between the sample and the refrigerator. The copper block
was the top layer of a stack that also included a 0.032'' G-10 (Grade
G-10/FR4 Garolite, McMaster-Carr, Elmhurst, IL, USA) spacer, a 0.5''
diameter piezoelectric actuator (EDO, West Salt Lake City, UT, USA),
and a 0.1'' G-10 spacer. Each layer of this stack was glued together
with Stycast$^{\textregistered}$ 2850 FT. The stack was screwed down
to a sample tilt stage piece which provided the desired angle between
cantilever and magnetic field. 32 AWG Copper wires were soldered to
either side of the piezoelectric actuator and to the center pins of
two SMA coaxial connectors. These connectors were connected to the
refrigerator's UT-85-BSS semi-rigid coaxial lines via jumpers of flexible
UT-085B-SS micro-coax (Janis, Wilmington, MA, USA) and $80\,$MHz
low pass filters (VLFX-80, Mini-Circuits, Brooklyn, NY, USA). At the
top of the refrigerator, these coaxial lines were connected to $1.9\,$MHz
low pass filters (SLP-1.9+, Mini-Circuits, Brooklyn, NY, USA).

\FloatBarrier

\subsubsection{\label{sub:CHExpSetup_CantileverElectronics}Electronics}

For measurements of the cantilever frequency, the arrangement of electronics
was adopted from \citep{albrecht1991frequency} as indicated in Fig.
\ref{fig:CHExpSetup_CantileverMeasurementSchematic}. The output of
the signal photodiode (with 2011 internal filters set to $300\,$Hz
and $30\,$kHz for most measurements) was fed into the input port
of lock-in 1 (7265 DSP Lock-in Amplifier, AMETEK Advanced Measurement
Technology, Oak Ridge, TN, USA). The signal monitor output of lock-in
1, the result of applying the lock-in's amplifier to the input signal,
was then passed through low and high pass filters (SIM 965 Analog
Filter, SIM900 Mainframe, Stanford Research Systems, Sunnyvale, CA,
USA) set to $\sim\pm10\%$ of the cantilever frequency and then into
the input of lock-in 2 (identical to lock-in 1). The signal monitor
of lock-in 2 was then connected to its own reference input. Lock-in
2's reference output, a square wave version of its reference input,
was connected to lock-in 1's reference input. Both lock-ins were operated
in external reference mode. The cantilever was driven in a phase-locked
loop by connecting the oscillator output of lock-in 2 to the piezoelectric
actuator on which the cantilever was mounted. The oscillator output
phase was chosen to maximize the amplitude of motion of the cantilever.
By setting lock-in 1 to measure the second harmonic of the input and
lock-in 2 to measure the first harmonic, the magnitude and phase of
the first two harmonics of the interferometer signal could be measured
simultaneously.

The frequency of the cantilever was recorded in one of two ways. First,
the reference output of lock-in 2 was fed into a frequency counter
(Agilent 53132A $225\,$MHz Universal Counter with 012 US oven option,
Agilent, Loveland, CO, USA). Second, the signal monitor output of
lock-in 2 was fed into a data acquisition (DAQ) board (NI PCI-6251,
National Instruments Corporation, Austin, TX, USA), which used the
oven-stabilized clock of the frequency counter as its timebase. The
digitized interferometer signal was then analyzed to determine its
frequency. An early implementation of the analysis software fit the
digitized signal to a sine wave and produced comparable results to
the frequency counter. A less computationally intensive and more informative
implementation took the Hilbert transform of the digitized signal
in order to obtain a trace of phase versus time, which could be differentiated
to give the real time frequency of the cantilever interferometer signal. 

The analysis of the cantilever interferometer signal described above
was performed using LabVIEW software (National Instruments Corporation,
Austin, TX, USA), which was also used to record all the readings of
all other electronic components of the experiment. Most other recordings
of instrument readings were made via GPIB connections. The other quantities
typically measured in this way were the temperature readings of the
various thermometers, the magnitude and phase of the input of both
lock-in amplifiers, the magnetic field strength, the reference photodiode
voltage (as measured by an Agilent 34410A 6$\unitfrac{1}{2}$ Digit
Multimeter, Agilent, Loveland, CO, USA), and the thermistor reading
of the laser's thermoelectric temperature controller (measured by
an analog to digital input of one of the lock-in amplifiers). Additionally,
LabVIEW routines controlled via GPIB the magnetic field strength,
the lock-in amplifier's oscillator output amplitude and phase, and
the laser's thermoelectric temperature controller setting.

\subsection{\label{sub:CHExpSetup_TransportSetUp}Transport measurement set-up}

Magnetoresistance measurements were performed on wires codeposited
with the aluminum persistent current rings onto the sample chips whose
fabrication was described in \ref{sub:CHExpSetup_PersistentCurrentFabrication}.
The original magentoresistance measurements of quasi-one dimensional
aluminum wires at low temperature were performed in the lab of Daniel
Prober at Yale University. In particular, the dissertations of Santhanam
\citep{santhanam1985localization}, Rooks \citep{rooks1987electron},
Wind \citep{wind1987electron}, and Chandrasekhar \citep{chandrasekhar1989electron}
(as well as personal communication from Dan himself) were useful references.

\begin{figure}
\begin{centering}
\includegraphics[width=0.7\paperwidth]{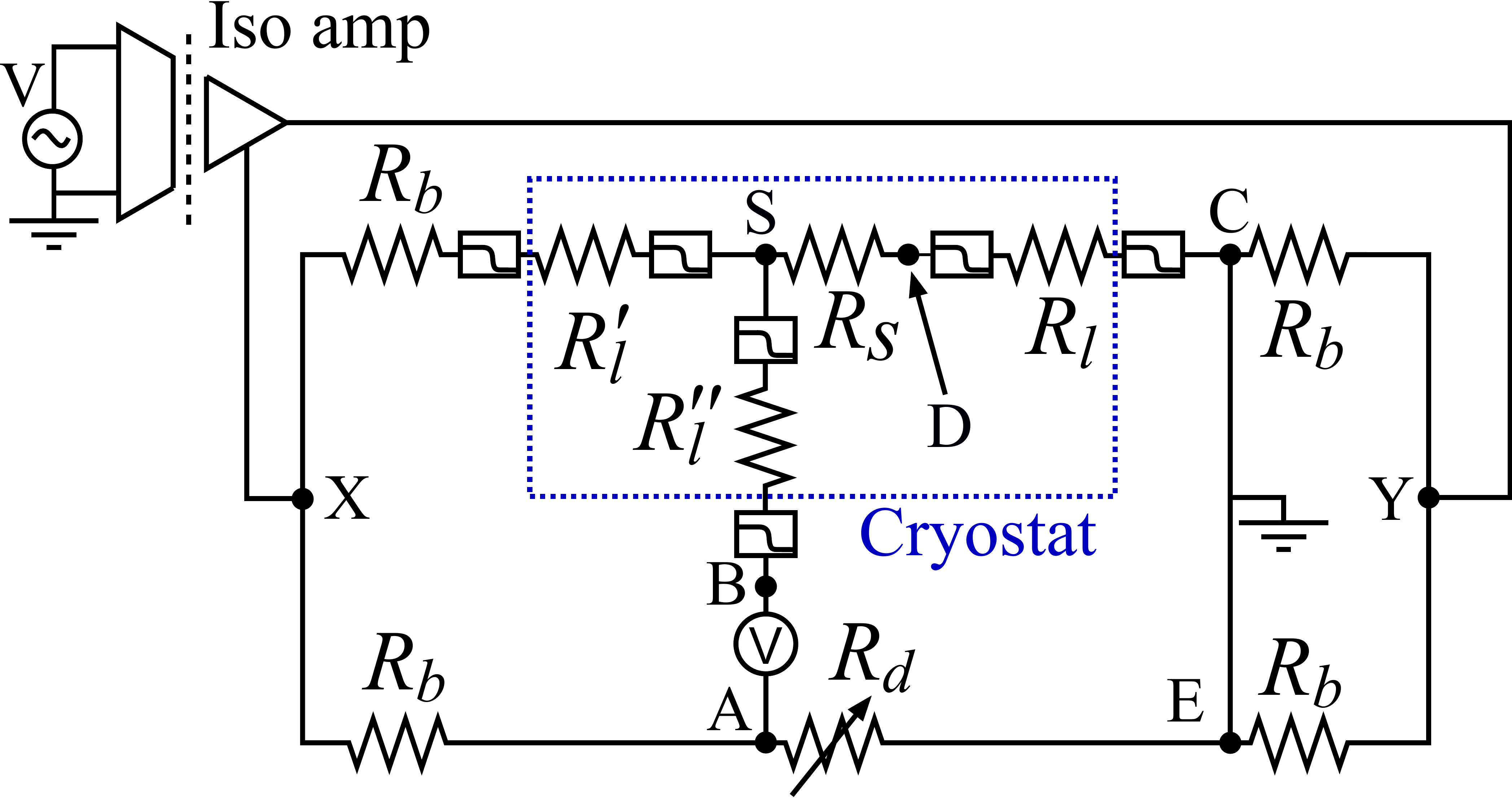}
\par\end{centering}

\caption[Circuit diagram of the four-point AC resistance bridge]{\label{fig:CHExpSetup_TransportBridgeSchematic}Circuit diagram of
the four-point AC resistance bridge. The figure depicts a {}``three
point'' resistance bridge driven by a lock-in amplifier (V) through
an isolation amplifier (Iso amp). The resistances for the bridge resistors
($R_{b}$), sample ($R_{s}$), decade ($R_{d}$), and sample leads
($R_{l}$ and primed equivalents) are indicated. Additionally, all
low-pass filters are indicated. The filters within the cryostat had
a $3\,$dB point of $80\,$MHz, while that of those outside it was
$1.9\,$MHz. Note that the resistances $R_{l}$ really encompass the
entire lead resistance on each side of these filters. The lock-in
performed a differential measurement between the decade (A) and the
sample (B). More discussion of the connections in the diagram is provided
in the main text.}
\end{figure}

All magnetoresistance measurements were performed using a four-point
AC resistance bridge modeled after the one described in the dissertations
of Rooks \citep{rooks1987electron} and Chandrasekhar \citep{chandrasekhar1989electron}.
A diagram of the bridge circuit is shown in Fig. \ref{fig:CHExpSetup_TransportBridgeSchematic}.
The bridge was excited and measured by the 7265 Signal Recovery lock-in
amplifier discussed in \ref{sub:CHExpSetup_CantileverElectronics}.
In order to avoid ground loops between voltage excitation and the
voltage measurement, a battery powered isolation amplifier (AD202JY
Isolation Amplifier, Analog Devices, Inc., Norwood, MA, USA) was inserted
between the lock-in's voltage output and the bridge circuit. The bridge
resistors were chosen to be well matched to a high tolerance and to
have good thermal stability (Vishay Intertechnology, Inc., Malvern,
PA, USA).%
\footnote{I believe we used VSMP series resistors with a 0.01\% tolerance and
a 0.1 $\text{ppm}/^{\circ}\text{C}$ temperature coefficient, but
I can not find the exact part number in my notes.%
} Each bridge resistor had a bridge resistance $R_{b}=47\,\text{k}\Omega$.%
\footnote{This choice of bridge resistance was the result of following the thesis
of Chandrasekhar a bit too closely. His samples had significantly
smaller resistance than ours so that his sample resistance was negligible
compared to the bridge resistance. Our sample resistance of around
$R_{s}\sim300\,\Omega$ was a bit less than 1\% of $R_{b}$, but a
bit higher bridge resistance of around $1\,\text{M}\Omega$ would
have been more strongly into the limit of $R_{s}\ll R_{b}$.%
} The bridge resistors were soldered to a small scrap piece of circuit
board which was screwed down to a large block of copper (approximately
$1/2"\times1"\times2"$) and enclosed in an aluminum box to ensure
thermal stability and to shield from stray electromagnetic radiation.
The refrigerator's coaxial lines were used for all transport measurements
with $80\,$MHz filters just before the sample on the sample stage
and $1.9\,$MHz just outside the refrigerator (in an arrangement identical
to that used for the connections to the piezoelectric actuator described
in \ref{sub:CHExpSetup_CantileverElectronics}). The sample resistance
$R_{s}$ was balanced by a decade resistor (GenRad 1433-X Decade Resistor,
General Radio Co., Concord, MA).

\begin{figure}
\begin{centering}
\includegraphics[width=0.6\paperwidth]{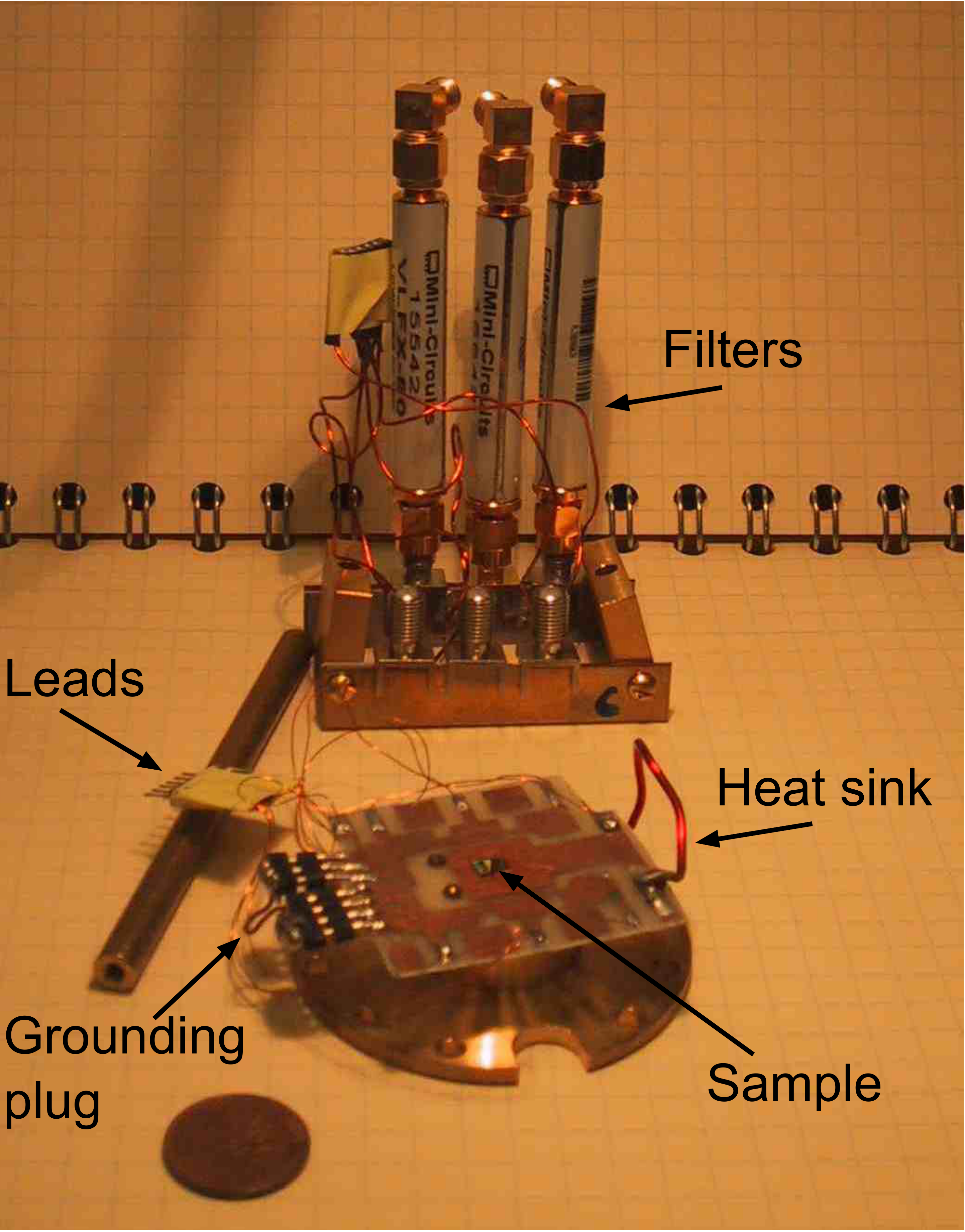}
\par\end{centering}

\caption[Transport measurement sample holder]{\label{fig:CHExpSetup_TransportSampleHolder}Transport measurement
sample holder. The sample chip is indicated on top of the ground plane,
which is electrically and thermally linked to the refrigerator by
the 16 AWG gauge copper wire shown (the other end of the wire is soldered
to a copper washer which is screwed down onto the refrigerator at
a point below the 3-helium cold plate. Each pad on the sample chip
is connected by a wire bond to one of the copper regions on the circuit
board. Each of these copper regions is connected to two sets of SIP
connectors (Series X518, Aries Electronics, Inc., Bristol, PA, USA),
one on the circuit board (grounding plug) and one separated from the
circuit board by 32 AWG gauge copper wire (leads). The circuit board
is shown screwed down to a stub of brass machined so that it can be
bolted to the top plate of the sample holder (not in figure). In the
background of the figure, the low pass filters, which hang below the
sample by brass rods like the one in the picture, can be seen, as
well as the plug for the sample leads.}
\end{figure}

A photograph of the sample chip holder is shown in Fig. \ref{fig:CHExpSetup_TransportSampleHolder}.
The sample chip was mounted onto a patterned circuit board. The circuit
board was a copper-coated dielectric onto which a circuit pattern
of our design was printed and then etched away to create six leads
and a ground plane.%
\footnote{We are indebted to the Devoret Lab, and Nicolas Bergeal in particular,
for supplying us with the necessary materials and assistance to fabricate
this board.%
} The sample itself was mounted with a cryogenic grease (Apiezon$^{\textregistered}$
N Cryogenic High Vacuum Grease, Manchester, UK) onto the ground plane,
which was connected directly to the refrigerator via a 16 AWG gauge
copper wire for heat sinking. The sample chip was then wire bonded
to the leads on the circuit board.%
\footnote{We are indebted to the Devoret and Schoelkopf labs for the use of
their wire bonder and for assistance in wire bonding.%
} Each chip had several wire samples fabricated onto it and in principal
four-point measurements of two wire samples could be performed in
one cool down using the six coaxial leads of the refrigerator. Let
A and B label the two bond pads on the same side of one wire sample
and C and D label the two bond pads on the same side of a second wire
sample. By connecting one coaxial lead to both A and C and another
coaxial lead to B and D, these leads could serve a dual role in measuring
the two wire samples. In practice, due to frequent lead blow outs
on the sample chips, we only measured one chip (and that chip only
had three leads available for that measurement as described below).

The circuit board leads were soldered to 32 AWG gauge copper wires
which were all soldered into one single-in-line (SIP) connector for
interfacing with the refrigerator wiring (all of the components connecting
the sample to the refrigerator's coaxial lines are shown in Fig. \ref{fig:CHExpSetup_TransportSampleHolder}
other than the micro-coax jumpers mentioned in \ref{sub:CHExpSetup_CantileverElectronics}).
Additionally, another set of SIP connectors was soldered to directly
to the circuit board leads. Another set of electrically connected
SIP connectors could be plugged into the circuit board SIP connectors
in order to ground the sample chip. This grounding plug was plugged
in during the wire bonding step. It was only removed after the other
set of leads had been connected to the fridge and then grounded (with
the heat sink connector also connected to the refrigerator and the
refrigerator itself tied to ground) in order avoid blowing out samples
by electrostatic discharge. The leads emerging from the refrigerator
were connected to a junction box. Inside the box, each lead was connected
to a switch which controlled whether the lead was grounded or electrically
connected to a connector on the other side of the box. When the sample
chip was first plugged into the refrigerator, all of these switches
were set to ground. Whenever an external wire was connected to or
disconnected from the sample, these switches were always set to ground
first. Despite all of these precautions, many sample chips were blown
out, and we ultimately settled on making a three lead measurement
of a single sample.

We now step briefly through the circuit depicted in Fig. \ref{fig:CHExpSetup_TransportBridgeSchematic}.
An excitation voltage (V) is supplied by the lock-in amplifier to
the isolation amplifier (Iso amp). In order to ensure accurate excitation
voltages, a voltage divider (not shown) was often included between
the lock-in and isolation amplifiers so that the lock-in output could
be operated at a higher setting. The frequency of the excitation was
typically around $200\,\text{Hz}$. The isolation amplifier output
is split off into the two arms of the bridge (X) passing through two
balanced bridge resistors $R_{b}$. The sample side of the bridge
then goes through two low pass filters and the refrigerator and sample
chip leads $R_{l}'$ (indicated schematically between the filters,
but in actuality distributed around both of them) before connecting
to the sample (S). Another lead attached to the other wire bond pad
on this same side of the sample (S, see Figs. \ref{fig:AppSampFab_TransportMaskDesign}
and \ref{fig:AppSampFab_WLSEM}) goes into the lock-in amplifier's
B input, while the lock-in's A input is connected to the other side
of the bridge. The other side of the sample (D) is then connected
(again through filters and leads) to a bridge resistor (C). On the
other side of the bridge, the bridge resistor is connected to the
decade (A), which in turn connects to a fourth bridge resistor (E).
These two bridge resistors on opposite sides of the decade and sample
from X are then tied together at Y where they are also connected to
the other side of the isolation amplifier. Additionally, the other
sides of these two bridge resistors (C and E) were both tied to the
lock-in's ground. 

The circuit described in this way is really a {}``three-point''
bridge. For a true four-point measurement, the sample-side connection
to ground should really be at point D connected to the sample through
a second sample bond pad lead (before the lead resistance $R_{l}$).
In the experiment, this fourth lead was unintentionally blown out,
so that only this three-point measurement arrangement was possible.

\FloatBarrier

\section{Calibrations and measurement procedures}

\subsection{\label{sub:ChExpSetup_CantDetectionMeasurement}Cantilever detection
calibrations and measurement procedures}

\subsubsection{\label{sub:CHExpSetup_CantileverFiberInterferometer}Model of the
cantilever-fiber interferometer}

\begin{figure}
\centering{}\includegraphics[width=0.7\paperwidth]{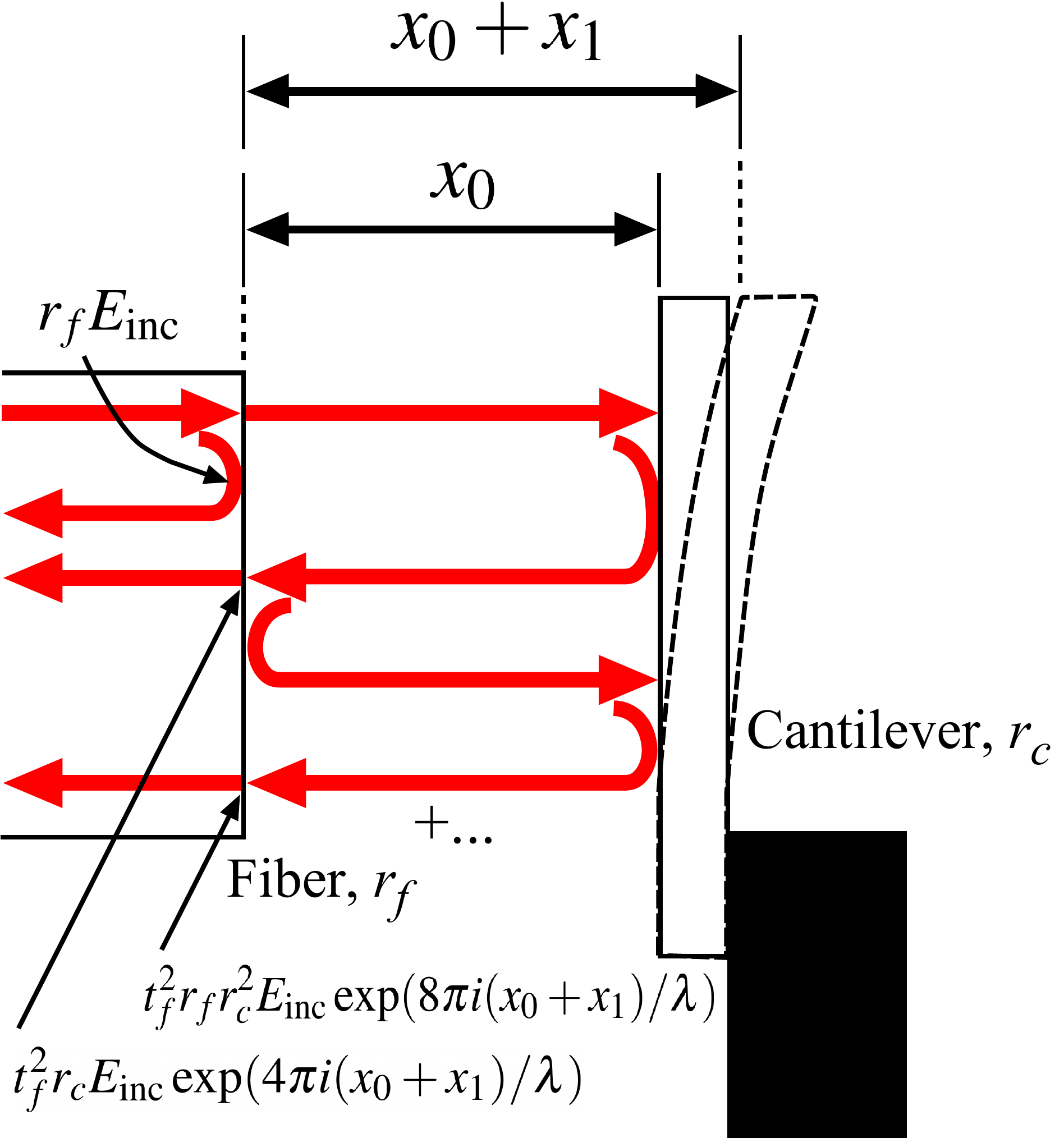}\caption[Cantilever-fiber interferometer diagram]{\label{fig:CHExpSetup_CantileverFiberInterferometer}Cantilever-fiber
interferometer diagram. The laser light with electric field amplitude
$E_{\text{inc}}$ enters the interferometer from the left and is partially
reflected off the end of the fiber (or Bragg reflector) which has
a reflection coefficient $r_{f}$. The rest of the laser light proceeds
to the cantilever from which it is either reflected back to the fiber
or transmitted through the cantilever and lost. The light reentering
the fiber has amplitude $t_{f}^{2}r_{c}E_{\text{inc}}$ and also picks
up a phase factor given by $2\pi$ times the round trip cavity distance,
$2(x_{0}+x_{1})$, divided by laser wavelength $\lambda$. Here $x_{0}$
is the cavity length with the cantilever at its equilibrium position
and $x_{1}$ is the displacement of the cantilever from equilibrium.
The second roundtrip of the light inside the cavity is also shown.
On each trip, some amount of the light is transmitted through the
cantilever (not shown). The loss factors discussed in the text are
omitted from the figure for simplicity.}
\end{figure}

We model the cantilever-fiber system as a Fabry-Perot interferometer
in the limit of low finesse. A diagram of the interferometer is shown
in Fig. \ref{fig:CHExpSetup_CantileverFiberInterferometer}. We call
the distance from the point of reflection in the fiber (either the
Bragg reflector or the cleaved fiber end) to the cantilever surface
at equilibrium $x_{0}$, and we call the cantilever displacement from
equilibrium $x_{1}$. We describe the incident laser light with a
complex electric field amplitude $E_{\text{inc}}\exp(2\pi ix/\lambda)$
at point $x$ where $E_{\text{inc}}$ is the magnitude of the electric
field amplitude of the incident light and $\lambda$ is its wavelength. 

At each interface, the light is partially reflected and partially
transmitted. Additionally, there can be loss associated with the light
propagation. With the materials involved, absorptive loss should be
negligible. However, a large degree of loss is still possible due
to misalignment of the optical components and due to the mismatch
of the beam profile and the fiber core shape. We call the electric
field amplitude reflection and transmission coefficients for the fiber
$r_{f}$ and $t_{f}$ respectively and the cantilever reflection coefficient
$r_{c}$ (see e.g. Ref. \citep{jackson1998classical} for a description
of how to treat a series of dielectric surfaces as one optical unit
characterized by its reflection and transmission coefficients). We
denote by $p_{ft,n}$, $p_{fr,n}$, and $p_{cr,n}$ the additional
reduction factors associated with the laser light being transmitted
into the fiber, reflected off of the fiber, and reflected off of the
cantilever on the $n^{\text{th}}$ trip. The index is necessary since
these factors can be different after different numbers of round trips
around the cavity (due to diffraction of the laser beam, for example).

As shown in Fig. \ref{fig:CHExpSetup_CantileverFiberInterferometer},
the ratio of the total light amplitude traveling back down the fiber
away from the cantilever to the incident light amplitude, $E_{\text{cant}}/E_{\text{inc}}$,
is found by summing the infinite series of successively higher numbers
of reflections as
\begin{align}
\frac{E_{\text{cant}}}{E_{\text{inc}}}=\, & r_{f}-t_{f}r_{c}p_{cr,1}t_{f}p_{ft,1}\exp\left(\frac{4\pi i\left(x_{0}+x_{1}\right)}{\lambda}\right)\nonumber \\
 & -t_{f}r_{c}p_{cr,1}r_{f}p_{fr,1}r_{c}p_{cr,2}t_{f}p_{ft,2}\exp\left(\frac{8\pi i\left(x_{0}+x_{1}\right)}{\lambda}\right)\nonumber \\
 & +\ldots\label{eq:CHExpSetup_EFullInterferometer}
\end{align}
where the factors have been written sequentially. Ignoring common
factors, the ratio of the third term to the second in Eq. \ref{eq:CHExpSetup_EFullInterferometer}
is $r_{c}r_{f}p_{cr,2}p_{ft,2}(p_{fr,1}/p_{ft,1})$. 

The amplitude reflection coefficient $r_{f}$ is typically around
$.2$ for a bare fiber and $.45$ for the Bragg reflectors in our
fibers, while the maximum possible value for the cantilever reflection
coefficient $r_{c}$ is $.85$.%
\footnote{This value is obtained by using $n\approx3.5$ for the refractive
index of silicon \citep{philipp1960optical} and treating the cantilever
as an Fabry-Perot etalon of optimal thickness \citep{hecht1987optics2nd}.%
} A typical value for $p_{cr,1}p_{ft,1}$ in either of our interferometer
arrangements is .19, and the value of $p_{cr,2}p_{ft,2}$ should be
smaller since misalignment and similar effects should be more pronounced
for higher orders of reflections. The ratio $(p_{fr,1}/p_{ft,1})$
should be less than one since misalignment/mismatch effects should
be more pronounced in reflection than transmission. Thus, we find
that the ratio of magnitudes between the third and second terms in
Eq. \ref{eq:CHExpSetup_EFullInterferometer} has an upper bound of
0.07 and should in fact be less than this figure. We make the low
finesse approximation by dropping all terms beyond the first two in
Eq. \ref{eq:CHExpSetup_EFullInterferometer} (all of which should
be at least a factor of $\sim0.07$ smaller), keeping only the first
reflection off of the cantilever. 

So far we have ignored the motion of the cantilever. When the cantilever
moves, it tilts in addition to changing its linear position and thus
changes the factor $p_{cr,1}p_{ft,1}$ which parametrizes the coupling
of light from the cantilever back into the fiber. For small cantilever
displacements, we can account for this {}``optical lever'' effect
by replacing $r_{c}p_{cr,1}p_{ft,1}$ by $r_{c}(1-\epsilon x_{1})$,
where we have combined the loss factors into $r_{c}$ for simplicity
of notation. The factor $\epsilon x_{1}$ is typically small for the
cantilever displacements used in the experiment. With these modifications,
we obtain the simplified form for the amplitude $E_{\text{cant}}$
of electric field returning from the cantilever,
\[
\frac{E_{\text{cant}}}{E_{\text{inc}}}=\, r_{f}-t_{f}^{2}r_{c}\left(1-\epsilon x_{1}\right)\exp\left(\frac{4\pi i\left(x_{0}+x_{1}\right)}{\lambda}\right).
\]
 We measure light intensity $P_{\text{cant}}$ which is proportional
to the square modulus of $E_{\text{cant}}$, satisfying
\begin{equation}
\frac{P_{\text{cant}}}{P_{\text{inc}}}=\, R_{f}+T_{f}^{2}R_{c}(1-2\epsilon x_{1})-2T_{f}\sqrt{R_{c}R_{f}}\left(1-\epsilon x_{1}\right)\cos\left(\frac{4\pi}{\lambda}(x_{0}+x_{1})\right)\label{eq:CHExpSetup_PInterferometerFull}
\end{equation}
 to leading order in $\epsilon$. In the previous equation we have
introduced the power reflectivities given by the square of the amplitude
coefficients, $R_{f}=r_{f}^{2}$, $T_{f}=t_{f}^{2}$, and $R_{c}=r_{c}^{2}$.
When $x_{0}\gg\lambda$ and $x_{1}$ is held constant, the cantilever
interferometer signal $P_{\text{cant}}$ should trace out a sinusoidal
curve over small changes in $\lambda$. Our observations of a highly
sinusoidal $P_{\text{cant}}$ (not shown) as a function of wavelength
justifies the use of the low finesse approximation.

Generally, in the experiment the cantilever signal is measured by
a lock-in amplifier, which monitors a single frequency component.
Therefore, a Fourier decomposition of the time dependence of $P_{\text{inc}}$
is appropriate. The time dependence enters the expression for $P_{\text{inc}}$
through the motion of the cantilever which we write as $x_{1}=x_{f,\max}\cos\omega t$
where $x_{f,\max}$ is the amplitude of motion of the cantilever at
the position $z_{f}$ addressed by the optical fiber and $\omega$
is the angular frequency of the cantilever's oscillation. We obtain
a harmonic decomposition by writing
\[
\cos\left(\frac{4\pi}{\lambda}(x_{0}+x_{1})\right)=\cos\left(\frac{4\pi}{\lambda}x_{0}\right)\cos\left(\frac{4\pi}{\lambda}x_{f,\max}\cos\omega t\right)-\sin\left(\frac{4\pi}{\lambda}x_{0}\right)\sin\left(\frac{4\pi}{\lambda}x_{f,\max}\cos\omega t\right)
\]
 and applying the Jacobi-Anger identity given in Eqs. \ref{eq:AppMath_JacAngCC}
and \ref{eq:AppMath_JacAngSC}. The first three terms are

\begin{align}
\left(\frac{P_{\text{cant}}}{P_{\text{inc}}}\right)_{0}= & R_{f}+T_{f}^{2}R_{c}-T_{f}\sqrt{R_{f}R_{c}}\cos\left(\frac{4\pi}{\lambda}x_{0}\right)J_{0}\left(\frac{4\pi}{\lambda}x_{f,\max}\right)\nonumber \\
 & +T_{f}\sqrt{R_{f}R_{c}}\sin\left(\frac{4\pi}{\lambda}x_{0}\right)\epsilon x_{f,\max}J_{1}\left(\frac{4\pi}{\lambda}x_{f,\max}\right)\label{eq:CHExpSetup_DCInterferometer}
\end{align}
\begin{align}
\left(\frac{P_{\text{cant}}}{P_{\text{inc}}}\right)_{1}= & 4T_{f}\sqrt{R_{f}R_{c}}\sin\left(\frac{4\pi}{\lambda}x_{0}\right)J_{1}\left(\frac{4\pi}{\lambda}x_{f,\max}\right)+2T_{f}^{2}R_{c}\epsilon x_{f,\max}\nonumber \\
 & +2T_{f}\sqrt{R_{f}R_{c}}\cos\left(\frac{4\pi}{\lambda}x_{0}\right)\epsilon x_{f,\max}\left(J_{0}\left(\frac{4\pi}{\lambda}x_{f,\max}\right)-J_{2}\left(\frac{4\pi}{\lambda}x_{f,\max}\right)\right)\label{eq:CHExpSetup_1stHarmInterferometer}
\end{align}
\begin{align}
\left(\frac{P_{\text{cant}}}{P_{\text{inc}}}\right)_{2}= & 4T_{f}\sqrt{R_{f}R_{c}}\cos\left(\frac{4\pi}{\lambda}x_{0}\right)J_{2}\left(\frac{4\pi}{\lambda}x_{f,\max}\right)\nonumber \\
 & -2T_{f}\sqrt{R_{f}R_{c}}\sin\left(\frac{4\pi}{\lambda}x_{0}\right)\epsilon x_{f,\max}\left(J_{1}\left(\frac{4\pi}{\lambda}x_{f,\max}\right)-J_{3}\left(\frac{4\pi}{\lambda}x_{f,\max}\right)\right)\label{eq:CHExpSetup_2ndHarmonicInterferometer}
\end{align}
where we have expanded 
\begin{equation}
\left(\frac{P_{\text{cant}}}{P_{\text{inc}}}\right)=\sum_{n}\left(\frac{P_{\text{cant}}}{P_{\text{inc}}}\right)_{n}\cos n\omega t.\label{eq:CHExpSetup_InterferometerExpansion}
\end{equation}
For $x_{0}=\lambda(n+1/4)/2$ with $n$ an integer, the first harmonic
$(P_{\text{cant}})_{1}$ of the interferometer signal is proportional
to $J_{1}(4\pi x_{f,\max}/\lambda)+Ax_{f,\max}$ for $A$ as given
by Eq. \ref{eq:CHExpSetup_DCInterferometer} and is proportional to
$x_{f,\max}$ for $x_{f,\max}\ll0.14\lambda$. We refer to such a
value of $x_{0}$ as an {}``optimal fringe position'' of the interferometer.

\begin{figure}

\begin{centering}
\includegraphics[width=0.6\paperwidth]{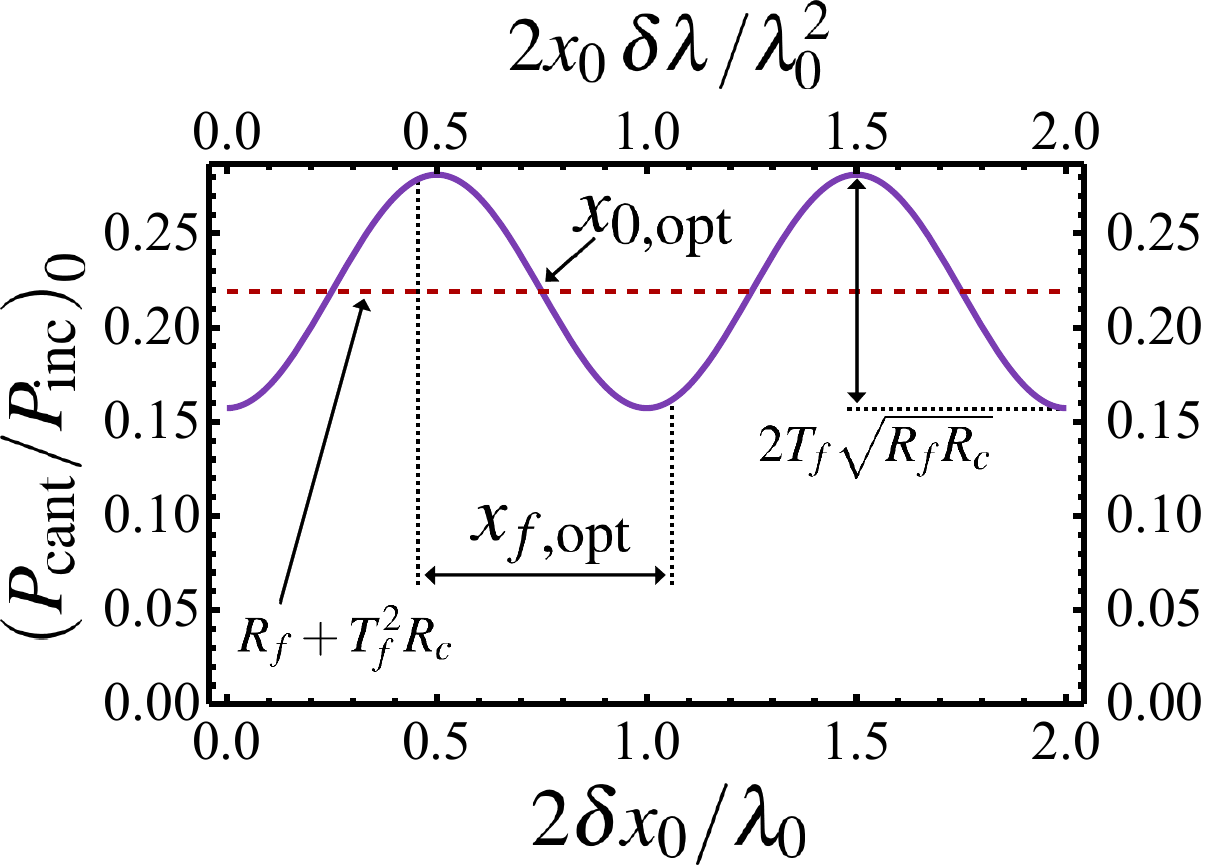}
\par\end{centering}

\caption[Fiber-cantilever interferometer fringe pattern]{\label{fig:CHExpSetup_FiberCantileverFringe}Fiber-cantilever interferometer
fringe pattern. The figure plots Eq. \ref{eq:CHExpSetup_DCInterferometer}
with $R_{f}=.2$, $T_{f}=.8$, $R_{c}=.03$, and $x_{0}=5300\lambda$.
The horizontal axis is parametrized in two equivalent ways. On the
bottom axis, the fringe pattern is measured in terms of a small detuning
$\delta x_{0}=x_{0}-5300\lambda$ away from a value of $x_{0}=5300\lambda$
that is an integral multiple of the fixed wavelength $\lambda$. On
the top axis, the wavelength $\lambda$ itself is detuned through
the parameter $\delta\lambda=\lambda-x_{0}/5300$ for fixed cavity
length $x_{0}$. In the limit of $x_{0}/\lambda\gg1$, these two parametrizations
are equivalent when $\delta x_{0}$ and $\delta\lambda$ are scaled
as indicated in the labels of these axes. Additionally, the figure
indicates $x_{0,\text{opt}}$, the optimal fringe position where the
first harmonic of the cantilever interferometer signal is maximal
(assuming $\epsilon\sim0$) as discussed in Section \ref{sub:CHExpSetup_WavelengthTuning}.
The range $x_{f,\text{opt}}=1.84\lambda/4\pi$ indicated in the figure
depicts the peak-to-peak amplitude $2x_{f,\max}$ of the cantilever
oscillation, centered at the optimal fringe position $x_{0,\text{opt}}$,
that maximizes the first harmonic signal of the interferometer (again
for $\epsilon\sim0$).}
\end{figure}

\FloatBarrier

\subsubsection{\label{sub:CHExpSetup_WavelengthTuning}Optimal interferometer fringe
position and laser wavelength tuning}

The constant value $(P_{\text{cant}}/P_{\text{inc}})_{0}$ of the
cantilever interferometer signal when plotted as a function of wavelength
$\lambda$ or cavity length $x_{0}$ is known as the interferometer
{}``fringe.'' In Fig. \ref{fig:CHExpSetup_FiberCantileverFringe},
the interferometer fringe pattern is plotted for the case of $x_{f,\max}=0$
and typical interferometer parameters for the Bragg reflector set-up.
In the limit $\epsilon\rightarrow0$, the condition $x_{0}=\lambda(n+1/4)/2$
maximizes the first harmonic of the cantilever interferometer signal
$(P_{\text{cant}}/P_{\text{inc}})_{1}$ and causes the second harmonic
$(P_{\text{cant}}/P_{\text{inc}})_{2}$ to be zero. With both the
Bragg reflector and the lens, the length of the interferometer was
$\sim8\,$mm, so $x_{0}\sim5300\lambda$. In this case, the required
change $\delta\lambda$ in wavelength necessary to change $x_{0}/\lambda$
by $1/2$ (and so to move from optimal fringe position $n$ to optimal
fringe position $n\pm1$) is given by 
\begin{eqnarray*}
\frac{1}{2} & = & \frac{x_{0}}{\lambda}-\frac{x_{0}}{\lambda+\delta\lambda}=\frac{x_{0}}{\lambda^{2}}\delta\lambda\\
\Longrightarrow & \delta\lambda & \approx.15\,\text{nm.}
\end{eqnarray*}
 With the thermoelectric cooler, the wavelength of the JDS Uniphase
laser was tunable between $1549\text{ nm}$ and $1554\text{ nm}$.
Thus the wavelength could be tuned to the optimal fringe position
for any cantilever displacement $x_{0}$.

With $\epsilon=0$, the peak response of the first harmonic $(P_{\text{cant}}/P_{\text{inc}})_{1}$
of the cantilever interferometer signal occurs at the first peak of
$J_{1}(4\pi x_{f,\max}/\lambda)$ where $x_{f,\max}\approx1.84\lambda/4\pi\approx227\,\text{nm}$.
Typical values of $x_{f,\max}$ during persistent current measurements
ranged from $10\,$nm to $227\,$nm, so that the arguments of each
of the Bessel functions in Eqs. \ref{eq:CHExpSetup_DCInterferometer},
\ref{eq:CHExpSetup_1stHarmInterferometer}, and \ref{eq:CHExpSetup_2ndHarmonicInterferometer}
were below the locations of all of the Bessel functions' non-trivial
zeros (i.e. those for non-zero argument). For $\epsilon=0$, the location
on the interferometer {}``fringe'' can be adjusted by tuning the
laser wavelength to maximize the first harmonic of the interferometer
signal either by tuning to the midpoint of the constant interferometer
response value $(P_{\text{cant}}/P_{\text{inc}})_{0}$, tuning to
the maximum of the first harmonic $(P_{\text{cant}}/P_{\text{inc}})_{1}$
directly, or tuning to the zero of the second harmonic $(P_{\text{cant}}/P_{\text{inc}})_{2}$.
In practice, tuning to the zero of the second harmonic is the most
practical because it can be measured more precisely (via a lock-in
measurement) than the constant interferometer reading and provides
feature more sensitive to laser wavelength than the first harmonic
reading which varies only to second order about its maximum value.

\begin{figure}

\begin{centering}
\includegraphics[width=0.7\paperwidth]{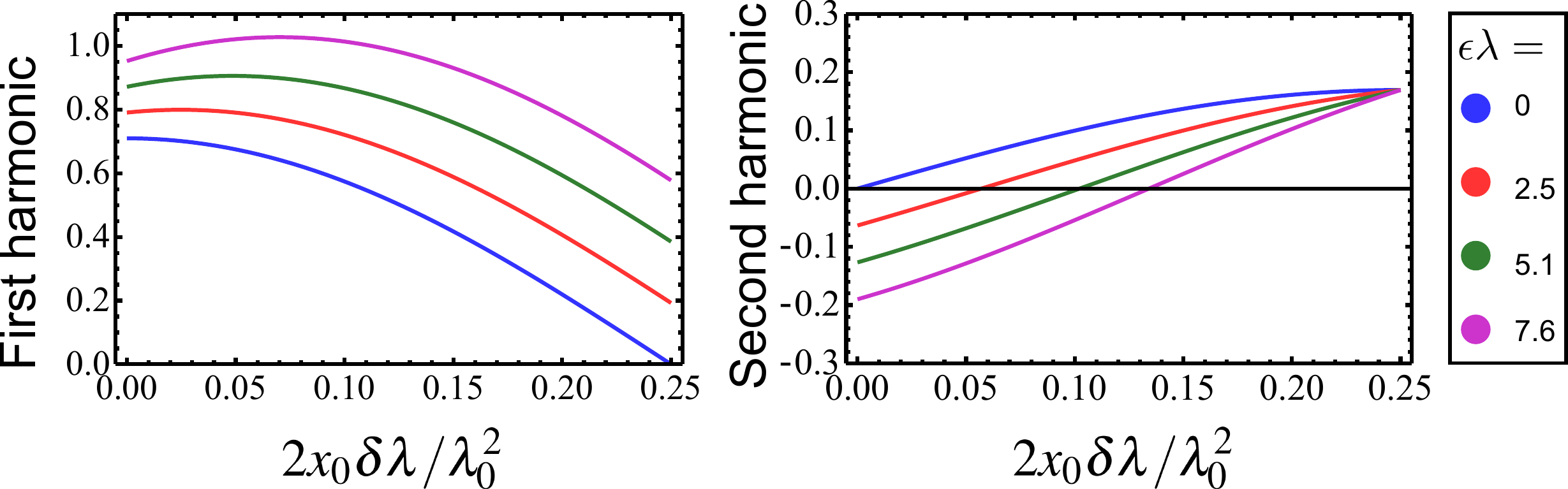}\caption[Effect of the optical lever term on the cantilever interferometer
signal]{\label{fig:CHExpSetup_OpticalLever}Effect of the optical lever term
on the cantilever interferometer signal. The figure shows the first
harmonic $(P_{\text{cant}}/P_{\text{inc}})_{1}$ and the second harmonic
$(P_{\text{cant}}/P_{\text{inc}})_{2}$ of the cantilever interferometer
signal as a function of wavelength $\delta\lambda=\lambda-\lambda_{0}$
for different values of the optical lever coefficient $\epsilon$.
The cavity length is set to $x_{0}=(5300+1/4)\lambda_{0}/2$ and the
amplitude $x_{f,\max}$ of motion of the cantilever is set to $x_{f,\max}=1.84\lambda/8\pi$,
half of the value giving the maximum response of the first harmonic
in the absence of the optical lever effect. The two harmonics have
been normalized by the maximum value of the first harmonic in the
absence of the optical lever effect. For $\epsilon=0$ and $\delta\lambda=0$,
the first harmonic is maximized while the second harmonic is zero.
As $\epsilon$ increases, the location of both these conditions moves
to the right. However, even for $\epsilon\lambda=7.6$, the first
harmonic is still within 10\% of its maximum at the zero of the second
harmonic }

\par\end{centering}

\end{figure}

The optical lever effect can modify the locations of the peaks and
zeros of the various harmonics of the cantilever interferometer signal.
Typical values for the optical lever coefficient $\epsilon$ ($\epsilon\lambda$)
were $4.1\times10^{-5}\,\text{nm}^{-1}$ (0.064) and $4.9\times10^{-3}\,\text{nm}^{-1}$
(7.6) for the Bragg reflector and lens set-ups respectively. The larger
optical lever effect for the lens set-up is due to the much longer
lever arm. The fiber to cantilever distance was only $\sim100\,\text{\ensuremath{\mu}m}$
for the Bragg reflector set-up whereas it was closer to $8\,\text{mm}$
for the lens set-up. However, the optical lever effect tends to shift
the maximum of the first harmonic and the zero of the second harmonic
of the cantilever interferometer signal in the same way as seen in
Fig. \ref{fig:CHExpSetup_OpticalLever} where we plot $(P_{\text{cant}}/P_{\text{inc}})_{1}$
and $(P_{\text{cant}}/P_{\text{inc}})_{2}$ as functions of $\lambda$
for various values of $\epsilon$ and for values of other parameters
similar to those of the experimental conditions. The location of the
zero of the second harmonic of the interferometer signal is nearly
insensitive to cantilever amplitude $x_{f,\max}$ for the experimentally
relevant range of drives. Thus, the zero of the second harmonic provides
a stable reference point for the wavelength that is close to the maximum
response of the first harmonic. We note that for the Bragg reflector
case the zero of the second harmonic is less than 1\% away from its
value in the absence of the optical lever effect. For large $\epsilon$,
it may be necessary to characterize the difference between operating
the system with $\lambda$ chosen to maximize the first harmonic and
with $\lambda$ chosen to minimize the second harmonic. In order to
use the second harmonic to stabilize the fringe position at a value
that provides a large first harmonic response, the second harmonic
would need to be tuned to a non-zero value.

\FloatBarrier

\subsubsection{\label{sub:CHExpSetup_PhaseScan}Determination of resonant phase
for the cantilever drive signal}

\begin{figure}

\centering{}\includegraphics[width=0.7\paperwidth]{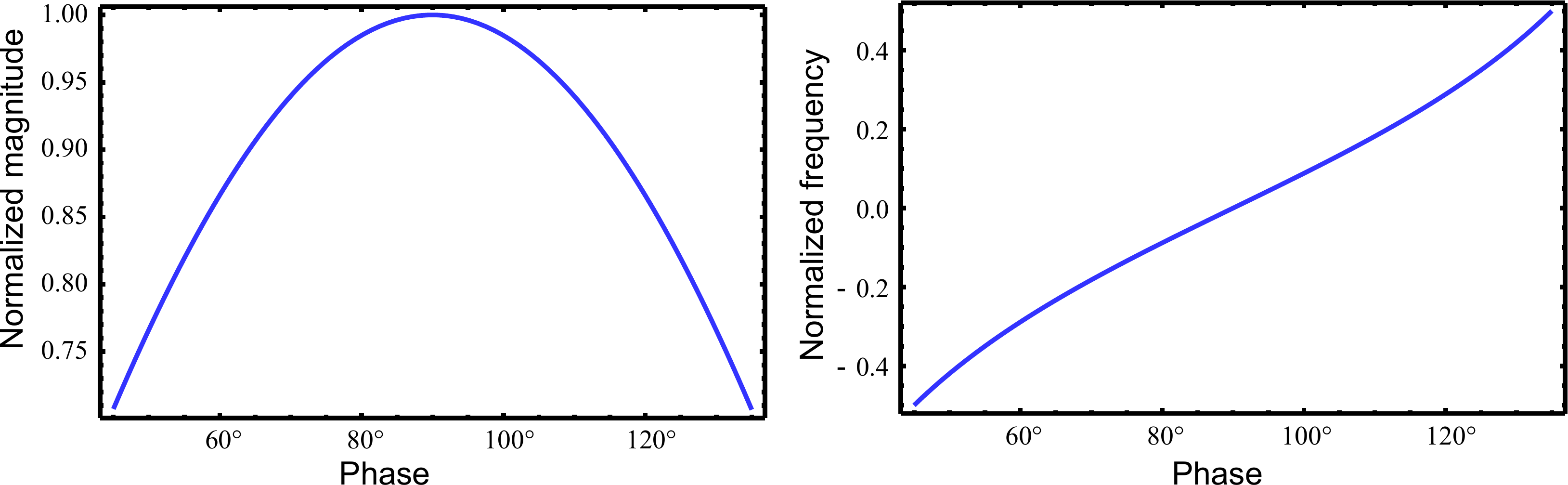}\caption[Normalized magnitude and frequency response of a simple harmonic oscillator]{\label{fig:CHExpSetup_SHOMagPhase}Normalized magnitude and frequency
response of a simple harmonic oscillator. The magnitude and frequency
of a damped, driven simple harmonic oscillator are plotted as a function
of the phase difference between the oscillator's motion and the driving
signal. The magnitude has been normalized so that it is unity on resonance.
The normalized frequency $Q\frac{\omega-\omega_{0}}{\omega_{0}}$
represents the difference between the oscillator's frequency $\omega$
and the resonant frequency $\omega_{0}$ multiplied by the oscillator's
quality factor $Q$ and divided by the resonant frequency. The phase
$\phi$ is found by writing the transfer function given in Eq. \ref{eq:GCantileverTransferFunction}
as $G\left(\omega\right)=\left|G(\omega)\right|\exp(i\phi)$ and solving
for $\phi$.}
\end{figure}

For the cantilever frequency measurements, the cantilever was driven
in a phase-locked loop using the arrangement described in \ref{sub:CHExpSetup_CantileverDetectionSetup}.
In this set-up, a lock-in amplifier drives the cantilever while in
external mode using the cantilever interferometer signal as its reference.
In this mode, the lock-in maintains a fixed phase between its reference
and the cantilever drive signal, effectively adjusting its output
frequency to do so. On resonance, a simple harmonic oscillator's motion
is $90^{\circ}$ out of phase with its drive as shown in Fig. \ref{fig:CHExpSetup_SHOMagPhase}.
However, due to extra phase shifts from the leads, filters, and amplifiers
in the phase-locked loop, the reference phase of the lock-in required
to drive the cantilever on resonance often differs from $90^{\circ}$.

In order to find the resonant phase, the magnitude of the first harmonic
of the cantilever interferometer signal was measured while the lock-in's
reference phase was scanned. The cantilever was driven sufficiently
weakly that the first harmonic of the interferometer signal was nearly
proportional to the cantilever displacement as mentioned at the end
of \ref{sub:CHExpSetup_CantileverFiberInterferometer}. In this case,
the magnitude of the first harmonic of the cantilever interferometer
signal traces out a curve similar to the one shown in Fig. \ref{fig:CHExpSetup_SHOMagPhase}.
The measured magnitude can be fit to the expected oscillator response
function to determine the resonant phase. In practice, it is sufficient
to determine the resonant phase by eye from a measured phase scan.
Because the frequency of the simple harmonic oscillator as a function
of phase is linear over tens of degrees (Fig. \ref{fig:CHExpSetup_SHOMagPhase}),
the phase-locked loop is relatively insensitive to the choice of the
reference phase.

\FloatBarrier

\subsubsection{\label{sub:CHExpSetup_CalibrationDriveMotion}Calibration of the
cantilever drive and interferometer signal}

The amount of displacement per voltage $V_{piezo}$ applied to the
piezoelectric actuator can be difficult to calibrate. Even when this
calibration is known, it is not straightforward to model the coupling
of this motion to the motion of a cantilever sitting on top of the
piezo. In principal, the interferometer signal could be converted
into a displacement from knowledge of the incident laser power and
wavelength and the relevant reflection and transmission coefficients.
However, the optical lever term complicates the conversion of this
signal into cantilever amplitude. The difficulties associated with
calibrating both the interferometer signal and the piezo voltage in
units of cantilever amplitude can be overcome by performing a cantilever
amplitude scan.

An amplitude scan is taken, for an experimental arrangement like the
one shown in Fig. \ref{fig:CHExpSetup_CantileverMeasurementSchematic},
by measuring the value of the first harmonic of the cantilever interferometer
signal at a series of drive voltages to the piezo. The drive frequency
must be the cantilever's resonant frequency. If the cantilever frequency
drifts significantly over time or has an amplitude dependence, the
amplitude scan must be performed with the lock-in maintaining a phase-locked
loop (which requires the resonant phase to be shifted by $180^{\circ}$
when the interferometer signal changes sign, a complication that can
be avoided by using the lock-in's internal reference when the cantilever
frequency is constant as a function of time and amplitude). Before
the scan begins, the laser wavelength must be tuned to the optimal
fringe position as described in \ref{sub:CHExpSetup_WavelengthTuning}
(or adjusting the constant value of the interferometer signal to its
midpoint).

\begin{figure}[h]

\centering{}\includegraphics[width=0.7\paperwidth]{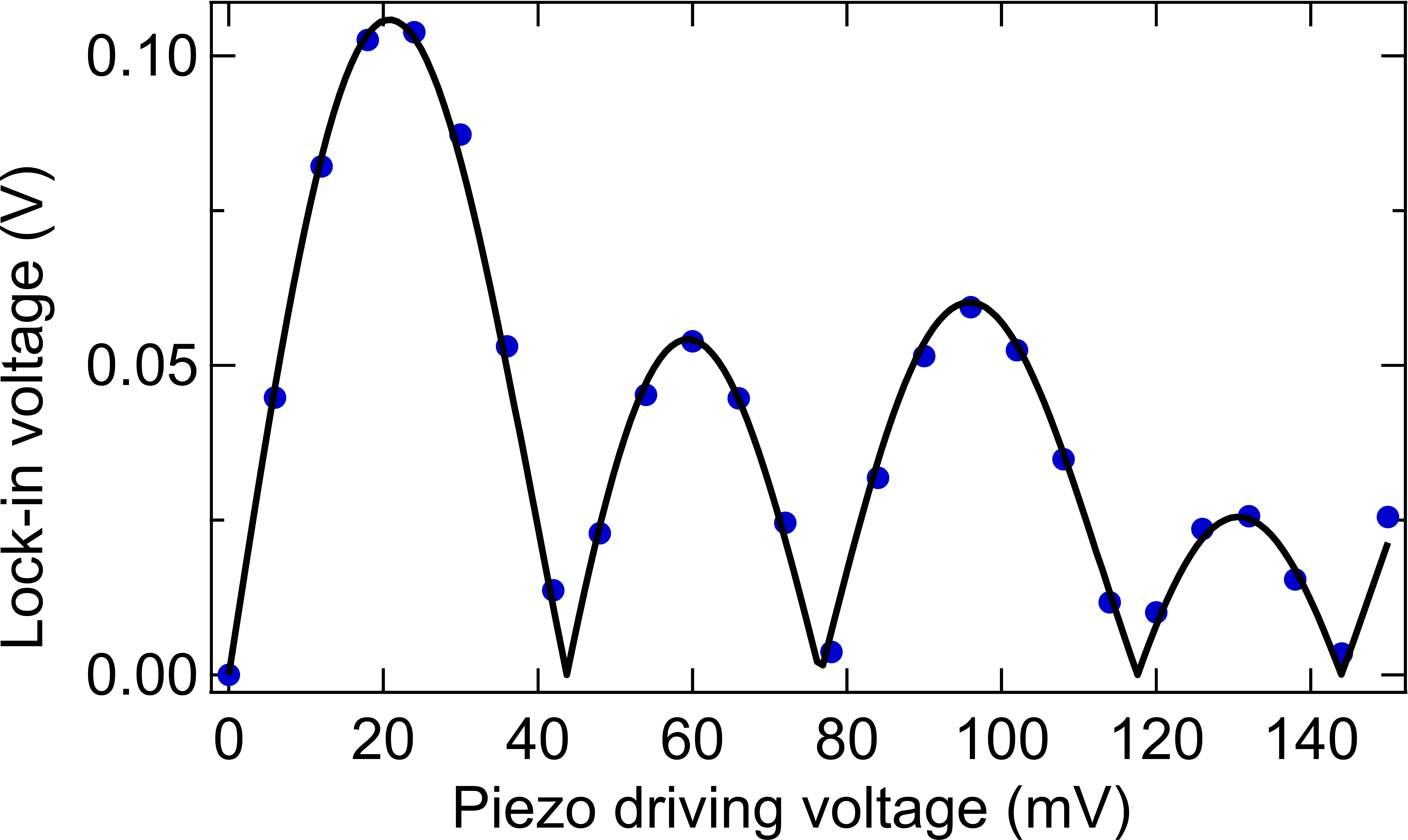}\caption[Cantilever first harmonic signal as measured by the lock-in as a function
of piezo driving voltage]{\label{fig:CHExpSetup_Ampscan}Cantilever first harmonic signal as
measured by the lock-in as a function of piezo driving voltage. The
figure shows the first harmonic $V_{1}$ of the cantilever interferometer
signal as measured by the lock-in as a function of piezo driving voltage
$V_{\text{piezo}}$ as well as a fit to Eq. \ref{eq:CHExpSetup_AmpscanFitFunction}.
The extracted fit coefficients are $V_{1,\max}=0.10\,$ V, $V_{\text{peak}}=20.6\,$
mV, and $\epsilon_{V}=1.17\times10^{-6}\,\text{V}^{-1}$. This amplitude
scan was taken with the cantilever shown in Fig. \ref{fig:CHExpSetup_BareArrows}
and was used to calibrate the Brownian motion measurements discussed
in \ref{sub:CHSensitivity_ThermometrySection}.}
\end{figure}

With the conditions described in the previous paragraph, the lock-in
first harmonic magnitude $V_{1}$ can be fit to a function of the
form
\begin{equation}
V_{1}=\frac{V_{1,\max}}{0.582}\left|J_{1}\left(1.841\frac{V_{\text{piezo}}}{V_{\text{peak}}}\right)+\epsilon_{V}V_{\text{piezo}}\right|\label{eq:CHExpSetup_AmpscanFitFunction}
\end{equation}
where, in the absence of the optical lever effect (i.e. for $\epsilon=0$),
$V_{1,\max}$ is the maximum value of $V_{1}$ and $V_{\text{peak}}$
the value of $V_{\text{piezo}}$ corresponding to that maximum (0.582
and 1.841 are approximations for the magnitude and location of the
first peak of $J_{1}$). Here we assume that the piezo motion and
consequently the cantilever motion are proportional to $V_{\text{piezo}}$.
By comparison with Eq. \ref{eq:CHExpSetup_1stHarmInterferometer},
we can write the calibration of the piezo drive as 
\[
\frac{x_{f,\max}}{V_{\text{piezo}}}=\frac{1.841\lambda}{4\pi V_{\text{peak}}}
\]
and the calibration of the lock-in magnitude at small cantilever amplitudes
as 
\[
\frac{dV_{1}}{dx_{f,\max}}\bigg|_{x_{f,\max}=0}=\frac{2\pi V_{0}}{.582\lambda}\left(1+\frac{2\epsilon_{V}V_{\text{peak}}}{1.841}\right).
\]
The ratio $x_{f,\max}/V_{\text{piezo}}$ allows the calculation of
the cantilever amplitude $x_{f,\max}$ from the voltage drive to the
piezo. The cantilever amplitude can also calculated from the measured
value of $V_{1}$ using 
\[
V_{1}=\frac{V_{1,\max}}{0.582}\left|J_{1}\left(\frac{4\pi x_{f,\max}}{\lambda}\right)+\frac{4\pi\epsilon_{V}V_{\text{peak}}x_{f,\max}}{1.841\lambda}\right|
\]
without reference to the piezo driving voltage. This method is useful
when the ratio $x_{f,\max}/V_{\text{piezo}}$ is not reliably constant
(such as when the cantilever quality factor is drifting in time).
The quantity $dV_{1}/dx_{f,\max}|_{x_{f,\max}=0}$ allows for the
lock-in voltage $V_{1}$ to be converted into displacement $x_{f,\max}$
when the cantilever amplitude is very small ($x_{f,\max}\ll\lambda/2\pi$
so that $V_{1}\propto x_{f,\max}$), as is the case during measurements
of the Brownian motion of an undriven cantilever (see Chapter \ref{cha:CHSensitivity}).
Assuming the distance $z_{f}$ of the laser spot from the cantilever
base is known, all of the calibrations can be expressed in terms of
the motion of the cantilever tip $x_{\max}$ by using $x_{\text{\ensuremath{\max}}}=x_{f,\max}/U_{m}(z_{f}/l)$
with $U_{m}$ as given by Eq. \ref{eq:CHTorsMagn_UmCantileverMode}.
We can write the conversion factor $\Gamma_{V\,\text{to}\, x}$ from
the lock-in reading $V_{1}$ to $x_{\text{\ensuremath{\max}}}$ (valid
for $x_{f,\max}\ll\lambda/2\pi$) as
\begin{equation}
\Gamma_{V\,\text{to}\, x}=\left(U_{m}\left(z_{f}/l\right)\frac{dV_{1}}{dx_{f}}\bigg|_{x_{f,\max}=0}\right)^{-1}\label{eq:CHExpSetup_VoltsToMetersXTip}
\end{equation}
and the conversion factor $g_{\text{piezo}}$ from piezo voltage $V_{\text{piezo}}$
to $x_{\max}$ as 
\begin{equation}
g_{\text{piezo}}\equiv\frac{x_{\max}}{V_{\text{piezo}}}=\frac{1.841\lambda}{4\pi V_{\text{peak}}}\frac{1}{U_{m}\left(z_{f}/l\right)}.\label{eq:ChExpSetup_gPiezoFactor}
\end{equation}
The optical lever coefficient $\epsilon$ is given by 
\[
\epsilon=\frac{8\pi V_{\text{peak}}\epsilon_{V}}{1.841\lambda T_{f}\sqrt{R_{c}/R_{f}}}.
\]
A typical amplitude scan and fit are shown in Fig. \ref{fig:CHExpSetup_Ampscan}.

\FloatBarrier

\subsubsection{\label{sub:CHExpSetup_CantileverMeasurementProcedure}Cantilever
measurement procedure}

We now give a brief step by step summary of the operations necessary
to go from new cantilever samples to persistent current (or aluminum
grain thermometry) measurements.
\begin{enumerate}
\item Preliminary steps at room temperature. 

\begin{enumerate}
\item Arrange all optical and electronic components as in Fig. \ref{sub:CHExpSetup_CantileverDetectionSetup}.
\item Mount new sample chip. 
\item Position fiber holder so that the laser is shining on a cantilever
at room temperature with a fiber to cantilever distance at least $500\,\mu$m
to allow for drift during the cool down (for the lens set up, the
lens should be positioned so that the laser spot is focused on a cantilever). 
\item Cool down refrigerator.
\end{enumerate}
\item Reposition the fiber holder.

\begin{enumerate}
\item Scan the fiber holder laterally to move the laser spot onto a cantilever
if it drifted off during the cool down.
\item Calibrate the step size in nanometers of the Attocube that moves the
fiber closer to and further away from the sample chip. This can be
done by monitoring the position on the interferometer fringe over
the course of many steps. The constant interferometer power or the
magnitude of the first harmonic response to shaking of the Attocube
itself can be monitored and fit to $P\sin(4\pi N\, dz/\lambda+\phi)$
where $N$ is the number of Attocube steps and $dz$ is displacement
per step (and $P$ and $\phi$ are the other fit parameters besides
$dz$).
\item Adjust the fiber to cantilever distance. With a cleaved fiber tip,
this adjustment is made by carefully moving the fiber towards the
cantilever until it just touches (indicated by the abrupt absence
of cantilever motion in the interferometer signal) and then retracting
the fiber tip by the desired amount. With the lens, the lens to cantilever
distance is adjusted to locate the point of maximum reflected power
and maximum depth of modulation of the reflected power when the wavelength
is tuned.
\item Determine laser spot position on the sample chip, and move to the
desired cantilever. Identification of chip position is usually determined
by scanning the laser spot laterally until the chip window is located.
The chip window is distinguished by being much wider than the cantilevers
and shaking less (as observed by the interferometer signal). Then
the laser spot is scanned back to the desired cantilever. The individual
cantilevers are easily distinguished from empty space by their modulation
of the interferometer signal.
\item Calibrate the step size of the Attocubes which move parallel to the
sample chip. This calibration is performed by noting the number of
steps required to move between two points separated by a known distance.
Typically, the number of steps required to move between two adjacent
cantilevers can be used for the direction perpendicular to the cantilever
beams, and the number of steps required to move from the cantilever
tip to the base (by moving to the tip, moving just off to the side
of the cantilever, and then moving back towards the base until it
appears again in the interferometer signal) is used for the direction
parallel to the cantilever beams.
\item Position the laser spot at a known distance $z_{f}$ from the base
of the cantilever.
\end{enumerate}
\item Calibrate and characterize cantilever motion.

\begin{enumerate}
\item Determine the resonant phase for the piezo driving voltage by performing
a phase scan as described in \ref{sub:CHExpSetup_PhaseScan}.
\item Tune the laser wavelength to the optimal fringe position as described
in \ref{sub:CHExpSetup_WavelengthTuning} and perform an amplitude
scan as described in \ref{sub:CHExpSetup_CalibrationDriveMotion}
to find the relationship between piezo drive voltage and cantilever
tip motion and between lock-in magnitude and cantilever tip motion.
\item Characterize the cantilever signal noise.

\begin{enumerate}
\item Determine the cantilever quality factor $Q$ from cantilever ringdown
measurements. The cantilever ringdown measurement procedure consists
of exciting the cantilever, setting the piezo driving voltage to zero,
and measuring the first harmonic of the cantilever interferometer
signal as a function of time. The portion of the signal corresponding
to $x_{f,\max}\ll\lambda/2\pi$ (where the interferometer signal is
proportional to the cantilever motion) can then be fit to Eq. \ref{eq:CantileverRingdown}
for $Q$.
\item Measure the power spectral density of the undriven cantilever's motion.
If this motion is larger than the thermal limit (Brownian motion),
external vibrations or heating may be contributing appreciably to
the motion of the cantilever.
\item Measure the power spectral density of the driven cantilever's frequency
(by using the Hilbert transform method of frequency measurement) at
various cantilever excitation amplitudes. This measurement allows
one to detect contributions to the uncertainty in the frequency measurement
beyond the unavoidable thermal contribution and to determine the optimal
conditions for the frequency measurement.
\end{enumerate}
\end{enumerate}
\item Measure the cantilever frequency as a function of magnetic field

\begin{enumerate}
\item Drive the cantilever in a phase-locked loop at the desired amplitude
(often chosen to give a certain amplitude of flux $\phi_{a}$ through
the ring as discussed in \ref{sec:CHTorsMagn_FiniteDriveSection};
alternatively chosen to minimize uncertainty in the frequency measurement
for the case of cantilevers which show increasing frequency noise
with amplitude).
\item \label{enu:CHExpSetup_PCMeasurementProcMeasureStep}Make a series
of measurements at constant magnetic field of the cantilever frequency
and any other parameters to be recorded.
\item \label{enu:CHExpSetup_PCMeasurementProcRampStep}Ramp the magnetic
field by the desired magnetic field step. 

\begin{enumerate}
\item While the magnetic field is ramped, a PID feedback loop is run in
LabVIEW to minimize the second harmonic of the cantilever interferometer
signal by tuning the laser wavelength via the thermoelectric temperature
controller. This feedback loop corrects for any shift in the interferometer
length $x_{0}$ due to tilting of the sample stage with magnetic field.
\item At the end of the magnetic field ramp, after the laser wavelength
has been tuned to the optimal fringe position, another PID feedback
loop can be run by LabVIEW to set the first harmonic of the cantilever
interferometer signal to the desired set point by adjusting the piezo
driving voltage. This feedback loop corrects for any change in the
calibration of the piezo driving voltage in units of cantilever amplitude
which can occur if, for example, the cantilever quality factor or
phase-locked loop resonant phase changes with time or magnetic field.
Because the piezo drive should not change much in time or with magnetic
field and because each adjustment to the piezo driving voltage must
be given a cantilever ringdown time to take full effect, this feedback
loop was not usually run at each magnetic field point and was instead
run when the first harmonic of the cantilever interferometer signal
moved outside of a specified tolerance window.
\end{enumerate}
\item Steps \ref{enu:CHExpSetup_PCMeasurementProcMeasureStep} and \ref{enu:CHExpSetup_PCMeasurementProcRampStep}
are repeated over the desired field range. In order to counteract
low frequency noise due to drift in the cantilever frequency as a
function of time, the number of measurements of the cantilever frequency
made at any one magnetic field value is usually kept to a small number
so that whole field range can be swept over several times. These sweeps
can then be averaged together.
\end{enumerate}
\end{enumerate}

\subsection{\label{sub:CHExpSetup_TransportCalibMeasurement}Transport calibrations
and measurement procedures}

\subsubsection{Mathematical relations for the resistance bridge}

In the magnetoresistance measurements, the lock-in is used to measure
the difference $V_{AB}$ between the voltage dropped across the sample
(plus a lead) $V_{ss}$ and that dropped across the decade resistor
$V_{ds}$. In Fig. \ref{fig:CHExpSetup_TransportBridgeSchematic},
$V_{ss}$ ({}``sample side'') and $V_{ds}$ ({}``decade side'')
are, respectively, the voltages from points B and A to ground. We
can write these voltages by using Ohm's law. If we call the voltage
across the entire bridge $V_{XY}$ (dropped between points X and Y
in Fig. \ref{fig:CHExpSetup_TransportBridgeSchematic}), the sample
side current $I_{ss}$ and decade side current $I_{ds}$ satisfy $I_{ss}=V_{XY}/(2R_{b}+R_{l}+R_{l}'+R_{s})$
and $I_{ds}=V_{XY}/(2R_{b}+R_{d})$, respectively. The voltage measured
by the lock-in is then
\begin{eqnarray}
V_{AB} & = & I_{ss}\left(R_{s}+R_{l}\right)-I_{ds}R_{d}\nonumber \\
 & = & V_{XY}\left(\frac{R_{s}+R_{l}}{2R_{b}+R_{l}+R_{l}'+R_{s}}-\frac{R_{d}}{2R_{b}+R_{d}}\right).\label{eq:CHExpSetup_VABfull}
\end{eqnarray}

In the experiment, the voltage $V_{AB}$ is measured while the sample
resistance $R_{s}$ is varied by sweeping the magnetic field. If we
write the sample resistance as $R_{s}=R_{s0}+\delta R_{s}$ where
$R_{s0}$ is the sample resistance at zero magnetic field, we can
invert the change in voltage $\delta V_{AB}=V_{AB}(R_{s0}+\delta R_{s})-V(R_{s0})$
to find the change in resistance $\delta R_{s}$ as
\begin{equation}
\delta R_{s}=\left(\frac{\delta V_{AB}}{V_{XY}}\right)\frac{\left(2R_{b}+R_{l}+R_{l}'\right)^{2}}{2R_{b}+R_{l}-\left(\delta V_{AB}/V_{XY}\right)\left(2R_{b}+R_{l}+R_{l}'\right)}.\label{eq:CHExpSetup_BridgeDeltaRsFull}
\end{equation}
Typical values for the difference resistances in bridge were $R_{b}=4.7\times10^{4}\,\Omega$,
$R_{l},R_{l}'=30\,\Omega$, and $R_{s}=300\,\Omega$. In the limit
of $R_{b}\gg R_{l},R_{l},R_{s}$, Eq. \ref{eq:CHExpSetup_BridgeDeltaRsFull}
becomes
\begin{equation}
\delta R_{s}\approx2R_{b}\frac{\delta V_{AB}}{V_{XY}}.\label{eq:CHExpSetup_BridgeDeltaRsApprox}
\end{equation}

\subsubsection{\label{sub:CHExpSetup_VXYCalSection}Calibration of the applied voltage
$V_{XY}$}

The output of the isolation amplifier used in the bridge circuit did
not have unity gain. In fact, the gain factor of its output to its
input was observed to drift slowly over time, possibly due to the
fact it was powered by batteries in order to minimize its output noise.
To calibrate the output of the isolation amplifier, {}``decade scans''
were taken each day.

\begin{figure}[h]

\centering{}\includegraphics[width=0.7\paperwidth]{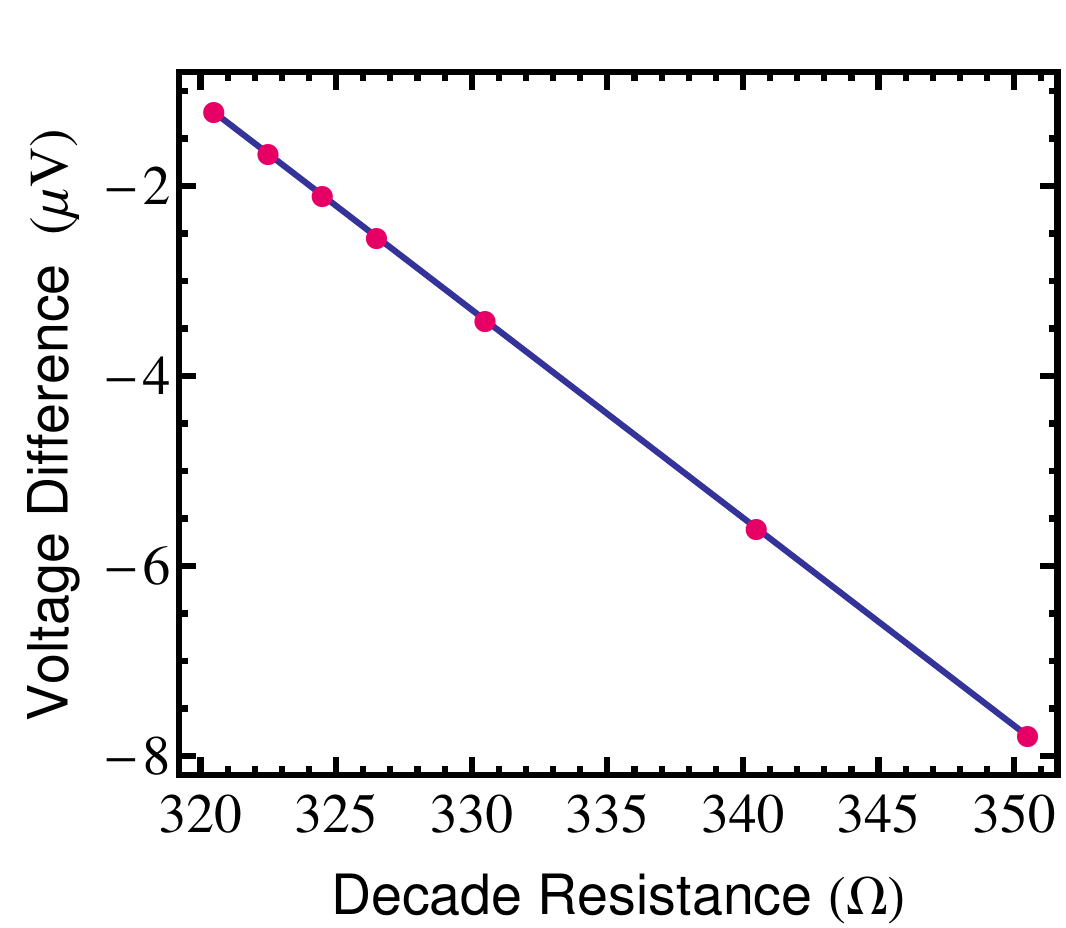}\caption[Example of a decade resistance scan]{\label{fig:CHExpSetup_VXYCal}Example of a decade resistance scan.
The voltage difference $V_{AB}$ is plotted against the decade resistance
$R_{d}$ along with a fit. For this particular scan, the bridge voltage
$V_{XY}$ was $-20.7\,$mV and the factor $w_{\text{offset}}$ was
$-3.34\times10^{-3}$.}
\end{figure}

A decade scan consisted of measuring the sample-decade voltage difference
$V_{AB}$ at a series of values of the decade resistance $R_{d}$.
Typically, $R_{d}$ was stepped up from the value used for the magnetoresistance
measurements over a range of about $30\,\Omega$. The measured values
of $V_{AB}$ were then fit to Eq. \ref{eq:CHExpSetup_VABfull} in
the form
\[
V_{AB}=V_{XY}\left(w_{\text{offset}}-\frac{R_{d}}{2R_{b}+R_{d}}\right)
\]
 for $V_{XY}$ and $w_{offset}$ using the known value of $R_{b}$.
This fitted value for $V_{XY}$ was then used with Eq. \ref{eq:CHExpSetup_BridgeDeltaRsFull}
(or Eq. \ref{eq:CHExpSetup_BridgeDeltaRsApprox}) to find $\delta R_{s}$
for the magnetic field sweeps. The data from one decade scan along
with the corresponding fit are shown in Fig. \ref{fig:CHExpSetup_VXYCal}.

\FloatBarrier

\subsubsection{\label{sub:CHExpSetup_LeadResistances}Determination of lead resistances
$R_{l}$ and $R_{l}'$}

Since a three-point rather than a four-point measurement was used,
it was helpful to determine the resistances of the leads attached
to either side of the sample labeled $R_{l}$ and $R_{l}'$ in Fig.
\ref{fig:CHExpSetup_TransportBridgeSchematic} (though to a good approximation
they could be ignored when finding $\delta R_{s}$ since $R_{b}\gg R_{l},R_{l}'$).
The lead resistance $R_{l}$ was found by noting the resistance of
the decade necessary to null out $V_{AB}$ while the sample was in
the superconducting state. The resistance $R_{l}'$ could then be
determined by measuring the resistance through $R_{l}'$, $R_{s}$,
and $R_{l}$ while the sample was superconducting and subtracting
the known value for $R_{l}$. The resistance of the sample can also
be inferred by measuring its superconducting transition and using
the inferred values for $R_{l}$ and $R'_{l}$ and solving Eq. \ref{eq:CHExpSetup_VABfull}
for $R_{s}$. The value of both lead resistances was about $34\pm5\,\Omega$
(where the imprecision is due to the uncertainty in determining the
end of the superconducting transition in Fig. \ref{fig:AppTransport_RfromSuperconTran}).

\subsubsection{Magnetoresistance measurement procedure}

We now give a brief outline of the steps for mounting and measuring
a transport sample.
\begin{enumerate}
\item Mount a sample chip (for details regarding the elements discussed
in all of these steps see \ref{sub:CHExpSetup_TransportSetUp})

\begin{enumerate}
\item The guiding principle behind the following steps is that an unknown
voltage should never be connected to the sample. The sample and any
connectors should always be grounded before making a new connection
in order to avoid a blow out.
\item Wire bond the sample. During part of the wire bonding process, the
sample may be ungrounded in order to measure its resistance and verify
that the wire bonding was successful. The end result though should
be that the sample is bonded to the circuit board with the grounding
plug in place.
\item Mount the sample chip holder onto the refrigerator.
\item Connect the sample chip holder leads to the refrigerator leads and
connect the heat sink/ground plane to the refrigerator.
\item Connect the leads coming out of the refrigerator to the junction box
and set all of the junction box switches to ground.
\item Remove grounding plug from sample chip holder.
\item Cool down the refrigerator.
\item Connect external electronics. Always ground the leads to the sample
(with the junction box switches) before changing external electronics
connections or settings.
\end{enumerate}
\item Measure Magnetoresistance

\begin{enumerate}
\item Perform a decade scan to calibrate the applied bridge voltage $V_{XY}$
(see \ref{sub:CHExpSetup_VXYCalSection}).
\item Set the decade resistance to null out $V_{AB}$.

\begin{enumerate}
\item When a four-point measurement is used, this decade setting gives the
sample resistance $R_{s}$. If only a three-point measurement is possible,
determine the lead resistances as described in \ref{sub:CHExpSetup_LeadResistances}. 
\item Due to stray capacitance in the leads, it is not possible to completely
null out $V_{AB}$ because tuning the decade resistance only adjusts
the real part of the bridge impedance. The following approaches can
be employed to minimize the effect of the stray capacitance on the
measurement:

\begin{enumerate}
\item Theoretically, it is possible to null out the difference in capacitance
between the two sides of the bridge using a variable capacitor. We
never attempted to do this in our measurements.
\item Alternatively, the reference phase of the lock-in can be adjusted
so that the lock-in's X quadrature voltage is entirely due to resistance
offsets in the two sides of the bridge and the Y quadrature contains
a constant voltage due to the constant capacitive offset. The voltage
in this quadrature should function as the voltage $V_{AB}$. We did
not use this approach in our measurements. 
\item In practice, we addressed this issue by tuning the decade resistance
$\sim20\,\Omega$ away from its balanced value. Because the capacitive
offset is small, it contributed negligibly to the magnitude of $V_{AB}$
at this bridge imbalance, so changes in voltage with magnetic field
could be treated as being proportional to changes in the sample resistance.
For this $20\,\Omega$ bridge imbalance, $V_{AB}$ was, for example,
$2.6\,\mu$V at $2.2\,$K and changes in $V_{AB}$ of $\sim5\,\text{nV}$
could be resolved with the lock-in. If higher resolution were needed,
it would be necessary to null out the capacitive offset so that the
gain of the lock-in could be increased (nulling out the offset voltage
being the whole point of the bridge circuit in the first place).
\end{enumerate}
\end{enumerate}
\item Measure magnetoresistance

\begin{enumerate}
\item Measure the voltage difference across the bridge several times at
one field point.
\item Ramp to the next field point (at a ramp rate of $\sim.1\,\text{mT}/\text{s}$).
\item Repeat over desired field range (this ramp used the same LabVIEW routine
as that used in the persistent current measurement procedure discussed
in \ref{sub:CHExpSetup_CantileverMeasurementProcedure}).
\end{enumerate}
\item Repeat the magnetoresistance measurements at several different excitation
voltages to verify that the signal does not depend on excitation.
This check is to make sure that the electrons are not being heated.
A good guide for the choice of excitation is that $eI_{ss}R_{s}\lesssim k_{B}T$
where $I_{ss}R_{s}$ is the voltage drop across the sample. When measuring
at many temperatures, it is a safe assumption that a sufficiently
low excitation voltage for temperature $T_{1}$ will also be sufficiently
low for $T_{2}>T_{1}$. Thus, it is advisable to perform temperature
measurements in order of increasing temperature so that this check
need not be made at each new temperature value. However, due to the
need for a high signal to noise ratio, it will be necessary to as
temperature increases.\end{enumerate}
\end{enumerate}

\chapter{\label{cha:CHSensitivity}Sensitivity of cantilever torsional magnetometry
to persistent currents}

\section{Sources of uncertainty in cantilever torsional magnetometry measurements}

There are many potential sources of uncertainty in the estimation
of the frequency of a cantilever excited in the phase-locked loop
circuit depicted in Fig. \ref{fig:CHExpSetup_CantileverMeasurementSchematic}.
Almost every component in the figure can in principal be assigned
a contribution to the measurement noise. These noise sources can be
broken up into four general classes: fluctuations of the cantilever's
position due to unavoidable applied forces, noise in the readout of
the interferometer signal, and the input voltage noise associated
with each piece of electronics, and the output noise of the driving
voltage sent to the piezoelectric actuator.

\subsection{\label{sub:CHSensitivity_ForceNoise}Fluctuating forces acting on
the cantilever}

When the cantilever is sufficiently well isolated from external vibrations,
the two fluctuating forces intrinsic to the cantilever detection arrangement
are those of the thermal noise associated with the cantilever's finite
temperature and the radiation pressure shot noise due to the backaction
of the readout laser on the cantilever \citep{albrecht1991frequency,sidles1995magnetic}.
We treat both noise sources as white and thus characterized by their
respective force power spectral densities, $S_{F,\text{th}}$ and
$S_{F,\text{RP}}$.%
\footnote{Unless otherwise specified, all power spectral densities discussed
in this text will be single-sided.%
} 

The magnitude of the thermal force noise can be computed from the
equipartition theorem \citep{pathria1996statistical}, which states
that at thermal equilibrium 
\begin{equation}
k\left\langle x^{2}\right\rangle =k_{B}T,\label{eq:CHSensitivity_EquipartitionTheorem}
\end{equation}
where $k$ and $x$ are the spring constant and displacement of the
cantilever tip, $\langle\ldots\rangle$ denotes thermodynamic ensemble
averaging, and $T$ is the cantilever temperature. The mean square
displacement is given by Parseval's theorem $\langle x^{2}\rangle=(1/2\pi)\int_{0}^{\infty}d\omega\, S_{x,\text{th}}(\omega)$
where $S_{x,\text{th}}(\omega)$ is the power spectral density of
the cantilever's displacement when the cantilever is driven by the
thermal force noise. Using the transfer function given in Eq. \ref{eq:GCantileverTransferFunction},
we can relate displacement to force as $S_{x,\text{th}}(\omega)=|G\left(\omega\right)|^{2}S_{F,\text{th}}$
and then evaluate the expression for the mean square displacement:
\begin{align*}
\left\langle x^{2}\right\rangle  & =\frac{S_{F,\text{th}}}{2\pi m_{\text{eff}}^{2}}\int_{0}^{\infty}d\omega\,\frac{1}{\left(\omega^{2}-\omega_{0}^{2}\right)^{2}+\omega^{2}\omega_{0}^{2}/Q^{2}}\\
 & =\frac{S_{F,\text{th}}}{2\pi m_{\text{eff}}^{2}}\frac{1}{2\omega_{0}^{3}}\int_{-\infty}^{\infty}dy\,\frac{1}{\left(y^{2}-1\right)^{2}+y^{2}/Q^{2}},
\end{align*}
where we have taken $y=\omega/\omega_{0}$. Using the quadratic formula,
the denominator of the integrand can be rewritten as $(y^{2}-r_{+}^{2})(y^{2}-r_{-}^{2})$
with the roots $r_{\pm}^{2}=1-1/2Q^{2}\pm i\sqrt{(1-1/4Q^{2})/Q^{2}}$.
In the limit $Q\gg1$, these roots become $r_{\pm}^{2}\approx1\pm i/Q$,
and $r_{\pm}\approx1\pm i/2Q$. We can rewrite the integral as 
\begin{align}
\left\langle x^{2}\right\rangle  & =\frac{S_{F,\text{th}}}{2\pi m_{\text{eff}}^{2}}\frac{1}{2\omega_{0}^{3}}\int_{-\infty}^{\infty}dy\,\frac{1}{\left(y-r_{+}\right)\left(y+r_{+}\right)\left(y-r_{-}\right)\left(y+r_{-}\right)}\label{eq:CHSensitivity_X2ContourPoles}
\end{align}
which can be evaluated by contour integration. The integral in Eq.
\ref{eq:CHSensitivity_X2ContourPoles} is equal to the integral for
standard semi-circular contour enclosing the upper half of the complex
plane that contains poles at $+r_{+}$ and $-r_{-}$. Calculating
the residues at these poles, we find
\begin{align*}
\left\langle x^{2}\right\rangle  & =\frac{S_{F,\text{th}}}{2\pi m_{\text{eff}}^{2}}\frac{2\pi i}{2\omega_{0}^{4}}\left(\frac{1}{2r_{+}\left(r_{+}^{2}-r_{-}^{2}\right)}+\frac{1}{2r_{-}\left(r_{+}^{2}-r_{-}^{2}\right)}\right)\\
 & \approx\frac{S_{F,\text{th}}}{2\pi m_{\text{eff}}^{2}}\frac{2\pi i}{2\omega_{0}^{4}}\left(\frac{Q}{2i}\right)\\
 & \approx\frac{S_{F,\text{th}}}{4m_{\text{eff}}^{2}}\frac{Q}{\omega_{0}^{3}}.
\end{align*}
This result combined with the Equipartition Theorem allows us to write
the thermal force spectral density as 
\begin{equation}
S_{F,\text{th}}=\frac{4kk_{B}T}{Q\omega_{0}}\label{eq:CHSensitivity_ThermalForcePSD}
\end{equation}
where we have used $k=m_{\text{eff}}\omega_{0}^{2}$. We can also
rewrite the relation $S_{x,\text{th}}(\omega)=|G(\omega)|^{2}S_{F,\text{th}}$
in the form
\begin{equation}
S_{x,\text{th}}\left(\omega\right)=\frac{4\left\langle x^{2}\right\rangle }{\omega_{0}Q}\frac{1}{\left(\left(\omega/\omega_{0}\right)^{2}-1\right)^{2}+\left(\omega/\omega_{0}Q\right)^{2}}.\label{eq:CHSensitivity_SxFitForm}
\end{equation}

The radiation pressure backaction force arises from the corpuscularity
of the readout laser beam which is composed many photons each carrying
a momentum $p=h/\lambda$. When a photon hits the cantilever and is
reflected, it imparts a net impulse of $2h/\lambda$ onto the cantilever.
A typical laser beam can be modeled as a series of propagating photons.
The distance between adjacent photons is random and follows a Poisson
distribution. Because the spacing between adjacent photons is random
and uncorrelated, the intensity of the laser beam (the number of photons
passing through a plane in space per unit time) possesses a fluctuating
component. These {}``shot noise'' fluctuations are also present
in the number of impulses imparted on the cantilever per unit time
and lead to a fluctuating component in the radiation pressure force
\citep{edelstein1978limitsto}. 

Consider a Poissonian stream of particles each having a value $q$
for some quantity (e.g. energy, electric charge, or momentum). If
we denote by $R$ the mean rate at which this quantity passes a particular
point in space per unit time (e.g. power, electric current, or force,
to match the previous examples for $q$), then the shot noise of the
stream can be expressed as the power spectral density $S_{R}$ of
the fluctuations in this rate. This power spectral density is white
and obeys \citep{devoret2006classlecture} 
\begin{equation}
S_{R}=2qR.\label{eq:CHSensitivity_ShotNoise}
\end{equation}
For radiation pressure, we can take $q$ to be the impulse kick per
photon $2h/\lambda$ and $R$ to be the rate of impulses on the cantilever,
in other words the force $F_{\text{RP}}$, given by the impulse kick
per photon $2h/\lambda$ times the rate of photons $\dot{N}$. We
write the rate of photons in terms of the incident laser power $P$
by dividing by the energy per photon $hc/\lambda$: $\dot{N}=\lambda P/hc$.
We have 
\begin{align}
S_{F,\text{RP}} & =2\left(\frac{2h}{\lambda}\right)\left(\frac{2h}{\lambda}\frac{\lambda P}{hc}\right)\nonumber \\
 & =\frac{8hP}{\lambda c}.\label{eq:CHSensitivity_RadiationPressure}
\end{align}
The highest power laser powers discussed in this text are $5\,\mu\text{W}$.
If we assume that the cantilever is perfectly reflective, Eq. \ref{eq:CHSensitivity_RadiationPressure}
gives a radiation pressure force noise of $\sqrt{S_{F,\text{RP}}}=8\times10^{-21}\,\text{N/\ensuremath{\sqrt{\text{Hz}}}}$.
This figure provides an upper bound on the radiation pressure force
noise experienced by a partially reflective cantilever. On the other
hand, using our refrigerator's base temperature of $T=300\,\text{mK}$
and the smallest value of $k/Q\omega_{0}$ of all the cantilevers
discussed in this text, Eq. \ref{eq:CHSensitivity_ThermalForcePSD}
gives a minimum thermal force noise of $\sqrt{S_{F,\text{th}}}=2\times10^{-18}\,\text{N/\ensuremath{\sqrt{\text{Hz}}}}$,
over two orders of magnitude larger than $\sqrt{S_{F,\text{RP}}}$.
We will drop $S_{F,\text{RP}}$ from further analysis of the uncertainty
in the cantilever frequency measurement.

\subsection{\label{sub:CHSensitivity_InterferometerNoise}Interferometer readout
noise}

Error in the interferometric detection of the cantilever's position
arises from fluctuations both in the laser used to monitor the cantilever
and in the detector used measure the laser signal. An ideal laser
source contains fluctuations due to the shot noise effect discussed
in the previous section. A real laser also possesses technical noise
in both its intensity and wavelength. Any optical detection circuit
based upon a photodiode contains several resistors and transistors
which possess some voltage and current noise. In this section, we
will compare these sources of uncertainty in units of optical intensity.
As we are considering small fluctuations, the optical lever effect
(see \ref{sub:CHExpSetup_CantileverFiberInterferometer}) will be
ignored in each derivation.

We use Eq. \ref{eq:CHSensitivity_ShotNoise} again to calculate the
power fluctuations due to shot noise. Power $P$ is the rate of passage
of energy and so plays the role of $R$ in Eq. \ref{eq:CHSensitivity_ShotNoise}.
The energy associated with a single photon of wavelength $\lambda$
can be written as $hc/\lambda$ \citep{hecht1987optics2nd}. Thus
the power fluctuations due to shot noise are $S_{P,\text{SN}}=2hcP/\lambda$.
The average power on the photodiode during the persistent current
measurement is $P=(R_{f}+T_{f}^{2}R_{c})P_{\text{inc}}$ (see Eq.
\ref{eq:CHExpSetup_PInterferometerFull}). Thus the laser power fluctuations
due to shot noise are
\begin{equation}
S_{P,\text{SN}}=\frac{2hc\left(R_{f}+T_{f}^{2}R_{c}\right)P_{\text{inc}}}{\lambda}.\label{eq:CHSensitivity_ShotNoiseLaserPower}
\end{equation}

In addition to the shot noise contribution, the intensity noise of
a laser can also have a component which is technical in nature. Laser
intensity fluctuations due to technical noise are often referred to
as {}``relative intensity noise'' and are proportional to the mean
laser power. Relative intensity noise arises from noise in the laser's
components such as fluctuations in the laser's driving current. Writing
the sum of the mean laser power and its fluctuations due to relative
intensity noise as $P+\delta P_{\text{RIN}}(t)=P(1+\gamma(t))$, the
power spectral density of the power fluctuations due to relative intensity
noise is $S_{P,\text{RIN}}=P^{2}S_{\gamma}$. Again using $P=(R_{f}+T_{f}^{2}R_{c})P_{\text{inc}}$,
we have
\begin{equation}
S_{P,\text{RIN}}=\left(R_{f}+T_{f}^{2}R_{c}\right)^{2}P_{\text{inc}}^{2}S_{\gamma}.\label{eq:CHExpSetup_RINPower}
\end{equation}

The change in the interferometer power $\delta P_{\lambda}$ due to
a fluctuation $\delta\lambda$ in the wavelength can be written as
$\delta P_{\lambda}=P_{\text{cant}}(\lambda+\delta\lambda)-P_{\text{cant}}(\lambda)$.
With the use of Eq. \ref{eq:CHExpSetup_PInterferometerFull}, we have
\begin{align}
\delta P_{\lambda} & =-2T_{f}\sqrt{R_{c}R_{f}}P_{\text{inc}}\left(\cos\left(\frac{4\pi}{\lambda+\delta\lambda}(x_{0}+x_{1})\right)-\cos\left(\frac{4\pi}{\lambda}(x_{0}+x_{1})\right)\right)\nonumber \\
 & \approx-2T_{f}\sqrt{R_{c}R_{f}}P_{\text{inc}}\bigg[\begin{array}[t]{c}
{\displaystyle \left(\cos\left(\frac{4\pi\delta\lambda}{\lambda^{2}}(x_{0}+x_{1})\right)-1\right)\cos\left(\frac{4\pi}{\lambda}(x_{0}+x_{1})\right)}\\
{\displaystyle +\sin\left(\frac{4\pi\delta\lambda}{\lambda^{2}}(x_{0}+x_{1})\right)\sin\left(\frac{4\pi}{\lambda}(x_{0}+x_{1})\right)}\bigg]
\end{array}\nonumber \\
 & \approx-2T_{f}\sqrt{R_{c}R_{f}}P_{\text{inc}}\frac{4\pi\delta\lambda}{\lambda^{2}}x_{0}\cos\left(\frac{4\pi}{\lambda}x_{1}\right)\label{eq:ChSensitivity_dPlambda}
\end{align}
where we have kept terms to first order in $\delta\lambda$. We have
also assumed that the interferometer is operated such that $4\pi x_{0}/\lambda=2\pi N+\pi/2$
and $x_{0}\gg x_{1}$. When the cantilever is driven, the factor involving
$x_{1}$ can have a time dependence. We find the upper bound of $\delta P_{\lambda}$
by setting the $\cos(4\pi x_{1}/\lambda)$ factor to unity. With the
relationship between $\delta P_{\lambda}$ and $\delta\lambda$ given
by Eq. \ref{eq:ChSensitivity_dPlambda}, the upper bound for the power
spectral density of laser intensity fluctuations can be written as
\begin{equation}
S_{P,\delta\lambda}=4T_{f}^{2}R_{c}R_{f}P_{\text{inc}}^{2}\left(\frac{4\pi x_{0}}{\lambda^{2}}\right)^{2}S_{\lambda}\label{eq:CHSensitivity_PowerWavelengthNoise}
\end{equation}
where $S_{\lambda}$ is the power spectral density of the laser's
wavelength fluctuations. 

Wavelength fluctuations are often specified by the full width at half
maximum $\Delta\nu$ of the laser frequency lineshape. The power spectral
density $S_{\nu}$ of laser frequency is given by $S_{\nu}=\Delta\nu/\pi$
\citep{2006application}. We can re-express $S_{\nu}$ in terms of
$\lambda$ by using $\Delta\nu=c\Delta\lambda/\lambda^{2}$ to write
\begin{equation}
S_{\lambda}=\frac{\lambda^{4}}{\pi^{2}cL_{c}},\label{eq:CHSensitivity_WavelengthFluc}
\end{equation}
where $L_{c}=\lambda^{2}/\pi\,\Delta\lambda$ is the coherence length.
When the wavelength of the light wave follows a lorentzian distribution
with a full width at half maximum $\Delta\lambda$, the coherence
length $L_{c}$ corresponds to the characteristic length scale over
which the autocorrelation of the laser's electric field amplitude
decays.

The total power spectral density $S_{P,\text{int}}$ of the interferometer
signal out of the photodetector is 
\[
S_{P,\text{int}}=S_{P,\text{SN}}+S_{P,\text{RIN}}+S_{P,\delta\lambda}+S_{P,pd}
\]
where $S_{P,pd}$ is the voltage noise of the photodetector converted
into units of optical power. On the highest gain setting (which was
used for all measurements), the 2011 photoreceiver has a noise equivalent
power specification of $\sqrt{S_{P,pd}}=200\,\text{fW/\ensuremath{\sqrt{\text{Hz}}}}$.
We compare this noise to the other noise figures discussed above by
using typical values for the operating input power $P_{\text{inc}}=20\,\text{nW}$,
the reflectivities $R_{f}=.2$ and $R_{c}=.03$, the cavity length
$x_{0}=8\,\text{mm}$, and the laser coherence length $L_{c}=10\,\text{cm}$.%
\footnote{The cavity length can be estimated from the period of oscillation
of the interferometer signal with laser wavelength. The coherence
length can be determined by comparing the depth of modulation of the
interferometer signal with the rf modulation on ($A$) and with the
rf modulation off ($B$). By assuming that, in the absence of rf modulation,
the coherence length is much longer than the cavity length $x_{0}$,
the coherence length with the rf modulation can be found from the
relation $A/B=\exp(-2x_{0}/L_{c})$. The rf modulation was on for
all persistent current measurements.%
} Eq. \ref{eq:CHSensitivity_ShotNoiseLaserPower} gives a noise power
due to shot noise of $\sqrt{S_{P,\text{SN}}}=34\,\text{fW/\ensuremath{\sqrt{\text{Hz}}}}$,
a factor of 6 below the detector's noise level. The relative intensity
noise specification for the JDS Uniphase laser is $\sqrt{S_{\gamma}}=10^{-8}\,/\text{\ensuremath{\sqrt{\text{Hz}}}}$
which corresponds to $\sqrt{S_{P,\text{RIN}}}=44\,\text{aW/\ensuremath{\sqrt{\text{Hz}}}}$,
negligible on the scale of the detector's noise.%
\footnote{This specification is valid between $20\,\text{MHz}$ and $1\,\text{GHz}$.
The frequencies of interest in the measurements discussed in this
text were well below $20\,\text{MHz}$ for which no relative intensity
noise specification is given. No change in the noise out of the photodetector
was observed when the laser turned on and off, so it is difficult
to assign a magnitude to low frequency laser intensity noise other
than to say that it is much less than $5\times10^{-4}\,/\sqrt{\text{\text{Hz}}}$,
which would result in a noise level equal to the photodetector's on
the photodetector output.%
} Eqs. \ref{eq:CHSensitivity_PowerWavelengthNoise} and \ref{eq:CHSensitivity_WavelengthFluc}
yield a noise of $46\,\text{fW/\ensuremath{\sqrt{\text{Hz}}}}$ for
wavelength fluctuations, a factor of four below the detector's noise.
So for the measurements discussed in this text, the photodetector's
noise was the dominant noise source of the interferometer readout
but was comparable in magnitude to the noise due to the laser.

For comparison with the force noise discussed in the previous section,
it is most convenient to express both the force noise and interferometer
noise in units of cantilever displacement. For the force noise, we
calculate the uncertainty in cantilever position at the cantilever's
resonance frequency $\omega=\omega_{0}$. Eqs. \ref{eq:CHSensitivity_EquipartitionTheorem}
and \ref{eq:CHSensitivity_SxFitFormWithOffset} give $S_{x,\text{th}}(\omega_{0})=4Qk_{B}T/\omega_{0}k$.
For $T=300\,\text{mK}$ and typical cantilever parameters $Q=10^{5}$,
$\omega_{0}=2\pi\times2\,\text{kHz}$, and $k=10^{-3}\,\text{N/m}$,
the cantilever displacement uncertainty is $\sqrt{S_{x,\text{th}}}=360\,\text{pm}/\sqrt{\text{Hz}}$.
To convert interferometer power into cantilever displacement, we use
$\delta P_{\text{cant}}\approx\delta x_{1}(dP_{\text{cant}}/dx_{1})$.
Assuming the interferometer fringe position is optimized with $4\pi x_{0}/\lambda=2\pi N+\pi/2$,
differentiating Eq. \ref{eq:CHExpSetup_PInterferometerFull} yields
\[
\frac{dP_{\text{cant}}}{dx_{1}}=\frac{8\pi}{\lambda}T_{f}\sqrt{R_{c}R_{f}}P_{\text{inc}}
\]
and thus
\begin{align*}
\sqrt{S_{x,\text{int}}} & =\frac{\lambda}{8\pi P_{\text{inc}}T_{f}\sqrt{R_{c}R_{f}}}\sqrt{S_{P,\text{int}}}\\
 & =10\,\text{pm}/\sqrt{\text{Hz}}.
\end{align*}
Within $\sim\omega_{0}/2\pi Q$ of the cantilever resonance frequency
$\omega_{0}$, the uncertainty in the cantilever position is dominated
by the noise $\sqrt{S_{x,\text{th}}}$ due to the thermal force. Away
from $\omega_{0}$, this contribution is suppressed and the interferometer
noise $\sqrt{S_{x,\text{int}}}$ dominates.

\subsection{\label{sub:CHSensitivity_ElectronicsNoise}Measurement electronics
noise}

The highest gain setting of the 2011 photoreceiver was $1.65\times10^{7}\,\text{V/W}$.
The photoreceiver's $200\,\text{fW}/\sqrt{\text{Hz}}$ noise level
thus corresponds to a voltage noise of $3.3\,\text{\ensuremath{\mu}V}/\sqrt{\text{Hz}}$.
This value is well above the $40\,\text{nV}/\sqrt{\text{Hz}}$ input
voltage noise of the lock-in amplifier (for $10\,\text{dB}$ gain
setting). Following the chain of electronics laid out in Fig. \ref{fig:CHExpSetup_CantileverMeasurementSchematic},
the photoreceiver signal was passed into lock-in 1 and then into a
set of filters. Lock-in 1 was typically operated at a gain of at least
12 (the gain factor for the $10\,\text{dB}$ setting), meaning that
photoreceiver's noise level into the filters was $40\,\text{\ensuremath{\mu}V}/\sqrt{\text{Hz}}$
or higher, safely above the SRS filter's $300\,\text{nV}/\sqrt{\text{Hz}}$
input voltage noise. After the filters, the signal was passed through
lock-in 2, which usually had a gain of 5 (for which the input noise
was $100\,\text{nV}/\sqrt{\text{Hz}}$). The amplified photoreceiver
noise was thus $200\,\text{\ensuremath{\mu}V}/\sqrt{\text{Hz}}$ out
of this lock-in. Finally, the signal was passed into the DAQ or the
frequency counter in order to perform the frequency measurement. The
DAQ input noise was $20\,\text{nV}/\sqrt{\text{Hz}}$ while that of
the frequency counter was $66\,\text{nV}/\sqrt{\text{Hz}}$, both
well below the amplified photoreceiver noise. Thus, the noise on the
photoreceiver output (including both the photoreceiver's noise and
that due to the cantilever's thermally driven motion) dominated the
noise of all subsequent electronic components in the circuit.

The voltage output of the lock-in amplifier used to drive the piezoelectric
actuator had an output voltage noise of $1\,\text{\ensuremath{\mu}V}/\sqrt{\text{Hz}}$.
Typically, the RMS amplitude of the voltage drive out of the lock-in
amplifier was kept fixed at close to $1\,\text{V}$. When a smaller
voltage drive was required, a voltage divider (composed of resistors
whose Johnson noise was of order $40\,\text{nV}/\sqrt{\text{Hz}}$)
was put between the lock-in output and the leads to the piezo.

\section{\label{sec:CHSensitivity_FrequencyError}Derivation of the error
in the frequency measurement of a cantilever driven in a phase-locked
loop}

Several different authors have calculated the uncertainty in the frequency
measurement of a resonator in a phase-locked loop \citep{yurke1995theoryof,hajimiri1998ageneral,albrecht1991frequency}.
Here we adopt the treatment of Yurke \emph{et al.} \citep{yurke1995theoryof}
to the detection set-up used in the persistent current measurements
(see Fig. \ref{fig:CHExpSetup_CantileverMeasurementSchematic}). The
total uncertainty $\sigma_{f,\text{tot}}$ in the measured cantilever
frequency has two major contributions: the fluctuations $\sigma_{f,\text{cant}}$
present in the driven cantilever motion itself and the fluctuations
$\sigma_{f,pd}$ due to noise added to the detection signal by the
photodetector.

\subsection{Fluctuations $\sigma_{f,\text{cant}}$ in the frequency of motion
of the driven cantilever}

We begin by rewriting Eq. \ref{eq:CHTorsMagnCantEquationMotionTime}
in the form
\begin{equation}
\ddot{X}+\frac{\omega_{0}}{Q}\dot{X}+\omega_{0}^{2}X=\frac{F_{d}\left(t\right)+F_{\text{th}}\left(t\right)}{m_{\textrm{eff}}}\label{eq:CHSensitivity_FreqErrorEqOfMotion}
\end{equation}
where $X$ is the displacement of the cantilever tip, $F_{d}$ is
force drive applied to the piezoelectric actuator by the feedback
circuit, and $F_{\text{th}}$ is the thermal force discussed in \ref{sub:CHSensitivity_ForceNoise}.
We write $X(t)$ in terms of its quadratures $X_{1}(t)$ and $X_{2}(t)$
as
\begin{equation}
X\left(t\right)=X_{1}\left(t\right)\cos\omega t+X_{2}\left(t\right)\sin\omega t.\label{eq:ChSensitivity_QuadraturesX1X2}
\end{equation}
The main frequency component of the driving force $F_{d}$ and the
cantilever's motion $X$ will be taken to be $\omega$. When $\omega$
is close to the resonant frequency $\omega_{0}$ and the quality factor
$Q$ is large, the quadratures $X_{1}(t)$ and $X_{2}(t)$ vary on
the time scale set by the ringdown time $\tau=2Q/\omega_{0}$, which
is much longer than the cantilever period $2\pi/\omega_{0}$. 

We combine the two quadratures into a single quantity by introducing
the complex amplitude 
\[
x(t)=X_{1}(t)-iX_{2}(t).
\]
The cantilever displacement can be written in terms of $x(t)$ as

\begin{align}
X(t) & =\left|x\left(t\right)\right|\cos\left(\omega t+\arg x_{0}\left(t\right)\right)\nonumber \\
 & =\frac{1}{2}\left(x\left(t\right)e^{i\omega t}+x^{*}\left(t\right)e^{-i\omega t}\right).\label{eq:CHSensitivity_FreqErrorXt}
\end{align}
Since the amplitude $x(t)$ varies slowly on the scale of the cantilever
period, it can be thought of as the $\omega$ Fourier component of
$X(t)$ evaluated over a time window that is centered at time $t$
and that is much smaller than the cantilever's ringdown time $\tau$.
Throughout this section, we use an upper-cased symbol to denote a
real-valued time dependent quantity (e.g. $X(t)$), the same upper-case
symbol with a 1 or 2 subscript to denote that quantity's two quadratures
(e.g. $X_{1}(t)$ and $X_{2}(t)$), and lower-case symbols for the
quantity's complex amplitude (e.g. $x(t)$). For instance, we can
write the driving force $F_{d}$ and the thermal force noise $F_{\text{th}}$
as 
\begin{align*}
F_{d}\left(t\right) & =F_{d,1}\cos\omega t+F_{d,2}\sin\omega t=\frac{1}{2}\left(f_{d}\left(t\right)e^{i\omega t}+f_{d}^{*}e^{-i\omega t}\right)
\end{align*}
and
\begin{align*}
F_{\text{th}}\left(t\right) & =F_{\text{th},1}\cos\omega t+F_{\text{th},2}\sin\omega t=\frac{1}{2}\left(f_{\text{th}}\left(t\right)e^{i\omega t}+f_{\text{th}}^{*}e^{-i\omega t}\right).
\end{align*}
We do not write out these definitions explicitly for every quantity.
We will set $\omega=\omega_{0}$ below.

The driving force $F_{d}$ is produced by a series of operations acting
on $X$. First, the cantilever position $X$, monitored interferometrically,
produces a voltage signal on the photodetector $V_{pd}(t)=G_{pd}(P_{\text{cant}}(t)+P_{N,\text{int}}(t))$,
where $G_{pd}$ is the gain of the photodetector in volts per watts,
$P_{\text{cant}}(t)$ is the signal from the photodetector and $P_{N,\text{int}}(t)$
is the noise in units of optical power out of the photodetector (see
\ref{sub:CHSensitivity_InterferometerNoise}). As discussed in \ref{sub:CHExpSetup_CantileverFiberInterferometer},
the interferometer signal $P_{\text{cant}}(t)$ has frequency components
at the frequency $\omega$ of cantilever motion and each of its harmonics.
Because of the bandpass filter in the feedback circuit, we keep only
the fundamental component. Because the first harmonic of $P_{\text{cant}}$
(see Eq. \ref{eq:CHExpSetup_InterferometerExpansion}) is in phase
with the cantilever position $X(t)$, the complex amplitude $p_{\text{cant}}(t)$
can be written as 
\begin{align*}
p_{\text{cant}}\left(t\right) & =P_{\text{inc}}\left(4T_{f}\sqrt{R_{f}R_{c}}J_{1}\left(\frac{4\pi}{\lambda}U\left|x\right|\right)+2T_{f}^{2}R_{c}\epsilon U\left|x\right|\right)\frac{x}{\left|x\right|}\\
 & \equiv R\left(\left|x\right|\right)x.
\end{align*}
Here we have assumed that the cantilever is operated at the optimal
fringe position (see \ref{sub:CHExpSetup_CantileverFiberInterferometer}).
For a cantilever of length $l$ excited in its $m^{th}$ flexural
mode, the factor $U=U_{m}(z_{f}/l)$ converts the displacement $x$
of the tip to that at the position $z_{f}$ monitored by the interferometer
(see Eq. \ref{eq:CHTorsMagn_UmCantileverMode}).

Continuing along the phase-locked loop diagram in Fig. \ref{fig:CHExpSetup_CantileverMeasurementSchematic},
we pass from the photodetector to the lock-in amplifier. We combine
the lock-in amplifications factors together with the photodetector's
gain in the quantity $G_{pd}$. As discussed in \ref{sub:CHSensitivity_ElectronicsNoise},
we can ignore the input noise of all electronic components following
the photodetector. The lock-in produces an output signal $V_{\text{piezo}}(t)$
whose frequency $\omega$ is set by the lock-in's internal feedback
circuit in order to main a fixed phase $\phi_{c}$, set externally
by the user, relative to the lock-in's input $V_{pd}(t)$.%
\footnote{In principle there could be additional phase shifts in the detection
circuit which we do not discuss (e.g. the phase shift due to the band-pass
filter). However, these shifts just result in an overall shift to
$\phi_{c}$ which can be negated by our freedom to set the value of
$\phi_{c}$.%
} The lock-in output's amplitude $V_{0}$ is also set by the user.
We model this behavior of the lock-in as that of an ideal phase shifter
and limiter plus a noise term $V_{N,d}(t)$, writing the complex amplitude
$v_{\text{piezo}}$ of the lock-in output as 
\begin{equation}
v_{\text{piezo}}=V_{0}e^{i\phi_{c}}\frac{G_{pd}\left(R\left(\left|x\right|\right)x+p_{N,pd}\right)}{\left|G_{pd}\left(R\left(\left|x\right|\right)x+p_{N,pd}\right)\right|}+v_{N,d}.\label{eq:ChSensitivity_vPiezoComplexAmp}
\end{equation}
Denoting by $\beta_{\text{piezo}}$ the amount of force exerted on
the cantilever per applied voltage to the piezoelectric actuator,
the complex amplitude $f_{d}$ of the driving force is
\begin{equation}
f_{d}=\beta_{\text{piezo}}V_{0}e^{i\phi_{c}}\frac{R\left(\left|x\right|\right)x+p_{N,\text{int}}}{\left|R\left(\left|x\right|\right)x+p_{N,\text{int}}\right|}+\beta_{\text{piezo}}v_{N,d}\label{eq:CHSensitivity_FreqErrorfdCmplxDriveAmp}
\end{equation}

In terms of the complex amplitudes, the equation of motion (Eq. \ref{eq:CHSensitivity_FreqErrorEqOfMotion})
for the cantilever driven in a phase-locked loop becomes
\begin{equation}
2i\omega_{0}\frac{dx}{dt}-\left(\omega_{0}^{2}+2\omega_{0}\,\delta\omega\right)x+i\frac{\omega_{0}^{2}}{Q}x+\omega_{0}^{2}x=\frac{f_{d}+f_{\text{th}}}{m_{\textrm{eff}}}\label{eq:CHSensitivity_FreqErrorComplexAmpEqOfMotion}
\end{equation}
where $\delta\omega=\omega-\omega_{0}\ll\omega_{0}$. We have dropped
terms containing the factors $\delta\omega/Q$, $\delta\omega\, dx/dt$,
$Q^{-1}\, dx/dt$ and $d^{2}x/dt^{2}$, each of which we assume to
be negligible due to the resonator's high $Q$ and thus long ringdown
time. We write 
\begin{equation}
x=\left(X_{0}+\delta X_{0}\left(t\right)\right)e^{i\phi_{0}+i\phi\left(t\right)}.\label{eq:CHSensitivity_FreqErrorFluctuatingTerms}
\end{equation}
Here, $X_{0}$ and $\phi_{0}$ the real valued steady-state (independent
of time) amplitude and phase of the complex amplitude $x$, and $\delta X_{0}(t)$
and $\phi(t)$ are the real valued fluctuations about these steady-state
values. We choose to shift the zero of $t$ so that $\phi_{0}=0.$
To leading order, we can write $x(t)\approx X_{0}+\delta X_{0}(t)+iX_{0}\phi(t)$.
The steady-state solution is given by setting $\delta X_{0}(t)$,
$\phi(t)$ and all noise terms to zero. The steady-state solution
can be written as
\begin{equation}
X_{0}=e^{i\phi_{c}}\frac{\beta_{\text{piezo}}V_{0}}{m_{\text{eff}}\left(\frac{i\omega_{0}^{2}}{Q}-2\omega_{0}\,\delta\omega\right)}.\label{eq:ChSensitivity_SteadyStateSolution}
\end{equation}
The steady-state amplitude $X_{0}$ and frequency offset $\delta\omega$
are determined by specifying $V_{0}$ and $\phi_{c}$.

Ultimately, we are seeking an expression for the fluctuations of the
phase-locked loop frequency when the loop is operated to drive the
cantilever on resonance. These fluctuations of the loop's frequency
are directly related to the fluctuations $\phi(t)$ in the loop's
phase. We obtain an expression for $\phi(t)$ when the cantilever
is driven resonantly by solving the equation of motion, Eq. \ref{eq:CHSensitivity_FreqErrorComplexAmpEqOfMotion},
with the resonance condition $\delta\omega=0$ and the expansion $x(t)\approx X_{0}+\delta X_{0}(t)+iX_{0}\phi(t)$.

We now solve for $\phi(t)$. For $\delta\omega=0$, the steady-state
solution, Eq. \ref{eq:ChSensitivity_SteadyStateSolution}, requires
that $\phi_{c}=\pi/2$ and $X_{0}=V_{0}Q\beta_{\text{piezo}}/m_{\text{eff}}\omega_{0}^{2}$.
With these relations, we can rewrite $f_{d}$ (Eq. \ref{eq:CHSensitivity_FreqErrorfdCmplxDriveAmp})
as 
\begin{align}
f_{d} & =i\frac{m_{\text{eff}}\omega_{0}^{2}}{Q}X_{0}\frac{R\left(\left|x\right|\right)x+p_{N,\text{int}}}{\left|R\left(\left|x\right|\right)x+p_{N,\text{int}}\right|}+\beta_{\text{piezo}}v_{N,d}\nonumber \\
 & \approx i\frac{m_{\text{eff}}\omega_{0}^{2}}{Q}X_{0}\left(1+i\phi+\frac{p_{N,\text{int}}-p_{N,\text{int}}^{*}}{2R\left(X_{0}\right)X_{0}}\right)+\beta_{\text{piezo}}v_{N,d}\nonumber \\
 & \approx i\frac{m_{\text{eff}}\omega_{0}^{2}}{Q}X_{0}\left(1+i\phi-i\frac{P_{N,\text{int},2}}{R\left(X_{0}\right)X_{0}}\right)+\beta_{\text{piezo}}v_{N,d}.\label{eq:ChSensitivity_fDriveComplexAmp}
\end{align}
where we have used the fact that for any real $\Gamma$ and complex
$\delta$ such that$\left|\delta\right|\ll\Gamma,$ 
\[
\frac{\Gamma+\delta}{|\Gamma+\delta|}\approx1+\frac{\delta-\delta^{*}}{2\Gamma}.
\]
Plugging this result for $f_{d}$ as well as $\delta\omega=0$ and
$x(t)\approx X_{0}+\delta X_{0}(t)+iX_{0}\phi(t)$ into the equation
of motion (Eq. \ref{eq:CHSensitivity_FreqErrorComplexAmpEqOfMotion}),
we have
\begin{multline}
2\left(\frac{d\left(\delta X_{0}\right)}{dt}+iX_{0}\frac{d\phi}{dt}\right)+\frac{\omega_{0}}{Q}\left(X_{0}+\delta X_{0}+iX_{0}\phi\right)=\\
\frac{\omega_{0}X_{0}}{Q}\left(1+i\phi-i\frac{P_{N,\text{int},2}}{R\left(X_{0}\right)X_{0}}\right)-\frac{i}{m_{\text{eff}}\omega_{0}}\left(\beta_{\text{piezo}}v_{N,d}+f_{\text{th}}\right).\label{eq:CHSensitivity_FreqErrorFluctuatingEqOfMotion}
\end{multline}
Taking the imaginary part of Eq. \ref{eq:CHSensitivity_FreqErrorFluctuatingEqOfMotion}
yields the following equation describing time evolution of the phase
fluctuations $\phi(t)$:
\begin{equation}
2\frac{d\phi}{dt}=-\frac{\omega_{0}}{Q}\left(\frac{P_{N,\text{int},2}}{R\left(X_{0}\right)X_{0}}\right)-\frac{1}{m_{\text{eff}}\omega_{0}X_{0}}\left(\beta_{\text{piezo}}V_{N,d,1}+F_{\text{th},1}\right).\label{eq:CHSensitivity_FreqErrordPhasedt}
\end{equation}

To find the time dependence of $\phi(t)$ we employ the Fourier transform
\[
\phi\left(\xi\right)=\int_{-\infty}^{\infty}dt\,\phi\left(t\right)e^{-2\pi i\xi t}
\]
and its inverse
\[
\phi\left(t\right)=\int_{-\infty}^{\infty}d\xi\,\phi\left(\xi\right)e^{2\pi i\xi t}.
\]
Each of $P_{N,\text{int},2}$, $V_{N,d,1}$, and $F_{\text{th},1}$
are quadrature amplitudes of uncorrelated fluctuating quantities which
we assume have the properties of white noise. In particular, for quadrature
amplitudes $A_{i}$ and $B_{j}$ of $A(t)$ and $B(t)$, we assume
\begin{equation}
\left\langle A_{i}\left(\xi\right)\right\rangle =0\label{eq:CHSensitivity_QuadratureAvg}
\end{equation}
\begin{equation}
\left\langle A_{i}\left(\xi\right)A_{i}\left(\xi'\right)\right\rangle =S_{A}\delta\left(\xi+\xi'\right)\label{eq:CHSensitivity_QuadratureCorr}
\end{equation}
\begin{equation}
\left\langle A_{i}\left(\xi\right)B_{j}\left(\xi'\right)\right\rangle =0.\label{eq:CHSensitivity_CrossQuadrature}
\end{equation}
where $S_{A}$ is the power spectral density of $A$ and $\langle\ldots\rangle$
denotes averaging over the ensemble of possible $A(t)$ and $B(t)$.
Taking the Fourier transform of Eq. \ref{eq:CHSensitivity_FreqErrordPhasedt},
we find
\begin{align}
\phi\left(\xi\right) & =-\frac{1}{4\pi i\xi}\left(\frac{\omega_{0}}{Q}\left(\frac{P_{N,\text{int},2}\left(\xi\right)}{R\left(X_{0}\right)X_{0}}\right)+\frac{1}{m_{\text{eff}}\omega_{0}X_{0}}\left(\beta_{\text{piezo}}V_{N,d,1}\left(\xi\right)+F_{\text{th},1}\left(\xi\right)\right)\right)\label{eq:CHSensitivity_PhiFourierTransform}
\end{align}
from which it follows that $\langle\phi(\xi)\rangle=0$. Using Eqs.
\ref{eq:CHSensitivity_CrossQuadrature} and \ref{eq:CHSensitivity_PhiFourierTransform},
we have 
\begin{equation}
\left\langle \phi\left(\xi\right)\phi\left(\xi'\right)\right\rangle =\frac{1}{16\pi^{2}\xi^{2}}\delta\left(\xi+\xi'\right)\left(\left(\frac{\omega_{0}}{QR\left(X_{0}\right)X_{0}}\right)^{2}S_{P,\text{int}}+\left(\frac{1}{m_{\text{eff}}\omega_{0}X_{0}}\right)^{2}\left(\beta_{\text{piezo}}^{2}S_{N,V,d}+S_{F,\text{th}}\right)\right)\label{eq:CHSensitivity_FreqErrorPhaseFreqCorrelation}
\end{equation}
where $S_{P,\text{int}}$ is the power spectral densities of the photodiode
signal in units of optical power, $S_{N,V,d}$ is the voltage power
spectral density of the lock-in output, and $S_{F,\text{th}}$ is
the power spectral density of the thermal force acting on the cantilever
(see \ref{sub:CHSensitivity_ForceNoise}). We will use this expression
for $\langle\phi(\xi)\phi(\xi')\rangle$ to find the fluctuations
in the phase-locked loop frequency momentarily.

We now consider the significance of the voltage noise $S_{N,V,d}$
of the lock-in output. As discussed in \ref{sub:CHSensitivity_ElectronicsNoise},
the feedback loop is typically operated with the voltage output of
the lock-in at $1\,\text{V}$. In most cases, the cantilever tip is
driven on resonance to an RMS amplitude of approximately 
\[
X_{0}=\frac{1}{\sqrt{2}}\frac{1}{0.7}\frac{1.84\lambda}{4\pi}\approx230\,\text{nm}
\]
by this $1\,\text{V}$ drive. This drive corresponds to the peak of
$J_{1}$ in the first harmonic response of the interferometer given
in Eq. \ref{eq:CHExpSetup_1stHarmInterferometer}. The factor of 0.7
in the expression above represents the scaling factor $U_{m}(z_{f}/l)$
for a typical detection position $z_{f}$. . On resonance, Eq. \ref{eq:GCantileverTransferFunction}
gives the conversion between force $F$ and displacement $X_{0}$
as $F=kX_{0}/Q$. For the typical values of $k=10^{-3}\,\text{N/m}$
and $Q=10^{5}$, the force per voltage is 
\begin{align*}
\beta_{\text{piezo}} & =\frac{F}{\left(1\,\text{V}\right)}\\
 & =\frac{kX_{0}}{Q\left(1\,\text{V}\right)}\\
 & =2.3\,\text{fN/V}.
\end{align*}
Thus, the $1\,\text{\ensuremath{\mu}V}/\sqrt{\text{Hz}}$ magnitude
of $\sqrt{S_{N,V,d}}$ (see \ref{sub:CHSensitivity_ElectronicsNoise})
corresponds to a force noise of $2.3\,\text{zN}/\sqrt{\text{Hz}}$,
much less than the typical magnitude for the thermal force noise of
$1\,\text{aN}$ (see \ref{sub:CHSensitivity_ForceNoise}). Therefore,
we now drop the $S_{N,V,d}$ term.

We now calculate the diffusion of the phase. Since $\langle\phi(\xi)\rangle=0,$
it follows that $\langle\phi(t)\rangle=0$. The typical fluctuation
$\delta\phi(\tau_{M})$ in phase accumulated over time $\tau_{M}$
is given by
\begin{align}
\delta\phi^{2}\left(\tau_{M}\right) & =\left\langle \left(\phi\left(t+\tau_{M}\right)-\phi\left(t\right)\right)^{2}\right\rangle \nonumber \\
 & =\left\langle \left(\int_{-\infty}^{\infty}d\xi\,\phi\left(\xi\right)e^{2\pi i\xi t}\left(e^{2\pi i\xi\tau_{M}}-1\right)\right)\left(\int_{-\infty}^{\infty}d\xi'\,\phi\left(\xi'\right)e^{2\pi i\xi't}\left(e^{2\pi i\xi'\tau_{M}}-1\right)\right)\right\rangle \nonumber \\
 & =\frac{S_{\phi}}{16\pi^{2}}\int_{-\infty}^{\infty}d\xi\,\frac{2-e^{2\pi i\xi\tau_{M}}-e^{-2\pi i\xi\tau_{M}}}{\xi^{2}}\label{eq:CHSensitivity_FreqErrordPhiSquaredSetup}
\end{align}
where we have used 
\begin{equation}
S_{\phi}=\omega_{0}^{2}\left(\frac{S_{P,\text{int}}}{\left(QR\left(X_{0}\right)X_{0}\right)^{2}}+\frac{S_{F,\text{th}}}{\left(kX_{0}\right)^{2}}\right).\label{eq:CHSensitivity_Sphi}
\end{equation}

The integral in Eq. \ref{eq:CHSensitivity_FreqErrordPhiSquaredSetup}
can be evaluated with contour integration. We use the standard semi-circular
contours with their flat sides along the real axis. The first two
terms in the numerator of the integrand of Eq. \ref{eq:CHSensitivity_FreqErrordPhiSquaredSetup}
will have a negligible contribution along the semicircular arc enclosing
the upper half plane as its radius is taken to infinity, while the
third term is negligible for such an arc enclosing the lower half
plane (which introduces an extra minus sign in the first line of the
equation below). To evaluate the integral, we replace the denominator
of the integrand $\xi^{2}$ by $\xi^{2}+\delta^{2}$ so that there
are poles at $\xi=\pm i\delta$. We can then perform the integral
using the calculus of residues to obtain
\begin{align*}
\delta\phi^{2}\left(\tau_{M}\right) & =\frac{S_{\phi}}{16\pi^{2}}2\pi i\left(\text{Res}\left[\frac{2-e^{2\pi i\xi\tau_{M}}}{\left(\xi-i\delta\right)\left(\xi+i\delta\right)},\, i\delta\right]-\text{Res}\left[\frac{-e^{-2\pi i\xi\tau_{M}}}{\left(\xi-i\delta\right)\left(\xi+i\delta\right)},-i\delta\right]\right)\\
 & =\frac{S_{\phi}}{16\pi^{2}}2\pi i\left(\frac{2-e^{-2\pi\delta\tau_{M}}}{2i\delta}-\frac{-e^{-2\pi\delta\tau_{M}}}{-2i\delta}\right)\\
 & =\frac{S_{\phi}}{8\pi}\left(\frac{1-e^{-2\pi\delta\tau_{M}}}{\delta}\right)\\
 & \approx\frac{S_{\phi}}{4}\tau_{M}\\
 & =\frac{\tau_{M}}{4}\omega_{0}^{2}\left(\frac{S_{P,\text{int}}}{\left(QR\left(X_{0}\right)X_{0}\right)^{2}}+\frac{S_{F,\text{th}}}{\left(kX_{0}\right)^{2}}\right)
\end{align*}
where in the next to last step we have taken $\delta\rightarrow0$. 

We now find the relation between a fluctuation $\delta\phi(\tau_{M})$
of the resonator's phase over time $\tau_{M}$ and the corresponding
fluctuation $\delta f$ of the resonator's frequency. Over time $\tau_{M}$,
a resonator with period $T$ undergoes $N=\tau_{M}/T$ oscillations.
If the total fluctuation of the phase over $\tau_{M}$ is $\delta\phi$,
then the fluctuation of the phase per oscillation is $\delta\phi/N$.
Typically a resonator oscillates at a constant frequency and the phase
\[
\phi=\frac{2\pi}{T}t
\]
evolves linearly in time $t$. A fluctuation $\delta\phi/N$ in the
phase over one period can be thought of as a fluctuation 
\[
\delta T=\frac{T}{2\pi}\left(\frac{\delta\phi}{N}\right)=\frac{T^{2}}{2\pi\tau_{M}}\delta\phi
\]
in the period $T$. The corresponding fluctuation $\delta f$ in frequency
is given by 
\begin{align}
\delta f & =\frac{1}{T}-\frac{1}{T+\delta T}\nonumber \\
 & \approx\frac{\delta T}{T^{2}}\nonumber \\
 & =\frac{\delta\phi}{2\pi\tau_{M}}.\label{eq:CHSensitivity_Phase2Freq}
\end{align}
Using Eq. \ref{eq:CHSensitivity_Phase2Freq} to convert the phase
fluctuation $\delta\phi(\tau_{M})$ into the fluctuation $\sigma_{f,\text{cant}}$
of the phase-locked loop frequency over $\tau_{M}$, we find
\begin{align}
\sigma_{f,\text{cant}}^{2}\left(\tau_{M}\right) & =\frac{\delta\phi^{2}\left(\tau_{M}\right)}{\left(2\pi\tau_{M}\right)^{2}}\nonumber \\
 & =\frac{f_{0}^{2}}{4\tau_{M}}\left(\frac{S_{P,\text{int}}}{\left(QR\left(X_{0}\right)X_{0}\right)^{2}}+\frac{S_{F,\text{th}}}{\left(kX_{0}\right)^{2}}\right)\nonumber \\
 & =\frac{1}{4\tau_{M}}\left({\displaystyle \frac{f_{0}^{2}S_{P,\text{int}}}{Q^{2}P_{\text{inc}}^{2}\left(4T_{f}\sqrt{R_{f}R_{c}}J_{1}\left(\frac{4\pi}{\lambda}UX_{0}\right)+2T_{f}^{2}R_{c}\epsilon UX_{0}\right)^{2}}+{\displaystyle \frac{2f_{0}k_{B}T}{\pi QkX_{0}^{2}}}}\right)\label{eq:CHSensitivity_FreqErrorFullExpression}
\end{align}
where we have used the abbreviation $U=U_{m}(z_{f}/l)$. We reiterate
now that $\sigma_{f,\text{cant}}$ is the typical fluctuation of the
actual frequency at which the cantilever oscillates. Throughout the
derivation, the cantilever's resonant frequency $f_{0}$ was assumed
to be constant. The actual frequency of motion of the cantilever shows
slight deviations from $f_{0}$ because the cantilever is driven by
fluctuating forces (both the thermal noise force and the force created
by noise in the phase-locked loop circuit being fed back to the piezoelectric
actuator).

\subsection{Frequency fluctuations $\sigma_{f,\text{int}}$ added to the cantilever
signal by detector noise}

So far, we have considered only the actual fluctuations of the phase
of the cantilever's motion, either caused by noise forces intrinsic
to the cantilever or transduced by the phase-locked loop circuit from
noise in the cantilever position detection. Noise in the detection
of the cantilever position can also lead to error in the inferred
value of the frequency. 

Typically, the cantilever frequency is measured by feeding the lock-in
output signal $V_{\text{piezo}}(t)$ into a frequency counter. Following
similar steps to those used deriving Eq. \ref{eq:ChSensitivity_fDriveComplexAmp},
the complex amplitude $v_{\text{piezo}}$ (see Eq. \ref{eq:ChSensitivity_vPiezoComplexAmp})
may be written as
\begin{align*}
v_{\text{piezo}} & =iV_{0}\frac{G_{pd}\left(R\left(\left|x\right|\right)x+p_{N,\text{int}}\right)}{\left|G_{pd}\left(R\left(\left|x\right|\right)x+p_{N,\text{int}}\right)\right|}\\
 & \approx iV_{0}\left(1+i\phi-i\frac{P_{N,\text{int},2}}{R\left(X_{0}\right)X_{0}}\right)
\end{align*}
From this expression, it can be seen that the fluctuating component
of the lock-in output's phase is the sum of the fluctuating phase
$\phi$ of the cantilever motion and another term, $-P_{N,\text{int},2}/R(X_{0})X_{0}$,
caused by the noise of the photodetector.%
\footnote{Following the observations of \ref{sub:CHSensitivity_ForceNoise},
we treat the noise of other electronic components, including the output
noise of the lock-in and the input noise of the frequency counter,
to be negligible compared to the noise of the photodetector.%
}

Using the approach of Ref. \citealp{yurke1995theoryof}, we take the
total measured fluctuating phase $\phi_{\text{tot}}$ to be a filtered
form of the lock-in output's fluctuating phase $\phi-P_{N,\text{int},2}/R(X_{0})X_{0}$:
\begin{align*}
\phi_{\text{tot}}\left(t\right) & =\frac{e^{-t/\tau_{F}}}{\tau_{F}}\int_{-\infty}^{t}dt'\, e^{t'/\tau_{F}}\left(\phi\left(t'\right)-\frac{P_{N,\text{int},2}\left(t'\right)}{R\left(X_{0}\right)X_{0}}\right)\\
 & =\frac{e^{-t/\tau_{F}}}{\tau_{F}}\int_{-\infty}^{\infty}d\xi\int_{-\infty}^{t}dt'\, e^{t'/\tau_{F}}e^{2\pi i\xi t'}\left(\phi\left(\xi\right)-\frac{P_{N,\text{int},2}\left(\xi\right)}{R\left(X_{0}\right)X_{0}}\right)\\
 & =\frac{e^{-t/\tau_{F}}}{\tau_{F}}\int_{-\infty}^{\infty}d\xi\frac{e^{t/\tau_{F}}e^{2\pi i\xi t}}{2\pi i\xi+1/\tau_{F}}\left(\phi\left(\xi\right)-\frac{P_{N,\text{int},2}\left(\xi\right)}{R\left(X_{0}\right)X_{0}}\right)\\
 & =\int_{-\infty}^{\infty}d\xi\frac{e^{2\pi i\xi t}}{1+2\pi i\xi\tau_{F}}\left(\phi\left(\xi\right)-\frac{P_{N,\text{int},2}\left(\xi\right)}{R\left(X_{0}\right)X_{0}}\right)
\end{align*}
where $\tau_{F}$ is the time constant of the filter. The average
measured phase fluctuation $\langle\phi_{\text{tot}}(t)\rangle=\langle\phi_{\text{tot}}(\xi)\rangle=0$
is zero because $\langle\phi(t)\rangle=\langle P_{N,\text{int},2}(\xi)\rangle=0$.
The total diffusion of the measured phase is 
\begin{align*}
\delta\phi_{\text{tot}}^{2}\left(\tau_{M}\right) & =\left\langle \left(\phi_{\text{tot}}\left(t+\tau_{M}\right)-\phi_{\text{tot}}\left(t\right)\right)^{2}\right\rangle \\
 & =\int_{-\infty}^{\infty}d\xi\int_{-\infty}^{\infty}d\xi'\,\left(\frac{e^{2\pi i\xi t}\left(e^{2\pi i\xi\tau_{M}}-1\right)}{1+2\pi i\xi\tau_{F}}\right)\left(\frac{e^{2\pi i\xi't}\left(e^{2\pi i\xi'\tau_{M}}-1\right)}{1+2\pi i\xi'\tau_{F}}\right)\\
 & \phantom{=\int_{-\infty}^{\infty}d\xi\int_{-\infty}^{\infty}d\xi'\,}\times\left\langle \left(\phi\left(\xi\right)-\frac{P_{N,\text{int},2}\left(\xi\right)}{R\left(X_{0}\right)X_{0}}\right)\left(\phi\left(\xi'\right)-\frac{P_{N,\text{int},2}\left(\xi'\right)}{R\left(X_{0}\right)X_{0}}\right)\right\rangle .
\end{align*}
The phase correlation $\langle\phi(\xi)\phi(\xi')\rangle$ was given
in Eq. \ref{eq:CHSensitivity_FreqErrorPhaseFreqCorrelation} and can
be abbreviated as 
\[
\left\langle \phi\left(\xi\right)\phi\left(\xi'\right)\right\rangle =\frac{1}{16\pi^{2}}S_{\phi}\left(\xi\right)\delta\left(\xi+\xi'\right).
\]
Following our assumption, Eq. \ref{eq:CHSensitivity_QuadratureCorr},
that each noise source is white, we can write 
\[
\langle P_{N,\text{int},2}(\xi)P_{N,\text{int},2}(\xi')\rangle=S_{P,\text{int}}(\xi)\delta(\xi+\xi').
\]
Using the expression for $\phi(\xi)$ given in Eq. \ref{eq:CHSensitivity_PhiFourierTransform},
the cross-correlation term can be evaluated as
\begin{align*}
\left\langle \phi\left(\xi\right)\frac{P_{N,\text{int},2}\left(\xi'\right)}{R\left(X_{0}\right)X_{0}}\right\rangle  & =\left\langle -\frac{1}{4\pi i\xi}\left(\frac{\omega_{0}}{Q}\left(\frac{P_{N,\text{int},2}\left(\xi\right)}{R\left(X_{0}\right)X_{0}}\right)+\frac{\beta_{\text{piezo}}V_{N,d,1}\left(\xi\right)+F_{\text{th},1}\left(\xi\right)}{m_{\text{eff}}\omega_{0}X_{0}}\right)\frac{P_{N,\text{int},2}\left(\xi'\right)}{R\left(X_{0}\right)X_{0}}\right\rangle \\
 & =-\frac{1}{4\pi i\xi}\frac{\omega_{0}}{Q}\frac{S_{P,\text{int}}\left(\xi\right)\delta\left(\xi+\xi'\right)}{\left(R\left(X_{0}\right)X_{0}\right)^{2}}
\end{align*}
With these results, the measured phase diffusion is
\begin{align*}
\delta\phi_{\text{tot}}^{2}\left(\tau_{M}\right) & =\int_{-\infty}^{\infty}d\xi\int_{-\infty}^{\infty}d\xi'\,\frac{-4e^{2\pi i\left(\xi+\xi'\right)\left(t+\tau_{M}/2\right)}\sin\left(\pi\xi\tau_{M}\right)\sin\left(\pi\xi'\tau_{m}\right)}{\left(1+2\pi i\xi\tau_{F}\right)\left(1+2\pi i\xi'\tau_{F}\right)}\times\ldots\\
 & \phantom{=\int_{-\infty}^{\infty}d\xi\int_{-\infty}^{\infty}d\xi'\,}\times\Bigg(\frac{1}{16\pi^{2}}\frac{\delta\left(\xi+\xi'\right)}{\xi^{2}}S_{\phi}+\frac{S_{P,\text{int}}}{\left(R\left(X_{0}\right)X_{0}\right)^{2}}\delta\left(\xi+\xi'\right)+\ldots\\
 & \phantom{=\int_{-\infty}^{\infty}d\xi\int_{-\infty}^{\infty}d\xi'\,\times\Bigg(}-\frac{1}{2}\frac{1}{2\pi i\xi}\frac{\omega_{0}}{Q}\frac{S_{P,\text{int}}}{R^{2}\left(X_{0}\right)}\delta\left(\xi+\xi'\right)\Bigg)\\
 & =\int_{-\infty}^{\infty}d\xi\,\frac{4\sin^{2}\left(\pi\xi\tau_{M}\right)}{1+4\pi^{2}\xi^{2}\tau_{F}^{2}}\left(\frac{S_{\phi}}{16\pi^{2}\xi^{2}}+\frac{S_{P,\text{int}}}{\left(R\left(X_{0}\right)X_{0}\right)^{2}}-\frac{1}{4\pi i\xi}\frac{\omega_{0}}{Q}\frac{S_{P,\text{int}}}{\left(R\left(X_{0}\right)X_{0}\right)^{2}}\right)\\
 & =\frac{S_{\phi}}{4}\int_{-\infty}^{\infty}d\xi\,\frac{1}{4\pi^{2}\xi^{2}}\frac{4\sin^{2}\left(\pi\xi\tau_{M}\right)}{1+4\pi^{2}\xi^{2}\tau_{F}^{2}}+\frac{S_{P,\text{int}}}{\left(R\left(X_{0}\right)X_{0}\right)^{2}}\int_{-\infty}^{\infty}d\xi\,\frac{4\sin^{2}\left(\pi\xi\tau_{M}\right)}{1+4\pi^{2}\xi^{2}\tau_{F}^{2}}\\
 & =\frac{\tau_{M}S_{\phi}}{2\pi}\int_{-\infty}^{\infty}dx\,\frac{1}{x^{2}}\frac{\sin^{2}\left(x/2\right)}{1+\left(x/\alpha\right)^{2}}+\frac{2S_{P,\text{int}}}{\pi\tau_{M}\left(R\left(X_{0}\right)X_{0}\right)^{2}}\int_{-\infty}^{\infty}dx\,\frac{\sin^{2}\left(x/2\right)}{1+\left(x/\alpha\right)^{2}}
\end{align*}
where we have used 
\[
\alpha=\tau_{M}/\tau_{F}.
\]
We dropped the integral of the cross-term (the $1/\xi$ term) because
it is anti-symmetric in $\xi$. 

Evaluating the integrals in the expression for $\delta\phi_{\text{tot}}^{2}(\tau_{M})$
with the help of Eqs. \ref{eq:AppMath_FilterInt1} and \ref{eq:AppMath_FilterInt2},
we find
\[
\delta\phi_{\text{tot}}^{2}\left(\tau_{M}\right)=\frac{S_{\phi}}{4}\left(\tau_{M}-\tau_{F}\left(1-\exp\left(-\frac{\tau_{M}}{\tau_{F}}\right)\right)\right)+\frac{S_{P,\text{int}}}{\left(R\left(X_{0}\right)X_{0}\right)^{2}}\frac{1}{\tau_{F}}\left(1-\exp\left(-\frac{\tau_{M}}{\tau_{F}}\right)\right).
\]
Using Eq. \ref{eq:CHSensitivity_Phase2Freq} to convert phase to frequency,
we find
\begin{align}
\sigma_{f,\text{tot}}^{2}\left(\tau_{M}\right) & =\frac{\delta_{\text{tot}}^{2}\left(\tau_{M}\right)}{\left(2\pi\tau_{M}\right)^{2}}\nonumber \\
 & =\frac{S_{\phi}}{16\pi^{2}\tau_{M}}\left(1-\frac{\tau_{F}}{\tau_{M}}\left(1-e^{-\frac{\tau_{M}}{\tau_{F}}}\right)\right)+\frac{S_{P,\text{int}}}{4\pi^{2}\left(R\left(X_{0}\right)X_{0}\right)^{2}}\frac{1}{\tau_{M}^{2}\tau_{F}}\left(1-e^{-\frac{\tau_{M}}{\tau_{F}}}\right).\label{eq:CHSensitivity_SigmaFreqFiltered}
\end{align}
This expression may also be written explicitly in terms of the different
noise spectral densities as
\begin{align}
\sigma_{f,\text{tot}}^{2}\left(\tau_{M}\right) & =\frac{f_{0}^{2}}{4\tau_{M}}\frac{S_{P,\text{int}}}{Q^{2}\left(R\left(X_{0}\right)X_{0}\right)^{2}}\left(1-\frac{\tau_{F}}{\tau_{M}}\left(1-\exp\left(-\frac{\tau_{M}}{\tau_{F}}\right)\right)\right)\nonumber \\
 & \phantom{=}+\frac{f_{0}^{2}}{4\tau_{M}}\frac{S_{F,\text{th}}}{\left(kX_{0}\right)^{2}}\left(1-\frac{\tau_{F}}{\tau_{M}}\left(1-\exp\left(-\frac{\tau_{M}}{\tau_{F}}\right)\right)\right)\nonumber \\
 & \phantom{=}+\frac{1}{4\pi^{2}}\frac{1}{\tau_{M}^{2}\tau_{F}}\frac{S_{P,\text{int}}}{\left(R\left(X_{0}\right)X_{0}\right)^{2}}\left(1-\exp\left(-\frac{\tau_{M}}{\tau_{F}}\right)\right).\label{eq:ChSensitivity_SigmaFTotal}
\end{align}

In the limit of long measurement times $\tau_{M}\gg\tau_{F}$, the
fluctuations $\sigma_{f,\text{tot}}$ in the observed frequency may
be written as
\[
\sigma_{f,\text{tot}}^{2}=\sigma_{f,\text{cant}}^{2}+\sigma_{f,\text{int}}^{2}
\]
with $\sigma_{f,\text{cant}}^{2}$ the actual fluctuations of the
resonator given in Eq. \ref{eq:CHSensitivity_FreqErrorFullExpression}
and
\[
\sigma_{f,\text{int}}^{2}=\frac{S_{P,\text{int}}}{4\pi^{2}\left(R\left(X_{0}\right)X_{0}\right)^{2}}\frac{1}{\tau_{M}^{2}\tau_{F}}
\]
the additional fluctuations in the detected frequency due to the noise
in the interferometer signal. Since $\sigma_{f,\text{cant}}^{2}\propto\tau_{M}^{-1}$
and $\sigma_{f,\text{int}}^{2}\propto\tau_{M}^{-2}$, there is a transitional
value $\tau_{M}^{*}$ beyond which $\sigma_{f,\text{tot}}^{2}$ is
dominated by the frequency fluctuations $\sigma_{f,\text{cant}}$
of the cantilever. For shorter times $\tau_{M}<\tau_{M}^{*}$, the
fluctuations $\sigma_{f,\text{int}}$ added by the detector dominate. 

Optimal filtering involves fixing $\alpha_{F}=\tau_{M}/\tau_{F}$
with $\alpha_{F}\apprge1$. Using this condition to eliminate $\tau_{F}$,
we obtain
\begin{align}
\sigma_{f,\text{tot}}^{2}\left(\tau_{M}\right) & =\frac{f_{0}^{2}}{4\tau_{M}}\frac{S_{P,\text{int}}}{Q^{2}\left(R\left(X_{0}\right)X_{0}\right)^{2}}\left(1-\frac{1}{\alpha_{F}}\left(1-\exp\left(-\alpha_{F}\right)\right)\right)\nonumber \\
 & \phantom{=}+\frac{f_{0}^{2}}{4\tau_{M}}\frac{S_{F,\text{th}}}{\left(kX_{0}\right)^{2}}\left(1-\frac{1}{\alpha_{F}}\left(1-\exp\left(-\alpha_{F}\right)\right)\right)\nonumber \\
 & \phantom{=}+\frac{1}{4\pi^{2}}\frac{\alpha_{F}}{\tau_{M}^{3}}\frac{S_{P,\text{int}}}{\left(R\left(X_{0}\right)X_{0}\right)^{2}}\left(1-\exp\left(-\alpha_{F}\right)\right).\label{eq:ChSensitivity_TotFreqUncertainty}
\end{align}
For optimal filtering, it remains the case that $\sigma_{f,\text{cant}}^{2}\propto\tau_{M}^{-1}$,
but now $\sigma_{f,\text{int}}^{2}\propto\tau_{M}^{-3}$. In either
case, we see that, by measuring for sufficiently long times $\tau_{M}$,
the fluctuations of the cantilever's phase can be made the dominant
source of fluctuations in the detected frequency.

\section{\label{sec:CHSensitivity_OptimalCantDimensions}Optimal cantilever
dimensions for measuring persistent currents}

We now comment briefly on the choice of adjustable parameters for
the persistent current measurements. When designing samples, we choose
parameters that maximize the sensitivity defined as 
\begin{equation}
\mathcal{S}_{pc}=\frac{\Delta f_{pc}}{\sqrt{(\sigma_{f,\text{tot}}(\tau_{M}))^{2}\tau_{M}}}\label{eq:ChSensitivity_Sensitivity}
\end{equation}
where $\Delta f_{pc}$ is the cantilever frequency shift due to the
persistent current given in Eq. \ref{eq:CHTorsMagn_FiniteAmpFreqShift}
and $\sigma_{f,\text{tot}}(\tau_{M})$ is the uncertainty in a frequency
measurement over time $\tau_{M}$ given in Eq. \ref{eq:ChSensitivity_SigmaFTotal}.
Because $\Delta f_{pc}$ depends non-linearly on several sample parameters
and $\sigma_{f,\text{tot}}(\tau_{M})$ trades off between two different
sources of uncertainty, determining the optimal sample parameters
is a complicated task. We do not have the space here to give this
topic full justice but will outline some of its most important facets.

We begin by rewriting results from previous chapters in forms convenient
for the present analysis. Eq. \ref{eq:CHTorsMagn_FiniteAmpFreqShift}
gave the frequency shift $\Delta f_{pc}$ due to the persistent current
in a ring mounted at the cantilever tip as 
\begin{equation}
\Delta f_{pc}=N^{\sigma}\frac{f_{0}}{2k}\frac{2\pi I_{1}}{\phi_{0}}\left(\pi r^{2}B\cos\theta_{0}\frac{\alpha}{l}\right)^{2}\mathrm{jinc}\left(2\pi\frac{\pi r^{2}B}{\phi_{0}}\cos\theta_{0}\frac{\alpha}{l}x_{\text{\ensuremath{\max}}}\right).\label{eq:CHSensitivity_OptDimDeltaFreqPC}
\end{equation}
Here we assume that the first harmonic $p=1$ of the current is measured.
Using Eq. \ref{eq:CHPCTh_IIFiniteTZSO} and assuming strong spin-orbit
scattering, we can write the magnitude $I_{1}$ of the first harmonic
of the typical persistent current%
\footnote{As discussed in \ref{sub:CHPCTh_AverageCurrent}, other persistent
current mechanisms, such the average current due to either interacting
or non-interacting electrons, have a similar exponential dependence
on $r$, $D$, and $T$, with the interacting electron case also having
the same $D/r^{2}$ prefactor.%
} as 
\begin{equation}
I_{1}\approx0.028\frac{eD}{r^{2}}\exp\left(-\frac{4k_{B}Tr^{2}}{\hbar D}\right)\label{eq:CHSensitivity_OptDimI1}
\end{equation}
where $r=L/2\pi$ is the radius of the ring. The number of rings on
the cantilever is given by $N$. The exponent $\sigma$ determines
how the total persistent current scales with $N$. When the sign of
the persistent current varies randomly from ring to ring, $\sigma=1/2$,
and, when all the rings have a persistent current of the same sign,
$\sigma=1$. The maximum value of $N$ for a given cantilever scales%
\footnote{If a sufficiently large section of the cantilever is covered with
rings, the cantilever mode shape factor $\alpha$ defined in Eq. \ref{eq:CHTorsMagn_Alpha}
may no longer be constant. It is then necessary to integrate over
the portion $C$ of the cantilever covered with rings as $(\Delta f_{pc})_{C}=\int_{C}dz\,\Delta f_{pc}(\alpha(z))/l_{C}$
where $z$ is the distance from the cantilever base and $l_{C}$ is
the length of the cantilever covered with rings. We ignore this mode
shape dependence in our analysis. For the fundamental flexural mode,
the factor $\alpha$ changes by only $\sim10\%$ over the $\sim40\%$
of the cantilever length closest to the tip.%
} with the cantilever width $w$ and length $l$ and ring radius $r$
as $N\propto wl/r^{2}$. The other parameters included in Eqs. \ref{eq:CHSensitivity_OptDimDeltaFreqPC}
and \ref{eq:CHSensitivity_OptDimI1} represent the same quantities
as they did in Eqs. \ref{eq:CHPCTh_IIFiniteTZSO} and \ref{eq:CHTorsMagn_FiniteAmpFreqShift}.

In the previous section, it was found that the fluctuations $\sigma_{f,\text{tot}}$
in the measured frequency were made up of two contributions, the fluctuations
$\sigma_{f,\text{cant}}$ of the cantilever frequency and additional
fluctuations $\sigma_{f,\text{int}}$ added to the measurement by
the detector. In the limit that the thermal noise contribution to
$\sigma_{f,\text{tot}}$ is negligible ($S_{F,\text{th}}\approx0$),
the ratio between these two contributions is
\[
\frac{\sigma_{f,\text{cant}}}{\sigma_{f,pd}}=\pi^{2}\tau_{M}\tau_{F}\frac{f_{0}^{2}}{Q^{2}}\left(\frac{1}{1-\exp\left(-\frac{\tau_{M}}{\tau_{F}}\right)}-\frac{\tau_{F}}{\tau_{M}}\right).
\]
In the limit of $\tau_{M}/\tau_{F}\gg1$, this ratio can be written
as 
\[
\frac{\sigma_{f,\text{cant}}}{\sigma_{f,\text{int}}}=\frac{\tau_{M}\tau_{F}}{\tau^{2}}
\]
where $\tau=Q/\pi f_{0}$ is the cantilever ringdown time. When $\tau_{M},\tau_{F}\gg\tau$,
the uncertainty $\sigma_{f,\text{int}}$ added to the measurement
by the detector is negligible in comparison to $\sigma_{f,\text{cant}}$. 

In principle, the contribution $\sigma_{f,\text{int}}$ can be made
a negligible contribution $\sigma_{f,\text{tot}}$ for any set of
sample parameters simply by measuring for a sufficiently long time
$\tau_{M}$. However, in the derivation of $\sigma_{f,\text{tot}}$
in the previous section, it was assumed that the cantilever resonant
frequency $f_{0}$ was constant in time. In practice, the resonant
frequency $f_{0}$ drifts slowly over time. When the measurement time
$\tau_{M}$ is sufficiently long, this random drift leads to an increase
in the total measured frequency uncertainty $\sigma_{f,\text{tot}}$.
We found that this frequency drift led to an optimal measurement time
of $\tau_{M}\approx5\,\text{s}$. For the samples listed in Table
\ref{tab:ChData_CLs}, the ringdown times $\tau\sim20\,\text{s}$
were greater than this optimal value of $\tau_{M}$, so $\sigma_{f,\text{int}}$
could not be neglected.

We now use the expressions reviewed so far in this section to write
the sensitivity $\mathcal{S}_{pc}$ in terms of parameters which can
be controlled during sample fabrication and during measurement. In
order to discuss the optimal choice of cantilever dimensions, we use
the relations given in Section \ref{sec:CHTorsMagn_CantileverSHO}
to write the cantilever frequency and spring constant in terms of
the cantilever dimensions as 
\begin{equation}
f_{0}=\beta_{m}\sqrt{\frac{E_{Y}}{12\rho}}\frac{t}{l^{2}}\label{eq:CHSensitivity_OptDimResFreq}
\end{equation}
\begin{equation}
k=\frac{\beta_{m}^{2}}{48}E_{Y}\frac{wt^{3}}{l^{3}}\label{eq:CHSensitivity_OptDimSpringK}
\end{equation}
where $E_{Y}$ and $\rho$ are the cantilever material's Young's modulus
and density and $\beta_{m}$ was defined implicitly by Eq. \ref{eq:CHTorsMagn_BetaMDef}
for cantilever mode $m$. With these relations, we find
\begin{equation}
\mathcal{S}_{pc}=A_{\mathcal{S}}\frac{\exp\left(-B_{\mathcal{S}}r^{2}T/D\right)r^{2-2\sigma}\left(\alpha B\cos\theta_{0}\right)^{2}\text{jinc}\left(2\pi r^{2}\alpha B\cos\theta_{0}x_{\max}/l\right)}{\sqrt{{\displaystyle \frac{w^{2-2\sigma}S_{P,\text{int}}}{R^{2}\left(x_{\max}\right)x_{\max}^{2}}\left(C_{\mathcal{S}}\frac{t^{6}}{Q^{2}l^{2+2\sigma}}+D_{\mathcal{S}}\frac{t^{4}l^{2-2\sigma}}{\tau_{M}\tau_{F}}\right)+E_{\mathcal{S}}w^{1-2\sigma}t^{2}l^{3-2\sigma}\frac{T}{Qx_{\max}^{2}}}}}\label{eq:CHSensitivity_SensitivityFull}
\end{equation}
where we have grouped constants into the terms $A_{\mathcal{S}}=0.020\, e$,
$B_{\mathcal{S}}=4k_{B}/\hbar$, $C_{\mathcal{S}}=(\beta_{m}^{2}E_{Y}/48)^{2}$,
$D_{\mathcal{S}}=(\beta_{m}^{2}E_{Y}\rho)/(192\pi^{2})$ and $E_{\mathcal{S}}=(k_{B}\beta_{m}\sqrt{E_{Y}\rho})/(4\pi\sqrt{3})$.
Recall that $R(x)$ was defined above by 
\[
R\left(x_{\max}\right)x_{\max}=P_{\text{inc}}\left(4T_{f}\sqrt{R_{f}R_{c}}J_{1}\left(\frac{4\pi}{\lambda}Ux_{\max}\right)+2T_{f}^{2}R_{c}\epsilon Ux_{\max}\right).
\]
For simplicity, we have assumed that $\tau_{F}/\tau_{M}\gg1$. The
symbol $x_{\max}$ has been used to denote amplitude of the cantilever
tip.%
\footnote{The parameter $x_{\max}$ is equivalent to $X_{0}$ from the previous
section. We changed notation in the previous section in order to follow
the convention of lower-case symbols for complex quantities and upper-case
symbols for the real-valued equivalents.%
}

We now consider the choice of the parameters listed in Eq. \ref{eq:CHSensitivity_SensitivityFull}
that maximizes the sensitivity $\mathcal{S}_{pc}$. The dependence
of $\mathcal{S}_{pc}$ on a few parameters is quite straightforward.
Under all conditions, the sensitivity is improved by increasing the
diffusion constant $D$ and the mechanical quality factor $Q$ and
decreasing the temperature $T$ and the cantilever thickness $t$.
However, when the frequency uncertainty is dominated by the detector
noise $\sigma_{f,\text{int}}$, $\mathcal{S}_{pc}$ is not affected
by $Q$. In this case, the sensitivity is improved by increasing the
measurement time $\tau_{M}$ until another noise source becomes dominant
(see the discussion above about the effect of resonant frequency drift
on the frequency uncertainty).

While in the previous paragraph we discussed $Q$ and $t$ as independent
variables, it is important to note that $Q$ is not necessarily an
independent parameter. Over the course of the persistent current experiment,
samples have been fabricated two times each with $t\sim110\,\text{nm}$
and $t\sim340\,\text{nm}$. For this limited number of fabrication
runs, it has been observed that $Q_{340}/Q_{110}\sim4-7$. Thus it
appears that $Q\propto t^{\delta}$ for $\delta>1$. It has previously
been reported that $\delta\approx1$ in silicon-nitride cantilevers
\citep{yasumura2000quality}. In the same study, $Q$ was found to
depend only weakly on cantilever length in single crystal silicon
cantilevers similar to the ones discussed in this text. As long as
$\delta\leq2$, the sensitivity is always improved by making $t$
as small as possible, especially when the uncertainty $\sigma_{f,\text{tot}}$
is limited by the interferometer noise $S_{P,\text{int}}$. We also
note that $Q$ typically decreases with $T$. This dependence reinforces
the trend noted above of increasing $\mathcal{S}_{pc}$ with decreasing
$T$.

The importance of the interferometer parameters $P_{\text{inc}}$,
$R_{f}$, and $R_{c}$ to $\mathcal{S}_{pc}$ depends on the nature
of the leading contribution to $S_{P,\text{int}}$. When the interferometer
noise $S_{P,\text{int}}$ is dominated by shot noise or the detector's
internal noise, the sensitivity increases with incident laser power
$P_{\text{inc}}$ until another noise contribution becomes dominant
or the laser begins to heat the cantilever. If $S_{P,\text{int}}$
is dominated by the laser's technical noise, either intensity or wavelength
fluctuations, the sensitivity is independent of $P_{\text{inc}}$.
Also when $S_{P,\text{int}}$ is dominant, the sensitivity depends
weakly on $R_{c}$ and $R_{f}$, generally increasing with $R_{c}$
and peaking for $R_{f}\approx.4$. The exact dependence of $\mathcal{S}_{pc}$
on $R_{c}$ and $R_{f}$ is contingent upon the dominant contribution
to $S_{P,\text{int}}$. Reflectivities of $R_{c}=.3$ and $R_{f}=.4$
achieve a factor of $\sim21$ improvement over $R_{c}=.01$ and $R_{f}=.01$
when the photodetector noise dominates, and a factor of $\sim4$ improvement
when relative intensity noise or shot noise dominates. No improvement
is seen when $S_{P,\text{int}}$ is dominated by wavelength fluctuations.
These factors of improvement assume that the same source of noise
remains dominant. Once one contribution to the noise is reduced to
the same level as another source, further improvements will result
in diminishing returns for $\mathcal{S}_{pc}$.

The choice of optimal cantilever width $w$ depends upon the nature
of the persistent current sample through the exponent $\sigma$. For
a single ring sample, we can set $\sigma=0$. Then the sensitivity
increases as the width decreased for all noise sources. For a current
of random sign, we have $\sigma=1/2$. For this value of $\sigma$,
$\mathcal{S}_{pc}$ increases with decreasing $w$ while the interferometer
noise dominates and then is independent of $w$ once the thermal force
noise dominates. For a current with a well defined sign, $\sigma=1$.
In this case $\mathcal{S}_{pc}$ increases with increasing $w$ until
the uncertainty due the interferometer noise dominates. Once it does,
$\mathcal{S}_{pc}$ becomes independent of $w$.

The dependence of $\mathcal{S}_{pc}$ on the remaining experimental
parameters is complicated by their appearance in the argument, $2\pi r^{2}\alpha B\cos\theta_{0}x_{\max}/l$,
of the $\text{jinc}$ function in Eq. \ref{eq:CHSensitivity_SensitivityFull}.
The function $\text{jinc}(x)$ is peaked at $x=0$ and drops to and
remains below $\sim10\%$ of its peak value once $x\apprge1/2$ (see
Fig. \ref{fig:CHTorsMagn_JincPlot}). Thus, $r$, $\alpha B\cos\theta_{0}$,
$x_{\max}$, and $l$ must be chosen so that the combination $2\pi r^{2}\alpha B\cos\theta_{0}x_{\max}/l$
remains less than $1/2$. 

Disregarding the $\text{jinc}$ factor, $\mathcal{S}_{pc}$ scales
as $(\alpha B\cos\theta_{0})^{2}$. This quadratic dependence of $\mathcal{S}_{pc}$
on $\alpha B\cos\theta_{0}$ is stronger than its dependence on $r^{2}$,
$x_{\max}$, or $l$ for all values of $\sigma$ and all noise limits.
Thus $\mathcal{S}_{pc}$ is always increased by scaling down $r^{2}$,
$x_{\max}$, or $l$ so that $\alpha B\cos\theta_{0}$ may be increased.
The maximum magnetic field $B$ is generally set by the specifications
of the equipment.%
\footnote{Our magnet was rated to produce fields up to $9\,\text{T}$. The maximum
value of $B$ used in the experiments discussed here was $8.4\,\text{T}$.
The magnet quenched multiple times above $8.4\,\text{T}$ during early
cooldowns. It was decided that it was not worth risking further quenching
to measure the persistent current between $8.4\,\text{T}$ and $9\,\text{T}$.
In principle, we could have continued measuring persistent currents
at larger magnetic field strengths.%
}

The cantilever mode factor $\alpha$ is maximal at the cantilever
tip and increases with the order of the flexural mode $m$. Thus optimal
sensitivity is achieved when persistent current rings are placed near
the cantilever tip and the highest possible order of flexural mode
is used. For arrays of rings, the fraction of the cantilever length
covered with rings must be decreased with increasing order of flexural
mode. Otherwise, the variation of $\alpha$ across the cantilever
length complicates the analysis of the frequency shift signal. The
portion of the cantilever over which $\alpha$ is constant scales
as $\alpha\propto\sim2^{m}$ while $\alpha$ at the cantilever tip
scales roughly as $\alpha\propto\sim2.8m$. This dependence leads
to the third order mode being optimal for $\sigma=1$ and the sixth
order mode being optimal for $\sigma=1/2$. 

In addition to its appearance in the expression for $\mathcal{S}_{pc}$,
the angle $\theta_{0}$ determines the degree of correlation of the
persistent current oscillations in applied magnetic field. The magnetic
field frequency $\beta_{1}=\pi r^{2}\sin\theta_{0}/\phi_{0}$ of the
first harmonic of the persistent current signal scales as $\sin\theta_{0}$.
As discussed in \ref{sub:CHPCTh_FluxThroughMetal}, the typical persistent
current oscillation is correlated on the magnetic field scale $\gamma B_{c,1}$,
where $B_{c,1}$ was given by Eq. \ref{eq:CHPCTh_BcpToroidalField}
and is independent of $\theta_{0}$. Assuming the geometrical factor
$\gamma$ depends only weakly on $\theta_{0}$, it is desirable to
make $\gamma B_{c,1}\beta_{1}$ as large as possible. For the typical
current, $\gamma B_{c,1}\beta_{1}$ is the number of oscillations
of the persistent current that are correlated with each other in field.
When this number becomes too small ($\apprle1$), the persistent current
signal no longer follows a sinusoidal form and becomes difficult to
distinguish from the frequency shift background. Additionally, for
the case of the average persistent current, $\gamma B_{c,1}\beta_{1}$
sets the number of oscillations observable before the current is suppressed.
A reasonable choice for the angle is $\theta_{0}\approx45^{\circ}$
because the $\cos^{2}\theta_{0}$ prefactor of $\mathcal{S}_{pc}$
decreases only by $1/2$ as $\theta_{0}$ is increased from $0^{\circ}$
to $45^{\circ}$, while $\gamma B_{c,1}\beta_{1}$ increases strongly.
Beyond $\theta_{0}=45^{\circ}$, the factor $\cos^{2}\theta_{0}$
drops towards zero as $\theta_{0}$ approaches $90^{\circ}$, while
$\gamma B_{c,1}\beta_{1}$ varies only weakly. In the experiment,
we initially used $\theta_{0}=6^{\circ}$ in order to maximize $\mathcal{S}_{pc}$
and measure our first persistent current signal. We then adjusted
$\theta_{0}$ to $45^{\circ}$ to obtain data with more oscillations
per correlation field $B_{c,1}$. 

Although the ring thickness $t_{r}$ and linewidth $w_{r}$ do not
appear in the expression for $\mathcal{S}_{pc}$, they do affect the
measurement through appearance in $B_{c,1}\propto(w_{r}t_{r})^{-1/2}$.
In order to maximize $\gamma B_{c,1}\beta_{1}$, the ring's cross-sectional
dimensions should be made as small as possible. However, once the
cross-sectional dimensions are reduced to roughly the same magnitude
as the bulk elastic mean free path $l_{e}$, surface scattering becomes
an important factor in the electron diffusion and the diffusion constant
$D$ is reduced. Since the current magnitude is suppressed exponentially
in $D$, the cross-sectional dimensions of the rings should not be
reduced to lengths much smaller than the value of the elastic mean
free path in the absence of surface scattering.

Outside of the $\text{jinc}$ factor, $\mathcal{S}_{pc}$ scales with
$r$ as $\exp(-B_{\mathcal{S}}r^{2}T/D)r^{2-2\sigma}$. When $\sigma\neq1,$
this factor is optimized for $r_{\text{opt}}=\sqrt{(1-\sigma)\hbar D/4k_{B}T}$.
Using typical values of $D=0.02\,\text{m}^{2}/\text{s}$ and $T=0.3\,\text{K}$,
the optimal radius is $252\,\text{nm}$ for an array of rings with
current of random sign ($\sigma=1/2$) and $357\,\text{nm}$ for a
single ring ($\sigma=0$). These dimensions are similar to the ones
used in the experiment. For an array of rings with well-defined sign
($\sigma=1$), the factor $\exp(-B_{\mathcal{S}}r^{2}T/D)r^{2-2\sigma}$
decreases monotonically with $r$ so that it is best to make the rings
as small as possible lithographically. Because the average current
is suppressed as magnetic flux penetrates the metal of the ring on
the field scale $\gamma B_{c,1}$, the minimum $r$ is set by the
value that gives the minimum acceptable number of observable oscillations.
As mentioned above this number is set by the factor $\gamma B_{c,1}\beta_{1}\propto r$.

The final two parameters $x_{\max}$ and $l$ have the most complicated
relationships to the sensitivity $\mathcal{S}_{pc}$. When the uncertainty
is thermally limited, the sensitivity increases with decreasing $l$
for all values of $\sigma$ (i.e. for single rings and arrays of rings
with persistent current of either random or well-defined sign). The
dependence on $l$ differs between the two contributions to the uncertainty
due to noise $S_{P,\text{int}}$ in the interferometer. When uncertainty
in the cantilever's frequency of motion cased by interferometer noise
(the $C_{\mathcal{S}}$ term of Eq. \ref{eq:CHSensitivity_SensitivityFull})
dominates, the sensitivity is improved with increasing $l$ for all
values of $\sigma$. This dependence on $l$ follows the opposite
trend as that followed when thermal noise dominates. When the uncertainty
added to the measurement by noise in the interferometer (the $D_{\mathcal{S}}$
term of Eq. \ref{eq:CHSensitivity_SensitivityFull}) dominates, the
sensitivity is independent of $l$ for an array of rings with well
defined sign ($\sigma=1$) and otherwise ($\sigma=0\:\text{or \,}1/2$)
increases with decreasing $l$. This trend (except for the $\sigma=1$
case) is the same as that followed when the sensitivity is limited
by thermal noise. Adjusting $l$ tunes the system from being thermally
limited and being limited by one of the two contributions related
to interferometer noise. Thus the optimal value of $l$ depends on
all of the other system parameters required to calculate the three
contributions to the frequency uncertainty. These parameters include
$P_{\text{inc}}$, $R_{c}$, $R_{f}$, $Q$, $\sigma$, among others. 

Generally, there is an optimal value of $x_{\max}$ for a given set
of values of the other parameters in the experiment. When the frequency
uncertainty $\sigma_{f,\text{tot}}$ is dominated by terms derived
from the noise $S_{P,\text{int}}$ in the interferometric detection,
the sensitivity scales as 
\[
\mathcal{S}_{pc}\propto R\left(x_{\max}\right)x_{\max}\text{jinc}\left(2\pi r^{2}\alpha B\cos\theta_{0}x_{\max}/l\right)
\]
When the optical lever effect is small, $R(x_{\max})x_{\max}$ typically
has a maximum value for realistic values of $x_{\max}$. The value
of $x_{\max}$ producing this peak increases as the laser spot is
moved toward the base of the cantilever where the cantilever motion
(i.e. the parameter $U$) is smaller. When the frequency uncertainty
is limited by $S_{P,\text{int}}$, it is preferable to position the
laser spot as close to the cantilever tip as possible so that the
argument of the $\text{jinc}$ factor can be minimized (i.e. by maximizing
$R(x_{\max})x_{\max}$ with the smallest possible value of $x_{\max}$).
When the frequency uncertainty is thermally limited, the sensitivity
is proportional to
\[
\mathcal{S}_{pc}\propto x_{\max}\text{jinc}\left(2\pi r^{2}\alpha B\cos\theta_{0}x_{\max}/l\right)
\]
Because $x_{\max}\text{jinc}(Ax_{\max})\propto J_{1}(Ax_{\max})$,
the optimal value of $x_{\max}$ in this case corresponds to the first
peak of $J_{1}\left(2\pi r^{2}\alpha B\cos\theta_{0}x_{\max}/l\right)$
at $x_{\max}=1.84l/2\pi r^{2}\alpha B\cos\theta_{0}$. If this value
of $x_{\max}$ occurs at a value above that optimizing $R(x_{\max})x_{\max}$
when the laser spot is at the cantilever tip and thus causes the interferometer
noise to dominate (i.e. because $R(x_{\max})x_{\max}$ is small),
the factor $R(x_{\max})x_{\max}$ can be optimized by moving the laser
spot closer to the cantilever base as mentioned above. As was the
case for $l$, the optimal choice of $x_{\max}$ depends on several
other parameters of the system.

There are a couple of caveats to our discussion of $l$ and $x_{\max}$.
In \ref{sec:CHTorsMagn_CantileverSHO}, we introduced the cantilever
as a simple harmonic oscillator with restoring force $F_{\text{restoring}}=-kx$.
For sufficiently large values of $x/l$, higher order corrections
$F_{\text{restoring}}$ become non-negligible. These non-linearities
must be included in the derivation of the uncertainty in the frequency
measurement given in \ref{sec:CHSensitivity_FrequencyError}. In general,
they lead to an increase in the uncertainty $\sigma_{f,\text{tot}}$.
Additionally, the laser spot takes up a finite amount of space on
the cantilever and must be kept a certain distance away from the rings
in order to avoid heating them. Thus, when the sensitivity increases
with decreasing $l$, there is a lower bound on $l$ of $\sim100\,\text{\ensuremath{\mu}m}$.%
\footnote{This figure is obtained by assuming that the laser spot is $10\,\text{\ensuremath{\mu}m}$
wide, that 40\% of the cantilever is covered with rings, and that
a $\sim50\,\text{\ensuremath{\mu}m}$ distance is kept between the
rings and the laser spot (with the edge of the laser spot touching
the edge of the cantilever base).%
}

To summarize, we have seen that it is generally desirable to produce
ring-cantilever samples with high diffusion constant $D$, cantilever
reflectivity $R_{c}$, and quality factor $Q$ and small thickness
$t$. The ring samples should be located near the tip of the cantilever.
The highest sensitivity in measuring these samples is achieved when
the magnetic field $B$ and incident laser power $P_{\text{inc}}$
are maximized and the temperature $T$ is minimized. The relationship
between the sensitivity to the cantilever width $w$, cantilever thickness
$l$, the sample orientation angle $\theta_{0}$, the cantilever mode
$m$, the fiber reflectivity $R_{f}$, and the cantilever tip amplitude
of motion $x_{\max}$ depends on the details of the measurement. The
sensitivity $\mathcal{S}_{pc}$ peaks for fixed values of these parameters
rather than scaling monotonically with them. The choice of the value
for other parameters such the orientation angle $\theta_{0}$ is less
straightforward.

Finally, we note that the persistent current measurement in a uniform
magnetic field is fundamentally a torque measurement. A torque $\tau$
experienced by a cantilever of length $l$ is equivalent to a force
of magnitude $F=\tau/l$ applied at the cantilever tip. This fact
explains why the sensitivity to persistent currents scales inversely
to cantilever length $l$ whereas in force detection the sensitivity
scales with $l$ \citep{sidles1995magnetic}. It is possible that
a transducer designed more specifically for a torque rather than a
force such as the ones described in Refs. \citealp{chabot1999microfabrication,lobontiu2006modeling,haiberger2007highlysensitive}
could achieve greater sensitivity than a cantilever.

\section{Experimental characterization of the sensitivity of the cantilever
detection apparatus}

\subsection{\label{sub:CHSensitivity_ThermometrySection}Cantilever and sample
thermometry measurements}

In \ref{sec:CHSensitivity_OptimalCantDimensions}, we saw that the
sensitivity $\mathcal{S}_{pc}$ of the persistent current measurement
depends strongly on the temperature of the electrons in the ring and
the temperature of the cantilever with the signal decreasing with
temperature while the noise increases. One might think that mounting
the cantilever chip on a sample stage in good thermal contact with
a good cryostat cooled to a low bath temperature $T_{b}$ would be
sufficient to ensure that $T_{n}$ and $T_{e}$ are also cooled down.
However, the requirements for fabricating a high sensitivity cantilever
is at odds with those of achieving good thermal contact with the cantilever's
base. In particular, a high cantilever mechanical quality factor $Q$
corresponds to a weak coupling of the cantilever to its external environment.
Moreover, the extreme cantilever aspect ratio and the insulating properties
of cantilever materials (such as silicon) contribute to the expectation
of poor thermal conductivity for a high sensitivity cantilever. Additionally,
the common detection mechanism of cantilever motion using a laser
introduces a potentially strong source of heating into the system.

When we began planning the persistent current experiment, we were
not aware any experimental investigation of both the temperature of
a cantilever's macroscopic degree of freedom $T_{n}$ and that of
its microscopic degrees of freedom (which should be equal to the temperature
of the electrons $T_{e}$ in a sample mounted on the end of the cantilever)
below $1\,$K. Thurber \emph{et al.} had previously performed careful
measurements of the temperature of the microscopic degrees of freedom
of a cantilever and their tracking of the bath temperature $T_{b}$
from $4\,$K to $16\,$K by measuring the paramagnetism of solid air
contamination on the cantilever \citep{thurber2003temperature}. Other
experiments had been performed to study various samples on the ends
of cantilevers at temperatures as low as $250\,$mK, though concerns
about possible discrepancies between the sample temperature and the
bath temperature were not addressed directly \citep{harris2001magnetization,schwarz2002sawtoothlike,harris2003damping,mamin2007nuclear}.
Measurements of the temperature of a resonator's macroscopic degree
of freedom by observing its Brownian motion had been performed at
temperatures as low as $56\,$mK in nano-electromechanical systems
\citep{lahaye2004approaching} and as low as $220\,$mK in single
crystal silicon cantilevers similar to those discussed in this text
\citep{mamin2001subattonewton}. It is worth noting that in the latter
measurement the bath temperature was $110\,$mK, so that decoupling
of the cantilever noise temperature $T_{n}$ from the bath temperature
$T_{b}$ was in fact observed. We decided to perform a preliminary
thermometry experiment of both $T_{e}$ and $T_{n}$ on the same sample
chip in order to assess the feasibility of the persistent current
measurement and to provide an early milestone along the daunting ascent
towards persistent current measurements \citep{bleszynski-jayich2008noisethermometry}.

Our cantilever thermometry measurements were performed on the commercial
atomic force microscope cantilevers discussed in \ref{sub:CHExpSetup_ThermometryCantilevers}.
The macroscopic temperature $T_{n}$ was measured via the cantilever's
Brownian motion, while the the microscopic temperature $T_{e}$ was
measured via observation of the superconducting transition of a macroscopic
aluminum grain attached to the end of the cantilever. The measurements
were not performed on the same cantilever but were performed on two
cantilevers on the same sample chip during the same cool down. Cantilever
detection was performed with a set-up similar to that described in
\ref{sub:CHExpSetup_CantileverDetectionSetup} using a bare cleaved
optical fiber kept at a distance of $\sim100\,\mu\text{m}$ from the
cantilever.

\subsubsection{\label{sub:CHSensitivity_MeasureBrownianMotion}Measurements of the
cantilever's noise temperature}

The cantilever's noise temperature $T_{n}$ was extracted from measurements
of the cantilever's undriven motion at a series of bath temperatures
$T_{b}$ and incident laser powers $P_{\text{inc}}$. The amplified
voltage signal from the photodiode (the lead connected to the {}``Frequency
counter'' in Fig. \ref{fig:CHExpSetup_CantileverMeasurementSchematic})
was fed into the DAQ where it was digitized into an array of $N$
voltage readings $v_{m}$ spaced equally in time by $\Delta t$. Many
such time series of the voltage were recorded, and for each one the
single-sided power spectral density 
\begin{equation}
S_{V}\left(f\right)=2\frac{\left|{\displaystyle \sum_{m=0}^{N}v_{m}e^{-2\pi if\,\Delta t\, m/N}}\right|^{2}}{N\Delta t}\label{eq:CHSensitivity_ThermometryVoltagePSD}
\end{equation}
was computed. These power spectral densities were then averaged together.
The resulting trace $S_{V}(f)$ was then converted into a cantilever
amplitude of motion power spectral density $S_{x}$ using the conversion
factor $\Gamma_{V\,\text{to}\, x}$ discussed in Section \ref{sub:CHExpSetup_CalibrationDriveMotion}
and defined in Eq. \ref{eq:CHExpSetup_VoltsToMetersXTip} to find
\begin{equation}
S_{x}\left(f\right)=\frac{\Gamma_{V\,\text{to}\, x}^{2}}{2g_{\text{lock-in}}^{2}}S_{V}\left(f\right)\label{eq:CHSensitivity_SxFromSVPSD}
\end{equation}
where $g_{\text{lock-in}}$ is the voltage gain between the lock-in's
input (which was used to define $\Gamma_{V\,\text{to}\, x}$) and
its output ({}``signal monitor'' on the 7265). Note that Eq. \ref{eq:CHSensitivity_ThermometryVoltagePSD}
gives the root-mean-square (RMS) voltage power spectral density and
that the lock-in reading used to calculate $\Gamma_{V\,\text{to}\, x}$
is also the RMS voltage. The conversion factor $\Gamma_{V\,\text{to}\, x}$
has been defined so that it converts from RMS voltage to peak (not
RMS) amplitude of motion at the cantilever tip. This choice of conversion
factor explains the extra factor of 2 in Eq. \ref{eq:CHSensitivity_SxFromSVPSD}.

The power spectral density was calculated using these short time trace
sections because the cantilever is quite susceptible to external vibrations.
Obvious outlier traces where the cantilever amplitude was several
times as large as that observed in the majority of the traces could
easily be identified and discarded. Additionally, because the undriven
cantilever motion is a random quantity, the fluctuations of $S_{x}(f)$
for any one value of $f$ in a single trace are as large as $S_{x}(f)$.
It is only by averaging many traces that the measured $S_{x}(f)$
resembles the theoretically expected form The effect of averaging
many traces could also be accomplished by binning the frequency components
of $S_{x}(f)$ from a single large time trace, but this method is
susceptible to the effects of external vibrations described above.

\begin{figure}
\centering{}\includegraphics[width=0.5\paperwidth]{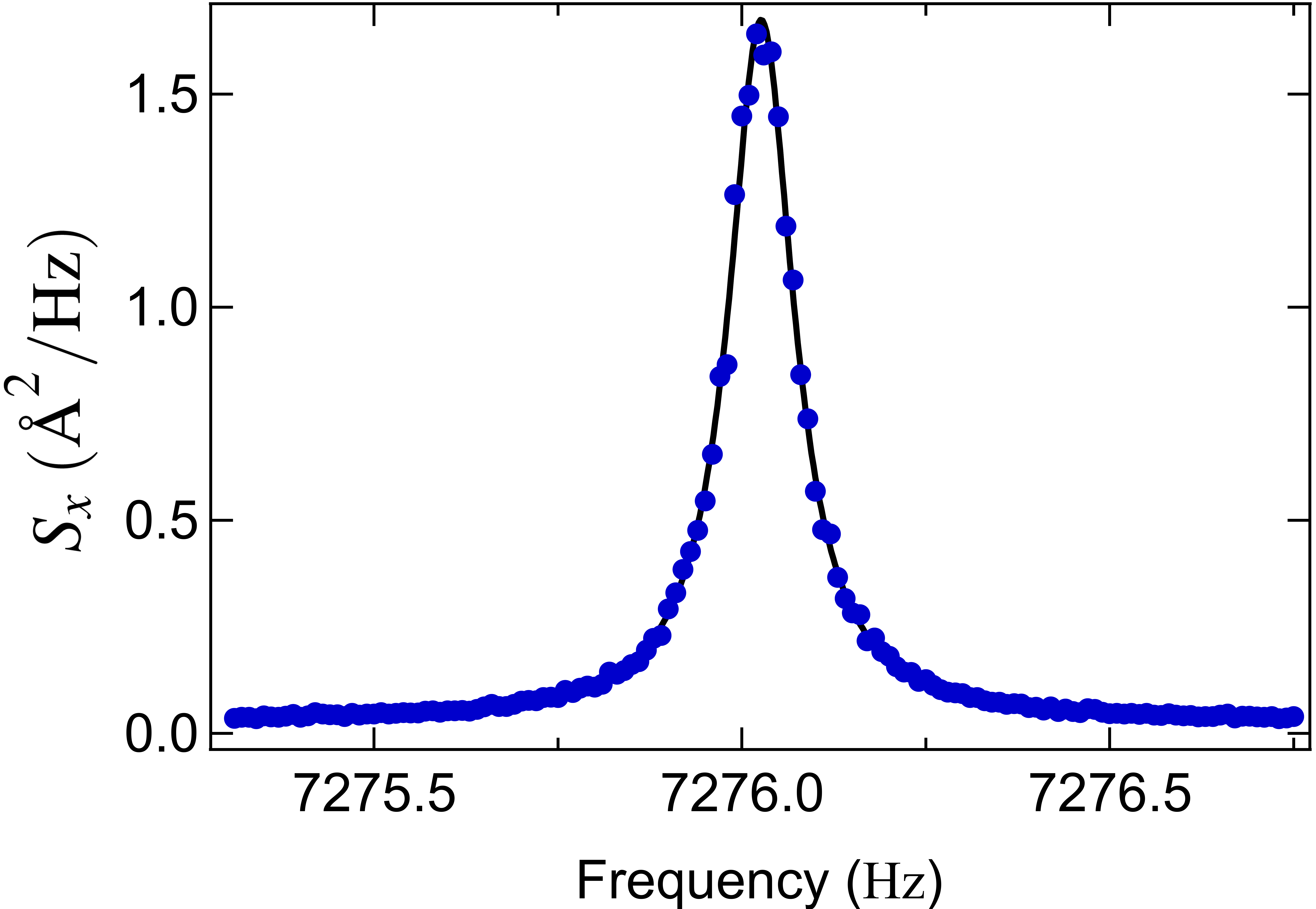}\caption[Power spectral density of cantilever motion]{\label{fig:CHSensitivity_BrownianPSDThermometry}Power spectral density
of cantilever motion. The data points shown were calculated from measurements
of the cantilever's undriven motion taken at $4.2\,\text{K}$ using
the procedure described in the text. The figure also shows a fit to
Eq. \ref{eq:CHSensitivity_SxFitFormWithOffset}. The extracted fitting
parameters were $\langle x^{2}\rangle=0.28\,\text{\AA}^{2}$, $f_{0}=7276\,\text{Hz}$,
$Q=6.8\times10^{4}$, and $S_{x,\text{int}}=0.029\,\text{\AA}^{2}/\text{Hz}$.
The magnitude of the background $S_{x,\text{int}}$ is a factor of
four greater than that expected for shot noise for the incident laser
power of $150\,\text{nW}$.}
\end{figure}

The measured power spectral density curves were fit using Eq. \ref{eq:CHSensitivity_SxFitFormWithOffset}
in the form 
\begin{equation}
S_{x}\left(f\right)=\frac{2\left\langle x^{2}\right\rangle }{\pi f_{0}Q}\frac{1}{\left(\left(f/f_{0}\right)^{2}-1\right)^{2}+\left(f/f_{0}Q\right)^{2}}+S_{x,\text{int}}\label{eq:CHSensitivity_SxFitFormWithOffset}
\end{equation}
where $S_{x,\text{int}}$ represents the constant offset in the power
spectral density due to the noise of the interferometric measurement
which was described in \ref{sub:CHSensitivity_InterferometerNoise}.
The free parameters in the fit were $\langle x^{2}\rangle$, $f_{0}$,
$Q$, and $S_{x,\text{int}}$. Fig. \ref{fig:CHSensitivity_BrownianPSDThermometry}
shows the power spectral density of cantilever displacement taken
at $4.2\,\text{K}$ along with a fit to Eq. \ref{eq:CHSensitivity_SxFitFormWithOffset}.%
\footnote{The data in this figure were originally published in \citep{bleszynski-jayich2008noisethermometry}.
The scale of the vertical axis was incorrect in the original paper
and has been corrected here.%
} Fits were also performed with $Q$ fixed to a value determined by
ringdown measurements as described in \ref{sub:CHExpSetup_CantileverMeasurementProcedure}.
These fits produced less consistent results than the ones for which
$Q$ was allowed to vary. The fit gives $\langle x^{2}\rangle$ directly.
Alternatively, after subtracting the baseline noise $S_{x,\text{int}}$,
the area under the data points of the $S_{x}(f)$ curve could be integrated
near the peak at $f_{0}$ to give $\langle x^{2}\rangle$. This procedure
gave similar results to the fits for $\langle x^{2}\rangle$.

\begin{figure}
\begin{centering}
\includegraphics[width=0.55\paperwidth]{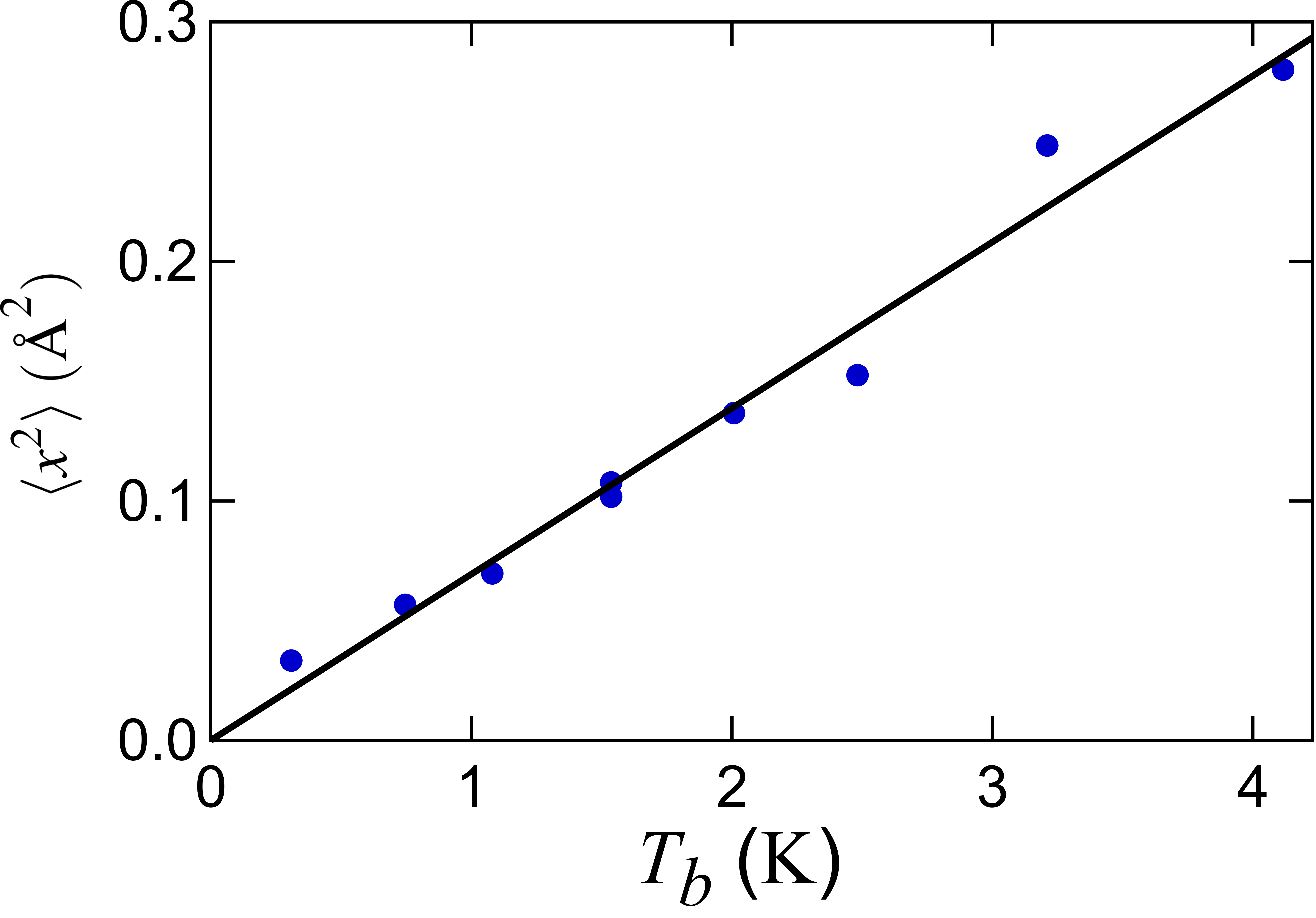}\caption[Mean square cantilever displacement versus refrigerator temperature]{\label{fig:CHSensitivity_xRMSvsT}Mean square cantilever displacement
versus refrigerator temperature. The data points represent the values
of $\langle x^{2}\rangle$ extracted from measurements such as the
one shown in Fig. \ref{fig:CHSensitivity_BrownianPSDThermometry}
at a series of refrigerator temperatures $T_{b}$. For each measurement,
the incident laser power was $150\,\text{nW}$. The line represents
a fit of the form $\langle x^{2}\rangle=k_{B}T/k+x_{\text{off}}^{2}$
for $k$ and $x_{\text{off}}^{2}$. The best fit parameters were $k=0.020\,\text{N/m}$
and $x_{\text{off}}^{2}=2.0\times10^{-3}\,\text{\AA}^{2}$.}

\par\end{centering}

\end{figure}

Measurements of $S_{x}(f)$ such as the one shown in Fig. \ref{fig:CHSensitivity_BrownianPSDThermometry}
were repeated at a series of refrigerator temperatures using $P_{\text{inc}}=150\,\text{nW}$
of power incident on the cantilever. The resulting trace of mean square
cantilever displacement $\langle x^{2}\rangle$ versus refrigerator
temperature $T_{b}$ is shown in Fig. \ref{fig:CHSensitivity_xRMSvsT},
as is a linear fit. According to the equipartition theorem given in
Eq. \ref{eq:CHSensitivity_EquipartitionTheorem}, the mean square
displacement $\langle x^{2}\rangle$ should be proportional to the
cantilever temperature with constant of proportionality $k_{B}/k$.
From Fig. \ref{fig:CHSensitivity_xRMSvsT}, we can conclude that the
cantilever temperature $T_{n}$ tracks the refrigerator temperature
down to its base temperature because the data matches the linear fit
down to the lowest temperature and because the $\langle x^{2}\rangle$-intercept
of the fit is consistent with zero within the uncertainty of the fit
due to the scatter in the data. The presence of an external heat source
preventing the cantilever from equilibrating with the refrigerator
would lead to a saturation of $\langle x^{2}\rangle$ at low temperature,
while a non-zero $\langle x^{2}\rangle$-intercept would indicate
an additional source vibrations to the thermal force noise. The Arrow$^{\text{TM}}$
TL8 cantilevers described in \ref{sub:CHExpSetup_ThermometryCantilevers}
had a specified spring constant of $0.03\,\text{N/m}$ with values
between 0.004 and $0.54\,\text{N/m}$ typical. The slope of the fit
in Fig. \ref{fig:CHSensitivity_xRMSvsT} corresponds to a spring constant
of $0.02\,\text{N/m}$, consistent with these specifications. The
spring constant can be calculated directly from the dimensions using
Eq. \ref{eq:CHTorsMagn_springKfromDimensions}. The uncertainty in
cantilever thickness (specified to be between $0.5$ and $2.5\,\text{\ensuremath{\mu}m}$)
results in the large uncertainty in the specification for the cantilever
spring constant, which scales as the cube of the thickness. From the
width and length measured in Fig. \ref{fig:CHExpSetup_BareArrows}
and the measured frequency of $7276\,\text{Hz}$ (Fig. \ref{fig:CHSensitivity_BrownianPSDThermometry}),
we estimate the cantilever thickness to be between 0.89 and $1.29\,\text{\ensuremath{\mu}m}$
using Eqs. \ref{eq:CHTorsMagn_FreqFromKandMeff}, \ref{eq:CHTorsMagn_springKfromDimensions},
and \ref{eq:CHTorsMagn_MeffFromDimensions}. These two values were
obtained defining the cantilever length to exclude and include the
triangular tip, respectively. The spring constants corresponding to
these dimensions are 0.036 and $0.11\,\text{N/m}$. We are not sure
of the origin of the discrepancy between these values and our extracted
value of $0.02\,\text{N/m}$ for $k$. 

In Ref. \citealp{mamin2001subattonewton}, a deviation of $T_{n}$
from $T_{b}$ was observed below $300\,\text{mK}$. The fact that
this deviation of $T_{n}$ from $T_{b}$ was observed in Ref. \citealp{mamin2001subattonewton}
and not by us can be explained by the fact that the cantilever used
in Ref. \citealp{mamin2001subattonewton} had a smaller cross-section
leading to increased phonon-boundary scattering \citep{asheghi2002thermal}.
Additionally, this deviation was observed only for $T_{b}\sim100\,\text{mK}$,
below the range of temperatures measured by us. Presumably, these
factors outweighed the fact that the cantilevers of Ref. \citealp{mamin2001subattonewton}
were fabricated from undoped silicon which should possess a larger
thermal conductivity and a lower optical absorption coefficient than
our doped silicon \citep{asheghi2002thermal,schmid1981optical}. We
discuss the transport of heat through the cantilever further in \ref{sub:CHSensitivity_HeatTransport}.

\FloatBarrier

\subsubsection{Measurements of the electron temperature of a metallic sample at
the end of a cantilever}

We use the superconducting transition of an aluminum grain mounted
on the end of the cantilever (Fig. \ref{fig:CHExpSetup_AlGrainArrow})
to determine the temperature $T_{e}$ of the cantilever's microscopic
degrees of freedom near its tip. In the presence of a static magnetic
field, a bulk superconductor develops surface currents which screen
the magnetic field from the superconductor's interior, a phenomenon
known as the Meissner effect \citep{tinkham2004introduction}. These
screening currents and the superconductor's corresponding magnetic
moment $\boldsymbol{\mu}$ are proportional to the applied magnetic
field $\boldsymbol{B}$. From Eq. \ref{eq:ChTorsMagn_EnergyMagMoment},
the energy $E$ of the superconductor in the magnetic field is proportional
to $B^{2}$. For an aspherical superconducting grain, we can write
\[
E=m_{0}B^{2}N\left(\theta\right)
\]
where $m_{0}$ is a constant of proportionality with units of $\text{A m}^{2}$,
$\theta$ is the angle between $\boldsymbol{\mu}$ and $\boldsymbol{B}$,
and $N(\theta)$ is a shape anisotropy factor \citep{stoner1948amechanism}. 

By Eq. \ref{eq:CHTorsMagn_FreqShiftdEdThetaGeneral}, it follows that
a cantilever mounted with a superconducting grain will exhibit a frequency
shift proportional to $B^{2}$. In the normal state, the grain should
produce a negligible frequency shift. According to the BCS model,
the critical field $B_{c}$ at which a superconductor transitions
to the normal state can be approximated within four percent by 
\begin{equation}
B_{c}\left(T_{e}\right)=B_{c}\left(0\right)\sqrt{1-\left(\frac{T_{e}}{T_{c}}\right)^{2}}\label{eq:CHSensitivity_BCSBc}
\end{equation}
where $T_{e}$ is the temperature of the superconductor and $T_{c}$
is the superconducting transition temperature in the absence of an
applied magnetic field \citep{bardeen1957theoryof,tinkham2004introduction}.
Thus, if $T_{c}$ is known, Eq. \ref{eq:CHSensitivity_BCSBc} can
be used to determine $T_{e}$ of a superconducting sample on the end
of a cantilever by measuring the cantilever frequency while sweeping
the magnetic field and noting the magnitude of field at which the
cantilever frequency stops changing quadratically with $B$ and becomes
independent of it.%
\footnote{We note here that Eq. \ref{eq:CHSensitivity_BCSBc} applies exactly
only for a long, skinny superconductor oriented with its long axis
parallel to the direction of the applied magnetic field. For a superconductor
with a realistic shape, demagnetization effects will lead to magnetic
flux penetration at a lower applied magnetic field than that given
in Eq. \ref{eq:CHSensitivity_BCSBc}. We do not attempt model these
intermediate states in our analysis here.%
}

\begin{figure}
\begin{centering}
\includegraphics[width=0.5\paperwidth]{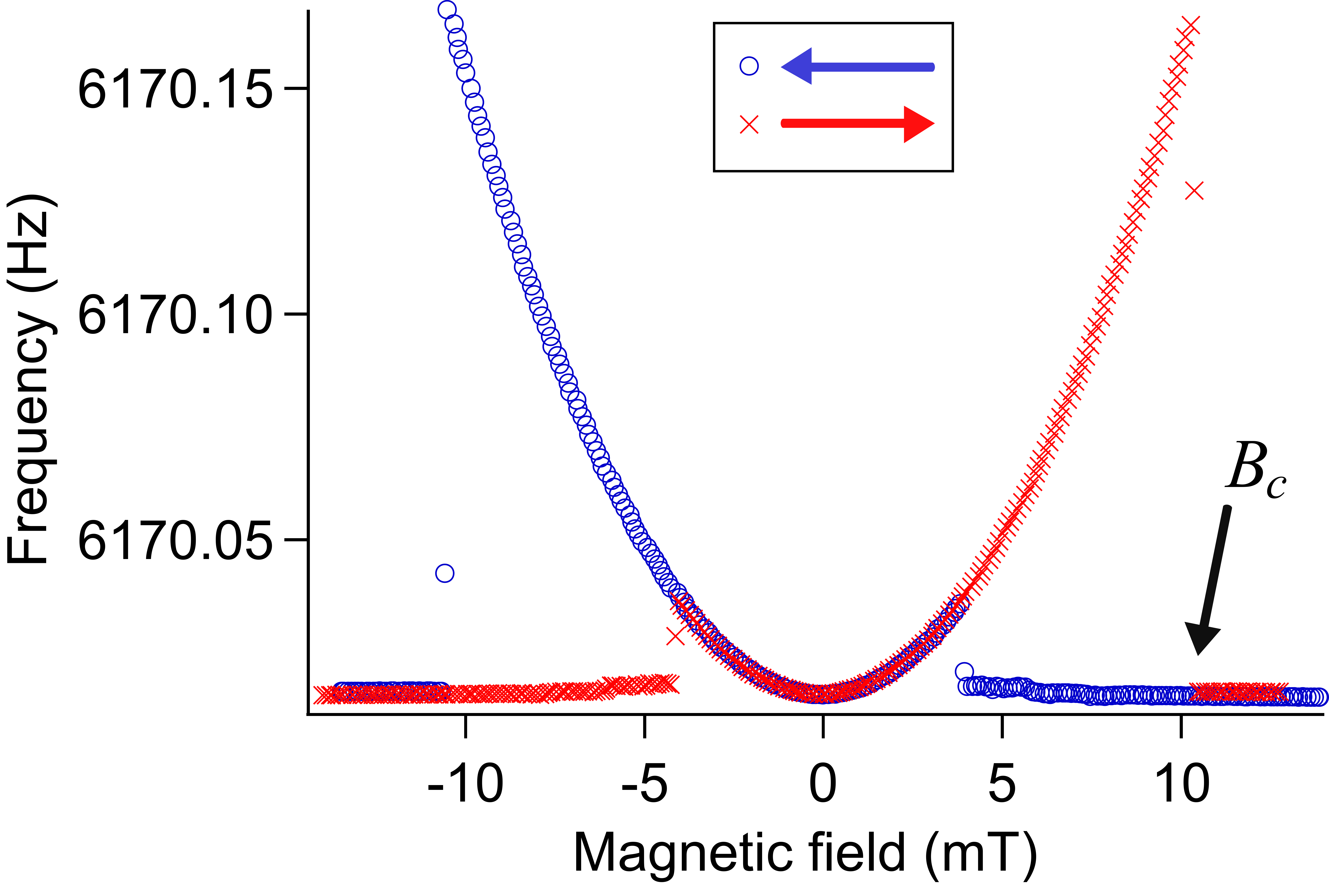}\caption[Resonant frequency versus magnetic field for the Al grain-mounted
cantilever]{\label{fig:CHSensitivity_FreqVsBAlGrain}Resonant frequency versus
magnetic field for the aluminum grain-mounted cantilever. As indicated
in the inset, the red \textsf{X}'s represent data taken as the magnetic
field was swept from negative to positive values (upward), while the
blue \textsf{O}'s correspond to a sweep in the opposite direction
(downward). The transition from the superconducting to the normal
state at $B=11.6\,\text{mT}$ is indicated as $B_{c}$ in the figure
for the upward sweep. The data shown were taken at a refrigerator
temperature $T_{b}=313\,\text{mK}$ and with an incident laser power
of $P_{\text{inc}}=25\,\text{nW}$. For both sweeps, the initial magnetic
field magnitude was larger than the largest magnetic field values
shown in the figure.}

\par\end{centering}

\end{figure}

At a series of refrigerator temperatures $T_{b}$, we made such measurements
of frequency as a function field for our aluminum grain-mounted cantilever.
We began driving the cantilever in a phase-locked loop as described
in \ref{sub:CHExpSetup_CantileverDetectionSetup}. For each measurement,
the incident laser power was $P_{\text{inc}}=25\,\text{nW}$. We initialized
the system by sweeping the magnetic field well above the critical
field $B_{c}$. We then swept the magnetic field continuously back
through zero and through the critical field with the opposite sign
of the initial field. Fig. \ref{fig:CHSensitivity_FreqVsBAlGrain}
shows two such measurements, one for each direction of magnetic field
sweep. As expected, the cantilever frequency exhibits a quadratic
dependence on magnetic field near zero field and a sharp transition
to independence from $B$ at higher fields. We interpret the region
of quadratic magnetic field dependence as the superconducting state
and the flat regions as the normal state. In both sweeps, the aluminum
grain transitions to the superconducting state well below $B_{c}$,
exhibiting supercooling \citep{faber1955thephase}. The transition
back to the normal state where the frequency drops abruptly is assumed
to occur at the critical field $B_{c}$. 

The magnetic field produced by the solenoid displayed a small amount
of hysteresis from sweep to sweep, which offset the frequency curves
in $B$. The hysteresis was corrected for by fitting plots of frequency
versus magnetic field with opposite sweep direction to second order
polynomials. The data were then shifted so that their minima both
located at zero magnetic field (see Fig. \ref{fig:CHSensitivity_FreqVsBAlGrain}).
The critical field $B_{c}$ was taken to be the average of the observed
$B_{c}$ for the upward and downward sweeps.

\begin{figure}

\begin{centering}
\includegraphics[width=0.6\paperwidth]{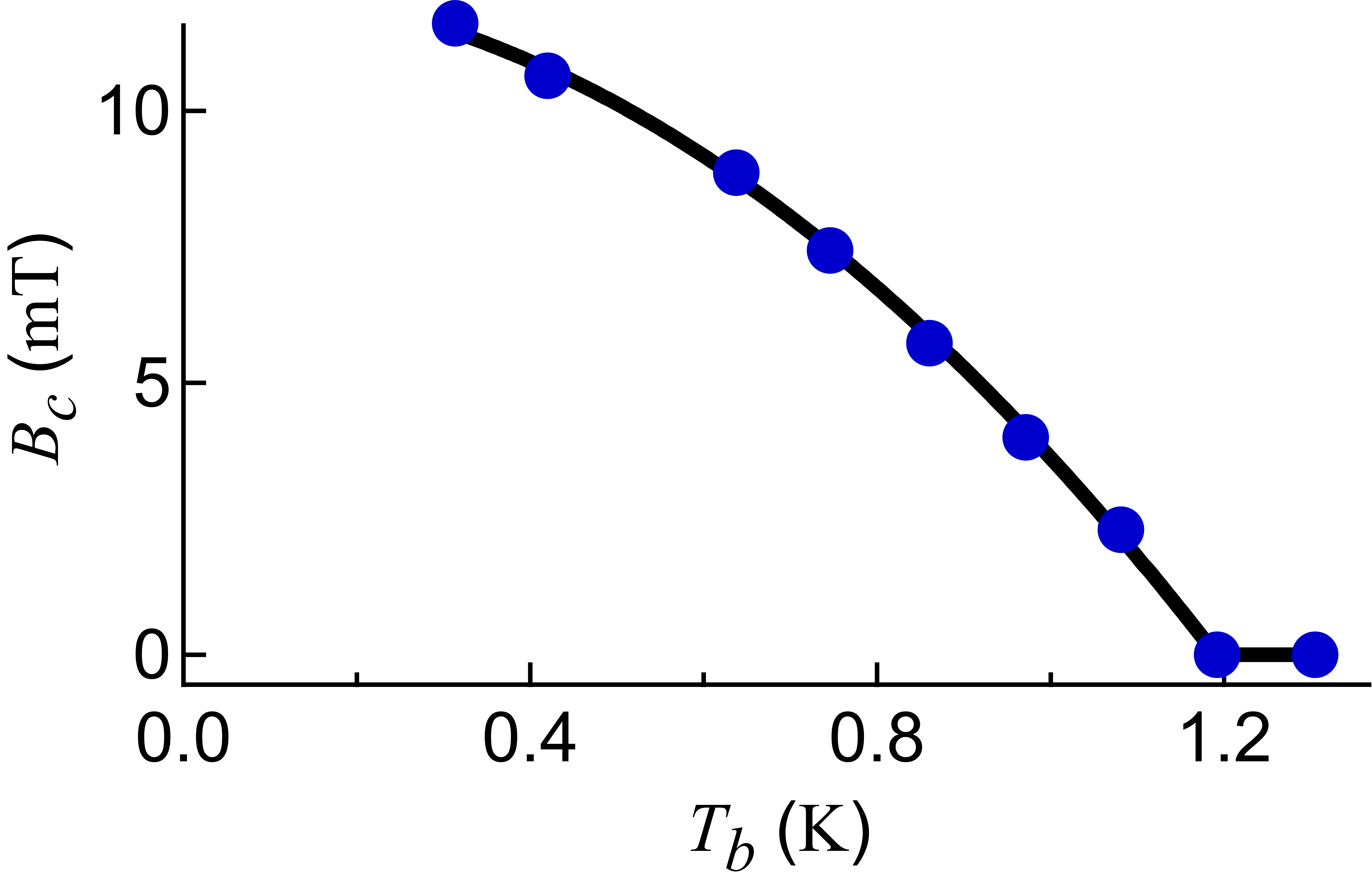}
\par\end{centering}

\caption[Superconducting critical field versus refrigerator temperature]{\label{fig:CHSensitivity_BcVsTBCS}Superconducting critical field
versus refrigerator temperature. The data points represent critical
fields extracted from measurements of the cantilever frequency like
the one shown in Fig. \ref{fig:CHSensitivity_FreqVsBAlGrain}. The
curve is a fit to the prediction of the BCS model for the critical
field of a superconductor. A discussion of the fit is given in the
text.}

\end{figure}

The measured critical field $B_{c}$ as a function of refrigerator
temperature $T_{b}$ is shown in Fig. \ref{fig:CHSensitivity_BcVsTBCS}.
These data were then fit to Eq. \ref{eq:CHSensitivity_BCSBc} for
$T_{c}$ and $B_{c}(0)$ using $T_{e}=T_{b}$. The fit, shown in Fig.
\ref{fig:CHSensitivity_BcVsTBCS}, produced the values $T_{c}=1.19\,\text{K}$
and $B_{c}(0)=12.3\,\text{mT}$. This value of $T_{c}$ agrees within
one percent of the bulk value of $1.18\,\text{K}$, while the critical
field is larger than the bulk value of $10.5\,\text{mT}$. The discrepancy
from the bulk value for $B_{c}(0)$ may be due to finite size effects,
the grain's aspherical shape, or trace impurities, each of which can
increase $B_{c}(0)$ without changing the dependence of $B_{c}$ on
temperature \citep{devoretprivate}. We interpret the excellent match
of the functional form of Eq. \ref{eq:CHSensitivity_BCSBc} with the
data as evidence that the temperature $T_{e}$ of the aluminum grain
at the cantilever tip follows the temperature $T_{b}$ of the refrigerator
down to $313\,\text{mK}$ for $P_{\text{inc}}=25\,\text{nW}$.

\subsubsection{\label{sub:CHSensitivity_HeatTransport}Effect of a localized heat
source on the cantilever temperature}

In order to test our conclusions from the previous sections that the
cantilever's noise temperature $T_{n}$ and the cantilever's microscopic
temperature $T_{e}$ could be deduced from measurements of the cantilever
Brownian motion $\langle x^{2}\rangle$ and the superconducting critical
field $B_{c}$ of the aluminum grain, we tested the effects of a local
heat source on the measurements of $T_{n}$ and $T_{e}$. In our case,
the detection laser, operated at higher intensity than in the measurements
discussed in the previous sections, acted as the local heat source
through the cantilever's small but finite absorption of the incident
optical power. The detection laser was directed at a point near the
cantilever tip so that for appreciable absorbed power the cantilever
no longer remains in equilibrium with the temperature $T_{b}$ of
the refrigerator at its base.

\begin{figure}

\centering{}\includegraphics[width=0.55\paperwidth]{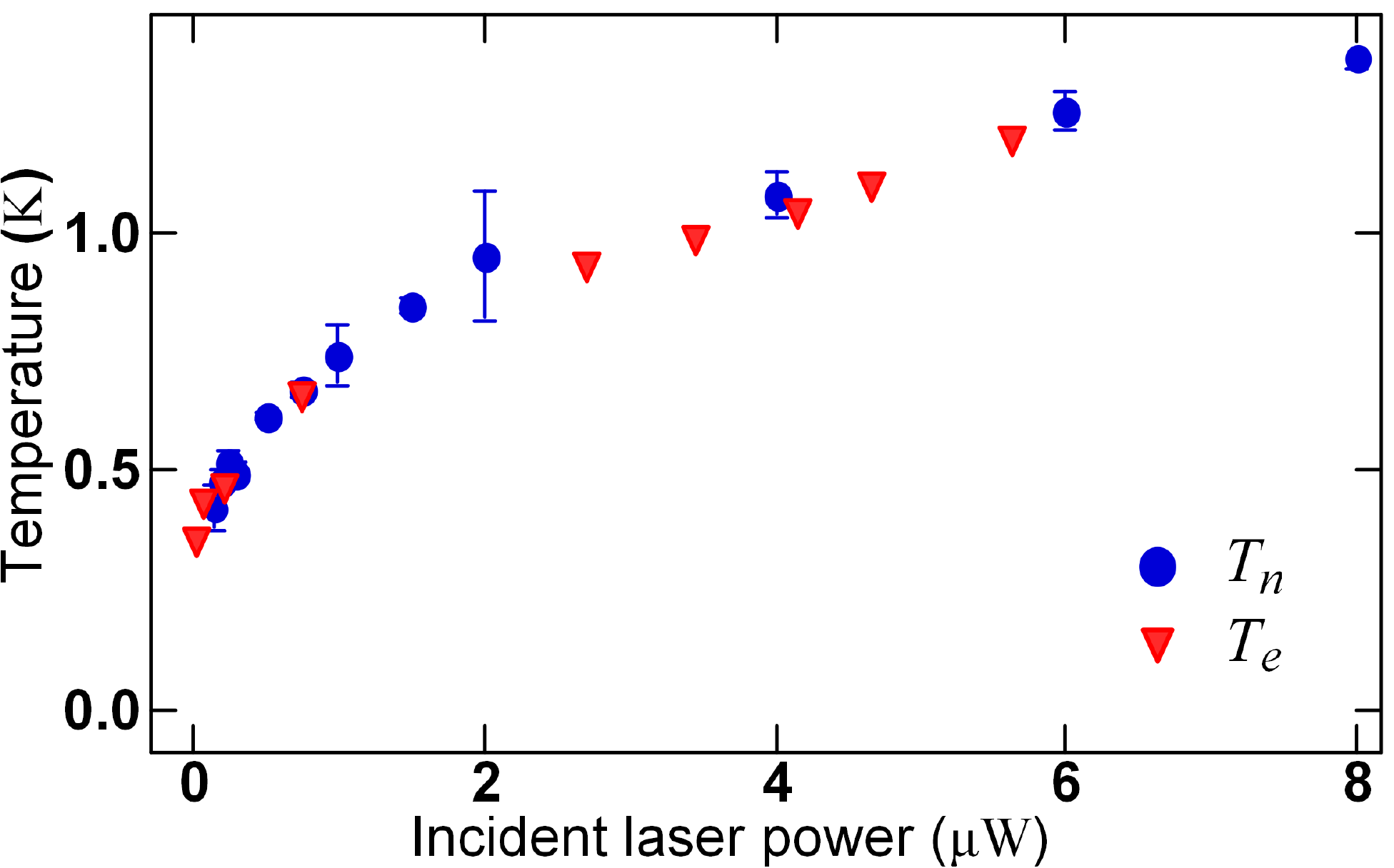}\caption[Measured cantilever temperature versus incident laser power]{\label{fig:CHSensitivity_LaserHeating}Measured cantilever temperature
versus incident laser power. Both $T_{n}$ (blue dots) and $T_{e}$
(red triangles), found by the method in the text, are plotted against
the incident laser power. As was the case for the measurements in
the previous sections, the measurements of $\langle x^{2}\rangle$
and $B_{c}$ were performed on different cantilevers on the same chip.
Further analysis of this data is given in the text and in Fig. \ref{fig:CHSensitivity_CantileverHeatTransport}.}
\end{figure}

For our measurements of both $\langle x^{2}\rangle$ and $B_{c}$,
the laser spot was directed at a point $z_{f}=400\,\text{\ensuremath{\mu}m}$
from the cantilever base on the $\sim500\,\text{\ensuremath{\mu}m}$
long cantilever. The refrigerator was kept at its base temperature
of $T_{b}=310\,\text{mK}$. At each incident power $P_{\text{inc}}$,
the cantilever motion $\langle x^{2}\rangle$ was measured and used
to infer $T_{n}$ by inverting the fit from Fig. \ref{fig:CHSensitivity_xRMSvsT}.
Similarly, the fit in Fig. \ref{fig:CHSensitivity_BcVsTBCS} was used
to convert the measured values of $B_{c}$ into $T_{e}$. Fig. \ref{fig:CHSensitivity_LaserHeating}
shows the extracted values of $T_{n}$ and $T_{e}$ versus the incident
laser power. 

Given that the silicon of the cantilever is an insulator, we assume
that, in the steady state, any heat $Q_{H}$ introduced to the cantilever
at the position of the laser spot must be conducted to the base by
phonons. According to Fourier's law \citep{ashcroft1976solidstate},
the flow $\boldsymbol{J}_{Q}$ of heat per unit time through a cross-section
of unit area satisfies
\begin{equation}
\boldsymbol{J}_{Q}=-\kappa\nabla T\label{eq:CHSensitivity_FouriersLaw}
\end{equation}
where $\kappa$ is the thermal conductivity. At low temperatures,
the phonon thermal conductivity of an insulator is proportional to
the specific heat at constant volume $c_{v}$, which itself is proportional
to $T^{3}$ in the low temperature limit \citep{ashcroft1976solidstate}.
We write $\kappa=bT^{3}$. By integrating $\boldsymbol{J}_{Q}$ over
the cross-section $w\times t$ of the cantilever, we obtain the rate
$\dot{Q}$ of total heat transfer through the cantilever. Using $z$
to denote the distance from the cantilever base, the rate of heat
transfer out of the cantilever ($-z$ direction) is $\dot{Q}_{\text{out}}=wtbT^{3}\, dT/dz$. 

In a material for which the fraction of optical power lost per unit
of distance is a constant, we can write the magnitude of optical power
as a function of distance $z$ traveled through the material as $P(z)=P(0)e^{-az}$,
where $a$ is the coefficient of absorption. For $az\ll1$, we have
$P(z)\approx P(0)-azP(0)$. Therefore, assuming all absorbed optical
power is converted into heat, we can write the rate of heat transfer
into the thin cantilever as $\dot{Q}_{\text{in}}=atP_{\text{inc}}$.

In the steady state, there is no build up of heat and $\dot{Q}_{\text{in}}=\dot{Q}_{\text{out}}$.
For the cantilever, we have $P_{\text{inc}}=(wb/a)T^{3}\, dT/dz$.
Integrating this expression from the cantilever base at $z=0$ to
the laser spot at $z=z_{f}$, we find
\begin{align*}
z_{f}P_{\text{inc}} & =\int_{0}^{z_{f}}dz\, T^{3}\frac{dT}{dz}\\
 & =\frac{wb}{a}\int_{T_{b}}^{T_{f}}dT\, T^{3}\\
 & =\frac{wb}{4a}\left(T_{f}^{4}-T_{b}^{4}\right)
\end{align*}
or
\begin{equation}
T_{f}^{4}-T_{b}^{4}=\frac{4az_{f}}{wb}P_{\text{inc}}\label{eq:CHSensitivity_HeatTransferEquation}
\end{equation}
where $T_{f}$ is the temperature of the cantilever at the spot of
the laser. Since there is no heat sink at the tip of the cantilever,
there is no heat flow from the laser spot towards the tip. Eq. \ref{eq:CHSensitivity_FouriersLaw}
then requires that the temperature from the laser spot to the tip
of the cantilever is constant and equal to $T_{f}$.

\begin{figure}
\centering{}\includegraphics[width=0.4\paperwidth]{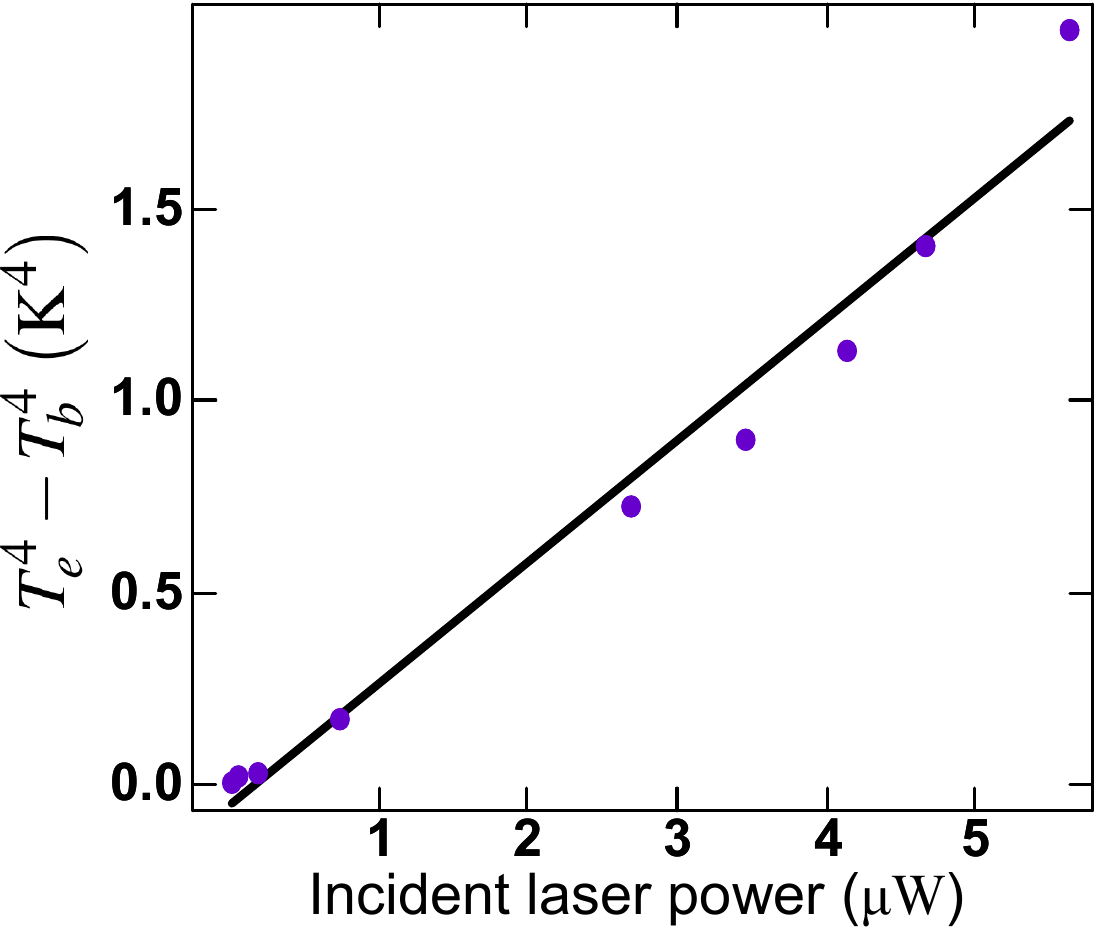}\caption[Comparison of observed cantilever heating to a phonon thermal conductivity
model]{\label{fig:CHSensitivity_CantileverHeatTransport}Comparison of observed
cantilever heating to a phonon thermal conductivity model. We plot
$T_{e}^{4}-T_{b}^{4}$ against $P_{\text{inc}}$, in order to compare
our data with Eq. \ref{eq:CHSensitivity_HeatTransferEquation} which
predicts the two quantities to be proportional to each other. We fit
the data to Eq. \ref{eq:CHSensitivity_HeatTransferEquation} with
$b$ as the only free parameter, obtaining $b=0.13\,\text{W\,\ K}^{-4}\text{m}^{-1}$
as described in the text.}
\end{figure}

In Fig. \ref{fig:CHSensitivity_CantileverHeatTransport}, we plot
$T_{e}^{4}-T_{b}^{4}$ against $P_{\text{inc}}$ using the data shown
in Fig. \ref{fig:CHSensitivity_LaserHeating}. The Arrow cantilevers
are doped such that their resistivity is specified to be on the order
of $0.025\,\Omega\,\text{cm}$. This resistivity corresponds to a
doping concentration of $\sim1-2\times10^{18}\,\text{cm}^{3}$ \citep{none2010resistivity}.
For a similar doping concentration, the absorption coefficient of
silicon has been measured to be $\sim20\,\text{cm}^{-1}$ at $4\,\text{K}$
\citep{schmid1981optical}. Using this value along with $z_{f}=400\,\text{\ensuremath{\mu}m}$
and $w=100\,\text{\ensuremath{\mu}m}$, we fit the data to Eq. \ref{eq:CHSensitivity_HeatTransferEquation}
and extract $b=0.13\,\text{W\,\ K}^{-4}\text{m}^{-1}$. This value
is a factor of four smaller than that found previously in similarly
doped silicon \citep{fortier1976effectof,asheghi2002thermal}. This
discrepancy could be due to uncertainty in the exact doping level
of the cantilever and also to finite size effects. We note that if
the thermal conductivity coefficient $b$ can be measured independently
that the analysis performed here gives a direct measure of the absorption
coefficient $a$. Usually $a$ is deduced from measurements of optical
transmission and reflection which do not distinguish between loss
due to absorption and loss due to diffusive scattering.

We do not analyze the data for $T_{n}$ because unlike $T_{e}$ this
quantity depends on the temperature of the entire cantilever, which
is no longer a constant. We note, however, that in Fig. \ref{fig:CHSensitivity_LaserHeating}
$T_{e}$ and $T_{n}$ display a similar dependence on $P_{\text{inc}}$.
This agreement between $T_{e}$ and $T_{n}$ can be explained by the
fact that when $T_{f}\apprge2T_{b}$ most of the drop temperature
in temperature along the cantilever is located near to the cantilever
base.%
\footnote{This can be seen by replacing $T_{f}$ and $z_{f}$ by $T\left(z\right)$
and $z$ in Eq. \ref{eq:CHSensitivity_HeatTransferEquation} and plotting
$T\left(z\right)$.%
} Additionally, as noted above, the portion of the cantilever from
the laser spot to the cantilever tip is at a constant temperature.
Thus, the portion of the cantilever from its midpoint to its tip,
the section responsible for most of the cantilever displacement, is
at a fairly constant temperature. Since the laser spot was located
at the same distance $z_{f}$ for both the measurements of $T_{n}$
and $T_{e}$, the temperature $T_{f}$ at $z_{f}$ should be the same
for a given $P_{\text{inc}}$ in both measurements. Thus, it is reasonable
that these inferred temperatures are in agreement.

\section{\label{sec:ChSensitivity_PCUncertainty}Characterization of uncertainty
in measurement of persistent current samples}

We conclude this chapter by presenting data from a few different measurements
characterizing the persistent current sensitivity $\mathcal{S}_{pc}$
of the cantilevers used for the measurements discussed in this text.
The parameters of these samples are collected in Tables \ref{tab:ChData_CLs}
and \ref{tab:ChData_Rings}. 

\begin{figure}
\centering{}\includegraphics[width=0.55\paperwidth]{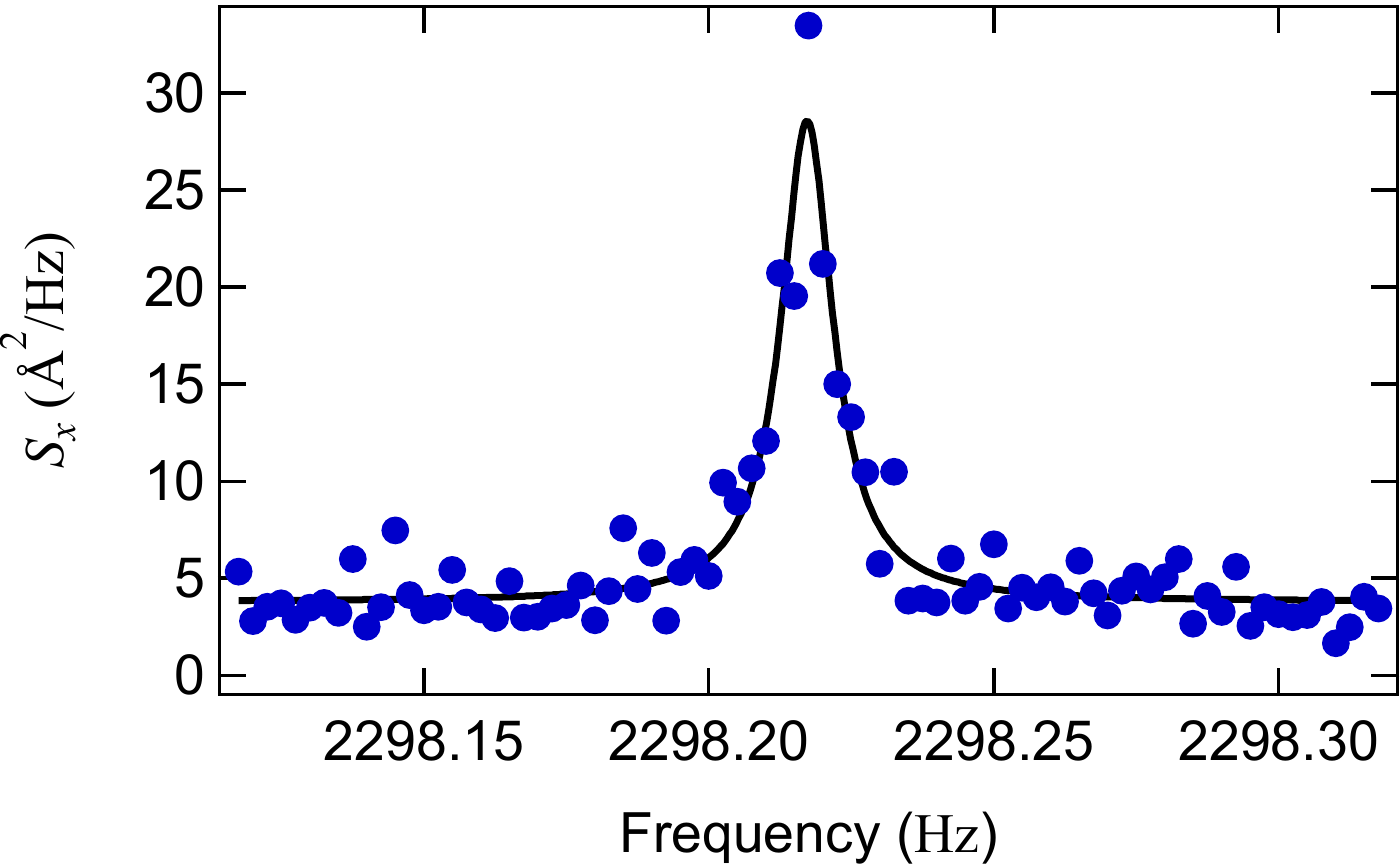}\caption[Power spectral density of cantilever displacement for sample CL14]{\label{fig:CHSensitivity_CLBrownian}Power spectral density of cantilever
displacement for sample CL14. The figure shows the power spectral
density $S_{x}$ obtained by the method described in \ref{sub:CHSensitivity_MeasureBrownianMotion}
as well as a fit to Eq. \ref{eq:CHSensitivity_SxFitFormWithOffset}.
The results of the fit are discussed in the text. This data was taken
with the Bragg reflector set-up with an incident laser power of $P_{\text{inc}}=1.9\,\text{nW}$.
The background noise level corresponds to a noise equivalent power
of $0.29\,\text{pW}/\sqrt{\text{Hz}}$ for the photoreceiver-amplifier
package.}
\end{figure}

In Fig. \ref{fig:CHSensitivity_CLBrownian}, the power spectral density
$S_{x}$ of cantilever displacement, obtained by the method outlined
in \ref{sub:CHSensitivity_MeasureBrownianMotion}, is shown for sample
CL14. Also shown is a fit to the form of $S_{x}$ given above in Eq.
\ref{eq:CHSensitivity_SxFitFormWithOffset} for a white thermal noise
force. The extracted best fit coefficients are $\langle x^{2}\rangle=2.5\times10^{-19}\,\text{m}^{2}/\text{Hz}$,
$f_{0}=2298.217\,\text{Hz}$, $Q=2.1\times10^{5}$, and $S_{x,\text{int}}=3.8\times10^{-20}\,\text{m}^{2}/\text{Hz}$.
Using the spring constant $k=1.08\times10^{-3}\,\text{N/m}$ deduced
from the cantilever's dimensions, the equipartition theorem (Eq. \ref{eq:CHSensitivity_EquipartitionTheorem})
gives a cantilever noise temperature of $T_{n}=k\langle x^{2}\rangle/k_{B}=334\,\text{mK}$,
in close agreement with the refrigerator temperature of $T_{b}=323\,\text{mK}$,
which was the base temperature for this set of persistent current
measurements. Thus we see that $T_{n}$ tracks $T_{b}$ down to the
base temperature of the refrigerator and that the thermal contribution
dominates the force noise acting on the cantilevers used in the persistent
current measurements.

In the next two figures, we display another characterization of the
cantilever's thermal motion, this time measured while the cantilever
was subjected to an external driving force. In the measurement discussed
above, the cantilever was not subjected to an external driving force.
In order to describe this measurement, we use the quadrature notation
introduced in Eqs. \ref{eq:ChSensitivity_QuadraturesX1X2} and \ref{eq:CHSensitivity_FreqErrorXt}.
We alter the notation slightly to write the cantilever tip displacement
$x(t)$ as 
\begin{align}
x\left(t\right) & =x_{\max}\left(t\right)\cos\left(\omega t+\phi\left(t\right)\right)\nonumber \\
 & =x_{1}\left(t\right)\cos\omega t+x_{2}\left(t\right)\sin\omega t.\label{eq:CHSensitivity_QuadAmplitudes}
\end{align}
For $x\ll\lambda/2\pi$, the photodetector voltage is proportional
to $x$, as explained in \ref{sub:CHExpSetup_CalibrationDriveMotion}.
In this case, the quadratures measured by lock-in 2 in Fig. \ref{fig:CHExpSetup_CantileverMeasurementSchematic}
are proportional to $x_{1}$ and $x_{2}$. The conversion from lock-in
voltage to cantilever tip displacement was given in section \ref{sub:CHExpSetup_CalibrationDriveMotion}. 

\begin{figure}

\begin{centering}
\includegraphics[width=0.7\paperwidth]{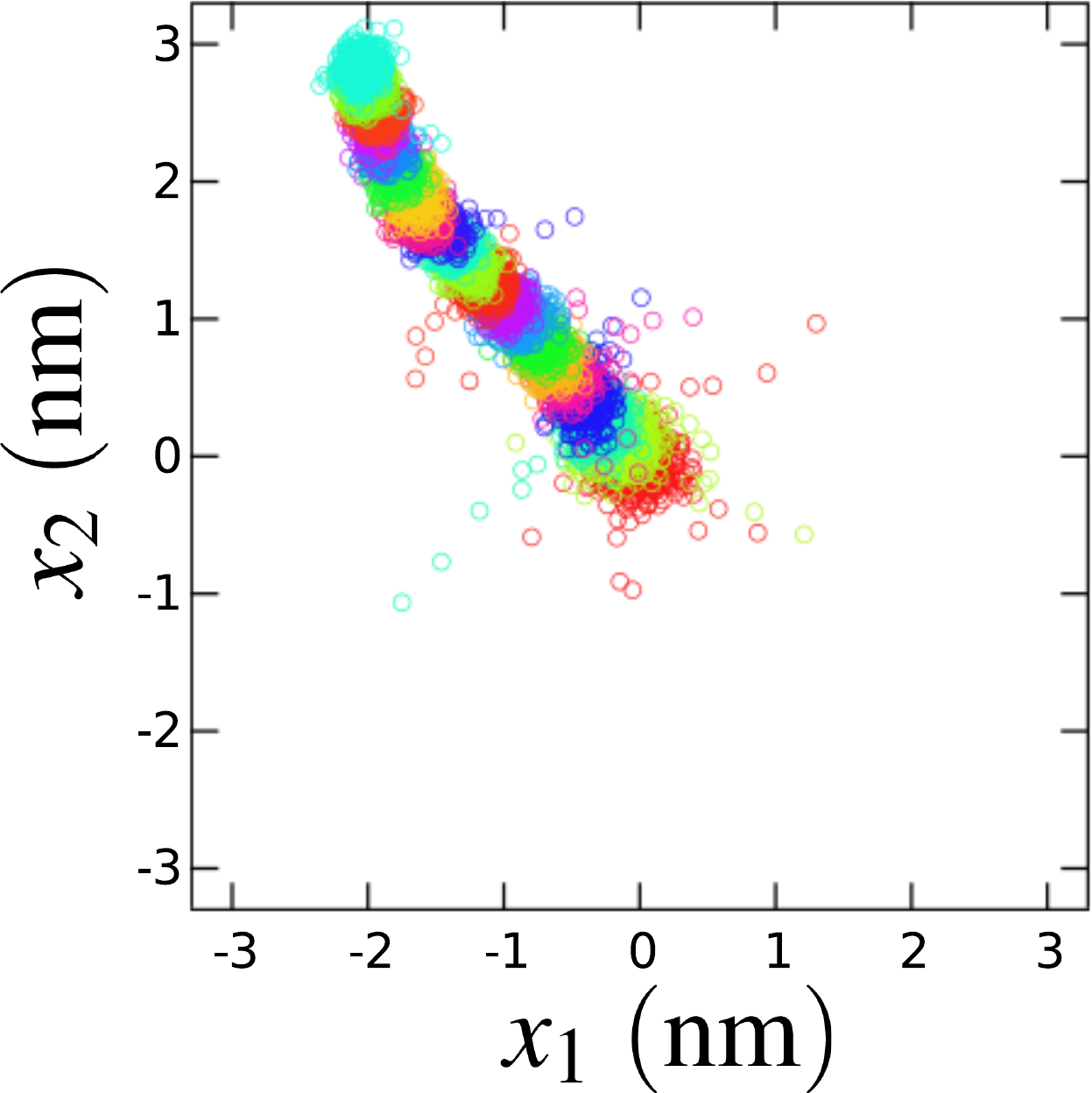}
\par\end{centering}

\caption[Quadratures of cantilever motion at a series of amplitudes of motion]{\label{fig:CHSensitivity_NoiseEllipses}Quadratures of cantilever
motion at a series of amplitudes of motion. Data points points represent
values of $x_{1}$ and $x_{2}$ measured by the method described in
the text for sample CL14 at $T=323\,\text{mK}$. The quadratures were
measured at a series of piezo drive voltages. Different values of
the driving voltage are indicated by differently colored data points
in the figure. The driving voltage was stepped to produce roughly
$0.18\,\text{nm}$ increments in the mean amplitude $\langle x_{\max}\rangle$.
At each drive setting, 400 data points were taken at approximately
$5\,\text{s}$ intervals. A linear relationship between $x_{1}$ and
$x_{2}$ is expected for a constant drive frequency. The slope of
this line is related to the phase between the piezo drive voltage
and the cantilever motion, but extra phase shifts in the measurement
chain make the exact value of this phase uninteresting. Deviations
from the linear relationship indicate a change in this phase and are
likely due to the temporal drift of the cantilever's resonant frequency
(see Fig. \ref{fig:CHExpSetup_SHOMagPhase}) which is not monitored
during the measurement. As expected, the scatter about $(\langle x_{1}\rangle,\langle x_{2}\rangle)$
is uniform and independent of $\langle x_{\max}\rangle$.}

\end{figure}

We measured the quadratures $x_{1}$ and $x_{2}$ while driving the
cantilever to a series small amplitudes $x_{\max}\ll\lambda/2\pi$.
For these measurements, lock-in 2 was not operated in a phase-locked
loop and instead took its reference from a stabilized clock signal
at the frequency measured to be the cantilever's resonant frequency
just before the measurement began.%
\footnote{The clock signal is provided by an arbitrary waveform generator (33220A
20 MHz Function / Arbitrary Waveform Generator, Agilent, Santa Clara,
CA) which is stabilized by the 10 MHz clock output of the frequency
counter described in \ref{sub:CHExpSetup_CantileverElectronics}.%
} The measured quadratures for drives between 0 and $3\,\text{nm}$
are shown for sample CL14 in Fig. \ref{fig:CHSensitivity_NoiseEllipses}.

When the lock-in's time constant is much less than the ringdown time
$\tau$ of the cantilever, the timescale over which the cantilever
amplitude can change appreciably, the lock-in voltage provides a real-time
estimate of $x_{1}(t)$ and $x_{2}(t)$. By measuring $x_{1}$ and
$x_{2}$ many times over a time span $t_{M}\gg\tau$, a statistically
significant sample size of the random component of the cantilever's
motion can be gathered. For CL14, the ringdown time $\tau$ was approximately
$30\,\text{s}$. The data in Fig. \ref{fig:CHSensitivity_NoiseEllipses}
was taken with a lock-in time constant of $100\,\text{ms}$. At each
value of the piezo driving voltage, the quadratures $x_{1}$ and $x_{2}$
were recorded approximately every $5\,\text{s}$ over a $2000\,\text{s}$
measurement time.

Because the cantilever equation of motion, Eq. \ref{eq:CHTorsMagnCantEquationMotionTime},
is linear, we can write $x_{i}(t)=x_{i,D}+x_{i,N}(t)$ for $i=1,2$
where $x_{i,D}$ is the quadrature amplitude due to the external resonant
driving force which is constant in time and $x_{i,N}(t)$ is the quadrature
amplitude due to the white fluctuating noise force. The time average
of the quadrature amplitudes is $\langle x_{i}\rangle=x_{i,D}$. The
standard deviation $\sigma(x_{i})$ of the quadrature amplitudes over
time is given by
\begin{align*}
\sigma^{2}\left(x_{i}\right) & =\left\langle \left(x_{i}-\left\langle x_{i}\right\rangle \right)^{2}\right\rangle \\
 & =\left\langle x_{i,N}^{2}\right\rangle .
\end{align*}
Since the $x_{i,N}$ are random and result from a white noise force
with no special phase reference, we can take $\langle x_{i,N}^{2}\rangle=\langle x_{1,N}^{2}\rangle=\langle x_{2,N}^{2}\rangle$.
It appears that the noise source producing the scatter in Fig. \ref{fig:CHSensitivity_NoiseEllipses}
has no phase reference since the data set associated with each cantilever
amplitude appears to have the same width in $x_{1}$ as in $x_{2}$.

To discuss the total scatter of the cantilever amplitude, it is convenient
to model the cantilever amplitude as a two component vector $\vec{x}_{\max}=x_{1}\hat{i}+x_{2}\hat{j}$.
We have already done this to some extent by plotting the quadrature
amplitudes together in the $x_{1}x_{2}$-plane in Fig. \ref{fig:CHSensitivity_NoiseEllipses}.
From the two lines in Eq. \ref{eq:CHSensitivity_QuadAmplitudes},
we can write 
\[
x_{1}=x_{\max}\cos\phi
\]
\[
x_{2}=-x_{\max}\sin\phi
\]
from which it follows that 
\[
x_{\max}^{2}=x_{1}^{2}+x_{2}^{2}.
\]
Thus, the vectorial notation is justified as $|\vec{x}_{\max}|=x_{\max}$.
With this $\vec{x}_{\max}$, we can define the time averages 
\begin{align*}
\left\langle \vec{x}_{\max}\right\rangle  & =\left\langle x_{1}\right\rangle \hat{i}+\left\langle x_{2}\right\rangle \hat{j}\\
 & =x_{1,D}\hat{i}+x_{2,D}\hat{j}
\end{align*}
and
\begin{align*}
\left\langle x_{\max}\right\rangle  & =\left|\left\langle \vec{x}_{\max}\right\rangle \right|\\
 & =\sqrt{\left\langle x_{1}\right\rangle ^{2}+\left\langle x_{2}\right\rangle ^{2}}\\
 & =\sqrt{x_{1,D}^{2}+x_{2,D}^{2}}.
\end{align*}
We can also define the typical size $\sigma(x_{\max})$ of the fluctuations
of $\vec{x}_{\max}$ by
\begin{align}
\sigma^{2}(x_{\max}) & =\left\langle \left|\vec{x}_{\max}-\left\langle \vec{x}_{\max}\right\rangle \right|^{2}\right\rangle \nonumber \\
 & =\left\langle \left|x_{1,N}\hat{i}+x_{2,N}\hat{j}\right|^{2}\right\rangle \nonumber \\
 & =\sigma^{2}\left(x_{1}\right)+\sigma^{2}\left(x_{2}\right).\label{eq:ChSensitivity_SigmaXmax}
\end{align}
In the absence of external drive, we have 
\begin{align*}
\left\langle x^{2}\right\rangle  & =\left\langle x_{\max}^{2}\right\rangle /2\\
 & =\sigma^{2}\left(x_{\max}\right)/2.
\end{align*}
Thus, by the Equipartition theorem, we have $\sigma(x_{\max})=\sqrt{2\langle x^{2}\rangle}=\sqrt{2k_{B}T/k}$.
Since the driving voltage introduces a negligible force noise compared
to the thermal force noise (see \ref{sec:CHSensitivity_FrequencyError}),
the typical deviation $\sigma(x_{\max})$ from the mean amplitude
vector $\langle\vec{x}_{\max}\rangle$ should be independent of amplitude
$\langle x_{\max}\rangle$.

\begin{figure}
\begin{centering}
\includegraphics[width=0.7\paperwidth]{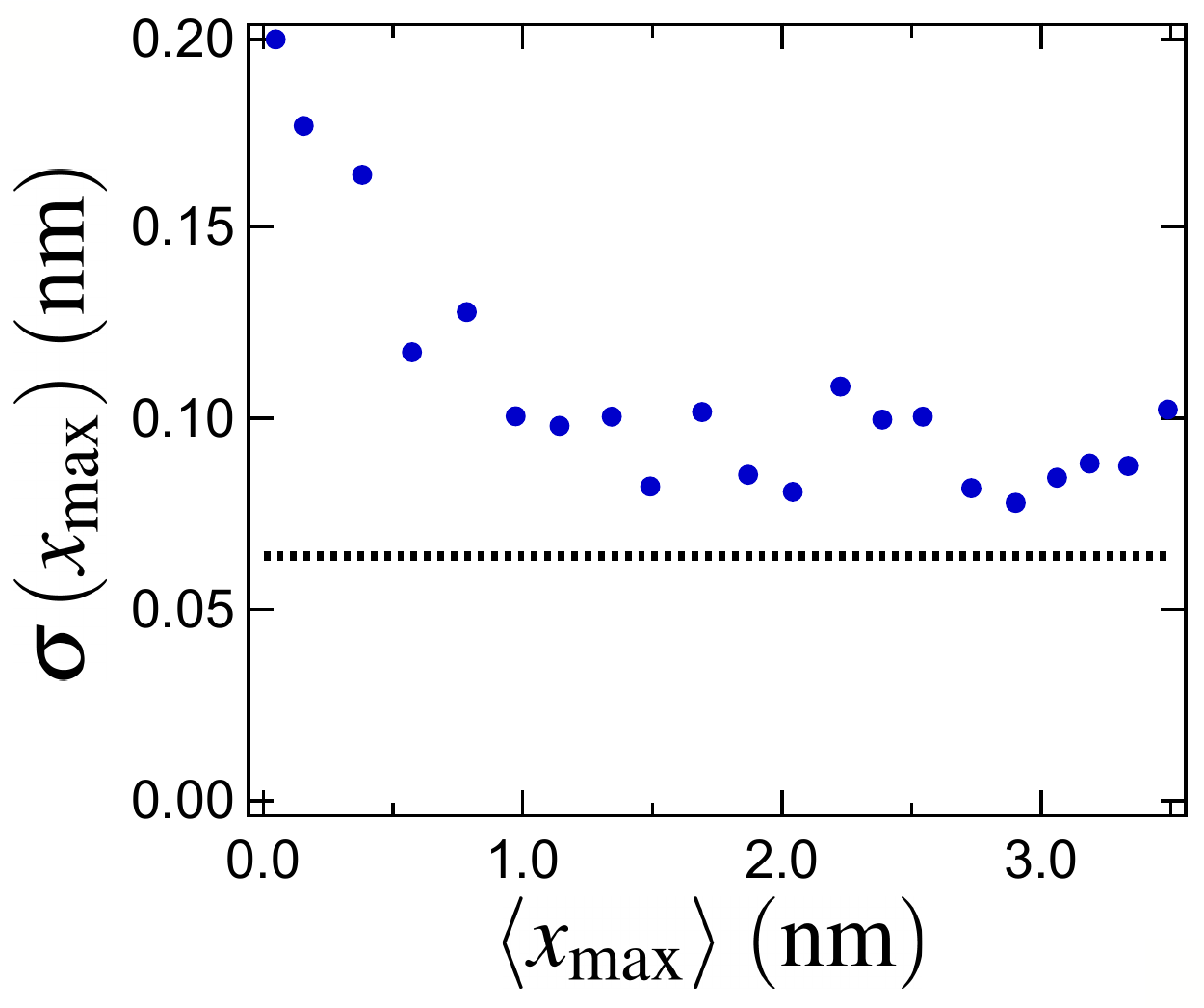}
\par\end{centering}

\caption[Standard deviation $\sigma(x_{\max})$ of cantilever amplitude versus
mean cantilever amplitude $\langle x_{\max}\rangle$]{\label{fig:CHSensitivity_sigmaXvsX}Standard deviation $\sigma(x_{\max})$
of cantilever amplitude versus mean cantilever amplitude $\langle x_{\max}\rangle$.
The data points represent the values of $\sigma(x_{\max})$ calculated
from the data in Fig. \ref{fig:CHSensitivity_NoiseEllipses} with
Eq. \ref{eq:ChSensitivity_SigmaXmax} plotted against the corresponding
values of $\langle x_{\max}\rangle=\sqrt{\langle x_{1}\rangle^{2}+\langle x_{2}\rangle^{2}}$
for each piezo drive setting. The dashed line indicates $\sigma(x_{\max})=\sqrt{2k_{B}T/k}$
for measured temperature $T=323\,\text{mK}$ and estimated spring
constant $k=1.08\times10^{-3}\,\text{N/m}$ of sample CL14.}
\end{figure}

In Fig. \ref{fig:CHSensitivity_sigmaXvsX}, we plot $\sigma(x_{\max})$
for each set of quadrature data shown in Fig. \ref{fig:CHSensitivity_NoiseEllipses}
as a function of the mean amplitude $\langle x_{\max}\rangle$ associated
with that set. We also mark $\sigma(x_{\max})=3.5\times10^{-9}\,\text{nm}$,
the value of the thermal limit expected for $T=323\,\text{mK}$ and
$k=1.08\times10^{-3}\,\text{N/m}$. Over most of the range of amplitudes,
$\sigma(x_{\max})$ is close to the thermal limit, indicating that
the piezo drive of the cantilever does not introduce any additional
force noise. The larger magnitude of $\sigma(x_{\max})$ at low drive
is not understood.

\begin{figure}

\begin{centering}
\includegraphics[width=0.7\paperwidth]{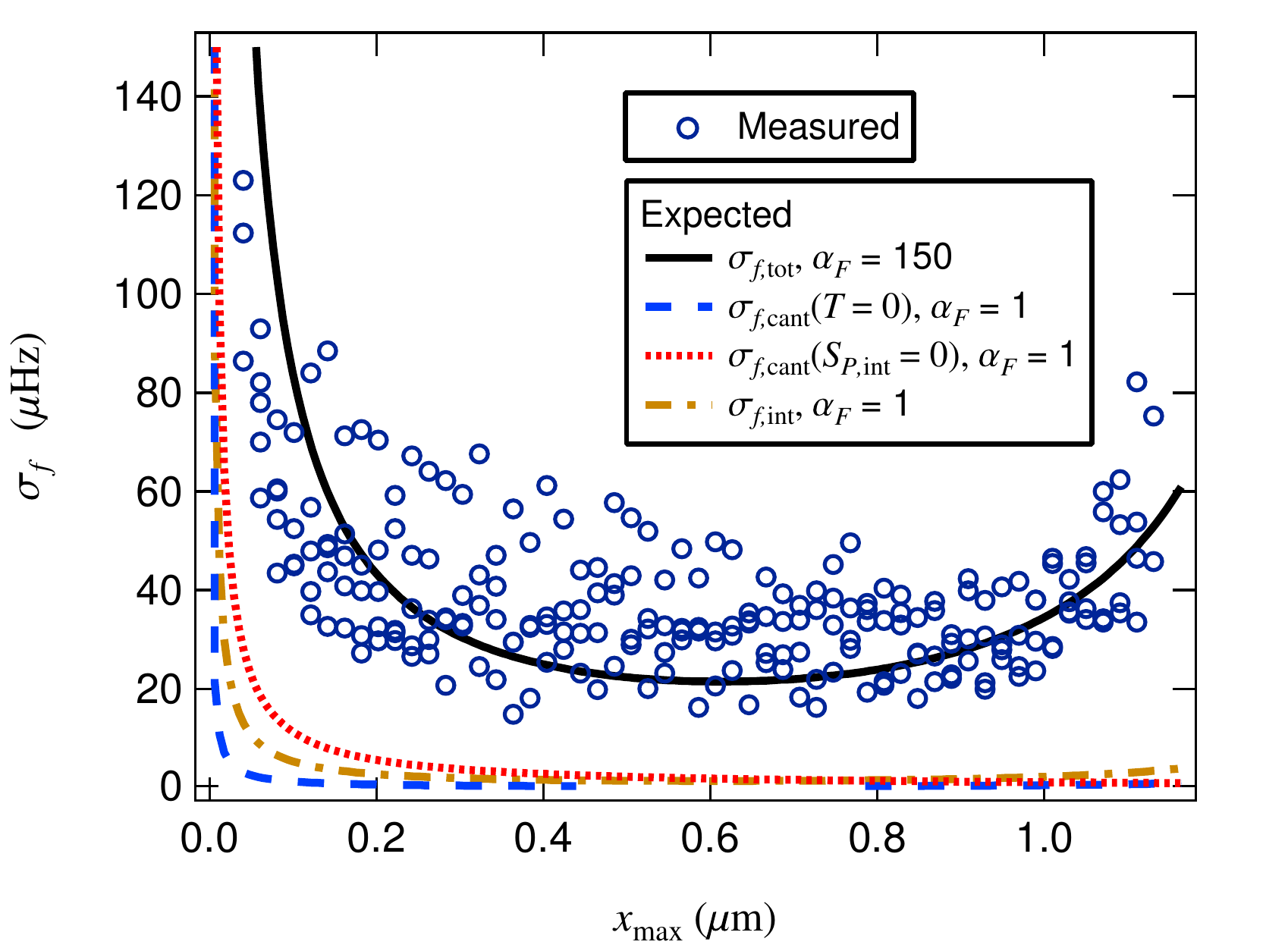}
\par\end{centering}

\caption[Measured frequency uncertainty as a function of cantilever amplitude]{\label{fig:CHSensitivity_dFvsxMax}Measured frequency scatter as
a function of cantilever amplitude. The measured $\sigma_{f}$ data
points (circles) represent the standard deviations of the cantilever
frequency calculated from sets of twenty measurements taken at each
value of the cantilever amplitude $x_{\max}$. The curves represent
the expected frequency uncertainty calculated from Eqs. \ref{eq:CHSensitivity_sigmaFthermal}
through \ref{eq:ChSensitivity_sigmaFtot}. For the parameters given
in the text and $\alpha_{F}=1$, it is expected that $\sigma_{f,\text{tot}}$
will be dominated by the thermal contribution $\sigma_{f,\text{cant}}(S_{P,\text{int}}=0)$
except at the highest drives where the response of the interferometer
drops to zero and the frequency uncertainty $\sigma_{f,\text{int}}$
added during the measurement by noise in the interferometer dominates.
The contribution $\sigma_{f,\text{cant}}(T=0)$ scales with $x_{\max}$
in the same way as $\sigma_{f,\text{int}}$ and is smaller than $\sigma_{f,\text{int}}$
by a factor of $\sim4$ for set of parameters given in the text and
$\alpha_{F}=1$. Also shown for reference is a curve of the expected
$\sigma_{f,\text{tot}}$ for $\alpha_{F}=150$. This curve, which
is dominated by the $\sigma_{f,\text{int}}$ contribution, matches
the data better the curves for $\alpha_{F}=1$ and is $\sim10$ times
larger than $\sigma_{f,\text{tot}}(\alpha_{F}=1)$.}

\end{figure}

Finally, in Fig. \ref{fig:CHSensitivity_dFvsxMax}, we present a measurement
of the noise $\sigma_{f}$ in the frequency measurement as a function
of cantilever amplitude $x_{\max}$. The data shown in the figure
were taken from measurements on sample CL17 of Table \ref{tab:ChData_CLs}.
For this measurement, the cantilever was driven in a phase-locked
loop, and the driving voltage to the piezo was stepped in small increments
so that the amplitude of motion observed by the fiber $x_{f,\max}$
spanned the full range of the first peak of $J_{1}(4\pi x_{f,\max}/\lambda)$
(see Eq. \ref{eq:CHExpSetup_1stHarmInterferometer} describing the
interferometer response). At each drive amplitude, the cantilever
frequency was measured twenty times using the frequency counter with
a gate time of $\tau_{M}=5\,\text{s}$.%
\footnote{The data shown in Fig. \ref{fig:CHSensitivity_dFvsxMax} were taken
during a measurement like the one shown in Fig. \ref{fig:ChData_DIA2AmpScan},
which plots the frequency shift versus cantilever amplitude. At each
cantilever amplitude, the frequency was measured twenty times and
averaged for Fig. \ref{fig:ChData_DIA2AmpScan}. The standard deviation
of those frequency measurements was calculated for Fig. \ref{fig:CHSensitivity_dFvsxMax}.
I note this fact to point out that the data shown in Fig. \ref{fig:CHSensitivity_dFvsxMax}
was taken from a measurement primarily intended to measure the persistent
current, not to characterize the measurement uncertainty. Some parameters
in the analysis below, namely $S_{P,\text{int}}$, $R_{c}$, and $Q$,
were inferred by me from other measurements taken at about the same
time as the measurement shown in Fig. \ref{fig:CHSensitivity_dFvsxMax}.
If these parameters had changed over time from the values I used,
some systematic uncertainty would be introduced to the analysis. It
is unlikely that any of these parameters would change by much more
than a factor of 4.%
} The data shown (circles) in Fig. \ref{fig:CHSensitivity_dFvsxMax}
represents the standard deviation of each of these twenty frequency
measurements.

Fig. \ref{fig:CHSensitivity_dFvsxMax} also displays curves representing
the frequency scatter expected from noise in the interferometric measurement
and from thermal force noise. Using Eq. \ref{eq:CHSensitivity_FreqErrorFullExpression},
we write the standard deviation of the frequency measurement due to
the noise of the interferometric measurement as

The curves distinguish three contributions to the total frequency
uncertainty $\sigma_{f,\text{tot}}$ given in Eq. \ref{eq:ChSensitivity_TotFreqUncertainty}:
the typical fluctuation $\sigma_{f,\text{cant}}(S_{P,\text{int}}=0)$
of the cantilever's frequency of motion in the absence of detector
noise given by
\begin{align}
\sigma_{f,\text{cant}}^{2}\left(S_{P,\text{int}}=0\right) & =\frac{f_{0}^{2}}{4\tau_{M}}\frac{S_{F,\text{th}}}{\left(kx_{\max}\right)^{2}}\left(1-\frac{1}{\alpha_{F}}\left(1-\exp\left(-\alpha_{F}\right)\right)\right)\nonumber \\
 & =\frac{1}{2\pi\tau_{M}}\frac{k_{B}Tf_{0}}{Qkx_{\max}^{2}}\left(1-\frac{1}{\alpha_{F}}\left(1-\exp\left(-\alpha_{F}\right)\right)\right),\label{eq:CHSensitivity_sigmaFthermal}
\end{align}
the typical fluctuation $\sigma_{f,\text{cant}}(T=0)$ of the cantilever's
frequency of motion in the absence of the thermal noise force given
by 
\begin{equation}
\sigma_{f,\text{cant}}^{2}\left(T=0\right)=\frac{f_{0}^{2}}{4\tau_{M}}\frac{S_{P,\text{int}}}{Q^{2}\left(R\left(x_{\max}\right)x_{\max}\right)^{2}}\left(1-\frac{1}{\alpha_{F}}\left(1-\exp\left(-\alpha_{F}\right)\right)\right),\label{eq:CHSensitivity_sigmaFcantT0}
\end{equation}
and the typical fluctuation $\sigma_{f,\text{int}}$ in the measured
frequency due to noise in the interferometer signal given by
\begin{equation}
\sigma_{f,\text{int}}^{2}=\frac{1}{4\pi^{2}}\frac{\alpha_{F}}{\tau_{M}^{3}}\frac{S_{P,\text{int}}}{\left(R\left(x_{\max}\right)x_{\max}\right)^{2}}\left(1-\exp\left(-\alpha_{F}\right)\right).\label{eq:ChSensitivity_sigmaFint}
\end{equation}
The total frequency uncertainty can be written as 
\begin{equation}
\sigma_{f,\text{tot}}=\sqrt{\sigma_{f,\text{cant}}^{2}\left(S_{P,\text{int}}=0\right)+\sigma_{f,\text{cant}}^{2}\left(T=0\right)+\sigma_{f,\text{int}}^{2}}.\label{eq:ChSensitivity_sigmaFtot}
\end{equation}
In these definitions, we have used $\alpha_{F}=\tau_{F}/\tau_{M}$
as was done in Section \ref{sec:CHSensitivity_FrequencyError}. In
addition to the dimensions, spring constant and resonant frequency
given for sample CL17 in Table \ref{tab:ChData_CLs}, we use the observed
parameters of $Q=1.1\times10^{5}$, $P_{\text{inc}}=24\,\text{nW}$,
$R_{f}=.23$, $R_{c}=0.009$, $\epsilon\approx6.7\times10^{-3}\,\text{nm}^{-1}$,
$S_{P,\text{int}}=0.29\,\text{pW}/\sqrt{\text{Hz}}$ and $U=U_{1}(249/449)=0.40$
and $T=323\,\text{mK}$ to plot the various contributions to the frequency
uncertainty. Curves representing $\sigma_{f,\text{cant}}(S_{P,\text{int}}=0)$,
$\sigma_{f,\text{cant}}(T=0)$, and $\sigma_{f,\text{int}}$ are each
plotted assuming ideal filtering for which $\alpha_{F}=1$. The measured
frequency scatter $\sigma_{f}$ is much larger than any of these curves
and seems to agree with the total frequency uncertainty $\sigma_{f,\text{tot}}$
calculated using the above parameters and $\alpha_{F}=150$. The total
frequency uncertainty $\sigma_{f,\text{tot}}$ calculated for $\alpha_{F}=150$
is $\sim10$ times larger than the curve of $\sigma_{f,\text{tot}}$
found by assuming $\alpha_{F}=1$.

The observed $35\,\mu\text{Hz}$ minimum for $\sigma_{f}$ was typical
of all samples in Table \ref{tab:ChData_CLs}. The measurement shown
in Fig. \ref{fig:CHSensitivity_dFvsxMax} was performed with the fiber
$z_{f}=249\,\text{\ensuremath{\mu}m}$ from the base of the $449\,\text{\ensuremath{\mu}m}$
long cantilever. For most persistent current measurements discussed
in this text, the fiber was positioned closer to the cantilever tip.
This positioning results in a larger value for $U$ and thus rescales
the dependence of $\sigma_{f,\text{int}}$ and $\sigma_{f,\text{cant}}(T=0)$
on $x_{\max}$. The net effect of this rescaling is that the values
of $x_{\max}$ indicated in Fig.\ref{fig:CHSensitivity_dFvsxMax}
are divided by $\sim2$ for $\sigma_{f,\text{int}}$ and $\sigma_{f,\text{cant}}(T=0)$.
The dependence of the measured $\sigma_{f}$ on $x_{\max}$ obeyed
a similar rescaling with fiber position, with the relatively flat
region of $\sigma_{f}$ occurring between $x_{\max}\approx100\,\text{nm}$
and $x_{\max}\approx500\,\text{nm}$ (rather than between $200\,\text{nm}$
and $1\,\text{\ensuremath{\mu}m}$ as in Fig. \ref{fig:CHSensitivity_dFvsxMax}).
During the persistent current measurements, the cantilever amplitude
$x_{\max}$ was on the order of $0.1\,\text{\ensuremath{\mu}m}$,
where the observed $\sigma_{f}\approx35\,\text{\ensuremath{\mu}Hz}$
was only a factor of $\sim3$ larger than the expected magnitude of
$\sigma_{f,\text{cant}}(S_{P,\text{int}}=0)$, the contribution due
to thermal force noise. Further study is necessary to understand the
deviation of $\sigma_{f}$ from the expected value in Fig. \ref{fig:CHSensitivity_dFvsxMax}.%
\footnote{Such study has been undertaken by my successors in the Harris Lab.
In preliminary analysis, the transition of $\sigma_{f,\text{tot}}$
from the regime dominated by $\sigma_{f,\text{int}}$ to that dominated
by $\sigma_{f,\text{cant}}(S_{P,\text{int}}=0)$ (the thermally limited
frequency uncertainty) has been observed by changing the measurement
time $\tau_{M}$. These measurements were performed on thinner, shorter
cantilevers than those listed in Table \ref{tab:ChData_CLs}. Decreasing
the cantilever thickness increases $\sigma_{f,\text{cant}}(S_{P,\text{int}}=0)$
while decreasing $\sigma_{f,\text{cant}}(T=0)$ and leaving $\sigma_{f,\text{int}}$
unchanged and thus overall increases the relative importance of the
$\sigma_{f,\text{cant}}(S_{P,\text{int}}=0)$ contribution. Additionally,
the shorter cantilevers had higher frequencies and lower quality factors
and ringdown times. Consequently shorter measurement times, necessary
to observe the cross-over from $\sigma_{f,\text{int}}$ to $\sigma_{f,\text{cant}}(S_{P,\text{int}}=0)$,
were possible. For these measurements, $\alpha_{F}$ was found to
be $\sim1$. Thermally limited frequency uncertainty was also observed
for cantilevers with lengths similar to those in Table \ref{tab:ChData_CLs}
(but with $t=110\,\text{nm}$).%
} However, the observed $\sigma_{f}$ was sufficiently low to measure
persistent currents with a signal to noise ratio over 30. These measurements
will be discussed in the next chapter.

\chapter{\label{cha:Data}Data from and analysis of persistent current measurements}

In this chapter, we first outline the procedure used to convert the
measured trace of cantilever frequency versus magnetic field into
persistent current versus magnetic field and then present the results
of measurements of persistent currents in four different samples.

\section{\label{sec:ChData_SigProc}Signal processing of the measured cantilever
frequency shift}

\subsection{\label{sub:ChData_SigProcDescription}Description of signal processing}

In \ref{sec:CHTorsMagn_FiniteDriveSection}, we found that a persistent
current signal of the form
\begin{equation}
I\left(B\right)=\text{Im}\left[\sum_{p}I_{p}e^{2\pi ip\beta_{1}B}e^{i\psi_{p}}\right]\label{eq:ChData_IofB}
\end{equation}
led to a cantilever frequency shift
\begin{equation}
\Delta f\left(B\right)=\text{Im}\left[FB^{2}\sum_{p}2\pi ip\beta_{1}I_{p}e^{2\pi ip\beta_{1}B}e^{i\psi_{p}}\mathrm{jinc}\left(2\pi p\beta_{1}GB\right)\right].\label{eq:ChData_DeltaF}
\end{equation}
In the preceding lines we have rewritten Eqs. \ref{eq:CHTorsMagnPCHarmonicExpansion}
and \ref{eq:CHTorsMagn_FiniteAmpFreqShift} with $\beta_{1}=A\sin\theta_{0}/\phi_{0}$,
\[
F=-\sqrt{N}\frac{f_{0}}{2k}\left(\frac{\alpha}{l}\right)^{2}\frac{A\cos^{2}\theta_{0}}{\sin\theta_{0}},
\]
and $G=\alpha x_{\max}/l\tan\theta_{0}$. The quantity $\beta_{1}$
is the magnetic field frequency whose period corresponds to a magnetic
flux of $\phi_{0}$ threading a ring of area $A$ lying on a plane
at an angle $\theta_{0}$ relative to the applied magnetic field $B$
(see Figs. \ref{fig:LabeledUnflexedCantilever} and \ref{fig:Flexed-cantilever-schematic}
for more clarification). If the applied magnetic field were threaded
entirely through the ring's hole, the persistent current oscillation
would possess non-zero coefficients $I_{p}$ only for magnetic field
frequencies $p\beta_{1}$ with $p$ an integer. As described in \ref{sub:CHPCTh_FluxThroughMetal},
the magnetic field penetrating the metal of the ring introduces a
finite range of correlation to the persistent current oscillation.
This finite correlation broadens the peaks in the magnetic field frequency
spectrum around each $p\beta_{1}$, allowing the coefficients $I_{p}$
to be non-zero even for non-integer $p$. In the definition of $F$,
we have included a factor of $\sqrt{N}$, where $N$ is the number
of rings on the cantilever. Following the discussion of \ref{sec:CHPCTh_DiffusiveRegime},
we expect that the amplitude $I_{p}$ should vary randomly from ring
to ring. Due to the finite magnetic field correlation, the phase $\psi_{p}$
should also grow stochastically with applied field and be random for
the values of magnetic applied in the measurements discussed below.
Thus the amplitude of the total current in the array should be random
with a typical magnitude $\sqrt{N}$ larger than that of a single
ring. Because of the inclusion of the $\sqrt{N}$ factor in $F$,
all quantities derived below representing current are scaled to correspond
to the typical single ring current.

In the limit of small cantilever amplitude ($\beta_{1}GB\ll1$ for
the range of $B$ of interest), the $\text{jinc}$ term is approximately
unity, and the $p^{th}$ complex Fourier coefficient $I_{p}e^{i\psi_{p}}$
of the persistent current is proportional to the $p^{th}$ complex
Fourier coefficient of $\Delta f\left(B\right)/B^{2}$. At finite
cantilever amplitude, inference of the persistent current $I\left(B\right)$
from the cantilever frequency shift $\Delta f\left(B\right)$ is complicated
by the $\text{jinc}(2\pi p\beta_{1}GB)$ term, which depends on both
the Fourier transform index $p$ and the magnetic field $B$. We will
now discuss two methods, which we will refer to as method A and method
B, for estimating the persistent current $I(B)$ from the frequency
shift $\Delta f(B)$ measured at finite cantilever amplitude.

For method A, we scale the entire trace $\Delta f(B)$ by a different
function of $B$ and $p$ for each value of the index $p$ and then
infer the $p^{th}$ component $I_{p}$ from the Fourier transform
of this scaled trace. Specifically, for a set of $M+1$ measurements
of $\Delta f(B)$ taken at regular intervals $\Delta B=(B_{\max}-B_{\min})/M$
between $B_{\min}$ and $B_{\max}$ (with $B_{\max}>B_{\min}>0$),
we calculate 
\begin{equation}
dI_{p}^{A}=\frac{1}{B_{\max}-B_{\min}}\sum_{n=0}^{M}\Delta B\,\frac{\Delta f\left(B_{\min}+n\Delta B\right)}{F\left(B_{\min}+n\Delta B\right)^{2}\text{jinc}\left(2\pi p\beta_{1}G\left(B_{\min}+n\Delta B\right)\right)}e^{-2\pi ip\beta_{1}\left(B_{\min}+n\Delta B\right)},\label{eq:ChData_dIpA}
\end{equation}
which is the discrete form of the Fourier transform
\[
\frac{1}{B_{\max}-B_{\min}}\int_{B_{\min}}^{B_{\max}}dB\,\frac{\Delta f\left(B\right)}{FB^{2}\text{jinc}\left(2\pi p\beta_{1}GB\right)}e^{-2\pi ip\beta_{1}B}.
\]
For $p\beta_{1}GB_{\max}\apprle0.4$ for which the $\text{jinc}$
function in the denominator does not pass through zero (see Fig. \ref{fig:CHTorsMagn_JincPlot})
and for $B_{\max}-B_{\min}\gg1/p\beta_{1}$ (with all the usual caveats
related to Fourier transform windowing and the Nyquist-Shannon sampling
theorem), the coefficient $dI_{p}^{A}$ satisfies
\[
dI_{p}^{A}\approx2\pi ip\beta_{1}I_{p}e^{i\psi_{p}},
\]
and thus we can define the approximate derivative of the current obtained
by method A:

\begin{align*}
\frac{\partial I^{A}}{\partial B} & =\text{Im}\left[\sum_{p}dI_{p}^{A}e^{2\pi ip\beta_{1}B}\right]\\
 & \approx\text{Im}\left[\sum_{p}2\pi ip\beta_{1}I_{p}e^{2\pi ip\beta_{1}B}e^{i\psi_{p}}\right]\\
 & \approx\frac{\partial I}{\partial B}.
\end{align*}
Finally, we can numerically integrate $\partial I^{A}/\partial B$
with respect to $B$ to obtain $I^{A}(B)$, the trace of persistent
current versus magnetic field as estimated from $\Delta f(B)$ by
method A.

Now we describe method B. The major difference between method A and
method B is that in method B the trace $\Delta f(B)$ is scaled by
a single function when performing Fourier analysis (rather than scaling
$\Delta f(B)$ by a different function for each Fourier index $p$
as was in done in method A). For method B, the trace $\Delta f(B)$
is divided by $FB^{2}\text{jinc}(2\pi\beta_{1}GB)$. In method A,
this factor is used to convert from $\Delta f\left(B\right)$ to $\partial I/\partial B$
for $p=1$ (i.e. for magnetic field frequency $\beta=\beta_{1}$).
As with the definition of $dI_{p}^{A}$ in method A, we define the
coefficient 
\[
dI_{p}^{B}=\frac{1}{B_{\max}-B_{\min}}\sum_{n=0}^{M}\Delta B\,\frac{\Delta f\left(B_{\min}+n\Delta B\right)}{F\left(B_{\min}+n\Delta B\right)^{2}\text{jinc}\left(2\pi\beta_{1}G\left(B_{\min}+n\Delta B\right)\right)}e^{-2\pi ip\beta_{1}\left(B_{\min}+n\Delta B\right)}
\]
 and then the inverse Fourier transform integral
\[
\frac{\partial I^{B}}{\partial B}=\text{Im}\left[\sum_{p}dI_{p}^{B}e^{2\pi ip\beta_{1}B}\right].
\]
We write these expressions simply to mirror the presentation of method
A given above. Because $\Delta f(B)$ is scaled by the same function
for all $p$ when calculating $dI_{p}^{B}$, we can also write 
\[
\frac{\partial I^{B}}{\partial B}=\frac{\Delta f\left(B\right)}{FB^{2}\text{jinc}\left(2\pi\beta_{1}GB\right)}.
\]
When the only contribution to $\Delta f(B)$ is from a persistent
current oscillation $I(B)$ with frequency components close to $\beta_{1}$,
we have 
\[
\frac{\partial I^{B}}{\partial B}\approx\frac{\partial I}{\partial B}.
\]
As with method A, we can numerically integrate $\partial_{B}I^{B}$
to obtain $I^{B}$.

The preceding steps are adequate for a measurement free from noise
and systematic error. In practice, it is necessary to remove a smooth
background from the $\Delta f(B)$ trace found by using either a polynomial
fit or a smoothing function prior to finding $I^{A}$ or $I^{B}$.
A second background subtraction is often desirable as a final step
to remove low frequency drift caused by the interaction of the integration
of $\partial_{B}I$ with the noise in the $\Delta f(B)$ trace (i.e.
the remaining low frequency noise that was not removed by the first
background subtraction step).

Typically, method A gives a more accurate estimation of the persistent
current than method B. However, the argument of the $\text{jinc}$
function in the denominator of the expression (Eq. \ref{eq:ChData_dIpA})
defining $dI_{p}^{A}$ will go to zero for some values of the Fourier
index $p$ which we call $p_{\text{zero}}$. Since the first zero
of the $\text{jinc}(x)$ is at $x\approx2\pi\times0.61$, 
\begin{equation}
p_{\text{zero}}\approx\frac{0.61}{\beta_{1}GB_{\min}}\label{eq:ChData_pzero}
\end{equation}
where $B_{\min}$ is the minimum magnitude of the magnetic field in
the trace being analyzed. For $p$ near these $p_{\text{zero}}$,
the coefficients $dI_{p}^{A}$ become unrealistically large. The reason
for this behavior is that for the chosen cantilever amplitude $x_{\max}$
(contained in the factor $G$) the cantilever frequency shift $\Delta f$
is simply not sensitive to the components $p$ of the persistent current
near these $p_{\text{zero}}$ (as can be seen from Eq. \ref{eq:ChData_DeltaF}
or Fig. \ref{fig:CHTorsMagn_JincPlot}). 

In order to avoid introducing a large component to the persistent
current with magnetic field frequency 
\begin{align}
\beta_{\text{zero}} & =p_{\text{zero}}\beta_{1}\nonumber \\
 & =\frac{0.61}{GB_{\min}},\label{eq:ChData_BetaZero}
\end{align}
we set all of the $dI_{p}^{A}$ coefficients with $p\ge p_{\text{zero}}$
to zero before finding $\partial_{B}I^{A}$. This procedure is essentially
a strong low-pass filtering of the data. All of the data discussed
below was analyzed in this way. However, we also analyzed the data
using method B which, though slightly less accurate, requires less
processing of the data. In this way, we verified that the persistent
current inferred by method A did not introduce or remove any significant
features to the persistent current trace $I^{A}(B)$ which were not
present in the frequency shift trace $\Delta f(B)$.

For most of the measurements of samples CL11 and CL15 at $\theta_{0}=6^{\circ}$
and sample CL14 at $45^{\circ}$, the lowest $p_{\text{zero}}$ was
close to $p=2$. Thus these measurements were not sensitive to the
$h/2e$ component of the persistent current oscillation. For these
combinations of sample and angle, we also performed measurements at
reduced cantilever amplitudes and found no evidence of the second
harmonic of the persistent current signal within the uncertainty of
our measurement. This result was not surprising as the expected magnitude
of the higher harmonics of the current based on Eq. \ref{eq:CHPCTh_IIFiniteTZSO}
was below our sensitivity for these samples.

In some instances, we plot the data in the form 
\[
I_{A,B}^{\prime}\left(B\right)\equiv\frac{1}{2\pi\beta_{1}}\frac{\partial I^{A,B}}{\partial B}
\]
and 
\begin{equation}
I_{p}^{\prime A,B}\equiv\frac{1}{2\pi\beta_{1}}dI_{p}^{A,B}\label{eq:ChData_dIPrimeScaling}
\end{equation}
in order to minimize the number of processing steps applied to the
raw data. The quantity $I_{A,B}^{\prime}(B)$ has units of current
and should have the same amplitude as $I(B)$ for features with magnetic
field frequencies close to $\beta_{1}$. The quantity $I_{p}^{\prime A,B}$
also has units of current.

\subsection{\label{sub:ChData_MethodAWalkThrough}Step-by-step walk-through of
signal processing}

We now walk through the conversion of measured cantilever frequency
to persistent current using method A on a typical data set. For this
data set, sample CL15 (see Tables \ref{tab:ChData_CLs} and \ref{tab:ChData_Rings})
was driven in a phase-locked loop as described in \ref{sub:CHExpSetup_CantileverDetectionSetup}
and \ref{sub:ChExpSetup_CantDetectionMeasurement}. The resonant frequency
of the cantilever was measured seven times by the frequency counter
with a gate time of 5 seconds while the magnetic field was held constant.
The magnetic field was then increased by $0.5\,\text{mT}$. This measure
and step process was repeated from $7.3\,\text{T}$ to $7.9\,\text{T}$. 

Prior to beginning the sweep of the magnetic field, the excitation
of the cantilever was calibrated following the procedure described
in \ref{sub:CHExpSetup_CalibrationDriveMotion}. During the scan,
the cantilever was driven with a tip amplitude $x_{\max}=31\,\text{nm}$.
The cantilever was mounted at an angle $\theta_{0}=45^{\circ}$ relative
to the magnetic field, which for CL15 corresponded to a magnetic field
frequency $\beta_{1}=94\,\text{T}^{-1}$ and a magnetic field period
$B_{1}=\beta_{1}^{-1}=10.7\,\text{mT}$. The temperature of the refrigerator
was $360\,\text{mK}$.

\begin{figure}
\begin{centering}
\includegraphics[width=0.6\paperwidth]{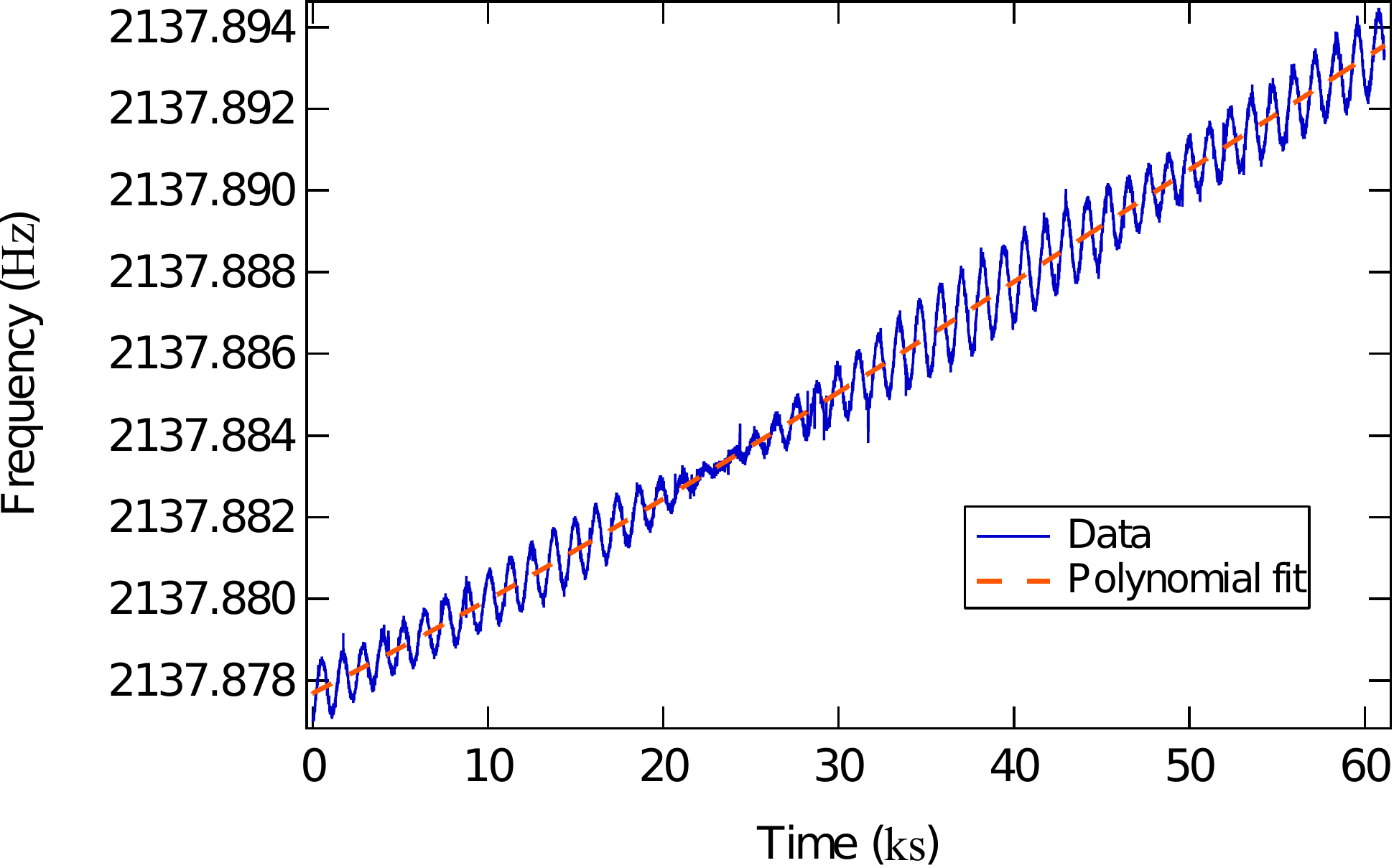}
\par\end{centering}

\caption[Frequency versus time for a typical data set]{\label{fig:ChData_SP1FreqvsTime}Frequency versus time for a typical
data set. The frequency of cantilever CL15 was measured seven times
at a fixed value of magnetic field with each measurement lasting approximately
five seconds. The magnetic field was then increased by $0.5\,\text{mT}$
over a time scale of $20$ seconds. Oscillations of the persistent
current can clearly be seen on top of a time-dependent drift of the
cantilever frequency. Also shown is a third order polynomial fit to
the frequency, which is then subtracted from the frequency data to
produce the curve in Fig. \ref{fig:ChData_SP2dFreqvsB}.}
\end{figure}

Figure \ref{fig:ChData_SP1FreqvsTime} shows the raw frequency data
versus measurement time as well as a third order polynomial fit. Oscillations
due to the persistent current are clearly present in the raw data
on top of a smooth frequency drift. Immediately after the data shown
in Fig. \ref{fig:ChData_SP1FreqvsTime} was taken, the direction of
the magnetic field step was reversed. While the oscillations were
also reversed, the nearly linear drift was unaffected. This observation
indicates that the drift of the cantilever's resonant frequency was
purely time-dependent and independent of magnetic field.%
\footnote{It was not always the case that the frequency drift was independent
of magnetic field. This field dependence could possibly be due to
a magnetic field dependence of the cantilever quality factor. Alternatively,
an electrostatic interaction between the cantilever and detection
fiber could lead to a magnetic field dependence of the cantilever
frequency if the magnetic field caused the sample stage to tilt and
thus the distance between the cantilever and fiber to change.%
}

\begin{figure}
\begin{centering}
\includegraphics[width=0.7\paperwidth]{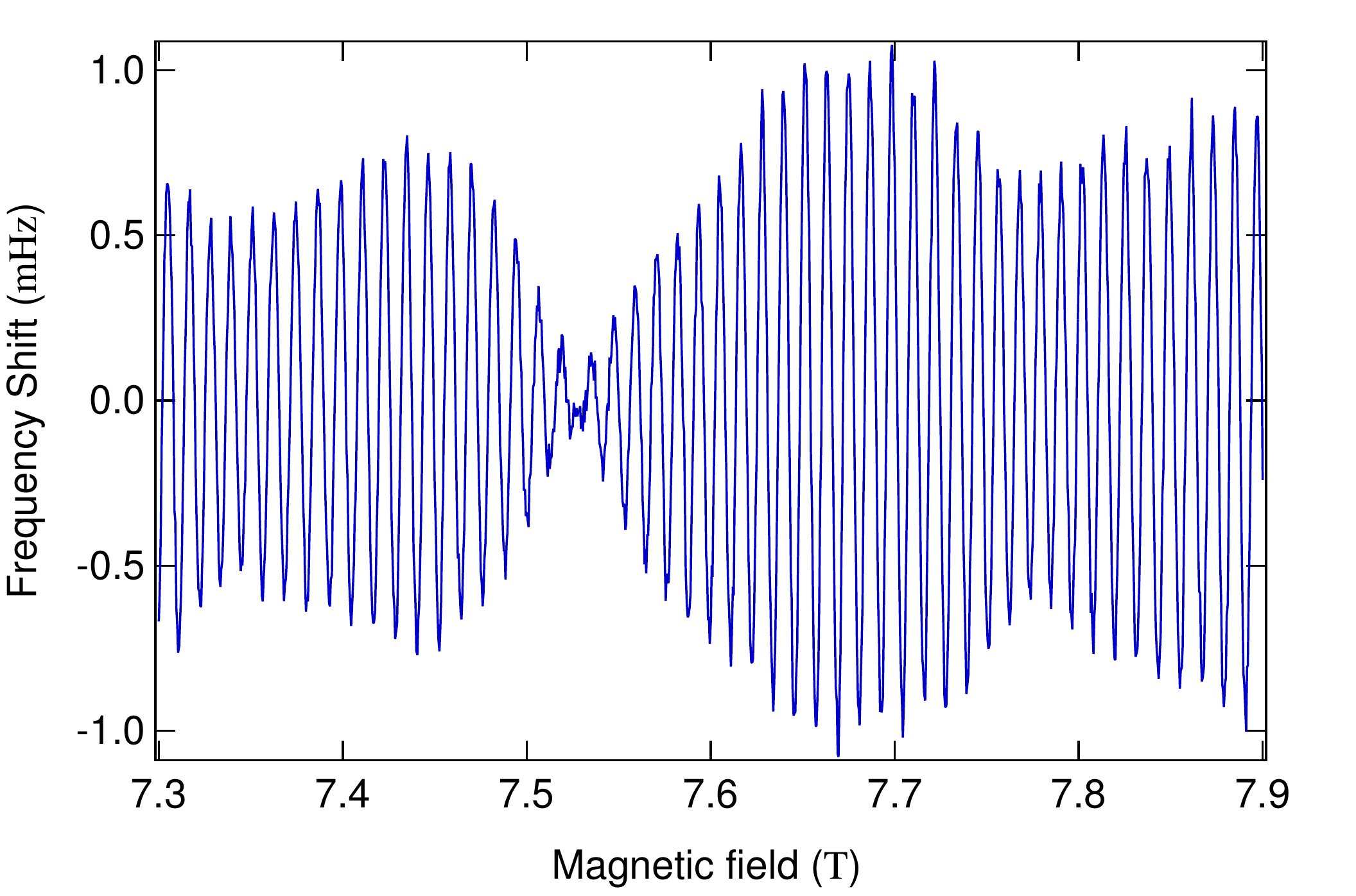}
\par\end{centering}

\caption[Frequency shift versus magnetic field for a typical data set]{\label{fig:ChData_SP2dFreqvsB}Frequency shift versus magnetic field
for a typical data set. The figure shows the frequency shift $\Delta f(B)$
obtained by removing the smooth background from the data shown in
Fig. \ref{fig:ChData_SP1FreqvsTime} and averaging together the data
points taken at the same value of magnetic field $B$. Oscillations
with a period close to the expected $\phi_{0}$ periodicity $B_{1}=10.7\,\text{mT}$
can be seen.}
\end{figure}

\begin{figure}
\begin{centering}
\includegraphics[width=0.7\paperwidth]{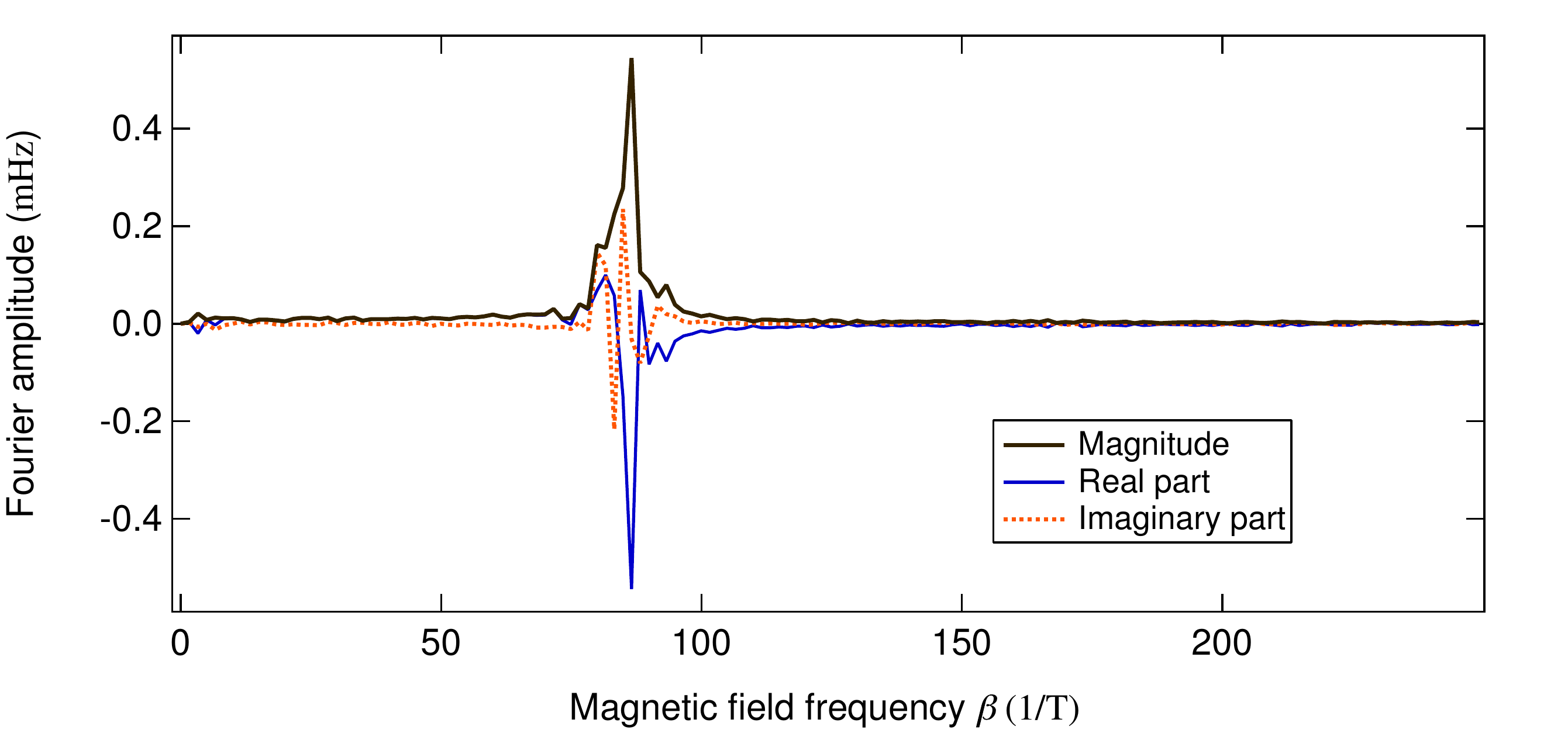}
\par\end{centering}

\caption[Frequency shift versus magnetic field frequency $\beta$ for a typical
data set]{\label{fig:ChData_SP3dFreqvsBeta}Frequency shift versus magnetic
field frequency $\beta$ for a typical data set. The Fourier spectrum
of the $\Delta f(B)$ trace shown in Fig. \ref{fig:ChData_SP2dFreqvsB}
is plotted for low values of the magnetic field frequency $\beta$.
A peak is located close to the expected magnetic field periodicity
of $\beta=94\,\text{T}^{-1}$. The peak has an irregular shape due
to the small number of correlation fields $B_{c,p=1}$ spanned by
the data in Fig. \ref{fig:ChData_SP2dFreqvsB}. The white fluctuating
background seen above $\beta=150\,\text{T}^{-1}$ maintains a roughly
constant level of fluctuations out to $\beta=10^{3}\,\text{T}^{-1}$.}
\end{figure}

Fig. \ref{fig:ChData_SP2dFreqvsB} shows the frequency shift $\Delta f(B)$,
the result of subtracting the polynomial fit from the data in Fig.
\ref{fig:ChData_SP1FreqvsTime} and then averaging together all of
the measurements recorded at the same value of the magnetic field.
A few errant data points from Fig. \ref{fig:ChData_SP1FreqvsTime},
selected for being several times the typical frequency scatter away
from the other frequency recordings at the same magnetic field, were
discarded before performing this averaging. Fig. \ref{fig:ChData_SP3dFreqvsBeta}
displays the discrete Fourier transform of the frequency shift trace
defined as 
\begin{equation}
\Delta f\left(\beta=\beta_{1}p\right)=\frac{1}{B_{\max}-B_{\min}}\int_{B_{\min}}^{B_{\max}}dB\,\Delta f\left(B\right)e^{-2\pi ip\beta_{1}B}.\label{eq:ChData_FourierTransform}
\end{equation}
A strong peak located close to the expected value of $\beta_{1}$
is visible in the spectrum.

\begin{figure}
\begin{centering}
\includegraphics[width=0.65\paperwidth]{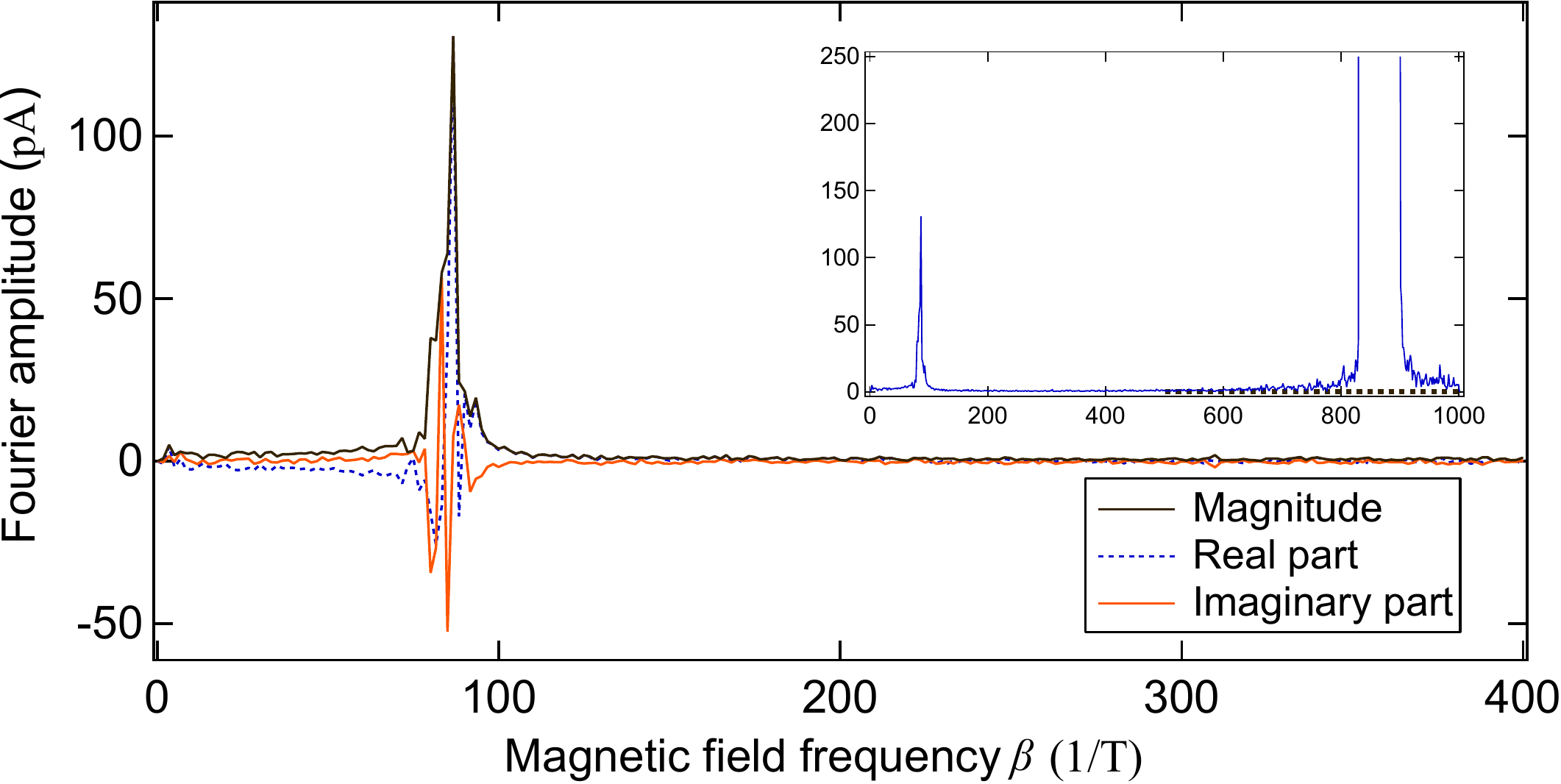}
\par\end{centering}

\caption[Fourier coefficients $dI_{p}^{\prime A}$ for a typical data set]{\label{fig:ChData_SP4dIpPrimeAvsBeta}Fourier coefficients $I_{p}^{\prime A}$
for a typical data set. The Fourier coefficients $dI_{p}^{A}$ were
calculated from the frequency shift $\Delta f(B)$ shown in Fig. \ref{fig:ChData_SP2dFreqvsB}
using Eq. \ref{eq:ChData_dIpA} and the values specified in the text
and in Tables \ref{tab:ChData_CLs} and \ref{tab:ChData_Rings} for
CL15. The traces shown in the figure represent the scaled form of
the coefficients $I_{p}^{\prime A}=dI_{p}^{A}/(2\pi\beta_{1})$ introduced
in Eq. \ref{eq:ChData_dIPrimeScaling}. This scaling converts the
current derivative coefficient $dI_{p}^{A}$ into a quantity $I_{p}^{\prime A}$
with units of current and similar magnitude to the current for $\beta\sim\beta_{1}$.
The coefficients $dI_{p}^{\prime A}$ are plotted versus $\beta=\beta_{1}p$.
The inset shows $I_{p}^{\prime A}$ over the full range of $\beta$
calculable from the data. Above $\beta=500\,\text{T}^{-1}$, and especially
between 829 and $904\mbox{ mT}^{-1}$, the values of the coefficients
$I_{p}^{\prime A}$ are enhanced due to an effect described in the
text.}
\end{figure}

Fig. \ref{fig:ChData_SP4dIpPrimeAvsBeta} shows the coefficients $I_{p}^{\prime A}$
calculated from Eqs. \ref{eq:ChData_dIpA} and \ref{eq:ChData_dIPrimeScaling}.
The $I_{p}^{\prime A}$ are the Fourier coefficients of the magnetic
field derivative $\partial_{B}I$ of the persistent current inferred
by method A. The coefficients $I_{p}^{\prime A}$ are plotted versus
magnetic field frequency $\beta=\beta_{1}p$. Because of the scaling
introduced in Eq. \ref{eq:ChData_dIPrimeScaling}, the $I_{p}^{\prime A}$
have units of current with the coefficients near $\beta=\beta_{1}$
possessing similar amplitudes to those of the corresponding Fourier
coefficients $I_{p}^{A}$ of the persistent current trace $I(B)$. 

In the inset of Fig. \ref{fig:ChData_SP4dIpPrimeAvsBeta}, the coefficients
$I_{p}^{\prime A}$ are shown over a wider range of magnetic field
frequency $\beta$. A large peak can be seen between $\beta\approx829\,\text{T}^{-1}$
and $\beta\approx904\,\text{T}^{-1}$. For these values of $\beta$,
the expression $\text{jinc}(2\pi p\beta_{1}GB)$ passes through zero
for $B=B_{\max}=7.9\,\text{T}$ and $B=B_{\min}=7.3\,\text{T}$ respectively.
Within this peak, the denominator of Eq. \ref{eq:ChData_dIpA} becomes
very small for some value of $B$ and thus causes the coefficients
$I_{p}^{\prime A}$ to be large. Because no such feature was present
in the Fourier transform of the frequency shift and no lower harmonics
of the persistent current other than the fundamental can be seen in
the spectrum, we can take this peak to be purely an artifact of the
signal processing procedure.

\begin{figure}
\begin{centering}
\includegraphics[width=0.7\paperwidth]{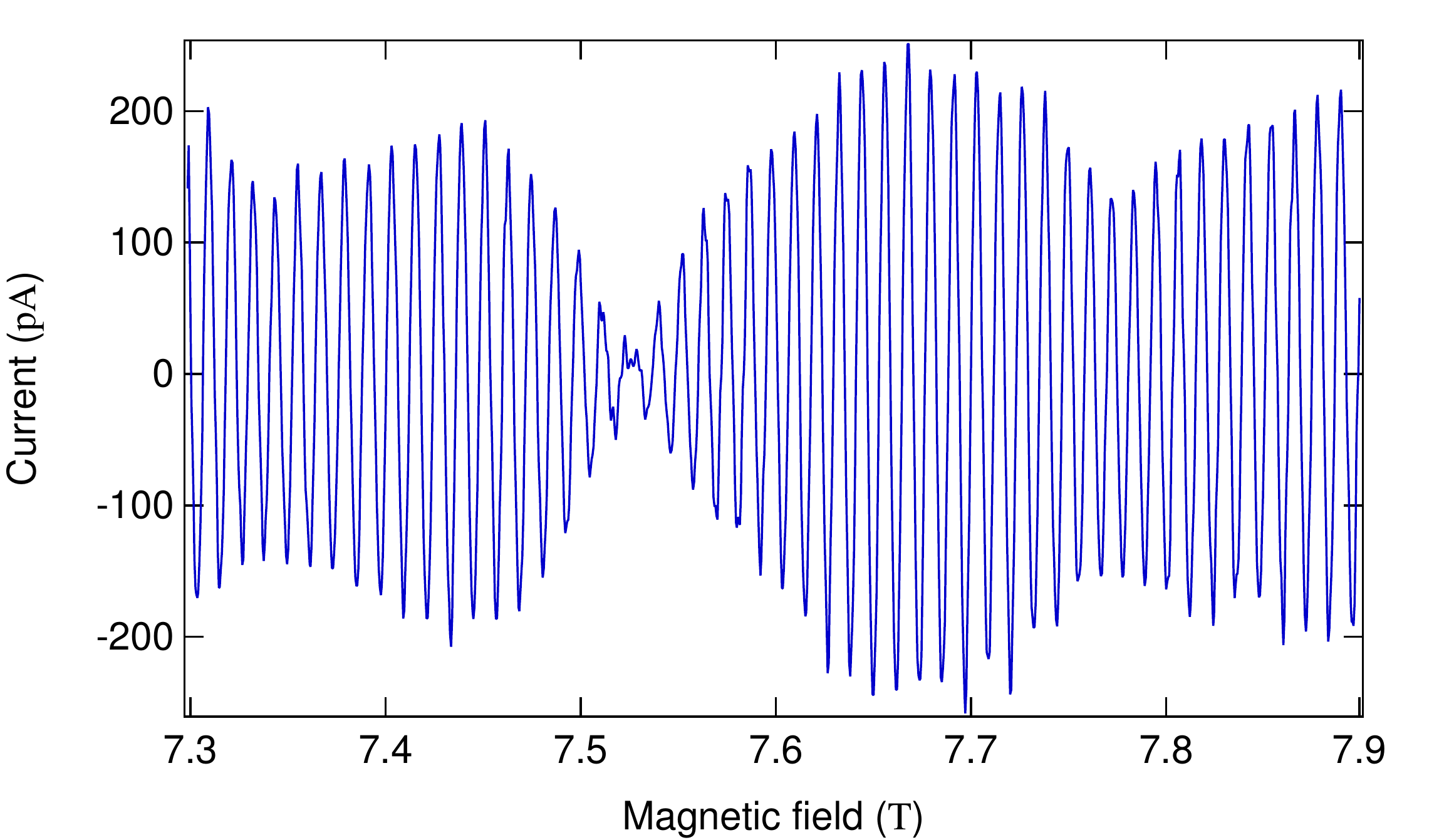}
\par\end{centering}

\caption[Derivative of persistent current $I^{\prime A}$ versus magnetic field
for a typical data set]{\label{fig:ChData_SP5IprimeAvsB}Derivative of persistent current
$I^{\prime A}$ versus magnetic field for a typical data set. The
curve shown is the inverse transform of the coefficients $I_{p}^{\prime A}$
shown in Fig. \ref{fig:ChData_SP4dIpPrimeAvsBeta} with the coefficients
corresponding to $\beta>500\,\text{T}^{-1}$ (black dashed line in
the inset of Fig. \ref{fig:ChData_SP4dIpPrimeAvsBeta}) set to zero.
This curve is the field derivative $\partial_{B}I^{A}$ scaled by
$(2\pi\beta_{1})^{-1}$ so that it has units of current.}
\end{figure}

\begin{figure}
\begin{centering}
\includegraphics[width=0.6\paperwidth]{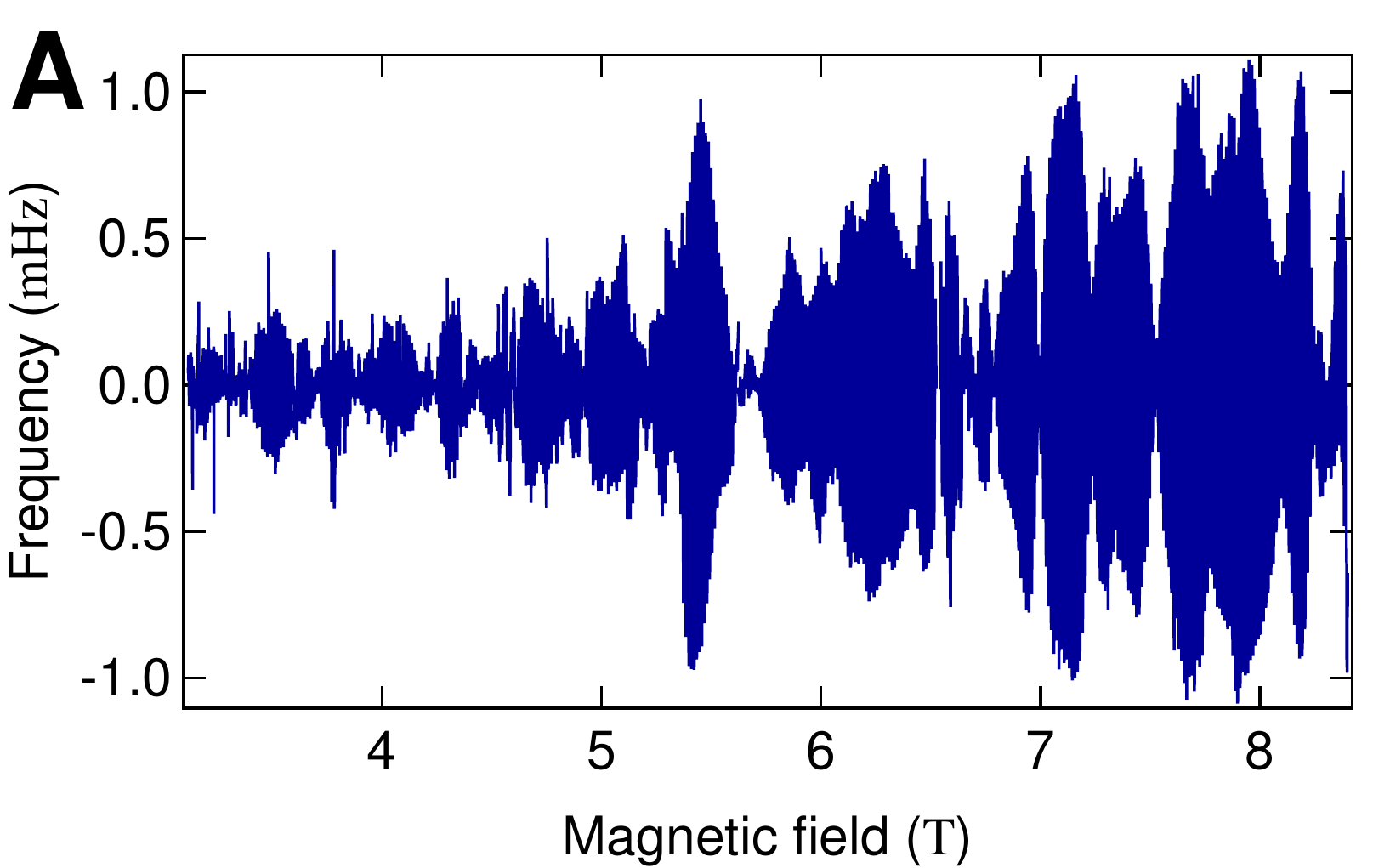}
\par\end{centering}

\begin{centering}
\includegraphics[width=0.6\paperwidth]{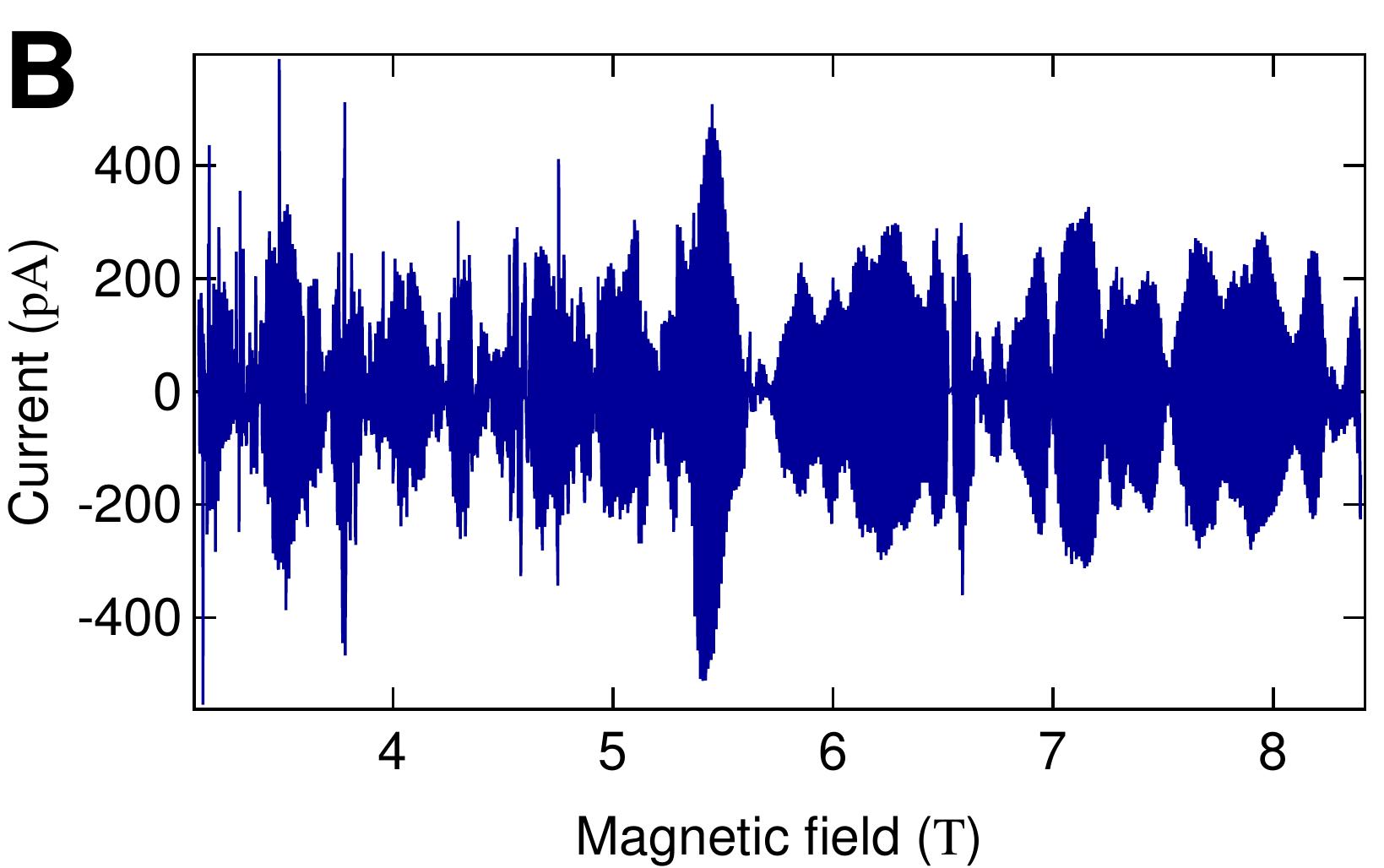}
\par\end{centering}

\caption[Frequency shift and inferred persistent current derivative $I^{\prime A}$
over a large range of magnetic field]{\label{fig:ChData_SP5bcBigFreqIprimevsB}Frequency shift and inferred
persistent current derivative $I^{\prime A}$ over a large range of
magnetic field. Panel A shows all of the frequency shift $\Delta f$
data taken on sample CL15 at $\theta_{0}=45^{\circ}$ and $T=360\,\text{mK}$.
The data shown was taken in a series of different magnetic field scans
at different points in time. A smooth background has been removed
from each $\Delta f(B)$ trace. The cantilever amplitude was held
constant during a single scan but varied in between scans so that
the argument of the $\text{jinc}$ factor in Eq. \ref{eq:ChData_DeltaF}
was relatively constant across the entire field range. Panel B shows
the inferred scaled current derivative $I^{\prime A}(B)$ calculated
by the method described in the text. The data between $7.3\,\text{T}$
and $7.9\,\text{T}$ match that shown in Fig. \ref{fig:ChData_SP5IprimeAvsB}
(on the scale of Fig. \ref{fig:ChData_SP5bcBigFreqIprimevsB}, the
individual oscillations of the persistent current can not be made
out). Despite the wide variation of the typical amplitude of the oscillations
in $\Delta f(B)$, the amplitude of the oscillations in $I^{\prime A}$
remain roughly constant over the entire range of magnetic field.}

\end{figure}

Fig. \ref{fig:ChData_SP5IprimeAvsB} shows the scaled magnetic field
derivative $I^{\prime A}=\partial_{B}I^{A}/(2\pi\beta_{1})$ of the
persistent current found by taking the inverse Fourier transform of
the spectrum shown in Fig. \ref{fig:ChData_SP4dIpPrimeAvsBeta}. Like
Fig. \ref{fig:ChData_SP4dIpPrimeAvsBeta}, the data in Fig. \ref{fig:ChData_SP5IprimeAvsB}
represents the derivative of the current but is scaled to have units
of current as described in \ref{sub:ChData_SigProcDescription}. For
the calculation of $I^{\prime A}$, all of the values of $dI_{p}^{A}$
corresponding to $\beta>500\,\text{T}^{-1}$ (the region marked by
the black dashed line in the inset of Fig. \ref{fig:ChData_SP4dIpPrimeAvsBeta})
were set to zero. This operation amounts to performing a low-pass
filter of the data. This region of high $\beta$ was removed because
it was quite clear that no observable persistent current signal was
present within it and because the white noise in the spectrum of $\Delta f$
(Fig. \ref{fig:ChData_SP3dFreqvsBeta}) was being significantly amplified
in this range (see discussion above) leading to the large peak in
the inset of Fig. \ref{fig:ChData_SP4dIpPrimeAvsBeta}. Because the
fractional change in magnetic field is fairly small over the range
shown, $I^{\prime A}(B)$ in Fig. \ref{fig:ChData_SP5IprimeAvsB}
looks qualitatively similar to $\Delta f(B)$ in Fig. \ref{fig:ChData_SP2dFreqvsB}.
The frequency shift $\Delta f(B)$ and inferred $I^{\prime A}$ from
measurements of sample CL15 over a wider range of magnetic field are
shown in Fig. \ref{fig:ChData_SP5bcBigFreqIprimevsB}.

\begin{figure}
\begin{centering}
\includegraphics[width=0.7\paperwidth]{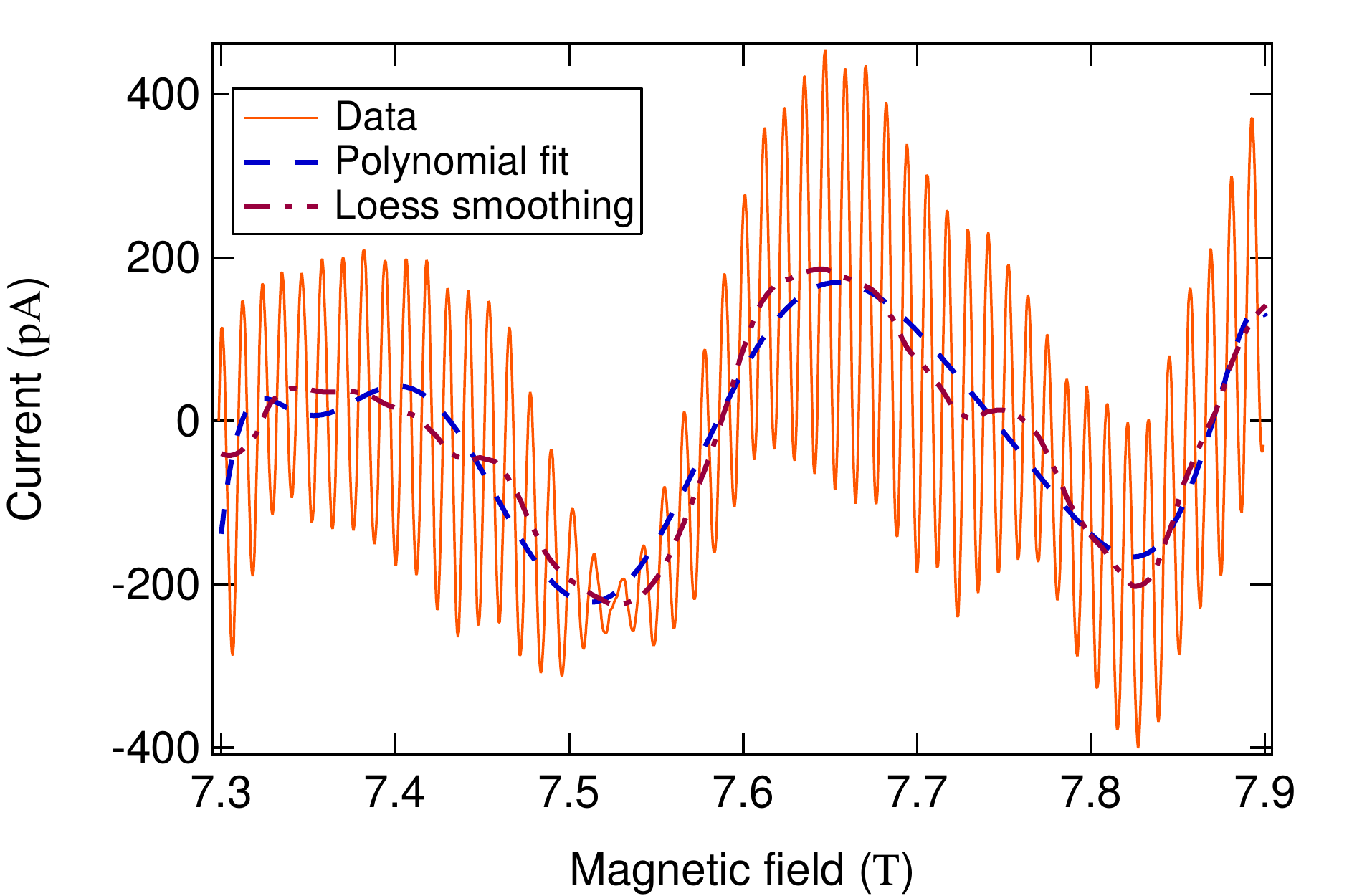}
\par\end{centering}

\caption[Persistent current $I^{A}$ versus magnetic field without background
subtraction for a typical data set]{\label{fig:ChData_SP6IAvsB}Persistent current $I^{A}$ versus magnetic
field without background subtraction for a typical data set. The curve
shown represents the persistent current $I^{A}(B)$ and is found by
numerically integrating the curve $I^{\prime A}(B)$ shown in Fig.
\ref{fig:ChData_SP5IprimeAvsB} (and multiplying by $2\pi\beta_{1}$
to remove the extra scale factor used in that figure). The process
of integration introduces large low frequency fluctuations to $I^{A}(B)$.
These fluctuations can also be seen in the Fourier spectrum of $I^{A}(B)$
shown in Fig. \ref{fig:ChData_SP7IAvsBeta}. Two approximations of
the low frequency fluctuations are also shown in the figure: a ninth
order polynomial fit (dashed line) and a lowest-order regression smoothing
with a window of $60\,\text{mT}$ (dot-dashed line).}
\end{figure}

\begin{figure}
\begin{centering}
\includegraphics[width=0.7\paperwidth]{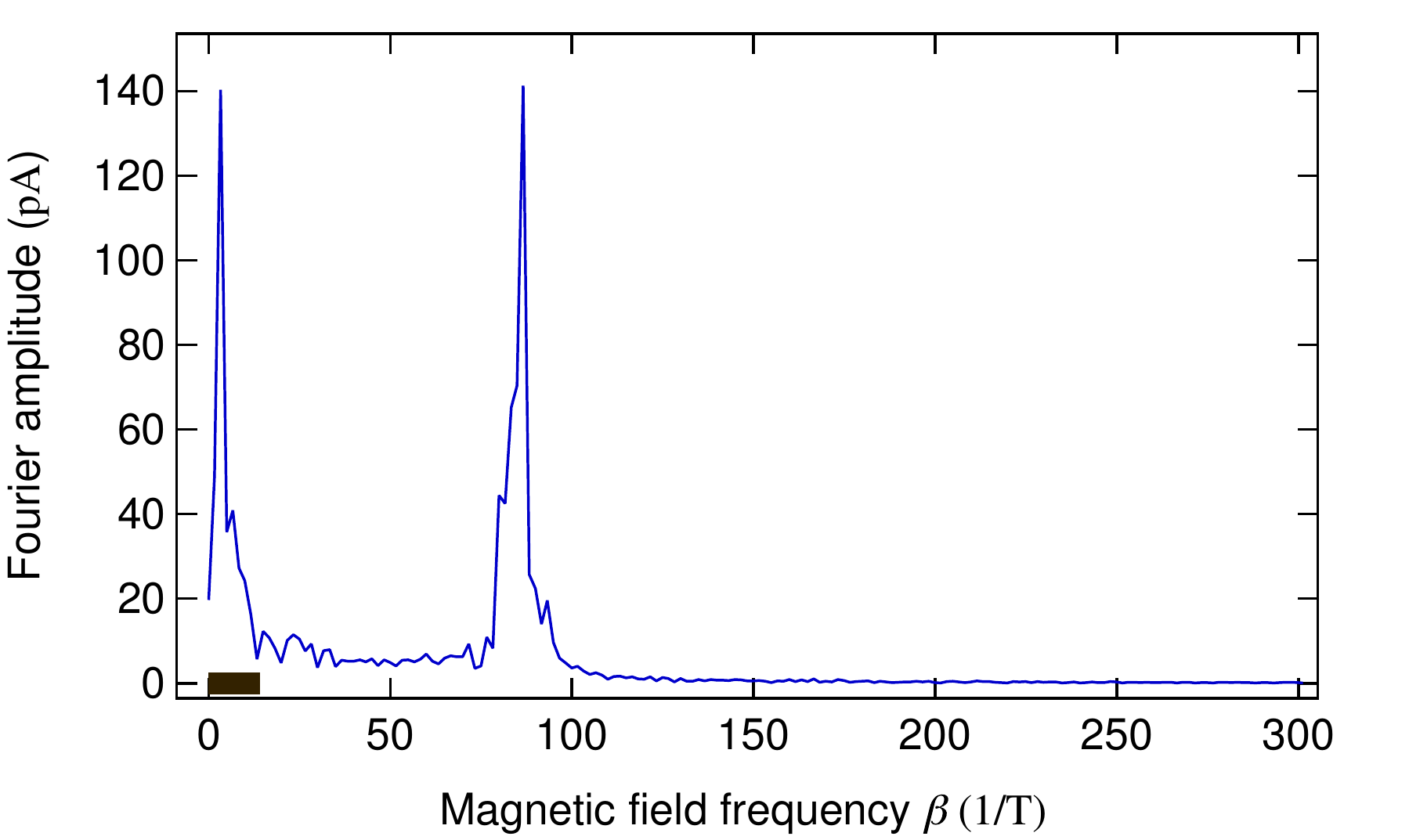}
\par\end{centering}

\caption[Fourier transform of the persistent current for a typical data set]{\label{fig:ChData_SP7IAvsBeta}Fourier transform of the persistent
current for a typical data set. The Fourier coefficients $I_{p}^{A}(\beta)$
of the current $I^{A}(B)$ plotted in Fig. \ref{fig:ChData_SP6IAvsB}
are shown versus magnetic field frequency $\beta=\beta_{1}p$. Low
frequency fluctuations enhanced by the integration process are indicated
by a black bar.}
\end{figure}

Fig. \ref{fig:ChData_SP6IAvsB} displays the integral with respect
to magnetic field of the data representing $I^{\prime A}=\partial_{B}I^{A}/(2\pi\beta_{1})$
in Fig. \ref{fig:ChData_SP5IprimeAvsB}. The scale factor $(2\pi\beta_{1})^{-1}$
used for $I^{\prime A}$ has been removed so that curve in Fig. \ref{fig:ChData_SP6IAvsB}
depicts the persistent current $I^{A}(B)$ as a function of magnetic
field. Numerical integration of the data has the undesirable side
effect of magnifying low frequency noise. This magnification can be
seen in Fig. \ref{fig:ChData_SP7IAvsBeta} which shows the Fourier
transform coefficients $I_{p}^{A}(\beta)$ of the current trace $I^{A}(B)$
shown in Fig. \ref{fig:ChData_SP6IAvsB}. The spectrum $I_{p}^{A}(\beta)$
in Fig. \ref{fig:ChData_SP7IAvsBeta} looks quite similar to the spectrum
$I_{p}^{\prime A}(\beta)$ shown \ref{fig:ChData_SP4dIpPrimeAvsBeta}
other than an added spike at low frequency (indicated by a black bar
in Fig. \ref{fig:ChData_SP7IAvsBeta}).

\begin{figure}
\begin{centering}
\includegraphics[width=0.7\paperwidth]{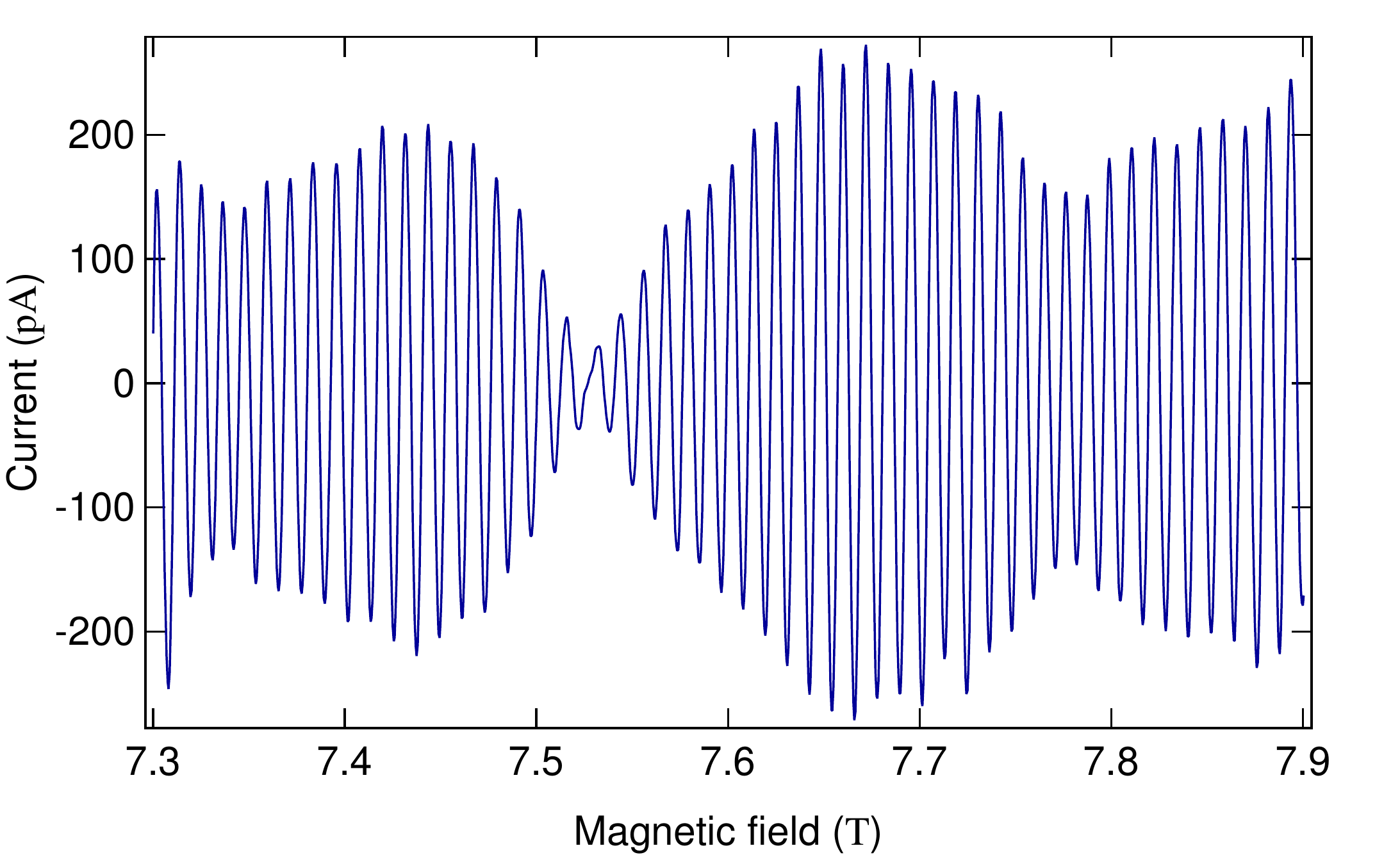}
\par\end{centering}

\caption[Persistent current versus magnetic field with smooth background removed
for a typical data set]{\label{fig:ChData_SP8IALPvsB}Persistent current versus magnetic
field with smooth background removed for a typical data set. The persistent
current $I^{A}(B)$ inferred from the data in Fig. \ref{fig:ChData_SP1FreqvsTime}
is plotted versus magnetic field $B$. This curve was obtained by
taking the difference of the persistent current trace and the lowest-order
regression smoothing trace both shown in Fig. \ref{fig:ChData_SP6IAvsB}.}
\end{figure}

The added low frequency noise can removed in several ways. In Fig.
\ref{fig:ChData_SP6IAvsB}, we show both a ninth order polynomial
fit to $I^{A}(B)$ and the result of a lowest-order regression smoothing
(LOESS) routine with a window of $60\,\text{mT}$ on $I^{A}(B)$.
Subtracting either of these curves from $I^{A}(B)$ removes most of
the low frequency fluctuations. A similar result is obtained by performing
a high-pass filter on $I^{A}(B)$. This filter can be realized by
setting to zero all of the low frequency components spanned by the
black bar in Fig. \ref{fig:ChData_SP7IAvsBeta} and then taking the
inverse transform. The difference between the current trace and the
LOESS smoothed trace is plotted in Fig. \ref{fig:ChData_SP8IALPvsB}
and represents the final form of the inferred current $I^{A}(B)$
from the measured cantilever frequency $f(B)$.

In calculating the current trace $I^{A}(B)$, we twice removed smooth
backgrounds and also integrated the data once. These operations correspond
to high and low pass filters respectively. We also low-pass filtered
the $dI_{p}^{A}$ to remove high frequency components of the noise
which were strongly magnified by the conversion of frequency shift
to current derivative. All of these filters were applied to ranges
of the magnetic field frequency $\beta$ spectrum well separated from
the persistent current features and should not affect the analysis
of the persistent current data. This conclusion can be seen by noting
that the main feature in the Fourier spectra of Figs. \ref{fig:ChData_SP3dFreqvsBeta},
\ref{fig:ChData_SP4dIpPrimeAvsBeta}, and \ref{fig:ChData_SP7IAvsBeta}
is largely unchanged throughout the signal processing routine. However,
these filtering steps are the reason that the trace of $I^{A}(B)$
in Fig. \ref{fig:ChData_SP8IALPvsB} appears slightly smoother than
the $\Delta f(B)$ trace in Fig. \ref{fig:ChData_SP2dFreqvsB}.

\FloatBarrier

\section{\label{sec:ChData_Diagnostics}Persistent current measurement diagnostics}

Before discussing our main experimental results regarding persistent
currents, we review several diagnostic measurements which were made
in order to check the validity of the persistent current signal processing
procedure outlined in \ref{sec:ChData_SigProc} and the validity of
the torsional magnetometry technique more generally.

\begin{figure}
\begin{centering}
\includegraphics[width=0.7\paperwidth]{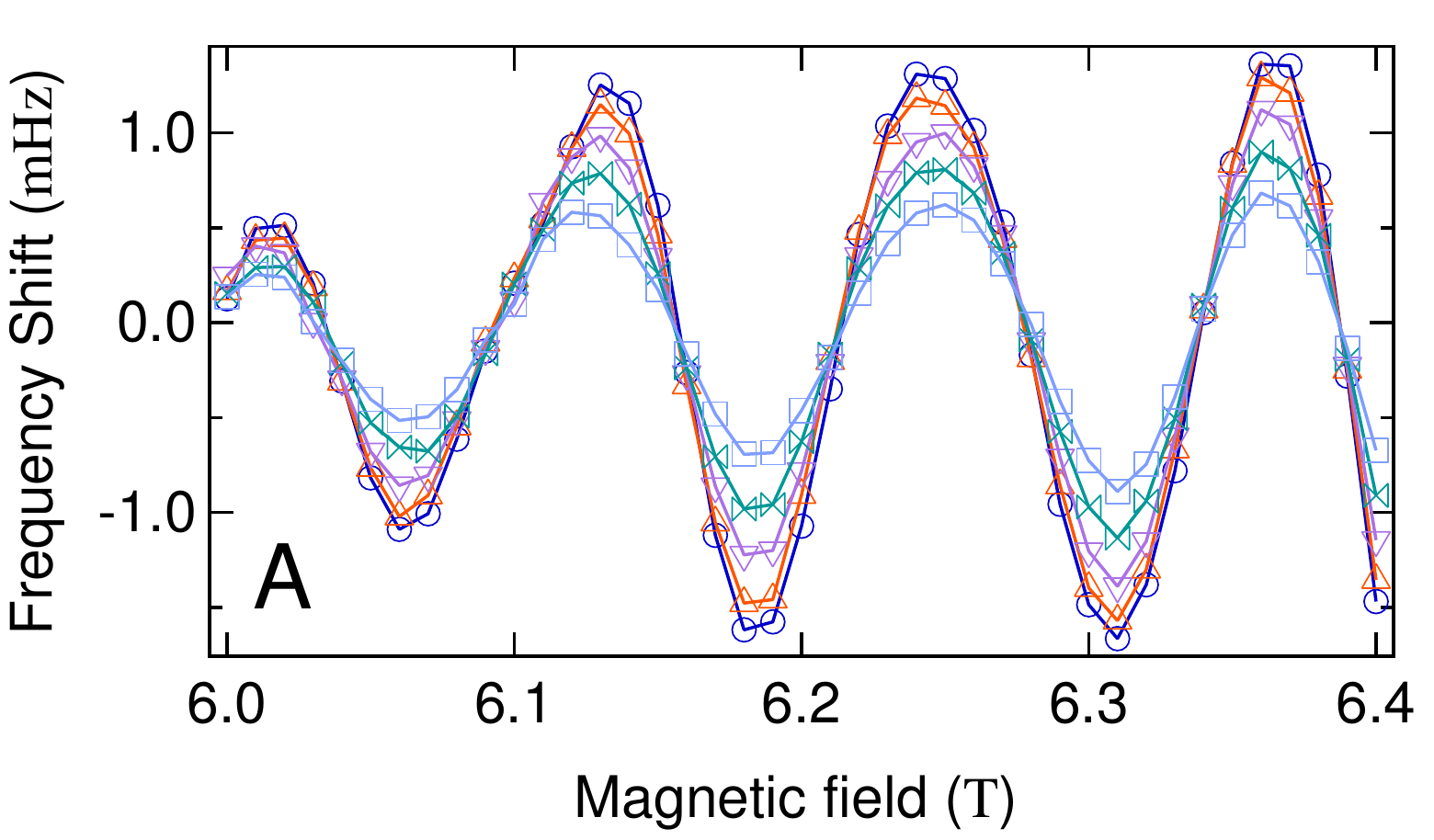}
\par\end{centering}

\begin{centering}
\includegraphics[width=0.7\paperwidth]{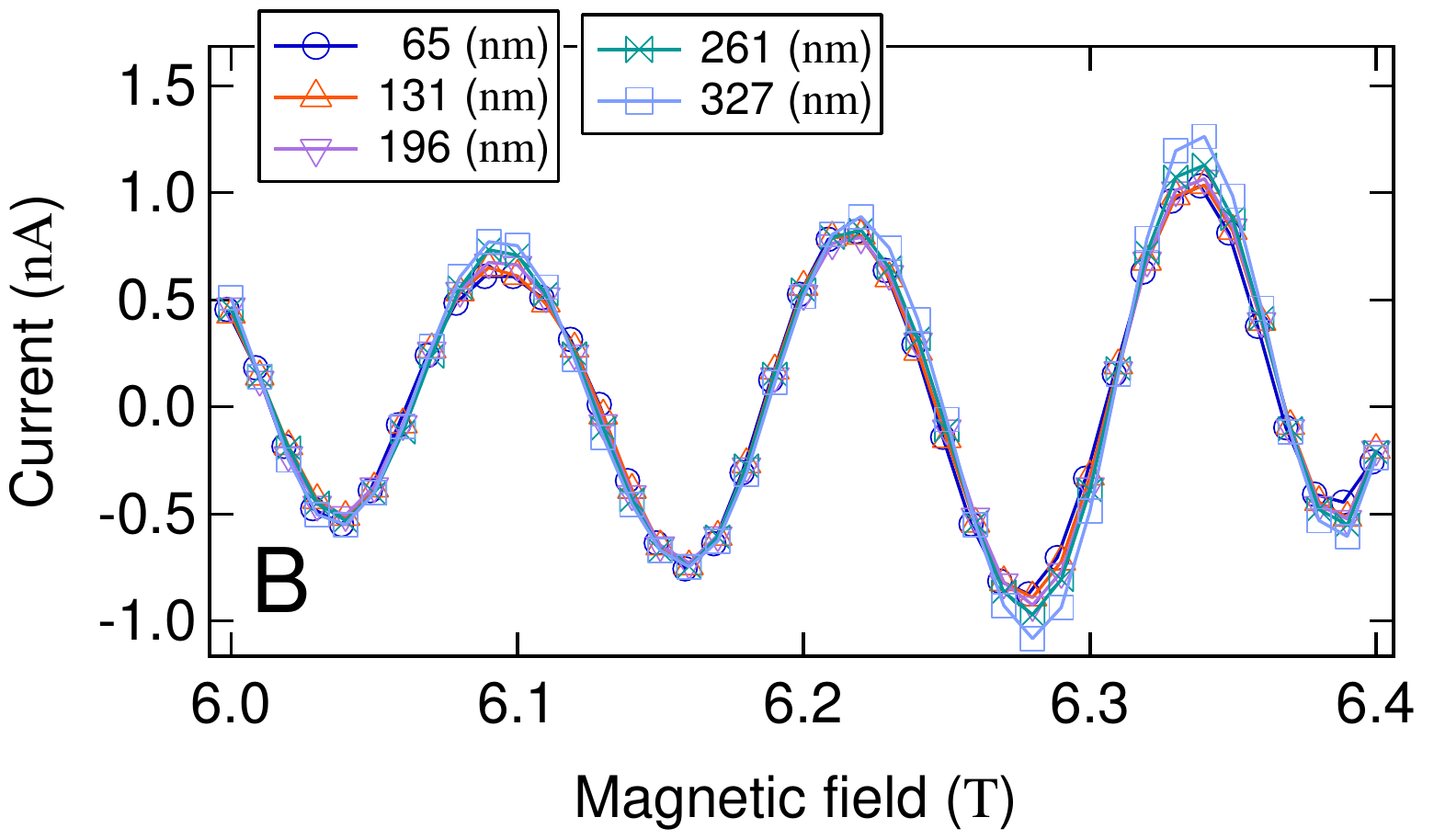}
\par\end{centering}

\caption[Frequency shift and inferred persistent current versus magnetic field
for a series of cantilever amplitudes]{\label{fig:ChData_DIA1DriveSeries}Frequency shift and inferred persistent
current versus magnetic field for a series of cantilever amplitudes.
The data shown were taken on sample CL17 at $T=323\,\text{mK}$. The
angle $\theta_{0}$ between the cantilever beam and the magnetic field
was $6^{\circ}$. The persistent current traces were calculated using
method A of \ref{sub:ChData_SigProcDescription}.}
\end{figure}

\begin{figure}
\begin{centering}
\includegraphics[width=0.6\paperwidth]{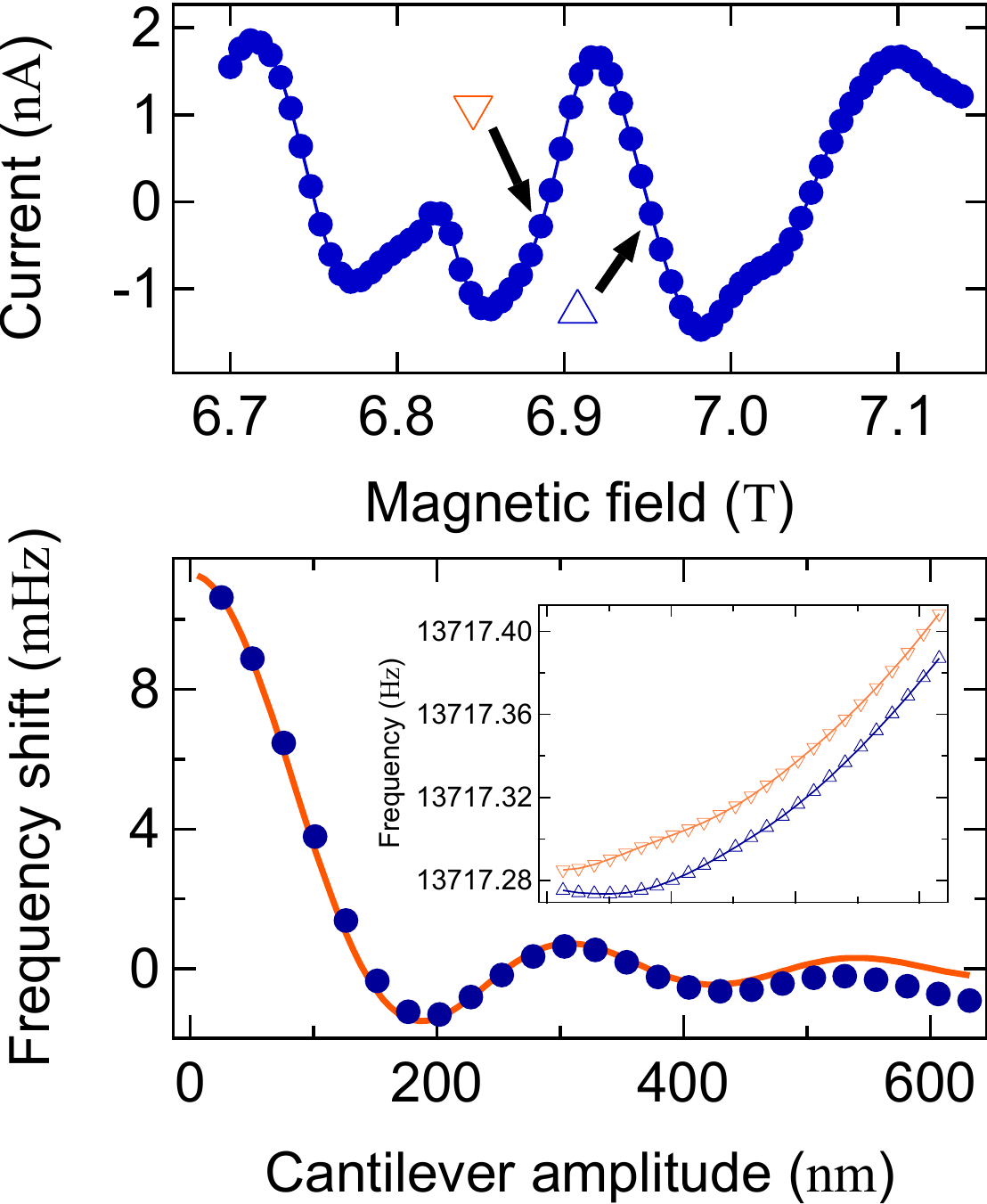}
\par\end{centering}

\caption[Frequency shift versus cantilever amplitude]{\label{fig:ChData_DIA2AmpScan}Frequency shift versus cantilever
amplitude. In the upper plot the full $I(B)$ trace is shown over
the magnetic field range of interest for the amplitude scan measurements
of the lower plot. To generate the data shown in the lower plot, the
resonant frequency of cantilever CL17 was measured at $B_{u}=6.885\,\text{T}$
and $B_{d}=6.95\,\text{T}$ (indicated by arrows in the upper plot)
as a function of cantilever amplitude $x_{\max}$. For these measurements,
the angle $\theta_{0}$ between the cantilever beam and the magnetic
field was $6^{\circ}$, the temperature $T$ was $323\,\text{mK}$,
and the cantilever was excited in its second flexural mode. The measured
frequency as a function cantilever amplitude at both field values
is shown in the inset of the lower plot with the frequency measured
at $B_{u}$ represented by downward pointing triangles and that measured
at $B_{d}$ by upward pointing triangles. The large lower plot shows
the difference $\Delta f_{pc}(x_{\max})=\Delta f(x_{\max},B_{d})-\Delta f(x_{\max},B_{u})$
(dots) between the measured frequency shifts at the two magnetic field
values. Also shown is a curve calculated using Eq. \ref{eq:ChData_dFreqvsXmax}.
The accuracy of the match between the curve and the data is discussed
in the text.}
\end{figure}

\begin{figure}
\begin{centering}
\includegraphics[width=0.6\paperwidth]{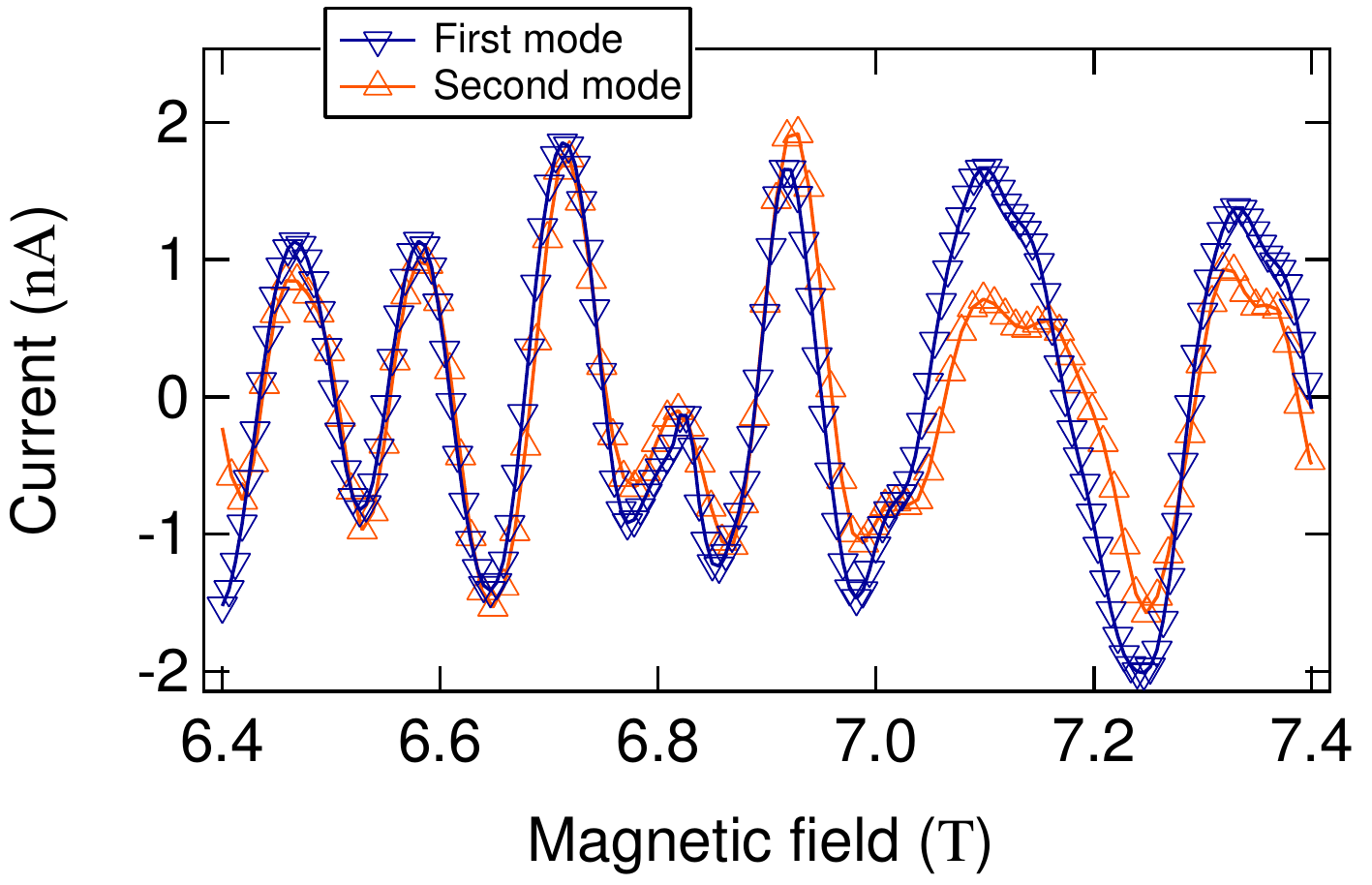}
\par\end{centering}

\caption[Persistent current measurements taken while exciting the first and
second flexural modes of the cantilever]{\label{fig:ChData_DIA3FlexuralMode}Persistent current measurements
taken while exciting the first and second flexural modes of the cantilever.
The two traces were calculated from measurements of the cantilever
frequency shift of sample CL17 using method A of \ref{sub:ChData_SigProcDescription}.
The measurements were taken with a temperature $T$ of $323\,\text{mK}$
and an angle $\theta_{0}$ of $6^{\circ}$ between the cantilever
beam and magnetic field.}

\end{figure}

Figs. \ref{fig:ChData_DIA1DriveSeries}, \ref{fig:ChData_DIA2AmpScan},
and \ref{fig:ChData_DIA3FlexuralMode} each display a different test
of Eq. \ref{eq:CHTorsMagn_FiniteAmpFreqShift} and Section \ref{sec:ChData_SigProc}.
Each of the figures displays data taken on sample CL17 of Tables \ref{tab:ChData_CLs}
and \ref{tab:ChData_Rings} at an angle of $\theta_{0}=6^{\circ}$.
In Fig. \ref{fig:ChData_DIA1DriveSeries}, panel A shows the frequency
shift $\Delta f(B)$ measured over a region of magnetic field $B$
for a series of different values of the cantilever amplitude $x_{\max}$,
while panel B displays the current $I^{A}(B)$ inferred by method
A of \ref{sec:ChData_SigProc} for value of $x_{\max}$. While the
features in the frequency shift data vary by over a factor of 2, the
traces representing the inferred value of the current show good agreement. 

Fig. \ref{fig:ChData_DIA2AmpScan} provides a more direct check of
Eq. \ref{eq:CHTorsMagn_FiniteAmpFreqShift} for the cantilever frequency
shift due to the persistent current by showing measurements of the
cantilever frequency as a function of cantilever amplitude at fixed
values of the magnetic field. For simplicity, we assume that the Fourier
series expansion of the persistent current given in Eq. \ref{eq:ChData_IofB}
contains only one non-zero term so that the current can be written
in the form
\[
I\left(B\right)=I_{p}\sin\left(2\pi p\beta_{1}B+\psi_{p}\right).
\]
Eq. \ref{eq:ChData_DeltaF} for the change in resonant frequency of
the cantilever due to the persistent current can then be written as
\[
\Delta f_{pc}\left(x_{\max},B\right)=FB^{2}2\pi p\beta_{1}I_{p}\cos\left(2\pi p\beta_{1}B+\psi_{p}\right)\mathrm{jinc}\left(2\pi p\beta_{1}\frac{\alpha}{l\tan\theta_{0}}x_{\max}B\right).
\]
The expected frequency shift should reach its maximum magnitude when
the current $I(B)$ crosses through zero and $\cos(2\pi p\beta_{1}B+\psi_{p})=\pm1$.

For Fig. \ref{fig:ChData_DIA2AmpScan}, the frequency shift $\Delta f$
was measured as a function of the cantilever amplitude $x_{\max}$
at two fixed values of the magnetic field $B$. These measurements
were made on sample CL17 with an angle $\theta_{0}$ of $6^{\circ}$
between the cantilever beam and the applied magnetic field. The cantilever
was excited in its second flexural mode in order to increase the argument
of the $\text{jinc}$ function in the expression for $\Delta f_{pc}$
without requiring larger values of $x_{\max}$. A comparison of measurements
performed using the cantilever's first and second flexural modes are
discussed below and shown in Fig. \ref{fig:ChData_DIA3FlexuralMode}.
A list of the cantilever parameters relevant to the calculation of
$\Delta f_{pc}$ that depend on mode index is given in the discussion
concerning Fig. \ref{fig:ChData_DIA3FlexuralMode}.

The two values of the magnetic field used in this measurement (indicated
by the arrows in the upper trace in Fig. \ref{fig:ChData_DIA2AmpScan})
were chosen to correspond to adjacent magnetic field positions $B_{u}$
and $B_{d}$ where the current trace $I(B)$ passed through zero with
an upward and downward slope respectively. The measured frequency
shifts $\Delta f(x_{\max},B_{u})$ and $\Delta f(x_{\max},B_{d})$
are shown in the inset of the lower plot of Fig. \ref{fig:ChData_DIA2AmpScan}.
Both traces show a similar trend of increasing frequency with cantilever
amplitude, possibly the result of an electrostatic interaction between
the cantilever and the nearby optical fiber tip or of a small mechanical
non-linearity of the cantilever.

In order to eliminate this background trend of the frequency shift
present at both $B_{u}$ and $B_{d}$, the net frequency shift $\Delta f_{\text{pc}}(x_{\max})$
was found by subtracting the frequency shift measured at the two magnetic
field values: 
\[
\Delta f_{\text{pc}}(x_{\max})=\Delta f(x_{\max},B_{d})-\Delta f(x_{\max},B_{u}).
\]
The measured frequency shift difference $\Delta f_{pc}$ is represented
by the dots in the lower plot of Fig. \ref{fig:ChData_DIA2AmpScan}.
Taking $B\equiv(B_{u}+B_{d})/2$, $I_{p}\equiv I_{p}(B_{u})\approx I_{p}(B_{d})$,
$\cos(2\pi p\beta_{1}B_{u}+\psi_{p})=+1$, and $\cos(2\pi p\beta_{1}B_{d}+\psi_{p})=-1,$
the expected frequency shift difference $\Delta f_{pc}(x_{\max})$
is
\begin{equation}
\Delta f_{pc}(x_{\max})=\frac{f_{0}}{k}\frac{2\pi p}{\phi_{0}}I_{pS}\left(AB\cos\theta_{0}\frac{\alpha}{l}\right)^{2}\mathrm{jinc}\left(2\pi p\frac{AB}{\phi_{0}}\cos\theta_{0}\frac{\alpha}{l}x_{\text{\ensuremath{\max}}}\right).\label{eq:ChData_dFreqvsXmax}
\end{equation}
A curve calculated using Eq. \ref{eq:ChData_dFreqvsXmax} with $I_{p}=4\,\text{nA}$
and $p=0.93$ is also shown in the lower plot of Fig. \ref{fig:ChData_DIA2AmpScan}
and agrees well with the measured $\Delta f_{pc}$ data. The $4\,\text{nA}$
value for $I_{p}$ is close to the $3.1\,\text{nA}$ peak-to-peak
amplitude observed in the trace $I(B)$ of current versus magnetic
field (upper plot of figure). The seven percent deviation of $p$
from unity is well within the range of expected fluctuations of the
persistent current oscillation frequency due to magnetic field penetrating
the ring's finite linewidth (see the discussion in \ref{sub:ChData_Quantitative}
below). If $p$ is fixed to unity inside the argument of the $\text{jinc}$
function in Eq. \ref{eq:ChData_dFreqvsXmax}, the same curve can be
generated by changing the ring radius from $308\,\text{nm}$ to $297\,\text{nm}$,
well within the ring's $115\,\text{nm}$ linewidth.

\begin{figure}
\begin{centering}
\includegraphics[width=0.7\paperwidth]{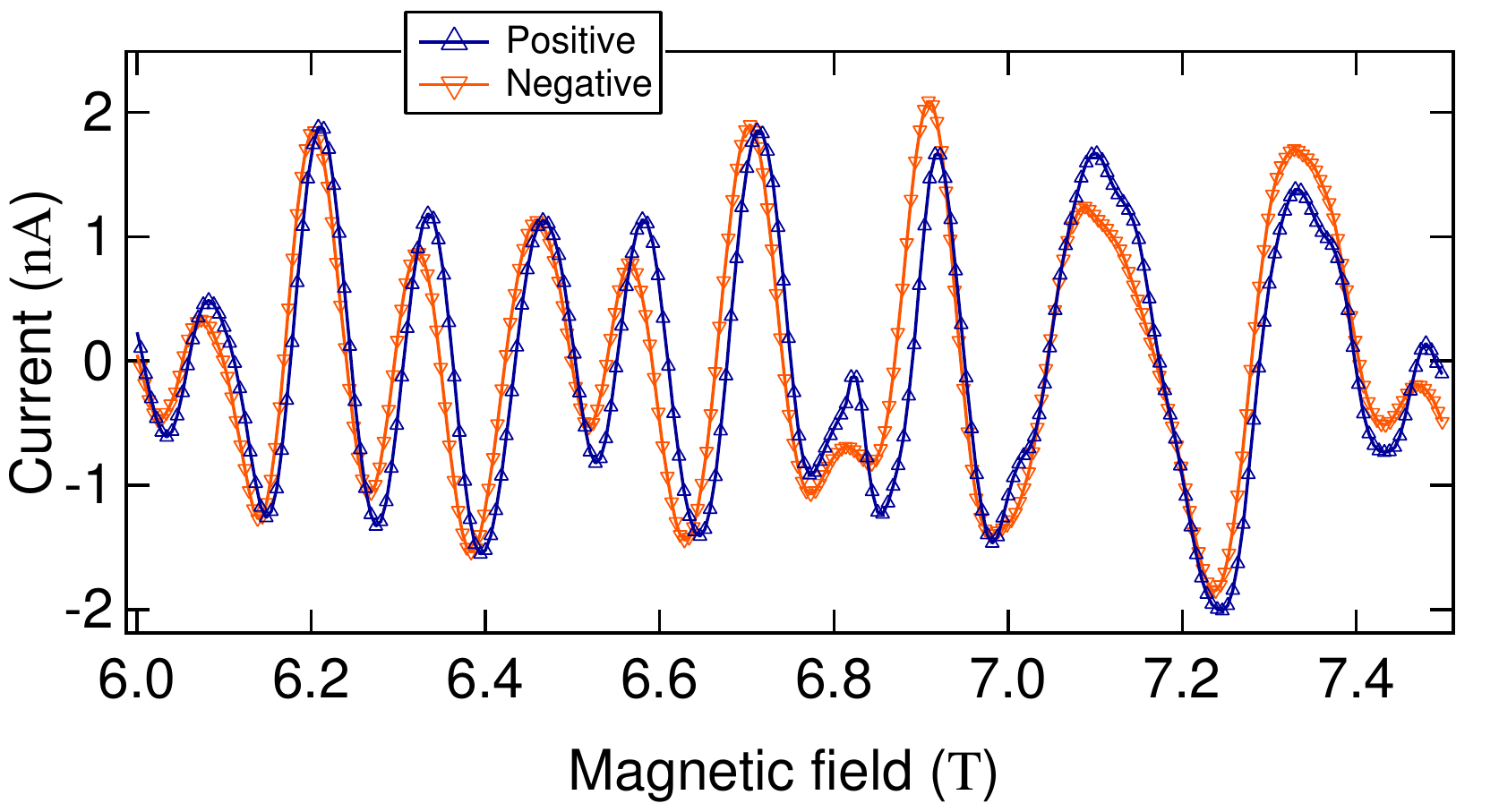}
\par\end{centering}

\caption[Persistent current versus magnetic field measured for both magnet
polarities]{\label{fig:ChData_DIA6MagnetPolarity}Persistent current versus magnetic
field measured for both magnet polarities. For the measurements with
negative magnet polarity, both the current and the magnetic field
axis have been reversed. The measurements shown were taken on sample
CL17 with a temperature $T$ of $323\,\text{mK}$ and an angle $\theta_{0}$
of $6^{\circ}$ between the cantilever beam and magnetic field.}
\end{figure}

Fig. \ref{fig:ChData_DIA3FlexuralMode} shows two traces of the current
$I(B)$ inferred from measurements of the frequency shift $\Delta f(B)$
taken while different flexural modes of the cantilever were excited.
For these two measurements, the fundamental flexural mode of the cantilever
with resonant frequency $f_{0}=2186\,\text{Hz}$ and the second order
flexural mode with resonant frequency $f_{0}=13,718\,\text{Hz}$ were
excited. Beyond the resonant frequency, the parameters in Eq. \ref{eq:CHTorsMagn_FiniteAmpFreqShift}
that varied with flexural mode were the spring constant $k$ ($1.34\times10^{-3}\,\text{N/m}$
for the fundamental mode, $5.27\times10^{-2}\,\text{N/m}$ for the
second order mode) and the mode shape derivative $\alpha$ (1.38 for
the fundamental mode, 4.78 for the second order mode). The two traces
shown in Fig. \ref{fig:ChData_DIA3FlexuralMode} exhibit good agreement
in terms of current magnitude and magnetic field features. The discrepancy
in the two traces near $7.2\,\text{T}$ could possibly be due to different
resonances in the sample holder in this region of magnetic field.
Similar variation between traces can be observed in Fig. \ref{fig:ChData_DIA6MagnetPolarity}
for which both measurements were performed while exciting the cantilever's
fundamental mode.

In addition to providing another confirmation of the accuracy of Eq.
\ref{eq:CHTorsMagn_FiniteAmpFreqShift}, the independence of the inferred
current from the cantilever flexural mode can also be viewed as independence
of the current from cantilever excitation frequency in the low kilohertz
regime relevant to the cantilever torsional magnetometry technique
discussed here. Thus the inferred persistent current magnitude can
be taken to be the equilibrium value. This result is not surprising
as almost all of the timescales related to the ring sample (the correlation
energy $E_{c}$, the electron phase coherence time $\tau_{\phi}$,
the electron elastic scattering time $\tau_{e}$, the decay time associated
with the ring inductance ($L/R$ where $L$ is the ring inductance
and $R$ the ring resistance), the temperature $T$, and the Zeeman
splitting $E_{Z}$ of the electron) are on order of one gigahertz
or higher. One notable exception, the single level spacing at the
Fermi energy, is still greater than $200\,\text{kHz}$ for each sample.

\begin{figure}
\begin{centering}
\includegraphics[width=0.7\paperwidth]{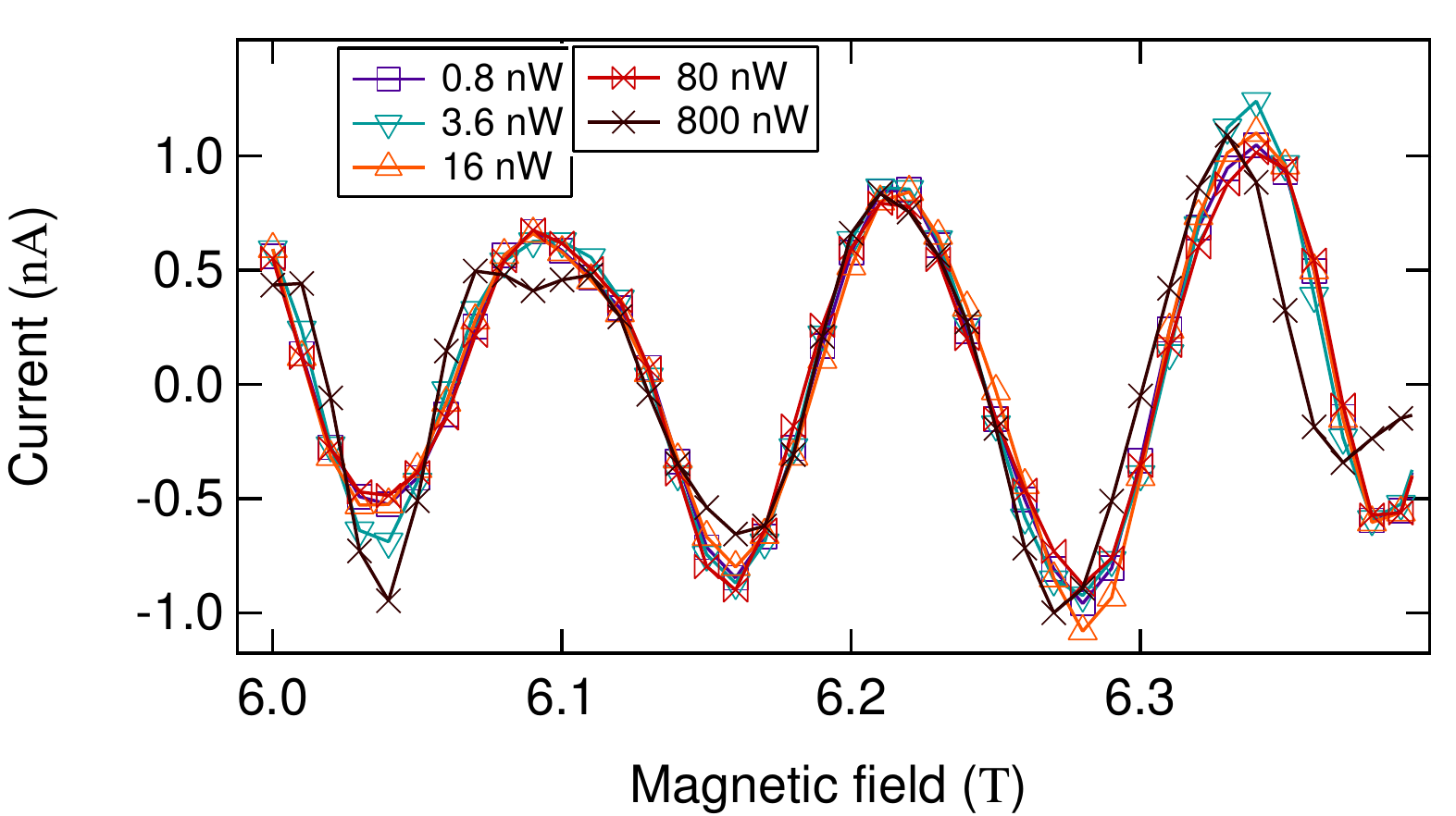}
\par\end{centering}

\caption[Persistent current versus magnetic field for a series of readout laser
powers]{\label{fig:ChData_DIA4LaserPower}Persistent current versus magnetic
field for a series of readout laser powers. The persistent current
is clearly unaffected by the incident power of the detection laser
for powers up to $80\,\text{nW}$. Only slight variations in the current
trace are visible for $800\,\text{nW}$. The measurements shown were
taken with a temperature $T$ of $323\,\text{mK}$ and an angle $\theta_{0}$
of $6^{\circ}$ between the cantilever beam and magnetic field.}
\end{figure}

\begin{figure}
\begin{centering}
\includegraphics[width=0.7\paperwidth]{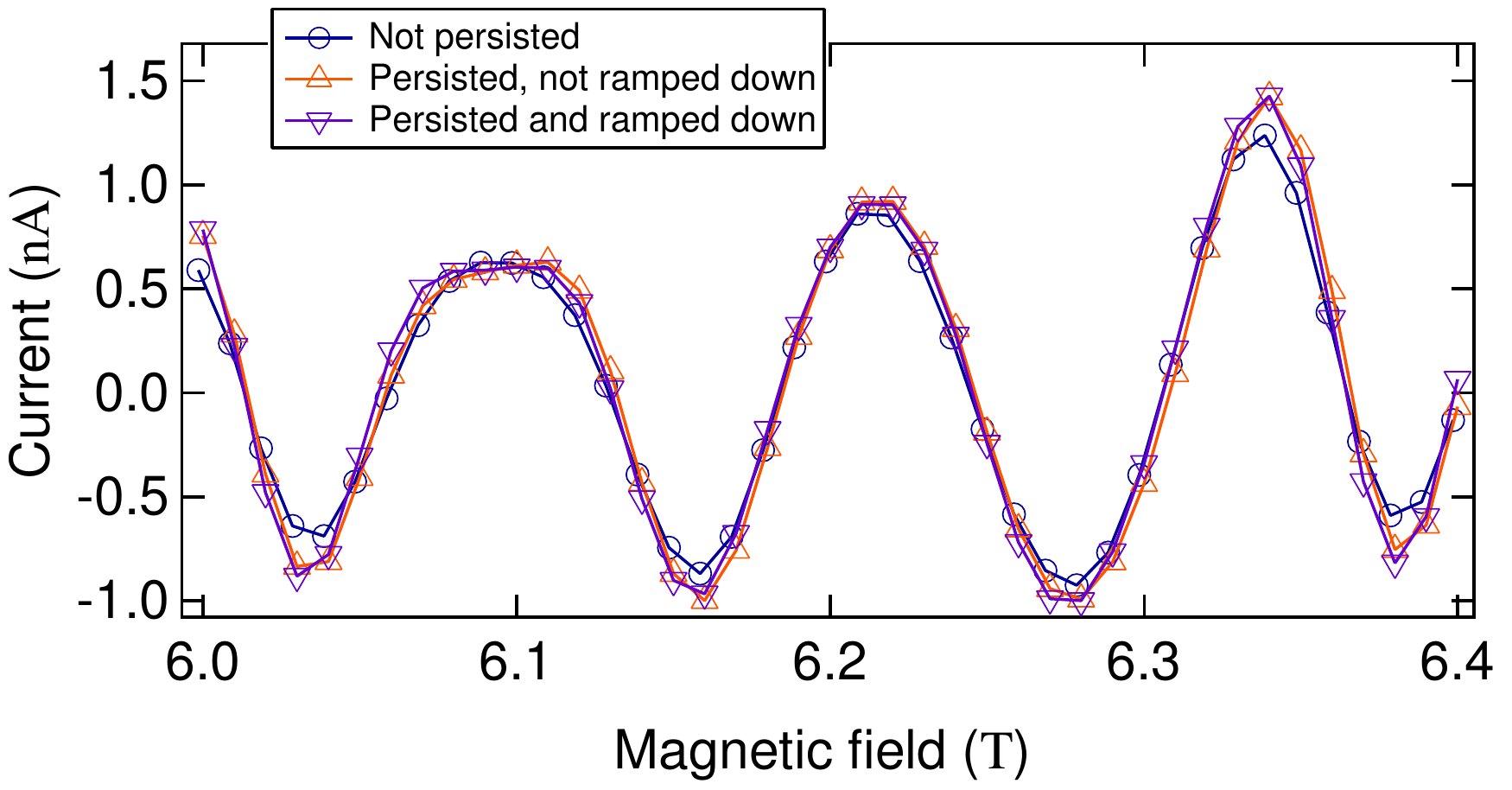}
\par\end{centering}

\caption[Persistent current versus magnetic field observed with the magnet
in different modes of operation]{\label{fig:ChData_DIA5MagnetState}Persistent current versus magnetic
field observed with the magnet in different modes of operation. Measurements
were performed on sample CL17 with the solenoid not persisted (persistent
switch heater on), the solenoid persisted but the leads not ramped
down (persistent switch heater off), and the solenoid persisted and
the leads ramped down (persistent switch heater off, no external current
sourced to magnet) as described in the text. The measurements shown
were taken with a temperature $T$ of $323\,\text{mK}$ and an angle
$\theta_{0}$ of $6^{\circ}$ between the cantilever beam and magnetic
field.}
\end{figure}

Figs. \ref{fig:ChData_DIA4LaserPower}, \ref{fig:ChData_DIA5MagnetState},
and \ref{fig:ChData_DIA6MagnetPolarity} show the results of measurements
performed to verify the independence of the inferred persistent current
trace $I(B)$ from various configurations of the experimental apparatus.
In Fig. \ref{fig:ChData_DIA4LaserPower}, the current traces $I(B)$
detected for a series of values of the optical power of the cantilever
detection laser are shown over a small range of magnetic field. These
measurements were performed at the refrigerator's base temperature
of $323\,\text{mK}$. For $80\,\text{nW}$ or less incident on the
cantilever, little variation in the current traces is observed, and
even for the trace taken with $800\,\text{nW}$ only a moderate deviation
is evident. Outside of Fig. \ref{fig:ChData_DIA4LaserPower}, the
incident laser power on the cantilever was kept to $5\,\text{nW}$
or less, a range for which the ring sample could be assumed to be
in thermal equilibrium with the refrigerator. Although we did not
thoroughly characterize the optical absorption of the samples listed
in Table \ref{tab:ChData_CLs}, we note that Fig. \ref{fig:ChData_DIA4LaserPower}
indicates less sensitivity to laser power than was observed in Fig.
\ref{fig:CHSensitivity_CantileverHeatTransport}. As the cantilevers
from these two sets of measurements had similar dimensions (see \ref{sub:CHExpSetup_ThermometryCantilevers}
and Table \ref{tab:ChData_CLs}), we surmise that lower level of doping
(resistivities of $\sim0.025\,\text{\ensuremath{\Omega}\,\ cm}$ for
the Arrow cantilevers and $\sim20\,\text{\ensuremath{\Omega}\ cm}$
for the persistent current samples) accounted for the decreased sensitivity
to incident laser power.

Fig. \ref{fig:ChData_DIA5MagnetState} presents measurements of the
persistent current for three different modes of operation of the solenoid
(see \ref{sub:ChSens_DewarFridge} for a description of the experimental
apparatus). The solenoid consisted of a large coil of superconducting
material closed electrically by a small section of superconducting
wire referred to as the {}``persistent switch.'' By operating a
nearby heater, the persistent switch could be driven normal, creating
a resistive {}``break'' of the closed superconducting loop. 

Most measurements of the persistent current were performed with the
persistent switch heater turned on and with the electrical current
in the solenoid sourced by a room temperature power supply. For Fig.
\ref{fig:ChData_DIA5MagnetState}, measurements of the cantilever
frequency were performed with the persistent switch heater turned
off but current from the external supply still flowing through the
leads down to the magnet. In order to change the magnetic field in
between measurements of the cantilever frequency, the persistent switch
heater was turned on and the current supplied by the external supply
ramped up or down. Additionally, measurements were performed with
the persistent switch heater turned off and the current in the magnet
leads ramped down to zero. For these measurements, all electrical
connections to the cryostat other than the leads to the piezoelectric
actuator and the leads to the magnet were disconnected. Little variation
in the current was observed for these three different modes of solenoid
operation. We did not attempt measurements in which the magnet leads
were physically disconnected from room temperature electronics. However,
we note that magnetic field fluctuations in the solenoid due to voltage
noise coupled to the magnet leads from the room temperature electronics
should be strongly suppressed due to the solenoid's large ($19\,\text{H}$)
inductance.

In Fig. \ref{fig:ChData_DIA6MagnetPolarity}, the persistent current
traces measured with the magnet operated in both polarities are compared.
For the trace representing the negative magnetic polarity, the sign
of both the current $I$ and the magnetic field $B$ are inverted.
The two traces show good agreement with each other. This result is
as expected since time reversal invariance requires that $I(-B)=-I(B)$.

We also measured the frequency shift signal due to the supercurrent
in the rings near $B\approx0\,\text{T}$. While the frequency shift
signal is roughly proportional to $B^{2}$ and thus small at low field,
the current in the superconducting state was over a thousand times
larger in magnitude than that in the normal state and was observable
at fields as low as $\sim15\,\text{mT}$. A rough analysis of the
measurements of the rings in the superconducting state is given in
Appendix \ref{cha:AppSC_}. The transition temperature of the rings
was found to be consistent with transport measurements of a co-deposited
wire (see Appendix \ref{cha:AppTransport_}) and close to the bulk
value for aluminum (see Fig. \ref{fig:CHSensitivity_BcVsTBCS} and
accompanying discussion). The magnetic field dependence and inferred
magnitude of the current in the superconducting state were similar
to those observed in previous measurements, indicating that the rings
were of good quality and that our modeling of the cantilever was accurate.

\FloatBarrier

\section{\label{sec:ChData_Quantitative}Persistent current measurements}

We now present our main experimental results and interpret them using
the theoretical picture developed in Chapter \ref{cha:CHMeso_}. The
measured samples are described in Tables \ref{tab:ChData_CLs} and
\ref{tab:ChData_Rings}. These samples were all contained on one sample
chip and thus could all be measured in the same cooldown. The measurements
described in this section were taken in two cooldowns, one with the
sample oriented with the angle $\theta_{0}$ between the cantilever
beam and the magnetic field at $6^{\circ}$ and one with $\theta_{0}=45^{\circ}$.
Persistent currents have also been observed in other cooldowns and
with other samples created in a different fabrication run.

\begin{table}
\begin{centering}
\begin{tabular}{|cccccc|}
\hline 
$\vphantom{{\displaystyle \sum_{a}^{a}}}$Sample & $l\,\text{(\ensuremath{\mu}m)}$ & $w\,\text{(\ensuremath{\mu}m)}$ & $f_{0}$ (Hz) & $k\,\text{(mN/m)}$ & $Q\,(\times10^{5})$\tabularnewline
\hline 
${\displaystyle \vphantom{\sum_{-}}}$CL11 & 459 & 40 & 2091 & 0.63 & 1.2\tabularnewline
${\displaystyle \vphantom{\sum_{-}}}$CL14 & 438 & 60 & 2298 & 1.08 & 1.6\tabularnewline
${\displaystyle \vphantom{\sum_{-}}}$CL15 & 454 & 60 & 2138 & 0.97 & 1.3\tabularnewline
CL17 & 449 & 80 & 2186 & 1.34 & 1.2\tabularnewline
\hline 
\end{tabular}
\par\end{centering}

\caption[Cantilever parameters of persistent current samples]{\label{tab:ChData_CLs}Cantilever parameters of persistent current
samples. The cantilever widths $w$ were measured optically, and the
cantilever resonant frequencies $f_{0}$ were measured using the detection
arrangement described in Chapter \ref{cha:CHExpSetup_}. The cantilever
thickness $t$ was measured to be $340\,\text{nm}$ using the instrument
mentioned in \ref{sec:APPsampFab_recipe}. The cantilever length $l$
and spring constant $k$ were calculated using the values from the
other measurements and Eqs. \ref{eq:CHTorsMagn_FreqFromKandMeff}
and \ref{eq:CHTorsMagn_springKfromDimensions}. The cantilever length
measured optically was not used due to uncertainty caused by over-etching
of the silicon handle layer on which the cantilever was mounted. The
length values given above are consistent with the values observed
optically. The listed cantilever quality factors $Q$ represent typical
measured values. The quality factor varied with temperature and over
time. More details about the cantilevers are given in \ref{sub:CHExpSetup_PersistentCurrentFabrication}
and \ref{sec:APPsampFab_recipe}.}
\end{table}

\begin{table}
\begin{centering}
\begin{tabular}{|cccccc|}
\hline 
$\vphantom{{\displaystyle \sum_{a}^{a}}}$Sample & $L\,\text{(\ensuremath{\mu}m)}$ & $w_{r}\,\text{(nm)}$ & $t_{r}\,\text{(nm)}$ & $N$ & $h_{A}\,\text{(\ensuremath{\mu}m)}$\tabularnewline
\hline 
${\displaystyle \vphantom{\sum_{-}}}$CL11 & 5.0 & 85 & 90 & 242 & 30\tabularnewline
${\displaystyle \vphantom{\sum_{-}}}$CL14 & 2.6 & 85 & 90 & 1 & n/a\tabularnewline
${\displaystyle \vphantom{\sum_{-}}}$CL15 & 2.6 & 85 & 90 & 990 & 18\tabularnewline
CL17 & 1.9 & 115 & 90 & 1680 & 20\tabularnewline
\hline 
\end{tabular}
\par\end{centering}

\caption[Ring specifications of persistent current samples]{\label{tab:ChData_Rings}Ring specifications of persistent current
samples. The mean circumference $L$ and linewidth $w_{r}$ of the
aluminum rings as measured by a scanning electron microscope are given
for each sample. The thickness $t_{r}$ of the deposited aluminum
was measured by an atomic force microscope. For each array, the amount
of distance $h_{A}$ spanned by the array along the cantilever's long
dimension is given, as is the number $N$ of rings in the array. Each
array was located at the tip of the cantilever with a $5\,\text{\ensuremath{\mu}m}$
margin from the three edges of the cantilever. The single ring was
also located $5\,\text{\ensuremath{\mu}m}$ from the cantilever tip
and was centered along the cantilever's width dimension. }
\end{table}

\subsection{\label{sub:ChData_Qualitative}Qualitative discussion}

\begin{figure}
\begin{centering}
\includegraphics[width=0.7\paperwidth]{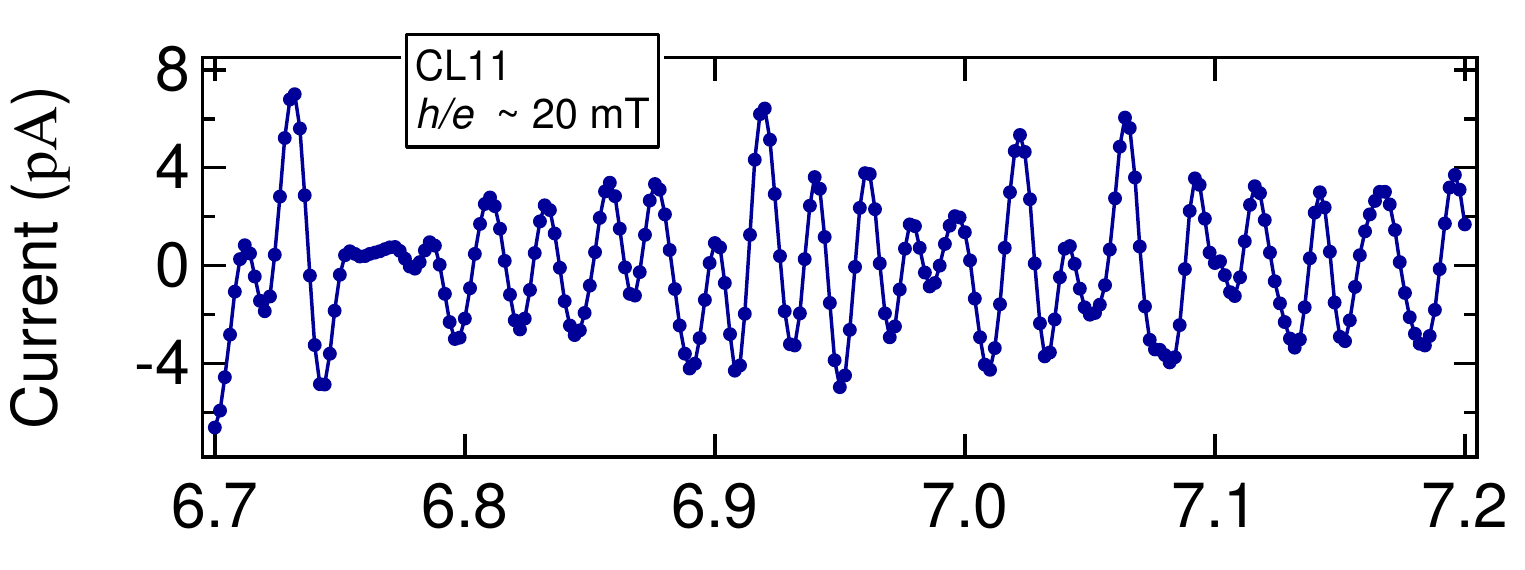}
\par\end{centering}

\begin{centering}
\includegraphics[width=0.7\paperwidth]{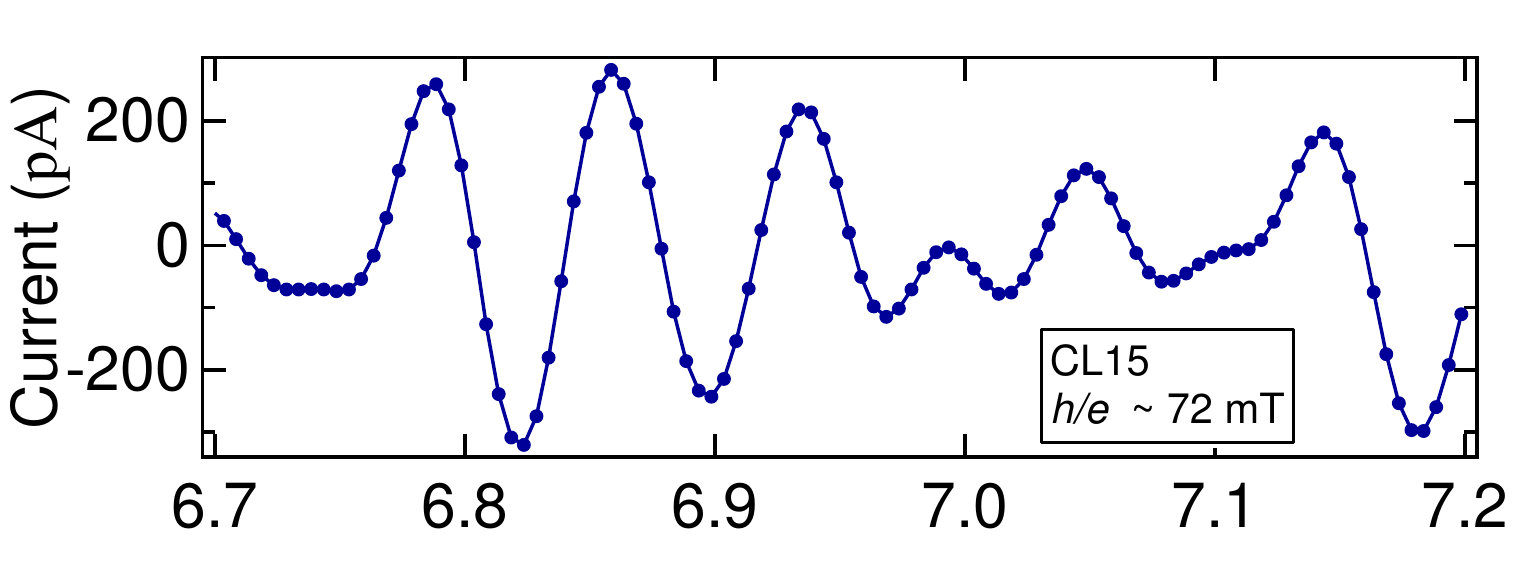}
\par\end{centering}

\begin{centering}
\includegraphics[width=0.7\paperwidth]{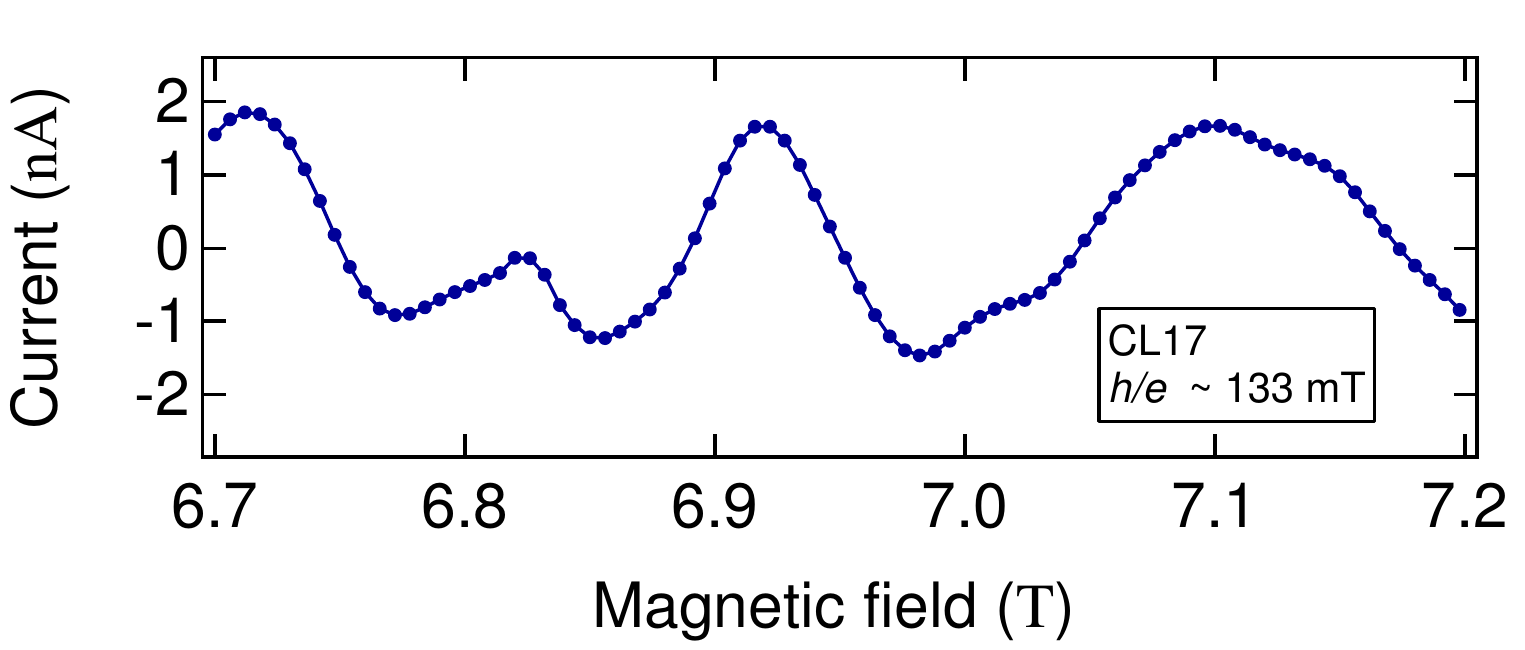}
\par\end{centering}

\caption[Comparison of persistent current measured for samples with three different
ring sizes]{\label{fig:ChData_DAT1_6DegIvsB}Comparison of persistent current
measured for samples with three different ring sizes. From top to
bottom, the observed current for samples CL11, CL15, and CL17 of Tables
\ref{tab:ChData_CLs} and \ref{tab:ChData_Rings} are shown in order
of decreasing ring size. Each trace is labeled with the expected magnetic
field scale $1/\beta_{1}$ for threading $h/e$ through the mean ring
radius with an angle $\theta_{0}$ of $6^{\circ}$ between the magnetic
field and the plane of the rings. As ring size decreases, both the
magnetic field scale and the current magnitude of typical features
increases. The measurements shown were taken at $T=323\,\text{mK}$.}
\end{figure}

In Fig. \ref{fig:ChData_DAT1_6DegIvsB}, measurements of rings with
three different circumferences $L$ are shown between $B=6.7\,\text{T}$
and $7.2\,\text{T}$ for $\theta_{0}=6^{\circ}$ and $T=323\,\text{mK}$.
The magnetic field period of the persistent current oscillation (the
magnetic field necessary to thread $h/e$ through the ring) is inversely
proportional to the ring's area and thus to $L^{2}$. Qualitatively,
this trend is borne out in Fig. \ref{fig:ChData_DAT1_6DegIvsB} where
the magnetic field scale of the typical feature is seen to increase
with decreasing $L$. Additionally, in Eq. \ref{eq:CHPCTh_IIFiniteTZSO}
it was predicted that the typical magnitude $I^{\text{typ}}(T)=\langle I^{2}(T)\rangle^{1/2}$
scales roughly as 
\[
I^{\text{typ}}\propto\frac{1}{R^{2}}\exp(-aL^{2}T/D)
\]
with $a$ a constant and $D$ the diffusion constant of the metal.
The data in Fig. \ref{fig:ChData_DAT1_6DegIvsB} is qualitatively
consistent with this relationship as the size scale of the persistent
current increases with decreasing $L$.

\begin{figure}
\begin{centering}
\includegraphics[width=0.7\paperwidth]{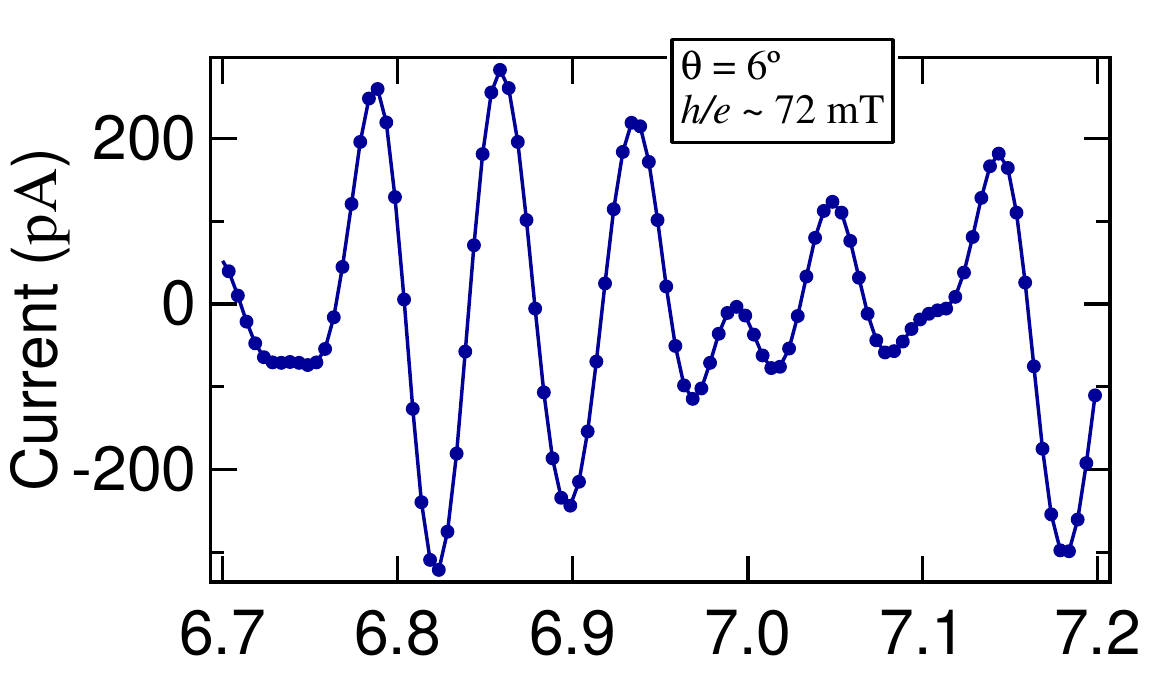}
\par\end{centering}

\begin{centering}
\includegraphics[width=0.7\paperwidth]{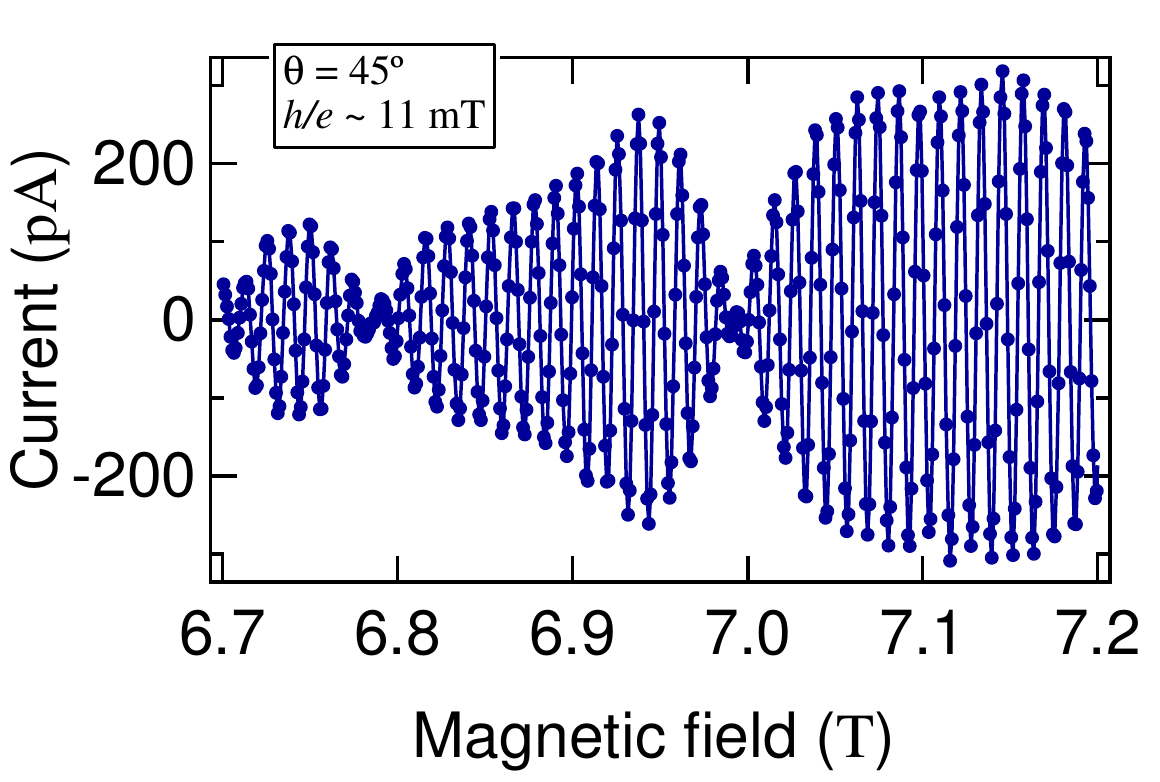}
\par\end{centering}

\caption[Comparison of persistent current observed in sample CL15 for two different
angles $\theta_{0}$]{\label{fig:ChData_DAT2_AngCL15}Comparison of persistent current
observed in sample CL15 for two different angles $\theta_{0}$. The
current inferred from measurements with $\theta_{0}=6^{\circ}$ and
$T=323\,\text{mK}$ (top graph) and $\theta_{0}=45^{\circ}$ and $T=365\,\text{mK}$
(bottom graph) are shown. The expected $h/e$ periodicity in terms
of applied magnetic field is given with each plot for the mean ring
radius and specified angle $\theta_{0}$.}
\end{figure}

\begin{figure}
\begin{centering}
\includegraphics[width=0.7\paperwidth]{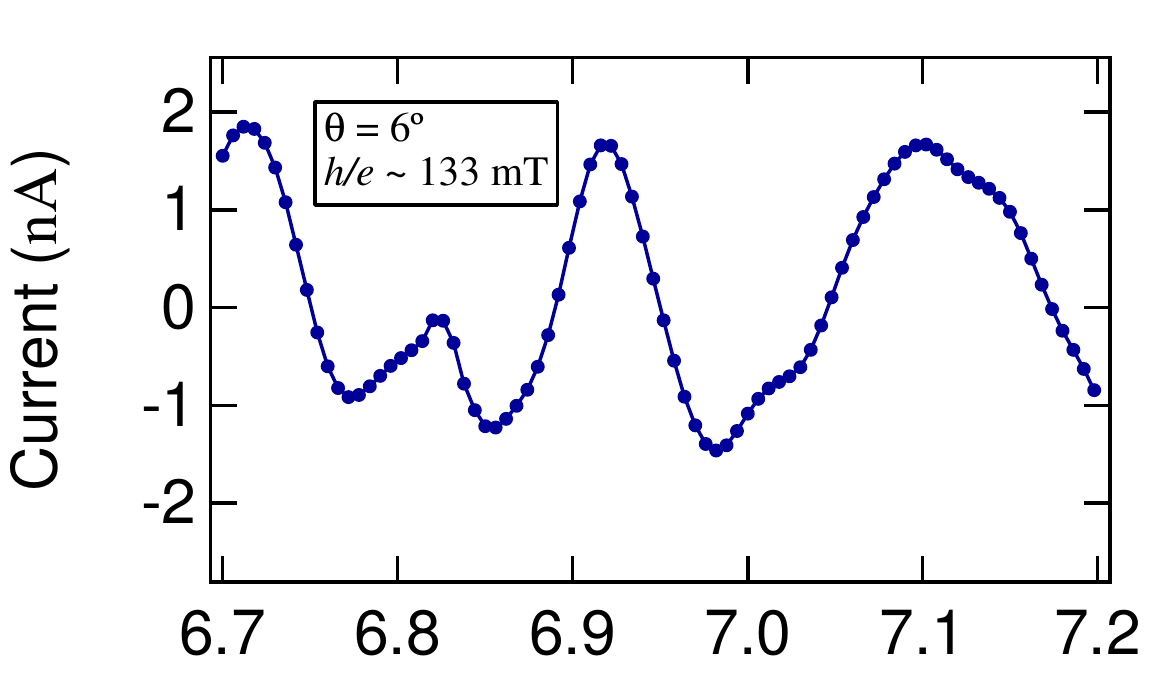}
\par\end{centering}

\begin{centering}
\includegraphics[width=0.7\paperwidth]{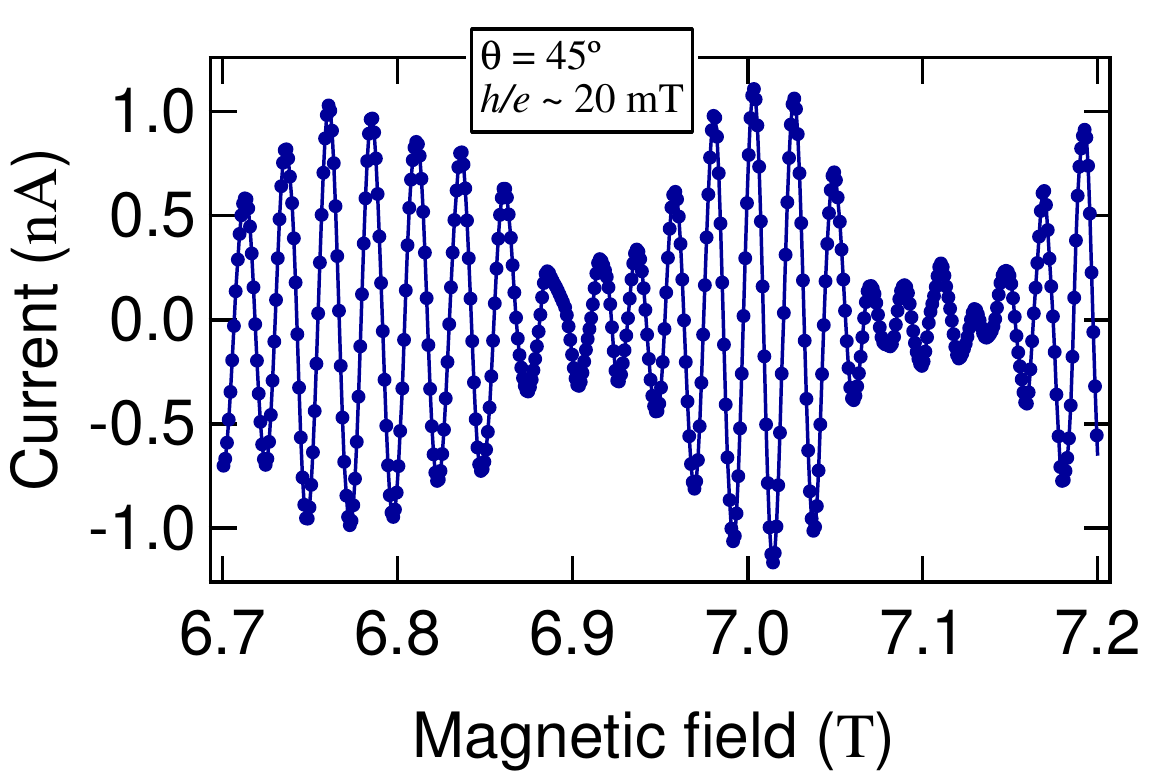}
\par\end{centering}

\caption[Comparison of persistent current observed in sample CL17 for two different
angles $\theta_{0}$]{\label{fig:ChData_DAT3_AngCL17}Comparison of persistent current
observed in sample CL17 for two different angles $\theta_{0}$. The
current inferred from measurements with $\theta_{0}=6^{\circ}$ and
$T=323\,\text{mK}$ (top graph) and $\theta_{0}=45^{\circ}$ and $T=365\,\text{mK}$
(bottom graph) are shown. The expected $h/e$ periodicity in terms
of applied magnetic field is given with each plot for the mean ring
radius and specified angle $\theta_{0}$.}
\end{figure}

In Figs. \ref{fig:ChData_DAT2_AngCL15} and \ref{fig:ChData_DAT3_AngCL17},
the current observed in samples CL15 and CL17 is shown for both $\theta_{0}=6^{\circ}$
and $\theta_{0}=45^{\circ}$ over the same $0.5\,\text{T}$ field
region as shown in Fig. \ref{fig:ChData_DAT1_6DegIvsB}. The magnetic
field scale $B_{h/e}$ associated with threading a flux quantum $h/e$
through the mean ring radius can be written as 
\[
B_{h/e}=\frac{1}{\beta_{1}}=\frac{4\pi\phi_{0}}{L^{2}\sin\theta_{0}}.
\]
Thus one expects the period of the oscillation to decrease as $\theta_{0}$
increases from $0^{\circ}$ to $90^{\circ}$. This trend is clearly
followed by the data in both Figs. \ref{fig:ChData_DAT2_AngCL15}
and \ref{fig:ChData_DAT3_AngCL17}. 

The effect of the angle $\theta_{0}$ on the persistent current was
addressed briefly in \ref{sub:CHPCTh_FluxThroughMetal} in which the
consequences of magnetic field penetrating the metal portions of the
ring was discussed. For the most part, Section \ref{sub:CHPCTh_FluxThroughMetal}
discussed the magnetic field penetrating the metal in terms of the
toroidal field model. It was found that the magnetic flux through
the metal introduced a finite scale $B_{c}$ of correlation to the
persistent current oscillation. It was also argued that this field
scale would need to be rescaled by a factor of order unity for a different
geometrical arrangement between the ring and the applied magnetic
field. That the observed persistent current oscillation has a finite
range of correlation is evident in Figs. \ref{fig:ChData_DAT2_AngCL15}
and \ref{fig:ChData_DAT3_AngCL17}, especially in the data taken with
$\theta_{0}=45^{\circ}$ for which more oscillations fit into the
$0.5\,\text{T}$ field range shown. The analysis of Section \ref{sub:CHPCTh_FluxThroughMetal}
predicted no change in the typical magnitude of the current, and indeed
the current magnitudes in Figs. \ref{fig:ChData_DAT2_AngCL15} and
\ref{fig:ChData_DAT3_AngCL17} are of the same order of magnitude
for both orientations.

In Section \ref{sub:ChData_SigProcDescription}, the frequency shift
was scaled by a factor of $\sqrt{N}$ to determine the current per
ring where $N$ is the number of rings on the cantilever. This choice
of scaling was justified by citing the fact that the persistent current
discussed in \ref{sec:CHPCTh_DiffusiveRegime} is random in sign.
Additionally, near the end of Section \ref{sub:CHPCTh_AvgInteraction}
it was argued that for moderate ring to ring variation the phase of
oscillation in magnetic field should be random from ring to ring in
the regime of magnetic field over which the measurements discussed
in this section were performed.%
\footnote{The effect of magnetic flux through the metal discussed in Section
\ref{sub:CHPCTh_FluxThroughMetal} also leads to a random phase. We
point out the other source of phase randomization because fewer assumptions
about the nature of the persistent current were required to derive
it.%
} 

The $\sqrt{N}$ scaling is justified as follows. The frequency of
a cantilever containing an array of $N$ rings is affected by the
total current $I_{\Sigma}$ of all of the rings in the array. The
quantity $I_{\Sigma}$ can be written as 
\[
I_{\Sigma}=\sum_{j=1}^{N}I_{j}
\]
where $I_{j}$ is the current in the $j^{th}$ ring. Writing the Fourier
series expansion with respect to magnetic field of the $j^{th}$ ring
as 
\[
I_{j}\left(B\right)=\text{Im}\left[\sum_{p}I_{j,p}e^{2\pi ip\beta_{1}B}e^{i\psi_{p}}\right],
\]
 the $p^{th}$ coefficient of the expansion
\[
I_{\Sigma}\left(B\right)=\text{Im}\left[\sum_{p}I_{\Sigma,p}e^{2\pi ip\beta_{1}B}e^{i\psi_{p}}\right]
\]
of the total current can be written as
\[
I_{\Sigma,p}=\sum_{j=1}^{N}I_{j,p}.
\]
For a cantilever with an array of rings, the coefficient $I_{p}$
in Eq. \ref{eq:CHTorsMagn_FiniteAmpFreqShift} relating the frequency
shift and the persistent current becomes $I_{\Sigma,p}$. The central
limit theorem of probability theory states that, when all of the $I_{j,p}$
satisfy $\langle I_{j,p}\rangle=0$ and $\sqrt{\langle I_{j,p}^{2}\rangle}=I_{p}^{\text{typ}}$,
the sum $I_{\Sigma,p}$ satisfies $\langle I_{\Sigma,p}\rangle=0$
and $\sqrt{\langle I_{\Sigma,p}^{2}\rangle}=\sqrt{N}I_{p}^{\text{typ}}$
in the limit of large $N$. According to the central limit theorem,
the typical magnitude of the current in the array is a factor of $\sqrt{N}$
larger than the single ring current. Therefore, when the total current
trace from an array of rings is rescaled by $\sqrt{N}$, the resulting
trace has the same typical magnitude as the trace from a single ring.
We note that more generally the central limit theorem requires that
in the limit of large $N$ the total current amplitude $I_{\Sigma,p}$
in an array should follow the normal distribution with zero mean and
a standard deviation of $\sqrt{N}I_{p}^{\text{typ}}$.

\begin{figure}
\begin{centering}
\includegraphics[width=0.7\paperwidth]{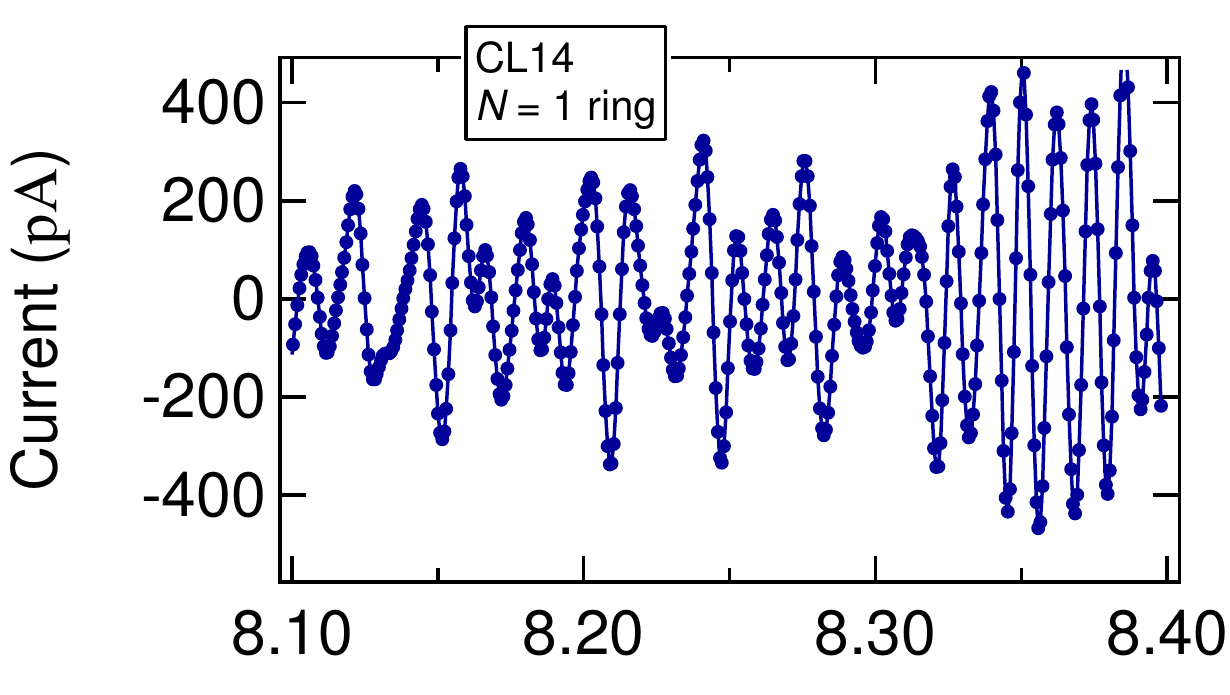}
\par\end{centering}

\begin{centering}
\includegraphics[width=0.7\paperwidth]{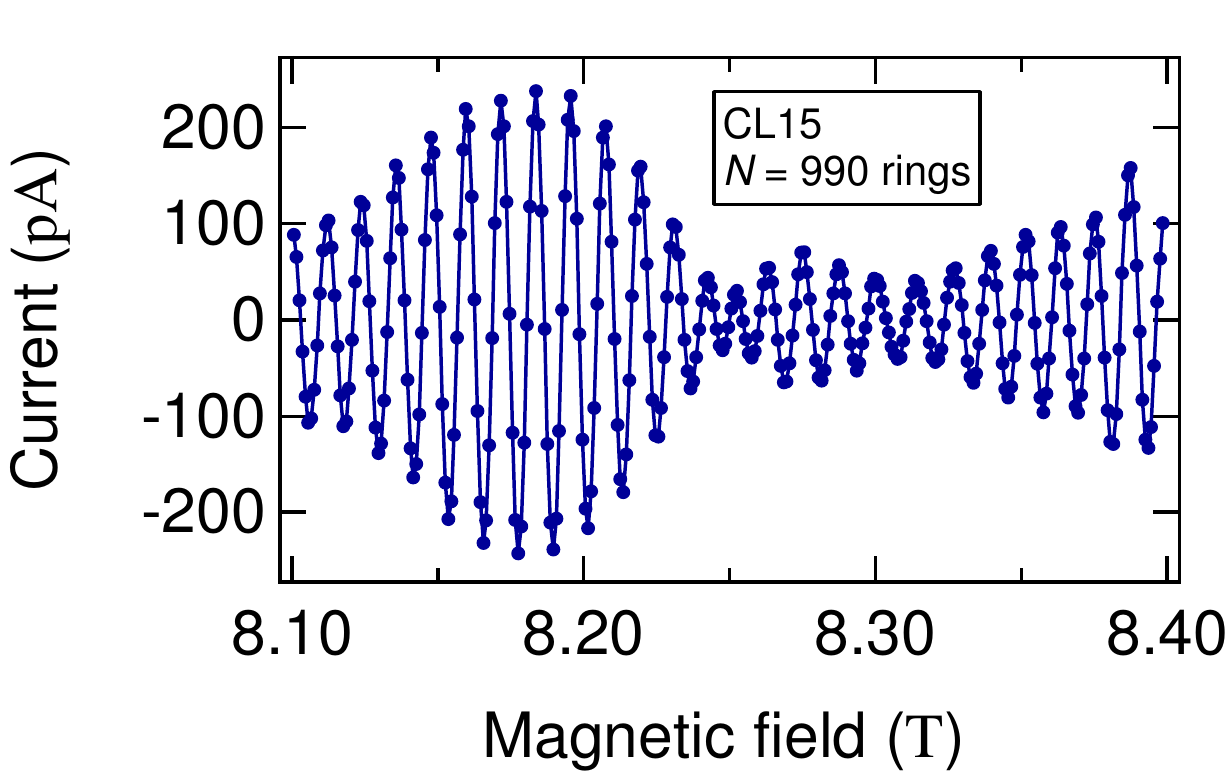}
\par\end{centering}

\caption[Comparison of the persistent current signal from a single ring with
the signal from an array of rings]{\label{fig:ChData_DAT4_RingNum}Comparison of the persistent current
signal from a single ring with the signal from an array of rings.
The inferred persistent current per ring for samples CL14 (upper plot)
and CL15 (lower plot) are shown for $\theta_{0}=45^{\circ}$ and $T=365\,\text{mK}$
between $8.1$ and $8.4\,\text{T}$. The cantilevers and rings of
both samples are nominally the same with the main difference that
sample CL14 had only a single ring while sample CL15 had an array
of 990 rings. With the signal scaled to represent the current per
ring as described in the text, both samples show similar current magnitudes
and magnetic field dependence.}
\end{figure}

Fig. \ref{fig:ChData_DAT4_RingNum} displays an experimental result
in support of the $\sqrt{N}$ scaling. In the figure, the current
observed for samples CL14 and CL15 are plotted between $8.1$ and
$8.4\,\text{T}$ for $\theta_{0}=45^{\circ}$ and $T=365\,\text{mK}$.
The two samples both contain rings with the same circumference $L=2.6\,\text{\ensuremath{\mu}m}$,
but CL15 had an array of 990 rings fabricated on its tip whereas sample
CL14 had only a single ring.%
\footnote{The chip also contained single ring samples with ring radii of $540\,\text{nm}$
and $793\,\text{nm}$. No current was observed in these samples, but,
given the diffusion constants (see below) and the scatter in the cantilever
frequency, the signal to noise ratio for these samples was expected
to be less than one.%
} Nevertheless, with the frequency shift from CL15 converted into a
current per ring using the $\sqrt{N}$ scaling, the current was observed
to be of the same order of magnitude, $\sim200\,\text{pA}$, for both
rings. A more quantitative comparison of the current magnitude observed
for each sample will be provided below.

\FloatBarrier

\subsection{\label{sub:ChData_Quantitative}Quantitative discussion}

In order to test the theoretical picture of Chapter \ref{cha:CHMeso_}
more directly, we measured each sample listed in Tables \ref{tab:ChData_CLs}
and \ref{tab:ChData_Rings} over as wide a range of magnetic field
and temperature as possible. The large magnetic field scans were all
performed at the refrigerator's base temperature with the sample chip
oriented at both $\theta_{0}=6^{\circ}$ and $\theta_{0}=45^{\circ}$
relative to the magnetic field. Plots of the inferred persistent current
versus magnetic field from these measurements are presented in \ref{sub:ChData_FullCurrentTraces}.
Measurements at different temperatures were performed over small regions
of magnetic field as described below.

The main result of Chapter \ref{cha:CHMeso_} that can be tested by
our measurements is Eq. \ref{eq:CHPCTh_IIFiniteTZSO} which specifies
the persistent current correlation function $\langle I(\phi,B_{M})I(\phi',B_{M}^{\prime})\rangle$
in the presence of finite temperature, spin-orbit scattering, and
Zeeman splitting. We do this by determining the magnitude $I_{p,\text{meas}}^{\text{typ}}$
of the typical current at a series of temperatures for each sample
and comparing this quantity to $I_{p}^{\text{typ}}(T,E_{Z},E_{SO})$
given in Eq. \ref{eq:ChPCTh_IpTypTEZESO}.%
\footnote{For clarity, we add the subscript {}``meas'' to the magnitude extracted
from the data rather than that expected from theory.%
} We now describe the procedure used to determine the measured typical
current $I_{p,\text{meas}}^{\text{typ}}$. In discussing our experimental
results, we treat $\phi$ as the magnetic flux threading the mean
radius of the ring and write $\phi/\phi_{0}=\beta_{1}B$ as was done
in Eq. \ref{eq:ChData_DeltaF}.

\subsubsection{Power spectral density of the persistent current}

According to \ref{sub:CHPCTh_FluxThroughMetal}, the magnetic field
$B$ of the experimental arrangement differs from the toroidal magnetic
field $B_{M}$ considered in that section through a geometrical factor
$\gamma$ of order unity. We define this factor by the relation $B=\gamma B_{M}$.
The finite range of correlation in applied magnetic field can then
be written as $\gamma B_{c,p}$ where $B_{c,p}$ was introduced in
\ref{sub:CHPCTh_FluxThroughMetal} as the correlation range in terms
of the toroidal field $B_{M}$ for the $p^{th}$ harmonic of the current.

The finite correlation in applied magnetic field of the persistent
current oscillation implies that measurements of the persistent current
made in one ring at values of $B$ separated by $\Delta B\gg\gamma B_{c,p}$
are uncorrelated and thus effectively independent measurements of
the persistent current for that ring. By measuring the current over
a field range $B_{0}\gg\gamma B_{c,p}$, we can build up a data set
of many independent measurements of the current. The uncertainty in
the calculated cumulants of such a finite, correlated data set has
been derived previously \citep{tsyplyatyev2003applicability} and
will be used below to find the uncertainty in our measurements.

The finite correlation in applied magnetic field $B$ of the persistent
current oscillation also leads to a broadening of the peak located
at $\beta=p\beta_{1}$ associated with the $p^{th}$ harmonic of the
current. We can rewrite the persistent current autocorrelation function
of Eq. \ref{eq:CHPCTh_IIFiniteTZSO} in terms of the applied field
$B$ as
\[
\left\langle I\left(B\right)I\left(B+B'\right)\right\rangle =\sum_{p=1}^{\infty}\left(I_{p}^{\text{typ}}\left(T,E_{Z},E_{SO}\right)\right)^{2}\cos\left(2\pi p\beta_{1}B'\right)K_{p}\left(\frac{B'}{\gamma B_{c,p}}\right)
\]
where $I_{p}^{\text{typ}}(T,E_{Z},E_{SO})$ was given in Eq. \ref{eq:ChPCTh_IpTypTEZESO}
and $K_{p}$ is the autocorrelation function 
\begin{equation}
K_{p}\left(x\right)=\frac{c_{p}^{T}\left(T,B_{c,p}x,E_{Z},E_{SO}\right)}{c_{p}^{T}\left(T,0,E_{Z},E_{SO}\right)}\label{eq:ChData_Kp}
\end{equation}
normalized so that $K_{p}(0)=1$ and rescaled in $B$ so that an argument
of $x=1$ corresponds to $B_{M}=B_{c,p}$. For $T=0$ and $E_{SO}\gg E_{c}$,
the Fourier transform of $K_{p}(x)$ is a peak centered at zero with
a half-width at half-maximum of $\Delta\sim1/6$. Consequently, the
Fourier transform of $\langle I(B)I(B+B')\rangle$ with respect to
$B'$ consists of a series of peaks located at $\beta=p\beta_{1}$
and possessing half widths at half maximum of 
\begin{equation}
\Delta\beta_{p}\approx\frac{1}{12\gamma B_{c,p}}\label{eq:ChData_dBetaP}
\end{equation}
for each positive integer $p$. From Fig. \ref{fig:CHPCTh_BcorrLSO},
it can be seen that the correlation scale $B_{c,p}$ increases slightly
with temperature and thus that $\Delta\beta_{p}$ decreases. We note
that the Fourier transform of $\langle I(B)I(B+B')\rangle$ is equal
to the power spectral density $S_{I}(\beta)$ of the persistent current.

\begin{figure}
\begin{centering}
\includegraphics[width=0.7\paperwidth]{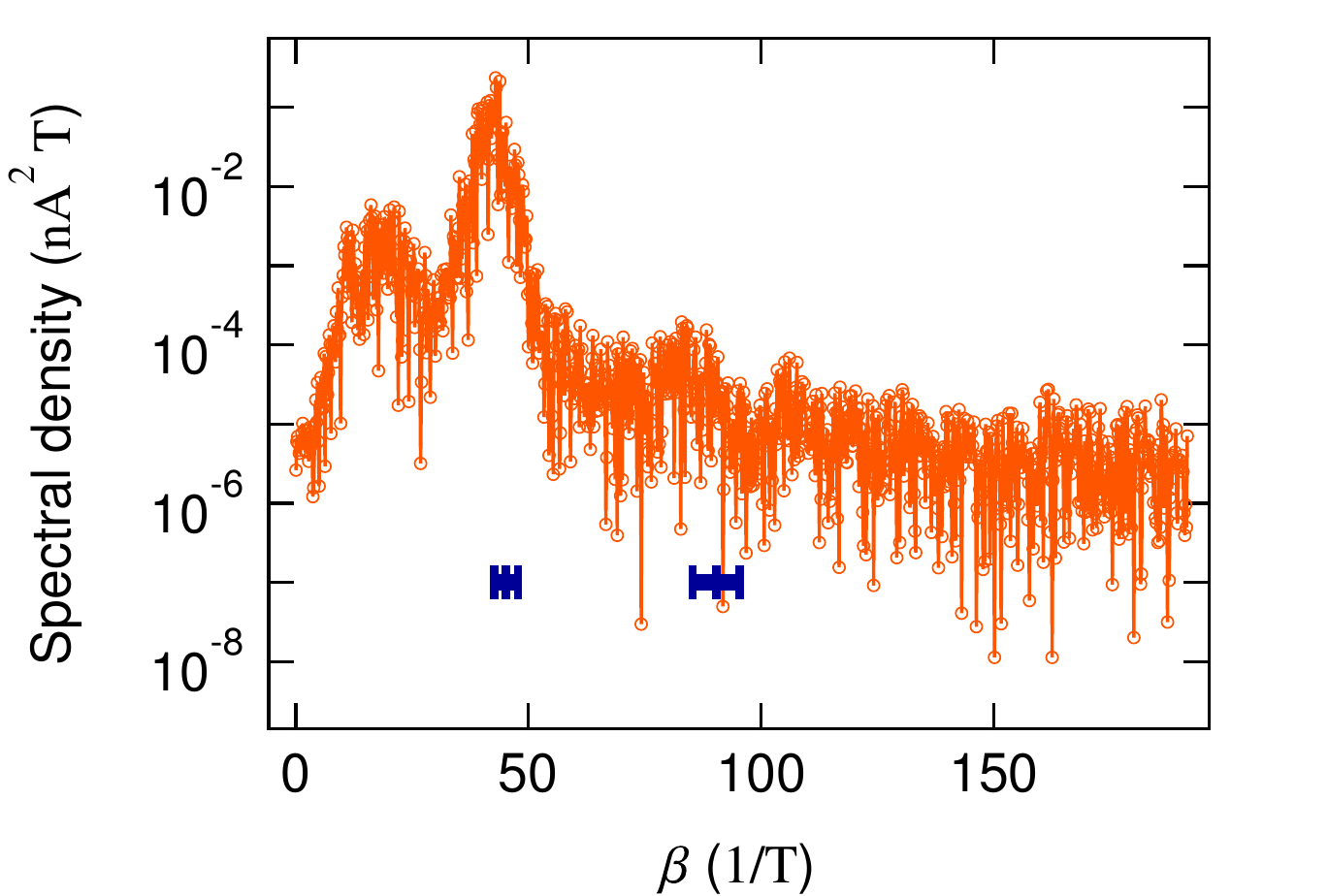}
\par\end{centering}

\caption[Persistent current power spectral density for sample CL17 at $45^{\circ}$]{\label{fig:ChData_DAT5_PSDCL17_45Deg}Persistent current power spectral
density for sample CL17 at $45^{\circ}$. Horizontal bars indicate
the expected location $\beta_{p}$ and width $\Delta\beta_{p}$ of
the peaks associated with the persistent current. The first harmonic
peak is clearly visible. As discussed in the text, a small second
harmonic peak is also present. The total current under the first harmonic
peak is $653\,\text{pA}$ while that under the second is $27\pm9\,\text{pA}$.
The data shown was taken at $T=365\,\text{mK}$.}
\end{figure}

\begin{figure}
\begin{centering}
\includegraphics[width=0.7\paperwidth]{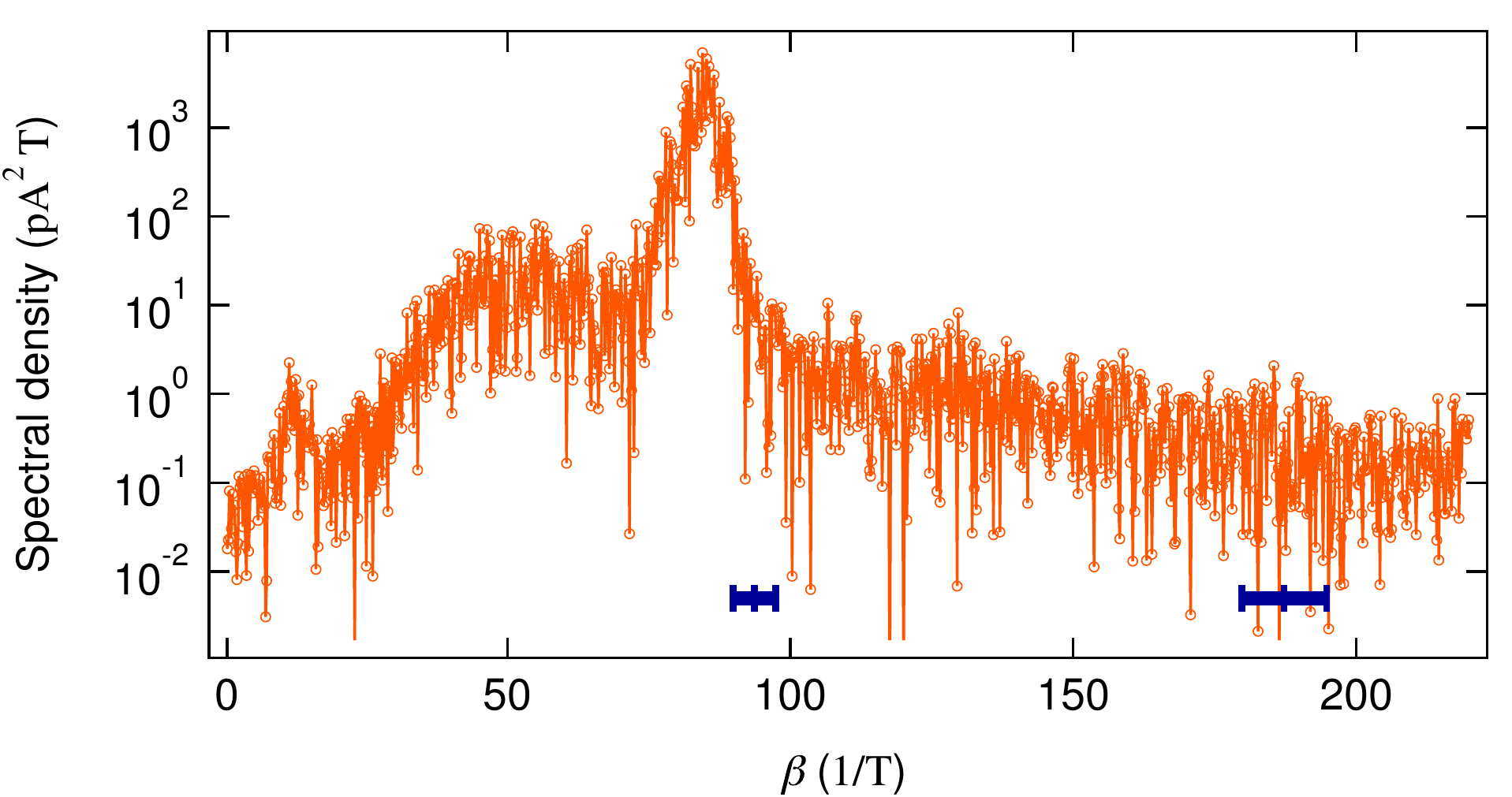}
\par\end{centering}

\caption[Persistent current power spectral density for sample CL15 at $45^{\circ}$]{\label{fig:ChData_DAT6_PSDCL15_45Deg}Persistent current power spectral
density for sample CL15 at $45^{\circ}$. Horizontal bars indicate
the expected location $\beta_{p}$ and width $\Delta\beta_{p}$ of
the peaks associated with the persistent current. The first harmonic
peak is clearly visible. No second harmonic peak is detected above
the level of the noise background. The total current under the first
harmonic peak is $131\,\text{pA}$. The data shown was taken at $T=365\,\text{mK}$.}
\end{figure}

\begin{figure}
\begin{centering}
\includegraphics[width=0.65\paperwidth]{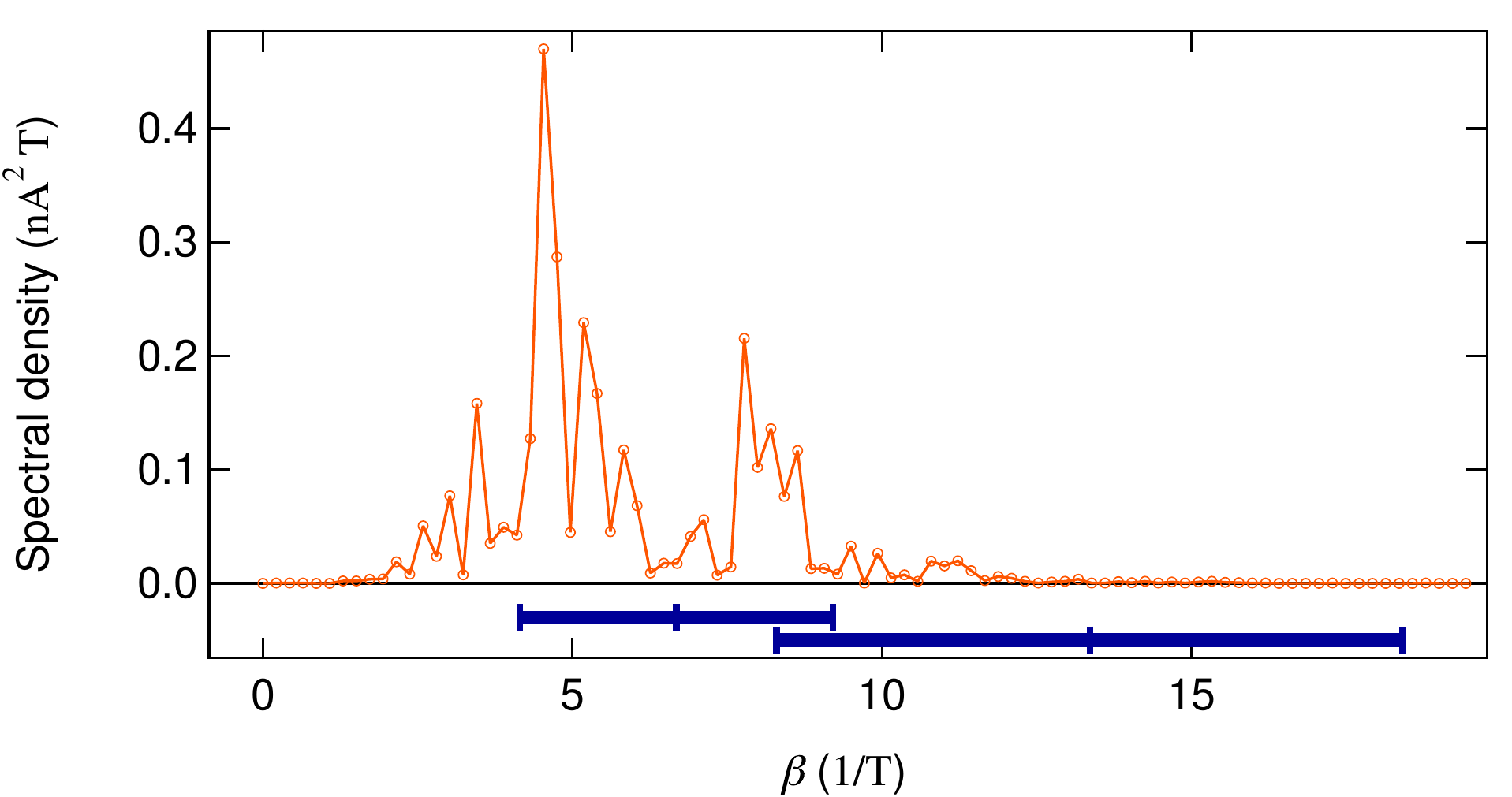}
\par\end{centering}

\caption[Persistent current power spectral density for sample CL17 at $6^{\circ}$]{\label{fig:ChData_DAT7_PSDCL17_6Deg}Persistent current power spectral
density for sample CL17 at $6{}^{\circ}$. All of the features present
in the spectrum are associated with the first harmonic of the persistent
current signal. Horizontal bars indicate the expected location $\beta_{p}$
and width $\Delta\beta_{p}$ of the peaks associated with the persistent
current. At this small angle $\theta_{0}$, the magnetic field frequencies
$\beta_{p}$ of the persistent current harmonics are not much larger
than the widths $\Delta\beta_{p}$ of the associated peaks. The second
harmonic, which was barely detectable at $\theta_{0}=45^{\circ}$,
is indistinguishable from the slightly larger background observed
for $\theta_{0}=6^{\circ}$. The total current associated with all
features is $693\,\text{pA}$. The data shown was taken at $T=323\,\text{mK}$.}
\end{figure}

\begin{figure}
\begin{centering}
\includegraphics[width=0.7\paperwidth]{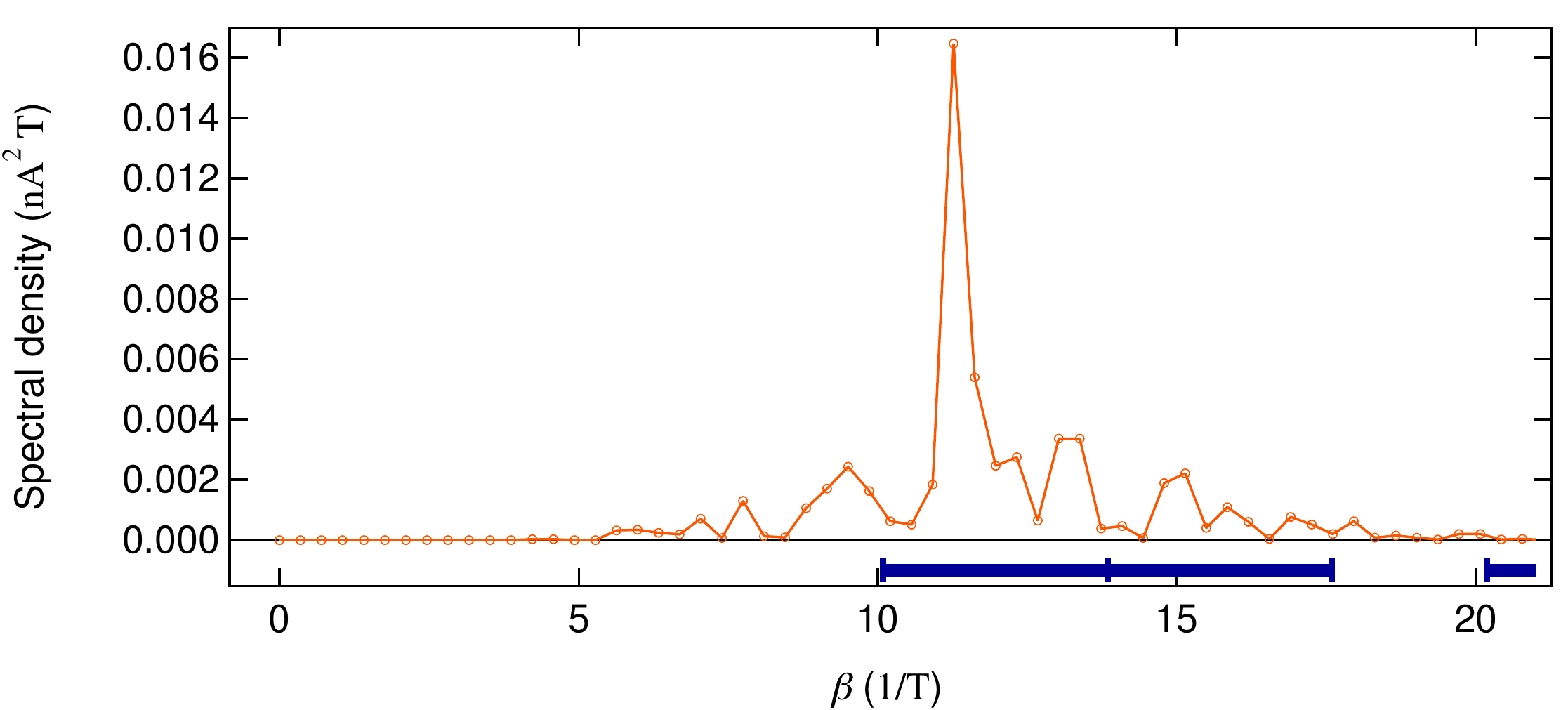}
\par\end{centering}

\caption[Persistent current power spectral density for sample CL15 at $6^{\circ}$]{\label{fig:ChData_DAT8_PSDCL15_6Deg}Persistent current power spectral
density for sample CL15 at $6^{\circ}$. A broad peak associated with
the persistent current is visible. Horizontal bars indicate the expected
location $\beta_{p}$ and width $\Delta\beta_{p}$ of the peaks associated
with the persistent current. Most of the region of the spectrum associated
with the second harmonic is above the cut-off frequency $\beta_{\text{zero}}=21\,\text{T}^{-1}$
beyond which the measurement was insensitive to the persistent current.
The large spike in the middle of the peak is presumably due to insufficient
averaging associated with the small number of measured statistically
independent sections of the persistent current signal. The total current
under the peak is $133\,\text{pA}$. The data shown was taken at $T=323\,\text{mK}$.}
\end{figure}

\begin{figure}
\begin{centering}
\includegraphics[width=0.7\paperwidth]{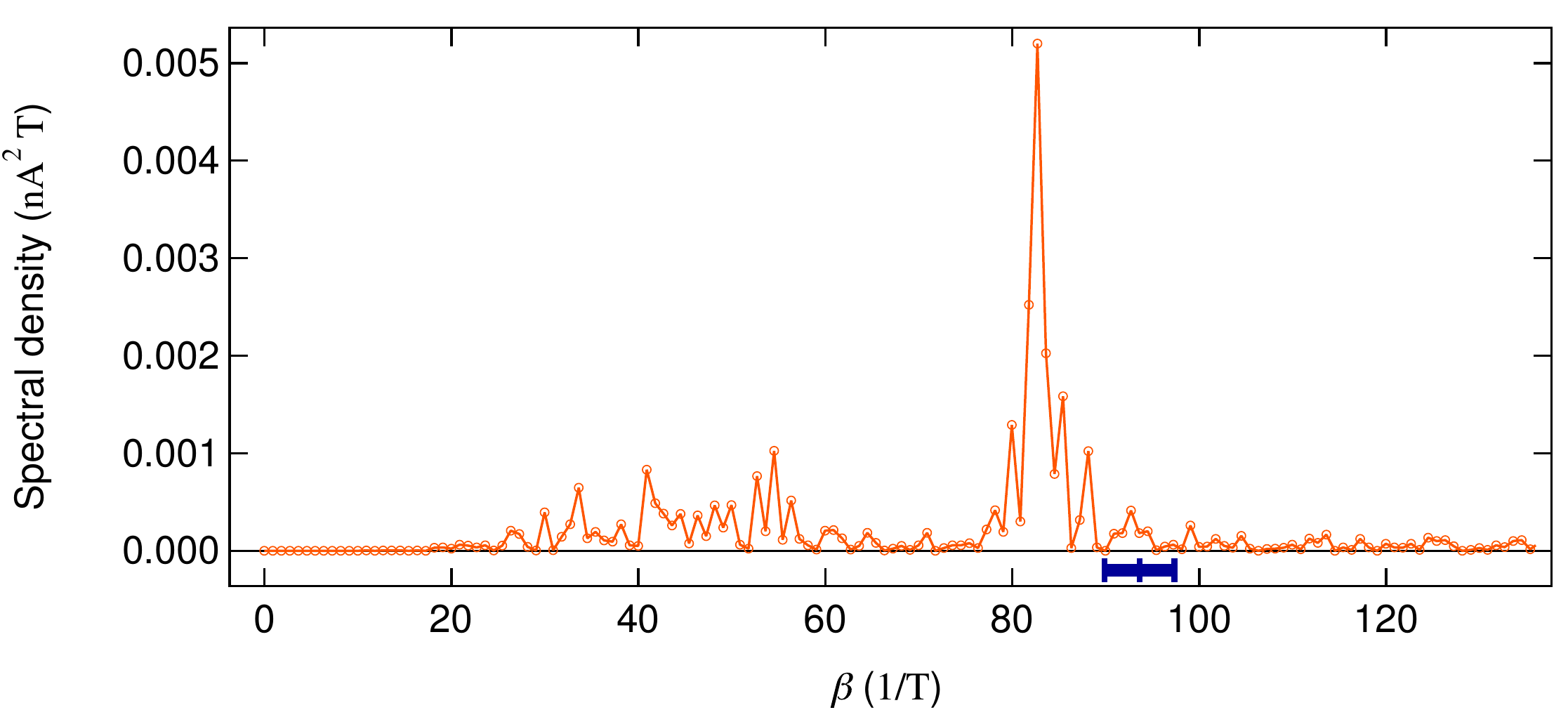}
\par\end{centering}

\caption[Persistent current power spectral density for sample CL14 at $45^{\circ}$]{\label{fig:ChData_DAT9_PSDCL14}Persistent current power spectral
density for sample CL14 at $45^{\circ}$. The horizontal bar indicates
the expected location $\beta_{p}$ and width $\Delta\beta_{p}$ of
the peak associated with the first harmonic of the persistent current
signal. A strong peak due to the persistent current is present at
values of $\beta$ just below the region covered by this bar. The
cut-off frequency $\beta_{\text{zero}}$ discussed in the text occurs
at $\beta=136\,\text{T}^{-1}$, below the region of the spectrum associated
with the second harmonic. A low, broad peak is located at lower values
of $\beta$ than the large peak near the expected location of the
first harmonic of the persistent current signal. This smaller peak
is the remnant of the low $\beta$ noise removed during the subtraction
of the smooth background. The total current under the peak is $116\,\text{pA}$.
The data shown was taken at $T=365\,\text{mK}$.}
\end{figure}

\begin{figure}
\begin{centering}
\includegraphics[width=0.7\paperwidth]{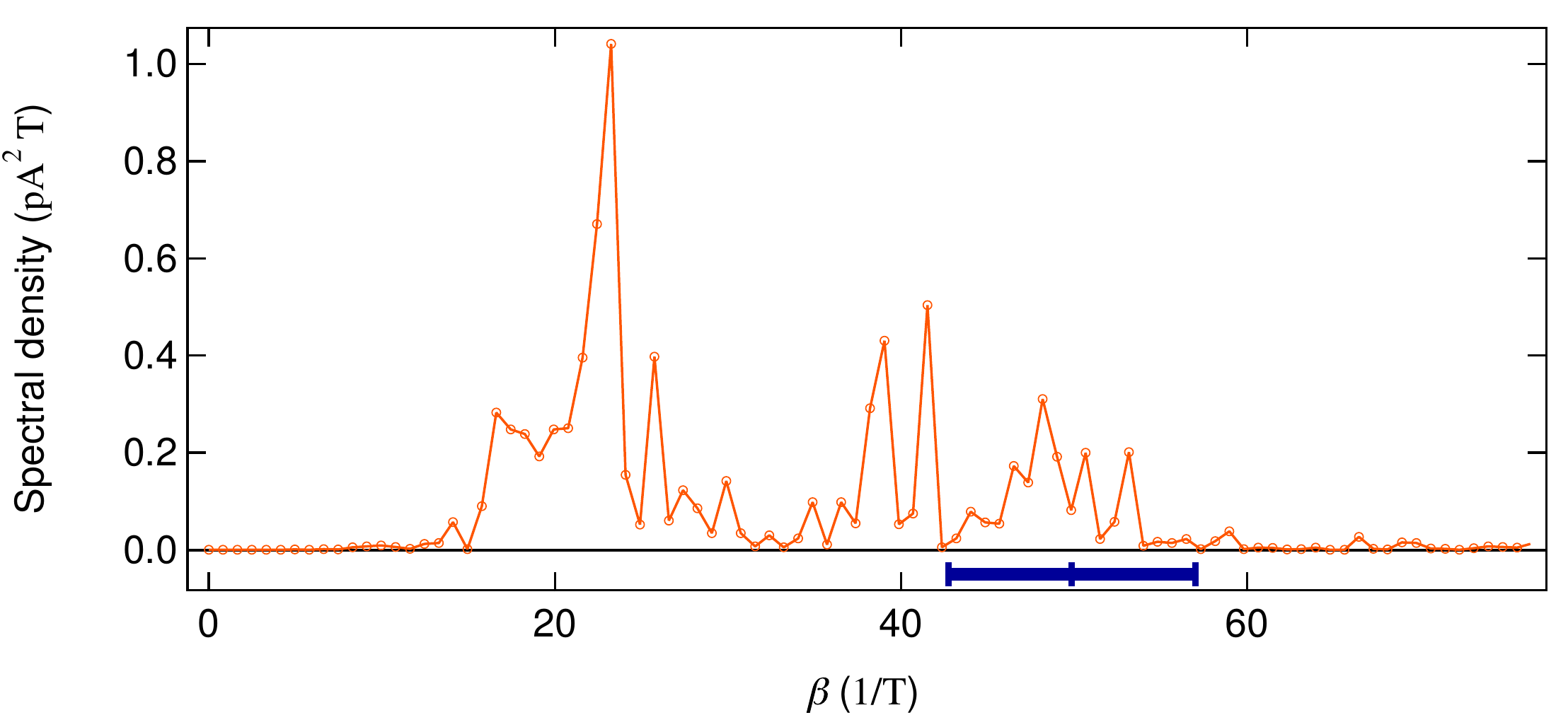}
\par\end{centering}

\caption[Persistent current power spectral density for sample CL11 at $6^{\circ}$]{\label{fig:ChData_DAT10_PSDCL11}Persistent current power spectral
density for sample CL11 at $6^{\circ}$. The horizontal bar indicates
the expected location $\beta_{p}$ and width $\Delta\beta_{p}$ of
the peak associated with the first harmonic of the persistent current
signal. The wide peak above the bar and the two slightly taller peaks
located just below it in $\beta$ are due to the persistent current.
The larger peak closer to $\beta=20\,\text{T}^{-1}$ is the shoulder
left by the removal of the smooth background from the low $\beta$
noise in the $\Delta f(B)$ and $I(B)$ traces (see discussion in
text and Fig. \ref{fig:ChData_DAT30_FreqPSD_CL11}). The cut-off frequency
$\beta_{\text{zero}}$ at which the measurement was insensitive to
the persistent current occurs at $\beta=82\,\text{T}^{-1}$, below
the region of the spectrum associated with the second harmonic. Total
current associated with the features which we attribute to the persistent
current is $1.7\,\text{pA}$. The data shown was taken at $T=323\,\text{mK}$.}
\end{figure}

In Figs. \ref{fig:ChData_DAT5_PSDCL17_45Deg} through \ref{fig:ChData_DAT10_PSDCL11},
we plot the power spectral densities $S_{I}(\beta)$ associated with
each of the measured persistent current traces (see Section \ref{sub:ChData_FullCurrentTraces}
for the complete traces). The power spectral densities are calculated
from the data by
\[
S_{I}\left(\beta\right)=\frac{2}{B_{0}}\left|\sum_{j=0}^{M-1}\Delta BI_{j}\exp\left(2\pi ij\beta\Delta B\right)\right|^{2}.
\]
Here the $\left\{ I_{j}\right\} $ are the $M$ values of the current
arranged in order of ascending magnetic field $B_{j}=B_{i}+j\Delta B$
with $B_{i}$ the initial value of $B$ and $\Delta B$ the magnetic
field spacing. We use $B_{0}=(M-1)\Delta B$ to indicate the range
of magnetic field spanned by the measurement. Also shown in each plot
are bars spanning the regions from $p\beta_{1}-\Delta\beta_{p}$ to
$p\beta_{1}+\Delta\beta_{p}$ for $p=1,2$ in order to indicate the
expected peak location and width of the first two harmonics of the
persistent current. For these bars, we use the zero-temperature value
for $\Delta\beta_{p}$ given in Eq. \ref{eq:ChData_dBetaP} with $\gamma$
set to 1.

In all of the power spectral density plots shown, peaks are visible
in the spectrum close to the bars associated with the first harmonic
of the persistent current oscillation. For the spectra with more well-defined
peaks, the peaks appear to be centered at $\sim90\%$ of the expected
value for $\beta_{1}$. This discrepancy in magnetic field frequency
could be due to an offset error in the values of the orientation angle
$\theta_{0}$ of the cantilever chip relative to the applied magnetic
field. An offset error of $\sim-6^{\circ}$ in $\theta_{0}$ would
be necessary to account for the observed discrepancy at a nominal
angle of $45^{\circ}$. Such an offset in $\theta_{0}$ would result
in an underestimation of the current magnitude by $\sim10\%$. However,
we estimate that the offset angle should not be larger than $1^{\circ}$.
We propose an alternative explanation of the discrepancy below. To
account for the observed discrepancy at a nominal angle of $6^{\circ}$,
an error of $\sim-0.5^{\circ}$is required and results in a negligible
error in the inferred persistent current magnitude.

Inaccuracy in defining the effective mean area $A$ of the ring from
its geometrical arrangement in the applied magnetic field could be
responsible for the discrepancy in the observed magnetic field frequency.
In \ref{sec:CHPCTh_DiffusiveRegime}, the finite linewidth of the
ring is taken into account by modeling the applied magnetic field
as an idealized Aharonov-Bohm flux plus a toroidal magnetic field.
It is possible that an accurate model for the experimental geometry
of a uniform magnetic field applied at angle to the plane of the ring
would calculate the flux $\phi_{\text{tot}}$ threading the ring using
a different area than the area corresponding to the mean radius of
the ring%
\footnote{By which we mean the area $\pi R^{2}\sin\theta_{0}$ covered when
a circle with a radius $R$ equal to the mean radius of the ring is
projected onto a plane perpendicular to the direction of the applied
magnetic field. %
} and thus a different expression for the expected frequency $\beta_{1}$.
Using $\beta_{1}=\pi R_{\text{eff}}^{2}\sin\theta_{0}/\phi_{0}$ with
the nominal value for $\theta_{0}$, the observed values of $\beta_{1}$
correspond to values for the effective radius $R_{\text{eff}}$ that
are $\sim5\%$ smaller than the nominal mean radii and thus still
within the nominal linewidths of the rings. It is unlikely that there
is a significant error in the dimensions of the rings as these dimensions
were measured with a scanning electron microscope.

The clearest peaks were obtained for samples CL15 and CL17 at $\theta_{0}=45^{\circ}$.
For these two measurements, the widest range ($B_{0}>5\,\text{T}$)
of magnetic field was studied and the greatest number $M_{\text{eff}}$
of statistically independent measurements (see discussion below for
a precise definition of $M_{\text{eff}}$) were obtained. The peaks
in the spectra of the other measurements appear uneven due to the
lower number of independent measurements. The spectra for samples
CL15 and CL17 at $\theta_{0}=45^{\circ}$ are plotted on a log scale
in Figs. \ref{fig:ChData_DAT5_PSDCL17_45Deg} and \ref{fig:ChData_DAT6_PSDCL15_45Deg}.
The suppression of the spectrum floor at low $\beta$ is the result
of the background subtraction steps (which act as high pass filters)
performed during the frequency to current conversion (see Section
\ref{sec:ChData_SigProc}). The downward sloping trend of $S_{I}(\beta)$
for higher values of $\beta$ is produced by the intervening integration
step that converts the current derivative $\partial_{B}I$ into the
current $I$. The cantilever frequency spectrum $S_{f}(\beta)$ at
high $\beta$ is flat.

During the discussion of the conversion from frequency shift $\Delta f(B)$
to current $I(B)$, it was mentioned (see e.g. Fig. \ref{fig:ChData_SP4dIpPrimeAvsBeta})
that the coefficient $dI_{p}^{A}$, which is related to the value
$S_{I}(\beta)$ of the power spectral density of the current at $\beta=p\beta_{1}$,
was set to zero for $p\apprge p_{\text{zero}}=0.6/\beta_{1}GB_{i}$
(for which the argument of the $\text{jinc}$ factor in Eq. \ref{eq:ChData_DeltaF}
goes to zero) where $B_{i}$ was the lowest value of $B$ in the $\Delta f(B)$
trace. For Figs. \ref{fig:ChData_DAT5_PSDCL17_45Deg} through \ref{fig:ChData_DAT10_PSDCL11},
only frequencies $\beta$ below this cut-off frequency $\beta_{\text{zero}}=p_{\text{zero}}\beta_{1}=0.6/GB_{i}$
are shown. In all cases, the feature in the spectrum located close
to the expected value of $\beta_{1}$ ends distinctly below this cut-off
frequency.

For the measurements performed at $\theta_{0}=6^{\circ}$ for which
the magnetic field frequency is lower, one might worry that subtractions
of a smooth background from the data (performed in going from Fig.
\ref{fig:ChData_SP1FreqvsTime} to Fig. \ref{fig:ChData_SP2dFreqvsB}
and from Fig. \ref{fig:ChData_SP6IAvsB} to Fig. \ref{fig:ChData_SP7IAvsBeta}
in the walk-through of method A given in \ref{sub:ChData_MethodAWalkThrough})
remove part of the persistent current signal. We do not believe this
to be the case. For samples CL15 and CL17, the signal to noise ratio
is quite large, comparable to that observed at $\theta_{0}=45^{\circ}$.
An indication of the repeatability of the signal for sample CL17 is
given by the figures of \ref{sub:ChData_MethodAWalkThrough} which
show comparisons of the current inferred from measurements made under
several different experimental arrangements. No feature appeared consistently
in the spectrum of the cantilever frequency $f$ at the low values
of $\beta$ removed by the background subtraction. 

Sample CL11 had a lower signal to noise ratio due to its weaker persistent
current. We believe the lowest peak in the spectrum $S_{I}(\beta)$
Fig. \ref{fig:ChData_DAT10_PSDCL11} is actually not due to the persistent
current. The drift of the cantilever frequency $f$ over time results
in an enhanced amount of noise in the low $\beta$ region of the spectrum
that falls off as $\beta^{-x}$ for $x>0$. Removal of the smooth
background reduces the magnitude of the spectral components at the
lowest $\beta$, resulting in a peak-like shoulder that, for sample
CL11, is of comparable size to the persistent current signal. Separation
of the persistent current signal from the background is aided by measuring
the current at different temperatures. At higher temperature, the
features at higher values of $\beta$ decay while the low $\beta$
shoulder does not. We discuss the signal from sample CL11 further
in Section \ref{sub:ChData_FullCurrentTraces}.

In the measurement of sample CL17 at $\theta_{0}=45^{\circ}$, we
were able to observe the second harmonic ($p=2$) of the current.
In Fig. \ref{fig:ChData_DAT5_PSDCL17_45Deg}, the second harmonic
peak is difficult to discern as it is only slightly larger than the
background. From Eqs. \ref{eq:CHPCTh_IIFiniteTZSO} and \ref{eq:CHTorsMagn_FiniteAmpFreqShift},
it can be seen that the frequency shift $\Delta f$ due to the $p^{th}$
harmonic of the persistent current is proportional to $p^{-0.5}\exp(-p^{2}k_{B}T/E_{c})$.
Thus the signal from higher harmonics is reduced compared to the fundamental
and more strongly suppressed by the effect of finite temperature.
Additionally, the correlation field $B_{c,p}$ associated with the
$p^{th}$ harmonic scales as $p^{-1}$. Thus the width $\Delta\beta_{p}$
of the $p^{th}$ peak in $S_{I}(\beta)$ grows linearly with $p$.
This effect contributes to the difficulty in observing the second
harmonic peak as the smaller persistent current power is spread out
over a wider range of $\beta$. When we measure the current over a
small range of magnetic field closer to the correlation field $B_{c,p=2}$,
we observe a stronger peak for the second harmonic of the current.
Such a measurement is shown below in Figs. \ref{fig:ChData_DAT12_TempSerIvsB}
and \ref{fig:ChData_DAT13_TempSerIvsBeta}. That this feature decays
much more strongly with temperature than the peak associated with
the first harmonic is a strong confirmation that it is indeed the
second harmonic of the persistent current oscillation.

\subsubsection{Determination of the temperature dependence from the measured persistent
current}

\begin{figure}
\begin{centering}
\includegraphics[width=0.7\paperwidth]{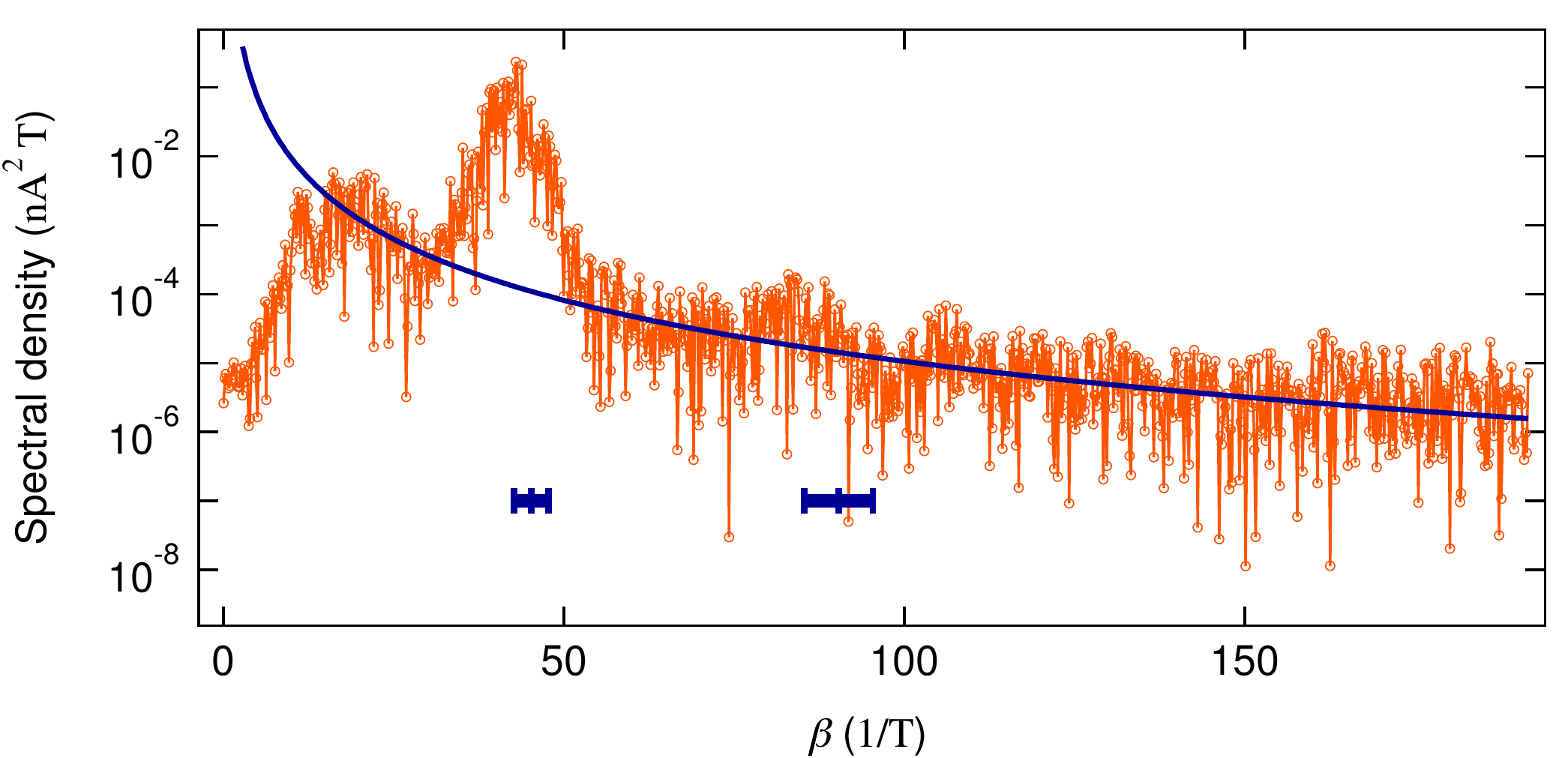}
\par\end{centering}

\caption[Fit to the background of the persistent current power spectral density
for sample CL17 at $45^{\circ}$]{\label{fig:ChData_DAT5b_PSDCL17_45Deg_BG}Fit to the background of
the persistent current power spectral density for sample CL17 at $45^{\circ}$.
The data (circles) and horizontal bars are the same as those shown
in Fig. \ref{fig:ChData_DAT5_PSDCL17_45Deg}. The solid line added
in this plot represents a fit $b(\beta)$ to the spectrum $S_{I}(\beta)$
close to, but outside of, the peak associated with the first harmonic
of the persistent current signal. The curve shown represents $b(\beta)=(8.1\times10^{-23}\,\text{A}^{2}\,\text{T})(\beta\times1\,\text{T})^{-2.9}$.
For each persistent current sample and orientation $\theta_{0}$,
the square of the typical current was determined by finding the area
under the peak in $S_{I}(\beta)$ minus the area under the background
$b(\beta)$ over the same region of $\beta$.}
\end{figure}

We determined the measured typical current $I_{p,\text{meas}}^{\text{typ}}(T_{b})$
at the refrigerator's base temperature $T_{b}$ by finding the area
under the peak of the power spectral densities $S_{I}(\beta)$ shown
in Figs. \ref{fig:ChData_DAT5_PSDCL17_45Deg} through \ref{fig:ChData_DAT10_PSDCL11}.
To do this, we first fit a curve $b(\beta)$ to each spectral density
$S_{I}(\beta)$ excluding the region dominated by the peak. The fitted
background curve $b(\beta)$ for $S_{I}(\beta)$ of sample CL17 at
$\theta_{0}=45^{\circ}$ is shown in Fig. \ref{fig:ChData_DAT5b_PSDCL17_45Deg_BG}.
The measured typical current $I_{p,\text{meas}}^{\text{typ}}(T_{b})$
is taken to be the square root of the area under $S_{I}(\beta)$ minus
the area under $b(\beta)$ in the vicinity of the peak:
\begin{align*}
\left(I_{p,\text{meas}}^{\text{typ}}(T_{b})\right)^{2} & =\int_{\beta_{p}^{-}}^{\beta_{p}^{+}}d\beta\,\left(S_{I}\left(\beta\right)-b\left(\beta\right)\right)\\
 & =\sum_{\beta_{j}=\beta_{p}^{-}}^{\beta_{p}^{+}}\Delta\beta\left(S_{I}\left(\beta_{j}\right)-b\left(\beta_{j}\right)\right)
\end{align*}
where $\Delta\beta=1/B_{0}$ is the point spacing in $\beta$ of the
power spectral density $S_{I}(\beta)$ calculated from the measured
current data. The bounds of the integration, $\beta_{p}^{-}$ and
$\beta_{p}^{+}$, are chosen to be outside of, but close to, the edges
of the peak. Their precise locations are not critical since the quantity
$S_{I}(\beta)-b(\beta)$ is zero on average outside of the peak. 

For each measurement, the background fit was of the form 
\[
b\left(\beta\right)=X\beta^{-Y}
\]
with $X$ and $Y$ fitting parameters. The value $I_{p,\text{meas}}^{\text{typ}}(T_{b})$
of the typical current inferred from each large magnetic field scan
is given in the caption of the corresponding power spectral density
plot (Figs. \ref{fig:ChData_DAT5_PSDCL17_45Deg} through \ref{fig:ChData_DAT10_PSDCL11}).
For each first harmonic, the error in the calculated typical current
$I_{p,\text{meas}}^{\text{typ}}(T_{b})$ due to fluctuations in the
background were a few percent or less and thus negligible compared
to the systematic uncertainty both due to error in the angle $\theta_{0}$
discussed above and due to the finite number $M_{\text{eff}}$ of
statistically independent measurements of the current discussed below.
The uncertainty in the second harmonic of the signal from sample CL17
due to the background measurement noise was about $33\%$.

\begin{figure}
\begin{centering}
\includegraphics[width=0.7\paperwidth]{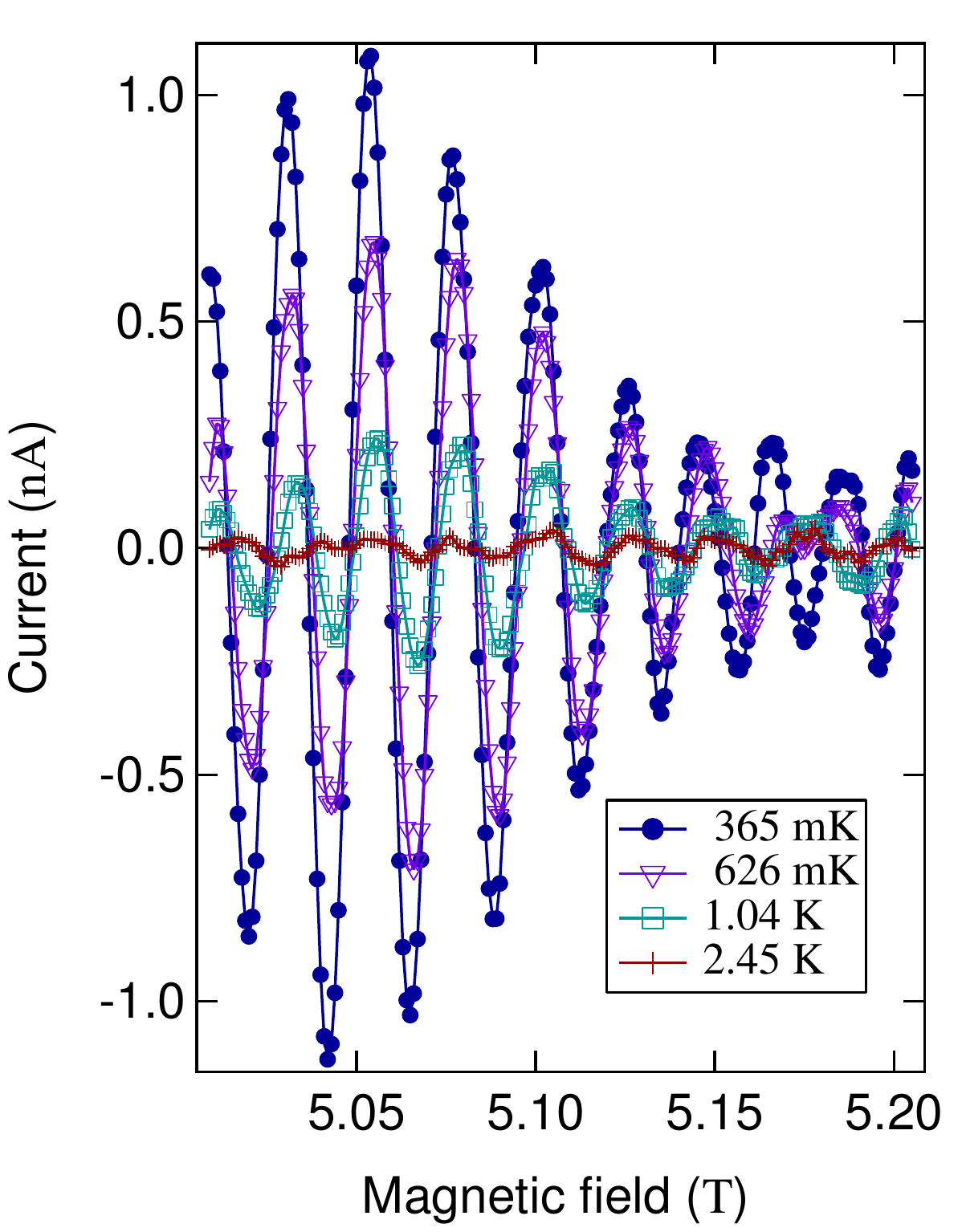}
\par\end{centering}

\caption[Persistent current versus magnetic field for a series of temperatures]{\label{fig:ChData_DAT12_TempSerIvsB}Persistent current versus magnetic
field for a series of temperatures. The traces shown were taken on
sample CL17 with $\theta_{0}=45^{\circ}$ and $T=0.365,$ 0.420, 0.626,
0.831, 1.04, and $2.45\,\text{K}$. Varying the temperature produces
an overall scaling of the data while largely leaving the shape of
the trace unchanged. Fourier transforms of the traces shown are plotted
in Fig. \ref{fig:ChData_DAT13_TempSerIvsBeta}.}
\end{figure}

\begin{figure}
\begin{centering}
\includegraphics[width=0.65\paperwidth]{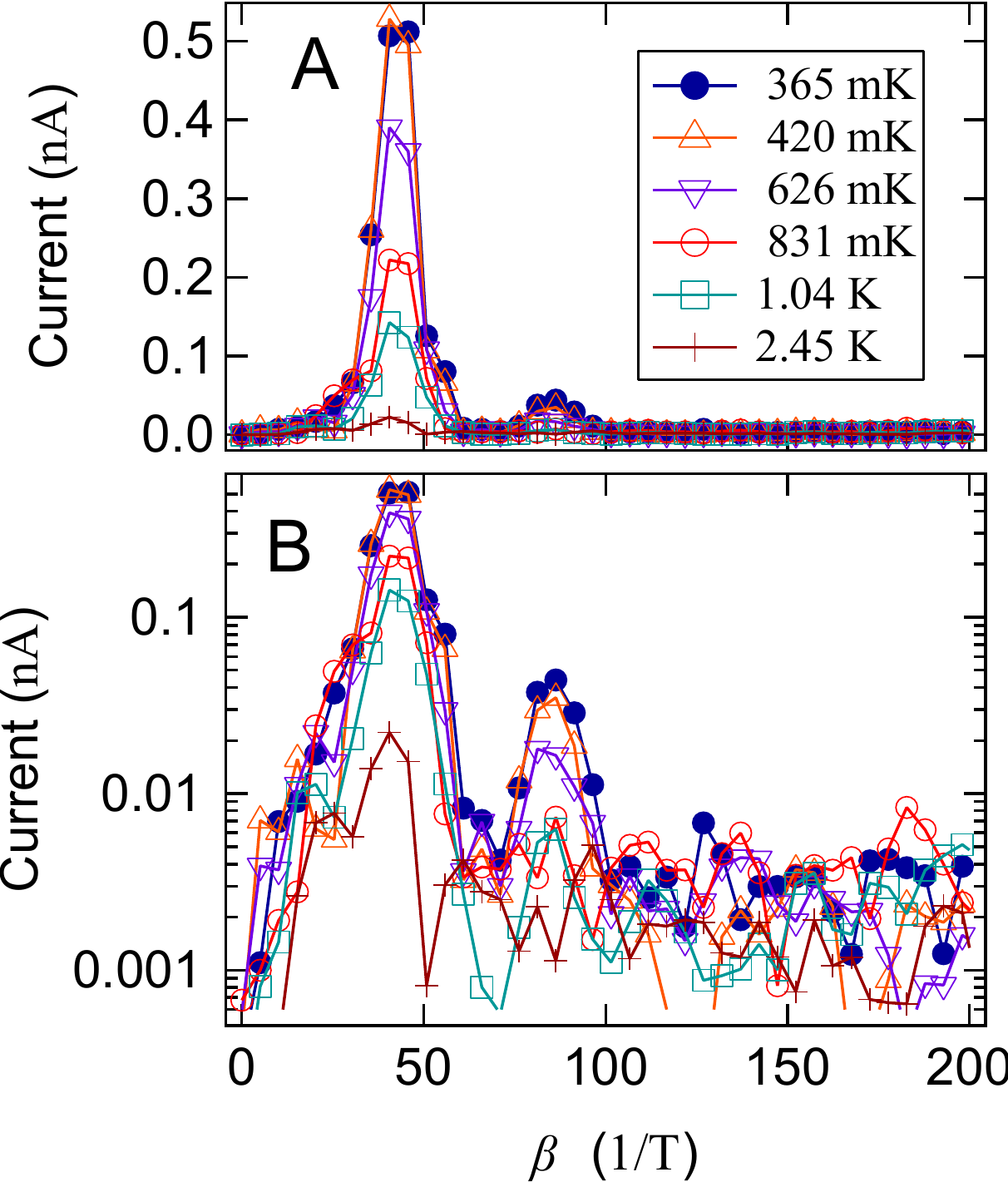}
\par\end{centering}

\caption[Fourier transform of the persistent current signal for a series of
temperatures]{\label{fig:ChData_DAT13_TempSerIvsBeta}Fourier transform of the
persistent current signal for a series of temperatures. The magnitude
$|I(\beta)|$ of the Fourier transform of the persistent current traces
measured for sample CL17 at $\theta_{0}$ is shown for $T=0.365,$
0.420, 0.626, 0.831, 1.04, and $2.45\,\text{K}$. Panels A and B show
the same data on linear and log scales respectively. Peaks in the
spectrum at $\beta\approx42\,\text{T}^{-1}$ and $\beta\approx84\,\text{T}^{-1}$
represent the first and second harmonics of the persistent current
signal respectively. The size of these peaks is used to calculate
the data shown in Fig. \ref{fig:ChData_DAT14_IvsT} for the temperature
dependence of the typical current. Several of the persistent current
traces $I(B)$ corresponding to the Fourier transforms above are shown
in Fig. \ref{fig:ChData_DAT12_TempSerIvsB}. The Fourier transform
$I(\beta)$ was calculated using the form given in Eq. \ref{eq:ChData_FourierTransform}
for the Fourier transform of $\Delta f(B)$. }
\end{figure}

In order to characterize the temperature dependence of the typical
current, we measured the persistent current of each sample over a
small region of magnetic field at a series of temperatures. Some of
the persistent current traces $I(B)$ measured for sample CL17 at
$\theta_{0}=45^{\circ}$ are shown in Fig. \ref{fig:ChData_DAT12_TempSerIvsB}.
The magnitude of the Fourier transform of all persistent current traces
measured for this sample are shown in Fig. \ref{fig:ChData_DAT13_TempSerIvsBeta}.
In both figures, the features of the persistent current show little
change with temperature other than an overall scaling of the current
amplitude. In the Fourier transform, a second peak associated with
the second harmonic of the persistent current signal is visible. As
expected from Eq. \ref{eq:CHPCTh_IIFiniteTZSO}, this peak decays
much more quickly than the first harmonic peak. The second harmonic
signal was only observable for sample CL17 and only at $\theta_{0}=45^{\circ}$.

\begin{figure}
\begin{centering}
\includegraphics[width=0.7\paperwidth]{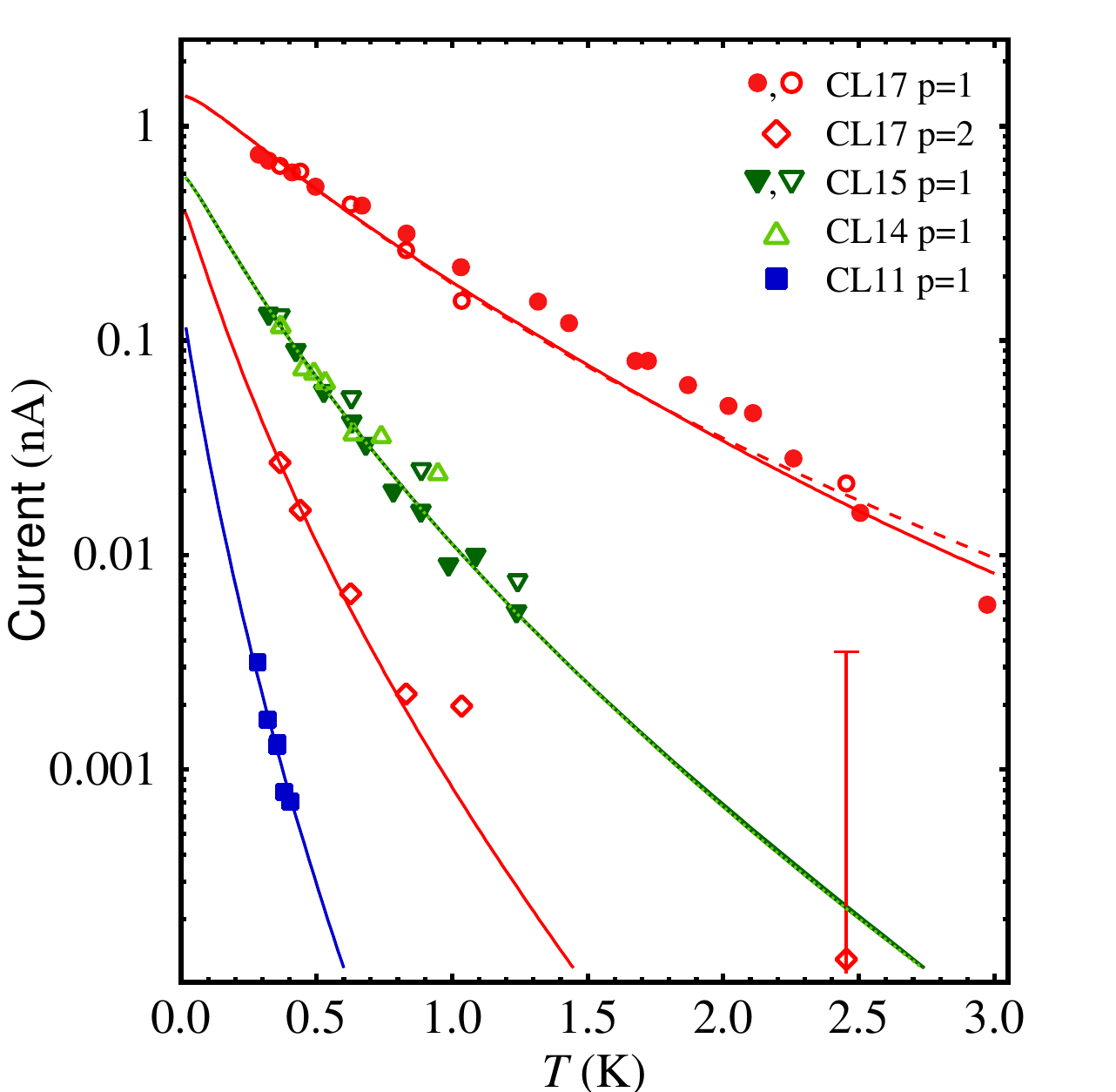}
\par\end{centering}

\caption[Typical magnitude of the persistent current versus temperature]{\label{fig:ChData_DAT14_IvsT}Typical magnitude of the persistent
current versus temperature. The data points show the inferred magnitude
$I_{p,M}^{\text{typ}}(T)$ of the current for each sample. Fits to
Eq. \ref{eq:ChData_IpTypFitFunction} are also shown and are described
in Section \ref{sub:ChData_TempFits}. Solid markers represent measurements
taken with an angle $\theta_{0}=6^{\circ}$ between the plane of the
rings and the applied magnetic field, while for hollow markers $\theta_{0}=45^{\circ}$.
For sample CL17, the magnitude of the second harmonic of the current
is represented by diamonds. For each other sample, only the first
harmonic was observable. The error bars associated with noise in the
cantilever frequency measurement were roughly the size of the markers
or smaller except where indicated explicitly. The value for the diffusion
constant $D$ extracted from the fits are given in Table \ref{tab:ChData_RingResults}.}
\end{figure}

For each measurement temperature $T$ in the temperature series, we
infer a current magnitude $I_{p,M}^{TS}(T)$ for the $p^{th}$ harmonic
of the persistent current signal from the size of the corresponding
peak in the Fourier spectrum.%
\footnote{We characterized the size of the peak by averaging the two or three
highest points defining the peak. An alternative method would be to
integrate the peak in the power spectral density. As the shape of
the peak was largely unchanged with temperature, the exact method
chosen should not be critical.%
} The typical current magnitude $I_{p,M}^{\text{typ}}(T)$ at temperature
$T$ is then found by scaling the temperature series magnitude by
$I_{p,M}^{\text{typ}}(T_{b})/I_{p,M}^{TS}(T_{b})$, a factor which
relates the magnitude $I_{p,M}^{TS}$ of the current measured over
the small region used for the temperature series to the typical magnitude
$I_{p,M}^{\text{typ}}$ of the current inferred from measurements
over a large range of magnetic field at the base temperature $T_{b}$.
That is, 
\[
I_{p,M}^{\text{typ}}\left(T\right)=I_{p,M}^{TS}\left(T\right)\left(\frac{I_{p,M}^{\text{typ}}\left(T_{b}\right)}{I_{p,M}^{TS}\left(T_{b}\right)}\right).
\]
The inferred typical current magnitudes $I_{p,M}^{\text{typ}}(T)$
for each sample, each angle $\theta_{0}$, each harmonic $p$, and
all temperatures are shown in Fig. \ref{fig:ChData_DAT14_IvsT} along
with fits described below.

This scaling procedure assumes that, as the temperature is varied,
the magnitude of the current measured in the small region of field
chosen for the temperature series maintains the same proportionality
to the typical current determined from a measurement over a large
region of magnetic field. From Eq. \ref{eq:CHPCTh_IIFiniteTZSO} and
Fig. \ref{fig:CHPCTh_BcorrLSO}, it can be seen that the magnetic
field scale of the correlation of the persistent current signal is
expected to increase with temperature. This increasing correlation
of the persistent current signal with temperature should lead to a
change in its shape and thus to a change in the proportionality between
the current magnitude within a particular small region of magnetic
field and its value averaged over a large region of magnetic field.
This effect produces a systematic error in our inferred magnitude
of the typical current.

For the range of temperatures covered by our measurements (see Fig.
\ref{fig:ChData_DAT14_IvsT}), the magnetic field scale $B_{1/2}(T)$
of the persistent current autocorrelation function, which we define
as the magnetic field scale $B_{1/2}$ satisfying $\langle I(\phi,B)I(\phi,B+B_{1/2})\rangle=(1/2)\langle I(\phi,B)I(\phi,B)\rangle$
(see Fig. \ref{fig:CHPCTh_BcorrLSO}), roughly doubles. The size $\Delta B_{TS}$
of the magnetic field region studied for each temperature series was
$\Delta B_{TS}\sim1.5B_{1/2}(T_{b})$, where $T_{b}$ was the base
temperature of the refrigerator. Thus at the highest temperature $T_{\max}$,
$\Delta B_{TS}\sim0.75B_{1/2}(T_{\max})$. In principle, one could
expect a systematic error of $\apprle33\%$ in the inferred $I_{p,M}^{\text{typ}}(T_{\max})$
due the change in $B_{1/2}(T)$ as $T$ increases from $T_{b}$ to
$T_{\max}$. In practice, we expect this error to be less for our
measurements as we did not observe significant changes in the shape
of the persistent current signal in our temperature series measurements.%
\footnote{This result could be related to the fact that, rather than a typical
region, we specifically chose regions where the current magnitude
was largest for the temperature series.%
} It would be interesting, but time consuming, to measure the persistent
current over a large range $B_{0}\gg B_{1/2}$ of magnetic field at
a series of temperatures to investigate changes in the correlation
scale of the persistent current signal with temperature.

For samples CL11 and CL17 at $\theta_{0}=6^{\circ}$, measurements
were taken at $T\approx285\,\text{mK}$, below the nominal base temperature
$T_{b}=323\,\text{mK}$. This lower temperature was achieved by filling
the 1K pot with helium, closing the needle valve connecting it to
the helium bath, and then pumping it down to the lowest possible temperature
($\sim1.8\,\text{K}$). We used this procedure sparingly because it
provided a low temperature only for a short duration (until the 1K
pot was empty) and introduced additional measurement noise due to
the vibrations of the pump. In the case of CL11 (largest rings and
thus smallest characteristic temperature), the increase in current
magnitude was large enough to offset the increase in noise, while
for CL17 the signal was large enough that the signal to noise ratio
was still large even with the added noise.

The inferred typical current magnitudes $I_{p,M}^{\text{typ}}(T)$
plotted in Fig. \ref{fig:ChData_DAT14_IvsT} represent measurements
for which several parameters were varied. The ring circumference was
varied by over a factor of two, going from $L=1.9\,\mu\text{m}$ (CL17)
to $5.0\,\mu\text{m}$ (CL11). Measurements were taken on two separate
cooldowns each corresponding to a different angle $\theta_{0}$ between
the magnetic field and the plane of the rings, with $\theta_{0}=6^{\circ}$
represented by solid markers and $\theta_{0}=45^{\circ}$ by hollow
markers. As expected, no dependence on the angle $\theta_{0}$ is
observed for the two samples, CL15 and CL17, measured in both orientations.
For $L=2.6\,\mu\text{m}$, the inferred typical current per ring for
a single ring (CL14) and an array of 990 rings (CL15) are both plotted
and show good agreement, justifying the $\sqrt{N}$ scaling of the
arrays discussed above. Finally, for sample CL17, the magnitudes of
both the first and second harmonics of the persistent current are
shown.

\subsubsection{\label{sub:ChData_TempFits}Analysis of the temperature dependence
of the persistent current}

In order to compare the data shown in Fig. \ref{fig:ChData_DAT14_IvsT}
to the theoretical picture discussed in \ref{sec:CHPCTh_DiffusiveRegime},
we fit the data for each sample using Eq. \ref{eq:ChPCTh_IpTypTEZESO}
for $I_{p}^{\text{typ}}(T,E_{Z},E_{SO})$. The quantity $I_{p}^{\text{typ}}(T,E_{Z},E_{SO})$
depends on $D$ and $L$ through the correlation energy $E_{c}$.
Each sample was fit individually, but for each sample the data from
both angles $\theta_{0}$ and harmonics $p$ were all used in a single
fit. The only parameter varied during the fitting routine was the
diffusion constant $D$. The sample circumference $L$ was fixed to
the value corresponding to the mean radius measured with a scanning
electron microscope and listed in Table \ref{tab:ChData_Rings}. The
spin-orbit scattering length $L_{SO}$ was held fixed to $1.1\,\mu\text{m}$,
the value obtained from transport measurements of a co-deposited wire
(see Appendix \ref{cha:AppTransport_}). To account for Zeeman splitting,
the actual fitting function took the form
\begin{equation}
I_{p,\text{fit}}^{\text{typ}}(T,D)=\sqrt{\frac{\int_{B_{M,\min}}^{B_{M,\max}}dB\,\left(I_{p}^{\text{typ}}\left(T,E_{Z}\left(B\right),E_{SO}\right)\right)^{2}}{B_{M,\max}-B_{M,\min}}}\label{eq:ChData_IpTypFitFunction}
\end{equation}
where $B_{M,\min}$ and $B_{M,\max}$ were the minimum and maximum
magnetic field values of the big magnetic field sweep (see \ref{sub:ChData_FullCurrentTraces})
used to determine $I_{p,M}^{\text{typ}}(T_{b})$. The quantity $I_{p,\text{fit}}^{\text{typ}}(T,D)$
given in Eq. \ref{eq:ChData_IpTypFitFunction} represents the average
of the typical square magnitude of the current for the region of magnetic
field over which the persistent current was measured.%
\footnote{It is not clear that this is the best way to account for the effect
of the Zeeman splitting. An alternative approach would be to use one
value for $E_{Z}(B)$ with $B$ chosen to be a value from within the
measurement range, such as the midpoint. In any case, different methods
for accounting for the Zeeman splitting produce only small changes
in the form of $I_{p,\text{fit}}^{\text{typ}}(T,D)$ for the range
of Zeeman splitting $E_{Z}$ and temperature spanned by the measurements
shown in Fig. \ref{fig:ChData_DAT14_IvsT}. In fact, different methods
of accounting for the finite Zeeman splitting in the form of $I_{p,\text{fit}}^{\text{typ}}(T,D)$
produce fits to the data with the same fitted diffusion constants
$D$, although the fit curves look slightly different (with differences
similar to the two curves in Fig. \ref{fig:ChData_DAT14_IvsT} fit
to the first harmonic data from sample CL17).%
}

The curves resulting from fits of the typical persistent current data
to Eq. \ref{eq:ChData_IpTypFitFunction} are shown in Fig. \ref{fig:ChData_DAT14_IvsT}.
For sample CL17, two curves are shown with the dashed curve corresponding
to $\theta_{0}=45^{\circ}$ and the solid curve to $\theta_{0}=6^{\circ}$.
The measurements for the two angles were made over regions of magnetic
field (and corresponding ranges of $E_{Z}$) different enough to produce
the slight separation between the curves at high temperature. Both
curves are part of one single fit to the data from sample CL17 using
Eq. \ref{eq:ChData_IpTypFitFunction}. Only one curve is shown for
samples CL14 and CL15. The same curve fit the data from both samples
independently.

The fits shown in Fig. \ref{fig:ChData_DAT14_IvsT} agree well with
the measured temperature dependence of the persistent current. The
fitting function has roughly an exponential form $I_{p}^{D}\exp(-T/T_{p})$
where both the amplitude $I_{p}^{D}=I_{p}^{\text{typ}}(T=0,E_{Z},E_{SO})$
and characteristic temperature $T_{p}$ are proportional to the only
fitting parameter, the diffusion constant $D$.%
\footnote{For the sake of clarity, we reiterate that for sample CL17 both harmonics
of the persistent current are fit simultaneously for one value of
the diffusion constant $D$. In this case, there are essentially two
characteristic temperatures, $T_{p=1}$ and $T_{p=2}$.%
} Table \ref{tab:ChData_RingResults} lists the values of the fitted
diffusion constant $D$ and of $I_{p}^{\text{typ}}(T=0,E_{Z},E_{SO})$
and $T_{p}$ calculated for the first harmonic of each sample using
fitted $D$.

\begin{table}
\begin{centering}
\begin{tabular}{|ccccccc|}
\hline 
$\vphantom{{\displaystyle \sum_{a}^{a}}}$Sample & $D\,\text{(m}^{2}/\text{s})$ & $I_{1}^{D}\,(\text{nA})$ & $T_{1}\,\text{(mK)}$ & $\gamma$ & $M_{\text{eff}}$ & $\eta(M_{\text{eff}})\,(\%)$\tabularnewline
\hline 
${\displaystyle \vphantom{\sum_{-}}}$CL11 & 0.0196 & 0.15 & 62 & (2.5) & (29) & (13)\tabularnewline
${\displaystyle \vphantom{\sum_{-}}}$CL14 & 0.0195 & 0.61 & 228 & 1.1 & 12 & 21\tabularnewline
${\displaystyle \vphantom{\sum_{-}}}$CL15 & 0.0195 & 0.61 & 228 & 1.1, (2.5) & 57, (14) & 9.4, (19)\tabularnewline
CL17 & 0.0234 & 1.43 & 513 & 0.83, (2.8) & 78, (20) & 8.0, (16)\tabularnewline
\hline 
\end{tabular}
\par\end{centering}

\caption[Extracted parameters for persistent current samples]{\label{tab:ChData_RingResults}Extracted parameters for persistent
current samples. The diffusion constants $D$ are the values inferred
from the fits to $I_{p,\text{fit}}^{\text{typ}}(T,D)$ (Eq. \ref{eq:ChData_IpTypFitFunction})
shown in Fig. \ref{fig:ChData_DAT14_IvsT}. The magnitude $I_{1}^{D}=I_{p=1}^{\text{typ}}(T=0,E_{Z}=\infty,E_{SO})$
of the persistent current in the zero-temperature, large Zeeman splitting
limit (see Eq. \ref{eq:ChPCTh_IpTypTEZESO}) and the characteristic
temperature $T_{1}$ (see Eq. \ref{eq:CHPCTh_TpDiffusive}) are calculated
for the first harmonic from the fitted value for $D$ and the value
of $L$ given in Table \ref{tab:ChData_Rings}. The spin-orbit scattering
length is fixed to $L_{SO}=1.1\,\mu\text{m}$, the value found in
transport measurements (see Table \ref{tab:AppTransport_WL115Properties}).
The geometrical factors $\gamma$ were found from fits to the autocorrelation
of the persistent current shown in Figs. \ref{fig:ChData_DAT15_Cor_CL17_45Deg}
through \ref{fig:ChData_DAT20_Cor_CL11}. The effective number $M_{\text{eff}}$
of independent realizations and the fractional standard error $\eta(M_{\text{eff}})$
in the typical current are calculated from $\gamma$ using Eqs. \ref{eq:ChData_NormalStandardError}
and \ref{eq:ChData_EffectiveNumberIndependent}. For these last three
quantities (see discussion in Section \ref{sub:ChData_StatUncertainty}),,
values in parentheses correspond to measurements at $\theta_{0}=6^{\circ}$
and those without parentheses to measurements at $\theta_{0}=45^{\circ}$.}
\end{table}

A further confirmation of the accuracy of the theoretical picture
of Section \ref{sub:CHPCTh_TypicalCurrent} (and Eq. \ref{eq:CHPCTh_IIFiniteTZSO}
in particular) is provided by a comparison of the fitted diffusion
constants $D$ from the different samples. The values of $D$ found
for samples CL11, CL14, and CL15 all agree within the 6\% error margin
which we discuss below. The value of $D$ obtained for sample CL17
was $\sim20\%$ larger than the values obtained for the other samples.
While a 20\% discrepancy is reasonable for different measurements
of the diffusion constant, we note that the ring linewidth $w_{r}=115\,\text{nm}$
of sample CL17 was somewhat larger than the linewidth $w_{r}=85\,\text{nm}$
of the other samples. Additionally, scanning electron microscopy of
other samples on the same chip (see Fig. \ref{fig:AppSampFab_linewidthSEM})
revealed that features with the wider linewidth of sample CL17 had
much more uniform sidewalls than features with the thinner linewidth
of the other samples. If boundary scattering contributed significantly
to electron diffusion in our samples, one would expect that the wider
sample with smoother sidewalls would possess a larger value of $D$.

Comparison with the diffusion constant $D_{\rho}$ found from transport
measurements of a co-deposited wire provides another check of the
accuracy of our analysis. In Appendix \ref{cha:AppTransport_} the
diffusion constant of transport sample WL115 is reported as $D_{\rho}=0.0259\pm0.0014\,\text{m}^{2}/\text{s}$.
Sample WL115 was co-deposited with the persistent current samples
and had nominally the same linewidth $w_{w}=115\,\text{nm}$ as sample
CL17. That the diffusion constants obtained from transport and persistent
current measurements agree to within $\sim10\%$, roughly the combined
experimental uncertainty for the two measurements, is a strong endorsement
for the theoretical picture reviewed in Section \ref{sub:CHPCTh_TypicalCurrent}.

\subsubsection{\label{sub:ChData_StatUncertainty}Analysis of persistent current
autocorrelation and estimation of statistical uncertainty of persistent
current magnitude}

We now discuss the sources of uncertainty in our measurements of the
persistent current and in the extraction of the diffusion constant
$D$. As discussed in chapters \ref{cha:CHMeso_} and \ref{cha:CHPrevWork},
the amplitude of the persistent current is expected to be a random
quantity and to follow the normal distribution with a mean of zero
and a standard deviation given by $I_{p}^{\text{typ}}(T,E_{Z},E_{SO})$
(see Eq. \ref{eq:ChPCTh_IpTypTEZESO}). For the arrays, we expect
that the current per ring (using $\sqrt{N}$ scaling) will follow
this distribution regardless of the distribution for the single ring
current because of the central limit theorem. With the inference of
any value found by averaging over measurements of a probabilistic
quantity, there is an accompanying uncertainty due to the finite number
of measurements entering into the statistical average. For $M$ measurements
of a quantity $x$ following the normal distribution with mean $\langle x\rangle=0$,
it can be shown that the fractional standard error $\eta(M)$ due
to finite sampling in the typical value $x^{\text{typ}}=\sqrt{\langle x^{2}\rangle}$
of $x$ is \citep{kendall1947advanced,castro1989anintroduction} 
\begin{equation}
\eta(M)\equiv\frac{\delta x^{\text{typ}}}{x^{\text{typ}}}=\sqrt{\frac{1}{2M}}.\label{eq:ChData_NormalStandardError}
\end{equation}

For the persistent current measurements, the form of the standard
error in the estimation of the typical current $I_{p,M}^{\text{typ}}$
is not as simple as Eq. \ref{eq:ChData_NormalStandardError}. The
finite correlation of the persistent current amplitude makes unclear
what number should be used for the number of measurements $M$. In
Ref. \citealp{tsyplyatyev2003applicability}, Tsyplyatyev \emph{et
al.} derive expressions for the standard error of the cumulants calculated
from a data set with a finite range of correlation. From their results
one finds that the fractional error in the typical magnitude (which
is related to the second cumulant when the mean is zero) satisfies
\begin{equation}
\frac{\delta I_{p}^{\text{typ}}}{I_{p}^{\text{typ}}}=\sqrt{\frac{\gamma B_{c,p}}{B_{0}}}\sqrt{\int_{0}^{\infty}dx\,\left(K_{p}\left(x\right)\right)^{2}}\label{eq:ChData_TypCurrentUncertainty}
\end{equation}
where $K_{p}(x)$ is the normalized and scaled correlation function%
\footnote{We can also write $K_{p}(x)$ in terms of the notation of \ref{sec:CHPCTh_DiffusiveRegime}
(see e.g. Eq. \ref{eq:CHPCTh_IIFiniteTZSO} and subsequent formulae)
as$K_{p}\left(x\right)=c_{p}^{T}\left(T,B_{c,p}x,E_{Z},E_{SO}\right).$%
}
\begin{equation}
K_{p}(x)=\frac{\left\langle I_{p}\left(B_{M}\right)I_{p}\left(B_{M}+B_{c,p}x\right)\right\rangle }{\left\langle I_{p}^{2}\left(B_{M}\right)\right\rangle },\label{eq:ChData_KpCorr}
\end{equation}
$B_{0}$ is the magnetic field range of the persistent current measurement
and $B_{c,p}$ is the toroidal magnetic field correlation scale given
in Eq. \ref{eq:CHPCTh_BcpToroidalField}. $\gamma$ is the geometrical
factor that, for a given geometrical arrangement of the ring in the
applied magnetic field, relates the toroidal field $B_{c,p}$ to a
quantity $\gamma B_{c,p}$ which is expressed in units of the applied
field. This expression is valid for a large measurement field range
$B_{0}\gg B_{c,p}$. At $T=0$, $\int_{0}^{\infty}dx\,(K_{p}(x))^{2}\approx1.8$,
so that 
\[
\frac{\delta I_{p}^{\text{typ}}}{I_{p}^{\text{typ}}}\approx\sqrt{1.8\frac{\gamma B_{c,p}}{B_{0}}}.
\]

We define an effective number of independent realizations $M_{\text{eff}}$
of the persistent current as
\begin{equation}
M_{\text{eff}}=\frac{1}{2\int_{0}^{\infty}dx\,\left(K_{p}\left(x\right)\right)^{2}}\frac{B_{0}}{\gamma B_{c,p}}\label{eq:ChData_EffectiveNumberIndependent}
\end{equation}
by equating Eq. \ref{eq:ChData_TypCurrentUncertainty} for the statistical
uncertainty in the typical persistent current magnitude with Eq. \ref{eq:ChData_NormalStandardError}
for the uncertainty in a finite number of measurements of a normally
distributed quantity.%
\footnote{Because the signal to noise ratio for the second harmonic signal observed
for sample CL17 was so low, we do not analyze its autocorrelation
function. Since we only consider the first harmonic, we do not index
$M_{\text{eff}}$ by $p$.%
} At $T=0$, $M_{\text{eff}}=B_{0}/3.6\gamma B_{c,p}$. From Fig. \ref{fig:CHPCTh_BcorrLSO},
it can be seen that the correlation of the persistent current is slightly
enhanced at finite temperature and spin-orbit scattering and thus
that $M_{\text{eff}}$ is reduced from its value at $T=0$.

\begin{figure}
\begin{centering}
\includegraphics[width=0.7\paperwidth]{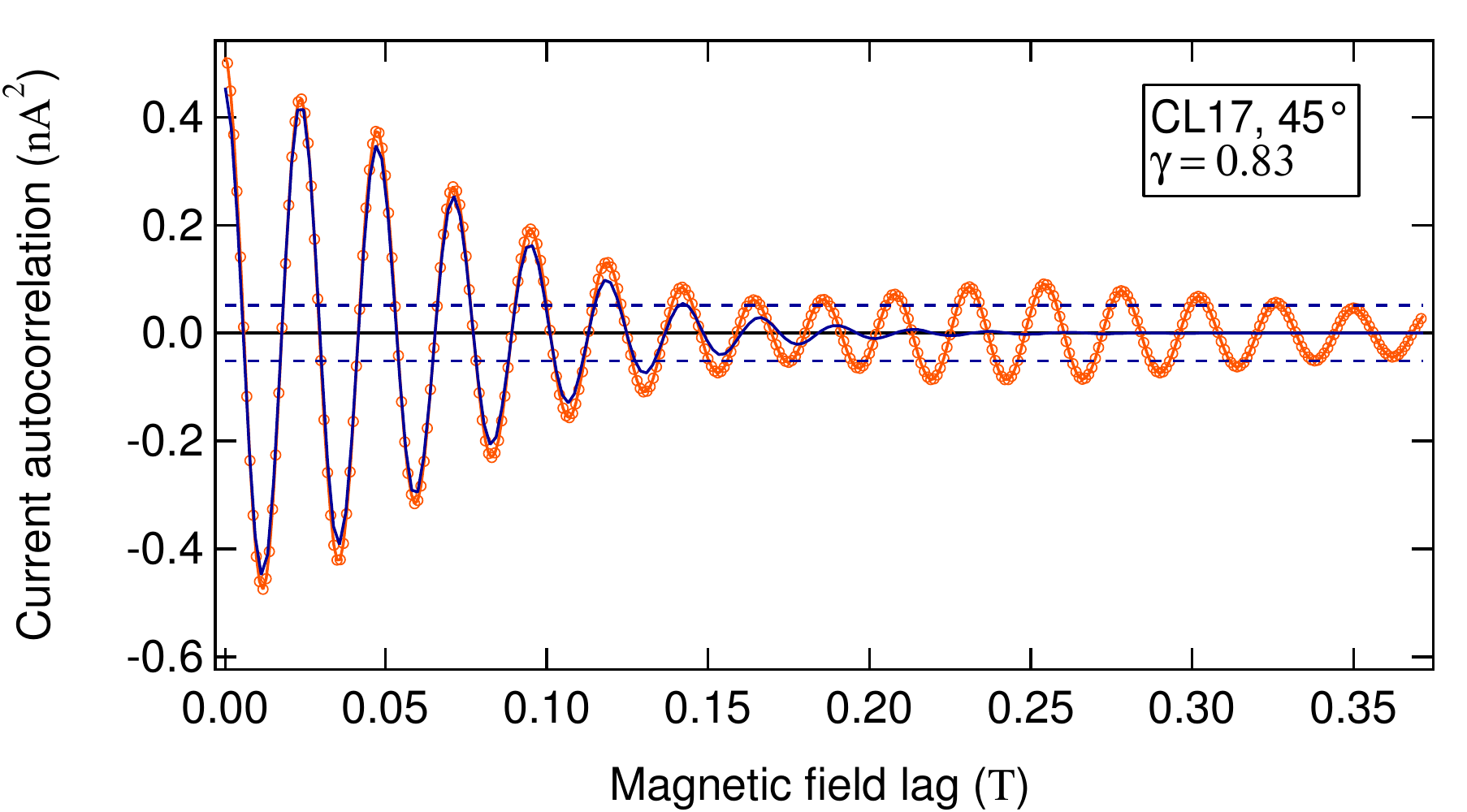}
\par\end{centering}

\caption[Persistent current autocorrelation for sample CL17 at $\theta_{0}=45^{\circ}$]{\label{fig:ChData_DAT15_Cor_CL17_45Deg}Persistent current autocorrelation
for sample CL17 at $\theta_{0}=45^{\circ}$. Circles represent the
autocorrelation of the data shown in Figs. \ref{fig:ChData_DAT21_IvsB_CL17_45DegP1}
through \ref{fig:ChData_DAT21_IvsB_CL17_45DegP3}. The curve is a
fit to Eq. \ref{eq:ChData_AutocorrelationFitFunction} as described
in the text with $\gamma=0.83$ and $p=0.83$. The estimated standard
error in the autocorrelation (dashed horizontal lines) is $\sqrt{2}\eta(M_{\text{eff}})=11\%$
of its value at zero magnetic field lag $B'$. Fluctuations of this
magnitude are present at large $B'$ where the autocorrelation is
expected to be small.}
\end{figure}

\begin{figure}
\begin{centering}
\includegraphics[width=0.7\paperwidth]{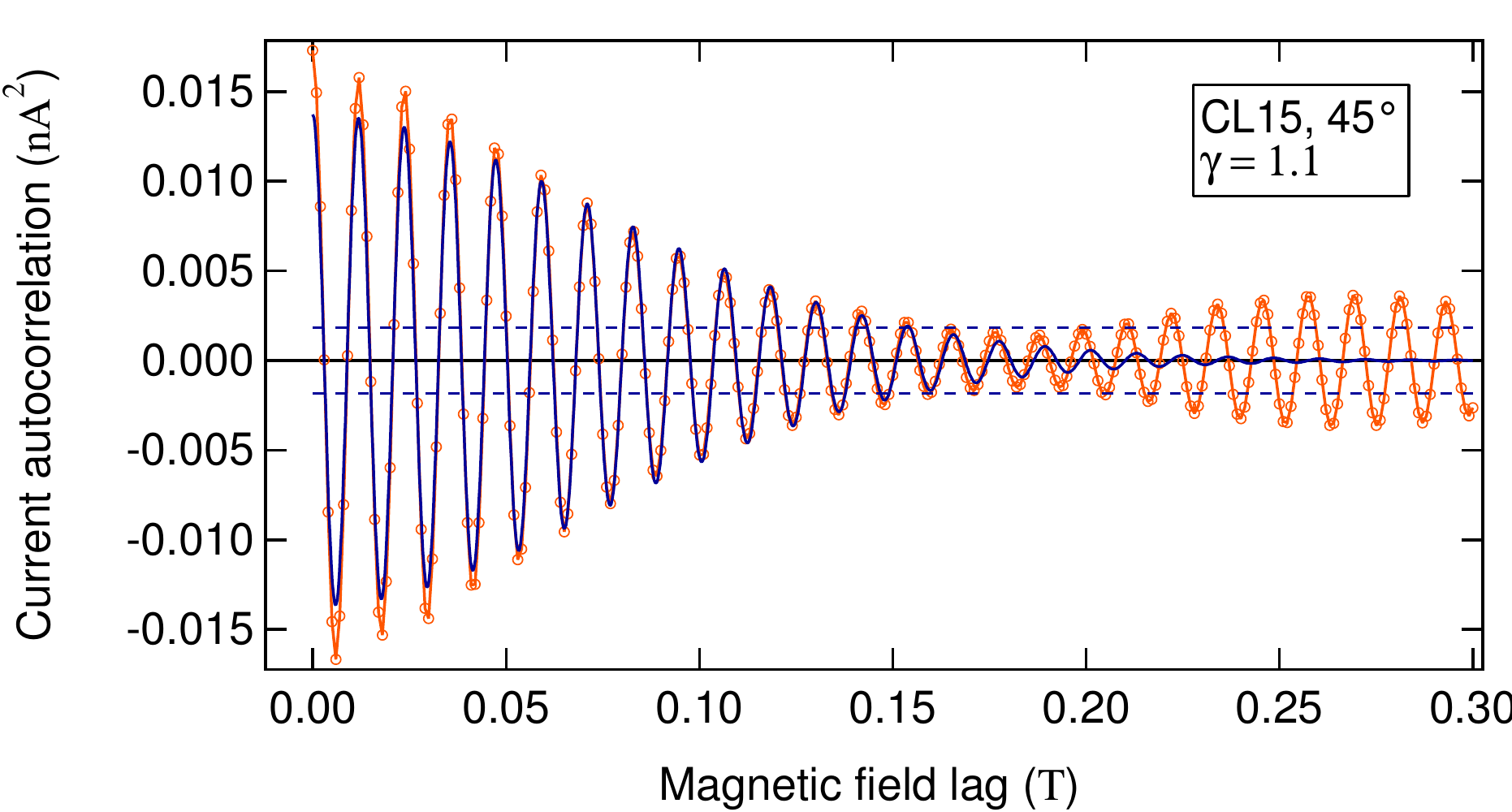}
\par\end{centering}

\caption[Persistent current autocorrelation for sample CL15 at $\theta_{0}=45^{\circ}$]{\label{fig:ChData_DAT16_Cor_CL15_45Deg}Persistent current autocorrelation
for sample CL15 at $\theta_{0}=45^{\circ}$. Circles represent the
autocorrelation of the data shown in Figs. \ref{fig:ChData_DAT22_IvsB_CL15_45DegP1}
through \ref{fig:ChData_DAT22_IvsB_CL15_45DegP4}. The curve is a
fit to Eq. \ref{eq:ChData_AutocorrelationFitFunction} as described
in the text with $\gamma=1.1$ and $p=0.90$. The estimated standard
error in the autocorrelation (dashed horizontal lines) is $\sqrt{2}\eta(M_{\text{eff}})=13\%$
of its value at zero magnetic field lag $B'$. Fluctuations of this
magnitude are present at large $B'$ where the autocorrelation is
expected to be small.}
\end{figure}

\begin{figure}
\begin{centering}
\includegraphics[width=0.7\paperwidth]{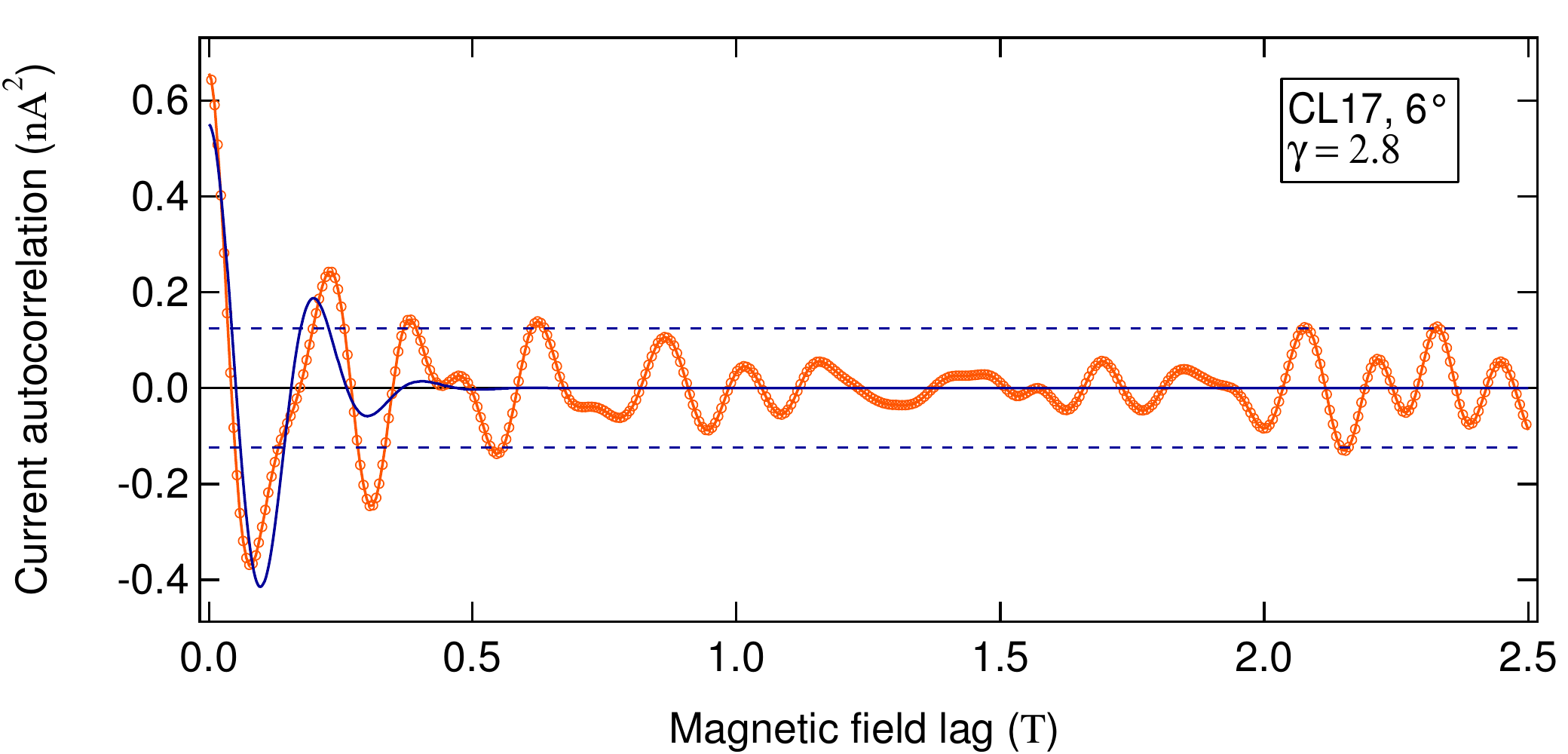}
\par\end{centering}

\caption[Persistent current autocorrelation for sample CL17 at $\theta_{0}=6^{\circ}$]{\label{fig:ChData_DAT17_Cor_CL17_6Deg}Persistent current autocorrelation
for sample CL17 at $\theta_{0}=6^{\circ}$. Circles represent the
autocorrelation of the data shown in Fig. \ref{fig:ChData_DAT23_IvsB_CL17_6Deg}.
The curve is a fit to Eq. \ref{eq:ChData_AutocorrelationFitFunction}
as described in the text with $\gamma=2.8$ and $p=0.63$. Forcing
$p=1$ in the expression for the magnetic field frequency $\beta=\pi R^{2}\sin\theta_{0}/\phi_{0}$
would require a radius $R$ of $244\,\text{nm}$ to match the frequency
$\beta$ for $p=0.63$ and $R$ the mean radius. This value is close
to the inner radius of $250\,\text{nm}$ for sample CL17. The estimated
standard error in the autocorrelation (dashed horizontal lines) is
$\sqrt{2}\eta(M_{\text{eff}})=23\%$ of its value at zero magnetic
field lag $B'$. Fluctuations of this magnitude are present at large
$B'$ where the autocorrelation is expected to be small. These fluctuations
show a wide range of frequencies, consistent with the large peak width
$\Delta\beta_{1}$ relative to the peak location $\beta_{1}$ in the
power spectral density (Fig. \ref{fig:ChData_DAT7_PSDCL17_6Deg}).}
\end{figure}

\begin{figure}
\begin{centering}
\includegraphics[width=0.7\paperwidth]{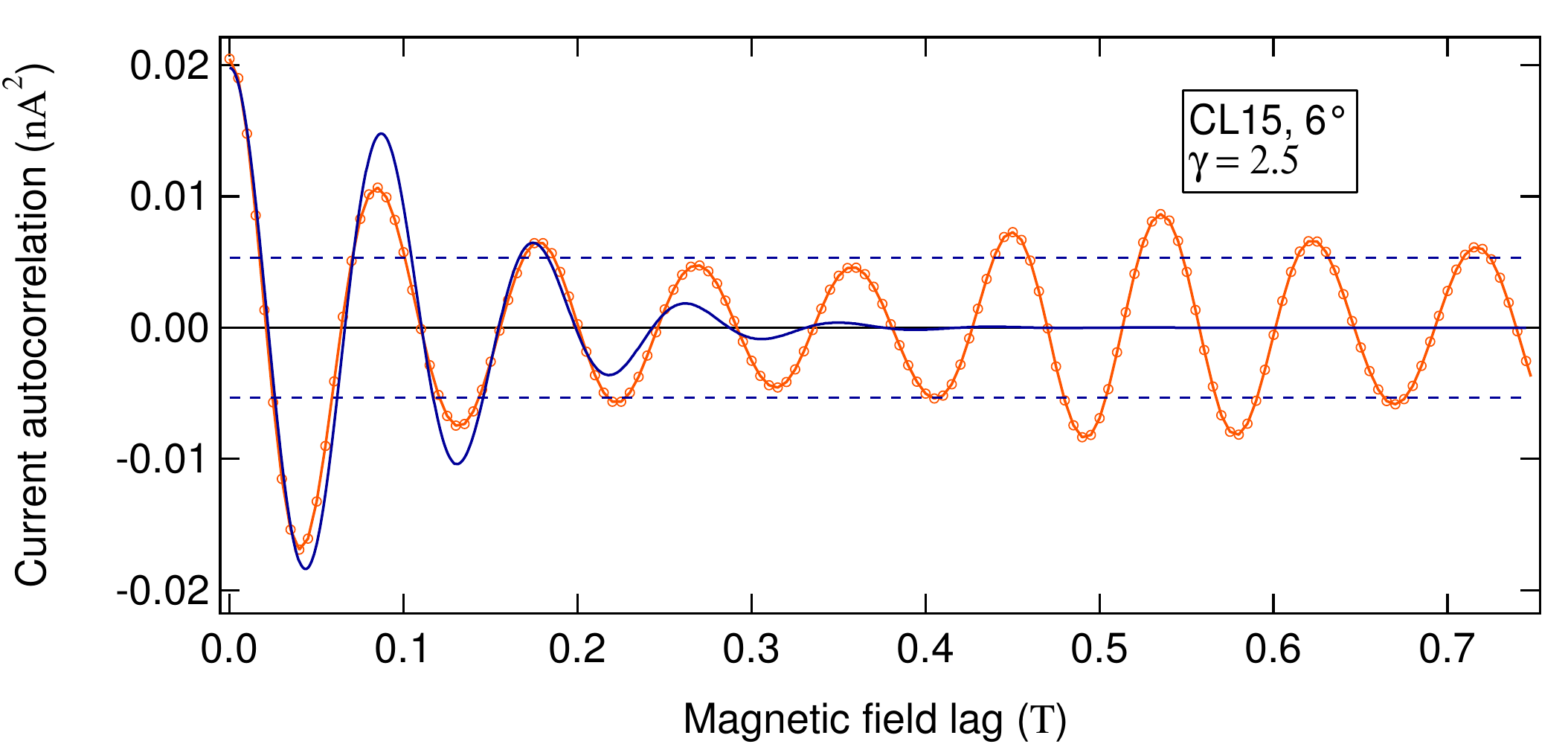}
\par\end{centering}

\caption[Persistent current autocorrelation for sample CL15 at $\theta_{0}=6^{\circ}$]{\label{fig:ChData_DAT18_Cor_CL15_6Deg}Persistent current autocorrelation
for sample CL15 at $\theta_{0}=6^{\circ}$. Circles represent the
autocorrelation of the data shown in Fig. \ref{fig:ChData_DAT24_IvsB_CL15_6Deg}.
The curve is a fit to Eq. \ref{eq:ChData_AutocorrelationFitFunction}
as described in the text with $\gamma=2.5$ and $p=0.82$. The estimated
standard error in the autocorrelation (dashed horizontal lines) is
$\sqrt{2}\eta(M_{\text{eff}})=27\%$ of its value at zero magnetic
field lag $B'$. Fluctuations of this magnitude are present at large
$B'$ where the autocorrelation is expected to be small.}
\end{figure}

\begin{figure}
\begin{centering}
\includegraphics[width=0.7\paperwidth]{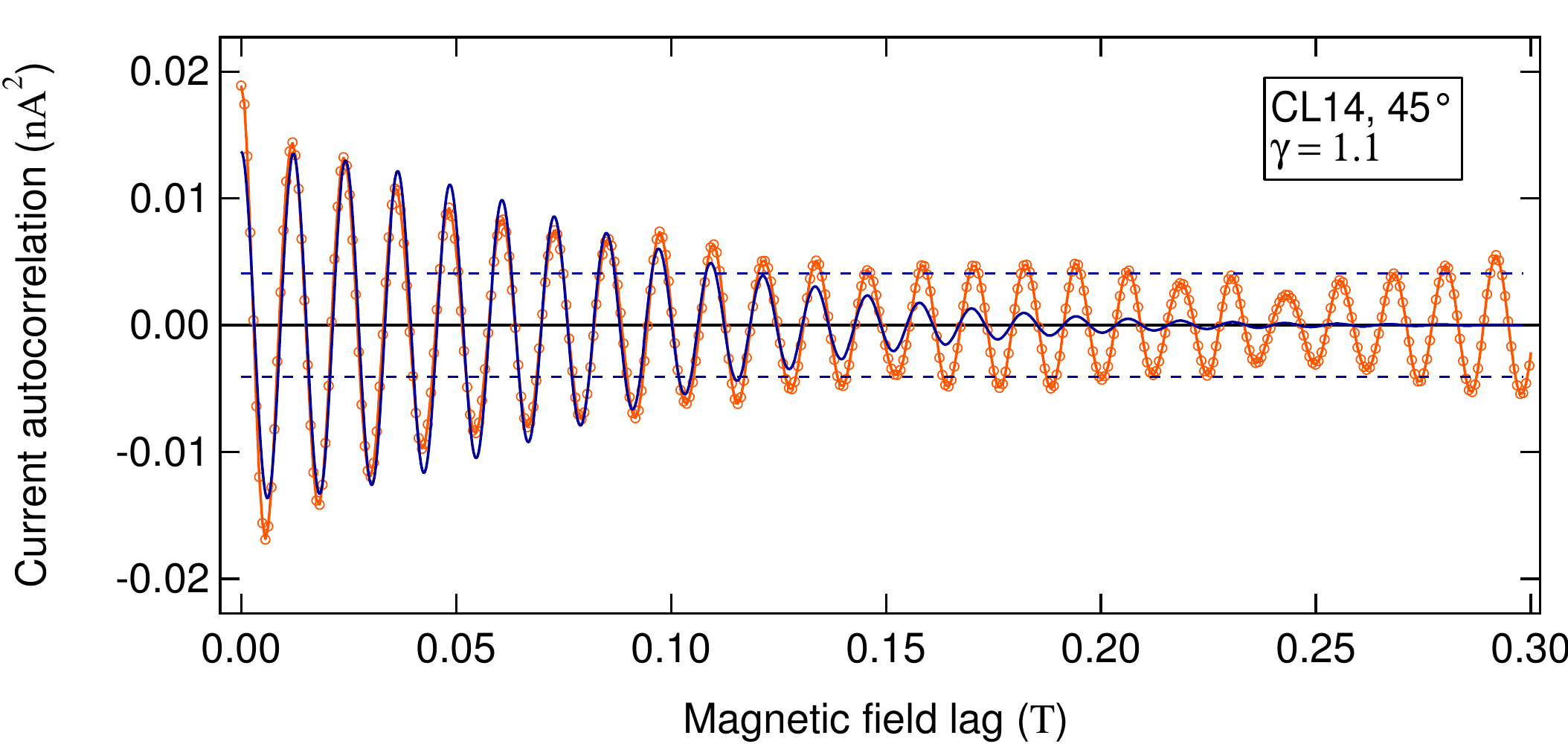}
\par\end{centering}

\caption[Persistent current autocorrelation for sample CL14]{\label{fig:ChData_DAT19_Cor_CL14}Persistent current autocorrelation
for sample CL14. Circles represent the autocorrelation of the data
shown in Fig. \ref{fig:ChData_DAT25_IvsB_CL14}. In calculating the
autocorrelation, the features with magnetic field frequency $\beta$
less than $60\,\text{T}^{-1}$ (see Fig. \ref{fig:ChData_DAT9_PSDCL14})
were removed. As discussed in the text, we attribute these features
to the slowly varying background and not to the persistent current.
The curve is a fit to Eq. \ref{eq:ChData_AutocorrelationFitFunction}
as described in the text with $\gamma=1.1$ and $p=0.88$. The estimated
standard error in the autocorrelation (dashed horizontal lines) is
$\sqrt{2}\eta(M_{\text{eff}})=30\%$ of its value at zero magnetic
field lag $B'$. Fluctuations of this magnitude are present at large
$B'$ where the autocorrelation is expected to be small.}
\end{figure}

\begin{figure}
\begin{centering}
\includegraphics[width=0.7\paperwidth]{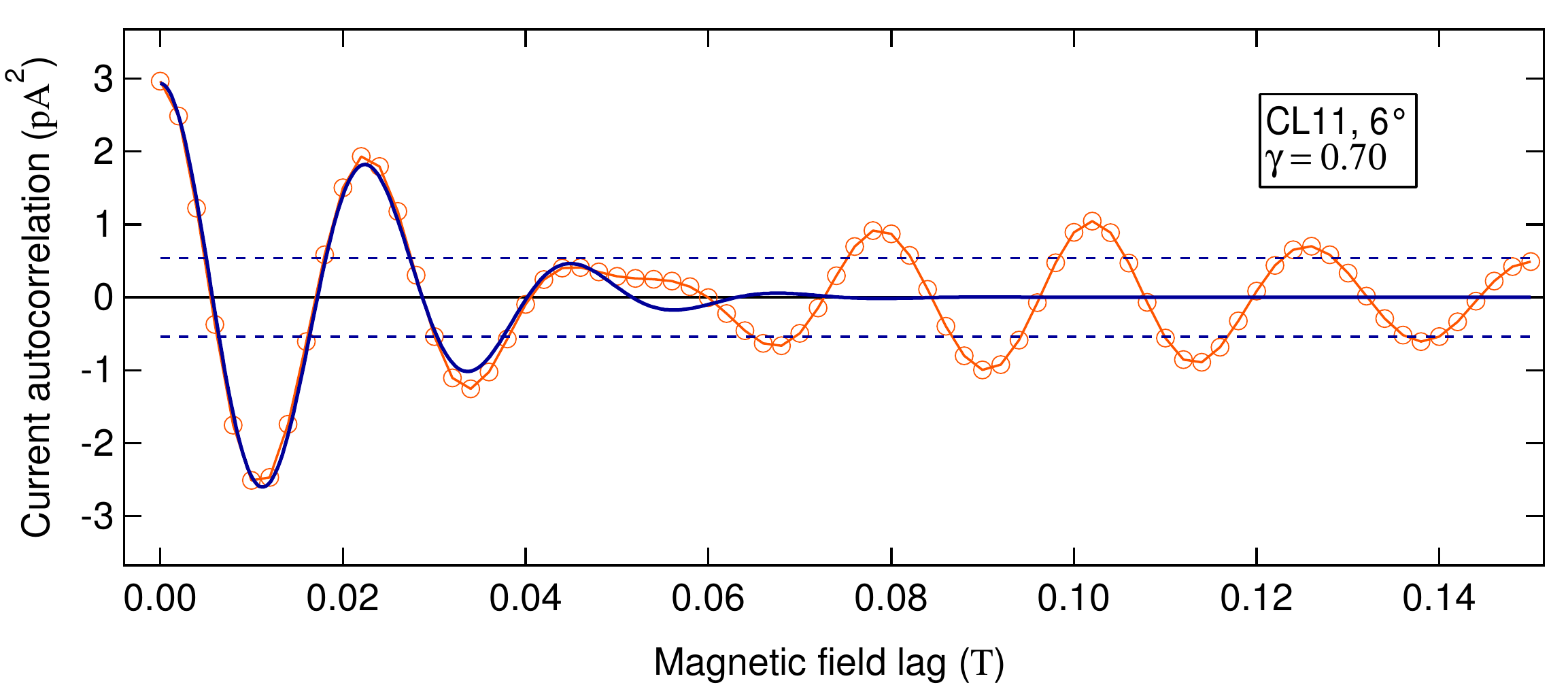}
\par\end{centering}

\caption[Persistent current autocorrelation for sample CL11]{\label{fig:ChData_DAT20_Cor_CL11}Persistent current autocorrelation
for sample CL11. Circles represent the autocorrelation of the data
shown in Fig. \ref{fig:ChData_DAT26_IvsB_CL11}. In calculating the
autocorrelation, the features with magnetic field frequency $\beta$
less than $30\,\text{T}^{-1}$ (see Fig. \ref{fig:ChData_DAT10_PSDCL11})
were removed. As discussed in the text, we attribute these features
to the slowly varying background and not to the persistent current.
The curve is a fit to Eq. \ref{eq:ChData_AutocorrelationFitFunction}
as described in the text with $\gamma=0.70$ and $p=0.90$. The estimated
standard error in the autocorrelation (dashed horizontal lines) is
$\sqrt{2}\eta(M_{\text{eff}})=18\%$ of its value at zero magnetic
field lag $B'$. Fluctuations of this magnitude are present at large
$B'$ where the autocorrelation is expected to be small.}
\end{figure}

In order to estimate $M_{\text{eff}}$ for each measurement of $I_{p,M}^{\text{typ}}$,
we analyze the autocorrelation of the large persistent current traces
shown in \ref{sub:ChData_FullCurrentTraces}. We calculate the autocorrelation
$\langle I(B)I(B+B')\rangle_{M}$ of the persistent current data using
\begin{align}
\left\langle I\left(B\right)I\left(B+j\Delta B\right)\right\rangle _{M} & =\frac{1}{P-j-1}\sum_{k=0}^{P-j-1}I\left(B_{\min}+k\Delta B\right)I\left(B_{\min}+\left(j+k\right)\Delta B\right)\nonumber \\
 & \approx\frac{1}{B_{0}}\int_{B_{\min}}^{B_{\min}+B_{0}}dB'\, I\left(B'\right)I\left(B'+j\Delta B\right)\label{eq:ChData_AutocorrelationCalculation}
\end{align}
where $B_{\min}$ is the minimum magnetic field measured, $P$ is
the total number of magnetic field values measured, and $\Delta B$
is the magnetic field spacing between measurements of the persistent
current. The resulting traces of the autocorrelation $\langle I(B)I(B+B')\rangle_{M}$
versus the magnetic field lag $B'$ are shown in Figs. \ref{fig:ChData_DAT15_Cor_CL17_45Deg}
through \ref{fig:ChData_DAT20_Cor_CL11}. The measurements of samples
CL17 (Fig. \ref{fig:ChData_DAT15_Cor_CL17_45Deg}) and CL15 (\ref{fig:ChData_DAT16_Cor_CL15_45Deg})
at $\theta_{0}=45^{\circ}$ resulted in the nicest looking autocorrelation
traces. As we shall see shortly, these two measurements achieved the
highest values of $M_{\text{eff}}$ and represent the best averaging
over the statistical distribution of the persistent current for our
measurements.

We analyze the autocorrelation data by fitting each trace to the first
harmonic component of Eq. \ref{eq:CHPCTh_IIFiniteTZSO} which we rewrite
in the form
\begin{equation}
\left\langle I\left(B\right)I\left(B+B'\right)\right\rangle =\left(I_{p=1}^{\text{typ}}\right)^{2}\cos\left(2\pi p\beta_{1}B'\right)c_{p=1}^{T}\left(T,\frac{B'}{\gamma},E_{Z}=\infty,E_{SO}\right).\label{eq:ChData_AutocorrelationFitFunction}
\end{equation}
In Eq. \ref{eq:ChData_AutocorrelationFitFunction}, we take the limit
of large Zeeman splitting, $E_{Z}\rightarrow\infty$, because finite
Zeeman splitting results in only small corrections to the autocorrelation
function for the temperatures at which the large magnetic field traces
were measured. For fitting the correlation function of the first harmonic,
the coefficient $p$ is fixed to 1 for both the persistent current
typical magnitude $I_{p=1}^{\text{typ}}$ and the normalized correlation
function $c_{p=1}^{T}$. However, in the argument of $\cos(2\pi p\beta_{1}B')$,
the coefficient $p$ is allowed to vary as it is critical for the
accuracy of the fit that the oscillating term have the correct period.
The other free parameter varied during the fitting routine is the
geometrical factor $\gamma$. The diffusion constant $D$ is held
fixed to its fitted value listed in Table \ref{tab:ChData_RingResults}.
The parameters $L$ and $\theta_{0}$ are fixed to their nominal values,
and $L_{SO}$ is fixed to the value found in transport measurements
(see Appendix \ref{cha:AppTransport_}).

The curves associated with the best fits of Eq. \ref{eq:ChData_AutocorrelationFitFunction}
to the autocorrelation data are plotted in Figs. \ref{fig:ChData_DAT15_Cor_CL17_45Deg}
through \ref{fig:ChData_DAT20_Cor_CL11}. The fitted values of the
geometrical factor $\gamma$ are shown in each figure and are collected
in Table \ref{tab:ChData_RingResults}. The fitted values of the coefficient
$p$ are listed in each figure caption. As expected, all of the fitted
values for $\gamma$ are of order unity. With one exception (sample
CL17 at $\theta_{0}=6^{\circ}$), the fitted values of $p$ were all
between 0.8 and 0.9, slightly below the expected value $p=1$ but
consistent with the location of the peaks in the power spectral densities
(Figs. \ref{fig:ChData_DAT5_PSDCL17_45Deg} through \ref{fig:ChData_DAT10_PSDCL11})
discussed earlier. Also shown in Table \ref{tab:ChData_RingResults}
are the values of $M_{\text{eff}}$ and $\eta(M_{\text{eff}})$ found
using Eqs. \ref{eq:ChData_NormalStandardError} and \ref{eq:ChData_EffectiveNumberIndependent}
and the values of $\gamma$ found from the fits to the persistent
current autocorrelation data.

At large magnetic field lag ($B'$ in Eq. \ref{eq:ChData_AutocorrelationFitFunction}),
the autocorrelation function $\langle I(B)I(B+B')\rangle$ decays
to zero. The autocorrelation calculated from each measured persistent
current trace (see Figs. \ref{fig:ChData_DAT15_Cor_CL17_45Deg} through
\ref{fig:ChData_DAT20_Cor_CL11}), however, continues to oscillate
at a finite, varying amplitude. Once the field lag $B'$ is large
enough that $\langle I(B)I(B+B')\rangle\ll\langle I^{2}(B)\rangle$,
the autocorrelation calculation essentially involves averaging the
product of $M_{\text{eff}}$ uncorrelated values of the persistent
current. Thus, the standard error $\sigma(\langle I(B)I(B+B')\rangle)$
in the autocorrelation due to the finite measurement size is 
\begin{align*}
\sigma\left(\left\langle I\left(B\right)I\left(B+B'\right)\right\rangle \right) & =\frac{\left\langle I^{2}\left(B\right)\right\rangle }{\sqrt{M_{\text{eff}}}}\\
 & =\sqrt{2}\eta\left(M_{\text{eff}}\right)\left\langle I^{2}\left(B\right)\right\rangle 
\end{align*}
where we have assumed that the measurement has $M_{\text{eff}}$ independent
realizations as calculated with Eq. \ref{eq:ChData_EffectiveNumberIndependent}.%
\footnote{This expression for the error in the autocorrelation is only valid
provided that $B'\ll B_{0}$. Otherwise, $M_{\text{eff}}$ must be
recalculated with $B_{0}$ replaced with $B_{0}-B'$ in Eq. \ref{eq:ChData_EffectiveNumberIndependent}
because only the overlapping magnetic field region of length $B_{0}-B'$
contributes to the persistent current autocorrelation calculated using
Eq. \ref{eq:ChData_AutocorrelationCalculation}. Also, it is possible
that my expression is a factor of $\sqrt{2}$ too small. This factor
does not change the qualitative conclusions drawn here.%
} The magnitude of the autocorrelation at large magnetic field lag
in Figs. \ref{fig:ChData_DAT15_Cor_CL17_45Deg} through \ref{fig:ChData_DAT20_Cor_CL11}
is consistent with $\sqrt{2}\eta(M_{\text{eff}})\langle I^{2}(B)\rangle$
for the values of $\eta(M_{\text{eff}})$ listed in Table \ref{tab:ChData_RingResults}.
The region of magnetic field lag $B'$ covered by each fit extended
from $B'=0$ to the value of $B'$ at which $\langle I(B)I(B+B')\rangle\approx\sqrt{2}\eta(M_{\text{eff}})\langle I^{2}(B)\rangle$.
That the phase of the autocorrelation begins to drift at large $B'$
(e.g. in Fig. \ref{fig:ChData_DAT15_Cor_CL17_45Deg}) is further indication
that the large magnitude of the autocorrelation at large $B'$ is
due to random fluctuations in the persistent current trace. 

The relatively large magnitude of the fluctuations in the autocorrelation
make an accurate analysis difficult. Other than the autocorrelation
of samples CL15 and CL17 at $\theta_{0}=45^{\circ}$, each measured
autocorrelation has an estimated error of over 25\% of its value $\langle I^{2}(B)\rangle$
at zero magnetic field lag $B'$, and thus the values for the geometrical
factor $\gamma$ extracted from fits to these autocorrelations must
be taken only as rough estimates. For samples CL15 and CL17 at $\theta_{0}=45^{\circ}$,
we estimate an uncertainty of $\sim25\%$ in the best fit values for
$\gamma$.%
\footnote{Unlike measurement noise which is uncorrelated from point to point,
the fluctuations in the autocorrelation due to finite sample size
are correlated in magnetic field lag $B'$. This correlation complicates
the estimation of the uncertainty in the fitted values for $\gamma$.
We took sections of the autocorrelation at large $B'$, shifted them
down to $B'=0$, and then added or subtracted them from the autocorrelation
data. By fitting these new forms of the autocorrelation, we got an
idea of the sensitivity of $\gamma$ to the typical fluctuations in
the autocorrelation.%
}

With this analysis of the persistent current autocorrelation, we have
estimated fractional uncertainties $\eta(M_{\text{eff}})$ ranging
from $8$ to 19\% for the magnitude of the typical persistent current
inferred from the large field traces. As these magnitudes were used
to scale all of the current versus temperature data shown in Fig.
\ref{fig:ChData_DAT14_IvsT}, the same fractional error is present
in each of these traces. This uncertainty leads to an over- or under-scaling
of the entire current versus temperature trace and does not introduce
any scatter to the data points. 

The scatter present in the current versus temperature plot is due
to the fluctuations of the cantilever frequency discussed in Chapter
\ref{cha:CHSensitivity}. Considering the persistent current data
in the magnetic field frequency $\beta$ domain, drift of the cantilever
frequency over a long time scale leads to increased noise at low $\beta$
(see e.g. Figs. \ref{fig:ChData_DAT10_PSDCL11} and \ref{fig:ChData_DAT30_FreqPSD_CL11}),
but otherwise the noise in the persistent current spectrum is fairly
flat and independent of temperature. This flat persistent current
background can be seen in Fig. \ref{fig:ChData_DAT13_TempSerIvsBeta}
which plots the Fourier transform of the persistent current for sample
CL17 at several different temperatures. For sample CL17, the background
noise was on the order of $30\,\text{pA}$. The background noise levels
were roughly $0.5$, 10, and $3\,\text{pA}$ for samples CL11, CL14,
and CL15 respectively. Statistical error in the fitting routine for
$D$ due to this background scatter was about 2\%.

The final possible error to consider in the analysis of the current
versus temperature data is the error in the nominal temperature of
the refrigerator. Based on the manufacturer's specifications for the
thermometer and comparison of its reading to fixed temperature points,
we estimate an error of about 7\% in the thermometer reading. As the
current's temperature dependence is roughly exponential, the error
in the temperature is much more significant in fitting the data than
the error in the current magnitude. 

Taking all of these sources of error into account (statistical uncertainty
in the current magnitude, scatter in the cantilever frequency, and
error in the thermometer calibration), we estimate an uncertainty
in the fitted values for the diffusion constant of 7\%. The uncertainty
in the analysis of the autocorrelation data and the extracted values
of $\gamma$ was discussed above. In that case, the uncertainty was
dominated by the standard error due to the finite sample size and
varied between measurements (see discussion above).

\FloatBarrier

\subsection{\label{sub:ChData_FullCurrentTraces}Complete persistent current
traces}

\begin{figure}
\begin{centering}
\includegraphics[width=0.6\paperwidth]{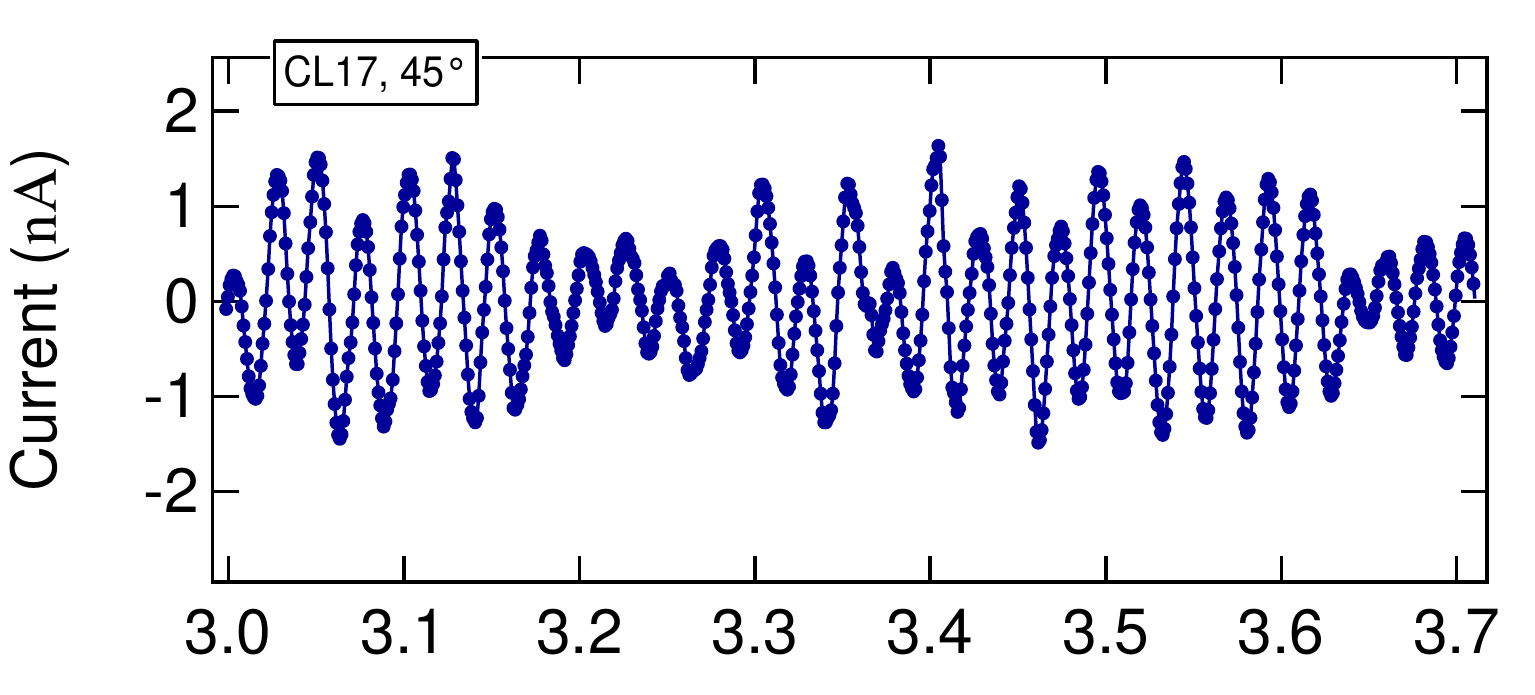}
\par\end{centering}

\begin{centering}
\includegraphics[width=0.6\paperwidth]{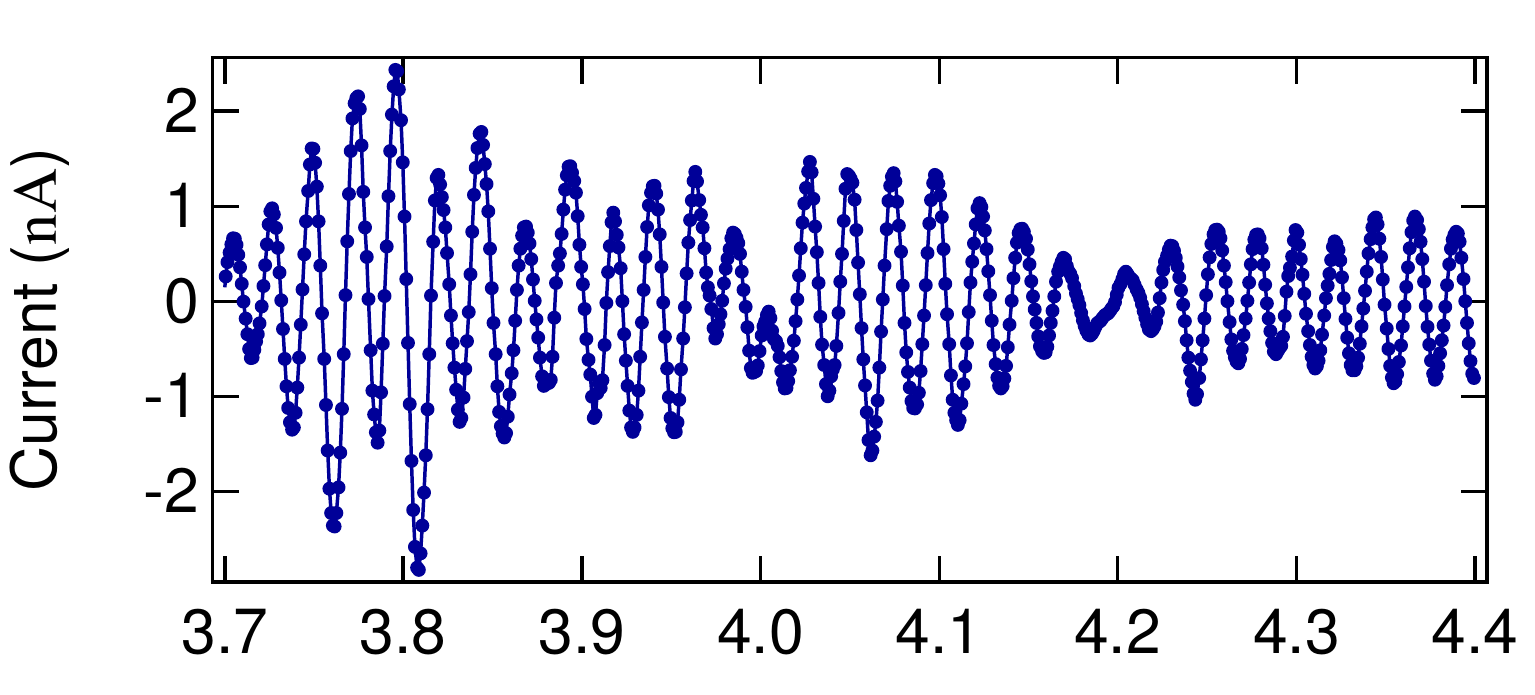}
\par\end{centering}

\begin{centering}
\includegraphics[bb=0bp 0bp 442bp 221bp,width=0.6\paperwidth]{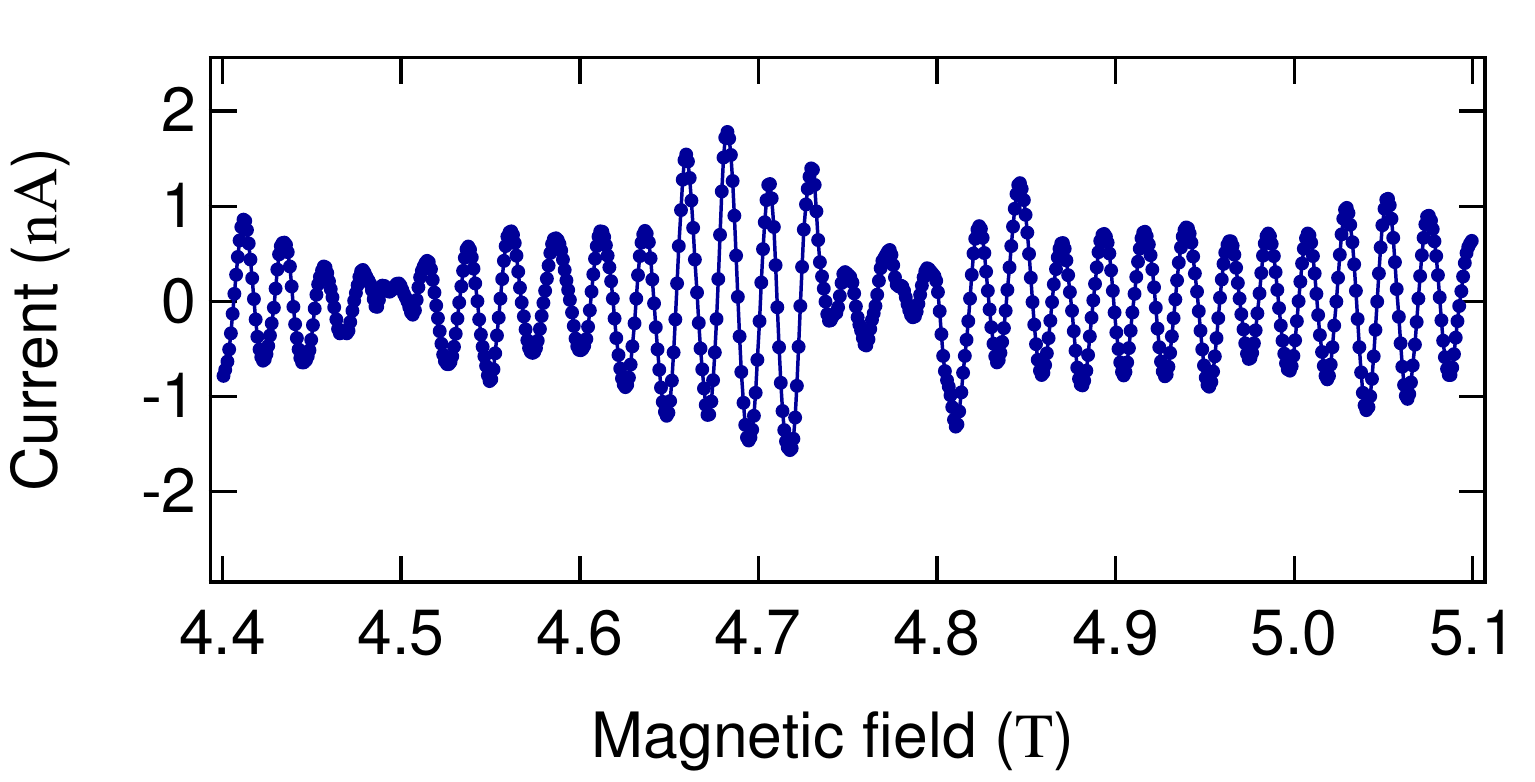}
\par\end{centering}

\caption[Full current versus magnetic field trace for CL17 at 45$^{\circ}$,
part 1]{\label{fig:ChData_DAT21_IvsB_CL17_45DegP1}Full current versus magnetic
field trace for CL17 at 45$^{\circ}$, part 1. The data shown were
recorded at $365\,\text{mK}$ and converted from frequency shift to
current using method A described in \ref{sec:ChData_SigProc}.}
\end{figure}

\begin{figure}
\begin{centering}
\includegraphics[width=0.6\paperwidth]{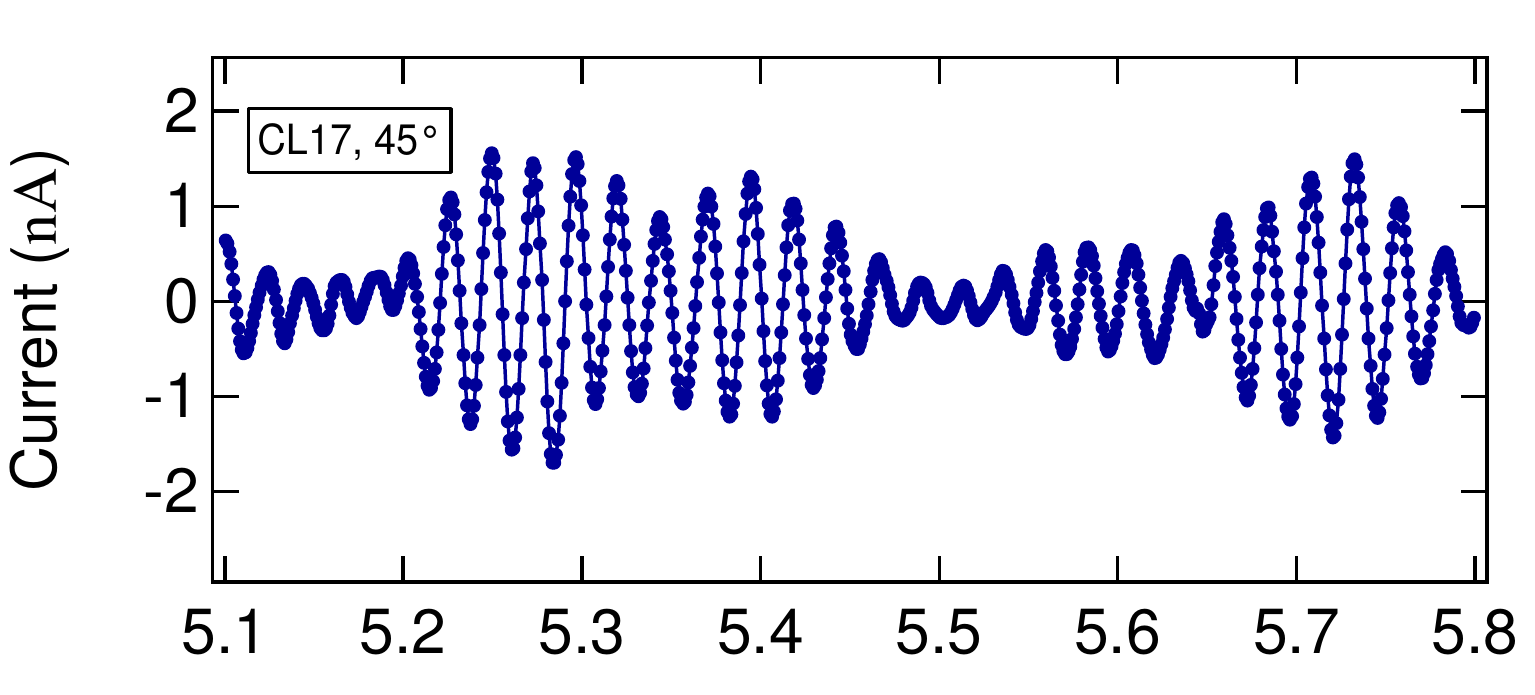}
\par\end{centering}

\begin{centering}
\includegraphics[width=0.6\paperwidth]{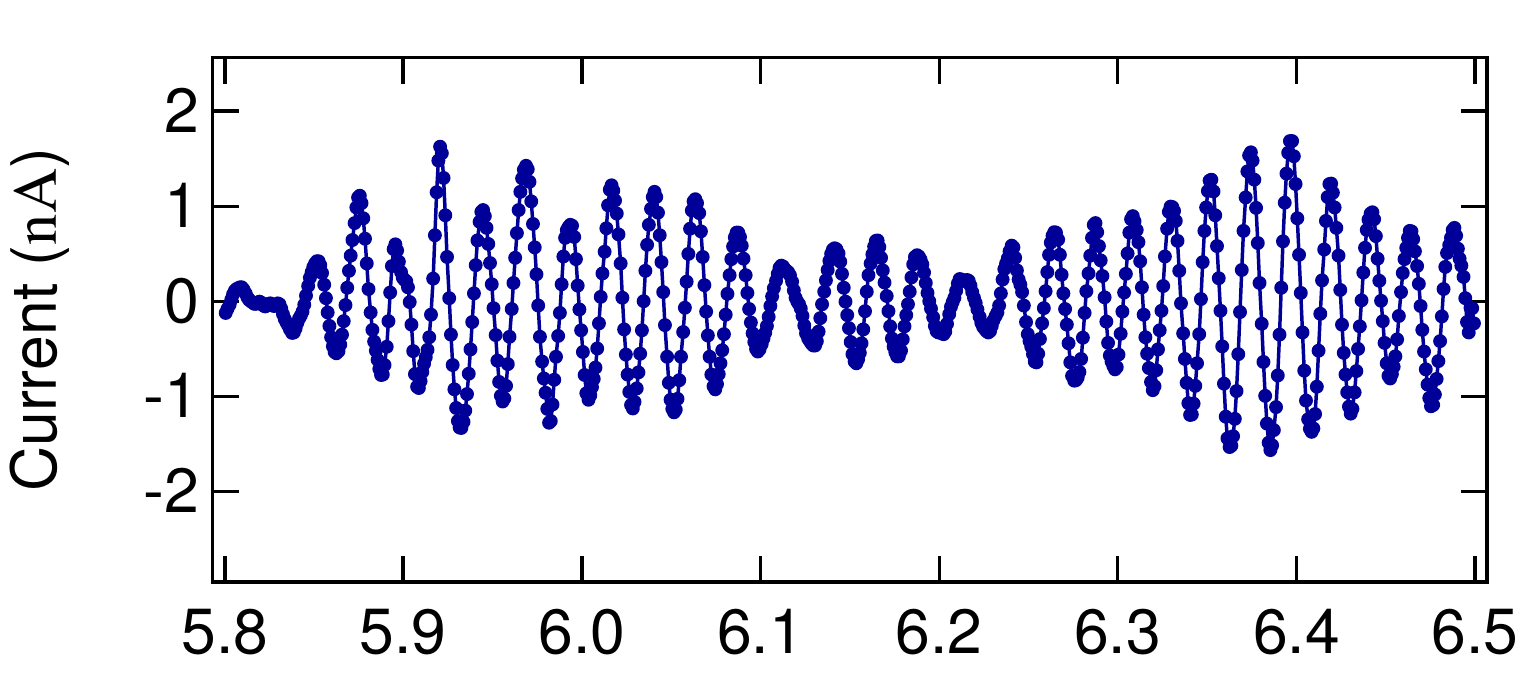}
\par\end{centering}

\begin{centering}
\includegraphics[width=0.6\paperwidth]{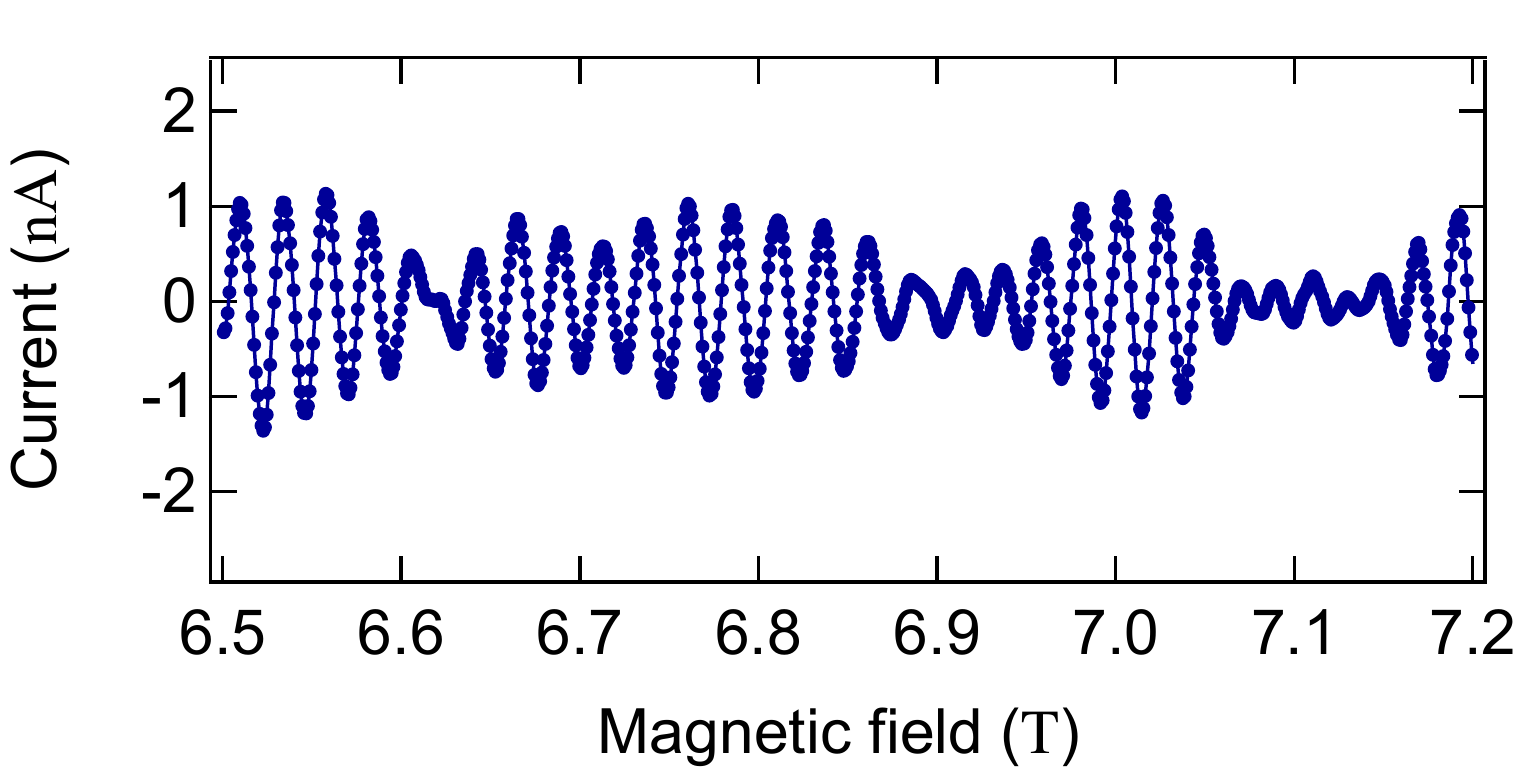}
\par\end{centering}

\caption[Full current versus magnetic field trace for CL17 at 45$^{\circ}$,
part 2]{\label{fig:ChData_DAT21_IvsB_CL17_45DegP2}Full current versus magnetic
field trace for CL17 at 45$^{\circ}$, part 2. The data shown were
recorded at $365\,\text{mK}$ and were converted from frequency shift
to current using method A described in \ref{sec:ChData_SigProc}.}
\end{figure}

\begin{figure}
\begin{centering}
\includegraphics[width=0.6\paperwidth]{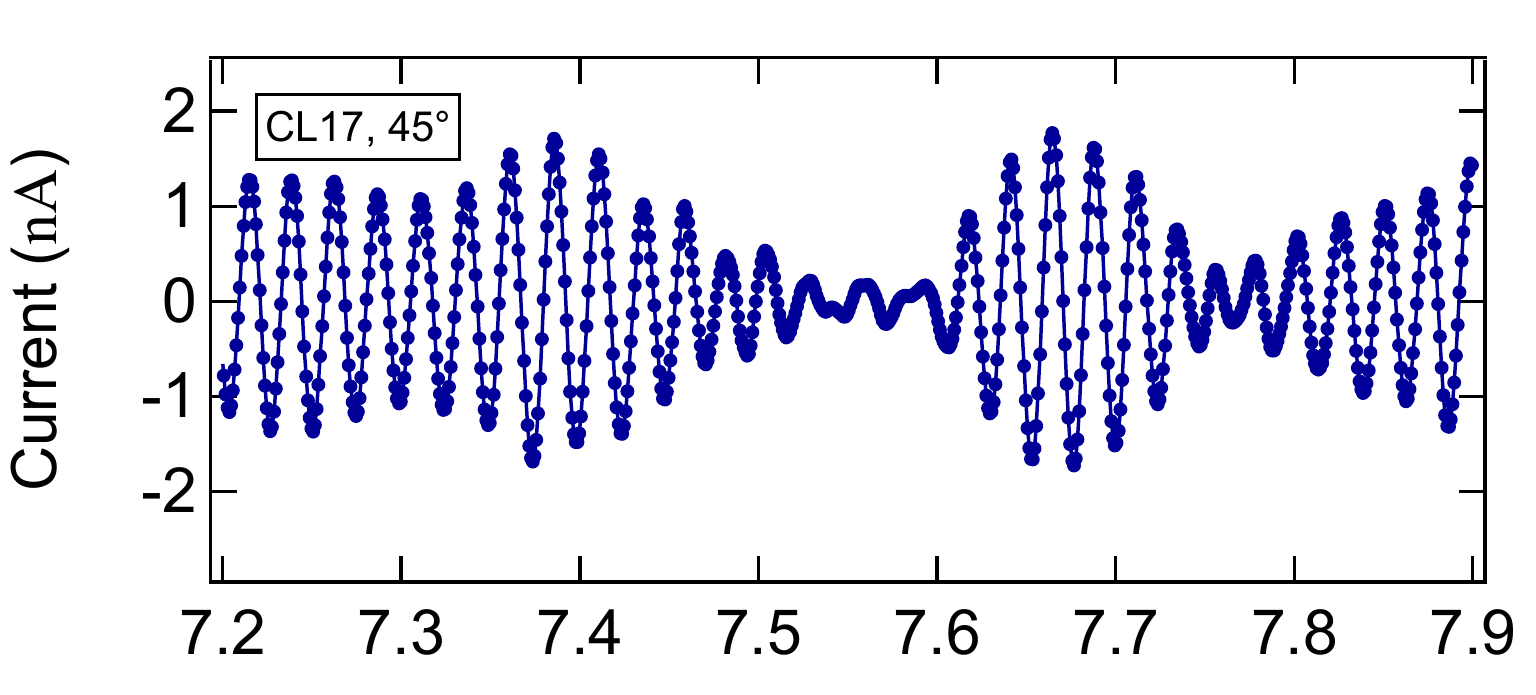}
\par\end{centering}

\begin{centering}
\includegraphics[width=0.6\paperwidth]{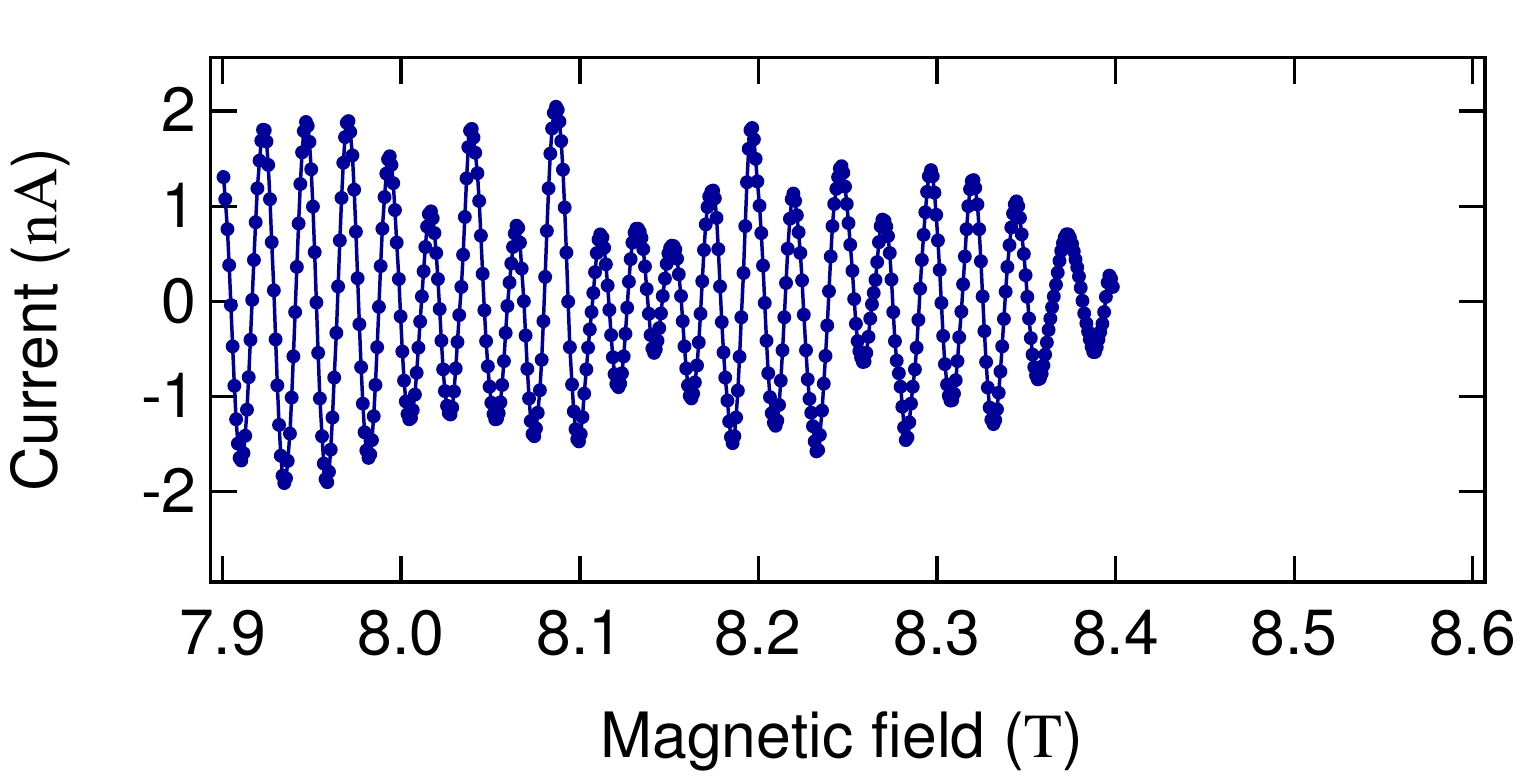}
\par\end{centering}

\caption[Full current versus magnetic field trace for CL17 at 45$^{\circ}$,
part 3]{\label{fig:ChData_DAT21_IvsB_CL17_45DegP3}Full current versus magnetic
field trace for CL17 at 45$^{\circ}$, part 3. The data shown were
recorded at $365\,\text{mK}$ and converted from frequency shift to
current using method A described in \ref{sec:ChData_SigProc}.}
\end{figure}

\begin{figure}
\begin{centering}
\includegraphics[width=0.6\paperwidth]{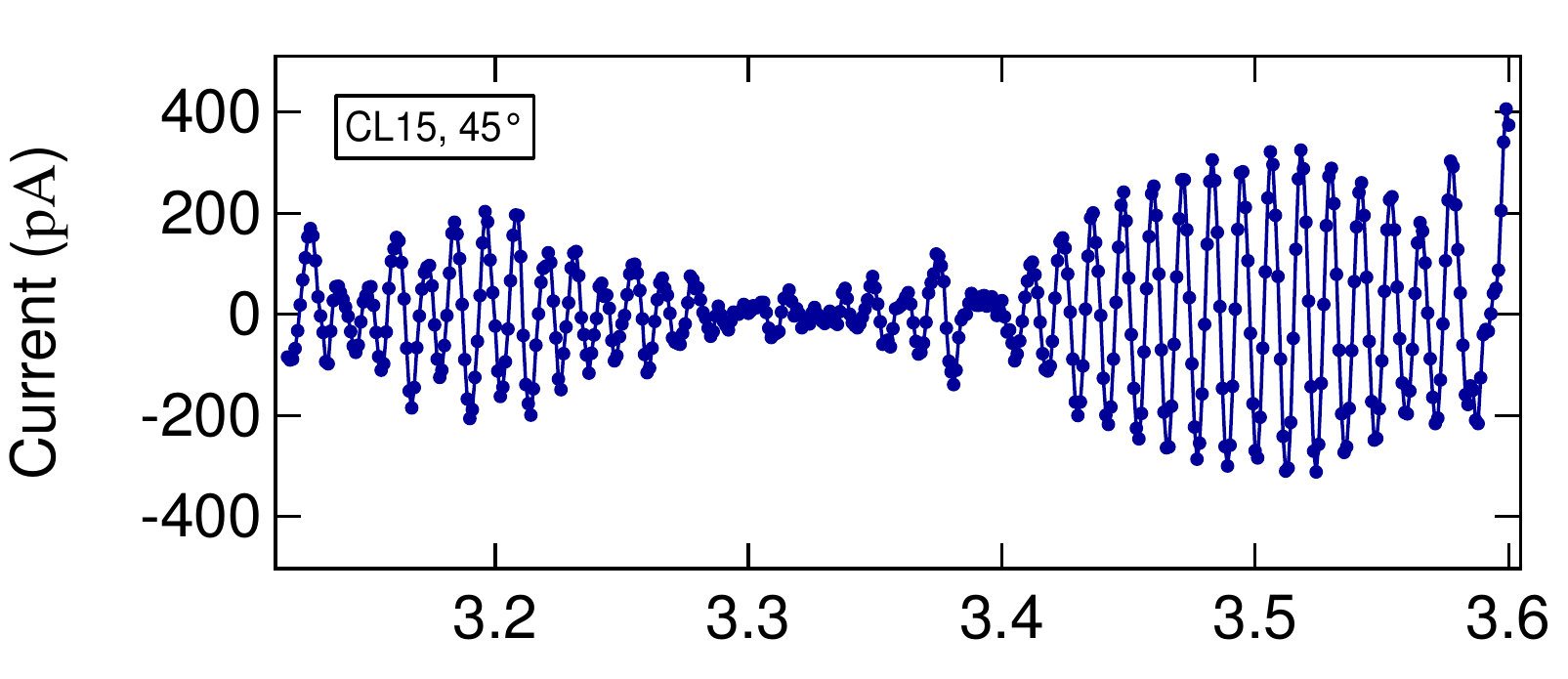}
\par\end{centering}

\begin{centering}
\includegraphics[width=0.6\paperwidth]{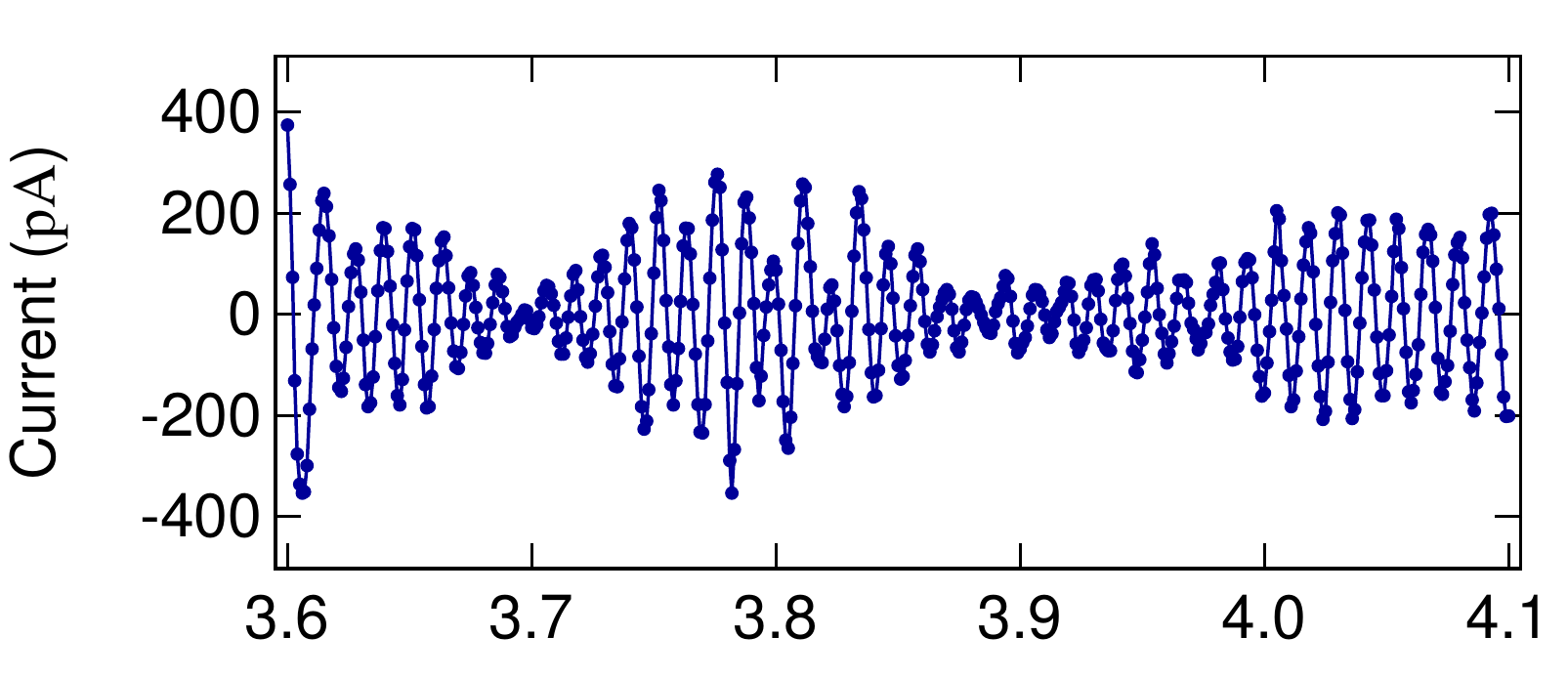}
\par\end{centering}

\begin{centering}
\includegraphics[width=0.6\paperwidth]{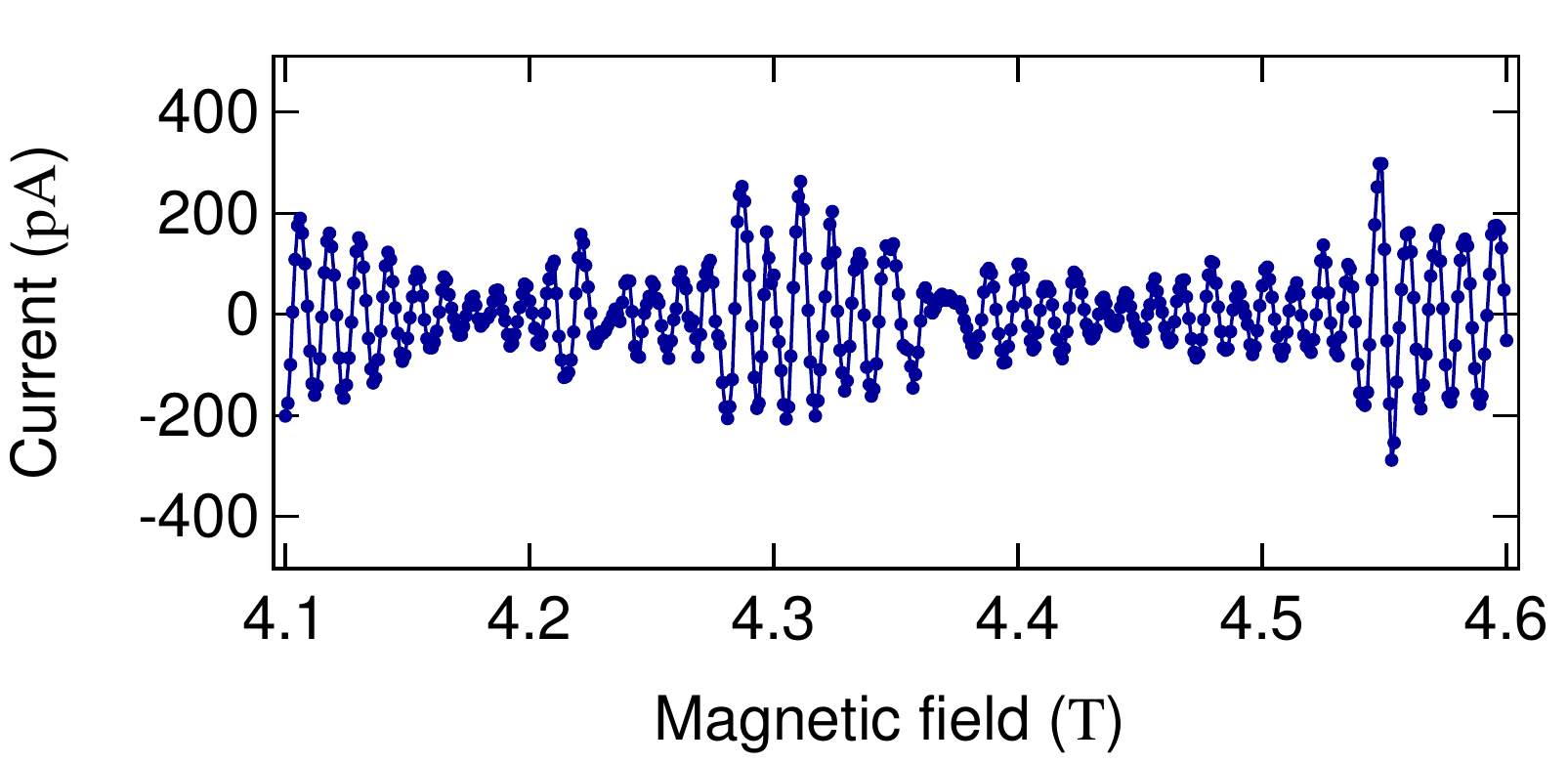}
\par\end{centering}

\caption[Full current versus magnetic field trace for CL15 at 45$^{\circ}$,
part 1]{\label{fig:ChData_DAT22_IvsB_CL15_45DegP1}Full current versus magnetic
field trace for CL15 at 45$^{\circ}$, part 1. The data shown were
recorded at $365\,\text{mK}$ and converted from frequency shift to
current using method A described in \ref{sec:ChData_SigProc}.}
\end{figure}

\begin{figure}
\begin{centering}
\includegraphics[width=0.6\paperwidth]{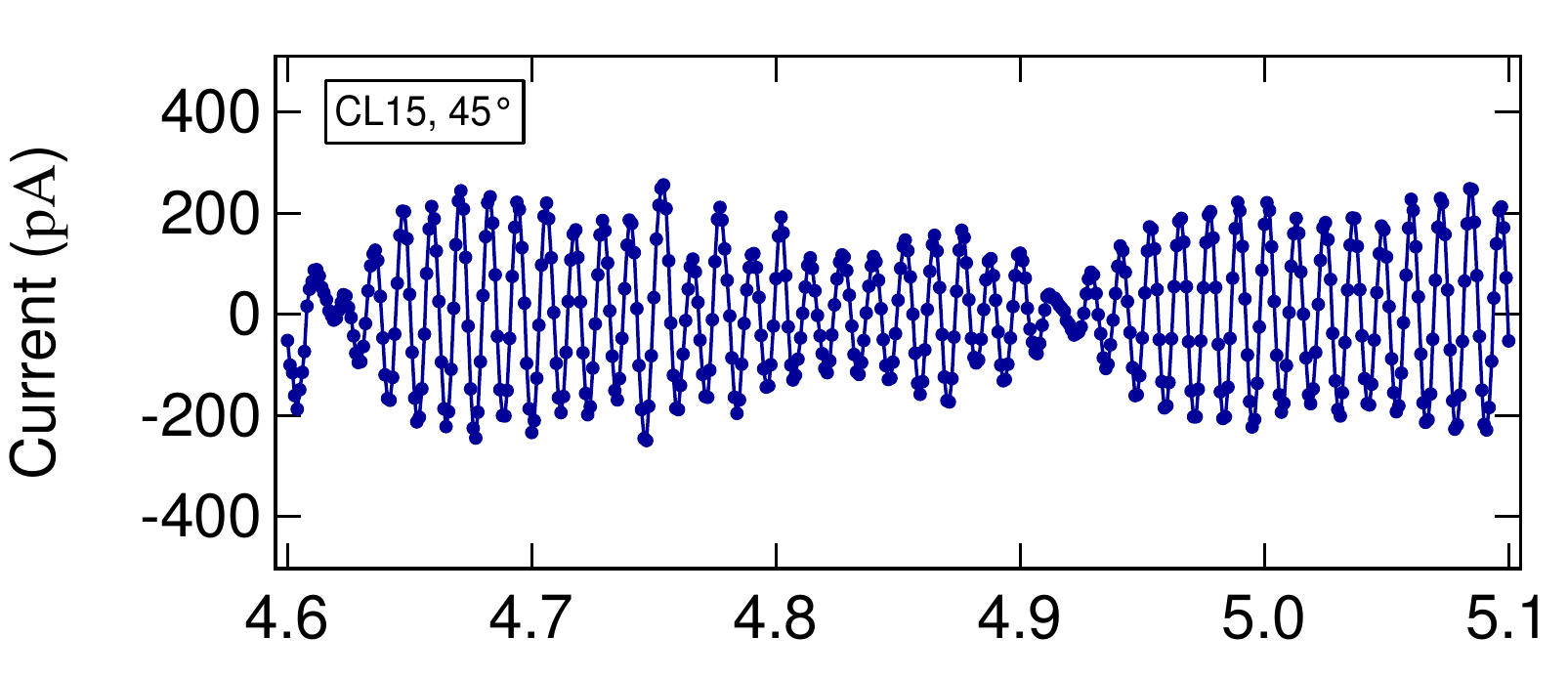}
\par\end{centering}

\begin{centering}
\includegraphics[width=0.6\paperwidth]{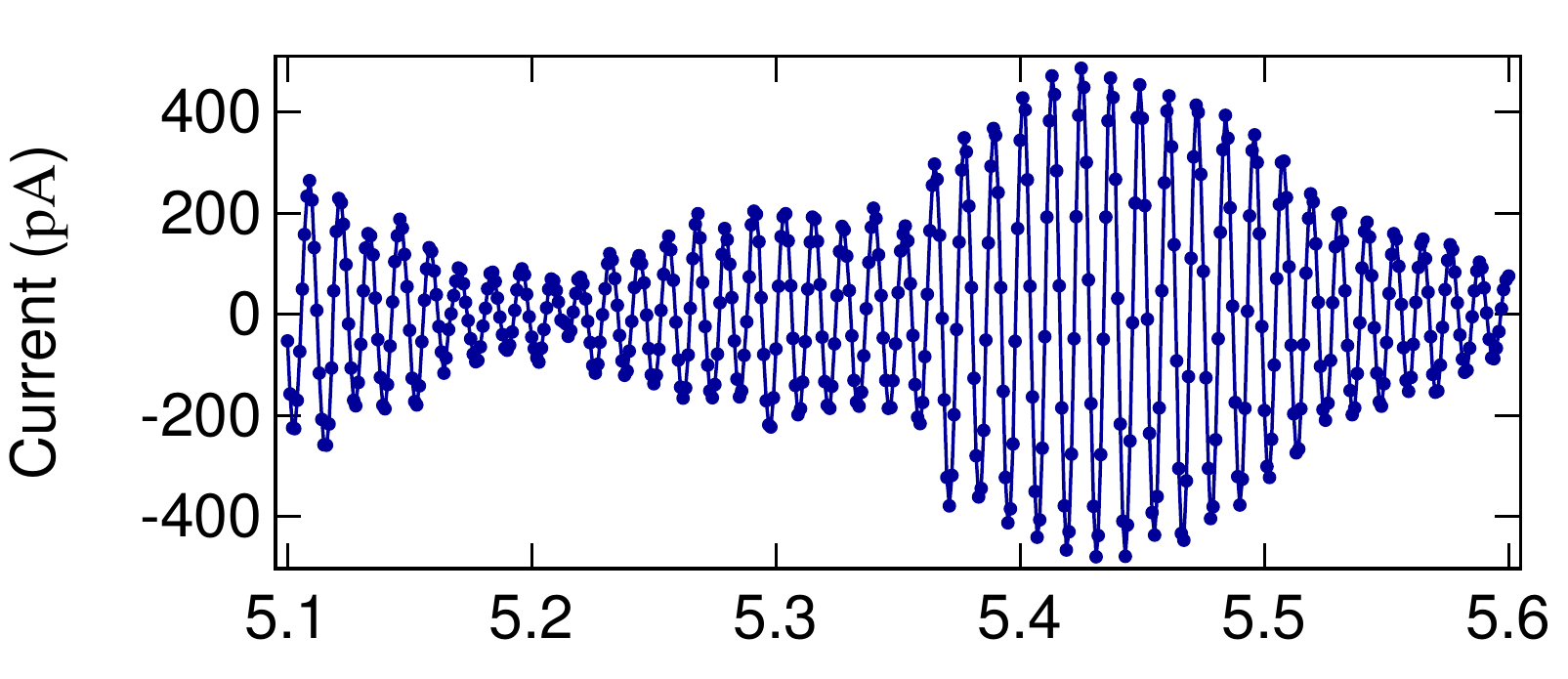}
\par\end{centering}

\begin{centering}
\includegraphics[width=0.6\paperwidth]{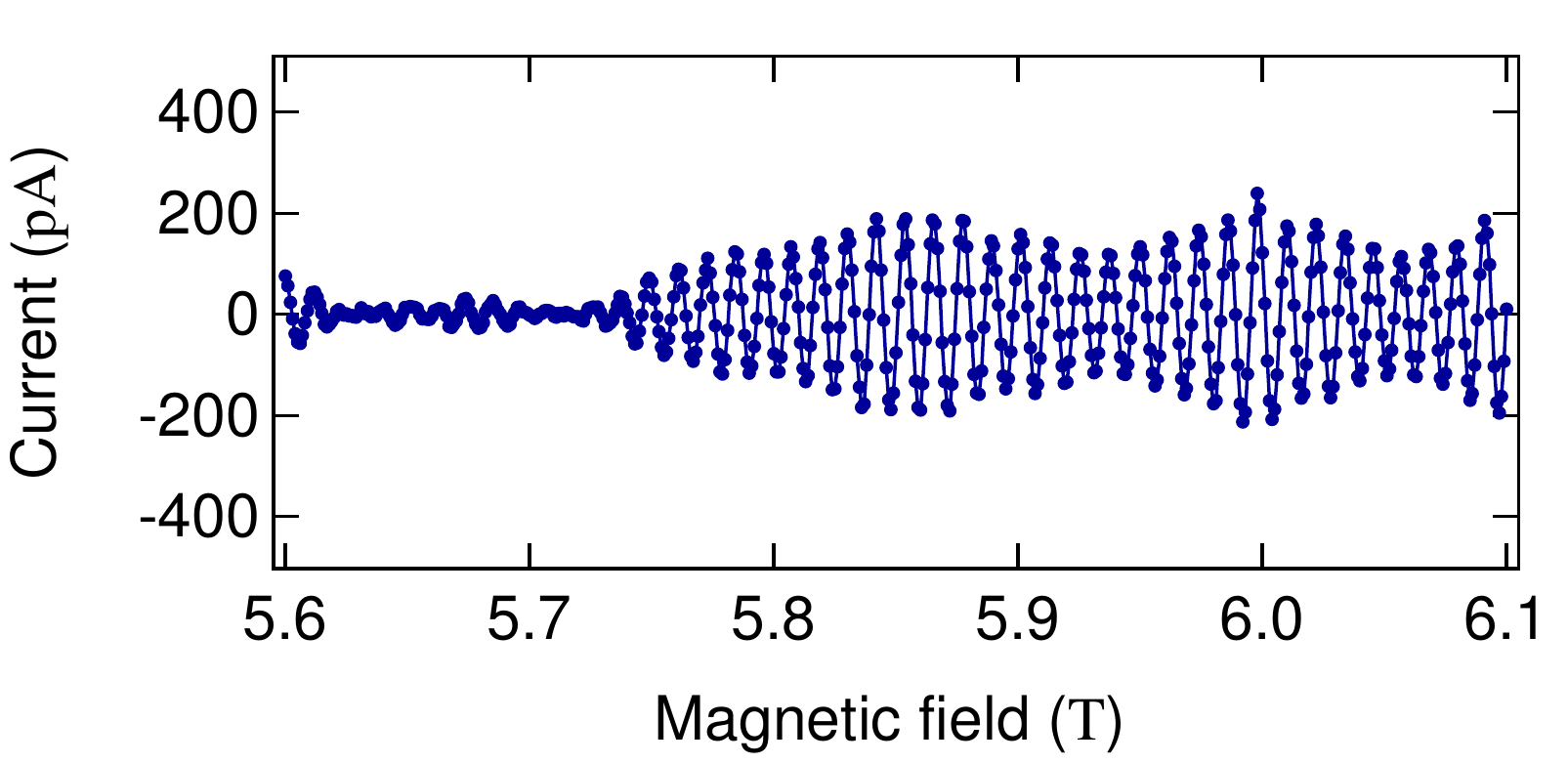}
\par\end{centering}

\caption[Full current versus magnetic field trace for CL15 at 45$^{\circ}$,
part 2]{\label{fig:ChData_DAT22_IvsB_CL15_45DegP2}Full current versus magnetic
field trace for CL15 at 45$^{\circ}$, part 2. The data shown were
recorded at $365\,\text{mK}$ and converted from frequency shift to
current using method A described in \ref{sec:ChData_SigProc}.}
\end{figure}

\begin{figure}
\begin{centering}
\includegraphics[width=0.6\paperwidth]{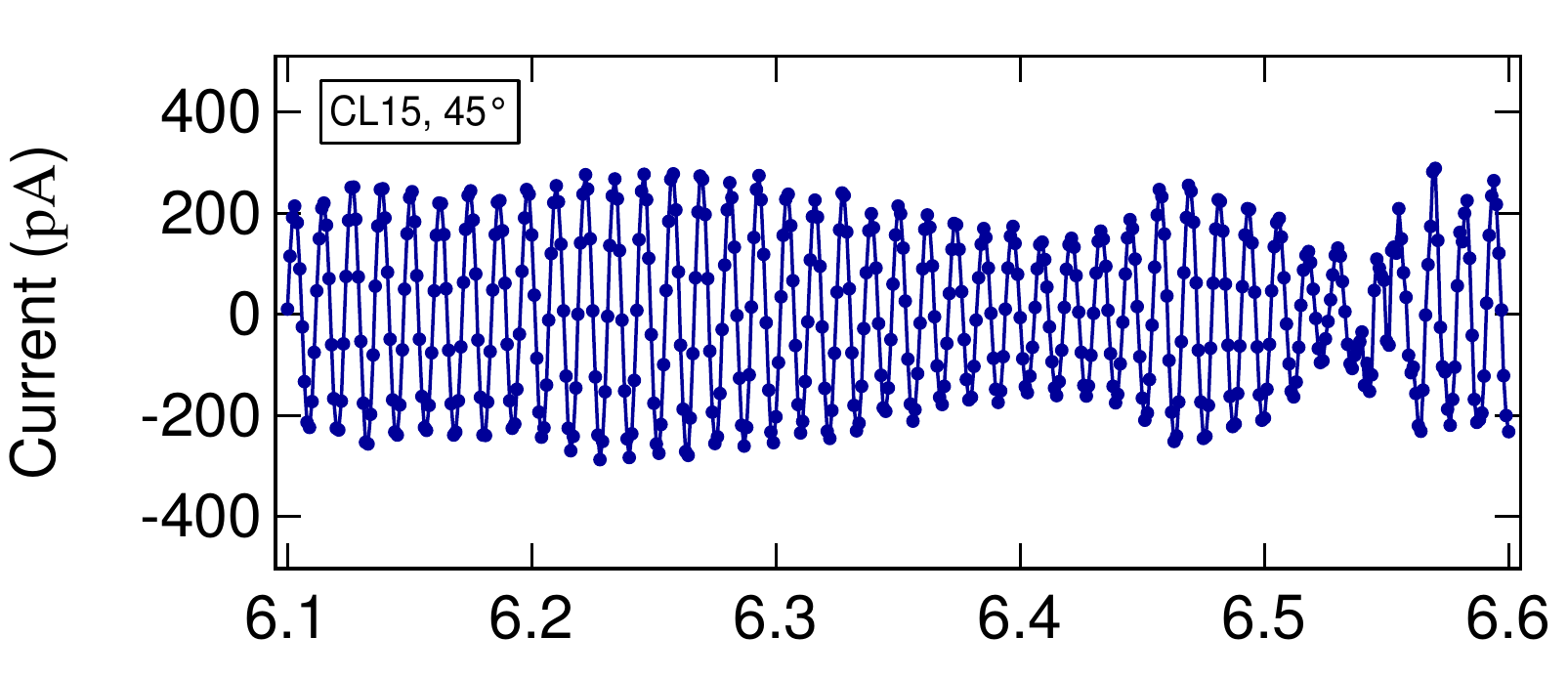}
\par\end{centering}

\begin{centering}
\includegraphics[width=0.6\paperwidth]{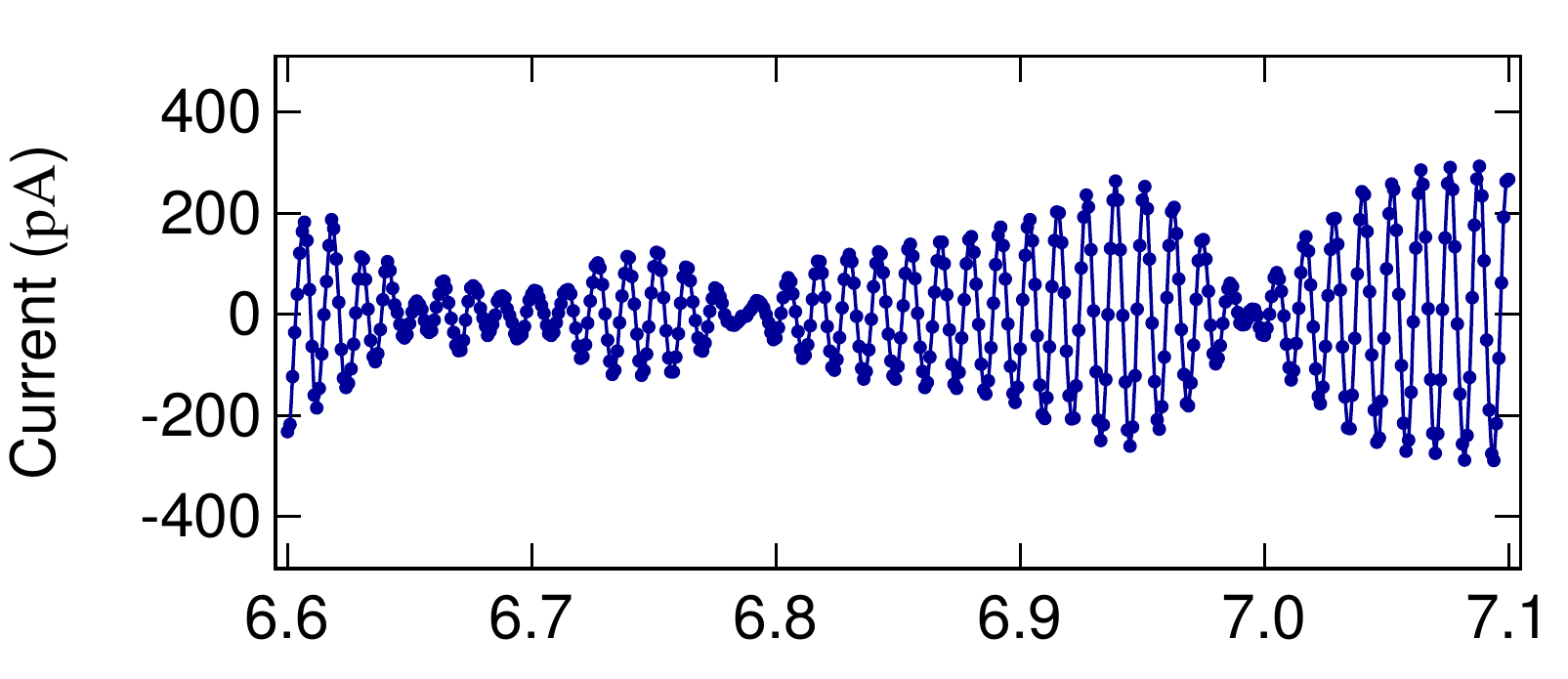}
\par\end{centering}

\begin{centering}
\includegraphics[width=0.6\paperwidth]{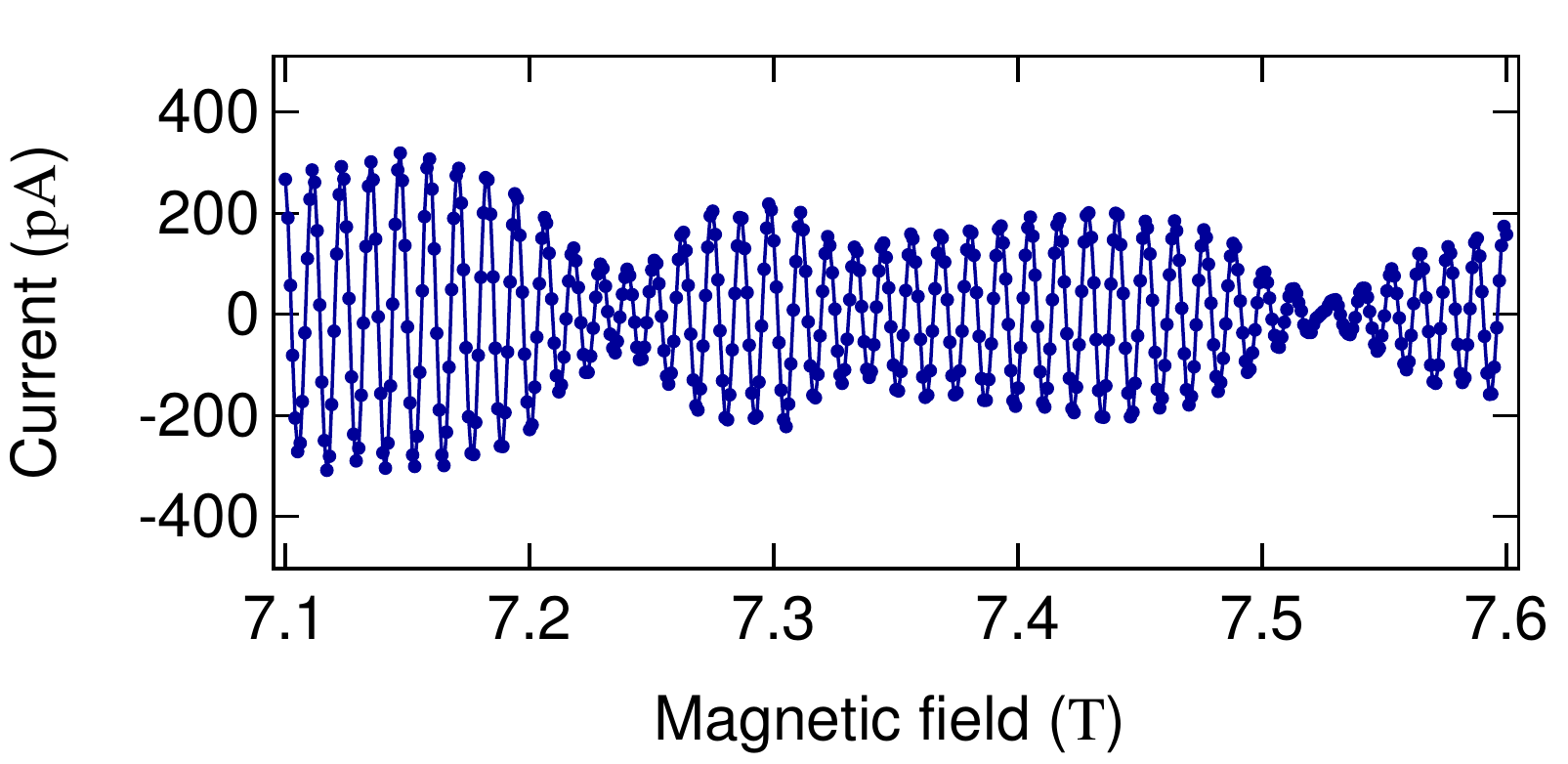}
\par\end{centering}

\caption[Full current versus magnetic field trace for CL15 at 45$^{\circ}$,
part 3]{\label{fig:ChData_DAT22_IvsB_CL15_45DegP3}Full current versus magnetic
field trace for CL15 at 45$^{\circ}$, part 3. The data shown were
recorded at $365\,\text{mK}$ and converted from frequency shift to
current using method A described in \ref{sec:ChData_SigProc}. Due
to intermittent noise in the measurement, the data between $6.528\,\text{T}$
and $6.537\,\text{T}$ was not usable. We approximated this section
by replicating the data from the oscillation just below $6.528\,\text{T}$.}
\end{figure}

\begin{figure}
\begin{centering}
\includegraphics[width=0.6\paperwidth]{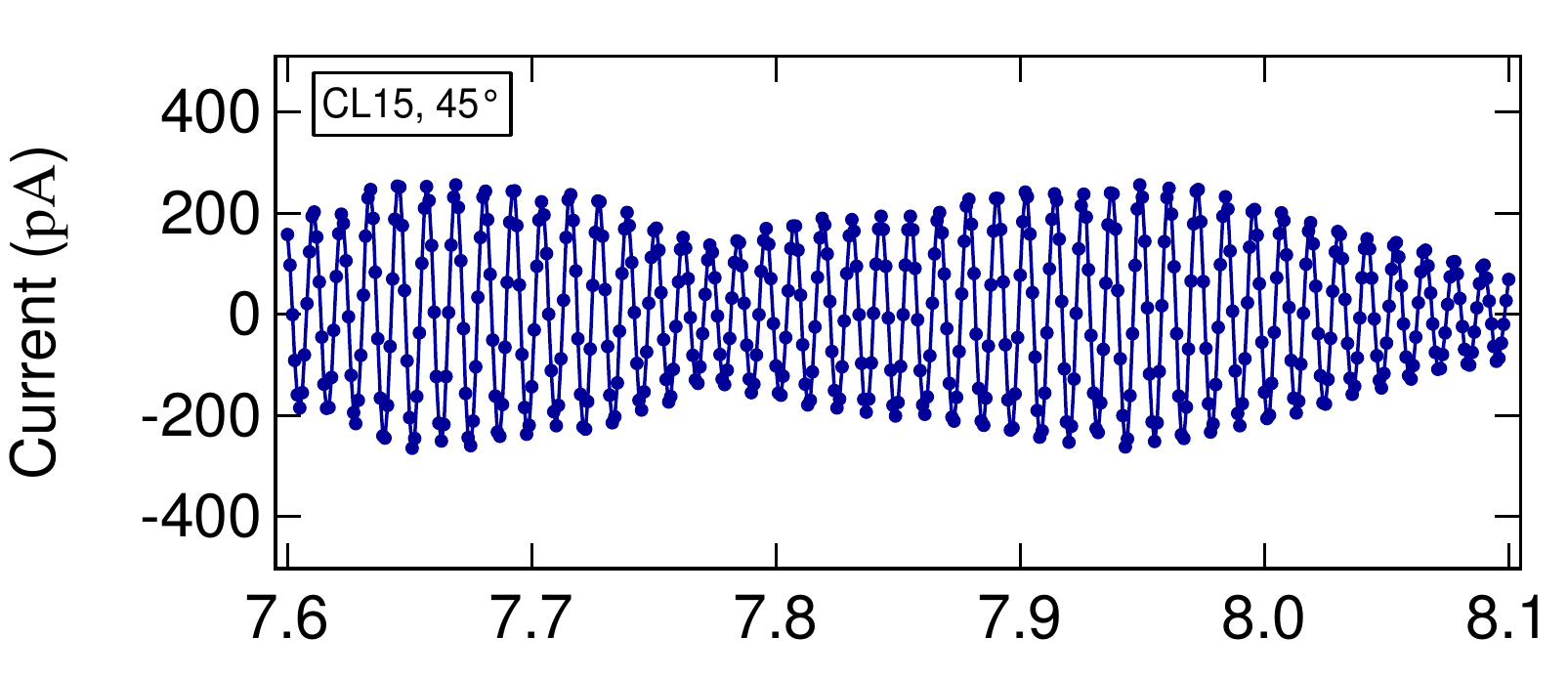}
\par\end{centering}

\begin{centering}
\includegraphics[width=0.6\paperwidth]{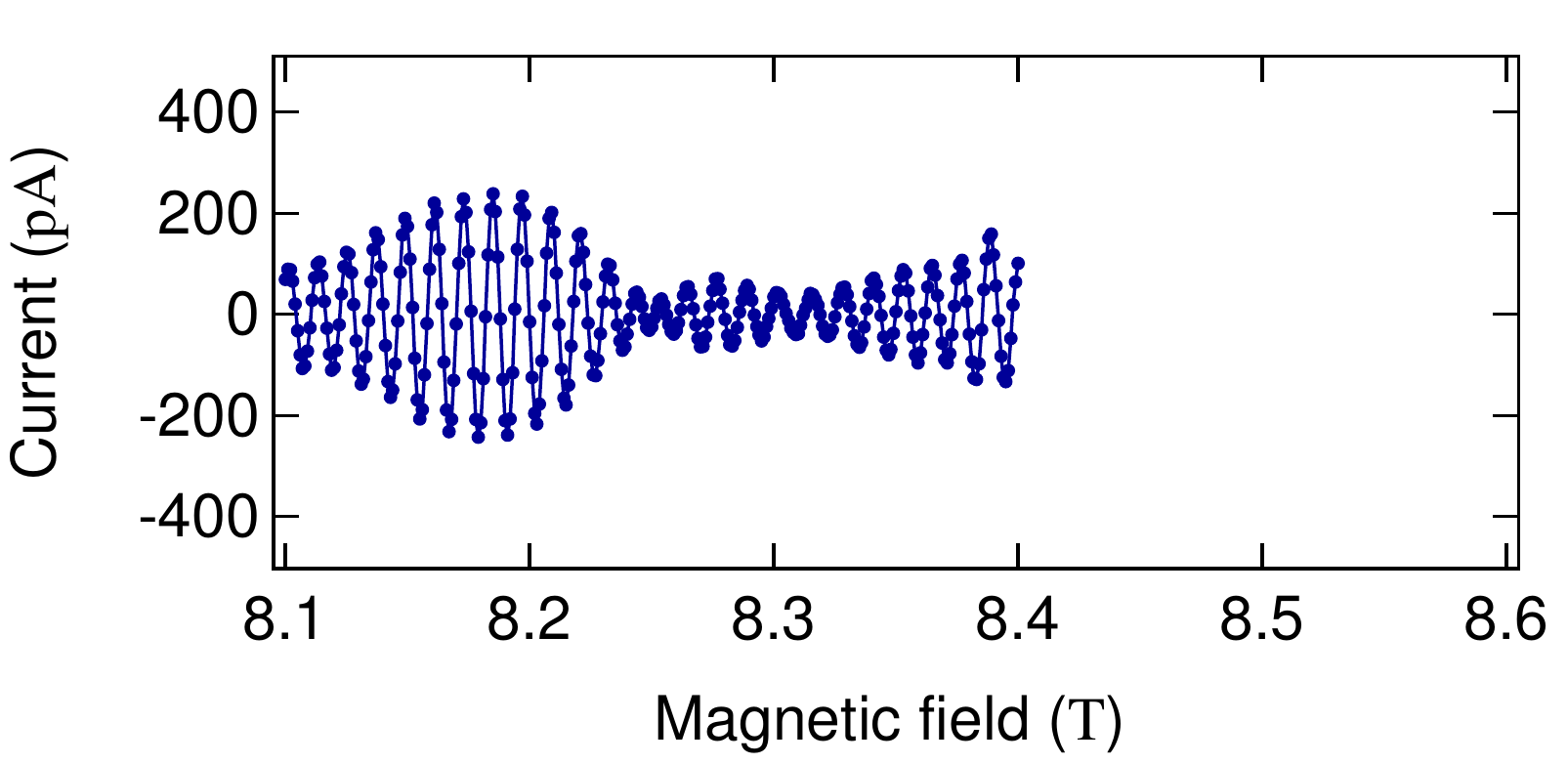}
\par\end{centering}

\caption[Full current versus magnetic field trace for CL15 at 45$^{\circ}$,
part 4]{\label{fig:ChData_DAT22_IvsB_CL15_45DegP4}Full current versus magnetic
field trace for CL15 at 45$^{\circ}$, part 4. The data shown were
recorded at $365\,\text{mK}$ and converted from frequency shift to
current using method A described in \ref{sec:ChData_SigProc}.}
\end{figure}

\begin{figure}
\begin{centering}
\includegraphics[width=0.7\paperwidth]{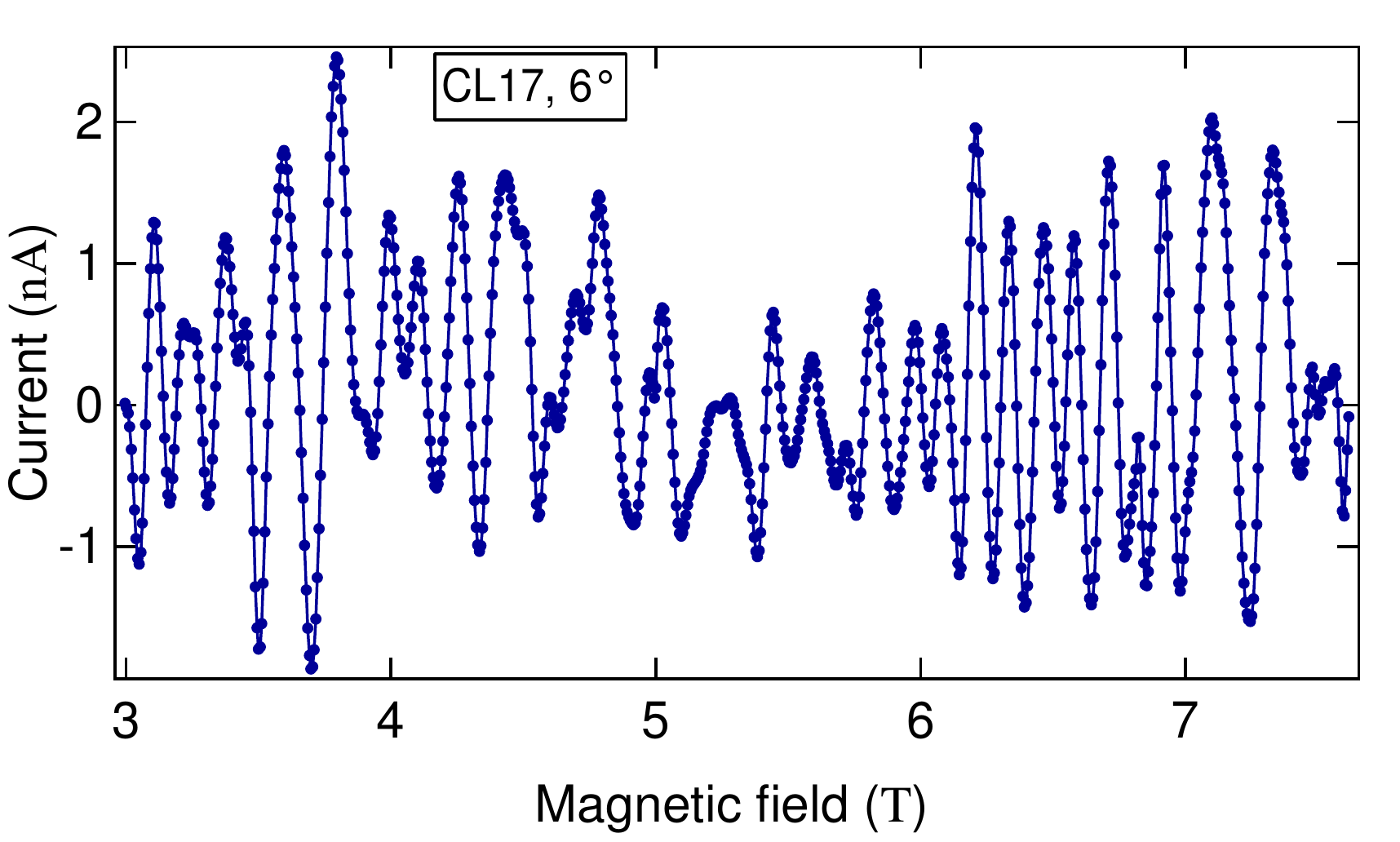}
\par\end{centering}

\caption[Full current versus magnetic field trace for CL17 at 6$^{\circ}$]{\label{fig:ChData_DAT23_IvsB_CL17_6Deg}Full current versus magnetic
field trace for CL17 at 6$^{\circ}$. The data shown were recorded
at $323\,\text{mK}$ and converted from frequency shift to current
using method A described in \ref{sec:ChData_SigProc}.}

\end{figure}

\begin{figure}
\begin{centering}
\includegraphics[width=0.7\paperwidth]{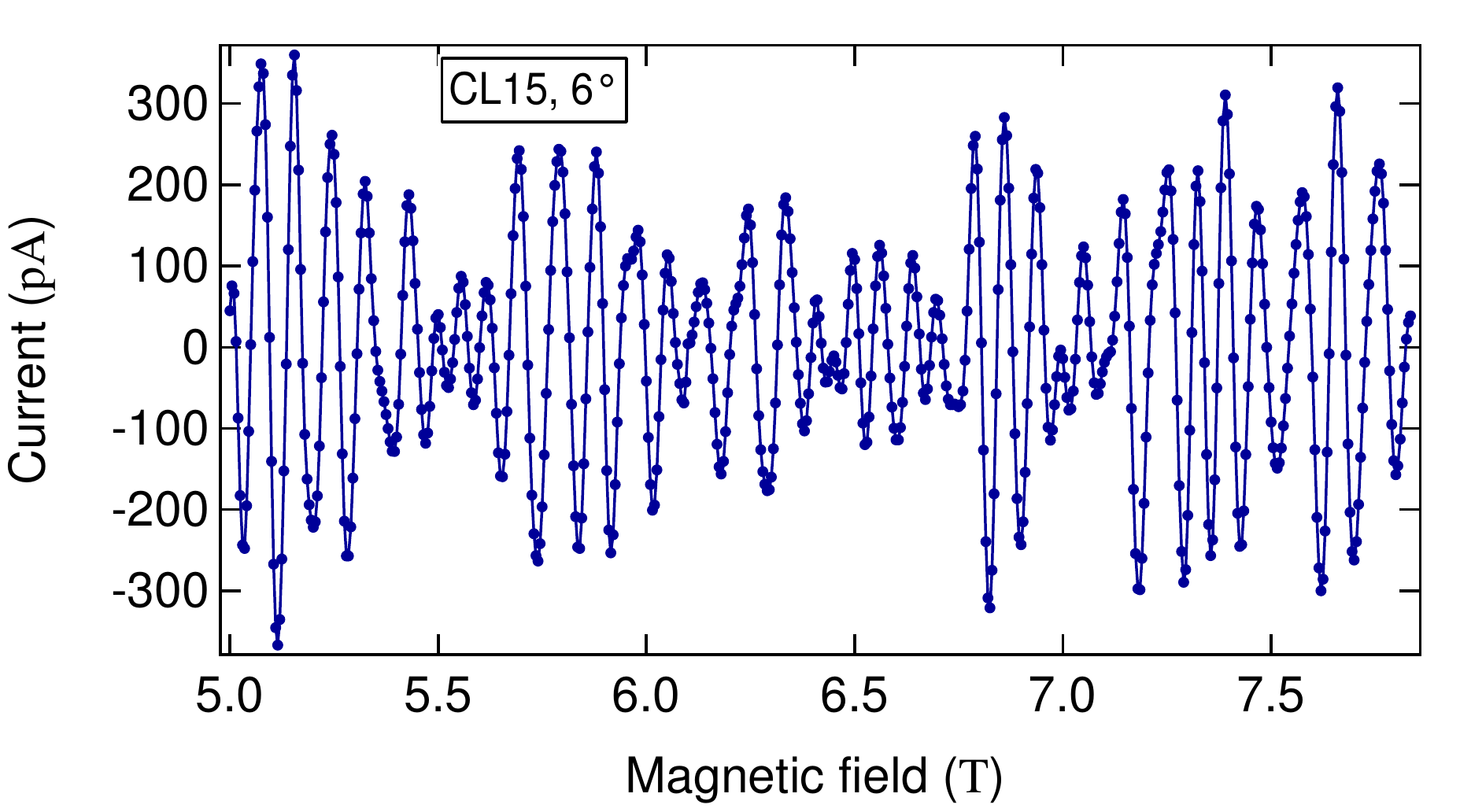}
\par\end{centering}

\caption[Full current versus magnetic field trace for CL15 at 6$^{\circ}$]{\label{fig:ChData_DAT24_IvsB_CL15_6Deg}Full current versus magnetic
field trace for CL15 at 6$^{\circ}$. The data shown were recorded
at $323\,\text{mK}$ and converted from frequency shift to current
using method A described in \ref{sec:ChData_SigProc}.}
\end{figure}

\begin{figure}
\begin{centering}
\includegraphics[width=0.6\paperwidth]{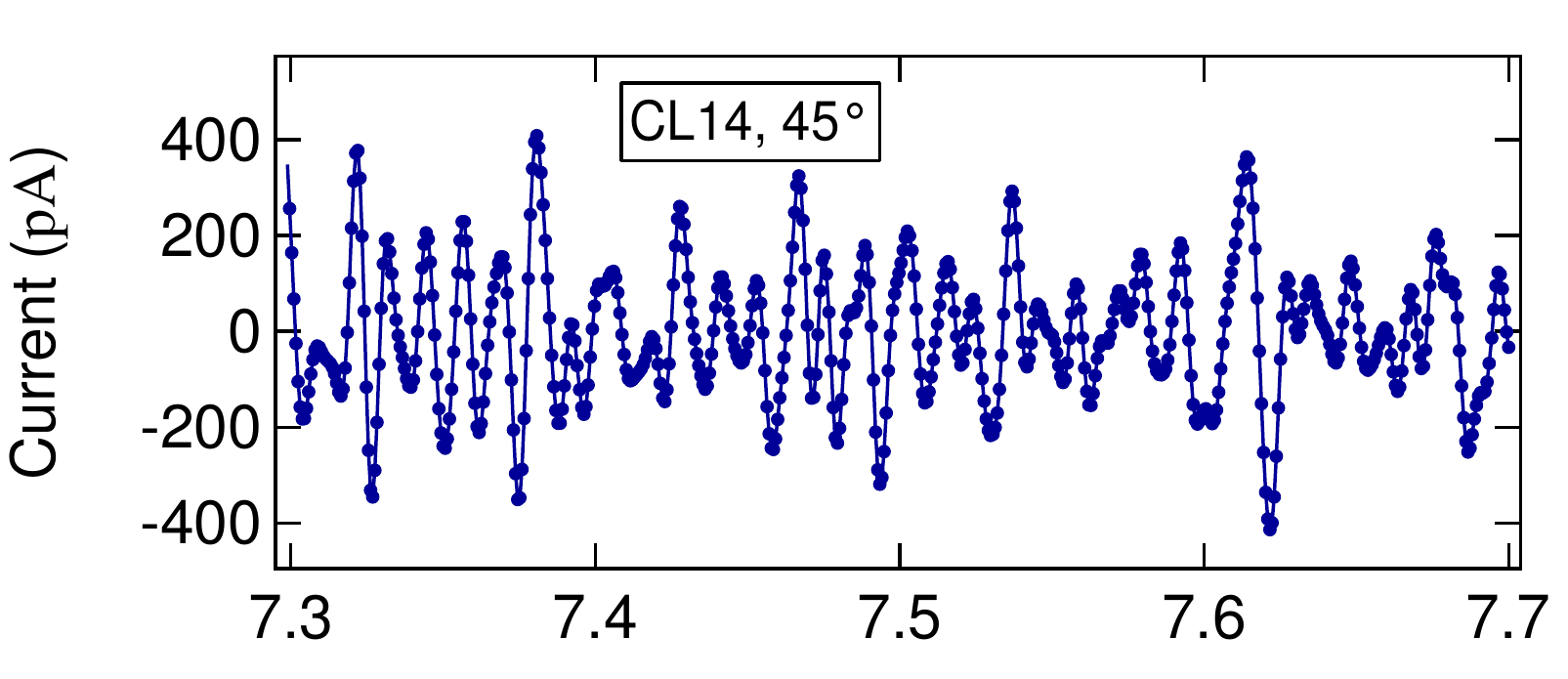}
\par\end{centering}

\begin{centering}
\includegraphics[width=0.6\paperwidth]{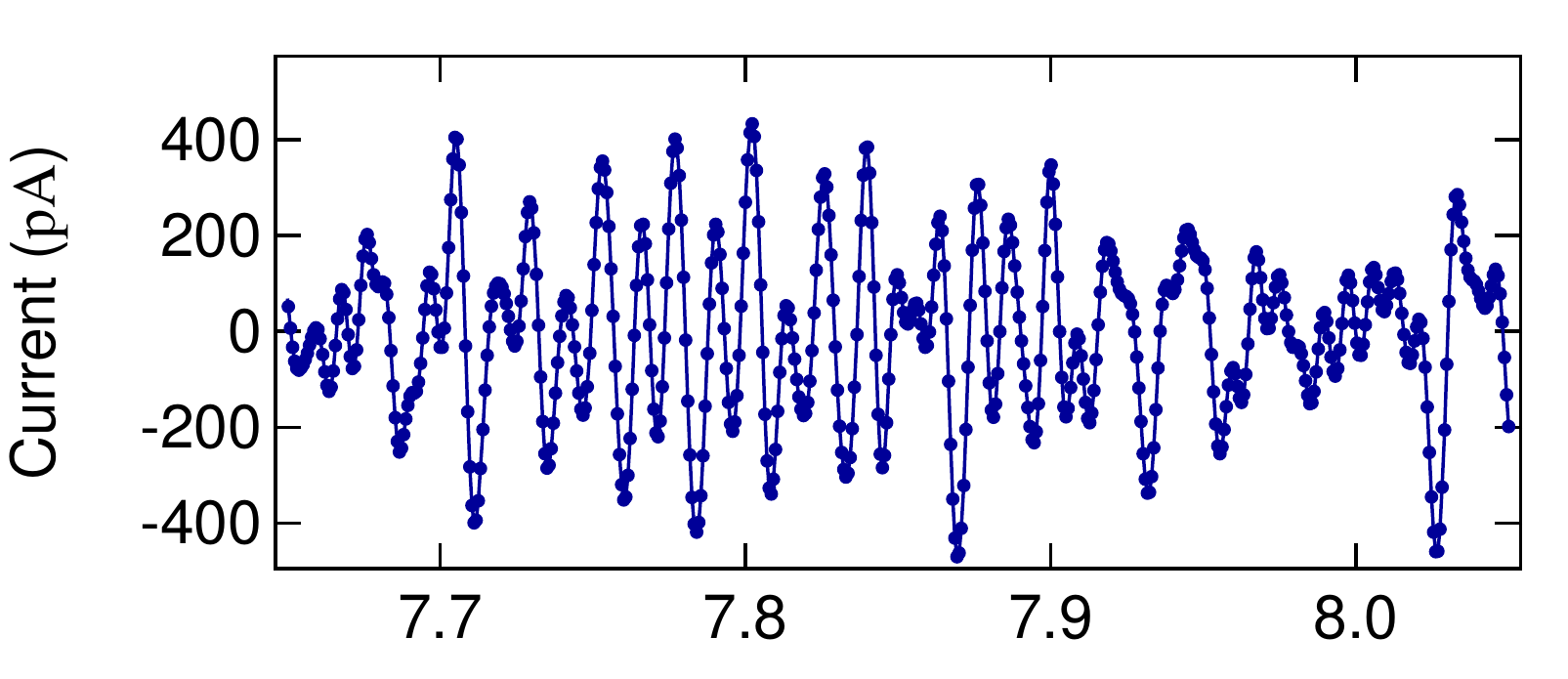}
\par\end{centering}

\begin{centering}
\includegraphics[width=0.6\paperwidth]{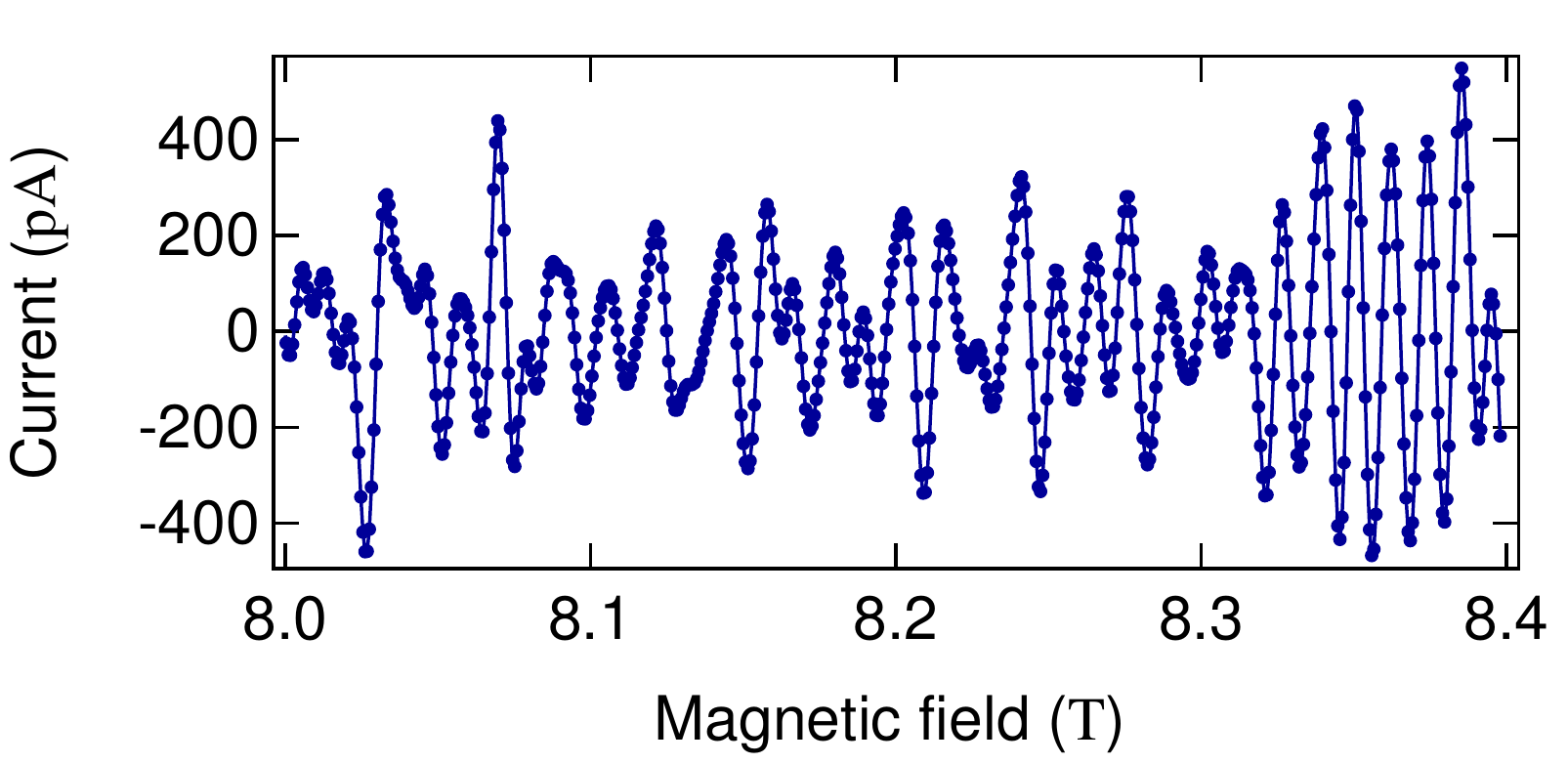}
\par\end{centering}

\caption[Full current versus magnetic field trace for CL14 at 45$^{\circ}$]{\label{fig:ChData_DAT25_IvsB_CL14}Full current versus magnetic field
trace for CL14 at 45$^{\circ}$. The data shown were recorded at $365\,\text{mK}$
and converted from frequency shift to current using method A described
in \ref{sec:ChData_SigProc}.}
\end{figure}

\begin{figure}
\begin{centering}
\includegraphics[width=0.7\paperwidth]{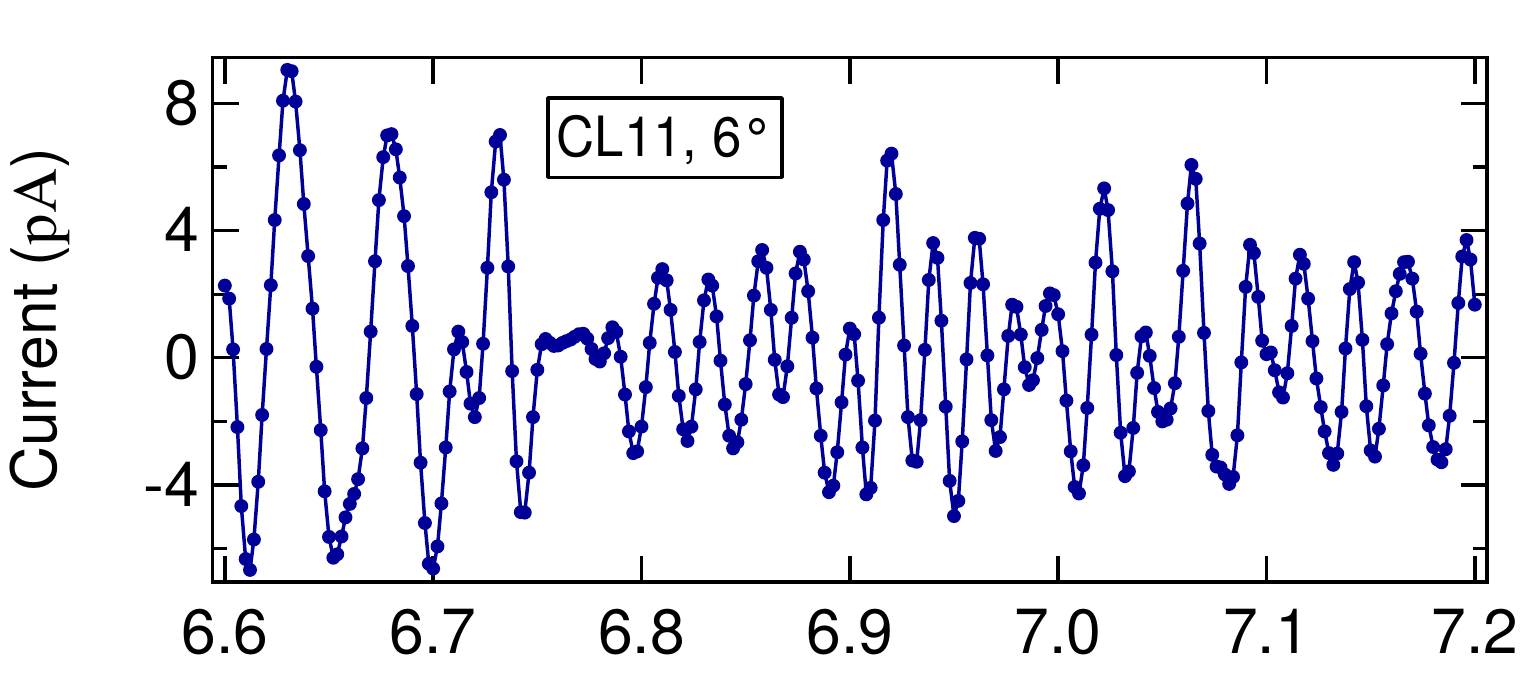}
\par\end{centering}

\begin{centering}
\includegraphics[width=0.7\paperwidth]{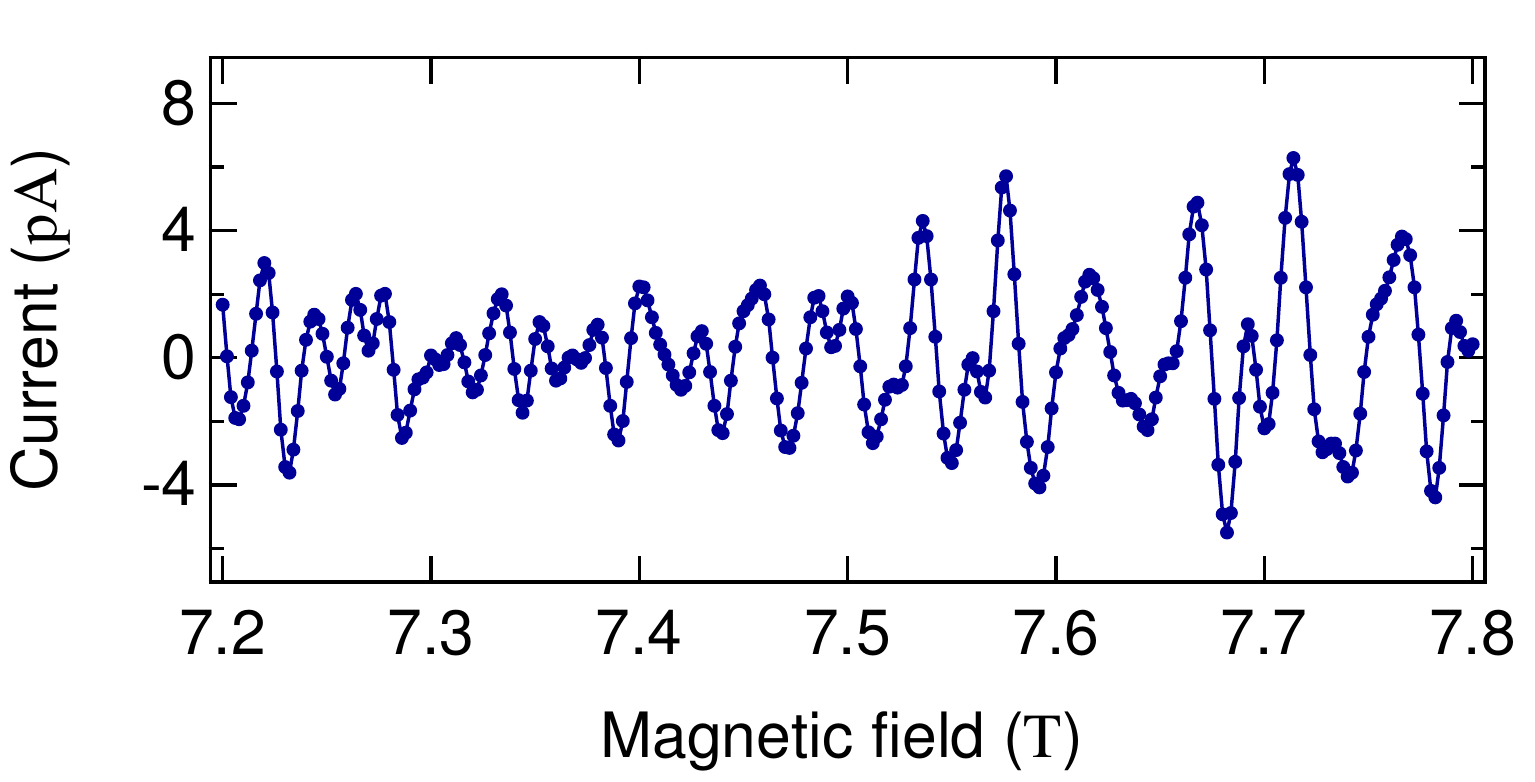}
\par\end{centering}

\caption[Full current versus magnetic field trace for CL11 at 6$^{\circ}$]{\label{fig:ChData_DAT26_IvsB_CL11}Full current versus magnetic field
trace for CL11 at 6$^{\circ}$. The data shown were recorded at $323\,\text{mK}$
and converted from frequency shift to current using method A described
in \ref{sec:ChData_SigProc}.}
\end{figure}

In this section, we present, in Figs. \ref{fig:ChData_DAT21_IvsB_CL17_45DegP1}
through \ref{fig:ChData_DAT26_IvsB_CL11}, the complete current versus
magnetic field traces which were analyzed in Sections \ref{sub:ChData_Qualitative}
and \ref{sub:ChData_Quantitative}. These traces were calculated,
using method A, from measurements of the cantilever frequency performed
at the refrigerator's base temperature $T_{b}$, which was $323\,\text{mK}$
for $\theta_{0}=6^{\circ}$ and $365\,\text{mK}$ for $\theta_{0}=45^{\circ}$.%
\footnote{We define the base temperature as the temperature of the refrigerator
when the helium-3 was condensed and all heat sources turned off. It
was possible to reach slightly lower temperatures by also pumping
on the 1K pot, but these temperatures could only be sustained for
short periods of time. The base temperature was different for the
two angles $\theta_{0}$ because different wiring arrangements led
to different heat loads on the refrigerator during the two cooldowns.%
}

\clearpage{}

As discussed in Sections \ref{sub:ChData_Qualitative} and \ref{sub:ChData_Quantitative},
the signal to noise ratios from samples CL11 (large rings) and CL14
(single ring) were smaller than those from samples CL15 and CL17,
which had bigger arrays of smaller rings. At high field, the absolute
signal to noise ratio for the latter two samples ranged from 30 to
40, making analysis quite straightforward. For samples CL11 and CL14,
the total signal to noise ratio for the large magnetic field scans
was between 5 and 10.

The weaker signal for samples CL11 and CL14 necessitated driving the
cantilevers hard so that the cut-off frequency $\beta_{\text{zero}}$
(see Eq. \ref{eq:ChData_BetaZero}), above which the cantilever frequency
shift is insensitive, was somewhat close to $\beta_{1}$, the expected
magnetic field frequency of the persistent current. In the data processing
steps of method A, it is necessary to filter out the components $\beta_{\text{zero}}$
to avoid introducing artifacts to the data. It was also necessary
to remove the smooth background which was of comparable magnitude
to the persistent current signal. Between these filtering steps and
the other data processing performed in converting the cantilever frequency
shift into a persistent current, one could be concerned that the persistent
current signals for samples CL11 and CL14 were mere artifacts of the
signal processing routine. 

\begin{figure}
\begin{centering}
\includegraphics[width=0.6\paperwidth]{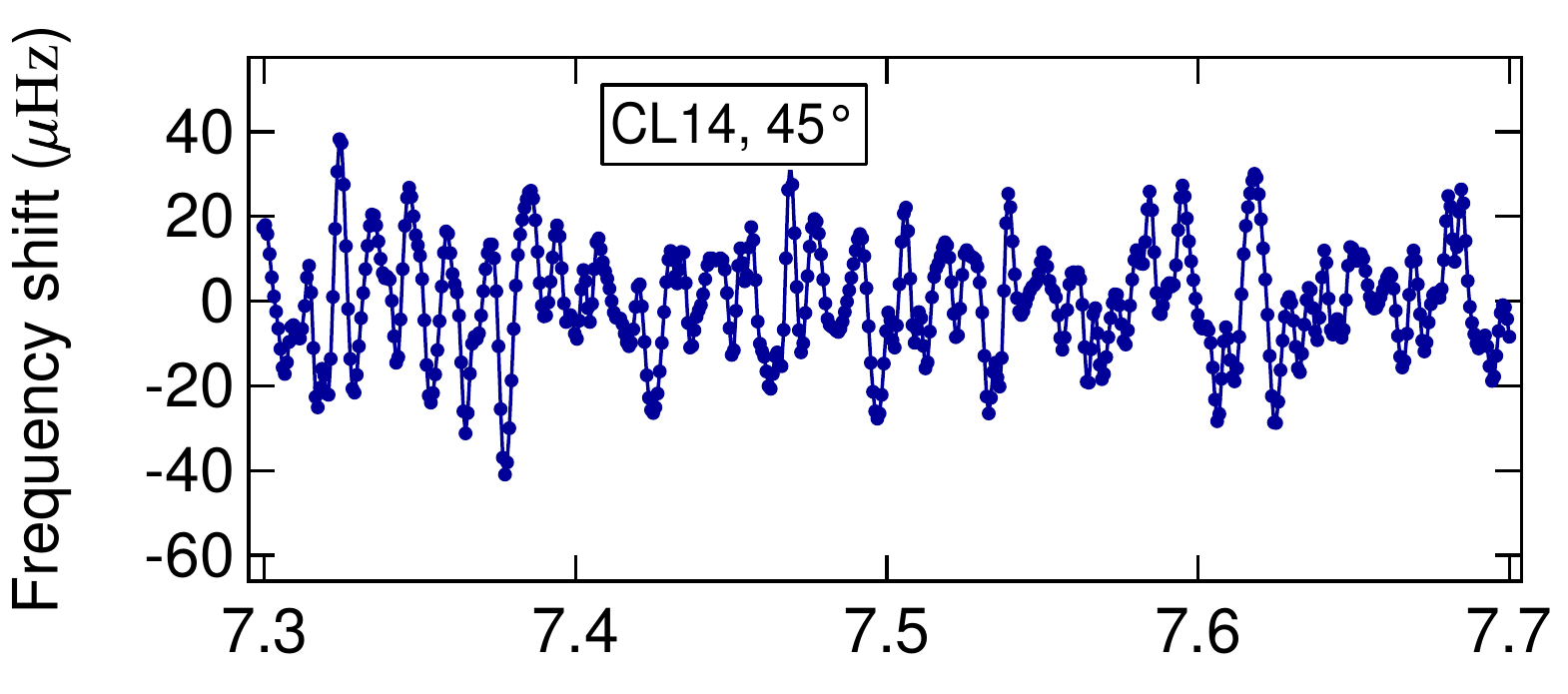}
\par\end{centering}

\begin{centering}
\includegraphics[width=0.6\paperwidth]{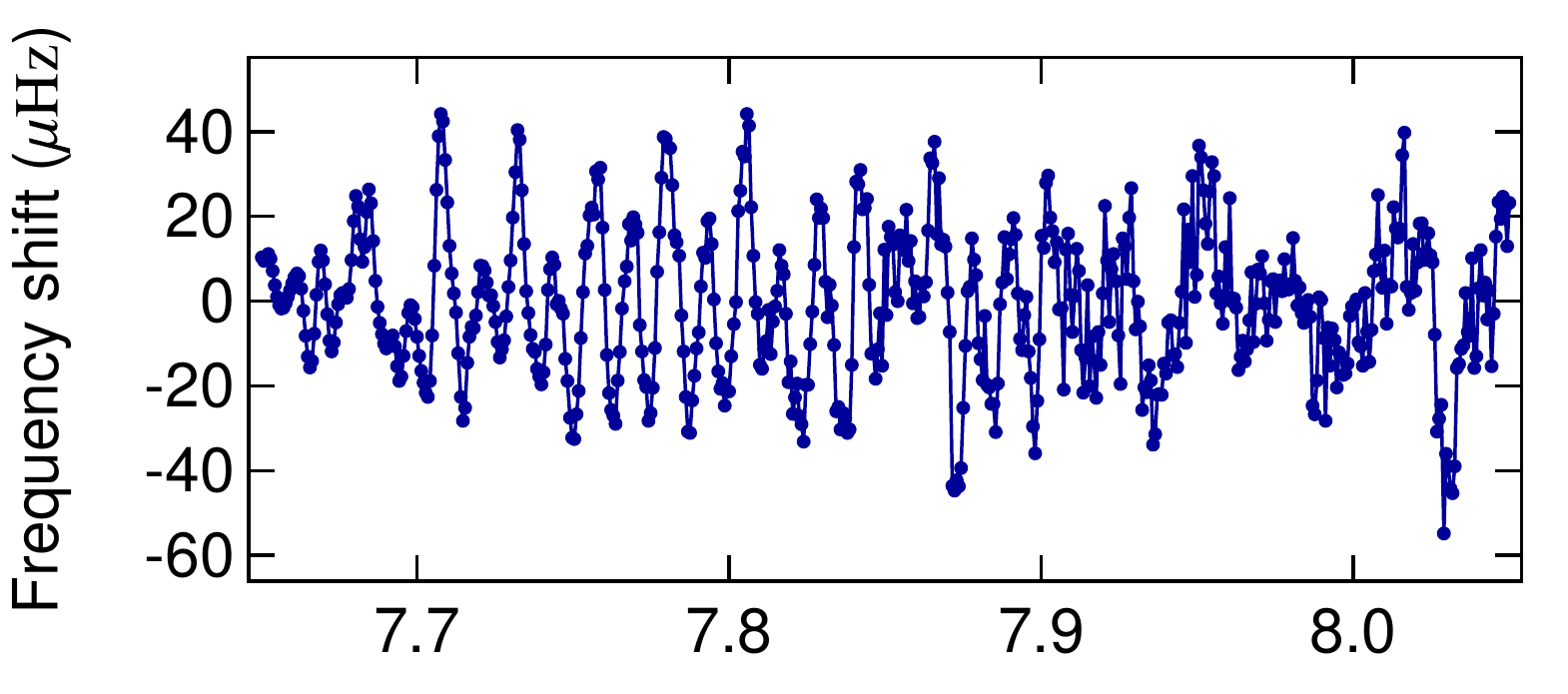}
\par\end{centering}

\begin{centering}
\includegraphics[width=0.6\paperwidth]{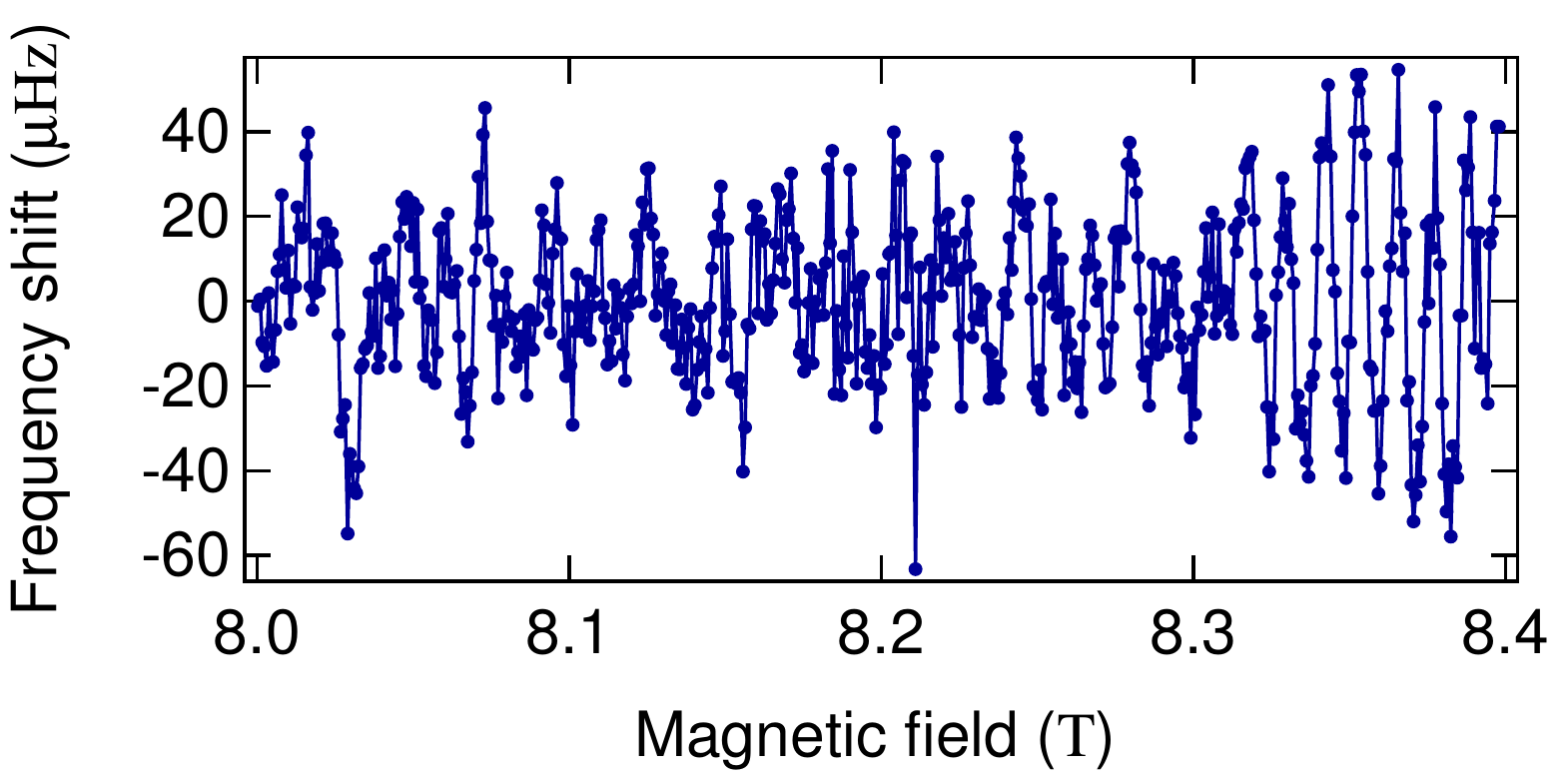}
\par\end{centering}

\caption[Full frequency shift versus magnetic field trace for CL14 at 45$^{\circ}$]{\label{fig:ChData_DAT27_FreqvsB_CL14}Full frequency shift versus
magnetic field trace for CL14 at 45$^{\circ}$. The data shown plot
the full data set of cantilever frequency shift versus magnetic field
for sample CL14 at $T_{b}=365\,\text{mK}$. A smooth background has
been subtracted from the cantilever frequency to remove its slow drift
in time. Otherwise, no manipulation of the data has been performed.
The size of the persistent current signal varies with magnetic field
with large amplitude oscillations visible near 7.35, 7.75, and $8.35\,\text{T}$.
The power spectral density of the trace is shown in Fig. \ref{fig:ChData_DAT28_FreqPSD_CL14}.}
\end{figure}

\begin{figure}
\begin{centering}
\includegraphics[width=0.7\paperwidth]{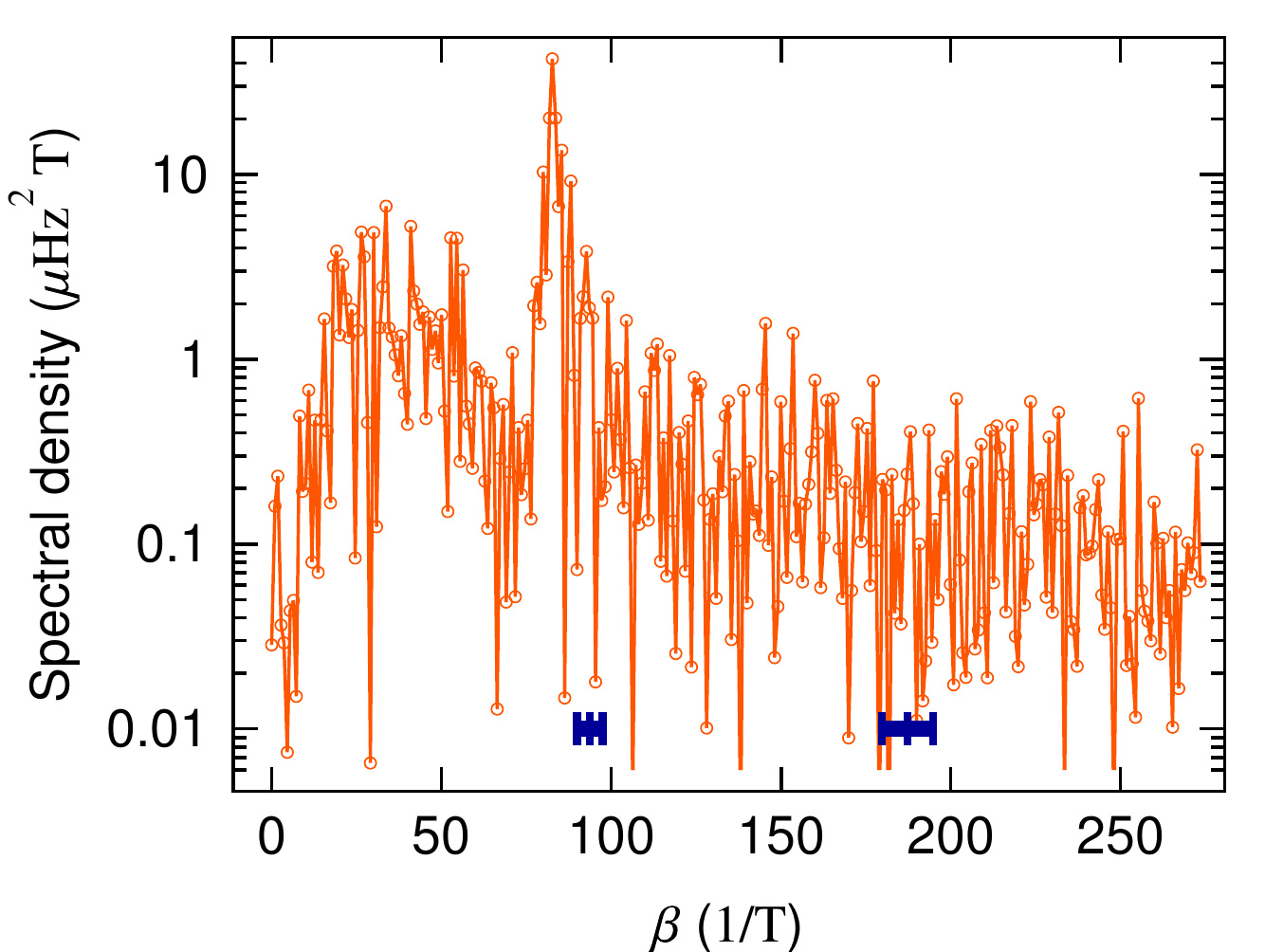}
\par\end{centering}

\caption[Spectral density of the measured frequency shift for CL14 at 45$^{\circ}$]{\label{fig:ChData_DAT28_FreqPSD_CL14}Spectral density of the measured
frequency shift for CL14 at 45$^{\circ}$. The data shown represent
the power spectral density of the frequency shift versus magnetic
field trace (Fig. \ref{fig:ChData_DAT27_FreqvsB_CL14}) for sample
CL14 taken with $\theta_{0}=45^{\circ}$ and $T=365\,\text{mK}$.
The horizontal bars represent the expected locations and widths of
the peaks in the spectrum expected for the first two harmonics of
the persistent current signal using the nominal experimental parameters.
A clear peak is visible just below the expected location of the first
harmonic.}
\end{figure}

\begin{figure}
\begin{centering}
\includegraphics[width=0.7\paperwidth]{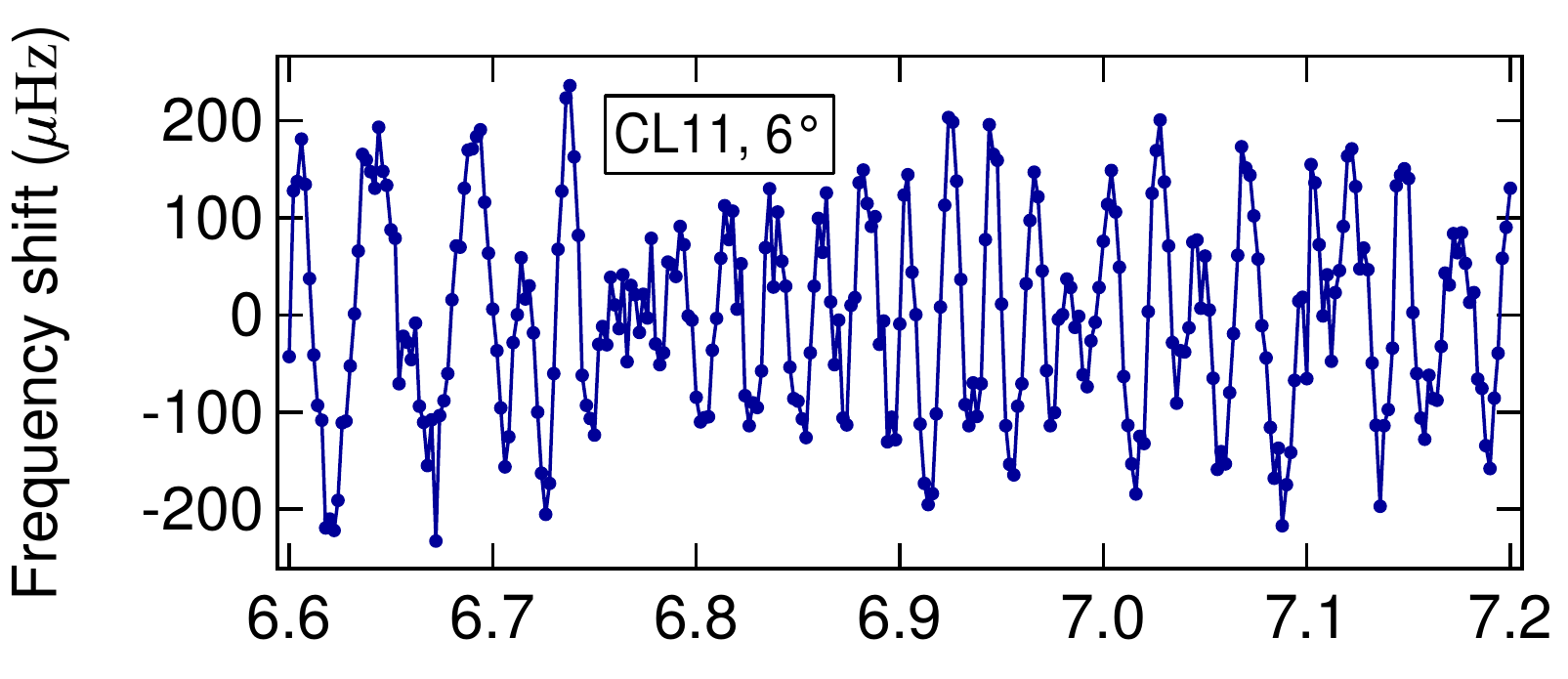}
\par\end{centering}

\begin{centering}
\includegraphics[width=0.7\paperwidth]{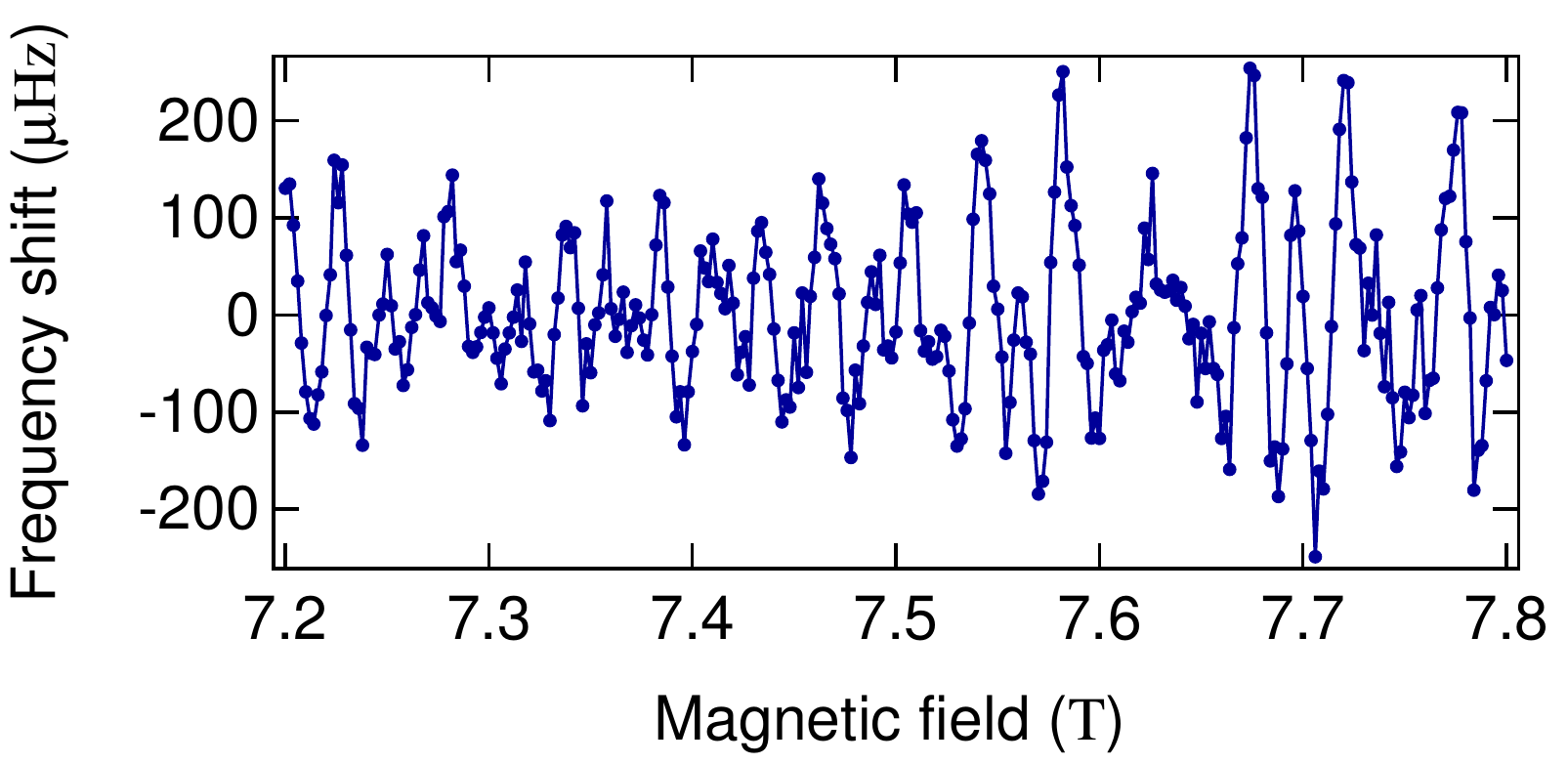}
\par\end{centering}

\caption[Full frequency shift versus magnetic field trace for CL11 at 6$^{\circ}$]{\label{fig:ChData_DAT29_FreqvsB_CL11}Full frequency shift versus
magnetic field trace for CL11 at 6$^{\circ}$. The data shown plot
the full data set of cantilever frequency shift versus magnetic field
for sample CL11 at $T_{b}=323\,\text{mK}$. A smooth background has
been subtracted from the cantilever frequency to remove its slow drift
in time. Otherwise, no manipulation of the data has been performed.
The size of the persistent current signal varies with magnetic field
with large amplitude oscillations visible near $6.9\,\text{T}$. The
power spectral density of the trace is shown in Fig. \ref{fig:ChData_DAT30_FreqPSD_CL11}.}
\end{figure}

\begin{figure}
\begin{centering}
\includegraphics[width=0.7\paperwidth]{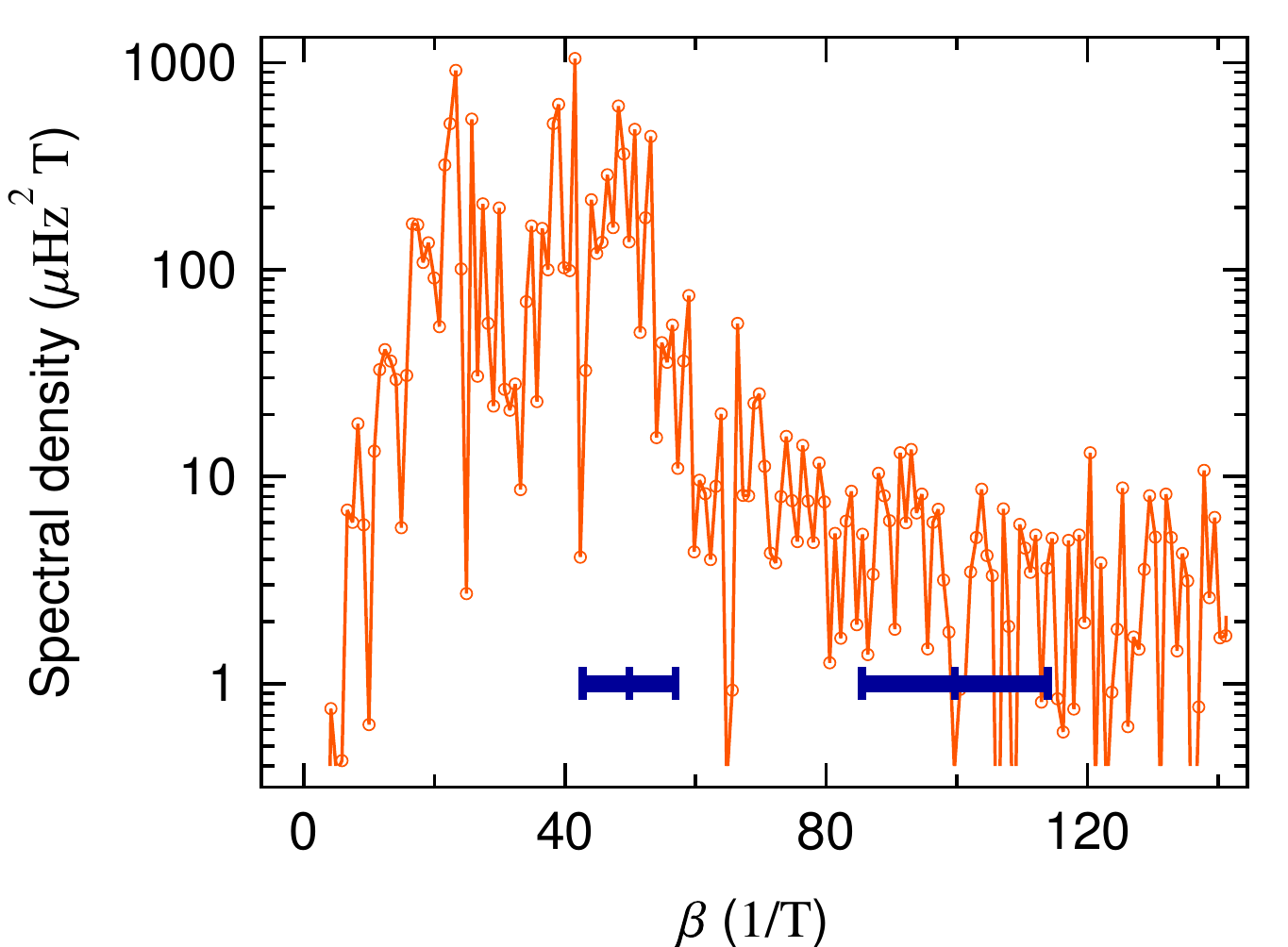}
\par\end{centering}

\caption[Spectral density of the measured frequency shift for CL11 at 6$^{\circ}$]{\label{fig:ChData_DAT30_FreqPSD_CL11}Spectral density of the measured
frequency shift for CL11 at 6$^{\circ}$. The data shown represent
the power spectral density of the frequency shift versus magnetic
field trace (Fig. \ref{fig:ChData_DAT29_FreqvsB_CL11}) for sample
CL11 taken with $\theta_{0}=6^{\circ}$ and $T=323\,\text{mK}$. The
horizontal bars represent the expected locations and widths of the
peaks in the spectrum expected for the first two harmonics of the
persistent current signal using the nominal experimental parameters.
The peak at $25\,\text{T}^{-1}$ was created by the subtraction of
the smooth background which suppresses the low $\beta$ components
of the spectrum. Above $25\,\text{T}^{-1}$, the remaining low $\beta$
noise results in a downward sloping trend in the spectrum. On top
of this trend, a peak is present between 35 and $60\,\text{T}^{-1}$,
roughly the expected location of the first harmonic of the persistent
current signal. We attribute this peak to the persistent current.}
\end{figure}

To dispel these concerns, we present in Figs. \ref{fig:ChData_DAT27_FreqvsB_CL14}
and \ref{fig:ChData_DAT29_FreqvsB_CL11} the raw frequency shift versus
magnetic field data for samples CL14 and CL11. The only processing
that has been performed on the data is the removal of a smooth, slowly
drifting background. In Figs. \ref{fig:ChData_DAT28_FreqPSD_CL14}
and \ref{fig:ChData_DAT30_FreqPSD_CL11}, we show the power spectral
densities of the two traces of frequency shift versus magnetic field.

In the spectrum of Fig. \ref{fig:ChData_DAT28_FreqPSD_CL14}, the
peak due to the persistent current in the single ring of sample CL14
can clearly be seen. The peak is located at $\sim80\,\text{T}^{-1}$.
Measurements taken at the same angle $\theta_{0}=45^{\circ}$ on another
sample (CL15) with rings of the same size found a peak at the same
location (see Fig. \ref{fig:ChData_DAT6_PSDCL15_45Deg}). The roll-off
at low $\beta$ due to the subtraction of the smooth background can
also be made out in Fig. \ref{fig:ChData_DAT6_PSDCL15_45Deg} at around
$\beta\approx25\,\text{T}^{-1}$, well below the location of the persistent
current peak.

The spectrum for sample CL11 (Fig. \ref{fig:ChData_DAT30_FreqPSD_CL11})
is similar to that of CL14 (Fig. \ref{fig:ChData_DAT28_FreqPSD_CL14}).
As with the spectrum for CL14, the low $\beta$ suppression due to
the subtraction of the smooth background occurs at $\sim25\,\text{T}^{-1}$.
Above this point, the spectrum begins to descend, following the usual
$S_{f}\propto\beta^{-1+\delta}$ behavior of low frequency noise.
This trend is visible from 25 to $35\,\text{T}^{-1}$ and above $60\,\text{T}^{-1}$.
Between $35$ and $60\,\text{T}^{-1}$, a broad peak is present on
top of this low frequency background in the spectrum. We attribute
this peak to the persistent current. As indicated by the horizontal
bar this peak is roughly in the expected location for $\theta_{0}=6^{\circ}$
and the dimensions of sample CL11.

\chapter{\label{cha:ChOutlook_}Outlook}

In the preceding chapters, I have described the magnetometer developed
in the Harris lab for the study of persistent currents in normal metal
rings and reported measurements of persistent currents in arrays of
aluminum rings. I would like to conclude by highlighting some simple
experiments that build upon the results reported here. Before doing
so, I will describe the place of those results within the context
of previous persistent current measurements.

We studied the dependence of the typical current%
\footnote{The typical current was defined in Chapter \ref{cha:Introduction_}
as the contribution to the persistent current which varies randomly
from ring to ring.%
} on several parameters that had not previously been varied and extended
the range of variation of several other parameters. We reported the
largest signal to noise ratio of all measurements of the typical current
and observed many more independent realizations $M_{\text{eff}}$
of the persistent current than all previous measurements combined.
We observed the typical current in an environment free from high frequency
electromagnetic radiation, avoiding a source of potential systematic
error present in previous measurements.%
\footnote{There is no published prediction for the effects of high frequency
radiation on the typical current. See Chapter \ref{cha:CHPrevWork}
for a discussion of related effects. In previous measurements, the
sample was exposed to high frequency radiation due to the Josephson
oscillations in the measurement SQUID.%
}

We performed the first measurements of the typical current in strong
magnetic fields. Previous measurements,%
\footnote{These experiments, Refs. \citealp{chandrasekhar1991magnetic,jariwala2001diamagnetic,bluhm2009persistent},
are reviewed in Chapter \ref{cha:CHPrevWork}. In this chapter, we
compare directly to the other measurements of the typical current.
Many of our statements about previous work apply equally well to measurements
of the average persistent current as well.%
} all of which used SQUIDs, had been restricted to fields $\apprle10\,\text{mT}$,
whereas our measurements spanned a magnetic field range 270 times
larger, achieving a maximum field of $8.4\,\text{T}$. We increased
the highest temperature at which the persistent current had been observed
by a factor of five to $2.5\,\text{K}$. We performed the first direct
study of the effect of ring size on the current's magnitude and temperature
dependence. We did this by measuring co-deposited rings of three significantly
different sizes. For each previous experiment, only one characteristic
temperature was observed. We reported the first definitive observation
of the second harmonic of the typical persistent current. Ref. \citealp{jariwala2001diamagnetic}
also reported a signal with $h/2e$ flux periodicity but could not
differentiate the contributions from the typical current and the average
current. We studied for the first time the dependence of the persistent
current on its orientation in the applied magnetic field by measuring
the same samples at two different angles with respect to magnetic
field. We measured aluminum rings (all previous measurements of the
typical persistent current studied gold rings).

We reported a greater absolute signal to noise ratio and a greater
sensitivity to persistent currents than previous measurements of the
typical current. Our largest signal to noise ratio was $\sim40$,
achieved with sample CL17 of Table \ref{tab:ChData_Rings}, while
the signal to noise ratio of previous measurements was typically $\apprle7$.
Because the persistent current signal scales differently with ring
size for the torsional magnetometry and SQUID measurements, it is
difficult to compare the persistent current sensitivity of measurements
made on rings of different sizes. Still, our measured current magnitude
of $\sim0.7\,\text{pA}$ for sample CL11 ($L=5\,\text{\ensuremath{\mu}m}$)
is much lower than the $\sim200\,\text{pA}$ minimum current magnitude
reported in other experiments on similar rings ($L=3.6\,\text{\ensuremath{\mu}m}$
for Ref. \citealp{bluhm2008magnetic}; $L=8\,\text{\ensuremath{\mu}m}$
for Ref. \citealp{jariwala2001diamagnetic}). Our signal to noise
ratio does not display as large of an improvement over previous measurements
as our persistent current sensitivity because we measured at higher
temperatures where the persistent current was suppressed. Operating
at lower temperatures would increase our signal to noise ratio, especially
for larger rings which have smaller characteristic temperatures. The
signal to noise ratio and persistent current sensitivity could also
be increased by covering a larger fraction of the cantilever with
rings and by making thermally limited measurements of the cantilever
frequency.%
\footnote{As discussed in \ref{sec:ChSensitivity_PCUncertainty}, the measurements
reported in Chapter \ref{cha:Data} did not reach the thermal limit.
Recently, thermally limited measurements have been performed in the
Harris lab with thinner ($110\,\text{nm}$) cantilevers. For these
measurements, the cantilever frequency was obtained using the Hilbert
transform technique mentioned in \ref{sub:CHExpSetup_CantileverElectronics}.%
}

The improvement in sensitivity was due in part to the significantly
larger magnetic field range accessible with torsional magnetometry.
The large number of observed persistent current oscillations aided
the task of distinguishing the persistent current signal from the
measurement's smooth background. Previous experiments were much more
sensitive to the measurement background. These measurements were performed
over a small range near zero magnetic field for which only a small
number of oscillations were observed and over which paramagnetic effects
led to a large background magnetization signal.

Beyond aiding in the analysis of the measurement, the large magnetic
field range enabled the study of new physics. We measured the persistent
current signal over many correlation fields $B_{c,p}$ and so were
able to confirm the predictions of \ref{sub:CHPCTh_FluxThroughMetal}
and Ref. \citealp{ginossar2010mesoscopic} for the effect of magnetic
flux piercing the metal on the autocorrelation of the persistent current
oscillation. In previous measurements, it was not possible to apply
a magnetic field much greater than the correlation field $B_{c,p}$.
With our measurement range $B_{0}\gg B_{c,p}$, we were able to acquire
$M_{\text{eff}}\approx210$ independent measurements of the persistent
current magnitude,%
\footnote{This number is the sum of all of the $M_{\text{eff}}$ listed in Table
\ref{tab:ChData_RingResults}.%
} a figure approximately seven times the total number of samples studied
previously. The large values of $M_{\text{eff}}$ for each sample
allowed for accurate estimates of the typical current magnitude.

There are several ways of making further use of cantilever torsional
magnetometry to study persistent currents. The simplest extension
of the work performed so far is to make additional measurements of
individual rings over a wide magnetic field range. Such measurements
would enable the study of the statistics of the persistent current
beyond its typical magnitude and test the prediction cited in Chapter
\ref{cha:CHPrevWork} that the persistent current magnitude is normally
distributed. Single ring measurements are required because, as discussed
in \ref{sec:ChData_Quantitative}, the central limit theorem requires
that the distribution of the total current in an array of rings tends
toward the normal distribution regardless of the underlying single
ring distribution. An example of the analysis that would be performed
on such single ring measurements is given in Appendix \ref{cha:AppCumul_}
for the measurements on the arrays of rings reported in Chapter \ref{cha:Data}.
The analysis in Appendix \ref{cha:AppCumul_} shows that the measured
set of values of the persistent current in the arrays is consistent
with a normal distribution. This result confirms the applicability
of the central limit theorem but does not provide much information
about the underlying single ring distribution.

The one single ring measurement discussed in Chapter \ref{cha:Data}
contained $M_{\text{eff}}\approx12$ independent samples of the persistent
current (see Table \ref{tab:ChData_RingResults} and accompanying
discussion). While the size of this data set was sufficient to infer
the magnitude of the typical current with reasonable statistical uncertainty,
a larger data set is necessary to study the properties of the current's
statistical distribution beyond its variance. By redesigning the sample
chip to contain many cantilevers with single rings, a much larger
number $M_{\text{eff}}$ of independent realizations of the persistent
current could be observed. The signal to noise ratio could also potentially
be improved by using cantilevers with dimensions chosen by following
the considerations reviewed in \ref{sec:CHSensitivity_OptimalCantDimensions}.
Together with Manuel Castellanos Beltran, I have fabricated such samples.
At the time of writing, measurements of these samples are underway
in the Harris lab.%
\footnote{To add to the credibility of the measurements discussed in Chapter
\ref{cha:Data}, I note that a persistent current signal has already
been measured in many additional single ring samples. These new samples
were part of a round of fabrication distinct from the one in which
the samples discussed in Chapter \ref{cha:Data} were created, indicating
good repeatability for the entire process of measuring persistent
currents with cantilever torsional magnetometry.%
}

The effects of high frequency electromagnetic radiation on the persistent
current could be studied by cantilever torsional magnetometry. To
study these effects on the typical current in a strong magnetic field,
the only necessary modification to the current experimental apparatus
is the addition of a radiation source coupled to the persistent current
sample. Many works investigating the non-equilibrium currents induced
by external radiation were mentioned in Chapter \ref{cha:CHPrevWork}.
However, none of these works discusses the typical current in a metal
ring. I am not aware of any published work studying the effect of
electromagnetic radiation on the typical persistent current in a normal
metal ring.

Cantilever torsional magnetometry could also be applied to the study
of the average persistent current at low magnetic field. The fact
that the frequency shift signal scales quadratically with magnetic
field was a great advantage in the study of the typical persistent
current because it meant that the persistent current was observable
over a large magnetic field range. However, one of the most intriguing
puzzles in the persistent current literature is the unexpected magnitude
and sign of the average current. This contribution to the persistent
current is only present at low magnetic field. 

With a modified cantilever design, it is possible to study this average
contribution to the persistent current as well. Because the total
current in the array scales as the ring number $N$, the signal to
noise ratio increases with cantilever size.%
\footnote{For the typical current, the signal to noise ratio decreases with
cantilever length and is independent of cantilever width when the
measurement is limited by the thermal motion of the cantilever. See
\ref{sec:CHSensitivity_OptimalCantDimensions}.%
} Using Eq. \ref{eq:CHPCTh_Iee} for the average current $I^{ee}$
due to electron-electron interactions and assuming that the thermal
motion of the cantilever produces the leading contribution to the
noise in the frequency measurement (see Eq. \ref{eq:CHSensitivity_FreqErrorFullExpression}),
one finds a sensitivity $\mathcal{S}_{pc}$ (defined in Eq. \ref{eq:ChSensitivity_Sensitivity})
of $\sim8.2\,(1/\sqrt{\text{\text{Hz}}})$ for a cantilever $400\,\text{\ensuremath{\mu}m}$
long, $2\,\text{mm}$ wide, and $110\,\text{nm}$ thick covered with
an array of $1.3\times10^{6}$ rings with radii of $250\,\text{nm}$
and cross-sections of $30\,\text{nm}$ by $30\,\text{nm}$. This figure
for the sensitivity was calculated for a temperature of $323\,\text{mK}$
and for gold rings each with a diffusion constant $D=0.014\,\text{m}^{2}/\text{s}$
(corresponding to an elastic mean free path of $l_{e}=30\,\text{nm}$)
and an effective electron-electron interaction coefficient $\lambda_{\text{eff}}=0.333$.
For these figures, the persistent current should be observable at
least between $\sim50\,\text{mT}$ and $\sim250\,\text{mT}$, a span
covering $\sim15$ oscillations.

Other modifications to the cantilever design and experimental apparatus
could potentially improve sensitivity to persistent currents. In \ref{sec:CHTorsMagn_deltaFZeroDrive},
it was noted that a measurement using a strong magnetic field gradient
could achieve a sensitivity comparable to that of the uniform field
measurements discussed in most of this text. Such a magnetic field
gradient measurement could be performed at low magnetic field and,
unlike the low magnetic field measurement described above, would require
a small array of rings so that each ring would experience a similar
magnetic field gradient strength. Also, as mentioned in \ref{sec:CHSensitivity_OptimalCantDimensions},
geometries other than the cantilevered beam could be more sensitive
to persistent currents \citep{chabot1999microfabrication,lobontiu2006modeling,haiberger2007highlysensitive}.

Finally, cantilever torsional magnetometry could used to study persistent
currents in other materials. Semiconductor rings in the ballistic
regime are of particular interest. Many predictions regarding the
persistent current in the ballistic regime have been made and remain
untested (see Chapter \ref{cha:CHPrevWork}). Another potential system
is a superconductor such as niobium with a large critical magnetic
field. If the critical field is large enough ($\sim1\,\text{T}$),
the sensitivity would be sufficient to observe persistent current
in the normal state across the transition into the superconducting
state.

\appendix

\chapter{Mathematical relations}

\section{Poisson summation formula}

\subsection{General formula}

The Poisson summation formula relates the infinite sum of a function
with evenly spaced arguments 
\begin{equation}
\sum_{n}f\left(\phi+n\phi_{0}\right)=F\left(\phi\right)\label{eq:AppMath_PoissonSumf}
\end{equation}
to its Fourier transform 
\[
\tilde{f}\left(p/\phi_{0}\right)=\int_{-\infty}^{\infty}d\phi\, f\left(\phi\right)e^{-2\pi ip\phi/\phi_{0}}.
\]
The formula is useful in the calculation of persistent currents where
sums over all the states in the ring often take the form of Eq. \ref{eq:AppMath_PoissonSumf}.
The identity is shown by noting that $F(\phi)$ is periodic in $\phi$
with period $\phi_{0}$ and can be expanded in a Fourier series:
\begin{align*}
F\left(\phi\right) & =\sum_{p}\left(\frac{1}{\phi_{0}}\int_{0}^{\phi_{0}}d\phi'\, F\left(\phi'\right)e^{-2\pi ip\phi'/\phi_{0}}\right)e^{2\pi ip\phi/\phi_{0}}\\
 & =\sum_{n}\sum_{p}\left(\frac{1}{\phi_{0}}\int_{0}^{\phi_{0}}d\phi'\, f\left(\phi'+n\phi_{0}\right)e^{-2\pi ip\phi'/\phi_{0}}\right)e^{2\pi ip\phi/\phi_{0}}\\
 & =\sum_{n}\sum_{p}\left(\frac{1}{\phi_{0}}\int_{-n\phi_{0}}^{-\left(n-1\right)\phi_{0}}d\phi'\, f\left(\phi'\right)e^{2\pi ipn}e^{-2\pi ip\phi'/\phi_{0}}\right)e^{2\pi ip\phi/\phi_{0}}\\
 & =\sum_{p}\sum_{n}\left(\frac{1}{\phi_{0}}\int_{-n\phi_{0}}^{-\left(n-1\right)\phi_{0}}d\phi'\, f\left(\phi'\right)e^{-2\pi ip\phi'/\phi_{0}}\right)e^{2\pi ip\phi/\phi_{0}}
\end{align*}
\begin{align}
F\left(\phi\right) & =\sum_{p}\left(\frac{1}{\phi_{0}}\int_{-\infty}^{\infty}d\phi'\, f\left(\phi'\right)e^{-2\pi ip\phi'/\phi_{0}}\right)e^{2\pi ip\phi/\phi_{0}}\nonumber \\
\sum_{n}f\left(\phi+n\phi_{0}\right) & =\frac{1}{\phi_{0}}\sum_{p}\tilde{f}\left(\frac{p}{\phi_{0}}\right)e^{2\pi ip\phi/\phi_{0}}.\label{eq:AppMath_PoissonSummationFormula}
\end{align}
The last line is the Poisson summation formula.

\subsection{Application to $\nu(\varepsilon,\phi)$ for the one-dimensional ring}

The density of states for a system with discrete energy levels such
as the one-dimensional ring considered in \ref{sec:CHPCTh_1DRing}
can be written in the form 
\[
\nu\left(\varepsilon\right)=\sum_{n}\delta\left(\varepsilon-\varepsilon_{n}\right)
\]
where the sum is over the discrete energy levels $\varepsilon_{n}$.

For the one-dimensional ring, the energy levels, given in Eq. \ref{eq:CHPCTh_EnergyLevel1DPerfectRing},
take the form
\[
\varepsilon_{n}\left(\phi\right)=\varepsilon_{0}\left(\phi+n\phi_{0}\right)
\]
with 
\[
\varepsilon_{0}\left(\phi\right)=\frac{h^{2}}{2mL^{2}}\left(\frac{\phi}{\phi_{0}}\right)^{2}.
\]
Thus, the density of states for the one-dimensional ring matches the
form of the Poisson summation formula in Eq. \ref{eq:AppMath_PoissonSummationFormula}
with $f(\phi)=\delta(\varepsilon-\varepsilon_{0}(\phi))$. We can
evaluate the Fourier transform as 
\begin{align*}
\tilde{f}\left(\frac{p}{\phi_{0}}\right) & =\int_{-\infty}^{\infty}d\phi\,\delta\left(\varepsilon-\varepsilon_{0}\left(\phi\right)\right)e^{-2\pi ip\phi/\phi_{0}}\\
 & =\int_{-\infty}^{\infty}d\phi\,\frac{mL^{2}\phi_{0}^{2}}{h^{2}\left|\phi\right|}\left(\delta\left(\phi-\sqrt{\frac{2mL^{2}\phi_{0}^{2}\varepsilon}{h^{2}}}\right)+\delta\left(\phi+\sqrt{\frac{2mL^{2}\phi_{0}^{2}\varepsilon}{h^{2}}}\right)\right)e^{-2\pi ip\phi/\phi_{0}}\\
 & =\frac{2mL^{2}\phi_{0}^{2}}{h^{2}}\sqrt{\frac{h^{2}}{2mL^{2}\phi_{0}^{2}\varepsilon}}\cos\left(\frac{2\pi p}{\phi_{0}}\sqrt{\frac{2mL^{2}\phi_{0}^{2}\varepsilon}{h^{2}}}\right)\\
 & =\sqrt{\frac{2mL^{2}\phi_{0}^{2}}{h^{2}\varepsilon}}\cos\left(2\pi p\sqrt{\frac{2mL^{2}\varepsilon}{h^{2}}}\right).
\end{align*}
In the second line we use the identity $\delta(g(x))=\sum_{i}\delta(x-x_{i})/|g'(x_{i})|$
where the $x_{i}$ are the values of $x$ for which $g(x)=0$. The
density of states is then
\begin{align}
\nu\left(\varepsilon,\phi\right) & =\frac{1}{\phi_{0}}\sum_{p}\sqrt{\frac{2mL^{2}\phi_{0}^{2}}{h^{2}\varepsilon}}\cos\left(2\pi p\sqrt{\frac{2mL^{2}\varepsilon}{h^{2}}}\right)e^{2\pi ip\phi/\phi_{0}}\label{eq:AppMath_PCDOS}\\
 & =\sum_{p>0}\sqrt{\frac{2mL^{2}}{h^{2}\varepsilon}}\cos\left(2\pi p\sqrt{\frac{2mL^{2}\varepsilon}{h^{2}}}\right)\cos\left(2\pi p\frac{\phi}{\phi_{0}}\right)\nonumber \\
 & \phantom{=}+\sqrt{\frac{2mL^{2}}{h^{2}\varepsilon}}.\nonumber 
\end{align}

\section{Fourier transform of sech$^{2}(t)$}

Here we derive the Fourier transform of $\text{sech}^{2}(t)$ 
\[
\mathcal{F}\left[\text{sech}^{2}\left(t\right),\omega\right]=\int_{-\infty}^{\infty}dt\,\text{sech}^{2}\left(t\right)e^{-i\omega t}
\]
by contour integration. We begin by noting 
\begin{align*}
\text{sech}^{2}\left(t+i\pi\right) & =\frac{4}{\left(e^{t}e^{i\pi}+e^{-t}e^{-i\pi}\right)^{2}}\\
 & =\frac{4}{\left(e^{t}+e^{-t}\right)^{2}}\\
 & =\text{sech}^{2}\left(t\right).
\end{align*}
We thus choose the contour $C$ bounded by $(-\infty,\infty)$, $(\infty,\infty+i\pi)$,
$(\infty+i\pi,-\infty+i\pi)$, and $(-\infty+i\pi,-\infty)$. The
two segments at $\infty$ give negligible contributions to the total
integral. Thus
\begin{align*}
\oint_{C}dz\,\text{sech}^{2}\left(z\right)e^{-i\omega z} & =\int_{-\infty}^{\infty}dt\,\text{sech}^{2}\left(t\right)e^{-i\omega t}+\int_{\infty}^{-\infty}dt\,\text{sech}^{2}\left(t+i\pi\right)e^{-i\omega t}e^{\pi\omega}\\
 & =\left(1-e^{\pi\omega}\right)\int_{-\infty}^{\infty}dt\,\text{sech}^{2}\left(t\right)e^{-i\omega t}.
\end{align*}
The function $\text{sech}^{2}(z)$ has poles at $z=i\pi/2+i\pi n$
for integer $n$. The only pole enclosed by $C$ is $z=i\pi/2$. To
find the residue, we find the Taylor expansion of the numerator $e^{-i\omega z}$
and the denominator $\cosh^{2}(z)$ about $z=i\pi/2$:
\[
e^{-i\omega z}\approx e^{\pi\omega/2}-i\omega e^{\pi\omega/2}\left(z-i\frac{\pi}{2}\right)+\mathcal{O}\left(\left(z-i\frac{\pi}{2}\right)^{2}\right)
\]
and
\begin{align*}
\cosh^{2}\left(z\right) & \approx\cosh^{2}\left(i\frac{\pi}{2}\right)+2\sinh\left(i\frac{\pi}{2}\right)\cosh\left(i\frac{\pi}{2}\right)\left(z-i\frac{\pi}{2}\right)\\
 & \phantom{\approx}+2\left(\sinh^{2}\left(i\frac{\pi}{2}\right)+\cosh^{2}\left(i\frac{\pi}{2}\right)\right)\frac{\left(z-i\frac{\pi}{2}\right)^{2}}{2}\\
 & \phantom{\approx}+8\sinh\left(i\frac{\pi}{2}\right)\cosh\left(i\frac{\pi}{2}\right)\frac{\left(z-i\frac{\pi}{2}\right)^{3}}{3!}+\mathcal{O}\left(\left(z-i\frac{\pi}{2}\right)^{4}\right)\\
 & \approx0+0-\left(z-i\frac{\pi}{2}\right)^{2}+0+\mathcal{O}\left(\left(z-i\frac{\pi}{2}\right)^{4}\right).
\end{align*}
These expansions allow us to find the Laurent series
\begin{align*}
\text{sech}^{2}\left(z\right)e^{-i\omega z} & =\frac{e^{\pi\omega/2}-i\omega e^{\pi\omega/2}\left(z-i\frac{\pi}{2}\right)+\mathcal{O}\left(\left(z-i\frac{\pi}{2}\right)^{2}\right)}{-\left(z-i\frac{\pi}{2}\right)^{2}+\mathcal{O}\left(\left(z-i\frac{\pi}{2}\right)^{4}\right)}\\
 & =-\frac{e^{\pi\omega/2}-i\omega e^{\pi\omega/2}\left(z-i\frac{\pi}{2}\right)+\mathcal{O}\left(\left(z-i\frac{\pi}{2}\right)^{2}\right)}{\left(z-i\frac{\pi}{2}\right)^{2}}\left(1+\mathcal{O}\left(\left(z-i\frac{\pi}{2}\right)^{2}\right)\right)\\
 & =-\frac{e^{\pi\omega/2}}{\left(z-i\frac{\pi}{2}\right)^{2}}+\frac{i\omega e^{\pi\omega/2}}{\left(z-i\frac{\pi}{2}\right)}+\mathcal{O}\left(\left(z-i\frac{\pi}{2}\right)^{0}\right).
\end{align*}
Thus we find 
\begin{align}
\int_{-\infty}^{\infty}dt\,\text{sech}^{2}\left(t\right)e^{-i\omega t} & =\frac{1}{1-e^{\pi\omega}}\oint_{C}dz\,\text{sech}^{2}\left(z\right)e^{-i\omega z}\nonumber \\
 & =\frac{2\pi i}{1-e^{\pi\omega}}\text{Res}\left[\text{sech}^{2}\left(z\right)e^{-i\omega z},i\frac{\pi}{2}\right]\nonumber \\
 & =-\frac{2\pi\omega e^{\pi\omega/2}}{1-e^{\pi\omega}}\nonumber \\
 & =\frac{\pi\omega}{\sinh\left(\frac{\pi\omega}{2}\right)}.\label{eq:AppMath_sechSquaredFourierTransform}
\end{align}
We note in particular that 
\begin{align*}
\lim_{\omega\rightarrow0}\int_{-\infty}^{\infty}dt\,\text{sech}^{2}\left(t\right)e^{-i\omega t} & =\lim_{\omega\rightarrow0}\frac{\pi\omega}{\sinh\left(\frac{\pi\omega}{2}\right)}\\
 & =2,
\end{align*}
so that, as noted in Section \ref{sub:CHPCTh_IdealRingFiniteTemperature},
\[
\int_{-\infty}^{\infty}d\varepsilon\,\frac{\text{sech}^{2}\left(\left(\varepsilon-\varepsilon_{F}\right)/2k_{B}T\right)}{4k_{B}T}=1
\]
while the integrand itself peaks at a value of $1/4k_{B}T$ at $\varepsilon=\varepsilon_{F}$
and thus becomes more and more sharply peaked as $T\rightarrow0$.

\section{Integral of $\text{sech}^{2}(\sigma+\frac{\varepsilon}{2})\text{sech}^{2}(\sigma-\frac{\varepsilon}{2})$
in current-current correlation function calculation$\;$}

In the calculation of the current-current function at finite temperature
in Section \ref{sub:CHPCTh_TypicalIFiniteT}, the following integral
is encountered:
\[
\int_{-\infty}^{\infty}d\sigma\,\text{sech}^{2}\left(\sigma+\frac{\varepsilon}{2}\right)\text{sech}^{2}\left(\sigma-\frac{\varepsilon}{2}\right).
\]
 Using
\begin{align*}
\cosh\left(\sigma+\frac{\varepsilon}{2}\right)\cosh\left(\sigma-\frac{\varepsilon}{2}\right) & =\frac{1}{4}\left(e^{\sigma+\varepsilon/2}+e^{-\sigma-\varepsilon/2}\right)\left(e^{\sigma-\varepsilon/2}+e^{-\sigma+\varepsilon/2}\right)\\
 & =\frac{1}{4}\left(e^{2\sigma}+e^{\varepsilon}+e^{-\varepsilon}+e^{-2\sigma}\right)
\end{align*}
and defining $x=e^{2\sigma}$ and $x_{0}=e^{\varepsilon}$, we can
rewrite the $\sigma$ integration as
\begin{align*}
\int_{-\infty}^{\infty}d\sigma\,\text{sech}^{2}\left(\sigma+\frac{\varepsilon}{2}\right)\text{sech}^{2}\left(\sigma-\frac{\varepsilon}{2}\right) & =\int_{-\infty}^{\infty}d\sigma\,\left(\frac{4}{e^{2\sigma}+e^{\varepsilon}+e^{-\varepsilon}+e^{-2\sigma}}\right)^{2}\\
 & =\int_{0}^{\infty}dx\,\frac{1}{2x}\left(\frac{4}{x+x_{0}+x_{0}^{-1}+x^{-1}}\right)^{2}\\
 & =8\int_{0}^{\infty}dx\,\frac{x}{\left(x^{2}+\left(x_{0}+x_{0}^{-1}\right)x+1\right)^{2}}.
\end{align*}
The quadratic formula gives the roots of the denominator as
\begin{align*}
x_{\pm} & =\frac{-\left(x_{0}+x_{0}^{-1}\right)\pm\sqrt{\left(x_{0}+x_{0}^{-1}\right)^{2}-4}}{2}\\
 & =\frac{-\left(x_{0}+x_{0}^{-1}\right)\pm\sqrt{\left(x_{0}^{-1}-x_{0}\right)^{2}}}{2}\\
 & =-x_{0}^{\pm1}.
\end{align*}
The integrand can then be rewritten as
\begin{align*}
\int_{-\infty}^{\infty}d\sigma\,\text{sech}^{2}\left(\sigma+\frac{\varepsilon}{2}\right)\text{sech}^{2}\left(\sigma-\frac{\varepsilon}{2}\right) & =8\int_{0}^{\infty}dx\,\frac{x}{\left(x^{2}+\left(x_{0}+x_{0}^{-1}\right)x+1\right)^{2}}\\
 & =8\int_{0}^{\infty}dx\,\frac{x}{\left(x-x_{+}\right)^{2}\left(x-x_{-}\right)^{2}}\\
 & =8\frac{\partial^{2}}{\partial x_{+}\partial x_{-}}\int_{0}^{\infty}dx\,\frac{x}{\left(x-x_{+}\right)\left(x-x_{-}\right)}.
\end{align*}
We can factor the denominator by solving for $A$ and $B$ in
\begin{align*}
\frac{x}{\left(x-x_{+}\right)\left(x-x_{-}\right)} & =\frac{A}{x-x_{+}}+\frac{B}{x-x_{-}}\\
 & =\frac{A\left(x-x_{-}\right)+B\left(x-x_{+}\right)}{\left(x-x_{+}\right)\left(x-x_{-}\right)}
\end{align*}
which requires
\begin{align*}
A+B & =1\\
Ax_{-}+Bx_{+} & =0
\end{align*}
and gives
\begin{align*}
A & =\frac{x_{+}}{x_{+}-x_{-}}\\
B & =\frac{-x_{-}}{x_{+}-x_{-}}.
\end{align*}
With this factorization, the integral can now be evaluated explicitly
to find
\begin{align*}
 & \int_{-\infty}^{\infty}d\sigma\,\text{sech}^{2}\left(\sigma+\frac{\varepsilon}{2}\right)\text{sech}^{2}\left(\sigma-\frac{\varepsilon}{2}\right)=\\
 & \phantom{\int_{-\infty}^{\infty}d\sigma\,\text{sech}^{2}}=8\frac{\partial^{2}}{\partial x_{+}\partial x_{-}}\int_{0}^{\infty}dx\,\left(\frac{A}{x-x_{+}}+\frac{B}{x-x_{-}}\right)\\
 & \phantom{\int_{-\infty}^{\infty}d\sigma\,\text{sech}^{2}}=8\frac{\partial^{2}}{\partial x_{+}\partial x_{-}}\left(\frac{x_{+}}{x_{+}-x_{-}}\ln\left(x-x_{+}\right)-\frac{x_{-}}{x_{+}-x_{-}}\ln\left(x-x_{-}\right)\right)\Bigg|_{0}^{\infty}\\
 & \phantom{\int_{-\infty}^{\infty}d\sigma\,\text{sech}^{2}}=8\frac{\partial}{\partial x_{-}}\Bigg(\frac{\ln\left(x-x_{+}\right)}{x_{+}-x_{-}}-\frac{x_{+}}{\left(x_{+}-x_{-}\right)^{2}}\ln\left(x-x_{+}\right)+\ldots\\
 & \phantom{\int_{-\infty}^{\infty}d\sigma\,\text{sech}^{2}}\phantom{=8\frac{\partial}{\partial x_{-}}\Bigg(}-\frac{x_{+}}{x_{+}-x_{-}}\frac{1}{x-x_{+}}+\frac{x_{-}}{\left(x_{+}-x_{-}\right)^{2}}\ln\left(x-x_{-}\right)\Bigg)\Bigg|_{0}^{\infty}
\end{align*}
\begin{align*}
 & \int_{-\infty}^{\infty}d\sigma\,\text{sech}^{2}\left(\sigma+\frac{\varepsilon}{2}\right)\text{sech}^{2}\left(\sigma-\frac{\varepsilon}{2}\right)=\\
 & \phantom{\int_{-\infty}^{\infty}d\sigma\,\text{sech}^{2}}=8\Bigg(\frac{\ln\left(x-x_{+}\right)}{\left(x_{+}-x_{-}\right)^{2}}-\frac{2x_{+}}{\left(x_{+}-x_{-}\right)^{3}}\ln\left(x-x_{+}\right)+\ldots\\
 & \phantom{\int_{-\infty}^{\infty}d\sigma\,\text{sech}^{2}}\phantom{=8\Bigg(}-\frac{x_{+}}{\left(x_{+}-x_{-}\right)^{2}}\frac{1}{x-x_{+}}+\frac{1}{\left(x_{+}-x_{-}\right)^{2}}\ln\left(x-x_{-}\right)+\ldots\\
 & \phantom{\int_{-\infty}^{\infty}d\sigma\,\text{sech}^{2}}\phantom{=8\Bigg(}-+\frac{2x_{-}}{\left(x_{+}-x_{-}\right)^{3}}\ln\left(x-x_{-}\right)-\frac{x_{-}}{\left(x_{+}-x_{-}\right)^{2}}\frac{1}{\left(x-x_{-}\right)}\Bigg)\Bigg|_{0}^{\infty}\\
 & \phantom{\int_{-\infty}^{\infty}d\sigma\,\text{sech}^{2}}=8\Bigg(-\frac{x_{+}+x_{-}}{\left(x_{+}-x_{-}\right)^{3}}\ln\left(x-x_{+}\right)+\frac{x_{+}+x_{-}}{\left(x_{+}-x_{-}\right)^{3}}\ln\left(x-x_{-}\right)+\ldots\\
 & \phantom{\int_{-\infty}^{\infty}d\sigma\,\text{sech}^{2}}\phantom{=8\Bigg(}-\frac{1}{\left(x_{+}-x_{-}\right)^{2}}\left(\frac{x_{+}}{x-x_{+}}+\frac{x_{-}}{x-x_{-}}\right)\Bigg)\Bigg|_{0}^{\infty}\\
 & \phantom{\int_{-\infty}^{\infty}d\sigma\,\text{sech}^{2}}=\frac{-8}{\left(x_{+}-x_{-}\right)^{2}}\left(\frac{x_{+}+x_{-}}{x_{+}-x_{-}}\ln\left(\frac{x-x_{+}}{x-x_{-}}\right)+\frac{x_{+}}{x-x_{+}}+\frac{x_{-}}{x-x_{-}}\right)\Bigg|_{0}^{\infty}\\
 & \phantom{\int_{-\infty}^{\infty}d\sigma\,\text{sech}^{2}}=\frac{8}{\left(x_{+}-x_{-}\right)^{2}}\left(\frac{x_{+}+x_{-}}{x_{+}-x_{-}}\ln\left(\frac{x_{+}}{x_{-}}\right)-2\right).
\end{align*}
Restoring $x_{\pm}=-e^{\pm\varepsilon}$, we have
\begin{align}
\int_{-\infty}^{\infty}d\sigma\,\text{sech}^{2}\left(\sigma+\frac{\varepsilon}{2}\right)\text{sech}^{2}\left(\sigma-\frac{\varepsilon}{2}\right) & =\frac{8}{\left(e^{\varepsilon}-e^{-\varepsilon}\right)^{2}}\left(\frac{e^{\varepsilon}+e^{-\varepsilon}}{e^{\varepsilon}-e^{-\varepsilon}}\left(2\varepsilon\right)-2\right)\nonumber \\
 & =4\frac{\varepsilon\cosh\varepsilon-\sinh\varepsilon}{\sinh^{3}\varepsilon}.\label{eq:AppMath_SechSechSigmaIntegral}
\end{align}
We note that the expression
\[
f_{2}\left(\varepsilon\right)=\frac{\varepsilon\cosh\varepsilon-\sinh\varepsilon}{\sinh^{3}\varepsilon}
\]
 can be written in the more compact form
\[
f_{2}\left(\varepsilon\right)=4\frac{\partial^{2}}{\partial\varepsilon^{2}}\left(\frac{\varepsilon}{1+e^{-2\varepsilon}}\right).
\]

\section{\label{sec:AppMath_gDsumExp}Summation form of the normalized temperature
dependence of the current-current correlation function}

In Eq. \ref{eq:CHPCTh_CurrCurrCorTempDependence} we introduce the
function
\begin{equation}
g_{D}\left(x,y\right)=\text{Re}\left(\int_{-\infty}^{\infty}d\varepsilon\,\left(\frac{\varepsilon\cosh\varepsilon-\sinh\varepsilon}{\sinh^{3}\varepsilon}\right)\left(1+\sqrt{x+iy\varepsilon}+\frac{\left(x+iy\varepsilon\right)}{3}\right)\exp\left(-\sqrt{x+iy\varepsilon}\right)\right)\label{eq:AppMath_gDIntegral}
\end{equation}
where $x$ and $y$ are real, positive numbers set by parameters of
the persistent current ring system. Note that despite the complex
components of the integrand the function $g_{D}(x,y)=g_{D}^{*}(x,y)$
is always real. The first factor of the integral 
\[
\lim_{\varepsilon\rightarrow0}\left(\frac{\varepsilon\cosh\varepsilon-\sinh\varepsilon}{\sinh^{3}\varepsilon}\right)=\frac{1}{3}
\]
is finite as $\varepsilon\rightarrow0$. However, for $\varepsilon\rightarrow2\pi iN$
with $N$ a non-zero integer, this factor has a third order pole.
Away from these poles, the entire integrand decays to zero as $|\varepsilon|\rightarrow\infty$
in the lower half plane due to the $\sinh^{3}\varepsilon$ term in
the denominator and the $\exp(-\sqrt{x+iy\varepsilon})$ in the numerator
because $\text{Re}(-\sqrt{x+iy\varepsilon})<0$ . The function $g_{D}(x,y)$
can thus be evaluated by contour integration with the contour following
the real axis and enclosing the lower half of the complex plane. Because
the integrand is a complicated function with third order poles, calculating
the residues by hand is quite tedious. Using the Series command of
Mathematica to evaluate the Laurent series at $i\pi N$, we find
\begin{align}
 & g_{D}\left(x,y\right)=\nonumber \\
 & \phantom{g_{D}}=-2\pi i\sum_{N=-1}^{-\infty}\text{Res}\left[\left(\frac{\varepsilon\cosh\varepsilon-\sinh\varepsilon}{\sinh^{3}\varepsilon}\right)\left(1+\sqrt{x+iy\varepsilon}+\frac{\left(x+iy\varepsilon\right)}{3}\right)\exp\left(-\sqrt{x+iy\varepsilon}\right),2\pi iN\right]\\
 & \phantom{g_{D}}=-2\pi i\sum_{N=-1}^{-\infty}\left(-i\frac{N\pi y^{2}}{24}\exp\left(-\sqrt{x-\pi Ny}\right)\right)\nonumber \\
 & \phantom{g_{D}}=\frac{\pi^{2}y^{2}}{12}\sum_{N=1}^{\infty}N\exp\left(-\sqrt{x+\pi Ny}\right).\label{eq:AppMath_gDTempFunction}
\end{align}
Typically, the summation in Eq. \ref{eq:AppMath_gDTempFunction} agrees
with numerical evaluation of the integral over the experimentally
relevant range $y>2.5$ when the first twenty terms are kept.

When Zeeman splitting is considered, the quantity $\varepsilon$ is
shifted to $\varepsilon+w$ where $w$ is a real number. Effectively,
this replaces $x$ in the expression for $g_{D}(x,y)$ by $x+iwy$.
This change does not affect the location or evaluation of the poles
of $g_{D}(x,y)$. However, it does make the integral in Eq. \ref{eq:AppMath_gDIntegral}
complex. We then can not drop the operation of taking the real part
of the integral. The more general result for complex $x$ is then
\[
g_{D}\left(x,y\right)=\frac{\pi^{2}y^{2}}{12}\sum_{N=1}^{\infty}N\text{Re}\left(\exp\left(-\sqrt{x+\pi Ny}\right)\right).
\]

\begin{figure}
\begin{centering}
\includegraphics[width=0.7\paperwidth]{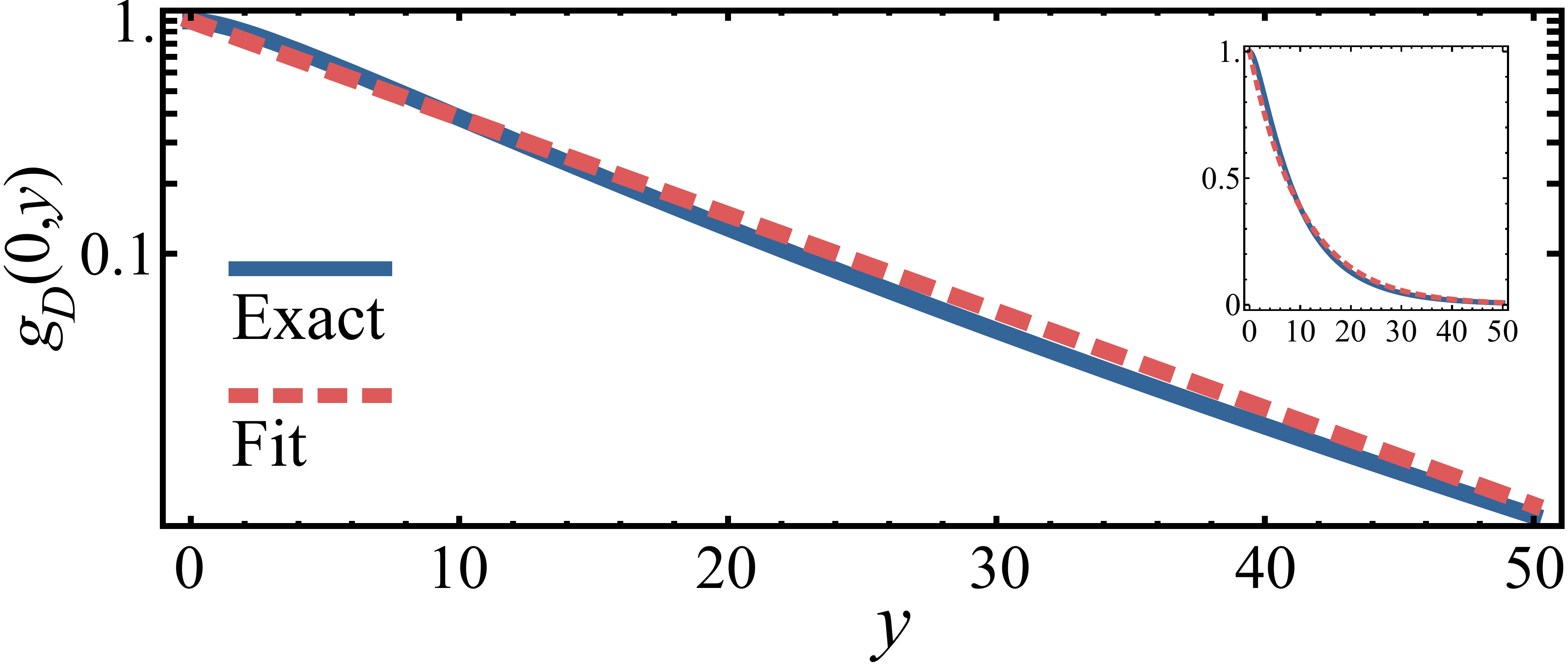}
\par\end{centering}

\caption[Exponential fit to $g_{D}(0,y)$]{\label{fig:AppMath_gDExponential}Exponential fit to $g_{D}(0,y)$.
The solid curve shows a numerical evaluation of the integral given
in Eq. \ref{eq:AppMath_gDIntegral} for $g_{D}(0,y)$. The dashed
curve shows the exponential $\exp(-0.095y)$ obtained from a best
fit for $a$ in $\exp(-ay)$ to the solid curve over the range $y<50$.}
\end{figure}

Finally, we note that for $x=0$, the function $g_{D}(x,y)$ is approximately
exponential in $y$ over the experimentally relevant parameter range
$y<50$, following the form
\[
g_{D}\left(0,y\right)\approx\exp\left(-0.096y\right)
\]
obtained by a numerical fit to $g_{D}(0,y)$ over this range. The
accuracy of this exponential approximation is demonstrated in Fig.
\ref{fig:AppMath_gDExponential}. A slightly better fit can be achieved
by allowing the prefactor of the exponential to deviate from 1. In
all analysis of data in this text, the exact form for $g_{D}\left(x,y\right)$
is used. The exponential approximation is only used when modeling
the dependence of the signal to noise ratio on the various properties
of the ring and cantilever system.

\FloatBarrier

\section{Trigonometric identities}

Many useful relations can be derived from algebraic manipulations
of Euler's formula,
\[
e^{i\theta}=\cos\theta+i\sin\theta.
\]
 The main ones used in this text are
\begin{eqnarray}
\cos(A+B) & = & \cos A\cos B-\sin A\sin B,\label{eq:AppMath_TrigCosAB}\\
\sin(A+B) & = & \sin A\cos B+\cos A\sin B,\label{eq:AppMath_TrigSinAB}\\
\cos^{2}\theta & = & \frac{1+\cos(2\theta)}{2},\label{eq:AppMath_TrigCosSquared}\\
\sin^{2}\theta & = & \frac{1-\cos(2\theta)}{2}.\label{eq:AppMath_TrigSinSquared}
\end{eqnarray}
 From these relations it is easily seen that 
\begin{eqnarray}
\int_{0}^{1}d\theta\,\sin\left(2\pi m\theta\right)\sin\left(2\pi n\theta\right) & = & \frac{\delta_{mn}}{2},\label{eq:AppMath_TrigIntegrals1}\\
\int_{0}^{1}d\theta\,\cos\left(2\pi m\theta\right)\cos\left(2\pi n\theta\right) & = & \frac{\delta_{mn}}{2},\label{eq:AppMath_TrigIntegrals}\\
\int_{0}^{1}d\theta\,\sin\left(2\pi m\theta\right)\cos\left(2\pi n\theta\right) & = & 0.\label{eq:AppMath_TrigIntegrals2}
\end{eqnarray}

\section{Jacobi-Anger expansion}

The Jacobi-Anger expansion provides a convenient way of expressing
nested sinusoidal functions in terms of a sum over sinusoidal functions
that are no longer nested. In Ref. \citep{arfken2001mathematical},
the Bessel functions $J_{n}$ are defined in terms of the generating
function $g(x,t)$ as
\begin{eqnarray}
g\left(x,t\right) & = & e^{(x/2)(t-1/t)}\label{eq:AppMath_gBesselGeneratingFunc}\\
 & = & \sum_{n=-\infty}^{\infty}J_{n}(x)t^{n}.\nonumber 
\end{eqnarray}
 Then, for $t=e^{i\theta}$, 
\[
g\left(x,e^{i\theta}\right)=e^{ix\sin\theta}=\sum_{n=-\infty}^{\infty}J_{n}(x)e^{in\theta}.
\]
By expanding the right-hand side of Eq. \ref{eq:AppMath_gBesselGeneratingFunc}
one can show that $J_{-n}(x)=(-1)^{n}J_{n}(x)$. The real and imaginary
parts of Eq. \ref{eq:AppMath_ImaginaryArgBesselGen} simplify to
\begin{align}
\cos(x\sin\theta) & =\sum_{n=-\infty}^{\infty}J_{n}(x)\cos\left(n\theta\right)\nonumber \\
 & =\sum_{n=-\infty}^{\infty}J_{2n}(x)\cos\left(2n\theta\right)\nonumber \\
 & =J_{0}(x)+2\sum_{n=1}^{\infty}J_{2n}(x)\cos\left(2n\theta\right)\label{eq:AppMath_JacAngCS}\\
\sin(x\sin\theta) & =\sum_{n=-\infty}^{\infty}J_{n}(x)\sin\left(n\theta\right)\nonumber \\
 & =2\sum_{n=1}^{\infty}J_{2n-1}(x)\sin\left(\left(2n-1\right)\theta\right).\label{eq:AppMath_JacAngSS}
\end{align}
 We obtain the comparable expressions for $\cos\theta$ by taking
$\theta\rightarrow\theta+\pi/2$ which gives
\begin{align}
\cos(x\cos\theta) & =\sum_{n=-\infty}^{\infty}(-1)^{n}J_{2n}(x)\cos\left(2n\theta\right)+\sum_{n=-\infty}^{\infty}(-1)^{n}J_{2n-1}(x)\sin\left((2n-1)\theta\right)\nonumber \\
 & =\sum_{n=-\infty}^{\infty}(-1)^{n}J_{2n}(x)\cos\left(2n\theta\right)\nonumber \\
 & =J_{0}(x)+2\sum_{n=1}^{\infty}(-1)^{n}J_{2n}(x)\cos\left(2n\theta\right)\label{eq:AppMath_JacAngCC}\\
\sin(x\cos\theta) & =\sum_{n=-\infty}^{\infty}(-1)^{n}J_{2n}(x)\sin\left(2n\theta\right)-\sum_{n=-\infty}^{\infty}(-1)^{n}J_{2n-1}(x)\cos\left((2n-1)\theta\right)\nonumber \\
 & =\sum_{n=-\infty}^{\infty}(-1)^{n+1}J_{2n-1}(x)\cos\left((2n-1)\theta\right)\nonumber \\
 & =2\sum_{n=1}^{\infty}(-1)^{n+1}J_{2n-1}(x)\cos\left(\left(2n-1\right)\theta\right).\label{eq:AppMath_JacAngSC}
\end{align}
Eqs. \ref{eq:AppMath_JacAngCS}, \ref{eq:AppMath_JacAngSS}, \ref{eq:AppMath_JacAngCC},
and \ref{eq:AppMath_JacAngSC} are collectively referred to as the
Jacobi-Anger expansion.

\section{Select integral identities}

\subsection{$\frac{\sin^{2}(x/2)}{1+(x/\alpha)^{2}}$}

By taking $x\rightarrow-x$ and reversing the direction of integration
for part of the integrand, we can write
\begin{align*}
\int_{-\infty}^{\infty}dx\,\frac{\sin^{2}\left(x/2\right)}{1+\left(x/\alpha\right)^{2}} & =\frac{1}{4}\int_{-\infty}^{\infty}dx\,\frac{2-e^{ix}-e^{-ix}}{1+\left(x/\alpha\right)^{2}}\\
 & =\frac{1}{4}\int_{-\infty}^{\infty}dx\,\frac{2-e^{ix}}{1+\left(x/\alpha\right)^{2}}-\frac{1}{4}\int_{\infty}^{-\infty}(-dx)\,\frac{e^{ix}}{1+\left(x/\alpha\right)^{2}}\\
 & =\frac{1}{4}\int_{-\infty}^{\infty}dx\,\frac{2-e^{ix}}{1+\left(x/\alpha\right)^{2}}-\frac{1}{4}\int_{-\infty}^{\infty}dx\,\frac{e^{ix}}{1+\left(x/\alpha\right)^{2}}\\
 & =\frac{1}{2}\int_{-\infty}^{\infty}dx\,\frac{1-e^{ix}}{1+\left(x/\alpha\right)^{2}}.
\end{align*}
The latter integral has poles at $\pm i\alpha$ and is easily evaluated
using the standard infinite semi-circular contour in the upper half
plane bordering the real axis:
\begin{align}
\int_{-\infty}^{\infty}dx\,\frac{\sin^{2}\left(x/2\right)}{1+\left(x/\alpha\right)^{2}} & =2\pi i\text{Res}\left[\frac{1}{2}\frac{\alpha^{2}\left(1-e^{ix}\right)}{\left(x+i\alpha\right)\left(x-i\alpha\right)},i\alpha\right]\nonumber \\
 & =\frac{\pi}{2}\alpha\left(1-e^{-\alpha}\right).\label{eq:AppMath_FilterInt1}
\end{align}

\subsection{\textmd{\normalsize $\frac{1}{x^{2}}\frac{\sin^{2}(x/2)}{1+(x/\alpha)^{2}}$}}

Using the same manipulations as in the previous section, we can rewrite
the integrand as
\[
\int_{-\infty}^{\infty}dx\,\frac{1}{x^{2}}\frac{\sin^{2}\left(x/2\right)}{1+\left(x/\alpha\right)^{2}}=\frac{1}{2}\int_{-\infty}^{\infty}dx\,\frac{1}{x^{2}}\frac{1-e^{ix}}{1+\left(x/\alpha\right)^{2}}.
\]
This integrand has second order pole on the real axis. We use the
standard trick of shifting this second order pole to $\pm\delta$
by changing $x^{-2}$ to $(x^{2}+\delta^{2})^{-1}$. The integral
can then be evaluated using the standard infinite semi-circular contour
in the upper half plane bordering the real axis
\begin{align*}
 & \int_{-\infty}^{\infty}dx\,\frac{1}{x^{2}+\delta^{2}}\frac{\sin^{2}\left(x/2\right)}{1+\left(x/\alpha\right)^{2}}=\\
 & \phantom{\int_{-\infty}^{\infty}dx\,}=2\pi i\left(\text{Res}\left[\frac{1}{2}\frac{1}{x^{2}+\delta^{2}}\frac{\alpha^{2}\left(1-e^{ix}\right)}{\alpha^{2}+x^{2}},i\alpha\right]+\text{Res}\left[\frac{1}{2}\frac{1}{x^{2}+\delta^{2}}\frac{\alpha^{2}\left(1-e^{ix}\right)}{\alpha^{2}+x^{2}},i\delta\right]\right)\\
 & \phantom{\int_{-\infty}^{\infty}dx\,}=\frac{\pi}{2}\left(\frac{1}{-\alpha^{2}+\delta^{2}}\right)\alpha\left(1-e^{-\alpha}\right)+\frac{\pi}{2}\frac{1}{\delta}\frac{\alpha^{2}\left(1-e^{-\delta}\right)}{\alpha^{2}+\delta^{2}}
\end{align*}
\begin{align*}
 & \int_{-\infty}^{\infty}dx\,\frac{1}{x^{2}+\delta^{2}}\frac{\sin^{2}\left(x/2\right)}{1+\left(x/\alpha\right)^{2}}=\\
 & \phantom{\int_{-\infty}^{\infty}dx\,}\approx-\frac{\pi}{2}\frac{1}{\alpha}\left(1-e^{-\alpha}\right)+\frac{\pi}{2}
\end{align*}
where in the last line we have taken the limit $\delta\rightarrow0$.
Thus we have
\begin{equation}
\int_{-\infty}^{\infty}dx\,\frac{1}{x^{2}}\frac{\sin^{2}\left(x/2\right)}{1+\left(x/\alpha\right)^{2}}=\frac{\pi}{2}\left(1-\frac{1}{\alpha}\left(1-e^{-\alpha}\right)\right).\label{eq:AppMath_FilterInt2}
\end{equation}

\chapter{\label{cha:AppGrFu_}Green functions in mesoscopics}

The best introduction to the Green function techniques used in mesoscopic
physics that I have found is the text of Akkermans and Montambaux
\citep{akkermans2007mesoscopic}. In this appendix, I will review
a few properties of Green functions relevant to the calculation of
persistent currents. Many of these results come from chapter three
of Akkermans and Montambaux. The main purpose of this appendix is
to introduce enough of the Green function formalism to describe what
the diffuson and cooperon are. The diffuson and cooperon are central
to the calculation of the persistent current measurements discussed
in this text.

\section{General properties}

For a system governed by the Hamiltonian $\hat{H}=\hat{H}_{0}+\hat{V}$
with $\hat{H}_{0}$ the Hamiltonian of a free particle, a complete
characterization amounts to solving the Schrödinger equation
\[
i\hbar\frac{\partial\psi}{\partial t}=\hat{H}\psi
\]
for $\psi(t)$. Formally, this equation can be solved%
\footnote{Here we are assuming that $\hat{V}$ is time independent.%
} by the time evolution Green function $\hat{G}(t)=\exp(-i\hat{H}t/\hbar)$
for which $\psi(t)=\hat{G}(t)\psi(0)$. It is often more convenient
in computations to make use of the retarded $\hat{G}^{R}$ and advanced
$\hat{G}^{A}$ Green functions
\begin{align*}
\hat{G}^{R}\left(t\right) & =-i\theta\left(t\right)\exp\left(-i\hat{H}t/\hbar\right)\\
\hat{G}^{A}\left(t\right) & =i\theta\left(-t\right)\exp\left(-i\hat{H}t/\hbar\right)
\end{align*}
where $\theta(t)$ is the Heaviside function ($\theta(t)=1$ for $t>0$
and 0 otherwise). Because of the Heaviside functions, these Green
functions are related to propagation forward and backward in time
respectively.

Consider the expectation value%
\footnote{This term is just used formally for the mathematical quantity under
consideration. The Green function does not represent a physical observable.%
} of $\hat{G}^{R,A}(t)$ for the $n^{th}$ eigenstate of the Hamiltonian
with eigenenergy $\varepsilon_{n}$,
\begin{align*}
\left\langle n\left|\hat{G}^{R,A}\left(t\right)\right|n\right\rangle  & =\mp i\theta\left(\pm t\right)\left\langle n\left|\exp\left(-i\hat{H}t/\hbar\right)\right|n\right\rangle \\
 & =\mp i\theta\left(\pm t\right)\exp\left(-i\varepsilon_{n}t/\hbar\right).
\end{align*}
Taking the Fourier transform with respect to $t$, we find
\begin{align*}
\left\langle n\left|\hat{G}^{R,A}\left(\varepsilon\right)\right|n\right\rangle  & =\int_{-\infty}^{\infty}dt\,\left\langle n\left|\hat{G}^{R,A}\left(t\right)\right|n\right\rangle e^{i\varepsilon t/\hbar}\\
 & =-i\int_{0}^{\pm\infty}dt\,\exp\left(i\left(\varepsilon-\varepsilon_{n}\right)t/\hbar\right)\\
 & =-\hbar\frac{\exp\left(i\left(\varepsilon-\varepsilon_{n}\right)t/\hbar\right)}{\varepsilon-\varepsilon_{n}}\bigg|_{0}^{\pm\infty}.
\end{align*}
This last expression contains an oscillating term that is not well-defined
at $\pm\infty$. This oscillating term can be eliminated if we give
each eigenenergy a small imaginary component $\varepsilon_{n}\rightarrow\varepsilon_{n}\mp i\gamma$.
In this case, 
\begin{align}
\left\langle n\left|\hat{G}^{R,A}\left(\varepsilon\right)\right|n\right\rangle  & =\lim_{t\rightarrow\infty}\hbar\frac{1-e^{-\gamma\left|t\right|/\hbar}\exp\left(i\left(\varepsilon-\varepsilon_{n}\right)t/\hbar\right)}{\varepsilon-\varepsilon_{n}\pm i\gamma}\nonumber \\
 & =\hbar\frac{1}{\varepsilon-\varepsilon_{n}\pm i\gamma}.\label{eq:AppGrFu_GRAnnEnergy}
\end{align}
One result of complex analysis is that 
\begin{align*}
\lim_{\gamma\rightarrow0}\frac{1}{\varepsilon-\varepsilon_{n}\pm i\gamma} & =\text{p.p.}\left(\frac{1}{\varepsilon-\varepsilon_{n}}\right)-i\pi\delta\left(\varepsilon-\varepsilon_{n}\right)
\end{align*}
where $\text{p.p.}$ represents the principal part. With this expression,
the density of states 
\begin{equation}
\nu\left(\varepsilon\right)=\sum_{n}\delta\left(\varepsilon-\varepsilon_{n}\right)\label{eq:AppGrFu_DOSDefiniton}
\end{equation}
can be written using Green functions as
\begin{align}
\nu\left(\varepsilon\right) & =\mp\frac{1}{\pi\hbar}\sum_{n}\text{Im}\left(\left\langle n\left|\hat{G}^{R,A}\left(\varepsilon\right)\right|n\right\rangle \right)\nonumber \\
 & =\mp\frac{1}{\pi\hbar}\text{Im}\left(\text{Tr}\left(\hat{G}^{R,A}\left(\varepsilon\right)\right)\right).\label{eq:AppGrFu_DOSImTrGRA}
\end{align}
In particular, we can take the trace over the position states to write
\begin{equation}
\nu\left(\varepsilon\right)=\mp\frac{1}{\pi\hbar}\text{Im}\left(\int d\boldsymbol{r}G^{R,A}\left(\boldsymbol{r},\boldsymbol{r},\varepsilon\right)\right).\label{eq:AppGrFu_DOSSpatialForm}
\end{equation}
The integral in this expression can be thought of as the spatial average
of the probability amplitude associated with all closed paths followed
by a particle with energy $\varepsilon$. 

Returning to Eq. \ref{eq:AppGrFu_GRAnnEnergy}, we note that the operator
$\hat{G}^{R,A}(\varepsilon)$ can be formally extended beyond the
energy eigenstate basis as
\begin{equation}
\hat{G}^{R,A}\left(\varepsilon\right)=\frac{\hbar}{\varepsilon-\hat{H}\pm i\gamma},\label{eq:AppGrFu_GRAenergy}
\end{equation}
 where the matrix element between any two states can be calculated
as 
\[
\left\langle k\left|\hat{G}^{R,A}\left(\varepsilon\right)\right|k'\right\rangle =\sum_{n}\left\langle k|n\vphantom{\hat{G}^{R,A}}\right\rangle \left\langle n\left|\hat{G}^{R,A}\left(\varepsilon\right)\right|n\right\rangle \left\langle n|k'\vphantom{\hat{G}^{R,A}}\right\rangle .
\]
In particular we can write the real space representation in the form
\begin{align}
G^{R,A}\left(\boldsymbol{r},\boldsymbol{r}',\varepsilon\right) & =\left\langle \boldsymbol{r}\left|\hat{G}^{R,A}\left(\varepsilon\right)\right|\boldsymbol{r}'\right\rangle \nonumber \\
 & =\sum_{n}\left\langle \boldsymbol{r}|n\vphantom{\hat{G}^{R,A}}\right\rangle \left\langle n\left|\hat{G}^{R,A}\left(\varepsilon\right)\right|n\right\rangle \left\langle n|\boldsymbol{r}'\vphantom{\hat{G}^{R,A}}\right\rangle \nonumber \\
 & =\sum_{n}\frac{\phi_{n}^{*}\left(\boldsymbol{r}'\right)\phi_{n}\left(\boldsymbol{r}\right)}{\varepsilon-\varepsilon_{n}\pm i\gamma}\label{eq:AppGrFu_GRArrEigenRepresentation}
\end{align}
where the $\phi_{n}(\boldsymbol{r})=\langle\boldsymbol{r}|n\rangle$
are the energy eigenfunctions. We note in passing that, if we define
the non-local density of states as 
\begin{equation}
\nu\left(\boldsymbol{r},\boldsymbol{r}',\varepsilon\right)=\sum_{n}\phi_{n}^{*}\left(\boldsymbol{r}'\right)\phi_{n}\left(\boldsymbol{r}\right)\delta\left(\varepsilon-\varepsilon_{n}\right),\label{eq:AppGrFu_NonLocalDOS}
\end{equation}
we can write
\[
\nu\left(\boldsymbol{r},\boldsymbol{r}',\varepsilon\right)=\mp\frac{1}{\pi\hbar}\text{Im}G^{R,A}\left(\boldsymbol{r},\boldsymbol{r}',\varepsilon\right)
\]
in analogy with the previous expression for $\nu(\varepsilon)$. From
Eqs. \ref{eq:AppGrFu_GRAnnEnergy} and \ref{eq:AppGrFu_GRAenergy},
it follows that the imaginary part of the operators $\hat{G}^{R,A}(\varepsilon)$
can be formally defined as 
\begin{equation}
\text{Im}\hat{G}^{R}\left(\varepsilon\right)=\frac{\hat{G}^{R}\left(\varepsilon\right)-\hat{G}^{A}\left(\varepsilon\right)}{2i}\label{eq:AppGrFu_ImGR}
\end{equation}
and
\begin{equation}
\text{Im}\hat{G}^{A}\left(\varepsilon\right)=\frac{\hat{G}^{A}\left(\varepsilon\right)-\hat{G}^{R}\left(\varepsilon\right)}{2i}.\label{eq:AppGrFu_ImGA}
\end{equation}
Denoting by $G_{0}$ the Green functions associated with $\hat{H}=\hat{H}_{0}$
and taking $\gamma\rightarrow0$, we can write
\begin{align*}
\left(\varepsilon-\hat{H}\right)\hat{G}^{R,A}\left(\varepsilon\right) & =\hbar\\
\hat{G}_{0}^{R,A}\left(\varepsilon\right)\left(\varepsilon-\hat{H}_{0}\right) & =\hbar.
\end{align*}
From these relations it follows that 
\begin{align*}
\hat{G}_{0}^{R,A}\left(\varepsilon\right) & =\hat{G}_{0}^{R,A}\left(\varepsilon\right)\frac{\left(\varepsilon-\hat{H}\right)}{\hbar}\hat{G}^{R,A}\left(\varepsilon\right)\\
 & =\hat{G}_{0}^{R,A}\left(\varepsilon\right)\frac{\left(\varepsilon-\hat{H}_{0}\right)}{\hbar}\hat{G}^{R,A}\left(\varepsilon\right)-\hat{G}_{0}^{R,A}\left(\varepsilon\right)\frac{\hat{V}}{\hbar}\hat{G}^{R,A}\left(\varepsilon\right)\\
 & =\hat{G}^{R,A}\left(\varepsilon\right)-\hat{G}_{0}^{R,A}\left(\varepsilon\right)\frac{\hat{V}}{\hbar}\hat{G}^{R,A}\left(\varepsilon\right),
\end{align*}
from which we obtain the recursive expression for $\hat{G}^{R,A}(\varepsilon)$
\begin{equation}
\hat{G}^{R,A}\left(\varepsilon\right)=\hat{G}_{0}^{R,A}\left(\varepsilon\right)+\hat{G}_{0}^{R,A}\left(\varepsilon\right)\frac{\hat{V}}{\hbar}\hat{G}^{R,A}\left(\varepsilon\right).\label{eq:AppGrFu_DysonExpansion}
\end{equation}
Generally, the potential $\hat{V}$ is taken to be a small perturbation
so that $\hat{G}^{R,A}$ and $\hat{G}_{0}^{R,A}$ are similar. In
that case, corrections of successively higher orders in $\hat{V}$
can be found for $\hat{G}^{R,A}$ by substituting the right-hand side
of Eq. \ref{eq:AppGrFu_DysonExpansion} for the factor of $\hat{G}^{R,A}$
in that same right-hand side. That is, 
\[
\hat{G}^{R,A}\approx\hat{G}_{0}^{R,A}+\hat{G}_{0}^{R,A}\frac{\hat{V}}{\hbar}\hat{G}_{0}^{R,A}+\hat{G}_{0}^{R,A}\frac{\hat{V}}{\hbar}\hat{G}_{0}^{R,A}\frac{\hat{V}}{\hbar}\hat{G}_{0}^{R,A}+\ldots
\]
This equation is known as the Dyson equation.

We assume that the perturbing potential is a function of position
and can be written as $\langle\boldsymbol{r}|\hat{V}|\boldsymbol{r}'\rangle=V(\boldsymbol{r})\delta(\boldsymbol{r}-\boldsymbol{r}')$
and that the function $V(\boldsymbol{r})$ varies randomly with each
individual realization of a conducting material due to microscopic
defects. The Dyson equation then has the following spatial representation
\begin{align}
G^{R,A}\left(\boldsymbol{r}\right) & \approx G_{0}^{R,A}\left(\boldsymbol{r}\right)+\frac{1}{\hbar}\int d\boldsymbol{r}_{1}\, G_{0}^{R,A}\left(\boldsymbol{r}_{1}\right)V\left(\boldsymbol{r}_{1}\right)G_{0}^{R,A}\left(\boldsymbol{r}-\boldsymbol{r}_{1}\right)\nonumber \\
 & \phantom{\approx}+\frac{1}{\hbar^{2}}\int d\boldsymbol{r}_{1}\int d\boldsymbol{r}_{2}\, G_{0}^{R,A}\left(\boldsymbol{r}_{1}\right)V\left(\boldsymbol{r}_{1}\right)G_{0}^{R,A}\left(\boldsymbol{r}_{2}-\boldsymbol{r}_{1}\right)V\left(\boldsymbol{r}_{2}\right)G_{0}^{R,A}\left(\boldsymbol{r}-\boldsymbol{r}_{2}\right)+\ldots\label{eq:AppGrFu_DysonEq}
\end{align}
A simple but accurate model for the disorder in a metal is Gaussian
white noise.%
\footnote{Gaussian noise is actually a sufficient assumption for the results
we discuss, but white noise is a reasonable assumption as well.%
} For Gaussian noise, all cumulants are zero except for the second.
We will not write down the general formula for the cumulants but will
only note that the result of all cumulants other than the second being
zero is that all of the odd moments of disorder (e.g. $\langle V(\boldsymbol{r}_{1})\rangle$,
$\langle V(\boldsymbol{r}_{1})V(\boldsymbol{r}_{2})V(\boldsymbol{r}_{3})\rangle$,
$\langle\prod_{p=1}^{2n+1}V(\boldsymbol{r}_{p})\rangle$, etc. where
$\langle\ldots\rangle$ denotes an average over disorder realization%
\footnote{Hopefully, the difference between quantum expectation value and disorder
average is clear from context. The other standard notation for averaging,
$\overline{\ldots}$, is already made use of in Chapter \ref{cha:CHMeso_}.
All expectation values below use unabbreviated {}``bra'' and {}``ket''
notation.%
}) are zero and all of the even moments can be written as a polynomial
consisting only of second moments $\langle V(\boldsymbol{r}_{1})V(\boldsymbol{r}_{2})\rangle$
(i.e. $\langle V(\boldsymbol{r}_{1})V(\boldsymbol{r}_{2})V(\boldsymbol{r}_{3})V(\boldsymbol{r}_{4})\rangle$
can be written as the sum of the products $\langle V(\boldsymbol{r}_{1})V(\boldsymbol{r}_{2})\rangle\langle V(\boldsymbol{r}_{3})V(\boldsymbol{r}_{4})\rangle$,
$\langle V(\boldsymbol{r}_{1})V(\boldsymbol{r}_{3})\rangle\langle V(\boldsymbol{r}_{2})V(\boldsymbol{r}_{4})\rangle$,
and $\langle V(\boldsymbol{r}_{1})V(\boldsymbol{r}_{4})\rangle\langle V(\boldsymbol{r}_{2})V(\boldsymbol{r}_{3})\rangle$).
For white noise, the disorder potential has no spatial correlation
so $\langle V(\boldsymbol{r}_{1})V(\boldsymbol{r}_{2})\rangle=\hbar^{2}B\delta(\boldsymbol{r}_{1}-\boldsymbol{r}_{2})$.
Thus the disorder averaged Green function (with dependence on $\varepsilon$
suppressed) is
\begin{align*}
\left\langle G^{R,A}\left(\boldsymbol{r}\right)\right\rangle  & =G_{0}^{R,A}\left(\boldsymbol{r}\right)+B\int d\boldsymbol{r}_{1}d\boldsymbol{r}_{2}\,\delta\left(\boldsymbol{r}_{1}-\boldsymbol{r}_{2}\right)G_{0}^{R,A}\left(\boldsymbol{r}_{1}\right)G_{0}^{R,A}\left(\boldsymbol{r}_{2}-\boldsymbol{r}_{1}\right)G_{0}^{R,A}\left(\boldsymbol{r}-\boldsymbol{r}_{2}\right)\\
 & \phantom{=}+B^{2}\int d\boldsymbol{r}_{1}d\boldsymbol{r}_{2}d\boldsymbol{r}_{3}d\boldsymbol{r}_{4}\,\delta\left(\boldsymbol{r}_{1}-\boldsymbol{r}_{2}\right)\delta\left(\boldsymbol{r}_{3}-\boldsymbol{r}_{4}\right)G_{0}^{R,A}\left(\boldsymbol{r}_{1}\right)G_{0}^{R,A}\left(\boldsymbol{r}_{2}-\boldsymbol{r}_{1}\right)G_{0}^{R,A}\left(\boldsymbol{r}_{3}-\boldsymbol{r}_{2}\right)\\
 & \phantom{=+B^{2}\int d\boldsymbol{r}_{1}d\boldsymbol{r}_{2}d\boldsymbol{r}_{3}d\boldsymbol{r}_{4}\,}\times G_{0}^{R,A}\left(\boldsymbol{r}_{4}-\boldsymbol{r}_{3}\right)G_{0}^{R,A}\left(\boldsymbol{r}-\boldsymbol{r}_{4}\right)\\
 & \phantom{=}+B^{2}\int d\boldsymbol{r}_{1}d\boldsymbol{r}_{2}d\boldsymbol{r}_{3}d\boldsymbol{r}_{4}\,\delta\left(\boldsymbol{r}_{1}-\boldsymbol{r}_{3}\right)\delta\left(\boldsymbol{r}_{2}-\boldsymbol{r}_{4}\right)G_{0}^{R,A}\left(\boldsymbol{r}_{1}\right)G_{0}^{R,A}\left(\boldsymbol{r}_{2}-\boldsymbol{r}_{1}\right)G_{0}^{R,A}\left(\boldsymbol{r}_{3}-\boldsymbol{r}_{2}\right)\\
 & \phantom{=+B^{2}\int d\boldsymbol{r}_{1}d\boldsymbol{r}_{2}d\boldsymbol{r}_{3}d\boldsymbol{r}_{4}\,}\times G_{0}^{R,A}\left(\boldsymbol{r}_{4}-\boldsymbol{r}_{3}\right)G_{0}^{R,A}\left(\boldsymbol{r}-\boldsymbol{r}_{4}\right)\\
 & \phantom{=}+B^{2}\int d\boldsymbol{r}_{1}d\boldsymbol{r}_{2}d\boldsymbol{r}_{3}d\boldsymbol{r}_{4}\,\delta\left(\boldsymbol{r}_{1}-\boldsymbol{r}_{4}\right)\delta\left(\boldsymbol{r}_{2}-\boldsymbol{r}_{3}\right)G_{0}^{R,A}\left(\boldsymbol{r}_{1}\right)G_{0}^{R,A}\left(\boldsymbol{r}_{2}-\boldsymbol{r}_{1}\right)G_{0}^{R,A}\left(\boldsymbol{r}_{3}-\boldsymbol{r}_{2}\right)\\
 & \phantom{=+B^{2}\int d\boldsymbol{r}_{1}d\boldsymbol{r}_{2}d\boldsymbol{r}_{3}d\boldsymbol{r}_{4}\,}\times G_{0}^{R,A}\left(\boldsymbol{r}_{4}-\boldsymbol{r}_{3}\right)G_{0}^{R,A}\left(\boldsymbol{r}-\boldsymbol{r}_{4}\right)\\
 & \phantom{=}+\mathcal{O}\left(B^{4}\right)
\end{align*}
which simplifies to
\begin{align*}
\left\langle G^{R,A}\left(\boldsymbol{r}\right)\right\rangle  & =G_{0}^{R,A}\left(\boldsymbol{r}\right)+B\int d\boldsymbol{r}_{1}\, G_{0}^{R,A}\left(\boldsymbol{r}_{1}\right)G_{0}^{R,A}\left(0\right)G_{0}^{R,A}\left(\boldsymbol{r}-\boldsymbol{r}_{1}\right)\\
 & \phantom{=}+B^{2}\int d\boldsymbol{r}_{1}d\boldsymbol{r}_{3}\, G_{0}^{R,A}\left(\boldsymbol{r}_{1}\right)G_{0}^{R,A}\left(0\right)G_{0}^{R,A}\left(\boldsymbol{r}_{3}-\boldsymbol{r}_{1}\right)G_{0}^{R,A}\left(0\right)G_{0}^{R,A}\left(\boldsymbol{r}-\boldsymbol{r}_{3}\right)\\
 & \phantom{=}+B^{2}\int d\boldsymbol{r}_{1}d\boldsymbol{r}_{2}\, G_{0}^{R,A}\left(\boldsymbol{r}_{1}\right)G_{0}^{R,A}\left(\boldsymbol{r}_{2}-\boldsymbol{r}_{1}\right)G_{0}^{R,A}\left(\boldsymbol{r}_{1}-\boldsymbol{r}_{2}\right)G_{0}^{R,A}\left(\boldsymbol{r}_{2}-\boldsymbol{r}_{1}\right)G_{0}^{R,A}\left(\boldsymbol{r}-\boldsymbol{r}_{2}\right)\\
 & \phantom{=}+B^{2}\int d\boldsymbol{r}_{1}d\boldsymbol{r}_{2}\, G_{0}^{R,A}\left(\boldsymbol{r}_{1}\right)G_{0}^{R,A}\left(\boldsymbol{r}_{2}-\boldsymbol{r}_{1}\right)G_{0}^{R,A}\left(0\right)G_{0}^{R,A}\left(\boldsymbol{r}_{1}-\boldsymbol{r}_{2}\right)G_{0}^{R,A}\left(\boldsymbol{r}-\boldsymbol{r}_{1}\right)\\
 & \phantom{=}+\mathcal{O}\left(B^{4}\right).
\end{align*}
This expression is simpler in momentum space as the correlation integrals
become simple products:
\begin{align*}
\left\langle G^{R,A}\left(\boldsymbol{k}\right)\right\rangle  & =\int d\boldsymbol{r}\,\left\langle G^{R,A}\left(\boldsymbol{r}\right)\right\rangle e^{-i\boldsymbol{k}\cdot\boldsymbol{r}}\\
 & =G_{0}^{R,A}\left(\boldsymbol{k}\right)+G_{0}^{R,A}\left(\boldsymbol{k}\right)\left(BG_{0}^{R,A}\left(\boldsymbol{r}=0\right)G_{0}^{R,A}\left(\boldsymbol{k}\right)\right)\\
 & \phantom{=}+G_{0}^{R,A}\left(\boldsymbol{k}\right)\left(BG_{0}^{R,A}\left(\boldsymbol{r}=0\right)G_{0}^{R,A}\left(\boldsymbol{k}\right)\right)^{2}\\
 & \phantom{=}+G_{0}^{R,A}\left(\boldsymbol{k}\right)\left(B^{2}\int d\boldsymbol{r}_{2}\, e^{-i\boldsymbol{k}\cdot\boldsymbol{r}_{2}}\left(G_{0}^{R,A}\left(\boldsymbol{r}_{2}\right)\right)^{2}G_{0}^{R,A}\left(-\boldsymbol{r}_{2}\right)G_{0}^{R,A}\left(\boldsymbol{k}\right)\right)\\
 & \phantom{=}+G_{0}^{R,A}\left(\boldsymbol{k}\right)\left(B^{2}G_{0}^{R,A}\left(\boldsymbol{r}=0\right)\int d\boldsymbol{r}_{2}\, G_{0}^{R,A}\left(\boldsymbol{r}_{2}\right)G_{0}^{R,A}\left(-\boldsymbol{r}_{2}\right)G_{0}^{R,A}\left(\boldsymbol{k}\right)\right)\\
 & \phantom{=}+\mathcal{O}\left(B^{4}\right).
\end{align*}
From this expression, it can (perhaps) be seen that $\langle G^{R,A}(\boldsymbol{k})\rangle$
is a geometric series of geometric series of terms involving successively
higher orders of $B$.%
\footnote{This conclusion might be best drawn from a diagrammatic framework
which we will not introduce here.%
} That is, we can write 
\[
\left\langle G^{R,A}\left(\boldsymbol{k}\right)\right\rangle =G_{0}^{R,A}\left(\boldsymbol{k}\right)+G_{0}^{R,A}\left(\boldsymbol{k}\right)\sum_{n=1}^{\infty}\left(\Sigma^{R,A}\left(\boldsymbol{k},\varepsilon\right)G_{0}^{R,A}\left(\boldsymbol{k}\right)\right)^{n}
\]
with $\Sigma^{R,A}(\boldsymbol{k},\varepsilon)=\sum_{n=1}^{\infty}\Sigma_{n}^{R,A}$
and
\begin{align}
\Sigma_{1}^{R,A}\left(\boldsymbol{k},\varepsilon\right) & =\sum_{n=1}^{\infty}\left(BG_{0}^{R,A}\left(\boldsymbol{r}=0\right)\right)^{n}\label{eq:AppGrFu_SelfEnergy}\\
\Sigma_{2}^{R,A}\left(\boldsymbol{k},\varepsilon\right) & =\sum_{n=1}^{\infty}\left(B^{2}\int d\boldsymbol{r}_{2}\, e^{-i\boldsymbol{k}\cdot\boldsymbol{r}_{2}}\left(G_{0}^{R,A}\left(\boldsymbol{r}_{2}\right)\right)^{2}G_{0}^{R,A}\left(-\boldsymbol{r}_{2}\right)\right)^{n}\nonumber \\
\Sigma_{3}^{R,A}\left(\boldsymbol{k},\varepsilon\right) & =\sum_{n=1}^{\infty}\left(B^{2}G_{0}^{R,A}\left(\boldsymbol{r}=0\right)\int d\boldsymbol{r}_{2}\, G_{0}^{R,A}\left(\boldsymbol{r}_{2}\right)G_{0}^{R,A}\left(-\boldsymbol{r}_{2}\right)\right)^{n}\nonumber 
\end{align}
with the series for $\Sigma_{n}^{R,A}$ with $n>3$ being series involving
$B^{m}$ for $m>2$. With this expansion for $\langle G^{R,A}(\boldsymbol{k})\rangle$,
we can write
\begin{align*}
\left\langle G^{R,A}\left(\boldsymbol{k}\right)\right\rangle  & =G_{0}^{R,A}\left(\boldsymbol{k}\right)+G_{0}^{R,A}\left(\boldsymbol{k}\right)\sum_{n=1}^{\infty}\left(\Sigma^{R,A}\left(\boldsymbol{k},\varepsilon\right)G_{0}^{R,A}\left(\boldsymbol{k}\right)\right)^{n}\\
 & =G_{0}^{R,A}\left(\boldsymbol{k}\right)\left(\sum_{n=0}^{\infty}\left(\Sigma^{R,A}\left(\boldsymbol{k},\varepsilon\right)G_{0}^{R,A}\left(\boldsymbol{k}\right)\right)^{n}\right)\\
 & =\frac{G_{0}^{R,A}\left(\boldsymbol{k}\right)}{1-\Sigma^{R,A}\left(\boldsymbol{k},\varepsilon\right)G_{0}^{R,A}\left(\boldsymbol{k}\right)}.
\end{align*}
Using the form of Eq. \ref{eq:AppGrFu_GRAnnEnergy} for $G_{0}^{R,A}(\boldsymbol{k})$
(since the free Hamiltonian $\hat{H_{0}}$ is diagonal in $\boldsymbol{k}$),
we have
\[
\left\langle G^{R,A}\left(\boldsymbol{k,}\varepsilon\right)\right\rangle =\frac{\hbar}{\varepsilon-\varepsilon\left(\boldsymbol{k}\right)-\hbar\Sigma^{R,A}\left(\boldsymbol{k},\varepsilon\right)}.
\]
From this expression, we can see that the real part of $\Sigma^{R,A}(\boldsymbol{k},\varepsilon)$
gives a shift to the energy of state $\boldsymbol{k}$ due to the
disorder average while the imaginary part of $\Sigma^{R,A}(\boldsymbol{k},\varepsilon)$
plays the role of $\gamma$.

The condition of weak disorder can be formulated as $k_{F}l_{e}\gg1$
where $k_{F}$ is the wave vector of electrons at the Fermi level
and $l_{e}$ is the elastic mean free path which can be conceptualized
as the average distance between collisions with the disorder potential
for a freely propagating electron. It is defined by the relation $l_{e}=v_{F}\tau_{e}$
where $v_{F}$ is the velocity of electrons at the Fermi level and
$\tau_{e}$ is the elastic scattering time (the average time between
collisions). It can be shown that for $n>1$ 
\[
\frac{\Sigma_{n}^{R,A}\left(\boldsymbol{k},\varepsilon\right)}{\Sigma_{1}^{R,A}\left(\boldsymbol{k},\varepsilon\right)}\propto\left(\frac{1}{k\left(\varepsilon\right)l_{e}}\right)^{m}
\]
with $m\geq1$. Since we are usually concerned with $\varepsilon\sim\varepsilon_{F}$,
these terms may be neglected in the limit of weak disorder. From Eq.
\ref{eq:AppGrFu_SelfEnergy}, we see that $\Sigma_{1}(\boldsymbol{k},\varepsilon)$,
usually referred to as the self-energy, does not depend on $\boldsymbol{k}$.
Since the real part of $\Sigma_{1}$ just gives a constant shift to
energy levels $\varepsilon(\boldsymbol{k})$, we will neglect it. 

To find the imaginary part of the self energy, we make the further
approximation of dropping the higher order terms in $B$, leaving
\begin{align*}
\Sigma_{1}^{R,A} & \approx BG_{0}^{R,A}\left(\boldsymbol{r}=0\right)\\
 & =B\left\langle \boldsymbol{r}=0\left|G_{0}^{R,A}\right|\boldsymbol{r}=0\right\rangle \\
 & =B\sum_{\boldsymbol{k},\boldsymbol{k}'}\left\langle \boldsymbol{r}=0|\boldsymbol{k}\right\rangle \left\langle \boldsymbol{k}\left|G_{0}^{R,A}\right|\boldsymbol{k}'\right\rangle \left\langle \boldsymbol{k}'|\boldsymbol{r}=0\right\rangle \\
 & =B\frac{1}{V_{d}}\sum_{\boldsymbol{k}}\left\langle \boldsymbol{k}\left|G_{0}^{R,A}\right|\boldsymbol{k}\right\rangle \\
 & =\frac{B}{V_{d}}\text{Tr}G_{0}^{R,A}
\end{align*}
where we have used the fact that $G_{0}^{R,A}$ is diagonal in $\boldsymbol{k}$
and $\langle\boldsymbol{r}=0|\boldsymbol{k}\rangle=V_{d}^{-1/2}$
where $V_{d}$ is the volume of the system. Using Eq. \ref{eq:AppGrFu_DOSImTrGRA},
we have
\begin{align}
\text{Im}\Sigma_{1}^{R,A}\left(\varepsilon\right) & =\pm\pi\hbar\frac{B}{V_{d}}\nu_{0}\left(\varepsilon\right)\label{eq:AppGrFu_SelfEnergyBVnu}
\end{align}
where $\nu_{0}(\varepsilon)$ is the density of states in the absence
of disorder. As a comparison, Fermi's golden rule (see e.g. Ref. \citep{shankar1994principles})
states that the rate of transition $R_{\boldsymbol{k}\rightarrow\boldsymbol{k}'}$
from state $\boldsymbol{k}$ to $\boldsymbol{k}'$ to leading order
in the perturbation $V$ is
\begin{align*}
R_{\boldsymbol{k}\rightarrow\boldsymbol{k}'} & =\frac{2\pi}{\hbar}\left|\left\langle \boldsymbol{k}\left|\hat{V}\right|\boldsymbol{k}'\right\rangle \right|^{2}\delta\left(\varepsilon\left(\boldsymbol{k}\right)-\varepsilon\left(\boldsymbol{k}'\right)\right).
\end{align*}
By summing up the transition rates to all possible states, we get
the total rate of decay which we refer to as the inverse lifetime
\begin{align*}
\tau^{-1}\left(\boldsymbol{k}\right) & =\int d\boldsymbol{k}'\,\left(\frac{V_{d}}{\left(2\pi\right)^{d}}\right)R_{\boldsymbol{k}\rightarrow\boldsymbol{k}'}\\
 & =\frac{2\pi}{\hbar}\int d\boldsymbol{k}'\,\left(\frac{V_{d}}{\left(2\pi\right)^{d}}\right)\left|\left\langle \boldsymbol{k}\left|\hat{V}\right|\boldsymbol{k}'\right\rangle \right|^{2}\delta\left(\varepsilon\left(\boldsymbol{k}\right)-\varepsilon\left(\boldsymbol{k}'\right)\right).
\end{align*}
Since 
\begin{align*}
\left\langle \boldsymbol{k}\left|\hat{V}\right|\boldsymbol{k}'\right\rangle  & =\int d\boldsymbol{r}\,\left\langle \boldsymbol{k}|\boldsymbol{r}\right\rangle \left\langle \boldsymbol{r}|\boldsymbol{k}'\right\rangle V\left(\boldsymbol{r}\right)\\
 & =\frac{1}{V_{d}}\int d\boldsymbol{r}\, e^{i\left(\boldsymbol{k}'-\boldsymbol{k}\right)\cdot\boldsymbol{r}}V\left(\boldsymbol{r}\right),
\end{align*}
the disorder average is
\begin{align*}
\left\langle \left|\left\langle \boldsymbol{k}\left|\hat{V}\right|\boldsymbol{k}'\right\rangle \right|^{2}\right\rangle  & =\left(\frac{1}{V_{d}}\right)^{2}\int d\boldsymbol{r}d\boldsymbol{r}'\, e^{i\left(\boldsymbol{k}'-\boldsymbol{k}\right)\cdot\left(\boldsymbol{r}-\boldsymbol{r}'\right)}\left\langle V\left(\boldsymbol{r}\right)V\left(\boldsymbol{r}'\right)\right\rangle \\
 & =\left(\frac{\hbar}{V_{d}}\right)^{2}B\int d\boldsymbol{r}d\boldsymbol{r}'\, e^{i\left(\boldsymbol{k}'-\boldsymbol{k}\right)\cdot\left(\boldsymbol{r}-\boldsymbol{r}'\right)}\delta\left(\boldsymbol{r}-\boldsymbol{r}'\right)\\
 & =\frac{\hbar^{2}B}{V_{d}}
\end{align*}
and thus the disorder average of the inverse lifetime is 
\begin{align}
\left\langle \tau^{-1}\left(\boldsymbol{k}\right)\right\rangle  & =\frac{2\pi\hbar B}{V_{d}}\int d\boldsymbol{k}'\,\left(\frac{V_{d}}{\left(2\pi\right)^{d}}\right)\delta\left(\varepsilon\left(\boldsymbol{k}\right)-\varepsilon\left(\boldsymbol{k}'\right)\right)\nonumber \\
 & =\frac{2\pi\hbar B}{V_{d}}\int_{0}^{\infty}dk'\,\left(\frac{V_{d}}{\left(2\pi\right)^{d}}\right)\frac{\delta\left(k-k'\right)}{d\varepsilon/dk}k^{\prime d-1}\int d\Omega_{d-1}\nonumber \\
 & =\frac{2\pi\hbar B}{V_{d}}\frac{dk}{d\varepsilon}\frac{d}{dk}\left(\frac{V_{d}}{\left(2\pi\right)^{d}}\frac{k^{d}}{d}\int d\Omega_{d-1}\right)\nonumber \\
 & =\frac{2\pi\hbar B}{V_{d}}\frac{dk}{d\varepsilon}\frac{dN}{dk}\nonumber \\
 & =\frac{2\pi\hbar B}{V_{d}}\nu_{0}\left(\varepsilon\right)\label{eq:AppGrFu_TauK}
\end{align}
where $\int d\Omega_{d-1}$ represents integration of the $d-1$ angular
degrees of freedom in $d$-dimensional space and $N$ is the number
of energy levels. We are usually interested in $\boldsymbol{k}$ near
the Fermi energy in which case $\langle\tau^{-1}(\boldsymbol{k})\rangle$
is the elastic scattering time $\tau_{e}$ mentioned above. For $\varepsilon$
near $\varepsilon_{F}$, Eqs. \ref{eq:AppGrFu_SelfEnergyBVnu} and
\ref{eq:AppGrFu_TauK} give
\[
\text{Im}\Sigma_{1}^{R,A}\left(\varepsilon\right)=\pm\frac{1}{2\tau_{e}}.
\]

We now consider some different forms of the Green functions. Since
momentum $\hat{P}=\hbar\hat{k}$ commutes with $\hat{H}_{0},$ the
Green function can be written as
\begin{align}
\left\langle G^{R,A}\left(\boldsymbol{k},\varepsilon\right)\right\rangle  & =\left\langle \boldsymbol{k}\left|\hat{G}^{R,A}\left(\varepsilon\right)\right|\boldsymbol{k}\right\rangle \nonumber \\
 & =\frac{\hbar}{\varepsilon-\frac{\hbar^{2}k^{2}}{2m}\pm\frac{i\hbar}{2\tau_{e}}},\label{eq:AppGrFu_GkDisorderAverage}
\end{align}
where in the first line on the right hand both the expectation value
and disorder average are taken. In real space, the Green function
becomes
\begin{align*}
G^{R,A}\left(\boldsymbol{r},\varepsilon\right) & =\left\langle \boldsymbol{r}=0\left|\hat{G}^{R,A}\left(\varepsilon\right)\right|\boldsymbol{r}\right\rangle \\
 & =\sum_{\boldsymbol{k}'}\left\langle \boldsymbol{r}=0|\boldsymbol{k}'\vphantom{\hat{G}^{R,A}}\right\rangle \left\langle \boldsymbol{k}'\left|\hat{G}^{R,A}\left(\varepsilon\right)\right|\boldsymbol{k}'\right\rangle \left\langle \boldsymbol{k}'|\boldsymbol{r}\vphantom{\hat{G}^{R,A}}\right\rangle \\
 & =\int d\boldsymbol{k}'\,\left(\frac{V_{d}}{\left(2\pi\right)^{d}}\right)\left(\frac{1}{\sqrt{V_{d}}}\right)G^{R,A}\left(\boldsymbol{k}',\varepsilon\right)\left(\frac{e^{-i\boldsymbol{k}'\cdot\boldsymbol{r}}}{\sqrt{V_{d}}}\right)\\
 & =\frac{1}{\left(2\pi\right)^{d}}\int d\boldsymbol{k}'\, G^{R,A}\left(\boldsymbol{k}',\varepsilon\right)e^{-i\boldsymbol{k}'\cdot\boldsymbol{r}}
\end{align*}
which is just the Fourier transform of $G^{R,A}(\boldsymbol{k},\varepsilon)\rangle$
($V_{d}$ is the system volume). Taking the disorder average of both
sides, this integral can be easily evaluated in one dimension by using
the calculus of residues with the standard contour following the real
axis and then enclosing the upper half plane. The result is
\begin{align*}
\int dk'\,\left\langle G^{R,A}\left(k',\varepsilon\right)\right\rangle e^{-ik'r} & =\int dk'\,\frac{2m}{\hbar^{2}}\frac{e^{-ik'r}}{k^{2}-k^{\prime2}\pm im/\hbar\tau_{e}}\\
 & \approx\frac{2m}{\hbar^{2}}2\pi i\text{Res}\left[\frac{e^{-ik'r}}{\left(k\pm i\frac{m}{2\hbar\tau_{e}k}-k'\right)\left(k\pm i\frac{m}{2\hbar\tau_{e}k}+k'\right)},\pm k+i\frac{m}{2\hbar\tau_{e}k}\right]\\
 & =\frac{2m}{\hbar^{2}}2\pi i\frac{1}{2}\frac{e^{\mp ikr}\exp\left(-\frac{m}{2\hbar\tau_{e}k}\right)}{\pm k+i\frac{m}{2\hbar\tau_{e}k}},
\end{align*}
which gives
\begin{align*}
\left\langle G_{0}^{R,A}\left(r,\varepsilon\right)\right\rangle  & =\pm\frac{m}{\hbar^{2}k}e^{\mp ikr}\left(\frac{\exp\left(-\frac{m}{2\hbar\tau_{e}k}r\right)}{1\mp i\frac{m}{2\hbar\tau_{e}k^{2}}}\right)
\end{align*}
where we have used $\varepsilon=\hbar^{2}k^{2}/2m$. A similar but
more lengthy calculation in three dimensions gives
\begin{align*}
\left\langle G_{0}^{R,A}\left(\boldsymbol{r},\varepsilon\right)\right\rangle  & =-\frac{m}{2\pi}\frac{e^{\mp ikr}}{r}\exp\left(-\frac{m}{2\hbar\tau_{e}k}r\right).
\end{align*}
At the Fermi level, $k$ becomes $k_{F}$ and $\hbar k_{F}/m$ is
the Fermi velocity $v_{F}$. Using the relation $l_{e}=v_{F}\tau_{e}$,
we can write the disorder averaged Green function as
\begin{equation}
\left\langle G_{0}^{R,A}\left(\boldsymbol{r},\varepsilon\right)\right\rangle =-\frac{m}{2\pi}\frac{e^{\mp ikr}}{r}\exp\left(-\frac{r}{2l_{e}}\right).\label{eq:AppGrFu_GDisorderAverage}
\end{equation}
The one-dimensional function $\langle G_{0}^{R,A}(r,\varepsilon)\rangle$
similarly has the factor of $\exp(-r/2l_{e})$.

We can use the form of $\langle G^{R,A}(\boldsymbol{k},\varepsilon)\rangle$
given in Eq. \ref{eq:AppGrFu_DOSImTrGRA} to find the disorder average
of the density of states as
\begin{align}
\left\langle \nu\left(\varepsilon\right)\right\rangle  & =\mp\frac{1}{\pi\hbar}\mbox{\ensuremath{\left\langle \text{Im}\left(\text{Tr}\left(\hat{G}^{R,A}\left(\varepsilon\right)\right)\right)\right\rangle }}\nonumber \\
 & =\mp\frac{1}{\pi\hbar}\sum_{\boldsymbol{k}'}\text{Im}\left(\frac{\hbar}{\varepsilon-\varepsilon\left(\boldsymbol{k}'\right)\pm i\hbar/2\tau_{e}}\right)\nonumber \\
 & =\frac{1}{\pi\hbar}\sum_{\boldsymbol{k}'}\left(\frac{\hbar\left(\hbar/2\tau_{e}\right)}{\left(\varepsilon-\varepsilon\left(\boldsymbol{k}'\right)\right)^{2}+\left(\hbar/2\tau_{e}\right)^{2}}\right)\nonumber \\
 & =\frac{1}{\pi\hbar}\sum_{\boldsymbol{k}'}\int d\eta\,\delta\left(\eta-\varepsilon\left(\boldsymbol{k}'\right)\right)\left(\frac{\hbar\left(\hbar/2\tau_{e}\right)}{\left(\varepsilon-\eta\right)^{2}+\left(\hbar/2\tau_{e}\right)^{2}}\right)\nonumber \\
 & =\frac{1}{\pi}\int d\eta\,\left(\frac{\hbar\left(\hbar/2\tau_{e}\right)}{\left(\varepsilon-\eta\right)^{2}+\left(\hbar/2\tau_{e}\right)^{2}}\right)\nu_{0}\left(\eta\right)\label{eq:AppGrFu_DOSLorentzianConvolution}
\end{align}
where we have used the expressions for $\nu$ given in Eqs. \ref{eq:AppGrFu_DOSDefiniton}
and \ref{eq:AppGrFu_DOSImTrGRA}.

\section{The diffuson and cooperon}

Having introduced the Green function and derived its form after averaging
over disorder, we now shift focus to the diffuson and cooperon. The
Green function allows the calculation of the wave function amplitude
for any set of parameters (position, momentum, time, energy, etc.)
given the wave function amplitude for some initial set of parameters.
The diffuson and cooperon represent contributions to the \emph{probability}
of measuring a particle with a given set of parameters given some
initial set of values for those parameters. Space does not allow for
a full derivation of these quantities (see Chapters 4 and 5 of Ref.
\citealp{akkermans2007mesoscopic} for an introduction; this section
is based on that text), but we will try to give enough description
of these quantities to give some physical intuition for the calculation
of the persistent current in the diffusive regime given in Section
\ref{sec:CHPCTh_DiffusiveRegime}.

For a Gaussian wavepacket $\psi(\boldsymbol{r},\varepsilon)$ of energy
$\varepsilon$ with energy width $\sigma_{\varepsilon}$ centered
at position $\boldsymbol{r}$ at time $t=0$, which can be specified
by 
\[
\psi\left(\boldsymbol{r}',t;\boldsymbol{r},\varepsilon\right)=\sum_{n}\chi_{n}^{*}\left(\boldsymbol{r}\right)\chi_{n}\left(\boldsymbol{r}\right)\exp\left(-i\frac{\varepsilon_{n}t}{\hbar}\right)\exp\left(-\frac{\left(\varepsilon_{n}-\varepsilon\right)^{2}}{4\sigma_{\varepsilon}^{2}}\right)
\]
where the $\chi_{n}$ and $\varepsilon_{n}$ are eigenstates and eigenenergies
of the exact Hamiltonian $\hat{H}$, it can be shown that 
\begin{equation}
P\left(\boldsymbol{r},\boldsymbol{r}',\omega\right)=\frac{V_{d}}{2\pi\nu_{0}\left(\varepsilon\right)}\left\langle G^{R}\left(\boldsymbol{r},\boldsymbol{r}',\varepsilon\right)G^{A}\left(\boldsymbol{r}',\boldsymbol{r},\varepsilon-\hbar\omega\right)\right\rangle \label{eq:AppGrFu_PomegaGeneral}
\end{equation}
where 
\[
P\left(\boldsymbol{r},\boldsymbol{r}',\omega\right)=\int dt\, P\left(\boldsymbol{r},\boldsymbol{r}',t\right)e^{-i\omega t}
\]
is the Fourier transform of 
\[
P\left(\boldsymbol{r},\boldsymbol{r}',t\right)=\left\langle \left|\psi\left(\boldsymbol{r}',t;\boldsymbol{r},\varepsilon\right)\right|^{2}\right\rangle ,
\]
the disorder-averaged probability of measuring the particle at position
$\boldsymbol{r}'$ at time $t>0$ \citep{akkermans2007mesoscopic}. 

The disorder average of the product $G^{R}G^{A}$ in Eq. \ref{eq:AppGrFu_PomegaGeneral}
is in principle a difficult quantity to calculate. It can be shown
that for three dimensions, when the product $\langle G^{R}G^{A}\rangle$
in Eq. \ref{eq:AppGrFu_PomegaGeneral} is replaced by $\langle G^{R}\rangle\langle G^{A}\rangle$,
that the Fourier transform of the resulting quantity $P_{0}(\boldsymbol{r},\boldsymbol{r}',\omega)$
is
\begin{align}
P_{0}\left(\boldsymbol{r},\boldsymbol{r}',t\right) & =\frac{V_{d}}{2\pi\nu_{0}\left(\varepsilon\right)}\left\langle G^{R}\left(\boldsymbol{r},\boldsymbol{r}',\varepsilon\right)\right\rangle \left\langle G^{A}\left(\boldsymbol{r}',\boldsymbol{r},\varepsilon-\hbar\omega\right)\right\rangle \nonumber \\
 & =\frac{\delta\left(\left|\boldsymbol{r}'-\boldsymbol{r}\right|-v\left(\varepsilon\right)t\right)}{4\pi\left|\boldsymbol{r}-\boldsymbol{r}'\right|^{2}}\exp\left(-\frac{t}{\tau_{e}}\right)\label{eq:AppGrFu_P0DrudeBoltzmann}
\end{align}
which describes a spherical plane wave originating at point $\boldsymbol{r}$
and traveling at speed $v(\varepsilon)$ while decaying on the timescale
$\tau_{e}$. From this result it is seen that replacing $\langle G^{R}G^{A}\rangle$
by $\langle G^{R}\rangle\langle G^{A}\rangle$ amounts to discarding
all contributions to $P(\boldsymbol{r},\boldsymbol{r}',t)$ in which
the particle scatters off the disorder potential. The timescale $\tau_{e}$
was introduced in the previous section as the typical timescale on
which a state $\boldsymbol{k}$ scatters off the disorder potential
into a state $\boldsymbol{k}'$ with $|\boldsymbol{k}'|=|\boldsymbol{k}|$.
The diffuson and cooperon represent the contributions to $P(\boldsymbol{r},\boldsymbol{r}',t)$
which involve scattering off the disorder potential and survive the
disorder averaging.

\begin{figure}
\begin{centering}
\includegraphics[width=0.45\paperwidth]{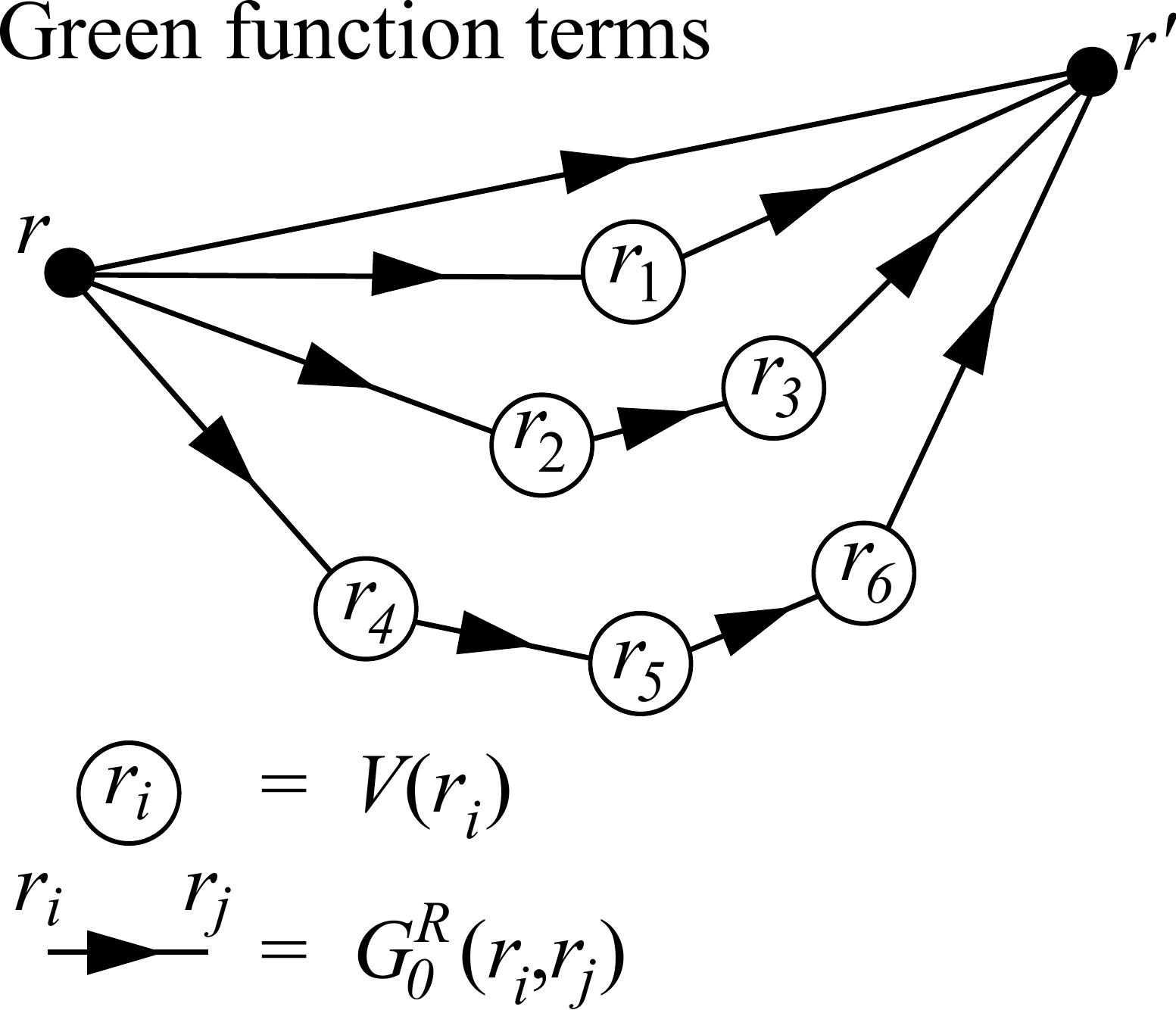}
\par\end{centering}

\caption[Contributions to the Green function in the presence of disorder]{\label{fig:AppGrFu_GreenFunctionTerms}Contributions to the Green
function in the presence of disorder. The different paths represent
integrands of first four terms of Eq. \ref{eq:AppGrFu_DysonEq}. The
total Green function $G^{R}(\boldsymbol{r},\boldsymbol{r}')$ is found
by summing the integral of each path over each $r_{i}$.}
\end{figure}

To see what other terms still contribute after disorder averaging,
we consider the form of $G^{R,A}$ in the presence of a disorder potential
$V$ as given by the Dyson equation (Eq. \ref{eq:AppGrFu_DysonEq}).
Each term in the Dyson equation represents a series of free propagations
broken up by scatterings off of the potential $V(\boldsymbol{r})$,
as illustrated in Fig. \ref{fig:AppGrFu_GreenFunctionTerms} for $G^{R}(\boldsymbol{r},\boldsymbol{r}')$.
The product $G^{R}G^{A}$ is the sum of terms composed of one set
of scatterings for $G^{R}$ (i.e. one path between $\boldsymbol{r}$
and $\boldsymbol{r}'$ in Fig. \ref{fig:AppGrFu_GreenFunctionTerms})
and one set of scatterings for $G^{A}$. For the white noise potential
which we have been considering, the potential $V$ has no spatial
correlation ($\langle V(\boldsymbol{r}_{i})V(\boldsymbol{r}_{j})\rangle=\delta(\boldsymbol{r}_{i}-\boldsymbol{r}_{j})$.
Thus, for $\langle G^{R}G^{A}\rangle$ only the cross terms of $G^{R}G^{A}$
which contain the same set of $V(\boldsymbol{r}_{i})$ will survive
the average over disorder.

\begin{figure}

\begin{centering}
\includegraphics[width=0.45\paperwidth]{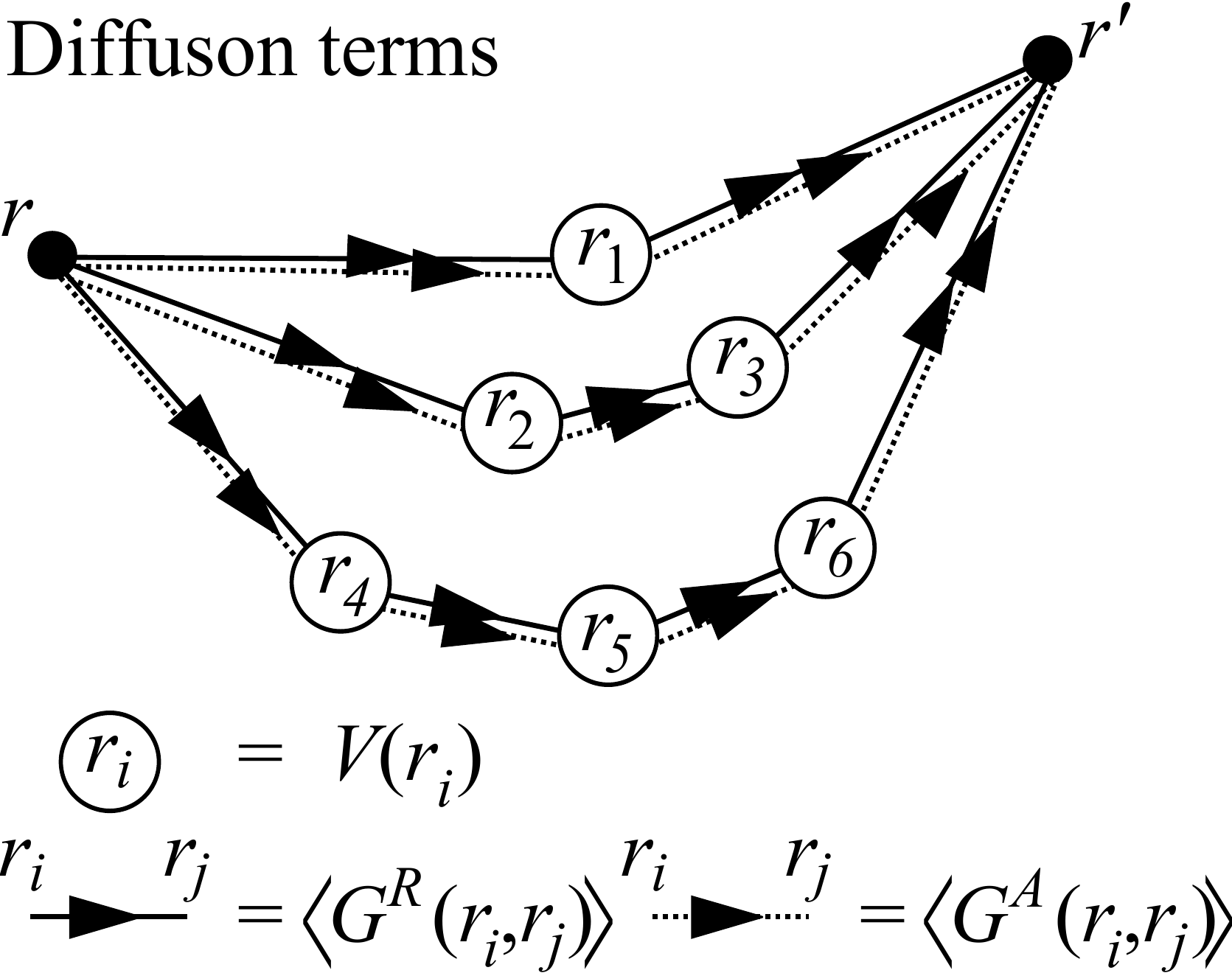}
\par\end{centering}

\caption[Contributions to the diffuson]{\label{fig:AppGrFu_DiffusonDrawing}Contributions to the diffuson.
One term of the contributions to the diffuson of each order in $V$
up to the third is shown in the figure. The diffuson represents all
terms in $\langle G^{R}G^{A}\rangle$ for which both the term from
$G^{R}$ and the term from $G^{A}$ follow the exact same path. The
accumulated phases are thus equal in magnitude and opposite in sign,
giving no net phase. In the actual calculation, the $G_{0}^{R,A}$'s
of Fig. \ref{fig:AppGrFu_GreenFunctionTerms} are replaced by $\langle G^{R,A}\rangle$.
To find the diffuson, each of these terms must be integrated over
all of the $\boldsymbol{r}_{i}$.}
\end{figure}

Next we consider the complex phases of the different terms contributing
to $G^{R,A}$. From Eq. \ref{eq:AppGrFu_GDisorderAverage}, we see
that the phase associated with $G_{0}^{R,A}(\boldsymbol{r}_{i},\boldsymbol{r}_{j},\varepsilon)$
is $\mp k(\varepsilon)|\boldsymbol{r}_{j}-\boldsymbol{r}_{i}|$. Since
$V(r)$ is taken to be real, the phases of the $G_{0}^{R,A}(\boldsymbol{r}_{i},\boldsymbol{r}_{j},\varepsilon)$
add up to give the total phase $\mp k(\varepsilon)L_{m}$ where $L_{m}$
is the length of the $m$th path between $r$ and $r'$. For paths
$m$ and $m'$ meeting the constraint stipulated above of scattering
off the potential $V$ at the same set of points $\{\boldsymbol{r}_{i}\}$,
with path $m$ corresponding to one term in $G^{R}$ and path $m'$
to one in $G^{A}$, the corresponding contribution to the phase of
$G^{R}G^{A}$ will generally be $\gg1$ unless the paths $m$ and
$m'$ follow the $\{\boldsymbol{r}_{i}\}$ in the exact same sequence
because the average distance between scatterers is $l_{e}$ and $k_{F}l_{e}\gg1$
in the weak disorder limit. The terms where $m$ and $m'$ are identical
do survive the disorder average. The sum of all such terms is collectively
known as the diffuson contribution to $\langle G^{R}G^{A}\rangle$.
Some of the terms contributing to the diffuson are shown in Fig. \ref{fig:AppGrFu_DiffusonDrawing}. 

The phase associated with $G^{R,A}$ does not depend on the direction
of propagation. Thus, the phase $k(L_{m}-L_{m'})$ will also be zero
when $m'$ represents the path through the same set of scattering
sites $\{r_{i}\}$ but in reverse order. As long as $\boldsymbol{r}$
is close to $\boldsymbol{r}'$, the terms corresponding to pairs of
reversed scattering paths will also survive averaging over disorder.
The sum of all such terms is known as the cooperon contribution to
$\langle G^{R}G^{A}\rangle$. One term contributing to the cooperon
is shown in Fig. \ref{fig:AppGrFu_DiffusonDrawing}. 

\begin{figure}
\begin{centering}
\includegraphics[width=0.45\paperwidth]{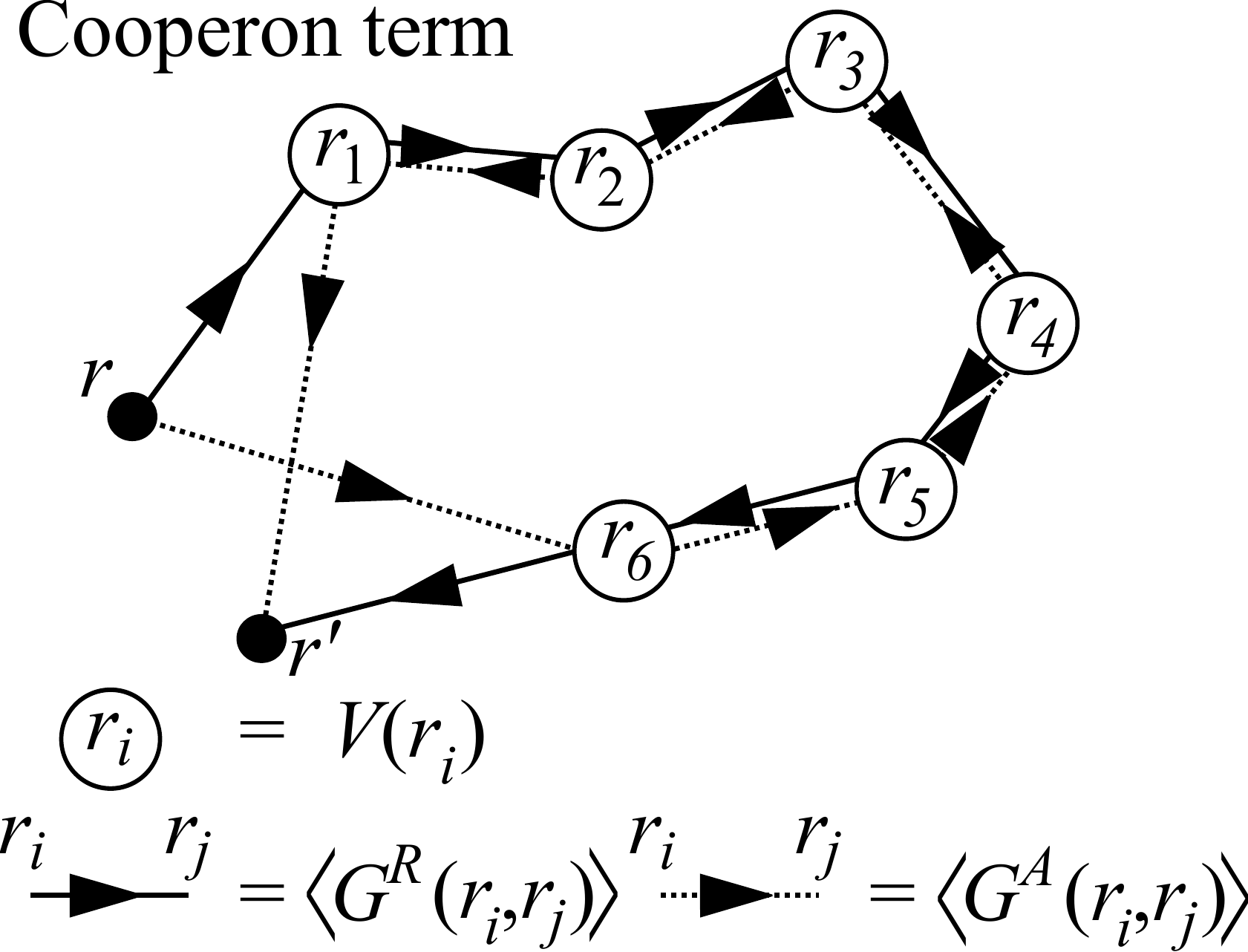}
\par\end{centering}

\caption[Contribution to the cooperon]{\label{fig:AppGrFu_CooperonDrawing}Contribution to the cooperon.
One sixth order contribution to the cooperon is shown. The path associated
with the $\langle G^{R}\rangle$'s through the scattering sites is
reversed from the path followed by the $\langle G^{A}\rangle$'s.
Just as with the diffuson terms in Fig. \ref{fig:AppGrFu_DiffusonDrawing},
the net phase of each term due to traveling through the sequence $\{\boldsymbol{r}_{i}\}$
is zero. However, here the connections to $\boldsymbol{r}$ and $\boldsymbol{r}'$
are not identical. The cooperon contributes significantly only when
$\boldsymbol{r}$ and $\boldsymbol{r}'$ are close. In the calculation
of the persistent current, only closed paths around the ring for which
$\boldsymbol{r}'=\boldsymbol{r}$ are important so the cooperon is
on equal footing with the diffuson.}
\end{figure}

The diffuson contribution to $\langle G^{R}G^{A}\rangle$ when multiplied
by $V_{d}/2\pi\nu_{0}(\varepsilon)$ leads to a contribution $P_{d}(\boldsymbol{r},\boldsymbol{r}',\omega)$
to $P(\boldsymbol{r},\boldsymbol{r}',\omega)$. By summing the terms
shown in Fig. \ref{fig:AppGrFu_DiffusonDrawing} and taking the disorder
average, this contribution can be written in the form
\begin{align}
P_{d}\left(\boldsymbol{r},\boldsymbol{r}',\omega\right) & =\frac{V_{d}}{2\pi\nu_{0}\left(\varepsilon\right)}\Bigg(B\int d\boldsymbol{r}_{1}\,\left\langle G^{R}\left(\boldsymbol{r},\boldsymbol{r}_{1}\right)\right\rangle \left\langle G^{A}\left(\boldsymbol{r}_{1},\boldsymbol{r}\right)\right\rangle \left\langle G^{R}\left(\boldsymbol{r}_{1},\boldsymbol{r}'\right)\right\rangle \left\langle G^{A}\left(\boldsymbol{r}',\boldsymbol{r}_{1}\right)\right\rangle \nonumber \\
 & \phantom{=}+B^{2}\int d\boldsymbol{r}_{1}d\boldsymbol{r}_{2}\,\left\langle G^{R}\left(\boldsymbol{r},\boldsymbol{r}_{1}\right)\right\rangle \left\langle G^{A}\left(\boldsymbol{r}_{1},\boldsymbol{r}\right)\right\rangle \left\langle G^{R}\left(\boldsymbol{r}_{1},\boldsymbol{r}_{2}\right)\right\rangle \left\langle G^{A}\left(\boldsymbol{r}_{2},\boldsymbol{r}_{1}\right)\right\rangle \times\ldots\nonumber \\
 & \phantom{=+B^{2}\int d\boldsymbol{r}_{1}d\boldsymbol{r}_{2}\,\left\langle G^{R}\left(\boldsymbol{r},\boldsymbol{r}_{1}\right)\right\rangle }\ldots\times\left\langle G^{R}\left(\boldsymbol{r}_{2},\boldsymbol{r}'\right)\right\rangle \left\langle G^{A}\left(\boldsymbol{r}',\boldsymbol{r}_{2}\right)\right\rangle +\ldots\nonumber \\
 & \phantom{=}+\ldots\Bigg),\label{eq:AppGrFu_DiffusonExpansion}
\end{align}
where all of the $\langle G^{R}\rangle$'s are at energy $\varepsilon$
and the $\langle G^{A}\rangle$'s are at energy $\varepsilon+\hbar\omega$.
The diffuson contribution $P_{d}$ also admits the simpler iterative
expression
\begin{equation}
P_{d}\left(\boldsymbol{r},\boldsymbol{r}',\omega\right)=\frac{2\pi\nu\left(\varepsilon\right)}{V_{d}}\int d\boldsymbol{r}_{1}d\boldsymbol{r}_{2}\, P_{0}\left(\boldsymbol{r},\boldsymbol{r}_{1},\omega\right)\Gamma_{d}\left(\boldsymbol{r}_{1},\boldsymbol{r}_{2}\right)P_{0}\left(\boldsymbol{r}_{2},\boldsymbol{r}',\omega\right)\label{eq:AppGrFu_PdDiffuson}
\end{equation}
with 
\begin{equation}
\Gamma_{d}\left(\boldsymbol{r}_{1},\boldsymbol{r}_{2}\right)=B\delta\left(\boldsymbol{r}_{1}-\boldsymbol{r}_{2}\right)+B\int d\boldsymbol{r}_{3}\,\Gamma_{d}\left(\boldsymbol{r}_{1},\boldsymbol{r}_{3}\right)\left(\frac{2\pi\nu_{0}\left(\varepsilon\right)}{V_{d}}P_{0}\left(\boldsymbol{r}_{3},\boldsymbol{r}_{2}\right)\right)\label{eq:AppGrFu_GammadDiffuson}
\end{equation}
where each term is also a function of $\omega$. In the diffusive
regime $\Gamma(\boldsymbol{r}_{1},\boldsymbol{r}_{3})$ changes slowly
on the scale of $l_{e}$ and that by Eq. \ref{eq:AppGrFu_P0DrudeBoltzmann}
$P_{0}$ decays exponentially on this scale, $\Gamma_{d}(\boldsymbol{r}_{1},\boldsymbol{r}_{3})$
can be expanded about $\boldsymbol{r}_{2}$ in a Taylor series, changing
Eq. \ref{eq:AppGrFu_GammadDiffuson} to a differential form. Using
the same assumptions, $\Gamma_{d}(\boldsymbol{r},\boldsymbol{r}')$
can be taken outside of the integral in Eq. \ref{eq:AppGrFu_PdDiffuson}
and can be seen to be proportional to $P_{d}(\boldsymbol{r},\boldsymbol{r}',\omega)$,
which thus also admits this differential form (see Ref. \citep{akkermans2007mesoscopic}
for the details of this calculation). The resulting differential \emph{diffusion}
equation for $P_{d}$ is
\begin{equation}
\left(-i\omega-D\nabla^{\prime2}\right)P_{d}\left(\boldsymbol{r},\boldsymbol{r}',\omega\right)=\delta\left(\boldsymbol{r}-\boldsymbol{r}'\right)\label{eq:AppGrFu_DiffusionEquation}
\end{equation}
where $D=v_{F}l_{e}/d$ is the diffusion constant and $\nabla^{\prime2}$
operates on $\boldsymbol{r}'$. In the limit of $\boldsymbol{r}'\rightarrow\boldsymbol{r}$
relevant to the calculation of persistent currents, the analogous
cooperon contribution $P_{c}(\boldsymbol{r},\boldsymbol{r}',\omega)$
is identical to $P_{d}(\boldsymbol{r},\boldsymbol{r}',\omega)$.

\FloatBarrier

\subsection{\label{sub:AppGrFu_DOScorrelation}Density of states correlation
function}

\begin{figure}
\begin{centering}
\includegraphics[width=0.56\paperwidth]{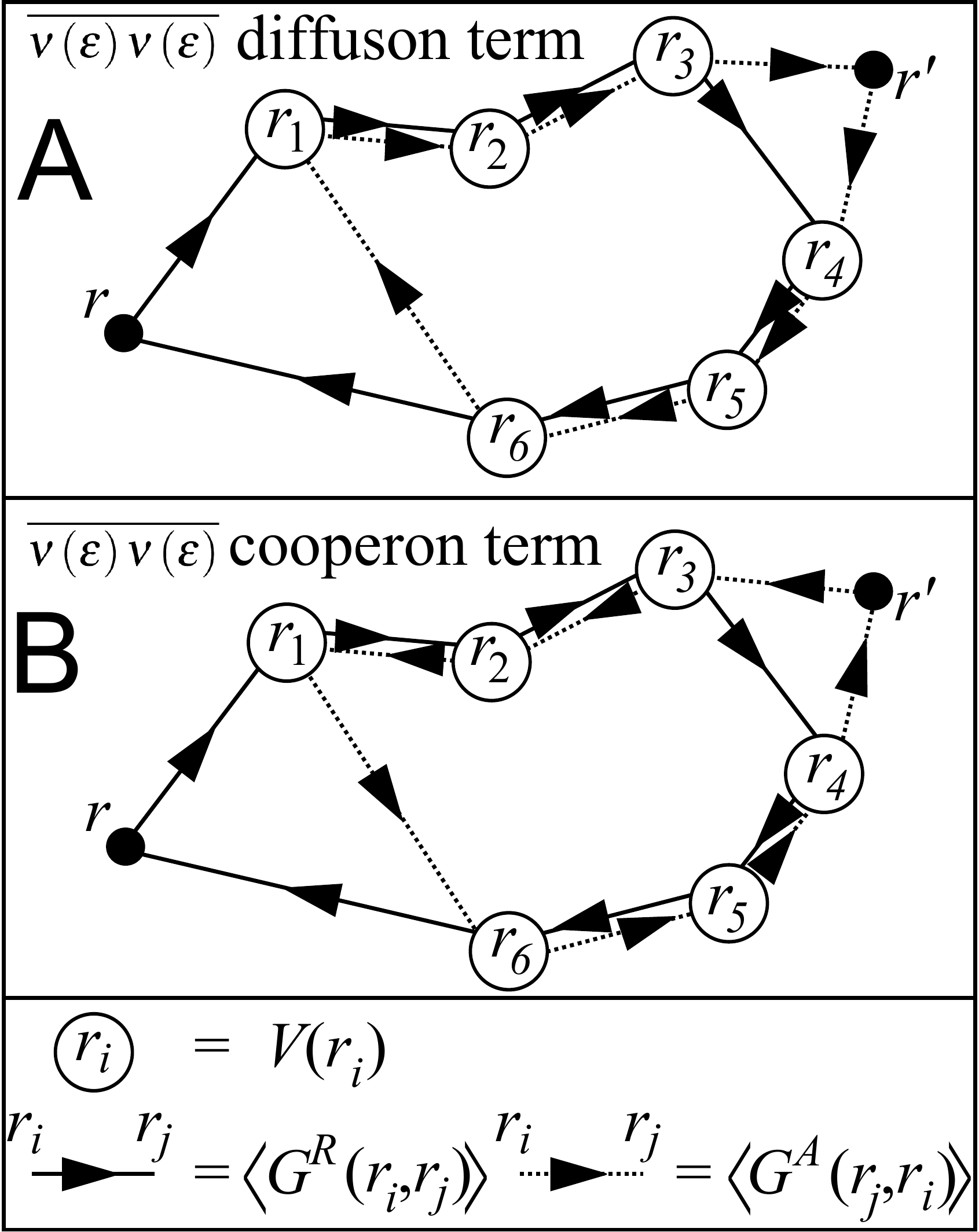}
\par\end{centering}

\caption[Long-range contributions of the diffuson and cooperon to the density
of states correlation function]{\label{fig:AppGrFu_DOSDiffusonCooperon}Long-range contributions
of the diffuson and cooperon to the density of states correlation
function. In panel A, the paths from $r_{1}$ to $r_{3}$ and from
$r_{4}$ to $r_{6}$ involve diffuson-like propagation, while in panel
B these paths involve cooperon-like propagation. Because of the short
range nature of $\langle G^{R}(\boldsymbol{r},\boldsymbol{r}_{i})\rangle$,
there must be scatterers close to points $\boldsymbol{r}$ and $\boldsymbol{r}'$
(e.g. in the drawings $\boldsymbol{r}$ must be close to $\boldsymbol{r}_{1}$
and $\boldsymbol{r}_{6}$ while $\boldsymbol{r}'$ must be close to
$\boldsymbol{r}_{3}$ and $\boldsymbol{r}_{4}$). In this case, summing
all the terms like those shown panels A and B results in contributions
$P_{d}(\boldsymbol{r},\boldsymbol{r}',\omega)P_{d}(\boldsymbol{r}',\boldsymbol{r},\omega)$
and $P_{c}(\boldsymbol{r},\boldsymbol{r}',\omega)P_{c}(\boldsymbol{r}',\boldsymbol{r},\omega)$
respectively.}
\end{figure}

In the calculation of the typical magnitude of the fluctuations of
the persistent current over different disorder configurations, the
quantity of interest is $\langle\nu(\varepsilon)\nu(\varepsilon-\hbar\omega)\rangle$.
By Eqs. \ref{eq:AppGrFu_DOSSpatialForm}, \ref{eq:AppGrFu_ImGR},
and \ref{eq:AppGrFu_ImGA}, this quantity can be written as
\begin{align}
 & \left\langle \nu\left(\varepsilon\right)\nu\left(\varepsilon-\hbar\omega\right)\right\rangle =\nonumber \\
 & \phantom{=}=\left(\frac{1}{2\pi\hbar}\right)^{2}\int d\boldsymbol{r}d\boldsymbol{r}'\,\left\langle \left(G^{R}\left(\boldsymbol{r},\boldsymbol{r},\varepsilon\right)-G^{A}\left(\boldsymbol{r},\boldsymbol{r},\varepsilon\right)\right)\left(G^{R}\left(\boldsymbol{r}',\boldsymbol{r}',\varepsilon-\hbar\omega\right)-G^{A}\left(\boldsymbol{r}',\boldsymbol{r}',\varepsilon-\hbar\omega\right)\right)\right\rangle .\label{eq:AppGrFu_DOSCorrelationInitial}
\end{align}
The procedure of identifying terms which survive averaging over disorder
follows the same principles as given above for $P_{d}(\boldsymbol{r},\boldsymbol{r}',\omega)$.
Restricting our focus to paths following the same (or reversed) sequence
of scatterers, the terms $G^{R}G^{R}$ and $G^{A}G^{A}$ become negligible
since each factor picks up a phase $kL_{m}$ giving a total phase
of $2kL_{m}\gg1$ which is essentially random. The other terms $G^{R}G^{A}$
contain contributions such as those shown in Fig. \ref{fig:AppGrFu_DOSDiffusonCooperon}
which survive disorder averaging. Summing up all such terms results
in
\begin{align}
 & \left\langle \nu\left(\varepsilon\right)\nu\left(\varepsilon-\hbar\omega\right)\right\rangle _{d,c}\nonumber \\
 & \phantom{\nu\left(\varepsilon\right)}=\left(\frac{1}{2\pi\hbar}\right)^{2}\sum_{\pm}\int d\boldsymbol{r}d\boldsymbol{r}'\,\left(P_{d}\left(\boldsymbol{r},\boldsymbol{r}',\pm\omega\right)P_{d}\left(\boldsymbol{r}',\boldsymbol{r},\pm\omega\right)+P_{c}\left(\boldsymbol{r},\boldsymbol{r}',\pm\omega\right)P_{c}\left(\boldsymbol{r}',\boldsymbol{r},\pm\omega\right)\right).\label{eq:AppGrFu_DOSCorrelationIntegral}
\end{align}
The correlation in energy of the density of states contains contributions
from all possible closed paths traversed by both diffusons and cooperons.
Although it is not especially illuminating, the diffuson/cooperon
contribution $\langle\nu(\varepsilon)\nu(\varepsilon-\hbar\omega)\rangle_{d,c}$
to the energy correlation function of the density of states can also
be viewed as the Fourier transform of the convolution $\int dt\, P_{d,c}(\boldsymbol{r},\boldsymbol{r}',t)P_{d,c}(\boldsymbol{r}',\boldsymbol{r},\tau-t)$
in time of the two halves of a closed path traversed by a diffuson
or cooperon. The connection of this component of the density of states
correlation function to the probability of completing a closed path
hints at its connection to the persistent current calculation as it
is closed paths around the ring which result in a current.

Denoting by $\phi_{n}(\boldsymbol{r})$ the eigenfunctions of $-D\nabla^{2}$
for the system under consideration with eigenvalues $E_{n}^{d,c}$,%
\footnote{Currently $E_{n}^{d,c}$ is the same for both the diffuson and cooperon.
They are different in the presence of a magnetic field as discussed
below.%
} we can write down the eigenfunction expansion, 
\[
P_{d,c}\left(\boldsymbol{r},\boldsymbol{r}',\omega\right)=\sum_{n}a_{n}\left(\boldsymbol{r}\right)\phi_{n}\left(\boldsymbol{r}'\right).
\]
Substituting this form into Eq. \ref{eq:AppGrFu_DiffusionEquation}
along with the representation $\delta(\boldsymbol{r}-\boldsymbol{r}')=\sum_{n}\phi_{n}^{*}(\boldsymbol{r})\phi_{n}(\boldsymbol{r}')$,
we have
\[
\sum_{n}\left(-i\omega-DE_{n}^{d,c}\right)a_{n}\left(\boldsymbol{r}\right)\phi_{n}\left(\boldsymbol{r}'\right)=\sum_{n}\phi_{n}^{*}\left(\boldsymbol{r}\right)\phi_{n}\left(\boldsymbol{r}'\right)
\]
from which we conclude
\begin{equation}
P_{d,c}\left(\boldsymbol{r},\boldsymbol{r}',\omega\right)=\sum_{n}\frac{\phi_{n}^{*}\left(\boldsymbol{r}\right)\phi_{n}\left(\boldsymbol{r}'\right)}{i\omega+DE_{n}^{d,c}}.\label{eq:AppGrFu_DiffCoopEigenExpansion}
\end{equation}
This result allows us to write 
\begin{align*}
\int d\boldsymbol{r}d\boldsymbol{r}'\, P_{d}\left(\boldsymbol{r},\boldsymbol{r}',\omega\right)P_{d}\left(\boldsymbol{r}',\boldsymbol{r},\omega\right) & =\sum_{n,n'}\int d\boldsymbol{r}d\boldsymbol{r}'\,\left(\frac{\phi_{n}^{*}\left(\boldsymbol{r}\right)\phi_{n}\left(\boldsymbol{r}'\right)}{i\omega+DE_{n}^{d,c}}\right)\left(\frac{\phi_{n'}^{*}\left(\boldsymbol{r}'\right)\phi_{n'}\left(\boldsymbol{r}\right)}{i\omega+DE_{n'}^{d,c}}\right)\\
 & =\sum_{n,n'}\int d\boldsymbol{r}\,\delta_{nn'}\left(\frac{\phi_{n}^{*}\left(\boldsymbol{r}\right)}{i\omega+DE_{n}^{d,c}}\right)\left(\frac{\phi_{n'}\left(\boldsymbol{r}\right)}{i\omega+DE_{n'}^{d,c}}\right)\\
 & =\sum_{n}\left(\frac{1}{i\omega+DE_{n}^{d,c}}\right)^{2}
\end{align*}
where we have used the orthonormality condition $\int d\boldsymbol{r}'\,\phi_{n}^{*}(\boldsymbol{r}')\phi_{n'}(\boldsymbol{r'})=\delta_{nn'}$.
With this result, we can rewrite the diffuson and cooperon contributions
to the density of states correlation function as
\begin{align}
\left\langle \nu\left(\varepsilon\right)\nu\left(\varepsilon-\hbar\omega\right)\right\rangle _{d,c} & =\left(\frac{1}{2\pi\hbar}\right)^{2}\sum_{\pm}\sum_{n}\left(\left(\frac{1}{\pm i\omega+DE_{n}^{d}}\right)^{2}+\left(\frac{1}{\pm i\omega+DE_{n}^{c}}\right)^{2}\right)\nonumber \\
 & =2\left(\frac{1}{2\pi\hbar}\right)^{2}\sum_{n}\text{Re}\left(\left(\frac{1}{i\omega+DE_{n}^{d}}\right)^{2}+\left(\frac{1}{i\omega+DE_{n}^{c}}\right)^{2}\right).\label{eq:AppGrFu_DOSCorrelationDiffCoop}
\end{align}

It also possible to create terms from the $G^{R}G^{A}$ product that
survive averaging over disorder for which $\boldsymbol{r}$ and $\boldsymbol{r}'$
are on the same side of the path through the scatterers such as those
shown in Fig. \ref{fig:AppGrFu_DOSDiffusonCooperon}. These terms
consist of a single diffuson or a single cooperon. They follow the
form (see Ref. \citealp{akkermans2007mesoscopic})
\[
\left\langle \nu\left(\varepsilon\right)\nu\left(\varepsilon-\hbar\omega\right)\right\rangle _{d,c}=\frac{\nu_{0}L^{d}}{\pi}\text{Re}\int d\boldsymbol{r}d\boldsymbol{r}'\,\left(g\left(\boldsymbol{r}-\boldsymbol{r}',\varepsilon\right)P_{d}\left(\boldsymbol{r}',\boldsymbol{r},\omega\right)+g\left(\boldsymbol{r}-\boldsymbol{r}',\varepsilon\right)P_{c}\left(\boldsymbol{r},\boldsymbol{r}',\omega\right)\right)
\]
where $L^{d}$ is the system volume and 
\[
g\left(\boldsymbol{r}\right)=\frac{\sin\left(k\left(\varepsilon\right)r\right)}{k\left(\varepsilon\right)r}\exp\left(-\frac{r}{2l_{e}}\right).
\]
Upon averaging over disorder, the free propagation between $\boldsymbol{r}$
and $\boldsymbol{r}'$ and the scattering sites decays exponentially
on the length scale $l_{e}$ resulting in the $g(\boldsymbol{r})$
factors in the expression. Due to the short range nature of $g(\boldsymbol{r})$,
these terms are not significant upon spatial integration and we neglect
them in the discussion below. They can be significant though when
calculating the local density of states correlation function, giving
a contribution
\begin{align}
 & \left\langle \nu\left(\boldsymbol{r},\boldsymbol{r},\varepsilon\right)\nu\left(\boldsymbol{r}',\boldsymbol{r}',\varepsilon-\hbar\omega\right)\right\rangle _{\text{SR}}\nonumber \\
 & \phantom{\nu\left(\boldsymbol{r},\boldsymbol{r},\varepsilon\right)}=\frac{\nu_{0}L^{d}}{\pi}\text{Re}\left(g\left(\boldsymbol{r}-\boldsymbol{r}',\varepsilon\right)P_{d}\left(\boldsymbol{r}',\boldsymbol{r},\omega\right)+g\left(\boldsymbol{r}-\boldsymbol{r}',\varepsilon\right)P_{c}\left(\boldsymbol{r},\boldsymbol{r}',\omega\right)\right).\label{eq:AppGrFu_DOSCorrShortRange}
\end{align}

\begin{figure}
\begin{centering}
\includegraphics[width=0.5\paperwidth]{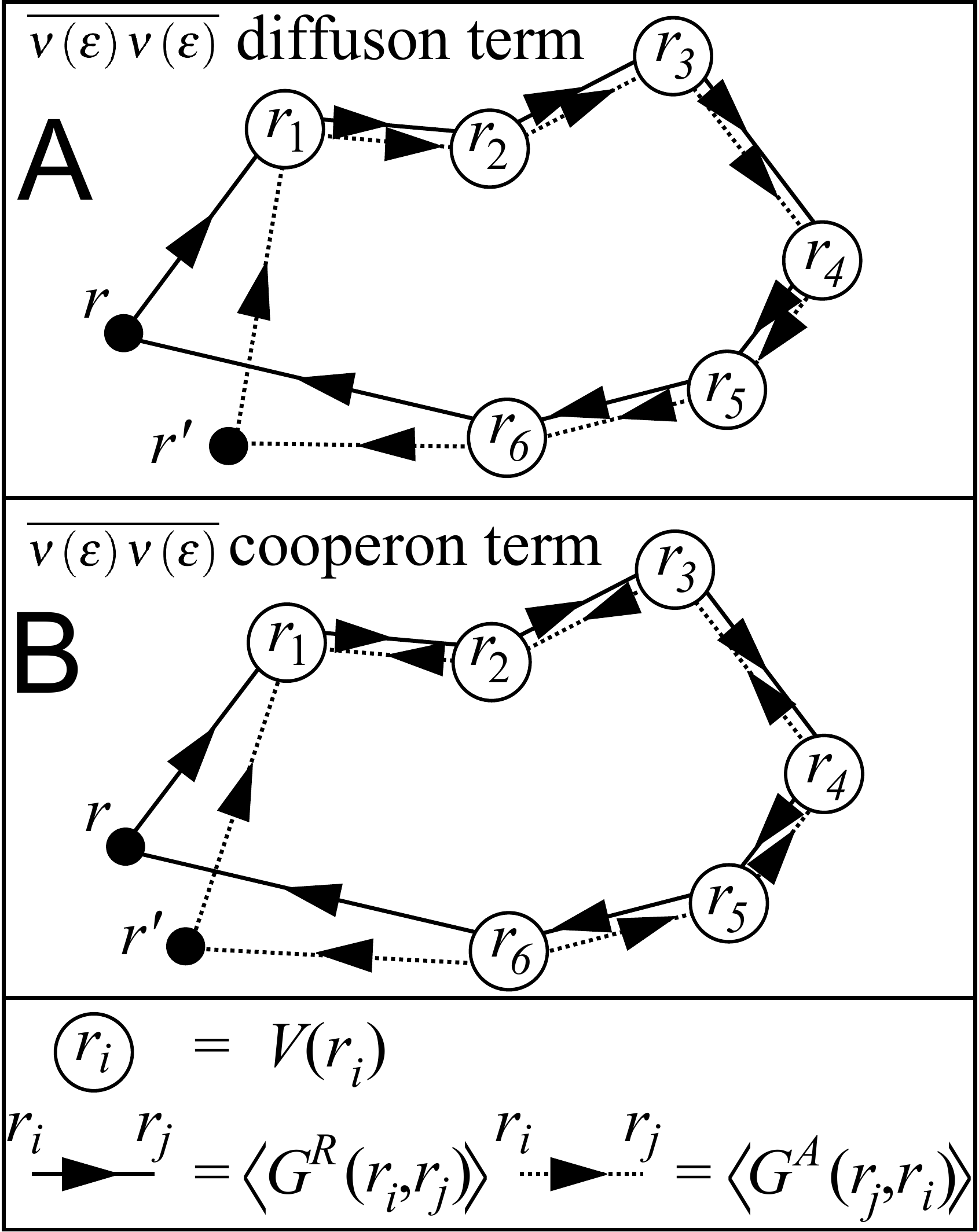}
\par\end{centering}

\caption[Short-range contributions of the diffuson and cooperon to the density
of states correlation function]{\label{fig:AppGrFu_DiffusonCooperonDOSShort}Short-range contributions
of the diffuson and cooperon to the density of states correlation
function. In panel A, $r_{1}$ and $r_{6}$ are connected by diffuson-like
propagation, while in panel B the connection is cooperon-like. In
contrast to Fig. \ref{fig:AppGrFu_DOSDiffusonCooperon}, $\boldsymbol{r}$
and $\boldsymbol{r}'$ must be close to each other due to the short
range nature of $\langle G^{R}(\boldsymbol{r},\boldsymbol{r}_{i})\rangle$.
Summing all of the terms like those shown in the figure leads to contributions
of the form $g(\boldsymbol{r}-\boldsymbol{r}')P_{d}(\boldsymbol{r},\boldsymbol{r}',\omega)$
and $g(\boldsymbol{r}-\boldsymbol{r}')P_{c}(\boldsymbol{r},\boldsymbol{r}',\omega)$
where $g(\boldsymbol{r}-\boldsymbol{r}')$ decays exponentially with
$|\boldsymbol{r}-\boldsymbol{r}'|/l_{e}$. }
\end{figure}

\FloatBarrier

\subsection{Magnetic field effects}

Now we address the question of how a finite magnetic field affects
the results of this appendix, in particular Eq. \ref{eq:AppGrFu_DOSCorrelationDiffCoop}.
In Section\ref{sub:CHPCTh_1DRingSingleLevelSolutions}, we discuss
the gauge transformation for a vector potential $\boldsymbol{A}$
that takes 
\[
\psi\left(\boldsymbol{r}\right)\rightarrow\psi'\left(\boldsymbol{r}\right)=\psi\left(\boldsymbol{r}\right)\exp\left(i\frac{e}{\hbar}\int_{\boldsymbol{r}_{0}}^{\boldsymbol{r}}d\boldsymbol{r}_{1}\cdot\boldsymbol{A}\left(\boldsymbol{r}_{1}\right)\right).
\]
With this transformation, $\psi'(\boldsymbol{r})$ represents the
same state as $\psi(\boldsymbol{r})$ but with $\boldsymbol{A}=0$
in the Hamiltonian. In that section, we argued that this result held
for regions where $\boldsymbol{B}=\nabla\times\boldsymbol{A}=0$.
However, the result can be extended to regimes where $\boldsymbol{A}$
varies slowly in space \citep{akkermans2007mesoscopic,aharonov1959significance}.
Using Eq. \ref{eq:AppGrFu_GRArrEigenRepresentation} and indicating
gauge transformed Green functions and eigenfunctions by $G'$ and
$\phi'$, we can relate the transformed Green function to the untransformed
one as 
\begin{align*}
G^{\prime R,A}\left(\boldsymbol{r},\boldsymbol{r}',\varepsilon\right) & =\sum_{n}\frac{\phi_{n}^{\prime*}\left(\boldsymbol{r}'\right)\phi'_{n}\left(\boldsymbol{r}\right)}{\varepsilon-\varepsilon_{n}\pm i\gamma}\\
 & =\sum_{n}\frac{\left(\exp\left(-i\frac{e}{\hbar}\int_{\boldsymbol{r}_{0}}^{\boldsymbol{r}'}d\boldsymbol{r}_{1}\cdot\boldsymbol{A}\left(\boldsymbol{r}_{1}\right)\right)\phi{}_{n}^{*}\left(\boldsymbol{r}'\right)\right)\left(\exp\left(i\frac{e}{\hbar}\int_{\boldsymbol{r}_{0}}^{\boldsymbol{r}}d\boldsymbol{r}_{1}\cdot\boldsymbol{A}\left(\boldsymbol{r}_{1}\right)\right)\phi{}_{n}\left(\boldsymbol{r}\right)\right)}{\varepsilon-\varepsilon_{n}\pm i\gamma}\\
 & =\exp\left(i\frac{e}{\hbar}\int_{\boldsymbol{r}'}^{\boldsymbol{r}}d\boldsymbol{r}_{1}\cdot\boldsymbol{A}\left(\boldsymbol{r}_{1}\right)\right)\sum_{n}\frac{\phi_{n}^{*}\left(\boldsymbol{r}'\right)\phi{}_{n}\left(\boldsymbol{r}\right)}{\varepsilon-\varepsilon_{n}\pm i\gamma}\\
 & =\exp\left(i\frac{e}{\hbar}\int_{\boldsymbol{r}'}^{\boldsymbol{r}}d\boldsymbol{r}_{1}\cdot\boldsymbol{A}\left(\boldsymbol{r}_{1}\right)\right)G{}^{R,A}\left(\boldsymbol{r},\boldsymbol{r}',\varepsilon\right).
\end{align*}
Thus, the magnetic field can be incorporated by giving the Green function
$G{}^{R,A}(\boldsymbol{r},\boldsymbol{r}',\varepsilon)$ a phase shift
of $i\frac{e}{\hbar}\int_{\boldsymbol{r}'}^{\boldsymbol{r}}d\boldsymbol{r}_{1}\cdot\boldsymbol{A}(\boldsymbol{r}_{1})$.
The diffuson $P_{d}(\boldsymbol{r},\boldsymbol{r}',\varepsilon)$
involves products $\langle G^{R}(\boldsymbol{r}_{i},\boldsymbol{r}_{j})\rangle\langle G^{A}(\boldsymbol{r}_{j},\boldsymbol{r}_{i})\rangle$
of Green functions with the $\{\boldsymbol{r}_{i}\}$ following a
path from $\boldsymbol{r}$ to $\boldsymbol{r}'$. In the persistent
current calculation, we are concerned with the quantity $\langle\nu(\varepsilon,B)\nu(\varepsilon-\hbar\omega,B')\rangle$,
which by Eq. \ref{eq:AppGrFu_DOSCorrelationInitial} can be seen to
involve $\langle G^{R}\rangle$ terms corresponding to $B$ and $\langle G^{A}\rangle$
terms corresponding to $B'$. Using $\boldsymbol{A}$ and $\boldsymbol{A}'$
as the corresponding vector potentials for $B$ and $B'$, we see
that 
\begin{align*}
 & \left\langle G^{\prime R,A}\left(\boldsymbol{r},\boldsymbol{r}',B\right)\right\rangle \left\langle G^{\prime R,A}\left(\boldsymbol{r}',\boldsymbol{r},B'\right)\right\rangle \\
 & \phantom{G^{\prime R,A}}=\exp\left(i\frac{e}{\hbar}\int_{\boldsymbol{r}'}^{\boldsymbol{r}}d\boldsymbol{r}_{1}\cdot\boldsymbol{A}\left(\boldsymbol{r}_{1}\right)\right)\exp\left(i\frac{e}{\hbar}\int_{\boldsymbol{r}}^{\boldsymbol{r}'}d\boldsymbol{r}_{1}\cdot\boldsymbol{A}'\left(\boldsymbol{r}_{1}\right)\right)\left\langle G^{R,A}\left(\boldsymbol{r},\boldsymbol{r}',0\right)\right\rangle \left\langle G^{R,A}\left(\boldsymbol{r}',\boldsymbol{r},0\right)\right\rangle \\
 & \phantom{G^{\prime R,A}}=\exp\left(i\frac{e}{\hbar}\int_{\boldsymbol{r}'}^{\boldsymbol{r}}d\boldsymbol{r}_{1}\cdot\left(\boldsymbol{A}\left(\boldsymbol{r}_{1}\right)-\boldsymbol{A}'\left(\boldsymbol{r}_{1}\right)\right)\right)\left\langle G^{R,A}\left(\boldsymbol{r},\boldsymbol{r}',0\right)\right\rangle \left\langle G^{R,A}\left(\boldsymbol{r}',\boldsymbol{r},0\right)\right\rangle 
\end{align*}
Thus, for vector potentials $\boldsymbol{A}$ and $\boldsymbol{A}'$,
the diffuson $P_{d}(\boldsymbol{r},\boldsymbol{r}',\varepsilon)$
picks up the extra phase $i\frac{e}{\hbar}\int_{\boldsymbol{r}'}^{\boldsymbol{r}}d\boldsymbol{r}_{1}\cdot(\boldsymbol{A}(\boldsymbol{r}_{1})-\boldsymbol{A}'(\boldsymbol{r}_{1}))$.
Similarly the cooperon involves terms of the form $\langle G^{R}(\boldsymbol{r}_{i},\boldsymbol{r}_{j})\rangle\langle G^{A}(\boldsymbol{r}_{i},\boldsymbol{r}_{j})\rangle$
and picks up a phase $i\frac{e}{\hbar}\int_{\boldsymbol{r}'}^{\boldsymbol{r}}d\boldsymbol{r}_{1}\cdot(\boldsymbol{A}(\boldsymbol{r}_{1})+\boldsymbol{A}'(\boldsymbol{r}_{1}))$.
We use the notation 
\[
\boldsymbol{A}_{\pm}=\boldsymbol{A}\pm\boldsymbol{A}'.
\]
When the Taylor expansion derivation leading to Eq. \ref{eq:AppGrFu_DiffusionEquation}
is performed in the presence of a magnetic field, the relation 
\[
\nabla P_{d,c}\left(\boldsymbol{r},\boldsymbol{r}',\varepsilon,\boldsymbol{A}_{\mp}\right)=\exp\left(i\frac{e}{\hbar}\int_{\boldsymbol{r}'}^{\boldsymbol{r}}d\boldsymbol{r}_{1}\cdot\boldsymbol{A}_{\mp}\right)\left(\nabla+i\frac{e}{\hbar}\boldsymbol{A}_{\mp}\right)P_{d,c}\left(\boldsymbol{r},\boldsymbol{r}',\varepsilon,0\right)
\]
results in the new diffusion equation
\begin{equation}
\left(-i\omega-D\left(\nabla'+i\frac{e}{\hbar}\boldsymbol{A}_{\mp}\right)^{2}\right)P_{d,c}\left(\boldsymbol{r},\boldsymbol{r}',\omega\right)=\delta\left(\boldsymbol{r}-\boldsymbol{r}'\right).\label{eq:AppGrFu_DiffusionEquationFieldDependent}
\end{equation}
With this result, we can write the field dependent form of the density
of states correlation function as
\begin{equation}
\left\langle \nu\left(\varepsilon,B\right)\nu\left(\varepsilon-\hbar\omega,B'\right)\right\rangle _{d,c}=2\left(\frac{1}{2\pi\hbar}\right)^{2}\sum_{\mp}\sum_{n}\text{Re}\left(\left(\frac{1}{i\omega+DE_{n}\left(B_{\mp}\right)}\right)^{2}\right)\label{eq:AppGrFu_DOSFieldDependent}
\end{equation}
where the $E_{n}^{d,c}(B_{\mp})$ are the eigenvalues in the expression
\begin{equation}
\left(\nabla'+i\frac{e}{\hbar}\boldsymbol{A}_{\mp}\right)^{2}P_{d,c}\left(\boldsymbol{r},\boldsymbol{r}',\omega\right)=E_{n}^{d,c}\left(B_{\mp}\right)P_{d,c}\left(\boldsymbol{r},\boldsymbol{r}',\omega\right).\label{eq:AppGrFu_DiffusonCooperonEigenvalues}
\end{equation}
The boundary conditions for $P_{d,c}$ are that 
\begin{equation}
\tilde{\boldsymbol{n}}\cdot\left(\nabla'+i\frac{e}{\hbar}\boldsymbol{A}_{\mp}\right)P_{d,c}\left(\boldsymbol{r},\boldsymbol{r}',\omega\right)=0\label{eq:AppGrFu_DiffusionBoundaryConditions}
\end{equation}
for reflecting boundaries with $\tilde{\boldsymbol{n}}$ the vector
normal to the surface of the boundary \citep{akkermans2007mesoscopic,ginossar2010mesoscopic}.

\subsection{\label{sub:AppGrFu_Spin}Spin effects}

We now address effects related to the electron spin. In the absence
of any effect to break the degeneracy of the two spin states $\uparrow$
and $\downarrow$, the spatial characterization of the two spin states
is identical and we can write
\[
\nu\left(\varepsilon\right)=\nu\left(\varepsilon,\uparrow\right)+\nu\left(\varepsilon,\downarrow\right)
\]
where both $\nu(\varepsilon,\uparrow)$ and $\nu(\varepsilon,\downarrow)$
are the just the density of states $\nu(\varepsilon)$ considered
in the preceding portion of this appendix. Now we consider applying
a magnetic field $\boldsymbol{B}$. In a magnetic field, the electron
Hamiltonian picks up a term 
\begin{equation}
\hat{H}_{Z}=\frac{g\mu_{B}}{2}\boldsymbol{\sigma}\cdot\boldsymbol{B},\label{eq:AppGrFu_HZ}
\end{equation}
and with its spin aligned (anti-aligned) with the field the electron
has a Zeeman energy
\begin{equation}
E_{Z}=\left(-\right)\frac{g\mu_{B}}{2}B\label{eq:AppGrFu_EZeeman}
\end{equation}
where $\boldsymbol{\sigma}$ is the vector of Pauli matrices, $g\approx2$
is the gyromagnetic ratio of the electron in the medium under consideration
and $\mu_{B}$ is the Bohr magneton. 

We now consider the correlation function of the density of states.
Reserving the $\varepsilon$ argument of $\nu(\varepsilon)$ for the
energy $\varepsilon_{s}$ associated with the electron's spatial degrees
of freedom but allowing $\varepsilon=\varepsilon_{s}+E_{Z}$ to denote
the total energy and taking $\uparrow$ to the be the spin direction
aligned with the magnetic field, we can write
\[
\nu\left(\varepsilon_{s},B,\uparrow\right)=\nu\left(\varepsilon-E_{Z},B,\uparrow\right)
\]
and
\[
\nu\left(\varepsilon_{s},B,\downarrow\right)=\nu\left(\varepsilon+E_{Z},B,\downarrow\right)
\]
where the expressions on the right-hand side correspond to the same
density of states expression discussed in previous sections where
spin degeneracy was assumed (the $\uparrow\downarrow$ label has only
been added for clarity). Using the fact that under the limits considered
the density of states correlation function $\langle\nu(\varepsilon)\nu(\varepsilon-\hbar\omega)\rangle$
depends only on the energy difference $\hbar\omega$, we can write
\begin{align*}
\left\langle \nu\left(\varepsilon,B\right)\nu\left(\varepsilon-\hbar\omega,B'\right)\right\rangle  & =\Big<\left(\nu\left(\varepsilon-E_{Z},B,\uparrow\right)+\nu\left(\varepsilon+E_{Z},B,\downarrow\right)\right)\times\ldots\\
 & \phantom{=\Big<}\ldots\left(\nu\left(\varepsilon-\hbar\omega-E_{Z},B',\uparrow\right)+\nu\left(\varepsilon-\hbar\omega+E_{Z},B',\downarrow\right)\right)\Big>\\
 & =\left\langle \nu\left(\varepsilon,B,\uparrow\right)\nu\left(\varepsilon-\hbar\omega,B',\uparrow\right)\right\rangle +\left\langle \nu\left(\varepsilon,B,\uparrow\right)\nu\left(\varepsilon-\hbar\omega+2E_{Z},B',\downarrow\right)\right\rangle +\ldots\\
 & \phantom{=}+\left\langle \nu\left(\varepsilon,B,\downarrow\right)\nu\left(\varepsilon-\hbar\omega-2E_{Z},B',\uparrow\right)\right\rangle +\left\langle \nu\left(\varepsilon,B,\downarrow\right)\nu\left(\varepsilon-\hbar\omega,B',\downarrow\right)\right\rangle 
\end{align*}
where we assumed that $E_{Z}^{\prime}\approx E_{Z}$ since we are
usually interested in magnetic field differences small on the scale
of the total field. From this result, we see that two terms, the $\uparrow\uparrow$
and $\downarrow\downarrow$ ones, match the expression found for the
spin degenerate case while the other two terms, $\uparrow\downarrow$
and $\downarrow\uparrow$, are shifted in energy by $\pm2E_{Z}$.
The effect of the Zeeman splitting is to separate in energy the orbital
energy levels for two spin orientations. When considering the correlation
of the density of states, the spin up levels maintain the same level
of correlation with themselves because they are all shifted by the
same energy and the disorder averaged correlation function is independent
of the absolute energy $\varepsilon$. The case is the same for the
spin down levels. The correlation between the two spin states, however,
changes as the correlated orbitals become shifted in energy relative
each by a total amount $2E_{z}$. In the presence of Zeeman splitting,
we can rewrite Eq. \ref{eq:AppGrFu_DOSFieldDependent} as
\begin{align}
 & \left\langle \nu\left(\varepsilon,B\right)\nu\left(\varepsilon-\hbar\omega,B'\right)\right\rangle _{d,c}=\nonumber \\
 & \phantom{==}=2\left(\frac{1}{2\pi\hbar}\right)^{2}\sum_{\mp}\sum_{n}\text{Re}\Bigg(2\left(\frac{1}{i\omega+DE_{n}\left(B_{\mp}\right)}\right)^{2}+\left(\frac{1}{i\omega+2i\frac{E_{Z}}{\hbar}+DE_{n}\left(B_{\mp}\right)}\right)^{2}+\nonumber \\
 & \phantom{===2\left(\frac{1}{2\pi\hbar}\right)^{2}\sum_{\mp}\sum_{n}\text{Re}\Bigg(}\ldots+\left(\frac{1}{i\omega-2i\frac{E_{Z}}{\hbar}+DE_{n}\left(B_{\mp}\right)}\right)^{2}\Bigg)\label{eq:AppGrFu_NuNuEZ}
\end{align}
with the eigenvalues $E_{n}(B_{\mp})$ unaffected by the Zeeman splitting.

Another important spin effect%
\footnote{We neglect a third important spin effect, the interaction of the electron
spin with magnetic impurities, because for the conditions relevant
to the measurements discussed in this text the magnetic field is strong
enough to polarize the magnetic impurities and weaken their ability
to interact with the electron spins.%
} is that of spin-orbit scattering.%
\footnote{As is the case with much of this appendix, this section is adapted
from Ref. \citealp{akkermans2007mesoscopic}.%
} Gauge invariance of the electromagnetic field requires that a pure
electric field $\boldsymbol{E}$ in one inertial reference frame becomes
a superposition of electric and magnetic fields in another reference
frame moving at velocity $\boldsymbol{v}$ relative to the first,
with the magnetic field given by $\boldsymbol{B}=-\boldsymbol{v}\times\boldsymbol{E}/c^{2}$.
An electron moving at velocity $\dot{\boldsymbol{r}}$ through a disorder
potential $V$ experiences an electric field given by $-\nabla V/e$
and thus a magnetic field $\dot{\boldsymbol{r}}\times\nabla V/ec^{2}$
in its rest frame. The spin of the electron couples to this magnetic
field and thus to the disorder potential. The Zeeman energy contribution
to the Hamiltonian (Eq. \ref{eq:AppGrFu_HZ}) for this effective magnetic
field takes the form
\[
\hat{H}_{SO}=\frac{g\mu_{B}}{2ec^{2}}\boldsymbol{\sigma}\cdot\left(\dot{\boldsymbol{r}}\times\nabla V\right),
\]
which for spin states $\alpha$ and $\beta$ has matrix elements of
the form
\begin{align*}
\left\langle \boldsymbol{k}'\beta\left|\hat{H}_{SO}\right|\boldsymbol{k}\alpha\right\rangle  & =\frac{g\mu_{B}}{2ec^{2}}\left\langle \beta\left|\boldsymbol{\sigma}\right|\alpha\right\rangle \cdot\left\langle \boldsymbol{k}'\left|\left(\dot{\boldsymbol{r}}\times\nabla V\right)\right|\boldsymbol{k}\right\rangle \\
 & =\frac{g\mu_{B}}{2ec^{2}}\left\langle \beta\left|\boldsymbol{\sigma}\right|\alpha\right\rangle \cdot\left(\frac{\hbar\boldsymbol{k}'}{m}\times i\boldsymbol{k}V\left(\boldsymbol{k}-\boldsymbol{k}'\right)\right)\\
 & =\left(i\frac{g\mu_{B}\hbar}{2emc^{2}}k'kV\left(\boldsymbol{k}-\boldsymbol{k}'\right)\right)\boldsymbol{\sigma}_{\beta\alpha}\cdot\left(\tilde{\boldsymbol{k}}'\times\tilde{\boldsymbol{k}}\right).
\end{align*}
The prefactor of this last term can be treated as new disorder potential
$V_{SO}$ which depends on the spin of the electrons in addition to
the wavevectors. 

The derivations of the diffuson, cooperon and density of states correlation
function of the preceding sections can be repeated adding this new
disorder potential to the original one which was independent of spin.
The major difference in the derivation is that now components of the
electron spins must be tracked. This means that, for example, in Fig.
\ref{fig:AppGrFu_CooperonDrawing} the solid and dashed lines entering
the $r_{i}$ must be assigned spin components $\alpha$ and $\beta$,
and the lines exiting must be assigned spin components $\gamma$ and
$\delta$, since the potential $V_{SO}$ can mix spin components.
The derivation including spin-orbit scattering is best performed in
reciprocal space, rather than real space which was used in the derivation
above. Rather than develop this framework to present the full derivation
(see e.g. Ref. \citealp{akkermans2007mesoscopic}), we simply state
the results. Eq. \ref{eq:AppGrFu_DiffusonCooperonEigenvalues} for
the eigenvalues of the diffuson%
\footnote{\label{fn:AppGrFu_CooperonZeemanSO}The cooperon term is slightly
different because the reversed paths have opposite wavevectors and
thus experience the opposite effective magnetic field form of the
spin-orbit interaction. The operator analogous to $H_{D,SO}$ is
\[
H_{C,SO}=\frac{2}{3D\tau_{SO}}\left[\begin{array}{cccc}
2 & 0 & 0 & 0\\
0 & 1 & -1 & 0\\
0 & -1 & 1 & 0\\
0 & 0 & 0 & 2
\end{array}\right].
\]
This term leads to the mixing of the $|\uparrow\downarrow\rangle$
and $|\downarrow\uparrow\rangle$ spin pairs and so to an interplay
with the Zeeman shifts experienced by these pairs. Incorporating the
Zeeman terms by changing the 1's on the diagonal to $1+(3D\tau_{SO}/2)2iE_{Z}$
and $1-(3D\tau_{SO}/2)2iE_{Z}$, the eigenvalues are found to be $4/3D\tau_{SO}$,
$4/3D\tau_{SO}$, and 
\[
\frac{2}{3D\tau_{SO}}\pm\sqrt{\left(\frac{2}{3D\tau_{SO}}\right)^{2}-4\left(\frac{E_{Z}}{\hbar D}\right)^{2}}.
\]
} is modified to include spin, taking the form
\[
\left(\left(\nabla'+i\frac{e}{\hbar}\boldsymbol{A}_{\mp}\right)^{2}+H_{D,SO}\right)\boldsymbol{P}_{d}\left(\boldsymbol{r},\boldsymbol{r}',\omega\right)=E_{n}^{d}\left(B_{\mp}\right)\boldsymbol{P}_{d}\left(\boldsymbol{r},\boldsymbol{r}',\omega\right)
\]
where the diffuson $\boldsymbol{P}_{d}(\boldsymbol{r},\boldsymbol{r}',\omega)$
is now a spinor and in the $\{|\uparrow\uparrow\rangle,|\uparrow\downarrow\rangle,|\downarrow\uparrow\rangle,|\downarrow\downarrow\rangle\}$
basis $H_{D,SO}$ takes the form
\[
H_{D,SO}=\frac{2}{3D\tau_{SO}}\left[\begin{array}{cccc}
1 & 0 & 0 & -1\\
0 & 2 & 0 & 0\\
0 & 0 & 2 & 0\\
-1 & 0 & 0 & 1
\end{array}\right]
\]
where $\tau_{SO}$ can be defined from the disorder average of $V_{SO}$
in a similar manner to $\tau_{e}$ being defined for $V$ in Eq. \ref{eq:AppGrFu_TauK}.
The unnormalized eigenvectors are $|\uparrow\uparrow\rangle+|\downarrow\downarrow\rangle$,
$|\uparrow\uparrow\rangle-|\downarrow\downarrow\rangle$, $|\uparrow\downarrow\rangle$,
and $|\downarrow\uparrow\rangle$, and the corresponding eigenvalues
are 0, $4/3D\tau_{SO}$, $4/3D\tau_{SO},$ and $4/3D\tau_{SO}$. Comparing
with the spin pairings for the Zeeman splitting, we see that the two
terms shifted by $\pm E_{Z}$ are eigenvectors of $H_{D,SO}$ both
with eigenvalue $4/3D\tau_{SO}$, while the two terms unaffected by
the Zeeman splitting are now split by the spin-orbit scattering. Incorporating
these eigenvalues, the full form for the diffuson contribution to
the density of states correlation function becomes
\begin{align}
 & \left\langle \nu\left(\varepsilon,B\right)\nu\left(\varepsilon-\hbar\omega,B'\right)\right\rangle _{d}=\nonumber \\
 & \phantom{==}=2\left(\frac{1}{2\pi\hbar}\right)^{2}\sum_{\mp}\sum_{n}\text{Re}\Bigg(\left(\frac{1}{i\omega+DE_{n}\left(B_{-}\right)}\right)^{2}+\left(\frac{1}{i\omega+\frac{4}{3\tau_{SO}}+DE_{n}\left(B_{-}\right)}\right)^{2}+\nonumber \\
 & \phantom{===2\left(\frac{1}{2\pi\hbar}\right)^{2}\sum_{\mp}\sum_{n}\text{Re}\Bigg(}\ldots+\left(\frac{1}{i\omega+2i\frac{E_{Z}}{\hbar}+\frac{4}{3\tau_{SO}}+DE_{n}\left(B_{-}\right)}\right)^{2}\nonumber \\
 & \phantom{===2\left(\frac{1}{2\pi\hbar}\right)^{2}\sum_{\mp}\sum_{n}\text{Re}\Bigg(}\ldots+\left(\frac{1}{i\omega-2i\frac{E_{Z}}{\hbar}+\frac{4}{3\tau_{SO}}+DE_{n}\left(B_{-}\right)}\right)^{2}\Bigg).\label{eq:AppGrFu_NuNuZSO}
\end{align}

\chapter{\label{app:AppCanonPert}Classical perturbation theory using action-angle
variables}

\section{Classical mechanics formalism}

In this appendix, we will outline the steps involved in calculating
the change in frequency of a periodic system due to a perturbation
using the action-angle variables in the Hamilton-Jacobi formalism.
To that end, we begin in this section by briefly reviewing the basic
concepts of the Hamilton-Jacobi formalism and action-angle variables.
For more detail, see \citep{goldstein2001classical}.%
\footnote{I recommend looking at the second edition as well as the third. The
second edition covers classical perturbation theory in one dimension
which is all that is needed for deriving the finite amplitude correction
to the persistent current signal and is much simpler notationally
than the higher dimensional case.%
}

\subsection{Hamiltonian mechanics}

In one dimension, the classical Hamiltonian $H=H(q,p)$ can be obtained
from the Lagrangian $L=T-V$ by the Legendre transformation $H=p\dot{q}-L$.
Here $q$ is a generalized coordinate with canonically conjugate momentum
$p=\partial_{\dot{q}}L$, and $T$ and $V$ are the kinetic and potential
energies of the system under consideration.%
\footnote{Although our actual perturbation ultimately arises from electrons
moving in a magnetic field, we will treat the perturbation solely
as a modification to the analytic function defining the cantilever's
potential energy landscape and so ignore modifications to the Lagrangian
and Hamiltonian relevant for the case of a charged particle in an
applied magnetic field. For simplicity, we also ignore the case of
$H$ depending explicitly on time.%
} The modified Hamilton's variational principle,
\[
\delta\intop_{t_{1}}^{t_{2}}dt\,\left(p\dot{q-H\left(q,p\right)}\right)=0,
\]
 leads to Hamilton's equations of motion,
\begin{eqnarray*}
\dot{q} & = & \frac{\partial H}{\partial p}\\
\dot{p} & = & -\frac{\partial H}{\partial q}.
\end{eqnarray*}

\subsection{Canonical transformations}

The key concept behind the Hamilton-Jacobi formalism and action-angle
variables is the canonical transformation. A canonical transformation
is a transformation from coordinates $(q,p)$ to coordinates $(\eta,J)$
such that the new coordinates satisfy Hamilton's equations of motion
\begin{eqnarray*}
\dot{\eta} & = & \frac{\partial K}{\partial J}\\
\dot{J} & = & -\frac{\partial K}{\partial\eta}
\end{eqnarray*}
 for some function $K(\eta,J)$. Generally, this kind of coordinate
transformation requires
\begin{equation}
p\dot{q}-H=J\dot{\eta}-K+\frac{dF}{dt}\label{eq:AppCanPertCanonicalTransformation}
\end{equation}
so that the modified Hamilton's principle holds in both coordinates
systems.%
\footnote{Because the end points in time $t_{1}$ and $t_{2}$ are fixed, $dF/dt$
does not change the variation.%
} Typically, $F$ is written as a function of one of the old coordinates
and one of the new coordinates so that it connects the two coordinate
systems. We will restrict ourselves to the case where $F$ has no
explicit time dependence ($\partial_{t}F=0$). 

The simplest form of $F$ is $F=F(q,\eta)$. In this case, Eq. \eqref{eq:AppCanPertCanonicalTransformation}
becomes
\[
p\dot{q}-H=J\dot{\eta}-K+\frac{\partial F}{\partial q}\dot{q}+\frac{\partial F}{\partial\eta}\dot{\eta}.
\]
 This relation only holds provided that
\begin{eqnarray*}
p & = & \frac{\partial F}{\partial q}\\
J & = & -\frac{\partial F}{\partial\eta}\\
K & = & H.
\end{eqnarray*}
Other pairs of coordinates can be used but require Legendre transformations
in order to follow the same form as $F(q,\eta)$. For the action-angle
variable analysis to follow, we will use the transformation function
$W(q,J)$, known as Hamilton's characteristic function in the literature,
for which $F=W(q,J)-J\eta$. With this transformation function, Eq.
\eqref{eq:AppCanPertCanonicalTransformation} takes the form
\[
p\dot{q}-H=J\dot{\eta}-K+\frac{\partial W}{\partial q}\dot{q}+\frac{\partial W}{\partial P}\dot{J}-J\dot{\eta}-\eta\dot{J}
\]
 which is satisfied when
\begin{eqnarray}
p & = & \frac{\partial W}{\partial q}\label{eq:AppCanonPertpWq}\\
\eta & = & \frac{\partial W}{\partial J}\label{eq:AppCanonPertEtaWJ}\\
K & = & H.
\end{eqnarray}

\subsection{Hamilton-Jacobi formalism}

In the Hamilton-Jacobi formalism, a transformation is chosen to make
the new Hamiltonian $K$ take a simple form. With our restricted assumptions
of one degree of freedom and no explicit time dependence in $H$,
the transformation we seek is one where $K=E(J)$ for some function
$E$ (with $E(J)$ constant because $J$ is). In this case where the
generalized coordinate $\eta$ does not appear in the Hamiltonian,
it is called \emph{cyclic}. 

Hamilton's equations of motion are easily solved in the case of cyclic
coordinates. The generalized momentum obeys
\[
\dot{J}=-\frac{\partial K}{\partial\eta}=0,
\]
 which requires $J$ to be a constant. The equation of motion for
the generalized position, 
\[
\dot{\eta}=\frac{\partial E}{\partial J}\equiv f_{0},
\]
 has the solution 
\begin{equation}
\eta=f_{0}t+\eta_{0}.\label{eq:AppCanonPertEtaTimeDependence}
\end{equation}
Given the initial conditions $(q_{0},p_{0})$ at $t=t_{0}$, the relation
$p=\partial_{q}W(q,J)$ can be inverted to give $J=J(q_{0},p_{0})$,
which can then be combined with the relation $\eta=\partial_{P}W(q,J)$
to give $\eta_{0}=\eta_{0}(q_{0},p_{0})$. In essence, this procedure
shifts the problem from solving the original equations of motion to
solving the Hamilton-Jacobi equation
\[
H\left(q,\frac{\partial W}{\partial q}\right)=E\left(J\right)
\]
for $W(q,J)$ (the constant of integration involved in going from
$\partial_{q}W$ to $W$ is trivial since only derivatives of $W$
are needed to obtain the equations of motion). The solutions of the
equations of motion for $(\eta,J)$ were derived above. The relations
$\eta=\partial_{J}W(q,J)$ and $p=\partial_{q}W(q,J)$ can be inverted
to give $q(\eta,J)$ and $p(\eta,J)$.

\subsection{Action-angle variables}

Action-angle variables are a particular choice of coordinates obtained
using the Hamilton-Jacobi procedure for the case of a periodic system.
By a {}``periodic system,'' we mean one where the solutions of the
equations of motion trace out closed curves in $(q,p)$ phase space.
The angle variable is the transformed position, while the action variable
is its canonically conjugate momentum. We define the action variable%
\footnote{This name comes from the usage of this integral in the form $\int dt\,\dot{q}p$
in the formulation of the principle of least action. It has the units
of angular momentum. The angle coordinate is so named because the
canonical conjugate of angular momentum is an angle.%
} by
\begin{equation}
J=\oint dq\, p\left(q,E\right)+A_{0}\label{eq:AppCanonPertActionDefinition}
\end{equation}
where $p(q,E)$ comes from inverting $H(q,p)=E$, and $A_{0}$ can
be any constant independent of $q$ and $p$. As usual, Hamilton's
characteristic function $W(q,J)$ connects the original coordinates
to the action-angle variables.

The usefulness of the action-angle variables is revealed by considering
the change in the angle variable over one period,
\begin{eqnarray*}
\Delta\eta & = & \oint dq\,\frac{\partial\eta}{\partial q}\\
 & = & \oint dq\,\frac{\partial^{2}W}{\partial q\partial J}\\
 & = & \frac{d}{dJ}\oint dq\,\frac{\partial W}{\partial q}\\
 & = & \frac{d}{dJ}\oint dq\, p\\
 & = & \frac{d}{dJ}\left(J-A_{0}\right)\\
 & = & 1
\end{eqnarray*}
where we have used Eqs. \eqref{eq:AppCanonPertpWq}, \eqref{eq:AppCanonPertEtaWJ},
and \eqref{eq:AppCanonPertActionDefinition} and the fact that $J$
is independent of $q$. From the equation of motion for $\eta$, Eq.
\eqref{eq:AppCanonPertActionDefinition}, we have $1=f_{0}\tau$ where
$\tau$ is the period of the system in units of time. Thus we see
that $f_{0}=\partial_{J}E$ is the frequency of the system. In principle,
this procedure allows one to determine the frequency of the system
without solving the equations of motion.

\subsection{Action-angle variables of the simple harmonic oscillator}

Now we will briefly set up the action-angle formalism for the case
of a simple harmonic oscillator for application to the study of the
cantilever's motion discussed in Chapter \eqref{cha:Cantilever-torsional-magnetometry}.
The Hamiltonian 
\[
H_{0}\left(q,p\right)=\frac{p^{2}}{2m}+\frac{1}{2}m\omega_{0}^{2}q^{2}=E
\]
can be rearranged as 
\[
p=\sqrt{2mE-m^{2}\omega_{0}^{2}q^{2}}
\]
with $m$ the mass of the oscillator. The action variable can then
be written as
\begin{eqnarray*}
J & = & \oint dq\, p\\
 & = & 4\int_{0}^{\sqrt{\frac{2E}{m\omega_{0}^{2}}}}dq\,\sqrt{2mE-m^{2}\omega_{0}^{2}q^{2}}\\
 & = & \frac{2\pi E}{\omega_{0}}.
\end{eqnarray*}
 where we have used the fact that $dq\, p>0$ to rewrite the integral
over a quarter of a period. We see immediately that the frequency
is $f_{0}=\partial_{J}E=\omega_{0}/2\pi$ as expected. We note that
$J$ can also be written in terms of the maximum displacement from
equilibrium $q_{\text{max}}$ as 
\[
J=\pi m\omega_{0}q_{\text{max}}^{2}.
\]
 Using the relation for the angle variable $\eta=f_{0}t+\eta_{0}$
found above in Eq. \ref{eq:AppCanonPertEtaTimeDependence}, the well
known solutions 
\begin{eqnarray*}
q\left(t\right) & = & \sqrt{\frac{2E}{m\omega_{0}^{2}}}\sin\left(2\pi f_{0}t+\phi_{0}\right)\\
p\left(t\right) & = & \sqrt{2mE}\cos\left(2\pi f_{0}t+\phi_{0}\right)
\end{eqnarray*}
 can be rewritten in terms of the action-angle variables as 
\begin{eqnarray}
q & = & \sqrt{\frac{J}{2\pi^{2}mf_{0}}}\sin2\pi\eta\label{eq:AppCanonPertqInTermsJEta}\\
p & = & \sqrt{2mf_{0}J}\cos2\pi\eta\nonumber 
\end{eqnarray}
 where we have set $\phi_{0}=2\pi\eta_{0}$.

\section{Perturbation theory with action-angle variables}

We will now review the calculation of the frequency of a periodic
system subject to a small perturbation. An analysis using action-angle
variables is particularly well suited to this task since the frequency
of the system has a particularly simple form. We write the Hamiltonian
in the original coordinate system as
\[
H\left(q,p,\varepsilon\right)=H_{0}\left(q,p\right)+\varepsilon H_{1}\left(q,p\right)
\]
where $H_{0}(q,p)$ is the Hamiltonian describing the unperturbed
periodic system and $\varepsilon$ is a dimensionless variable parametrizing
the strength of the perturbation. The unperturbed Hamiltonian $H_{0}$
can be transformed into $K(\eta,J)$ where $\eta$ and $J$ are the
angle and action variables described in the preceding section. For
small perturbations the system remains periodic and the perturbed
Hamiltonian $H$ can be transformed to $\mathcal{K}(\zeta,j)$ where
$\zeta$ and $j$ are now the angle and action variables for the system
described by $H$. The transformation $(q,p)\rightarrow(\eta,J)$
is just a set of analytic relations between variables independent
of the form of the Hamiltonian and so remains canonical when the perturbation
is added. By the construction of $(\eta,J)$, $q(\eta,J)$ and $p(\eta,J)$
are periodic in $\eta$ with period 1. However, when $\varepsilon\neq0$,
$(\eta,J)$ are no longer action-angle variables. It is thus no longer
guaranteed that $J$ is a constant nor that $\eta$ is linear in time.

In the presence of the perturbation, the Hamiltonians of both the
old and new action-angle variables are functions of $\varepsilon$
and can be expanded as 
\[
K\left(\eta,J,\varepsilon\right)=K_{0}\left(J\right)+\varepsilon K_{1}\left(\eta,J\right)+\varepsilon^{2}K_{2}\left(\eta,J\right)+\ldots
\]
 and
\[
\mathcal{K}\left(j,\varepsilon\right)=\mathcal{K}_{0}\left(j\right)+\varepsilon\mathcal{K}_{1}\left(j\right)+\varepsilon^{2}\mathcal{K}_{2}\left(j\right)+\ldots
\]
where $\mathcal{K}$ is independent of $\zeta$ because it was constructed
to be the Hamiltonian for the action-angle variables of the perturbed
system. The two sets of coordinates are related by a canonical transformation
which we can write as $W(\eta,j)$ in keeping with the notation developed
in the previous section. This function also admits an expansion in
$\varepsilon$:
\[
W\left(\eta,j,\varepsilon\right)=\eta j+\varepsilon W_{1}\left(\eta,j\right)+\varepsilon^{2}W_{2}\left(\eta,j\right)+\ldots
\]
 where the form of the first term has been constructed so that $W$
gives the trivial transformation $J\rightarrow j$, $\eta\rightarrow\zeta$
when $\varepsilon=0$. 

Now we will solve for the first correction to the frequency $\partial_{j}\mathcal{K}_{1}$.
We make use of the fact that $K$ and $\mathcal{K}$ must be equal
for each order of $\varepsilon.$ Using $J\approx j+\varepsilon\partial_{\eta}W_{1}(\eta,j)$,
we can rewrite $K$ to first order in $\varepsilon$ as
\begin{eqnarray*}
K\left(\eta,j,\varepsilon\right) & \approx & K_{0}\left(j+\varepsilon\partial_{\eta}W_{1}\left(\eta,j\right)\right)+\varepsilon K_{1}\left(\eta,j+\varepsilon\partial_{\eta}W_{1}\left(\eta,j\right)\right)\\
 & \approx & K_{0}(j)+\varepsilon\left[\partial_{j}K_{0}\left(j\right)\partial_{\eta}W_{1}\left(\eta,j\right)+K_{1}\left(\eta,j\right)\right].
\end{eqnarray*}
 Equating powers of $\varepsilon$, we have 
\begin{eqnarray}
\mathcal{K}_{0}\left(j\right) & = & K_{0}\left(j\right)\nonumber \\
\mathcal{K}_{1}\left(j\right) & = & \partial_{j}K_{0}\left(j\right)\partial_{\eta}W_{1}\left(\eta,j\right)+K_{1}\left(\eta,j\right).\label{eq:AppCanonPertHamiltonianPertExpansionTerms}
\end{eqnarray}
To carry the analysis further, we need to consider closely the properties
of $W_{1}(\eta,j)$. When the system moves through one period, the
angle variables must satisfy $\zeta\rightarrow\zeta+1$ and $\eta\rightarrow\eta+1$
for all values of $\varepsilon$. Since 
\[
\zeta=\eta+\varepsilon\partial_{j}W_{1}\left(\eta,j\right)+\ldots\,,
\]
it must hold that $\partial_{j}W_{m}$ is periodic in $\eta$ with
period 1 for each $m$. Similarly, although $J$ is no longer constant
in the presence of the perturbation, it can still be written in terms
of $q$ and $p$. The relation 
\begin{equation}
J=j+\varepsilon\partial_{\eta}W_{1}+\ldots\label{eq:AppCanonPertJjActionExpansion}
\end{equation}
 thus implies that $\partial_{\eta}W_{m}$ is also periodic in $\eta$
with period 1 for each $m$. The Fourier expansion 
\begin{equation}
\partial_{\eta}W_{m}=\sum_{n}C_{n}^{m}(j)e^{2\pi in\eta}\label{eq:AppCanonPertWFourierExpansion}
\end{equation}
 can be integrated to give
\[
W_{m}=\sum_{n\neq0}\frac{C_{n}^{m}(j)}{2\pi in}e^{2\pi in\eta}+D^{m}(j)+C_{0}^{m}(j)\eta.
\]
 Taking the derivative with respect to $j$, we find
\[
\partial_{j}W_{m}=\sum_{n\neq0}\frac{\partial_{j}C_{n}^{m}(j)}{2\pi in}e^{2\pi in\eta}+\partial_{j}D^{m}(j)+\partial_{j}C_{0}^{m}(j)\eta.
\]
For $\partial_{j}W_{m}$ to be periodic in $\eta$, the quantity $\partial_{j}C_{0}^{m}$
must be zero, and thus $C_{0}^{m}$ must be independent of $\eta$
and $j$. From Eqs. \ref{eq:AppCanonPertJjActionExpansion} and \ref{eq:AppCanonPertWFourierExpansion},
we see that all of the $C_{0}^{m}$ effectively shift $J$ by a constant
that depends only on $\varepsilon$. The action variable $J$ as defined
in Eq. \ref{eq:AppCanonPertActionDefinition} includes an arbitrary
offset $A$. We are thus free to shift $J$ appropriately to set all
$C_{0}^{m}=0$.

With all $C_{0}^{m}=0$, we can now evaluate Eq. \ref{eq:AppCanonPertHamiltonianPertExpansionTerms}
by average both sides over a period. This averaging gives 
\begin{eqnarray}
\mathcal{K}_{1}\left(j\right) & = & \int_{0}^{1}d\eta\left(\partial_{j}K_{0}\left(j\right)\partial_{\eta}W_{1}\left(\eta,j\right)+K_{1}\left(\eta,j\right)\right)\nonumber \\
 & = & \int_{0}^{1}d\eta K_{1}\left(\eta,j\right).\label{eq:AppCanonPertK1average}
\end{eqnarray}
To first order, the correction to the frequency of the system is thus
\begin{equation}
\delta f_{0}\approx\varepsilon\frac{\partial}{\partial j}\int_{0}^{1}d\eta K_{1}\left(\eta,j\right).\label{eq:AppCanonPertFreqShiftFormula}
\end{equation}
From Eqs. \ref{eq:AppCanonPertHamiltonianPertExpansionTerms} and
\ref{eq:AppCanonPertK1average}, it is possible to solve for $\partial_{\eta}W_{1}$.
To find the second order correction $\mathcal{K}_{2}$, all that is
needed is to carry out the Taylor expansions to second order and use
the values of $\mathcal{K}_{1}(j)$ and $\partial_{\eta}W_{1}(\eta,j)$
from the first order calculation. Successively higher order terms
can be found by iterating this procedure.

\chapter{\label{app:AppSampFab}Persistent current cantilever-with-ring sample
fabrication}

The actual samples discussed in this text were fabricated at the Cornell
NanoScale Facility (CNF) at Cornell University, Ithaca, NY, USA except
for the aluminum deposition which took place in the Devoret/Schoelkopf
thermal evaporator in Becton Engineering and Applied Science Center,
Yale University, New Haven, CT, USA. The details of the recipe were
developed largely by Ania Jayich and Rob Ilic with guidance from Jack
Harris and some input from myself. Some preliminary work was performed
at Yale with the assistance of Luigi Frunzio and use of the Devoret
lab's FEI scanning electron microscope (SEM) (FEI Company, Hillsboro,
OR, USA). 

Below is a detailed recipe for the single crystal silicon cantilevers
with integrated rings reported on in this text.%
\footnote{Ring/wire feature sizes obtained by this recipe were 80-130 nm. In
order to achieve smaller feature sizes, a thinner e-beam resist is
likely needed. To obtain a smooth coat of a thinner resist it would
be preferable to perform the e-beam lithography before the frontside
photolithography for the cantilevers. We attempted one run of this
alternate recipe. Because we did not achieve the desired feature sizes
(we got $\sim$100 nm rather than the target 30 nm, possibly due to
overexposure with the e-beam), I will not include the alternate recipe
in this thesis. Further refinements to our recipe are necessary to
achieve these smaller features. When aiming for the smallest possible
feature sizes with e-beam lithography, it is common practice to write
on suspended membranes to minimize electron backscattering off the
substrate. It is worth considering reversing the major steps in the
recipe below, performing the backside etch first, then writing the
frontside e-beam pattern and depositing the metal, and defining the
cantilevers with frontside photolithography last.%
} Samples were made over several iterations with slight tweaks to the
recipe each time. The recipe as presented represents a recommended
fabrication procedure for the production of future samples rather
than a step-by-step record of any one previous fabrication run. All
numbers in the recipe below should be treated as starting points in
future fabrication processes. When using a new tool for the first
time or an old tool after a long lapse, it is highly recommended to
consult with the tool manager about recent tool performance and calibrations
and to adjust the recipe accordingly. Also, when using a chemical
unfamiliar to you, always check for its container compatibility before
pouring it!

\section{\label{sec:APPsampFab_recipe}Cantilever-with-ring fabrication recipe}
\begin{figure}

\begin{centering}
\includegraphics[width=0.7\paperwidth]{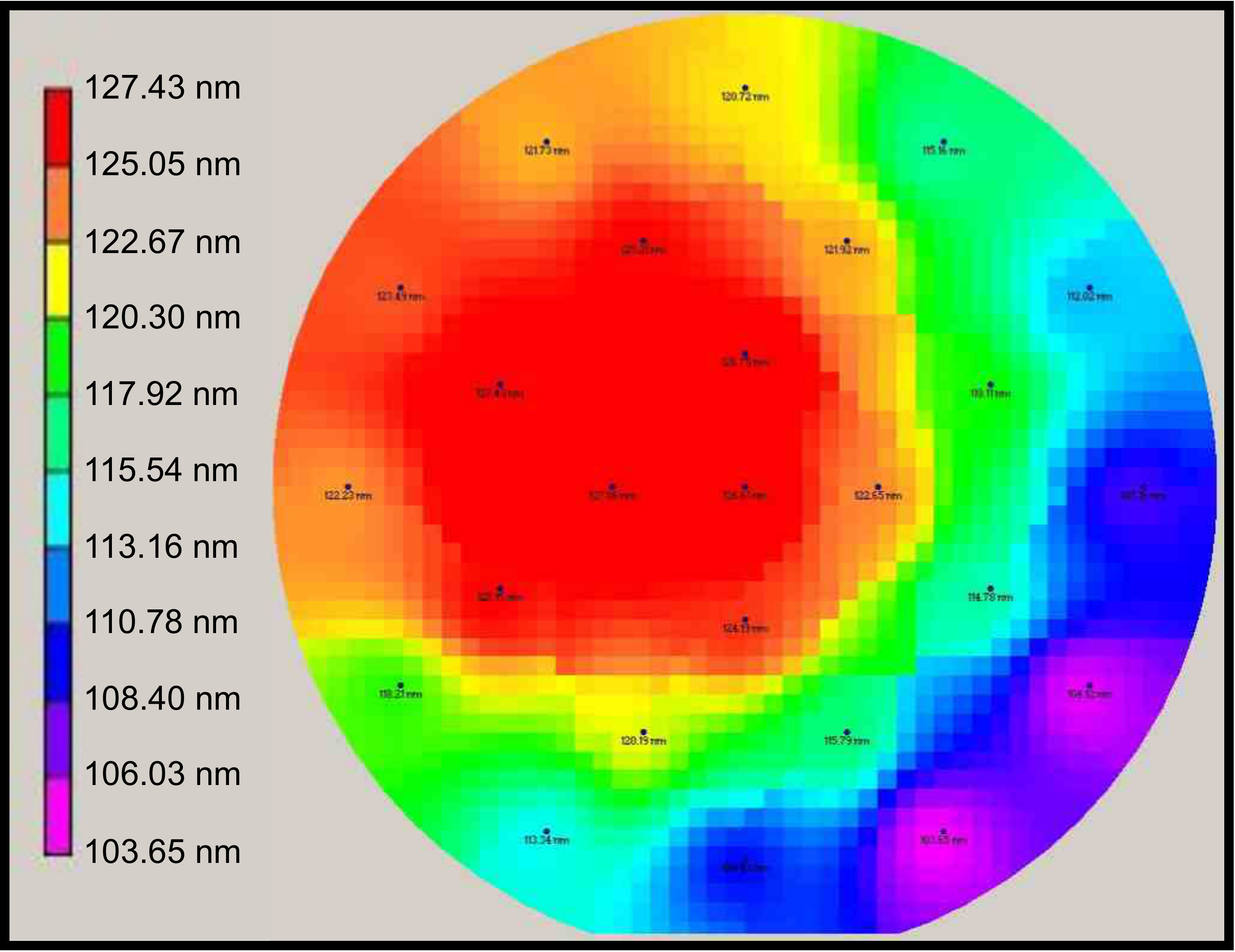}
\par\end{centering}

\begin{centering}
\caption[Wafer map displaying typical SOI wafer uniformity]{\label{fig:AppSampFab_waferMap}Wafer map displaying typical Soitec
SOI wafer uniformity. The wafer shown began with a $340\,$nm device
layer and was thinned down to $\sim114\,$nm. The $25\,$nm spread
in thickness is typical for these wafers and can actually be a bit
smaller in the wafers prior to thinning. Wafers obtained from other
suppliers that were thinned down from an initial thickness of $1\,\mu$m
had much greater variations in their final thicknesses.}

\par\end{centering}

\end{figure}

\begin{figure}
\begin{centering}
\includegraphics[width=0.7\paperwidth]{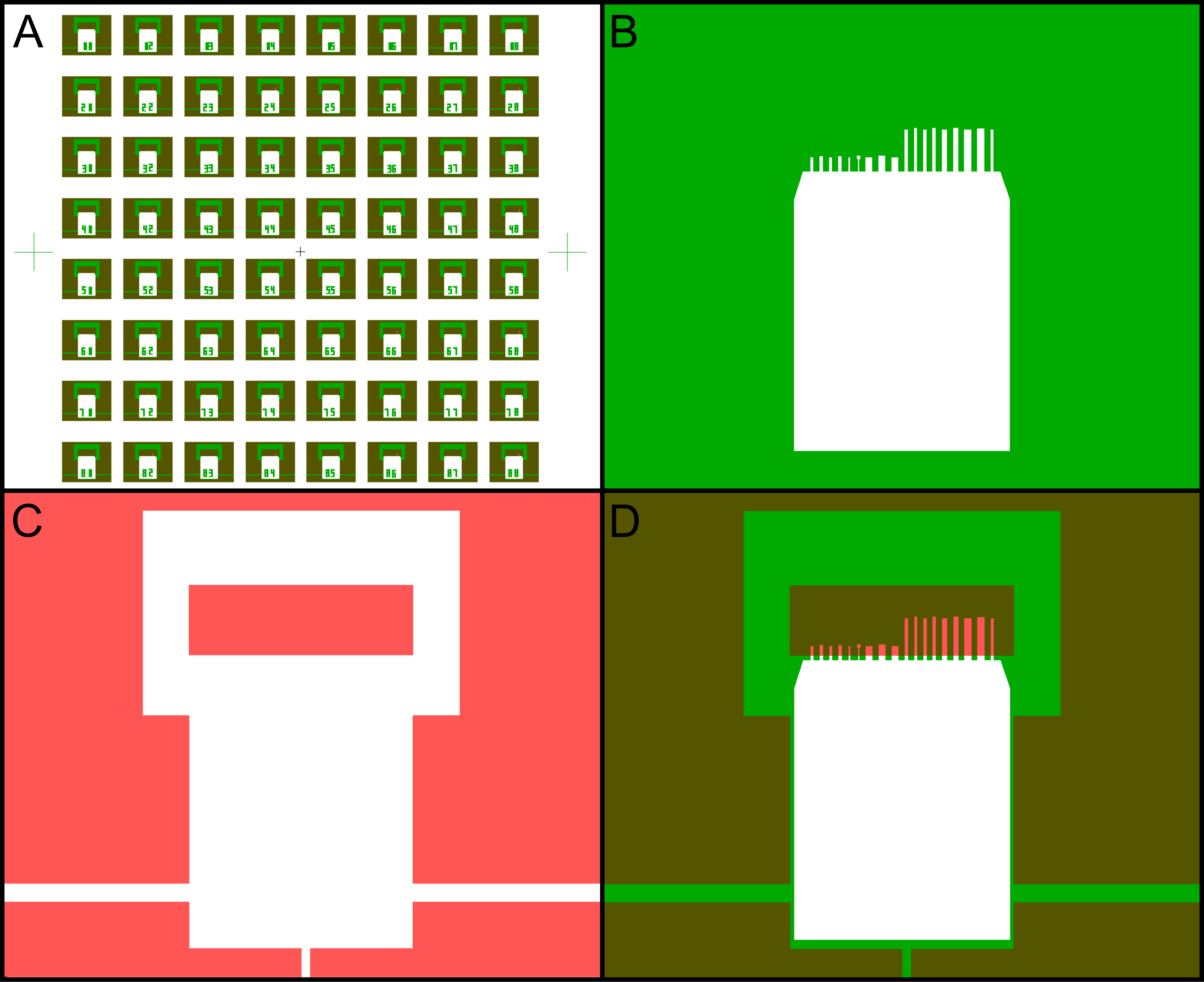}\caption[Images of cantilever sample mask designs]{\label{fig:AppSampFab_LEditMask}Images of cantilever sample mask
designs. Panel A shows the wafer die with both frontside and backside
patterns overlaid in a manner similar to panel D. Global alignment
marks are visible on the two sides of the pattern (see the note accompanying
step \ref{enu:AppSampFab_AlignmentMarks} of Section \ref{sec:APPsampFab_recipe}
for a suggestion for better alignment mark design). Panel B shows
the frontside pattern of one sample chip. The colored area is etched
away during frontside fabrication. The samples discussed in Chapter
\ref{cha:Data} were part of a chip made with this pattern. Panel
C shows the backside pattern of each sample chip. Again, the colored
area is etched away (during the deep reactive ion etch step). The
beams near the bottom of the chip hold the chip to the wafer after
this etch step. Panel D shows the frontside and backside etch masks
overlaid. In later iterations of the sample mask, cantilevers were
added to the top of the chip window to make more efficient use of
space and increase the number of samples available in one cooldown.
The cantilever bases on the frontside mask were also pulled further
down the chip to make the design less susceptible to overetching of
the window during the deep reactive ion etch.}

\par\end{centering}

\end{figure}

\begin{enumerate}
\item Photolithography mask creation

\begin{enumerate}
\item Create mask designs for all three stages of lithography (frontside
photolithography, backside photolithography, electron beam (e-beam)
lithography) by computer.

\begin{enumerate}
\item Lay out mask designs with a computer aided design (CAD) program (See
Fig. \ref{fig:AppSampFab_LEditMask}). We used the L-Edit Pro software
(Tanner EDA, www.tannereda.com, Monrovia, CA, USA).
\item \label{enu:AppSampFab_AlignmentMarks}Put global alignment marks on
the mask of each stage of lithography. These marks should line up
on each mask so that the sequential lithography steps can be aligned
to each other. We used two crosses, located on opposite sides of the
wafer. %
\footnote{A convenient choice for alignment marks on the photolithography masks
is a large window to be etched from the mask with a large unetched
cross centered inside of it. This configuration provides a large transparent
viewing area for hunting for marks on different layers when performing
alignment. We never did this during any of my fabrication runs. Instead
we had large transparent crosses for alignment marks. Such marks provide
only a narrow viewing window for searching for marks through the mask.
It is advisable to consult with the managers of the lithography tools
requiring alignment when choosing the dimensions and geometry of alignment
marks.%
}
\item Export masks to appropriate format for mask writing tools. In our
case, this format was the semiconductor standard GDSII.
\end{enumerate}
\item Write photolithography masks with mask writer. 

\begin{enumerate}
\item We used five inch chrome photomasks provided by the CNF. These masks
are five inch squares of glass about an eighth of an inch thick with
a thin film of chrome evaporated on one side and a layer of photoresist
on top of the chrome.
\item We used two mask writing tools at the CNF, the DWL 66 laser lithography
system (Heidelberg Instruments Mikrotechnik GmbH, Heidelberg, Germany)
and the GCA Mann 3600F pattern generator (this is an old machine.
GCA / D. W. Mann is no longer in business. Its intellectual properties
have been sold several times and are currently controlled by Ultratech,
San Jose, CA, USA). The tools have slightly different capabilities,
but our mask features were realizable in both. The deciding factor
in which mask writer was used was tool availability.
\item Many companies offer mask writing services. This step could be outsourced
if no mask writer is available.
\end{enumerate}
\item Develop mask photoresist and etch chrome.

\begin{enumerate}
\item At the CNF, these two steps can be done using the Steag-Hamatech HMP
900 mask processing system (HamaTech APE GmbH \& Co. KG, Sternenfels,
Germany). The Hamatech system contains a spinner and a chemical sprayer
and is preprogrammed with recipes for chrome photomask development
and etching.
\item For photoresist, AZ 300 MIF (AZ Electronic Materials USA Corp., Somerville,
NJ, USA) developer is used. The main working component of the developer
is tetramethylammonium hydroxide (TMAH).
\item For chrome etching, CR-14 Chromium Etchant (Cyantek Corporation, Fremont,
CA, USA) is used. The main active ingredients of the etchant are ceric
ammonium nitrate and acetic acid.
\end{enumerate}
\end{enumerate}
\item Preliminary preparation of silicon wafers

\begin{enumerate}
\item Obtain four inch diameter silicon-on-insulator wafers to fabricate
the persistent current samples. 

\begin{enumerate}
\item Some wafers used in early stages of the experiment were obtained from
Shin-Etsu (SEH America, Inc., Vancouver, WA, USA), but all of the
samples in which normal state persistent currents were actually observed
were obtained from Soitec (Soitec USA, Inc., Peabody, MA, USA).
\item Specifically, we used Soitec's Unibond wafer, part number G4P-022-01.
These wafers are manufactured with a 340 nm top silicon layer, a 1
$\mu$m thick buried silicon dioxide layer, and 450 $\mu$m silicon
handle layer. Soitec's Unibond wafer production process, which involves
bonding a silicon wafer weakened at well-defined depth by hydrogen
implantation to a handle wafer, produces thin but highly uniform top
silicon layers, well suited to wafer-scale parallel production of
cantilevers of $\sim$100 nm thickness.
\item Soitec's Unibond wafers are not mass produced. As a small-volume customer,
we were restricted in our purchases of SOI wafers to whatever was
leftover in Soitec's inventory. In one instance, we could only obtain
six inch diameter wafers. We had those wafers resized to four inch
diameter by MPE (Micro Precision Engineering, Greenville, TX, USA). 
\end{enumerate}
\item Clean wafer. We used a standard MOS clean recipe required at the CNF
before using any MOS compatible tool in the facility (such as the
furnace described in the oxidation step). The MOS clean recipe consisted
of the following steps.

\begin{enumerate}
\item Submerge wafer in 6:1:1 DI water:H$_{2}$O$_{2}$:NH$_{4}$OH bath
for 10 minutes.
\item Submerge wafer in 6:1:1 DI water:H$_{2}$O$_{2}$:HCl bath for 10
minutes.
\end{enumerate}
\item Thin down frontside silicon layer to desired cantilever thickness.
This step is not necessary if the wafers begin with an acceptable
thickness (the main experimental results reported in this text used
340 nm thick cantilevers which were not thinned down). A wafer map
of the final device layer silicon thickness (see Fig. \ref{fig:AppSampFab_waferMap})
should be recorded regardless of whether the wafer needs etching.

\begin{enumerate}
\item Oxidize wafer in thermal oxidation furnace.

\begin{enumerate}
\item $\sim$44\% of thermally grown oxide grows down into the existing
silicon. So to remove 44 nm of Si, 100 nm of SiO$_{\text{2}}$ must
be grown and etched.
\item Put wafers into furnace facing frontside-to-frontside and backside-to-backside
when processing multiple wafers. This arrangement helps to maintain
frontside polish through the oxidation process.
\item For removal of large amounts of silicon, we used the CNF's CMOS Wet
Oxidation Furnace to perform a wet oxide etch with HCl at 1000$^{\circ}$C.
These parameters grow SiO$_{2}$ at an approximate rate of 5 to 9
nm/min. Note that this rate is tool specific and nonlinear in time.
\item For removal of fine amounts of silicon, we used the CNF's CMOS Dry
Oxide Furnace to perform a dry oxide etch with HCl at 1000$^{\circ}$C.
These parameters grow SiO$_{2}$ at an approximate rate of 1 nm/min.
Note that this rate is tool specific and nonlinear in time.
\end{enumerate}
\item Etch oxide in HF for about three minutes (100 nm/min etch rate).
\item Check remaining silicon thickness. We used a microscope equipped with
the F40 optical film thickness measurement instrument (Filmetrics,
San Diego, CA, USA).
\item Iterate steps i-iii until desired frontside silicon thickness is achieved.
\end{enumerate}
\item Deposit 1.2 $\mu$m of oxide on backside of wafer.

\begin{enumerate}
\item We used the CNF's GSI PECVD (plasma-enhanced chemical vapor deposition)
tool (Ultradep, Group Sciences, Inc., San Jose, CA; this company went
out of business several years ago). The tool's standard oxide recipe
uses a temperature of 400$^{\circ}$C and a gas mix of SiH$_{4}$,
N$_{2}$O, and N$_{2}$ and produces oxide with 290 MPa compressive
stress. This recipe's nominal deposition rate is 260 nm/min.
\item The purpose of this step is to produce an extra mask layer for the
backside etch step. It is not strictly necessary if the backside photoresist
is baked sufficiently.
\item Measure the deposited oxide thickness if possible and note it for
when the oxide will be etched during the backside wafer processing.
\end{enumerate}
\end{enumerate}
\item Frontside cantilever definition (all steps performed to frontside
of wafer)

\begin{enumerate}
\item Spin $\sim$7 mL MicroPrime MP-P20 photoresist primer (Shin-Etsu MicroSi,
Phoenix, AR, USA) at 4000 rpm for 30 seconds on frontside of wafer.
\item Spin $\sim$14 mL Megaposit SPR220-3.0 photo resist (Rohm and Haas
Electronic Materials LLC, Marlboro, MA, USA) at 4000 rpm for 60 seconds
on frontside of wafer.%
\footnote{\begin{enumerate}
\item Make sure the lid is closed for all spinning steps as the composition
of the photoresist can be different when spun with the lid open, leading
to inconsistent photolithography results.\end{enumerate}
} 
\item Bake for 90 seconds at 115$^{\circ}$C on hotplate. (At the CNF, we
used the BLE-150 hotplate which has a lid and the ability to lift
the wafer off the plate when the timer is up). 
\item Expose the photoresist using the frontside mask (if this is the first
mask used on the wafer, alignment is not critical) with 12 mW/cm$^{2}$
for 5 seconds. At the CNF, we used the EV620 mask aligner (EV Group
GmbH, St. Florian am Inn, Austria) in soft contact mode.
\item Bake for 90 seconds at 115$^{\circ}$C on hotplate.
\item Develop photoresist with AZ 300 MIF for 60 seconds. At the CNF, we
used the STEAG-Hamatech HMP 900 system with a double puddle process
(recipe 6). If developing by hand, spray clean with de-ionized (DI)
water and blow dry with clean N$_{2}$ gas.
\item Examine developed photoresist in an optical microscope. If the resist
does not match the expected pattern of the mask, remove it by spinning
the wafer with acetone and try repeating steps (a) through (g).
\item Etch frontside silicon using a CF$_{4}$ reactive ion etch (RIE).

\begin{enumerate}
\item We used the Oxford PlasmaLab 80+ RIE System (Oxford Instruments, Tubney
Woods, Abingdon, Oxfordshire, UK).
\item For all RIE work, a 10 minute oxygen clean is recommended prior to
putting samples in the tool and after every $\sim$30 minutes of tool
use. The standard oxygen clean recipe for the Oxford 80 used 30 sccm
of O$_{2}$ at 60 mtorr and 150 W of RF power.
\item \label{enu:AppSampFab_CF4recipe}The standard CF$_{4}$ etch recipe
for the Oxford 80 used 30 sccm of CF$_{4}$ at 40 mtorr and 150 W
of RF power. Other tool parameters were an 80 V DC bias and -10$^{\circ}$
for the chilled house water. The nominal tool etch rate was $\sim$40
nm/min. We etched for 20 minutes to be sure the silicon was totally
gone. It is advisable to check with the tool manager about the standard
etch recipe and etch rate when using a new tool. Be careful about
overetching because the CF$_{4}$ etch can etch both Si and SiO$_{2}$.
\item We used pieces of quartz to pin the wafer in place and keep it from
sliding around on the etcher's electrode.
\end{enumerate}
\item Examine etch in optical microscope. A thin film (such as one composed
of residual silicon) should have strong color, usually a bright purple,
blue, or red. If possible, check film thicknesses with a film measurement
tool. If it appears that the silicon is not totally etched, submit
the wafer to further reactive ion etching.
\item Remove the remaining photoresist and clean the wafer.

\begin{enumerate}
\item Spin off photoresist with acetone.
\item Consider cleaning the wafer with a stronger cleaning agent. At the
CNF, we would either use the STEAG-Hamatech HMP 900 to run a wafer
cleaning recipe which spun hot Pirhana Nano-strip solution (Cyantek
Corporation, Fremont, CA, USA) on the wafer. This solution is primarily
sulfuric acid with small amounts of peroxymonosulfuric acid and hydrogen
peroxide. Alternatively, we used the two stage hot photoresist bath
which had tanks of AZ 300T photoresist stripper (AZ Electronic Materials
USA Corp., Somerville, NJ, USA), which is composed of 1-Methyl-2-pyrrolidone
(NMP), 1,2-Propanediol, and TMAH.
\end{enumerate}
\end{enumerate}
\item Electron beam lithography (all steps performed to frontside of wafer)%
\footnote{We use a bilayer e-beam resist for improved lift-off. The lower resist
layer is removed more easily by the e-beam so that an undercut is
formed below the top resist layer. The top resist layer defines the
feature size during evaporation. During lift-off, solvent is able
to enter the space between the lower resist layer and the evaporated
metal and so remove the remaining resist more easily than if the resist
were flush with the metal.%
}

\begin{enumerate}
\item Spin 5.5\% poly(methylmethacrylate) (PMMA) in Anisole (NANO$^{\text{TM}}$
495K A5.5\% PMMA Positive Radiation Sensitive Resist, Microchem, Newton,
MA, USA) at 2000 rpm for 60 seconds.
\item Bake wafer on hotplate at 170$^{\circ}$C for 10 minutes.
\item Spin 2\% PMMA in methyl isobutyl ketone (MIBK) (NANO$^{\text{TM}}$
950K M2\% PMMA Positive Radiation Sensitive Resist, Microchem, Newton,
MA, USA) at 4000 rpm for 60 seconds.
\item Bake wafer at 170$^{\circ}$C for 10 minutes
\item Write e-beam pattern with electron beam writing tool.

\begin{enumerate}
\item At the CNF, we used the JEOL JBX-9300FS Electron Beam Lithography
System (JEOL Ltd., Tokyo, Japan).
\item For fine features (namely the rings and transport measurement wires),
we used a 1.6 nA beam current and a dosage of 1200 $\mu$C/cm$^{2}$
(aperture 3 on the JEOL e-beam tool). However, our features always
ended up overexposed with fabricated linewidths about 40\% greater
than expected.
\item For big features (bond pads and leads to transport samples), we used
a 140 nA beam current and 2000 $\mu$C/cm$^{2}$ (aperture 8 on the
JEOL e-beam tool).
\item More detail might be desired regarding the e-beam writing procedure.
However, the e-beam tool is sophisticated and expensive, requiring
extensive training to be operated autonomously. During my time working
on the persistent current experiment, none of the members of the Harris
Lab achieved full independence with the tool. Some of the finer points
of the e-beam writer operation were handled by Rob Ilic and Daron
Westly of the CNF.
\end{enumerate}
\item Develop e-beam resist by dipping wafer in 1:3 MIBK:IPA (a mixture
of solvents) for 2 minutes. Use very slight agitation for $\sim$20
seconds. (Precise solvent specifications: MIBK (4-methyl-2-pentanone,
Sigma-Aldrich, Saint Louis, MO, USA) and IPA (2-propanol, Mallinckrodt
Baker, Phillipsburg, NJ, USA).
\item Descum wafer in barrel etcher (P2000 Branson International Plasma
Corp. (long out of business), Hayward, CA, USA) for 2 minutes at 150
W.
\end{enumerate}
\item Metal deposition and lift-off

\begin{enumerate}
\item Deposit metal.%
\footnote{We deposited aluminum at Yale using the Devoret/Schoelkopf PLASSYS
electron beam evaporator. This evaporator was chosen because its application
was the creation of Josephson tunnel junctions for qubits with long
coherence times. Long coherence times are linked to cleanliness of
the aluminum, which we also wanted for our persistent current samples.
The variety of materials allowed in the evaporator was highly limited
(mainly just aluminum and titanium though copper, gold, and possibly
a couple others had been used in the past), and none of them were
magnetic. Previous work has shown a link between electron phase coherence
and magnetic impurities at low magnetic field \citep{pierre2002dephasing,pierre2003dephasing},
and it has been predicted that magnetic impurities could have an effect
on persistent currents (se e.g. \citep{schwab1996impurity,schwab1997persistent,eckern2002persistent}
or Chapter \ref{cha:CHPrevWork} for more detail). We also tested
the fabrication of gold rings though we never measured them. The gold
evaporation was outsourced to Jose Aumentado, NIST, Boulder, CO, USA.
We made some preliminary lift-off tests and had mild success with
the same lift-off procedure described here for aluminum. Typically
99\% of the ring centers lifted off but that might not be good enough
when creating arrays of hundreds of thousands of rings. Also, we observed
that some of the gold rings lifted off of the silicon entirely. At
least one more wafer should be tested for lift-off using the procedure
given here as the starting point for finding one that works with gold
rings.%
} For the experiments discussed in the text 99.999\% pure aluminum
(Alfa Aesar, Ward Hill, MA, USA) was used.%
\footnote{Concentrations of some impurities as provided by Alfa Aesar: 0.5 ppm
Fe, 0.07 ppm Mn, 0.017 ppm Cr, <0.002 ppm Ni. The overall impurity
concentration was specified as 10$\pm$5 ppm.%
} Here is a detailed procedure for our aluminum deposition with the
PLASSYS electron beam evaporator (PLASSYS-BESTEK, Marolles-en Hurepoix,
France):

\begin{enumerate}
\item Mount wafer. The PLASSYS we used was design for three inch wafers.
We used double sided 1 mil Kapton$^{\textregistered}$ tape (KaptonTape.com,
Torrance, CA, USA) to secure our four inch wafers to the sample mount.
There was just barely enough clearance for the four inch wafer in
the evaporator.
\item Pump down sample load-lock and then open valve to evaporation chamber.
We usually pumped on the load-lock for at least four hours before
opening the valve to the electron beam evaporation chamber. A typical
chamber pressure was 3$\times$10$^{-8}$ torr after opening the valve.
\item Run titanium sweep of chamber. A typical chamber pressure after running
a Ti sweep was 1$\times$10$^{-8}$ torr.
\item Evaporate aluminum at $\sim$1 nm/s.
\item Wait 5 minutes for aluminum to cool.
\item Treat aluminum with static oxidation using 3 torr of O$_{2}$ for
10 minutes. The purpose of this step is to oxidize the aluminum with
a clean, controlled source of oxygen. If this step is skipped, the
aluminum will still oxidize $\sim$3-5 nm in from the surface, but
the oxidation will be uncontrolled and other impurities could enter
the aluminum as well.
\end{enumerate}
\item Lift off e-beam resist.

\begin{enumerate}
\item Put wafer in a solution of $\sim$90\% methylene chloride (MeCl)%
\footnote{Some people feel very strongly about not using MeCl, and it is banned
in some facilities. We had much greater success with MeCl than we
did with pure acetone for the aluminum lift-off.%
} (Dichloromethane, Sigma-Aldrich, Saint Louis, MO, USA) and \textasciitilde{}10\%
acetone (Mallinckrodt Baker, Inc., Phillipsburg, NJ, USA).

\begin{enumerate}
\item Holding the wafer upside down allows gravity to aid in the lift-off
process.
\item We machined a Teflon ring with lip to hold the wafer in place upside
down without touching the frontside of the wafer.
\item Use a screw-top lid for the MeCl container because it evaporates quickly.
Do not tighten the lid all the way because the evaporating MeCl can
make the lid very difficult to loosen.
\end{enumerate}
\item Let sample sit for 5 minutes. Then ultrasound container for 20 seconds.
\item Let sample sit for 1 hour. Then ultrasound container for 20 seconds.
This step can be repeated several times.
\item Let sample sit for \textasciitilde{}8 hours (typically overnight).
\item Ultrasound container for 3 minutes just before removing wafer.
\item Spray wafer with IPA while removing it from MeCl. MeCl evaporates
quickly and leaves an unwanted residue.
\item Rinse wafer in a fresh container of IPA. Ultrasound this container
for 20 seconds.
\item Blow wafer dry with clean N$_{2}$ gas.
\item Examine wafer in optical microscope.%
\footnote{\begin{enumerate}
\item Even with feature sizes smaller than the diffraction limit, it is
possible to distinguish rings for which the center has lifted off
from rings for which it has not (if you see two sorts of rings, then
one of those sorts is probably rings that have not lifted off properly).
If it appears that not all rings have lifted off, the wafer can be
returned to MeCl and further ultrasounding can be attempted. If the
wafer can be imaged before being dried off, that would be worth considering.
We had inconsistent results when trying to re-lift off the rings after
drying them off.\end{enumerate}
} It is possible to image metal features at this point with an SEM
(see Fig. \ref{fig:AppSampFab_linewidthSEM}) and an atomic force
microscope (AFM) (see Fig. \ref{fig:AppSampFab_AFMimage}) to determine
their linewidths and thickness. These measurements can also be delayed
until step \ref{enu:AppSampFab_LastRecipeStep}.
\end{enumerate}
\end{enumerate}
\item Backside wafer photolithography

\begin{enumerate}
\item Spin \textasciitilde{}7 mL MicroPrime MP-P20 photoresist primer at
4000 rpm for 30 seconds on frontside of wafer.%
\footnote{Good advice: always test the spinner at low speed before spinning
and always be aware of the location of the stop button when starting
a high speed spin. At this point in the fabrication procedure, a lot
of time has been invested into the wafer and the frontside metal is
currently exposed. Extra care is called for here.%
}

\begin{enumerate}
\item Spin Megaposit SPR220-7.0 photo resist (Rohm and Haas Electronic Materials
LLC, Marlboro, MA, USA) at 3000 rpm for 60 seconds on frontside of
wafer.%
\footnote{This resist is really thick. The easiest way to spin it on the wafer
is to pour the resist onto the wafer directly from the bottle. Pour
enough to create a circle of about 2 inch diameter in the center of
the wafer. When using this technique, always clean the edge of the
bottle top before and after pouring the resist out to prevent contamination
of the resist with dried resist residue.%
} The purpose of this resist is to protect the deposited metal during
subsequent processing steps.
\item Bake for 3 minutes at 115$^{\circ}$C using hotplate (again, at the
CNF, we used the BLE-150 hotplate).
\item Spin \textasciitilde{}7 mL MicroPrime MP-P20 photoresist primer at
4000 rpm for 30 seconds on backside of wafer.
\item Spin Megaposit SPR220-7.0 photo resist at 3000 rpm for 60 seconds
on backside of wafer. (See note on previous use of SPR220-7.0 above).
\item Bake wafer for 2 minutes at 115$^{\circ}$C. Do not put the wafer
directly onto the hotplate with photoresist on both sides! Put the
wafer facedown onto the polished face of a spare silicon wafer and
then bake with the clean side of this wafer in contact with the hotplate.
After baking, the wafers will need to be pried apart. Intentionally
misaligning the wafer flats when putting them together can help with
prying the wafers apart.
\item Expose backside of wafer with 12 mW/cm$^{2}$ for 14 seconds using
backside alignment to line up the backside mask to the features on
the front of the wafer. At the CNF, we used the EV620 mask aligner.
\item Allow the photoresist to rehydrate for at least 2 hours.%
\footnote{The wafer must be kept out of UV light at this time. If storing in
a wafer box with screw-top lid, keep the lid a little bit loose.%
}
\item Bake for 2 minutes at 115$^{\circ}$C. Use a same procedure as previous
bake step.
\item Develop photoresist in dish of AZ 300 MIF for 2 minutes. Agitate the
wafer gently by hand over this time.
\item Spray wafer with DI water and blow dry with clean N$_{2}$ gas.
\item Bake wafer in oven for 8 hours at 90$^{\circ}$C. At the CNF, we put
the wafer in a standard quartz wafer cassette to hold it in the oven.
\end{enumerate}
\end{enumerate}
\item Backside wafer etching

\begin{enumerate}
\item If PECVD oxide was deposited on the backside of the wafer:

\begin{enumerate}
\item Reactive ion etch the backside of the wafer with CF$_{4}$ in 20 minute
intervals to remove oxide (see CF$_{4}$ recipe in step \ref{enu:AppSampFab_CF4recipe}). 

\begin{enumerate}
\item Run 5-10 minute O$_{2}$ cleans in between etches (plus one before
beginning etching of course) with the wafer removed from the etcher.
\item Any color on the etched areas means that the oxide has not been fully
etched. Repeat previous steps until all oxide is removed.
\end{enumerate}
\item Check depth of windows with profilometer if desired. At the CNF, we
used a P-10 surface profiler (KLA-Tencor, Milpitas, CA, USA).
\end{enumerate}
\item Deep reactive ion etch (DRIE) the backside features (windows underneath
the cantilevers) through the handle down to the oxide. 

\begin{enumerate}
\item NOTE: at the CNF, we used the Unaxis ICP 770 deep silicon etcher (OC
Oerlikon (formerly Unaxis), Pfäffikon, Schwyz, Switzerland) and the
following detailed steps were developed based on that tool's performance. 
\item NOTE: DRIE typically employs a Bosch process which consists of a loop
of deposition of a passivation layer and a standard reactive ion etch.
The etch step is somewhat directional so that during each loop the
passivation layer is etched away from the bottom of the trench before
it is etched away from the sides. Silicon is then etched away from
the bottom. Then the passivation layer is deposited again before being
completely removed from the sides. In this way, a highly directional
etch is achieved. Ideally, the etch is highly selective to silicon
over silicon dioxide so that the oxide can serve as an etch stop and
make up for non-uniformity in etch rate across the wafer. At the CNF,
we used the standard etch recipe saved as 0trench in the Unaxis software.
Typical recipe performance was 0.7 $\mu$m per loop and 4 loops/min.
This recipe loop had three steps with the following parameters:

\begin{enumerate}
\item Deposition: 5 seconds of 70 sccm C$_{4}$F$_{8}$, 2 sccm SF$_{6}$,
100 sccm Ar at 24 mtorr with RF1 = 0.1 W and RF2 = 850 W (RF powers).
\item Etch 1: 2 seconds of 2 sccm C$_{4}$F$_{8}$, 70 sccm SF$_{6}$, 40
sccm Ar at 23 mtorr with RF1 = 8 W and RF2 = 850 W.
\item Etch 2: 5 seconds of 2 sccm C$_{4}$F$_{8}$, 100 sccm SF$_{6}$,
40 sccm Ar at 23 mtorr with RF1 = 8 W and RF2 = 850 W (RF powers).
\end{enumerate}
\item Run 20-50 loops of DRIE recipe on a junk silicon wafer. If the wafer
looks black, there is a problem with the tool that should be addressed.
Otherwise, proceed.
\item Etch \textasciitilde{}100 $\mu$m into the wafer (\textasciitilde{}150
loops of 0trench).
\item Check etch depth with profilometer. Check several regions of the wafer.
The etch rates of the center of the wafer and the perimeter might
be fairly different.
\item Etch the rest of silicon handle layer. This process may be broken
up into steps to allow for further checks of the etch rate or to work
around tool availability.
\item Inspect wafer in an optical microscope. A thin remaining layer of
silicon can appear light colored but opaque (see Fig. \ref{fig:AppSampFab_DRIEsilicon}).
The photoresist can appear to have features but is more cloudy and
transparent than the silicon.
\end{enumerate}
\end{enumerate}
\item Cantilever release

\begin{enumerate}
\item Break off some chips with tweezers. I prefer carbon coated tweezers
(758TW0000, Techni-Tool, Worcester, PA, USA) to minimize damage to
the sample chips.
\item Etch Bosch polymer residue on sample chips.

\begin{enumerate}
\item Lay down a strip of double-sided Kapton tape on the edge of a cleaved
piece of a silicon wafer. 
\item Rough up the exposed side of the tape with tweezers so it is not excessively
sticky. It can be difficult to remove the silicon chips from fresh
Kapton tape without damaging them.
\item Stick some sample chips%
\footnote{The exact number depends on the level of confidence in the subsequent
steps. As a first pass, I would recommend about 1/6 of the chips be
used in the first batch. This batch should favor the least desirable
chips (e.g. those suspected of damage or possibly on the edge of area
of metal deposition) but should also sample different areas of the
wafer. Breaking things up prevents one disaster from ruining all of
the work done up to this point.%
} cantilever-side up onto the tape with the cantilever window portion
of the chips hanging off the edge of the silicon piece (this arrangement
makes it easy to pull the samples off the tape afterwards).
\item Weakly tape the silicon pieces to a silicon wafer which you can pin
down in the RIE chamber with quartz pieces. Do not tape the pieces
directly to the RIE chamber electrode.
\item RIE the samples (after the standard O$_{2}$ clean of the chamber)
with 5 sccm CF$_{4}$, 30 sccm O$_{2}$ at 60 mtorr and 150 W of RF
power for 3-4 minutes.
\item Remove the samples from the silicon pieces and then retape them to
the silicon pieces upside down, again with the cantilever windows
hanging off of the sides (this time this arrangement is important
since the you do not want the fragile topside membrane of the cantilever
windows to touch anything).%
\footnote{A convenient way I devised for flipping the sample chips was to set
them right side up in a row on a flat surface and then to press the
silicon piece tape first down onto the sample chips so that they would
all stick to the silicon piece with the windows off the edge. I always
ended up damaging some chips when I tried the alternative of flipping
the chips one by one and sticking them to the tape upside down.%
}
\item Again tape the silicon pieces to a wafer and secure it in the RIE
chamber.
\item RIE the samples (after the standard O$_{2}$ clean of the chamber)
with 5 sccm CF$_{4}$, 30 sccm O$_{2}$ at 60 mtorr and 150 W of RF
power for 5-6 minutes.
\item The preceding etches can be varied from batch to batch of sample chips.
During early fabrication runs, it was discovered that the Bosch process
of the DRIE left a bit of residue which made it difficult for the
buffered oxide etchant (BOE) to etch the oxide and this RIE step solved
this problem. If performance of the DRIE tool changes, this step might
need to be adjusted.
\end{enumerate}
\item Etch silicon dioxide membrane with 6:1 BOE with surfactant.

\begin{enumerate}
\item Prepare the BOE with surfactant (Buffered etch 6:1 w/OHS; semi grade,
Fujifilm Electronic Materials U.S.A., Inc., North Kingstown, RI, USA). 

\begin{enumerate}
\item Pour \textasciitilde{}0.5 L into a plastic container (NOT GLASS!).
\item Add a few drops of Triton X surfactant (Triton X-100, Sciencelab.com,
Inc., Houston, TX, USA).
\item Add magnetic stirrer and put container on magnetic stirrer base stand.
Agitate the liquid steadily but weakly enough not to cause bubbles/turbulence.
\item Put stand capable of holding up the silicon pieces with the taped
sample chips over the magnetic stirrer.
\item Make sure Triton X has been dissolved into BOE. The Triton X requires
some agitation to dissolve.
\end{enumerate}
\item Put one silicon piece with sample chips taped to it onto stand in
the BOE with the backside of the sample chips facing up.%
\footnote{It is recommended to do one silicon piece at a time in the BOE as
the situation can become hectic when multiple pieces are involved.
Handling the silicon pieces with the weakly attached sample chips
in the BOE using tweezers, gloves, mask and apron can be tricky. Try
practicing with an empty silicon piece first to get a feel for how
the silicon piece will slide around on the stand in the BOE. Also,
practice taking the silicon piece out of the BOE quickly. Etch times
can get stretched out due to difficulty in grabbing the silicon pieces
quickly. Finally, be gentle with the silicon pieces in the BOE. The
tape should be only weakly holding the sample chips. It is possible
for them to fall off of the silicon pieces and get sucked into the
magnetic stirrer if handled too roughly. If this happens frequently,
do not rough up the Kapton tape so much with the tweezers (also, consider
fresh tape between Bosch etches).%
}
\item Blow out air bubble from sample chip windows with pipette. Air bubbles
will prevent BOE from etching the oxide membrane.
\item For $1\,\mu$m oxide layer, keep chips in BOE for 20 minutes.%
\footnote{\begin{enumerate}
\item The nominal etch rate for BOE is $100\,$nm/min. However, poor circulation
in windows necessitates a longer etch time. Also, by the time unetched
oxide (see Fig. \ref{fig:AppSampFab_unetchedOxide}) can be identified,
it is too late to perform further etching so it is advisable to overetch
a little bit. On the other hand, it is possible for the BOE to get
under the photoresist and keep etching sideways under the cantilever,
possibly to ill effect. If there is evidence of strong undercutting
in an early batch, the BOE time can be reduced in subsequent batches.\end{enumerate}
}
\item Rinse chips in three to four stages of $\sim0.5\,$L DI water.
\item Turn chips right side up in the last stage of water (or an earlier
stage of water).
\item Remove chips from silicon piece in the last stage of water.%
\footnote{I like to do this after flipping the whole silicon piece over to avoid
mishandling of individual sample chips. It is pretty easy to pull
the sample chips off the silicon piece after it has been flipped over.
It is recommended to remove the chips now because they will fall off
on their own in the next step.%
}
\end{enumerate}
\item Put chips face up in warm (at least 80$^{\circ}$C%
\footnote{This temperature was the highest allowed at the CNF for 1165.%
}) 1165 (Microposit$^{\text{TM}}$ Remover 1165, Rohm and Haas Electronics
Materials LLC, Marlborough, MA, USA) to remove protective frontside
photoresist until resist stops visibly dissolving.
\item Transfer some chips to a dish of IPA (to rinse) and then to a dish
of DI water. Examine the chips.%
\footnote{Looking at a dish of water under the optical microscope might be against
the cleanroom rules. If it is, do not blame me if you get in trouble
for it.%
} If chips look okay, proceed. If the cantilevers appear compromised
in some way (e.g. there is residual oxide or the cantilevers are warped),
make note of the deficiencies and adjust the release procedure in
future batches to try to correct them.
\item Return sample chips to warm 1165 and leave them there for \textasciitilde{}1
hour.
\item Rinse the chips in 2-3 stages of IPA.
\item Use a critical point dryer (CPD) to dry off the cantilevers.

\begin{enumerate}
\item Without critical point drying, the cantilevers will be deformed by
surface tension when allowed to dry in ambient conditions (typically,
the cantilevers are pulled down and become stuck to the chip).
\item At the CNF, we used the tousimis$^{\textregistered}$ Automegasamdri$^{\textregistered}$-915B,
series B supercritical point dryer (Tousimis, Rockville, MD, USA).
\item We used custom Teflon pieces to hold the sample chips in the CPD.
These pieces were Teflon rectangles a few millimeters thick with one
through-hole drilled into each. A square groove forming a shallow
step (less than the thickness of the silicon wafer) was milled off
the edge of each Teflon piece. A Teflon piece could then be screwed
down onto a sample chip with the sample chip under the stepped groove. 
\item Thinner cantilevers suffered some attrition in the critical point
dryer. We tried to position the cantilevers away from holes in the
sample holder boat and away from entry and exit holes in the critical
point dryer in order to shield the cantilevers from turbulence.
\end{enumerate}
\end{enumerate}
\item \label{enu:AppSampFab_LastRecipeStep}Examine chips in optical microscope
(see Fig. \ref{fig:AppSampFab_OpticalSampChip}), scanning electron
microscope (see Figs. \ref{fig:AppSampFab_linewidthSEM}, \ref{fig:AppSampFab_SEMchipAngle},
\ref{fig:AppSampFab_SEMchipOverhead} and \ref{fig:AppSampFab_SEMarrayAngle}),
and atomic force microscope (see Fig. \ref{fig:AppSampFab_AFMimage}).
We avoided imaging samples with the SEM that we wanted to use to study
persistent currents, but we do not think the SEM would have any serious
negative effects on the persistent current samples.
\end{enumerate}

\section{Supplementary figures related to sample fabrication}

In this section, we present some drawings and images related to fabrication
of the cantilever-with-ring samples relevant to the discussions in
Chapter \ref{cha:CHExpSetup_} and the preceding section of this appendix.

\begin{figure}

\begin{centering}
\includegraphics[width=0.5\paperwidth]{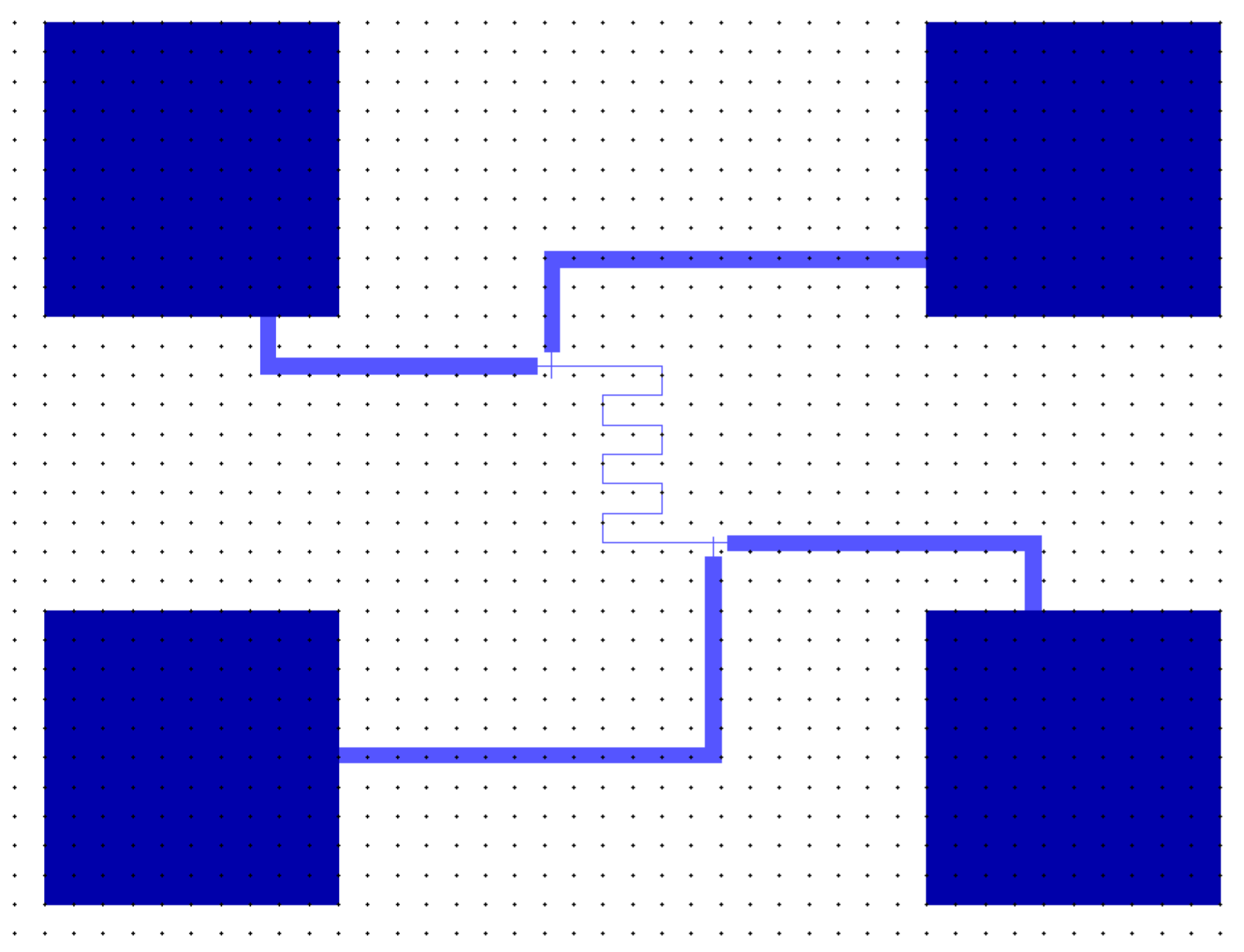}\caption[Image of transport sample mask design]{\label{fig:AppSampFab_TransportMaskDesign}Image of transport sample
mask design. The large squares are $100\,\mu$m on each side and were
used for wire bonding to the transport sample (the thin wire in the
center of the image). The large leads were $5\,\mu$m wide. The sample
wire shown has a linewidth of $80\,$nm and a length of $\sim280\,\mu$m.
This mask design was used for the measurements discussed in \ref{cha:AppTransport_}.}

\par\end{centering}

\end{figure}

\begin{figure}

\begin{centering}
\includegraphics[width=0.6\paperwidth]{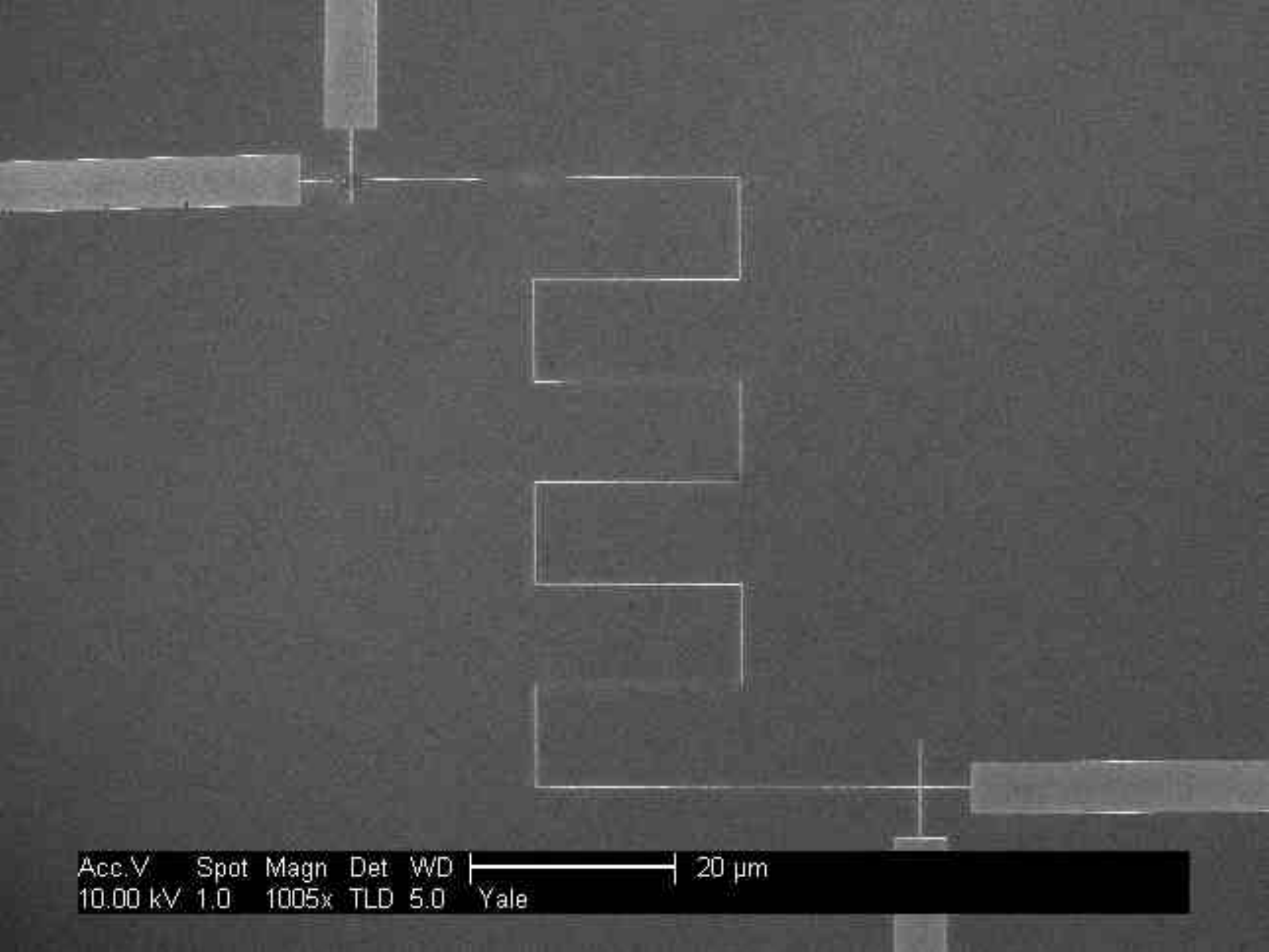}\caption[Scanning electron micrograph of transport sample]{\label{fig:AppSampFab_WLSEM}Scanning electron micrograph of transport
sample WL115. The figure shows the sample discussed in \ref{cha:AppTransport_}
with dimensions given in Table \ref{tab:AppTransport_WL115Properties}.
The sample mask design is shown in Fig. \ref{fig:AppSampFab_TransportMaskDesign}.
Magnified images of the wire are shown in Fig. \ref{fig:AppSampFab_linewidthSEM}.}

\par\end{centering}

\end{figure}

\begin{figure}

\centering{}\includegraphics[width=0.7\paperwidth]{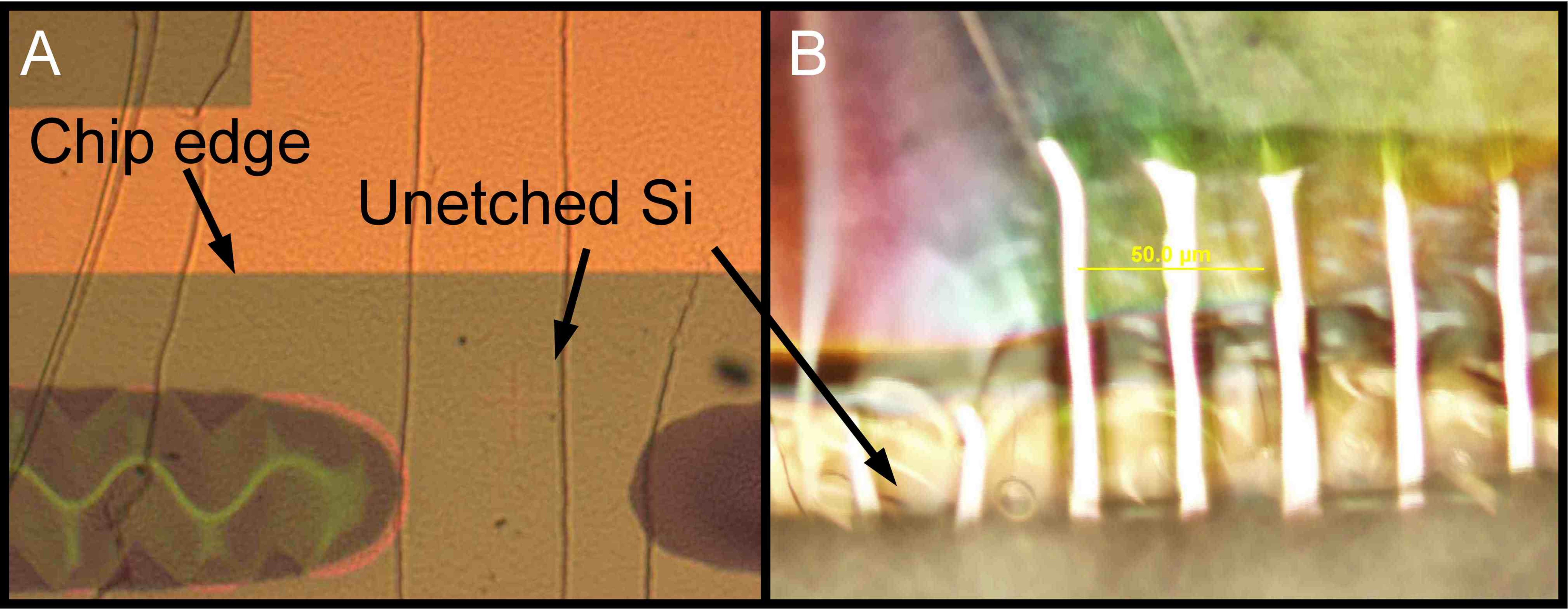}\caption[Optical microscope images of underetched silicon surrounding two sample
chips]{\label{fig:AppSampFab_DRIEsilicon}Optical microscope images of underetched
silicon surrounding two sample chips. The figure show images taken
during the deep reactive ion etch step of sample fabrication. Panel
A shows an image taken from the frontside of the wafer while panel
B displays the view from the backside. The frontside photoresist can
distort the shape of the sample chip, but from both sides of the chip
evidence of underetched silicon (opaque areas indicated in figure)
can be made out.}
\end{figure}

\begin{figure}

\centering{}\includegraphics[width=0.7\paperwidth]{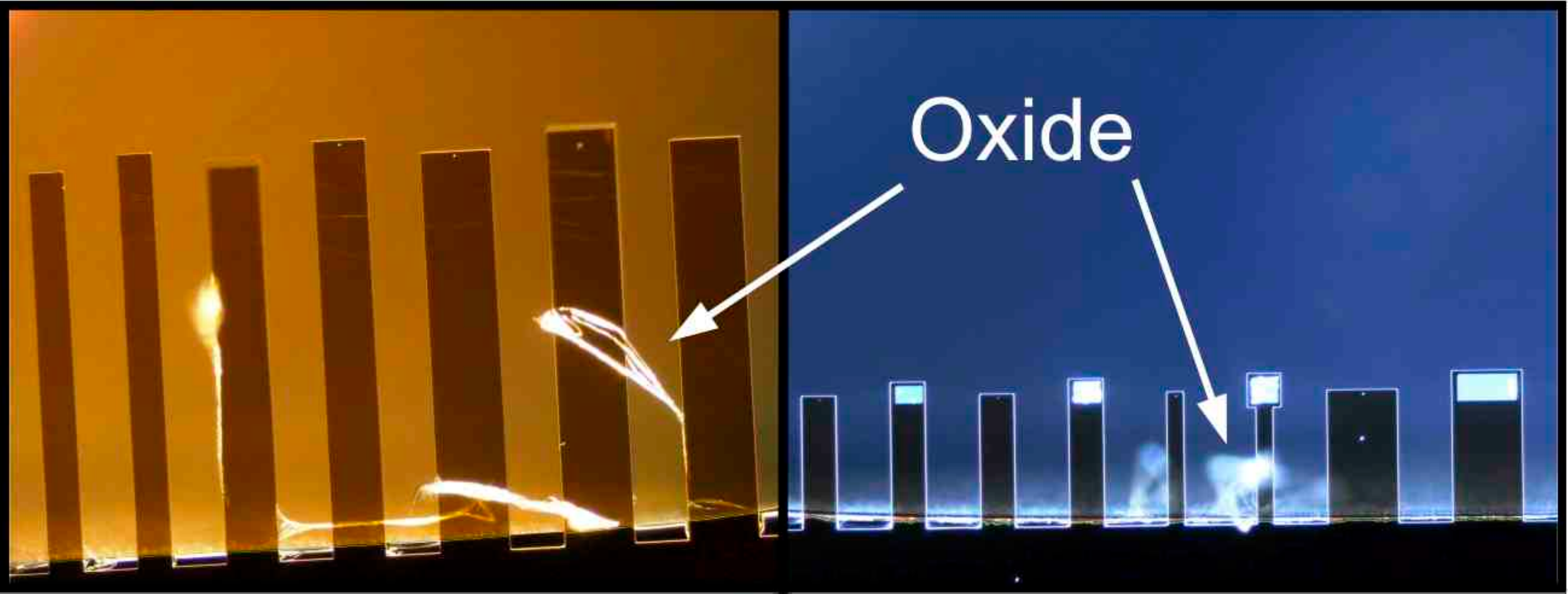}\caption[Optical microscope images of residual oxide film on two cantilever
sample chips]{\label{fig:AppSampFab_unetchedOxide}Optical microscope images of
residual oxide film on two cantilever sample chips. The oxide renders
the affected cantilevers unusable. The residual oxide film can be
minimized by blowing out any air bubbles in the sample chip windows
during the BOE etch step and etching for a sufficient amount of time.}
\end{figure}

\begin{figure}
\begin{centering}
\includegraphics[width=0.7\paperwidth]{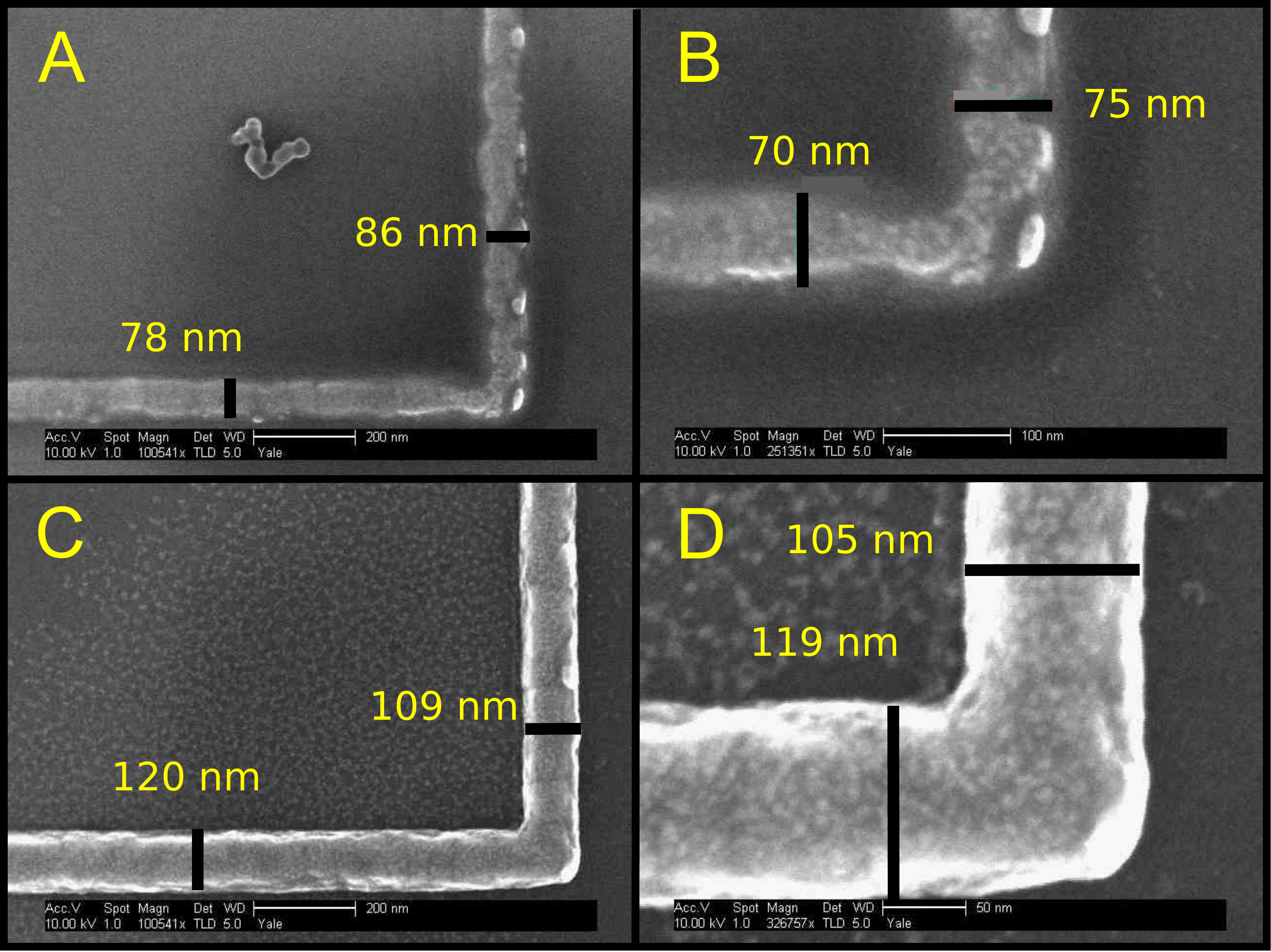}
\par\end{centering}

\caption[Images of aluminum wires displaying characteristic linewidths of measured
samples.]{\label{fig:AppSampFab_linewidthSEM}Images of aluminum wires displaying
characteristic linewidths of measured samples. Panels A and B show
SEM images at two different magnifications of a corner of an aluminum
meander with linewidth $85\pm15\,$nm (see scale bars in figure).
The target linewidth programmed into the e-beam tool for this sample
was $65\,$nm, the same as samples CL11, CL14, and CL15 in Table \ref{tab:ChData_Rings}.
Panels C and D show SEM images at two different magnifications of
a corner of an aluminum meander (sample WL115 in Table \ref{tab:AppTransport_WL115Properties})
with linewidth $116\pm6\,$nm (see scale bars in figure). The target
linewidth programmed into the e-beam tool for this sample was $80\,$nm,
the same as sample CL17 in Table \ref{tab:ChData_Rings}. We note
also that the wider linewidth sample appears to have a more uniform
surface. This surface uniformity might be related to the observed
overexposure in both samples.}
\end{figure}

\begin{figure}
\centering{}\includegraphics[width=0.7\paperwidth]{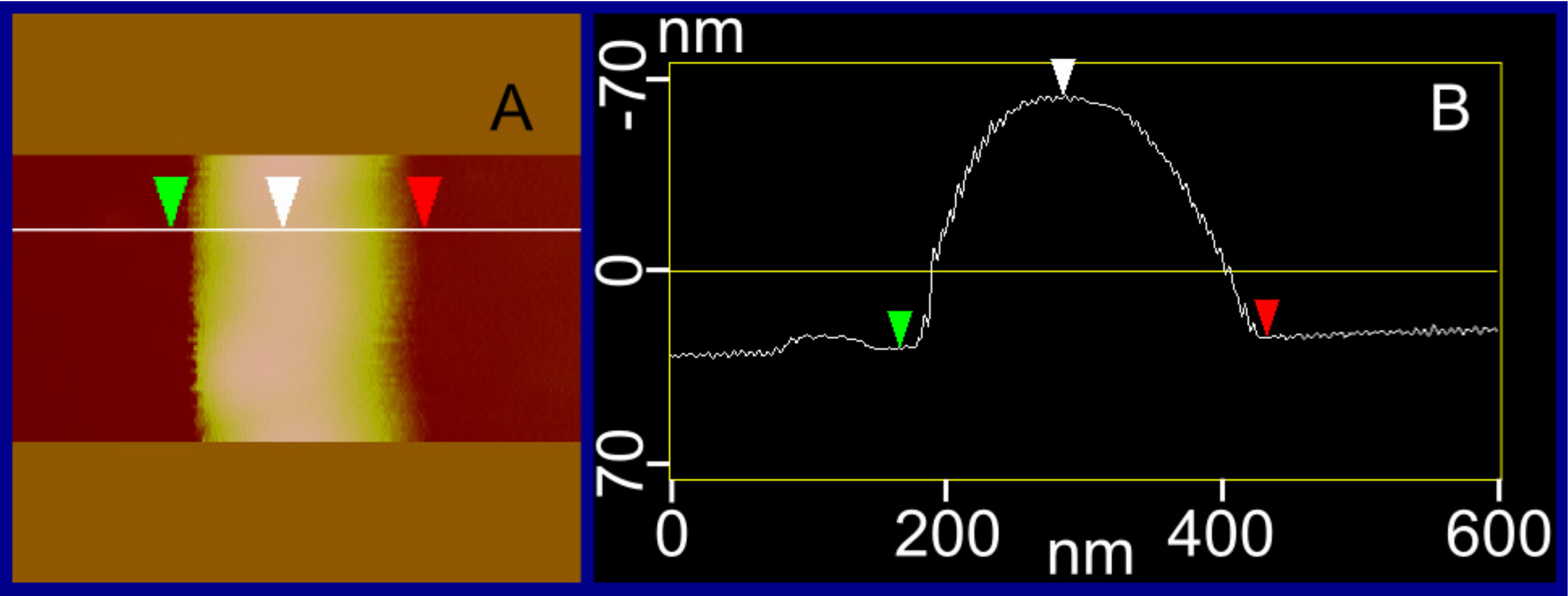}\caption[Atomic force microscope analysis of persistent current sample thickness]{\label{fig:AppSampFab_AFMimage}Atomic force microscope analysis
of persistent current sample thickness. Panel A shows a surface map
of a section of the wire (sample WL115 in Table \ref{tab:AppTransport_WL115Properties})
displayed in panels C and D of Fig. \ref{fig:AppSampFab_linewidthSEM}.
Panel B plots the line drawn through the sample in panel A. The height
difference between the left and middle markers is $91.6\,$nm, and
the difference between the right and middle markers is $87.6\,$nm.
Overall, the sample height is estimated to be $90\pm2\,$nm.}
\end{figure}

\begin{figure}
\centering{}\includegraphics[width=0.7\paperwidth]{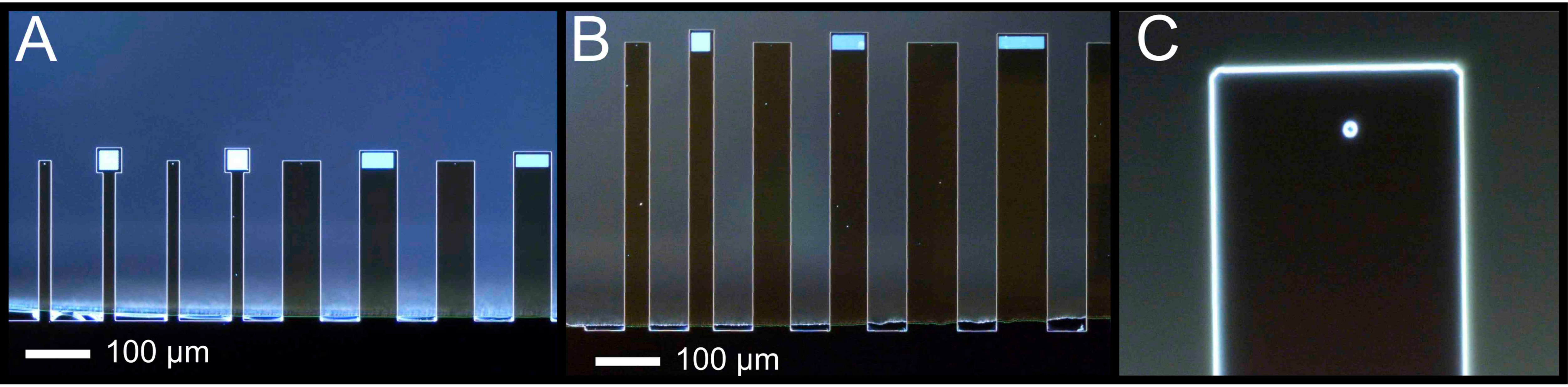}\caption[Dark field optical microscope images of cantilevers]{\label{fig:AppSampFab_OpticalSampChip}Dark field optical microscope
images of cantilevers. In dark-field, the arrays and single rings
(panel C) are clearly visible. The cantilever in panel C has a width
of $40\,\mu$m and supports a ring of diameter $1.5\,\mu$m and $115\,$nm
linewidth.}
\end{figure}

\begin{figure}

\centering{}\includegraphics[width=0.7\paperwidth]{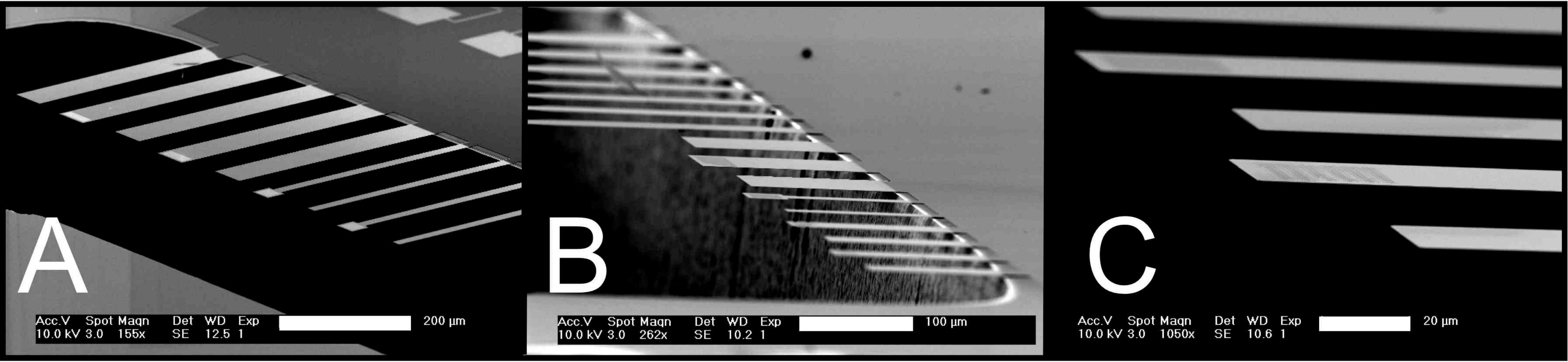}\caption[Scanning electron micrographs of sample chip: side views]{\label{fig:AppSampFab_SEMchipAngle}Scanning electron micrographs
of sample chip: side views. The three panels display successive magnifications
of a sample chip viewed from the side at different angles. These images
allow the extreme aspect ratio of the cantilevers to be appreciated.
The scale bars for panels A, B, and C are $200\,\mu$m, $100\,\mu$m,
and $20\,\mu$m respectively.}
\end{figure}

\begin{figure}

\begin{centering}
\includegraphics[width=0.7\paperwidth]{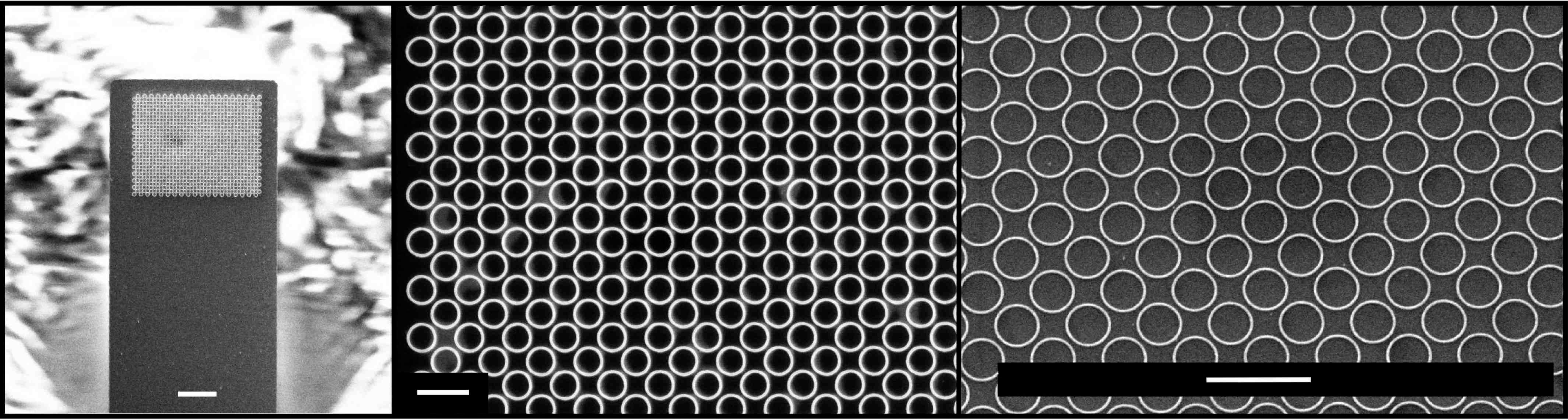}\caption[Overhead scanning electron micrographs of cantilever with array of
rings]{\label{fig:AppSampFab_SEMchipOverhead}Overhead scanning electron
micrographs of cantilever with array of rings. Each panel was taken
from a different sample. From left to right, the scale bars are $10\,\mu$m,
$2\,\mu$m, and $5\,\mu$m.}

\par\end{centering}

\end{figure}

\begin{figure}

\centering{}\includegraphics[width=0.7\paperwidth]{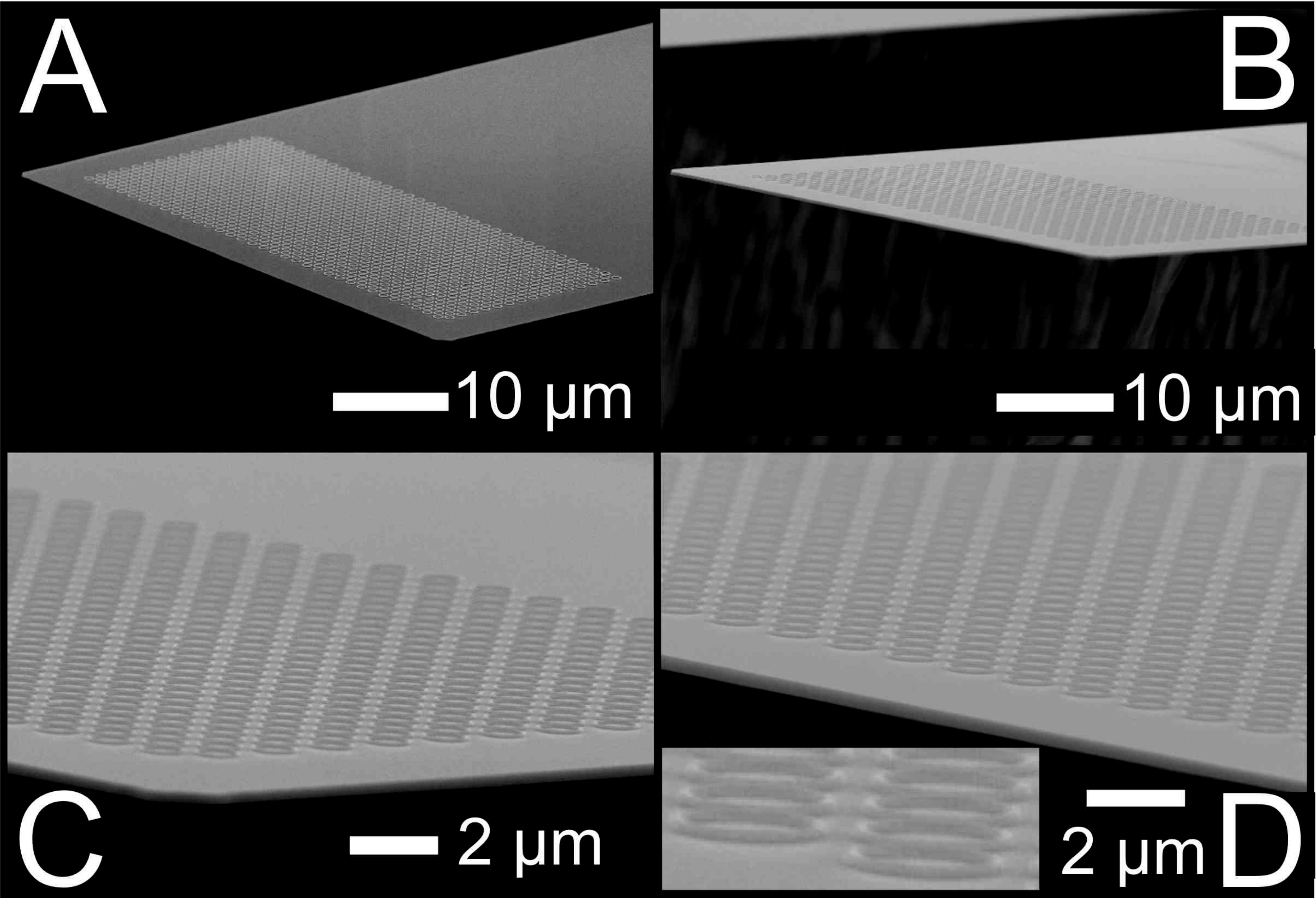}\caption[Angled scanning electron micrographs of array of rings on end of cantilever]{\label{fig:AppSampFab_SEMarrayAngle}Angled scanning electron micrographs
of array of rings on end of cantilever. In the lower panels, the thickness
of the cantilever can be discerned. The inset in panel D shows a region
of the main image in panel D at three times stronger magnification.}
\end{figure}

\chapter{\label{cha:AppTransport_}Transport characterization of persistent
current samples}

The persistent current samples consisted of small isolated aluminum
rings fabricated on the ends of pliable cantilevers less than $1\,\text{\ensuremath{\mu}m}$
thick. Because it would be nearly impossible to perform transport
measurements on these rings directly, we also included wires in the
same same e-beam mask defining the rings and so deposited these wires
onto the chip simultaneously with the rings as described in \ref{sub:CHExpSetup_PersistentCurrentFabrication}
and \ref{app:AppSampFab}. These wires allowed us to obtain values
of the diffusion constant $D$, the electron phase coherence length
$L_{\phi}$ and spin-orbit scattering length $L_{SO}$ for comparison
with the results from the persistent current measurements. 

All measurements were performed using the bridge circuit and following
the procedure detailed in Sections \ref{sub:CHExpSetup_TransportSetUp}
and \ref{sub:CHExpSetup_TransportCalibMeasurement}. The same sample
which we refer to as WL115 (see Table \ref{tab:AppTransport_WL115Properties}
and Fig. \ref{fig:AppSampFab_WLSEM}), was used for all measurements. 

\begin{center}
\begin{table}
\centering{}%
\begin{tabular}{|ccccccc|}
\hline 
Name & $L\,(\text{\ensuremath{\mu}m)}$ & $w\,(\text{nm})$ & $t\,(\text{nm})$ & $D_{\rho}\,(\text{cm}^{2}/\text{s})$ & $D_{B_{c}}\,(\text{cm}^{2}/\text{s})$ & $L_{SO}\,(\text{\ensuremath{\mu}m)}$\tabularnewline
\hline 
WL115 & $289$ & $115\pm5$ & $90\pm2$ & $259\pm14$ & $122\pm5$ & $1.10\pm0.25$\tabularnewline
\hline 
\end{tabular}\caption[Dimensions and properties of sample WL115]{\label{tab:AppTransport_WL115Properties}Dimensions and properties
of sample WL115. The table gives the wire's length $L$, linewidth
$w$, and thickness $t$. Also listed are the diffusion constant calculated
from the wire's resistance $D_{\rho}$ and from its superconducting
critical field $D_{B_{c}}$. The measured wire resistance was $R=285\pm7\,\Omega$,
which corresponds to a resistivity of $\rho=1.02\pm0.06\times10^{-8}\,\text{\ensuremath{\Omega}\,\ cm}$.
The final entry gives the wire's spin-orbit scattering length $L_{SO}$,
found from low field magnetoresistance measurements in the normal
state. The wire's electron phase coherence length $L_{\phi}$ is not
shown because it varies with temperature. At $T=2\,\text{K}$, $L_{\phi}$
was observed to be about $5\,\text{\ensuremath{\mu}m}$. The measurements
of $D_{\rho}$, $D_{B_{c}}$, $L_{SO}$, and $L_{\phi}$ are described
in the subsequent sections.}
\end{table}

\par\end{center}

\section{Transport measurements of the diffusion constant}

\subsection{Resistance measurement}

\begin{figure}

\begin{centering}
\includegraphics[width=0.7\paperwidth]{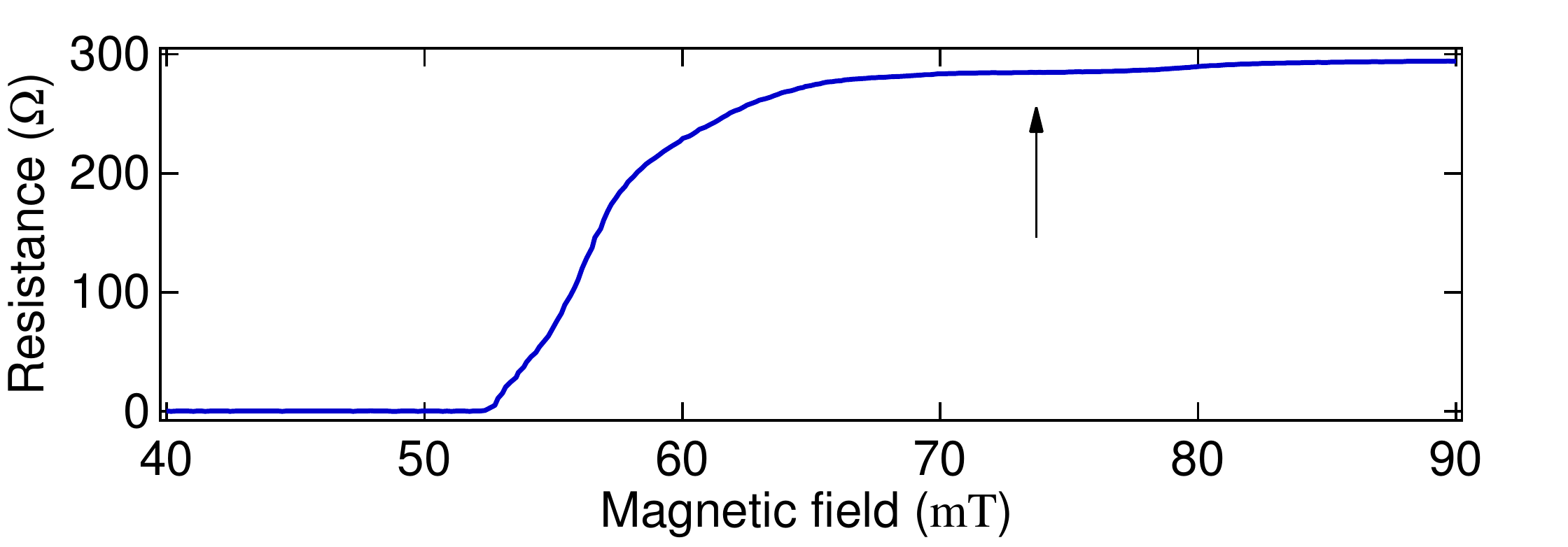}
\par\end{centering}

\caption[Magnetic field sweep through the superconducting transition of sample
WL115]{\label{fig:AppTransport_RfromSuperconTran}Magnetic field sweep through
the superconducting transition of sample WL115. Due to finite size
effects, the wire's superconducting critical field is enhanced and
the superconducting to normal transition is broadened as a function
of magnetic field. We take the normal state resistance $R$ to be
the value of the resistance at $\sim74\,\text{mT}$ where resistance
as a function of magnetic field is relatively flat. We assign a $7\,\Omega$
uncertainty to this value of resistance to account for the ambiguity
in choosing the value of the magnetic field for which the sample is
fully in the normal state. The sample resistance continues to vary
with magnetic field in the normal state presumably due to classical
magnetoresistance. The temperature during this measurement was $365\,\text{mK}$,
well below the superconducting transition temperature $T_{c}\approx1.2\,\text{K}$.
The magnetic field was swept up from zero field through the superconducting
transition. Due to the loss of a sample lead as described in \ref{sub:CHExpSetup_TransportSetUp},
the sample resistance had to be inferred from the change in resistance
across the superconducting transition rather than from a direct four
point measurement.}
\end{figure}

Sample WL115\textquoteright{}s resistivity $\rho$ was obtained by
measuring the total change in resistance $R$ of the sample at 360
mK as the magnetic field was swept through the wire\textquoteright{}s
superconducting critical field. From Fig. \ref{fig:AppTransport_RfromSuperconTran}
we take $R=286\pm7\,\Omega$ with the relatively large uncertainty
for a resistance measurement due to the broadening of the superconducting
transition. Using the dimensions of Table \ref{tab:AppTransport_WL115Properties}
the corresponding resistivity $\rho=\frac{wt}{L}R$ is $(1.02\pm0.06)\times10^{-8}\,\text{\ensuremath{\Omega}\,\ m}$.

The diffusion constant $D$ can be calculated using the Einstein relation

\begin{equation}
\rho^{-1}=e^{2}\eta D\label{eq:AppTransport_DiffusionConstant}
\end{equation}
with $e$ the electron charge and $\eta$ the electron density of
states per unit volume at the Fermi level. The density of states can
be written in terms of the free electron density $n$ and the Fermi
energy $\varepsilon_{F}$ as $\eta=3n/2\varepsilon_{F}$. With $n=1.81\times10^{29}\,\text{m}^{-3}$
and $\varepsilon_{F}=11.5\,\text{eV}$ for aluminum \citep{ashcroft1976solidstate},
the wire\textquoteright{}s measured resistivity corresponds to a diffusion
constant of $D_{\rho}=0.0259\pm0.014\,\text{m}^{2}/\text{s}$. To
avoid confusion in the following sections, we denote this value of
$D$ as $D_{\rho}$. 

\subsection{Superconducting critical field measurement}

The wire\textquoteright{}s superconducting critical field $B_{c}$
was measured as a function of temperature $T$ near the wire\textquoteright{}s
superconducting transition temperature $T_{c}$. In the Ginzburg Landau
framework valid for a dirty superconductor near $T_{c}$, the superconducting
critical field for a thin wire lying on a plane normal to the applied
magnetic field can be written as
\begin{equation}
B_{c}\left(T\right)=\frac{\sqrt{12h}}{\pi ew\sqrt{D}}\sqrt{k_{B}\left(T_{c}-T\right)}\label{eq:AppTransport_BcOfT}
\end{equation}
where $h$ is Planck\textquoteright{}s constant and $k_{B}$ is the
Boltzmann constant \citep{tinkham2004introduction}. For a superconductor
with known linewidth $w$, a measurement of $B_{c}$ as a function
of $T$ allows one to determine both $D$ and $T_{c}$.

\begin{figure}

\begin{centering}
\includegraphics[width=0.7\paperwidth]{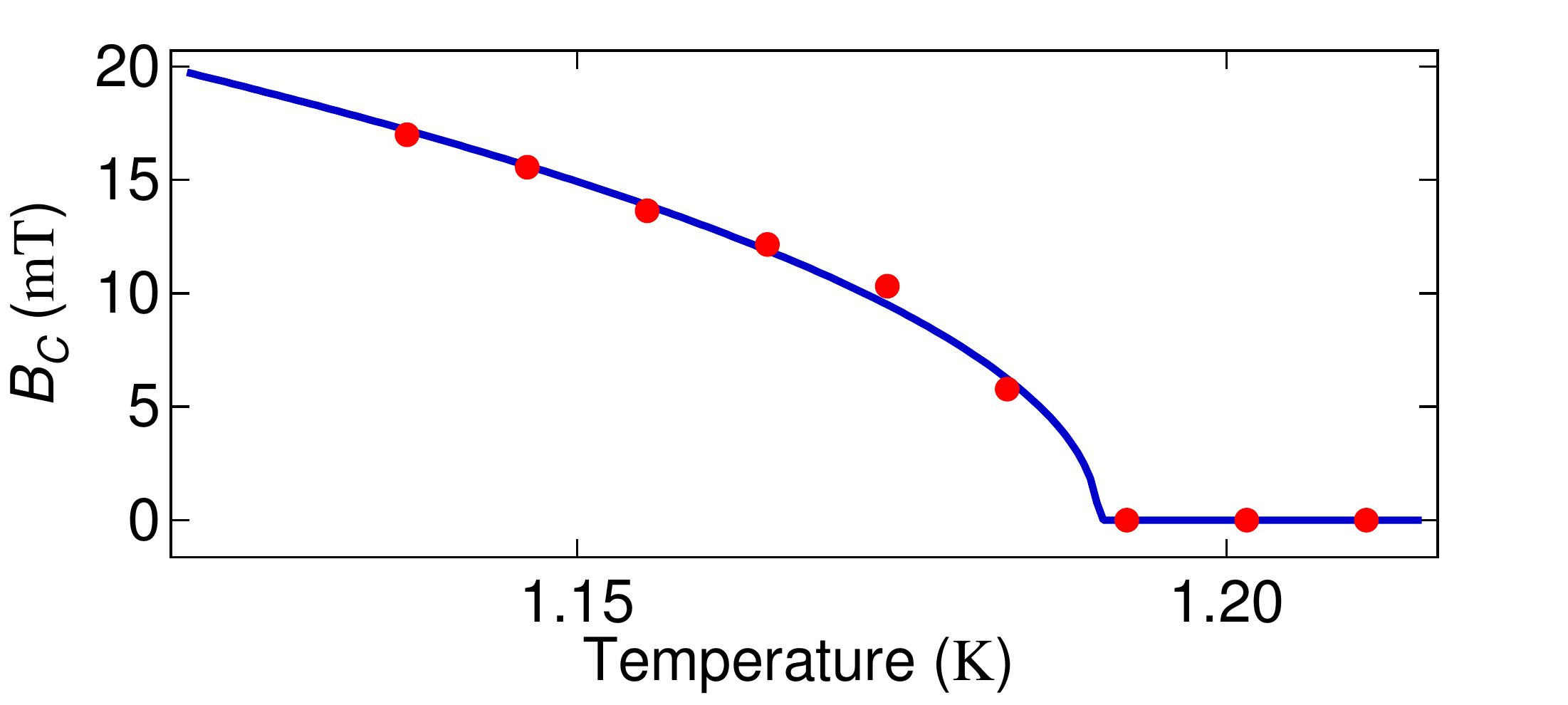}
\par\end{centering}

\caption[Superconducting critical field of transport sample versus temperature]{\label{fig:AppTransport_BcvsT}Superconducting critical field of
transport sample versus temperature. The superconducting critical
field was extracted from measurements of resistance as a function
of magnetic field (similar to the one shown in Fig. \ref{fig:AppTransport_RfromSuperconTran})
as described in the text. A fit to Eq. \ref{eq:AppTransport_BcOfT}
is also shown. Analysis of the results of the fit is provided in the
text.}
\end{figure}

The superconducting critical field was measured by sweeping the magnetic
field at different sample temperatures and observing the change in
resistance from the normal to the superconducting state. The superconducting
critical field was taken to be the field at which the wire resistance
reached a fixed fraction $\gamma_{R}$ of the normal state resistance.
Fig. \ref{fig:AppTransport_BcvsT} shows the extracted values of $B_{c}$
when $\gamma_{R}=0.1$. A different choice for $\gamma_{R}$ simply
offsets all of the data points in Fig. \ref{fig:AppTransport_BcvsT}
by a constant amount. From Eq. \ref{eq:AppTransport_BcOfT}, it can
be seen that such a shift in $T$ will result in an equal shift in
the inferred value of $T_{c}$ but will not affect the determination
of $D$. Choosing $\gamma_{R}=0.5$ produces a shift of $+20\,\text{mK}$
in $T_{c}$ relative to the value found for $\gamma_{R}=0.1$. In
addition to the measured $B_{c}(T)$, Fig. \ref{fig:AppTransport_BcvsT}
shows a fit to Eq. \ref{eq:AppTransport_BcOfT}. The extracted fit
parameters are $T_{c}=1.19\,\text{K}$ and $D_{B_{c}}=0.0122\pm0.0005\,\text{m}^{2}/\text{s}$.
We denote this value of $D$ as $D_{B_{c}}$.

The fitted value of $D_{B_{c}}$ is a factor of two smaller than that
which was found for $D_{\rho}$. We note that Eq. \ref{eq:AppTransport_BcOfT}
is applicable only when the electrons\textquoteright{} elastic scattering
length $l_{e}$ is much smaller than the wire\textquoteright{}s transverse
dimensions $w$ and $t$. If we use the value of $D_{\rho}$ determined
above for the diffusion constant $D$ and $v_{F}=2.0\times10^{6}\,\text{m}/\text{s}$
as the Fermi velocity of aluminum, we find a value of $l_{e}=3D/vF=40\,\text{nm}$.
This indicates that $l_{e}\thicksim w,\, d$, and hence that Eq. \ref{eq:AppTransport_BcOfT}
is not valid. As a result we do not consider $D_{B_{c}}$ to have
a physical meaning and include it here only for completeness. Discrepancies
between $D_{B_{c}}$ and $D_{\rho}$ due to the breakdown of Eq. \ref{eq:AppTransport_BcOfT}
have been noted previously \citep{lafarge1993twoelectron}. 

\FloatBarrier

\section{\label{sec:AppTransport_phaseCoherence}Determination of electron
phase coherence length and spin-orbit scattering length from low field
magnetoresistance measurements}

The electron phase coherence and spin orbit scattering lengths were
extracted from measurements of the magnetoresistance of sample WL115
at temperatures above $T_{c}$. As mentioned in \ref{sub:CHExpSetup_TransportSetUp},
these measurements were first performed in aluminum wires two decades
ago at Yale and have been reviewed previously \citep{santhanam1987localization}.%
\footnote{All of the equations in the next section are derived from relations
given in this paper.%
}

\subsection{Theoretical predictions for the magnetoresistance of thin wires}

The coherent interference of time reversed trajectories leads to an
increase in the probability for a quasiparticle to return to its original
position and thus to an increase in electrical resistance $R$, a
phenomenon known as weak localization. The presence of a magnetic
field $B$ suppresses weak localization by breaking time reversal
symmetry and allows a direct measure of electron phase coherence through
the resulting magnetoresistance. Spin orbit scattering can also modify
the spin components of time reversed paths and thus the weak localization
contribution to conductivity. 

The analytic form for the weak localization correction to the resistance
$R$ in a magnetic field $B$ is given by
\begin{align}
\frac{\delta R^{WL}}{R} & \equiv\frac{R\left(B\right)-R\left(B=0\right)}{R\left(B=0\right)}\nonumber \\
 & =\frac{3}{2}f_{1}\biggl(B,b\left(L_{2}\right)\biggr)-\frac{1}{2}f_{1}\biggl(B,b\left(L_{\phi}\right)\biggr)\label{eq:AppTransport_dRRWeakLocalization}
\end{align}
where $L_{\phi}$ is electron phase coherence length and 
\[
L_{2}=\frac{1}{\sqrt{L_{\phi}^{-2}+\frac{4}{3}L_{SO}^{-2}}}
\]
with $L_{SO}$ is the spin orbit scattering length. The function $f_{1}$
is given by%
\footnote{This definition of $f_{1}$ differs by a constant offset from the
definition given in Ref. \citealp{santhanam1987localization}. This
offset makes $f_{1}(0,B_{1})=0$.%
}
\[
f_{1}\biggl(B,B_{1}\biggr)=R_{\square}\frac{e^{2}}{\pi\hbar}\left(\frac{b\left(w\right)}{B_{1}}\right)^{1/2}\left(\left(1+\frac{B^{2}}{48b\left(w\right)B_{1}}\right)^{-1/2}-1\right)
\]
with $R_{\square}=\rho/t$ the sheet resistance per square unit of
the wire. The field scale $b\left(w\right)$ is given by
\[
b\left(l\right)=\frac{\hbar}{4el^{2}}
\]
where $l$ is in units of length. This form for the weak localization
correction to the magnetoresistance is derived from a perturbative
calculation and is valid for $B<12b(w)$ (approximately $300\,\text{mT}$
for sample WL115) \citep{santhanam1987localization}. 

Just above $T_{c}$, superconducting fluctuations result in a small,
temperature dependent population of Cooper pairs, which reduce the
resistance of the metal. Further above $T_{c}$, in the temperature
range relevant to our magnetoresistance measurements, Cooper pairs
from superconducting fluctuations are too short-lived to contribute
directly to the conductivity. However, after a Cooper pair decays,
the two electron quasiparticle wave functions are still correlated
and provide a contribution to the conductivity, known as the Maki-Thompson
contribution, which behaves similarly to the direct contribution of
a Cooper pair as long as the electrons maintain phase coherence. Because
all Cooper pairs are composed of electrons in the singlet state, spin
orbit scattering does not affect the Maki-Thompson contribution to
the conductivity. 

The correlation between quasiparticles formed by the decay of a Cooper
pair has a theoretically similar description to the cooperon which
describes weak localization and thus both effects have similar analytic
forms for their contributions to the magnetoresistance. Specifically,
the Maki-Thompson correction to the resistance obeys
\begin{align}
\frac{\delta R^{MT}}{R} & \equiv\frac{R\left(B\right)-R\left(B=0\right)}{R\left(B=0\right)}\nonumber \\
 & =-\beta\left(\frac{T}{T_{c}}\right)\, f_{1}\biggl(B,b\left(L_{\phi}\right)\biggr)\label{eq:AppTransport_dRRMakiThompson}
\end{align}
where $\beta\left(t\right)$ is a function introduced by Larkin which
diverges logarithmically as $t\rightarrow1$. Eq. \ref{eq:AppTransport_dRRMakiThompson}
is valid provided that $\hbar D/L_{\phi}^{2}\ll k_{B}T\ln(T/T_{c})$
and $B\ll(k_{B}T/4De)\ln(T/T_{c})$.

\subsection{Measurement and analysis of low field magnetoresistance measurements}

\begin{figure}
\begin{centering}
\includegraphics[width=0.7\paperwidth]{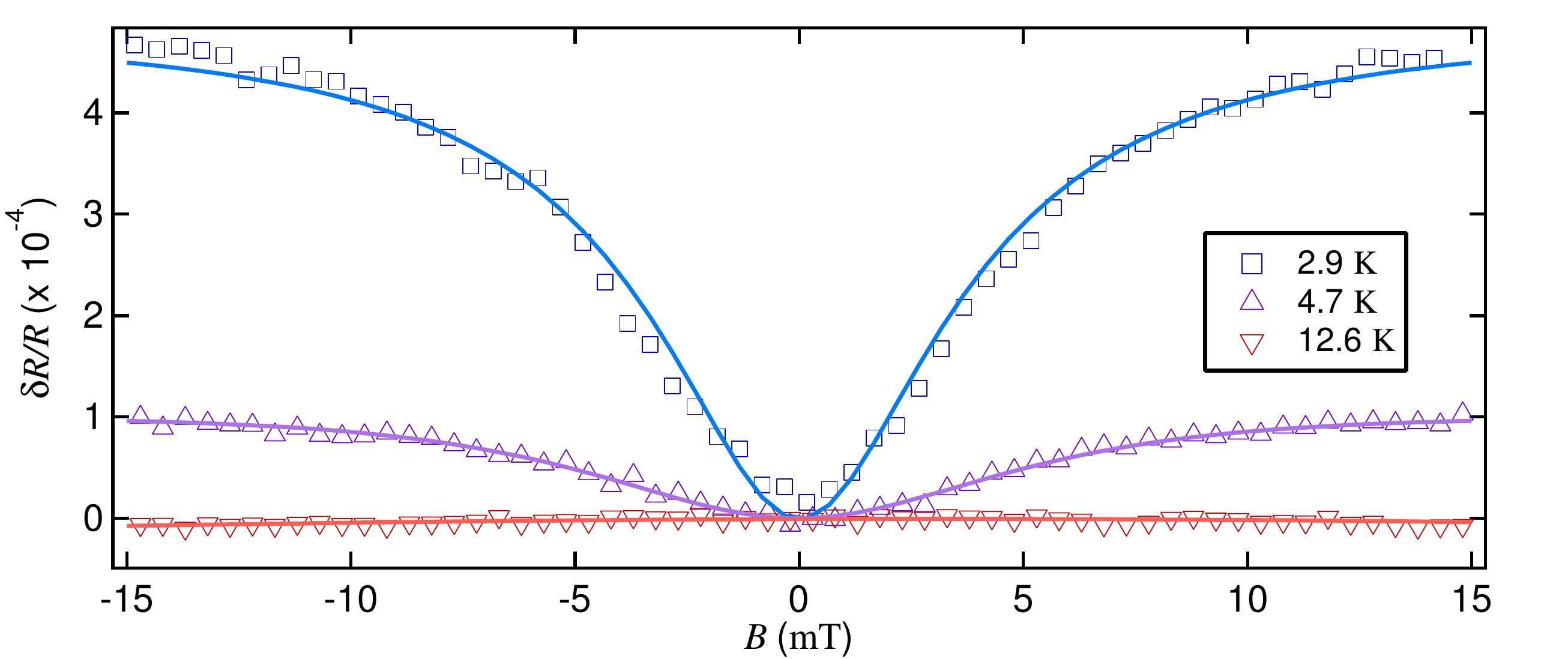}
\par\end{centering}

\caption[Magnetoresistance of sample WL115 at several temperatures]{\label{fig:AppTransport_MagnetoResistance}Magnetoresistance of sample
WL115 at several temperatures. Superconducting fluctuations (as well
as spin-orbit scattering) result in a reduction in the sample resistance
at zero magnetic field which is evident in the data from $2.9\,\text{K}$
and $4.7\,\text{K}$. Weak localization causes an increase in resistance
at zero magnetic field. The weak localization correction to the resistance
dominates once the temperature is far enough from $T_{c}$ for superconducting
fluctuations to be sufficiently suppressed ($12.6\,\text{K}$ data).
The curves in the figure shows fits to the sum of Eqs. \ref{eq:AppTransport_dRRWeakLocalization}
and \ref{eq:AppTransport_dRRMakiThompson} for $L_{\phi}$ with $L_{SO}$
fixed to $1.1\,\text{\ensuremath{\mu}m}$. Analysis of the fits is
provided in the text. Each fit curve and data set has been shifted
as described in the text.}
\end{figure}

Magnetoresistance measurements were made at a series of wire temperatures
above $T_{c}$ between $1.6$ and $12.6\,\text{K}$. Because of the
limitations on the validity of Eq. \ref{eq:AppTransport_dRRMakiThompson},
measurements could only be made at relatively high temperatures compared
to those for which we measured persistent currents. Fig. \ref{fig:AppTransport_MagnetoResistance}
shows magnetoresistance measurements with fits to the sum of Eqs.
\ref{eq:AppTransport_dRRWeakLocalization} and \ref{eq:AppTransport_dRRMakiThompson}
for three different temperatures. The exact form of the fitting function
used was
\[
\frac{\delta R}{R}=\frac{\delta R^{WL}\left(B-B_{o}\right)}{R}+\frac{\delta R^{MT}\left(B-B_{o}\right)}{R}-\frac{\delta R_{o}}{R}
\]
where $B_{o}$ and $\delta R_{o}$ correct for possible trapped flux
in the solenoid applying the magnetic field and for the imbalance
between the two sides of the resistance bridge at $B=0$. Besides
$B_{o}$ and $\delta R_{o}$, the only free parameter was $L_{\phi}$.
Some preliminary analysis also varied $L_{SO}$ as described below.
In Fig. \ref{fig:AppTransport_MagnetoResistance}, each data set and
fit curve are shifted by $B_{o}$ in $B$ and $\delta R_{o}/R$ in
$\delta R/R$ so that the extreme point of each fit curve is located
at the origin.

Note the first condition, $\hbar D/L_{\phi}^{2}\ll k_{B}T\ln(T/T_{c})$,
for the validity of Eq. \ref{eq:AppTransport_dRRMakiThompson} is
satisfied for T as low as $\sim1.35\,\text{K}$ where $\hbar D/L_{\phi}^{2}\approx100k_{B}T\ln(T/T_{c})$.
However, the second condition, $B\ll(k_{B}T/4De)\ln(T/T_{c})$, restricts
the valid range of magnetic fields to relatively small values over
the range of temperatures relevant to our measurements. For $1.6\,\text{K}$,
the second condition should be valid for fields much less than 0.43
mT. By $4\,\text{K}$, the field range is restricted to $\ll6\,\text{mT}$
and by $10\,\text{K}$, $\ll25\,\text{mT}$. Due to the small size
of the magnetoresistance signal for field ranges of less than $\thicksim3\,\text{mT}$,
all magnetoresistance data were fit to Eq. \ref{eq:AppTransport_dRRMakiThompson}
over a field range of $15\,\text{mT}$, as indicated in Fig. \ref{fig:AppTransport_MagnetoResistance}.
The extracted $L_{\phi}$ does depend weakly on the size of the fit
range. For example, at $2.4\,\text{K}$ where the expression is valid
for $B\ll2\,\text{mT}$, the fitted $L_{\phi}$ increases by $20\%$
as the fit range is increased from $3\text{\,\ mT}$ to $30\,\text{mT}$.
Previous efforts have produced modifications to Eq. \ref{eq:AppTransport_dRRMakiThompson}
which allow it to be applied to wider ranges of magnetic field and
temperature, but no modifications have addressed the range of temperatures
and magnetic fields relevant to our magnetoresistance measurements
\citep{gordon1985superconductingfluctuation,gordon1986electron}.

\begin{figure}[h]
\begin{centering}
\includegraphics[width=0.7\paperwidth]{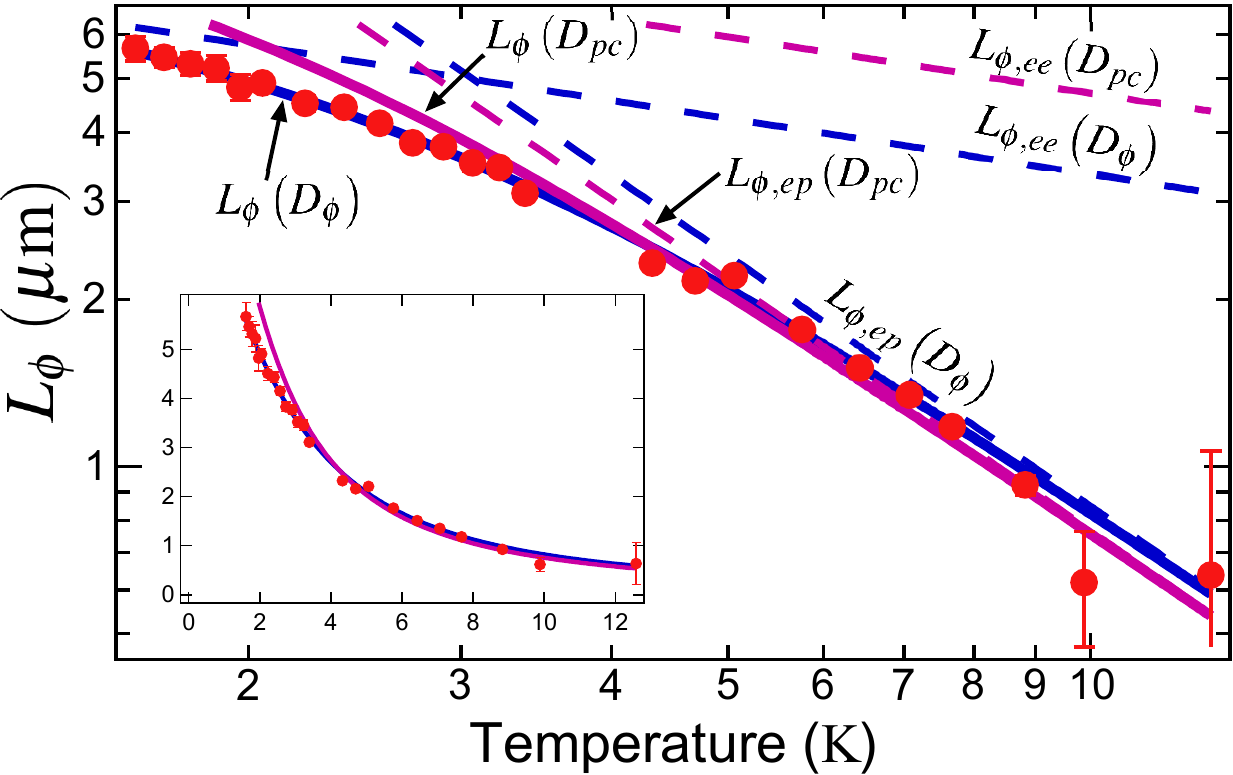}
\par\end{centering}

\caption[Fitted electron phase coherence length for sample WL115 versus temperature]{\label{fig:AppTransport_LphivsT}Fitted electron phase coherence
length for sample WL115 versus temperature. The data points were extracted
from fits to the magnetoresistance as described in the text. There
are also several curves shown. The curves labeled as functions of
$D_{\phi}$ were obtained from fits in which the diffusion constant
$D$ was a fitting parameter (as was $A_{ep}$), while those labeled
as functions of $D_{pc}$ were calculated from fits in which $D$
was fixed to the value $D_{pc}=0.024\,\text{m}^{2}/\text{s}$. The
dashed curves labeled $L_{\phi,ep}$ and $L_{\phi,ee}$ represent
the values of $L_{\phi}$ calculated from the extracted parameters
of these fits assuming $L_{\phi,ep}=\sqrt{D\tau_{ep}}$ and $L_{\phi,ee}=\sqrt{D\tau_{ee}}$,
respectively. We note that the non-linearity of the data points as
a function of $T$ on the log-log scale indicates that we observe
the cross-over from the regime where $L_{\phi}$ is dominated by the
electron-electron interaction to that dominated by the electron-phonon
interaction. A more detailed description and analysis of the fitting
procedure is provided in the text. The inset shows the data on a linear
scale, as well as the fitted curves $L_{\phi}(D_{\phi})$ and $L_{\phi}(D_{pc})$.}
\end{figure}

From fits to the magnetoresistance data in which both $L_{\phi}$
and $L_{SO}$ are varied, we verified that spin-orbit scattering contributes
significantly to the magnetoresistance only at higher temperatures
where the Maki-Thompson contribution is small. For temperatures above
4 K, the magnetoresistance data were fit with both $L_{\phi}$ and
$L_{SO}$ as fitting parameters. In this range, $L_{SO}$ was measured
to be $1.10\pm0.25\,\text{\ensuremath{\mu}m}$ and observed to be
independent of temperature. Following this analysis, the data for
the whole temperature range was fit with $L_{SO}$ fixed to $1.1\,\text{\ensuremath{\mu}m}$
 and $L_{\phi}$ as the only free parameter. In Fig. \ref{fig:AppTransport_LphivsT},
the values of $L_{\phi}$ found in this way are plotted versus temperature. 

The electron phase coherence length is limited to a finite value due
to scattering processes in which electrons change energy. For the
temperature regime of our measurements, the processes expected to
be dominant are electron-phonon and electron-electron scattering.
Scattering from magnetic impurities should be negligible for our high
purity aluminum film (see Appendix \ref{app:AppSampFab} for a description
of the aluminum source). The electron-phonon phase scattering rate
$\tau_{ep}^{-1}$ follows the form 
\begin{equation}
\tau_{ep}^{-1}=A_{ep}T^{3}\label{eq:AppTransport_AepElectronPhonon}
\end{equation}
The electron phonon coefficient has previously been reported as $A_{ep}=9.1\times10^{6}\,\text{s}^{-1}\text{K}^{-3}$
for bulk aluminum \citep{santhanam1987localization,lawrence1978calculation}.
For the conditions of our measurement, the electron-electron phase
scattering rate $\tau_{ee}^{-1}$ was observed in Ref. \citep{wind1986onedimensional,pierre2003dephasing}
to be dominated by the process of multiple collisions with small energy
transfers and to follow the form%
\footnote{There appears to be some disagreement in the literature over the numerical
prefactor multiplying $D$ in this expression. Compare Refs. \citealp{altshuler1985electronelectron,wind1986onedimensional,aleiner1999wavesin,pierre2003dephasing}.
The expression used here produces the largest value of $D$.%
} 
\begin{align}
\tau_{ee}^{-1} & =A_{ee}T^{2/3}\nonumber \\
 & =\left(\frac{R_{\square}e^{2}k_{B}}{2\sqrt{2}\hbar^{2}}\frac{\sqrt{D}}{w}\right)^{2/3}T^{2/3}.\label{eq:AppTransport_AeeElectronElectron}
\end{align}

The total electron phase breaking rate $\tau_{\phi}^{-1}$ is the
sum of these two rates, $\tau_{\phi}^{-1}=\tau_{ep}^{-1}+\tau_{ee}^{-1}$.
The blue lines in Fig. \ref{fig:AppTransport_LphivsT} show a fit
to the measured $L_{\phi}(T)$ using Eqs. \ref{eq:AppTransport_AepElectronPhonon}
and \ref{eq:AppTransport_AeeElectronElectron} and the relation $L_{\phi}=\sqrt{D\tau_{\phi}}$
with $A_{ep}$ and $D$ as the free parameters. We denote this fitted
value of $D$ as $D_{\phi}$. The fitted values are $A_{ep}=(1.21\pm0.07)\times10^{7}\,\text{s}^{-1}\text{K}^{-3}$
and $D_{\phi}=0.0044\pm0.0002\,\text{m}^{2}/\text{s}$. 

While $D_{\phi}$ differs from $D_{\rho}$ and $D_{pc}$, the value
of $D$ extracted from the persistent current measurements, we note
that $L_{\phi}$ is roughly proportional to $D^{1/2}$ and so provides
a relatively weak constraint on $D$. To illustrate this point, Fig.~\ref{fig:AppTransport_LphivsT}
also show fits (purple lines) with $D_{\phi}$ fixed to $0.024\,\text{m}^{2}/\text{s}$
(the value of $D_{pc}$ for the persistent current sample with comparable
linewidth to WL115) and $A_{ep}$ as the only fitting parameter. The
best fit value of $A_{ep}$ was $\left(4.09\pm0.04\right)\times10^{7}\,\text{s}^{-1}\text{K}^{-3}$
in this case. As can be seen from Fig. \ref{fig:AppTransport_LphivsT},
$D_{\phi}=D_{pc}$ provides reasonable agreement with the data, producing
slightly higher values of $L_{\phi}$ at the lowest temperatures.
We note that using the bulk value for $A_{ep}$ rather than our fitted
values would also result in larger values for $L_{\phi}$. These two
predictions of larger magnitudes of $L_{\phi}$ are consistent with
the findings of Refs. \citep{gordon1985superconductingfluctuation,gordon1986electron}
that fitting the magnetoresistance data over a larger magnetic field
range than that specified for Eq. \ref{eq:AppTransport_dRRMakiThompson}
results in lower magnitudes for the extracted $L_{\phi}$. Thus, the
values in Fig.~\ref{fig:AppTransport_LphivsT} can be taken as rough
lower bounds on $L_{\phi}(T)$. As our primary intent in measuring
$L_{\phi}(T)$ is to demonstrate that $L_{\phi}(T)$ is greater than
the circumference of the persistent current rings $L=1.9\,\text{to\,}5.0\,\text{\ensuremath{\mu}m}$,
such an interpretation is sufficient for our needs.

\chapter{\label{cha:AppSC_}Measurements of persistent currents in the superconducting
state}

We measured persistent currents with the rings in the superconducting
state as well as the normal state. Measurements of the superconducting
rings provided a method of checking the quality of the deposited aluminum
and the angle between the cantilever.

We analyze the measurements of the superconducting state using the
Ginzburg-Landau theory for superconductivity \citep{tinkham2004introduction}.
As the analysis serves mainly a diagnostic role, we make use of some
approximations that simplify the results notationally. A more accurate
numerical analysis is possible. The cases of a one dimensional ring
\citep{tinkham2004introduction} and a ring with finite linewidth
\citep{zhang1997susceptibility} in a magnetic field perpendicular
to the ring plane have been considered previously. Here we modify
these results to account for the angle between the applied magnetic
field and the plane of the ring.

In the Ginzburg-Landau theory, the free energy density of the superconductor
is written as
\[
f=f_{n0}+\alpha\left|\psi\right|^{2}+\frac{\beta}{2}\left|\psi\right|^{4}+\frac{1}{4m}\left|\left(\frac{\hbar}{i}\nabla+2e\boldsymbol{A}\right)\psi\right|^{2}
\]
with
\[
\alpha=-\frac{\hbar^{2}}{4m\xi^{2}}
\]
and
\[
\beta=\frac{\mu_{0}e^{2}\hbar^{2}}{2}\frac{\lambda_{\text{eff}}^{2}}{\xi^{2}}.
\]
Here $f_{n0}$ is the free energy density of the normal state, $\psi$
is the complex order parameter, $m$ and $e$ are the mass and charge
of the electron, $\phi_{0}$ is the normal state flux quantum $h/e$,
$\mu_{0}$ is the magnetic permeability of the superconductor, $\lambda_{\text{eff}}$
is the effective penetration depth (see Eq. 3.119 of Ref. \citealp{tinkham2004introduction}),
and $\xi$ is the Ginzburg-Landau coherence length. If the value of
$\psi$ would lead to $f>f_{n0}$, then $\psi$ is replaced with 0
and the superconductor is taken to be in the normal state. The total
free energy $F=\int d^{3}\boldsymbol{r}\, f$ of the superconductor
is found by integrating $f$ over the volume of the superconductor. 

The current density $\boldsymbol{J}$ of the superconductor is given
by
\begin{equation}
\boldsymbol{J}=-\frac{e}{m}\left|\psi\right|^{2}\left(\hbar\nabla\left(\arg\psi\right)+2e\boldsymbol{A}\right).\label{eq:AppSC_CurrentDensity}
\end{equation}
For a superconductor surrounded by an insulator, the boundary condition
ensuring that no current leaves the superconductor is 
\begin{equation}
\tilde{\boldsymbol{n}}\cdot\left(\frac{\hbar}{i}\nabla-2e\boldsymbol{A}\right)\psi=0\label{eq:AppSC_BC}
\end{equation}
where $\tilde{\boldsymbol{n}}$ is a unit vector normal to the surface
of the superconductor. To determine the current in a superconductor,
one finds the order parameter $\psi$ which satisfies the boundary
conditions and minimizes the free energy $F$. Then this $\psi$ is
plugged into Eq. \ref{eq:AppSC_CurrentDensity} above.

We now consider a superconducting ring with radius $R$, thickness
$t$, and linewidth $w$ centered at the origin so that its axis of
rotational symmetry is the $z$-axis. A magnetic field $\boldsymbol{B}$
is applied at an angle $\theta$ relative to the $z$-axis in the
$xz$-plane. We decompose the magnetic field into perpendicular $\boldsymbol{B}_{\perp}=B\sin\theta\tilde{\boldsymbol{z}}$
and in-plane $\boldsymbol{B}_{M}=B\cos\theta\tilde{\boldsymbol{x}}$
components. 

The boundary conditions (Eq. \ref{eq:AppSC_BC}) complicate the form
of $\psi$. To simplify the derivation, we will treat $\boldsymbol{B}_{\perp}$
exactly but will use the toroidal field model introduced in Section
\ref{sub:CHPCTh_FluxThroughMetal} to account for $\boldsymbol{B}_{M}$.
Following Eq. \ref{eq:CHPCTh_VectorAABflux}, we write the vector
potential $\boldsymbol{A}_{\perp}$ associated with $\boldsymbol{B}_{\perp}$
in cylindrical coordinates $(r,a,z)$ (with $a$ the azimuthal coordinate)
as
\[
\boldsymbol{A}_{\perp}=\frac{B_{\perp}r}{2}\tilde{\boldsymbol{a}}.
\]
For the toroidal field, we imagine breaking the ring and unbending
it to form a rectangular bar of length $2\pi R$. Then we replace
the rectangular cross-section with a circular one of radius $\sqrt{wt/\pi}$.
The volume $2\pi wtR$ of this cylinder is then the same as that of
the ring. The vector potential $\boldsymbol{A}_{M}$ associated with
$\boldsymbol{B}_{M}$ can be written as 
\[
\boldsymbol{A}_{M}=\frac{B_{M}r_{M}}{2}\tilde{\boldsymbol{a}}_{M}
\]
where $(r_{M},a_{M},z_{M})$ are the cylindrical coordinates corresponding
to the cylinder created by transforming the ring in the manner just
described with the $z_{M}$-axis running down the center of the cylinder. 

With these conventions, both $\boldsymbol{A}_{\perp}$ and $\boldsymbol{A}_{M}$
are everywhere parallel to the surface of the ring. We can then use
the form of the order parameter found previously in Refs. \citealp{tinkham2004introduction}
and \citealp{zhang1997susceptibility}, namely 
\[
\psi=\left|\psi\right|e^{ina}
\]
where $n$ is an integer so that $\psi$ is single valued in $a$.
Because $\tilde{\boldsymbol{n}}\cdot\nabla\psi=0$ over the surface
of the ring, the boundary conditions (Eq. \ref{eq:AppSC_BC}) are
satisfied. Integrating the free energy density over the ring with
this form for $\psi$, one finds
\begin{align*}
F & \approx F_{n0}+V\left(\alpha\left|\psi\right|^{2}+\frac{\beta}{2}\left|\psi\right|^{4}\right)\\
 & \phantom{\approx}+\frac{\left|\psi\right|^{2}}{4m}\left(2\pi t\left(\hbar^{2}n^{2}\left(\frac{w}{R}+\frac{w^{3}}{12R^{3}}\right)+\hbar neB_{\perp}Rw+e^{2}B_{\perp}^{2}\left(R^{3}w+\frac{1}{4}Rw^{3}\right)\right)+e^{2}\gamma^{2}B_{M}^{2}w^{2}t^{2}R\right)
\end{align*}
where $F_{n0}$ is the integral of $f_{n0}$ over the ring and, as
in Section \ref{sub:ChData_StatUncertainty}, we multiply $B_{M}$
by a geometrical factor $\gamma$ to account for the difference between
the toroidal field model and the field configuration of the experiment.
In performing the integration, it was assumed that $w\ll R$. in order
to drop higher order terms 

To find the value of $|\psi|$ that minimizes $F$, we solve $\partial_{|\psi|}F=0$.
Using the abbreviated form 
\[
F=C+D\left|\psi\right|^{2}+E\left|\psi\right|^{4},
\]
the free energy is minimized for 
\[
\left|\psi\right|^{2}=-\frac{D}{2E}.
\]
With this form of $|\psi|^{2}$, the free energy can be written as
\[
F=C-\frac{D^{2}}{4E}.
\]
Explicitly, the coefficients $C$, $D$ and $E$ are
\[
C=F_{n0}+V\frac{B^{2}}{2\mu_{0}}
\]
\[
D=V\left(\alpha+\frac{\hbar^{2}}{4mR^{2}}\left(\left(n+\frac{\phi_{\perp}}{\phi_{0}/2}\right)^{2}+\frac{w^{2}}{4R^{2}}\left(\frac{n^{2}}{3}+\frac{\phi_{\perp}^{2}}{(\phi_{0}/2)^{2}}\right)+\frac{\pi}{2}\frac{\gamma^{2}B_{M}^{2}R^{2}wt}{(\phi_{0}/2)^{2}}\right)\right)
\]
\[
E=V\frac{\beta}{2}
\]
where $\phi_{\perp}=\pi R^{2}B_{\perp}$ is the flux threading the
mean radius of the ring. One may then evaluate $F$ for each value
of $n$ to determine the value that minimizes $F$. Integrating $\tilde{\boldsymbol{a}}\cdot\boldsymbol{J}$
(see Eq. \ref{eq:AppSC_CurrentDensity}) over the ring cross-section
gives the current in the ring as 
\begin{align*}
I & =\frac{e\hbar}{m}\frac{1}{\beta}\frac{wt}{R}\left(n\left(1+\frac{w^{2}}{12R^{2}}\right)+\frac{\phi_{\perp}}{\phi_{0}/2}\right)\\
 & \phantom{=}\times\left(\alpha+\frac{\hbar^{2}}{4mR^{2}}\left(\left(n+\frac{\phi_{\perp}}{\phi_{0}/2}\right)^{2}+\frac{w^{2}}{4R^{2}}\left(\frac{n^{2}}{3}+\frac{\phi_{\perp}^{2}}{(\phi_{0}/2)^{2}}\right)+\frac{\pi}{2}\frac{\gamma^{2}B_{T}^{2}R^{2}wt}{(\phi_{0}/2)^{2}}\right)\right).
\end{align*}
In the limit of $w\ll R$ and $B_{T}\rightarrow0$, the free energy
and thus current become periodic in $\phi_{\perp}$ with period $\phi_{0}/2$.
Additionally, for $\xi\ll R$ so that $\alpha$ dominates the second
line in the expression for $I$, the current takes on a sawtooth shape
similar to $I_{N+0}$ of Fig. \ref{fig:PCTh_CurrentSimple}.

In Section \ref{sec:CHTorsMagn_deltaFZeroDrive}, the frequency shift
of the cantilever was found to be proportional to the second derivative
of the energy with respect to angle $\theta$. For the superconducting
ring the frequency shift is then
\begin{align}
\Delta f_{sc} & =N\frac{f_{0}}{2k}\left(\frac{\alpha_{m}\left(z_{r}\right)}{l}\right)^{2}\frac{\partial^{2}F}{\partial\theta^{2}}\nonumber \\
 & =-N\frac{(\phi_{0}/2)^{2}V}{16\pi^{2}\mu_{0}\lambda_{\text{eff}}^{2}}\frac{f_{0}}{2k}\left(\frac{\alpha_{m}\left(z_{r}\right)}{l}\right)^{2}\nonumber \\
 & \phantom{=-}\times\frac{\partial^{2}}{\partial\theta^{2}}\left(-\frac{1}{\xi}+\frac{\xi}{R^{2}}\left(\left(n+\frac{\phi_{\perp}}{\phi_{0}/2}\right)^{2}+\frac{w^{2}}{4R^{2}}\left(\frac{n^{2}}{3}+\frac{\phi_{\perp}^{2}}{(\phi_{0}/2)^{2}}\right)+\frac{\pi}{2}\frac{\gamma^{2}B_{M}^{2}R^{2}wt}{(\phi_{0}/2)^{2}}\right)\right)^{2}\label{eq:AppSC_DeltaFSC}
\end{align}
where $N$ is the number of rings in the array and $f_{0}$, $k$,
$\alpha_{m}(z_{r})$, and $l$ are properties of the cantilever defined
in Section \ref{sec:CHTorsMagn_deltaFZeroDrive}. To evaluate this
expression, one must restore the dependence on $\theta$ for $\phi_{\perp}=\pi R^{2}B\sin\theta$
and $B_{M}=B\cos\theta$. We do not write the full form for the frequency
shift $\Delta f_{sc}$ because it is very long. When measuring a superconducting
sample, the unknown parameters in the expression for $\Delta f_{sc}$
are $\gamma$, $n$, $\lambda_{\text{eff}}$, and $\xi$. The geometrical
factor $\gamma$ adjusts the overall shape of the curve $\Delta f_{sc}(B)$
and affects the location of the transition to the normal state. The
winding number $n$ must always be an integer and jumps discontinuously
as a function of magnetic field $B$, roughly at half integral values
of $\phi_{\perp}/(\phi_{0}/2)$ where the value of $n$ minimizing
$F$ changes. The penetration depth $\lambda_{\text{eff}}$ scales
the entire $\Delta f_{sc}$ curve. The coherence length $\xi$ sets
the critical field at which superconductivity is quenched.

A frequency shift due to the rings' superconductivity was observed
at low magnetic field for each sample listed in Tables \ref{tab:ChData_CLs}
and \ref{tab:ChData_Rings}. For measurements of superconducting rings,
the frequency shift signal $\Delta f_{sc}$ increases with ring radius
$R$. For this reason, we present representative measurements of the
superconducting state from a sample with the largest ring size. For
simplicity, we consider sample CL10 which contained a single ring
with the same dimensions as the rings of sample CL11. The dimensions
of sample CL10 are given in Tables \ref{tab:AppSC_CL10Cantilever}
and \ref{tab:AppSC_CL10Rings}.

\begin{table}
\begin{centering}
\begin{tabular}{|cccccc|}
\hline 
$\vphantom{{\displaystyle \sum_{a}^{a}}}$Sample & $l\,\text{(\ensuremath{\mu}m)}$ & $w\,\text{(\ensuremath{\mu}m)}$ & $f_{0}$ (Hz) & $k\,\text{(mN/m)}$ & $Q\,(\times10^{5})$\tabularnewline
\hline 
${\displaystyle \vphantom{\sum_{-}}}$CL10 & 439 & 40 & 2287 & 1.1 & 1.2\tabularnewline
\hline 
\end{tabular}
\par\end{centering}

\caption[Cantilever parameters of superconducting ring sample]{\label{tab:AppSC_CL10Cantilever}Cantilever parameters of superconducting
ring sample. The cantilever width $w$ was measured optically, and
the cantilever resonant frequency $f_{0}$ was measured using the
detection arrangement described in Chapter \ref{cha:CHExpSetup_}.
The cantilever thickness $t$ was measured to be $340\,\text{nm}$
using the instrument mentioned in Section \ref{sec:APPsampFab_recipe}.
The cantilever length $l$ and spring constant $k$ were calculated
using the values from these measurements and Eqs. \ref{eq:CHTorsMagn_FreqFromKandMeff}
and \ref{eq:CHTorsMagn_springKfromDimensions}. The cantilever length
measured optically was not used due to uncertainty caused by over-etching
of the silicon handle layer on which the cantilever was mounted. The
listed cantilever quality factor $Q$ is a typical value. The quality
factor varied with temperature and over time. More details about the
cantilevers are given in Sections \ref{sub:CHExpSetup_PersistentCurrentFabrication}
and \ref{sec:APPsampFab_recipe}.}
\end{table}

\begin{table}
\begin{centering}
\begin{tabular}{|ccccc|}
\hline 
$\vphantom{{\displaystyle \sum_{a}^{a}}}$Sample & $L\,\text{(\ensuremath{\mu}m)}$ & $w_{r}\,\text{(nm)}$ & $t_{r}\,\text{(nm)}$ & $N$\tabularnewline
\hline 
${\displaystyle \vphantom{\sum_{-}}}$CL10 & 5.0 & 85 & 90 & 1\tabularnewline
\hline 
\end{tabular}
\par\end{centering}

\caption[Superconducting ring specifications]{\label{tab:AppSC_CL10Rings}Superconducting ring specifications.
The mean circumference $L$ and linewidth $w_{r}$ of the aluminum
ring were measured by a scanning electron microscope. The thickness
$t_{r}$ of the deposited aluminum was measured by an atomic force
microscope. The ring was located $5\,\text{\ensuremath{\mu}m}$ from
the cantilever tip and was centered along the cantilever's width dimension. }
\end{table}

\begin{figure}
\begin{centering}
\includegraphics[width=0.7\paperwidth]{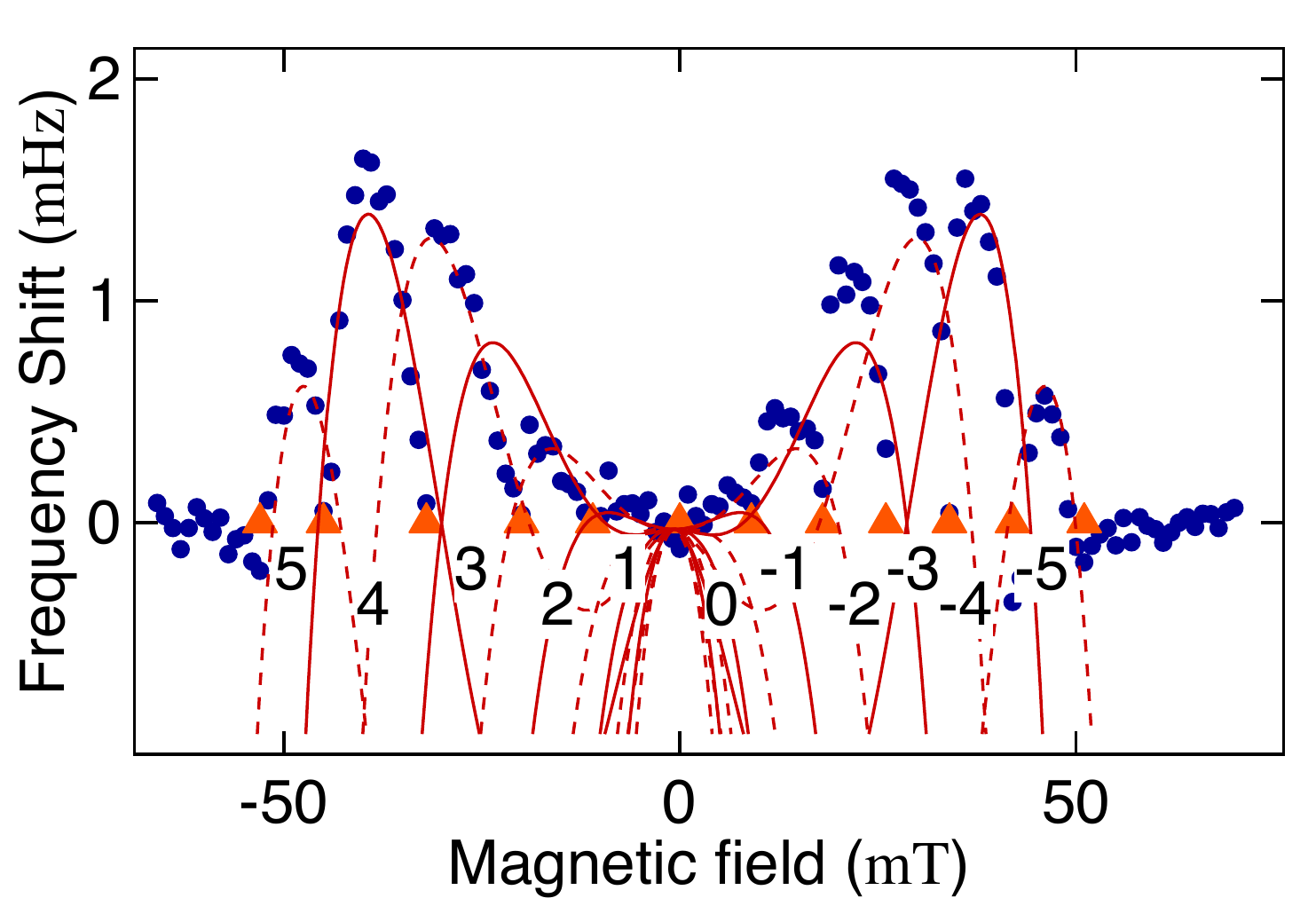}
\par\end{centering}

\caption[Frequency shift due to superconductivity in sample CL10]{\label{fig:AppSC_dFvsBAllN}Frequency shift due to superconductivity
in sample CL10. The cantilever frequency was measured at $\theta_{0}=6^{\circ}$
and $T=1.05\,\text{K}$ while sweeping the magnetic field from $B<-60\,\text{mT}$
to $B>60\,\text{mT}$. The frequency shift (dots) was found by subtracting
a fourth order polynomial to the trace of frequency versus magnetic
field. The polynomial coefficients were determined by fitting the
cantilever frequency in the magnetic field regions where the frequency
shift was expected to be negligible ($|B|\apprle5\,\text{mT}$ and
$|B|\apprge55\,\text{mT}$). A series of jumps is visible in the measured
frequency shift. These jumps occur at points in $B$ (triangles) where
the winding number $n$ changes by 1. The magnetic field regions between
jumps are labeled by the value of $n$ that best matches the observed
frequency shift. The curves representing $\Delta f_{sc}(B,n)$ are
plotted for each of these values of $n$ and for the values of $\lambda_{\text{eff}}$,
$\xi$, and $\gamma$ found with the fit shown in Fig. \ref{fig:AppSC_dFvsBFit}.
For even $n$, $\Delta f_{sc}$ is drawn with a solid curve, while
for odd $n$ the curve is dashed.}

\end{figure}

The frequency shift observed for sample CL10 at $\theta_{0}=6^{\circ}$
and $T=1.05\,\text{mK}$ is shown in Fig. \ref{fig:AppSC_dFvsBAllN}.
This measurement was taken as the magnetic field was stepped up from
below $-60\,\text{mT}$ to above $60\,\text{mT}$. The frequency shift
exhibits regular jumps at values of the magnetic field at which the
winding number $n$ increases by 1 as the state of the ring jumps
to a state with a lower free energy $F(B,n)$. In Fig. \ref{fig:AppSC_dFvsBAllN},
a triangle is located along the $\Delta f_{sc}=0$ line at the magnetic
field value of each jump. The frequency shift is clearly asymmetric
about $B=0\,\text{T}$, indicating that the ring experiences some
metastability and that the winding number $n$ does not change at
exactly the value of $B$ at which $F(B,n)=F(B,n-1)$.%
\footnote{We focus on a single ring sample in part because it is possible that
the locations of the jumps in $n$ could vary from ring to ring in
the array samples.%
}

We use Eq. \ref{eq:AppSC_DeltaFSC} to analyze the frequency shift
$\Delta f_{sc}$. In order to account for the observed metastability
of the superconducting state, the value of $n$ was held fixed over
each region of the magnetic field between the observed jumps in $\Delta f_{sc}$.
At each jump in $\Delta f_{sc}$, $n$ was incremented by 1. The particular
values of $n$ used were chosen to provide the best match to the data.
In practice, the set of values of $n$ was always symmetric about
$n=0$ (i.e. if $+n$ were present, so was $-n$). The region of field
between each jump in $\Delta f_{sc}$ is labeled by the appropriate
value for $n$ in Fig. \ref{fig:AppSC_dFvsBAllN}. The frequency shift
curves $\Delta f_{sc}(B,n)$ associated with each $n$ are also shown
in Fig. \ref{fig:AppSC_dFvsBAllN} for a particular set of $\lambda_{\text{eff}}$,
$\xi$, and $\gamma$.

\begin{figure}
\begin{centering}
\includegraphics[width=0.7\paperwidth]{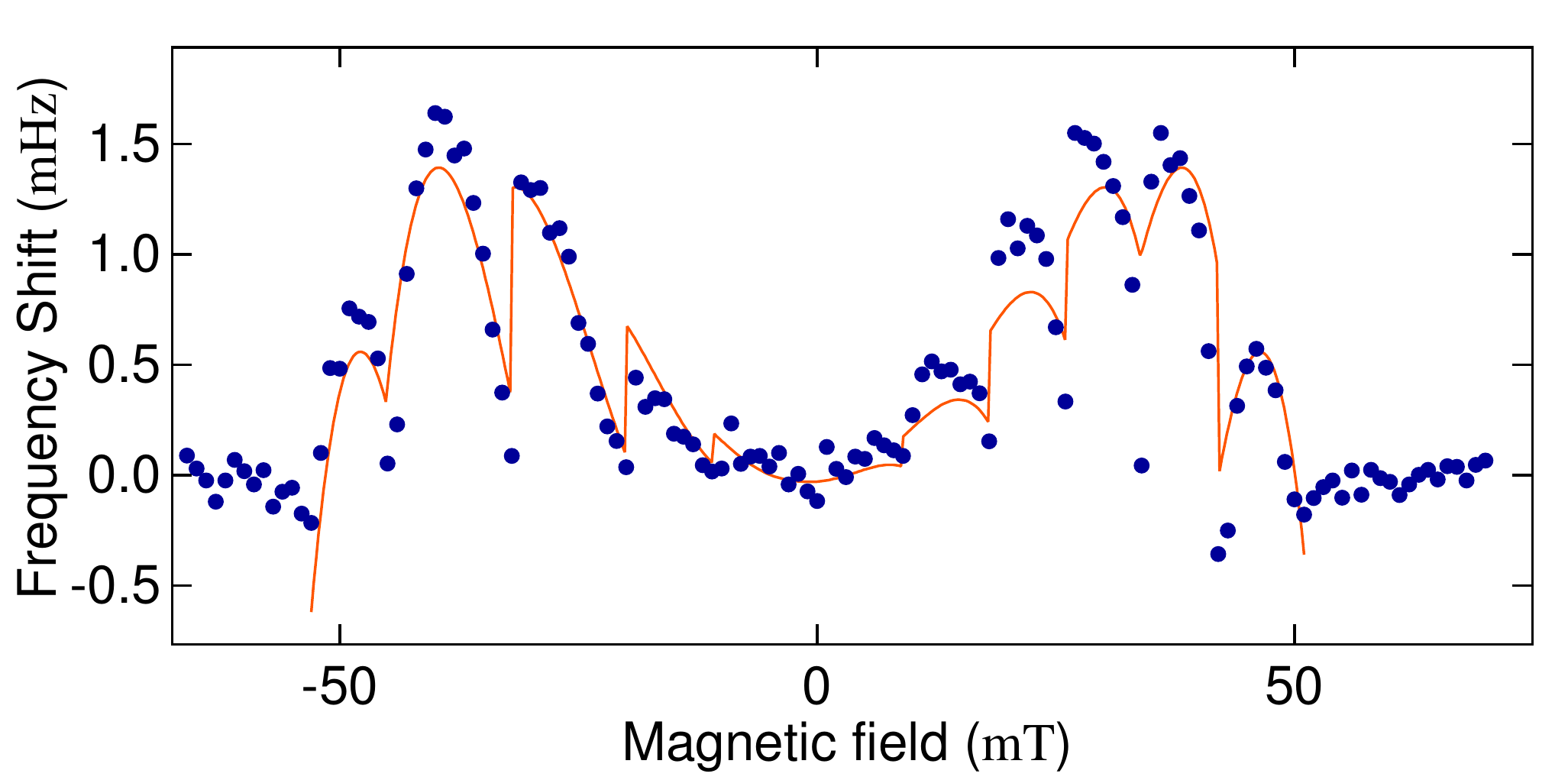}
\par\end{centering}

\caption[Fit to frequency shift of sample CL10 at low magnetic field]{\label{fig:AppSC_dFvsBFit}Fit to frequency shift of sample CL10
at low magnetic field. The same measured frequency shift (dots) as
shown in Fig. \ref{fig:AppSC_dFvsBAllN} is reproduced here. A fit
(solid curve) to Eq. \ref{eq:AppSC_DeltaFSC} is also shown. During
the fitting routine, the value of $n$ was held to integer values
that varied with magnetic field region as indicated in Fig. \ref{fig:AppSC_dFvsBAllN}.
The geometric factor $\gamma$ was held to 1 while $\lambda_{\text{eff}}$
and $\xi$ were varied. The fitting routine also allowed for overall
offsets in the frequency shift ($-30\,\text{\ensuremath{\mu}Hz}$)
and the magnetic field ($-0.7\,\text{mT}$). The best fit coefficients,
$\lambda_{\text{eff}}=145\,\text{nm}$ and $\xi=366\,\text{nm}$,
are in reasonable agreement with previous measurements of aluminum
rings. The deviations of the observed frequency shift from the fitted
curve could be due to corrections to the Ginzburg-Landau model away
from $T_{c}$.}
\end{figure}

In Fig. \ref{fig:AppSC_dFvsBFit}, the measured frequency shift for
sample CL10 at $\theta_{0}=6^{\circ}$ and $T=1.05\,\text{mK}$ is
shown again along with a fit to Eq. \ref{eq:AppSC_DeltaFSC}. During
the fitting routine, the value of $n$ was fixed to the values shown
in Fig. \ref{fig:AppSC_dFvsBAllN} as discussed above. At $\theta_{0}=6^{\circ}$,
the frequency shift $\Delta f_{sc}$ has a similar dependence on the
coherence length $\xi$ and the geometrical factor $\gamma$. To achieve
a reliable fit, the geometrical factor was fixed to $\gamma=1$ because
it is expected to be of order unity. The fitted superconductor parameters
were $\lambda_{\text{eff}}=145\,\text{nm}$ and $\xi=366\,\text{nm}$.
These values are in reasonable agreement with those found in previous
measurements of superconducting aluminum rings \citep{zhang1997susceptibility}.

\begin{figure}
\begin{centering}
\includegraphics[width=0.7\paperwidth]{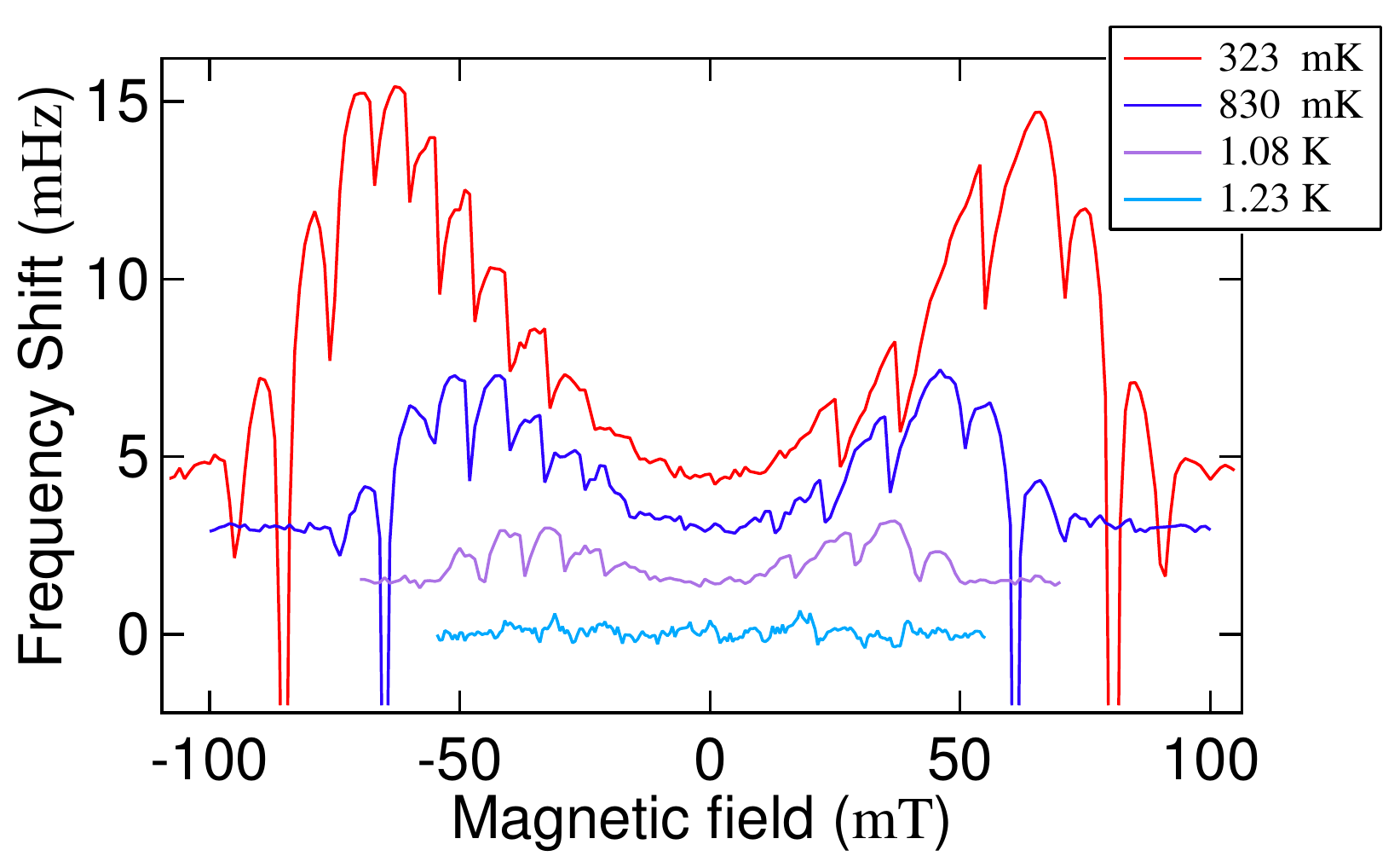}
\par\end{centering}

\caption[Frequency shift observed for sample CL10 in the superconducting state
at several temperatures]{\label{fig:AppSC_CL10_6Deg_dFvsB}Frequency shift observed for sample
CL10 in the superconducting state at several temperatures. All curves
were recorded at $\theta_{0}=6^{\circ}$ by first sweeping the magnetic
field above $100\,\text{mT}$ and then stepping it down while recording
the cantilever frequency. Smooth backgrounds were removed by the method
described for the measurement shown in Fig. \ref{fig:AppSC_dFvsBAllN}.
The curves are each offset from each other by $1.5\,\text{mT}$. In
descending order the curves represent measurements taken at $T=0.32$,
$0.83$, $1.08$, and $1.23\,\text{K}$. As expected, the strength
of the superconducting signal and the magnitude of the critical magnetic
field decrease with temperature and vanish near $T\approx1.2\,\text{K}$.}
\end{figure}

In Fig. \ref{fig:AppSC_CL10_6Deg_dFvsB}, the frequency shift observed
for sample CL10 at $\theta_{0}=6^{\circ}$ is shown at a series of
temperatures between $300\,\text{mK}$ and $1.2\,\text{K}$. The strength
of the signal decreases with temperature and appears to vanish near
$T\approx1.18\,\text{K}$ as expected for aluminum (see Fig. \ref{fig:CHSensitivity_BcVsTBCS}).
Similar curves were observed for each sample listed in Tables \ref{tab:ChData_CLs}
and \ref{tab:ChData_Rings}. The magnetic field scale of the jumps
in the frequency shift was observed to change with ring size, with
the smallest rings exhibiting only one or two jumps before going normal.

\chapter{\label{cha:AppCumul_}Cumulants of the persistent current in arrays
of rings}

We present a brief analysis of the statistical distribution of the
persistent current based on the measurements of samples CL15 and CL17
at $\theta_{0}=45^{\circ}$. The sample parameters are given in Tables
\ref{tab:ChData_CLs} and \ref{tab:ChData_Rings}. The full data sets
for these measurements are shown in Section \ref{sub:ChData_FullCurrentTraces}.
As discussed in Section \ref{sub:ChData_Qualitative}, we measure
the total current summed over the many rings making up the array on
each of these samples. By the central limit theorem, we expect the
total current in each sample to follow the normal distribution regardless
of the underlying distribution of the persistent current for a single
ring. Thus, the analysis presented here is mostly a check of the accuracy
of our measurement. To test the predictions regarding the single ring
distribution discussed in Section \ref{sec:ChPrevWork_theory}, more
measurements of samples with single rings must be performed.

We analyze the two quadratures of the persistent current signal. The
quadratures are found by making use of the Hilbert transform. To calculate
the Hilbert transform $I_{H}(B)$ of a signal $I(B)$, one finds the
Fourier transform 
\[
I\left(\beta\right)=\int_{-\infty}^{\infty}dB\, I\left(B\right)e^{-2\pi i\beta B},
\]
 defines 
\[
I_{H}\left(\beta\right)=\begin{cases}
I\left(\beta\right) & \beta>0\\
0 & \beta=0\\
-I\left(\beta\right) & \beta<0
\end{cases},
\]
and then finds the inverse Fourier transform
\[
I_{H}\left(B\right)=\int_{-\infty}^{\infty}d\beta\, I_{H}\left(\beta\right)e^{2\pi i\beta B}.
\]

When a signal $I(B)=I_{0}\sin(2\pi\beta_{0}B)$ is a monotone oscillation,
the Hilbert transform $I_{H}(B)=I_{0}\cos(2\pi\beta_{0}B)$ is also
a monotone oscillation with the same frequency but shifted in phase.
In this case, the amplitude of the oscillation can be found by taking
$\sqrt{I^{2}(B)+I_{H}^{2}(B)}=I_{0}$. Similarly, a signal possessing
a slowly varying amplitude $I_{0}(B)$ and containing a finite range
of frequencies (such as the range of frequencies exhibited by the
peaks in the spectra shown in Figs. \ref{fig:ChData_DAT5_PSDCL17_45Deg}
and \ref{fig:ChData_DAT6_PSDCL15_45Deg} for samples CL17 and CL15)
can be written as $I(B)=I_{0}(B)\sin(2\pi\beta_{0}B+\phi(B))$. As
long as $I_{0}(B)$ and $\phi(B)$ vary slowly enough, the Hilbert
transform satisfies $I_{H}(B)\approx I_{0}(B)\cos(2\pi\beta_{0}B+\phi(B))$,
so that the instantaneous amplitude still satisfies 
\begin{equation}
I_{0}(B)=\sqrt{I^{2}(B)+I_{H}^{2}(B)}.\label{eq:AppCumul_MagInstantaneous}
\end{equation}
The instantaneous phase $\phi(B)$ can also be found by 
\begin{equation}
\phi(B)=\arg(I(B)+iI_{H}(B))-2\pi\beta_{0}B.\label{eq:AppCumul_PhaseInstantaneous}
\end{equation}
For a more thorough review of the properties of the Hilbert transform,
see Ref. \citealp{huang1998theempirical}.

\begin{figure}
\begin{centering}
\includegraphics[width=0.7\paperwidth]{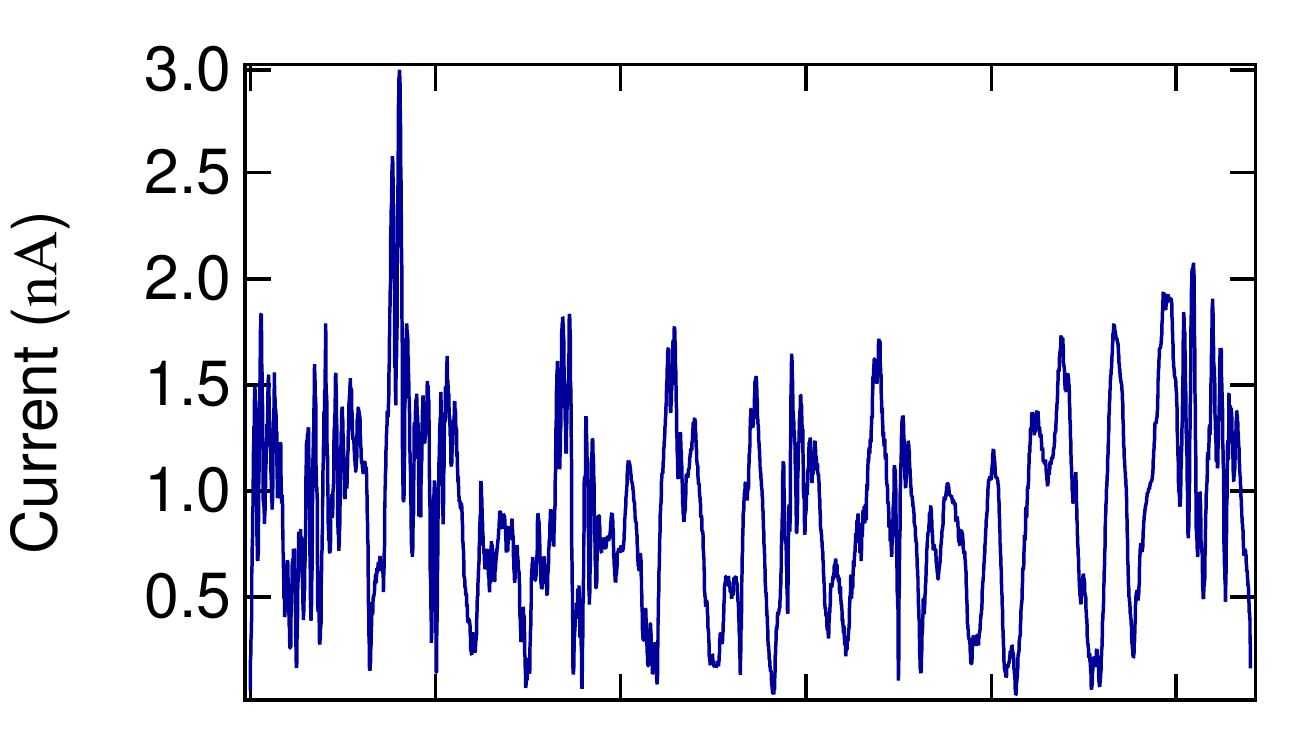}
\par\end{centering}

\begin{centering}
\includegraphics[width=0.7\paperwidth]{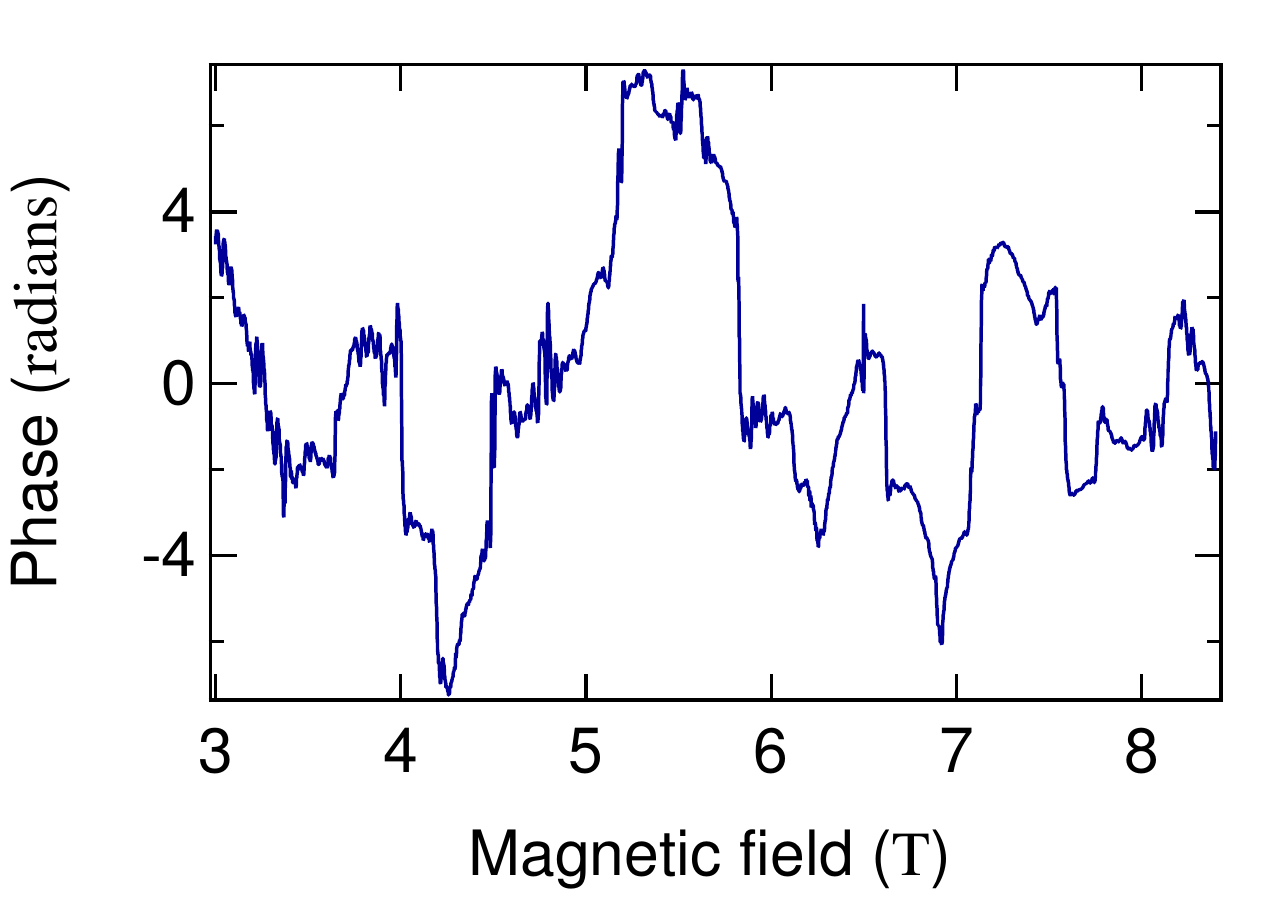}
\par\end{centering}

\caption[Magnitude and phase of the persistent current signal]{\label{fig:AppCumul_MagPhase}Magnitude and phase of the persistent
current signal. The current magnitude $I_{0}(B)$ (top graph) and
phase $\phi(B)$ (bottom graph) were calculated as described in the
text from the current versus magnetic field trace shown in Figs. \ref{fig:ChData_DAT21_IvsB_CL17_45DegP1}
through \ref{fig:ChData_DAT21_IvsB_CL17_45DegP3}. This trace was
measured with sample CL17 at $\theta_{0}=45^{\circ}$ and $T=365\,\text{mK}$.}
\end{figure}

To analyze the persistent current signals of samples CL15 and CL17,
we found the Hilbert transform of the current versus magnetic field
trace and then calculated the instantaneous current magnitude $I_{0}(B)$
and phase $\phi(B)$ using Eqs. \ref{eq:AppCumul_MagInstantaneous}
and \ref{eq:AppCumul_PhaseInstantaneous}. The magnitude and phase
found for sample CL17 are plotted versus magnetic field in Fig. \ref{fig:AppCumul_MagPhase}.
In calculating the phase, the function $\arg(I(B)+iI_{H}(B))$ was
first unwrapped by adding $2\pi$ whenever the phase jumped discontinuously
from $+\pi$ to $-\pi$. A line was fit and subtracted from this unwrapped
trace to determine $\phi(B)$. For both samples, the slope of the
fitted line was approximately $2\pi\beta_{0}$ where $\beta_{0}$
is the location of the center of the peak in the Fourier spectrum
(see Figs. \ref{fig:ChData_DAT5_PSDCL17_45Deg} and \ref{fig:ChData_DAT6_PSDCL15_45Deg}).

\begin{figure}
\begin{centering}
\includegraphics[width=0.7\paperwidth]{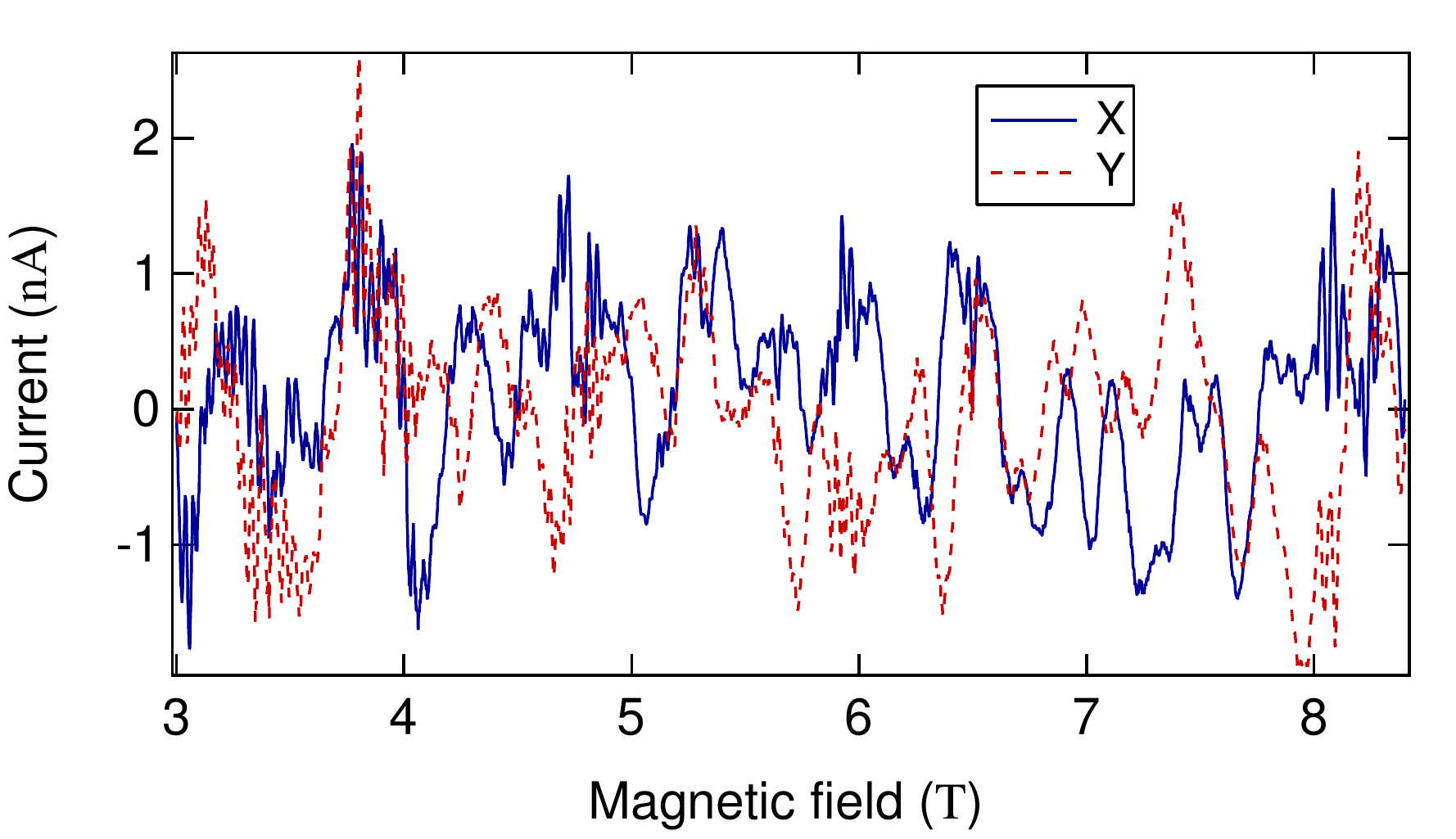}
\par\end{centering}

\caption[Quadratures of the persistent current signal]{\label{fig:AppCumul_Quadratures}Quadratures of the persistent current
signal. The quadratures $X_{I}$ (solid line) and $Y_{I}$ (dashed
line) were calculated from the magnitude and phase shown in Fig. \ref{fig:AppCumul_MagPhase}
using the definitions given in Eqs. \ref{eq:AppCumul_X} and \ref{eq:AppCumul_Y}.}
\end{figure}

\begin{figure}
\begin{centering}
\includegraphics[width=0.7\paperwidth]{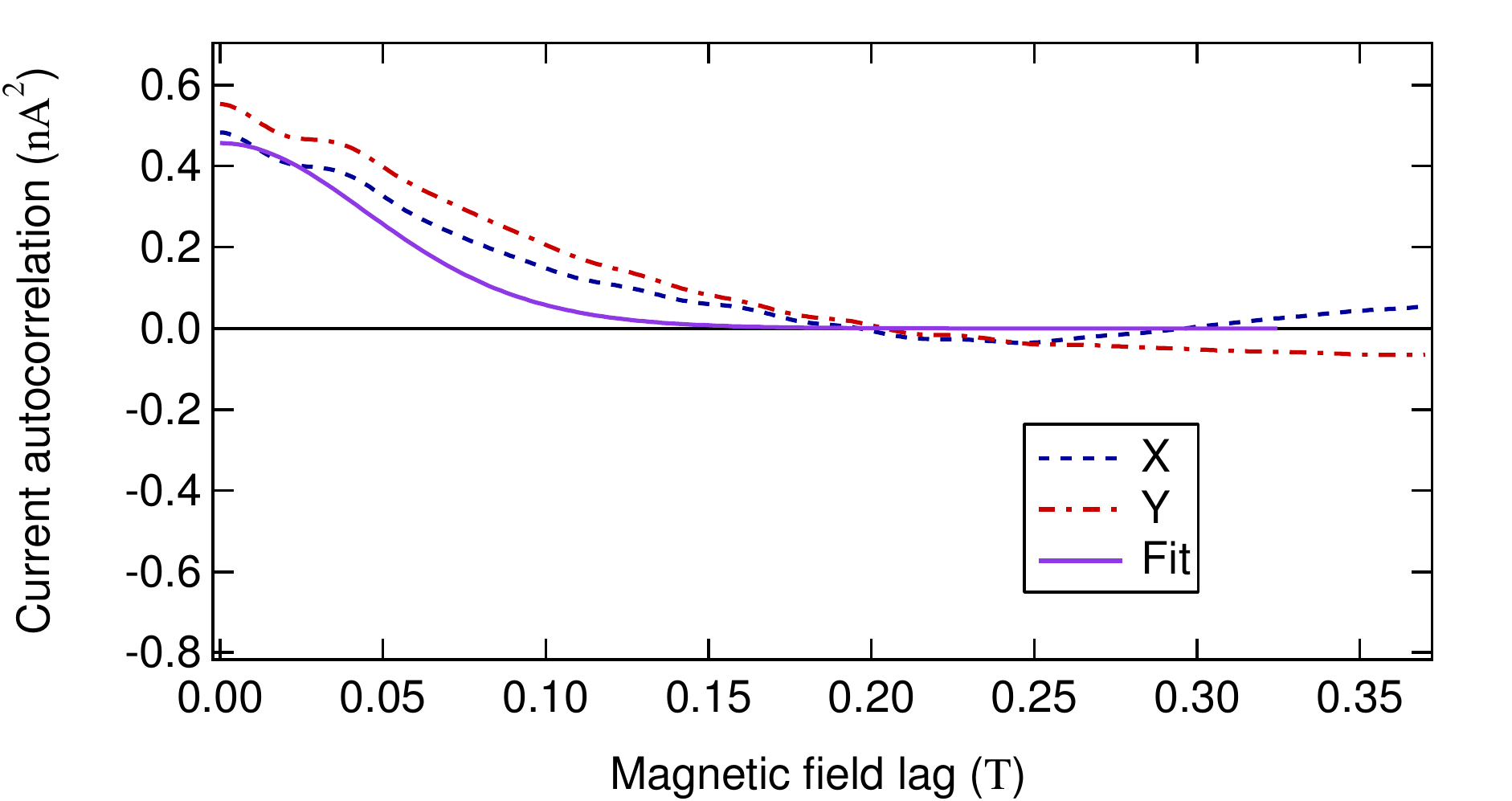}
\par\end{centering}

\caption[Autocorrelation of the persistent current quadratures]{\label{fig:AppCumul_QuadratureAutocorrelation}Autocorrelation of
the persistent current quadratures. The autocorrelation of the $X_{I}$
(dashed line) and $Y_{I}$ (dot-dashed line) quadratures from Fig.
\ref{fig:AppCumul_Quadratures} were calculated using Eq. \ref{eq:ChData_AutocorrelationCalculation}.
Also shown is the solid line representing the envelope of the fit
to the oscillating autocorrelation of the persistent current signal
given in Fig. \ref{fig:ChData_DAT15_Cor_CL17_45Deg}. Each quadrature
is correlated on roughly the same scale within the level of uncertainty
associated with the finite magnetic field range of the measurement.
For the autocorrelation, one expects an error of $2\eta(M_{\text{eff}})$,
which for this sample is $16\%$ (see Table \ref{tab:ChData_RingResults}).}
\end{figure}

From the current magnitude $I_{0}(B)$ and phase $\phi(B)$, the quadratures
of the persistent current calculation can be defined as 
\begin{align}
X_{I}(B) & =I_{0}(B)\cos(\phi(B))\label{eq:AppCumul_X}\\
Y_{I}(B) & =I_{0}(B)\sin(\phi(B)).\label{eq:AppCumul_Y}
\end{align}
Fig. \ref{fig:AppCumul_Quadratures} plots the quadratures associated
with the magnitude and phase shown in Fig. \ref{fig:AppCumul_MagPhase}
for sample CL17. In Fig. \ref{fig:AppCumul_QuadratureAutocorrelation},
the autocorrelation, calculated using Eq. \ref{eq:ChData_AutocorrelationCalculation},
is shown for both quadratures of the signal from sample CL17. As expected,
the magnitude and magnetic field range of this correlation agrees
with those observed in Fig. \ref{fig:ChData_DAT15_Cor_CL17_45Deg}
for the current versus magnetic field trace $I(B)$. Additionally,
the cross-correlation of the two quadratures, defined by 
\begin{equation}
\left\langle X_{I}\left(B\right)Y_{I}\left(B+j\Delta B\right)\right\rangle _{M}=\frac{1}{P-\left|j\right|-1}\sum_{k=0}^{P-\left|j\right|-1}X_{I}\left(B_{\min}+k\Delta B\right)Y_{I}\left(B_{\min}+\left(j+k\right)\Delta B\right),\label{eq:AppCumul_CrossCor}
\end{equation}
shows no significant correlation. In the definition, for both traces
$X_{I}$ and $Y_{I}$, $B_{\text{min}}$ is the minimum value of $B$,
$\Delta B$ is the spacing in $B$ between successive points, and
$P$ is the total number of points. 

\begin{figure}
\begin{centering}
\includegraphics[width=0.7\paperwidth]{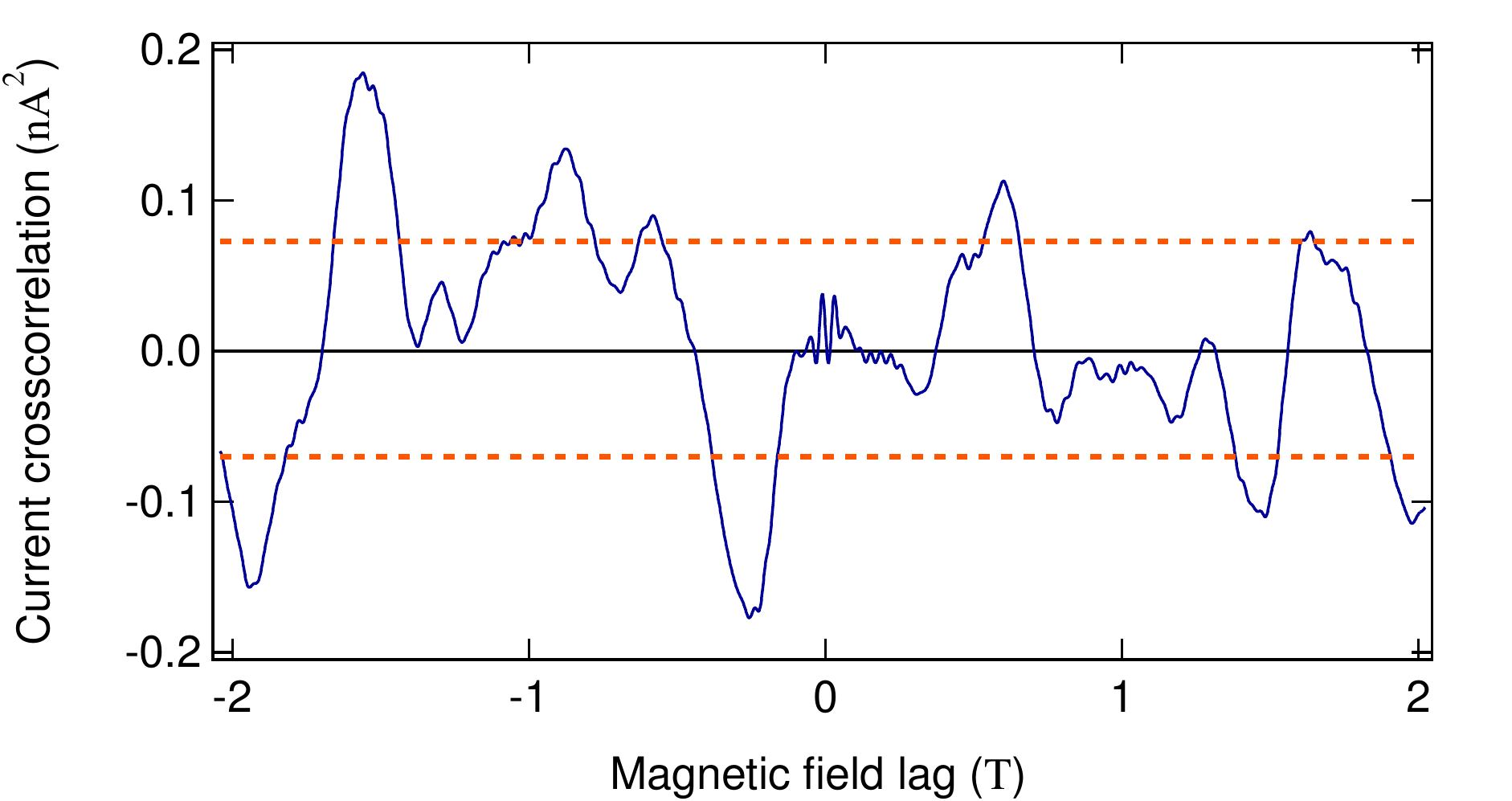}
\par\end{centering}

\caption[Cross-correlation of the persistent current quadratures]{\label{fig:AppCumul_QuadratureCrosscorrelation}Cross-correlation
of the persistent current quadratures. The cross correlation (solid
line) of the two quadratures shown in Fig. \ref{fig:AppCumul_QuadratureAutocorrelation}
was calculated using Eq. \ref{eq:AppCumul_CrossCor}. The horizontal
dashed lines represent the uncertainty $2\eta(M_{\text{eff}})(I^{\text{typ}})^{2}$
expected for the finite range of magnetic field measured and the observed
correlation of the persistent current signal. The lack of a significant
deviation from the horizontal lines indicates that the two quadratures
are uncorrelated.}
\end{figure}

To assess the distribution of the quadrature amplitudes we calculate
the cumulants of the quadratures. The $j^{th}$ moment $\mu_{j}$
of a set of numbers $\{r_{k}\}$ is defined as the average of the
$j^{th}$ power of those numbers
\[
\mu_{j}=\frac{1}{N}\sum_{k=1}^{N}r_{k}^{j}.
\]
The cumulants $\kappa_{j}$ are another set of numbers related to
the moments. The cumulants can be easier to work with analytically
in some instances but are not as easy to write down as the moments.
When the mean $\mu_{1}=0$, the cumulants $\kappa_{j}$ can be defined
in terms of the moments $\mu_{j}$ by the relation
\begin{equation}
\exp\left(\sum_{j=1}^{\infty}\frac{t^{j}\kappa_{j}}{j!}\right)=1+\sum_{j=1}^{\infty}\frac{\mu_{j}t^{j}}{j!}.\label{eq:AppCumul_CumulantDef}
\end{equation}
For a normal distribution with $\mu_{1}=0$ and $\mu_{2}=(I^{\text{typ}})^{2}$,
the probability $P(I)$ of measuring an amplitude $I$ is 
\[
P(I)=\frac{1}{\sqrt{2\pi}I^{\text{typ}}}\exp\left(-\frac{I^{2}}{2\left(I^{\text{typ}}\right)^{2}}\right).
\]
For such a distribution, $\kappa_{2}=(I^{\text{typ}})^{2}$ and $\kappa_{j}=0$
for $j\neq2$. More information about moments and cumulants is given
in Ref. \citealp{kendall1947advanced}. 

As discussed in Section \ref{sub:ChData_Quantitative}, the inference
of any quantity from a set of measurements with a random contribution
has some error related to the finite number of measurements averaged
together to calculate that quantity. When the quantity being measured
is correlated from measurement to measurement, as is the case for
the persistent current measured as a function of magnetic field, the
determination of this error is not straightforward. In Ref. \citealp{tsyplyatyev2003applicability},
Tsyplyatyev \emph{et al}. give the error $\delta\kappa_{j}$ in the
$j^{th}$ cumulant $\kappa_{j}$ for such a set of correlated data
as 
\[
\delta\kappa_{j}=\sqrt{\kappa_{2}^{j}\, j!\frac{B_{c}}{B_{0}}\int_{-\infty}^{\infty}dx\,\left(K_{p}\left(x\right)\right)^{j}}
\]
where $B_{0}$ is the range of the measurement field $B$, $B_{c}$
is a characteristic field scale of the correlation and $K_{p}(x)$
is the normalized and scaled correlation function for the measured
quantity. The correlation function $K_{p}(x)$ was discussed further
in Section \ref{sub:ChData_Quantitative} and given explicitly for
the persistent current in Eq. \ref{eq:ChData_Kp}.

\begin{figure}
\begin{centering}
\includegraphics[width=0.7\paperwidth]{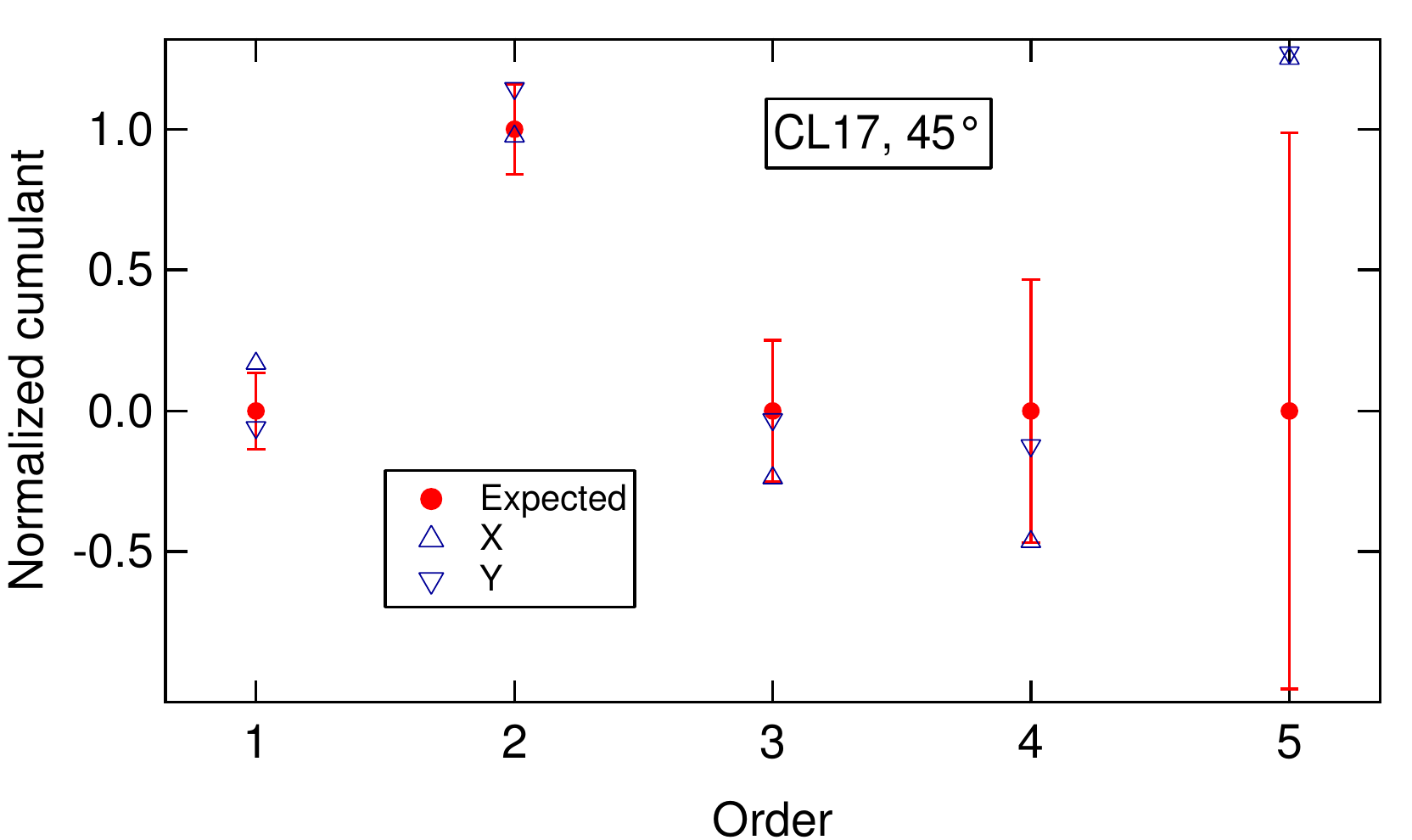}
\par\end{centering}

\begin{centering}
\includegraphics[width=0.7\paperwidth]{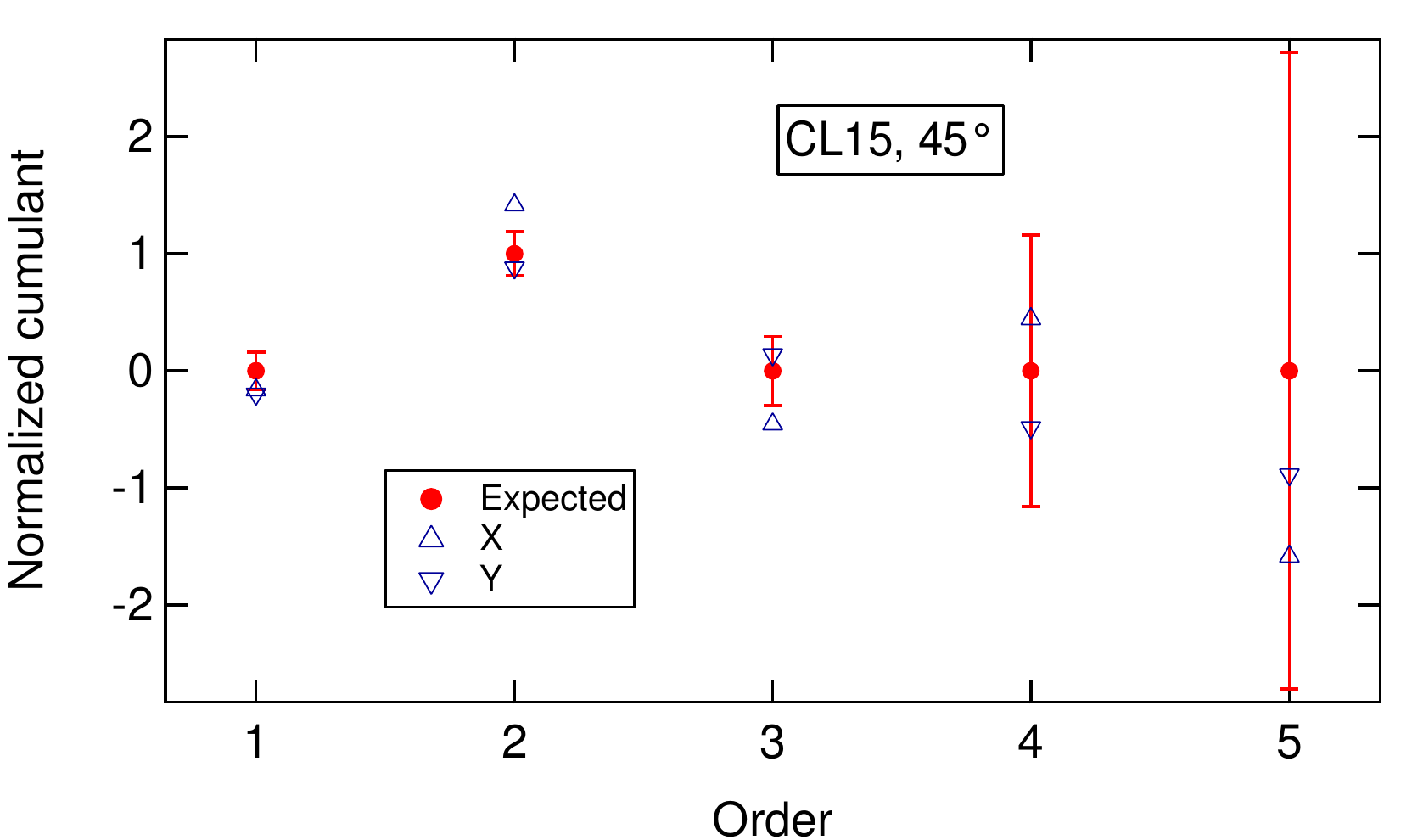}
\par\end{centering}

\caption[Cumulants of the persistent current for samples CL15 and CL17]{\label{fig:AppCumul_Cumulants}Cumulants of the persistent current
for samples CL15 and CL17. The normalized cumulants $\kappa_{j}^{N}$
are shown for the measurements of both sample CL17 (top graph) and
CL15 (bottom graph) at $\theta_{0}=45^{\circ}$ and $T=365\,\text{mK}$.
The cumulants were calculated for both the $X_{I}$ (upward pointing
triangles) and $Y_{I}$ (downward pointing triangles) quadratures
and then divided by $(I^{\text{typ}})^{j}$, the $j$th power of the
typical current expected from analysis of the temperature dependence
of the current (see the discussion in the text). The quadratures of
sample CL17 are shown in Fig. \ref{fig:AppCumul_Quadratures}. The
normalized cumulants $\kappa_{j}^{N}$ (dots) expected for the normal
distribution with zero mean are also shown above, as are the corresponding
normalized uncertainties $\delta\kappa_{j}^{N}$ (vertical bars) given
by Eq. \ref{eq:AppCumul_dKappa}. Most of the measured cumulants fall
within the expected uncertainties, indicating that the measured values
of the persistent current are consistent with the normal distribution.}
\end{figure}

The measured cumulants for samples CL15 and CL17 are shown in Fig.
\ref{fig:AppCumul_Cumulants}. The plotted cumulants have been normalized
as 
\[
\kappa_{j}^{N}=\frac{\kappa_{j}}{\left(I^{\text{typ}}\right)^{j}}
\]
where $\kappa_{j}$ is the unnormalized cumulant and $I^{\text{typ}}$
is the typical current given by Eq. \ref{eq:ChPCTh_IpTypTEZESO} for
the measurement temperature $T=365\,\text{mK}$ and the value of the
diffusion constant found from a fit to the temperature dependence
of the persistent current (see Table \ref{tab:ChData_RingResults}).
Also shown are bars representing the normalized error 
\begin{align}
\delta\kappa_{j}^{N} & =\frac{\delta\kappa_{j}}{\left(I^{\text{typ}}\right)^{j}}\nonumber \\
 & =\sqrt{j!\frac{B_{c}}{B_{0}}\int_{-\infty}^{\infty}dx\,\left(K_{p}\left(x\right)\right)^{j}}\label{eq:AppCumul_dKappa}
\end{align}
appropriate for a normal distribution with zero mean and $\kappa_{2}=(I^{\text{typ}})^{2}$.
These bars are centered at $\kappa_{j}^{N}=1$ for $j=2$ and $\kappa_{j}^{N}=0$
for $j\neq2$, the positions expected for a normal distribution. The
cumulants are plotted up to fifth order where $\delta\kappa_{j}^{N}\approx1$.
The measured cumulants agree well the values expected for the normal
distribution given the uncertainty specified by Eq. \ref{eq:AppCumul_dKappa}.
We reiterate that this agreement only confirms the applicability of
the central limit theorem to our arrays of rings and consequently
the accuracy of our data analysis procedure. To test the claim that
the persistent current in an individual ring follows the normal distribution,
larger data sets of individual rings must be measured.

\bibliographystyle{IEEEtranPCmod}
\bibliography{PC_Bibliography}

\end{document}